ORGANISATION EUROPÉENNE POUR LA RECHERCHE NUCLÉAIRE
**CERN** EUROPEAN ORGANIZATION FOR NUCLEAR RESEARCH

# Workshop on CP Studies and Non-Standard Higgs Physics

## May 2004 – December 2005

Edited by

Sabine Kraml[1], Georges Azuelos[2,3], Daniele Dominici[4], John Ellis[1], Gerald Grenier[5], Howard E. Haber[6], Jae Sik Lee[7], David J. Miller[8], Apostolos Pilaftsis[9] and Werner Porod[10]

[1] CERN, Geneva, Switzerland.
[2] Université de Montréal, Montreal, Canada.
[3] TRIUMF, Vancouver, Canada.
[4] Università di Firenze and INFN, Firenze, Italy.
[5] IPNL, Université Lyon-1, Villeurbanne, France.
[6] University of California, Santa Cruz, USA.
[7] Seoul National University, Seoul, Korea.
[8] University of Glasgow, Glasgow, UK.
[9] University of Manchester, Manchester, UK.
[10] IFIC-CSIC, Valencia, Spain.

**Conveners**

*The CP-violating two-Higgs doublet model:* G. Grenier, H. E. Haber, M. Krawczyk
*The MSSM with CP phases:* M. Boonekamp, M. Carena, S. Y. Choi, J. S. Lee, M. Schumacher
*SUSY models with an extra singlet:* S. Baffioni, J. Gunion, D. Miller, A. Pilaftsis, D. Zerwas
*The MSSM with R-parity violation:* M. Besançon, W. Porod
*Extra Gauge groups:* P. Langacker, A. Raspereza, S. Riemann
*Little Higgs models:* T. Gregoire, H. Logan, B. McElrath
*Large extra dimensions:* D. Dominici, S. Ferrag
*Randall-Sundrum model:* A. De Roeck, S. Ferrag, J. L. Hewett, T. G. Rizzo
*Higgsless models:* B. Lillie, J. Terning
*Strongly interacting Higgs sector:* G. Azuelos, W. Kilian, T. Han
*Technicolour:* G. Azuelos, F. Sannino
*Higgs Triplets:* J. F. Gunion, C. Hays

**Organizing Committee**

S. Y. Choi (Chonbuk)
J. Conway (Davis)
R. M. Godbole (Bangalore)
J. F. Gunion (Davis)
J. Ellis (CERN)
J. L. Hewett (SLAC)
S. Kraml (CERN)
M. Krawczyk (Warsaw)
M. Mangano (CERN)
D. J. Miller (Glasgow)
Y. Okada (KEK)
M. Oreglia (Chicago)

**Meetings**

14–15 May 2004 at CERN
2–3 December 2004 at CERN
24–25 March 2005 at SLAC
14–16 December 2005 at CERN

# Abstract

There are many possibilities for new physics beyond the Standard Model that feature non-standard Higgs sectors. These may introduce new sources of CP violation, and there may be mixing between multiple Higgs bosons or other new scalar bosons. Alternatively, the Higgs may be a composite state, or there may even be no Higgs at all. These non-standard Higgs scenarios have important implications for collider physics as well as for cosmology, and understanding their phenomenology is essential for a full comprehension of electroweak symmetry breaking. This report discusses the most relevant theories which go beyond the Standard Model and its minimal, CP-conserving supersymmetric extension: two-Higgs-doublet models and minimal supersymmetric models with CP violation, supersymmetric models with an extra singlet, models with extra gauge groups or Higgs triplets, Little Higgs models, models in extra dimensions, and models with technicolour or other new strong dynamics. For each of these scenarios, this report presents an introduction to the phenomenology, followed by contributions on more detailed theoretical aspects and studies of possible experimental signatures at the LHC and other colliders.

# Authors

E. Accomando[1], A. G. Akeroyd[2], E. Akhmetzyanova[3], J. Albert[4], A. Alves[5], N. Amapane[1], M. Aoki[2], G. Azuelos[6,7], S. Baffioni[8], A. Ballestrero[1], V. Barger[9], A. Bartl[10], P. Bechtle[11], G. Bélanger[12], A. Belhouari[1], R. Bellan[1], A. Belyaev[13], P. Beneš[14], K. Benslama[15], W. Bernreuther[16], M. Besançon[17], G. Bevilacqua[1], M. Beyer[18], M. Bluj[19], S. Bolognesi[1], M. Boonekamp[20], F. Borzumati[21,22], F. Boudjema[12], A. Brandenburg[16], T. Brauner[14], C. P. Buszello[23], J. M. Butterworth[24], M. Carena[25], D. Cavalli[26], G. Cerminara[1], S. Y. Choi[27,28], B. Clerbaux[29], C. Collard[8], J. A. Conley[11], A. Deandrea[30], S. De Curtis[31], R. Dermisek[32], A. De Roeck[33], G. Dewhirst[34], M. A. Díaz[35], J. L. Díaz-Cruz[36], D. D. Dietrich[37], M. Dolgopolov[3], D. Dominici[31], M. Dubinin[38], O. Eboli[5], J. Ellis[33], N. Evans[39], L. Fano[40], J. Ferland[6], S. Ferrag[41,42], S. P. Fitzgerald[23], H. Fraas[43], F. Franke[43], S. Gennai[44,45], I. F. Ginzburg[46], R. M. Godbole[47], T. Grégoire[48], G. Grenier[30], C. Grojean[33,49], S. B. Gudnason[37], J. F. Gunion[50], H. E. Haber[51], T. Hahn[52], T. Han[9], V. Hankele[53], C. Hays[54], S. Heinemeyer[55], S. Hesselbach[43], J. L. Hewett[11], K. Hidaka[56], M. Hirsch[57], W. Hollik[52], D. Hooper[25], J. Hošek[14], J. Hubisz[25], C. Hugonie[58], J. Kalinowski[59], S. Kanemura[60], V. Kashkan[1], T. Kernreiter[10], W. Khater[61], V. A. Khoze[62,63], W. Kilian[64,28], S. F. King[39], O. Kittel[65], G. Klämke[53], J. L. Kneur[58], C. Kouvaris[37], S. Kraml[33], M. Krawczyk[59], P. Krstonošić[28], A. Kyriakis[66], P. Langacker[67], M. P. Le[11], H.-S. Lee[68], J. S. Lee[69], M. C. Lemaire[20], Y. Liao[70], B. Lillie[71,72], V. Litvin[73], H. E. Logan[74], B. McElrath[50], T. Mahmoud[29], E. Maina[1], C. Mariotti[1], P. Marquard[75], A. D. Martin[62], K. Mazumdar[76], D. J. Miller[42], P. Miné[8], K. Mönig[77], G. Moortgat-Pick[33], S. Moretti[39], M. M. Mühlleitner[33], S. Munir[39], R. Nevzorov[39], H. Newman[73], P. Nieżurawski[78], A. Nikitenko[34], R. Noriega-Papaqui[79], Y. Okada[2,80], P. Osland[81], A. Pilaftsis[82], W. Porod[57], H. Przysiezniak[83], A. Pukhov[38], D. Rainwater[84], A. Raspereza[28], J. Reuter[28] S. Riemann[77], S. Rindani[85], T. G. Rizzo[11], E. Ros[57], A. Rosado[79], D. Rousseau[86], D. P. Roy[76], M. G. Ryskin[62,63], H. Rzehak[87], F. Sannino[37], E. Schmidt[18], H. Schröder[18], M. Schumacher[65], A. Semenov[88], E. Senaha[89], G. Shaughnessy[9], R. K. Singh[47], J. Terning[50], L. Vacavant[90], M. Velasco[91], A. Villanova del Moral[57], F. von der Pahlen[43], G. Weiglein[62], J. Williams[23,92], K. Williams[62], A. F. Żarnecki[78], D. Zeppenfeld[53], D. Zerwas[86], P. M. Zerwas[28], A. R. Zerwekh[93], J. Ziethe[16]

[1] Università di Torino, Facoltà di Fisica, Via P. Giuria 1, Torino 10100, Italy
[2] KEK, 1-1 Oho, Tsukuba, Ibaraki 305-0801, Japan
[3] Samara State University, 443011 Samara, Russia
[4] California Institute of Technology, Pasadena, CA 91125, USA
[5] Instituto de Física, Universidade de São Paulo, Caixa Postal 66318, 05315-970 São Paulo, Brazil
[6] Département de Physique, Université de Montréal, Montréal, Qué., H3C 3J7, Canada
[7] TRIUMF, 4004 Wesbrook Mall, Vancouver, B.C., V6T 2A3, Canada
[8] Laboratoire Leprince-Ringuet (LLR), Ecole Polytechnique, IN2P3-CNRS, F-91128 Palaiseau Cedex, France
[9] Department of Physics, University of Wisconsin, Madison, WI 53706, USA
[10] Institut für Theoretische Physik, Universität Wien, A-1090 Vienna, Austria
[11] Stanford Linear Accelerator Center (SLAC), P.O. Box 20450, Stanford, CA 94309, USA
[12] LAPTH, 9 Chemin de Bellevue, B.P. 110, F-74941 Annecy-le-Vieux, France
[13] Department of Physics and Astronomy, Michigan State University, East Lansing, MI 48824, USA
[14] Department of Theoretical Physics, Nuclear Physics Institute, 25068 Řež, Czech Republic
[15] Physics Department, Nevis Laboratories, Columbia University, 136 S Broadway, Irvington, NY 10533, USA
[16] Institut für Theoretische Physik, RWTH Aachen, D-52056 Aachen, Germany
[17] Centre d'Etudes de Saclay, Orme des Merisiers, F-91191 Gif-sur-Yvette Cedex, France
[18] Institute of Physics, University of Rostock, D-18051 Rostock, Germany
[19] Soltan Institute for Nuclear Studies, Hoża 69, 00-681 Warsaw, Poland
[20] DAPNIA/SPP, CEA-Saclay, F-91191 Gif-sur-Yvette Cedex, France
[21] International Center for Theoretical Physics (ICTP), I-34000 Trieste, Italy
[22] Scuola Internazionale Superiore di Studi Avanzati (SISSA), I-34000 Trieste, Italy
[23] Cambridge University, Dept. of Physics, Cavendish Laboratory, Madingley Road, Cambridge CB3 0HE, UK
[24] Physics and Astronomy Department, University College London, Gower St, London WC1E 6BT, UK
[25] Fermi National Accelerator Laboratory (FNAL), P.O. Box 500, Batavia, IL 60510-0500, USA
[26] INFN and Dipartimento di Fisica, Università di Milano, I-20133 Milano, Italy
[27] Department of Physics, Chonbuk National University, Jeonju 561-756, Korea
[28] Deutsches Elektronen-Synchrotron DESY, D-22603 Hamburg, Germany
[29] Université Libre de Bruxelles (ULB), CP230 Blvd. du Triomphe, B-1050 Brussels, Belgium
[30] Institut de Physique Nucléaire de Lyon, CNRS/IN2P3, Université Lyon 1, F-69622 Villeurbanne, France
[31] Dipartimento di Fisica, Univiversità di Firenze, and INFN, I-50019 Sesto Fiorentino (Firenze), Italy
[32] Institute for Advanced Study, Princeton, NJ 08540, USA
[33] CERN, CH-1211 Geneva 23, Switzerland
[34] Imperial College, Department of Physics, Prince Consort Road, London SW7 2BW, UK
[35] Departamento de Física, Pontificia Universidad Católica de Chile, Av. Vicuña Mackenna 4860, Santiago, Chile
[36] FCFM, BUAP. Apdo. Postal 1364, C.P. 72000 Puebla, Pue., México
[37] The Niels Bohr Institute, Blegdamsvej 17, DK-2100 Copenhagen, Denmark
[38] D.V. Skobeltsyn Institute of Nuclear Physics, Moscow State University, 119992 Moscow, Russia
[39] School of Physics and Astronomy, University of Southampton, Southampton SO17 1BJ, UK
[40] INFN and Università degli Studi di Perugia, Dipartimento di Fisica, via Pascoli, 06100 Perugia, Italy
[41] Department of Physics, University of Oslo, P.O. Box 1048 Blindern, NO-0316 Oslo, Norway
[42] Department of Physics and Astronomy, University of Glasgow, Glasgow G12 8QQ, UK
[43] Institut für Theoretische Physik und Astrophysik, Universität Würzburg, Am Hubland, D-97074 Würzburg, Germany
[44] INFN Pisa, Largo Ponte Corvo 3, Pisa 56126, Italy
[45] Centro Studi Enrico Fermi, Compendio Viminale, Roma 00184, Italy
[46] Sobolev Institute of Mathematics, Novosibirsk 630090, Russia
[47] Centre for High Energy Physics, Indian Institute of Science, Bangalore 560012, India
[48] Boston University, Dept. of Physics, 590 Commonwealth Ave., Boston, MA 02215, USA
[49] Service de Physique Théorique, CEA Saclay, F-91191 Gif-sur-Yvette, France
[50] Department of Physics, University of California at Davis, Davis, CA 95616, USA
[51] Santa Cruz Institute for Particle Physics, University of California, Santa Cruz, CA 95064, USA
[52] MPI für Physik (Werner-Heisenberg-Institut), Föhringer Ring 6, D-80805 München, Germany

[53] Institut für Theoretische Physik, Universität Karlsruhe, P.O. Box 6980, D-76128 Karlsruhe, Germany
[54] University of Oxford, Oxford OX1 3RH, UK
[55] Depto. de Física Teórica, Universidad de Zaragoza, 50009 Zaragoza, Spain
[56] Department of Physics, Tokyo Gakugei University, Koganei, Tokyo 184-8501, Japan
[57] Instituto de Física Corpuscular / C.S.I.C., Edificio Institutos de Paterna, Apartado 22085, E-46071 Valencia, Spain
[58] Laboratoire Physique Théorique et Astroparticules, Univ. Montpellier II, F-34095 Montpellier, France
[59] Institute of Theoretical Physics, Warsaw University, Hoża 69, 00-681 Warsaw, Poland
[60] Department of Physics, Osaka University, Toyonaka, Osaka 560-0043, Japan
[61] Department of Physics, Birzeit University, Birzeit, West Bank, Palestine
[62] Department of Physics and Institute for Particle Physics Phenomenology, University of Durham, Durham DH1 3LE, UK
[63] Petersburg Nuclear Physics Institute, Gatchina, St. Petersburg, 188300, Russia
[64] Fachbereich Physik, University of Siegen, D-57068 Siegen, Germany
[65] Physikalisches Institut der Universität Bonn, Nussallee 12, D-53115 Bonn, Germany
[66] Institute of Nuclear Physics, NCSR "Demokritos", Athens, Greece
[67] Department of Physics and Astronomy, University of Pennsylvania, Philadelphia, PA 19104, USA
[68] Department of Physics, University of Florida, Gainesville, FL 32608, USA
[69] Center for Theoretical Physics, School of Physics, Seoul National University, Seoul 151-747, Korea
[70] Department of Physics, Nankai University, Tianjin 300071, China
[71] High Energy Physics Division, Argonne National Laboratory, Argonne, IL 60439, USA
[72] Enrico Fermi Institute, University of Chicago, Chicago, IL 60637, USA
[73] Department of Physics, California Institute of Technology, MS356-48, Pasadena, CA 91125, USA
[74] Ottawa Carleton Institute for Physics, Carleton University, Ottawa K1S 5B6, Canada
[75] Institut für Theoretische Teilchenphysik, Universität Karlsruhe, D-76128 Karlsruhe, Germany
[76] Tata Institute of Fundamental Research, Homi Bhabha Rd., Mumbai 400005, India
[77] DESY, Platanenallee 6, D-15738 Zeuthen, Germany
[78] Institute of Experimental Physics, Warsaw University, Hoża 69, 00-681 Warsaw, Poland
[79] Instituto de Física, BUAP. Apdo. Postal J-48, C.P. 72570 Puebla, Pue., México
[80] Graduate University for Advanced Studies, Tsukuba, Ibaraki 305-0801, Japan
[81] Department of Physics and Technology, University of Bergen, Allegt. 55, N-5007 Bergen, Norway
[82] School of Physics and Astronomy, University of Manchester, Manchester M13 9PL, UK
[83] LAPP, 9 Chemin de Bellevue, B.P. 110, F-74941 Annecy-le-Vieux, France
[84] Department of Physics and Astronomy, University of Rochester, Rochester, NY 14627, USA
[85] Physical Research Lab, Navrangpura, Ahmedabad 380009, Gujarat, India
[86] LAL Orsay, Université de Paris-Sud et IN2P3-CNRS, F-91898 Orsay Cedex, France
[87] Paul Scherrer Institut, Würenlingen und Villigen, CH-5232 Villigen PSI, Switzerland
[88] Joint Institute for Nuclear Research (JINR), 141980, Dubna, Russia
[89] Department of Physics, National Central University, Chungli, Taiwan 320, R.O.C.
[90] CPPM, CNRS/IN2P3/Univ. Méditerranée, 163 Av. de Luminy, Case 902, F-13288 Marseille Cedex 09, France
[91] Northwestern University, Evanston, IL 60201, USA
[92] Physics and Electronics Department, Rhodes University, Grahamstown, 6140, South Africa
[93] Instituto de Física, Universidad Austral de Chile, Casilla 567, Valdivia, Chile

# Contents

## 1 INTRODUCTION

*John Ellis and Sabine Kraml*

The next step in high-energy particle physics will be the exploration of the TeV energy scale, an adventure that starts with the LHC and is to be continued with other colliders operating in a similar energy range. The energy scale of the LHC is largely determined by the need to complete the successful Standard Model and, in particular, to understand the origin of the elementary particle masses. Even though the Standard Model is very successful, having survived stringent high-energy tests at the SLC, LEP, HERA and the Tevatron, in particular, it is known to be incomplete. The existence of a Higgs sector is essential for the breaking of the electroweak gauge symmetry, enabling the $W$ and $Z$ gauge bosons and the matter fermions to acquire masses. The presence of a Higgs boson is the only way to avoid having the scattering amplitudes for massive particles grow indefinitely, leading to unrenormalizable divergences in loop diagrams.

General theoretical arguments based on unitarity and lattice calculations imply that an elementary Higgs boson should have a mass less than about a TeV. The Standard Model's successes in all its experimental tests to date implies that the dangerous loop diagrams must indeed be cut off by some undetected ingredient resembling a Higgs boson, with a mass that is likely to be no larger than a few hundred GeV. The discovery of the Standard Model Higgs boson or some equivalent substitute is therefore one of the primary objectives of the experimental programmes at the LHC and other TeV-scale colliders. Its discovery is expected to lead to a flowering of new physics, as illustrated in Fig. 1.

This is because there are many reasons to suspect that the simplest Higgs sector postulated in the original formulations of the Standard Model is unlikely to be the complete solution to the origin of particle masses. It may be supplemented by additional new physics beyond the Standard Model, or the Higgs sector may be more complicated, or it may be replaced by some very different dynamics serving a similar purpose. An elementary Higgs field and its associated Higgs boson are subject to quantum-mechanical instabilities induced by loop diagrams that threaten to subject the electroweak mass scale to large corrections. It is, in principle, possible to maintain a low electroweak scale despite these large corrections, but this would appear to require unnatural fine tuning of the model parameters. One of the favoured solutions to this naturalness problem is to postulate the appearance of supersymmetry at or below the TeV scale. Even the minimal supersymmetric extension of the Standard Model (MSSM) requires two doublets of Higgs fields and hence five physical Higgs bosons. However, there is still no experimental evidence for supersymmetry, and the first collider able to provide evidence for its relevance to particle physics is likely to be the LHC.

The Higgs sectors of the Standard Model and its minimal supersymmetric extension have been discussed extensively in connection with the experimental programmes of the LHC and other TeV-scale colliders. These define the standard options in Higgs physics. The prospects for discovering and characterizing the MSSM at these colliders have also been discussed in some detail. The purpose of this report is to explore the other petals of the Higgs flower shown in Fig. 1, by assembling studies of non-standard Higgs models within and beyond the framework of supersymmetry. Particular emphasis is placed on Higgs scenarios that violate CP. In addition to the problem of mass, one of the most puzzling aspects of the Standard Model is flavour physics, and particularly the violation of CP symmetry. The Standard Model accommodates CP violation quite economically via the Kobayashi-Maskawa mechanism, but it does not explain its origin. In principle, CP violation could also be present in the strong interactions, as a result of non-perturbative effects, but this has not been seen. An attractive option for suppressing this strong CP violation is to complicate the minimal Higgs sector of the Standard Model by postulating an axion. On the other hand, non-minimal Higgs sectors introduce in general additional sources of CP violation beyond the Kobayashi-Maskawa mechanism. This is not necessarily unwelcome, since some additional source of CP violation would in any case be required in order to explain the cosmological baryon asymmetry on the basis of elementary particle interactions. Cosmological baryogenesis could be achieved at any temperature scale at or beyond the electroweak scale. One particularly attractive option is to generate the cosmological baryon asymmetry at the electroweak scale by means of a non-minimal Higgs sector, such as that in the MSSM. The MSSM already contains a plethora of possible CP-violating phases, which may have interesting signatures within and beyond the Higgs sector of the theory. We place particular emphasis on the possible implications of these phases for Higgs phenomenology at colliders, including the LHC, International Linear Collider (ILC) and its photon-photon collider option.

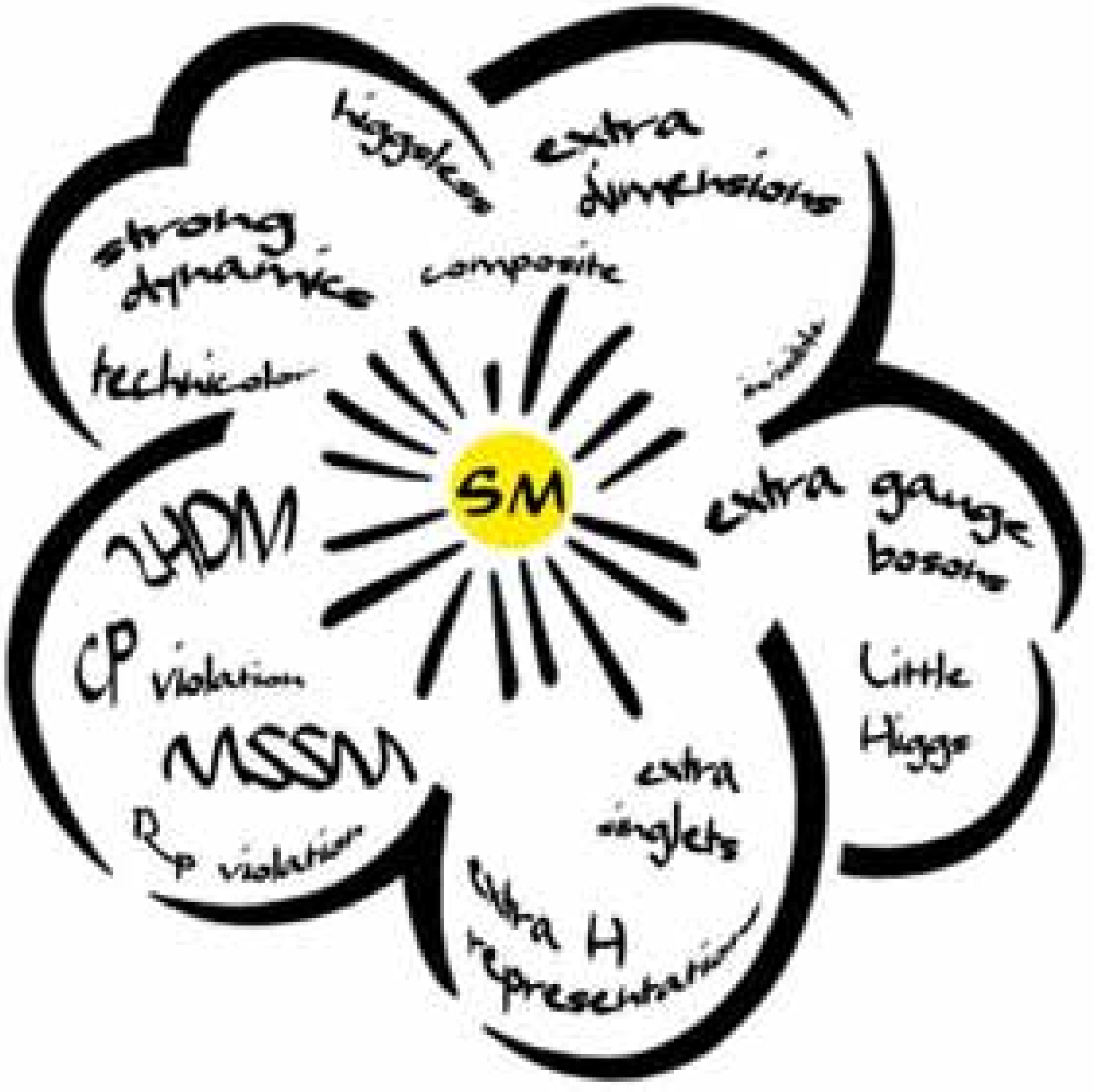

Fig. 1: The flowering of the Higgs physics that is expected to bloom at the TeV scale.

Even more possibilities for CP and flavour violation are offered by modifications of the MSSM in which R parity is violated. These introduce many novel Yukawa-like interactions that possess, in general, additional CP-violating phases. We give particular emphasis in this report to the possible mixing of Higgs bosons with sleptons and the corresponding phenomenological signatures. Yet another possibility is to augment the MSSM framework, for example by postulating its extension to include a singlet superfield that expands the Higgs sector of the theory. One of the motivations for such an addition is to avoid introducing *a priori* a Higgs mixing parameter with a magnitude similar to the electroweak scale, instead replacing it with a vacuum expectation value generated dynamically. In addition to enriching the possibilities for Higgs phenomenology at colliders, such scenarios also have interesting cosmological implications, e.g., for the nature of cold dark matter.

A more radical expansion of the field content of the Standard Model is to postulate an extension of the gauge group. Various such extensions have been considered in previous studies of collider physics. Among the motivations for such models are grand-unified and string models that contain supplementary U(1) gauge groups, and left-right symmetric models. Aspects of these have been studied previously: the new thrust here is to consider in more detail the phenomenology of Higgs bosons in such scenarios. An extra stimulus to such models has recently been provided by little Higgs models. Their central idea is to interpret the Higgs boson of the Standard Model as a pseudo-Goldstone boson of a higher electroweak gauge symmetry after its breakdown into the SU(2) $\times$ U(1) of the Standard Model. In such little Higgs models, the low scale of the electroweak vacuum relative to the breaking of the larger gauge symmetry is protected by the pseudo-Goldstone character of the Higgs boson. This protection is provided in specific models by additional matter, gauge and Higgs fields with relatively low masses that might be accessible to TeV-scale colliders. We discuss here the possible phenomenology of such extra degrees of freedom, as well as the phenomenology of the little Higgs boson itself.

A different set of possible extensions of the Standard Model are those with extra spatial dimensions. Such theories have been around for many decades, but recently gained motivation from string theory. This apparently requires such extra dimensions, though they might well be much smaller than the inverse-TeV scale. However, it has been realized that the phenomenological constraints on at least some extra-dimensional scenarios are quite weak, laying them open to experimental tests at colliders. In particular, they provide options for invisible Higgs decays and for other sources of missing transverse energy. One particularly interesting possibility is that an extra dimension is warped. An important new scalar degree of freedom in such a model is the radion, which has several potential interfaces with Higgs physics. Higgs-radion mixing must be taken into account, since several of the radion decay modes mimic those of a conventional Higgs boson, such as those to $\gamma\gamma$ and $ZZ^{(*)}$, and radion decays into pairs of Higgs bosons are also of potential interest.

The imminent exploration of the Higgs sector by the LHC and other colliders has prompted new questions whether Higgsless models are viable. In their original four-dimensional formulations, they lead to strong $WW$ scattering at relatively low energies, and run into related problems with the precision electroweak data. However, these difficulties may be alleviated by postulating an extension to five dimensions, where electroweak symmetry may be broken by appropriate boundary conditions. From a theoretical point of view, the absence of a Higgs boson would be a very interesting outcome from the LHC, even if experimentalists might be disappointed. However, they should be encouraged by the fact that, even in such Higgsless models, there are possible experimental probes of the mechanism of electroweak symmetry breaking.

Strongly-interacting Higgs sectors arise in a number of other scenarios, in addition to Higgsless models. One general approach to these is provided by effective Lagrangian techniques modeled on those used in QCD at low energies. As well as probing strong $WW$ scattering and possible massive resonances via the production of pairs of weak gauge bosons, it may also be possible to study anomalous quartic gauge-boson couplings via triple weak-boson production. There is also an interesting class of models in which the elementary Higgs field of the Standard Model is replaced by a composite field in a theory of new strong ‘technicolour’ interactions. Models in which the technicolour dynamics is closely modeled on that in QCD have problems with precision electroweak data and the generation of fermion masses. However, the first problem may be mitigated in ‘walking’ technicolour models whose dynamics is not related to that of QCD by simple rescaling. The fermion mass problem may be solved in extended technicolour models, which offer interesting possibilities for light composite Higgs bosons as well as predicting complex strong dynamics at higher energies.

A final class of scenarios to consider is that with higher-dimensional Higgs representations. These arise in generic little Higgs scenarios of the type mentioned above, but may also arise in other models. These may give rise to distinctive signatures due to doubly-charged Higgs bosons, as well as interesting effects in the physics of neutral and singly-charged Higgs bosons.

This brief summary gives an impressionistic survey of the different non-standard Higgs scenarios that should be considered in preparations for collider experiments. It demonstrates that one should not allow one’s attention to be dominated by the single weakly-interacting Higgs boson of the Standard Model, nor even by its modest extension to the MSSM. One should keep in mind, in particular, the possibility that there may be a close link between the Higgs sector and CP violation, and one should be open to the possible appearance of non-standard Higgs representations such as singlets and triplets, as well as novel decay patterns, including invisible modes. Revealing the full details of the underlying mechanism of electroweak symmetry breaking may be considerably more complex than in the Standard

Model or the MSSM.

This report is the product of a Workshop which extended from May 2004 to December 2005, with significant support from the CERN Theory Division and elsewhere. It consists of chapters discussing each of the principal non-standard Higgs scenarios mentioned above and shown in the Higgs flower. Each chapter starts with a pedagogical introduction to the corresponding scenario, which is followed by a set of individual contributions describing specific studies made in the context of the Workshop. In addition to many phenomenological studies, this report reviews several studies made of LHC capabilities using detailed simulations of the ATLAS and CMS detectors. It is encouraging that, although many of the non-standard Higgs scenarios were not considered in the designs of ATLAS and CMS, nevertheless they have excellent capabilities for revealing such scenarios. This indicates that the ATLAS and CMS detector designs and trigger concepts are sufficiently robust to respond to new challenges. Thus one may also hope that they will also be sensitive to Nature's choice for new physics, even if it extends beyond the options considered here. This volume also contains many studies of non-standard Higgs signatures at linear $e^+e^-$ colliders. It is also encouraging that the ILC, in particular, also offers excellent prospects for exploring more aspects of non-standard Higgs scenarios, thanks to its very clean experimental conditions.

The LHC will soon start revealing what physics lies at the TeV scale, and in particular what the Higgs sector holds in store for us. We do not know in advance whether it will reveal a single elementary Standard Model Higgs boson, something more complicated, or even a Higgsless model. One must approach LHC physics in general, and Higgs physics in particular, with an open mind. The Higgs sector may be not only the completion of the Standard Model, but also the first window on physics beyond it. In addition to answering one of the key open questions in the Standard Model, for example, it provides some of the key motivation for supersymmetry and is deeply implicated in the linked problems of flavour and CP violation. This volume provides a hitchhiker's guide to these and other possible aspects of non-standard Higgs physics at the LHC and other colliders.

## 2 The CP-Violating Two-Higgs Doublet Model

### 2.1 Theory review

*Howard E. Haber and Maria Krawczyk*

The Standard Model (SM) of electroweak physics is an $\mathrm{SU}(2)_\mathrm{L}\times\mathrm{U}(1)$ gauge theory coupled to quarks, leptons and one complex hypercharge-one, $\mathrm{SU}(2)_\mathrm{L}$ doublet of scalar fields. Due to the form of the scalar potential, one component of the complex scalar field acquires a vacuum expectation value, and the $\mathrm{SU}(2)_\mathrm{L}\times\mathrm{U}(1)$ electroweak symmetry is spontaneously broken down to the $\mathrm{U}(1)_\mathrm{EM}$ gauge symmetry of electromagnetism. Hermiticity requires that the parameters of the SM scalar potential are real. Consequently, the resulting bosonic sector of the electroweak theory is CP-conserving.

The SM, with its minimal Higgs structure, provides an extremely successful description of observed electroweak phenomena. Nevertheless, there are a number of motivations to extend the Higgs sector of this model by adding a second complex doublet of scalar fields [1–10]. Perhaps the best motivated of these extended models is the minimal supersymmetric extension of the Standard Model (MSSM) [11–13], which requires a second Higgs doublet (and its supersymmetric fermionic partners) in order to preserve the cancellation of gauge anomalies. The Higgs sector of the MSSM is a two-Higgs-doublet model (2HDM), which contains two chiral Higgs supermultiplets that are distinguished by the sign of their hypercharge. The theoretical structure of the MSSM Higgs sector is constrained by the supersymmetry, leading to numerous relations among Higgs masses and couplings. In particular, as in the case of the SM, the tree-level MSSM Higgs sector is CP-conserving. However, the supersymmetric relations among Higgs parameters are modified by loop-corrections due to the effects of supersymmetry-breaking that enter via the loops. Thus, the Higgs-sector of the (radiatively-corrected) MSSM can be described by an effective field theory consisting of the most general CP-violating two-Higgs-doublet model.

The 2HDM Lagrangian contains eight real scalar fields. After electroweak symmetry breaking, three Goldstone bosons ($G^\pm$ and $G^0$) are removed from the spectrum and provide the longitudinal modes of the massive $W^\pm$ and $Z$. Five physical Higgs particles remain: a charged Higgs pair ($H^\pm$) and three neutral Higgs bosons. If experimental data reveals the existence of a Higgs sector beyond that of the SM, it will be crucial to test whether the observed scalar spectrum is consistent with a 2HDM interpretation. In order to be completely general within this framework, one should allow for the most general CP-violating 2HDM when confronting the data. Any observed relations among the general 2HDM parameters would surely contribute to the search for a deeper theoretical understanding of the origin of electroweak symmetry breaking.

#### *2.1.1 The general Two-Higgs-Doublet Model (2HDM)*

The 2HDM is governed by the choice of the Higgs potential and the Yukawa couplings of the two scalar-doublets to the three generations of quarks and leptons. Let $\Phi_1$ and $\Phi_2$ denote two complex hypercharge-one, $\mathrm{SU}(2)_L$ doublet scalar fields. The most general gauge-invariant renormalizable Higgs scalar potential is given by

$$\begin{aligned}\mathcal{V} \;=\;& m_{11}^2\Phi_1^\dagger\Phi_1 + m_{22}^2\Phi_2^\dagger\Phi_2 - [m_{12}^2\Phi_1^\dagger\Phi_2 + \mathrm{h.c.}] \\ & + \tfrac{1}{2}\lambda_1(\Phi_1^\dagger\Phi_1)^2 + \tfrac{1}{2}\lambda_2(\Phi_2^\dagger\Phi_2)^2 + \lambda_3(\Phi_1^\dagger\Phi_1)(\Phi_2^\dagger\Phi_2) + \lambda_4(\Phi_1^\dagger\Phi_2)(\Phi_2^\dagger\Phi_1) \\ & + \left\{\tfrac{1}{2}\lambda_5(\Phi_1^\dagger\Phi_2)^2 + \left[\lambda_6(\Phi_1^\dagger\Phi_1) + \lambda_7(\Phi_2^\dagger\Phi_2)\right]\Phi_1^\dagger\Phi_2 + \mathrm{h.c.}\right\}\,, \qquad (2.1)\end{aligned}$$

where $m_{11}^2$, $m_{22}^2$, and $\lambda_1,\cdots,\lambda_4$ are real parameters. In general, $m_{12}^2$, $\lambda_5$, $\lambda_6$ and $\lambda_7$ are complex.

2.1.1.1 Covariant notation with respect to scalar field redefinitions

In writing Eq. (2.1), we have implicitly chosen a basis in the two-dimensional "flavor" space of scalar fields. To allow for other basis choices, it will be convenient to rewrite Eq. (2.1) in a covariant form with respect to global U(2) transformations, $\Phi_a \to U_{a\bar{b}}\Phi_b$ (and $\Phi^\dagger_{\bar{a}} \to \Phi^\dagger_{\bar{b}} U^\dagger_{b\bar{a}}$), where the $2 \times 2$ unitary matrix $U$ satisfies $U^\dagger_{b\bar{a}} U_{a\bar{c}} = \delta_{b\bar{c}}$. In our index conventions, replacing an unbarred index with a barred index is equivalent to complex conjugation (for further details see section 2.3). Thus, Eq. (2.1) can be expressed in U(2)-covariant form as [10, 14]:

$$\mathcal{V} = Y_{a\bar{b}}\Phi^\dagger_{\bar{a}}\Phi_b + \tfrac{1}{2} Z_{a\bar{b}c\bar{d}}(\Phi^\dagger_{\bar{a}}\Phi_b)(\Phi^\dagger_{\bar{c}}\Phi_d)\,, \tag{2.2}$$

where the indices $a$, $\bar{b}$, $c$ and $\bar{d}$ run over the two-dimensional Higgs "flavor" space and $Z_{a\bar{b}c\bar{d}} = Z_{c\bar{d}a\bar{b}}$. Hermiticity of $\mathcal{V}$ implies that $Y_{a\bar{b}} = (Y_{b\bar{a}})^*$ and $Z_{a\bar{b}c\bar{d}} = (Z_{b\bar{a}d\bar{c}})^*$. Explicitly, the coefficients of the quadratic terms are

$$\begin{aligned} Y_{11} &= m^2_{11}\,, & Y_{12} &= -m^2_{12}\,, \\ Y_{21} &= -(m^2_{12})^*\,, & Y_{22} &= m^2_{22}\,, \end{aligned} \tag{2.3}$$

and the coefficients of the quartic terms are

$$\begin{aligned} Z_{1111} &= \lambda_1\,, & Z_{2222} &= \lambda_2\,, \\ Z_{1122} &= Z_{2211} = \lambda_3\,, & Z_{1221} &= Z_{2112} = \lambda_4\,, \\ Z_{1212} &= \lambda_5\,, & Z_{2121} &= \lambda^*_5\,, \\ Z_{1112} &= Z_{1211} = \lambda_6\,, & Z_{1121} &= Z_{2111} = \lambda^*_6\,, \\ Z_{2212} &= Z_{1222} = \lambda_7\,, & Z_{2221} &= Z_{2122} = \lambda^*_7\,. \end{aligned} \tag{2.4}$$

Under the global U(2) transformation, the tensors $Y$ and $Z$ transform covariantly: $Y_{a\bar{b}} \to U_{a\bar{c}} Y_{c\bar{d}} U^\dagger_{d\bar{b}}$ and $Z_{a\bar{b}c\bar{d}} \to U_{a\bar{e}} U^\dagger_{f\bar{b}} U_{c\bar{g}} U^\dagger_{h\bar{d}} Z_{e\bar{f}g\bar{h}}$. Indices can only be summed over using the U(2)-invariant tensor $\delta_{a\bar{b}}$.

The advantage of introducing the U(2)-covariant notation is that one can immediately identify U(2)-invariant quantities as *basis-independent*; such quantities do not depend on the original choice of the $\Phi_1$–$\Phi_2$ basis. In particular, any physical observable must be independent of the basis choice and hence can be identified as some U(2)-invariant quantity. For example, the well-known $\tan\beta$ parameter of the general 2HDM is *not* a physical quantity [14–16].

2.1.1.2 Counting the degrees of freedom

The 2HDM scalar potential depends on six real parameters and four complex parameters, for a total of fourteen degrees of freedom. However, these parameters depend on the choice of the $\Phi_1$–$\Phi_2$ basis. In order to determine the number of physical degrees of freedom, one must take into account the possibility that unphysical degrees of freedom can be removed by redefining the scalar fields via the global U(2) "flavor" transformations. However, note that the global U(2) group can be decomposed as U(2) $\cong$ SU(2)$\times$U(1), where the global hypercharge U(1) transformation has *no* effect on the scalar potential parameters. In contrast, the scalar potential parameters will be modified by a general SU(2)-"flavor" transformation. Since an SU(2) transformation is specified by three parameters, three degrees of freedom can be removed by a redefinition of the scalar fields. Thus, the scalar potential provides eleven physical degrees of freedom that govern the properties of the 2HDM scalar sector [14, 15, 17].

2.1.1.3 Discrete symmetries and the 2HDM potential

The general 2HDM is not phenomenologically viable over most of its parameter space. In particular, if we allow for the most general Higgs-fermion Yukawa couplings, the model exhibits tree-level Higgs-mediated flavor-changing neutral currents (FCNCs), which may contradict the experimental bounds on FCNCs. This can be ameliorated by either avoiding the untenable regions of parameter space or by introducing additional structure into the model. For example, in the Higgs sector of the MSSM, tree-level Higgs-mediated FCNCs are absent due to the supersymmetric structure of the Higgs-fermion Yukawa couplings. Tree-level Higgs-mediated FCNCs can also be eliminated by invoking appropriate discrete symmetries [18]. Here, we focus on discrete symmetries imposed on the scalar fields. Consider a discrete $Z_2$ symmetry realized for some choice of basis: $\Phi_1 \to \Phi_1$, $\Phi_2 \to -\Phi_2$. This discrete symmetry implies that $m_{12}^2 = \lambda_6 = \lambda_7 = 0$. A basis-independent characterization of this discrete symmetry has been given in [14, 19]. In practice, the discrete symmetry must also be extended to the fermion sector. By specifying the transformation properties of the fermions with respect to the discrete symmetry, one can constrain the form of the Higgs-fermion Yukawa interactions. In fact, removing the possibility of dangerous FCNC effects can also be achieved if the symmetry of the $Z_2$ discrete transformation of the Higgs potential is *softly* broken; *i.e.*, there exists a basis in which $\lambda_6 = \lambda_7 = 0$ but $m_{12}^2 \neq 0$ [15, 17]. A basis-independent characterization of the softly-broken discrete symmetry can also be given [14]. Finally, hard-breaking of the discrete $Z_2$ symmetry corresponds to the case in which no basis exists in which $\lambda_6 = \lambda_7 = 0$. Additional implications of the broken $Z_2$ symmetry can be found in section 2.4.

2.1.1.4 The scalar field vacuum expectation values

Electroweak symmetry breaking arises if the minimum of the scalar potential occurs for nonzero expectation values of the scalar fields. The condition for extrema of the scalar potential

$$\left.\frac{\partial \mathcal{V}}{\partial \Phi_1}\right|_{\substack{\Phi_1=\langle\Phi_1\rangle,\\ \Phi_2=\langle\Phi_2\rangle}} = 0, \qquad \left.\frac{\partial \mathcal{V}}{\partial \Phi_2}\right|_{\substack{\Phi_1=\langle\Phi_1\rangle,\\ \Phi_2=\langle\Phi_2\rangle}} = 0 \tag{2.5}$$

yields the vacuum expectation values (vevs) $\langle\Phi_{1,2}\rangle$. The scalar fields will develop non-zero vevs if the mass matrix constructed from the quadratic squared-mass parameters of the Higgs potential $(m_{ij}^2)$ has at least one negative eigenvalue. By employing an appropriate weak isospin and U$(1)_Y$ transformation, it is always possible to write the scalar field vevs in the following form

$$\langle\Phi_1\rangle = \frac{1}{\sqrt{2}}\begin{pmatrix} 0 \\ v_1 \end{pmatrix}, \qquad \langle\Phi_2\rangle = \frac{1}{\sqrt{2}}\begin{pmatrix} u \\ v_2 e^{i\xi} \end{pmatrix}, \tag{2.6}$$

where $v_1$ and $v_2$ are real and positive, and $0 \leq \xi < 2\pi$. Depending on the parameters of Higgs potential, the extremum for $u \neq 0$ describes either saddle point or a minimum of the potential, called the *charged vacuum*, where the U$(1)_{\rm EM}$ symmetry is spontaneously broken [15, 20–22]. The vacuum solution with $u = 0$ preserves the U$(1)_{\rm EM}$ symmetry; it corresponds to a local minimum of potential if its parameters are such that the physical Higgs squared-masses are non-negative. In this case, one can show that the energy of the charged vacuum is larger than energy of the U$(1)_{\rm EM}$ preserving vacuum [20, 22].

Henceforth, we assume that the global minimum of the scalar potential respects the U$(1)_{\rm EM}$ gauge symmetry. In this case $u = 0$ and it is convenient to write:

$$v_1 \equiv v\cos\beta\,, \qquad v_2 \equiv v\sin\beta\,, \tag{2.7}$$

where $v^2 \equiv v_1^2 + v_2^2 = (\sqrt{2}G_F)^{-1/2} = (246~\text{GeV})^2$ and $0 \leq \beta \leq \pi/2$.

One is always free to rephase $\Phi_2$ in order to set $\xi = 0$. In the following, we shall always work in a basis in which the two neutral Higgs field vevs are real and positive (corresponding to a *real vacuum*). The scalar minimum conditions (2.5) then yield:

$$m_{11}^2 = m_{12}^2\, t_\beta - \tfrac{1}{2}v^2\left[\lambda_1 c_\beta^2 + (\lambda_3 + \lambda_4 + \lambda_5 s_\beta^2 + (2\lambda_6 + \lambda_6^*) s_\beta c_\beta + \lambda_7 s_\beta^2 t_\beta\right] \tag{2.8}$$

$$m_{22}^2 = (m_{12}^2)^*\, t_\beta^{-1} - \tfrac{1}{2}v^2\left[\lambda_2 s_\beta^2 + (\lambda_3 + \lambda_4 + \lambda_5^* c_\beta^2 + \lambda_6^* c_\beta^2 t_\beta^{-1} + (\lambda_7 e^{i\xi} + 2\lambda_7^* e^{-i\xi}) s_\beta c_\beta\right], \tag{2.9}$$

where $s_\beta = \sin\beta$, $c_\beta = \cos\beta$ and $t_\beta = \tan\beta$. Since $m_{11}^2$ and $m_{22}^2$ are both real, the imaginary part of either Eq. (2.8) or Eq. (2.9) yields one independent equation:

$$\mathrm{Im}\,(m_{12}^2) = \tfrac{1}{2}v^2\left[\mathrm{Im}\,(\lambda_5)s_\beta c_\beta + \mathrm{Im}\,(\lambda_6)c_\beta^2 + \mathrm{Im}\,(\lambda_7)s_\beta^2\right] . \tag{2.10}$$

The quantities

$$\delta \equiv \frac{\mathrm{Im}\,(m_{12}^2)}{v^2 s_\beta c_\beta}\,, \qquad \eta \equiv \frac{\mathrm{Re}\,(m_{12}^2)}{v^2 s_\beta c_\beta}\,, \tag{2.11}$$

will be useful in our discussion of the Higgs mass eigenstates and the mixing of CP-even and CP-odd states. Note that $\mathrm{Re}\,(m_{12}^2)$ is not determined by the scalar potential minimum conditions.

#### 2.1.1.5 Theoretical constraints on the Higgs potential parameters

The parameters of Higgs potential are constrained by various conditions. To have a *stable vacuum*, the potential must be positive at large quasi–classical values of the magnitudes of the scalar fields for an arbitrary direction in the $(\Phi_1, \Phi_2)$ plane. These are the *positivity constraints* [23–26]. The *minimum constraints* are the conditions ensuring that the extremum is a minimum for all directions in $(\Phi_1, \Phi_2)$ space, except for the direction of the Goldstone modes. It is realized when the squared-masses of the five physical Higgs bosons are all positive.

The tree-level amplitudes for the scattering of longitudinal gauge bosons at high energy can be related via the equivalence theorem [27] to the corresponding amplitudes in which the longitudinal gauge bosons are replaced by Goldstone bosons. The latter can be computed in terms of quartic couplings $\lambda_i$ that appear in the Higgs potential. By imposing *tree-level unitarity constraints* on these amplitudes, one can derive upper bounds on the values of certain combinations of Higgs quartic couplings [28–34].

The *perturbativity condition* for a validity of a tree approximation in the description of interactions of the lightest Higgs boson may be somewhat less restrictive than the unitarity constraints. For example, by requiring that one-loop corrections to Higgs self-couplings are small compared to tree-level couplings, one expects that $|\lambda_i|/16\pi^2 \ll 1$.

Unitarity constraints for the 2HDM were first derived for the potential without a hard violation of the discrete $Z_2$ symmetry and for the CP conserving case (*e.g.*, see [32]). Extension to the CP-violating case can be found in [33], and for the case of hard discrete $Z_2$ symmetry violation in [34].

### *2.1.2 Conditions for Higgs sector CP-violation*

Higgs sector CP-violation may be either explicit or spontaneous. Explicit CP conservation[1] or violation refers respectively to the consistent or inconsistent CP transformation properties of the various terms that appear in the Lagrangian. If the scalar Lagrangian is explicitly CP-conserving, but the vacuum state of the theory violates CP, then one says that CP is spontaneously broken [1, 10, 35]. The observable consequences of Higgs sector CP-violation (either explicit or spontaneous) include the mixing of neutral Higgs states of opposite CP quantum numbers and/or the existence of (direct) CP-violating Higgs interactions.

The CP state mixing and the direct CP-violation in the gauge/Higgs interactions are determined by the properties of the scalar Lagrangian (and the corresponding vacuum state). These CP-violating effects are absent if and only if there exists a basis in which the two neutral Higgs vacuum expectation values and the scalar potential parameters are simultaneously real [36, 37]. Given an arbitrary potential, the existence or non-existence of such a basis may be difficult to determine directly. For this problem, the basis-independent methods are invaluable. In particular, a set of basis-independent conditions can be found to test for the CP-invariance of the scalar sector. Following [38], we introduce three U(2)-invariant quantities [37]:

$$-\tfrac{1}{2}v^2 J_1 \equiv \widehat{v}_{\bar{a}}^* Y_{a\bar{b}} Z^{(1)}_{b\bar{d}} \widehat{v}_d \,, \tag{2.12}$$

$$\tfrac{1}{4}v^4 J_2 \equiv \widehat{v}_{\bar{b}}^* \widehat{v}_{\bar{c}}^* Y_{b\bar{e}} Y_{c\bar{f}} Z_{e\bar{a}f\bar{d}} \widehat{v}_a \widehat{v}_d \,, \tag{2.13}$$

$$J_3 \equiv \widehat{v}_{\bar{b}}^* \widehat{v}_{\bar{c}}^* Z^{(1)}_{b\bar{e}} Z^{(1)}_{c\bar{f}} Z_{e\bar{a}f\bar{d}} \widehat{v}_a \widehat{v}_d \,, \tag{2.14}$$

where $\langle\Phi_a^0\rangle \equiv v\widehat{v}_a/\sqrt{2}$, and $\widehat{v}_a$ is a unit vector in the complex two-dimensional Higgs flavor space. Then, the scalar sector is CP-conserving (*i.e.*, no explicit nor spontaneous CP-violation is present) if $J_1$, $J_2$ and $J_3$ defined in Eqs. (2.12)–(2.14) are real.[2] If the scalar potential is CP-violating, then the CP state mixing depends only on $\operatorname{Im} J_2$ [16, 39], whereas CP-violation in the gauge/Higgs boson interactions is governed by all three quantities $\operatorname{Im} J_k$, $k = 1, 2, 3$.

[1] Since CP is violated in the SM via the CKM mixing of the quarks, it is generally unnatural to demand that the Higgs sector of the 2HDM explicitly conserve CP. Nevertheless, one can naturally impose a CP-conserving Higgs sector by employing an appropriate discrete symmetry. In the MSSM, the Higgs sector is CP-conserving at tree-level (due to the supersymmetry), although one finds CP-violation arising at one-loop due to supersymmetry-breaking effects.

### 2.1.2.1 Explicit CP-conservation

The general 2HDM scalar potential explicitly violates the CP symmetry. An explicitly CP-conserving scalar potential requires the existence of a $\Phi_1$–$\Phi_2$ basis in which all the Higgs potential parameters are real. Such a basis will henceforth be called a *real basis*. However, given an arbitrary potential, the existence or non-existence of a real basis may be difficult to discern, as already noted. In Ref. [37], the necessary and sufficient *basis-independent* conditions for an explicitly CP-conserving scalar potential have been established, in terms of the following four potentially complex invariants:

$$I_{Y3Z} \equiv \operatorname{Im}(Z^{(1)}_{a\bar{c}} Z^{(1)}_{e\bar{b}} Z_{b\bar{e}c\bar{d}} Y_{d\bar{a}}) \,, \tag{2.15}$$

$$I_{2Y2Z} \equiv \operatorname{Im}(Y_{a\bar{b}} Y_{c\bar{d}} Z_{b\bar{a}d\bar{f}} Z^{(1)}_{f\bar{c}}) \,, \tag{2.16}$$

$$I_{6Z} \equiv \operatorname{Im}(Z_{a\bar{b}c\bar{d}} Z^{(1)}_{b\bar{f}} Z^{(1)}_{d\bar{h}} Z_{f\bar{a}j\bar{k}} Z_{k\bar{j}m\bar{n}} Z_{n\bar{m}h\bar{c}}) \,, \tag{2.17}$$

$$I_{3Y3Z} \equiv \operatorname{Im}(Z_{a\bar{c}b\bar{d}} Z_{c\bar{e}d\bar{g}} Z_{e\bar{h}f\bar{q}} Y_{g\bar{a}} Y_{h\bar{b}} Y_{q\bar{f}}) \,, \tag{2.18}$$

where $Z^{(1)}_{a\bar{d}} \equiv \delta_{b\bar{c}} Z_{a\bar{b}c\bar{d}}$.

The conditions for a CP-conserving scalar potential depend on the invariant quantity [14, 19]:

$$Z \equiv 2\,\mathrm{Tr}\ [Z^{(1)}]^2 - (\mathrm{Tr}\ Z^{(1)})^2 = (\lambda_1 - \lambda_2)^2 + 4|\lambda_6 + \lambda_7|^2 \,, \tag{2.19}$$

Note that if $Z$ vanishes, then Eq. (2.19) implies that $\lambda_1 = \lambda_2$ and $\lambda_7 = -\lambda_6$ for *all* basis choices. Two distinct cases are possible. If $Z \neq 0$, then the necessary and sufficient conditions for an explicitly CP-conserving 2HDM scalar potential are given by $I_{Y3Z} = I_{2Y2Z} = I_{6Z} = 0$. (A similar result has also been obtained in [40].) In this case $I_{3Y3Z} = 0$ is automatically satisfied. If $Z = 0$, then the aforementioned first three invariants automatically vanish, in which case the necessary and sufficient condition for an explicitly CP-conserving 2HDM scalar potential is given by $I_{3Y3Z} = 0$. Explicit expressions for the imaginary parts of the four CP-odd invariants above can be found in [37]. The significance of the four conditions above from a group-theoretical perspective has been recently discussed in [19, 41].

Finally, we note that the imposition of the discrete $Z_2$ symmetry $\Phi_1 \to \Phi_1$, $\Phi_2 \to -\Phi_2$ implies that the scalar potential is CP-conserving. Since $\lambda_5$ is the only nonzero complex parameter in the basis where the discrete symmetry is manifest, it is a simple matter to rephase one of the scalar doublets to render $\lambda_5$ real. Explicit CP-violation can arise if the $\Phi_1 \to \Phi_1$, $\Phi_2 \to -\Phi_2$ discrete $Z_2$ symmetry breaking is either hard or soft. In the latter case, *e.g.*, CP violation is a consequence of a nontrivial relative phase in the complex parameters $m_{12}^2$ and $\lambda_5$.

[2] One can show that the reality of the $J_k$ is equivalent to the invariant conditions given in Eq. (2.75) of section 2.3 [14, 38].

2.1.2.2 Spontaneous CP-violation

If the scalar Lagrangian is explicitly CP-conserving but the Higgs vacuum is CP-violating, then CP is spontaneously broken. However, both spontaneous and explicit CP-violation yield similar CP-violating phenomenology. To distinguish between the two, one would need to discover CP-violation in the Higgs sector and prove that the fundamental scalar Lagrangian is CP-conserving. In principle, such a distinction is possible. For example, suppose one could verify that $I_{Y3Z} = I_{2Y2Z} = I_{6Z} = I_{3Y3Z} = 0$, whereas at least one of three invariants $J_1$, $J_2$ and $J_3$ possesses a non-zero imaginary part. In this case, the CP-symmetry in the Higgs sector is spontaneously broken.[3] In practice, distinguishing between explicit and spontaneous CP-violation by experimental observations and analysis seems extremely difficult.

Spontaneous CP-violation cannot arise in the presence of the $\Phi_1 \to \Phi_1$, $\Phi_2 \to -\Phi_2$ discrete $Z_2$ symmetry. In particular, in this case the scalar potential minimum condition implies that it is possible to transform to a real basis in which the two neutral vacuum expectation values are real.

### *2.1.3 The Higgs mass spectrum*

2.1.3.1 CP violation and mixing of states

We introduce the following field decomposition

$$\Phi_1 = \begin{pmatrix} \varphi_1^+ \\ \dfrac{v_1 + \varphi_1 + i\chi_1}{\sqrt{2}} \end{pmatrix}, \qquad \Phi_2 = \begin{pmatrix} \varphi_2^+ \\ \dfrac{v_2 + \varphi_2 + i\chi_2}{\sqrt{2}} \end{pmatrix}. \tag{2.20}$$

Then the corresponding scalar squared-mass matrix can be transformed to the block diagonal form by a separation of the massless charged and neutral Goldstone boson fields, $G^\pm$ and $G^0$, and the charged Higgs boson fields $H^\pm$:

$$G^\pm = \cos\beta\,\varphi_1^\pm + \sin\beta\,\varphi_2^\pm\,, \tag{2.21}$$

$$G^0 = \cos\beta\;\chi_1 + \sin\beta\;\chi_2\,. \tag{2.22}$$

The physical charged Higgs boson is orthogonal to $G^\pm$:

$$H^\pm = -\sin\beta\,\varphi_1^\pm + \cos\beta\,\varphi_2^\pm\,. \tag{2.23}$$

The mass of the charged Higgs boson is easily obtained:

$$M_{H^\pm}^2 = \left[\eta - \frac{1}{2}(\lambda_4 + \mathrm{Re}\,\lambda_5 + \mathrm{Re}\,\lambda_{67})\right] v^2\,, \tag{2.24}$$

where $\lambda_{67} \equiv \lambda_6 \cot\beta + \lambda_7 \tan\beta$ and $\eta$ is defined in Eq. (2.11). The physical neutral Higgs bosons are mixtures of the two CP-even fields $\varphi_1$, $\varphi_2$ and a CP-odd field

$$A = -\sin\beta\;\chi_1 + \cos\beta\;\chi_2\,, \tag{2.25}$$

that is orthogonal to $G^0$. Consequently, in the general 2HDM, the physical neutral Higgs bosons are states of indefinite CP.

In the $\{\varphi_1, \varphi_2, A\}$ basis, the real symmetric squared-mass matrix $\mathcal{M}^2$ for neutral sector is obtained:

$$\mathcal{M}^2 = \begin{pmatrix} M_{11}^2 & M_{12}^2 & M_{13}^2 \\ M_{12}^2 & M_{22}^2 & M_{23}^2 \\ M_{13}^2 & M_{23}^2 & M_{33}^2 \end{pmatrix}. \tag{2.26}$$

[3] One would also have to prove the absence of explicit CP-violation in the Higgs-fermion couplings. The relevant basis-independent conditions have been given in [10, 38].

Diagonalizing the matrix $\mathcal{M}^2$ by using an orthogonal transformation $R$ we obtain the physical neutral states $h_{1,2,3}$, with corresponding squared-masses $M_i^2$ that are the eigenvalues of the matrix $\mathcal{M}^2$:

$$\begin{pmatrix} h_1 \\ h_2 \\ h_3 \end{pmatrix} = R \begin{pmatrix} \varphi_1 \\ \varphi_2 \\ A \end{pmatrix} , \quad \text{with} \quad R\mathcal{M}^2 R^T = \text{diag}(M_1^2,\, M_2^2,\, M_3^2)\,. \tag{2.27}$$

The diagonalizing matrix $R$ can be written as a product of three rotation matrices $R_i$, corresponding to rotations by three angles $\alpha_i \in (0, \pi)$ about the $z$, $y$ and $x$ axes, respectively:

$$R = R_3 R_2 R_1 = \begin{pmatrix} c_1\, c_2 & c_2\, s_1 & s_2 \\ -c_1\, s_2\, s_3 - c_3\, s_1 & c_1\, c_3 - s_1\, s_2\, s_3 & c_2\, s_3 \\ -c_1\, c_3\, s_2 + s_1\, s_3 & -c_1\, s_3 - c_3\, s_1\, s_2 & c_2\, c_3 \end{pmatrix} . \tag{2.28}$$

Here, we define $c_i = \cos\alpha_i$, $s_i = \sin\alpha_i$ and adopt the convention for masses that $M_1 \le M_2 \le M_3$.

One can first diagonalize the upper left $2 \times 2$ block of the matrix $\mathcal{M}^2$. This partial diagonalization [15] results in the neutral, CP-even Higgs fields which we denote as $h$ and $(-H)$,

$$H = \cos\alpha\, \varphi_1 + \sin\alpha\, \varphi_2, \qquad h = -\sin\alpha\, \varphi_1 + \cos\alpha\, \varphi_2\,, \tag{2.29}$$

where $\alpha \equiv \alpha_1 - \pi/2$ is the mixing angle that renders the $2 \times 2$ CP-even submatrix diagonal.[4] At this stage the CP–odd field $A$ remains unmixed:

$$\begin{pmatrix} h \\ -H \\ A \end{pmatrix} = R_1 \begin{pmatrix} \varphi_1 \\ \varphi_2 \\ A \end{pmatrix} , \quad \text{with} \quad R_1 \mathcal{M}^2 R_1^T = \mathcal{M}_1^2 \equiv \begin{pmatrix} M_h^2 & 0 & M_{13}'^2 \\ 0 & M_H^2 & M_{23}'^2 \\ M_{13}'^2 & M_{23}'^2 & M_A^2 \end{pmatrix} , \tag{2.30}$$

where

$$M_A^2 = \left[\eta - \text{Re}\,(\lambda_5 - \tfrac{1}{2}\lambda_{67})\right] v^2\,, \tag{2.31}$$

$$M_{h,H}^2 = \tfrac{1}{2}\left[M_{11} + M_{22} \mp \sqrt{(M_{11} - M_{22})^2 + 4M_{12}^2}\,\right] . \tag{2.32}$$

The off-diagonal squared-masses $M_{13}'^2$ and $M_{23}'^2$ are given by

$$M_{13}'^2 = c_1 M_{13}^2 + s_1 M_{23}^2 = -\tfrac{1}{2}\left[2\delta\cos(\beta + \alpha) - \,\text{Im}\,\tilde\lambda_{67}\cos(\beta - \alpha)\right] v^2, \tag{2.33}$$

$$M_{23}'^2 = -s_1 M_{13}^2 + c_1 M_{23}^2 = \tfrac{1}{2}\left[2\delta\sin(\beta + \alpha) + \,\text{Im}\,\tilde\lambda_{67}\sin(\beta - \alpha)\right] v^2\,, \tag{2.34}$$

where $\tilde\lambda_{67} \equiv \lambda_6 \cot\beta - \lambda_7 \tan\beta$ and $\delta$ is defined in Eq. (2.11).

In the general CP-violating 2HDM, the states $h$, $H$ and $A$ are useful intermediaries, which do not directly correspond to physical objects. In the case of CP conservation (realized for $M_{13}'^2 = M_{23}'^2 = 0$), the fields $h$, $H$ and $A$ represent physical Higgs bosons: $h_1 = h$, $h_2 = -H$, $h_3 = A$. If at least one of the off diagonal terms differs from zero, an additional diagonalization is necessary, and the mass eigenstates, which are now admixtures of CP–even and CP–odd states, violate the CP symmetry. In this case we express the physical Higgs boson states $h_{1,2,3}$ as linear combinations of $h$, $H$, $A$:

$$\begin{pmatrix} h_1 \\ h_2 \\ h_3 \end{pmatrix} = R_3 R_2 \begin{pmatrix} h \\ -H \\ A \end{pmatrix} \quad \text{with} \quad R\,\mathcal{M}^2\, R^T = R_3 R_2 \mathcal{M}_1^2 R_2^T R_3^T = \begin{pmatrix} M_1^2 & 0 & 0 \\ 0 & M_2^2 & 0 \\ 0 & 0 & M_3^2 \end{pmatrix} . \tag{2.35}$$

[4] The appearance of the minus sign in $-H$ and the shift by $\pi/2$ in the definition of $\alpha$ is needed in order to match the standard convention used for CP-conserving case [8].

The following mass sum rule holds:

$$M_1^2 + M_2^2 + M_3^2 = M_h^2 + M_H^2 + M_A^2 = M_{11}^2 + M_{22}^2 + M_{33}^2 \,. \qquad (2.36)$$

In general, the Higgs mass-eigenstates $h_i$ [Eq. (2.27)] are not states of definite CP parity since they are mixtures of fields $\varphi_{1,2}$ and $A$, which possess opposite CP parities. Such CP-state mixing is absent if and only if $M_{13}^2 = M_{23}^2 = 0$. In particular, for $\sin 2\beta \neq 0$, the absence of CP-state mixing implies that $\mathrm{Im}\,\tilde{\lambda}_{67} = 0$ and $\delta \propto \mathrm{Im}\,(m_{12}^2) = 0$. In this latter case, $h$, $H$ and $A$ are the physical Higgs bosons, with masses given by eqs. (2.31) and (2.32), and $\alpha_2 = \alpha_3 = 0$.

2.1.3.2 Various cases of CP mixing

We consider a number of possible interesting patterns of CP-even/CP-odd scalar state mixing [15]:

• If $\boldsymbol{\varepsilon_{13} \equiv |M_{13}'^2/(M_A^2 - M_h^2)| \ll 1}$, then $\alpha_2 \approx 0$ and the Higgs boson $h_1$ practically coincides with the lighter CP-even state, $h$. In addition, the CP-violating couplings of $h$ are very small, typically of $\mathcal{O}(\varepsilon_{13})$. The diagonalization of the residual $\langle 23 \rangle$ corner of the squared-mass matrix (2.30) using the rotation matrix $R_3$ yields the mass eigenstates $h_2$ and $h_3$. These are superpositions of $H$ and $A$ with a potentially large mixing angle $\alpha_3$:

$$\tan 2\alpha_3 \approx \frac{-2M_{23}'^2}{M_A^2 - M_H^2} \,. \qquad (2.37)$$

If $M_A \approx M_H$, then the CP-violating state mixing can be strong even at small but nonzero $|M_{23}'^2|/v^2$. For large values of $M_H \approx M_A$ the proper widths of $H$ and $A$ become large and the $H$ and $A$ mass peaks strongly overlap. Here, one should include a (complex) matrix of Higgs polarization operators [42,43].

• If $\boldsymbol{\varepsilon_{23} \equiv |M_{23}'^2/(M_A^2 - M_H^2)| \ll 1}$, then $\alpha_3 \approx 0$ and the Higgs boson $h_2$ practically coincides with the heavier CP-even state, $-H$. Similarly to the previous case, the diagonalization of the $\langle 13 \rangle$ part of squared-mass matrix (2.30), using the rotation matrix $R_2$ yields the mass eigenstates $h_1$ and $h_3$. These are superpositions of $h$ and $A$ states, which can strongly mix with large mixing angle $\alpha_2$:

$$\tan 2\alpha_2 \approx \frac{-2M_{13}'^2}{M_A^2 - M_h^2} \,. \qquad (2.38)$$

As in the previous case, if $M_A \approx M_h$, the CP-violating state mixing can be strong even at small $M_{13}'^2/v^2$.

• The case of *weak CP-violating state mixing* combines both cases above. That is $\varepsilon_{13}, \varepsilon_{23} \ll 1$, which imply that $\alpha_2, \alpha_3 \approx 0$, in which case the CP–even states $h$, $H$ are weakly mixed with the CP–odd state $A$. The corresponding physical Higgs masses are given by

$$M_1^2 \simeq M_h^2 - s_2^2(M_A^2 - M_h^2), \qquad M_2^2 \simeq M_H^2 - s_3^2(M_A^2 - M_H^2), \qquad (2.39)$$

with $M_3^2$ given by the sum rule (2.36). In the particular case of *soft-violation of the discrete $Z_2$ symmetry* we also have

$$s_2 \simeq \delta\, \frac{\cos(\beta+\alpha)}{M_A^2 - M_h^2} v^2, \qquad s_3 \simeq -\delta\, \frac{\sin(\beta+\alpha)}{M_A^2 - M_H^2} v^2. \qquad (2.40)$$

• The case of the *intense coupling regime* with $M_A \approx M_h \approx M_H$ [44] may also yield strong CP-violating state mixing even when both $\delta$ and $\mathrm{Im}\,\tilde{\lambda}_{67}$ are small.

### *2.1.4 Higgs boson couplings*

In the investigation of phenomenological aspects of 2HDM it is useful to introduce *relative couplings*, defined as the couplings of each neutral Higgs boson $h_i$ $(i = 1, 2, 3)$ to gauge bosons $W^+W^-$ or $ZZ$,

Higgs bosons $H^+H^-$ and $h_jh_k$, quarks $\bar{q}q$ ($q = u, d$) and charged leptons $\ell^+\ell^-$, normalized to the corresponding couplings of the SM Higgs boson:

$$\chi^{(i)} = g_j^{(i)}/g_j^{\rm SM}\,, \qquad j = W^\pm, Z, H^\pm, u, d, \ell \ldots\,, \tag{2.41}$$

where $g_j^{(i)}$ denotes the $jjh_i$ coupling. Note that for bosonic $j$, the relative couplings are real. In the case of neutral Higgs boson ($h_i$) couplings to fermions pairs $f\bar{f}$, the Yukawa couplings take the form

$$-\mathcal{L}_Y = \bar{f}(g_{Ri} + ig_{Ii}\gamma_5)f\, h_i = \bar{f}_L(g_{Ri} + ig_{Ii})f_R\, h_i + \text{h.c.}\,, \tag{2.42}$$

where the right and left-handed fermion fields are defined as usual: $f_R \equiv P_R f$ and $f_L \equiv P_L f$, with $P_{R,L} \equiv \frac{1}{2}(1 \pm \gamma_5)$. Hence, we shall compute the Higgs–fermion relative coupling in Eq. (2.41) by employing the *complex* couplings $g_i = g_{Ri} + ig_{Ii}$.

One can also make use of basis-independent techniques to obtain expressions for Higgs couplings to gauge bosons, Higgs bosons and fermions that are invariant under U(2) field redefinitions of the two complex scalar doublet fields [16]. Further details of this procedure and a complete collection of 2HDM couplings can be found in section 2.3.

#### 2.1.4.1 Bosonic sector

The gauge bosons $V$ ($W$ and $Z$) couple only to the CP–even fields $\varphi_1$, $\varphi_2$. In terms of the relative couplings defined in Eq. (2.41), the couplings of gauge bosons to the physical Higgs bosons $h_i$ are:

$$\chi_V^{(i)} = \cos\beta\, R_{i1} + \sin\beta\, R_{i2}, \qquad V = W \text{ or } Z. \tag{2.43}$$

In particular, in the case of weak CP-violating state mixing considered above, we obtain

$$\chi_V^{(1)} \simeq \sin(\beta - \alpha), \qquad \chi_V^{(2)} \simeq -\cos(\beta - \alpha), \qquad \chi_V^{(3)} \simeq -s_2 \sin(\beta - \alpha) + s_3 \cos(\beta - \alpha). \tag{2.44}$$

The cubic and quartic Higgs self-couplings as functions of the Higgs potential parameters and the elements of mixing matrix were obtained in [15, 16, 45–47]. In the case of soft $Z_2$ symmetry violation in the CP-conserving case, these latter results simplify. The Higgs self-couplings can be expressed in terms of the Higgs masses and the mixing angles $\alpha$ and $\beta$. Moreover, if the Higgs-fermion Yukawa interactions are of type-II [as defined below Eq. (2.46)], the trilinear couplings can be given in terms of the Higgs masses, the relative couplings to gauge bosons and quarks, and the parameter $\eta$ [15]. As an important example, in the case of weak CP-violating state mixing and soft $Z_2$ symmetry-violation, the coupling of the neutral scalar $h_i$ to a charged Higgs boson pair (normalized to $2M_{H^\pm}^2/v$) can be expressed in terms of the relative neutral Higgs couplings to the gauge bosons and fermions as follows:

$$\chi_{H^\pm}^{(i)} = \left(1 - \frac{M_i^2}{2M_{H^\pm}^2}\right)\chi_V^{(i)} + \frac{M_i^2 - \eta v^2}{2M_{H^\pm}^2}\, \text{Re}\,(\chi_u^{(i)} + \chi_d^{(i)}). \tag{2.45}$$

Deviations of the cubic Higgs boson self-couplings from the corresponding Standard Model value would also provide insight into the dynamics of the 2HDM. In particular, as emphasized in section 2.6, there is a strong correlation between the loop-corrected $hhh$ coupling and successful electroweak baryogenesis (that makes critical use of the CP-violation from the Higgs sector).

#### 2.1.4.2 Fermion–Higgs boson Yukawa couplings

The Higgs couplings to fermions are model dependent. The most general structure for the Higgs-fermion Yukawa couplings, often referred to as the type-III model [48, 49], is given in the generic basis by:

$$-\mathcal{L}_{\rm Y} = \overline{Q_L^0}\widetilde{\Phi}_1\Gamma_1 U_R^0 + \overline{Q_L^0}\Phi_1\Delta_1 D_R^0 + \overline{Q_L^0}\widetilde{\Phi}_2\Gamma_2 U_R^0 + \overline{Q_L^0}\Phi_2\Delta_2 D_R^0 + \text{h.c.}\,, \tag{2.46}$$

where $\widetilde{\Phi}_i \equiv i\sigma_2\Phi_i^*$, $Q_L^0$ is the weak isospin quark doublet, and $U_R^0$, $D_R^0$ are weak isospin quark singlets. Here, $Q_L^0$, $U_R^0$, $D_R^0$ denote the interaction basis states, which are vectors in the quark flavor space, and $\Gamma_1, \Gamma_2, \Delta_1, \Delta_2$ are Yukawa coupling matrices in quark flavor space.[5] We have omitted the leptonic couplings in Eq. (2.46); these follow the same pattern as the down-type quark couplings.

In some models, not all the terms in Eq. (2.46) are present at tree-level [50]. For example, in a type-I model (2HDM-I) [51], there exists a basis where $\Gamma_2 = \Delta_2 = 0$.[6] Similarly, in a type-II model (2HDM-II) [52], there exists a basis where $\Gamma_1 = \Delta_2 = 0$. The vanishing of certain Higgs-fermion couplings at tree-level can be enforced by imposing a discrete $Z_2$ symmetry under which $\Phi_1 \to \Phi_1$, $\Phi_2 \to -\Phi_2$, and the fermion fields are either invariant or change sign according to whether one wishes to preserve either the type-I or type-II Higgs-fermion couplings while eliminating the other possible terms in Eq. (2.46). Another well-known example is the MSSM Higgs sector, which exhibits a type-II Higgs-fermion coupling pattern that is enforced by supersymmetry.

The fermion–Higgs boson Yukawa couplings can be derived from Eq. (2.46) (see, *e.g.*, chapter 22 of [10]). Without loss of generality, we choose a basis corresponding to a real vacuum (*i.e.*, $\xi = 0$). The fermion mass eigenstates are related to the interaction eigenstates by bi-unitary transformations:

$$P_L U = V_L^U P_L U^0 \,, \qquad P_R U = V_R^U P_R U^0 \,, \qquad P_L D = V_L^D P_L D^0 \,, \qquad P_R D = V_R^D P_R D^0 \,, \tag{2.47}$$

and the Cabibbo-Kobayashi-Maskawa matrix is defined as $K \equiv V_L^U V_L^{D\,\dagger}$. It is also convenient to define "rotated" linear combinations of the Yukawa coupling matrices:

$$\kappa^U \equiv V_L^U(\Gamma_1 c_\beta + \Gamma_2 s_\beta)V_R^{U\,\dagger} \,, \qquad \rho^U \equiv V_L^U(-\Gamma_1 s_\beta + \Gamma_2 c_\beta)V_R^{U\,\dagger} \,, \tag{2.48}$$

$$\kappa^D \equiv V_L^D(\Delta_1 c_\beta + \Delta_2 s_\beta)V_R^{D\,\dagger} \,, \qquad \rho^D \equiv V_L^D(-\Delta_1 s_\beta + \Delta_2 c_\beta)V_R^{D\,\dagger} \,. \tag{2.49}$$

The quark mass terms are identified by replacing the scalar fields with their vacuum expectation values. The unitary matrices $V_L^U$, $V_L^D$, $V_R^U$ and $V_R^D$ are chosen so that $\kappa^D$ and $\kappa^U$ are diagonal with real non-negative entries. These quantities are proportional to the *diagonal* quark mass matrices:

$$M_D = \frac{v}{\sqrt{2}}\kappa^D \,, \qquad M_U = \frac{v}{\sqrt{2}}\kappa^U \,. \tag{2.50}$$

In a general model, the matrices $\rho^D$ and $\rho^U$ are independent *complex* non-diagonal matrices.

It is convenient to rewrite Eq. (2.46) in terms of the CP-even Higgs fields $H$ and $h$ and the CP-odd fields $A$ (and the Goldstone boson $G^0$). The end result is:

$$\begin{aligned}
-\mathcal{L}_Y = {} & \frac{1}{v}\,\overline{D}\left[M_D s_{\beta-\alpha} + \frac{v}{\sqrt{2}}(\rho^D P_R + \rho^{D\,\dagger} P_L)c_{\beta-\alpha}\right] Dh + \frac{i}{v}\overline{D}M_D\gamma_5 D G^0 \\
& + \frac{1}{v}\,\overline{D}\left[M_D c_{\beta-\alpha} - \frac{v}{\sqrt{2}}(\rho^D P_R + \rho^{D\,\dagger} P_L)s_{\beta-\alpha}\right] DH + \frac{i}{\sqrt{2}}\overline{D}(\rho^D P_R - \rho^{D\,\dagger} P_L)DA \\
& + \frac{1}{v}\,\overline{U}\left[M_U s_{\beta-\alpha} + \frac{v}{\sqrt{2}}(\rho^U P_R + \rho^{U\,\dagger} P_L)c_{\beta-\alpha}\right] Uh - \frac{i}{v}\overline{U}M_D\gamma_5 U G^0 \\
& + \frac{1}{v}\,\overline{U}\left[M_U c_{\beta-\alpha} - \frac{v}{\sqrt{2}}(\rho^U P_R + \rho^{U\,\dagger} P_L)s_{\beta-\alpha}\right] UH - \frac{i}{\sqrt{2}}\overline{U}(\rho^U P_R - \rho^{U\,\dagger} P_L)UA \\
& + \left\{\overline{U}\left[K\rho^D P_R - \rho^{U\,\dagger} K P_L\right] DH^+ + \frac{\sqrt{2}}{v}\,\overline{U}\left[K M_D P_R - M_U K P_L\right] DG^+ + \text{h.c.}\right\} \,,
\end{aligned} \tag{2.51}$$

[5] We have reversed the lettering conventions for these coupling matrices as compared to [10] since $\Delta$ is more naturally associated with the coupling to down-type quarks.

[6] A type-I model can also be defined as a model in which $\Gamma_1 = \Delta_1 = 0$ in some basis. Clearly, the two definitions are equivalent, since the difference in the two conditions is simply an interchange of $\Phi_1$ and $\Phi_2$ which can be viewed as a change of basis.

where $s_{\beta-\alpha} = \sin(\beta-\alpha)$ and $c_{\beta-\alpha} = \cos(\beta-\alpha)$. In the most general CP-violating 2HDM, the physical Higgs fields are linear combinations of $h$, $H$ and $A$. As advertised, since $\rho^D$ and $\rho^U$ are non-diagonal, Eq. (2.51) exhibits tree-level Higgs-mediated FCNCs.[7] See section 2.5 for a study of the implications of flavor-changing fermion–Higgs boson couplings for a variety of neutral current processes.

The fermion–Higgs boson Yukawa couplings simplify considerably in type-I and type-II models. In particular, $\rho^D$ and $\rho^U$ are no longer independent parameters. For example, in a one-generation type-II model, $\Gamma_1 = \Delta_2 = 0$, which implies that [14]

$$\tan\beta = \frac{-\rho^D}{\kappa^D} = \frac{\kappa^U}{\rho^U}\,. \tag{2.52}$$

These two equations are consistent, since the type-II condition is equivalent to $\kappa^U\kappa^D + \rho^U\rho^D = 0$. Moreover, using Eqs. (2.50) and (2.52), it follows that:

$$\rho^D = -\frac{\sqrt{2}m_d}{v}\tan\beta\,, \qquad \rho^U = \frac{\sqrt{2}m_u}{v}\cot\beta\,. \tag{2.53}$$

Inserting this result into Eq. (2.51) yields the well-known Feynman rules for the type-II Higgs-quark interactions. For example, in the case of weak CP-violating state mixing, one finds the expected form for the relative couplings of the neutral Higgs bosons to the up and down-type quarks:

$$\chi_d^{(1)} = -\frac{\sin\alpha}{\cos\beta} = s_{\beta-\alpha} - \tan\beta\, c_{\beta-\alpha}\,, \qquad \chi_u^{(1)} = \frac{\cos\alpha}{\sin\beta} = s_{\beta-\alpha} + \cot\beta\, c_{\beta-\alpha}\,, \tag{2.54}$$

$$-\chi_d^{(2)} = \frac{\cos\alpha}{\cos\beta} = c_{\beta-\alpha} + \tan\beta\, s_{\beta-\alpha}\,, \qquad -\chi_u^{(2)} = \frac{\sin\alpha}{\sin\beta} = c_{\beta-\alpha} - \cot\beta\, s_{\beta-\alpha}\,, \tag{2.55}$$

$$\chi_d^{(3)} = -i\tan\beta\,, \qquad \chi_u^{(3)} = -i\cot\beta\,. \tag{2.56}$$

Note the extra minus sign in $\chi_i^{(2)}$ which arises due to the identification of $h_2 \simeq -H$ in this limiting case.

A similar analysis can be given for models of type-I. In the same CP-conserving limiting case considered above, $\chi_u^{(i)}$ is identical to the corresponding type-II values given above, but $\chi_d^{(i)} = \chi_u^{(i)\,*}$.

#### 2.1.4.3 The decoupling limit and implications for a SM-like Higgs boson

Suppose that all the coefficients of the quartic terms are held fixed [with values that are not allowed to exceed $\mathcal{O}(1)$]. Then, in the limit that $M_{H^\pm} \gg v = 246$ GeV, we find that one neutral Higgs boson has mass of $\mathcal{O}(v)$, while the other two neutral Higgs bosons have mass of $\mathcal{O}(M_{H^\pm})$. In this *decoupling limit*, one can formally integrate out the heavy Higgs states from the theory [53–58]. The resulting Higgs effective theory yields precisely the SM Higgs sector up to corrections of of $\mathcal{O}(v^2/M_{H^\pm}^2)$. Thus, the properties of the light neutral Higgs boson of the model, $h_1$, are nearly identical to those of the CP-even SM Higgs boson. Note that the CP-violating couplings of the lightest neutral Higgs boson to the fermions, gauge bosons and to itself are suppressed by a factor of $\mathcal{O}(v^2/M_{H^\pm}^2)$. In contrast, the two heavy neutral Higgs bosons will generally be significant admixtures of the CP-even and CP-odd states $H$ and $A$.

In the approach to the decoupling limit, $c_{\beta-\alpha} \simeq \mathcal{O}(v^2/M_{H^\pm}^2)$ [58]. Then, Eqs. (2.44) and (2.54) yields $\chi_V^{(1)} \simeq \chi_d^{(1)} \simeq \chi_u^{(1)} \simeq 1$, as expected. The flavor structure of the Higgs-quark interactions in the decoupling limit is also noteworthy. Eq. (2.51) yields approximately flavor-diagonal $\overline{Q}Qh_1$ couplings, since the contribution of the non-diagonal $\rho^Q$ is suppressed by $c_{\beta-\alpha}$. The heavier neutral Higgs bosons possess unsuppressed flavor non-diagonal Yukawa interactions, and thus can mediate FCNCs at tree-level. Of course, such FCNC effects would be suppressed by a factor of $\mathcal{O}(v^2/M_{H^\pm}^2)$ due to the heavy masses of the exchanged Higgs bosons. The existence of a decoupling limit depends on the possibility of taking $M^2_{H^\pm}$ arbitrarily large while holding the parameters $\lambda_i$ fixed. Using Eq. (2.24), it follows that the approach to the decoupling limit corresponds to the region of 2HDM parameter space where $\eta \gg |\lambda_i|$. This implies that no decoupling limit exists in a 2HDM with an exact discrete $Z_2$ symmetry.

[7] Note that even in the case of a CP-conserving Higgs potential, where $h$, $H$ and $A$ are physical mass eigenstates, Eq. (2.51) exhibits CP-violating Yukawa couplings proportional to the *complex* matrices $\rho^D$ and $\rho^U$.

The presence of a SM-like Higgs boson (which is defined as a neutral scalar that possesses tree-level couplings which are nearly identical to those of the SM Higgs boson) is consistent with a 2HDM with parameters near the decoupling limit. However, a SM-Higgs boson can arise in non-decoupling regions of the 2HDM parameter space. As an example, in the CP-conserving limit with $c_{\beta-\alpha} \simeq 1$ and $\cot\beta\, s_{\beta-\alpha} \ll 1$, it follows that the heavier CP-even state $H$ strongly resembles a SM-like Higgs boson. Other examples of a SM-like Higgs boson in a non-decoupling regime can be found in [58, 59].

The decoupling limit is also a regime in which all but the lightest Higgs boson are very heavy and nearly mass-degenerate. However, large Higgs masses (often with significant mass splittings) can also arise in a non-decoupling parameter regime in which the $\lambda_i$ are large. In this case, the heavy Higgs boson masses are bounded from above by imposing unitarity constraints on the $\lambda_i$. These unitarity constraints, which have been obtained for the CP-conserving case in [32], can be more severe in the CP-violating case [33,34]. For example, the unitarity limit constrains the parameter $|\lambda_5|$ while the Higgs mass formulae depend on $\mathrm{Re}\,\lambda_5$. In general, reasonably large $H$,$H^\pm$ and $A$ masses (up to about 600 GeV), consistent with the unitarity constraints, can be obtained for very large or very small $\tan\beta$ and reasonably small values of $\eta \approx (M_h/v)^2$, as well as for $\tan\beta \approx 1$ with $\eta \approx 0$ [15].

Finally, we note that in a non-decoupling parameter regime, the loop effects due to virtual exchange of heavy Higgs boson states do not decouple. Thus, in this parameter regime, one can also deduce upper bounds for heavy Higgs masses (or equivalently a bound on the departure from the decoupling limit). As an example, the non-decoupling effects of charged and neutral Higgs boson one-loop contributions to leptonic $\tau$-decays can yield an upper limit on the charged Higgs boson mass [60].

#### 2.1.4.4 Pattern relations and sum rules

The orthogonality of the mixing matrix $R$ allows one to obtain a number of relations [15, 61–63] among the relative couplings of neutral Higgs particles to the gauge bosons and fermions. For simplicity, we restrict the following analysis to the case of one generation of quarks (and leptons). Consider the following ratio of tree-level relative neutral Higgs couplings in the general CP-violating 2HDM:

$$\mathcal{R} \equiv \frac{\chi_V^{(i)} - \chi_d^{(i)*}}{\chi_u^{(i)} - \chi_V^{(i)}}\,, \tag{2.57}$$

where $i$ labels the Higgs mass eigenstates. One can easily verify that Eq. (2.57) holds separately for each value of $i = 1, 2, 3$. Moreover, if the denominator of Eq. (2.57) vanishes, then the numerator must vanish as well and vice versa. The neutral Higgs boson relative couplings $\chi_j$ also satisfy a *vertical sum rule* [46, 64, 65]:

$$\sum_{i=1}^{3} (\chi_j^{(i)})^2 = 1 \qquad (j = V, d, u)\,. \tag{2.58}$$

Note that Eqs. (2.57) and (2.58) also holds for the corresponding relative couplings of the definite CP scalar states $h$, $H$ and $A$.

In models with type-I and type-II Higgs–fermion Yukawa couplings, additional tree-level *pattern relations* and sum rules are respected. This is not surprising, given that the type-I and type-II conditions impose extra relations among the Higgs-fermion couplings. For example, Eq. (2.57) can be extended to the following result [15]:

$$\mathcal{R} = \frac{\chi_V^{(i)} - \chi_d^{(i)*}}{\chi_u^{(i)} - \chi_V^{(i)}} = \frac{1 - |\chi_d^{(i)}|^2}{|\chi_u^{(i)}|^2 - 1} = \frac{\mathrm{Im}\,\chi_d^{(i)}}{\mathrm{Im}\,\chi_u^{(i)}}\,, \tag{2.59}$$

where $\mathcal{R}$ is independent of the index $i$ that labels the neutral Higgs state. A brief computation shows that in type-I models, $\mathcal{R} = -1$, whereas in type-II models, $\mathcal{R} = \tan^2\beta$. In writing Eq. (2.59), we implicitly assumed that the denominators do not vanish. For example, Eq. (2.59) can also be applied to the couplings of the neutral Higgs states of definite CP: $h$, $H$ and $A$. However, in the CP-conserving limit, $\chi_u^{(i)}$ and $\chi_d^{(i)}$ are real for $i = h$ and $H$, so for these states the ratio of imaginary parts in Eq. (2.59) should be removed.

From Eq. (2.59), one can derive a *horizontal sum rule* for the neutral Higgs boson couplings [65]:

$$\mathcal{R}\,|\chi_u^{(i)}|^2 + |\chi_d^{(i)}|^2 = 1 + \mathcal{R}\,. \tag{2.60}$$

Taken together, the vertical and horizontal sum rules guarantee that the cross section to produce each neutral Higgs boson $h_i$ (or $h, H, A$) of the 2HDM-I or 2HDM-II, in the processes involving the Yukawa interaction, cannot be lower than the corresponding cross section for the production of the SM Higgs boson with the same mass [65].

The following linear relation among neutral Higgs boson relative couplings [15] is also a consequence of Eq. (2.59)

$$(1+\mathcal{R})\chi_V^{(i)} = \chi_d^{(i)\,*} + \mathcal{R}\,\chi_u^{(i)}\,. \tag{2.61}$$

Models with type I and II Higgs-fermion Yukawa couplings can be distinguished by the following pattern relation, which only holds in the 2HDM-II: [62, 63]:

$$(\chi_u^{(i)} + \chi_d^{(i)})\chi_V^{(i)} - \chi_u^{(i)}\chi_d^{(i)} = 1\,. \tag{2.62}$$

Of course, Eqs. (2.60)–(2.62) can also be applied to the neutral Higgs states of definite CP: $h$, $H$ and $A$.

## 2.2 Overview of phenomenology

*Gérald Grenier, Howard E. Haber and Maria Krawczyk*

We present a brief tour of the phenomenological and experimental investigations of the Higgs of the general 2HDM sector [9, 66, 67] at existing colliders (LEP[8] and TEVATRON), and at colliders now under construction (LHC) and under development (the ILC and the associated Photon Linear Collider (PLC) [68]). Results from the LHC and the ILC/PLC can provide useful synergies for CP studies of the general 2HDM, as discussed in [69, 70] and illustrated in section 2.14. The possibility of higher energy lepton colliders such as CLIC [71] and $\mu^+\mu^-$ collider [72, 73] have also been considered, and these facilities provide additional opportunities for CP studies of the Higgs sector.

### *2.2.1 Present limits on Higgs boson masses and couplings*

Due to the complexity of the general 2HDM parameter space, there are no completely model-independent limits on Higgs boson masses and couplings. However, the absence of a Higgs boson discovery at LEP and the Tevatron places numerous constraints on the 2HDM parameters.

In the decoupling limit of the 2HDM (see section 2.1.4.3), where the properties of the lightest neutral Higgs boson approach those of the SM Higgs boson, the mass limits of the SM Higgs apply: $M_h > 114.4$ GeV [74]. Less definitive results exist away from the decoupling limit, where significant deviations of the properties of the lightest neutral Higgs boson from those of the SM Higgs boson can be realized [58, 63]. Within the context of the MSSM, numerous mass limits have been quoted depending on a number of underlying theoretical assumptions. Many of these limits are described in Section 3.2. Here, we briefly focus on some of the more model-independent limits that have been obtained at LEP.

[8] Although the LEP collider shut down in 2000, there are still ongoing analyses of data.

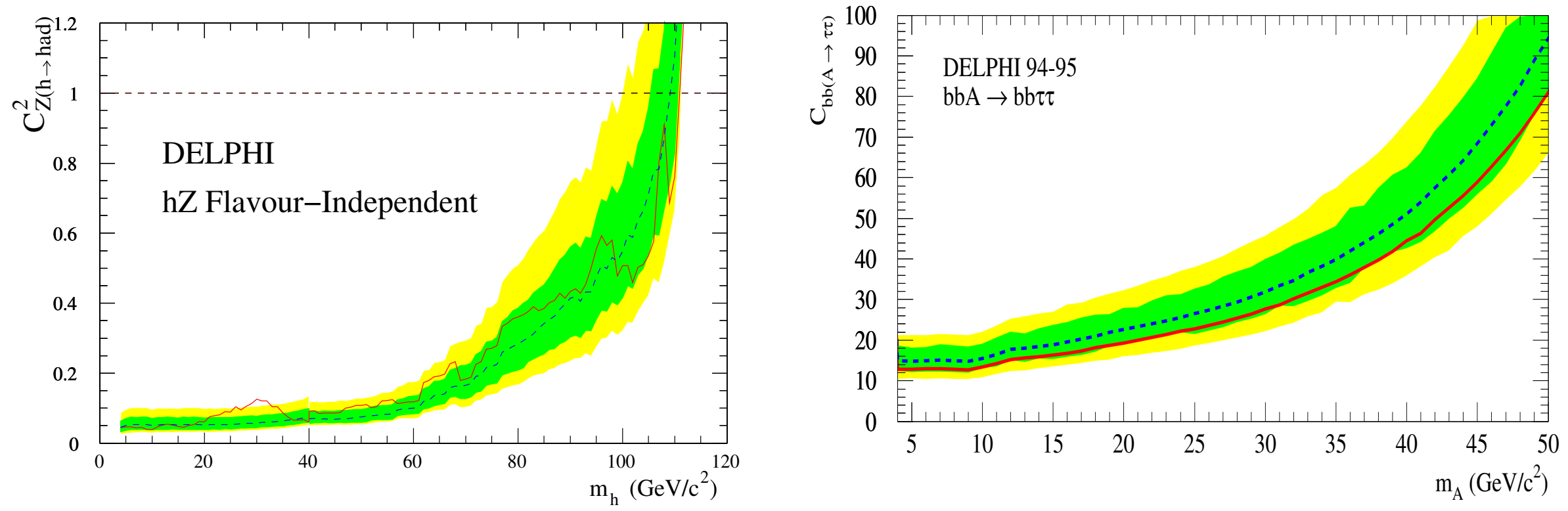

Fig. 2.1: Model independent upper limits for the CP-even and CP-odd Higgs boson masses as a function of their couplings to vector bosons and down-type fermions. The allowed parameter regimes (at 95% CL) lie below the solid lines. In the left panel, the squared relative coupling $\chi_V^2$ is shown as function of $M_h$ [75]. In the right panel, the relative coupling $\chi_d$ is shown as function of $M_A$ [76]. The relative couplings are defined in section 2.1.4 (note that the vertical axes employ different symbols for these relative couplings).

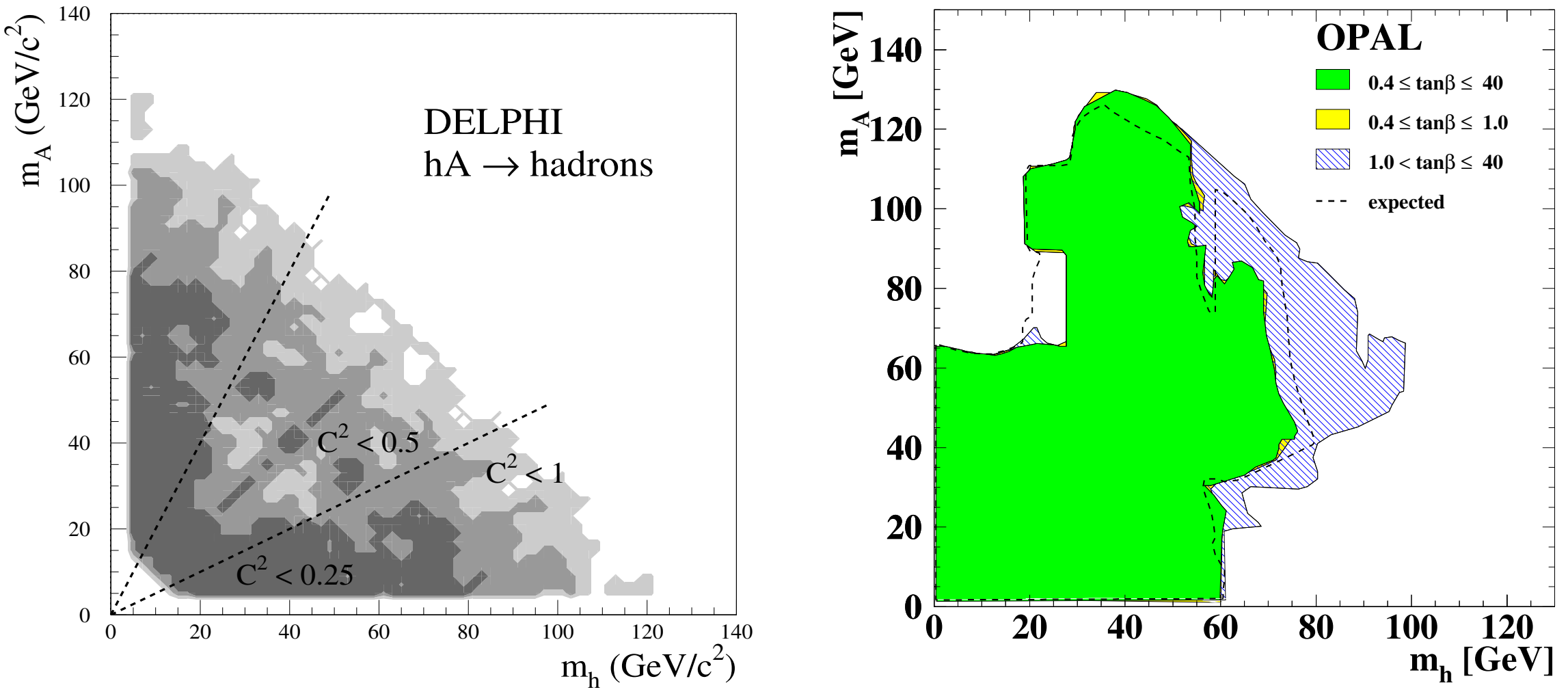

Fig. 2.2: In the left panel, constraints on the parameter $C^2 \equiv \cos^2(\beta - \alpha)$ are exhibited as a function of the Higgs masses $M_h$ and $M_A$, based on the non-observation of $e^+e^- \longrightarrow hA$ (assuming purely hadronic final states). Similar results are reinterpreted in the right panel in the CP conserving type-II 2HDM. The shaded area denotes the excluded regions of the $(M_h, M_A)$ plane, independently of the CP-even neutral Higgs boson mixing angle [77,78].

If the neutral Higgs boson coupling to gauge bosons is suppressed, then $(\chi_V^h)^2 = \sin^2(\beta{-}\alpha) \ll 1$, and the upper bound on the Higgs mass (derived from the non-observation of $e^+e^- \to Zh$) is significantly reduced from the corresponding SM Higgs boson mass limit [75, 79], as shown in the left panel of Fig. 2.1. In the CP-conserving 2HDM, the cross section for $e^+e^- \to hA$ is proportional $\cos^2(\beta - \alpha)$ and depends on the corresponding masses $M_h$ and $M_A$. The constraints on $\cos^2(\beta - \alpha)$ deduced from the non-observation of $hA$ production at LEP yields an exclusion limit in the $M_h$–$M_A$ plane shown in Fig. 2.2 [77, 78]. Note that these exclusion plots cannot exclude the possible existence of one very light neutral Higgs boson. For a CP-odd Higgs boson (which does not couple to gauge bosons at tree-level), important constraints on the Yukawa coupling $\chi_d$ as a function of the Higgs mass are derived from searches for $e^+e^- \to b\bar{b}A$ (where $A \to \tau^+\tau^-$). The absence of an observed $b\bar{b}\tau^+\tau^-$ signal above the SM background yields the exclusion plot shown in the right panel of Fig. 2.1 [76]. These limits depend on the enhancement of the $b\bar{b}A$ coupling (which for type-II Yukawa couplings is proportional to $\tan\beta$). A similar search was also performed for four $b$ final states resulting from $b\bar{b}A(\to b\bar{b})$ production and for four tau final states resulting from $\tau^+\tau^-A(\to \tau^+\tau^-)$. The same four-fermion signatures can also result from the production of the CP-even state $h$ [76].

The phenomenology of the charged Higgs boson of the 2HDM depends on fewer model parameters compared to that of the neutral Higgs bosons. Mass limits (at 95% CL) exist for the CP-conserving 2HDM-I and 2HDM-II and are given by 78.6 GeV [80] and 76.7 GeV [81], respectively. These limits are independent of $\tan\beta$. All other limits involving the charged Higgs boson mass depend on $\tan\beta$. For example, the CDF Collaboration reports [82] no charged Higgs bosons have been observed in top quark decays at the Tevatron. This data excludes certain regions of the $M_{H^\pm}$–$\tan\beta$ plane.

Virtual charged Higgs exchange can affect low-energy processes and place constraints on the 2HDM parameters. The most powerful constraint of this type can be obtained from the observed rate for $b\to s\gamma$, which is consistent with the predictions of the Standard Model. Thus, the contributions via loops due to new physics must be rather small. In the 2HDM, there is an extra contribution due to a charged Higgs boson loop. The significance of this contribution depends on the structure of the Higgs-fermion couplings. For example, there is almost no constraint in the CP-conserving 2HDM-I. In contrast, in the 2HDM-II, the experimental observation of $b\to s\gamma$ implies a 95% CL lower limit of $M_{H^\pm}\gtrsim 320$ GeV [83,84]. However, such a limit must be interpreted with care, since virtual effects originating from other new physics sources can cancel the charged Higgs contribution, thereby significantly relaxing (or removing entirely) the charged Higgs mass limit.

A number of other observables can also provide useful constraints on 2HDM parameters. For example, a lower bound on $M_{h^\pm}/\tan\beta$ can be obtained in precision measurements of semi-leptonic $b$ decays [85] and leptonic $\tau$-decays [60] by constraining the size of the tree-level charged Higgs boson exchange contributions. Constraints on Higgs masses and couplings have been obtained for type-II [60, 86] and for type III Higgs-fermion couplings [87–89]. Additional examples of this type can be found in section 2.5. Global fits using different electroweak observables such as $\rho$, $R_b$ and $b\to s\gamma$ [90] (and $(g-2)_\mu$ in [91]), have been made for the 2HDM-II, and these can significantly constrain the allowed regions of the parameter space.

### *2.2.2 Probing the CP nature of the neutral Higgs bosons*

If phenomena consistent with the 2HDM are discovered, it will be important to discover the form of the 2HDM that is realized in nature. One critical step in this program is the determination of the CP-properties of the three neutral Higgs bosons. If CP is conserved, then one can associate definite CP-quantum numbers with the three states (*i.e*, the CP-even $h$ and $H$ and the CP-odd $A$). If CP is violated, there is mixing among these states of definite CP, the corresponding mass-eigenstates $h_1$, $h_2$ and $h_3$ are states of indefinite CP. In this latter case, one of the main tasks of the experimental Higgs studies is to confirm this mixing and determine the corresponding mixing angles. If the 2HDM parameters lie in the vicinity of the decoupling regime, then to a very good approximation $h_1\simeq h$ is a SM-like CP-even Higgs boson. However, even in this case, there may be significant CP-violating effects involving the heavy neutral Higgs bosons, manifested by large mixing between the $H$ and $A$. In the decoupling regime, the mass splitting between the heavy neutral Higgs states tends to be small (of order $M_Z^2/M_{H^\pm}^2$). Consequently, the mixing between $H$ and $A$ in the CP-violating case and the overlapping of the $H$ and $A$ resonances in the CP-conserving case can lead to similar phenomenological effects. Moreover, in the case of nearly mass-degenerate scalar states, one should include the effects of the non-zero width in scalar mass matrix [42, 43], leading to the phenomena of resonant Higgs bosons production. A study of this type is presented in Section 3.12.

In order to determine the CP-properties of the neutral scalar states, one must study the various Higgs couplings to gauge bosons and fermions that govern the Higgs production and decay processes. In general, the CP-indefinite neutral states $h_i$ exhibit both scalar and pseudoscalar couplings to fermion pairs. In contrast, $W^+W^-$ and $ZZ$ couple dominantly to the CP-even component of the neutral Higgs bosons. This latter coupling enters at tree-level, whereas the coupling to the CP-odd component is a one-loop generated effect [92]. Nevertheless, in a completely model-independent study of Higgs decay to massive vector boson pairs, one may wish to allow for the possibility of a non-negligible CP-odd component that couples to vector bosons, as in [93]. For example, Lorentz invariance dictates that the most general interaction vertex for the coupling of a neutral Higgs boson $h_j$ to $ZZ$ is given by [94–96]:

$$\Gamma_{\mu\nu}(k_1\,,\,k_2) = \frac{ig}{m_Z\cos\theta_W}\left[a_j m_Z^2 g_{\mu\nu} + b_j(k_{2\mu}k_{1\nu} - k_1\cdot k_2 g_{\mu\nu}) + c_j\epsilon_{\mu\nu\alpha\beta}k_1^\alpha k_2^\beta\right]\,, \qquad (2.63)$$

where the incoming momenta $k_1$ and $k_2$ correspond to the fields $Z_\mu$ and $Z_\nu$, respectively, $a_j$ and $b_j$ are CP-even form factors and $c_j$ is a CP-odd form factor. (We assume a convention where $\epsilon_{0123} = 1$.) The form factors depend on Lorentz invariant combinations of the external momenta. For the CP-violating 2HDM at tree level, $b_j = c_j = 0$ and $a_j = R_{j1}\cos\beta + R_{j2}\sin\beta$, where $R$ is the mixing matrix defined in Eq. (2.27). The form factors $b_j$ and $c_j$ (and modifications of $a_j$) are generated by radiative loop corrections. These corrections can generate both dispersive (real) and absorptive (imaginary) contributions to the form factors (the latter corresponds to the possibility of on-shell intermediate states). Contributions to $b_j$ and $c_j$ can also be generated due to new physics at the TeV-scale. After integrating out the effects of the high-scale physics, effective local (dimension-five) operators of the form $Z_{\mu\nu}Z^{\mu\nu}h_j$ and $\epsilon_{\mu\nu\alpha\beta}Z^{\mu\nu}Z^{\alpha\beta}h_j$ can be generated in the low-energy effective Lagrangian that result in contributions to $b_j$ and $c_j$, respectively [97,98], where $Z_{\mu\nu} \equiv \partial_\mu Z_\nu - \partial_\nu Z_\mu + ig\cos\theta_W(W_\mu^+W_\nu^- - W_\nu^+W_\mu^-)$ is the field strength tensor of the $Z$ boson. The hermiticity of these two operators implies that the contributions of high-scale physics to the form factors must be real.

The discovery of a CP-violating signal in the $ZZh_j$ interaction requires a detection of an interference effect between the CP-even and CP-odd form factors of Eq. (2.63). If the tree-level term $a_j$ dominates, the direct observation of CP-violation will be difficult. In contrast, there are no tree-level couplings of the neutral Higgs bosons to gluon pairs or photon pairs, so that the loop coupling to the CP-even and CP-odd components of the neutral Higgs bosons can be competitive. One can also infer the existence of CP-violation indirectly (in the context of the 2HDM) if there is significant CP-mixing among the three neutral states. In this case, the observation of $ZZh_j$ couplings for all three states $h_j$ would provide strong evidence for the CP-mixing of the scalar states.

For a neutral Higgs boson with a mass below 140 GeV the most detailed information of its CP properties can be obtained from its decay into $\tau^+\tau^-$ [99–101] (a tool is provided in [102]). For larger Higgs masses, one must employ the decays into $W^+W^-$ and/or $ZZ$ [93], if the corresponding branching ratios are suitably large (in the decoupling regime only $h_1$ can have significant couplings to $W^+W^-$ and $ZZ$). Finally, for Higgs masses above 350 GeV, one can employ the $t\bar{t}$ decay mode [103]. In each case, the CP properties can be determined by studying the angular distributions of the various final state decay products. Additional CP-odd observables can be constructed by considering the properties of the Higgs production process. The production of a neutral Higgs boson in association with hadronic jets (*e.g*, as in gauge boson fusion at the LHC [104]) or in association with $t\bar{t}$ [105, 106] have been investigated. We note that if Higgs production via diffractive processes is observable at the LHC, then the azimuthal angular distribution of the tagged protons can be used to study the CP-properties of the produced Higgs boson. For further details, see [107, 108] and Section 3.8.

At the ILC and CLIC, the Higgs bosons are produced in pairs or in association with other particles. At the PLC and at a $\mu^+\mu^-$ collider, one can produce a single neutral Higgs boson through a resonant $s$-channel process via $\gamma\gamma$ fusion and $\mu^+\mu^-$ annihilation, respectively. For example, in $\gamma\gamma$ fusion to a neutral Higgs boson, one can make use of the polarization of the photon beams to study in detail the CP-properties of the resonant neutral Higgs states [109]. The interference between a Higgs signal and the SM background can also provide information that enables one to determine the CP property of the Higgs boson. Specific examples will be cited below in the discussion of CP studies at the ILC and PLC.

### 2.2.3 SM-like Higgs boson scenario

The Higgs boson direct search limits and the global electroweak fits to the precision electroweak data from LEP, SLC and the Tevatron provides a strong hint for the existence of a SM-like Higgs boson with a mass in the range of 115 GeV—207 GeV [110]. Precision measurements of Higgs couplings and quantum numbers (spin and CP) at future colliders are needed to confirm the nature of such a particle. A SM-like Higgs boson can appear in both the CP-conserving and the CP-violating 2HDM in the decoupling regime as previously noted [58, 63]. In this case, the other Higgs boson masses of the model must be significantly larger than $M_Z$.

For special choices of the 2HDM parameters, a SM-like Higgs boson can also appear in a non-decoupling regime [58, 59, 90] of the model parameter space. In this latter case, a SM-like Higgs boson would possess tree-level couplings that approximately match those of the SM Higgs boson (although opposite to the SM signs of couplings are possible). However in contrast to the decoupling regime, the masses of the non-SM-like neutral and charged Higgs bosons in the non-decoupling regime need not be particularly heavy. It may be possible to distinguish a SM-like Higgs boson from the Higgs boson of the Standard Model due to the effects of one-loop induced couplings. For example, the couplings of a SM-like Higgs boson to two gluons and to two photons can deviate from those of the corresponding SM Higgs couplings, due to the contribution of a charged Higgs boson loop [59, 62].

In the non-decoupling scenario just considered, a very light $h_1$ may exist with suppressed couplings to gauge bosons, whereas the heavier $H$ (in the CP-conserving model) or $h_2$ in the CP-violating model corresponds to a SM-like Higgs boson. In such a case, even a very light $h$ or $A$ (with the mass of the SM-like $H$ above 114 GeV) is not excluded by LEP data in the CP conserving 2HDM-II as shown in section 2.2.1. In the MSSM with large loop-induced CP-violating Higgs Yukawa interactions, a benchmark scenario named CPX [111] has been provided for further studies of this scenario. The OPAL collaboration has reanalyzed its data in the light of the CPX scenario [112]. Combined LEP results and ATLAS prospects for this benchmark are provided in Sections 3.2 and 3.3.

In [113] very large non-decoupling effects were found in the CP conserving 2HDM due to loop corrections in Higgs self-couplings (for small $\eta$). These can yield deviations as large as 100% from the Standard Model prediction, even when all the other couplings of the lightest Higgs boson to gauge bosons and fermions are in good agreement with the Standard Model.

### 2.2.4 Tri-mixing scenarios

In some portion of the 2HDM parameter space, the three neutral Higgs bosons can be close in mass and have similar coupling strengths to the $Z$. This has been called the three-way mixing regime (in the CP-conserving case sometimes referred to as the intense coupling regime [44]).

A good experimental Higgs mass resolution is important to probe this scenario at the LHC. Here the mass resolution expected for the SM Higgs searches, which is between $\sim$ 300 MeV and 2 GeV [114, 115], should be sufficient. The total cross-section for Higgs production is divided up (roughly equally) among the three neutral Higgs bosons. One must check that the three $h_iZZ$ couplings satisfy the vertical sum rule Eq. (2.58). This parameter regime seems very challenging at the Tevatron but might be easier to probe at the LHC. Detailed analyses of such scenarios were also performed for the ILC [116] and PLC [117]. A similar two-mixing scenario was considered in [43], for the PLC.

### 2.2.5 CP studies of the Higgs sector at the LHC

After the initial discovery of a (candidate) neutral Higgs boson at the LHC, it will be important to verify whether the properties of this state are consistent with those of the SM Higgs boson, or whether an extension of the SM Higgs sector is required. The CP properties can be determined at the LHC by studying angular distributions of the Higgs decay products. The $gg \to h_i \to f\bar{f}$ process has been considered for the $t\bar{t}$ final state in [103, 118] and analyses for $\tau^+\tau^-$ and $t\bar{t}$ final states are presented in sections 2.7 and 2.8. The decay $h_i \to VV$ with $V = W^\pm$ or $Z$, followed by leptonic decays of the vector boson $V$ has been considered in [93, 119]. The sensitivity of the four-lepton channel (for $h_i \to ZZ$) to CP-violating observables is examined in sections 2.12 and 2.11 and in section 2.13 for the $e^+e^-\mu^+\mu^-$ final states.

The CP properties can also be determined by studying angular distributions of particles produced in association with the $h_i$. The distribution of azimuthal angles of two light accompanying jets in $h_ijj$ production via gluon-gluon and vector boson fusion are studied in section 2.10 as a probe for the CP properties of $h_i$. In the case of gluon fusion, this distribution might be diluted by higher order corrections [120]. In the $gg \to h_it\bar{t}$ production, CP-sensitive variables can be build as studied in section 2.9. In this same production, certain weighted cross section integrals described in [121] can provide a determination of the CP nature of a light $h_i$. A similar study was performed in [122] for partonic processes involving gluons and light quarks accompanying by two jets, for neutral Higgs bosons lighter than 200 GeV.

Charged Higgs boson production and decays can also be useful for probing CP properties of the Higgs sector at LHC. The associated production of $H^\pm h_1$ with $H^\pm \to W^\pm h_1$ and $h_1 \to b\bar{b}$ yields events with four $b$-quarks, a charged lepton and missing transverse energy. There is virtually no Standard Model background, and the corresponding signal can be as large as 45 events for an integrated luminosity of 30 fb$^{-1}$ at the LHC [123]. For $m_{H^\pm} < m_t$, the $t\bar{t}$ pair production, with one top quark decaying into $bH^\pm$ and $H^\pm \to W^\pm h_1 \to W^\pm b\bar{b}$ (and the other top quark decays into $bW^\pm$), yields a signal of four $b$ quarks and two $W$ bosons. It is relatively free of Standard Model background and can result in roughly 5000 events for a luminosity of 30 fb$^{-1}$ at the LHC [124]. In some regions of the 2HDM parameter space, the study of $H^\pm W^\mp$ production might allow one to distinguish between Higgs sector of the MSSM and a general 2HDM [125]. Evidence for CP violation can be revealed in asymmetries in the associated production of a charged Higgs boson and the top quark [126]. Asymmetries in charged Higgs boson decay into $t\bar{b}, \bar{t}b$ can also be used to probe CP violation [127], whereas measuring the same asymmetries in $H^\pm \to \tau^\pm\nu_\tau$ decay is more challenging [128].

In a general 2HDM, one expects (at some level) the existence of FCNC processes and lepton-flavor-violating (LFV) processes in the leptonic sector, mediated by tree-level neutral Higgs boson exchange. For example, $h_i$ can decay into 2 charged leptons of different flavors. Typical branching ratios compatible with current experimental data for LFV $h_i$ decay can be found in [129]. The most promising decay is $h_1 \to \tau^\pm\mu^\mp$. Measurements of the muon anomalous magnetic moment [130] favor regions of the parameter space where the $h_1 \to \tau^\pm\mu^\mp$ decay can be seen at both the LHC and Tevatron [131].

### *2.2.6 CP studies of the Higgs sector at the ILC*

We briefly survey some of the main aspects of the CP study at a high energy linear $e^+e^-$ collider (ILC) with $\sqrt{s} = 500$—1000 GeV, assuming an integrated luminosity of 100 $fb^{-1}$ and a longitudinally polarized electron beam (up to 90%). The main neutral Higgs production mechanisms at the ILC [132,133] are governed by the Higgs interaction with gauge bosons: $e^+e^- \to Zh_i$ via Higgs-strahlung, $e^+e^- \to \nu\bar{\nu}h_i$ via $W^+W^-$ fusion and $e^+e^- \to e^+e^-h_i$ via $ZZ$-fusion. The gauge boson fusion processes become the dominant Higgs production processes at large $\sqrt{s}$. Neutral and charged Higgs bosons can also be produced in pairs: $e^+e^- \to h_ih_j$ $(i \neq j)$ via $s$-channel $Z$-exchange and $e^+e^- \to H^+H^-$ via $s$-channel $\gamma$ and $Z$-exchange. Note that in the decoupling limit, the two heaviest neutral Higgs bosons ($h_2$ and $h_3$) and $H^\pm$ are heavy and roughly mass degenerate. Thus, the pair (or associated) production of a pair of heavy states is kinematically possible only if $\sqrt{s}$ is larger than twice the mass of the heavy Higgs states.

The Higgs-strahlung cross section $e^+e^- \to Zh_i$ depends on whether the $h_i$ is CP-even, CP-odd, or a mixture [94, 95, 97, 134, 135]. For a CP-even (CP-odd) Higgs boson, the $Z$ is longitudinally (transversally) polarized. The spin, parity and charge conjugation quantum numbers, $J^{PC}$, of the Higgs boson can potentially be determined independently of the model by studying the threshold dependence and angular distribution of the Higgs and $Z$ boson [136]. The angular distribution of the fermions in the $Z \to f\bar{f}$ reflects the CP nature of the state $h_i$ [95,97,134,135]. A full simulation was performed in [137] for the TESLA design, with promising results. It should be noted that in the analyses cited above, generic CP-odd couplings of the Higgs boson to gauge boson pairs were assumed. In practice, such couplings are expected to be quite suppressed, as they are necessarily absent at tree-level and thus can only appear at one loop (in contrast to the tree-level CP-even couplings), as discussed in section 2.2.2.

The Yukawa couplings of the neutral Higgs bosons can be studied in the associated production processes $e^+e^- \to f\bar{f}h_i$, where $(f = b, t, \tau, \mu)$. The Yukawa process and the Higgs production processes $e^+e^- \to Zh_i$ and $e^+e^- \to h_ih_j$ $(i \neq j)$ are complementary in the search for at least one Higgs boson of the CP-violating 2HDM [65, 138]. In all these processes, correlations between the production and decay (with polarized initial beams) yield numerous observables that are sensitive to the spin and CP properties of the produced Higgs bosons. For example, the sensitivity to CP violation in the Yukawa coupling to $b$ quarks, was studied in [139]. In this analysis the process $e^+e^- \to b\bar{b}\nu\bar{\nu}$ was considered for a Higgs boson with a mass of 120 GeV that decays primarily into $b\bar{b}$, and the interference between the Higgs signal and SM background was exploited to determine the CP properties of the Higgs boson. The process $e^+e^- \to t\bar{t}h_i$ alone may be sufficient to provide a reasonable determination of the $t\bar{t}$ and $ZZ$ couplings and the CP-properties of the produced $h_i$ [140]. The Yukawa interaction is also responsible for singly-produced charged Higgs bosons via $e^-e^+ \to b\bar{c}H^+$, $\tau\bar{\nu}H^+$. These production processes provide a test of the chirality of the charged Higgs boson Yukawa couplings [141].

The Higgs self-couplings are very difficult to ascertain at the LHC. At the ILC, these couplings can be measured with limited accuracy in the processes $e^+e^- \to Zh_ih_j$ and $e^+e^- \to \nu\bar{\nu}h_ih_j$ [113,142, 143]. These processes depend both on the Higgs self couplings and the $VVh_i$ and $VVh_ih_j$ couplings.

### *2.2.7 CP studies of the Higgs sector at the PLC*

By shining intense laser light on the electron (and positron) beam, one can convert the initial beam of the ILC into a photon beam via Compton backscattering. This provides a mechanism for using the ILC as an electron-photon or photon-photon collider. In the photon linear collider (PLC) mode of the ILC, the photon beams are produced with energies up to 80% of the electron-positron center-of-mass energy and with a luminosity similar to that of the original colliding $e^+e^-$ beams. Moreover, it is possible to produce highly polarized photon beams (the degree of polarization depends on the polarization of the laser light employed in the Compton backscattering that produces the photon beam, and polarization of the electron beam). At the PLC, the neutral Higgs boson can be produced resonantly in the $s$-channel, leading to a higher mass reach than the parent $e^+e^-$ collider. Moreover, the polarization of the photon beams can be selected to form an (approximately pure) CP-odd or CP-even initial state. This provides an ideal laboratory for studying the CP properties of the neutral Higgs boson. For example, CP-violating asymmetries in neutral Higgs boson processes can be constructed even without information on the final states.

The spin and parity of the Higgs boson can be measured in a model-independent way at the PLC using the $h_i \to ZZ$ $(ZZ^*)$ decay channel [93], and the angular distributions of the fermions from the decays $Z \to f\bar{f}$. The detection of Higgs sector CP-violating effects can be ascertained by studying a variety of final states. By taking into account interference effects of the Higgs signal and background, one can extract both the Higgs partial width $\Gamma_{\gamma\gamma}$ and the phase of the Higgs amplitude, $\phi_{\gamma\gamma}$, for the $W^+W^-$ final state [144] and for the $t\bar{t}$ final state [145]. Other analyses of CP violation in $\gamma\gamma \to h_i \to W^+W^-$ have been given in [146]; realistic simulations for the 2HDM with a SM-like Higgs boson and in a model independent approach were performed in [144, 147] and are discussed in section 2.14. In particular, the simultaneous simulation of ZZ and $W^+W^-$ final states is crucial in determining the CP properties of the neutral Higgs bosons with masses in the range of 200—300 GeV. Various analyses related to heavy neutral Higgs boson production in $\gamma\gamma \to ZZ, ZH$, which also make use of the interference effects with the SM background, are given in [148]. The process $\gamma\gamma \to h_i \to t\bar{t}$ also provides an ideal setting for studying the CP properties of the neutral Higgs boson. Model independent studies of this channel are given in [145, 149–151].

Both linearly and circularly polarized photon beams are necessary in order to measure polarization asymmetries and to establish the CP property of the heavy $h_i$. Experimental signatures of CP-violating mixing of the heavy CP-even and CP-odd neutral Higgs states ($H/A$) at the PLC with (linear and circular) polarized beams were studied in [43] in the decoupling regime including effect of non-zero Higgs widths. Resonant loop-induced CP violation in Higgs-strahlung, in tri-mixing and two-mixing scenarios, was studied in [117]. The CP asymmetries in the production and decays of pairs of muons, taus, $b$ and $t$ quarks were used in this analysis. Although both analyses cited above were carried out for the case of the MSSM Higgs sector, the main results should hold for the more general (CP-violating) 2HDM.

The Yukawa couplings of the Higgs boson can be explored at the PLC by making use of another mechanism. The fermionic fusion, from the splitting of two photons into pairs of fermions, may lead to the production of neutral or charged Higgs bosons without strong suppression at high energies. This splitting leads to a collinear enhancement $\log(M_h/m_f)$ (where $M_h$ is the mass of the produced Higgs boson), which can be interpreted as the generation of "partonic densities" in the photon. The $\tau\tau$ fusion was used e.g. in [152] as a method to determine the Yukawa coupling of neutral Higgs boson. Likewise, single charged Higgs boson production in $\gamma\gamma \to b\bar{c}H^+, \tau\bar{\nu}H^+$, via $\gamma \to c\bar{c}$ and $\gamma \to b\bar{b}$, may be useful to discriminate models [141]. It is possible to determine the chirality of Yukawa couplings $H^+bc, H^+\tau\nu$ by choosing the polarization of the colliding photon beams.

Finally, $\gamma\gamma \to h_ih_j$ (via box and triangle loops with gauge bosons and fermions) can be used to determine Higgs self-couplings [143], with larger sensitivity than in the "parent" $e^+e^-$ collider.

### *2.2.8 CP studies of the Higgs sector at a multi-TeV lepton collider*

In the decoupling limit, $h_1$ is nearly indistinguishable from the SM-Higgs boson. The heavier Higgs bosons $h_2$, $h_3$ and $H^\pm$ (which are degenerate in mass up to corrections of order $m_Z^2/m_{H^\pm}$) may be too heavy to be studied directly at the LHC and ILC. In this case, a multi-TeV lepton collider will be required to fully explore the Higgs sector and provide a comprehensive study of the heavy Higgs states, in particular of the Higgs sector CP-violation. Thus, we briefly survey the potential for CP studies at a high luminosity multi-TeV $e^+e^-$ collider such as CLIC [71] and a $\mu^+\mu^-$ collider [72, 73].

At CLIC, the heavy Higgs states are produced in pairs via $e^+e^- \to h_ih_j$ ($j = 2, 3$) and $e^+e^- \to H^+H^-$ (note that single production of a heavy Higgs state: $e^+e^- \to h_1h_j$ ($j = 2, 3$) is suppressed in the decoupling limit [53, 58]). The dominant decay of the charged Higgs boson is $H^+ \to t\bar{b}$. An asymmetry between the $H^+$ and $H^-$ partial decay rates into $t\bar{b}$ and $\bar{t}b$ would be a signal of CP-violation. A simulation performed in [153] suggests that a 10% asymmetry could be detected as a $3\sigma$ effect with 5 ab$^{-1}$ of data at CLIC with $\sqrt{s} = 3$ TeV. No analogous study has yet been performed for the $h_ih_j$ final state (see [154] for an analysis of the discovery potential for $e^+e^- \to HA$ at CLIC).

At a $\mu^+\mu^-$ collider, the $s$-channel single production of a heavy neutral Higgs boson [155, 156] provides a new discovery mode as compared to the $e^+e^-$ colliders. This production mechanism is feasible at a $\mu^+\mu^-$ collider due to the mass enhancement in the Higgs coupling to $\mu^+\mu^-$ relative to $e^+e^-$. In addition, the charged Higgs boson can also be singly produced via $\mu^+\mu^- \to h_j \to H^\pm W^\mp$. The superb energy resolution of the $\mu^+\mu^-$ would permit the separation of the heavy neutral Higgs boson $s$-channel resonances, even though these states are nearly degenerate in mass. It is demonstrated in [157] that the $s$-channel production rates for the heavy neutral Higgs states and the transverse-polarization asymmetries are complementary in diagnosing Higgs sector CP-violation. Additional CP-violating observables can be studied by examining the heavy Higgs decays into a pair of third generation quarks or leptons [158]. The importance of the muon beam polarization for Higgs CP studies in $\mu^+\mu^- \to h_j$ ($j = 2, 3$) is emphasized in [159]. In singly produced charged Higgs bosons in association with the $W$, an asymmetry in the production rates for $H^\pm W^\mp$ also provides a signal of CP violation [160].

## 2.3 Basis-independent treatment of Higgs couplings in the CP-violating 2HDM

*Howard E. Haber*

In the most general two-Higgs-doublet model (2HDM), there is no distinction between the two complex hypercharge-one SU(2) doublet scalar fields, $\Phi_a$ ($a = 1, 2$). Thus, any two orthonormal linear combinations of these two fields can serve as a basis for the Lagrangian. All physical observables of the model must be basis-independent. For example, $\tan\beta \equiv v_2/v_1$ [see Eq. (2.7)] is basis-dependent and thus cannot be a physical parameter of the model [14–16]. Basis independent techniques have been exploited to great advantage in [14, 19, 37, 40, 41] in the study of the CP-violating structure of the 2HDM (and extend the results originally obtained in [38, 39].) In addition, the importance of the scalar-doublet field redefinitions (and rephasing transformations) have been emphasized, and some of their implications for 2HDM phenomenology have been explored in [15]. In this paper, we employ the basis-independent formalism to obtain an invariant description of all 2HDM couplings.

### 2.3.1 Basis-independent formalism for the 2HDM

The fields of the 2HDM consist of two identical complex hypercharge-one, SU(2) doublet scalar fields $\Phi_a(x) \equiv (\Phi_a^+(x)\,,\, \Phi_a^0(x))$, where $a = 1, 2$ can be considered a Higgs "flavor" index. The most general redefinition of the scalar fields (which leaves the form of the canonical kinetic energy terms invariant) corresponds to a global U(2) transformation, $\Phi_a \to U_{a\bar{b}}\Phi_b$ [and $\Phi_{\bar{a}}^\dagger \to \Phi_{\bar{b}}^\dagger U_{b\bar{a}}^\dagger$]. Here, it is convenient to introduce unbarred and barred indices with a summation convention in which only barred–unbarred index pairs of the same letter are summed. The basis-independent formalism consists of writing all equations involving the Higgs sector fields in a U(2)-covariant fashion. Basis-independent quantities can then be identified as invariant scalars under U(2). The U(2)-invariants are easily identified as products of tensor quantities with all barred and unbarred index pairs summed with no flavor indices left over.

The scalar potential can be written in U(2)-covariant form [10, 14] in terms of the tensors $Y_{a\bar{b}}$ and $Z_{a\bar{b}c\bar{d}}$ as shown in Eq. (2.2). The vacuum of the theory is assumed to respect the electromagnetic $\mathrm{U}(1)_{\mathrm{EM}}$ gauge symmetry. The $\mathrm{U}(1)_{\mathrm{EM}}$-conserving vacuum expectation value can be written as:

$$\langle\Phi_a\rangle = \frac{v}{\sqrt{2}}\begin{pmatrix} 0 \\ \widehat{v}_a \end{pmatrix}, \quad (a = 1, 2), \quad \text{with} \quad \widehat{v}_a \equiv e^{i\eta}\begin{pmatrix} c_\beta \\ s_\beta\, e^{i\xi} \end{pmatrix}, \tag{2.64}$$

where $v \equiv 2m_W/g = 246$ GeV, $c_\beta \equiv \cos\beta$, $s_\beta \equiv \sin\beta$ and $\widehat{v}_a$ is a vector of unit norm. The overall phase $\eta$ is arbitrary. By convention, we take $0 \le \beta \le \pi/2$ and $0 \le \xi < 2\pi$.

Under a U(2)-flavor transformation, $\widehat{v}_a \to U_{a\bar{b}}\widehat{v}_b$. The unit vector $\widehat{v}_a$ can also be considered to be an eigenvector of unit norm of the Hermitian matrix $V_{a\bar{b}} \equiv \widehat{v}_a\widehat{v}_{\bar{b}}^*$. Since $V$ is Hermitian, it possesses a second eigenvector of unit norm that is orthogonal to $\widehat{v}_a$. We denote this eigenvector by $\widehat{w}_a$, which satisfies $\widehat{v}_{\bar{b}}^*\widehat{w}_b = 0$. The most general solution for $\widehat{w}_a$, up to an overall multiplicative phase factor, is:

$$\widehat{w}_b \equiv \widehat{v}_{\bar{a}}^*\epsilon_{ab} = e^{-i\eta}\begin{pmatrix} -s_\beta\, e^{-i\xi} \\ c_\beta \end{pmatrix}. \tag{2.65}$$

The inverse relation to Eq. (2.65) is easily obtained: $\widehat{v}_{\bar{a}}^* = \epsilon_{\bar{a}\bar{b}}\,\widehat{w}_b$. Above, we have introduced two Levi-Civita tensors with $\epsilon_{12} = -\epsilon_{21} = 1$ and $\epsilon_{11} = \epsilon_{22} = 0$. However, $\epsilon_{ab}$ and $\epsilon_{\bar{a}\bar{b}}$ are not proper tensors with respect to the full flavor-U(2) group (although these are invariant SU(2)-tensors). That is, $\widehat{w}_a$ does not transform covariantly with respect to the full flavor-U(2) group. If $U = e^{i\psi}\widehat{U}$, with $\det\widehat{U} = 1$ (and consequently $\det U = e^{2i\psi}$), it is simple to check that under a U(2) transformation

$$\widehat{v}_a \to U_{a\bar{b}}\widehat{v}_b \qquad \text{implies that} \qquad \widehat{w}_a \to (\det U)^{-1}\, U_{a\bar{b}}\,\widehat{w}_b\,. \tag{2.66}$$

Henceforth, we shall define a pseudotensor as a tensor that transform covariantly with respect to the flavor-SU(2) subgroup but whose transformation law with respect to the full flavor-U(2) group is only covariant modulo an overall nontrivial phase equal to some integer power of $\det U$. Thus, $\widehat{w}_a$ is a pseudovector. However, we can use $\widehat{w}_a$ to construct proper tensors. For example, the Hermitian matrix $W_{a\bar{b}} \equiv \widehat{w}_a \widehat{w}^*_{\bar{b}} = \delta_{a\bar{b}} - V_{a\bar{b}}$ is a proper second-ranked tensor.

One can write a set of independent scalar quantities constructed out of $Y_{a\bar{b}}$, $Z_{a\bar{b}c\bar{d}}$, $v_a$ and $w_a$. There are six independent invariant quantities:

$$\begin{aligned} Y_1 &\equiv \mathrm{Tr}\,(YV)\,, & Y_2 &\equiv \mathrm{Tr}\,(YW)\,, \\ Z_1 &\equiv Z_{a\bar{b}c\bar{d}}\, V_{b\bar{a}} V_{d\bar{c}}\,, & Z_2 &\equiv Z_{a\bar{b}c\bar{d}}\, W_{b\bar{a}} W_{d\bar{c}}\,, \\ Z_3 &\equiv Z_{a\bar{b}c\bar{d}}\, V_{b\bar{a}} W_{d\bar{c}}\,, & Z_4 &\equiv Z_{a\bar{b}c\bar{d}}\, V_{b\bar{c}} W_{d\bar{a}}\,, \end{aligned} \tag{2.67}$$

and four independent pseudo-invariant quantities:

$$\begin{aligned} Y_3 &\equiv Y_{a\bar{b}}\, \widehat{v}^*_{\bar{a}}\, \widehat{w}_b\,, & Z_5 &\equiv Z_{a\bar{b}c\bar{d}}\, \widehat{v}^*_{\bar{a}}\, \widehat{w}_b\, \widehat{v}^*_{\bar{c}}\, \widehat{w}_d\,, \\ Z_6 &\equiv Z_{a\bar{b}c\bar{d}}\, \widehat{v}^*_{\bar{a}}\, \widehat{v}_b\, \widehat{v}^*_{\bar{c}}\, \widehat{w}_d\,, & Z_7 &\equiv Z_{a\bar{b}c\bar{d}}\, \widehat{v}^*_{\bar{a}}\, \widehat{w}_b\, \widehat{w}^*_{\bar{c}}\, \widehat{w}_d\,, \end{aligned} \tag{2.68}$$

that depend linearly on $Y$ and $Z$. Note that the invariants are manifestly real, whereas the pseudo-invariants are potentially complex. Using Eq. (2.66), it follows that under a flavor-U(2) transformation specified by $U$, the pseudo-invariants $Y_3$, $Z_5$, $Z_6$ and $Z_7$ transform as:

$$[Y_3, Z_6, Z_7] \to (\det U)^{-1} [Y_3, Z_6, Z_7] \qquad \text{and} \qquad Z_5 \to (\det U)^{-2} Z_5\,. \tag{2.69}$$

Thus, Eqs. (2.67) and (2.68) correspond to thirteen invariant real degrees of freedom (ten magnitudes and three relative phases) prior to imposing the scalar potential minimum conditions:

$$Y_1 = -\tfrac{1}{2} Z_1 v^2\,, \qquad\qquad Y_3 = -\tfrac{1}{2} Z_6 v^2\,. \tag{2.70}$$

This leaves eleven independent real degrees of freedom (one of which is the vacuum expectation value $v$) that specify the 2HDM parameter space.

### *2.3.2 (Pseudo)-invariants and the Higgs bases*

Once the scalar potential minimum is determined, which defines $\widehat{v}_a$, one class of basis choices is uniquely selected. Suppose we begin in a generic $\Phi_1$–$\Phi_2$ basis. We define new Higgs doublet fields:

$$\boldsymbol{H}_1 = (H_1^+\,,\, H_1^0) \equiv \hat{v}^*_{\bar{a}} \Phi_a\,, \qquad\qquad \boldsymbol{H}_2 = (H_2^+\,,\, H_2^0) \equiv \hat{w}^*_{\bar{a}} \Phi_a = \epsilon_{\bar{b}\bar{a}} \widehat{v}_b \Phi_a\,. \tag{2.71}$$

With respect to U(2) transformations, $\boldsymbol{H}_1$ is an invariant field and $\boldsymbol{H}_2$ is a pseudo-invariant field that transforms as $\boldsymbol{H}_2 \longrightarrow (\det U)\boldsymbol{H}_2$. The latter phase freedom defines a class of Higgs bases. The definitions of $\boldsymbol{H}_1$ and $\boldsymbol{H}_2$ imply that

$$\langle H_1^0 \rangle = v/\sqrt{2}\,, \qquad\qquad \langle H_2^0 \rangle = 0\,. \tag{2.72}$$

Hence, all Higgs bases are characterized by: $\widehat{v} = (1, 0)$ and $\widehat{w} = (0, 1)$. Using Eqs. (2.67) and (2.68), one identifies $Y_1$, $Y_2$, $Y_3$ and $Z_1$, $Z_2$,…,$Z_7$ as the coefficients of the 2HDM scalar potential in any Higgs basis.

Explicitly, the scalar field doublets in the Higgs basis can be parameterized as follows:

$$\boldsymbol{H}_1 = \begin{pmatrix} G^+ \\ \frac{1}{\sqrt{2}}\left(v + \phi_1^0 + iG^0\right) \end{pmatrix}, \qquad \boldsymbol{H}_2 = \begin{pmatrix} H^+ \\ \frac{1}{\sqrt{2}}\left(\phi_2^0 + ia^0\right) \end{pmatrix}, \tag{2.73}$$

where $G^\pm$ are the charged Goldstone bosons, $G^0$ is the CP-odd neutral Goldstone boson, and $H^\pm$ are the charged Higgs bosons with mass: $M^2_{H^\pm} = Y_2 + \frac{1}{2} Z_3 v^2$. Since $\widehat{v}$ is a vector and $\widehat{w}$ is a pseudovector, it follows that $G^\pm$ is an invariant field and $H^\pm$ is a pseudo-invariant field that transforms as:

$$H^\pm \to (\det U)^{\pm 1}\, H^\pm \tag{2.74}$$

with respect to flavor-U(2) transformations. The neutral Higgs mass-eigenstates are linear combinations of $\phi_1^0$, $\phi_2^0$ and $a^0$. CP-violation due to the mixing of neutral scalar CP-eigenstates and/or direct CP-violation in the bosonic interactions of the gauge/Higgs bosons are *absent* if and only if [14, 38, 39]:

$$\mathrm{Im}\,[Z_5^* Z_6^2] = \mathrm{Im}\,[Z_6 Z_7^*] = \mathrm{Im}\,[Z_5^*(Z_6+Z_7)^2] = 0\,. \tag{2.75}$$

### 2.3.3 The physical Higgs mass-eigenstates

It is simplest to perform the diagonalization of the neutral scalar squared-mass matrix in the Higgs basis. As in Section 2.1, we denote the neutral mass-eigenstate Higgs fields by $h_1$, $h_2$ and $h_3$. The angles $\theta_{ij}$ parameterize the rotation matrix that converts the neutral Higgs basis fields $\phi_1^0$, $\phi_2^0$ and $a^0$ into the mass-eigenstate fields $h_k$. Details of the diagonalization procedure can be found in [16].[9] The end result is:

$$\begin{pmatrix} h_1 \\ h_2 \\ h_3 \end{pmatrix} = \begin{pmatrix} q_{11} & \frac{1}{\sqrt{2}} q_{12}^*\, e^{i\theta_{23}} & \frac{1}{\sqrt{2}} q_{12}\, e^{-i\theta_{23}} \\ q_{21} & \frac{1}{\sqrt{2}} q_{22}^*\, e^{i\theta_{23}} & \frac{1}{\sqrt{2}} q_{22}\, e^{-i\theta_{23}} \\ q_{31} & \frac{1}{\sqrt{2}} q_{32}^*\, e^{i\theta_{23}} & \frac{1}{\sqrt{2}} q_{32}\, e^{-i\theta_{23}} \end{pmatrix} \begin{pmatrix} \sqrt{2}\,\mathrm{Re}\,H_1^0 - v \\ H_2^0 \\ H_2^{0\,\dagger} \end{pmatrix}\,, \tag{2.76}$$

where

$$\begin{aligned} q_{11} &= c_{13}c_{12}\,, & q_{21} &= c_{13}s_{12}\,, & q_{31} &= s_{13}\,, & q_{41} &= i\,, \\ q_{12} &= -s_{12} - ic_{12}s_{13}\,, & q_{22} &= c_{12} - is_{12}s_{13}\,, & q_{32} &= ic_{13}\,, & q_{42} &= 0\,, \end{aligned} \tag{2.77}$$

with $c_{ij} \equiv \cos\theta_{ij}$ and $s_{ij} \equiv \sin\theta_{ij}$. We have also defined $q_{4\ell}$ ($\ell = 1, 2$) for later use. In particular,

$$q_{k\ell} \to q_{k\ell}\,, \qquad \text{and} \qquad e^{i\theta_{23}} \to (\det U)^{-1} e^{i\theta_{23}}\,, \tag{2.78}$$

under a U(2) transformation. That is, the $q_{k\ell}$ are invariants, which implies that $\theta_{12}$ and $\theta_{13}$ are U(2)-invariant angles, whereas $e^{i\theta_{23}}$ is a pseudo-invariant. Note that since $H_1$ and $e^{i\theta_{23}}H_2$ are U(2)-invariant fields, it follows that the $h_k$ are invariant fields, as expected. We shall also define $Z_5 \equiv |Z_5|e^{2i\theta_5}$ and $Z_{6,7} \equiv |Z_{6,7}|e^{i\theta_{6,7}}$, in which case the $\phi_n \equiv \theta_n - \theta_{23}$ ($n = 5, 6, 7$) are U(2)-invariant angles.

If Im $(Z_5^* Z_6^2) = 0$, then the neutral scalar squared-mass matrix can be transformed into block diagonal form, which contains the squared-mass of a CP-odd neutral mass-eigenstate Higgs field $A$ and a $2 \times 2$ sub-matrix that yields the squared-masses of two CP-even neutral mass-eigenstate Higgs fields $h$ and $H$. The analytic form of this diagonalization is simple and yields the well-known results of the CP-conserving 2HDM . If Im $(Z_5^* Z_6^2) \neq 0$, then the neutral scalar mass-eigenstates do not possess definite CP quantum numbers, and the three invariant mixing angles $\theta_{12}$, $\theta_{13}$ and $\phi_6 \equiv \theta_6 - \theta_{23}$ are non-trivial.

The angles $\theta_{13}$ and $\phi_6$ are determined modulo $\pi$ from [16]:

$$\tan\theta_{13} = \frac{\mathrm{Im}\,(Z_5\, e^{-2i\theta_{23}})}{2\ \mathrm{Re}\,(Z_6\, e^{-i\theta_{23}})}\,, \qquad \tan 2\theta_{13} = \frac{2\ \mathrm{Im}\,(Z_6\, e^{-i\theta_{23}})}{Z_1 - A^2/v^2}\,, \tag{2.79}$$

where $A^2 \equiv Y_2 + \frac{1}{2}[Z_3 + Z_4 - \mathrm{Re}\,(Z_5 e^{-2i\theta_{23}})]v^2$. These equations exhibit multiple solutions (modulo $\pi$) corresponding to different orderings of the $h_k$ masses. Likewise, the angle $\theta_{12}$ is determined from:

$$\tan 2\theta_{12} = \frac{2\cos 2\theta_{13}\ \mathrm{Re}\,(Z_6\, e^{-i\theta_{23}})}{c_{13}\left[c_{13}^2(A^2/v^2 - Z_1) + \cos 2\theta_{13}\ \mathrm{Re}\,(Z_5\, e^{-2i\theta_{23}})\right]}\,. \tag{2.80}$$

For a given solution of $\theta_{13}$ and $\phi_6$, Eq. (2.80) yields two solutions for $\theta_{12}$ modulo $\pi$, corresponding to the two possible relative mass orderings of $h_1$ and $h_2$.

[9] This procedure differs somewhat from the one presented in section 2.1.3.1. In the latter, the diagonalization of the scalar squared-mass matrix is carried out in a generic basis. The advantage of performing this computation in the Higgs basis is that it allows one to easily identify the (pseudo)-invariant quantities that relate the neutral scalar interaction- and mass-eigenstates.

The neutral Higgs boson masses $M_k \equiv M_{h_k}$ can be expressed in terms of $Z_1$, $Z_6$ and the $\theta_{ij}$:

$$M_1^2 = \left[Z_1 - \frac{s_{12}}{c_{12}c_{13}} \operatorname{Re}(Z_6\, e^{-i\theta_{23}}) + \frac{s_{13}}{c_{13}} \operatorname{Im}(Z_6\, e^{-i\theta_{23}})\right] v^2\,, \tag{2.81}$$

$$M_2^2 = \left[Z_1 + \frac{c_{12}}{s_{12}c_{13}} \operatorname{Re}(Z_6\, e^{-i\theta_{23}}) + \frac{s_{13}}{c_{13}} \operatorname{Im}(Z_6\, e^{-i\theta_{23}})\right] v^2\,, \tag{2.82}$$

$$M_3^2 = \left[Z_1 - \frac{c_{13}}{s_{13}} \operatorname{Im}(Z_6\, e^{-i\theta_{23}})\right] v^2\,. \tag{2.83}$$

Eqs. (2.79) and (2.81)–(2.83) can be used to obtain [16]:

$$s_{13}^2 = \frac{(Z_1 v^2 - M_1^2)(Z_1 v^2 - M_2^2) + |Z_6|^2 v^4}{(M_3^2 - M_1^2)(M_3^2 - M_2^2)}\,, \qquad c_{13}^2 s_{12}^2 = \frac{(Z_1 v^2 - M_1^2)(M_3^2 - Z_1 v^2) - |Z_6|^2 v^4}{(M_2^2 - M_1^2)(M_3^2 - M_2^2)}\,, \tag{2.84}$$

and

$$\sin 2\theta_{12} = \frac{2\,|Z_6| v^2 \cos\phi_6}{c_{13}(M_2^2 - M_1^2)}\,, \qquad \tan 2\phi_6 = \frac{\operatorname{Im}(Z_5^* Z_6^2)}{\operatorname{Re}(Z_5^* Z_6^2) + \dfrac{|Z_6|^4 v^2}{M_3^2 - Z_1 v^2}}\,. \tag{2.85}$$

These results uniquely determine the invariant angles (modulo $\pi$) for a given $h_k$ mass ordering.

Using Eqs. (2.73) and (2.76), one obtains the following U(2)-covariant expression for the scalar fields in a generic basis in terms of mass-eigenstate fields:

$$\Phi_a = \begin{pmatrix} G^+ \widehat{v}_a + H^+ \widehat{w}_a \\ \dfrac{v}{\sqrt{2}} \widehat{v}_a + \dfrac{1}{\sqrt{2}} \displaystyle\sum_{k=1}^{4} \left( q_{k1} \widehat{v}_a + q_{k2} e^{-i\theta_{23}} \widehat{w}_a \right) h_k \end{pmatrix}\,, \tag{2.86}$$

where $h_4 \equiv G^0$ and the $q_{k\ell}$ have been given in Eq. (2.77).

### 2.3.4 *Higgs boson couplings*

The gauge boson–Higgs boson interactions are governed by the following interaction Lagrangians:

$$\begin{aligned} \mathscr{L}_{VVH} = &\left( g m_W W_\mu^+ W^{\mu -} + \frac{g}{2c_W} m_Z Z_\mu Z^\mu \right) \operatorname{Re}(q_{k1}) h_k \\ &+ e m_W A^\mu (W_\mu^+ G^- + W_\mu^- G^+) - g m_Z s_W^2 Z^\mu (W_\mu^+ G^- + W_\mu^- G^+)\,, \end{aligned} \tag{2.87}$$

$$\begin{aligned} \mathscr{L}_{VVHH} = &\left[ \tfrac{1}{4} g^2 W_\mu^+ W^{\mu -} + \frac{g^2}{8c_W^2} Z_\mu Z^\mu \right] \operatorname{Re}(q_{j1}^* q_{k1} + q_{j2}^* q_{k2})\, h_j h_k \\ &+ \left[ e^2 A_\mu A^\mu + \frac{g^2}{c_W^2} \left(\tfrac{1}{2} - s_W^2\right)^2 Z_\mu Z^\mu + \frac{2ge}{c_W} \left(\tfrac{1}{2} - s_W^2\right) A_\mu Z^\mu \right] (G^+ G^- + H^+ H^-) \\ &+ \left\{ \left( \tfrac{1}{2} e g A^\mu W_\mu^+ - \frac{g^2 s_W^2}{2c_W} Z^\mu W_\mu^+ \right) (q_{k1} G^- + q_{k2}\, e^{-i\theta_{23}} H^-) h_k + \text{h.c.} \right\}, \end{aligned} \tag{2.88}$$

$$\begin{aligned} \mathscr{L}_{VHH} = &\frac{g}{4c_W} \operatorname{Im}(q_{j1} q_{k1}^* + q_{j2} q_{k2}^*) Z^\mu h_j \overleftrightarrow{\partial}_\mu h_k + \frac{ig}{c_W} \left(\tfrac{1}{2} - s_W^2\right) Z^\mu (G^+ \overleftrightarrow{\partial}_\mu G^- + H^+ \overleftrightarrow{\partial}_\mu H^-) \\ &- \tfrac{1}{2} g \left\{ i W_\mu^+ \left[ q_{k1} G^- \overleftrightarrow{\partial}^\mu h_k + q_{k2} e^{-i\theta_{23}} H^- \overleftrightarrow{\partial}^\mu h_k \right] + \text{h.c.} \right\}, \\ &+ ie A^\mu (G^+ \overleftrightarrow{\partial}_\mu G^- + H^+ \overleftrightarrow{\partial}_\mu H^-) \end{aligned} \tag{2.89}$$

where $s_W \equiv \sin\theta_W$, $c_W \equiv \cos\theta_W$ and there is an implicit sum over the repeated indices $j, k = 1, \ldots, 4$ (with $h_4 = G^0$). Since $e^{-i\theta_{23}} H^-$ is an invariant field, Eqs. (2.87)–(2.89) are indeed U(2)-invariant.

Likewise, one can work out the cubic and quartic Higgs boson self-couplings [16]:

$$\begin{aligned}
\mathscr{L}_{3h} = -\tfrac{1}{2} v\, h_j h_k h_\ell \Big[ & q_{j1} q^*_{k1} \,\mathrm{Re}\,(q_{\ell 1}) Z_1 + q_{j2} q^*_{k2} \,\mathrm{Re}\,(q_{\ell 1})(Z_3 + Z_4) + \mathrm{Re}\,(q^*_{j1} q_{k2} q_{\ell 2} Z_5\, e^{-2i\theta_{23}}) \\
& + \mathrm{Re}\left( [2q_{j1} + q^*_{j1}] q^*_{k1} q_{\ell 2} Z_6\, e^{-i\theta_{23}} \right) + \mathrm{Re}\,(q^*_{j2} q_{k2} q_{\ell 2} Z_7\, e^{-i\theta_{23}}) \Big] \\
& - v\, h_k G^+ G^- \Big[ \mathrm{Re}\,(q_{k1}) Z_1 + \mathrm{Re}\,(q_{k2}\, e^{-i\theta_{23}} Z_6) \Big] - v\, h_k H^+ H^- \Big[ \mathrm{Re}\,(q_{k1}) Z_3 + \mathrm{Re}\,(q_{k2}\, e^{-i\theta_{23}} Z_7) \Big] \\
& - \tfrac{1}{2} v\, h_k \Big\{ G^- H^+\, e^{i\theta_{23}} \Big[ q^*_{k2} Z_4 + q_{k2}\, e^{-2i\theta_{23}} Z_5 + 2\,\mathrm{Re}\,(q_{k1}) Z_6\, e^{-i\theta_{23}} \Big] + \mathrm{h.c.} \Big\}, \qquad (2.90)
\end{aligned}$$

and

$$\begin{aligned}
\mathscr{L}_{4h} = -\tfrac{1}{8} h_j h_k h_l h_m \Big[ & q_{j1} q_{k1} q^*_{\ell 1} q^*_{m1} Z_1 + q_{j2} q_{k2} q^*_{\ell 2} q^*_{m2} Z_2 + 2 q_{j1} q^*_{k1} q_{\ell 2} q^*_{m2} (Z_3 + Z_4) \\
& + 2\,\mathrm{Re}\,(q^*_{j1} q^*_{k1} q_{\ell 2} q_{m2} Z_5\, e^{-2i\theta_{23}}) + 4\,\mathrm{Re}\,(q_{j1} q^*_{k1} q^*_{\ell 1} q_{m2} Z_6\, e^{-i\theta_{23}}) + 4\,\mathrm{Re}\,(q^*_{j1} q_{k2} q_{\ell 2} q^*_{m2} Z_7\, e^{-i\theta_{23}}) \Big] \\
& - \tfrac{1}{2} h_j h_k G^+ G^- \Big[ q_{j1} q^*_{k1} Z_1 + q_{j2} q^*_{k2} Z_3 + 2\,\mathrm{Re}\,(q_{j1} q_{k2} Z_6\, e^{-i\theta_{23}}) \Big] \\
& - \tfrac{1}{2} h_j h_k H^+ H^- \Big[ q_{j2} q^*_{k2} Z_2 + q_{j1} q^*_{k1} Z_3 + 2\,\mathrm{Re}\,(q_{j1} q_{k2} Z_7\, e^{-i\theta_{23}}) \Big] \\
& - \tfrac{1}{2} h_j h_k \Big\{ G^- H^+\, e^{i\theta_{23}} \Big[ q_{j1} q^*_{k2} Z_4 + q^*_{j1} q_{k2} Z_5\, e^{-2i\theta_{23}} + q_{j1} q^*_{k1} Z_6\, e^{-i\theta_{23}} + q_{j2} q^*_{k2} Z_7\, e^{-i\theta_{23}} \Big] + \mathrm{h.c.} \Big\} \\
& - \tfrac{1}{2} Z_1 G^+ G^- G^+ G^- - \tfrac{1}{2} Z_2 H^+ H^- H^+ H^- - (Z_3 + Z_4) G^+ G^- H^+ H^- - \tfrac{1}{2} Z_5 H^+ H^+ G^- G^- \\
& - \tfrac{1}{2} Z_5^* H^- H^- G^+ G^+ - G^+ G^- (Z_6 H^+ G^- + Z_6^* H^- G^+) - H^+ H^- (Z_7 H^+ G^- + Z_7^* H^- G^+), \qquad (2.91)
\end{aligned}$$

where there is an implicit sum over the repeated indices $j$, $k$, $\ell$, $m = 1, 2, 3, 4$ (with $h_4 = G^0$). Using Eq. (2.74) and noting that the combinations $Z_5 e^{-2i\theta_{23}}$, $Z_6 e^{-i\theta_{23}}$ and $Z_7 e^{-i\theta_{23}}$ are U(2)-invariant quantities, it follows that the cubic and quartic Higgs boson self-couplings are also U(2)-invariant.

Expressions for the cubic and quartic Higgs self-couplings in the CP-violating 2HDM have also been obtained in terms of generic basis parameters in [15, 47, 161], and an application of these results to the CPX scenario [111] of the minimal supersymmetric extension of the Standard Model (MSSM) is given in Section 3.6. Indeed, the effective Lagrangian of the MSSM Higgs sector *is* a general CP-violating 2HDM when one-loop radiative corrections (which are sensitive to supersymmetry-breaking effects and new CP-violating phases) are taken into account. The relative simplicity of the Higgs self-couplings given in Eqs. (2.90) and (2.91) illustrates the power of the basis-independent techniques.

The Higgs couplings to quarks are governed by the Yukawa Lagrangian given in Eq. (2.46) In terms of the quark mass-eigenstate fields, Eq. (2.46) can be expressed in U(2)-covariant form:.[10]

$$-\mathscr{L}_{\rm Y} = \overline{Q}_L \widetilde{\Phi}_{\bar a} \eta^U_a U_R + \overline{Q}_L \Phi_a \eta^{D\,\dagger}_{\bar a} D_R + \mathrm{h.c.}\,, \qquad (2.92)$$

where $\widetilde{\Phi}_{\bar a} \equiv i\sigma_2 \Phi^*_{\bar a}$, and $\eta^U_a \equiv (V^U_L \Gamma_1 V^{U\,\dagger}_R\,,\, V^U_L \Gamma_2 V^{U\,\dagger}_R)$ and $\eta^D_a \equiv (V^D_R \Delta_1 V^{D\,\dagger}_L\,,\, V^D_R \Delta_2 V^{D\,\dagger}_L)$. We employ the standard notation: $\psi_{R,L} \equiv P_{R,L}\psi$ with $P_{R,L} \equiv \frac{1}{2}(1 \pm \gamma_5)$. The unitary matrices $V^U_{L,R}$ and $V^D_{L,R}$ relate the quark interaction-eigenstate and quark mass-eigenstate fields via the bi-unitary transformations of Eq. (2.47). One can rewrite Eq. (2.92) in terms of Higgs basis scalar fields:

$$-\mathscr{L}_{\rm Y} = \overline{Q_L}(\widetilde{\boldsymbol{H}}_1 \kappa^U + \widetilde{\boldsymbol{H}}_2 \rho^U) U_R + \overline{Q}_L (\boldsymbol{H}_1 \kappa^{D\,\dagger} + \boldsymbol{H}_2 \rho^{D\,\dagger}) D_R + \mathrm{h.c.}\,, \qquad (2.93)$$

[10]To obtain the Higgs couplings to leptons, let $Q \to L$ and $D \to E$, and omit $U$ (right-handed neutrinos are not included).

where

$$\kappa^Q \equiv \widehat{v}_{\bar{a}}^* \eta_a^Q = \sqrt{2} M_Q / v \,, \qquad\qquad \rho^Q \equiv \widehat{w}_{\bar{a}}^* \eta_a^Q \,. \tag{2.94}$$

Under a U(2) transformation, $\kappa^Q$ is invariant, whereas $\rho^Q$ is a pseudo-invariant that transforms as:

$$\rho^Q \to (\det U) \rho^Q \,. \tag{2.95}$$

By construction, $\kappa^U$ and $\kappa^D$ are proportional to the (real non-negative) diagonal quark mass matrices, as indicated in Eq. (2.50). That is, we have chosen the unitary matrices $V_L^U$, $V_R^U$, $V_L^D$ and $V_R^D$ such that $M_D$ and $M_U$ are diagonal matrices with real non-negative entries. In the general 2HDM, the $\rho^Q$ are arbitrary complex $3 \times 3$ matrices.

In order to determine the interactions of the Higgs (and Goldstone) bosons with the quark mass eigenstates, one can bypass the intermediate step involving the Higgs basis by inserting Eq. (2.86) into Eq. (2.92) to obtain:

$$\begin{aligned} -\mathscr{L}_Y = {} & \frac{1}{v} \overline{D} \Big\{ M_D (q_{k1} P_R + q_{k1}^* P_L) + \frac{v}{\sqrt{2}} \Big[ q_{k2} \, [e^{i\theta_{23}} \rho^D]^\dagger P_R + q_{k2}^* \, e^{i\theta_{23}} \rho^D P_L \Big] \Big\} D h_k \\ & + \frac{1}{v} \overline{U} \Big\{ M_U (q_{k1} P_L + q_{k1}^* P_R) + \frac{v}{\sqrt{2}} \Big[ q_{k2}^* \, e^{i\theta_{23}} \rho^U P_R + q_{k2} \, [e^{i\theta_{23}} \rho^U]^\dagger P_L \Big] \Big\} U h_k \\ & + \Big\{ \overline{U} \Big[ K [\rho^D]^\dagger P_R - [\rho^U]^\dagger K P_L \Big] D H^+ + \frac{\sqrt{2}}{v} \overline{U} \left[ K M_D P_R - M_U K P_L \right] D G^+ + \text{h.c.} \Big\} \,, \end{aligned} \tag{2.96}$$

where $k = 1, \dots 4$. Since $e^{i\theta_{23}} \rho^Q$ and $[\rho^Q]^\dagger H^+$ are U(2)-invariant, it follows that Eq. (2.96) is a basis-independent expression of the Higgs-quark interactions.

The Higgs-quark couplings are generically CP-violating as a result of the complexity of the $q_{k2}$ and the fact that the matrices $e^{i\theta_{23}} \rho^Q$ are not generally Hermitian or anti-Hermitian. Consequently, the neutral Higgs bosons ($h_1$, $h_2$ and $h_3$) are typically states of indefinite CP quantum number (whereas $h_4 \equiv G^0$ is always a pure CP-odd state). Basis-independent conditions for the CP-invariance of the neutral Higgs boson couplings to quark pairs are obtained by requiring that Eq. (2.75) is satisfied and [16]:

$$Z_5 [\rho^Q]^2 \,, \quad Z_6 \rho^Q \,, \quad \text{and} \quad Z_7 \rho^Q \quad \text{are Hermitian matrices} \qquad (Q = U, D) \,. \tag{2.97}$$

In this case, the only remaining source of CP-violation in the 2HDM is the unremovable phase in the CKM matrix $K$ that enters via the charged current interactions mediated by either $W^\pm$ or $H^\pm$ exchange.

The Higgs-quark couplings also generate Higgs-boson-mediated flavor-changing neutral currents at tree-level by virtue of the fact that the $\rho^Q$ are not diagonal (in the quark mass basis). Thus, for a phenomenologically acceptable theory, the off-diagonal elements of $\rho^Q$ must be small.

### *2.3.5 Conclusions*

In the most general (CP-violating) 2HDM, physical observables do not depend on the choice of scalar field definitions (or basis). Employing the U(2) freedom of field redefinitions, one can write down the Higgs couplings of the 2HDM in a form that is manifestly basis independent. The U(2)-invariant forms for the Higgs boson couplings have been explicitly presented in this paper. In particular, the parameter $\tan\beta$, which depends on the choice of basis, does not appear in any of the Higgs boson (or Goldstone boson) couplings. In specialized versions of the 2HDM, additional theoretical assumptions are introduced that may implicitly select a preferred basis. For example, one can impose a discrete symmetry on the Lagrangian that takes a simple form in some particular basis. The type-I and type-II models discussed in Section 2.1.4.2 provide examples of such a scenario. In this case, $\tan\beta$ is promoted to a physical parameter, and one can express $\tan\beta$ directly in terms of U(2)-invariant quantities [14–16].

The basis-independent formalism provides a powerful approach for connecting physical observables that can be measured in the laboratory with fundamental invariant parameters of the 2HDM. This will permit the development of two-Higgs doublet model-independent analyses of data in Higgs studies at the LHC, ILC and beyond.

## 2.4 Symmetries of 2HDM and CP violation

*Ilya F. Ginzburg and Maria Krawczyk*

This contribution is based on the results published in [15] and some new results that have recently been obtained. The main aspects of the results of [15] are included in Section 2.1. Here we present alternative treatments of some problems and add new results, some of which were reported in [162].

The spontaneous electroweak symmetry breaking (EWSB) via the Higgs mechanism is described by the Lagrangian

$$\mathcal{L} = \mathcal{L}_{gf}^{SM} + \mathcal{L}_H + \mathcal{L}_Y \quad with \quad \mathcal{L}_H = T - V\,. \tag{2.98}$$

Here $\mathcal{L}_{gf}^{SM}$ describes the $SU(2)\times U(1)$ Standard Model interaction of gauge bosons and fermions, $\mathcal{L}_Y$ describes the Yukawa interactions of fermions with Higgs scalars and $\mathcal{L}_H$ is the Higgs scalar Lagrangian. Higgs potential. In our calculations we use the 2HDM Higgs potential as specified in Eq. (2.1); however, we insert explicit negative signs in the terms proportional to $m_{11}^2$ and $m_{22}^2$ terms. In this latter convention, if $m_{12}^2 = 0$ then EWSB is generated for positive values of $m_{11}^2$ and $m_{22}^2$. The most general renormalizable kinetic term is

$$T = (D_\mu\phi_1)^\dagger(D^\mu\phi_1) + (D_\mu\phi_2)^\dagger(D^\mu\phi_2) + \left[\varkappa(D_\mu\phi_1)^\dagger(D^\mu\phi_2) + h.c.\right]. \tag{2.99}$$

### 2.4.1 Reparametrization group

The "flavor" basis transformations discussed in Section 2.1 are described by a unitary group give by the direct product of the 3-parameter $SU(2)$ reparametrization (RPa) group and a $U(1)$ group, describing an overall phase freedom of the Lagrangian. This entire group operates on the space of fields while the RPa group operates also in the space of Lagrangians (with coordinates given by its parameters). The parameters of the Lagrangian can be determined in principle from measurements only with an accuracy up to the RPa freedom. All observable quantities (at least in principle) are invariants of the RPa group (IRPa). These are, for example, masses of observable Higgs bosons, i.e. eigenvalues of the mass matrix, and eigenvalues of the Higgs-Higgs scattering matrices. The transformations $\phi_k \to e^{-i\rho_k}\phi_k$ form a rephasing transformation (RPh) group, which is a subgroup of the RPa group with a single parameter $\rho = \rho_2 - \rho_1$.

The method described in Section 2.1 allows one to obtain a large series of (generally not independent) IRPa's [37]. An alternate method is based on the irreducible representations of SU(2) RPa group as discussed [19]. In this paper some basic objects, related to these irreducible representations, were determined: 3 scalars $A^k$, 2 vectors $L_i$ and $M_i$ and tensor $a_{ij}$ $(i,\, j = x,\, y,\, z)$. After a simple reorganization of scalars $A^k$, we get

$$A^I = \lambda_1 + \lambda_2 + 2\lambda_3, \quad A^{II} = \lambda_3 - \lambda_4, \quad A^{III} = m_{11}^2 + m_{22}^2\,, \tag{2.100a}$$

$$(L_x, L_y, L_z) = \frac{1}{\sqrt{2}}\left(-Re(\lambda_6+\lambda_7),\; Im(\lambda_6+\lambda_7),\; -(\lambda_1-\lambda_2)/2\right), \tag{2.100b}$$

$$(M_x, M_y, M_z) = \frac{1}{\sqrt{2}}\left(Re\; m_{12}^2,\; -Im\; m_{12}^2,\; (m_{11}^2 - m_{22}^2)/2\right)\,, \tag{2.100c}$$

$$a_{ij} = \frac{1}{2}\begin{pmatrix} Re\,\lambda_5 - b & -Im\,\lambda_5 & Re\,(\lambda_6-\lambda_7) \\ -Im\,\lambda_5 & -Re\,\lambda_5 - b & Im\,(\lambda_7-\lambda_6) \\ Re\,(\lambda_6-\lambda_7) & Im\,(\lambda_7-\lambda_6) & 2b \end{pmatrix}, \tag{2.100d}$$

with $b = (\lambda_1 + \lambda_2 - 2\lambda_3 - 2\lambda_4)/6$. Introducing the vectors $L_i^I = a_{ij}L_j$ and $L_i^{II} = a_{ij}L_j^I$ , a complete set of 11 independent invariants of RPa transformations can be naturally chosen as follows

$$\begin{gathered} I_1 = A^I, \quad I_2 = A^{II}\,, \quad I_3 = L_iL_i, \\ I_4 = Tr(a^2) \equiv Tr\,(a_{ij}a_{jk})\,, \quad I_5 = Tr(a^3) = Tr\,(a_{ij}a_{jk}a_{k\ell})\,, \\ I_6 = a_{ij}L_iL_j \equiv L_iL_i^I\,, \quad I_7 = \varepsilon_{ijk}L_iL_j^IL_k^{II}, \\ I_8 = A^{III}\,, \quad I_9 = M_iM_i\,, \quad I_{10} = L_iM_i\,, \quad I_{11} = \varepsilon_{ijk}L_iL_j^IM_k\,. \end{gathered} \tag{2.101}$$

### 2.4.2 $Z_2$ symmetry

The CP violation and the flavour changing neutral currents (FCNC) are absent for a 2HDM Lagrangian with a $Z_2$ symmetry, which inhibits the $\phi_1 \leftrightarrow \phi_2$ transitions [18]. The Lagrangian is invariant under the interchange $\phi_1 \leftrightarrow \phi_1, \phi_2 \leftrightarrow -\phi_2$ or $\phi_1 \leftrightarrow -\phi_1, \phi_2 \leftrightarrow \phi_2$.

• The case of exact $Z_2$ symmetry is described by the Lagrangian Eq. (2.98) with potential Eq. (2.1), where $\lambda_6 = \lambda_7 = m_{12}^2 = 0$, and a kinetic term Eq. (2.99) with $\varkappa = 0$.

• In the case of soft violation of $Z_2$ symmetry, one adds to the $Z_2$ symmetric Lagrangian a term of operator dimension two, $m_{12}^2(\phi_1^\dagger\phi_2) + h.c.$, with a generally complex $m_{12}^2$ (and $\lambda_5$) parameter. This type of violation respects the $Z_2$ symmetry at short distances (much shorter than a cut-off $1/M$) to all orders in the perturbative series, i.e. the amplitudes for $\phi_1 \leftrightarrow \phi_2$ transitions disappear at virtuality $k^2 \sim M^2 \to \infty$. That is why we call this a "soft" violation.

Let our physical system be described by the Lagrangian $\mathcal{L}_s$ with an exact or softly violated $Z_2$ symmetry. The general RPa transformation converts $\mathcal{L}_s$ to a hidden soft $Z_2$ violating form $\mathcal{L}_{hs}$, with $\lambda_6$, $\lambda_7 \neq 0$, $\varkappa = 0$. The 14 parameters of this type of Lagrangian are constrained since they can be obtained from 9 independent parameters of an initial Lagrangian $\mathcal{L}_s$ (+ 3 RPa group parameters); the nondiagonal $\varkappa$ kinetic term does not arise from the loop corrections. For such a physical system the preferable RPa representation is given by $\mathcal{L}_s$.

The criteria whether the soft $Z_2$ symmetry of the potential is hidden can be easily obtained from Eq. (2.101) (the invariant condition is provided in [14]). If this case is realized, one can consider the RPa representation with an explicit soft $Z_2$ symmetry ($\lambda_6 = \lambda_7 = 0$) and with a real $\lambda_5$ (this can be achieved by a suitable RPh transformation). In this representation, $a_{ij}$ is a diagonal matrix, and the vector $L_i \equiv (0,\, 0,\, L)$ has only a $z$-component. Therefore the vectors $L_i^I$ and $L_i^{II}$ also have only $z$-components. Hence, the invariant $I_7 = 0$. A straightforward calculation gives in addition: $I_3 = L^2$, $I_4 = (\lambda_5^2 + 3b^2)/2$, $I_5 = 3b(b^2 - \lambda_5^2)/4$, $I_6 = bL^2$. Therefore, these four invariants obey the relation $I_5I_3/I_6 + 3I_4/2 = 3(I_6/I_3)^2$. These relations are written for invariants. Hence, the necessary and sufficient conditions for soft $Z_2$ symmetry violation are

$$I_7 = 0\,, \qquad I_3^2 I_4 I_6 + (2/3) I_3^3 I_5 - 2I_6^3 = 0\,. \tag{2.102}$$

• In the general case, the terms of an operator of dimension four, with generally complex parameters $\lambda_6$, $\lambda_7$ and $\varkappa$, are added to the Lagrangian with a softly violated $Z_2$ symmetry. This case covers both the opportunity of a hidden soft $Z_2$ symmetry violation and of a true hard violation of the $Z_2$ symmetry of the Lagrangian, which cannot be transformed to an exact or softly violated $Z_2$ symmetry form by any RPa transformation. For the true hard violation case, the $Z_2$ symmetry is broken at both large and short distances in any scalar ("flavor") basis.

The mixed kinetic terms Eq. (2.99) can be eliminated by the nonunitary transformation (rotation + renormalization), e.g.

$$(\phi_1', \phi_2') \to \left( \frac{\sqrt{\varkappa^*}\phi_1 + \sqrt{\varkappa}\phi_2}{2\sqrt{|\varkappa|(1+|\varkappa|)}} \pm \frac{\sqrt{\varkappa^*}\phi_1 - \sqrt{\varkappa}\phi_2}{2\sqrt{|\varkappa|(1-|\varkappa|)}} \right). \tag{2.103}$$

However, in the presence of the $\lambda_6$ and $\lambda_7$ terms in the potential Eq. (2.1), the renormalization of the quadratically divergent, non-diagonal two-point functions leads anyway to mixed kinetic terms (e.g. from loops with $\lambda_6^*\lambda_{1,3-5}$ and $\lambda_7^*\lambda_{2-5}$). This means that $\varkappa$ becomes nonzero at higher orders in perturbative theory (and *vice versa* a mixed kinetic term generates counterterms with $\lambda_{6,7}$). Therefore all of these terms should be included in Lagrangian on the same footing, i.e. the treatment of the hard violation of $Z_2$ symmetry without $\varkappa$ terms is inconsistent. (This term does not appear if the parameters $\lambda_i$ are constrained by the relations given in Eq. (2.102).) In the case of true hard violation of the $Z_2$ symmetry, the parameter $\varkappa$ is running like the $\lambda$ parameters. Therefore, the diagonalization of Eq. (2.103) is scale dependent, and the Lagrangian remains off–diagonal in the fields $\phi_{1,2}$ even at very short distances in any RPa representation. Such a theory seems to be unnatural.

Although we present in [15] and here the relations for the case of hard violation of the $Z_2$ symmetry at $\varkappa = 0$, the loop corrections create a $\varkappa$ terms and can change the results significantly. Such a treatment of the case with true hard violation of the $Z_2$ symmetry is as incomplete as in other papers that consider the "most general 2HDM potential". Note, however that there is no consensus whether the parameters $\varkappa$ are independent parameters.

### *2.4.3 Ground state after EWSB. Criterium for CP conservation*

The extrema of the potential define the vacuum expectation values (v.e.v.'s) $\langle\phi_{1,2}\rangle$ of the fields $\phi_{1,2}$ via equations Eq. (2.5), $\partial V/\partial\phi_i|_{\phi_i=\langle\phi_i\rangle} = 0$. The physical reality can be described only by a potential with a ground state obeying the condition for $U(1)$ symmetry of electromagnetism[11]:

$$\langle\phi_1\rangle = \frac{1}{\sqrt{2}}\begin{pmatrix}0\\ v_1\end{pmatrix}, \quad \langle\phi_2\rangle = \frac{1}{\sqrt{2}}\begin{pmatrix}0\\ v_2 e^{i\xi}\end{pmatrix}, \tag{2.104}$$

The rephasing transformation of fields allows one to eliminate the phase difference $\xi$ and leads to the corresponding changes of the coefficients of Lagrangian. This real vacuum Lagrangian is used in [15] and in Section 2.1.

A standard decomposition of the fields $\phi_{1,2}$ in the component fields Eq. (2.20) at $\varkappa = 0$ retains a diagonal form of the kinetic terms for the fields $\varphi_i^+$, $\chi_i$, $\varphi_i$. The mass-squared matrix for the component fields has a block diagonal form with separate blocks, corresponding to massless Goldstone boson fields, charged Higgs boson fields $H^\pm$ and a $3\times 3$ matrix for the neutral fields Eq. (2.26), for two scalars $\eta_1$, $\eta_2$ and a pseudoscalar $A = -\chi_1 sin\beta + \chi_2 cos\beta$.

The possible mixing of the scalar and pseudoscalar states, which yields physical Higgs states $h_i$ having no definite CP parity, generates CP violation in Higgs sector. Therefore, a signature for CP conservation in the Higgs sector is given by the zero values of squared-mass matrix elements Eq. (2.26) responsible for this mixing (i.e. $M_{13} = M_{23} = 0$ in the real vacuum basis). The vanishing of $M_{13}$ and $M_{23}$ can be realized if in such a basis $m_{12}^2$ and $v_1^2\lambda_6 - v_2^2\lambda_7$ (and also $\lambda_5$ – see Eq. (2.10)) are real. The set of arbitrary bases can be obtained from the set of real vacuum bases by the rephasing of fields. Therefore the sufficient condition for CP conservation in the Higgs sector can be written in a basis independent form as

$$Im\lambda_5^*(m_{12}^2)^2 = 0\,, \quad Im(\lambda_6^* + \lambda_7^*)m_{12}^2 = 0\,, \quad Im\lambda_6^*\lambda_7 = 0. \tag{2.105}$$

Each of these quantities is not RPa invariant but these forms are very simple. (For the soft $Z_2$ violating potential, this condition becomes both necessary and sufficient; it is simply: $Im\,\lambda_5^*(m_{12}^2)^2 = 0$, cf. [163]).

The imaginary part of the quantity $v_1^2\lambda_6 - v_2^2\lambda_7$ can be equal to 0 (which is necessary for CP conservation) even for complex $\lambda_6$ and $\lambda_7$. The RPa transformation from one real vacuum basis to another depends on two independent parameters (3 parameters of a general RPa transformation with one parameter restoring the real basis). One can use these parameters to eliminate the imaginary parts of $\lambda_6$ and $\lambda_7$ separately. Using in addition Eq. (2.10) one can conclude that in the case of CP violation in the Higgs sector there exists a Higgs basis in which all the coefficients of the potential are real (which is necessary condition for CP conservation).

Generally, these v.e.v.'s can be complex even in the case of a Lagrangian with real parameters. Therefore, the coefficients of the real vacuum Lagrangian can be complex even in the case where all coefficients of the incident Higgs potential are real. Hence, the existence of an RPa representation of the potential with all real coefficients forms a necessary but not a sufficient condition for CP conservation.

[11] Detail analysis of two possible vacuum solutions, including charged vacuum [20], is presented in [162], [15].

The RPa invariant form of this necessary condition is presented in Section 2.1 [38, 40, 164], [14, 37], and with invariants (2.101) in [19].

### 2.4.4 Tree level unitarity constraints

The quartic terms in the Higgs potential (with $\lambda_i$) lead, in the tree approximation, to the s–wave Higgs-Higgs and $W_L W_L$ and $W_L H$, etc. scattering amplitudes for different elastic channels. These amplitudes should not overcome the unitary limit for this partial wave – that is the tree-level unitarity constraint. Such a constraint was obtained first in [18] for the minimal SM, with one Higgs field. For the 2HDM with soft $Z_2$ violation and CP conservation, they were derived in [32]. In the general CP nonconserving case, the corresponding constraints were obtained in [34].

Since in the Higgs–Higgs scattering the total hypercharge $Y$ and weak isospin $\sigma$ are conserved, one can consider separate S matrices, $S_{Y\sigma}$, for the different quantum numbers of the initial state. The unitarity constraint means that the eigenvalues of these $S_{Y\sigma}$ are less than 1 [34], where

$$16\pi S_{Y=2,\sigma=1} = \begin{pmatrix} \lambda_1 & \lambda_5 & \sqrt{2}\lambda_6 \\ \lambda_5^* & \lambda_2 & \sqrt{2}\lambda_7^* \\ \sqrt{2}\lambda_6^* & \sqrt{2}\lambda_7 & \lambda_3+\lambda_4 \end{pmatrix}, \quad 16\pi S_{Y=2,\sigma=0} = \lambda_3 - \lambda_4, \tag{2.106a}$$

$$16\pi S_{Y=0,\sigma=1} = \begin{pmatrix} \lambda_1 & \lambda_4 & \lambda_6 & \lambda_6^* \\ \lambda_4 & \lambda_2 & \lambda_7 & \lambda_7^* \\ \lambda_6^* & \lambda_7^* & \lambda_3 & \lambda_5^* \\ \lambda_6 & \lambda_7 & \lambda_5 & \lambda_3 \end{pmatrix}, \tag{2.106b}$$

$$16\pi S_{Y=0,\sigma=0} = \begin{pmatrix} 3\lambda_1 & 2\lambda_3+\lambda_4 & 3\lambda_6 & 3\lambda_6^* \\ 2\lambda_3+\lambda_4 & 3\lambda_2 & 3\lambda_7 & 3\lambda_7^* \\ 3\lambda_6^* & 3\lambda_7^* & \lambda_3+2\lambda_4 & 3\lambda_5^* \\ 3\lambda_6 & 3\lambda_7 & 3\lambda_5 & \lambda_3+2\lambda_4 \end{pmatrix}. \tag{2.106c}$$

The eigenvalues of these matrices can be found as roots of equations of the 3-rd or 4-th degree. It is useful to start the diagonalization from the corners of the above matrices, corresponding to the fixed values of the $Z_2$ parity. This particular diagonalization transforms $S_{Y\sigma}$ to a form with diagonal elements that are coincident with the eigenvalues found in [32] (for soft $Z_2$ violation without CP violation) with the sole change of $\lambda_5 \to |\lambda_5|$.

Next, one can use the following observation: For an Hermitian matrix $\mathcal{M} = ||M_{ij}||$ with maximal and minimal eigenvalues $\Lambda_+$ and $\Lambda_-$, all diagonal matrix elements $M_{ii}$ lie between them, $\Lambda_+ \geq M_{ii} \geq \Lambda_-$. By virtue of this fact, the mentioned corrected constraints from [32] form necessary conditions for unitarity. These constraints are enhanced due to the $\lambda_6$, $\lambda_7$ terms that govern the hard violation of the $Z_2$ symmetry.

### 2.4.5 Couplings to fermions

The general form of Yukawa interaction couples a 3-family vector of the left-handed quark isodoublets $Q_L$ with 3-family vectors of the right-handed field singlets $d_R$ and $u_R$ and Higgs fields $\phi_i$ Eq. (2.46).

If some fermion field singlet is coupled to both scalar fields $\phi_1$ and $\phi_2$, the counterterms corresponding to the one-loop propagator corrections to the Higgs Lagrangian contain operators of dimension 4, which violate the $Z_2$ symmetry in a hard way. They contribute to the renormalization of the parameters $\varkappa$, $\lambda_6$ and $\lambda_7$. Therefore, to have only a soft violation of $Z_2$ symmetry (to prevent $\phi_1 \leftrightarrow \phi_2$ transitions at short distances), one demands that [18, 165] each right-handed fermion couples to only one scalar field, either $\phi_1$ or $\phi_2$. The case $\Gamma_2 = \Delta_1 = 0$ with diagonal $\Gamma_1$, $\Delta_2$ corresponds to the well known Model II, while $\Gamma_2 = \Delta_2 = 0$ corresponds to Model I. Note that a general RPa transformation makes these properties of the Lagrangian hidden. The widespread Model II, with many useful relations for it partially obtained in [15] and first collected together there, is described in Section 2.1.

In this analysis we use in principle measurable relative couplings—that is, ratios of the couplings of each neutral Higgs boson $h_i$ to the corresponding SM couplings Eq. (2.41). We present here, for completeness, only the case of model I Yukawa interactions. In this model, all right handed fermions are coupled to one Higgs field, say $\phi_1$. The general RPa transformation makes this property hidden, changing simultaneously $\tan\beta$. The parameter $\beta$ corresponding to the Model I form of the Lagrangian will be labeled with a subscript I. For this form of the Lagrangian we have ($i = 1, 2, 3$):

$$\text{Model I}: \qquad \chi_u^{(i)} = \chi_d^{(i)} \equiv \chi_f^{(i)} = \frac{[R_{i2} - i\cos\beta_I\, R_{i3}]}{\sin\beta_I}. \tag{2.107}$$

In this case, among the methods presented for Model II, only one method succeeds in determining $\beta_I$ via observable quantities, namely [see Eq. (2.58)]

$$\frac{1}{\sin^2\beta_I} = \sum_{i=1}^{3} (\,\mathrm{Re}\,\chi_u^{(i)})^2. \tag{2.108}$$

### 2.4.6 *A natural set of parameters of 2HDM*

It is natural to assume that the 2HDM that describes physical reality allows for the existence among the reparametrization equivalent Lagrangians the one in which the fields $\phi_k$ do not mix at small distances (mixed kinetic term does not appear). This would correspond to the 2HDM with an exact or softly violated $Z_2$ symmetry. We assume such choice in this section. Besides, it is natural to assume that the CP symmetry in the Higgs sector is violated only weakly at least for the lightest Higgs boson $h_1$. This assumption together with rephasing invariance offers the basis for the selection of *the natural set of parameters of 2HDM*.

The Eq. (2.30) shows that the CP symmetry for the lightest Higgs boson is violated weakly if and only if $|M'_{13}| \ll |M_A^2 - M_h^2|$. In view of Eq. (2.34), for *the real vacuum Lagrangian* at $\beta + \alpha \neq \pi/2$ this condition can be rewritten in the form

$$v^2 |\,\mathrm{Im}\, m_{12}^2| \ll v_1 v_2 |M_A^2 - M_h^2|\,. \tag{2.109}$$

For all other rephasing equivalent Lagrangians, this condition Eq. (2.109) contains both $\mathrm{Im}\, m_{12}^2$ and $\mathrm{Re}\, m_{12}^2$. Therefore, for *the natural set of parameters of 2HDM* we require that both $|\,\mathrm{Im}\, m_{12}^2|(v^2/v_1 v_2)$ and $|\,\mathrm{Re}\, m_{12}^2|(v^2/v_1 v_2)$ are small for all rephasing equivalent Lagrangians. In the case of soft violation of $Z_2$ symmetry, the same requirements is transmitted to $\mathrm{Im}\,\lambda_5$ and $\mathrm{Re}\,\lambda_5$. Therefore, we define *a natural set of parameters* as follows

$$|\eta|,\ |\lambda_5| \ll |\lambda_{1-4}|\,. \tag{2.110}$$

For the natural set of parameters of the 2HDM, the breaking of the $Z_2$ symmetry is governed by a small parameter $\eta$. Due to the existence of a limit where the $Z_2$ symmetry holds, a small soft $Z_2$ violation in the Higgs Lagrangian and the Yukawa interactions also remains small beyond the tree level. In this respect, we use term *natural* in the same sense as in [166]. (Note also that the non-diagonal Yukawa coupling matrices $\Gamma_1$ and $\Delta_2$ (leading to FCNCs) are unnatural in this very sense).

In accordance with Eq. (2.26), for the natural set of parameters $M_A$ cannot be too large. This parameter regime is not ruled out by the data; in the CP conserving case see e.g. Section 2.2.1.

### 2.5 Textures and the Higgs boson-fermion couplings

*Justiniano L. Díaz-Cruz, Roberto Noriega-Papaqui and Alfonso Rosado-Sánchez*

The 2HDM [8, 167] has a potential problem with flavour changing neutral currents (FCNC) mediated by the Higgs bosons, which arise when one quark of type u or d is allowed to couple to both Higgs doublets. The possible solutions to this problem of the 2HDM involve an assumption about the Yukawa Lagrangian of the model. The specific choices for the Yukawa matrices $\Gamma_{1,2}$ and $\Delta_{1,2}$ define the versions of the 2HDM known as I, II or III, which involve certain mechanism to eliminate the otherwise unbearable FCNC problem or at least to keep it under control. In this paper we are interested in studying the 2HDM-III, where the FCNC problem is ameliorated by assuming a certain texture for the Yukawa couplings. However, the original six-texture ansatz that leads to the popular Cheng-Sher ansatz [168] seems disfavored by current data on the CKM mixing angles. More recently, mass matrices with four-texture ansatz have been considered, and are found in better agreement with the observed data [169,170]. In this paper we investigate how the form of the $ff'\phi^0$ couplings, get modified when one replaces the six-texture matrices by the four-texture one. We also discuss some implications for rare quark and lepton decays, as well as the phenomenology of the Higgs bosons [129].

#### 2.5.1 *The fermion sector of the 2HDM-III with four-texture mass matrices*

We will assume that both Yukawa matrices $\Gamma_{1,2}$ and $\Delta_{1,2}$ have the four-texture form and are Hermitian; following the conventions of [169], the quark mass matrix is written as:

$$M_q = \begin{pmatrix} 0 & C_q & 0 \\ C_q^* & \tilde{B}_q & B_q \\ 0 & B_q^* & A_q \end{pmatrix}. \tag{2.111}$$

when $\tilde{B}_q \to 0$ one recovers the six-texture form. We also consider the hierarchy:
$\mid A_q \mid \gg \mid \tilde{B}_q \mid, \mid B_q \mid, \mid C_q \mid$, which is supported by the observed fermion masses in the SM.
Because of the hermicity condition, both $\tilde{B}_q$ and $A_q$ are real parameters, while the phases of $C_q$ and $B_q$, $\Phi_{B_q,C_q}$, can be removed from the mass matrix $M_q$ by defining: $M_q = P_q^\dagger \tilde{M}_q P_q$, where $P_q = diag[1, e^{i\Phi_{C_q}}, e^{i(\Phi_{B_q}+\Phi_{C_q})}]$, and the mass matrix $\tilde{M}_q$ includes only the real parts of $M_q$. The diagonalization of $\tilde{M}_q$ is then obtained by an orthogonal matrix $O_q$, such that the diagonal mass matrix is: $\bar{M}_q = O_q^T \tilde{M}_q O_q$. Expanding in powers of $z_f = m_2^f/m_3^f$, where $m_{2,3}^f$ denote the masses for $2^{nd}$ and $3^{rd}$ generations, the Yukawa Lagrangian can then be expressed in terms of the mass-eigenstates for the neutral ($h^0, H^0, A^0$) and charged Higgs bosons ($H^\pm$). The interactions of the neutral Higgs bosons with the d-type and u-type quarks ($u, u' = u, c, t$ and $d, d' = d, s, b$) are expressed as follows.

$$\begin{aligned}
\mathcal{L}_Y^q &= \frac{g}{2}\bar{D}\left[\left(\frac{m_d}{m_W}\right)\frac{\cos\alpha}{\cos\beta}\delta_{dd'} + \frac{\sin(\alpha-\beta)}{\sqrt{2}\cos\beta}\left(\frac{\sqrt{m_d m_{d'}}}{m_W}\right)\tilde{\chi}_{dd'}\right]D'H^0 \\
&+\frac{g}{2}\bar{D}\left[-\left(\frac{m_d}{m_W}\right)\frac{\sin\alpha}{\cos\beta}\delta_{dd'} + \frac{\cos(\alpha-\beta)}{\sqrt{2}\cos\beta}\left(\frac{\sqrt{m_d m_{d'}}}{m_W}\right)\tilde{\chi}_{dd'}\right]D'h^0 \\
&+\frac{ig}{2}\bar{D}\left[-\left(\frac{m_d}{m_W}\right)\tan\beta\,\delta_{dd'} + \frac{1}{\sqrt{2}\cos\beta}\left(\frac{\sqrt{m_d m_{d'}}}{m_W}\right)\tilde{\chi}_{dd'}\right]\gamma^5 D'A^0 \\
&+\frac{g}{2}\bar{U}\left[\left(\frac{m_u}{m_W}\right)\frac{\sin\alpha}{\sin\beta}\delta_{uu'} - \frac{\sin(\alpha-\beta)}{\sqrt{2}\sin\beta}\left(\frac{\sqrt{m_u m_{u'}}}{m_W}\right)\tilde{\chi}_{uu'}\right]U'H^0 \\
&+\frac{g}{2}\bar{U}\left[\left(\frac{m_u}{m_W}\right)\frac{\cos\alpha}{\sin\beta}\delta_{uu'} - \frac{\cos(\alpha-\beta)}{\sqrt{2}\sin\beta}\left(\frac{\sqrt{m_u m_{u'}}}{m_W}\right)\tilde{\chi}_{uu'}\right]U'h^0 \\
&+\frac{ig}{2}\bar{U}\left[-\left(\frac{m_u}{m_W}\right)\cot\beta\,\delta_{uu'} + \frac{1}{\sqrt{2}\sin\beta}\left(\frac{\sqrt{m_u m_{u'}}}{m_W}\right)\tilde{\chi}_{uu'}\right]\gamma^5 U'A^0.
\end{aligned} \tag{2.112}$$

The corresponding Lagrangian for the charged lepton sector is obtained following a similar procedure, and can be read from [171]. Unlike the Cheng-Sher ansatz, the parameters $\tilde{\chi}_{ff'}$ $(f \neq f')$ are now complex. While the diagonal elements $\tilde{\chi}_{ff}$ are real, the phases in the off-diagonal elements are essentially unconstrained by present low-energy data. These phases modify the pattern of flavour violation (FV) in the Higgs sector. However, because of the Hermiticity of the Yukawa matrices, the three-level CP-properties of $h^0/H^0$ and $A^0$ remain valid i.e. the couplings $h^0(H^0)f\bar{f}$ are pure scalar, while the coupling $A^0 f\bar{f}$ is proportional to $\gamma_5$. Further, in our prescription the FV couplings satisfy some relations, such as: $|\tilde{\chi}_{\mu\tau}| = |\tilde{\chi}_{e\tau}|$ and $|\tilde{\chi}_{sb}| = |\tilde{\chi}_{db}|$, which simplifies the parameter analysis. Henceforth, we denote $|\tilde{\chi}_{ff'}|$ as $\chi_{ff'}$. On the other hand, by considering the effective Lagrangian for the couplings of the charged leptons to the neutral Higgs fields one can also relate our results with the SUSY-induced 2HDM-III. Thus, our result will cover (for specific choices of parameters) the general expectations for the corrections arising in the MSSM.

### 2.5.2 *Bounds on the FV Higgs parameters*

Constraints on the lepton flavour violation (LFV)-Higgs interaction will be obtained by studying LFV transitions, which include the 3-body modes ($l_i \to l_j l_k \bar{l}_k$), radiative decays ($l_i \to l_j + \gamma$), as well as the $\mu - e$ conversion in nuclei. On the other side, constraints on the Higgs boson-quark interaction can be obtained by studying FCNC transitions. In particular, we consider the radiative decay $b \to s\,\gamma$ and the decay $B_s^0 \to \mu^-\mu^+$, which together with LFV bounds derived in [171] constrain the parameter space of 2HDM-III, and determine possible Higgs boson signals that may be detected at future colliders.

#### 2.5.2.1 LFV three-body decays

To evaluate the LFV leptonic couplings, we calculate the decays $l_i \to l_j l_k \bar{l}_k$, including the contribution from the three Higgs bosons ($h^0$, $H^0$ and $A^0$). In particular, for the decay $\tau^- \to \mu^-\mu^+\mu^-$ we obtain the following expression for the branching ratio:

$$\begin{aligned} Br(\tau^- \to \mu^-\mu^+\mu^-) &= \frac{5}{3}\,\frac{\tau_\tau}{2^{12}\,\pi^3}\,\frac{m_\mu^3\,m_\tau^6}{v^4}\left\{\frac{\cos^2(\alpha-\beta)\,\sin^2\alpha}{m_{h^0}^4} + \frac{\sin^2(\alpha-\beta)\,\cos^2\alpha}{m_{H^0}^4}\right. \\ &\left. -2\,\frac{\cos(\alpha-\beta)\,\sin(\alpha-\beta)\,\cos\alpha\,\sin\alpha}{m_{h^0}^2\,m_{H^0}^2} + \frac{\sin^2\beta}{m_{A^0}^4}\right\}\frac{\chi_{\mu\tau}^2}{\cos^4\beta} \qquad (2.113) \end{aligned}$$

here $\tau_\tau$ corresponds to the life time of the $\tau$ lepton (we have also assumed $\chi_{\mu\mu} \ll 1$). Using $Br(\tau^- \to \mu^-\mu^+\mu^-) < 1.9 \times 10^{-7}$ [172], we get an upper bound on $\chi_{\mu\tau}$ ($(\chi_{\mu\tau})_{u.b.}^{\tau\to 3\mu}$) (see Table 2.1).

#### 2.5.2.2 Radiative decay $\mu \to e\gamma$

The B.R. of $\mu^+ \to e^+\gamma$ at one loop level is given by [173]

$$\begin{aligned} Br(\mu^+ \to e^+\gamma) &= \frac{\alpha_{em}\tau_\mu m_e m_\mu^4 m_\tau^4}{2^{12}\pi^4 v^4\cos^4\beta}\,\chi_{\mu\tau}^2\,\chi_{e\tau}^2\left\{\frac{\cos^4(\alpha-\beta)}{m_{h^0}^4}\left|\ln\frac{m_3^2}{m_{h^0}^2}+\frac{3}{2}\right|^2\right. \\ &+2\frac{\cos^2(\alpha-\beta)\sin^2(\alpha-\beta)}{m_{h^0}^2 m_{H^0}^2}\left|\ln\frac{m_3^2}{m_{h^0}^2}+\frac{3}{2}\right|\left|\ln\frac{m_3^2}{m_{H^0}^2}+\frac{3}{2}\right| \\ &\left.+\frac{\sin^4(\alpha-\beta)}{m_{H^0}^4}\left|\ln\frac{m_3^2}{m_{H^0}^2}+\frac{3}{2}\right|^2+\frac{1}{m_{A^0}^4}\left|\ln\frac{m_3^2}{m_{A^0}^2}+\frac{3}{2}\right|^2\right\}. \qquad (2.114) \end{aligned}$$

We make use of $Br(\mu^+ \to e^+\gamma) < 1.2 \times 10^{-11}$ [174] to constrain $\chi_{\mu\tau}(= \chi_{e\tau})$ (see Table 2.1).

2.5.2.3 Radiative decay $\tau \to \mu\gamma$

The B.R. of $\tau \to \mu\,\gamma$ at one loop level (assuming $\chi_{\tau\tau} \ll 1$) is given by [173]

$$\begin{aligned} Br(\tau \to \mu\,\gamma) &= \frac{3}{5}\,\frac{\alpha_{em}\, m_\mu\, m_\tau^3}{16\,\pi\,\cos^4\beta}\,\chi_{\mu\tau}^2 \left\{ \frac{\sin^2\alpha\,\cos^2(\alpha-\beta)}{m_{h^0}^4} \left|\ln\frac{m_\tau^2}{m_{h^0}^2}+\frac{3}{2}\right|^2 \right. \\ &+ \frac{\cos^2\alpha\,\cos^2(\alpha-\beta)+\sin^2\alpha\,\sin^2(\alpha-\beta)}{m_{h^0}^2\, m_{H^0}^2} \left|\ln\frac{m_\tau^2}{m_{h^0}^2}+\frac{3}{2}\right| \left|\ln\frac{m_\tau^2}{m_{H^0}^2}+\frac{3}{2}\right| \\ &+ \left. \frac{\cos^2\alpha\,\sin^2(\alpha-\beta)}{m_{H^0}^4} \left|\ln\frac{m_\tau^2}{m_{H^0}^2}+\frac{3}{2}\right|^2 + \frac{\sin^2\beta}{m_{A^0}^4} \left|\ln\frac{m_\tau^2}{m_{A^0}^2}+\frac{3}{2}\right|^2 \right\}. \end{aligned} \tag{2.115}$$

We constrain $\chi_{\mu\tau}$ by using $Br(\tau \to \mu + \gamma) < 3.1 \times 10^{-7}$ [175, 176] (see Table 2.1).

2.5.2.4 $\mu - e$ conversion

The formulas of the conversion branching ratios for the LFV muon electron process in nuclei at large $\tan\beta$, in the aluminum and lead targets, are approximately given by

$$Br(\mu^- Al(Pb) \to e^- Al(Pb)) \simeq 0.18(2.5) \times 10^{-3}\,\frac{m_e\, m_\mu^6\, m_p^2\, \tan^6\beta\,\cos^2\beta}{2\,v^4\, m_{H^0}^4\,\omega_{capt}}\,\chi_{e\mu}^2, \tag{2.116}$$

where $\omega_{capt}$ is the rate for muon capture in the nuclei [177]. $\omega_{capt} = 0.7054 \times 10^6\, s^{-1}$ and $\omega_{capt} = 13.45 \times 10^6\, s^{-1}$ in the $Al$ and the $Pb$ nuclei, respectively [178]. We get an upper bound on $\chi_{e\mu}$ ($(\chi_{e\mu})_{u.\,b.}^{\mu\mathcal{N}\to e\mathcal{N}}$) for $Al$ and $Pb$ (see Table 2.1), by using $Br(\mu^-\mathcal{N} \to e^-\mathcal{N}) < 6.1 \times 10^{-13}$ [179].

2.5.2.5 Radiative decay $b \to s\,\gamma$

We will make an estimation of the contribution due to the FV $ff'\phi^0$ couplings to the standard model branching ratio of $b \to s\,\gamma$ as follows

$$\Delta Br(b \to s\,\gamma) = \Delta\Gamma(b \to s\,\gamma) \times (\sum_{l=e,\mu,\tau} \Gamma(b \to c\,l\,\bar{\nu}_l))^{-1} \tag{2.117}$$

Such contribution to the branching ratio of $b \to s\,\gamma$ at one loop level is then given by [173]

$$\Delta Br(b \to s\,\gamma) = \frac{\alpha_{em}\, m_s\, m_b^3\, \cos^2(\alpha-\beta)}{16\,\pi\, m_{h^0}^4\, |V_{cb}|^2\, \cos^4\beta}\,\chi_{sb}^2 \left| -\sin\alpha + \frac{\cos(\alpha-\beta)}{\sqrt{2}}\,\tilde{\chi}_{bb} \right|^2 \left|\ln\frac{m_b^2}{m_{h^0}^2}+\frac{3}{2}\right|^2. \tag{2.118}$$

We make use of the good agreement between the experimental value for $Br(b \to s\,\gamma) = (3.3 \pm 0.4) \times 10^{-4}$ and the theoretical value obtained for $Br(b \to s\,\gamma) = (3.29 \pm 0.33) \times 10^{-4}$ in the context of the SM [174] to constrain any new contribution to $Br(b \to s\,\gamma)$, namely $\Delta Br(b \to s\,\gamma) \leq 10^{-5}$, and hence to bound $\chi_{sb}(= \chi_{db})$ (see Table 2.1).

2.5.2.6 $B_s^0 \to \mu^-\mu^+$ decay

The width of the decay $B_s^0 \to \mu^-\mu^+$ at the tree level is given as [180]

$$\Gamma(B_s^0 \to \mu^-\mu^+) = \frac{G_F^2\,\eta_{QCD}^2\, m_B^3\, f_B^2\, m_s\, m_b\, m_\mu^2\, \cos^2(\alpha-\beta)}{128\,\pi\, m_{h^0}^4\, \cos^4\beta}\,\chi_{sb}^2 \left| -\sin\alpha + \frac{\cos(\alpha-\beta)}{\sqrt{2}}\,\tilde{\chi}_{\mu\mu} \right|^2, \tag{2.119}$$

where $G_F = 1.16639 \times 10^{-5}\, GeV^{-2}$, $\eta_{QCD} \approx 1.5$, $m_B \simeq 5\, GeV$, and $f_B = 180\, MeV$. We make use of $\Gamma(B_s^0 \to \mu^-\mu^+) < 8.7 \times 10^{-19}\, GeV$ [180, 181] to constrain $\chi_{sb}(= \chi_{db})$ (see Table 2.1).

Table 2.1: Upper bounds on $\chi_{\mu\tau}$, $\chi_{e\mu}$ and $\chi_{sb}$ as functions of $\tan\beta$, for $\alpha = \beta$, $\alpha = \beta - \pi/4$, $\alpha = \beta - \pi/3$, taking $m_{h^0} = 120$ GeV, $m_{H^0} = 300$ GeV and $m_{A^0} = 300$ GeV and $\chi_{\mu\mu} = 0$ $\chi_{\tau\tau} = 0$. Upper bound $(\chi_{e\mu})^{\mu\mathcal{N}\to e\mathcal{N}}_{u.\,b.}$ as a function of $\tan\beta$ for $Al$, $Pb$ and assuming $m_{H^0} = 300$ GeV.

| upper bound | $\beta - \alpha$ | $\tan\beta = 10$ | $\tan\beta = 20$ | $\tan\beta = 30$ | $\tan\beta = 40$ | $\tan\beta = 50$ |
|---|---|---|---|---|---|---|
| $(\chi_{\mu\tau})^{\tau\to 3\mu}_{u.\,b.}$ | 0 | 8.1 | 2.1 | $9.0 \times 10^{-1}$ | $5.1 \times 10^{-1}$ | $3.3 \times 10^{-1}$ |
| | $\pi/4$ | $1.6 \times 10^{1}$ | 3.8 | 1.7 | $9.2 \times 10^{-1}$ | $5.9 \times 10^{-1}$ |
| | $\pi/3$ | $2.5 \times 10^{1}$ | 5.9 | 2.6 | 1.5 | $9.2 \times 10^{-1}$ |
| $(\chi_{\mu\tau})^{\mu\to e\gamma}_{u.\,b.}$ | 0 | $1.5 \times 10^{-1}$ | $7.4 \times 10^{-2}$ | $5.0 \times 10^{-2}$ | $3.7 \times 10^{-2}$ | $3.0 \times 10^{-2}$ |
| | $\pi/4$ | $1.9 \times 10^{-1}$ | $9.4 \times 10^{-2}$ | $6.3 \times 10^{-2}$ | $4.7 \times 10^{-2}$ | $3.8 \times 10^{-2}$ |
| | $\pi/3$ | $2.2 \times 10^{-1}$ | $1.1 \times 10^{-1}$ | $7.4 \times 10^{-2}$ | $5.6 \times 10^{-2}$ | $4.5 \times 10^{-2}$ |
| $(\chi_{\mu\tau})^{\tau\to\mu\gamma}_{u.\,b.}$ | 0 | 1.6 | $4.0 \times 10^{-1}$ | $1.8 \times 10^{-1}$ | $9.9 \times 10^{-2}$ | $6.4 \times 10^{-2}$ |
| | $\pi/4$ | 2.7 | $6.6 \times 10^{-1}$ | $2.9 \times 10^{-1}$ | $1.7 \times 10^{-1}$ | $1.1 \times 10^{-1}$ |
| | $\pi/3$ | 3.8 | $9.3 \times 10^{-1}$ | $4.1 \times 10^{-1}$ | $2.3 \times 10^{-1}$ | $1.5 \times 10^{-1}$ |
| $(\chi_{e\mu})^{\mu Al\to eAl}_{u.\,b.}$ | | $1.2 \times 10^{-1}$ | $3.1 \times 10^{-2}$ | $1.4 \times 10^{-2}$ | $7.6 \times 10^{-3}$ | $4.9 \times 10^{-3}$ |
| $(\chi_{e\mu})^{\mu Pb\to ePb}_{u.\,b.}$ | | $1.4 \times 10^{-1}$ | $3.6 \times 10^{-2}$ | $1.6 \times 10^{-2}$ | $8.9 \times 10^{-3}$ | $5.7 \times 10^{-3}$ |
| $(\chi_{sb})^{b\to s\gamma}_{u.\,b.}$ | 0 | $8.1 \times 10^{-2}$ | $2.1 \times 10^{-2}$ | $9.1 \times 10^{-3}$ | $5.1 \times 10^{-3}$ | $3.3 \times 10^{-3}$ |
| | $\pi/4$ | $1.8 \times 10^{-1}$ | $4.3 \times 10^{-2}$ | $1.9 \times 10^{-2}$ | $1.1 \times 10^{-2}$ | $6.7 \times 10^{-3}$ |
| | $\pi/3$ | $3.9 \times 10^{-1}$ | $8.9 \times 10^{-2}$ | $3.9 \times 10^{-2}$ | $2.2 \times 10^{-2}$ | $1.4 \times 10^{-2}$ |
| $(\chi_{sb})^{B^0_s\to\mu\mu}_{u.\,b.}$ | 0 | 1.1 | $2.7 \times 10^{-1}$ | $1.2 \times 10^{-1}$ | $6.6 \times 10^{-2}$ | $4.2 \times 10^{-2}$ |
| | $\pi/4$ | 2.4 | $5.6 \times 10^{-1}$ | $2.5 \times 10^{-1}$ | $1.4 \times 10^{-1}$ | $8.6 \times 10^{-2}$ |
| | $\pi/3$ | 5.1 | 1.2 | $5.0 \times 10^{-1}$ | $2.8 \times 10^{-1}$ | $1.8 \times 10^{-1}$ |

### *2.5.3 Higgs boson decays in the 2HDM-III*

One of the distinctive characteristic of the SM Higgs boson is the fact that its coupling to other particles is proportional to the mass of that particle, which in turn determines the search strategies proposed so far to detect it at future colliders. In particular, the decay pattern of the Higgs boson is dominated by the heaviest particle allowed to appear in its decay products. When one considers extensions of the SM it is important to study possible deviations from the SM decay pattern as it could provide a method to discriminate among the different models [182, 183]. Within the context of the 2HDM-III, not only modification of the Higgs boson couplings are predicted, but also the appearance of new channels with FV, both in the quark and leptonic sectors [129, 184, 185].

To explore the characteristics of Higgs boson decays in the 2HDM-III, we will focus on the lightest CP-even state ($h^0$), which could be detected first at LHC. The light Higgs boson-fermion couplings are given by Eq. (2.112), where we have separated the SM from the corrections that appear in a 2HDM-III. In fact, we have also separated the factors that arise in the 2HDM-III too. We notice that the correction to the SM result, depends on $\tan\beta$, $\alpha$ (the mixing angle in the neutral CP-even Higgs sector) and the factors $\tilde{\chi}_{ff'}$ that induce FCNC transitions (for $f \neq f'$) and further corrections to the SM vertex. In our analysis, we will include the decay widths for all the modes that are allowed kinematically for a Higgs boson with a mass in the range $80\,GeV < m_{h^0} < 160\,GeV$. Namely, we study the branching ratios for the decays $h^0 \to b\bar{b}$, $c\bar{c}$, $\tau\bar{\tau}$, $\mu\bar{\mu}$ and the FV $h^0 \to b\bar{s}(s\bar{b})$, $\tau\bar{\mu}(\mu\bar{\tau})$, as well as the decays into pairs of gauge bosons with one real an the other one virtual, i.e. $h^0 \to WW^*, ZZ^*$ [8, 167]. Overall, our results show that the usual search strategies to look for the SM Higgs boson in this mass range, may need to be modified in order to cover the full parameter space of the 2HDM-III (see Fig. 2.3).

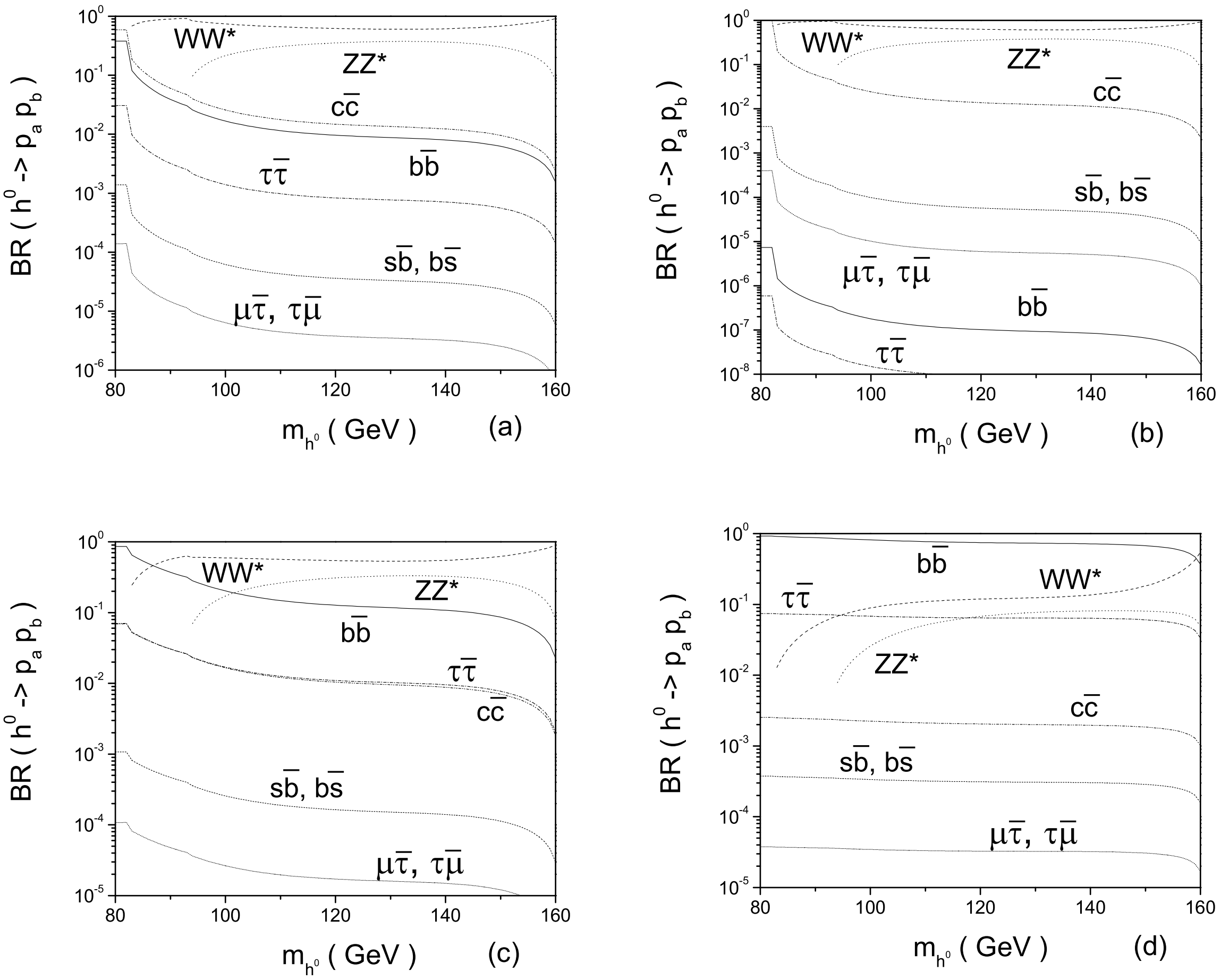

Fig. 2.3: B.R. for all the relevant decay modes that are allowed kinematically for $80\,GeV < m_{h^0} < 160\,GeV$; taking $\alpha = \beta - 3\pi/8$ and assuming $\tilde{\chi}_{ff'} = 0.1$ for $f = f'$ and $f \neq f'$. For: (a) $\tan\beta = 2$; (b) $\tan\beta = 2.61$; (c) $\tan\beta = 5$; (d) $\tan\beta = 15$.

### *2.5.4 Conclusions*

We have studied in this paper the $ff'\phi^0$ couplings that arise in the 2HDM-III, using a Hermitian four-texture form for the fermionic Yukawa matrix. Because of this, although the $ff'\phi^0$ couplings are complex, the three-level CP-properties of $h^0, H^0$ (even) and $A^0$ (odd) remain valid.

We have derived bounds on the parameters of the model, using current experimental bounds on LFV and FCNC transitions. One can say that the present bounds on the couplings $\tilde{\chi}_{ff'}$'s still allow the possibility to study interesting direct FV Higgs boson signals at future colliders.

In particular, the LFV couplings of the neutral Higgs bosons, can lead to new discovery signatures of the Higgs boson itself. For instance, the branching fraction for $h^0 \to \tau\bar{\mu}(\bar{\tau}\mu)$ can be as large as $10^{-5}$, while $Br(h \to b\bar{s}(\bar{b}s))$ is also about $10^{-4}$. These LFV Higgs modes complement the modes $B^0 \to \mu\mu$, $\tau \to 3\mu$, $\tau \to \mu\gamma$ and $\mu \to e\gamma$, as probes of FV in the 2HDM-III, which could provide key insights into the form of the Yukawa mass matrix sector.

## 2.6 Electroweak baryogenesis and quantum corrections to the Higgs potential

*Shinya Kanemura, Yasuhiro Okada and Eibun Senaha*

The connection between cosmology and particle physics is important to understand what the Universe is made of. The baryon asymmetric Universe observed today is one of the outstanding problems in cosmology. The asymmetry is characterized by the ratio of the baryon number density to the entropy density, $n_B/s \sim 10^{-10}$ [186], which remains constant during the expansion of the Universe if there is neither baryon number violation nor entropy production.

In order to construct such baryon asymmetry from the initially baryon symmetric Universe, three ingredients are required [187]: (a) baryon number violation, (b) $C$ and $CP$ violation, and (c) departure from thermal equilibrium. In the electroweak theories, these conditions can in principle be satisfied (*electroweak baryogenesis*). The condition (a) is fulfilled by the sphaleron process in Standard Model at high temperature. The sphaleron is an unstable classical solution of the $SU(2)$ gauge-Higgs system which corresponds to a saddle point connecting different topological vacua. Frequent baryon number violation processes occur near and above the critical temperature by the transition associated with a change of the topological number, which is called the sphaleron process, although it is completely negligible at zero temperature. On the other hand the Standard Model cannot satisfy the other two conditions under the current experimental data. One is that the electroweak phase transition is not first order for experimental lower bounds of the Higgs boson mass, $m_h > 114$ GeV [74], so that the condition (c) cannot be fulfilled. The other difficulty is that the magnitude of the $CP$ violation which is originated from the Kobayashi-Maskawa matrix is too small to generate the sufficient baryon asymmetry during the phase transition. Therefore, the extension of the Standard Model Higgs sector and the additional sources of the $CP$ violation are required. There are many attempts to explain the baryon asymmetry in the extension of the Standard Model. For reviews on electroweak baryogenesis, see Refs. [188–193].

Here, we study electroweak baryogenesis in the two Higgs doublet model [194–203] and the minimal supersymmetric standard model [204–221] focusing on its connections to collider phenomenology. In particular, we discuss relationship between the strength of the first-order electroweak phase transition and the quantum corrections to the trilinear coupling of the lightest Higgs boson [222]. Similar discussions on the Higgs self-coupling in the electroweak baryogenesis scenario can be found in Refs. [223, 224].

First we consider the two Higgs doublet model with the softly-broken discrete symmetry. The Higgs potential at the tree-level is given by Eq. (2.1) with $\lambda_6 = \lambda_7 = 0$. Though $m_{12}^2$ and $\lambda_5$ can be complex, one of the two becomes real by the redefinition of the either Higgs field. As mentioned above, the $CP$ violation plays a crucial role in the generation of the baryon asymmetry. In particular, the difference between the $CP$ violating phase in the symmetric phase and that in the broken phase at finite temperature gives a significant effect on the total amount of the baryon asymmetry. In order to calculate the magnitude of such $CP$ violating phases, the equation of motion for the Higgs bubble wall has to be solved at the critical temperature. In the previous studies, it was found that there is a solution in which the $CP$ violation can enhance only during the phase transition while it can become small at zero temperature enough to escape the experimental constraints of the electric dipole moment [201–203]. Here, we assume such a scenario so that we neglect the $CP$ violating phase as the first approximation. Furthermore, to simplify our analysis we consider the phase transition in the direction of $\langle\Phi_1\rangle = \langle\Phi_2\rangle = (0\ \varphi)^T/2$, which corresponds to $m_1 = m_2$, $\lambda_1 = \lambda_2$, in other words, $\sin(\alpha-\beta) = -1$ and $\tan\beta = 1$ [199–201].

The one-loop contributions to the effective potentials at zero and finite temperatures [225] are respectively given by

$$\mathcal{V}_1(\varphi) = n_i \frac{m_i^4(\varphi)}{64\pi^2}\left(\log\frac{m_i^2(\varphi)}{Q^2} - \frac{3}{2}\right), \quad \mathcal{V}_1(\varphi, T) = \frac{T^4}{2\pi^2}\Big[\sum_{j=\text{bosons}} n_j I_B(a_j^2) + n_t I_F(a_t^2)\Big], \quad (2.120)$$

with

$$I_{B,F}(a_i^2) = \int_0^\infty dx\, x^2 \log\left(1 \mp e^{-\sqrt{x^2+a_i^2}}\right), \quad a_i(\varphi) = \frac{m_i(\varphi)}{T}, \tag{2.121}$$

where $\mathcal{V}_1(\varphi)$ is regularized by using the $\overline{\mathrm{DR}}$-scheme, $Q$ is a renormalization scale, $m_i(\varphi)$ is the field dependent mass of the particle $i$, and $n_i$ is the degree of the freedom of $i$; i.e., $n_W = 6$, $n_Z = 3$ for gauge bosons ($W^\pm$, $Z$), $n_t = -12$ for the top quark ($t$) and $n_h = n_H = n_A = 1$, $n_{H^\pm} = 2$ for the five physical Higgs bosons ($h, H, A, H^\pm$).

Qualitative features of the phase transition can be understood from the effective potential (2.120) by the following high temperature expansion. When $m_\Phi^2 \gg m_h^2$, $M^2$ ($\Phi \equiv H, A, H^\pm$, $M^2 \equiv v^2\eta$), the field dependent masses of the heavy Higgs bosons can be written as $m_\Phi^2(\varphi) \simeq m_\Phi^2\varphi^2/v^2$. At high temperatures, the Higgs potential can be expanded in powers of $\varphi$ [226, 227].

$$\mathcal{V}_{\text{eff}}(\varphi, T) \simeq D(T^2 - T_0^2)\varphi^2 - ET|\varphi|^3 + \frac{\lambda_T}{4}\varphi^4 + \cdots, \tag{2.122}$$

with

$$T_0^2 = \frac{1}{D}\left(\frac{1}{4}m_h^2 - 2Bv^2\right), \tag{2.123}$$

$$B = \frac{1}{64\pi^2 v^4}\left(6m_W^4 + 3m_Z^4 - 12m_t^4 + m_H^4 + m_A^4 + 2m_{H^\pm}^4\right), \tag{2.124}$$

$$D = \frac{1}{24v^2}\left(6m_W^2 + 3m_Z^2 + 6m_t^2 + m_H^2 + m_A^2 + 2m_{H^\pm}^2\right), \tag{2.125}$$

$$E = \frac{1}{12\pi v^3}\left(6m_W^3 + 3m_Z^3 + m_H^3 + m_A^3 + 2m_{H^\pm}^3\right), \tag{2.126}$$

$$\lambda_T = \frac{m_h^2}{2v^2}\left[1 - \frac{1}{8\pi^2 v^2 m_h^2}\left\{6m_W^4 \log\frac{m_W^2}{\alpha_B T^2} + 3m_Z^4 \log\frac{m_Z^2}{\alpha_B T^2} - 12m_t^4 \log\frac{m_t^2}{\alpha_F T^2} + m_H^4 \log\frac{m_H^2}{\alpha_B T^2} + m_A^4 \log\frac{m_A^2}{\alpha_B T^2} + 2m_{H^\pm}^4 \log\frac{m_{H^\pm}^2}{\alpha_B T^2}\right\}\right], \tag{2.127}$$

where $\log\alpha_B = 2\log 4\pi - 2\gamma_E \simeq 3.91$, $\log\alpha_F = 2\log\pi - 2\gamma_E \simeq 1.14$, and $\gamma_E$ is the Euler constant. The first order phase transition is possible due to the appearance of the cubic term which originates from the bosonic loops at finite temperature. From Eq. (2.122), the critical temperature $T_c$ is expressed by

$$T_c^2 = \frac{T_0^2}{1 - E^2/(\lambda_{T_c} D)}. \tag{2.128}$$

At $T_c$, the effective potential $\mathcal{V}_{\text{eff}}$ has two degenerate minima at

$$\varphi = 0, \qquad \varphi_c = \frac{2ET_c}{\lambda_{T_c}}. \tag{2.129}$$

In order not to wash out the created baryon number density after the electroweak phase transition, we have to require that the sphaleron process should be sufficiently suppressed. The most reliable condition has been obtained from the lattice simulation study [228, 229]. It is expressed as

$$\frac{\varphi_c}{T_c} = \frac{2E}{\lambda_{T_c}} > 1. \tag{2.130}$$

Due to the contributions of the heavy Higgs bosons in the loop, the first order phase transition can be strong enough to satisfy Eq. (2.130). The high temperature expansion makes it easy to see the phase transition analytically. However, it breaks down when the masses of the particles in loops become larger than $T_c$. In the following, we therefore calculate $T_c$ and $\varphi_c$ numerically.

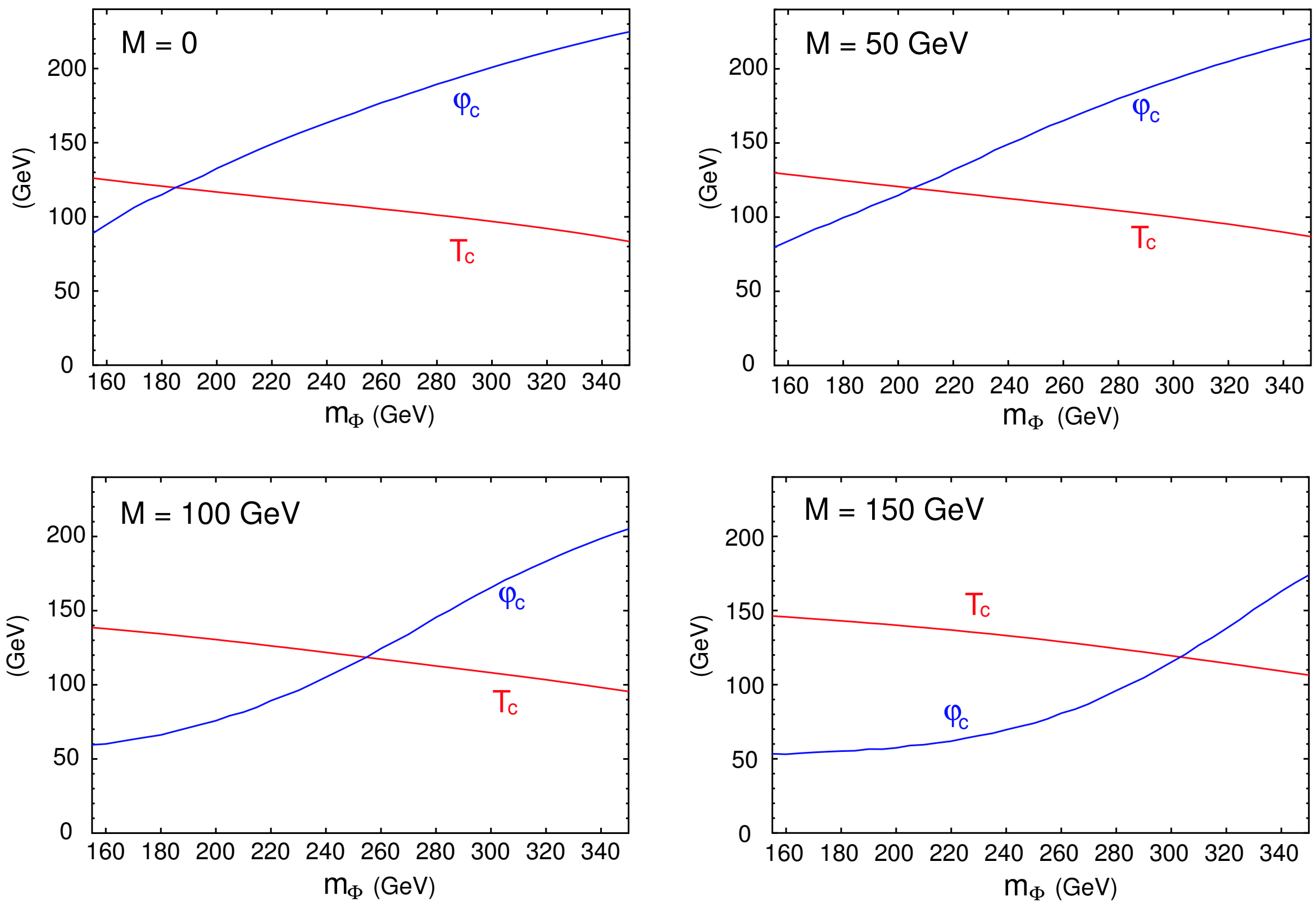

Fig. 2.4: The Higgs vacuum expectation value $\varphi_c$ at the critical temperature $T_c$ as a function of the heavy Higgs boson mass $m_\Phi$ ($m_\Phi = m_H = m_A = m_{H^\pm}$) for $M = 0$, 50, 100 and 150 GeV. Other parameters are fixed as $\sin(\alpha - \beta) = -1$, $\tan\beta = 1$ and $m_h = 120$ GeV.

In order to see phenomenological consequences of our scenario for successful electroweak baryogenesis, we study the trilinear coupling of the lightest Higgs boson (the $hhh$ coupling) at the zero temperature in the parameter region where the phase transition is strongly first order. The leading contribution of the heavy Higgs bosons and the top quark to the $hhh$ coupling can be extracted from the one-loop calculation by [113, 230]

$$\lambda^{\text{eff}}_{hhh}(\text{2HDM}) \simeq \frac{3m_h^2}{v}\left[1 + \frac{m_H^4}{12\pi^2 m_h^2 v^2}\left(1 - \frac{M^2}{m_H^2}\right)^3 + \frac{m_A^4}{12\pi^2 m_h^2 v^2}\left(1 - \frac{M^2}{m_A^2}\right)^3 + \frac{m_{H^\pm}^4}{6\pi^2 m_h^2 v^2}\left(1 - \frac{M^2}{m_{H^\pm}^2}\right)^3 - \frac{m_t^4}{\pi^2 m_h^2 v^2}\right]. \quad (2.131)$$

It is easily seen that the effects of the heavy Higgs boson loops are enhanced by $m_\Phi^4$ ($\Phi = H, A, H^\pm$) when $M^2$ is zero. These effects do not decouple even in the large mass limit $m_\Phi \to \infty$ and yield the large deviation of the $hhh$ coupling from the Standard Model prediction. In this case, $m_\Phi$ is bounded from above by perturbative unitarity ($m_\Phi < 550$ GeV) [31,32,34,231]. We note that when such nondecoupling loop effects due to the extra heavy Higgs bosons are large on the $hhh$ coupling, the coefficient $E$ of the cubic term in Eq. (2.122) becomes correspondingly large. Therefore there is a strong correlation between the large quantum correction to the $hhh$ coupling and successful electroweak baryogenesis.

We calculate the effective potential (2.120) varying the temperature $T$ and determine the critical temperature $T_c$ of the first order phase transition and the expectation value $\varphi_c$ at $T_c$. In the plots of Fig. 2.4, $T_c$ and $\varphi_c$ are shown as a function of the mass of the heavy Higgs boson $m_\Phi$ for $M = 0,\ 50,\ 100$ and 150 GeV. We take $m_h = 120$ GeV. For the heavy Higgs boson mass, we assume $m_H = m_A = m_{H^\pm} (\equiv m_\Phi)$ to avoid the constraint on the $\rho$ parameter from the LEP precision data [232]. We also take into account the ring summation for the contribution of the Higgs bosons to the effective potential at finite temperature to improve our calculation [225, 233–238]. In the case of $M = 0$, it is found that $\varphi_c = T_c \simeq 120$ GeV at $m_\Phi \simeq 185$ GeV, and the condition (2.130) is satisfied for $m_\Phi > 185$ GeV. One can also find that the condition (2.130) can still be satisfied for $M = 150$ GeV, if the masses of the heavy Higgs bosons are greater than about 300 GeV.

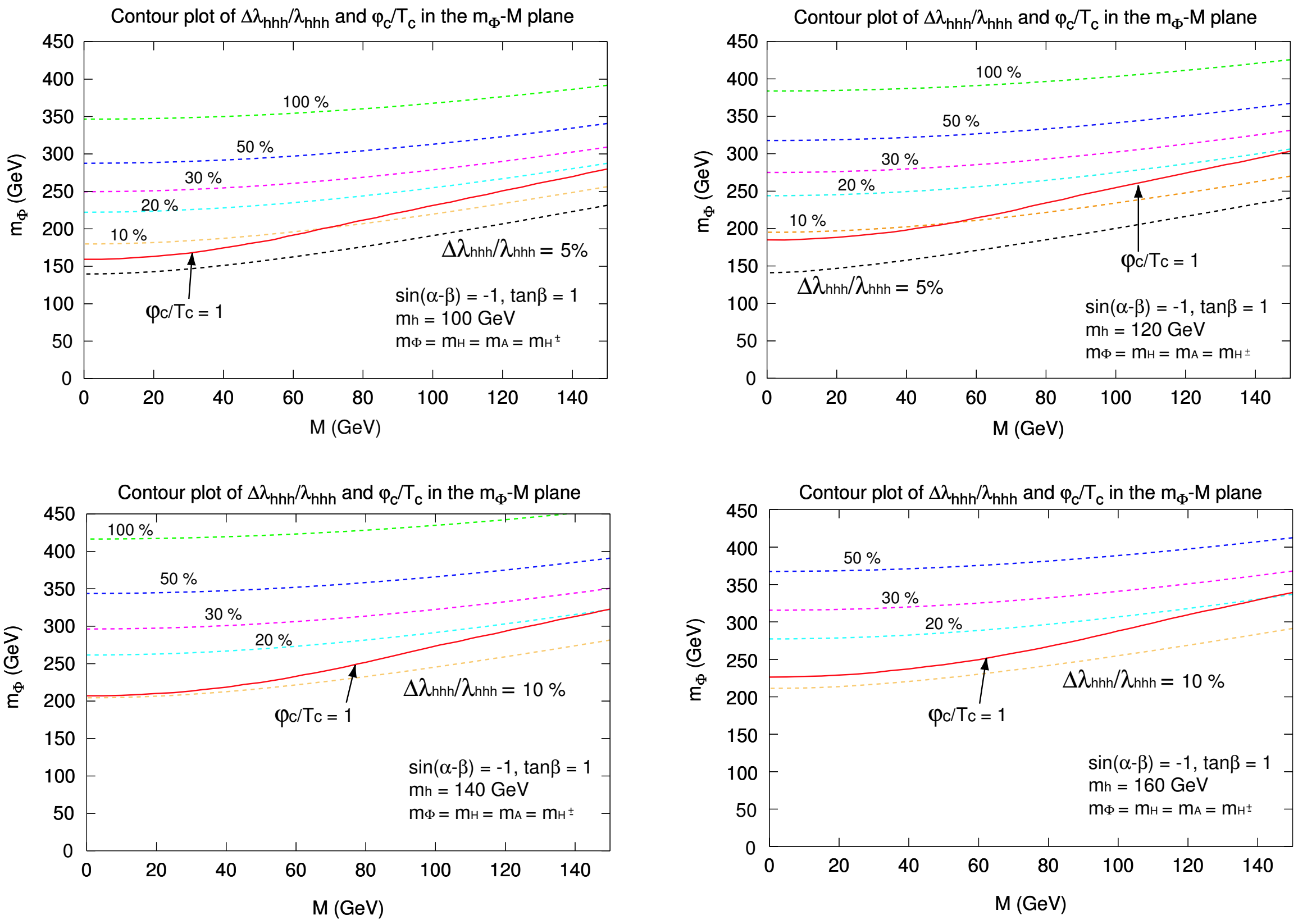

Fig. 2.5: Contours of the radiative correction of the triple Higgs boson coupling constant overlaid with the line $\varphi_c/T_c = 1$ in the $m_\Phi$-$M$ plane for $m_h$=100, 120, 140 and 160 GeV. Other parameters are the same as those in Fig. 2.4. The above the critical line, the phase transition is strong enough for the successful electroweak baryogenesis scenario.

In Fig. 2.5, we show the parameter region where the necessary condition of electroweak baryogenesis in Eq. (2.130) is satisfied in the $m_\Phi$-$M$ plane for $m_h = 100,\ 120,\ 140$ and 160 GeV. For $m_h = 120$ GeV, we can see that the phase transition becomes strong enough for successful baryogenesis when the masses of the heavy Higgs bosons are larger than about 200 GeV. For the larger values of $M$ or $m_h$, the greater $m_\Phi$ are required to satisfy the condition (2.130). In this figure we also plot the contour of the magnitude of the deviation in the $hhh$ coupling from the Standard Model value. We define the deviation $\Delta\lambda_{hhh}^{\rm 2HDM}/\lambda_{hhh}^{\rm eff}({\rm SM})$ by $\Delta\lambda_{hhh}^{\rm 2HDM} \equiv \lambda_{hhh}^{\rm eff}({\rm 2HDM}) - \lambda_{\rm hhh}^{\rm eff}({\rm SM})$. We calculated the deviation from the full one-loop results, which give a better approximation than the formula given in Eq. (2.131) [113, 230]. We can easily see that the magnitude of the deviation is significant ($> 10\%$) in the parameter region where electroweak baryogenesis is possible. Such magnitude of the deviation can be detected at future collider experiments [142, 239–241].

Next we discuss a scenario of electroweak baryogenesis in the minimal supersymmetric standard model. The strong first order phase transition can be induced by the loop effect of the light stop in the finite temperature effective potential [204]. We examine the loop effect of the light stop on the $hhh$ coupling in this scenario. In the following, we only consider the finite and zero temperature effective potentials using high temperature expansion to understand the qualitative feature. As we have done in the case of the two Higgs doublet model, we consider the relationship between the magnitude of the phase transition and the deviation of the $hhh$ coupling from the Standard Model value. The combined result is approximately expressed as

$$\frac{\Delta\lambda_{hhh}(\mathrm{MSSM})}{\lambda_{hhh}(\mathrm{SM})} \simeq \frac{2v^4}{m_t^2 m_h^2}(\Delta E_{\tilde{t}_1})^2, \tag{2.132}$$

where $m_h$ is the one-loop renormalized mass of the lightest Higgs boson and $\Delta E_{\tilde{t}_1}$ is the contribution of the light stop loop to the cubic term in the finite temperature effective potential. From the condition (2.130), the deviation in the $hhh$ coupling from the Standard Model value is estimated to be $\sim 6\%$ for $m_h = 120$ GeV. In the minimal supersymmetric standard model, the condition of the sphaleron decoupling also leads to the large deviation of the $hhh$ coupling from the Standard Model prediction at zero temperature.

In summary, we have discussed electroweak baryogenesis with special emphasis on its connections to the collider phenomenology. If the electroweak phase transition is strong enough for electroweak baryogenesis, the triple Higgs boson coupling can deviate from the Standard Model value. The magnitude of the deviation can be larger than 10% level in the two Higgs doublet model. Such magnitude of the deviation can be detected at future colliders.

## 2.7 Neutral Higgs bosons with (in)definite CP: decay distributions for $\tau^+\tau^-$ and $t\bar{t}$ final states

*Werner Bernreuther, Arnd Brandenburg and Jörg Ziethe*

This contribution deals with the question of how to determine the parity, respectively the CP property of a neutral Higgs boson. While the Standard Model Higgs boson is parity-even, SM extensions predict also parity-odd state(s) or, if the (effective) Higgs potential violates CP, states of undefined CP parity with Yukawa couplings both to scalar and pseudoscalar quark and lepton currents. Higgs sector CP violation (CPV) is, especially in view of its potentially enormous impact on the physics of the early universe, a fascinating speculation which can be investigated at the upcoming generation of colliders in several ways. The decays $h \to \tau^-\tau^+$ and/or $h \to t\bar{t}$ are particularly suited, provided that sufficiently large event numbers are available. The analysis presented here is based on the proposals and investigations of [99, 103] for the tau and of [99, 103, 242–245] for the top channel. Other investigations include [101, 246–250].

The following applies to any neutral Higgs boson $h_j$ with *flavor-diagonal* couplings to quarks and leptons $f$ (with mass $m_f$)

$$\mathcal{L}_Y = -(\sqrt{2}G_F)^{1/2}\sum_{j,f} m_f(a_{jf}\bar{f}f + b_{jf}\bar{f}i\gamma_5 f)\,h_j\ , \tag{2.133}$$

where $a_{jf}$ and $b_{jf}$ are the reduced scalar and pseudoscalar Yukawa couplings, respectively, which depend on the parameters of the scalar potential and on the type of model. In the SM $a_f = 1$ and $b_f = 0$. In models with two Higgs doublets there are three physical neutral Higgs fields $h_j$ in the mass basis. In the type II models the Yukawa couplings to top quarks and $\tau$ leptons are (see sections 2.1 and 3.1): $a_{jt} = R_{2j}/\sin\beta$, $b_{jt} = -R_{3j}\cot\beta$, $a_{j\tau} = R_{1j}/\cos\beta$, $b_{j\tau} = -R_{3j}\tan\beta$, where $\tan\beta = v_2/v_1$, and $(R_{ij})$ is a $3\times 3$ orthogonal matrix that describes the mixing of the neutral spin-zero states. At the Born level only the CP $= +1$ component of $h_j$ couples to $W^+W^-$ and to $ZZ$. If Higgs sector CP violation (CPV) is negligibly small then the fields $h_j$ describe two scalar states $h, H$ and a pseudoscalar $A$. In the following $\phi$ denotes, as in section 2.1 above, any of these Higgs bosons. We assume that the differences between the mass of the Higgs particle $\phi$ under consideration and the masses of the other neutral Higgs states are larger than the experimental resolution.

### 2.7.1 $\tau$ and top spin observables

The observables discussed below for determining the CP quantum number of a neutral Higgs boson in the decay channels $\phi \to \tau^-\tau^+$ and/or $\phi \to t\bar{t}$ may be applied to any Higgs production process. At the LHC this includes the gluon and gauge boson fusion processes $gg \to \phi$ and $q_i q_j \to \phi q_i' q_j'$, respectively, and associated production of a light Higgs boson, $t\bar{t}\phi$ or $b\bar{b}\phi$ with $\phi \to \tau^-\tau^+$. Likewise they can be used in future Higgs search at an ILC, or in Higgs production with envisaged high energetic muon or photon collisions, $\mu^-\mu^+$, $\gamma\gamma \to \phi \to f\bar{f}$. In the following we consider the semi-inclusive reactions

$$i \to \phi + X \to f(\mathbf{k_f}, \alpha) + \bar{f}(\mathbf{k_{\bar{f}}}, \beta) + X\,, \tag{2.134}$$

where $i$ is some initial state, $f = \tau^-, t$, $\mathbf{k_f}$ and $\mathbf{k_{\bar{f}}} = -\mathbf{k_f}$ are the 3-momenta of $f$ and $\bar{f}$ in the $f\bar{f}$ zero-momentum frame (ZMF), and $\alpha, \beta$ are spin labels. We make use of the fact that, at colliders, polarization and spin correlation effects are both measurable and reliably predictable for tau leptons and top quarks.

Let's assume that experiments at the LHC will discover a neutral boson resonance in the channel $gg \to \phi \to \tau^-\tau^+X$. The spin of $\phi$ may be inferred from the polar angle distribution of the tau leptons. Suppose the outcome of this is that $\phi$ is a spin-zero (Higgs) particle. One would next like to determine its Yukawa coupling(s), and specifically like to know whether $\phi$ is a scalar, a pseudoscalar, or a mixture of both, i.e., a state of undefined CP quantum number. For answering this question several CP-even and -odd observables involving the spins of $f, \bar{f}$ apply, and we emphasize that all of them should be used. It was shown in [99] that the correlation resulting from projecting the spin of $f$ onto the spin of $\bar{f}$,

$$\mathcal{O}_1 = \mathbf{s}_f \cdot \mathbf{s}_{\bar{f}}\,, \tag{2.135}$$

is the best choice for discriminating between a $CP = \pm 1$ state. Here $\mathbf{s}_f, \mathbf{s}_{\bar{f}}$ denote the $f, \bar{f}$ spin operators. This is easy to understand in simple quantum mechanical terms. Consider a reaction $i \to \phi \to f\bar{f}$ where $\phi$ production and decay factorizes. If $\phi$ is a scalar ($J^{PC} = 0^{++}$) then $f\bar{f}$ is in a ${}^3P_0$ state, and an elementary calculation yields $\langle \mathbf{s}_f \cdot \mathbf{s}_{\bar{f}} \rangle = 1/4$. If $\phi$ is a pseudoscalar ($J^{PC} = 0^{-+}$) then $f\bar{f}$ is in a ${}^1S_0$ state and $\langle \mathbf{s}_f \cdot \mathbf{s}_{\bar{f}} \rangle = -3/4$, which is strikingly different from the scalar case. These values do not depend on the mass of $\phi$, provided $m_\phi > 2m_f$. For general couplings (2.133) one gets $\langle \mathcal{O}_1 \rangle = (a_f^2\beta_f^2 - 3b_f^2)/(4a_f^2\beta_f^2 + 4b_f^2)$ [99]. In Fig. 2.6 (left) the expectation value $\langle \mathcal{O}_1 \rangle$ is shown for $\phi \to \tau^-\tau^+X$ as a function of the ratio $r_\tau = b_\tau/(a_\tau + b_\tau)$, taking $a_\tau, b_\tau > 0$ for definiteness, for arbitrary Higgs mass $m_\phi \gtrsim 100$ GeV. The figure applies also to $\phi \to t\bar{t}X$ (with $r_\tau \to r_t$) if $m_\phi$ is markedly above the $t\bar{t}$ threshold. For small $t$ quark velocities $\beta_t$ the resulting plot is distorted, as compared with Fig. 2.6, between the fixed points 1/4 and -3/4. The QED corrections to this observable, respectively the order $\alpha_s$ QCD corrections in the case of $f = t$ are very small [99].

We note in passing that the CP-even spin-spin correlation in the helicity basis, $\langle (\hat{\mathbf{k}}_\mathbf{f} \cdot \mathbf{s}_f)(\hat{\mathbf{k}}_{\bar{\mathbf{f}}} \cdot \mathbf{s}_{\bar{f}}) \rangle$, is insensitive to the CP quantum number of $\phi$ [99].

If $\gamma_{CP}^f \equiv -a_f b_f \neq 0$ the Yukawa interactions of $\phi$ break CP. This leads to CP-violating effects in the reactions (2.134). For an unpolarized initial state $i$ a general kinematic analysis of (2.134) yields the following [103, 243, 245]. If C-violating interactions do not matter in (2.134) then $\mathcal{L}_Y$ (which is C-invariant, but P- and CP-violating) induces two types of CPV effects in the $f\bar{f}$ state: a CP-odd spin-spin correlation and a CP-odd polarization asymmetry which correspond to the observables

$$\mathcal{O}_2 = \hat{\mathbf{k}}_\mathbf{f} \cdot (\mathbf{s}_f \times \mathbf{s}_{\bar{f}})\,, \quad \mathcal{O}_3 = \hat{\mathbf{k}}_\mathbf{f} \cdot (\mathbf{s}_f - \mathbf{s}_{\bar{f}})\,. \tag{2.136}$$

Here $\hat{\mathbf{k}}_\mathbf{f} = \mathbf{k_f}/|\mathbf{k_f}|$ in the $f\bar{f}$ ZMF. (A priori two more terms can appear in the squared matrix element of (2.134). They are obtained by replacing $\hat{\mathbf{k}}_\mathbf{f} \to \hat{\mathbf{p}}$ in (2.136), where $\hat{\mathbf{p}}$ is the direction of one of the colliding beams in $i$. However, for resonant $\phi$ production only the observables (2.136) are of interest.) The CP-odd and T-odd[12] variable $\mathcal{O}_2$ measures a correlation of the spins of the $f$ and $\bar{f}$ transverse to their directions of flight. A non-zero expectation value is generated already at tree level, $\langle \mathcal{O}_2 \rangle = \gamma^f_{CP}\beta_f/(a_f^2\beta_f^2 + b_f^2)$ [103], which can be as large as 0.5 in magnitude! In Fig. 2.6 (right) $\langle \mathcal{O}_2 \rangle$ is shown for $\phi \to \tau^-\tau^+ X$ as a function of $r_\tau$, for Higgs masses $m_\phi \gtrsim$ 100 GeV. The figure applies also to $\phi \to t\bar{t}X$ (with $r_\tau \to r_t$) for sufficiently heavy $\phi$. The QED corrections to this observable, respectively the order $\alpha_s$ QCD corrections in the case of $f = t$, are also very small [99]. The variable $\mathcal{O}_3$ measures an asymmetry in the longitudinal polarization of the $f$ and $\bar{f}$. As it is CP-odd but T-even, a non-zero $\langle \mathcal{O}_3 \rangle$ requires $\gamma^f_{CP} \neq 0$ and a non-zero absorptive part of the respective scattering amplitude. This variable is relevant for heavy Higgs $\to f\bar{f}$, e.g., for $gg \to \phi \to t\bar{t}$ (see below), but not for light Higgs $\to \tau\tau$.

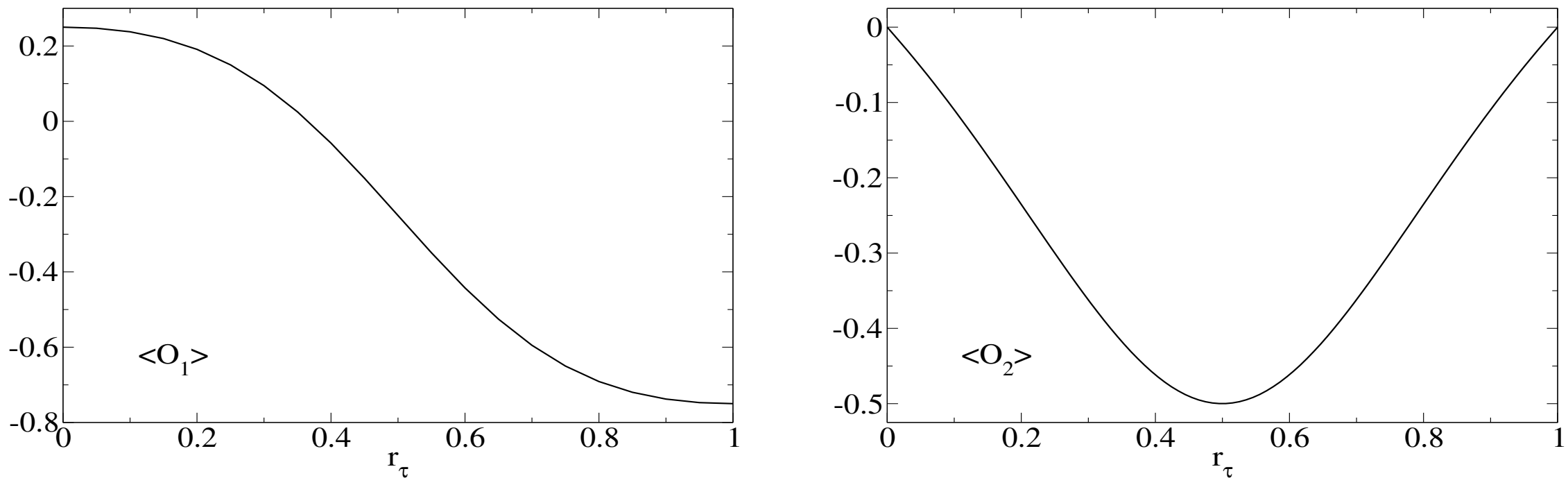

Fig. 2.6: Expectation value of $\langle \mathcal{O}_1 \rangle$ (left) and $\langle \mathcal{O}_2 \rangle$ (right) for $\phi \to \tau^-\tau^+ X$ as a function of $r_\tau$ [99].

If besides (2.133) and QCD also C-violating interactions (e.g. the standard weak interactions) matter for the reactions (2.134) then there can be, in principle, another CPV effect, namely $\langle \hat{\mathbf{n}} \cdot (\mathbf{s}_f - \mathbf{s}_{\bar{f}}) \rangle \neq 0$ [245]. Here $\hat{\mathbf{n}}$ denotes the normal to the $i \to f\bar{f}$ scattering plane. Below we consider the reactions $gg \to \phi \to f\bar{f}$ within QCD. In this case this CPV polarization effect is absent. This holds true also for reactions (2.134) where the production and decay of $\phi$ factorizes.

### *2.7.2 Distributions for the decay products of $\tau^+\tau^-$ and $t\bar{t}$*

The polarization and spin-correlation effects (2.135), (2.136) induced in the $f\bar{f}$ sample lead, through the parity-violating weak decays of the $\tau$ leptons and top quarks, to specific angular distributions and correlations in the respective final state. We consider here

$$i \to \phi + X \to f(\mathbf{k_f}) + \bar{f}(\mathbf{k_{\bar{f}}}) + X \to a(\mathbf{q_1}) + \bar{b}(\mathbf{q_2}) + X\,, \qquad (2.137)$$

where $a, \bar{b}$ denotes a charged particle or a jet from the decays $f \to a + \cdots$, $\bar{f} \to \bar{b} + \cdots$. The 3-momenta of $f$ and $\bar{f}$ in (2.137) refer as above to the $f\bar{f}$ ZMF, while the momenta $\mathbf{q_1}$ and $\mathbf{q_2}$ refer to the $f$ and $\bar{f}$ rest frames, respectively. For $f = \tau, t$ these frames and momenta can be reconstructed using kinematic constraints (c.f., e.g. [115, 251, 252]).
For the tau lepton one may take into account the decay channels $\tau^- \to \pi^-\nu_\tau$, $\rho^-\nu_\tau$, $a_1^-\nu_\tau$, $\ell^-\bar{\nu}_\ell\nu_\tau$, which comprise about 81 % of all tau decays. That is, $B(\tau^-\tau^+ \to a\bar{b}X) \simeq$ 66 % for $a, b = \pi, \rho, a_1, \ell$. Here we need to recall only the $\tau$-spin analyzing power of these particles, that is, the coefficient $c_a$ in the distribution $\Gamma_a^{-1} d\Gamma_a/d\cos\theta = (1 + c_a \cos\theta)/2$ of the decay $\tau^- \to a + \cdots$, where $\cos\theta$ is the angle between the $\tau$ spin vector and the direction of $a$ in the $\tau$ rest frame (c.f., e.g., [252]). They are collected in Table 2.2.

According to the SM the top quark decays into $Wb$ almost 100 % of the time, which leads to the CKM allowed semi- and nonleptonic final states, $t \to b\ell\nu_\ell,\ bq\bar{q}'$, $q\bar{q}'$ = $u\bar{d}$, $c\bar{s}$. Again we need here only the $t$-spin analyzing power $c_a$ of particle/jet $a$ in the decay $t \to a + \cdots$. Table 2.2 contains the values of the $c_a$ at tree level (c.f., for instance, [253]) and to order $\alpha_s$, which were computed for the semi-and non-leptonic decays in [254] and in [255], respectively. For the non-leptonic channels, $j_<$ and $j_>$ denote the least energetic and most energetic non-b jet defined by the Durham clustering algorithm.

[12] Here T-even/odd refers to a naive T transformation, i.e., reversal of momenta and spins only.

Table 2.2: Spin-analyzing power for tau decays and top quark decays in the SM.

| $\tau^- \to$ | $\pi^-$ | $\rho^-$ | $a_1^-$ | $\ell^-$ |
|---|---|---|---|---|
| $c_a$: | 1.0 | 0.46 | 0.12 | −0.33 |
| $t \to$ | $\ell^+$ | $b$ | $j_<$ | $j_>$ |
| $c_a$ (LO): | 1 | −0.41 | 0.51 | 0.2 |
| $c_a$ (NLO): | 0.999 | −0.39 | 0.47 | |

Table 2.3: Coefficient $D_{ab}$ in (2.138) for some final states in $\phi \to \tau\tau$.

| $\tau\tau \to$ | $\pi\pi$ | $\rho\rho$ | $\ell\ell'$ | $\pi\rho$ | $\pi\ell$ | $\rho\ell$ |
|---|---|---|---|---|---|---|
| $\phi\,(0^{++})$: | 0.33 | 0.07 | 0.04 | 0.15 | −0.11 | −0.05 |
| $\phi\,(0^{-+})$: | −1 | −0.21 | −0.11 | −0.46 | 0.33 | 0.15 |

Within 2HDMs the decays of the top quark will be mediated also by charged Higgs exchange. However the branching ratio $B(b \to s\gamma)$ implies that $H^+$ is much heavier than the top quarks, see section 2.2. Thus for the important channel $t \to \ell + \cdots$, $\ell = e, \mu$ the impact of $H^+$ exchange on the $c_\ell$ can be neglected. In any case, the results below can be straightforwardly extended if new top decay modes and/or decay mechanisms should be discovered.

The $\cos\theta$ distributions for the antiparticle decays $\bar{f} \to \bar{b} + \cdots$ are proportional to $(1 - c_b\,\cos\theta)$, assuming CP invariance. Violation of this relation requires that the respective decay amplitude has a CP-violating absorptive part [256]. In 2HMDs the one-loop corrections to the $tWb$ vertex generate such a term [257], but its effect on the $c_a$ of the top quark is negligible in the context of this report.

Let's now come to the analogue of the $\mathcal{O}_i$ at the level of the final states $a, \bar{b}$. The spin correlation $\langle \mathcal{O}_1 \rangle$ leads to a non-isotropic distribution in $\cos\varphi_{ab}$, where $\varphi_{ab} = \angle(\mathbf{q_1}, \mathbf{q_2})$. If no phase space cuts are applied – modulo cuts on the invariant mass $M_{f\bar{f}}$ of the fermion pair – this opening angle distribution is of the form [99, 244]:

$$\frac{1}{\sigma_{ab}}\frac{d\sigma_{ab}}{d\cos\varphi_{ab}} = \frac{1}{2}\left(1 - D_{ab}\,\cos\varphi_{ab}\right)\,, \quad D_{ab} = \frac{4}{3}\,c_a c_b\,\langle \mathbf{s}_f \cdot \mathbf{s}_{\bar{f}} \rangle\,. \tag{2.138}$$

### 2.7.3 $\tau$ decay channels

For $\phi \to \tau\tau$ we have listed in Table 2.3 the coefficients $D_{ab}$ for some of the final states mentioned above. As the charged pion in $\tau \to \pi\nu_\tau$ is the best $\tau$-spin analyzer, this channel discriminates most strikingly between a scalar and a pseudoscalar Higgs boson. In the case of a pseudoscalar $\phi$ the pion momenta $\mathbf{q_1}$, $\mathbf{q_2}$ are predominantly parallel, while for a scalar $\phi$ they tend to be antiparallel. As this channel has a small branching ratio, $B(\tau\tau \to \pi\pi) \simeq 0.01$, the other channels also matter.

The analogue of the CP-odd spin observables $\mathcal{O}_{2,3}$ are [99, 245]:

$$Q_2 \;=\; (\hat{\mathbf{k}}_f - \hat{\mathbf{k}}_{\bar{f}}) \cdot (\hat{\mathbf{q}}_2 \times \hat{\mathbf{q}}_1)/2\;, \quad Q_3 \;=\; \hat{\mathbf{k}}_f \cdot \hat{\mathbf{q}}_1 - \hat{\mathbf{k}}_{\bar{f}} \cdot \hat{\mathbf{q}}_2\;, \tag{2.139}$$

where $\hat{\mathbf{k}}_f$ and $\hat{\mathbf{k}}_{\bar{f}} = -\hat{\mathbf{k}}_f$ are defined as above in the $f\bar{f}$ ZMF. Measurement of (2.139) requires the determination of the signs of the charges of $a$ and $\bar{b}$ while this is not necessary for (2.138). The average of $Q_2$ should be computed for events (2.137) plus the charge conjugated events $\bar{a}b$, while the average of $\langle Q_3 \rangle$ is to be computed for diagonal channels $a\bar{a}$. Concerning $\mathcal{O}_3$ one may take advantage of larger event samples, as exemplified in the case of top quarks in (2.142) below. Asymmetries corresponding to (2.139) are:

$$A(Q) = \frac{N_{ab}(Q>0) - N_{ab}(Q<0)}{N_{ab}} , \tag{2.140}$$

where $N_{ab}$ is the number of events in the reaction (2.137). They should be experimentally more robust because only the signs of $Q_i$ have to be measured. If no phase-space cuts, besides cuts on $M_{f\bar{f}}$, are imposed then [245]

$$\begin{aligned} \langle Q_2 \rangle_{ab} &= \frac{4}{9} c_a c_b \langle \mathcal{O}_2 \rangle , \quad \langle Q_3 \rangle_{aa} = \frac{2}{3} c_a^2 \langle \mathcal{O}_3 \rangle , \\ A(Q_2) &= \frac{9\pi}{16} \langle Q_2 \rangle_{ab} , \quad A(Q_3) = \langle Q_3 \rangle_{aa} . \end{aligned} \tag{2.141}$$

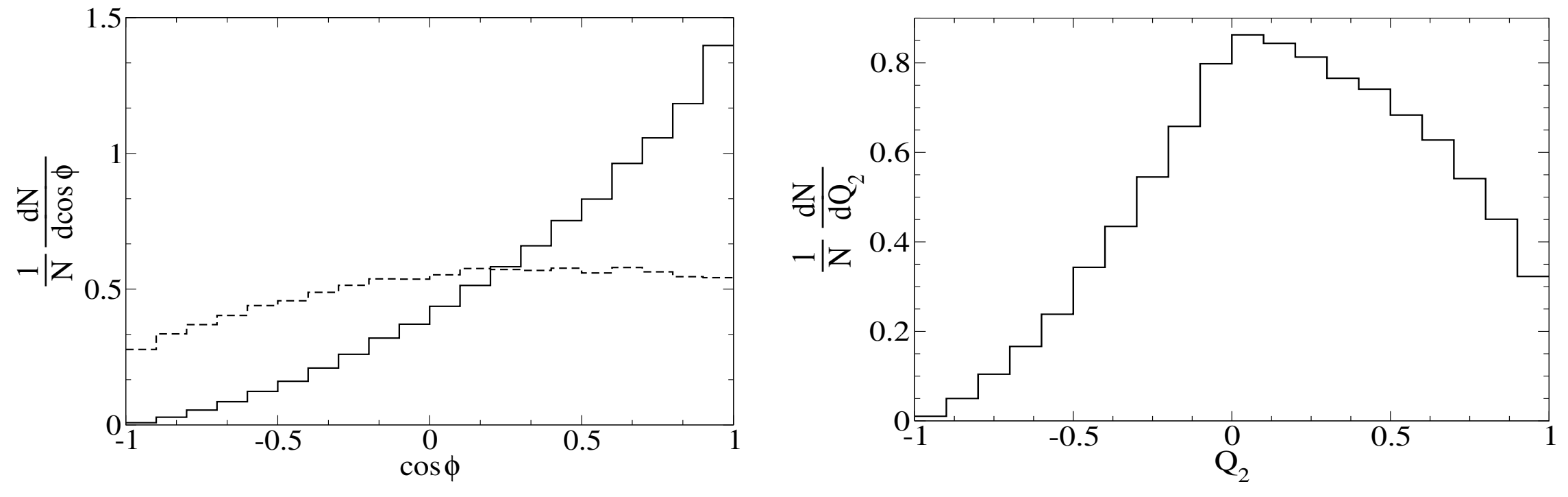

Fig. 2.7: Production of $\phi$ with $m_\phi$= 200 GeV at the LHC via gluon fusion and decay $\phi \to \tau\tau \to \pi\pi$. Left: opening angle distribution for a scalar (dashed) and a pseudoscalar (solid). Right: distribution of $Q_2$ for a Higgs boson with $a_\tau = -b_\tau$.

Let's apply $Q_2$ to $\phi \to \tau\tau$. (As already mentioned above $\langle Q_3 \rangle$ is in general small in this channel.) The observable $Q_2$ measures the distribution of the signed normal vector of the plane spanned by $\mathbf{q_1}, \mathbf{q_2}$ with respect to the $\tau^-$ direction of flight. If $\gamma_{CP}^\tau \neq 0$ then this distribution is asymmetric. If $\phi$ were an ideal mixture of a CP-even and -odd state, $|a_\tau| = |b_\tau|$, the asymmetry corresponding to $Q_2$ would take the value $|A(Q_2)| = 0.4$ in the $\pi\pi$ and $|A(Q_2)| = 0.06$ in the $\rho\ell$ channel, etc. Notice that the sign of $\langle Q_2 \rangle_{ab}$ resp. of $A(Q_2)$ measures the relative sign of the Yukawa couplings $a_\tau$ and $b_\tau$.

How are these results modified by cuts? We have analyzed this for the production of a Higgs boson $\phi$ via gluon fusion at the LHC, and we report here only on the $\pi\pi$ channel: $gg \to \phi \to \tau^-\tau^+ X \to \pi^-\pi^+ X$. Backgrounds are due to the irreducible $Z \to \tau\tau$ and the $t\bar{t}$, $b\bar{b}$ and $W + jet$ processes (c.f., for instance, [115,258]). We apply the cuts $E_T^{a,b} \geq$ 40 GeV, $|\eta| <$2.5. We take $m_\phi$ = 200 GeV and require 120 GeV $\leq M_{\tau\tau} \leq$ 280 GeV. Figure 2.7 (left) shows the opening angle distribution for a scalar ($b_\tau = 0$) and a pseudoscalar ($a_\tau = 0$) state. Because of the applied cuts the shapes of the distributions differ from (2.138), but the two cases are, nevertheless, clearly distinguishable. One can use $D_{ab} \equiv -3\langle \cos\varphi_{ab} \rangle$ as an unbiased estimator. We get $D_{\pi\pi} = -0.32$ (scalar) and $-1.37$ (pseudoscalar), which is to be compared with the respective values of Table 2.3. Thus only a few $\pi\pi$ events are required to decide whether $\phi$ is essentially a parity-even or -odd state. With (2.139) one can further check whether or not $\phi$ is a CPV mixture. In Fig. 2.7 (right) the distribution of $Q_2$ is plotted for the case of a Higgs boson with "maximal" CPV in its couplings to tau leptons, $a_\tau = -b_\tau$. This gives $\langle Q_2 \rangle_{\pi\pi} = 0.19$, which is a bit below the "no-cut" value 2/9, obtained from (2.141). We estimate that about 45 $\pi\pi$ events would establish this as a 5 $\sigma$ effect. The opening angle distribution and $Q_2$ can be evaluated in analogous fashion for the other tau decay channels which should, of course, also be taken into account to accumulate statistics.

### 2.7.4 Top decay channels

Finally we discuss heavy Higgs bosons $\phi$ with mass $m_\phi > 2m_t$ that strongly couple to top quarks. Of particular interest here is the case of a pseudoscalar, as $A \not\to W^+W^-, ZZ$ in lowest order, or a heavy scalar with strongly suppressed couplings to the weak gauge bosons. If $\tan\beta$ is of order 1, the top-quark Yukawa coupling(s) will be large and $\phi \to t\bar{t}$ is the dominant decay mode. For the investigation of the CP nature of $\phi$ with the observables (2.138) and (2.139), (2.140) in this mode the dilepton and the lepton + jets channels are suitable which, for $\ell = e, \mu$, comprise about 4/81 and 8/27, respectively, of all $t\bar{t}$ decays in the SM. In order to search for a longitudinal polarization asymmetry $\mathcal{O}_3$ it is useful to divide the lepton + jets sample into two classes: $\mathcal{A}: \; t\bar{t} \to \ell^+ + \cdots$, and $\bar{\mathcal{A}}: \; t\bar{t} \to \ell^- + \cdots$. For these events one can use [245]

$$\mathcal{E} \;=\; \langle \hat{\mathbf{k}}_f \cdot \hat{\mathbf{q}}_1 \rangle_{\mathcal{A}} - \langle \hat{\mathbf{k}}_{\bar{f}} \cdot \hat{\mathbf{q}}_2 \rangle_{\bar{\mathcal{A}}} \tag{2.142}$$

and a further asymmetry involving the above triple correlation.

For reactions (2.137) where $\phi$ production and decay factorizes to good approximation we get the following [99]: If no phase space cuts are applied – modulo cuts on $M_{t\bar{t}}$ – the opening angle distribution is of the form (2.138) with $D_{\ell\ell'} = 1.33\langle \mathbf{s}_t \cdot \mathbf{s}_{\bar{t}} \rangle$ in the dilepton channel and $D_{\ell j_<} = 0.66\langle \mathbf{s}_t \cdot \mathbf{s}_{\bar{t}} \rangle$ in the lepton + jets channel if $j_<$ is used as top-spin analyzer in the non-leptonic top decay modes. The expectation value of $\mathcal{O}_1$ was given above and takes the values 0.25 and $-0.75$ for a P-even and -odd Higgs boson $\phi$, respectively. In addition the formulae (2.141) apply. For a CPV Higgs boson a non-zero $\langle Q_3 \rangle$ and $\mathcal{E}$ are generated by the one-loop QCD corrections. With about 4000 $\phi \to t\bar{t}$ events the CP nature of $\phi$ could be established, in this ideal situation, for a large range of the coupling ratio $r_t$ with 5 $\sigma$ sensitivity when (2.138), (2.141), and (2.142) are used in combination [99].

At the LHC the main production reaction is expected to be gluon fusion, for which these results do not apply. The amplitude of $g\,g \to \phi \to t\bar{t} \to$ final state interferes with the amplitude of the QCD-induced non-resonant $t\bar{t}$ background, $g\,g \to t\bar{t} \to$ final state, and this interference is not negligible, even in the vicinity of $\sqrt{s} \sim m_\phi$, because the resonance is not narrow. The interference generates a peak-dip structure in the $t\bar{t}$ invariant mass distribution $M_{t\bar{t}}$ [244, 259]. Statistically significant signals are possible in the mass range 350 GeV $\lesssim m_\phi \lesssim$ 500 GeV, depending on the strength of the Yukawa couplings and on the width of $\phi$ [115, 244, 251, 259]. Needless to say, this is a difficult channel which requires very good $M_{t\bar{t}}$ resolution and a precise knowledge of the background contributions to the $M_{t\bar{t}}$ distribution[13].

If experiments will find a signal of a heavy neutral spin-zero boson $\phi$ in the $t\bar{t}$ channel, the above observables can of course be used in this case, too, to investigate its CP properties. The opening angle distribution (2.138) was investigated in [244] in the dilepton channel with the irreducible $t\bar{t}$ background included. This background dilutes the striking difference between the shapes of the distributions for a scalar and a pseudoscalar $\phi$ exhibited above. It depends critically on the Yukawa couplings, mass, and width of $\phi$ whether or not a statistically significant effect is obtained. In order to preserve the discriminating power of this distribution it should be determined only for events with $M_{t\bar{t}} = m_\phi - \Delta$, where $\Delta$ is of the order of 40 GeV [244]. For the $t\bar{t}$ background the distribution (2.138) was computed to NLO QCD in [261].

CPV (resonant and non-resonant) $\phi$ exchange at one loop was computed for $q\bar{q}, gg \to t\bar{t}$ within 2HDM in [103, 243] and confirmed by [118, 262]. The expectation values of the observables (2.139), (2.140), and (2.142) were analyzed in [245] for the dilepton and the lepton + jets channels. When evaluated for events with $M_{t\bar{t}} = m_\phi - \Delta$ it was found that CP effects of a few percent are possible. Observables composed of final state momenta in the laboratory frame yield smaller CP effects [243]. In [263] the CP asymmetry $\Delta_{LR} = [N(t_L\bar{t}_L) - N(t_R\bar{t}_R)]/(\mathrm{all}\, t\bar{t})$, which corresponds to $\langle \mathcal{O}_3 \rangle$, was computed within 2HDM for light $\phi$ exchange in $q\bar{q}, gg \to t\bar{t}$ production, and found to be $\Delta_{LR} \sim 0.1$ %.

Heavy Higgs production by weak gauge boson fusion at the LHC or at an ILC should also be an option to explore the $t\bar{t}$ decay channel. If high-energetic left- and right-circularly polarized photon beams will be available in the future then the respective production cross sections $\gamma\gamma \to \phi$ can be measured, and a difference would constitute a clean signal of Higgs sector CPV [109]. For unpolarized $\gamma\gamma$ collisions, the reactions $\gamma\gamma \to \phi \to \tau\tau,\ t\bar{t}$ may be employed to investigate the CP nature of $\phi$. In [149] CP observables were analyzed and computed within 2HDM for the $\ell$ + jets final states of the $\phi \to t\bar{t}$ channel.

[13]The matrix elements of [244] are contained in the MC generator TOPREX [260].

Table 2.4: $\phi \to \tau\tau$ event numbers $N_1$ and $N_2$ required to determine the CP-even and -odd correlation $D_{ab}$ and $\langle Q_2 \rangle_{ab}$ with 3 $\sigma$ significance, as a function of $r_\tau = b_\tau/(a_\tau + b_\tau)$.

| $r_\tau$ | 0 | 0.1 | 0.2 | 0.5 | 0.6 | 0.7 – 1.0 | |
|---|---|---|---|---|---|---|---|
| $N_1$: | $9 \times 10^3$ | $10^4$ | $1.2 \times 10^4$ | $1.2 \times 10^4$ | $3 \times 10^3$ | $10^3$ | |
| $r_\tau$: | 0.15 | 0.2 | 0.3 | 0.4 – 0.6 | 0.7 | 0.8 | 0.85 |
| $N_2$ : | $1.2 \times 10^4$ | $6 \times 10^3$ | $3 \times 10^3$ | $1.5 \times 10^3$ | $3 \times 10^3$ | $6 \times 10^3$ | $1.2 \times 10^4$ |

### *2.7.5 Conclusions*

In conclusion we have discussed, for $\phi \to \tau\tau$ and $\phi \to t\bar{t}$, a set of observables for determining the CP parity of a neutral Higgs boson $\phi$ and, in particular, for investigating whether or not there is CPV in the Higgs sector. The $\tau$ decay channel is clearly most suited to explore the nature of a light or heavy $\phi$, and the above correlations and asymmetries, applied in combination to the various charged final states, should provide powerful tools already at the LHC. Table 2.4 summarizes our results for the $\tau$ decay mode: $N_1$ and $N_2$ are the $\phi \to \tau\tau$ event numbers required to measure the CP-even and -odd correlation $D_{ab}$ and $\langle Q_2 \rangle_{ab}$ with 3 $\sigma$ significance as a function of $r_\tau = b_\tau/(a_\tau + b_\tau)$, using the $\tau$ decay channels discussed above. The numbers apply to light and heavy $\phi$. The opening angle distribution (2.138) is sensitive in the ranges $0 \le r_\tau \lesssim 0.2$ (scalar-like $\phi$) and $0.7 \lesssim r_\tau \le 1.0$ (pseudoscalar-like $\phi$). Assuming that at least $10^4$ $\phi \to \tau\tau$ events will be recorded at the LHC, Higgs sector CP violation can be established if the ratio of the Yukawa couplings lies in the range $0.2 \lesssim r_\tau \lesssim 0.8$.

At the LHC a heavy $\phi$ is expected to be observable in the $t\bar{t}$ channel only under favorable circumstances, i.e., for a restricted parameter range of various SM extensions. We found that the CP-odd correlations and asymmetries (2.141), (2.142), applied to the dilepton and lepton + jets channels and evaluated in appropriate mass bins, deviate from zero with $\gtrsim 3$ $\sigma$ for a Higgs boson with mass in the range 300 GeV$\lesssim m_\phi \lesssim$ 500 GeV and reduced Yukawa couplings $|a_t b_t| \gtrsim 0.1$. In any case the above observables may be applied to dileptonic and single-lepton $t\bar{t}$ events, irrespective of a significant resonance signal. Moreover, if $\phi \to t\bar{t}$ should be seen at a future high luminosity $e^+e^-$ and/or photon collider the variables above will also show their discriminating power.

## 2.8 CP-violating top Yukawa couplings in the 2HDM

*Wafaa Khater and Per Osland*

The Two-Higgs-Doublet Model is a simple extension of the Standard Model that can provide additional CP violation [1,36,264–266]. However, the model is rather constrained, it is not a priori obvious that the allowed parameter regions provide CP violation that could be of experimental interest at the LHC. The top Yukawa coupling is of particular interest, since it will become accessible at the LHC. It is interesting to establish how the Higgs sector can be explored via this coupling.

The process

$$pp \to t\bar{t} + X \tag{2.143}$$

has been studied in considerable detail [263, 267], in particular by Bernreuther and Brandenburg [243,

245] who identified the different kinematical structures appearing in the CP-violating part of the interaction, and evaluated them in a generic Two-Higgs-Doublet Model.

At very high energies, the dominant contribution to the process (2.143) is from the gluon-gluon initial state,

$$gg \to t\bar{t} \tag{2.144}$$

as indicated in Fig. 2.8. Also, among various observables proposed by Bernreuther and Brandenburg, we focus [118] on one that requires the decay to electrons (or muons):

$$t \to l^+ \nu_l b, \quad \bar{t} \to l^- \bar{\nu}_l \bar{b}. \tag{2.145}$$

In the process of producing $t\bar{t}$ via gluon fusion, the CP violation can arise at the one-loop level, via neutral Higgs exchange involving the $t$ and $\bar{t}$ lines, provided the top Yukawa coupling exhibits both scalar and pseudo-scalar terms as given in Eq. (2.146). Such a coupling induces correlations among the $t$ and $\bar{t}$ momenta and their spins. The most interesting of these correlations are the CP-odd ones which are transferred to the $t$ and $\bar{t}$ decay products, e.g., to the energies and momenta of the electron and the positron.

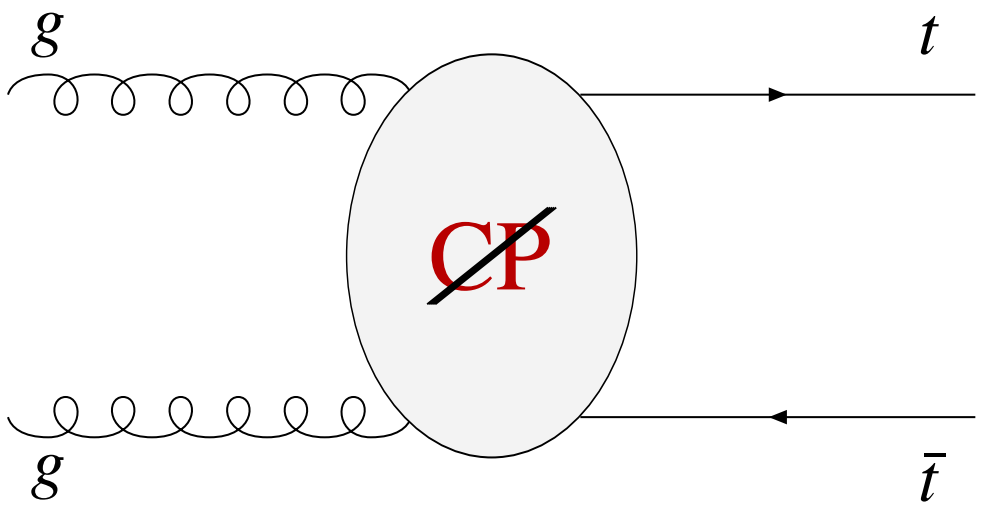

Fig. 2.8: The underlying $g + g \to t + \bar{t}$ reaction.

A necessary condition for having CP-violating Yukawa couplings, is that the mass matrix corresponding to the three neutral Higgs bosons not be block diagonal in the weak basis, i.e., in terms of the real and imaginary parts of the doublet fields $\Phi_1$ and $\Phi_2$. In the notation of [118] (see also section 2.1), this requires one or more among $\lambda_5$, $\lambda_6$ and $\lambda_7$ to be complex. The simplest case is to take $\lambda_5$ to be complex. Actually, the model considered in [118] takes $\lambda_6 = \lambda_7 = 0$. The $Z_2$ symmetry is thus respected by the quartic terms, and Flavour-Changing Neutral Couplings are naturally suppressed [18].

The resulting model can be parameterized in different ways. Let the top Yukawa coupling for a particular Higgs boson $H_j$ be written as

$$H_j t\bar{t}: \qquad [a + i\gamma_5 \tilde{a}] \qquad (j = 1, 2, 3) \tag{2.146}$$

or as

$$\frac{m_t}{v}\left[g_L \frac{1-\gamma_5}{2} + g_R \frac{1+\gamma_5}{2}\right]. \tag{2.147}$$

Then, a crucial quantity is the asymmetry between the left- and right-handed parts of the coupling

$$\gamma_{CP} = -a\tilde{a} = -i\,\frac{m_t^2}{4v^2}(g_L^2 - g_R^2). \tag{2.148}$$

In the Model II for Yukawa couplings, where only $\Phi_2$ couples to up-type quarks, and only $\Phi_1$ to down-type quarks, the couplings $a$ and $\tilde{a}$ are simply given in terms of elements of the rotation matrix that diagonalizes the mass-squared matrix $\mathcal{M}^2$ of the neutral Higgs bosons. This rotation matrix R is defined in (2.27) and (2.28). Relative to the SM coupling, the Yukawa couplings can then be written as

$$H_j t\bar{t}: \qquad \frac{1}{\sin\beta}\,[R_{j2} - i\gamma_5 \cos\beta R_{j3}], \tag{2.149}$$

with $R$ the rotation matrix as defined in (2.28)

Unless the couplings are suppressed, the dominant contribution to the CP violation will come from diagrams involving exchange of the *lightest* Higgs boson, $H_1$. According to the above discussion, this contribution will be proportional to $R_{12}R_{13}\cos\beta/\sin^2\beta$. Thus, in order to maximize the CP violation in $t\bar{t}$ production, we are interested in low values of $\tan\beta$, and large values of $|R_{12}R_{13}|$. The latter requirement means large $|\sin\alpha_1|$ and large $|\sin 2\alpha_2|$.

### 2.8.1 *A CP-violating observable*

Among various CP-violating observables proposed by Bernreuther and Brandenburg, the quantity

$$A_1 = E_+ - E_- \tag{2.150}$$

was found to be rather promising [118]. Here, $E_+$ and $E_-$ are the energies of the positron and electron of Eq. (2.145), defined in the laboratory frame.

In order to have a significant observation, the expectation value $\langle A_1\rangle$ must compare favourably with the statistical fluctuations, which behave like $\sqrt{N}$, where $N$ is the number of events. In order to assess this, it is convenient to consider the "signal to noise" ratio [243],

$$\frac{S}{N} = \frac{\langle A_1\rangle}{\sqrt{\langle A_1^2\rangle - \langle A_1\rangle^2}}. \tag{2.151}$$

The analytical expression for $\langle A_1\rangle$ is entirely determined by the coefficients of the CP-odd correlations between the momenta and the spins of the $t\bar{t}$ pair [268]. This explicitly shows that the CP-violation originating at the production level of the top pair manifests itself in the kinematics of their decay products.

### 2.8.2 *Results*

One can specify the Two-Higgs-Doublet Model in terms of the potential, plus additional parameters. We found it convenient to follow a different approach. In order to more easily identify regions of large CP violation, we take as input parameters those which are more directly related to the observables. Thus, we take the angles of the rotation matrix and the lowest masses as part of the input:

$$\text{Input:}\quad \tan\beta,\quad \alpha_1,\quad \alpha_2,\quad \alpha_3,\quad M_1,\quad M_2,\quad M_{H^\pm},\quad \mathrm{Re}\, m_{12}^2. \tag{2.152}$$

With this input, the mass of the heaviest Higgs boson, $M_3$, is determined, as well as the coefficients of the potential, $\lambda_1$, $\lambda_2$, $\lambda_3$, $\lambda_4$, and $\lambda_5$.

CP violation requires the mass-squared matrix not to be block diagonal. This, in turn, requires $\sin\alpha_2 \neq 0$, and/or $\sin\alpha_3 \neq 0$. However, only a small part of the 8-dimensional parameter space (2.152) yields viable models, when various physical constraints are taken into account. The constraints are of different kinds, the most important of which are: (i) the potential must satisfy positivity and unitarity (constraints most easily expressed in terms of the $\lambda$s), (ii) the spectrum must be compatible with the LEP searches (which essentially constrains a function of the lightest Higgs mass and its coupling to the $Z$ boson), and (iii) the charged Higgs must be compatible with constraints from direct searches at LEP [232] and $b \to s\gamma$ [83].

For a range of parameters, with the lightest Higgs mass of the order of 100 to 150 GeV, $\tan\beta = 0.5$ and the charged Higgs mass at 300 GeV, the "signal-to-noise" ratio was found to be of the order of $10^{-3}$. Thus, a number of semileptonically decaying $t\bar{t}$ events in excess of $10^6$ will be required in order to measure a significant CP-violating signal, for "optimal" parameters. This should be possible, after a few years of running at high luminosity [269, 270].

We note from Eqs. (2.148) and (2.149) that in the limit of three degenerate Higgs masses, the CP violation in the top Yukawa coupling vanishes, due to the orthogonality of the rotation matrix $R$. Also, due to the constraints inherent in the model (with $\lambda_6 = \lambda_7 = 0$), if two of the Higgs boson masses approach each other ($M_1 \to M_2$ or $M_2 \to M_3$) then also the third one will approach this value, and the CP violation will again vanish. Thus, "large" CP violation is only possible if one Higgs boson is fairly light, and the other two are heavy and non-degenerate.

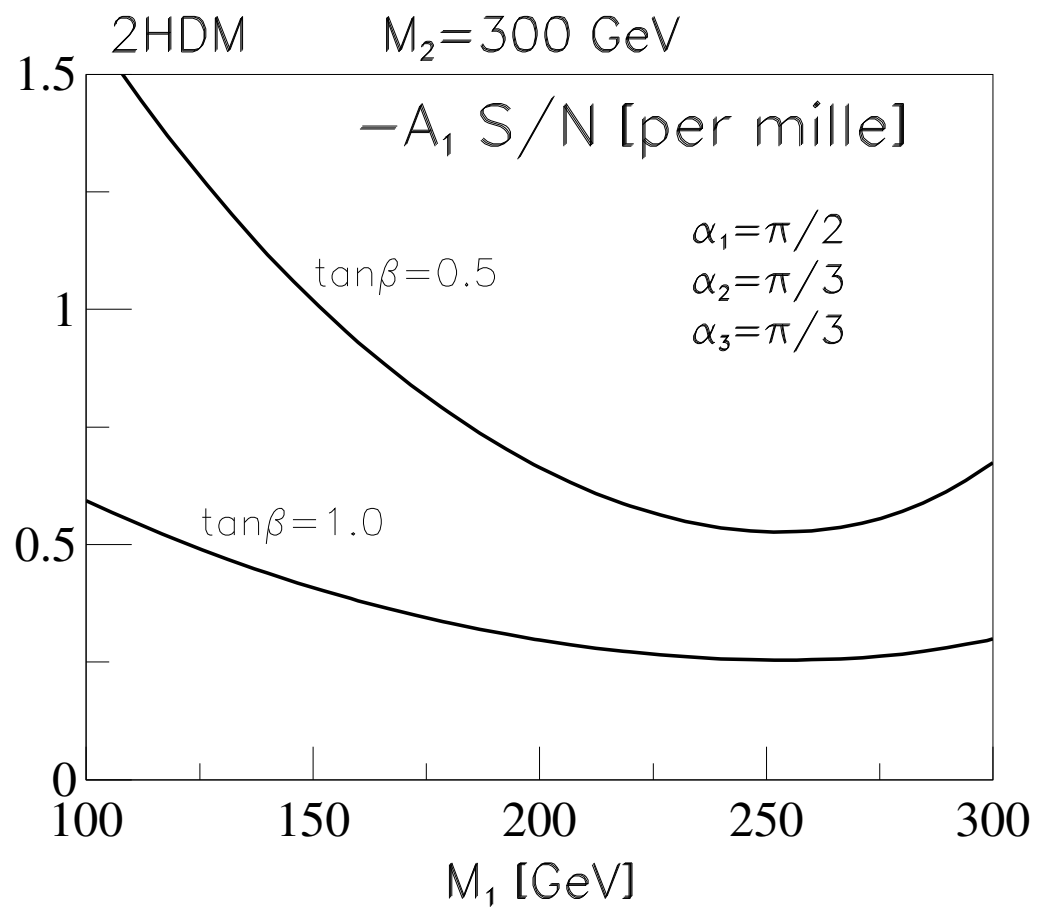

Fig. 2.9: Sensitivity of $A_1$ vs. lightest Higgs mass, $M_1$, for $M_2 = 300$ GeV, $M_3 = 500$ GeV, $M_{H^\pm} = 500$ GeV (from [271]).

In a recent update of this study [271], more constraints have been imposed on the model. As a result, the allowed regions of the parameter space shrink, and the detection of CP violation within the 2HDM thus becomes more challenging than found earlier, see Fig. 2.9.

## 2.9 Higgs CP measurement via $t\bar{t}\phi$ partial reconstruction at the LHC

*Justin Albert, Mikhail Dubinin, Vladimir Litvin, and Harvey Newman*

In the Standard Model, if the Higgs mass is below 140 GeV, the "golden" channels $\phi \to ZZ^* \to 4\ell$ and $\phi \to WW^* \to 2\ell 2\nu$ have small branching fraction, thus the mode $\phi \to \gamma\gamma$ begins to become more favorable for discovery. However, the latter does not, in general, encode information on Higgs properties such as CP and spin. In order for information on Higgs CP and spin to be obtained from such decays, the Higgs must be produced in association with two or more particles, such as $b\bar{b}$ or $t\bar{t}$. From an angular analysis of such processes, one can obtain information, in a model-independent way, on the Higgs spin and CP [272].

We consider here the process $gg \to t\bar{t}\phi$. This process has a relatively small cross-section in the SM (see the left plot of Fig. 2.10), however it has comparatively quite low background. In order to increase the size of the sample of $t\bar{t}\phi$ events, we reconstruct just one of the $t$ *or* $\bar{t}$, *but not both*, *i.e.* a **partial reconstruction** of $t(\bar{t})\phi$, as compared with a **full reconstruction** of *both* the $t$ *and* $\bar{t}$, as well as the Higgs. For an efficiency for top reconstruction of 20%, partial reconstruction increases the efficiency, relative to full reconstruction, from $(20\%)^2 = 4\%$ to $2 * 20\% - (20\%)^2 = 36\%$, nearly an order of magnitude. This could potentially introduce backgrounds of the form $t\phi + X$, however events that contain both top and Higgs are dominated by $t\bar{t} + X$, so this technique does not add significant irreducible background.

One may use both $\phi \to \gamma\gamma$ and $\phi \to b\bar{b}$ channels for this process. We consider here the $gg \to t(\bar{t})\phi, \phi \to \gamma\gamma$ channel, which has less background, although a far lower branching fraction, than $\phi \to b\bar{b}$. We select the $\phi \to \gamma\gamma$ in a similar manner as for the CMS inclusive $\phi \to \gamma\gamma$ analysis [273], and then add to that a top, reconstructed as a $b$-tagged jet and a high-$p_T$ ($> 40$ GeV) lepton. The events themselves, and thus the background channels, are a subset of the inclusive $\phi \to \gamma\gamma$ analysis. Backgrounds are dominated in this case by $t\bar{t}\gamma\gamma$, $b\bar{b}\gamma\gamma$, and $Z\gamma\gamma$ processes. As Higgs discovery and mass measurement would likely be performed by the inclusive Higgs analysis prior to this analysis to determine Higgs spin and CP, a selection on the Higgs mass can dramatically reduce these main "irreducible" backgrounds.

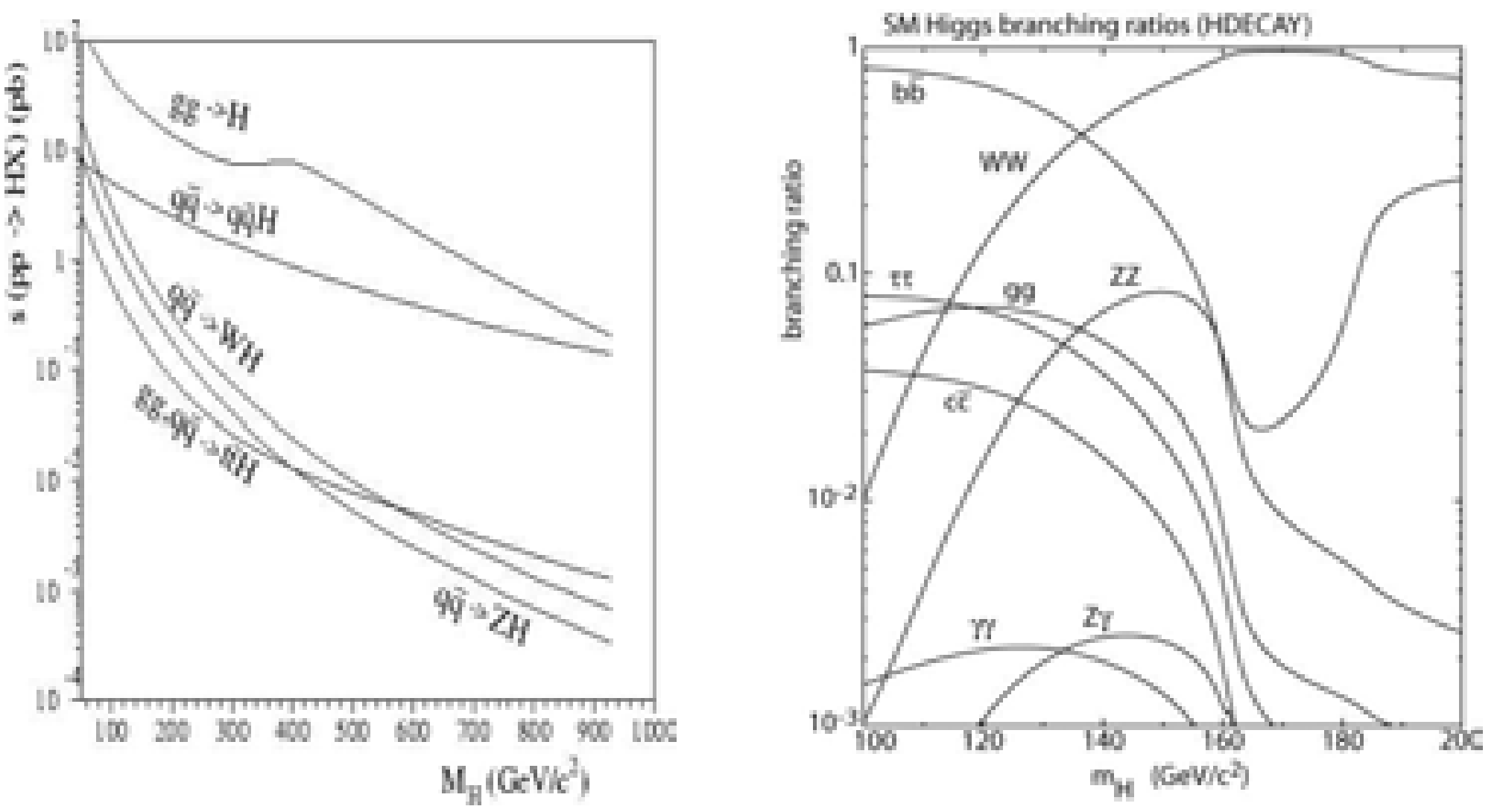

Fig. 2.10: Standard Model cross-sections (left) of processes containing a Higgs at the LHC as a function of Higgs mass, and branching fractions (right) of the Higgs, again as a function of Higgs mass.

In order to consider the $gg \to t(\bar{t})\phi$ process, the Yukawa Lagrangian can be divided into CP-even and CP-odd components:

$$\mathcal{L} = \bar{t}(c + id\gamma_5)t\phi, \tag{2.153}$$

where $0 \leq c \leq 1$ parametrizes the CP-even contribution and $d = 1 - c$ parametrizes the CP-odd fraction [272]. In the existing approaches to the MSSM with explicit CP violaton in the Higgs sector [47, 274, 275] the parameters $c$ and $d$ are expressed by means of the matrix elements $R_{ij}$ of the Higgs boson mixing matrix (see Eq. 2.27). For the lightest mass eigenstate $h_1$ we have $c = k\,(R_{21}\sin\alpha + R_{11}\cos\alpha)$ and $d = -k\,(R_{31}\cos\beta)$ where $k = -m_{top}/(\sin\beta\, v^2)$ and $\alpha$,$\beta$ are the standard mixing angles of the CP-even/odd states. However, in the following we are not going to use any particular model of explicit CP-violation but simply consider $c$ and $d$ as the model-independent weights parametrizing the CP-even and the CP-odd components in the Yukawa Lagrangian [272].

Gunion and He define 6 CP-sensitive variables, as follows [272]:

$$\begin{aligned}
a_1 &= \frac{(\vec{p}_t \times \hat{n}) \cdot (\vec{p}_{\bar{t}} \times \hat{n})}{|(\vec{p}_t \times \hat{n}) \cdot (\vec{p}_{\bar{t}} \times \hat{n})|}, \quad & a_2 &= \frac{p_t^x p_{\bar{t}}^x}{|p_t^x p_{\bar{t}}^x|} \\
b_1 &= \frac{(\vec{p}_t \times \hat{n}) \cdot (\vec{p}_{\bar{t}} \times \hat{n})}{p_t^T p_{\bar{t}}^T}, \quad & b_2 &= \frac{(\vec{p}_t \times \hat{n}) \cdot (\vec{p}_{\bar{t}} \times \hat{n})}{|\vec{p}_t||\vec{p}_{\bar{t}}|} \\
b_3 &= \frac{p_t^x p_{\bar{t}}^x}{p_t^T p_{\bar{t}}^T}, \quad & b_4 &= \frac{p_t^z p_{\bar{t}}^z}{|\vec{p}_t||\vec{p}_{\bar{t}}|},
\end{aligned} \tag{2.154}$$

where $\hat{n}$ is a unit vector in the $+z$ direction along the collision axis. Using the partial reconstruction technique, the information from the second top momentum must be replaced with the momentum of the reconstructed Higgs (or potentially with the unreconstructed [missing] momentum, or some combination. Here we simply use the momentum of the reconstructed Higgs for the replacement.) As shown in Fig. 2.11, the partial reconstruction has a similar overall per-event CP sensitivity to the full reconstruction, along with the much higher efficiencies. We denote the partial-reconstruction versions of the original Gunion-He variables with Greek letters: $(a_i, b_j) \to (\alpha_i, \beta_j)$.

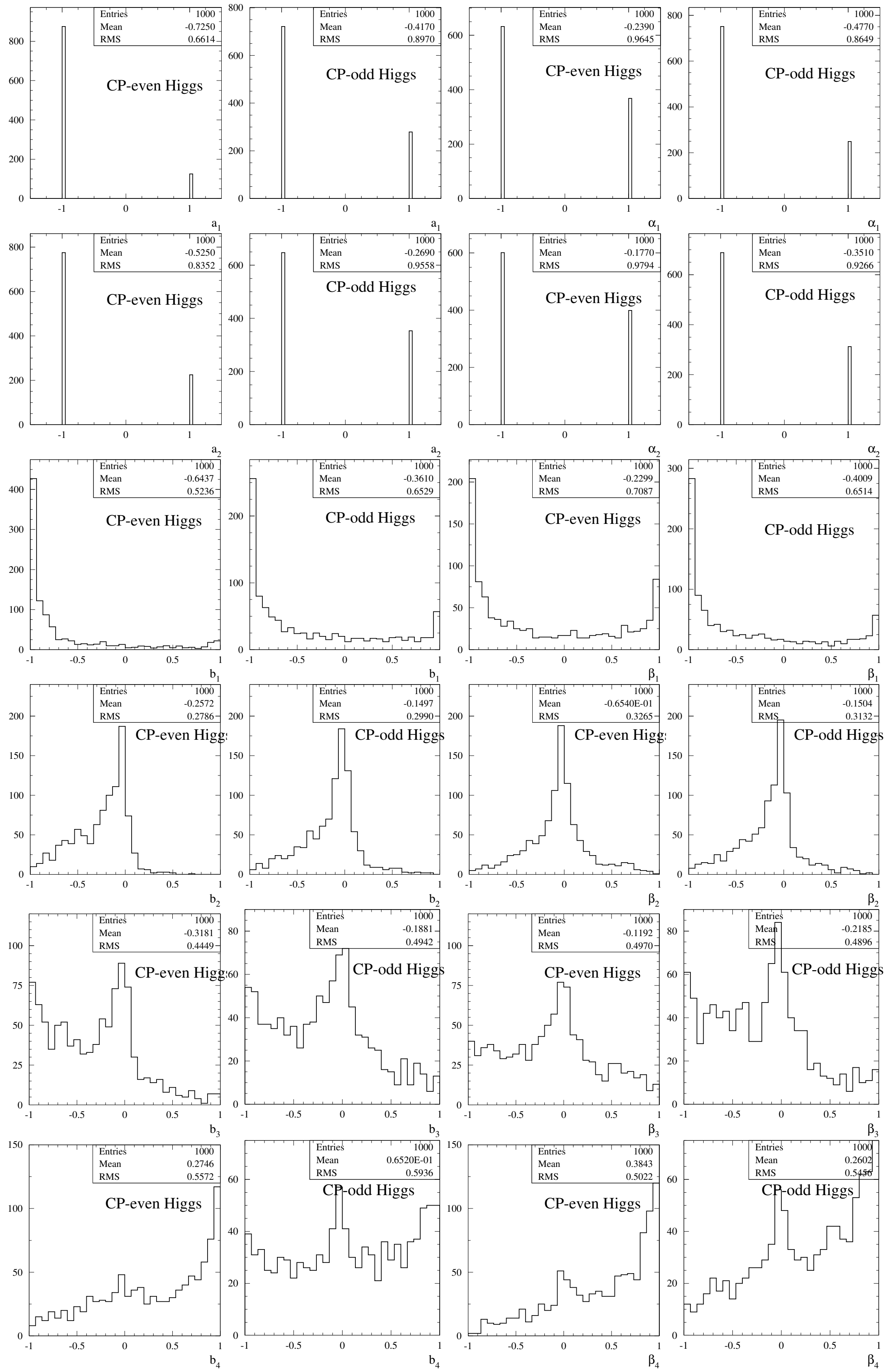

Fig. 2.11: Distributions of (as rows from top) the $a_1$, $a_2$, $b_1$, $b_2$, $b_3$, and $b_4$ variables. Within each row, the leftmost plot shows the distribution for CP-even Higgs full reconstruction, left middle shows CP-odd with full reconstruction, right middle shows CP-even Higgs partial reconstruction, and rightmost shows CP-odd Higgs partial reconstruction.

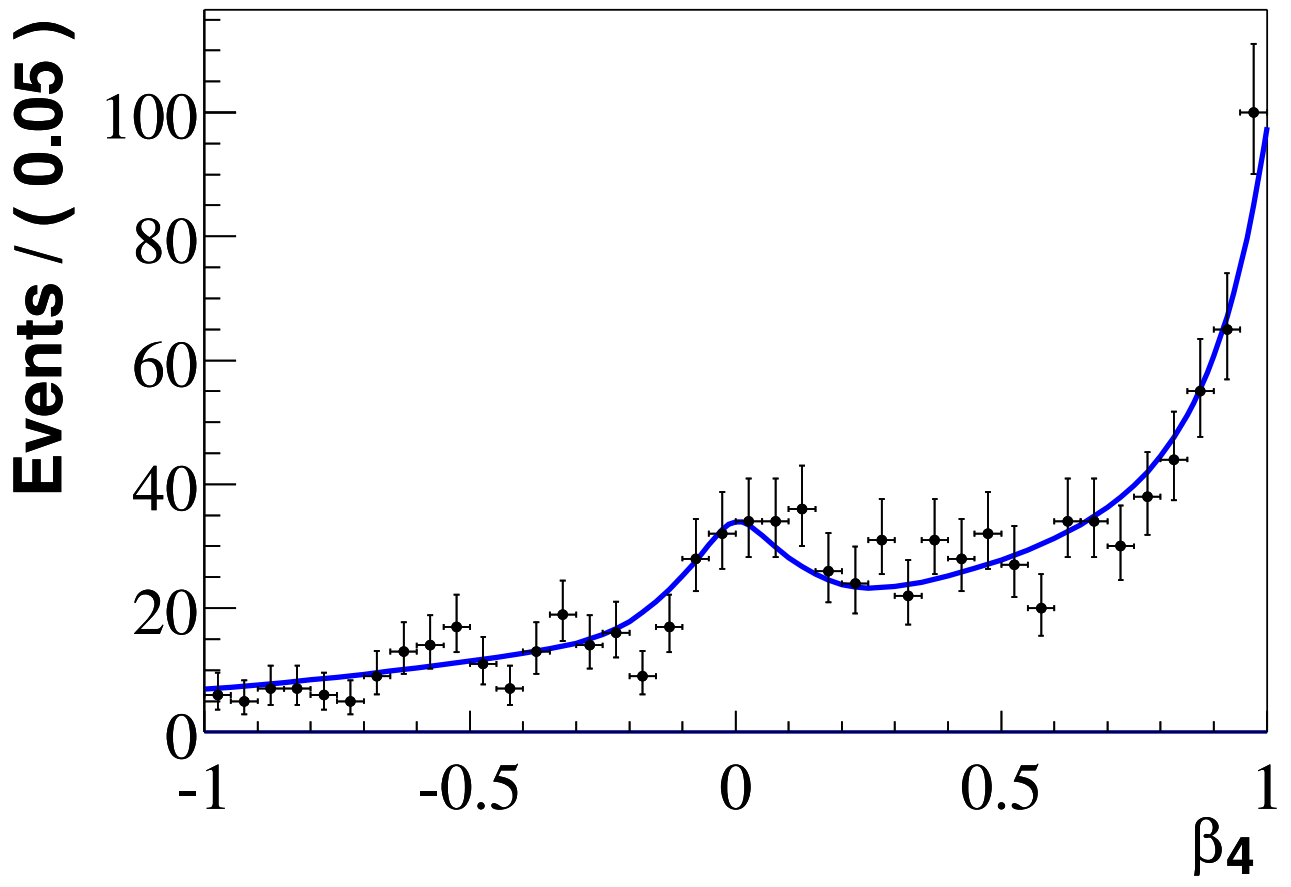

Fig. 2.12: Projection of Higgs CP fit onto the $\beta_4$ variable. Monte Carlo data are points with error bars (statistical only). The line shows the projection of the fit to a sum of CP-even and CP-odd components (as well as misreconstructed background).

In order to extract the CP-even and CP-odd fractions of the Higgs from reconstructed $t(\bar{t})\phi$ events, we have implemented an unbinned maximum-likelihood fit, combining the information from each of the CP-sensitive variables. For each event $i$ and hypothesis $j$ (CP-even signal, CP-odd signal, background) we define the probability density function (PDF) as

$$\mathcal{P}_j^i = \mathcal{P}_j(\alpha_1^i)\mathcal{P}_j(\alpha_2^i)\mathcal{P}_j(\beta_1^i)\mathcal{P}_j(\beta_2^i)\mathcal{P}_j(\beta_3^i)\mathcal{P}_j(\beta_4^i), \tag{2.155}$$

accounting for correlations between the 6 variables. The likelihood function is

$$\mathcal{L} = \frac{e^{-\sum Y_j}}{N!}\prod_{i=1}^{N}\sum_j Y_j \mathcal{P}_j^i \, , \tag{2.156}$$

where $Y_j$ is the yield of events of hypothesis $j$ and $N$ is the number of events in the sample.

We fit sets of 50, 100, 500, and 1000 partially-reconstructed $t(\bar{t})\phi$ events (corresponding to approximately 40, 80, 400, and 800 fb$^{-1}$ of integrated luminosity respectively), in each case with the $\phi$ generated as being 50% CP-odd and 50% CP-even. The resulting uncertainties on the CP (*i.e.* parameters $c$ and $d$ of Eq. 2.153) are $\pm 0.5$ for the 50-event case, $\pm 0.3$ for 100 events, $\pm 0.2$ for 500 events, and $\pm 0.1$ for 1000 events. Fig. 2.12 shows a projection of the maximum-likelihood fit onto the $\beta_4$ variable, as compared with the data (the points with error bars). The CP-even component has a gentler exponential slope, and smaller central Gaussian fraction, than the CP-odd component for this variable.

To improve measured uncertainties on the Higgs CP and spin, performing a combined analysis of this $gg \to t(\bar{t})\phi,\ \phi \to \gamma\gamma$ channel together with related channels such as $gg \to t(\bar{t})\phi,\ \phi \to b\bar{b}$; $gg \to b(\bar{b})\phi,\ \phi \to \gamma\gamma$; and vector boson fusion Higgs production (for a light Higgs) and $\phi \to ZZ^* \to 4\ell$ and $\phi \to WW^* \to 2\ell 2\nu$ (for a heavier Higgs) is the most promising direction.

### 2.10 Higgs + 2 jets as a probe for CP properties

*Vera Hankele, Gunnar Klämke, and Dieter Zeppenfeld*

At the LHC, one would like to experimentally determine the CP nature of any previously discovered (pseudo)scalar resonance. Such measurements require a complex event structure in order to provide the distributions and correlations which can distinguish between CP-even and CP-odd couplings. This can either be done by considering decays, e.g. $H \to ZZ \to l^+l^-l^+l^-$ and the correlations of the decay leptons [93, 119], (see sections 2.12, 2.11 and 2.13), or one can study correlations arising in the production process. Here the azimuthal angle correlations between the two additional jets in $Hjj$ events have emerged as a promising tool [98]. In the following we consider the prospects for using $\Phi jj$ events at the LHC, where $\Phi$ stands for a CP even boson, H, a CP odd state, A, or a mixture of the two. Two production processes are considered. The first is vector boson fusion (VBF), i.e. the electroweak process $qQ \to qQ\Phi$ (and crossing related ones) where $\Phi$ is radiated off a $t$-channel electroweak boson. The second is gluon fusion where $\Phi$ is produced in QCD dijet events, via the insertion of a heavy quark loop which mediates $gg \to \Phi + 0, 1, 2$ gluons.

The CP properties of a scalar field are defined by its couplings and here we consider interactions with fermions as well as gauge bosons. Within renormalizable models the former are given by the Yukawa couplings

$$\mathcal{L}_Y = y_f \bar{\psi} H \psi + \tilde{y}_f \bar{\psi} A i \gamma_5 \psi \,, \tag{2.157}$$

where $H$ (and $A$) denote (pseudo)scalar fields which couple to fermions $f = t,\ b,\ \tau$ etc. In our numerical analysis we consider couplings of SM strength, $y_f = \tilde{y}_f = m_f/v = y_{SM}$. Via these Yukawa couplings, quark loops induce effective couplings of the (pseudo)scalar to gluons which, for (pseudo)scalar masses well below quark pair production threshold, can be described by the effective Lagrangian

$$\mathcal{L}_{\text{eff}} = \frac{y_f}{y_{SM}} \cdot \frac{\alpha_s}{12\pi v} \cdot H\, G^a_{\mu\nu}\, G^{a\,\mu\nu} + \frac{\tilde{y}_f}{y_{SM}} \cdot \frac{\alpha_s}{16\pi v} \cdot A\, G^a_{\mu\nu}\, G^a_{\rho\sigma} \varepsilon^{\mu\nu\rho\sigma} \,. \tag{2.158}$$

Similar to the $\Phi gg$ coupling, Higgs couplings to $W$ and $Z$ bosons will also receive contributions from heavy particle loops which can be parameterized by the effective Lagrangian

$$\mathcal{L}_5 = \frac{f_{\text{e}}}{\Lambda_5}\, H\, \vec{W}_{\mu\nu} \vec{W}^{\mu\nu} + \frac{f_{\text{o}}}{\Lambda_5}\, A\, \vec{W}_{\mu\nu} \vec{W}_{\rho\sigma}\, \frac{1}{2} \varepsilon^{\mu\nu\rho\sigma} \,. \tag{2.159}$$

For most models, one expects a coupling strength of order $f_i/\Lambda_5 \sim \alpha/(4\pi v)$ for these dimension 5 couplings and, hence, cross section contributions to vector boson fusion processes which are suppressed by factors $\alpha/\pi$ (for interference effects with SM contributions) or $(\alpha/\pi)^2$ compared to those mediated by the tree level $HVV$ $(V = W,\ Z)$ couplings of the SM. However, together with the tree level couplings, the effective Lagrangian of Eq. (2.159) has the virtue that it parameterizes the most general $\Phi VV$ coupling which can contribute in the vector boson fusion process $qQ \to qQ\Phi$ and, thus, it is a convenient tool for phenomenological discussions and for quantifying, to what extent certain couplings can be excluded experimentally. Neglecting terms which vanish upon contraction with the conserved quark currents, the most general tensor structure for the fusion vertex $V^\mu(q_1)V^\nu(q_2) \to \Phi$ is given by

$$T^{\mu\nu}(q_1, q_2) = a_1(q_1, q_2)\, g^{\mu\nu} + a_2(q_1, q_2)\ [q_1 \cdot q_2 g^{\mu\nu} - q_2^\mu q_1^\nu] + a_3(q_1, q_2)\ \varepsilon^{\mu\nu\alpha\beta} q_{1\alpha} q_{2\beta} \,. \tag{2.160}$$

Here the $a_i(q_1, q_2)$ are scalar form factors, which, in the low energy limit, are given by the effective Lagrangian of Eq. (2.159). One obtains, e.g. for the $W^+W^-\Phi$ coupling, $a_2 = -2f_e/\Lambda_5$ and $a_3 = 2f_o/\Lambda_5$, while $a_1 = 2m_W^2/v$ is the SM vertex.

The CP-even and CP-odd couplings of Eqs. (2.158,2.159) lead to characteristic azimuthal angle correlations of the two jets in $\Phi jj$ production processes. Normalized distributions of the azimuthal angle between the two jets, $\triangle\phi_{jj}$, are shown in Fig. 2.13 for vector boson fusion (left panel) and for gluon fusion processes (right panel) leading to $\Phi jj$ events: A CP-odd coupling suppresses the cross section for planar events because the epsilon tensor contracted with the four linearly dependent momentum vectors of the incoming and outgoing partons disappears. For a CP-even coupling the dip, instead, appears at 90 degrees [98, 277]. Unfortunately, when both CP-even and CP-odd couplings are present simultaneously, the two $\triangle\phi_{jj}$ distributions simply add, i.e. one does not observe interference effects. The dip-structure which is present for pure couplings is, thus, washed out.

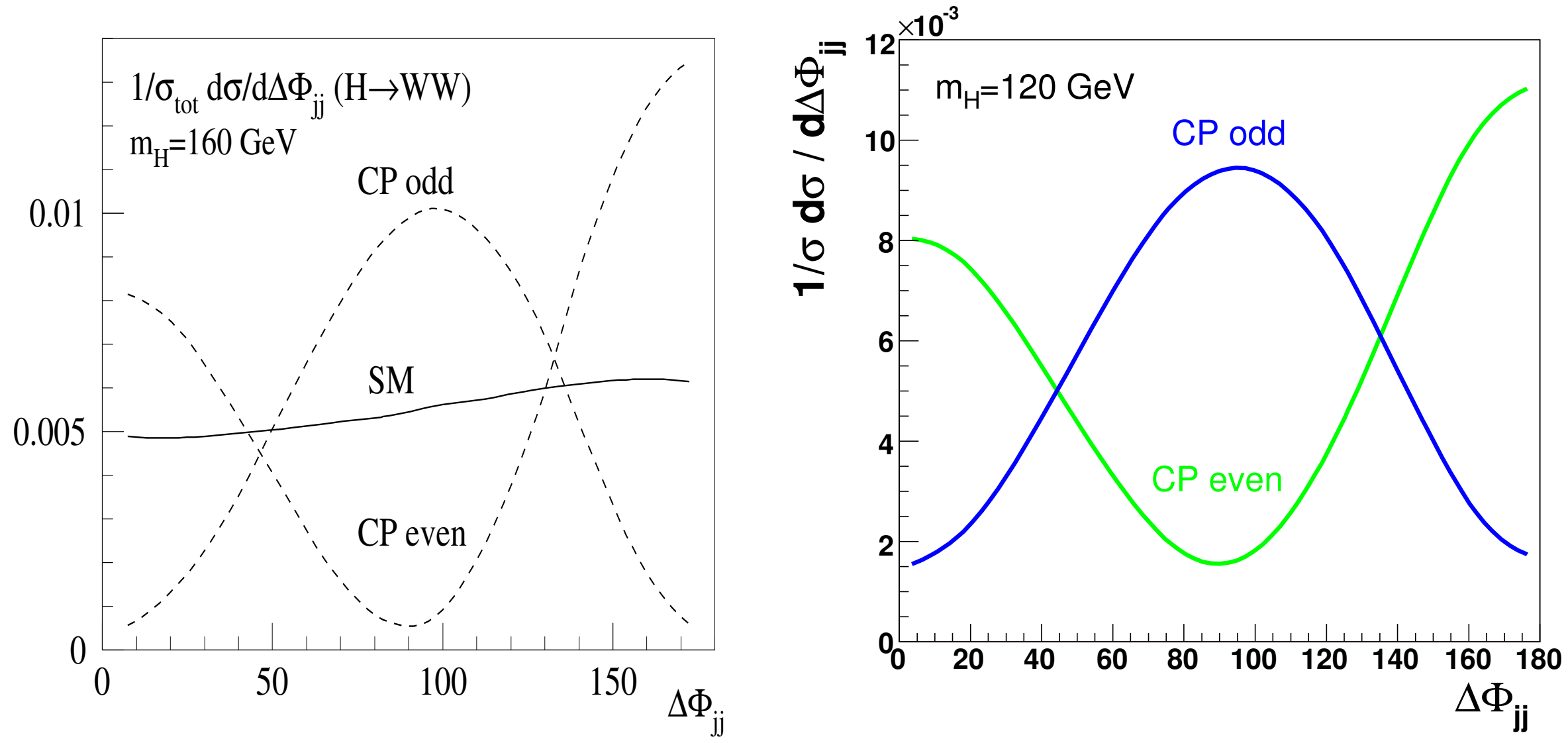

Fig. 2.13: *Left:* Normalized distributions of the azimuthal angle between the two tagging jets, for the $\Phi \to WW \to e\mu\not{p}_T$ signal in vector boson fusion at $m_\Phi = 160$ GeV, from Ref. [98]. Curves, after cuts as in Ref. [276], are for the SM and for single D5 operators as given in Eq. (2.159), i.e. they each assume a single nonzero coupling $a_i$ of Eq. (2.160). *Right:* The same for Higgs production in gluon fusion at $m_\Phi = 120$ GeV. Curves are for CP-even (i.e. SM) and CP-odd $\Phi tt$ couplings.

This behavior is demonstrated in Fig. 2.14. For CP-even and CP-odd couplings of the same strength, i.e. $f_e = f_o$, the azimuthal angle distribution is very similar to the SM case. However, in order to test the presence of anomalous couplings in such cases, other jet distributions can be used, e.g. transverse momentum distributions. The $\triangle\phi_{jj}$ distribution is quite insensitive to variations of form factors, NLO corrections and the like [278]. On the other hand, $p_T$ distributions depend strongly on form factor effects. We study these effects for a particular parameterization of the momentum dependence:

$$a_2(q_1, q_2) = a_3(q_1, q_2) \sim M^2\, C_0\,(\,q_1, q_2, M\,)\,, \tag{2.161}$$

where $C_0$ is the familiar Passarino-Veltman scalar three-point function [279]. This ansatz is motivated by the fact that the $C_0$ function naturally appears in the calculation of one-loop triangle diagrams, where the mass scale M is given by the mass of the heavy particle in the loop. As can be seen in the right panel of Fig. 2.14, even for a mass scale M of the order of 50 GeV the anomalous couplings produce a harder $p_T$ distribution of the tagging jets than is expected for SM couplings. Thus it is possible to experimentally distinguish EW vector boson fusion as predicted in the SM from loop induced $WW\Phi$ or $ZZ\Phi$ couplings by the shape analysis of distributions alone.

Let us now consider the gluon fusion processes where, for $\Phi tt$ couplings of SM strength, one does expect observable event rates from the loop induced effective $\Phi gg$ couplings [277]. In order to assess the visibility of the CP-even vs. CP-odd signatures of the azimuthal jet correlations at the LHC, we consider Higgs + 2 jet production with the Higgs decaying into a pair of $W$-bosons which further decay leptonically, $\Phi \to W^+W^- \to \ell^+\ell^-\nu\bar{\nu}$. We only consider electrons and muons ($\ell = e^\pm$, $\mu^\pm$) in the final state. The Higgs-mass is set to $m_\Phi = 160$ GeV. From previous studies on Higgs production in vector boson fusion [276] the main backgrounds are known to be top-pair production i.e. $pp \to t\bar{t},\ t\bar{t}j,\ t\bar{t}jj$ [280]. The three cases distinguish the number of $b$ quarks which emerge as tagging jets. The $t\bar{t}$ case corresponds to both bottom-quarks from the top-decays being identified as forward tagging jets, for $t\bar{t}j$ production only one tagging jet arises from a $b$ quark, while the $t\bar{t}jj$ cross section corresponds to both tagging jets arising from massless partons. Further backgrounds arise from QCD induced $W^+W^-$ + 2 jet production and electroweak $W^+W^-jj$ production. These backgrounds are calculated as in Refs. [281] and [282], respectively. In the EW $W^+W^-jj$ background, Higgs production in VBF is included, i.e. the VBF Higgs signal is considered as a background to the observation of $\Phi jj$ production in gluon fusion. We do not consider backgrounds from $Zjj,\ Z \to \tau\tau$ and from $b\bar{b}jj$ production because they have been shown to be small in the analyses of Refs. [276, 283].

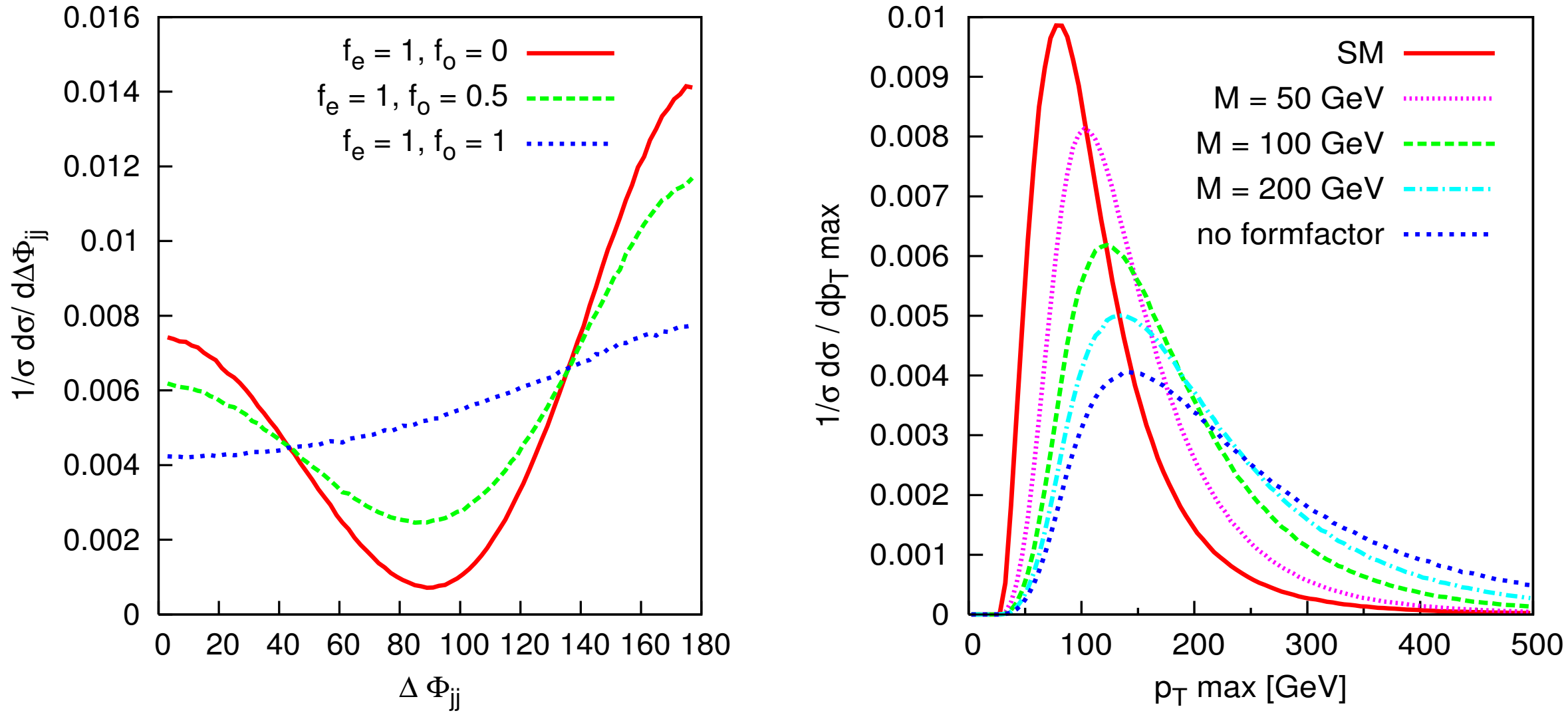

Fig. 2.14: Normalized distributions of the tagging jets in EW vector boson fusion with anomalous couplings and for a Higgs mass of $m_\Phi = 120$ GeV. Typical VBF cuts of $p_{Tj} > 30\,\text{GeV},\ |\eta_j| < 4.5,\ |\eta_{j_1} - \eta_{j_2}| > 4.0,\ m_{jj} >$ 600 GeV are applied. *Left:* Azimuthal angle distribution between the two tagging jets, for different strengths of the operators of Eq. (2.159). *Right:* Transverse momentum distribution of the hardest tagging jet for $f_e = f_o = 1$ and a form factor as in Eq. (2.161). The "no formfactor" curve corresponds to the limit $M \to \infty$, i.e. a constant $a_i$.

The inclusive cuts in Eq. (2.162) reflect the requirement that the two tagging jets and two charged leptons are observed inside the detector, and are well-separated from each other.

$$\begin{aligned} &p_{Tj} > 30\,\text{GeV}, \qquad |\eta_j| < 4.5, \qquad |\eta_{j_1} - \eta_{j_2}| > 1.0 \\ &p_{T\ell} > 10\,\text{GeV}, \qquad |\eta_\ell| < 2.5, \qquad \Delta R_{j\ell} > 0.7 \end{aligned} \tag{2.162}$$

The resulting cross sections for these cuts are shown in Table 2.5. The signal cross section of 121 fb (which includes the branching ratios into leptons) is quite sizeable. The QCD $WWjj$ cross section is about 3 times higher whereas the VBF process reaches 2/3 of the signal rate. The worst source of background arises from the $t\bar{t}$ processes, with a total cross section of more than 17 pb.
In order to improve the signal to background ratio the following selection cuts are applied:

$$\begin{aligned} &p_{T\ell} > 30\,\text{GeV}, \qquad m_{\ell\ell} < 75\,\text{GeV}, \qquad \Delta R_{\ell\ell} < 1.1 \\ &m_T^{WW} < 170\,\text{GeV}, \qquad m_{\ell\ell} < 0.5 \cdot m_T^{WW}\ . \end{aligned} \tag{2.163}$$

Table 2.5: Signal rates and background cross sections for $m_\Phi = 160\,\text{GeV}$. Results are given for the inclusive cuts of Eq. (2.162), the additional selection cuts of Eq. (2.163) and b-quark identification as discussed in the text, and with the additional $\Delta\eta_{jj}$ cut of Eq. (2.166) which improves the sensitivity to the CP nature of the $\Phi tt$ coupling. The events columns give the expected number of events for $\mathcal{L}_{int} = 30\,\text{fb}^{-1}$.

| | inclusive cuts | selection cuts | | selection cuts + Eq. (2.166) | |
|---|---|---|---|---|---|
| process | $\sigma$ [fb] | $\sigma$ [fb] | events / 30 $\text{fb}^{-1}$ | $\sigma$ [fb] | events / 30 $\text{fb}^{-1}$ |
| GF $pp \to \Phi + jj$ | 121.2 | 39.2 | 1176 | 13.1 | 393 |
| VBF $pp \to W^+W^- + jj$ | 75.2 | 20.8 | 624 | 17.4 | 521 |
| $pp \to t\bar{t}$ | 6832 | 29.6 | 888 | 2.0 | 60 |
| $pp \to t\bar{t} + j$ | 9712 | 56.4 | 1692 | 15.6 | 468 |
| $pp \to t\bar{t} + jj$ | 1200 | 8.8 | 264 | 3.2 | 97 |
| QCD $pp \to W^+W^- + jj$ | 364 | 15.2 | 456 | 3.9 | 116 |

Here, the transverse mass of the dilepton-$\not{p}_T$ system is defined as [276]

$$m_T^{WW} = \sqrt{(\not{E}_T + E_{T,\ell\ell})^2 - (\vec{p}_{T,\ell\ell} + \vec{\not{p}}_T)^2} \tag{2.164}$$

in terms of the invariant mass of the two charged lepton and the transverse energies

$$E_{T,\ell\ell} = (p_{T,\ell\ell}^2 + m_{\ell\ell}^2)^{1/2}, \qquad \not{E}_T = (\not{p}_T^2 + m_{\ell\ell}^2)^{1/2}. \tag{2.165}$$

In addition to these cuts we make use of a b-veto to reduce the large top-background. We reject all events where at least one jet is identified as a b-jet. Using numbers from Ref. [284], we assume b-tagging efficiencies in the range of $60\% - 75\%$ (depending on b-rapidity and transverse momentum) and an overall mistagging probability of $10\%$ for light partons.
With the selection cuts (2.163) and the b-veto the backgrounds can be strongly suppressed. Table 2.5 shows the resulting cross sections and the expected number of events for an integrated luminosity of $\mathcal{L}_{int} = 30\,\text{fb}^{-1}$. The signal rate is reduced by a factor of 3 but the backgrounds now have cross sections of the same order as the signal. The largest background still comes from the $t\bar{t}$ processes, especially $t\bar{t}+1j$. For $30\,\text{fb}^{-1}$ we get about 1000 signal events on top of 4000 background events. This corresponds to a purely statistical significance of the gluon fusion signal of $S/\sqrt{B} \approx 18$ and a sufficient number of events to analyze the azimuthal jet correlations.

Figure 2.15 shows the expected $\triangle\phi_{jj}$ distribution for $30\,\text{fb}^{-1}$. Plotted are signal events on top of the various backgrounds. An additional cut on the rapidity gap between the jets

$$|\eta_{j_1} - \eta_{j_2}| > 3.0 \tag{2.166}$$

has been applied. It enhances the shape of the distribution that is sensitive to the nature of the $\Phi tt$ coupling. Clearly visible, the distribution for the CP-even coupling has a slight minimum at $\triangle\phi_{jj} = 90°$ whereas for the CP-odd case there is a pronounced maximum. In order to quantify this, we define the fit-function

$$f(\triangle\phi) = C \cdot (1 + A \cdot \cos 2\triangle\phi + B \cdot \cos \triangle\phi) \tag{2.167}$$

with free parameters $A$, $B$, $C$. The fit is shown as black curves in Fig. 2.15. The parameter $A$ is now a measure for the $\triangle\phi_{jj}$ asymmetry, i.e. whether there is a CP-even or CP-odd $\Phi tt$ coupling. The fitted values are $A = 0.064 \pm 0.035$ for the CP-even and $A = -0.157 \pm 0.034$ for the CP-odd case, while $A_B = -0.039 \pm 0.040$ for the sum of all backgrounds. Defining a significance $s$ as

$$s = \frac{(A_{S+B} - A_B)}{\Delta A_{S+B}}, \tag{2.168}$$

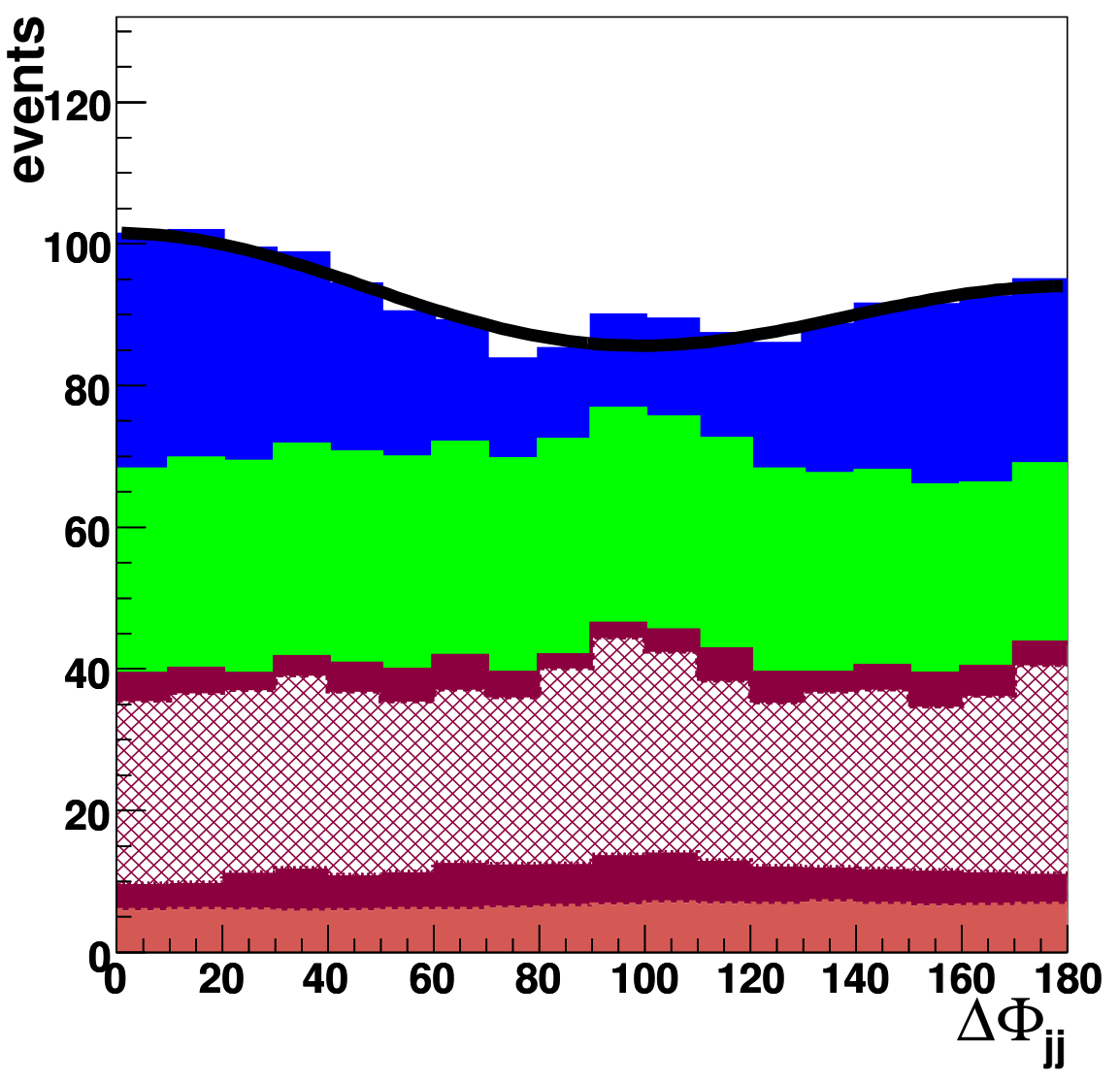

Fig. 2.15: Distribution of the azimuthal angle between the tagging jets in $\Phi jj$ events for a CP-even (*left*) and a CP-odd (*right*) $\Phi tt$ coupling. Shown are expected signal and background events per 10 degree bin for $\Phi \to W^+W^- \to \ell^+\ell^-\nu\bar{\nu}$ and $\mathcal{L}_{int} = 30\,\text{fb}^{-1}$ for the cuts of Eqs. (2.162, 2.163, 2.166) and an applied b-veto. Processes from top to bottom: gluon fusion (signal), VBF, $t\bar{t}$, $t\bar{t}j$, $t\bar{t}jj$, QCD $WWjj$. $m_\Phi = 160$ GeV is assumed.

we get $s = 3.0$ and $s = -3.4$ for the CP-even and CP-odd case, respectively. Thus, a distinction of a CP-odd and CP-even $\Phi tt$ coupling is possible at a $6\sigma$ level for the considered process and a Higgs mass of 160 GeV. This implies that, at least for favorable values of the Higgs boson mass, (i) an effective separation of VBF and gluon fusion sources of $\Phi jj$ events is possible and (ii) the CP nature of the $\Phi tt$ coupling of Eq. (2.157) can be determined at the LHC.

## 2.11 CP-violating Higgs bosons decaying via $H \to ZZ \to 4$ leptons at the LHC

*Rohini M. Godbole, David J. Miller, Stefano Moretti and Margarete M. Mühlleitner*

In this contribution, we study the decay of a Higgs boson to a pair of real and/or virtual $Z$ bosons which subsequently decay into pairs of fermions, $H \to ZZ \to (f_1\bar{f}_1)(f_2\bar{f}_2)$, where $f_1$ and $f_2$ are distinguishable. This channel is particularly important at the LHC for Higgs masses $M_H > 2M_Z$, where the $Z$ bosons are produced on-shell, but is also of use for smaller Higgs boson masses where one of the $Z$ bosons must be virtual [285].

To do a model-independent analysis we examine the most general vertex for a spin-0 boson coupling to two $Z$ bosons, including possible CP violation, which can be written as

$$\frac{ig}{m_Z \cos\theta_W}\left[a\, g_{\mu\nu} + b\,(k_{2\mu}k_{1\nu} - k_1\cdot k_2 g_{\mu\nu}) + c\,\epsilon_{\mu\nu\alpha\beta}{k_1}^{\alpha}{k_2}^{\beta}\right], \tag{2.169}$$

with $k_1$ and $k_2$ the four-momenta of the two $Z$ bosons, and $\theta_W$ the weak-mixing angle, c.f. Eq. (2.63) in the introduction. The form factors $b$ and $c$ may be complex, but since an overall phase will not effect the observables studied here, we are free to adopt a convention where $a$ is real. These form factors can arise from radiative loop corrections or from new physics at the TeV scale, i.e. from higher dimensional operators [98], and may themselves be functions of the momenta. The terms associated with $a$ and $b$ are CP-even, while that associated with $c$ is CP-odd. $\epsilon_{\mu\nu\alpha\beta}$ is totally antisymmetric with $\epsilon_{0123} = 1$. CP violation will be realized if at least one of the CP-even terms is present (i.e. either $a \neq 0$ and/or $b \neq 0$) and $c$ is non-zero. In the following, for the sake of simplicity, we will always assume $b = 0$.

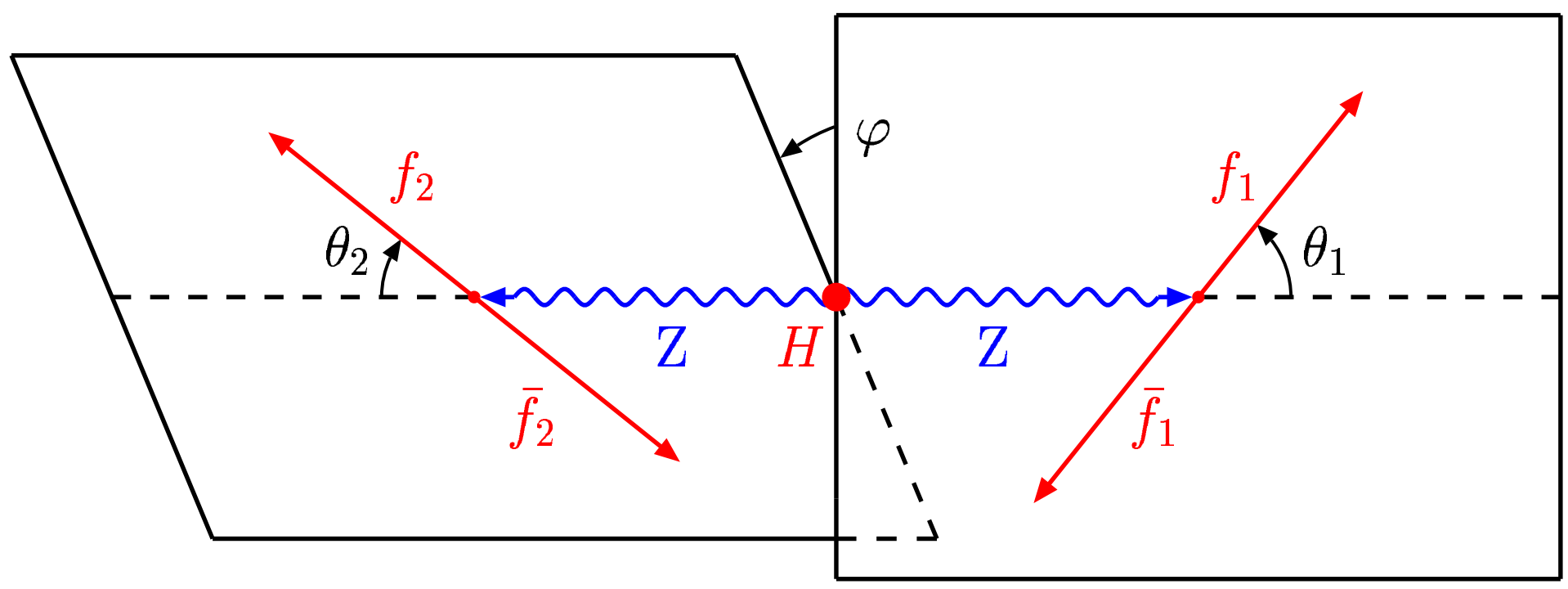

Fig. 2.16: The definition of the polar angles $\theta_i$ $(i = 1, 2)$ and the azimuthal angle $\varphi$ for the sequential decay $H \to Z^{(*)}Z \to (f_1\bar{f}_1)(f_2\bar{f}_2)$ in the rest frame of the Higgs boson.

For $b = 0$ this differs from the vertex of Refs. [93, 136] in the CP-odd term by a factor of 2, and differs from that of Refs [119, 286, 287] and section 2.12 in the choice of $m_Z$ as a normalization factor instead of $m_H$. For further related studies relevant to the LHC also see Refs. [98, 288, 289] and section 2.13; for those relevant to $e^+e^-$ colliders see Refs. [94–97, 134, 151]; for a study at a photon collider see Ref. [147] and section 2.14.

The Standard Model at tree-level is recovered for $a = 1$ and $b = c = 0$, which is obviously CP conserving. Nevertheless, it is interesting to ask if the LHC will be sensitive to any exotic new physics which might provide a CP violating $HZZ$ vertex of this form.

### *2.11.1 The distributions sensitive to CP violation*

In order to fully test for the occurance of CP violation in the $HZZ$ vertex it is helpful to find asymmetries which probe the real and imaginary parts of $c$. The real part of $c$ is probed by any observable which is CP odd and $\tilde{\mathrm{T}}$ odd (where $\tilde{\mathrm{T}}$ denotes pseudo-time-reversal, which reverses particle momenta and spin but does not interchange initial and final states), while the imaginary part is probed by any observable which is CP odd and $\tilde{\mathrm{T}}$ even. The nonvanishing of the CP$\tilde{\mathrm{T}}$ odd coefficients is related to the presence of absorptive parts in the amplitude [290].

An observable sensitive to $\mathrm{Im}\,(c)$ can be found by looking at the polar angular distributions of the process. We denote the polar angles of the fermions $f_1, f_2$ in the rest frames of the $Z$ bosons by $\theta_1$ and $\theta_2$ and define,

$$O_1 \equiv \cos\theta_1 = \frac{(\vec{p}_{\bar{f}_1} - \vec{p}_{f_1}) \cdot (\vec{p}_{\bar{f}_2} + \vec{p}_{f_2})}{|\vec{p}_{\bar{f}_1} - \vec{p}_{f_1}||\vec{p}_{\bar{f}_2} + \vec{p}_{f_2}|} \tag{2.170}$$

where $\vec{p}_f$ are the three-vectors of the corresponding fermions with $\vec{p}_{f_1}$ and $\vec{p}_{\bar{f}_1}$ in their parent $Z$'s rest frame but $\vec{p}_{f_2}$ and $\vec{p}_{\bar{f}_2}$ in the Higgs rest frame, see Fig. 2.16. The angular distribution in $\theta_i$ $(i = 1, 2)$ for a CP-odd state is $\sim (1 + \cos^2\theta_i)$, corresponding to transversely polarized $Z$ bosons, which is very distinct from the purely CP-even distribution proportional to $\sin^2\theta_i$ for longitudinally polarized $Z$ bosons in the large Higgs mass limit [97, 134]. $\mathrm{Im}\,(c) \neq 0$ will introduce a term linear in $\cos\theta_i$ leading to a forward-backward asymmetry. The distribution for $\cos\theta_1$ is shown in Fig. 2.17 for a Higgs mass of 200 GeV and a purely scalar, purely pseudoscalar and CP-mixed scenario. The asymmetry is absent if CP is conserved (for both CP-odd and CP-even states) but is non-zero if $\mathrm{Im}\,(c) \neq 0$ while simultaneously $a \neq 0$. This may then act as a definitive signal of CP violation in this vertex. However, note that this observable requires one to distinguish between $f_1$ and $\bar{f}_1$.

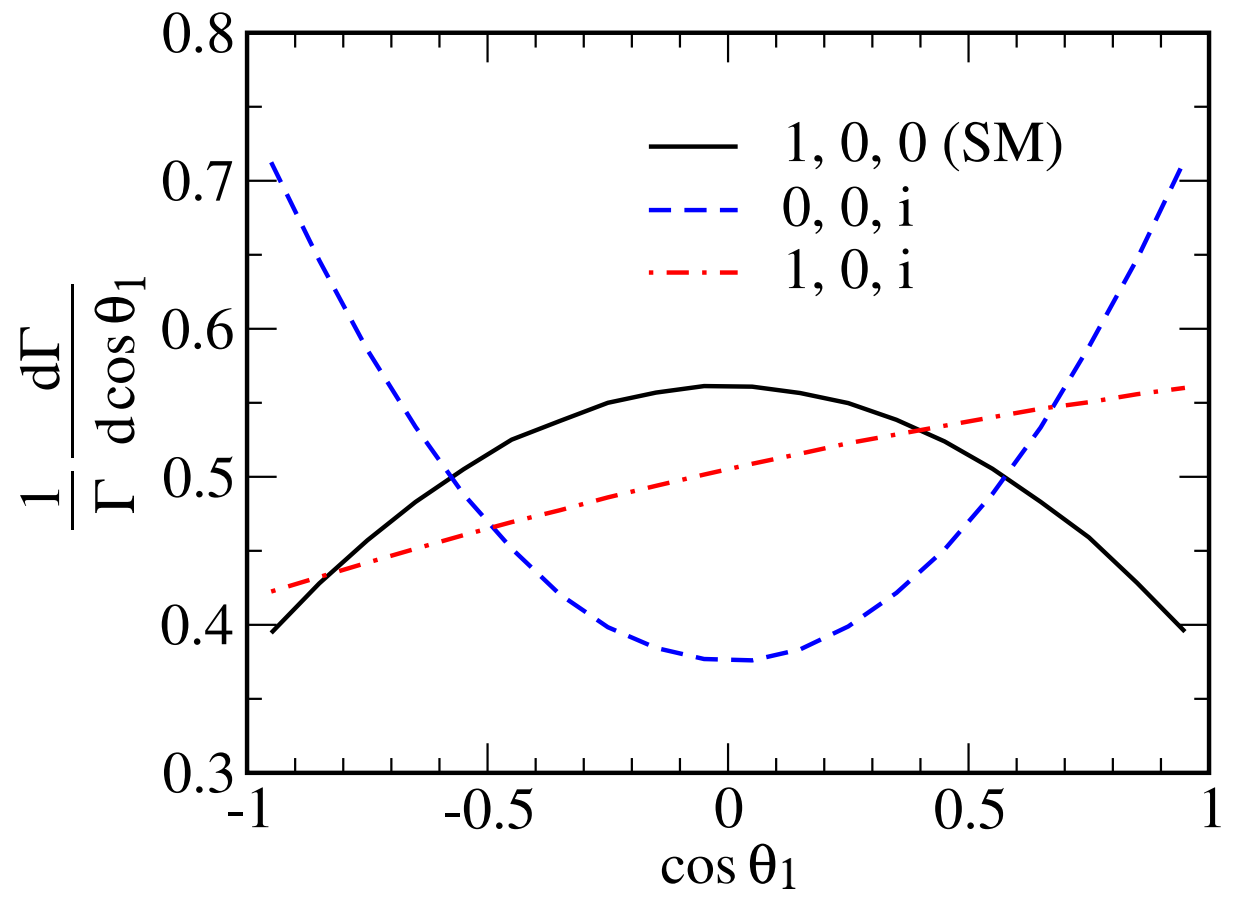

Fig. 2.17: The normalized differential width for $H \to ZZ \to (f_1\bar{f}_1)(f_2\bar{f}_2)$ with respect to the cosine of the fermion's polar angle. The solid (black) curve shows the SM ($a = 1$, $b = c = 0$) while the dashed (blue) curve is a pure CP-odd state ($a = b = 0$, $c = i$). The dot-dashed (red) curve is for a state with a CP violating coupling ($a = 1$, $b = 0$, $c = i$). One can clearly see an asymmetry about $\cos\theta_1 = 0$ for the CP violating case.

To quantify this we define an asymmetry by

$$\mathcal{A}_1 = \frac{\Gamma(\cos\theta_1 > 0) - \Gamma(\cos\theta_1 < 0)}{\Gamma(\cos\theta_1 > 0) + \Gamma(\cos\theta_1 < 0)}. \tag{2.171}$$

In the case of no CP violation $\mathcal{A}_1 = 0$, whereas any significant deviation from zero will be a sign that CP is violated. Fig. 2.18 *(left)* shows the value of $\mathcal{A}_1$ for a Higgs mass of 200 GeV as a function of the ratio $\mathrm{Im}\,(c)/a$. The value $\mathrm{Im}\,(c)/a = 0$ corresponds to the purely scalar state whereas $\mathrm{Im}\,(c)/a \to \infty$ to the purely CP-odd case. It is clear from Eq. (2.171) that $\mathcal{A}_1$ is sensitive only to the relative size of the couplings since any factor will cancel in the ratio. We find that the asymmetry is maximal for $\mathrm{Im}\,(c)/a \sim 1.4$ with a value of about $0.077$.

In order to get a first rough estimate whether this asymmetry can be measured at the LHC we calculate the significance with which a particular CP violating coupling would manifest at the LHC. In the purely SM case, we assumed that 100 fb$^{-1}$ provide 180 signal events containing $H \to ZZ \to 4$ leptons after cuts to remove background [285] (all production channels). We then divide this number by two to provide an estimate for $H \to ZZ \to e^+e^-\mu^+\mu^-$ (since we need to distinguish the leptons) and scaled this number up to 300 fb$^{-1}$ (i.e. giving 270 events). The number of events for the CP violating case has been obtained by multiplying the number of SM events by the ratio of CP violating to SM cross sections. We are therefore assuming the SM value for the CP even coefficient, i.e. $a = 1$. For simplicity we assume the charge of the particles is unambiguously determined.

Fig. 2.18 *(right)* shows the significance as a function of $\mathrm{Im}\,(c)$, calculated according to $\mathcal{A}_1\sqrt{N}$ where $N$ is the number of expected events. The maximum of the curve is slighlty shifted to higher values of $\mathrm{Im}\,(c)/a$ compared to Fig. 2.18 *(left)* due to the increasing Higgs decay rate with rising pseudoscalar coupling. The curve shows that, even in a best case scenario, the signficance is always less than $3\sigma$, so evidence for CP violation cannot be obtained in this channel without more luminosity. However, since one does not need to distinguish $f_2$ and $\bar{f}_2$ one could also consider using jets instead of muons, i.e. $H \to ZZ \to l^+l^-jj$, to increase the statistics. This process deserves further study.

To probe $\mathrm{Re}\,(c)$ we require an observable which is CP odd and $\tilde{\mathrm{T}}$ odd, so we choose to define,

$$O_2 \equiv \frac{(\vec{p}_{\bar{f}_1} - \vec{p}_{f_1}) \cdot (\vec{p}_{\bar{f}_2} \times \vec{p}_{f_2})}{|\vec{p}_{\bar{f}_1} - \vec{p}_{f_1}||\vec{p}_{\bar{f}_2} \times \vec{p}_{f_2}|}. \tag{2.172}$$

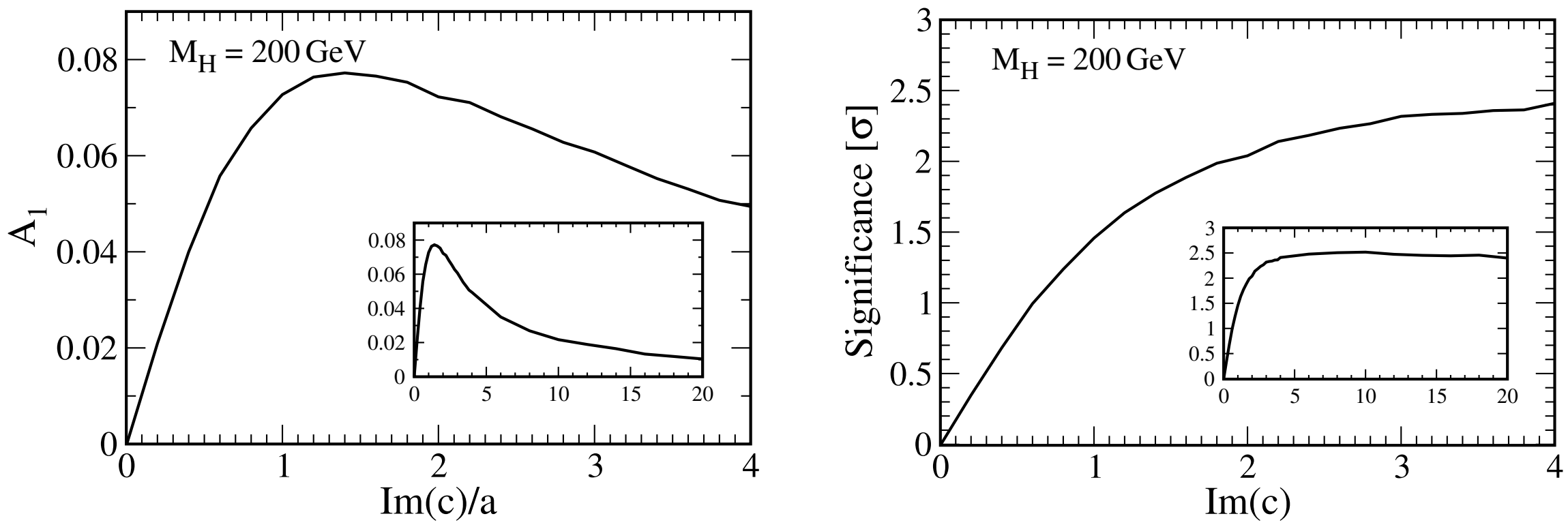

Fig. 2.18: *Left:* The asymmetry given by Eq. (2.171) as a function of the ratio $\mathrm{Im}\,(c)/a$, for a Higgs boson of mass 200 GeV. *Right:* The number of standard deviations the asymmetry deviates from zero as a function of $\mathrm{Im}\,(c)$. The inserts show the same quantities for a larger range of $\mathrm{Im}\,(c)$.

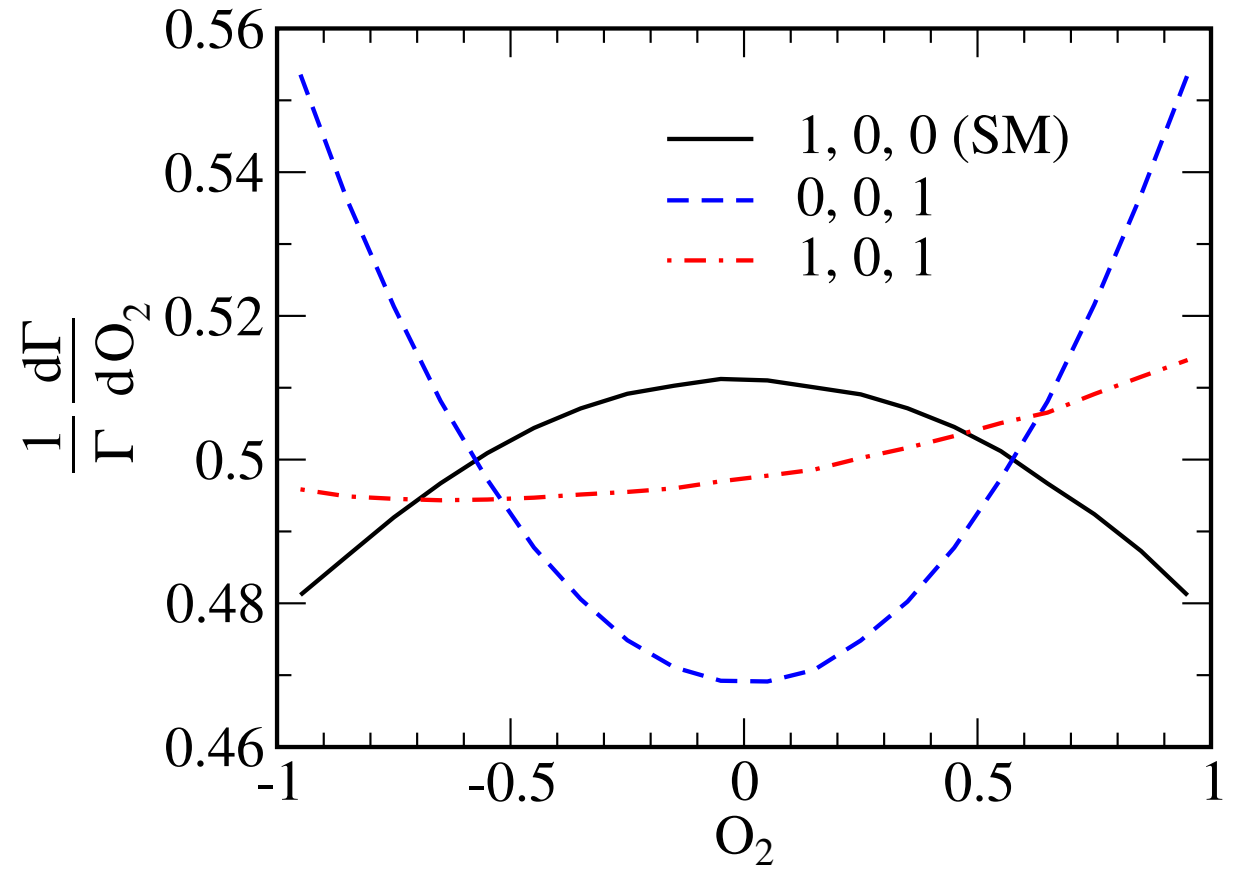

Fig. 2.19: The normalized differential width for $H \to ZZ \to (f_1\bar{f}_1)(f_2\bar{f}_2)$ with respect to the observable $O_2$ (see text). The solid (black) curve shows the SM ($a = 1$, $b = c = 0$) while the dashed (blue) curve is a pure CP-odd state ($a = b = 0$, $c = 1$). The dot-dashed (red) curve is for a state with a CP violating coupling ($a = 1$, $b = 0$, $c = 1$). Again one sees an asymmetry about zero for the CP violating case.

The dependence of the differential width on this observable is plotted in Fig. 2.19 but while an asymmetry is indeed present, it is very small and will be difficult to see in practice. The corresponding asymmetry is

$$\mathcal{A}_2 = \frac{\Gamma(O_2 > 0) - \Gamma(O_2 < 0)}{\Gamma(O_2 > 0) + \Gamma(O_2 < 0)}, \tag{2.173}$$

which is plotted in Fig. 2.20 *(left)* as a function of $\mathrm{Re}\,(c)/a$. The significance (as calculated for $\mathcal{A}_1$ above) is shown in Fig. 2.20 *(right)*. The significance is always very small, and it is difficult to see how this could provide useful information. In this case one cannot exploit the decay of Higgs bosons to jets since one must also distinguish $f_2$ and $\bar{f}_2$.

Another distribution sensitive to CP violation is the azimuthal angular distribution $d\Gamma/d\varphi$ where $\varphi$ denotes the angle between the planes of the fermion pairs stemming from the $Z$ boson decays, cf. Fig. 2.16. Whereas the purely SM case shows a distribution

$$\frac{d\Gamma}{d\varphi} \sim 1 + A\cos\varphi + B\cos 2\varphi, \tag{2.174}$$

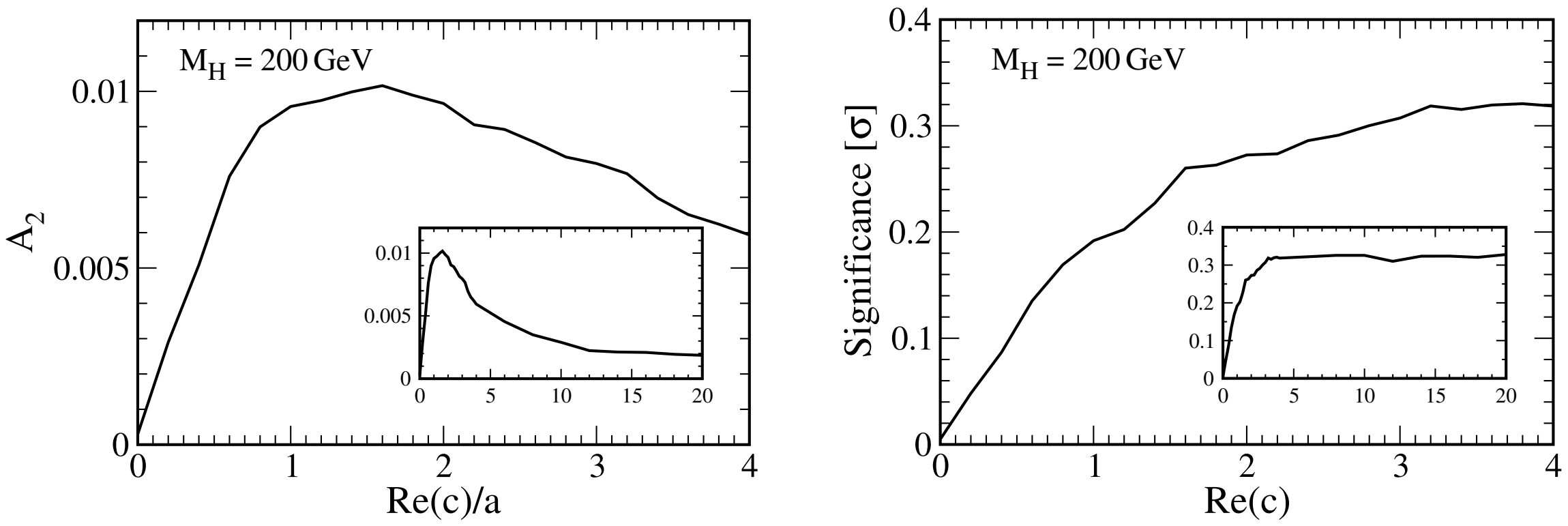

Fig. 2.20: *Left:* The asymmetry given by Eq. (2.173) as a function of the ratio $\mathrm{Re}\,(c)/a$, for a Higgs boson of mass 200 GeV. *Right:* The number of standard deviations the asymmetry deviates from zero as a function of $\mathrm{Re}\,(c)$. The inserts show the same quantities for a larger range of $\mathrm{Re}\,(c)$.

where the coefficients $A$ and $B$ are functions of the Higgs and $Z$ boson mass (see Ref. [93]), in the purely pseudoscalar case

$$\frac{d\Gamma}{d\varphi} \sim 1 - \frac{1}{4}\cos 2\varphi. \tag{2.175}$$

In the CP violating case we must include contributions from both the scalar and pseudoscalar couplings which will alter this behaviour. Knowing the Higgs mass from previous measurements, any deviation from the predicted distribution in the scalar/pseudoscalar case will be indicative of CP violation. This can be inferred from Fig.2.21 which shows the azimuthal angular distribution for $M_H = 200\,\mathrm{GeV}$ in the SM case, for a CP-odd Higgs boson and two CP violating cases. The purely CP-odd curve will always show the same behaviour independently of the value of $c$ since the curves are normalized to unit area. Therefore a special value of $c$ could not fake the flattening of the curve appearing in the CP violating examples. This flattening even leads to an almost constant distribution in $\varphi$ for the case $c/a = 1$. It should be kept in mind, though, that this method cannot be applied for large Higgs masses where the $\varphi$ dependence disappears in the SM. One must also beware of degenerate Higgs bosons of opposite CP; since one cannot distinguish which Higgs boson is in which event, one must add their contributions together, possibly mimicking the effect seen above.

This procedure is similar to that of Sections 2.12 and 2.13 where log-likelihood functions were constructed and minimised to extract the coefficients in the vertex or yield exclusion contours.

The next step will be to study in a more realistic simulation how well the $\varphi$ distribution can be fitted at the LHC and hence to which extent CP violation can be probed in the azimuthal angular distribution.

### *2.11.2 Summary and Outlook*

We have studied the decays of Higgs bosons into a pair of $Z$ bosons, which subsequently decay into leptons, for a general $HZZ$ coupling at the LHC. We examined CP violating asymmetries which probe the real and imaginary couplings of the CP-odd term. We found that the asymmetries produced are small and will not provide evidence of CP violation at the LHC without higher luminosity. However, it may be possible to exploit other channels, such as Higgs decays to leptons and jets, to increase significances. We also examined the dependence on the azimuthal angle between the lepton planes, which is similarly indicative of CP violation. Further studies of this azimuthal angle and the extension to arbitrary higher "Higgs" spin will be the subject of future work.

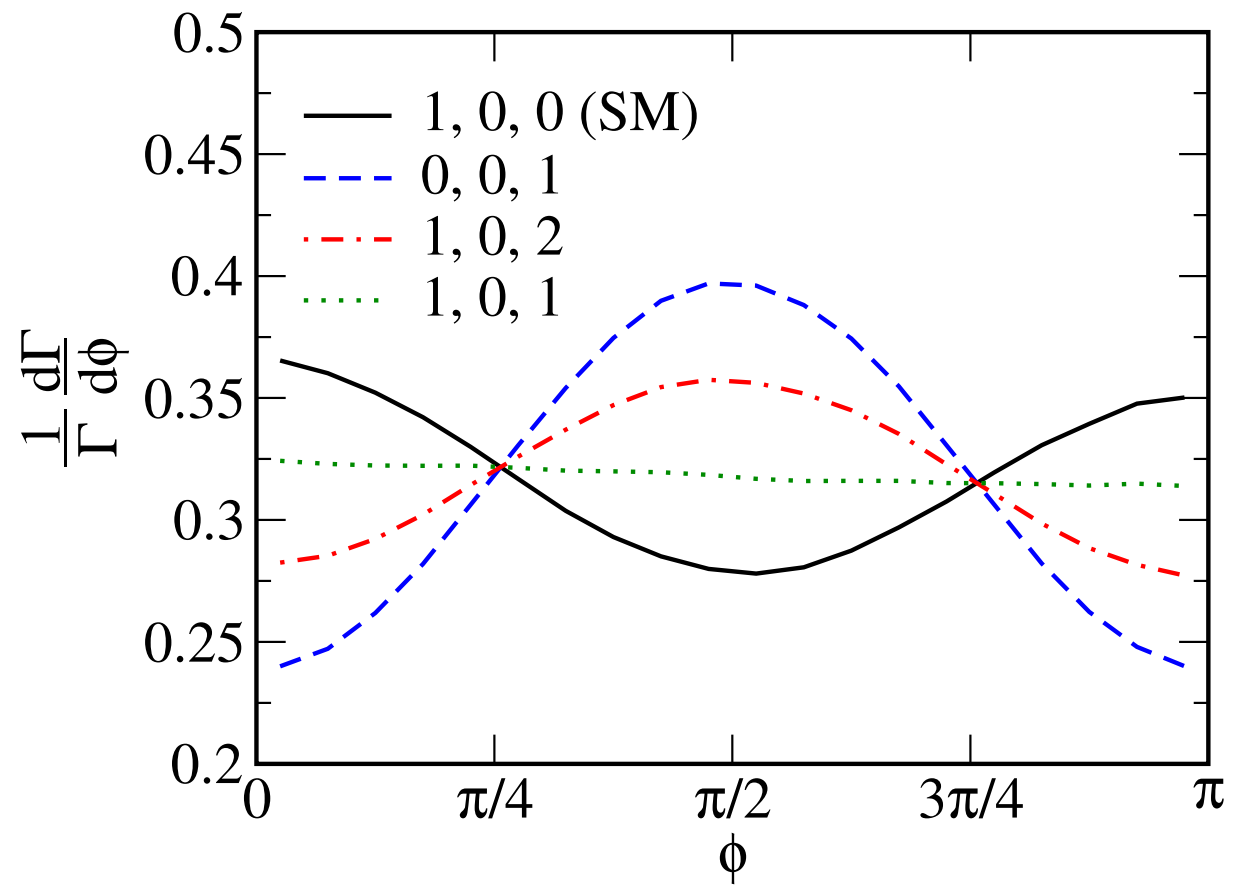

Fig. 2.21: The normalized differential width for $H \to Z^{(*)}Z \to (f_1\bar{f}_1)(f_2\bar{f}_2)$ with respect to the azimuthal angle $\varphi$. The solid (black) curve shows the SM ($a = 1$, $b = c = 0$) while the dashed (blue) curve is a pure CP-odd state ($a = b = 0$, $c = 1$). The dot-dashed (red) curve and the dotted (green) curve are for states with CP violating couplings $a = 1, b = 0$ with $c = 2$ and $c = 1$, respectively.

### 2.12 Testing the spin and CP properties of a SM-like Higgs boson at the LHC

*Claus P. Buszello and Peter Marquard*

To confirm the properties of a Higgs-like particle found at the LHC, we study the angular distributions of the final state particles in the decay $H \to ZZ \to 4\ell$. To this end we consider hypothetical couplings of the Higgs with momentum $k$ to $Z$ bosons with momenta $p, q$ reflecting the different spin/CP states. We use a parametrisation of the couplings as follows

$$\mathcal{L}_{scalar} = \mathbf{X} g_{\mu\nu} + \mathbf{Y} k_\mu k_\nu / m_H^2 + i\mathbf{P}\epsilon_{\mu\nu\rho\sigma} p^\rho q^\sigma / m_H^2 \tag{2.176}$$

for the spin 0 Higgs and

$$\mathcal{L}_{vector} = \mathbf{X}(g_{\rho\mu} p_\nu + g_{\rho\nu} q_\mu) + \mathbf{P}(i\epsilon_{\mu\nu\rho\sigma} p^\sigma - i\epsilon_{\mu\nu\rho\sigma} q^\sigma) \tag{2.177}$$

for the spin 1 case where $\epsilon_{1234} = i$. This parametrisation is discussed in detail in [119]. The scalar couplings in Section 2.11 differ slightly from these by the choice of the masses used to normalize the non-SM contributions. We choose $m_H$ over $m_Z$ as this is more convenient if one wants to use the same parameterisation for $HZZ$ and $HWW$ vertices. We then study the distributions of the final state leptons performing a one and a multi-dimensional analysis.

The analysis presented here is divided in two parts. First, we only consider pure states (i.e all but one of the parameters **X**,**Y** and **P** are zero). The analysis of the feasibility of the exclusion of the pure states is based on a the fast parameterised ATLAS detector simulation [291] . Next, we consider the exclusion of admixtures of the CP-even and CP-odd non-SM contributions. This analysis - detailed in [292] - is the first one that takes all interference terms into account and is based on the same cuts, efficiencies and momentum resolutions as the first part. The event samples for these studies were generated using a new matrix element generator written by the authors implementing the complete couplings including mixtures given above. It generates the decay $H \to ZZ(*) \to 4\ell$ with two on-shell $Z$ bosons in the narrow width approximation above the $ZZ$-threshold and one on-shell and one off-shell $Z$ below. In the following we give the main results of these analyses. The full results and details can be found in [119, 292]. Another one dimensional analysis of the pure states has also been performed in [93] and a PYTHIA based study of CP properties at CMS can be found in Section 2.13. A similar analysis can be carried out in other cases, where the Higgs vector boson vertex is present. In fact, in the $ZZ \to 4\ell$ decay the angular correlations are suppressed compared to $W$ decays or $ZZ \to 2\ell 2q$. This makes exploiting WBF and the Higgs decay to $W$ pairs so interesting. In that case, one can use the forward jets and the leptons from the $W$ decay to determine the spin-parity of the Higgs (see e.g. [287] and Section 2.10).

### *2.12.1 Analysis and results*

We study essentially two distributions. One is the distribution of the cosine of the polar angle, $\cos\theta$, of the decay leptons relative to the $Z$ boson. Because the heavy Higgs decays mainly into longitudinally polarised vector bosons the cross-section $d\sigma/d\cos\theta$ should show a maximum around $\cos\theta$ =0. The other is the distribution of the angle $\phi$ between the decay planes of the two $Z$ bosons in the rest frame of the Higgs boson. This distribution depends on the details of the Higgs decay mechanism. Within the Standard Model, a behaviour roughly like $1 + \beta\ \cos 2\phi$ is expected. This last distribution is flattened in the decay chain $H \to ZZ \to 4\ell$, because of the small vector coupling of the leptons, in contrast to the decay of the Higgs Boson into $W$ pairs or decay of the $Z$ into quarks. The angles under investigation are shown in Fig. 2.16.

I Pure states

The plane-correlation can be parametrised as

$$F(\phi) = 1 + \alpha \cdot \cos\phi + \beta \cdot \cos 2\phi \tag{2.178}$$

In all four cases discussed here, there is no $\sin\phi$ or $\sin 2\phi$ contribution. For the Standard Model Higgs, $\alpha$ and $\beta$ depend on the Higgs mass while they are constant over the whole mass range in the other cases. The polar angle distribution can be described by

$$G(\theta) = T \cdot (1 + \cos^2\theta) + L \cdot \sin^2\theta \tag{2.179}$$

reflecting the longitudinal or transverse polarisations of the $Z$ boson. We define the ratio

$$R := \frac{L - T}{L + T} \tag{2.180}$$

of transversal and longitudinal polarisation.

Figure 2.22 (left) shows the expected values and errors for the parameter R, using an integrated luminosity of 100 fb$^{-1}$. It is clearly visible that for masses above 250 GeV the measurement of this parameter allows the various non-SM hypotheses for the spin and CP-state of the "Higgs Boson" considered here to be unambiguously excluded. For a Higgs mass of 200 GeV only the pseudoscalar is excluded. Fig. 2.22 (right) shows the expected values and errors for $\alpha$ and $\beta$ for a 200 GeV Higgs and an integrated luminosity of 100 fb$^{-1}$.

The parameter $\alpha$ can be used to distinguish between a spin 1 and the SM Higgs particle, but its use is statistically limited. The same applies to the parameter $\beta$. Measuring $\beta$, which is zero for spin 1 and $> 0$ in the SM case, can contribute only very little to the spin measurement even if $m_H$ is in the range where $\beta$, in the SM case, is close to its maximum value. Nevertheless, $\beta$ can be useful to rule out a CP-odd spin 0 particle.

The significance of the parameter $\alpha$ can be improved by exploiting the correlation between the sign of $\cos\theta$ for the two $Z$ Bosons and $\phi$. In Fig. 2.23, we plot the parameters separately for $\mathrm{sign}(\cos\theta_1) = \mathrm{sign}(\cos\theta_2)$ and $\mathrm{sign}(\cos\theta_1) = -\mathrm{sign}(\cos\theta_2)$. As can be seen, the difference in $\alpha$ becomes bigger for $J = 1$ and CP-even. For higher masses $\alpha$ and $\beta$ of the SM Higgs approach 0; thus only $\alpha$ can be used to measure the spin. This is fortunately compensated by the measurement of R.

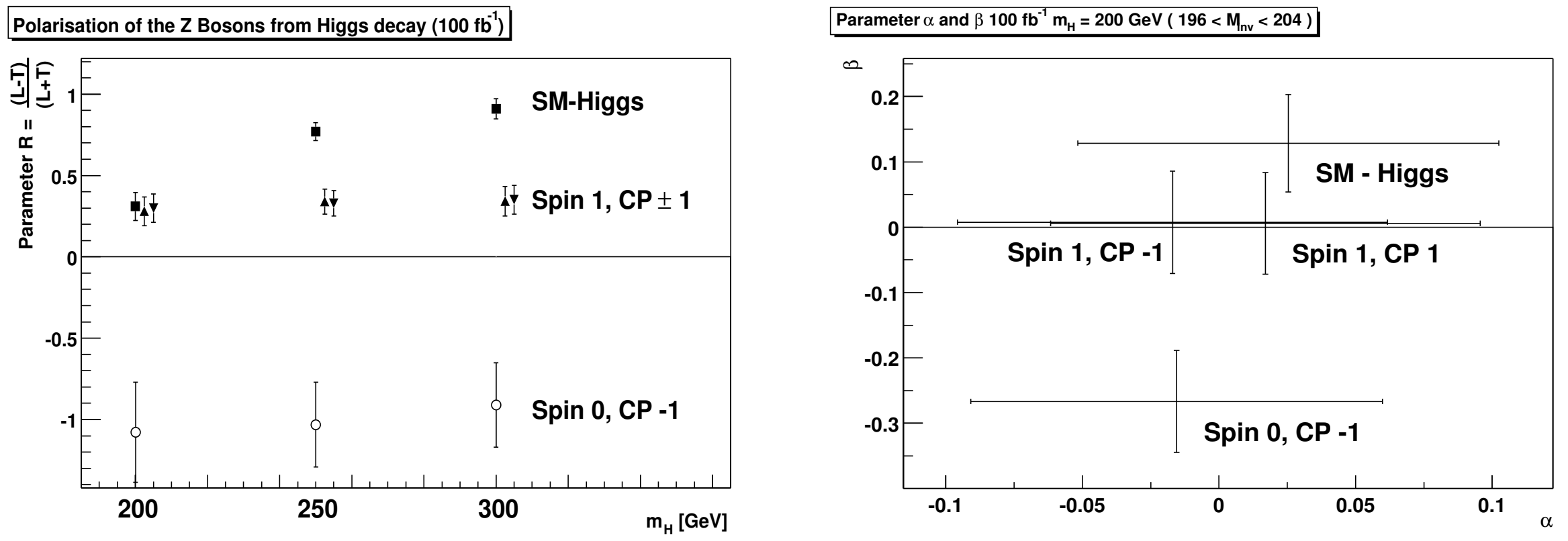

Fig. 2.22: The parameter $R$ for different Higgs massses (left) and $\alpha$ and $\beta$ (right) for $m_H$ = 200 GeV using 100 fb$^{-1}$. The error scales with the integrated luminosity as expected.

Fig. 2.25 shows the significance, i. e. the difference of the expected non-SM value and SM value divided by the expected error of the SM Higgs. We add up the significance for $\alpha$ and $\beta$ for the like-signed und unlike-signed $cos(\theta)$ combinations and plot the resulting significance together with the one from the polar angle measurement in Fig. 2.25 (left). For higher Higgs masses the decay plane angle correlation contributes almost nothing, but the polarisation leads to a good measurement of the parameters spin and CP-eigenvalue. For full luminosity (300 fb$^{-1}$) the significance can simply be multiplied by $\sqrt{3}$ assuming stable detector performance. This is especially interesting for a Higgs mass of 200 GeV. The spin 1, CP-even hypothesis can then be ruled out with a significance of 6.4$\sigma$, while for the spin 1, CP-odd case the significance is still only 3.9$\sigma$.

In principle, the same analysis can be done for Higgs masses below the $ZZ$ threshold. In practice this is complicated by the fact, that the cross-section for $H \to ZZ^*$ is a lot smaller and is further reduced by additional combined impact parameter and isolation cuts needed to suppress the $t\bar{t}$ and $Zb\bar{b}$ backgrounds. Due to this reduction in statistics, the decay plane correlation doesn't yield any useful results, and we limit the discussion to the polar angle and the spin 0 case. Furthermore we will always use an integrated luminosity of 300 fb$^{-1}$, the maximum foreseen for each of the LHC experiments. We use the number of signal and background events published in the ATLAS TDR [115].

The distortion of the polar angle distribution is sizeable, and we have to introduce a statistical correction. The correction reproducing the SM values properly will not necesarily correct the non-SM values back to the theoretical values. In Fig. 2.24 (left) we present the expected values of R after applying the correction to the distributions. The exclusion significance is shown in Fig. 2.25 (top right).

An additional distribution that is only available below the threshold, is the distribution of the off-shell $Z$ mass. Fig. 2.24 (right) shows this distribution for a Higgs of 150 GeV and the three different spin 0 couplings. The distributions are more robust against the cuts than the polar angle distribution. We generate 300 data samples of the expected number of signal and background events including all cuts for the three different hypothesis, and calculate the $\chi^2$ to the SM-distribution (again after all cuts are applied) for each one of them. The means of these values along with the corresponding confidence levels are plotted against the various Higgs masses in Fig. 2.25.

II Mixed states

In order to measure possible CP-violation in the Higgs to vector boson coupling, we consider the full matrix element including the mixed terms (**PX**, **PY** and **YX**). As in the one dimensional case, we assume the discovery of a signal in the $H \to ZZ$ channel. Since a significant deviation from the expected number of events would rule out a SM Higgs in a trivial way, we further assume, that the number of events seen is compatible with a SM Higgs. A deviation in the number of events would not allow to pinpoint the coupling structure anyway, as it would be a possible combination of effects in production and branching ratios. Instead, we use the angular correlations of the decay products to test for small non-SM contributions to the SM coupling. To give a better physical interpretation to the notion of a small coupling, we rescaled $\mathbf{Y}$ and $\mathbf{P}$ to $\mathbf{Y}'$ and $\mathbf{P}'$ such that now for the widths of the pure states $\Gamma_{P'} = \Gamma_{Y'} = \Gamma_X$. The exact scaling factors can be found in Table 2.6. In this study we demonstrate how CP violation in the $H \to ZZ$ coupling could be ruled out. Figure 2.26 and 2.27 show the exclusion significance for **Y'** and **P'** admixtures to an SM Higgs. By turning this around we can interpret a measurement of **P'** and **Y'** outside these boundaries as proof of a non-SM Higgs coupling to vector bosons.

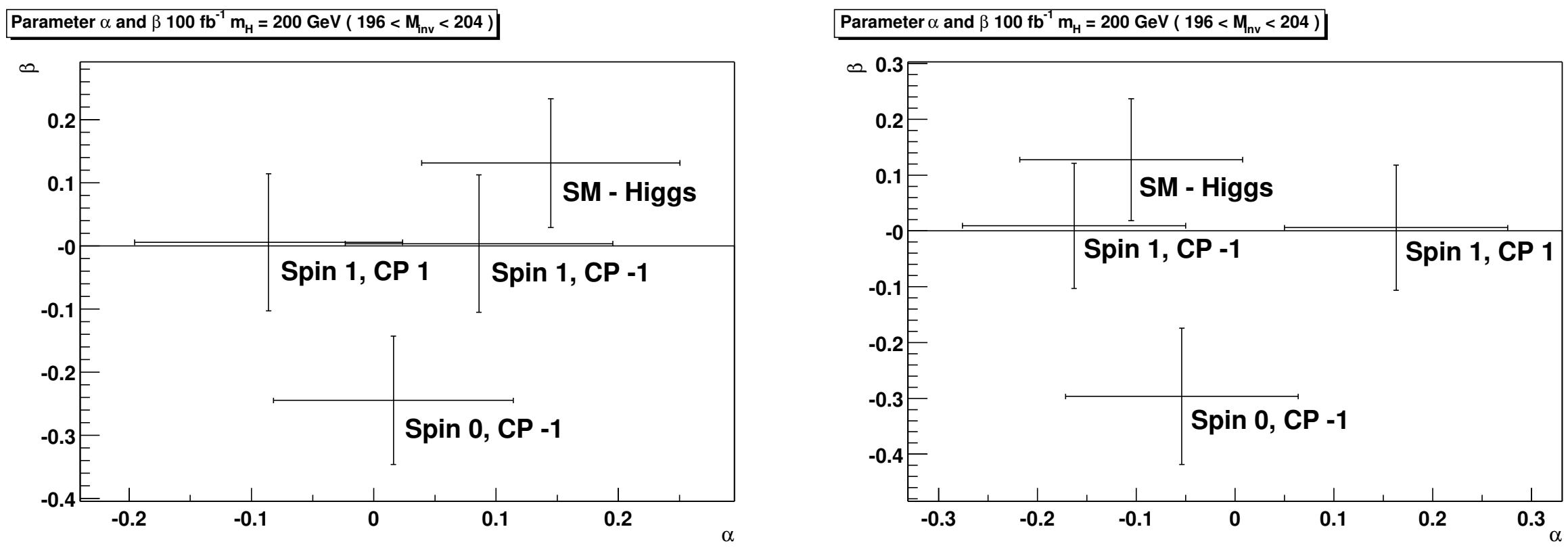

Fig. 2.23: The parameter $\alpha$ depends on the signs of the $\cos(\theta)$ of the two $Z$ bosons. The events where the signs are equal are used for the left plot, those where the signs are different are used for the right plot.

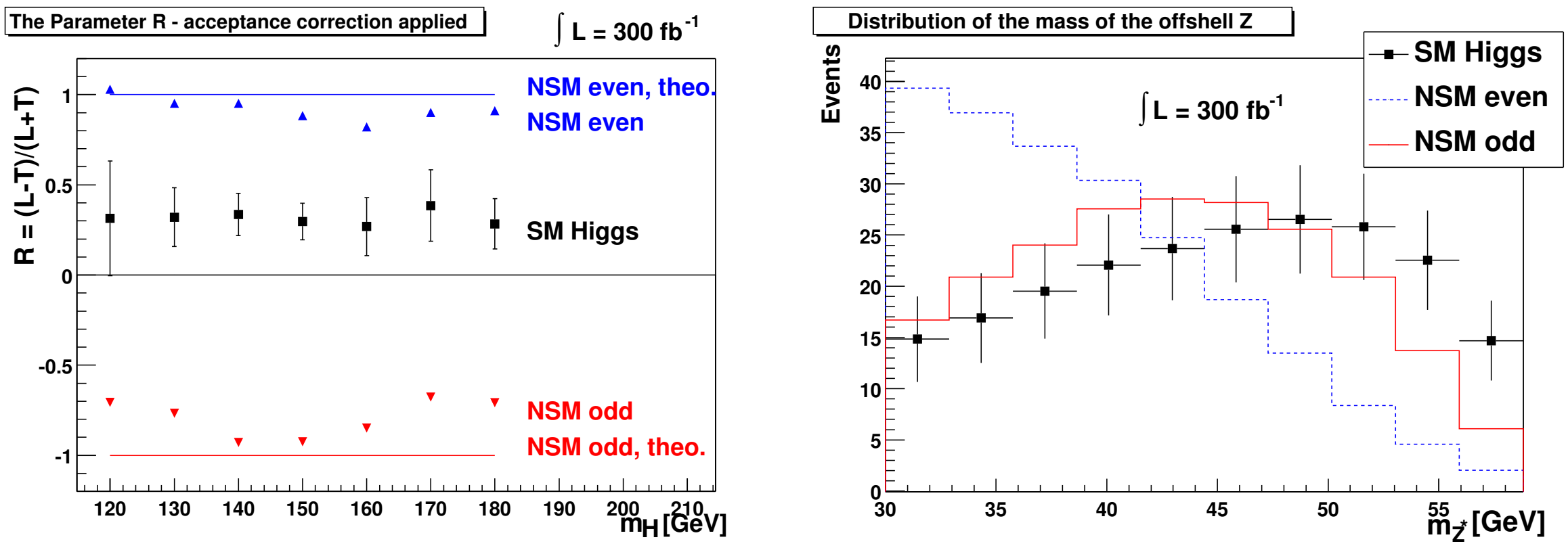

Fig. 2.24: Left: The expected values for the parameter R after reconstruction and signal selection and with a correction for detector effects applied, so that the SM Higgs values are recovered. The errorbars reflect the expected statistical error for the SM case using 300 fb$^{-1}$. Right: The off-shell $Z$ mass distribution for a 150 GeV Higgs. NSM even refers to states with **Y**=1, **P**=0, **X**=0. NSM odd refers to **Y**=0, **P**=1, **X**=0.

We use the full information from the three fold differential cross-section by constructing the following likelihood function:

$$L(\mathbf{Y},\mathbf{P}) = \sum_{k\in\text{events}} \log \frac{|\mathcal{M}|^2(\phi^k,\theta_1^k,\theta_2^k,\mathbf{P},\mathbf{Y},\mathbf{X}=1)}{\int |\mathcal{M}|^2(\phi,\theta_1,\theta_2,\mathbf{P},\mathbf{Y},\mathbf{X}=1)d\phi d\cos\theta_1 d\cos\theta_2} \tag{2.181}$$

where $|\mathcal{M}|^2$ is the squared matrix element evaluated at leading order. The value of $\mathbf{X}$ is always fixed to the SM value of 1, since we want to measure small contributions from non-standard couplings. By maximising the likelihood we expect to find a value of zero for $\mathbf{P}$ and $\mathbf{Y}$. In order to demonstrate the potential of measuring these parameters with ATLAS we show contour plots of the expected exclusion limits (see Fig. 2.26 and 2.27). The full luminosity of 300 fb$^{-1}$ has been used for all plots. The background has been statistically subtracted where the distribution of the background considered in this study was computed with PYTHIA [293]. The distortion of the signal is not negligible, but since the contributions of the non standard model couplings are small the distortions don't vary much. Therefore the expected likelihood distributions are affected only slightly by the detector effects. We do not include any corrections for this effect, which is visible as a small shift of the maximum in positive Y' direction. The plots were achieved by fitting to the whole dataset and as a check to many small samples with the expected number of events (pseudo-experiments). The results from the two methods agree perfectly. A remarkable feature of the contour-plots is the V-form in the $\mathbf{Y} - \mathbf{P}$ plane. This form is understandable, because some combinations of $\mathbf{Y}$ and $\mathbf{P}$ couplings behave very similar to the standard model coupling $\mathbf{X}$. Therefore, neglecting the $\mathbf{Y}$ term in the determination of CP violating contributions could lead to wrong results.

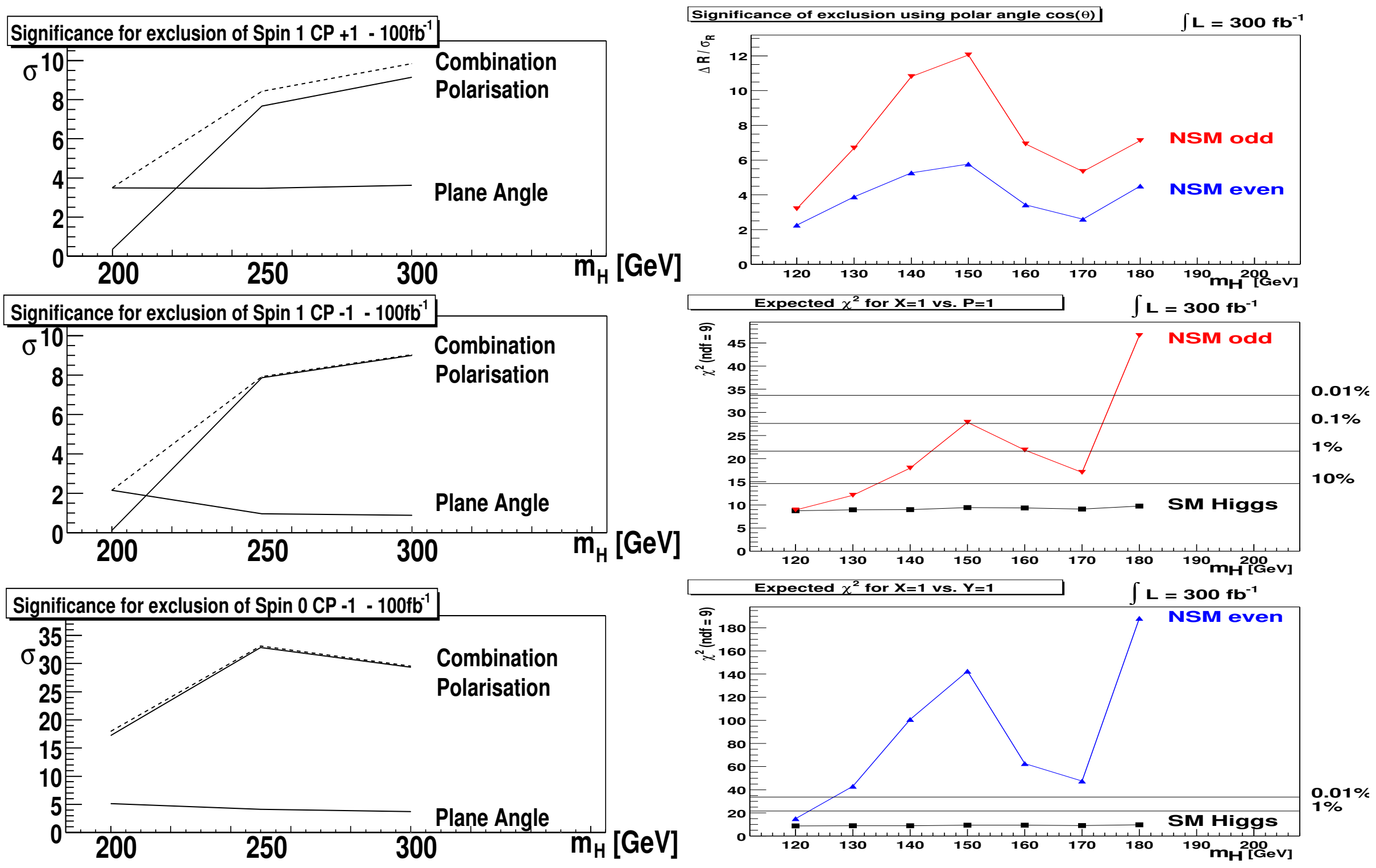

Fig. 2.25: Left: The overall significance for the exclusion of the non standard spin and CP-eigenvalue. The significance from the polar angle measurement and the decay-plane-correlation are plotted separately. Right: The exclusion significance for the non-SM cases for various Higgs masses. The top figure shows the exclusion using the polarisation of the $Z$. The middle and bottom ones show the exclusion from the distribution of the off-shell $Z$ mass distribution for the pseudoscalar and the scalar non-SM couplings. NSM even refers to states with **Y**=1, **P**=0, **X**=0. NSM odd refers to **Y**=0, **P**=1, **X**=0.

Table 2.6: Ratio of the roots of the total widths of the pure states for various Higgs masses $m_H$. These ratios can be used to scale the constants **P** and **Y** such, that the non standard model couplings are of the same strength as the standard model coupling.

| $\mathbf{m_H}$ **[GeV]** | 130 | 140 | 150 | 160 | 170 | 180 | 200 | 250 | 300 |
|---|---|---|---|---|---|---|---|---|---|
| $\sqrt{\frac{\Gamma_Y}{\Gamma_{SM}}}$ | 0.093 | 0.106 | 0.116 | 0.092 | 0.106 | 0.066 | 0.102 | 0.284 | 0.368 |
| $\sqrt{\frac{\Gamma_P}{\Gamma_{SM}}}$ | 0.106 | 0.117 | 0.125 | 0.123 | 0.126 | 0.102 | 0.146 | 0.156 | 0.121 |

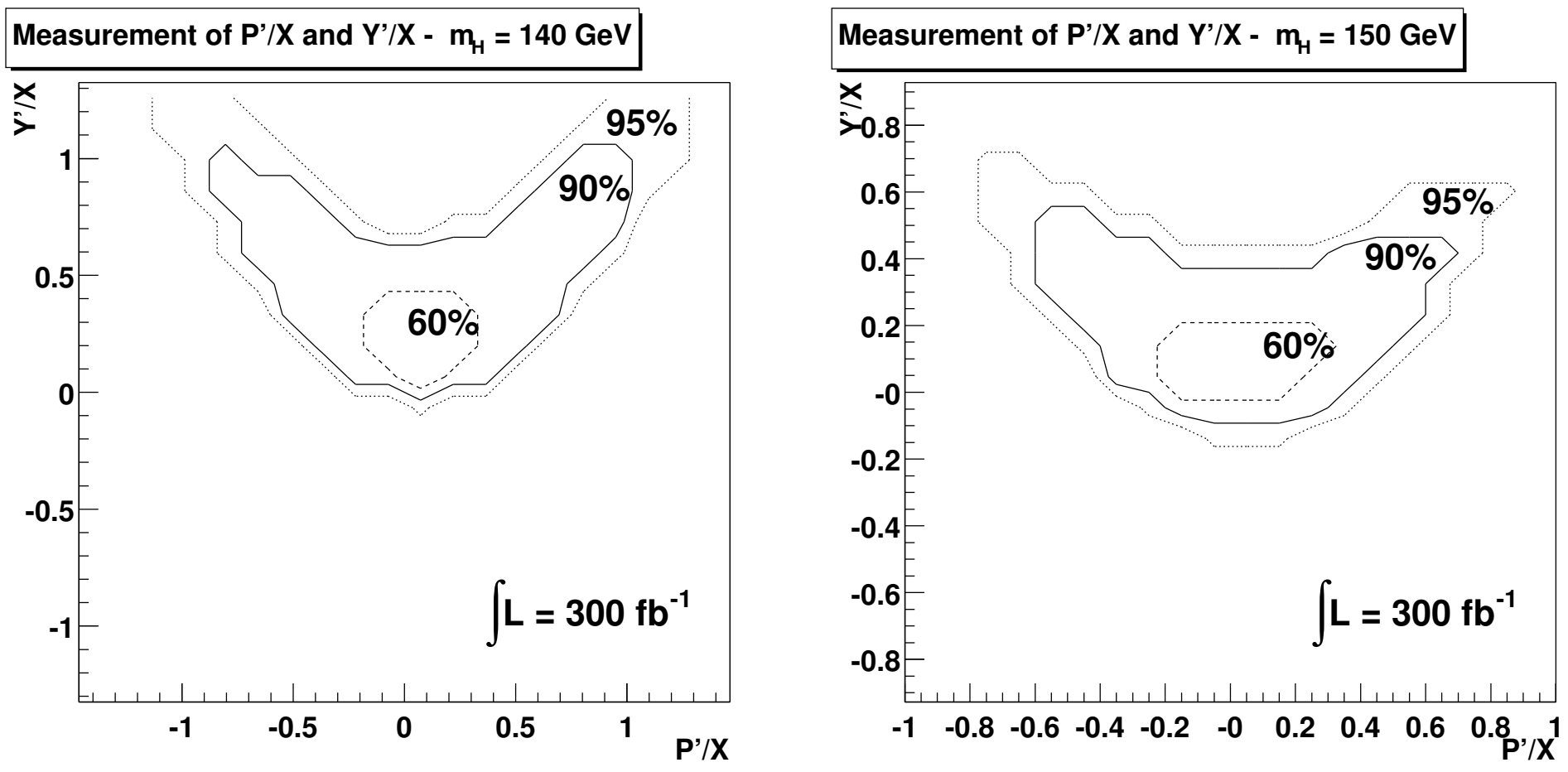

Fig. 2.26: Expected exclusion significance of **P'/X** and **Y'/X** for masses of the Higgs of 140 GeV and 150 GeV. The quality of the measurement is mainly limited by statistics.

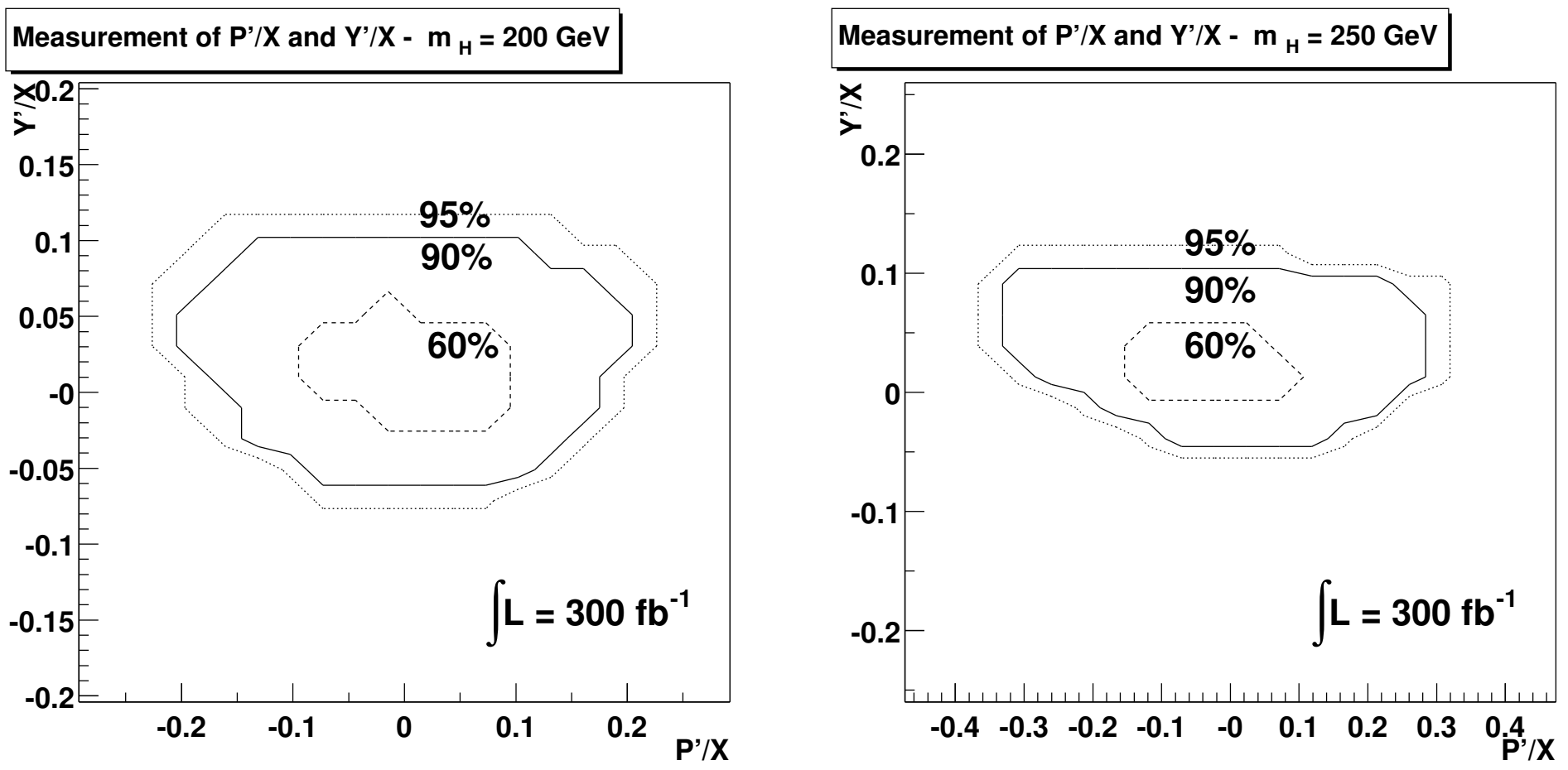

Fig. 2.27: Expected exclusion significance of **P'/X** and **Y'/X** for masses of the Higgs of 200 GeV and 250 GeV. The much higher number of events allows for a much better measurement of the coupling structure above the $ZZ$ threshold.

### *2.12.2 Conclusions*

We have shown with our analyses that the angular correlations of the decay products of the $Z$ bosons can be used to distiguish the SM Higgs-boson from hypothetical particles with different spin and CP quantum numbers. Furthermore, we have demonstrated how and to what extent CP violation in the scalar Higgs decay to Z-pairs can be studied and excluded. The methods discussed work well for a Higgs-boson with a mass above the $Z$ boson pair production threshold. Even small contributions of CP-even and CP-odd non SM couplings can be excluded in this case. Below, the analysis is statistically limited.

### 2.13 Study of the CP properties of the Higgs boson in the $\Phi \to ZZ \to 2e2\mu$ process in CMS

*Michał Bluj*

We study a possible measurement of the CP-parity of the Higgs boson $\Phi$ at the LHC, using the CMS detector. We consider a „golden channel" $\Phi \to ZZ \to 2e2\mu$ and angular correlations of leptons. The most general $\Phi VV$ coupling ($V = W^{\pm}, Z^0$) for a spin-0 Higgs boson looks as follows [93, 119, 288, 294]:

$$C^{J=0}_{\Phi VV} = \kappa \cdot g^{\mu\nu} + \frac{\zeta}{m_V^2} \cdot p^{\mu} p^{\nu} + \frac{\eta}{m_V^2} \cdot \epsilon^{\mu\nu\rho\sigma} k_{1\rho} k_{2\sigma}, \tag{2.182}$$

where $k_1$, $k_2$ are four-momenta of vector bosons $V$ and $p \equiv k_1 + k_2$ is four-momentum of the Higgs boson. In this analysis a simplified version of above $\Phi VV$ coupling (Eq. 2.182) is considered with a scalar and a pseudoscalar contributions only (i.e. $\kappa, \eta \neq 0$ and $\zeta = 0$). To study deviations from the Standard Model $\Phi ZZ$ coupling we take $\kappa = 1$[14]. The differential cross-section for the $\Phi \to Z_1 Z_2 \to (\ell_1 \bar{\ell}_1)(\ell_2 \bar{\ell}_2)$ process consists now of three terms: a scalar one (denoted by $H$), a pseudoscalar one $\sim \eta^2$ (denoted by $A$) and the interference term violating CP $\sim \eta$ (denoted by $I$):

$$d\sigma(\eta) \sim H + \eta\, I + \eta^2 A. \tag{2.183}$$

This way the Standard-Model scalar ($\eta = 0$) and the pseudoscalar (in the limit $|\eta| \to \infty$) contributions could be recovered. It is convenient to introduce a new parameter $\xi$, defined by $\tan\xi \equiv \eta$, with values between $-\pi/2$ and $\pi/2$. Expressions for $H$, $A$ and $I$ can be found in article [288].

To study the CP-parity of the Higgs boson we use two angular distributions. The first one is a distribution of the angle $\varphi$ (called a plane or an azimuthal angle) between the planes of two decaying $Z$s, in the Higgs boson rest frame[15]. The second one is a distribution of the polar angle $\theta$, in the $Z$ rest frame, between the momentum of the negatively charged lepton and the direction of motion of the $Z$ boson in the Higgs boson rest frame (Fig. 2.16).

#### *2.13.1 MC samples*

The Higgs-boson signal samples were generated using `PYTHIA` [295] for three masses of the Higgs boson ($m_\Phi = 200,\ 300,\ 400$ GeV). Generated events were required to contain $e^+e^-$ and $\mu^+\mu^-$ pairs within the detector acceptance region ($p_t^e > 5$ GeV, $|\eta^e| < 2.7$ and $p_t^\mu > 3$ GeV, $|\eta^\mu| < 2.5$). The analysis was performed for the scalar, pseudoscalar and CP-violating states (the latter for $\tan\xi$=$\pm$0.1, $\pm$0.4, $\pm$1, $\pm$4). Samples for the scalar, pseudoscalar and $\tan\xi = \pm 1$ states contain 10 000 events, while each of remaining samples contains 5 000 events. The predicted production cross-sections: $\sigma_\Phi$, $\sigma_\Phi \cdot BR(\Phi \to 4\ell)$ and $\sigma_\Phi \cdot \epsilon \cdot BR(\Phi \to 4\ell)$, where $\epsilon$ is the preselection efficiency for a the signal, are summarized in Table 2.7. We assume the Standard Model cross-section [296] and the Standard Model branching ratio [297] for each value of the $\xi$ parameter (independently on the CP-parity of the Higgs boson). A dependence of the analysis' results on the assumed cross-section is discussed in Section 2.13.4. The following background processes were considered:

Table 2.7: Production cross-sections of the signal. Errors are statistical only.

| mass (GeV) | $\sigma_\Phi$ (fb) | $\sigma_\Phi \cdot BR(\Phi \to 4\ell)$ (fb) | $\sigma_\Phi \cdot \epsilon \cdot BR(\Phi \to 4\ell)$ (fb) |
|---|---|---|---|
| 200 | $17.86 \cdot 10^3$ | 38.75 | $7.65 \pm 0.09$ |
| 300 | $9.41 \cdot 10^3$ | 24.03 | $5.08 \pm 0.06$ |
| 400 | $8.71 \cdot 10^3$ | 20.15 | $4.45 \pm 0.05$ |

[14] The $\Phi VV$ coupling with $\kappa = 1$ and arbitrary $\eta$ is implemented in the `PYTHIA` generator.

[15] The negatively charged leptons were used to fix planes' orientations.

Table 2.8: Production cross-sections and number of used events for background processes. Errors are statistical only.

| process | $\sigma_{bkg}$ (fb) | $\sigma_{bkg} \cdot BR$ (fb) | $\sigma_{bkg} \cdot \epsilon \cdot BR$ (fb) | # events |
|---|---|---|---|---|
| $ZZ/\gamma^*$ | $28.9 \cdot 10^3$ | 730.27 | $39.75 \pm 0.34$ | 20k |
| $t\bar{t}$ | $840 \cdot 10^3$ | $87.2 \cdot 10^3$ | $775.08 \pm 4.84$ | 48k |
| $Zb\bar{b}$ | $525 \cdot 10^3$ | $9.49 \cdot 10^3$ | $116.38 \pm 3.22$ | 5k |

Table 2.9: Selected cross-section for signal and background at chosen stages of the selection. All values in fb; errors are statistical only.

| level of selection | signal | background $ZZ/\gamma^*$ | background $t\bar{t}$ | background $Zb\bar{b}$ |
|---|---|---|---|---|
| selection for $m_\Phi = 200$ GeV | | | | |
| trigger | 6.45±0.09 | 30.30±0.30 | 305.04±3.11 | 81.17±2.69 |
| reco. $e^+e^-$ $\mu^+\mu^-$ | 5.46±0.08 | 22.57±0.26 | 164.04±2.29 | 32.77±1.73 |
| $Z$s' mass | 3.89±0.07 | 12.57±0.19 | 0.09±0.06 | <0.03 |
| $\Phi$s mass | 3.43±0.06 | 1.84±0.07 | <0.02 | <0.03 |
| selection for $m_\Phi = 300$ GeV | | | | |
| trigger | 4.34±0.06 | 30.30±0.30 | 305.04±3.11 | 81.17±2.69 |
| reco. $e^+e^-$ $\mu^+\mu^-$ | 3.74±0.05 | 22.57±0.26 | 164.04±2.29 | 32.77±1.73 |
| $Z$s' mass | 2.69±0.05 | 7.32±0.15 | 0.13±0.07 | 0.05±0.07 |
| $\Phi$s mass | 2.10±0.04 | 0.82±0.05 | <0.02 | <0.03 |
| selection for $m_\Phi = 400$ GeV | | | | |
| trigger | 3.84±0.06 | 30.30±0.30 | 305.04±3.11 | 81.17±2.69 |
| reco. $e^+e^-$ $\mu^+\mu^-$ | 3.35±0.06 | 22.57±0.26 | 164.04±2.29 | 32.77±1.73 |
| $Z$s' mass | 2.46±0.05 | 5.35±0.13 | 0.09±0.06 | <0.03 |
| $\Phi$s mass | 2.02±0.04 | 0.66±0.05 | <0.02 | <0.03 |

1. $ZZ/\gamma^* \to 2e2\mu$ (irreducible background). The leading order cross-section for the $q\bar{q} \to ZZ/\gamma^*$ process calculated using `MCFM` program [298] is equal to 18.7 pb. The next-to-leading order contribution as well as the contribution from the $gg \to ZZ/\gamma^*$ process, with estimated cross-section of about 20% of the $q\bar{q} \to ZZ/\gamma^*$ cross-section at the leading order, was included as a four-lepton-mass dependent K-factor. The K-factor is in average equal to 1.55 in four-lepton-mass range between 30 and 750 GeV, for example K=1.46, 1.66, 1.90 for $m_{4\ell}$=200, 300, 400 GeV, respectively.
2. $t\bar{t} \to W^+W^-b\bar{b} \to 2e2\mu X$. The $t\bar{t}$ cross-section is equal to 840 pb [299].
3. $Zb\bar{b} \to 2e2\mu X$. The $Zb\bar{b}$ cross-section at the next-to-leading order, determined using `MCFM` program [300–302] for $p_t^b > 1$ GeV, $|\eta^b| < 2.5$ and $81 < m_{Z^*} < 101$ GeV, is equal to 525 pb.

The information about the generated background samples are summarized in Table 2.8.
The minimum-bias pile-up events for the low LHC luminosity were added to each signal and background sample.

### *2.13.2 Selection*

We use selection criteria (for four isolated leptons) developed in the Standard-Model Higgs boson searches at CMS for the $H \to ZZ \to 2e2\mu$ process [303]. Values of the selection cuts depend on the Higgs boson mass. The selected cross-section, at chosen stages of the selection, for the signal and the background for three masses of the Higgs boson $m_\Phi = 200$, 300, 400 GeV are shown in Table 2.9. Fig. 2.28 shows the invariant mass of four reconstructed leptons before and after the off-line selection (i.e. after lepton reconstruction and after cut on two $Z$s' masses, respectively) for the Higgs boson sig-nal, with $m_\Phi$=300 GeV, and for the background. The reconstructed angular distributions after the final selection for the signal (with mass $m_\Phi$=300 GeV) for various values of the parameter $\xi$, and for the background are shown in Fig. 2.29. Shape of the angular distributions for the background slightly depends on the Higgs-mass-dependent selection. This effect is taken into account in our analysis.

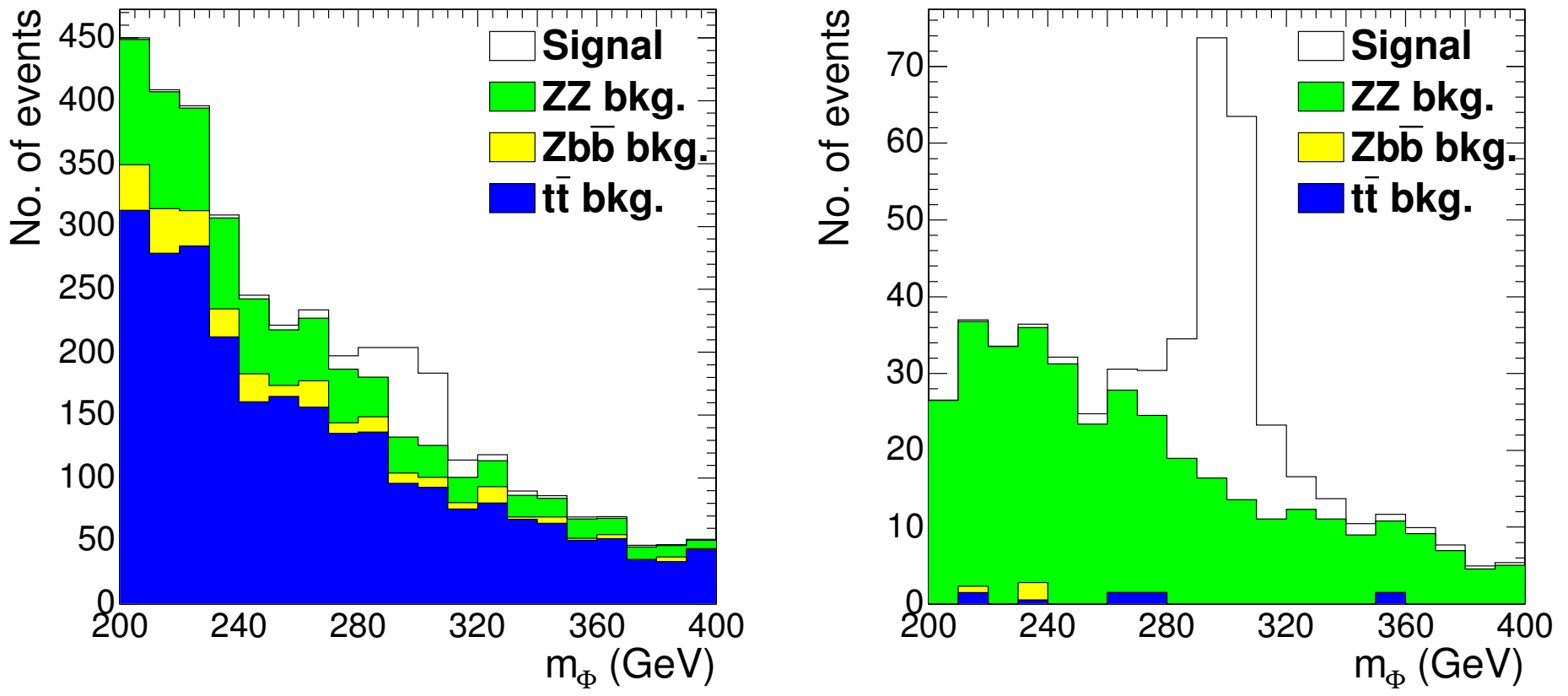

Fig. 2.28: Invariant mass distributions of four leptons before (left) and after (right) the off-line selection (normalized to 60 fb$^{-1}$). The signal of the Higgs boson with $m_\Phi$=300 GeV (empty histogram) and the background (filled histograms).

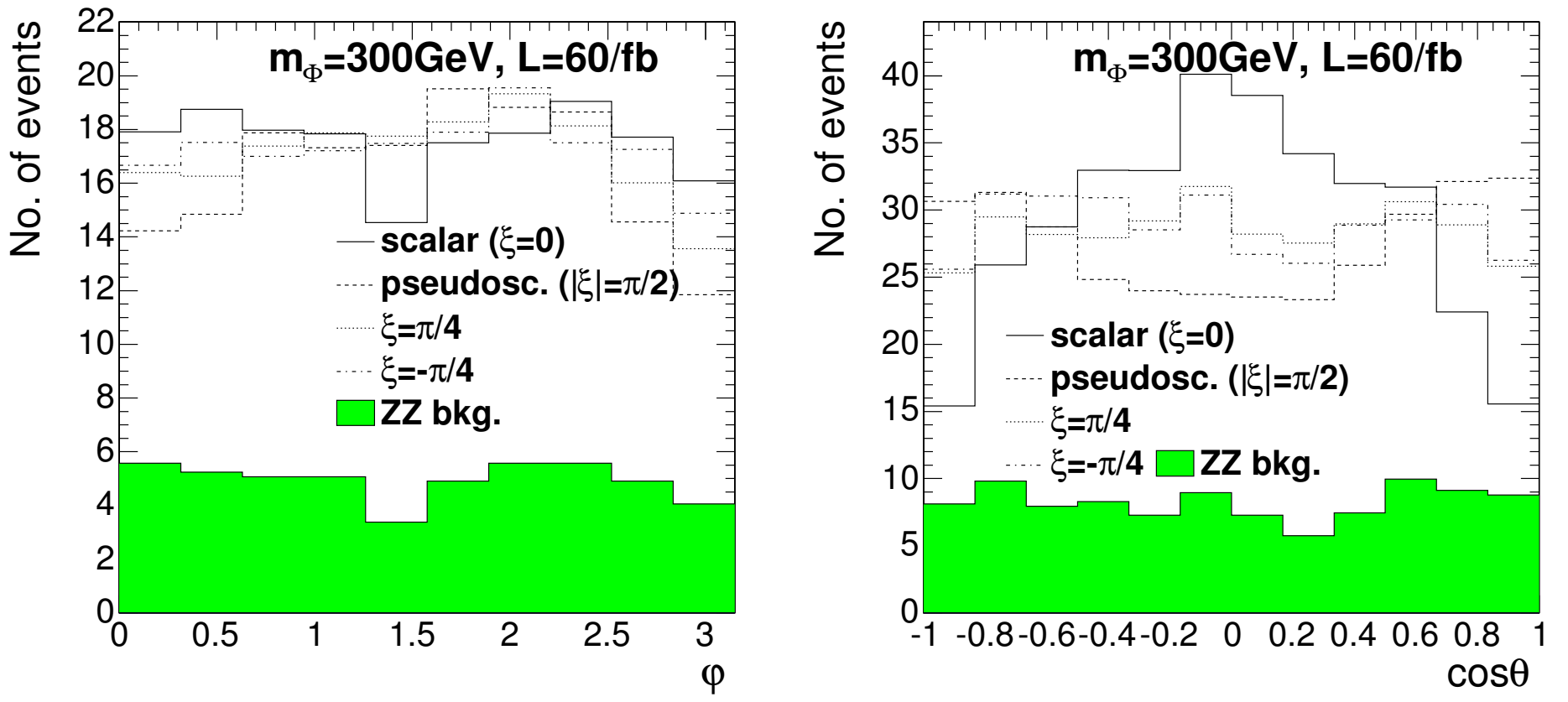

Fig. 2.29: The $\varphi$-distributions (left) and the $\theta$-distributions (right) for various values of the parameter $\xi$ after final selection (normalized to 60 fb$^{-1}$). The signal for $m_\Phi$=300 GeV and $\xi = 0$ (scalar), $\xi = -\pi/4$, $\xi = +\pi/4$ and $|\xi| = \pi/2$ (pseudoscalar), respectively (empty histograms). The $ZZ$ background - filled histogram.

### 2.13.3 *Determination of the parameter $\xi$*

The parameter $\xi$ was determined by maximization of the likelihood function $\mathcal{L}(\xi,R)$, which was constructed, for both the signal and the background, from the angular distributions and invariant mass distribution of four leptons. The function depends on two parameters: $\xi$ describing CP of the Higgs boson, and $R$ describing a fraction of the signal in the data sample. The function has the following form:

$$\mathcal{L}(\xi,R) \equiv 2 \sum_{x_i \in data} \log \mathcal{Q}(\xi,R;\, x_i), \text{ where } \mathcal{Q}(\xi,R;\, x_i) \equiv R \cdot \mathcal{PDF}_S(\xi;\, x_i) + (1-R) \cdot \mathcal{PDF}_B(x_i). \tag{2.184}$$

$\mathcal{PDF}_B(x_i)$ and $\mathcal{PDF}_S(\xi;\, x_i)$ are Probability Density Functions for the background and the signal respectively; $\{x_i\}$ are values of the measured quantities (angles and invariant mass) in the data event $i$.

They are products of probability densities $\mathcal{P}^M$, $\mathcal{P}^\varphi$, $\mathcal{P}^{\cos\theta_{1,2}}$ of four leptons invariant mass and angles $\varphi$ and $\cos\theta_{1,2}$ i.e. $\mathcal{PDF} \equiv \mathcal{P}^M\mathcal{P}^\varphi\mathcal{P}^{\cos\theta_1}\mathcal{P}^{\cos\theta_2}$. $\mathcal{P}^M$, $\mathcal{P}^\varphi$, $\mathcal{P}^{\cos\theta_{1,2}}$ are obtained by the Monte Carlo technique, using the normalized histograms of given quantities after the final selection.
A part of the function $\mathcal{Q}$, which describes angular distributions of the signal depends on the parameter $\xi$. From Eq. (2.183) we obtain:

$$\mathcal{P}(\xi) \equiv (\mathcal{P}_S^\varphi \cdot \mathcal{P}_S^{\cos\theta_1} \cdot \mathcal{P}_S^{\cos\theta_2})(\xi) \equiv (\mathcal{H} + \tan\xi \cdot \mathcal{I} + \tan^2\xi \cdot a^2\mathcal{A})/(1 + a^2\tan^2\xi), \qquad (2.185)$$

where: $\mathcal{H} \equiv \mathcal{P}_H^\varphi \cdot \mathcal{P}_H^{\cos\theta_1} \cdot \mathcal{P}_H^{\cos\theta_2}$ and $\mathcal{A} \equiv \mathcal{P}_A^\varphi \cdot \mathcal{P}_A^{\cos\theta_1} \cdot \mathcal{P}_A^{\cos\theta_2}$ are probability densities obtained by the Monte Carlo technique for the scalar (H) and the pseudoscalar (A), respectively. The parameter $a^2$ is a (mass dependent) relative strength of the pseudoscalar and scalar couplings. For example $a^2$=0.51, 1.65, 1.79 for $m_\Phi$=200, 300, 400 GeV, respectively. $\mathcal{I}$ is a normalized product of angular distributions for the CP-violating term. Since $\mathcal{I}$ is not positive, and its integral is equal to zero, it is not possible to simulate it separately. The $\mathcal{I}$ contribution can be obtained indirectly from the combined probability density for the signal with a non-zero value of the parameter $\xi$. For example by introducing $\mathcal{P}_+ \equiv \mathcal{P}(\pi/4) = (\mathcal{H}+\mathcal{I}+a^2\mathcal{A})/(1+a^2)$ and $\mathcal{P}_- \equiv \mathcal{P}(-\pi/4) = (\mathcal{H}-\mathcal{I}+a^2\mathcal{A})/(1+a^2)$ we have $\mathcal{I} = \frac{1+a^2}{2}(\mathcal{P}_+ - \mathcal{P}_-)$. Finally we obtain:

$$\mathcal{P}(\xi) \equiv [\mathcal{H} + \tan\xi \cdot \frac{1+a^2}{2} \cdot (\mathcal{P}_+ - \mathcal{P}_-) + \tan^2\xi \cdot a^2\mathcal{A}]/(1 + a^2\tan^2\xi). \qquad (2.186)$$

### *2.13.4 Results*

After selection all background contributions but $ZZ/\gamma^* \to 2e2\mu$ are negligible, therefore only such events were used to construct the probability density function for the background. We use the $ZZ/\gamma^*$ sample containing 15 000 events at the generator level. Signal probability density functions were constructed using samples of scalar Higgs boson (H), pseudoscalar (A) and $\mathcal{P}_+$, $\mathcal{P}_-$ samples, each containing 8 000 events at the generator level. Likelihood functions were constructed independently for three masses of the Higgs boson ($m_\Phi$=200, 300, 400 GeV).

For each value of parameter $\xi$ and for each Higgs-boson mass we made 200 pseudoexperiments for the integrated luminosity $\mathcal{L}$=60 fb$^{-1}$ (3 years of LHC at low luminosity). For each pseudoexperiment we randomly selected events from the signal and background samples to form a test sample[16]. The number of selected events was given by a Poisson probability distribution with mean defined by the process cross-section and the examined luminosity. Then to obtain a value of the parameter $\xi$, we performed a maximization of the likelihood function $\mathcal{L}(\xi,\mathcal{R})$ for the test sample. The expected and reconstructed values of the parameter $\xi$ (with its uncertainty), obtained for three masses of the Higgs boson are shown in Fig. 2.30.

In our analysis the Standard-Model signal cross-section and branching ratio were used as a reference. However, both of them may change for other Higgs models. An influence of a possible suppression (enhancement) factor $C^2$ of the Standard Model signal on the reconstructed $\xi$ were studied and we found that its value slightly depends on size of suppression (enhancement). On the other hand, the uncertainty of $\xi$ is approximately $\sim 1/C$ (i.e. it depends on square-root of number of events, what one can expect), namely:

$$\Delta\xi(\xi, C^2) \equiv \frac{\sigma_0(\xi)}{\sqrt{C^2}}. \qquad (2.187)$$

A value of $\sigma_0(\xi)$ (a precision factor) can be determined from the fit. Taking this into account we parametrize a relative error of the difference between $\xi$ and $\xi_0$ as follows:

$$\frac{\sigma_{|\xi-\xi_0|}(C^2)}{|\xi-\xi_0|} = \frac{\Delta\xi(\xi, C^2)}{|\xi-\xi_0|}. \qquad (2.188)$$

[16] Samples used to select events, contain 2 000 and 5 000 events for the signal and the background, respectively. The samples do not contain events used to construct probability densities for the likelihood function.

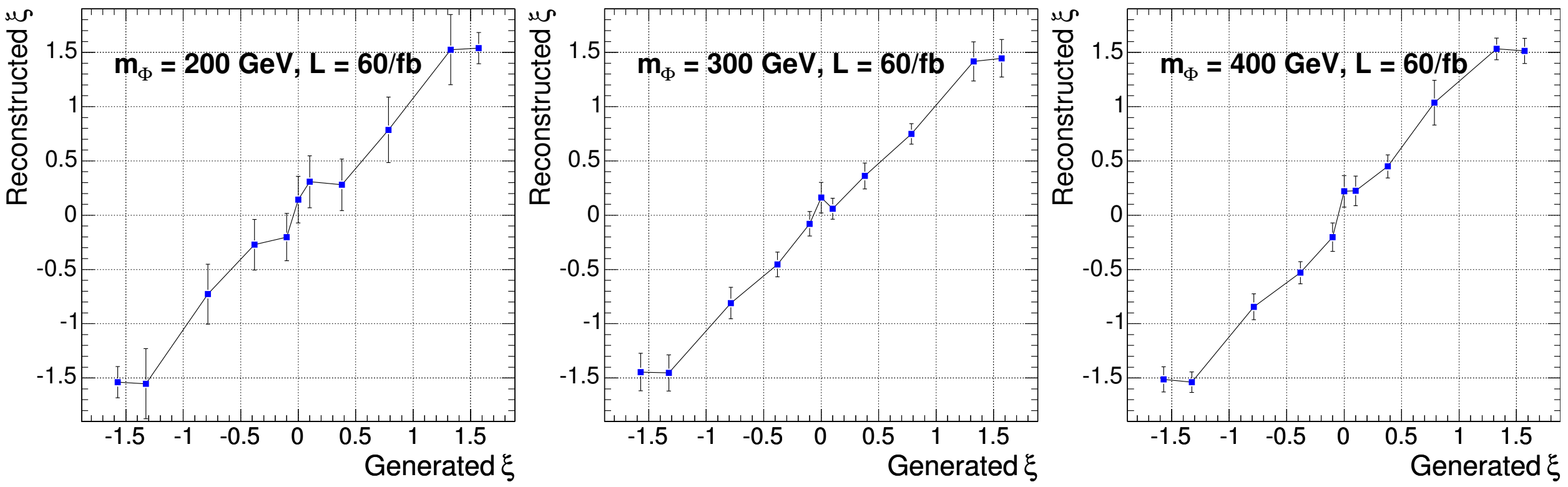

Fig. 2.30: Reconstructed value of the parameter $\xi$ as function of the generated value of the parameter $\xi$, for $\mathcal{L}$=60 fb$^{-1}$, for Higgs boson mass $m_\Phi$=200, 300, 400 GeV. Uncertainties correspond to one standard deviation.

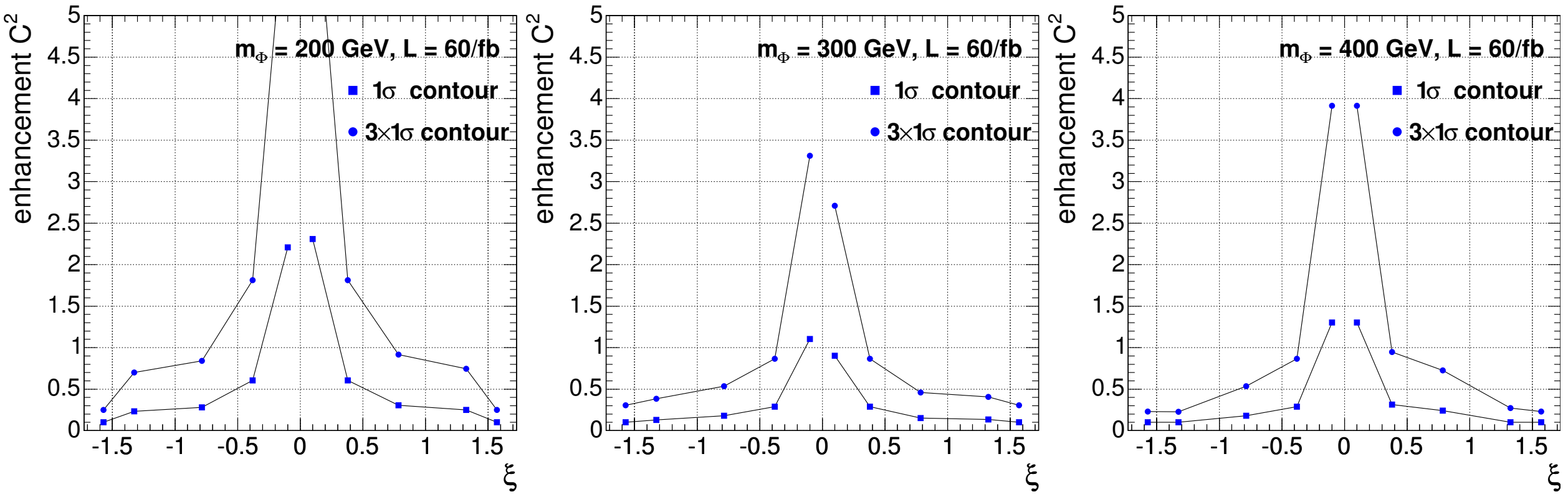

Fig. 2.31: Exclusion contours for scalar Higgs boson as a function of the enhancement factor $C^2$ for the Higgs boson masses $m_\Phi$=200, 300, 400 GeV (from left to right). Results were obtained for 60 fb$^{-1}$.

The requirement of exclusion of $\xi \neq \xi_0$ at the level of "$N$ sigmas" could be written as $\sigma_{|\xi - \xi_0|}(C^2) = |\xi - \xi_0|$

$$C^2(N) = N^2 \frac{\sigma_0^2(\xi)}{(\xi - \xi_0)^2}. \qquad (2.189)$$

The exclusion contours for $N$=1, 3 and for $\xi_0 = 0$ (scalar) are shown in Fig. 2.31.

### 2.13.5 *Summary*

A possibility of a measurement of the CP-properties of the Higgs boson $\Phi$ in the $\Phi \to ZZ \to 2e2\mu$ process at LHC with CMS detector was studied. It was shown that using angular correlations of the Higgs boson decay products (leptons) the measurement of the parameter $\xi$, describing a general $\Phi ZZ$ coupling, will be feasible. Precision of this measurement is sufficient for determination of the CP-parity of the Higgs boson, particularly it is sufficient to distinguish scalar from pseudoscalar.

## 2.14 Higgs-boson CP properties from decays to WW and ZZ at the Photon Linear Collider

*Piotr Nieżurawski, Aleksander Filip Żarnecki and Maria Krawczyk*

The process of resonant Higgs boson production at the Photon Linear Collider (PLC), due to $h\gamma\gamma$ coupling, is in the Standard Model sensitive to the Higgs boson couplings to both, the gauge-bosons and up-type fermions. Moreover, as the phases of the two dominant contributions to the $\gamma\gamma \to h$ amplitude, from $W^\pm$ and top loops, differ, the process turns out to be very sensitive to the possible effects of the CP violation.

In Ref. [304] we performed a realistic simulation of the Standard Model Higgs-boson production at the PLC for $W^+W^-$ and $ZZ$ decay channels, for Higgs-boson masses above 150 GeV. From the combined analysis of $W^+W^-$ and $ZZ$ invariant mass distributions the $\gamma\gamma$ partial width of the Higgs boson, $\Gamma_{\gamma\gamma}$, can be measured with an accuracy of 3 to 8% and the phase of $\gamma\gamma \to h$ amplitude, $\phi_{\gamma\gamma}$, with an accuracy between 30 and 100 mrad. In Ref. [144] we extended this analysis to the generalized Standard Model-like scenario $B_h$ of the Two Higgs Doublet Model II, 2HDM(II), with and without CP-violation. We also considered a general 2HDM (II) with CP violation, and found that only the combined analysis of LHC, ILC and PLC measurements allows for a precise determination of the Higgs-boson couplings and of CP-violating $H$–$A$ mixing angle [305, 306]. Finally, we considered model with a generic, CP-violating Higgs-boson couplings to vector bosons [93, 136, 307], which leads to different angular distributions for a scalar- and pseudoscalar-type of couplings. From a combined analysis of the invariant mass distributions and angular distributions of the $W^+W^-$ and $ZZ$ decay-products the CP-parity of the observed Higgs state can be determined independently on a production mechanism [147].

In this contribution we summarize selected results of [144, 147, 305, 306], related to the determination of the Higgs-boson CP properties at the PLC.

### 2.14.1 *Event simulation*

In analyses we use the `CompAZ` parametrization [308] of the realistic luminosity spectra for a Photon Linear Collider at TESLA [309, 310] and assume that the centre-of-mass energy of colliding electron beams, $\sqrt{s_{ee}}$, is optimized for the production of a Higgs boson with given mass. We consider the mass range between 200 and 350 GeV, where $W^+W^-$ and $ZZ$ decays are expected to dominate. All results presented in this paper were obtained for an integrated luminosity corresponding to one year of the PLC running, as given by [309, 310], i.e. from 600 fb$^{-1}$ for $\sqrt{s_{ee}} = 305$ GeV (optimal beam energy choice for $M = 200$ GeV) to about 1000 fb$^{-1}$ for $\sqrt{s_{ee}} = 500$ GeV (for $M = 350$ GeV).

Analyses described in this work were performed in two steps. In the first step we use samples of events generated with `PYTHIA` 6.152 [295] to estimate selection efficiency, as well as resolutions of the angular variable and of the invariant-mass reconstruction for $\gamma\gamma \to W^+W^-/ZZ$ events, as a function of the $\gamma\gamma$ centre-of-mass energy, $W_{\gamma\gamma}$ . We consider the direct vector-bosons production in $\gamma\gamma$ interactions (background) as well as the signal $\gamma\gamma \to h \to W^+W^-/ZZ$ and the interference between the signal and the background. To take into account effects which are not implemented in `PYTHIA` (photon beam polarization, interference term contribution, direct $\gamma\gamma \to ZZ$ production) we exploit the standard method used in various experimental analyses called a reweighting procedure. To each generated event a weight is attributed given by the ratio of the differential cross-section for a vector-boson production in the polarized photon interactions [311–314] to the `PYTHIA` differential cross section for given event. The fast simulation program `SIMDET` version 3.01 [315] is used to model the TESLA detector performance.

For the $W^+W^-$ events only $qq\bar{q}\bar{q}$ decay channel is considered, as without knowing the exact beam-photon energies, which is a case for the Photon Linear Collider, the semileptonic $W^\pm$ decays can not be fully reconstructed. For the $ZZ$ events, only $l\bar{l}q\bar{q}$ decay channel is considered, with one $Z$ decaying into $e^+e^-$ or $\mu^+\mu^-$. Selection of the leptonic channel is crucial for a suppression of the background from the direct $\gamma\gamma \to W^+W^-$ events.

The invariant-mass resolutions obtained from a full simulation of $W^+W^-$ and $ZZ$ events (based on the PYTHIA and SIMDET programs), have been parametrized as a function of the $\gamma\gamma$ centre-of-mass energy, $W_{\gamma\gamma}$ . This parametrization can then be used to obtain the parametric description of the expected invariant mass distributions, for $\gamma\gamma \to W^+W^-$ and $\gamma\gamma \to ZZ$ events, avoiding the time consuming event generation procedure. Resolutions expected in the reconstruction of angular variables are very good and the measurement errors can be safely neglected. The measured angular distributions are mainly affected by the detector acceptance and the corresponding selection cuts used in the analysis. The corresponding acceptance corrections have also been parametrized as a function of the relevant angular variables. For arbitrary model, and for arbitrary model parameters, we calculate the expected angular and invariant mass distributions for $ZZ$ and $W^+W^-$ events by convoluting the corresponding cross-section formula with the analytic photon-energy spectra CompAZ [308]. To take into account detector effects, we convolute these distributions further with the function parameterising the invariant-mass resolution and the acceptance function, which takes into account the angular- and jet-selection cuts. This approach has been developed in [304].

### *2.14.2 Generic model*

Following the analysis described in [93,136,307] we consider a generic model with a direct CP violation, i.e. with tensor couplings of a Higgs boson, $\Phi$, to $ZZ$ and $W^+W^-$ given by:

$$\begin{aligned} g_{\Phi ZZ} &= ig\frac{M_Z}{\cos\theta_W}\left(\lambda_H \cdot g^{\mu\nu} + \lambda_A \cdot \varepsilon^{\mu\nu\rho\sigma}\,\frac{(p_1+p_2)_\rho\,(p_1-p_2)_\sigma}{M_Z^2}\right) , \\ g_{\Phi WW} &= igM_W\left(\lambda_H \cdot g^{\mu\nu} + \lambda_A \cdot \varepsilon^{\mu\nu\rho\sigma}\,\frac{(p_1+p_2)_\rho\,(p_1-p_2)_\sigma}{M_W^2}\right) , \end{aligned} \tag{2.190}$$

where $p_1$ and $p_2$ are the 4-momenta of the vector bosons. The $\lambda_H$-terms have a structure of the CP-even SM Higgs boson coupling,[17] whereas the one with $\lambda_A$ corresponds to a general CP-odd coupling for the spin-0 boson. Coefficients $\lambda_H$ and $\lambda_A$ can be parametrized by:

$$\begin{aligned} \lambda_H &= \lambda \cdot \cos\Phi_{CP} , \\ \lambda_A &= \lambda \cdot \sin\Phi_{CP} . \end{aligned} \tag{2.191}$$

The couplings of the Standard Model Higgs boson are reproduced for $\lambda = 1$ and $\Phi_{CP} = 0$ (i.e. $\lambda_H = 1$ and $\lambda_A = 0$). Below we will limit ourselves to $\lambda \approx 1$ and $|\Phi_{CP}| \ll 1$ region, corresponding to a small deviation from the respective Standard Model coupling. However, we do not make any assumptions concerning Higgs-boson couplings to the fermions and we allow for deviations from SM predictions in $\Gamma_{\gamma\gamma}$ and $\phi_{\gamma\gamma}$. Therefore our results do not depend on the Higgs-boson production mechanism and our approach can be considered as a model-independent one.

The angular distributions of the secondary $W^+W^-$ and $ZZ$ decay products turn out to be very sensitive to the CP properties of the Higgs-boson [93, 136, 307]. Angular variables which can be used in the analysis are defined in Fig. 2.32 (see also Fig. 2.16). To test CP-properties of the Higgs-bosons the distributions of the polar angles $\Theta_1$ and $\Theta_2$ as well as the $\Delta\phi$ distribution, where $\Delta\phi$ is the angle between two $Z$- or two $W$-decay planes, are used. Here we propose to consider, instead of the two-dimensional distribution in ($\cos\Theta_1$, $\cos\Theta_2$), the distribution in a new variable, defined as

$$\zeta = \frac{\sin^2\Theta_1 \cdot \sin^2\Theta_2}{(1+\cos^2\Theta_1)\cdot(1+\cos^2\Theta_2)} . \tag{2.192}$$

[17] Other possible CP-even tensor structure, $\sim (p_1+p_2)^\mu (p_1+p_2)^\nu$, give the angular distributions similar to that of the SM Higgs boson and therefore we will not consider this case separately. See also Section 2.12.

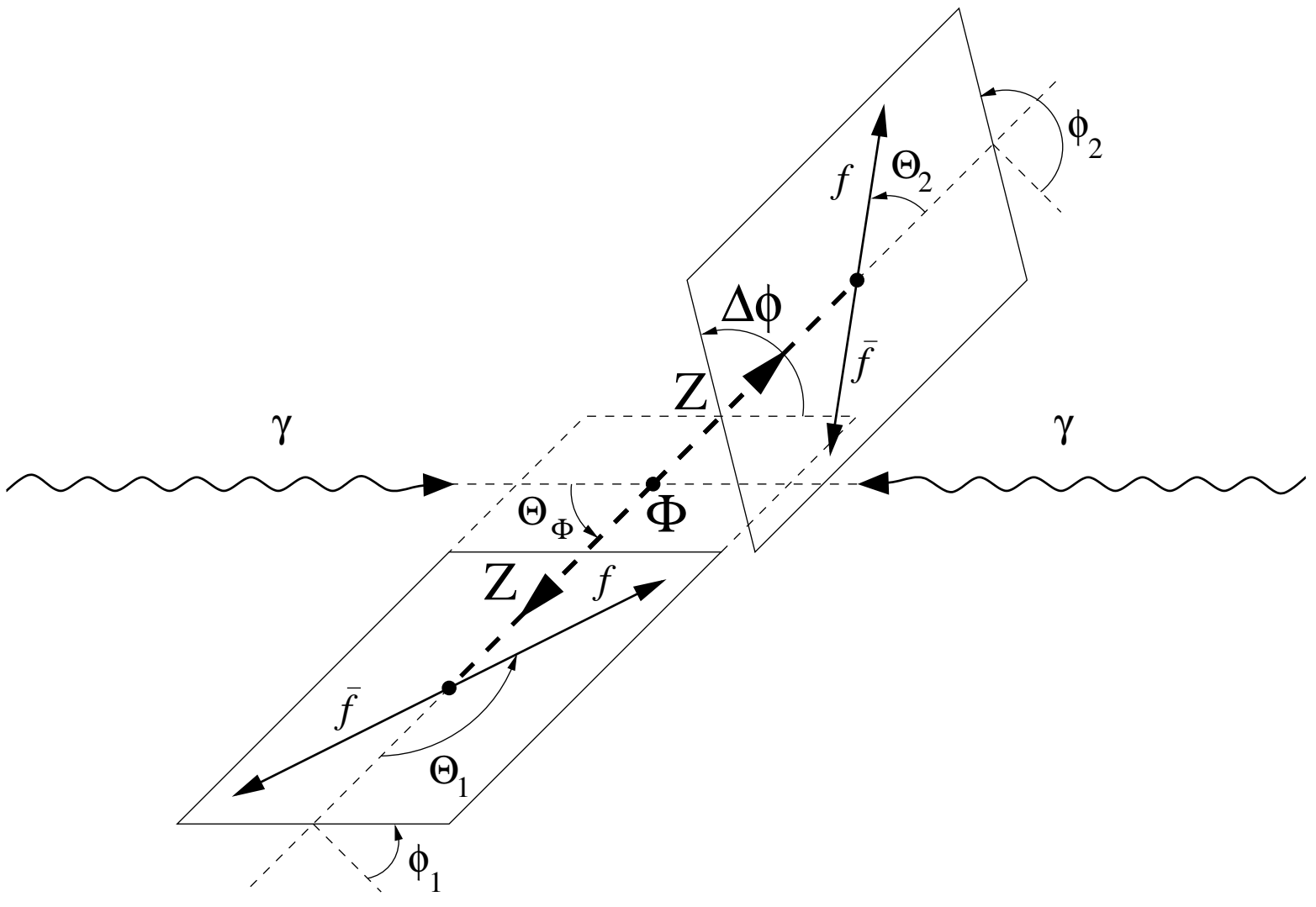

Fig. 2.32: The definition of the polar angles $\Theta_\Phi$, $\Theta_1$ and $\Theta_2$, and the azimuthal angles $\phi_1$ and $\phi_2$ for the process $\gamma\gamma \to \Phi \to ZZ \to 4\ f$. $\Delta\phi$ is the angle between two Z decay planes, $\Delta\phi = \phi_2 - \phi_1$. All polar angles are calculated in the rest frame of the decaying particle.

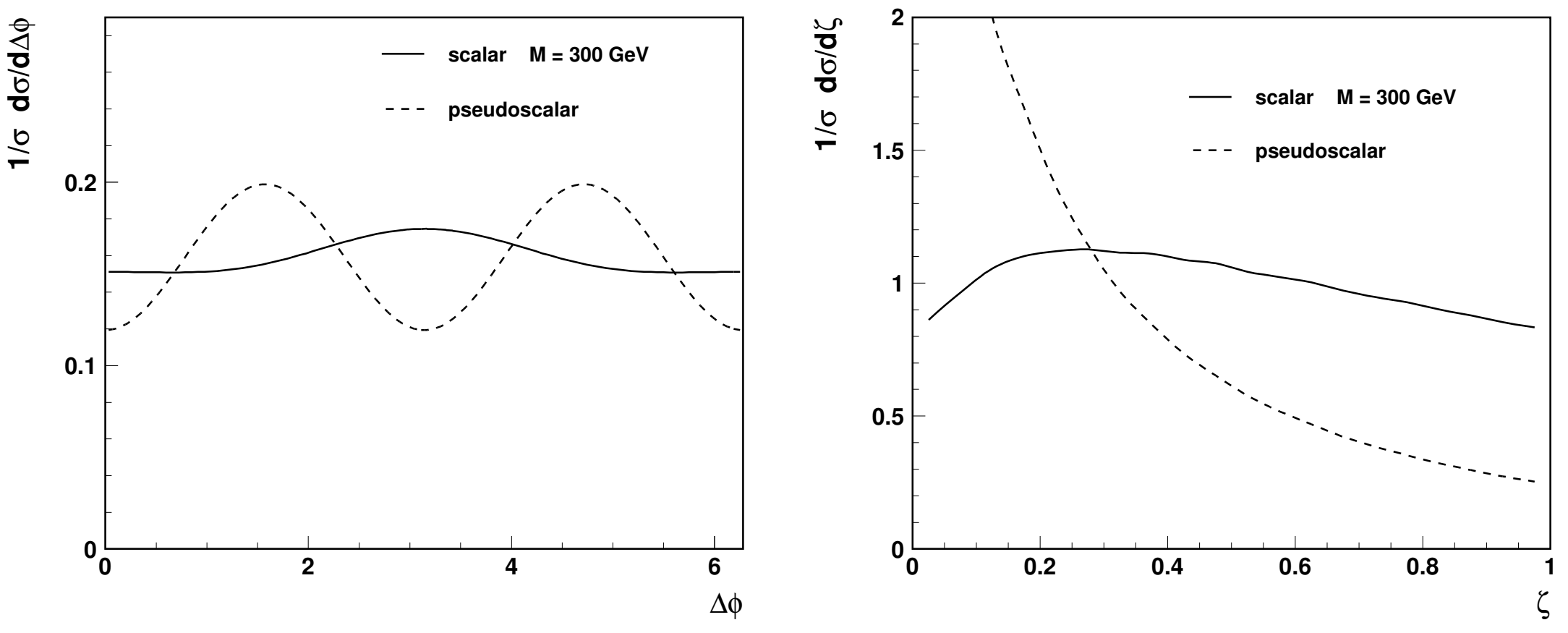

Fig. 2.33: Normalised angular distributions in $\Delta\phi$ (left plot) and $\zeta$ (right plot), expected for scalar (solid line) and pseudoscalar (dashed line) Higgs boson decays $H, A \to ZZ \to l^+l^-jj$, for the Higgs boson mass of 300 GeV.

The $\zeta$-variable corresponds to the ratio of the angular distributions expected for the decay of a scalar and a pseudoscalar (in a limit $M_\Phi \gg M_Z$) [93, 136, 307]. It proves to be very useful and complementary to the $\Delta\phi$ variable.

The angular distributions in $\Delta\phi$ and $\zeta$, expected for decays of a scalar $H$ ($\Phi_{CP} = 0$) and a pseudoscalar $A$ ($\Phi_{CP} = \frac{\pi}{2}$) Higgs boson with mass of 300 GeV, $\Phi \to ZZ \to l^+l^-jj$, are compared in Fig. 2.33. Both distributions clearly distinguish between decays of scalar and pseudoscalar Higgs boson; so it's possible to distinguish the CP-even and CP-odd states without taking into account the production mechanism. We point out the usefulness of the $\zeta$ distribution.

For the measurement of the $\Delta\phi$ and $\zeta$ distributions we introduce an additional cut on the reconstructed $ZZ$ or $W^+W^-$ invariant mass and, for $W^+W^-$ events only, the cut on the Higgs-boson decay angle $\Theta_\Phi$, to suppress large background from the nonresonant $W^+W^-$ production. The cuts were optimised for the smallest relative error in the signal cross-section measurement.

The expected precision in the measurements of the $\Delta\phi$- and of the $\zeta$-distributions, for $\gamma\gamma \to ZZ \to l^+l^-jj$ events is illustrated in Fig. 2.34. The reconstructed $\Delta\phi$ values range from 0 to $\pi$, since we are not able to distinguish between quark and antiquark jet. Calculations were performed for the primary electron-beam energy of 152.5 GeV and the Higgs-boson mass of 200 GeV. The results are compared with the expectation for $\Phi_{CP} = 0$ (as in SM) and $\Phi_{CP} = \frac{\pi}{2}$. We see, that even after taking into account the beam spectra, detector effects, selection cuts and background influence, the differences between shapes of the angular distributions for the scalar and pseudoscalar couplings are still significant. Therefore we should be able to constrain Higgs-boson couplings from the shape of the distributions, even if the overall normalisation related to the Higgs-boson production mechanism is not known.

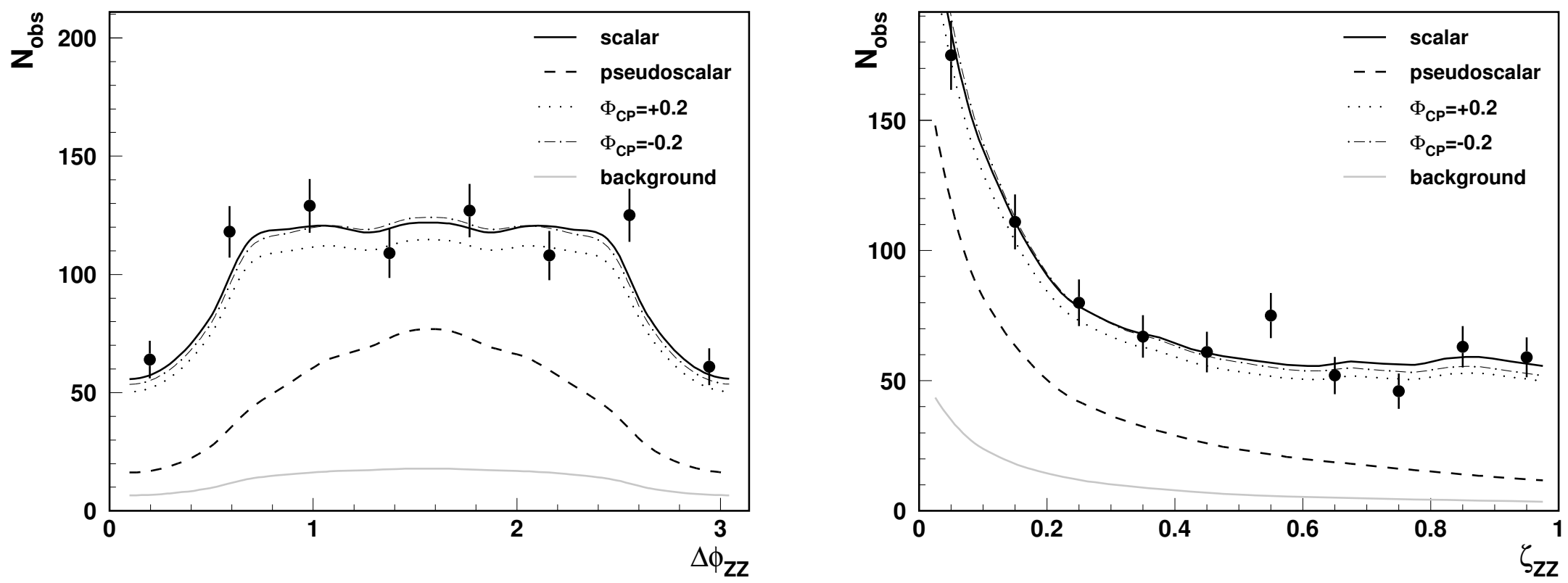

Fig. 2.34: Measurement of the angle $\Delta\phi_{ZZ}$ between two $Z$-decay planes (left plot) and of the variable $\zeta_{ZZ}$ calculated from the polar angles of the $Z \to l^+l^-$ and $Z \to jj$ decays (right plot) for $ZZ \to l^+l^-jj$ events. Error bars indicate the statistical precision of the measurement after one year of PLC running at nominal luminosity, for the scalar Higgs boson with mass of 300 GeV. The solid and dashed lines correspond to the predictions of the model with pure scalar ($\Phi_{CP} = 0$) and pseudoscalar ($\Phi_{CP} = \frac{\pi}{2}$) Higgs-boson couplings, whereas dotted and dash-dotted lines correspond to CP violating couplings with $\Phi_{CP} = \pm 0.2$. The gray line represents the SM background of non-resonant $ZZ$ production.

Each of the considered angular distributions discussed above can be fitted with the model expectations, given in terms of the parameters $\lambda$ and $\Phi_{CP}$ describing Higgs-boson couplings to gauge bosons, the parameters $\Gamma_{\gamma\gamma}$ and $\phi_{\gamma\gamma}$ describing the production mechanism, and an overall normalisation. We calculate the expected statistical errors on the parameters $\lambda$ and $\Phi_{CP}$, from the combined fit to angular distributions measured for the $ZZ$ and $W^+W^-$ decays, and to the invariant mass distributions. Results are shown in Fig. 2.35. The two photon width of the Higgs boson $\Gamma_{\gamma\gamma}$, the phase $\phi_{\gamma\gamma}$ and normalisations of both samples are allowed to vary in the fit, so the results are independent on the production mechanism. One observes that for Higgs-boson masses below 250 GeV, better constrains are obtained from the measurement of $W^+W^-$ events, whereas for masses above 300 GeV smaller errors are obtained from the $ZZ$ events. The error on $\Phi_{CP}$ expected from the combined fit is below 50 mrad in the whole considered mass range. The corresponding error on $\lambda$ is about 0.05.

### *2.14.3 SM-like Two Higgs Doublet Model*

Here we consider the CP violation in the Standard-Model scenario of the 2HDM. This is a generalization of a CP conserving scenario $B_h$, introduced in [15,63,144]. In the following we consider the CP-violating solution $B_h$, with a weak CP violation through a small mixing between $H$ and $A$ states.

In this scenario the Yukawa couplings of $h$ ($h \sim h_1$) are equal (up to a sign) to the corresponding SM Higgs-boson couplings. Then, it follows from Eqs. (2.43) and (2.54)–(2.56) that the coupling of $h$ to gauge bosons as well as the corresponding Yukawa and gauge boson couplings of $H$ and $A$ bosons are uniquely determined by $\tan\beta$, as shown in Table 2.10 (for relative couplings). Note that the tensor structure of all couplings is the same as in the Standard Model.

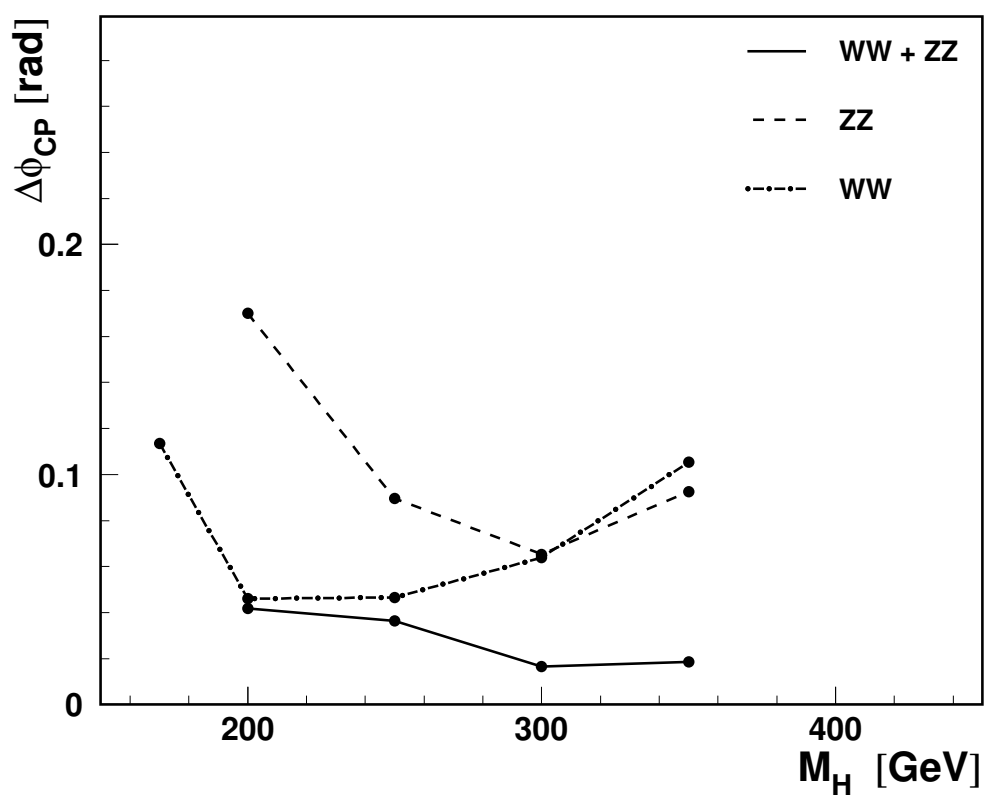

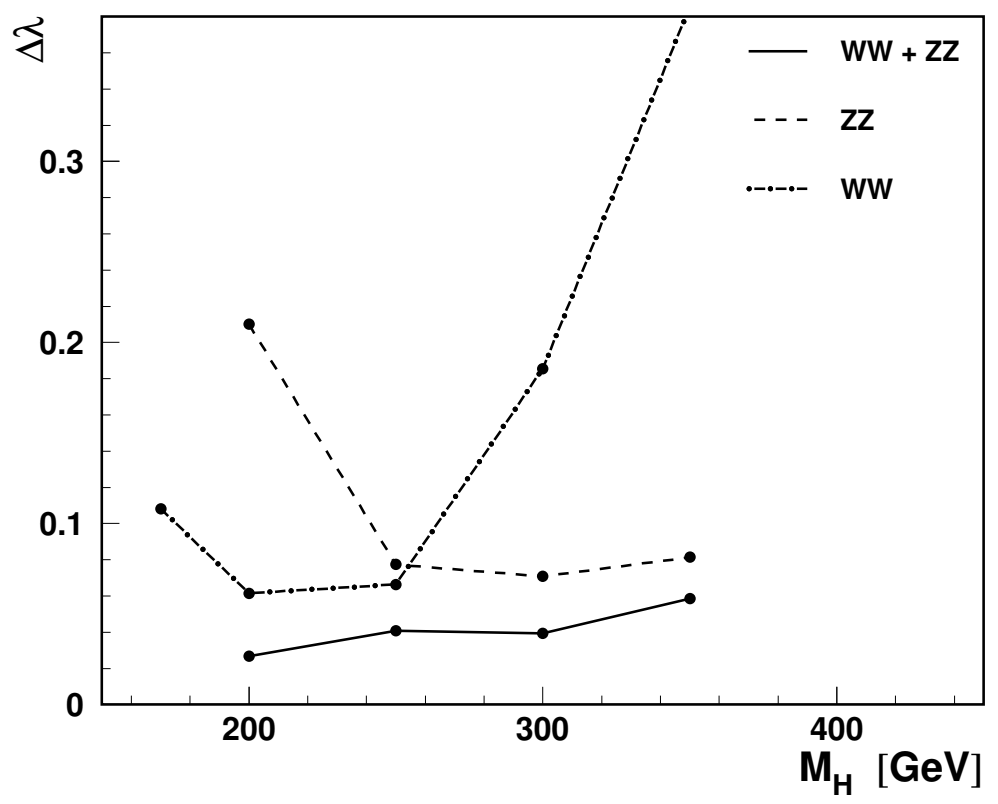

Fig. 2.35: Statistical error in the determination of $\Phi_{CP}$ (left plot) and $\lambda$ (right plot), expected after one year of Photon Linear Collider running, as a function of the Higgs-boson mass $M_\Phi$. Combined fits were performed to the considered angular distributions and invariant mass distributions for $ZZ$ events and $W^+W^-$ events. Results were obtained assuming small deviations from Standard Model predictions, i.e. $\lambda \approx 1$ and $\Phi_{CP} \approx 0$. The two photon width of the Higgs boson $\Gamma_{\gamma\gamma}$, the phase $\phi_{\gamma\gamma}$ and normalisations of both samples are allowed to vary in the fit.

The couplings of the lightest mass-eigenstate $h_1$ (with mass 120 GeV) are expected to correspond to the couplings of the SM-like $h$ boson, whereas couplings of $h_2$ and $h_3$ states can be described as the superposition of $H$ and $A$ couplings. For the relative basic couplings we have:

$$\begin{aligned} \chi_X^{h_1} &\approx \chi_X^h , \\ \chi_X^{h_2} &\approx \chi_X^H \cdot \cos\Phi_{HA} + \chi_X^A \cdot \sin\Phi_{HA} , \\ \chi_X^{h_3} &\approx \chi_X^A \cdot \cos\Phi_{HA} - \chi_X^H \cdot \sin\Phi_{HA} , \end{aligned} \tag{2.193}$$

where $X$ denotes a fermion or a vector boson, $X = u,\ d,\ V$ and $\Phi_{HA}$ is the $H - A$ mixing angle characterizing a weak CP violation.

We study the feasibility of $\Phi_{HA}$ determination from the combined measurement of the invariant-mass distributions[18] in the $ZZ$ and $W^+W^-$ decay-channels for the Higgs-boson mass-eigenstate $h_2$. From such measurement the $\gamma\gamma$ partial width, $\Gamma_{\gamma\gamma} \times BR(h \to W^+W^-/ZZ)$, and the phase of the $\gamma\gamma h$ amplitude, $\phi_{\gamma\gamma}$, can be extracted. Results obtained for $h_2$ with mass $M_{h_2} = 300$ GeV are presented in Fig. 2.36, for $M_{h_1} = 120$ GeV and $M_{H^\pm} = 800$ GeV. Error contours ($1\sigma$) on the measured deviation from the Standard Model predictions are shown for $\Phi_{HA} = 0$, i.e. when CP is conserved, and for the CP violation with $\Phi_{HA} = \pm 0.3$ rad. Even a small CP-violation can significantly influence the measured two-photon width and two-photon phase allowing to determine precisely both the CP-violating mixing angle $\Phi_{HA}$ and the parameter $\tan\beta$.

[18] It should be stressed that in the considered case of CP violation via $H - A$ mixing, contrary to the generic model studied in Section 2.14.2, only the invariant mass distributions are sensitive to the mixing angle $\Phi_{HA}$.

Table 2.10: Couplings of the neutral Higgs-bosons to up- and down-type fermions, and to vector bosons, relative to the Standard Model couplings, for the considered solution $B_h$ of the SM-like 2HDM (II).

| | $h$ | $H$ | $A$ |
|---|---|---|---|
| $\chi_u$ | $-1$ | $-\frac{1}{\tan\beta}$ | $-i\,\gamma_5\,\frac{1}{\tan\beta}$ |
| $\chi_d$ | $+1$ | $-\tan\beta$ | $-i\,\gamma_5\,\tan\beta$ |
| $\chi_V$ | $\cos(2\beta)$ | $-\sin(2\beta)$ | $0$ |

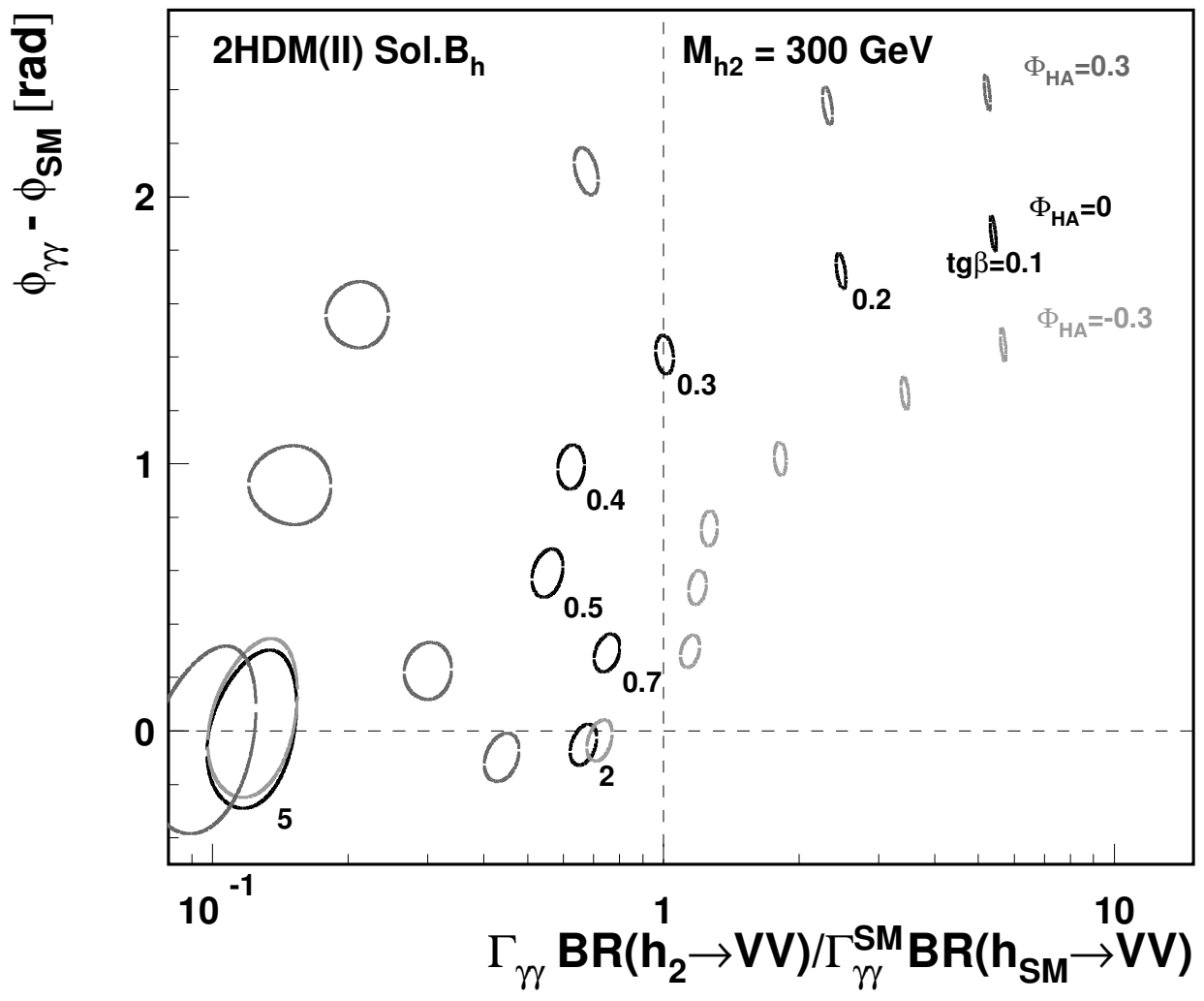

Fig. 2.36: The deviation from the SM predictions for the SM-like 2HDM II (sol. $B_h$) with CP-violation, for the heavy Higgs-boson $h_2$ with mass 300 GeV. A light Higgs-boson has mass $M_{h_1} = 120$ GeV. Three values of $H - A$ mixing angle $\Phi_{HA} = -0.3, 0, 0.3$ are considered.

As a large sample of events is expected at PLC, especially in the $\gamma\gamma \to W^+W^-$ channel, systematic uncertainties have to be taken into account, as they can significantly influence the final precision. In case of scenario $B_h$ with CP violation, a possible correlations between $\Phi_{HA}$ and $\tan\beta$ has to be considered if both parameters are to be constrained from the fit to the data. In this analysis the systematic uncertainties from following sources were considered: the total integrated $\gamma\gamma$ luminosity, shape of the luminosity spectra, energy and mass scale of the detector, reconstructed mass resolution, and in addition the Higgs-boson mass and width from other measurements. In order to take these uncertainties into account we include additional parameters in the fit. Variations of these parameters allow us to account for possible deviations of the invariant-mass distributions, from the nominal model expectation due to the systematic uncertainties.

The total error in the determination of the $H - A$ mixing angle $\Phi_{HA}$, as a function of $\tan\beta$ value, is presented in Fig. 2.37, for four values of heavy Higgs-boson mass $M_{h_2}$, between 200 and 350 GeV. The simultaneous fit of $\tan\beta$ and $\Phi_{HA}$ to the observed $W^+W^-$ and $ZZ$ mass spectra is considered assuming light Higgs-boson mass of 120 GeV, charged Higgs-boson mass of 800 GeV, and no $H - A$ mixing ($\Phi_{HA} = 0$). The error on $\Phi_{HA}$ is below $\sim$100 mrad for $\tan\beta \leq 1$ and increases rapidly for high $\tan\beta$ values.

### *2.14.4 Two Higgs Doublet Model*

In the CP violating 2HDM (II), couplings of the neutral Higgs-bosons to up- and down-type quarks (and leptons), and to vector bosons can be expressed in terms of two mixing angles, $\alpha$ and $\beta$, as discussed in Section 2.1.4. In the following we will consider production and decays of the heavy Higgs-boson $H$. Instead of parameters of the model, angles $\alpha$ and $\beta$, we will use its basic relative couplings $\chi_V^H$ and $\chi_u^H$, to parametrize cross sections and branching ratios. Moreover, couplings of the other neutral Higgs-bosons $h$ and $A$ are also uniquely defined by $\chi_V^H$ and $\chi_u^H$. As in Section 2.14.3 we consider a scenario with a weak CP violation, where the couplings of the lightest mass-eigenstate $h_1$ correspond to the couplings of $h$ boson, whereas relative couplings of mass-eigenstates $h_2$ and $h_3$ can be described as the superposition of $H$ and $A$ couplings (see Eq. 2.193). We study the feasibility of constraining the value of the mixing angle $\Phi_{HA}$ from the measurements of the heavy Higgs-boson $H$ (i.e. Higgs-boson mass-eigenstate $h_2$ for $\Phi_{HA} = 0$) production.

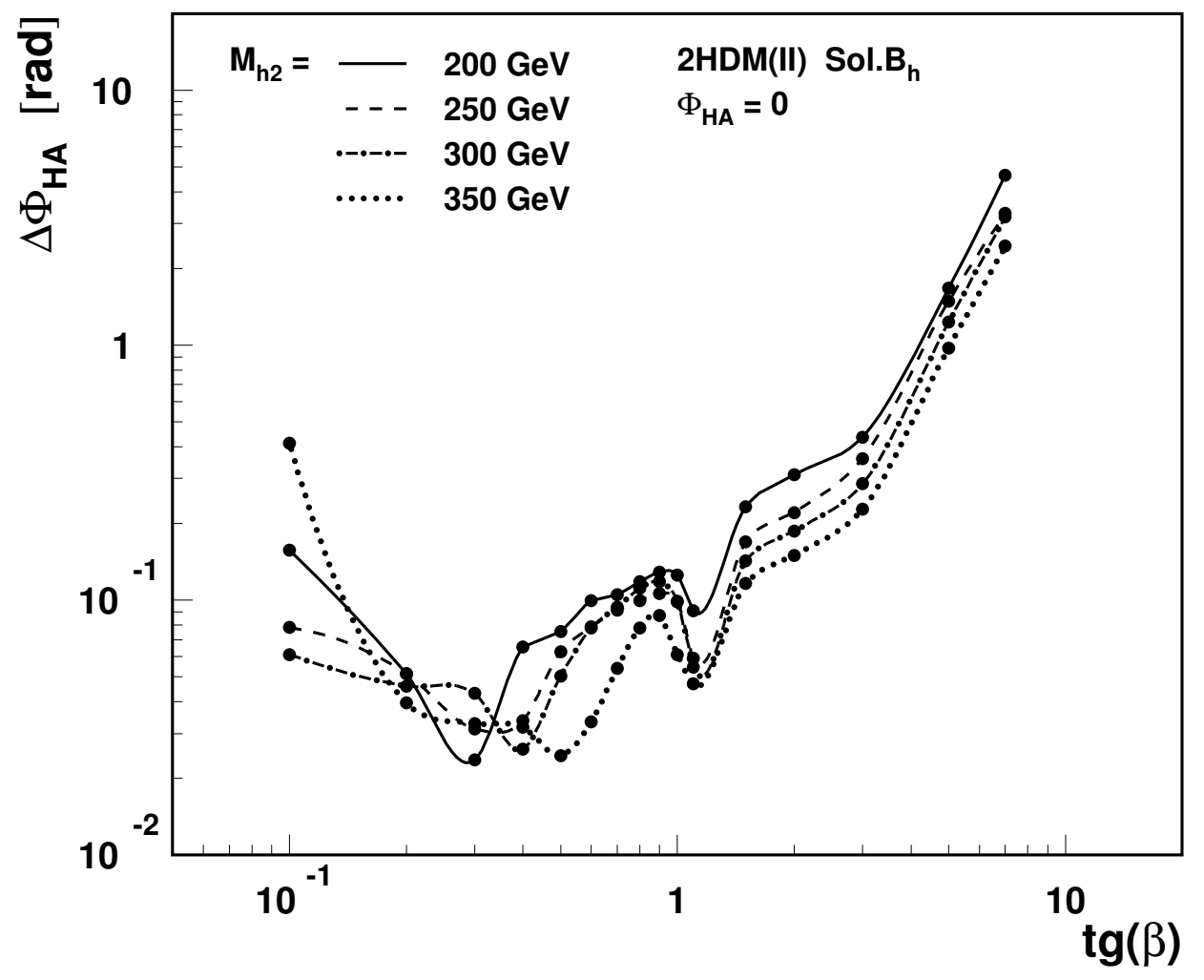

Fig. 2.37: The total error in the determination of the $H - A$ mixing angle $\Phi_{HA}$, as a function of $\tan\beta$ value, for four values of heavy Higgs-boson mass $M_{h_2}$. The simultaneous fit of $\tan\beta$ and $\Phi_{HA}$ to the observed $W^+W^-$ and $ZZ$ mass spectra is considered for the SM-like 2HDM II (sol. $B_h$), with light Higgs-boson mass of 120 GeV, charged Higgs-boson mass of 800 GeV, and no $H - A$ mixing ($\Phi_{HA} = 0$), Eq. 2.193. Systematic uncertainties related to the luminosity spectra, Higgs boson mass and total width, energy scale and mass resolution are taken into account.

The Photon Linear Collider by itself can not uniquely determine the Higgs-boson couplings in case of 2HDM (II) with CP-violation. Therefore, we consider determination of the heavy scalar Higgs-boson properties from the combined analysis of LHC, ILC and Photon Linear Collider data. Fig. 2.38 shows the expected Higgs-boson production rates times the $W^+W^-/ZZ$ branching ratios, at the LHC, ILC and PLC, as a function of $\chi_V$ and $\chi_u$. Cross section measurements at these machines are complementary, as they are sensitive to different combinations of the Higgs boson couplings. LHC, ILC and PLC measurements are also complementary in providing an evidence for a weak CP violation, as shown in Fig. 2.39.

An expected $h_2$ production rates for $h_2 \to W^+W^-/ZZ$ at LHC, ILC and PLC, are shown as a function of $\chi_u$ (LHC) or $\chi_V$ (ILC and PLC) and the CP-violating $H - A$ mixing angle $\Phi_{HA}$. For $\Phi_{HA} \approx 0$ LHC and ILC measurements weakly depend on the mixing angle $\Phi_{HA}$, as the cross section is dominated by one of the basic couplings, and there is no direct dependence on the coupling phase. At the Photon Linear Collider both couplings as well as their relative phase are important and the cross section is sensitive to the $H - A$ mixing angle (and its sign) even for small $\Phi_{HA}$.

In the simulation of LHC and ILC measurements we use approach similar to the method used for PLC, described in Section 2.14.1. We use results of [316] for the expected invariant mass distribution of the Higgs-boson signal ($pp \to H \to ZZ \to 4l$) and Standard Model background events at LHC, scaled to integrated luminosity of 300 $fb^{-1}$. For the Higgs-boson production via Higgs-strahlung and $WW$-fusion at ILC, for $\sqrt{s} = 500$ GeV and the integrated luminosity of 500 $fb^{-1}$, we use results of [317]. In both cases the signal distributions are obtained from a simple convolution of the Breit-Wigner mass distribution for the Higgs-boson with a detector resolution function. With such an assumption we can scale the SM signal expectations presented in [316, 317] to any scenario of the 2HDM (II).

For each simulated set of the LHC, ILC and PLC data, the Higgs-boson couplings and CP-violating $H$-$A$ mixing angle were used as the free parameters in the simultaneous fit of the expected distributions to all observed $W^+W^-$ and $ZZ$ mass spectra. To take into account systematic uncertainties additional parameters are added to the fit, as in Section 2.14.3. For LHC we assume $10\%$ systematic uncertainty in the normalization of the background and $20\%$ total systematic uncertainty in the expected signal rate [318]. For ILC the uncertainties in both signal and background normalization are assumed to be 5%. For PLC we take into account 5% uncertainty in the signal and $10\%$ uncertainty in the background normalization, as well as $10\%$ uncertainty in the parameters describing the shape of the luminosity spectra. The Higgs-boson mass is also used as a free parameter in the combined fit, since there will be no other measurements to constrain its value.

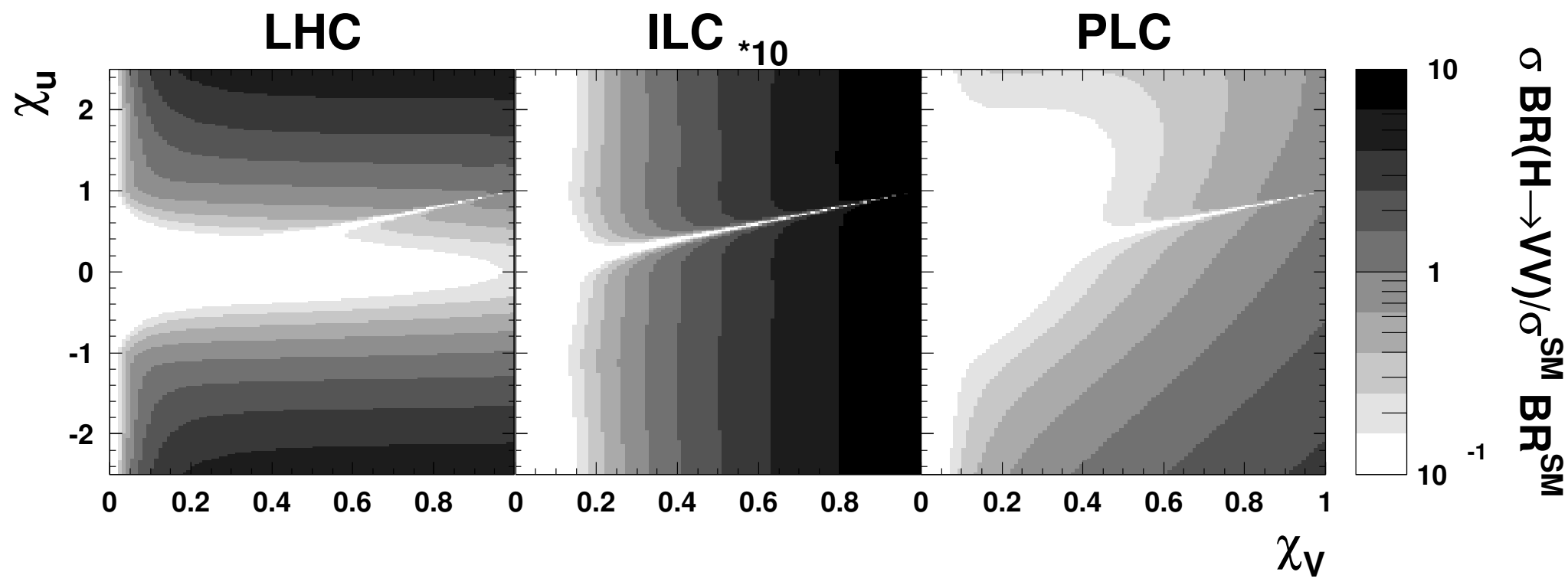

Fig. 2.38: Expected Higgs-boson $H$ production rates times $W^+W^-/ZZ$ branching ratios, relative to SM predictions, as a function of basic relative couplings to vector bosons ($\chi_V$) and up fermions ($\chi_u$). Higgs-boson production at LHC, ILC and PLC is studied for $M_H = 250$ GeV. For ILC the plotted ratio is multiplied by factor 10.

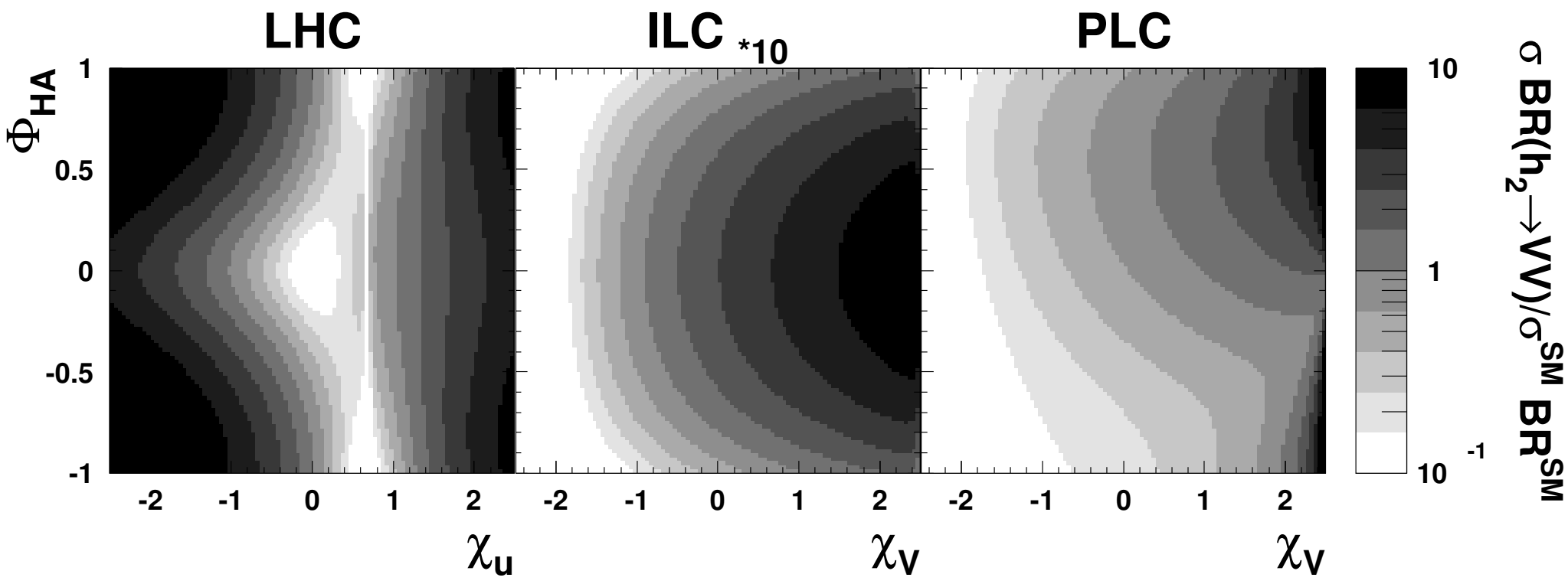

Fig. 2.39: Expected Higgs-boson $h_2$ production rates times $W^+W^-/ZZ$ branching ratios, relative to SM predictions, as a function of basic relative coupling to vector bosons ($\chi_V$) or up fermions ($\chi_u$), and the $H - A$ mixing angle $\Phi_{HA}$. Higgs-boson production at LHC, ILC and PLC is studied for $M_{h_2} = 250$ GeV. For ILC the plotted ratio is multiplied by factor 10.

In Fig. 2.40 the expected total error on the $H - A$ mixing angle $\Phi_{HA}$, calculated assuming weak CP violation ($\Phi_{HA} \approx 0$), is shown as a function of the couplings $\chi_V$ and $\chi_u$, for different heavy Higgs boson masses from 200 to 350 GeV. An average error on $\Phi_{HA}$ is about 150 mrad, although in most of the considered parameter space it can be measured with accuracy better than 100 mrad. The corresponding errors on the couplings $\chi_V$ and $\chi_u$, averaged over the same parameter range are equal to 0.03 and 0.13, respectively. No significant variations with the Higgs boson mass are observed.

The final step in verifying the coupling structure of the model is the comparison of the direct heavy neutral Higgs boson measurements with constraints on the model parameters resulting from other measurements in the Higgs sector. Assuming no CP violation, constraints on the couplings $\chi_V^H$ and $\chi_u^H$, used to parametrize the model obtained from heavy Higgs-boson (with mass of 250 GeV) and light Higgs-boson (with mass of 120 GeV) measurements at LHC, ILC and PLC, are compared in Fig. 2.41. Measurements of the light Higgs-boson production result in constraints on the $H$ couplings, comparable with the precision of the direct measurements (indicated by the ellipse).

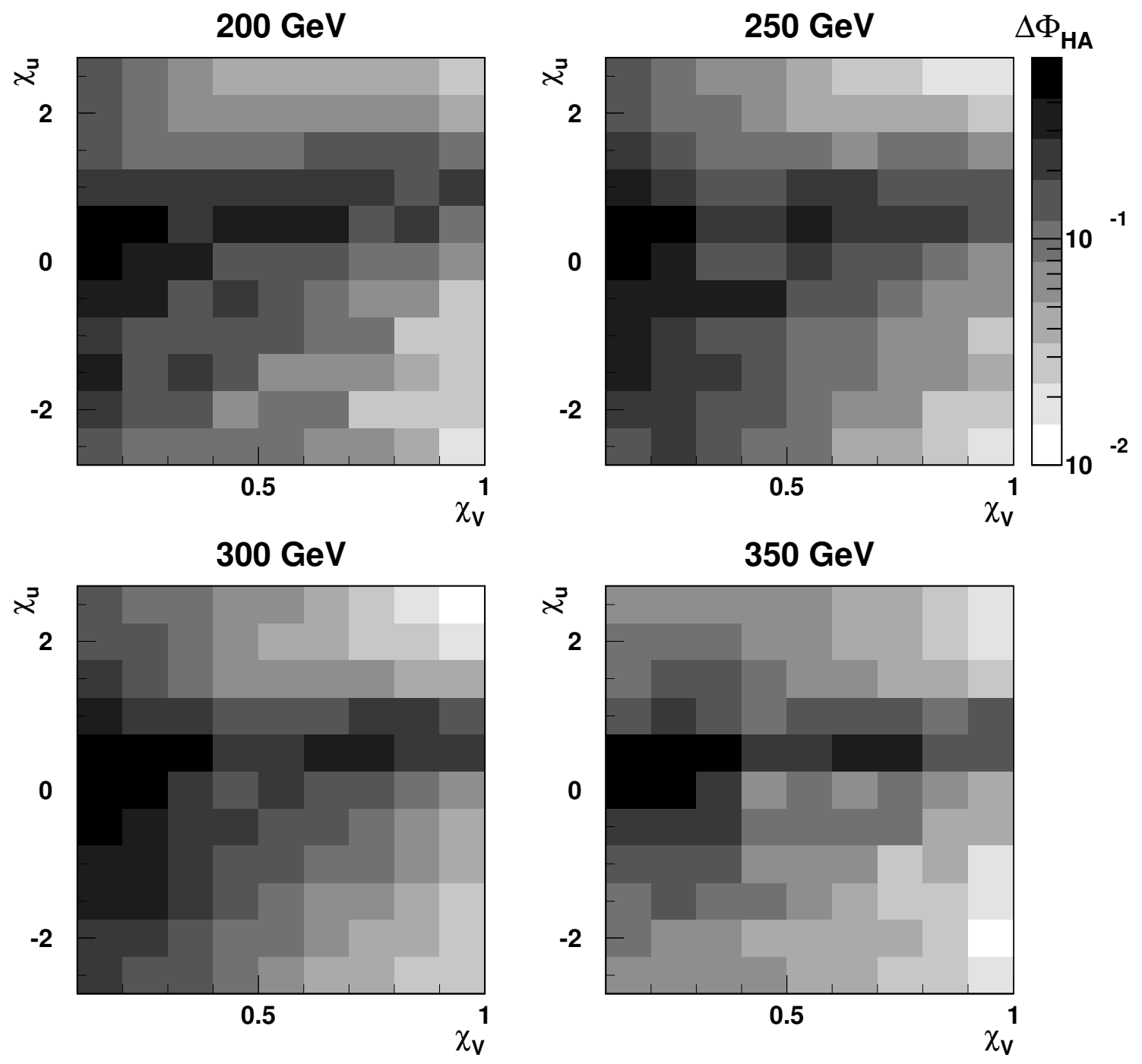

Fig. 2.40: Expected total errors on the $H-A$ mixing angle $\Phi_{HA}$, from combined fit to the invariant mass distributions measured at LHC, ILC and PLC, for $\Phi_{HA}=0$ and different heavy Higgs boson $h_2$ masses, as indicated in the plot.

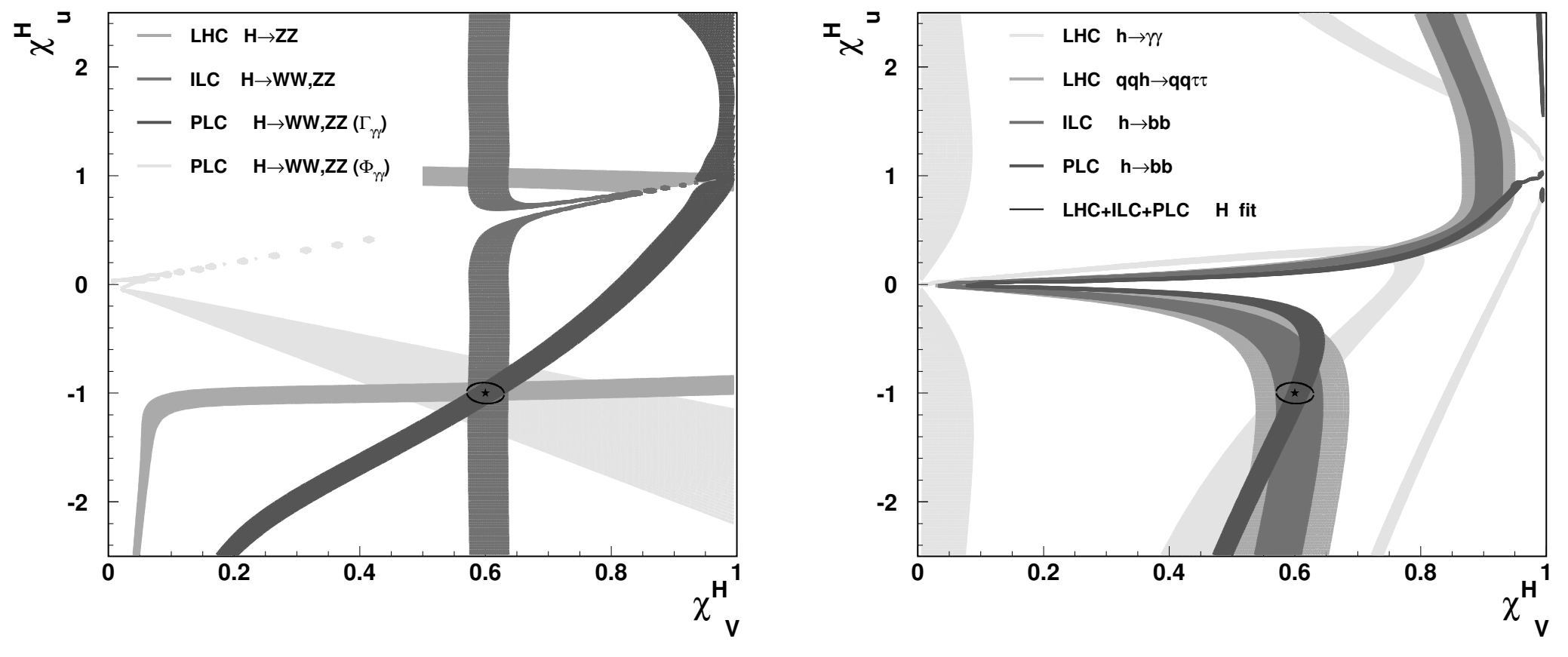

Fig. 2.41: A complementarity of LHC, ILC and PLC measurements in the determination of the 2HDM (II) parameters. Bands show values of the basic heavy Higgs-boson couplings to vector bosons ($\chi_V^H$) and up-type fermions ($\chi_u^H$) consistent (on $1\sigma$ statistical error level) with heavy Higgs-boson (left plot) and light Higgs-boson (right plot) measurements at LHC, ILC and PLC, assuming CP conservation. Model with $\chi_V^H=0.6$, $\chi_u^H=-1$ (star) and $H$ mass of 300 GeV is considered, while the mass of $h$ is set to 120 GeV.

*2.14.5 Summary*

An opportunity of measuring Higgs-boson CP properties at the Photon Linear Collider has been studied in detail for Higgs boson masses between 200 and 350 GeV, using realistic luminosity spectra and detector simulation. We considered three different models with CP violation. For a generic model with the CP violating Higgs tensor couplings to gauge bosons, the angle describing CP violation can be determined with accuracy of about 50 mrad in a model independent way. In the so called solution $B_h$ of the Standard Model-like 2HDM (II), the $H - A$ mixing angle describing the weak CP violation can be determined to about 100 mrad, for low $\tan\beta$. For the Two Higgs Doublet Model, only the combined analysis of LHC, ILC and PLC measurements allows for the determination of the CP-violating mixing angle $\Phi_{HA}$. In most of the considered parameter space, $\Phi_{HA}$ can be measured to better than 100 mrad. Our results demonstrate that the Photon Linear Collider will be an unique place for a precise determination of the CP properties of the neutral Higgs boson.

# 3 The Minimal Supersymmetric Standard Model with CP Phases

## 3.1 Introduction

*Maarten Boonekamp, Marcela Carena, Seong Youl Choi, Jae Sik Lee and Markus Schumacher*

One of the most theoretically appealing realizations of the Higgs mechanism for mass generation is provided by supersymmetry (SUSY). The minimal supersymmetric extension of the standard model (MSSM) has a number of interesting field–theoretic and phenomenological properties, if SUSY is softly broken such that super–particles acquire masses not greatly exceeding 1 TeV. Specifically, within the MSSM, the gauge hierarchy can be made technically natural [1–6]. Unlike the SM, the MSSM exhibits quantitatively reliable gauge-coupling unification at the energy scale of the order of $10^{16}$ GeV [7–14]. Furthermore, the MSSM provides a successful mechanism for cosmological baryogenesis via a strongly first-order electroweak phase transition [15–29], and provides viable candidates for cold dark matter [30–37].

The MSSM makes a crucial and definite prediction for future high-energy experiments, that can be directly tested at the Tevatron and/or the LHC. It guarantees the existence of (at least) one light neutral Higgs boson with mass bounded from above at $\mathcal{O}(140\ \mathrm{GeV})$ [38–44]. This rather strict upper bound on the lightest Higgs boson mass is in accord with global analyses of the electroweak precision data, which point towards a relatively light SM Higgs boson, with $M_{H_{\rm SM}} \lesssim 186$ GeV at the 95 % confidence level [45]. Furthermore, because of the decoupling properties of heavy superpartners, the MSSM predictions for the electroweak precision observables can easily be made consistent with all the experimental data [46, 47].

An important and interesting phenomenological feature of the MSSM Higgs sector is that loop effects mediated dominantly by third-generation squarks may lead to sizeable violations of the tree-level CP invariance of the MSSM Higgs potential, giving rise to significant Higgs scalar–pseudoscalar transitions [48, 49], in particular. As a consequence, the three neutral Higgs mass eigenstates $H_{1,2,3}$, labeled in order of increasing mass such that $M_{H_1} \leq M_{H_2} \leq M_{H_3}$, have no definite CP parities, but become mixtures of CP-even and CP-odd states. In this case, the conventional CP-odd Higgs mass $M_A$ is no longer a physical parameter. Instead, the charged Higgs mass is still physical and can be used as an input.

Much work has been devoted to studying in greater detail this radiative Higgs–sector CP violation in the framework of the MSSM [50–61]. In the MSSM with explicit CP violation, the upper bound on the lightest Higgs boson mass is almost identical to the one obtained in the CP conserving case [50]. The couplings of the Higgs bosons to the SM gauge bosons and fermions, to their supersymmetric partners and to the Higgs bosons themselves may be considerably modified from those predicted in the CP-conserving case. Consequently, radiative CP violation in the MSSM Higgs sector can significantly affect the production rates and decay branching fractions of the Higgs bosons. In particular, the drastic modification of the couplings of the $Z$ boson to the two lighter Higgs bosons $H_1$ and $H_2$ might enable a relatively light Higgs boson with a mass $M_{H_1}$ even less than about 70 GeV to have escaped detection at LEP 2 [62]. The upgraded Tevatron collider and the LHC will be able to cover a large fraction of the MSSM parameter space, including the challenging regions with a light Higgs boson without definite CP parity [62–77]. Furthermore, complementary and accurate explorations of the CP-noninvariant MSSM Higgs sector may be carried out using high-luminosity $e^+e^-$ [78–82] and/or $\gamma\gamma$ colliders [83–93]. In addition, a complete determination of the CP properties of the neutral Higgs bosons is possible at muon colliders by exploiting polarized muon beams [94–102].

This introductory section is devoted to a short description of the key aspects and important experimental implications of the MSSM Higgs sector with radiatively–induced CP violation.

### *3.1.1 CP phases in the MSSM*

Any phenomenologically viable SUSY model requires us to introduce terms which break SUSY *softly*, without spoiling the supersymmetric mechanism solving the hierarchy problem. There are three kinds of soft SUSY breaking terms in the framework of the MSSM:

– The gaugino mass terms:

$$\frac{1}{2}\left(M_3\,\widetilde{g}^a\widetilde{g}^a + M_2\,\widetilde{W}^i\widetilde{W}^i + M_1\,\widetilde{B}\widetilde{B} + \text{h.c.}\right)\,, \tag{3.1}$$

where $M_3$ is a gluino mass parameter of the gauge group SU(3)$_c$ and $M_2$ and $M_1$ are wino and bino mass parameters of the gauge groups SU(2)$_L$ and U(1)$_Y$, respectively.

– The trilinear $A$ terms:

$$\widetilde{u}_R^*\,h_u\,A_u\,\widetilde{Q}H_2 - \widetilde{d}_R^*\,h_d\,A_d\,\widetilde{Q}H_1 - \widetilde{e}_R^*\,h_e\,A_e\,\widetilde{L}H_1 + \text{h.c.}\,, \tag{3.2}$$

where $\widetilde{Q}$ and $\widetilde{L}$ are SU(2)$_L$ doublet squark and slepton fields and $\widetilde{u}_R$, $\widetilde{d}_R$, and $\widetilde{e}_R$ are SU(2)$_L$ singlet fields.

– The scalar mass terms:

$$\begin{aligned}&\widetilde{Q}^\dagger\,M^2_{\widetilde{Q}}\,\widetilde{Q} + \widetilde{L}^\dagger\,M^2_{\widetilde{L}}\,\widetilde{L} + \widetilde{u}_R^*\,M^2_{\widetilde{u}}\,\widetilde{u}_R + \widetilde{d}_R^*\,M^2_{\widetilde{d}}\,\widetilde{d}_R + \widetilde{e}_R^*\,M^2_{\widetilde{e}}\,\widetilde{e}_R\\ &+ m_2^2 H_2^* H_2 + m_1^2 H_1^* H_1 - (m_{12}^2 H_1 H_2 + \text{h.c.})\,.\end{aligned} \tag{3.3}$$

One crucial observation is that all the massive parameters appearing in the soft SUSY breaking terms can be complex with non-trivial CP-violating phases. Together with the phase of the Higgsino mass parameter $\mu$ of the term $-\mu H_1 H_2$ in the superpotential, all the physical observables depend on the CP phases of the combinations $\text{Arg}[M_i\,\mu\,(m_{12}^2)^*]$ and $\text{Arg}[A_f\,\mu\,(m_{12}^2)^*]$ [103, 104]. We have taken the convention of $\text{Arg}(m_{12}^2) = 0$ keeping the explicit dependence of $\mu$. These new CP phases would lead to various interesting phenomena and, moreover, reopen the possibility of explaining the baryon asymmetry of the Universe in the framework of MSSM [16, 18, 20–22, 25–27, 29, 105, 106].

### *3.1.2 Loop-induced CP violation in the Higgs sector*

Through the radiative corrections, the CP-violating mixing among the CP-even $\phi_{1,2}$ and CP-odd $a$ states is induced [48–54]. Due to large Yukawa couplings, the third generation scalar quarks contribute most significantly at one-loop level. The size of the mixing is proportional to

$$\frac{3}{16\pi^2}\,\frac{\text{Im}\,(A_f\mu)}{m^2_{\tilde{f}_2} - m^2_{\tilde{f}_1}} \tag{3.4}$$

with $f = t, b$. At two-loop level, the gluino mass parameter becomes relevant, for example, through the possibly important threshold corrections to the top- and bottom-quark Yukawa couplings. More CP phases become relevant by including other radiative corrections than those from the stop and sbottom sectors [55–57].

A. Mass spectra and couplings

The most comprehensive calculation of the CP-violating mixing and Higgs-boson mass spectrum in full consideration of the dependence on CP phases can be found in Refs. [53, 58] and [60, 61]. The Higgs-boson pole masses are calculated and all leading two-loop logarithmic corrections are incorporated in the one-loop RG–improved diagrammatic approach.

Due to the loop-induced CP-violating mixing, the neutral Higgs bosons do not have to carry any definite CP parities and the mixing among them is described by 3×3 real orthogonal matrix $O$ instead of 2×2 one with a rotation angle $\alpha$. The matrix $O$ relates the Electroweak states to the mass eigenstates as:

$$(\phi_1\,,\phi_2\,,a)^T = O\,(H_1\,,H_2\,,H_3)^T\,. \tag{3.5}$$

We find the relation $O = R^T$ in which the rotation matrix $R$ is given by Eq. (2.27) in Section 2.1. The Higgs-boson couplings to the SM and SUSY particles could be modified significantly due to the CP violating mixing. The most eminent example is the Higgs-boson coupling to a pair of vector bosons, $g_{H_iVV}$, which is responsible for the production of Higgs bosons at $e^+e^-$ colliders:

$$\mathcal{L}_{HVV} = gM_W\left(W^+_\mu W^{-\mu} + \frac{1}{2c_W^2}Z_\mu Z^\mu\right)\sum_{i=1}^{3} g_{H_iVV}H_i\,, \tag{3.6}$$

where $g_{H_iVV} = c_\beta O_{\phi_1 i} + s_\beta O_{\phi_2 i}$ which is normalized to the SM value and given by the weighted sum of the CP-even components of the $i$-th Higgs mass eigenstate. Compared to the CP-conserving case, it's possible for the lightest Higgs boson to develop significant CP-odd component and its coupling to a pair of vector bosons becomes vanishingly small. The interactions of a pair of Higgs bosons to a vector boson are given by

$$\begin{aligned}\mathcal{L}_{HHZ} &= \frac{g}{4c_W}\sum_{i,j=1}^{3} g_{H_iH_jZ}\,Z^\mu(H_i\,i\overleftrightarrow{\partial}_\mu\,H_j)\,,\\ \mathcal{L}_{HH^\pm W^\mp} &= -\frac{g}{2}\sum_{i=1}^{3} g_{H_iH^+W^-}\,W^{-\mu}(H_i\,i\overleftrightarrow{\partial}_\mu\,H^+) + \text{h.c.}\,,\end{aligned} \tag{3.7}$$

where

$$g_{H_iH_jZ} = \text{sign}[\det(O)]\;\varepsilon_{ijk}\,g_{H_kVV} \quad \text{and} \quad g_{H_iH^+W^-} = c_\beta\,O_{\phi_2 i} - s_\beta\,O_{\phi_1 i} - iO_{ai} \tag{3.8}$$

leading to the following sum rules:

$$\sum_{i=1}^{3} g^2_{H_iVV} = 1 \quad \text{and} \quad g^2_{H_iVV} + |g_{H_iH^+W^-}|^2 = 1 \quad \text{for each } i\,. \tag{3.9}$$

The effective Lagrangian governing the interactions of the neutral Higgs bosons with quarks and charged leptons is

$$\mathcal{L}_{H_i\bar{f}f} = -\sum_{f=u,d,l}\frac{gm_f}{2M_W}\sum_{i=1}^{3} H_i\,\bar{f}\left(g^S_{H_i\bar{f}f} + ig^P_{H_i\bar{f}f}\gamma_5\right)f\,. \tag{3.10}$$

At the tree level, $(g^S_{H_i\bar{f}f}, g^P_{H_i\bar{f}f}) = (O_{\phi_1 i}/c_\beta, -O_{ai}\tan\beta)$ and $(g^S_{H_i\bar{f}f}, g^P_{H_i\bar{f}f}) = (O_{\phi_2 i}/s_\beta, -O_{ai}\cot\beta)$ for $f = (l, d)$ and $f = u$, respectively. We observe that all neutral Higgs bosons can couple to both scalar and pseudoscalar fermion bilinear currents simultaneously ($\bar{f}f$ and $\bar{f}\gamma_5 f$, respectively) in the presence of CP-violating mixing. In the case of third-generation quarks, the couplings depend on the threshold corrections induced by the exchanges of gluinos and charginos which modify the relations between quark masses and the corresponding Yukawa couplings as:

$$\begin{aligned}h_b &= \frac{\sqrt{2}m_b}{v\cos\beta}\,\frac{1}{1 + (\delta h_b/h_b) + (\Delta h_b/h_b)\tan\beta}\,,\\ h_t &= \frac{\sqrt{2}m_t}{v\sin\beta}\,\frac{1}{1 + (\delta h_t/h_t) + (\Delta h_t/h_t)\cot\beta}\,.\end{aligned} \tag{3.11}$$

We note that the corrections do not decouple in the limit of large SUSY breaking parameters and the dominant contributions to $h_b$ are $\tan\beta$-enhanced. The relation between $m_\tau$ and the Higgs–tau-lepton Yukawa coupling $h_\tau$ is also modified, but the corrections are expected to be smaller than those to $h_b$. The corrections depend on the combinations of $\mu M_3$ and $\mu A_t$, stop and sbottom masses, etc. We refer to, for example, Ref. [58] for details of them.

The couplings of the charged Higgs bosons to quarks are of the form $\mathcal{L}_{H^\pm t\bar{b}} = \bar{b}\,(g^L_{H^- t\bar{b}} P_L + g^R_{H^- t\bar{b}} P_R)\, tH^- + \text{h.c.}$ with the couplings given by

$$g^L_{H^- t\bar{b}} \simeq \frac{\sqrt{2}\,m_b}{v}\tan\beta - \frac{\Delta h_b}{\cos\beta} \qquad \text{and} \qquad g^R_{H^- t\bar{b}} \simeq \frac{\sqrt{2}\,m_t}{v}\cot\beta - \frac{(\Delta h_t)^*}{\sin\beta}\,. \tag{3.12}$$

An explicit computation of the CP-violating $H^- t\bar{b}$ vertex can be found in Ref. [58].

For large values of the charged Higgs boson mass and for heavy supersymmetric particles, the expressions of the lightest neutral Higgs boson coupling to fermions reduce to those of the (CP-conserving) SM Higgs boson, as expected for the decoupling limit. In contrast, the two heavy neutral Higgs bosons are still admixtures of CP-even and CP-odd eigenstates; hence, CP-violating effects are still present in the heavy neutral Higgs sector. However, due to the high degeneracy in mass of the heavy scalar sector (especially in the decoupling limit [107]), CP–violating effects may be difficult to observe without precision measurements of the heavy neutral Higgs properties.

The so called CPX scenario has been defined as a benchmark point for studying the CP-violating Higgs-mixing phenomena [54]. In this scenario, the parameters have been fixed as follows:

$$\begin{aligned} &M_{\tilde{Q}_3} = M_{\tilde{U}_3} = M_{\tilde{D}_3} = M_{\tilde{L}_3} = M_{\tilde{E}_3} = M_{\rm SUSY}\,, \\ &|\mu| = 4\,M_{\rm SUSY}\,, \ \ |A_{t,b,\tau}| = 2\,M_{\rm SUSY}\,, \ \ |M_3| = 1\ \ \text{TeV}\,. \end{aligned} \tag{3.13}$$

The parameter $\tan\beta$, the charged Higgs-boson pole mass $M_{H^\pm}$, and the common SUSY scale $M_{\rm SUSY}$ can be varied. For CP phases, taking $\Phi_\mu = 0$ convention and a common phase for $A$ terms $\Phi_A = \Phi_{A_t} = \Phi_{A_b} = \Phi_{A_\tau}$, we have two physical phases to vary: $\Phi_A$ and $\Phi_3 = \text{Arg}(M_3)$.

Corrections to the MSSM Higgs boson sector have been evaluated in several approaches. At the one loop level the complete result for radiative corrections to the masses and mixing angles in the MSSM Higgs sector is known [38–44]. Concerning the two-loop effects, their computation is quite advanced and has now reached a stage such that all the presumably dominant contributions are known [108–129] with a remaining theoretical uncertainty on the light CP-even Higgs boson mass which is estimated to be below $\sim$ 3 GeV [47, 130]. The results of the radiative correction calculations have been implemented into public codes.

The code `CPsuperH` [131] is based on the renormalization group (RG) improved effective potential approach [109–111, 132–140] and it implements the results obtained in Refs. [53, 58, 141–143], see Section 3.4. The program `FeynHiggs` [144–148] is based on the results obtained in the Feynman-diagrammatic (FD) approach [113–115, 129, 130], see Section 3.5. For the MSSM with real parameters the two codes can differ by up to $\sim$ 4 GeV for the light CP-even Higgs boson mass, mostly due to subleading two-loop corrections that are included only in `FeynHiggs`. For the MSSM with complex parameters the phase dependence at the two-loop level is included in a more advanced way in `CPsuperH`, but, on the other hand, `CPsuperH` does not contain all the subleading one-loop contributions that are included in `FeynHiggs`. The plots of this Introduction have been obtained by use of `CPsuperH`.

Figure 3.1 shows the Higgs-boson pole masses $M_{H_i}$ and the couplings squared $g^2_{H_i VV}$ as functions of $\Phi_A$ for the CPX scenario. When $M_{H^\pm} = 120$ GeV (left frames), around $\Phi_A = 90^\circ$, we observe $H_1$ becomes light with vanishingly small couplings. In other words, it becomes lighter than 50 GeV and behaves almost like a CP-odd state. In this case, $H_1$ production rate at LEP is very low and $H_2$ dominantly decays into a pair of the lightest Higgs bosons which subsequently decays into 4 $b$ quarks.

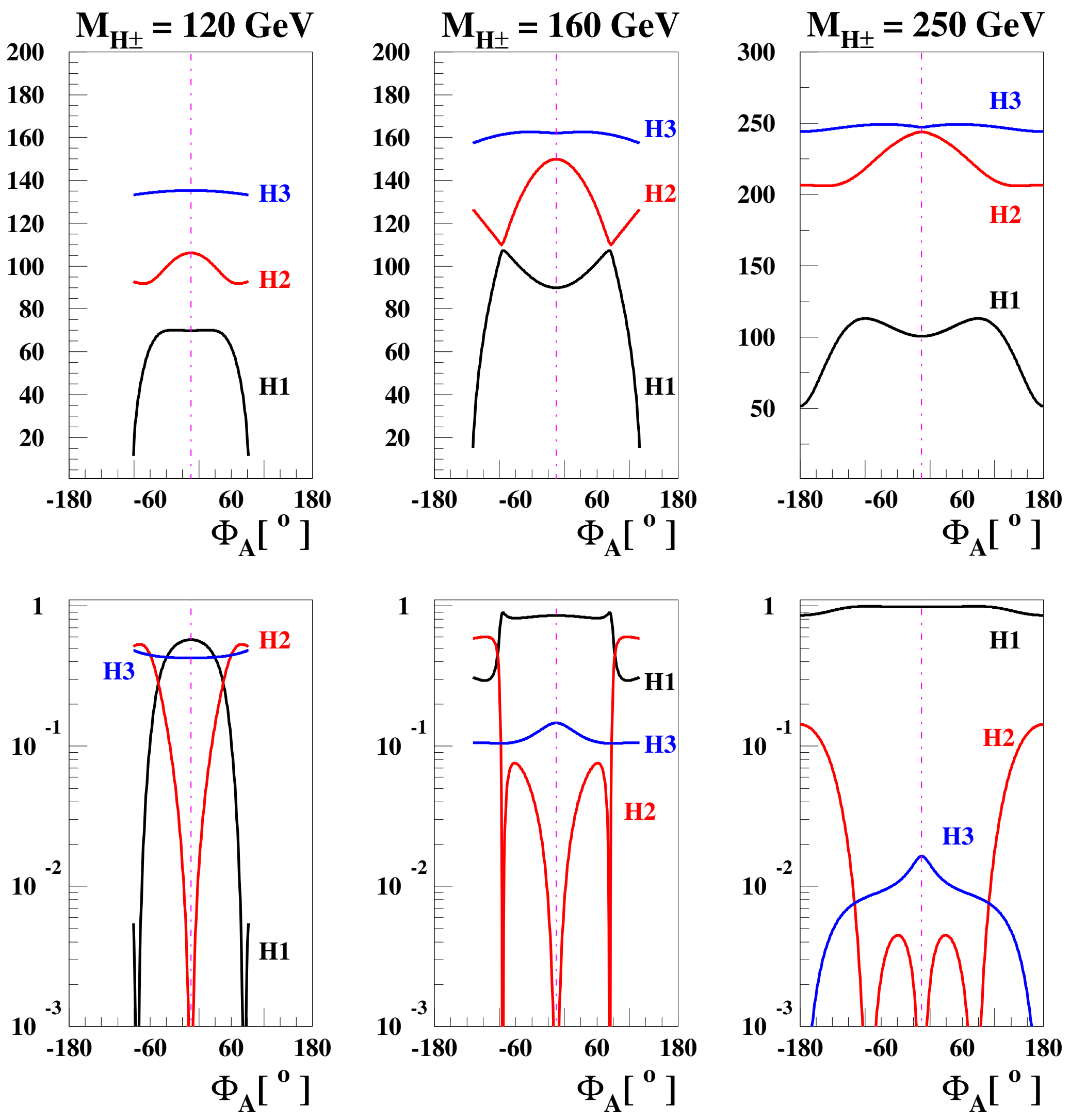

Fig. 3.1: The Higgs-boson masses $M_{H_i}$ (upper frames) in GeV and $g^2_{H_iVV}$ (lower frames) as functions of $\Phi_A$ for the CPX scenario when $\tan\beta = 4$, $\Phi_3 = 0°$, and $M_{\rm SUSY} = 0.5$ TeV. Three values of the charged Higgs-boson pole mass have been taken: 120 GeV (left frames), 160 GeV (middle frames), and 250 GeV (right frames).

This makes the Higgs detection at LEP difficult and the region with $M_{H_1} \leq 50$ GeV and $\tan\beta = 4$–8 has not been excluded yet [149, 150]. In the middle frames with $M_{H^\pm} = 160$ GeV, we observe a resonant-mixing behavior between $H_1$ and $H_2$ around $\Phi_A = 90°$. The lightest Higgs becomes SM-like Higgs boson and decouples from the 3×3 mixing when the charged Higgs boson becomes heavy, see the right frames of Fig. 3.1. Nevertheless, there still can be significant mixing between the two heavier neutral mass eigenstates due to their highly–degenerate masses.

The Higgs-boson decay patterns strongly depend on the CP-violating mixing. For this we show the branching fractions and decay widths of the MSSM Higgs bosons in Figs. 3.2 and 3.3. In the CP-conserving case (Fig. 3.2), the decay channels $H_2 \to H_1H_1, WW, ZZ$ and $H_3 \to H_1Z$ are forbidden. In the CP-violating case (Fig. 3.3), on the other hand, all the decay channels are open for the heavier Higgs bosons. We note that, in the CP-violating case, both heavier Higgs bosons $H_2$ and $H_3$ dominantly decay into the lightest Higgs-boson pairs where the decay widths drastically increase. Also there, the decay width of the charged Higgs boson increases and it mainly decays into $W^\pm H_1$.

The phenomenological implication of the CP-violating couplings of the charged Higgs boson to quarks can be found in Ref. [68].

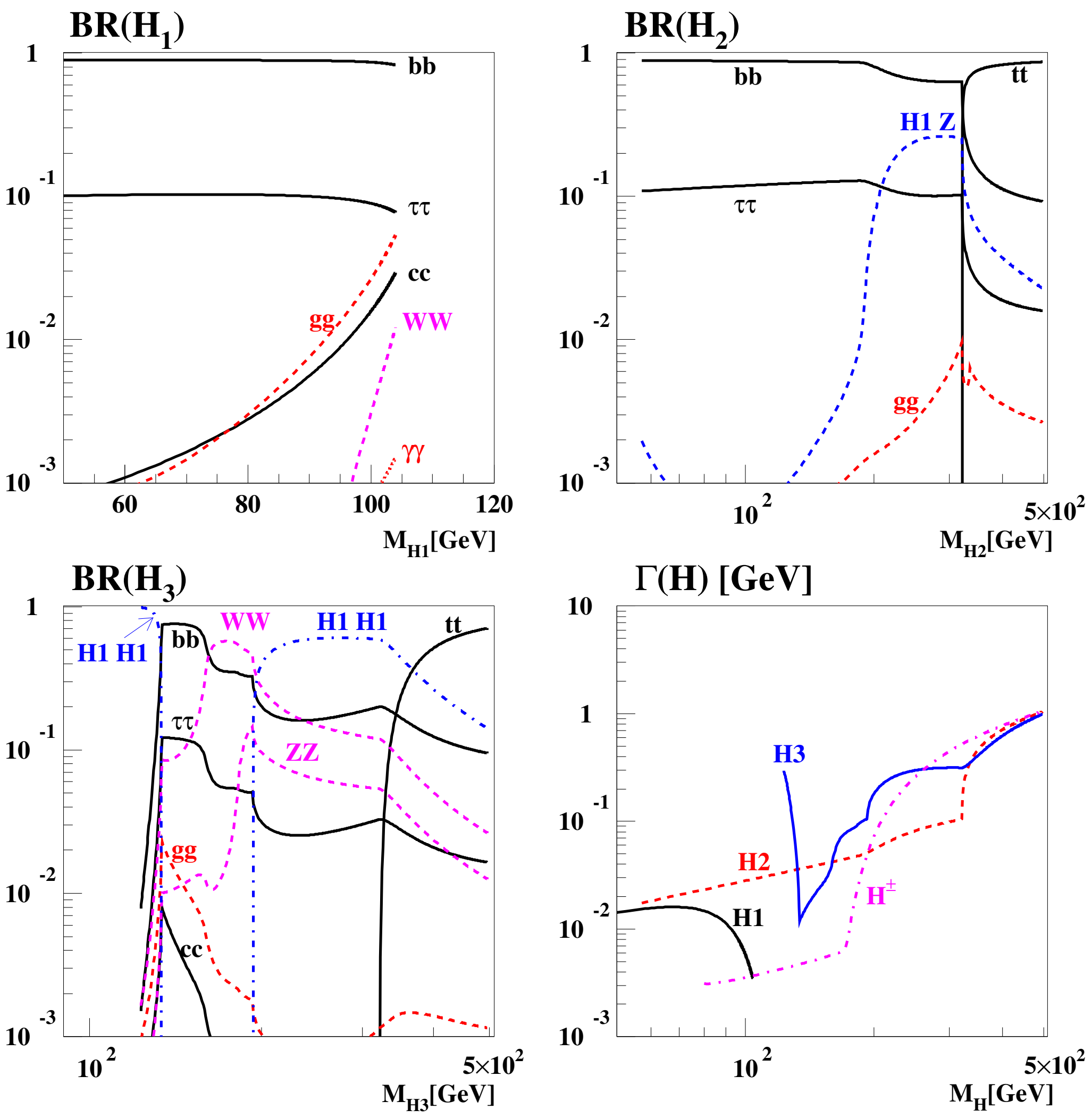

Fig. 3.2: The branching fractions and decay widths of the MSSM Higgs bosons for the CPX scenario with $\tan\beta = 4$ and $M_{\rm SUSY} = 0.5$ TeV as functions of their masses. Here, we are taking the CP-conserving case ($\Phi_A = \Phi_3 = 0°$). See Fig. 3.3 for the CP-violating case.

B. Low–energy constraints

Low-energy observables provide indirect constraints on the soft SUSY breaking parameters. The observables are particularly useful for identifying the favoured range of parameter space when the SM predictions for them are strongly suppressed and/or precise experimental measurements of them have been performed. Such observables include EDMs, $(g-2)_\mu$, $\mathrm{BR}(b \to s\gamma)$, $\mathcal{A}_{\rm CP}(b \to s\gamma)$, $\mathrm{BR}(B \to K l\, l)$ and $\mathrm{BR}(B_{s,d} \to l^+ l^-)$.

Currently, the EDM of the thallium atom provides one of the best constraints on the CP-violating phases, depending on the SUSY scale. The main contributions to the atomic EDM of $^{235}$Tl come from two terms. One of them is the electron EDM $d_e$ and the other is the coefficient $C_S$ of a CP-odd electron-nucleon interaction. The coefficient $C_S$ is essentially given by the gluon-gluon-Higgs couplings and the two-loop Higgs-mediated electron EDM [151, 152] is given by the sum of contributions from third-generation quarks and squarks and charginos. As can be seen for example in Ref. [153], the two kinds of dominant contributions could cancel each other allowing narrow region compatible with the EDM constraint but with sizable CP phases. If no cancellations take place, the allowed CP-phases are highly constrained, in particular for large values of $\tan\beta$ and small values of the CP-odd Higgs mass. However, they may be large enough to allow for the possibility of electroweak baryogenesis in the MSSM [154–156]. The Thallium EDM constraint can be evaded more easily by assuming cancellations between the two-loop and possible one-loop contributions. This shows that the possibility of large CP phases which can induce significant CP-violating mixing in the Higgs sector cannot be excluded *a priori*.

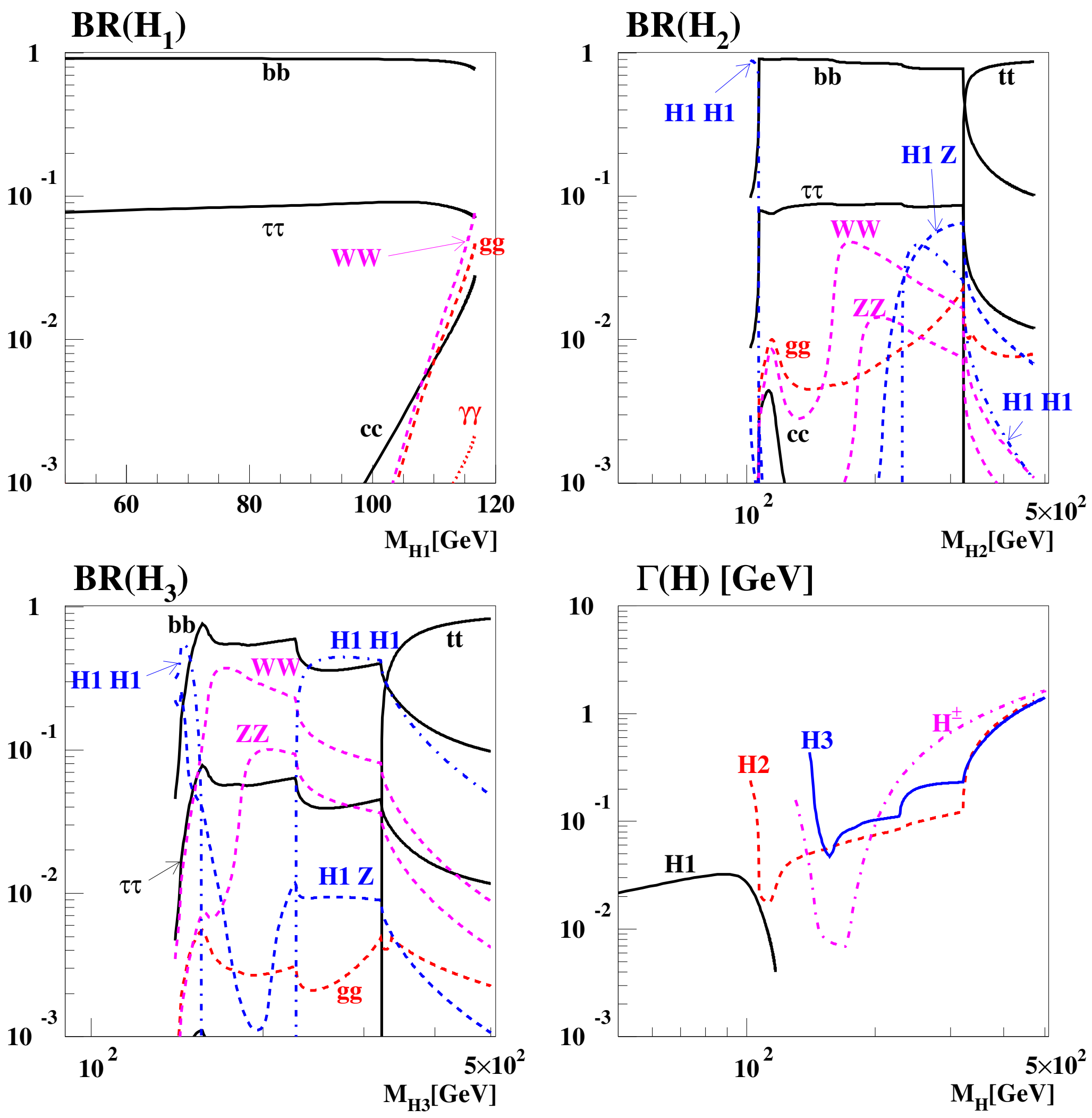

Fig. 3.3: The same as in Fig. 3.2 but with $\Phi_A = \Phi_3 = 90°$.

### *3.1.3 Coupled channel analysis*

In the presence of non-trivial CP violating phases, the MSSM neutral Higgs bosons have a tendency to show strong mixing among them, with small mass differences comparable to their widths. When the charged Higgs boson is heavy, the two heavier Higgs bosons mix significantly. On the other hand, all three neutral Higgs bosons show strong three-way mixing acquiring significant CP-even and CP-odd components when the charged Higgs boson is light and, especially, when the values of $\tan\beta$ is large. In this case, when considering Higgs boson production at colliders, each Higgs boson can not be treated separately and all three neutral Higgs boson should be considered as a coupled system. The characteristic feature of the coupled–channel analysis lies in the off-diagonal absorptive parts in the inverse of the full 3×3 propagator matrix [157–159]. For an explicit example, see Ref. [160] that shows the effects of including the off-diagonal absorptive parts in the Higgs-boson propagators based on a scenario in which all three neutral Higgs bosons are nearly degenerate in mass, around 120 GeV, and with widths of the order of 1–3 GeV.

#### *3.1.4 Experimental signatures*

A. The Large Electron-Positron Collider

At LEP, CP violating scenarios are probed *via* a re-interpretation of the usual (CP-conserving) MSSM or 2HDM searches [149, 161, 162]. Compared to the MSSM, the mass and coupling constraints are relaxed, leading to an often richer mixture of final states. The results of Refs. [149, 161, 162] exploit the earlier searches for the $e^+e^- \to$ hA, hZ processes (in CP-conserving nomenclature), now accounting for the possible presence of more than two neutral Higgs bosons in the spectrum, and for generally diluted cross-sections and modified branching fractions.

The considered modes include the familiar $H_i \to bb, \tau\tau$ [163], but also gluonic or photonic decays (so-called flavour-blind and fermiophobic searches, respectively) [164–168]. Details concerning the analyses and the statistical combination procedure are given in Section 3.2.

B. The Large Hadron Collider

At the LHC, via gluon fusion, we probe loop-induced Higgs-boson couplings to two gluons. The diagrams with top and bottom quarks inside loops can induce both scalar $S_i^g$ and pseudoscalar $P_i^g$ form factors simultaneously for a specific $H_i$ when CP is violated. We refer to Sec. 3.4 for specific forms of the form factors. Also, note that the CP-odd component of each Higgs boson can contribute to the scalar form factor in the presence of non-vanishing CP phases.

The $s$-channel production cross section of a neutral Higgs boson $H_i$ in $gg$ fusion is given by

$$\sigma(pp \to H_i) = K \frac{\alpha_s^2}{256\pi v^2} \left(|S_i^g|^2 + |P_i^g|^2\right) \tau \frac{d\mathcal{L}^{gg}}{d\tau}, \tag{3.14}$$

where the factor $K \approx$ 1.5-1.7 for QCD corrections and the Drell-Yan variable $\tau = M_{H_i}^2/s$ with $s$ being the invariant hadron collider energy squared. The gluon-gluon luminosity $\tau d\mathcal{L}^{gg}/d\tau \sim 500$ when $M_{H_i} \sim 100$ GeV at the LHC and $\alpha_s^2/256\pi v^2 \sim 0.1$ pb.

Even the absolute values of the scalar and pseudoscalar form factors depend strongly on the CP phases [62–74], we need observables which vanish in the CP-conserving limit to establish CP-violating Higgs mixing at the LHC. The product $\mathrm{Re\,e}(S_i^g P_i^g)/(|S_i^g|^2 + |P_i^g|^2)$ might be measured by examining the azimuthal angular distribution of the tagged forward protons [75]. On the other hand, when the Higgs bosons decay into $\tau$ leptons, one can construct CP-odd observables if the polarizations of the $\tau$ leptons can be determined reasonably. Even the production rates are known to be low in general, the exclusive double diffractive process, thanks to a clean environment due to the large rapidity gap and a good Higgs-mass resolution of the order of 1 GeV, may offer unique possibilities for exploring Higgs physics in ways that would be difficult or even impossible in inclusive Higgs production at the LHC [76].

The inclusion of supersymmetric threshold corrections to the $b$-quark mass [124–126, 128, 169–176] has significant consequences in scenarios with large CP-mixing effects in the Higgs sector. Depending on the size of $\mathrm{Arg}(A_{t,b}\,\mu)$, $\mathrm{Arg}(M_3\,\mu)$, and the details of the spectrum, the lightest sbottom squark becomes tachyonic and, possibly, the $b$-quark Yukawa coupling nonperturbative for values of $\tan\beta$ ranging from intermediate up to large or very large [77]. In this case, the main production mechanism of Higgs bosons at the LHC is not gluon fusion but $b$-quark fusion. Inability of distinguishing gluon fusion from $b$-quark fusion leads us to consider $W^+W^-$ fusion into Higgs bosons and subsequent decays of Higgs bosons into tau leptons [158]. It would be a promising channel for studying signature of Higgs-sector CP violation at the LHC.

Finally, we note that the existing MSSM Higgs CP studies at the LHC are mostly at the parton level and still in need of a detailed experimental validation including detector simulations.

C. The International Linear Collider

At the international linear collider (ILC), the neutral Higgs bosons are produced via Higgs couplings to vector boson pairs, $g_{H_iVV}$, and vector-boson couplings to Higgs boson pairs, $|g_{H_iH_jZ}| = |\epsilon_{ijk}\, g_{H_kVV}|$. This relation, together with the sum rule Eq. (3.9), leads to the selection rule that only two CP-even Higgs bosons can appear in the Higgs–strahlung process and only two pairs of CP-even and CP-odd Higgs bosons can be produced in CP-invariant MSSM framework. In other words, if one observe three Higgs bosons in Higgs–strahlung and/or all the three pairs of Higgs bosons, $H_1H_2$, $H_2H_3$, and $H_1H_3$, in pair productions, this is a signal of CP violation in the MSSM framework [78,79]. But in the non-minimal supersymmetric extension(s) of the SM, there can be additional Higgs singlet(s) and/or doublets(s) implying the observation made above does not necessarily mean a signal of CP violation.

Higgs-boson production via the Higgs–strahlung process $e^+e^- \rightarrow H_i\, Z$, where the $Z$ boson decays into electron or muon pairs, offers a unique environment for determining the masses and widths of the neutral Higgs bosons by the recoil-mass method [80–82]. Thanks to the excellent energy and momentum resolution of electrons and muons coming from the $Z$-boson decay, the recoil mass against the $Z$ boson, $p^2 = s - 2 \cdot \sqrt{s} \cdot E_Z + M_Z^2$, can be reconstructed with a precision as good as 1 GeV. Here $s$ and $E_Z$ are the the collider centre-of-mass energy squared and the energy of the $Z$ boson, respectively. It is shown [153] that the production lineshape of a coupled system of neutral Higgs bosons decaying into $b\bar{b}$ quarks is sensitive to the CP-violating parameters. When the Higgs bosons decay into $\tau^-\tau^+$, two CP asymmetries can be defined using the longitudinal and transverse polarizations of the $\tau$ leptons.

D. Photon Linear Collider

By means of Compton back–scattering of laser light, almost the entire energy of electrons/positrons at ILC can be transferred to photons [83] so that $e\gamma$ and $\gamma\gamma$ processes can be studied for energies close to the ILC energy scale [84, 177, 178]. The luminosities are expected to be about one third of the $e^+e^-$ luminosity in the high energy regime. Especially, the photon–photon collision is an ideal option to look for the signatures of neutral Higgs bosons. The $\gamma\gamma$ formation allows us to generate heavy Higgs bosons [179, 180] in a wedge centered around medium $\tan\beta$ values, in which neither the LHC nor the ILC give access to the spectrum of heavy Higgs bosons. Various options of choosing the photon polarization, circular and linear, allow unique experimental analyses of the properties and interactions of Higgs bosons. For example, more than 20 independent observables, half of which are CP-odd, can be constructed by exploiting the controllable photon beam polarization and the possibly measurable final–state fermion polarizations.

In the narrow-width approximation, the $s$-channel Higgs-boson production $\gamma\gamma \rightarrow H_i$ can be expressed in the simple form,

$$\sigma(\gamma\gamma \rightarrow H_i) = \frac{\alpha_{\rm em}^2}{32\pi v^2} \left(|S_i^\gamma|^2 + |P_i^\gamma|^2\right) \delta(1 - M_{H_i}^2/s)\,, \tag{3.15}$$

where $s$ is the c. m. energy squared of two colliding photons and $\alpha_{\rm em}^2/32\pi v^2 \sim 4$ fb. For the scalar $S_i^\gamma$ and pseudoscalar $P_i^\gamma$ form factors, we again refer to Section 3.4. The $s$-channel production of neutral Higgs bosons and its decays into several final states have been studied by many authors taking account of possible interference effects with the tree-level $t$- and $u$-channel continuum amplitudes [85–92,159]. Recently, a comprehensive study has been done taking into account $\mu^+\mu^-$, $\tau^+\tau^-$, $\bar{b}b$ and $\bar{t}t$ final states [93]. Some signatures of the resonant CP–violating Higgs mixing due to near degeneracy of heavy Higgs bosons in the decoupling limit [107] have been investigated as well [159], cf. Section 3.12.

E. Muon colliders and other experimental probes

The main physics advantage of the muon collider is that the larger Yukawa coupling of muons in many cases admits copious production of Higgs bosons as $s$-channel resonances. Moreover, with controllable energy resolution and beam polarizations, the muon collider provides a powerful probe of the Higgs sector CP violation [94–100, 181]. Several detailed studies considering fermion and/or sfermion final states can be found, for example, in Refs. [101, 102].

### 3.2 Search for CP-violating neutral Higgs bosons in the MSSM at LEP

*Philip Bechtle*

In this section, the results from the combination of the Higgs boson searches of the four LEP collaborations [149, 182–184] at $\sqrt{s} = 91$–$209$ GeV in model-independent cross-section limits on various MSSM-Higgs-like topologies and in exclusion of CP-violating MSSM benchmark scenarios are presented. Because of the different Higgs boson production and decay properties outlined in the previous sections, the experimental exclusions published so far for the CP-conserving MSSM scenario are partly invalidated by CP-violating effects.

#### *3.2.1 Higgs boson searches at LEP*

In the CP-conserving MSSM, the two dominant production mechanisms Higgsstrahlung ($e^+e^- \to hZ$, $\sigma_{hZ} \propto \sin^2(\beta-\alpha)$) and pair production ($e^+e^- \to hA$, $\sigma_{hA} \propto \cos^2(\beta-\alpha)$) are complementary and ensure the coverage of the whole kinematically accessible plane of Higgs boson masses, because for large $\cos^2(\beta-\alpha)$ the two Higgs bosons $h$ (CP-even) and $A$ are close to each other in mass.

In the CP-violating MSSM, the experimental coverage of the mass plane is lost, since first all three Higgs bosons can be produced in Higgsstrahlung (and hence the direct complementary of two modes is lost), and second, because there can be large mass differences between $M_{H_1}$ and $M_{H_2}$ over the whole parameter space. Additionally, the cascade decay $H_2 \to H_1H_1$ is dominant in large areas of the parameter space. Hence, the coverage of non-diagonal pair production mechanisms and the coverage of cascade decays is crucial for the experimental access to the CP-violating models. This is shown in Fig. 3.4. In (a), the 95 % confidence level (CL) exclusion limits on $\sigma \times BR$ in the process $e^+e^- \to H_2Z \to H_1H_1Z \to b\bar{b}b\bar{b}Z$ relative to the nominal 2HDM cross-section is shown. For $M_{H_2}$ up to 105 GeV and all $M_{H_1}$, models which predict a $\sigma \times BR$ value of more than 40 % of the SM cross-section can be excluded. In (b), the coverage for the process $e^+e^- \to H_2H_1 \to H_1H_1H_1 \to b\bar{b}b\bar{b}b\bar{b}$ is shown, with limits relative to the nominal pair-production cross-section with $\cos^2(\beta-\alpha) = 1$.

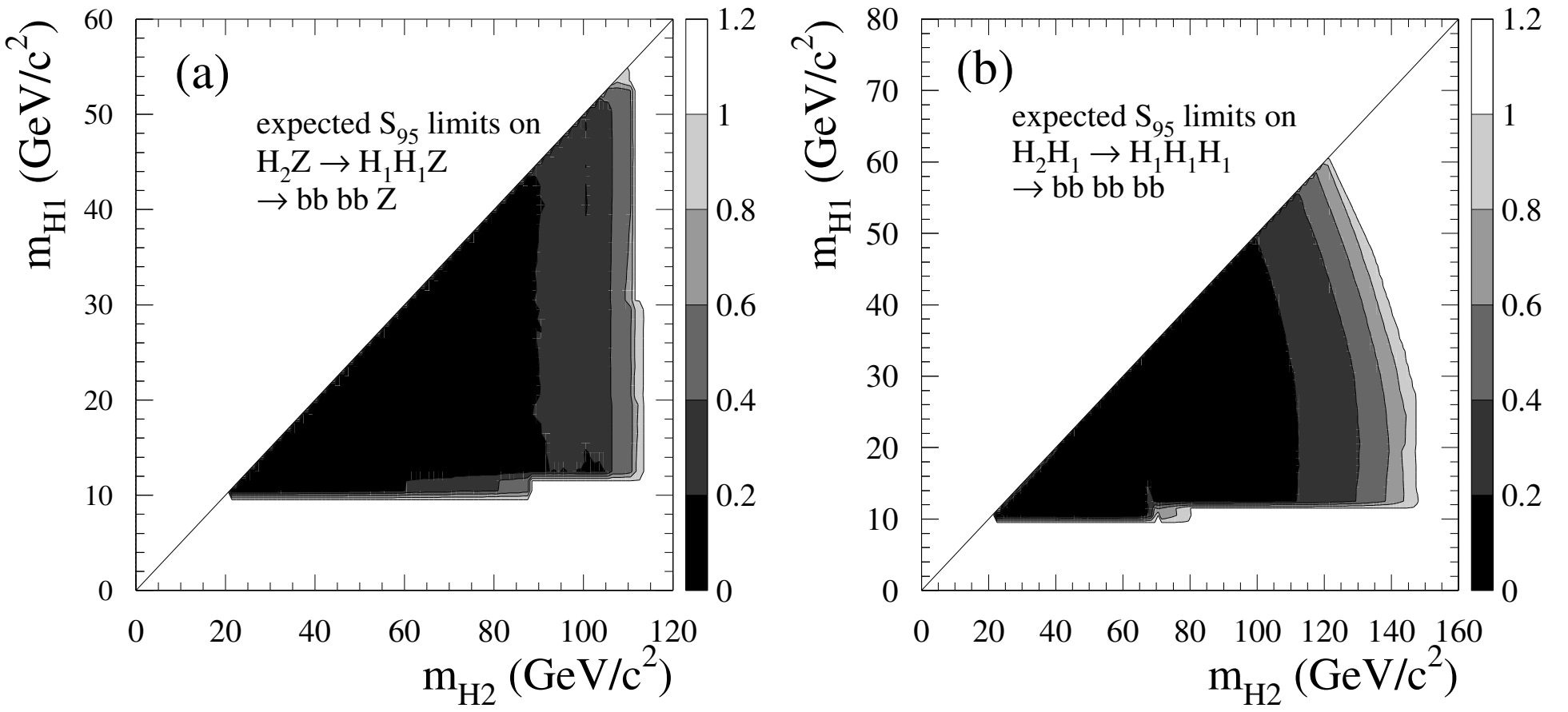

Fig. 3.4: Model independent limits on $\sigma \times BR$ relative to the nominal 2HDM cross-sections for $\sin^2(\beta-\alpha) = 1$ in (a) and $\cos^2(\beta-\alpha) = 1$ in (b). The scale on the right side of the figures shows the fraction of the nominal cross-section which can be excluded at the 90 % CL.

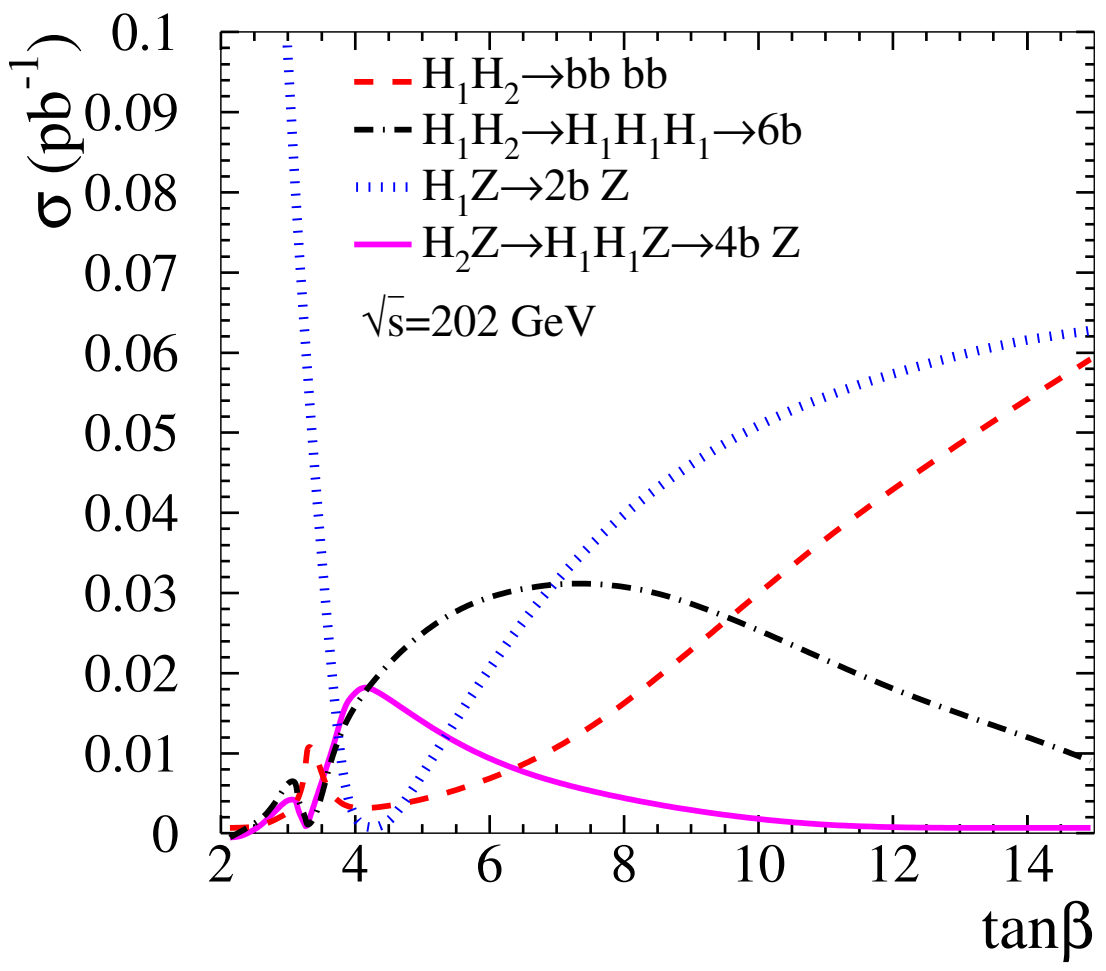

Fig. 3.5: Model predictions for $\sigma \times BR$ values of dominant Higgs boson production mechanisms in the CPX scenario for $30\,\mathrm{GeV} < M_{H_1} < 40\,\mathrm{GeV}$ at the average center-of-mass energy of LEP in the last two years of data-taking.

### *3.2.2 MSSM models with additional CP violation*

The benchmark model used in the combination of the LEP data is the CPX scenario [54]. It is characterized by large mixing between CP-even and CP-odd states in the mass eigenstates. The CP-even/CP-odd mixing $\mathcal{M}^2_{SP}$ is characterized by

$$\mathcal{M}^2_{SP} \propto \frac{m_t^4}{v^2} \frac{\mathrm{Im}\,(A_{t,b}\mu)}{M^2_{SUSY}}. \tag{3.16}$$

Therefore, large values of the top quark mass $m_t$, the Higgsino mixing parameter $\mu$ and the imaginary part of the trilinear couplings in the stop and sbottom sector $A_{t,b}$ ($\arg A_{t,b,\tau} = \Phi_3 = 90°$), coupled with a not too large scale of the squark masses $M_{SUSY}$ is chosen. Effects of the variation of these parameters are studied. Detailed calculations on the two-loop order [60, 144] or on the one-loop renormalization-improved order [54] are used to calculate the model predictions.

The resulting predictions for selected processes in the CPX scenario are shown in Fig. 3.5 for lightest Higgs masses of $30\,\mathrm{GeV} < M_{H_1} < 40\,\mathrm{GeV}$. For low $\tan\beta \sim 2$, the SM-like production mechanism $H_1 Z \to (b\bar{b}, \tau^+\tau^-)Z$ is dominant and has a large production cross-section. For intermediate $\tan\beta \sim 4$, however, all production cross-sections are reduced with respect to the area at $\tan\beta \sim 2$, since the kinematically accessible $H_1$ decouples from the $Z$ because it becomes entirely CP-odd, hence no Higgsstrahlung occurs, and since $M_{H_2} \approx 110\,\mathrm{GeV}$ is close to the kinematic limit. Additionally, the experimentally more difficult cascade decay $H_2 \to H_1 H_1$ becomes dominant. For large $\tan\beta$ the production cross-sections increase and finally $H_1 H_2 \to b\bar{b}b\bar{b}$ becomes the dominant mode at $\tan\beta > 15$.

### *3.2.3 Interpretation of the LEP data in the CPV MSSM*

The statistical combination of all Higgs boson searches from all four LEP collaborations uses the modified frequentist approach as implemented in [185, 186]. The result of this combination shows no statistically significant excesses of the data over the expected background. Hence, limits on the parameter space are computed [162]. These limits are shown in Fig. 3.6, 3.7 and 3.8 for the CPX scenario. In each case, the full set of MSSM parameters is fixed to the values chosen for the scenario (as given in [54, 162]), apart from $\tan\beta$ and the charged Higgs boson mass $M_{H^\pm}$, which are scanned. The result is then shown in the $\tan\beta, M_{H_1}$ projection.

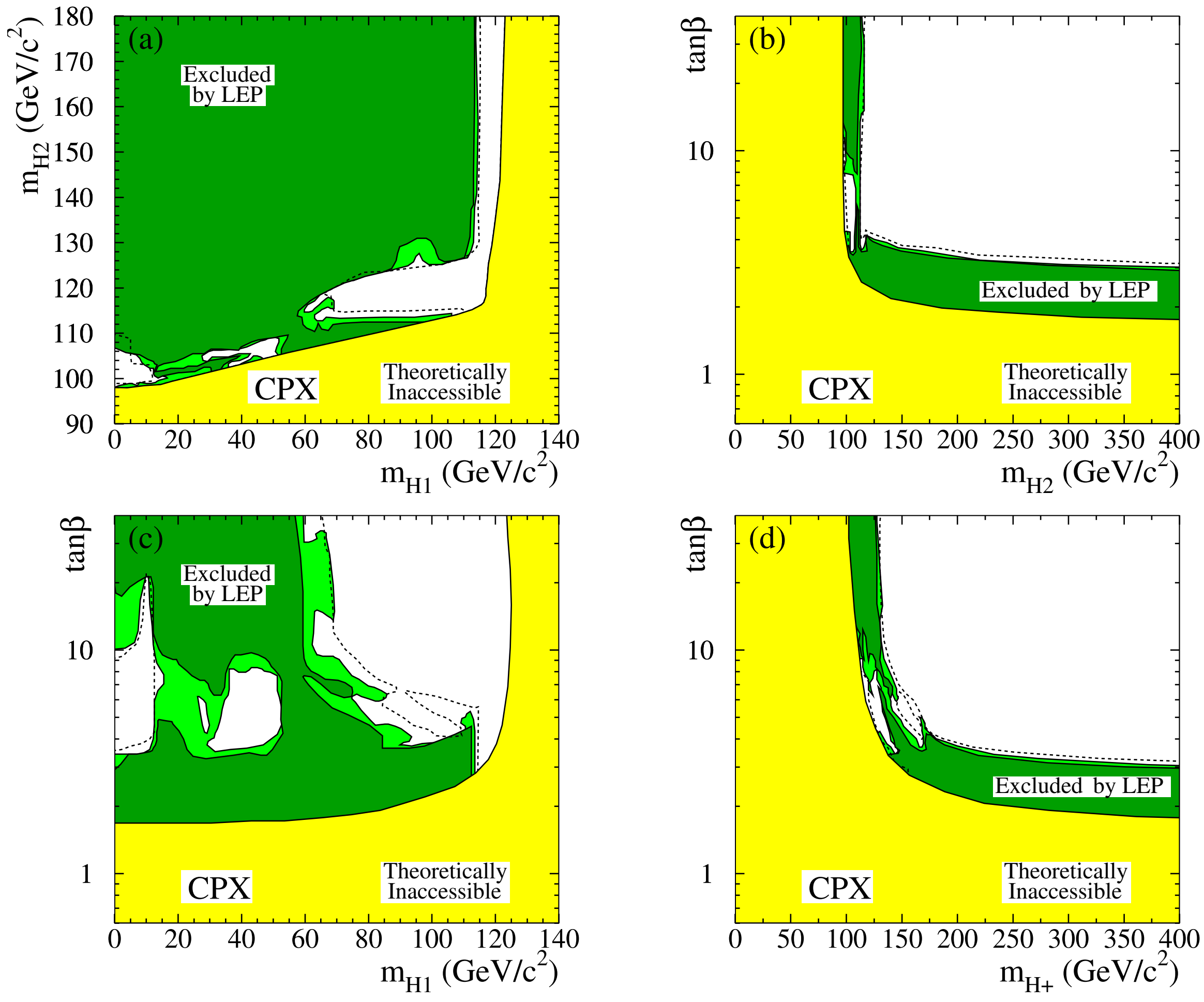

Fig. 3.6: Exclusion areas in the (a) $(M_{H_2}, M_{H_1})$, (b) $(\tan\beta, M_{H_2})$, (c) $(\tan\beta, M_{H_1})$ and (d) $(\tan\beta, M_{H^\pm})$ planes in the CPX scenario for $m_t = 174.3\,\text{GeV}$. Theoretically inaccessible regions are shown in yellow, experimentally excluded areas in light green ($CL = 95\,\%$) and dark green ($CL = 99.7\,\%$).

Fig. 3.6 shows the excluded region in the CPX scenario for four different projections and $m_t = 174.3\,\text{GeV}$. The reduction of production cross-sections for intermediate $\tan\beta$ described in Section 3.2.2 causes unexcluded regions for low values of the lightest Higgs boson mass $M_{H_1}$. No absolute limit on $M_{H_1}$ can be set. In Fig. 3.7 the results in the CPX scenario are shown for different top quark masses $m_t$. The present experimental value of $m_t = 172.7\,\text{GeV}$ [187] lies between the values used for Fig. 3.7 (a) and the nominal CPX scenario shown in Fig. 3.6 (c). For larger values of $m_t$ the unexcluded region increases, since $m_t$ strongly influences the mixing of the mass eigenstates (see (3.16)) and increases the mass splitting between $M_{H_1}$ and $M_{H_2}$, hence further decreasing the production cross-sections of $ZH_2$ states for intermediate $\tan\beta$.

The effect of unexcluded regions in the parameter space for low $M_{H_1}$ is clearly connected to the CP-violating imaginary phase of the trilinear couplings $A_{t,b}$. This is shown in Fig. 3.8. Only for large phases (and hence large mixings in (3.16)) the effect of large inaccessible regions is strong.

### *3.2.4 Conclusions*

The results from neutral Higgs bosons searches in the context of the MSSM described in this paper are based on data collected by the four LEP collaborations, ALEPH, DELPHI, L3 and OPAL at $\sqrt{s} =$

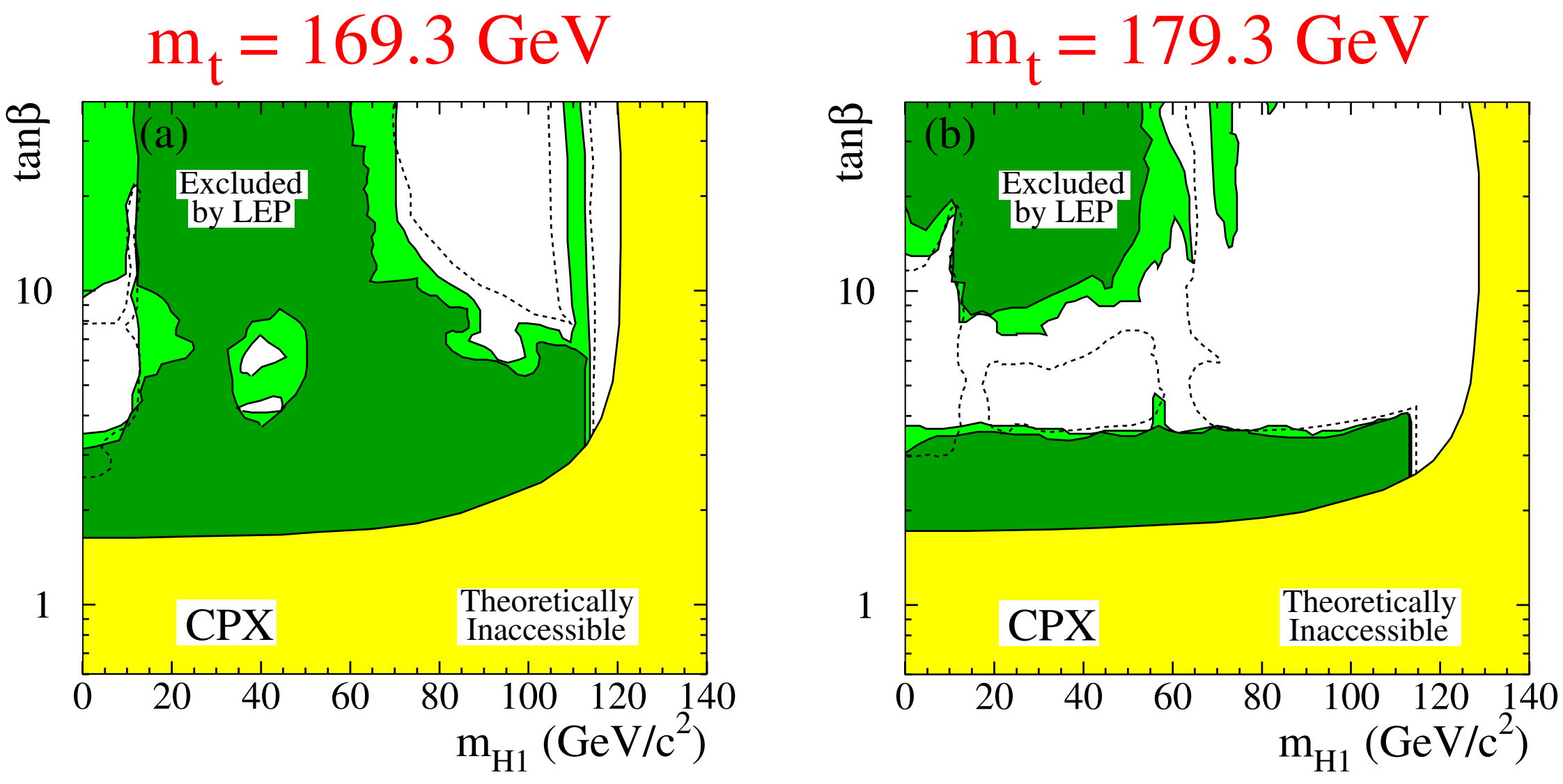

Fig. 3.7: Exclusion areas in the $(\tan\beta, M_{H_1})$ plane in the CPX scenario for $m_t = 169.3\,\mathrm{GeV}$ and $179.3\,\mathrm{GeV}$. For the corresponding exclusion areas for $m_t = 174.3\,\mathrm{GeV}$ please see Fig. 3.6 (c). Theoretically inaccessible regions are shown in yellow, experimentally excluded areas in light green ($CL = 95\,\%$) and dark green ($CL = 99.7\,\%$). The phase is set to $\arg A = \Phi_3 = 90°$.

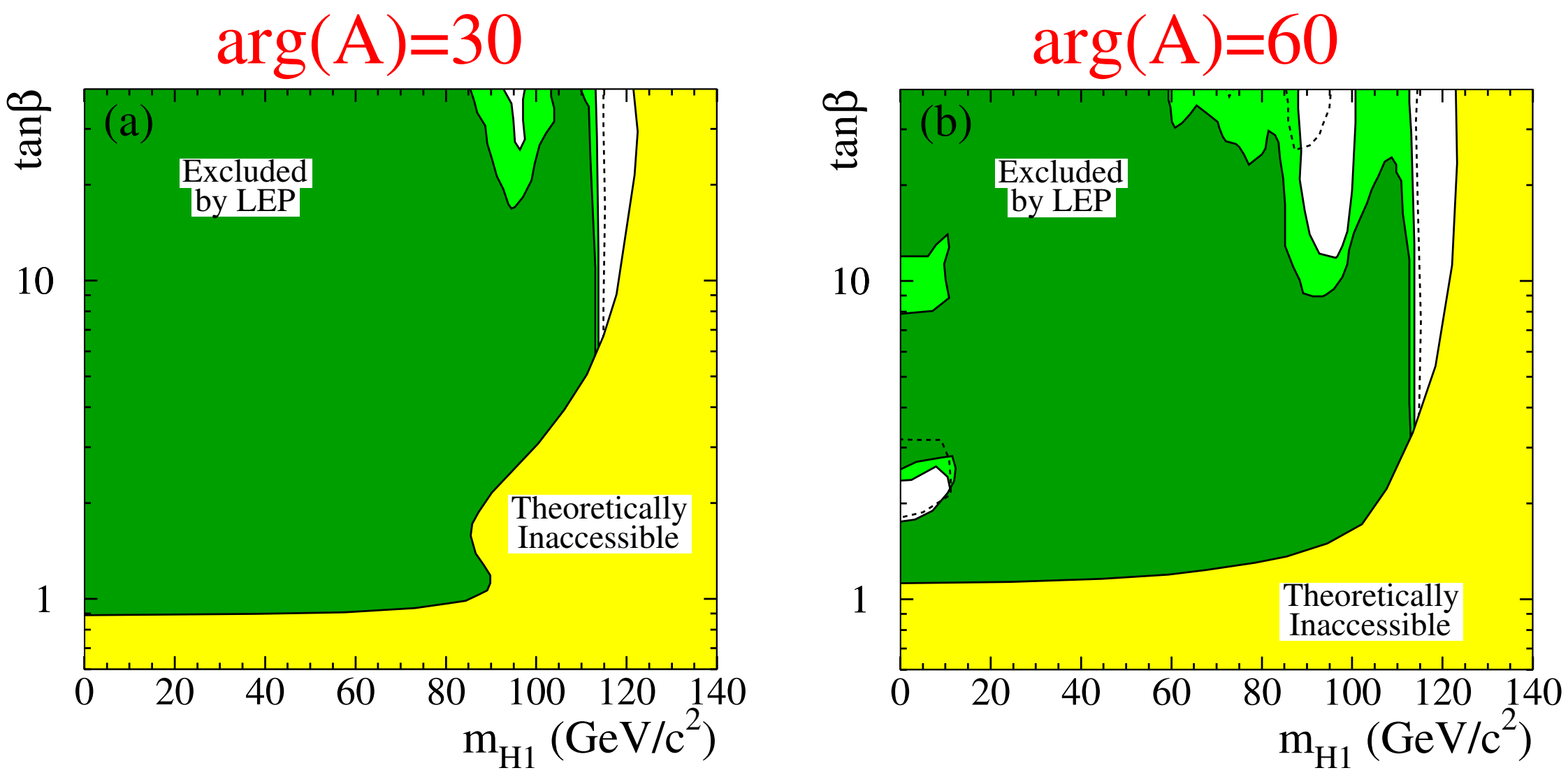

Fig. 3.8: Exclusion areas in the $(\tan\beta, M_{H_1})$ plane for different values of the phase $\arg A = \Phi_3 = 30°, 60°$ of the trilinear coupling parameters in the stop and sbottom sector. a value of $m_t = 174.3\,\mathrm{GeV}$ is chosen and the unvaried parameter values are identical to those of the CPX scenario. For the corresponding plot of the nominal CPX phase of $\arg A = \Phi_3 = 90°$ see Fig. 3.6 (c).

$91 - 209\,\mathrm{GeV}$. No significant excess of data over the expected backgrounds has been found. From these results, upper bounds are derived for the cross sections of a number of Higgs event topologies. These upper bounds cover a wide range of Higgs boson masses and are typically much lower than the largest cross sections predicted within the MSSM framework. In the CP-violating benchmark scenario CPX and the variants which have been studied, the combined LEP data show large unexcluded domains, down to the smallest masses; hence, no absolute limits can be set for the Higgs boson masses. On the other hand, $\tan\beta$ can be restricted to values larger than 2.9 for $m_t = 174.3$ GeV. While the excluded mass domains vary considerably with $m_t$, the bound in tan is barely sensitive to the precise choice of the top quark mass.

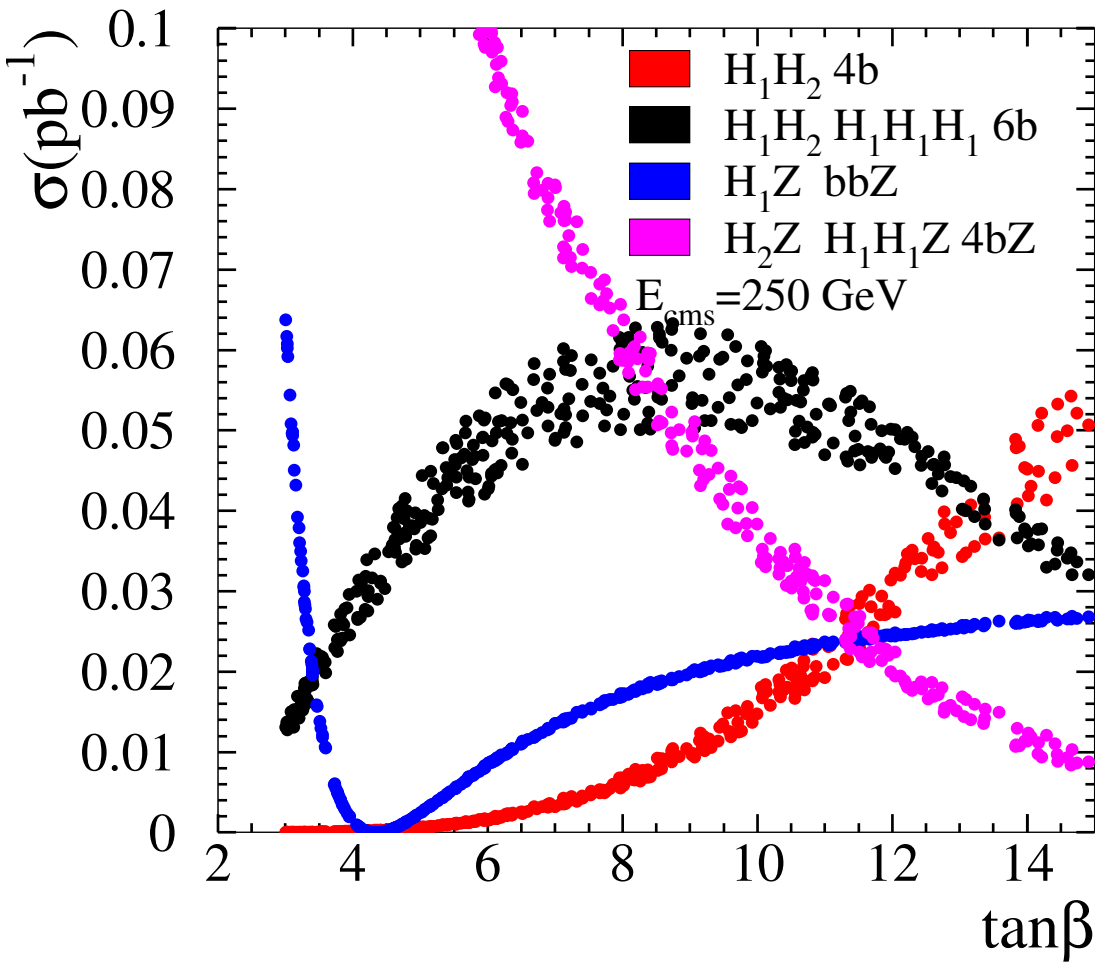

Fig. 3.9: Model predictions for $\sigma \times BR$ values of dominant Higgs boson production mechanisms in the CPX scenario for 30 GeV $< M_{H_1} <$ 40 GeV at $\sqrt{s} = 250$ GeV.

Fig. 3.9 shows a selection of the same cross-sections as in Fig. 3.5, this time for a center-of-mass energy of 250 GeV instead of 202 GeV. At higher energies, the $ZH_2$ Higgsstrahlung channel is kinematically open and no suppression of the cross-sections at intermediate $\tan\beta$ can be seen. Therefore it is expected that in this model the ILC will have the same access to the Higgs sector as in a CP conserving MSSM model.

## 3.3 The ATLAS discovery potential for Higgs bosons in the CPX scenario

*Markus Schumacher*

The investigation of the discovery potential for Higgs bosons of the MSSM at the LHC has so far concentrated on the CP conserving case with real SUSY breaking parameters. Here we discuss a preliminary investigation of the discovery potential of the ATLAS experiment for Higgs bosons in the CPX scenario [58] of the CP violating MSSM. The soft SUSY breaking parameters have been fixed according to Eq. 3.13 with the common SUSY scale chosen to be $M_{\mathrm{SUSY}}$ = 500 GeV, the SU(2) gaugino mass parameter $M_2$ = 500 GeV and $M_1$, the U(1) gaugino mass parameter, is derived from $M_2$ using the GUT relation. The mass of the top quark used in this study is 175 GeV. The two parameters $\tan\beta$ and $M_{H^\pm}$, which determine the Higgs sector at Born level, have been scanned between 1 to 40 and 50 to 1000 GeV, respectively.

### *3.3.1 Experimental peculiarities*

The results of the published ATLAS Monte Carlo (MC) studies, shown in Table 3.1, are used for the investigation of the discovery potential. The key performance figures for e.g. lepton identification and isolation, b-tagging, $\tau$ identification, trigger efficiencies and mass resolutions have been obtained from studies using a full simulation of the ATLAS detector. The number of expected signal and background events have been estimated then using a fast simulation of the ATLAS detector.

In order to evaluate the discovery potential of a search channel in a specific $(\tan\beta, M_{H^\pm})$ parame-ter point, the number of expected signal and background events after all selection cuts need to be derived for this parameter set. This is done in the following way:

#### 3.3.1.1 Masses, coupling and branching ratios

The masses of the Higgs bosons, their coupling strength and branching ratios are calculated with `FeynHiggs 2.1` [145]. Preliminary checks have also been performed using an alternative program `CPsuperH` [131]. The differences in the predictions among the two programs are significant for certain areas of the parameter space e.g. a shift in the mass of $H_1$ of about 5 GeV can be achieved. This results in a change of the contribution of a particular search channel, however the basic conclusions stay the same. In this report results are only shown which were obtained with `FeynHiggs`.

#### 3.3.1.2 Production cross sections

Leading order cross sections are used for all production processes. All cross sections are calculated using the CTEQ5L parton distribution functions [188]. For the production of neutral Higgs bosons via gluon fusion, weak boson fusion[1] and heavy quarks ($ttH_i$ and $bbH_i$) the SM like cross sections are calculated using the programs from reference [189][2] and then applying the appropriate correction factors to obtain the MSSM cross section values as detailed below:

$$\text{Gluon fusion}: \quad \sigma^{GGF}_{MSSM} = \frac{\Gamma_{H_i \to \text{gluons}}(MSSM)}{\Gamma_{H \to \text{gluons}}(SM)} \times \sigma^{GGF}_{SM} \tag{3.17}$$

$$\text{Weak Boson Fusion}: \quad \sigma^{WBF}_{MSSM} = \left(\frac{g_{H_iVV}(MSSM)}{g_{HVV}(SM)}\right)^2 \times \sigma^{WBF}_{SM} \tag{3.18}$$

$$\begin{aligned} \text{ffH}_\text{i}: \quad \sigma^{ffH_i}_{MSSM} &= \left(\frac{g^S_{H_iff}(MSSM)}{g^S_{Hff}(SM)}\right)^2 \times \sigma^{Hff}_{SM} \\ &+ \left(\frac{g^P_{H_iff}(MSSM)}{g^S_{ffH}(SM)}\right)^2 \times \sigma^{ffA}_{2HDM,tan\beta=1} \end{aligned} \tag{3.19}$$

Here $g_{H_iVV}$ denote the coupling to weak gauge bosons and $g^P_{H_iff}$ and $g^S_{H_iff}$ the scalar and pseudoscalar coupling to fermions, respectively. The SM values are equal to unity in our convention. For heavy charged Higgs bosons ($M^{\pm}_H >$ 180 GeV) production via the process $gb \to tH^{\pm}$ is considered. For a light Higgs boson ($M^{\pm}_H <$ 170 GeV) the production in the decay of a top quark in a charged Higgs boson and $b$ quark are investigated. The intermediate mass region is for now excluded from the evaluation of the discovery potential. A discussion of a proper handling of this transition region may be found in [190–192] and is awaiting a detailed experimental MC study.

The cross section $\sigma_{Pseudo-SM}$ for $gb \to H^{\pm}t$ as a function of the charged Higgs boson mass is calculated with the program of [193]. Here "pseudo" means that the factor $(M_b \tan\beta)^2 + (M_t/\tan\beta)^2$, which enters the cross section, is set to one. The cross section for each parameter point is then scaled according to:

$$\sigma_{MSSM} = [(M_b(\ in\ GeV) \tan\beta)^2 + (M_t(\ in\ GeV)/\tan\beta)^2)]\sigma_{Pseudo-SM} \tag{3.20}$$

Here $M_b$ and $M_t$ denote the running quark masses at the scale $(M_{H^{\pm}} + M_t)/4$ as recommended in [190, 193].

[1] $H_i$ here and in the following denotes a general neutral Higgs boson mass eigenstate. Only its CP even component couples to W and Z boson. Associated production with weak gauge bosons ($W(Z)H_i$) is not considered as a discovery channel at the LHC.

[2] The codes for `HIGLU, VV2H` and `HQQ` are accessible via `http://people.web.psi.ch/spira/proglist.html`.

Table 3.1: Channels contributing to the ATLAS discovery potential. Shown are the production mechanisms, decay channels, mass ranges considered and references for the analysis.

| Production process | Decay mode | Mass range (GeV) | Reference |
|---|---|---|---|
| Inclusive | $H_i \to \gamma\gamma$ | 60 to 400 | [196] |
| Gluon Fusion | $H_i \to ZZ \to 4$ leptons | 100 to 450 | [196] |
| Gluon Fusion | $H_i \to WW \to l\nu l\nu$ | 140 to 200 | [196] |
| Weak Boson Fusion | $H_i \to WW \to l\nu\nu(qql\nu)$ | 110 (130) to 250 | [197] |
| Weak Boson Fusion | $H_i \to \tau\tau \to ll4\nu, lhad.3\nu$ | 110 to 180 | [197] |
| $ttH_i$ | $H_i \to bb$ | 70 to 150 | [198] |
| Gluon Fusion/$bbH_i$ | $H_i \to \mu\mu$ | 70 to 1000 | [199, 200] |
| Gluon Fusion/$bbH_i$ | $H_i \to \tau\tau \to l\ had.(had.\ had.)$ | 110(450) to 1000 | [201, 202] ( [203]) |
| Gluon Fusion | $H_i \to H_j H_k \to \gamma\gamma bb$ | 100(40) to 360(130) | [196] |
| Gluon Fusion | $H_i \to H_j Z \to bbll$ | 100(40) to 360(130) | [196] |
| Gluon Fusion | $H_i \to tt$ | 350 to 500 | [196] |
| $tt \to H^{\pm}bWb$ | $H^{\pm} \to \tau\nu, W \to qq(l\nu)$ | 70 to 170 | [204] ( [196]) |
| $gb \to H^{\pm}t$ | $H^{\pm} \to \tau\nu$ | 180 to 1000 | [192] |

The cross section for charged Higgs boson production in top quark decay is calculated in the following way. A leading order top quark pair production cross section of 492 pb is used and multiplied with the $t \to H^{\pm}b$ branching ratio as obtained from PYTHIA [194, 195] depending on $\tan\beta$ and the charged Higgs boson mass.

#### 3.3.1.3 Signal and background rates

The expected background rates are independent from the MSSM parameter point and their values for a given Higgs boson mass are taken directly from the published ATLAS MC studies. The expected signal rate $N_{signal}$ is calculated according to:

$$N_{signal} = \sigma_{MSSM} \times BR \times \mathcal{L} \times \epsilon \quad , \tag{3.21}$$

where $\mathcal{L}$ denotes the integrated luminosity and $\sigma_{MSSM} \times BR$ the product of cross section times branching ratio for the particular MSSM parameter. The signal efficiency $\epsilon$ for a given Higgs boson mass is taken from published detailed ATLAS MC studies. However two type of corrections are applied to the efficiencies as discussed below.

Most experimental MC studies have been performed for a light SM like Higgs boson or for the heavy Higgs bosons for particular choice of the MSSM parameters. These parameter choices include e.g. the following assumptions: (i) the natural total decay width of the Higgs bosons is negligible compared to the mass resolution, (ii) the mass degeneracy between e.g. $\mathcal{H}_2$ and $\mathcal{H}_3$ is perfect i.e. the masses are exactly the same, which means the signal contributions can be simply added. During the scan of the MSSM parameter space these assumption might no be fulfilled for all points. Deviations from the above assumptions have been corrected for in the following way.

The effect of an increased total decay width leading to a broadening of the reconstructed mass peak and therefore to a reduced signal efficiency when using the standard mass window cuts are taken into account by multiplying the signal efficiency with the following correction factor $K_{width}$:

$$K_{width} = \frac{\int_{M-\Delta M}^{M+\Delta M} BW(\Gamma_{total}^{MSSM}) \otimes G(\sigma_M)}{\int_{M-\Delta M}^{M+\Delta M} BW(\Gamma_{total}^{MC\ study}) \otimes G(\sigma_M)} \tag{3.22}$$

The integration borders $M - \Delta M$ and $M + \Delta M$ are the mass window cuts in the standard MC study. $\Gamma_{total}^{MC\ study}$ denotes the total decay width assumed in the ATLAS MC study, $\Gamma_{total}^{MSSM}$ the total decay width in the particular MSSM point and $\sigma_M$ the mass resolution for the signal. $BW(\Gamma_{total})$ is the Breit-Wigner distribution which is folded ($\otimes$) with the Gaussian mass resolution $G(\sigma_M)$. The correction factor is calculated and applied for each individual search channel and each MSSM parameter point separately. In the case that $\Gamma_{total}^{MSSM}$ is smaller than $\Gamma_{total}^{MCstudy}$ no correction factor, which would be larger than 1, is applied.

Depending on the MSSM parameter point the masses of two of the neutral Higgs bosons might be closer than the expected mass resolution or the mass window cut applied. Then the partially overlapping signal will lead to an increased discovery potential. Examples of such channels are: production via weak boson fusion, associated production with b-quarks and Higgs boson production with decay to a pair of top quarks. In such cases the signal rate of the two contributing mass states are combined in the following way. Consider that the masses of the two bosons are $M_1$ and $M_2$. The expected signal and background rate for the Higgs boson with mass $M_1$ have been evaluated. In addition the signal rate of the boson with mass $M_2$ leaking into the mass window around $M_1$ is evaluated and added to the expected signal rate. The same procedure is repeated interchanging the role of the two Higgs bosons. For both cases the significance for observation of a signal excess is evaluated and the one yielding the larger value is retained.

Two luminosity scenarios are distinguished: (i) low luminosity running at $\mathcal{L} = 10^{33}$ cm$^{-2}$s$^{-1}$ yielding an integrated luminosity of 10 fb$^{-1}$ per year, (ii) high luminosity running at $\mathcal{L} = 10^{34}$ cm$^{-2}$s$^{-1}$ yielding an integrated luminosity of 100 fb$^{-1}$ per year. At high luminosity running several performance numbers are degraded e.g. $b$-tagging performance, mass resolutions etc. This change in the detector performance is taken into account when deriving the signal and background rates and from those the discovery potential.

### 3.3.2 ATLAS discovery potential in the CPX benchmark scenario

The evaluation of the discovery potential is based on Poissonian statistics requiring that the probability of a background fluctuation to the number of expected signal+background events is less than $2.85\times10^{-7}$. In the case of combining different final states the likelihood ratio method [205] is applied. The results are shown for integrated luminosities of 30 fb$^{-1}$ and 300 fb$^{-1}$. In the latter case 30 fb$^{-1}$ collected during low luminosity running and 270 fb$^{-1}$ collected at high luminosity running are assumed. The weak boson channels, charged Higgs boson channels for the Higgs boson mass below the top quark mass and the decay $H_i \to \tau\tau$ have only been studied for low luminosity running. Hence all results for these channels are only shown for an integrated luminosity of 30 fb$^{-1}$.

The discovery potential for the lightest neutral Higgs boson $H_1$ after collecting an integrated luminosity of 30 and 300 fb$^{-1}$ for the two projections $(M_{H^\pm}, \tan\beta)$ and $(M_{H_1}, \tan\beta)$ is shown in Fig. 3.10. In the $(M_{H^\pm}, \tan\beta)$ projection the discovery potential for 30 fb$^{-1}$ is dominated by the weak boson fusion channels which cover a large area left over by the LEP experiments [206]. Additional small areas of parameter space are covered by associated production with b and top quarks. With 300 fb$^{-1}$ the coverage of the latter two channels is increased and furtheron a large fraction of parameter space is covered also by the decay into a pair of photons and a pair of Z bosons. Masses below 60 GeV have been not studied up to now. In the weak boson fusion channels only masses above 110 GeV have been investigated. Therefore a significant area in the $(M_{H_1}, \tan\beta)$ plane has not been investigated yet.

The discovery potential for the heavier neutral Higgs boson $H_2$ and $H_3$ and the charged Higgs bosons in the projection $(M_{H^\pm}, \tan\beta)$ are shown in Fig. 3.11. For the heavy neutral Higgs bosons the area of large $\tan\beta$ is covered by the associated production with b quarks. Areas of low $M_{H^\pm}$ are covered by the production of $H_{2/3}$ in weak boson fusion with subsequent decay into tau leptons and the associated production with top quarks. The area of small $\tan\beta$, which to large extent has already excluded by the LEP searches, is covered by one ore more of several search channels as indicated in the

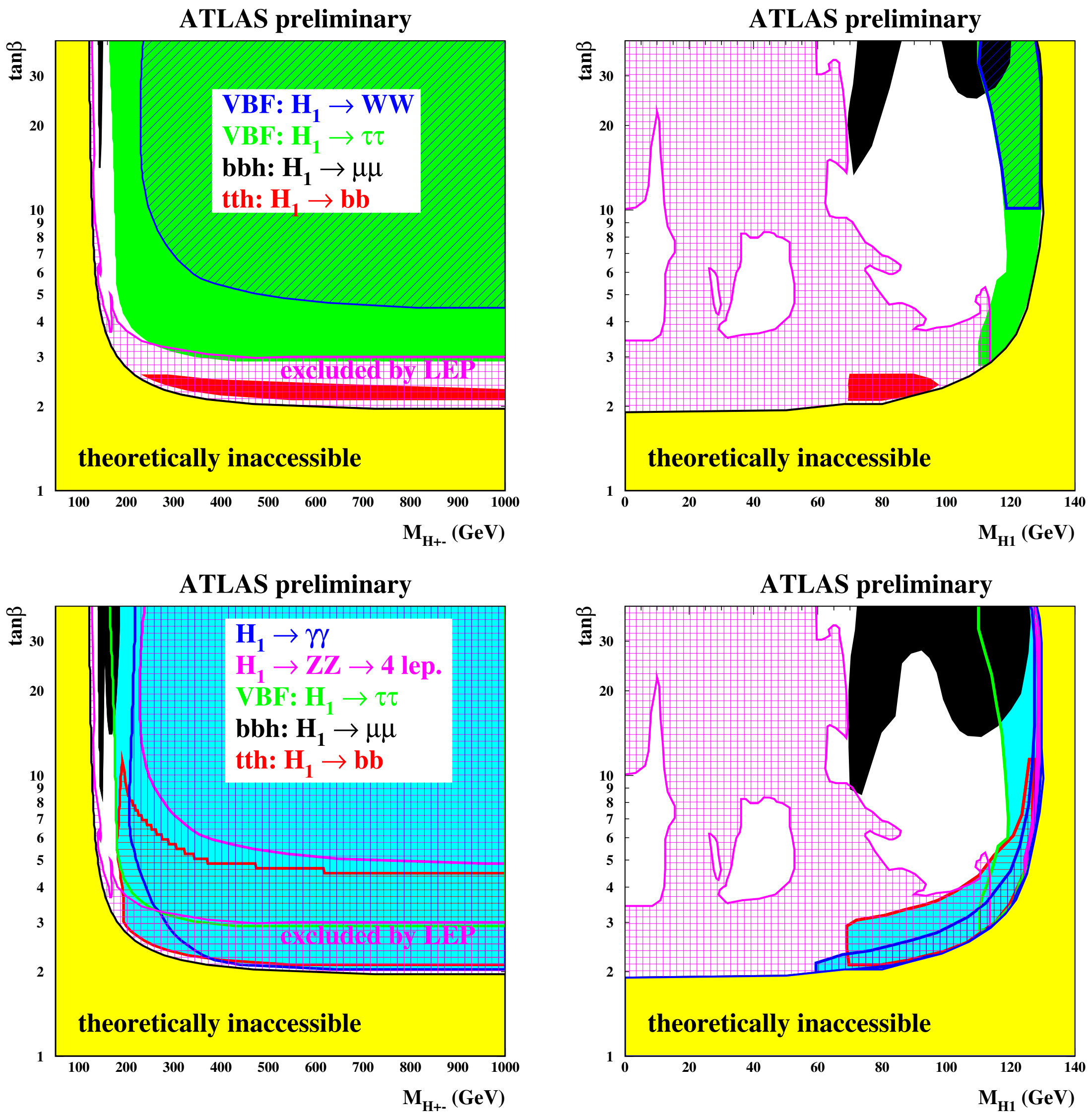

Fig. 3.10: Top row: discovery potential for the lightest neutral Higgs boson $H_1$ after collecting 30 fb$^{-1}$ In the black area observation via $bbH_1, H_1 \to \mu\mu$ is expected, in the medium grey (green) area via WBF $H_1 \to \tau\tau$, in the single hatched (blue) area via WBF, $H_1 \to WW$ and in dark grey (red) area via $ttH_1, H_1 \to bb$.
Bottom row: discovery potential for the lightest neutral Higgs boson $H_1$ after collecting 300 fb$^{-1}$, for the WBF channels only 30 fb$^{-1}$ are used. In the black area observation via $bbH_1, H_1 \to \mu\mu$ is expected, and in the horizontally lined (red) area at low $\tan\beta$ via $ttH_1, H_1 \to bb$, in the horizontally lined (magenta) area at large $\tan\beta$ via $H_1 \to ZZ \to 4$ leptons, in the vertically lined (blue) area via $H_1 \to \gamma\gamma$, and in the remaining solid area surrounded by the green line via WBF, $H_1 \to \tau\tau$.
The light grey (yellow) area is theoretically inaccessible. The cross hatched (magenta) area is excluded by the LEP experiments [206].

figure. At intermediate $\tan\beta$ and charged Higgs boson masses above 700 GeV no discovery potential is found with the current analysis and an integrated luminosity of 300 fb$^{-1}$. Light charged Higgs bosons below the top quark mass are expected to be observed in the tau lepton decay mode produced in top quark pair productions. Heavy charged Higgs bosons can be discovered via $gb \to tH^{\pm}$ and subsequent decay to $\tau\nu$ for large values of $\tan\beta$.

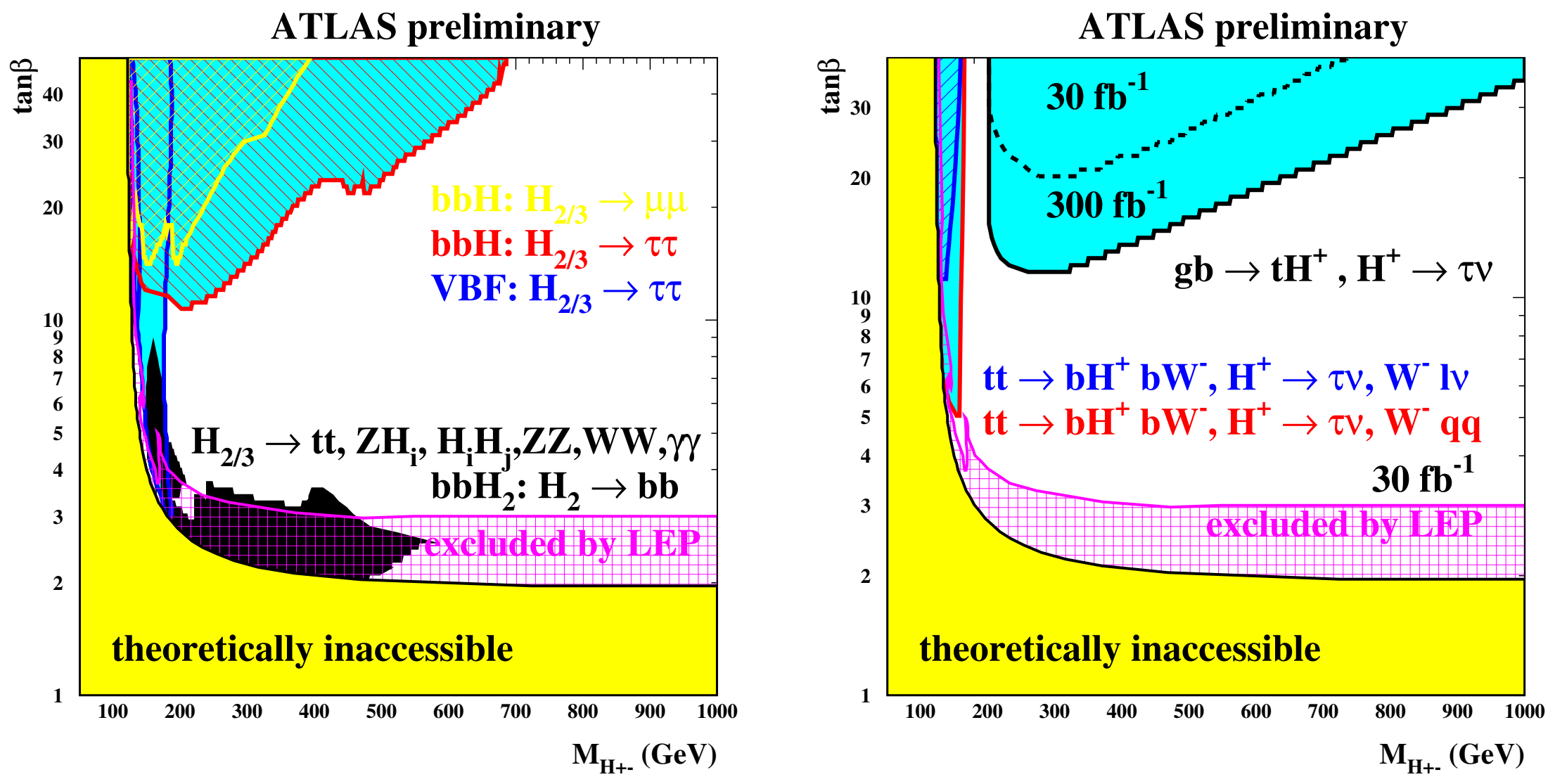

Fig. 3.11: Left: discovery potential for the heavy neutral Higgs bosons. In the black area at small $\tan\beta$ one or both Higgs boson are expected to be observed in the decays $H_{2/3} \to tt, ZH_i, H_iH_j, ZZ, WW, \gamma\gamma$ or in $ttH_{2/3}$ production with $H_{2/3} \to bb$ after collecting 300 fb$^{-1}$. In the area surrounded by the dark gray (blue) line the WBF channel with $H \to \tau\tau$ contributes, shown for 30 fb$^{-1}$ only. In the left bottom to top right hatched (yellow) area discovery is expected via $bbH_{2/3}, H_{2/3} \to \mu\mu$ after collecting 300 fb$^{-1}$ and in the right bottom to left top hatched (red) area discovery is expected via $bbH_{2/3}, H_{2/3} \to \tau\tau$ after collecting 30 fb$^{-1}$.
Right: discovery potential for the charged Higgs bosons. In the area at large $\tan\beta$ with $M_{H^\pm} >$180 GeV discovery is expected via $gb \to tH^\pm, H^\pm \to \tau\nu$. The dashed and solid line show the expected sensitivity after collecting 30 and 300 fb$^{-1}$, respectively. For $M_{H^\pm} <$170 GeV observation is expected via charged Higgs bosons produced in top decays with $H^\pm \to \tau\nu$. The expected sensitivity for the leptonic and hadronic decay of the W boson after collecting 30 fb$^{-1}$ is shown in the blue hatched and red solid area, respectively.
The cross hatched (magenta) area is excluded by the LEP experiments [206]. The light grey (yellow) area is theoretically inaccessible.

The overall ATLAS discovery potential after collecting an integrated luminosity of 300 fb$^{-1}$ for the two projections $(M_{H^\pm}, \tan\beta)$ and $(M_{H_1}, \tan\beta)$ are shown in Fig. 3.12. Almost the whole parameter space in the projection $(M_{H^\pm}, \tan\beta)$ is covered by the observation of at least one Higgs boson (see Fig. 3.12 top left). In the intermediate $\tan\beta$ regime only the lightest neutral Higgs boson $H_1$ is expected to be observable. The zoom in for small charged Higgs boson masses and moderate $\tan\beta$ in the projection $(M_{H^\pm}, \tan\beta)$ shows a small yet uncovered and not yet by LEP excluded region (see Fig. 3.12 top right). The same uncovered region is visible in the $(M_{H_1}, \tan\beta)$ projection (see Fig. 3.12 bottom). Using `FeynHiggs 2.1` for the calculations and assuming a top quark mass of 175 GeV the uncovered region corresponds to the following mass values: $M_{H_1} < 50$ GeV, $105 < M_{H_2} < 115$ GeV, $140 < M_{H_3} < 180$ GeV and $130 < M_{H^\pm} < 170$ GeV. Preliminary studies with `CPsuperH` and a different top quark mass indicate that the size and location of the uncovered region depends on the calculation used, but that this region is existing in all investigations performed. So far the LHC collaborations have not investigated the discovery potential for such a light Higgs boson.

### *3.3.3 Conclusions*

The discovery potential of the ATLAS experiment for Higgs bosons in the CPX scenario of the CP violating MSSM based on current MC studies has been discussed. Almost all of the model parameter space is covered by the observation of at least one Higgs boson. However a “small” corner of the model parameter space is uncovered by the current MC studies. This area corresponds to small values for the lightest Higgs boson of less than 50 GeV, which is not yet excluded by the LEP searches and the sensitivity of the ATLAS experiment has not yet been studied. The most promising channel for simultaneous observation of the charged and the lightest neutral Higgs boson in this area of parameter space seems to be top quark pair production with the following decay chain: $t \to bW \to bl\nu$, $t \to bH^{\pm} \to bWH_1 \to bqqbb$ [73]. An ATLAS MC study in this channel is in preparation.

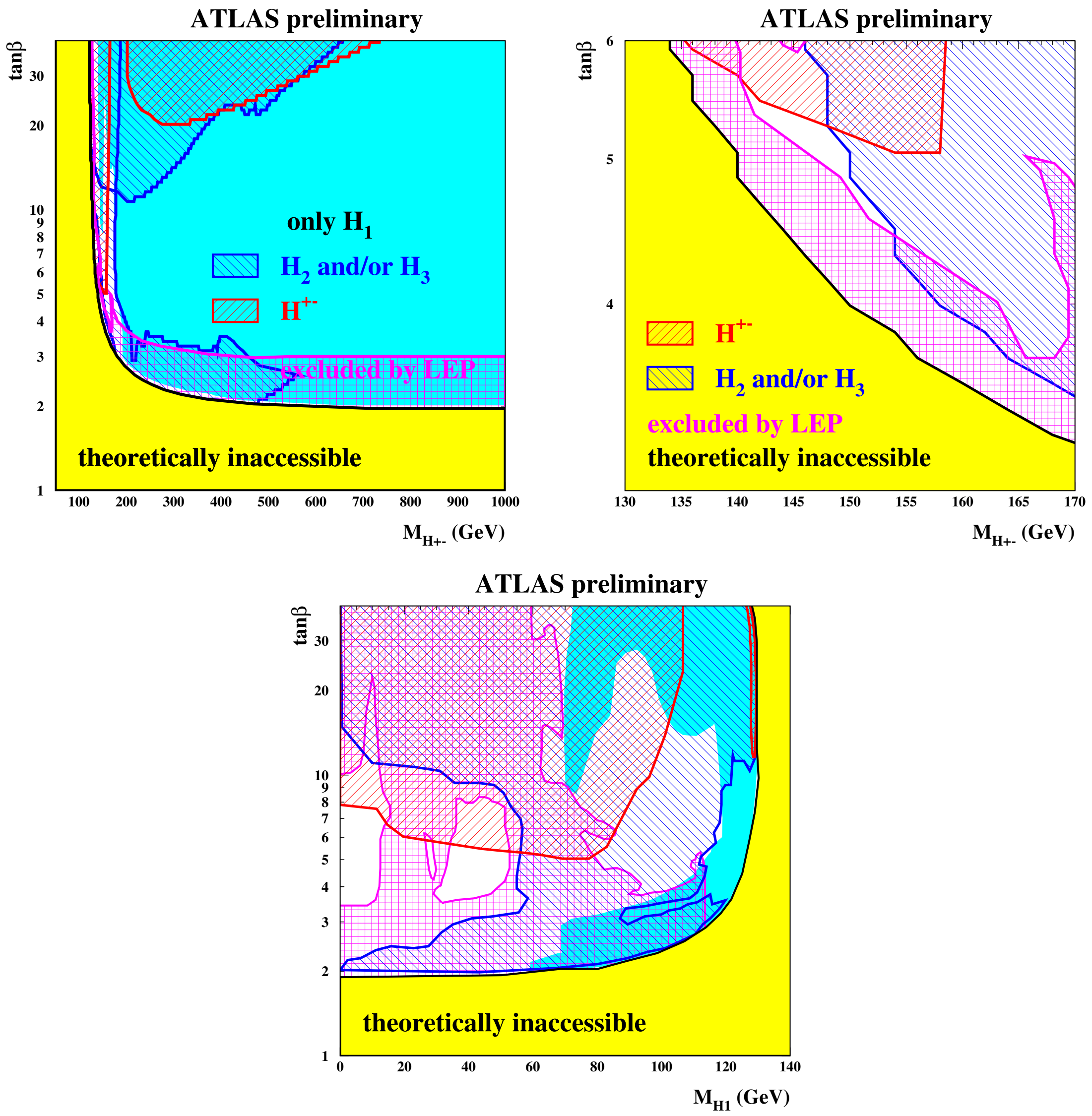

Fig. 3.12: Overall ATLAS discovery potential for Higgs bosons after collecting 300 fb$^{-1}$. The lightest neutral Higgs boson $H_1$ is expected to be seen in the solid medium gray (cyan) area. One or both of the heavier neutral Higgs bosons $H_2$ and $H_3$ are expected to be observed in in the right top to left bottom hatched (blue ) area. The charged Higgs bosons $H^{\pm}$ are expected to be observed in in the right bottom to left top hatched (red) area. The cross hatched (magenta) area is excluded by the LEP experiments [206]. The light grey (yellow) area is theoretically inaccesible.

### 3.4 Higgs phenomenology with CPsuperH

*John Ellis, Jae Sik Lee and Apostolos Pilaftsis*

The `Fortran` code `CPsuperH` [131] is a powerful and efficient computational tool for understanding quantitatively phenomenological subjects within the framework of the MSSM with explicit CP violation. It calculates the mass spectrum and decay widths of the neutral and charged Higgs bosons in the most general MSSM including CP-violating phases. In addition, it computes all the couplings of the neutral Higgs bosons $H_{1,2,3}$ and the charged Higgs boson $H^+$. The program is based on the results obtained in Refs. [141–143] and the most recent renormalization-group-improved effective-potential approach, which includes dominant higher-order logarithmic and threshold corrections, $b$-quark Yukawa-coupling resummation effects, and Higgs-boson pole-mass shifts [53, 58]. The masses and couplings of the charged and neutral Higgs bosons are computed at a similar high-precision level. Even in the CP-conserving case, `CPsuperH` is unique in computing the neutral and charged Higgs-boson couplings and masses with equally high levels of precision, and is therefore a useful tool for the study of MSSM Higgs phenomenology at present and future colliders.

#### *3.4.1 Introduction to* `CPsuperH`

The tarred and gzipped program file `CPsuperH.tgz` can be downloaded from [3]:

`http://www.hep.man.ac.uk/u/jslee/CPsuperH.html`

Typing `tar -xvzf CPsuperH.tgz` will create a directory called CPsuperH containing files : `0LIST, ARRAY, COMMON, cpsuperh.f, fillpara.f, fillhiggs.f, fillcoupl.f, fillgambr.f, makelib, compit, and run`. A library file `libcpsuperh.a` will be created by `./makelib` from the four Fortran files of `fillpara.f`, `fillhiggs.f`, `fillcoupl.f`, and `fillgambr.f` and the shell-script `compit` compiles `cpsuperh.f` linked with the library file. Input values are supplied by the shell-script file `run`. Straightforwardly, type '`./makelib`' and '`./compit`' followed by '`./run`':

| Run `CPsuperH`: `./makelib` $\longrightarrow$ `./compit` $\longrightarrow$ `./run` |
|---|

and then one can see some outputs depending on input values. For a full description of the input parameters `SMPARA_H(IP)`, `SSPARA_H(IP)`, `IFLAG_H(NFLAG)`, see Ref. [131].

In `CPsuperH`, the main numerical output is stored in arrays. The masses of the three neutral Higgs bosons, labelled in order of increasing mass such that $M_{H_1} \leq M_{H_2} \leq M_{H_3}$, are stored in `HMASS_H(3)`. Since the neutral pseudoscalar Higgs bosons mixes with the neautral scalars in the presence of CP violation, the charged Higgs boson mass $M_{H^\pm}$ is used as an input parameter. The array `OMIX_H(3,3)` yields the $3 \times 3$ Higgs mixing matrix, $O_{\alpha i}$: $(\phi_1, \phi_2, a)^T_\alpha = O_{\alpha i}(H_1, H_2, H_3)^T_i$. All the couplings of the neutral and charged Higgs bosons are stored in `NHC_H(NC,IH)` and `CHC_H(NC)`, respectively. These include Higgs couplings to leptons, quarks, neutralinos, charginos, stops, sbottoms, staus, tau sneutrinos, gluons, photons, and massive vector bosons. The array `SHC_H(NC)` contains Higgs-boson self-couplings. We note that the masses and mixing matrices of the stops, sbottoms, staus, charginos, and neutralinos are also calculated and stored in corresponding arrays. For the decay widths and branching fractions, the arrays `GAMBRN(IM,IWB,IH)` and `GAMBRC(IM,IWB)` are used for the neutral and charged Higgs bosons. For a full description, we refer again to Ref. [131]. The masses and branching fractions of the Higgs bosons and its couplings to a pair of vector bosons obtained by use of `CPsuperH` are shown in Sec. 3.1.

[3] Some new features appearing in this write-up, for example, the propagator matrix `DH(3,3)` and some low-energy observables, will be implemented in the forthcoming version of `CPsuperH`.

### 3.4.2 *Collider signatures*

To analyze CP-violating phenomena in the production, mixing and decay of a coupled system of multiple CP-violating MSSM neutral Higgs bosons at colliders, we need a "full" $3 \times 3$ propagator matrix $D(s)$, given by [158]

$$D(\hat{s}) = \hat{s} \begin{pmatrix} \hat{s} - M_{H_1}^2 + i\,\mathrm{Im}\,\mathrm{m}\widehat{\Pi}_{11}(\hat{s}) & i\,\mathrm{Im}\,\mathrm{m}\widehat{\Pi}_{12}(\hat{s}) & i\,\mathrm{Im}\,\mathrm{m}\widehat{\Pi}_{13}(\hat{s}) \\ i\,\mathrm{Im}\,\mathrm{m}\widehat{\Pi}_{21}(\hat{s}) & \hat{s} - M_{H_2}^2 + i\,\mathrm{Im}\,\mathrm{m}\widehat{\Pi}_{22}(\hat{s}) & i\,\mathrm{Im}\,\mathrm{m}\widehat{\Pi}_{23}(\hat{s}) \\ i\,\mathrm{Im}\,\mathrm{m}\widehat{\Pi}_{31}(\hat{s}) & i\,\mathrm{Im}\,\mathrm{m}\widehat{\Pi}_{32}(\hat{s}) & \hat{s} - M_{H_3}^2 + i\,\mathrm{Im}\,\mathrm{m}\widehat{\Pi}_{33}(\hat{s}) \end{pmatrix}^{-1}, \tag{3.23}$$

where $\hat{s}$ is the center-of-mass energy squared, $M_{H_{1,2,3}}$ are the one-loop Higgs-boson pole masses, and the absorptive parts of the Higgs self-energies $\mathrm{Im}\,\mathrm{m}\widehat{\Pi}_{ij}(\hat{s})$ receive contributions from loops of fermions, vector bosons, associated pairs of Higgs and vector bosons, Higgs-boson pairs, and sfermions. The calculated propagator matrix has been stored in the array `DH(3,3)`.

The so called tri-mixing scenario has been taken for studying the production, mixing and decay of a coupled system of the neutral Higgs bosons at colliders. This scenario is different from the CPX scenario and characterized by large value of $\tan\beta = 50$ and the light charged charged Higgs boson $M_{H^\pm}^{\rm pole} = 155$ GeV. All the three-Higgs states mix significantly in this scenario in the presence of CP-violating mixing. Without CP violation, only two CP-even states mix. For details of the scenario, see Refs. [76, 93, 153, 158].

A. LHC

At the LHC, the matrix element for the process $g(\lambda_1)g(\lambda_2) \to H \to f(\sigma)\bar{f}(\bar{\sigma})$ can conveniently be represented by the helicity amplitude

$$\mathcal{M}^{gg}(\sigma\bar{\sigma};\lambda_1\lambda_2) = \frac{\alpha_s\, g_f\, \sqrt{\hat{s}}\,\delta^{ab}}{4\pi v}\langle\sigma;\lambda_1\rangle_g \delta_{\sigma\bar{\sigma}}\delta_{\lambda_1\lambda_2}\,, \tag{3.24}$$

where $a$ and $b$ are indices of the SU(3) generators in the adjoint representation and $\sigma$, $\bar{\sigma}$, and $\lambda_{1,2}$ denote the helicities of fermion, antifermion, and gluons, respectively. The amplitude $\langle\sigma;\lambda\rangle_g$ is defined as

$$\langle\sigma;\lambda\rangle_g \equiv \sum_{i,j=1,2,3} [S_i^g(\sqrt{\hat{s}}) + i\lambda P_i^g(\sqrt{\hat{s}})]\, D_{ij}(\hat{s})\,(\sigma\beta_f g^S_{H_j\bar{f}f} - i g^P_{H_j\bar{f}f})\,, \tag{3.25}$$

where

$$\begin{aligned} S_i^g(\sqrt{\hat{s}}) &= \sum_{f=b,t} g_f g^S_{H_i\bar{f}f}\frac{v}{m_f}F_{sf}(\tau_f) - \sum_{\tilde{f}_j=\tilde{t}_1,\tilde{t}_2,\tilde{b}_1,\tilde{b}_2} g_{H_i\tilde{f}_j^*\tilde{f}_j}\frac{v^2}{4m^2_{\tilde{f}_j}}F_0(\tau_{\tilde{f}_j})\,, \\ P_i^g(\sqrt{\hat{s}}) &= \sum_{f=b,t} g_f g^P_{H_i\bar{f}f}\frac{v}{m_f}F_{pf}(\tau_f)\,, \end{aligned} \tag{3.26}$$

where $\tau_x = \hat{s}/4m_x^2$ and $\hat{s}$-dependent $S_i^g$ and $P_i^g$ are scalar and pseudoscalar form factors [4], respectively. The Higgs-boson couplings to quarks $g^{S,P}_{H_i\bar{f}f}$ and squarks $g_{H_i\tilde{f}_j^*\tilde{f}_j}$, and the explicit forms of the functions $F_{sf,pf,0}$ are coded in `CPsuperH` [131]. When $f = \tau, t, \chi^0, \chi^\pm$, etc, one can construct CP asymmetries in the longitudinal and/or transverse polarizations of the final fermions which can be observed at the LHC. For other production mechanisms such as $b$-quark and weak-boson fusions, the CP asymmetries can be defined similarly as in the case of gluon fusion.

When $\tan\beta$ is large and $M_{H^\pm} \sim 150$ GeV, $b$-quark fusion is a dominant production mechanism and the CP asymmetries in gluon fusion can be diluted. In this case, the most promising channel for probing Higgs-sector CP violation may be the weak-boson fusion process and subsequent decays into tau leptons: $W^+W^- \to H_{1,2,3} \to \tau^+\tau^-$. The cross section $\sigma[pp(W^+W^-) \to H \to \tau^+\tau^- X]$ lies between 0.2 and 0.6 pb and the CP asymmetry is large for a wide range of CP phases, see the two upper frames of Fig. 3.13.

[4] These $\hat{s}$-dependent gluon-gluon-Higgs couplings are stored in arrays `SGLUE(3)` and `PGLUE(3)`.

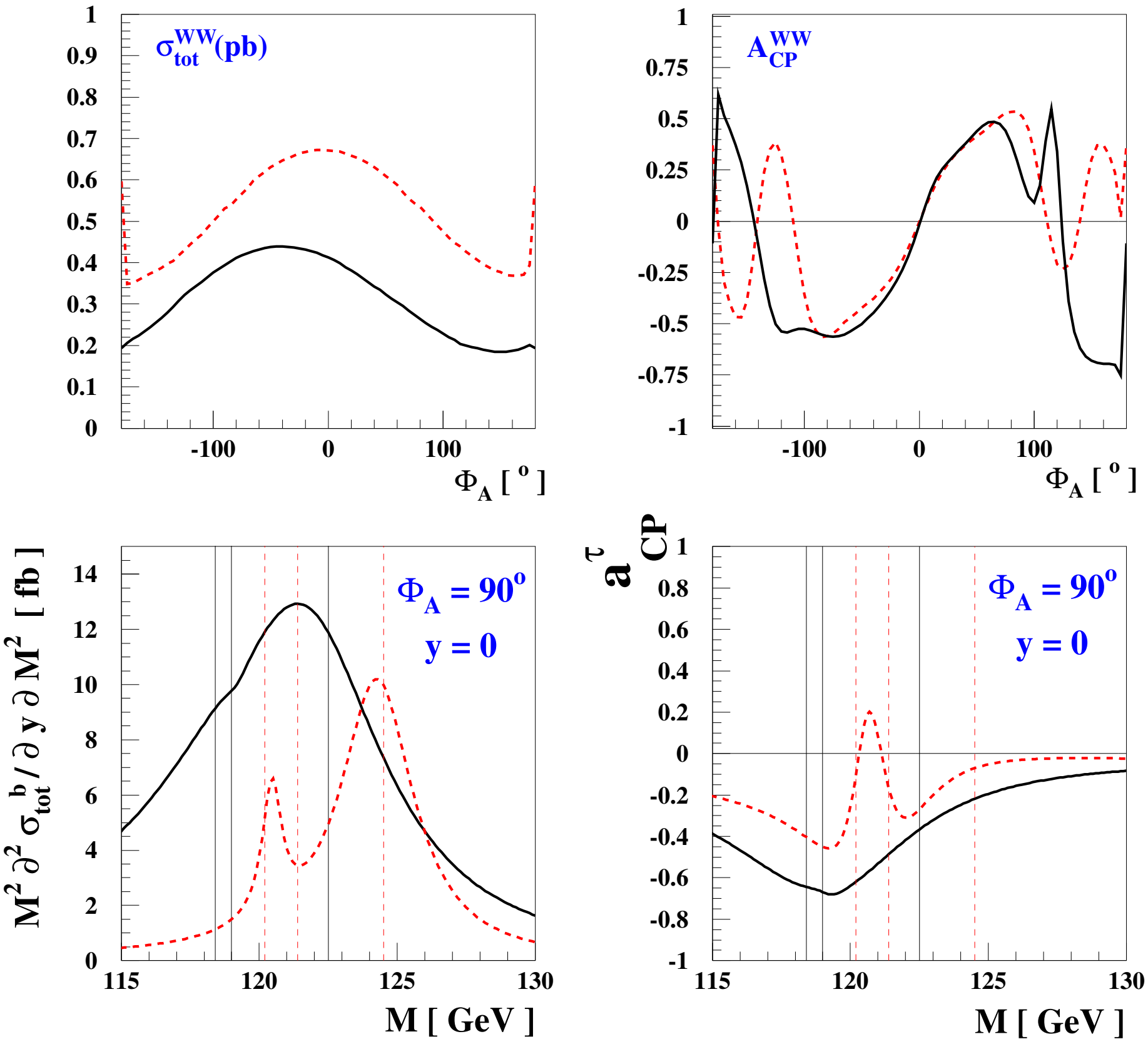

Fig. 3.13: The cross section $\sigma^{WW}_{\rm tot}[pp(WW) \to \tau^+\tau^- X]$ (upper-left panel) at the LHC and its associated total CP asymmetry $\mathcal{A}^{WW}_{\rm CP} \equiv [\sigma^{WW}_{RR} - \sigma^{WW}_{LL}]/[\sigma^{WW}_{RR} + \sigma^{WW}_{LL}]$ (upper-right panel) as functions of $\Phi_A = \Phi_{A_t} = \Phi_{A_b} = \Phi_{A_\tau}$, where $\sigma^{WW}_{RR(LL)} \equiv \sigma(pp\,(WW) \to \tau^+_{R(L)}\tau^-_{R(L)}X)$. We have considered a tri-mixing scenario with $\Phi_3 = -10^\circ$ (dotted lines) and $-90^\circ$ (solid lines). For details of the scenario and the CP asymmetry, see [158]. The lower frames are for Higgs bosons produced in diffractive collisions at the LHC. The lower-left frame shows the hadron-level cross sections when the Higgs bosons decay into $b$ quarks, as functions of the invariant mass $M$ with $\Phi_A = 90^\circ$ and rapidity $y = 0$ . The vertical lines indicate the three Higgs-boson pole-mass positions. The lower-right frame shows the CP-violating asymmetry when the Higgs bosons decay into $\tau$ leptons. See [76] for details.

Higgs-boson production in an exclusive diffractive collision $p + p \to p + H_i + p$ offers unique possibilities for exploring Higgs physics in ways that would be difficult or even impossible in inclusive Higgs production [76]. In spite of the low and theoretically uncertain luminosity of the process, what makes diffraction so attractive compared to the inclusive processes are the clean environment due to the large rapidity gap and the good Higgs-mass resolution of the order of 1 GeV which may be achievable by precise measurements of the momenta of the outgoing protons in detectors a long way downstream from the interaction point. It may be possible to disentangle nearly-degenerate Higgs bosons by examining the production lineshape of the coupled system of neutral Higgs bosons, see lower-left frame of Fig. 3.13. Moreover, the CP-odd polarization asymmetry can be measured when the polarization information of Higgs decay products is available, see the lower-right frame of Fig. 3.13. For more studies of the diffractive production and decay of Higgs bosons in the MSSM scenario with CP violation, see Section 3.8.

B. ILC

A future $e^+e^-$ linear collider, such as the projected ILC, will have the potential to probe the Higgs sector with higher precision than the LHC. At the ILC, the Higgs-boson coupling to a pair of vector bosons, $g_{H_iVV} = c_\beta O_{\phi_1 i} + s_\beta O_{\phi 2i}$, plays a crucial role. There are three main processes for producing the neutral Higgs bosons: Higgs strahlung, $WW$ fusion, and pair production. The cross section of each process is given by

$$\begin{aligned}
\sigma(e^+e^- \to ZH_i) &= g^2_{H_iVV}\, \sigma_{\rm SM}(M_{H_{\rm SM}} \to M_{H_i})\,, \\
\sigma(e^+e^- \to \nu\nu H_i) &= g^2_{H_iVV}\, \sigma^{WW}_{\rm SM}(M_{H_{\rm SM}} \to M_{H_i})\,, \\
\sigma(e^+e^- \to H_iH_j) &= g^2_{H_iH_jV}\, \frac{G_F^2 M_Z^4}{6\pi s}(v_e^2 + a_e^2)\frac{\lambda^{3/2}(1, M^2_{H_i}/s, M^2_{H_j}/s)}{(1 - M_Z^2/s)^2}\,,
\end{aligned} \tag{3.27}$$

where $g_{H_iH_jV} = {\rm sign}[\det(O)]\epsilon_{ijk}\, g_{H_kVV}$, $v_e = -1/4 + s^2_W\,, a_e = 1/4$ and $\sigma^{(WW)}_{\rm SM}$ denotes the corresponding production cross section of the SM Higgs boson. As is well known, the $WW$ fusion cross section grows as $\ln(s)$ compared to the Higgs strahlung and becomes dominant for large center-of-mass energy $\sqrt{s}$. In the decoupling limit, $M_{H^\pm} \gtrsim 200$ GeV, the couplings for heavier Higgs bosons $g_{H_{2,3}VV}$ are suppressed, see Fig. 3.1 of Sec. 3.1. In this case, for the production of $H_2$ and $H_3$, the pair production mechanism is active since $|g_{H_2H_3V}| = |g_{H_1VV}| \sim 1$. When $M_{H^\pm} \lesssim 200$ GeV, the excellent energy and momentum resolution of electrons and muons coming from measurements of the $Z$ boson in Higgs strahlung may help to resolve a coupled system of neutral Higgs bosons by analyzing the production lineshape, see the two upper panels of Fig. 3.14.

As noted in the previous section, if one observes three Higgs bosons in Higgsstrahlung and $WW$ fusion and/or all three pairs of Higgs bosons in pair production, this can be interpreted as a signal of CP violation in the MSSM framework. However, such an interpretation relies on the hypothesis that there exist no additional singlet or doublet Higgs fields. To confirm the existence of genuine CP violation, one needs to measure other observables such as CP asymmetries. In this light, the final fermion spin-spin correlations in Higgs decays into tau leptons, neutralinos, charginos, and top quarks need to be investigated. See the lower panels of Fig. 3.14 for an example when Higgs bosons are are produced in Higgs strahlung and decay into $\tau$ leptons. The two CP asymmetries $a^\tau_L$ and $a^\tau_T$ are defined in terms of the longitudinal and transverse polarizations of final $\tau$ leptons, see [153] for details.

C. $\gamma$LC

The two-photon collider option of the ILC, the $\gamma$LC, offers unique capabilities for probing CP violation in the MSSM Higgs sector, because one may vary the initial-state polarizations as well as measure the polarizations of some final states in Higgs decays [93]. The amplitude contributing to $\gamma(\lambda_1)\gamma(\lambda_2) \to H \to f(\sigma)\bar{f}(\bar{\sigma})$ is given by

$$\mathcal{M}_H = \frac{\alpha\, m_f \sqrt{\hat{s}}}{4\pi v^2}\langle\sigma;\lambda_1\rangle_H \delta_{\sigma\bar{\sigma}}\delta_{\lambda_1\lambda_2}, \tag{3.28}$$

where the reduced amplitude

$$\langle\sigma;\lambda\rangle_H = \sum_{i,j=1}^{3}[S_i^\gamma(\sqrt{\hat{s}}) + i\lambda P_i^\gamma(\sqrt{\hat{s}})]\, D_{ij}(\hat{s})\, (\sigma\beta_f g^S_{H_j\bar{f}f} - i g^P_{H_j\bar{f}f}), \tag{3.29}$$

is a quantity given by the Higgs-boson propagator matrix Eq. (3.23) combined with the production and decay vertices. The one-loop induced complex couplings of the $\gamma\gamma H_i$ vertex, $S_i^\gamma(\sqrt{\hat{s}})$ and $P_i^\gamma(\sqrt{\hat{s}})$, get dominant contributions from charged particles such as the bottom and top quarks, tau leptons, $W^\pm$

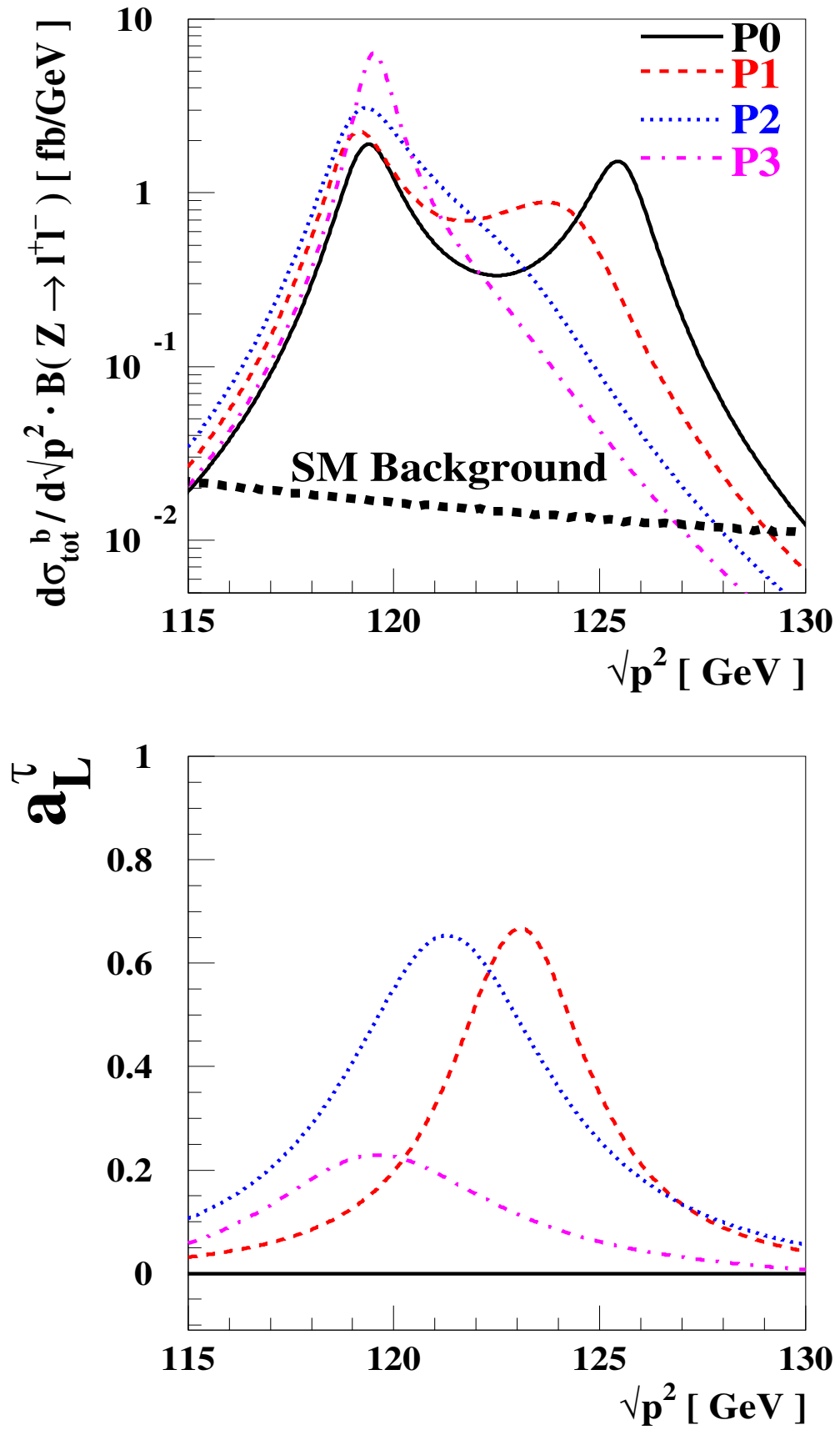

Fig. 3.14: The differential total cross section $d\hat{\sigma}^f_{\rm tot}(e^-e^+ \to ZH_i \to Z\bar{f}f)/d\sqrt{p^2}$ multiplied by $B(Z \to l^+l^-) = B(Z \to e^+e^-) + B(Z \to \mu^+\mu^-)$ as functions of the invariant mass of the Higgs decay products $\sqrt{p^2}$ in units of fb/GeV when $f = b$ (upper-left panel) and $f = \tau$ (upper-right panel). The CP-conserving two-way mixing (**P0**) and three CP-violating tri-mixing (**P1**-**P3**) scenarios have been taken. The lower two frames show the CP-violating asymmetries when Higgs bosons decay into tau leptons. See [153] for details.

bosons, charginos, third-generation sfermions and charged Higgs bosons [5] :

$$\begin{aligned}
S_i^{\gamma}(\sqrt{\hat{s}}) &= 2\sum_{f=b,t,\tilde{\chi}^{\pm}_{1,2}} N_C\, Q_f^2\, g_f g^S_{H_i\bar{f}f}\,\frac{v}{m_f}F_{sf}(\tau_f) - \sum_{\tilde{f}_j=\tilde{t}_{1,2},\tilde{b}_{1,2},\tilde{\tau}_{1,2}} N_C\, Q_f^2 g_{H_i\tilde{f}_j^*\tilde{f}_j}\frac{v^2}{2m^2_{\tilde{f}_j}}F_0(\tau_{\tilde{f}_j}) \\
&\quad - g_{H_iVV}F_1(\tau_W) - g_{H_iH^+H^-}\frac{v^2}{2M^2_{H^\pm}}F_0(\tau_{H^\pm})\,, \\
P_i^{\gamma}(\sqrt{\hat{s}}) &= 2\sum_{f=b,t,\tilde{\chi}^{\pm}_{1,2}} N_C\, Q_f^2\, g_f g^P_{H_i\bar{f}f}\,\frac{v}{m_f}F_{pf}(\tau_f)\,,
\end{aligned} \tag{3.30}$$

where $\tau_x = \hat{s}/4m_x^2$, $N_C = 3$ for (s)quarks and $N_C = 1$ for staus and charginos, respectively. For the explicit forms of $F_1$ and couplings, see [131].

One advantage of $\gamma$LC over the $e^+e^-$ option at the ILC is that one can construct CP asymmetries even when Higgs bosons decay into muons and $b$ quarks, by exploiting the controllable beam polarizations of the colliding photons, see Fig. 3.15.

[5] The arrays `SPHO(3)` and `PPHO(3)` are used for the $\hat{s}$-dependent $\gamma$-$\gamma$-Higgs couplings.

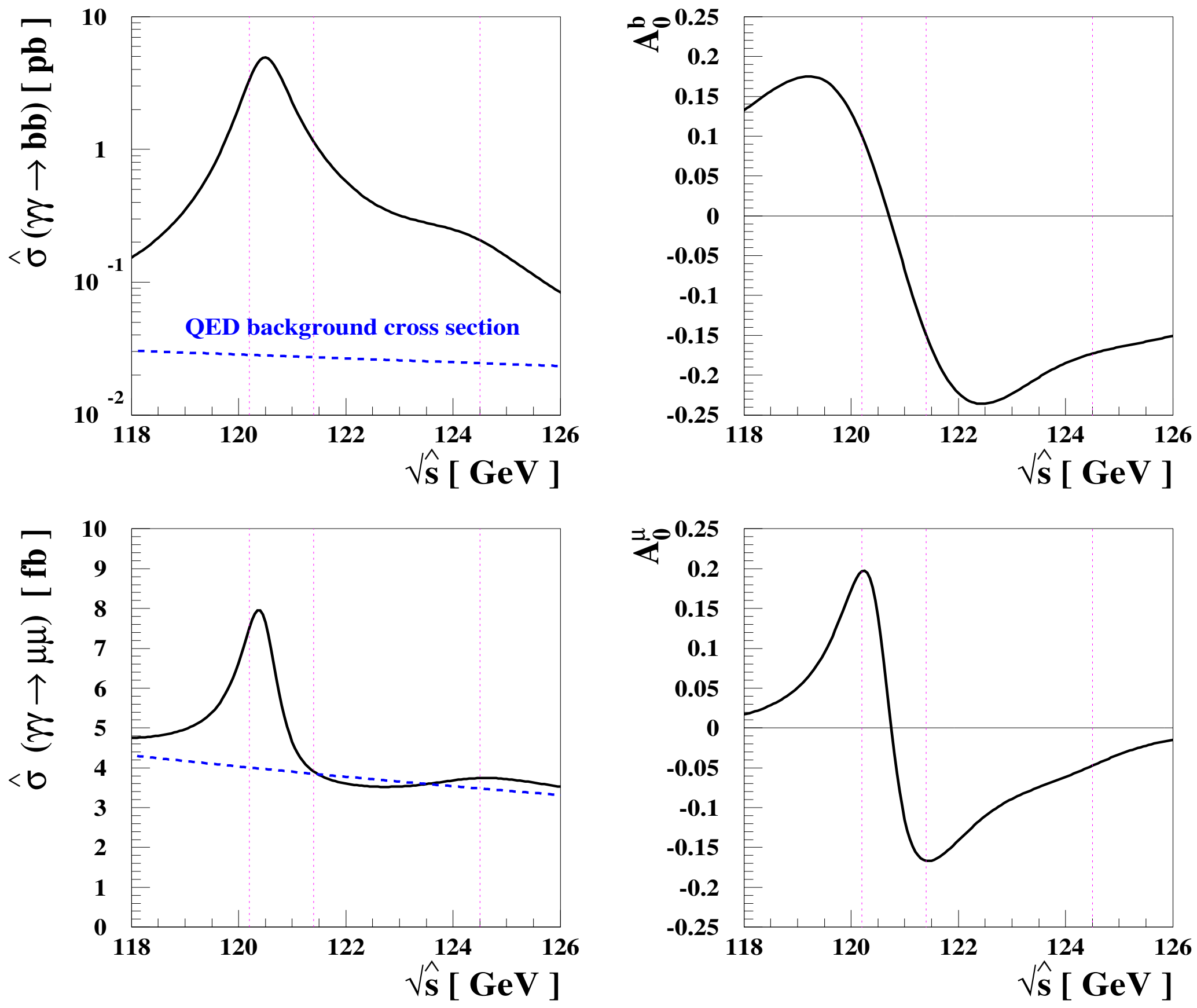

Fig. 3.15: The cross sections (left column) and the CP asymmetries $\mathcal{A}_0^f$ (right column) for the processes $\gamma\gamma \to \bar{b}b$ (upper) and $\gamma\gamma \to \mu^+\mu^-$ (lower). The QED continuum contributions to the cross sections are also shown. The tri-mixing scenario with $\Phi_3 = -10^\circ$ and $\Phi_A = 90^\circ$ has been considered, making the angle cuts $\theta^b_{\rm cut} = 280$ mrad and $\theta^\mu_{\rm cut} = 130$ mrad. The three Higgs masses are indicated by vertical lines. See [93] for details.

One can investigate all possible spin-spin correlations in the final states such as tau leptons, neutralinos, charginos, top quarks, vector bosons, etc., with the goal of complete determination of CP-violating Higgs-boson couplings to them. The cases of tau-lepton and top-quark final states are demonstrated in [93].

For the complete determination of CP-violating Higgs-boson couplings to SM as well as Supersymmetric particles, a muon collider is even better than the $\gamma$LC. At a muon collider, it is possible to control the energy resolution and polarizations of both the muon and the anti-muon. Compared to the $\gamma$LC case, the center-of-mass frame is known and it has much better resolving power for a nearly-degenerate system of Higgs bosons [100].

### *3.4.3 Low-energy observables*

Low-energy observables such as EDMs, $(g-2)_\mu$, $\mathrm{BR}(b \to s\gamma)$, $\mathcal{A}_{\rm CP}(b \to s\gamma)$, $\mathrm{BR}(B \to K l l)$, $\mathrm{BR}(B_{s,d} \to l^+ l^-)$, etc. provide indirect constraints on the soft SUSY breaking parameters. Specifically, we show in [153] that it is straightforward to obtain the expressions of the coefficient $C_S$ and the Higgs-mediated $d_e$ for the thallium EDM by use of couplings calculated by `CPsuperH`. Fig. 3.16 shows that one can implement the thallium EDM constraint on the CP-violating phases and demonstrate that they leave open the possibility of large CP-violating effects in Higgs production at the ILC.

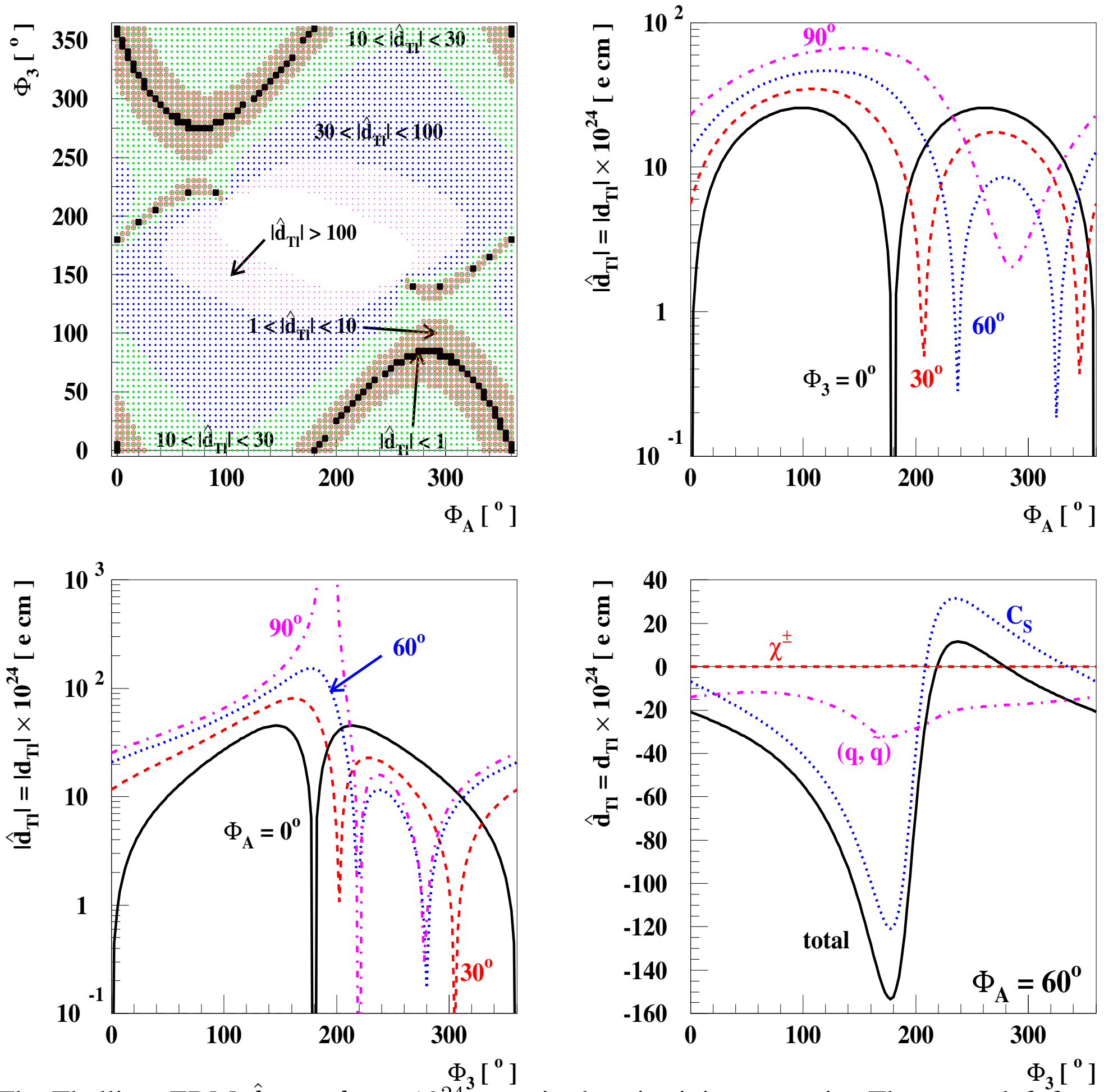

Fig. 3.16: The Thallium EDM $\hat{d}_{\rm Tl} \equiv d_{\rm Tl} \times 10^{24}\, e\, cm$ in the tri-mixing scenario. The upper-left frame displays $|\hat{d}_{\rm Tl}|$ in the $(\Phi_A, \Phi_3)$ plane. The unshaded region around the point $\Phi_3 = \Phi_A = 180^o$ is not theoretically allowed. The different shaded regions correspond to different ranges of $|\hat{d}_{\rm Tl}|$, as shown: specifically, $|\hat{d}_{\rm Tl}| < 1$ in the narrow region denoted by filled black squares. In the upper-right frame, we show $|\hat{d}_{\rm Tl}|$ as a function of $\Phi_A$ for several values of $\Phi_3$. In the lower-left frame, we show $|\hat{d}_{\rm Tl}|$ as a function of $\Phi_3$ for four values of $\Phi_A$. In the lower-right frame, we show the $C_S$ (dotted line) and $d_e$ (dash-dotted line) contributions to $\hat{d}_{\rm Tl}$ separately as functions of $\Phi_3$ when $\Phi_A = 60^o$. As shown by the dashed line, the chargino contribution is negligible. See [153] for details.

The code `CPsuperH` will be extended in near future to include CP-violating effective FCNC Higgs-boson interactions to up- and down-type quarks [207–210]. The determination of these effective interactions may be further improved in the framework of an effective potential approach, where the most significant subleading contributions to the couplings can be consistently incorporated. At large values of $\tan\beta$, Higgs-mediated interactions contribute significantly to the $B$-meson observables mentioned above and so may offer novel constraints on the parameter space of constrained versions of the MSSM, such as the scenario of minimal flavour violation.

### 3.5 Higgs phenomenology in the Feynman-diagrammatic approach / FeynHiggs

*Thomas Hahn, Sven Heinemeyer, Wolfgang Hollik, Heidi Rzehak, Georg Weiglein and Karina Williams*

In this contribution we present recent higher-order corrections to the Higgs boson masses and decay widths involving complex phases that have been obtained in the Feynman diagrammatic (FD) approach. A precise prediction for the masses of the Higgs bosons, their couplings, and their production and decay processes in terms of the relevant SUSY parameters is necessary in order to determine the discovery and exclusion potential of the Tevatron [211–214], and for physics at the LHC [196, 215–219] and the ILC [220–225].

In the following we discuss the two-loop corrections of $\mathcal{O}(\alpha_t\alpha_s)$ to the Higgs boson masses and mixings. We give a short description of the calculation and show numerical examples, comparing the phase dependence at the one-loop and the two-loop level. We furthermore present the vertex corrections to the decay $H_2 \to H_1 H_1$, which plays an important role for the Higgs search in the CP-violating MSSM [162]. We discuss some details of the calculation and present the numerical results, showing the impact of higher-order corrections and the dependence on the complex phase. The new results discussed in this contribution are currently implemented into the Fortran code `FeynHiggs`. We provide a brief description of this code, including a summary of the evaluated observables and instructions for its installation and use.

#### *3.5.1 Higher-order corrections to the Higgs boson masses and mixings*

The current status can be summarized as follows: after the first more general investigations [48, 49], one-loop calculations have been performed in the effective potential (EP) approach [51, 52], and leading two-loop contributions have been incorporated with the renormalisation-group (RG) improved one-loop EP method [50, 53]. These results have been restricted to the corrections coming from the (s)fermion sector and some leading logarithmic corrections from the gaugino sector. Within the FD approach the leading one-loop corrections have been calculated in Ref. [59], and the complete one-loop result has been obtained in Refs. [60, 61, 148]. Within the FD approach the two-loop corrections in the $t/\tilde{t}$ sector had so far been restricted to the MSSM with real parameters [113–115, 118]. The FD result in the MSSM with real parameters contains subleading two-loop corrections that go beyond the result obtained in the EP/RG approach, leading to a shift in the lightest Higgs boson mass of about 4 GeV [132]. It is clearly desirable to extend the FD two-loop result to the CP-violating MSSM.

In this section we present the $\mathcal{O}(\alpha_t\alpha_s)$ corrections to Higgs boson masses and mixings including the full phase dependence at the two-loop level.

##### 3.5.1.1 Calculation of two-loop corrections

In order to compute the Higgs boson masses and mixings up to $\mathcal{O}(\alpha_t\alpha_s)$ the determinant of the inverse propagator matrix $\Gamma$ has to be evaluated,

$$\Gamma(k^2) = k^2 \mathbb{1} - \begin{pmatrix} (M_H^{(0)})^2 - \hat{\Sigma}_{HH}(k^2) & -\hat{\Sigma}_{hH}(k^2) & -\hat{\Sigma}_{AH}(k^2) \\ -\hat{\Sigma}_{hH}(k^2) & (M_h^{(0)})^2 - \hat{\Sigma}_{hh}(k^2) & -\hat{\Sigma}_{Ah}(k^2) \\ -\hat{\Sigma}_{AH}(k^2) & -\hat{\Sigma}_{Ah}(k^2) & (M_A^{(0)})^2 - \hat{\Sigma}_{AA}(k^2) \end{pmatrix}. \tag{3.31}$$

The Higgs masses are given by the roots of det($\Gamma$). The tree-level masses are denoted by $M^{(0)}$. The renormalized self-energies, $\hat{\Sigma}_{\phi\phi}(k^2)$ with $\phi = H, h, A$, contain a one-loop and a two-loop part,

$$\hat{\Sigma}_{\phi\phi} = \hat{\Sigma}^{(1)}_{\phi\phi} + \hat{\Sigma}^{(2)}_{\phi\phi}\,. \tag{3.32}$$

In our calculation we have evaluated the dominant part of the two-loop self-energies, i.e. the contributions of $\mathcal{O}(\alpha_t\alpha_s)$, taking into account the full complex phase dependence. To extract this dominant part the generic self-energy diagrams (see Fig. 3.17) and the corresponding diagrams with counterterm insertions (see Fig. 3.18) have been evaluated applying the approximation of vanishing electroweak gauge couplings and vanishing external momenta. The Feynman diagrams have been generated with the package `FeynArts` [226–229] and the tensor reduction has been performed with the package `TwoCalc` [230].

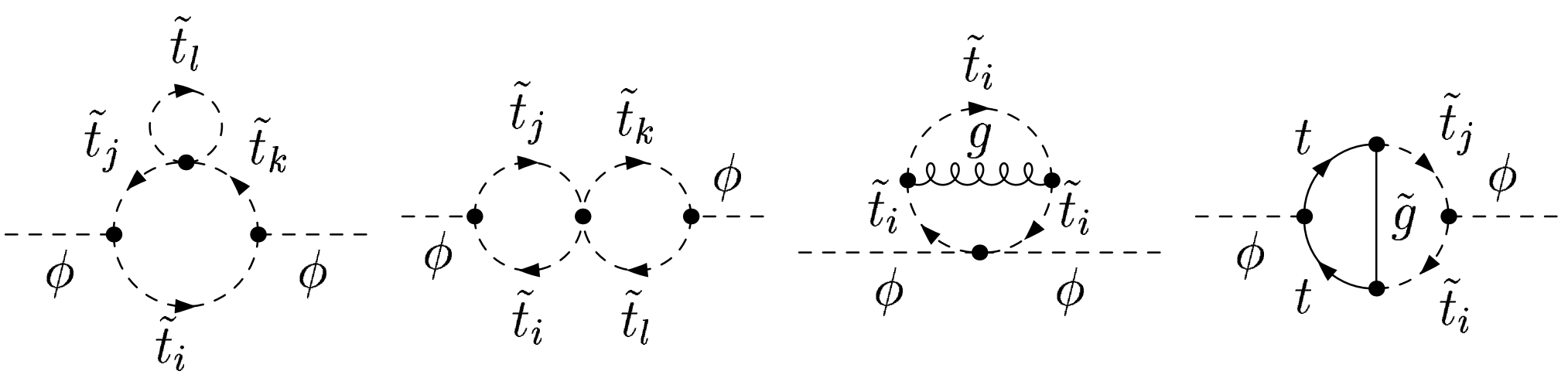

Fig. 3.17: Sample of two-loop diagrams for the Higgs-boson self-energies ($\phi = h, H, A;\ i, j, k, l = 1, 2$).

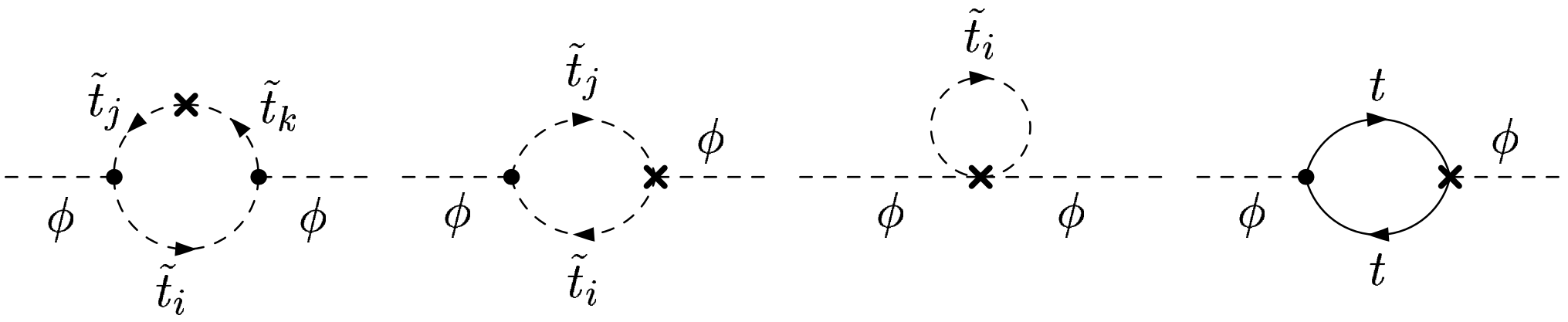

Fig. 3.18: Sample diagrams for the Higgs self-energies with counterterm insertion ($\phi = h, H, A;\ i, j, k = 1, 2$).

In the calculation of the renormalised self-energies the input parameter $M_{H^\pm}$ enters and has to be defined at the two-loop level. We use the on-shell renormalisation for the charged Higgs boson,

$$\mathrm{Re}\hat{\Sigma}_{H^+H^-}(M_{H^\pm}^2) = 0\ , \tag{3.33}$$

where as explained above the external momentum is neglected in the two-loop contribution. The on-shell condition ensures that $M_{H^\pm}$ corresponds to the physical (pole) mass. Also the SM gauge bosons are renormalized on-shell. The tadpole coefficients must vanish in order not to shift the vacuum expectation values,

$$T_\phi + \delta T_\phi = 0 \quad (\phi = h, H, A)\ . \tag{3.34}$$

The counterterm expressions enter the the renormalised Higgs self-energies in Eq. (3.31).

The parameters of the $\tilde{t}$ sector have to be defined at the one-loop level. The top quark mass, $m_t$, as well as the two $\tilde{t}$ masses, $m_{\tilde{t}_1}$ and $m_{\tilde{t}_2}$, are defined as pole masses. The mixing is fixed by (generalizing the condition used in Ref. [231])

$$\widetilde{\mathrm{Re}}\hat{\Sigma}_{\tilde{t}_{12}}(m_{\tilde{t}_1}^2) + \widetilde{\mathrm{Re}}\hat{\Sigma}_{\tilde{t}_{12}}(m_{\tilde{t}_2}^2) = 0\ , \tag{3.35}$$

where $\widetilde{\mathrm{Re}}$ gives the real part of the loop functions and does not act on complex parameters.

#### 3.5.1.2 Numerical results

At the two-loop level the phases of the $\tilde{t}$ sector, $\Phi_{A_t}$ and $\Phi_\mu$, and of the gluino mass parameter, $\Phi_3$, enter the prediction of the Higgs boson masses and mixings. The phase of the Higgs mixing parameter, $\Phi_\mu$, is tightly constrained by the measurements of the electric dipole moments [232] (we use the convention where $\Phi_{M_2} = 0$), and we therefore do not consider non-zero values of this phase.

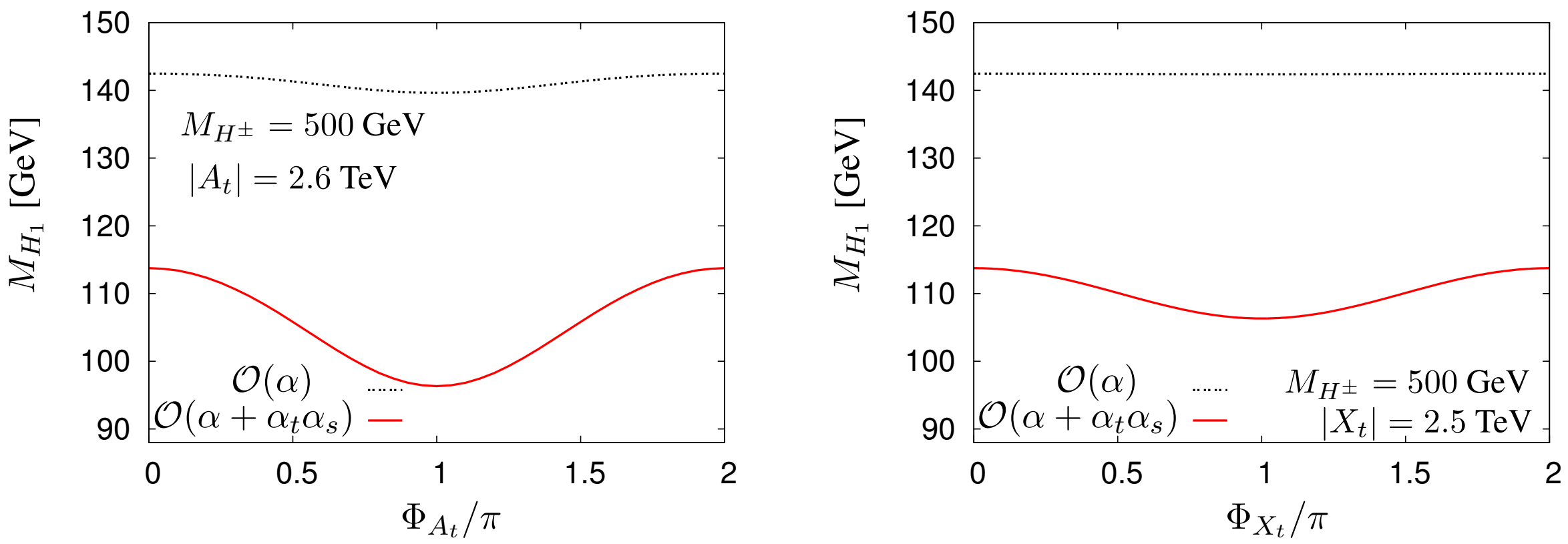

Fig. 3.19: Dependence of the mass of the lightest Higgs boson, $M_{H_1}$, on the phase $\Phi_{A_t}$ (left) and the phase $\Phi_{X_t}$ (right). The mass prediction is shown including $\mathcal{O}(\alpha)$ (one-loop) and $\mathcal{O}(\alpha + \alpha_t\alpha_s)$ (two-loop) corrections.

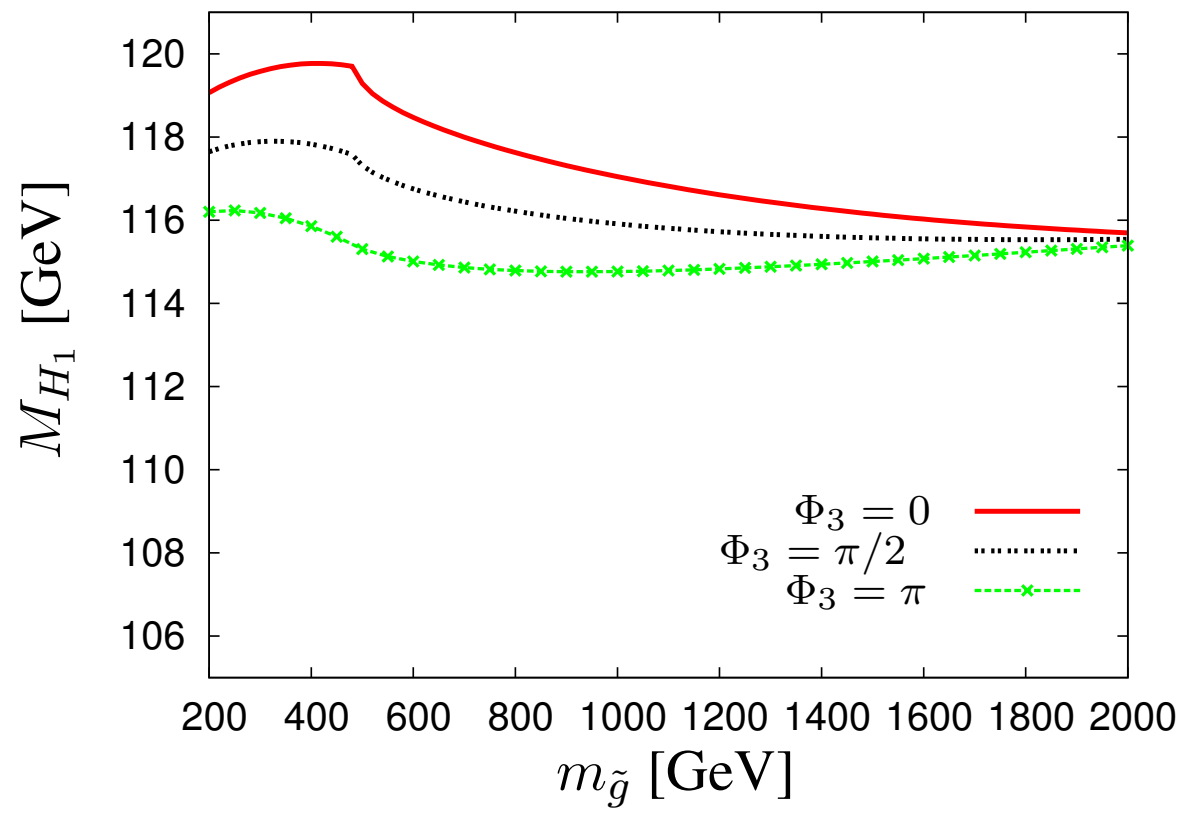

Fig. 3.20: Dependence of $M_{H_1}$ on $m_{\tilde{g}}$ for $\Phi_3 = 0, \pi/2, \pi$ with $M_{\rm SUSY} = 500$ GeV.

The following default values for the parameters have been used unless indicated otherwise: $m_t = 174.3$ GeV, $M_{H^\pm} = 500$ GeV, $\tan\beta = 10$, $\mu = 1000$ GeV, $M_1 = 5/3\tan^2\theta_W M_2$, $M_2 = 500$ GeV, $M_3 = 1000$ GeV, $M_{\rm SUSY} = 1000$ GeV (the soft SUSY-breaking parameters in the diagonal entries in the sfermion mass matrices), $|A_f| = 1000$ GeV (the trilinear sfermion-Higgs couplings), and $\Phi_{A_f} = 0$.

In Fig. 3.19 two numerical examples are shown. In the left plot the dependence of $M_{H_1}$ on the phase of the trilinear coupling, $\Phi_{A_t}$, is shown, taking into account the one-loop and the one- and two-loop contributions, respectively. The phase dependence is much stronger in the case of the two-loop corrected mass. This is related in particular to the two-loop contributions with gluino exchange, see Fig. 3.17. Varying $\Phi_{A_t}$ also changes the amount of $\tilde{t}$ mixing, $X_t := A_t - \mu^* \cot\beta$, and hence also the values of the $\tilde{t}$ masses. In the right plot of Fig. 3.19 the phase of the squark mixing $\Phi_{X_t}$ is varied, which keeps the $\tilde{t}$ masses constant. The parameters are chosen such that for vanishing phases the Higgs masses in both plots in Fig. 3.19 are equal. In the right plot the phase dependence is negligible for the one-loop mass but still sizeable in the two-loop case. This behaviour is related to the fact that the one-loop result in the MSSM is symmetric w.r.t. changing the sign of $X_t$, while the FD two-loop result contains contributions proportional to odd powers of $X_t$ that amount to several GeV in $M_{H_1}$, see e.g. Ref. [132].

In Fig. 3.20 we show the dependence of $M_{H_1}$ on $m_{\tilde{g}} \equiv |M_3|$ for three different values of the gluino phase, $\Phi_3 = 0, \pi/2, \pi$. The plot shows that both the variation of $m_{\tilde{g}}$ (see also Ref. [115]) and the impact of the complex phase $\Phi_3$ can lead to shifts in the prediction for $M_{H_1}$ of several GeV. The dependence on $\Phi_3$ is most pronounced in the threshold region seen in Fig. 3.20, where $m_{\tilde{g}} \approx m_{\tilde{t}_1} + m_t$.

The numerical examples shown above demonstrate that the effects of complex phases at the two-loop level are relevant for a precise prediction of the Higgs masses and thus for confronting SUSY theories with present and future experimental results from the Higgs searches. The implementation of the new two-loop corrections into the program `FeynHiggs` is in progress.

### *3.5.2 Higher-order corrections to the decay $H_2 \to H_1 H_1$*

Due to the two Higgs-doublet structure of the MSSM, the on-shell decay of a heavier Higgs boson to two lighter Higgs bosons is possible. For real parameters this can be $h \to AA$ in a small parameter region with very light $M_A$ [233] or $H \to hh$ for large values of $M_A$. The former decay leads to small unexcluded parameter regions in the $M_A$–$\tan\beta$ plane from LEP Higgs searches [162] (especially in the "no-mixing" scenario [234]).

Within the CP-violating MSSM, where all three neutral Higgs bosons can mix, the decays $H_2, H_3$ – $H_1 H_1$ can be important. In the parameter region of the CPX scenario [54] probed by the LEP Higgs searches the decay $H_2 \to H_1 H_1$ can be large, leading to unexcluded areas in the $M_{H^\pm}$–$\tan\beta$ parameter plane for $\tan\beta \sim 4$ and $M_{H_1}$ values of $\sim$ 40 GeV [162]. A precise prediction of this decay in the CP-violating MSSM is crucial in order to translate the experimental limits into reliable bounds on the SUSY parameter space.

In the following we present results for the leading vertex corrections to the decay $H_2 \to H_1 H_1$, obtained in the FD approach. The results for the genuine vertex contributions are combined with the propagator corrections for the external Higgs bosons (evaluated with `FeynHiggs` [60, 115, 130, 144, 147, 148, 235], see Section 3.5.3).

#### 3.5.2.1 Calculation

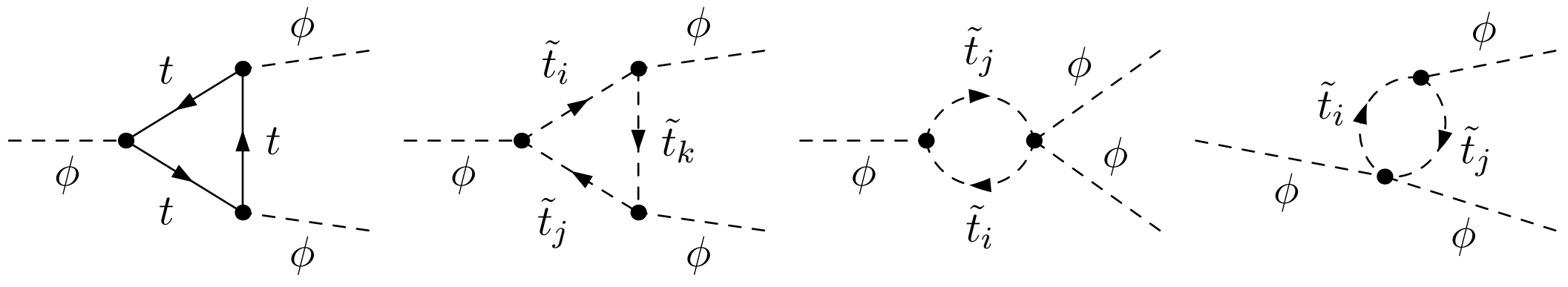

Fig. 3.21: Leading vertex corrections to the decay $H_2 \to H_1 H_1$, involving $t/\tilde{t}$ loops. ($\phi = h, H, A$; $i, j, k = 1, 2$)

The first step in the calculation is the evaluation of the leading one-loop vertex contributions. They consist of the Yukawa-enhanced terms (i.e. those proportional to $m_t^4$) of the diagrams with $t/\tilde{t}$ loops depicted in Fig. 3.21. In order to extract the leading contributions it is sufficient to neglect the gauge couplings and the external momentum. The contributions obtained in this way form a UV-finite subclass. The diagrams were evaluated using the packages `FeynArts` [226–229] and `FormCalc` [236].

The vertex corrections are supplemented with the external propagator corrections, evaluated up to the two-loop level [59–61, 115, 130, 147, 148]. These contributions are incorporated using the the elements of the matrix $U$, see Eq. (3.37). The elements of the mixing matrix and the masses $M_{H_1}$, $M_{H_2}$ of the external particles were obtained from `FeynHiggs` (see Section 3.5.3). Accordingly, the amplitude

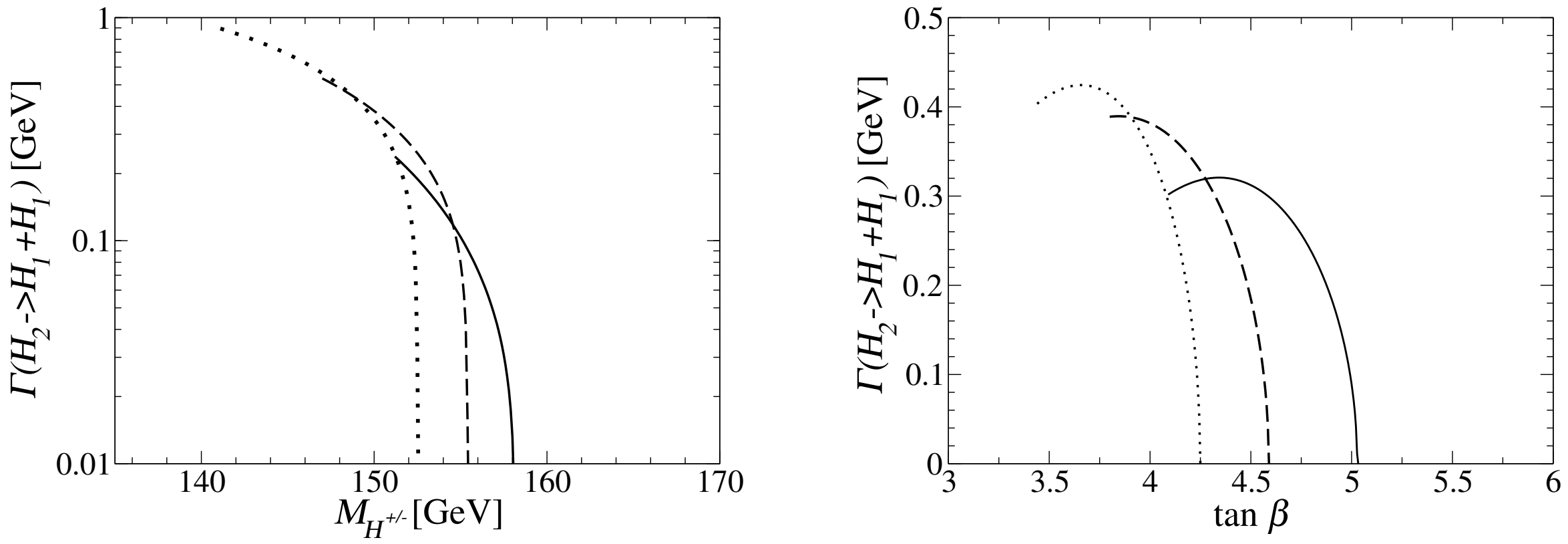

Fig. 3.22: Decay width $\Gamma(H_2 \to H_1 H_1)$ in the CPX scenario (using the CPX values as on-shell parameters) as function of $M_{H^\pm}$ for $\tan\beta = 4$ (left) and as function of $\tan\beta$ for $M_{H^\pm}$ = 150 GeV (right). The genuine vertex contributions are supplemented with the external propagator corrections evaluated with `FeynHiggs` incorporating the one-loop $t, \tilde{t}, b, \tilde{b}$ contributions for vanishing external momentum (dotted lines), the full one-loop result (dashed lines), and the two-loop result (solid lines).

for the decay can be written as:

$$\begin{aligned} A(H_2 \to H_1 H_1) \;=\;& U_{21}U_{11}^2\Gamma_{hhh} + U_{13}^2 U_{23}\Gamma_{AAA} + \left(U_{21}U_{13}^2 + 2U_{11}U_{23}U_{13}\right)\Gamma_{hAA} \\ &+ \left(U_{22}U_{13}^2 + 2U_{12}U_{23}U_{13}\right)\Gamma_{HAA} + \left(U_{23}U_{11}^2 + 2U_{13}U_{21}U_{11}\right)\Gamma_{hhA} \\ &+ (2U_{12}U_{13}U_{21} + 2U_{11}U_{13}U_{22} + 2U_{11}U_{12}U_{23})\Gamma_{hHA} \\ &+ \left(U_{23}U_{12}^2 + 2U_{13}U_{22}U_{12}\right)\Gamma_{HHA} + \left(U_{22}U_{11}^2 + 2U_{12}U_{21}U_{11}\right)\Gamma_{hhH} \\ &+ \left(U_{21}U_{12}^2 + 2U_{11}U_{22}U_{12}\right)\Gamma_{hHH} + U_{12}^2 U_{22}\Gamma_{HHH}, \end{aligned} \tag{3.36}$$

with

$$U = O^T \begin{pmatrix} -s_\alpha & c_\alpha & 0 \\ c_\alpha & s_\alpha & 0 \\ 0 & 0 & 1 \end{pmatrix}, \tag{3.37}$$

where $O$ is defined in Eq. (3.5), and $\Gamma$ denotes the genuine one-loop vertex contributions. $\alpha$ is the angle diagonalizing the CP-even Higgs-boson mass matrix at tree-level.

#### 3.5.2.2 Numerical results

Results for the decay width $\Gamma(H_2 \to H_1 H_1)$ are shown in Figs. 3.22, 3.23. As described above, our numerical results for $\Gamma(H_2 \to H_1 H_1)$ are obtained by supplementing the genuine one-loop vertex contributions with the external propagator corrections according to Eq. (3.36). These external propagator corrections are evaluated with the program `FeynHiggs` incorporating different sets of higher-order contributions. The dotted lines in Figs. 3.22, 3.23 indicate the result where only the one-loop $t, \tilde{t}, b, \tilde{b}$ contributions for vanishing external momentum are taken into account, the dashed lines correspond to the full one-loop result for the propagator corrections, while the full lines indicate the results incorporating also the two-loop propagator corrections.

Fig. 3.22 shows the prediction for the decay width $\Gamma(H_2 \to H_1 H_1)$ as function of $M_{H^\pm}$ (left plot) and as a function of $\tan\beta$ (right plot). The parameters are those of the CPX scenario as defined in Eq. (3.13), with $M_{\rm SUSY} = 500$ GeV and $\Phi_{A_t} = \pi/2$. Once two-loop corrections are taken into account, the renormalisation scheme for the parameters in the stop sector needs to be specified. For simplicity, we interpret $|A_t|$ and $M_{\rm SUSY}$ as on-shell parameters in Fig. 3.22. The figure illustrates the fact that the decay $H_2 \to H_1 H_1$ can be important in the CPX scenario for $\tan\beta \approx 4$ and relatively low $M_{H^\pm}$. The dependence on both $M_{H^\pm}$ and $\tan\beta$ is very pronounced in this region. While the three implementations of the external propagator corrections lead to the same qualitative behaviour of $\Gamma(H_2 \to H_1 H_1)$, they give rise to a sizeable shift in $M_{H^\pm}$ and $\tan\beta$.

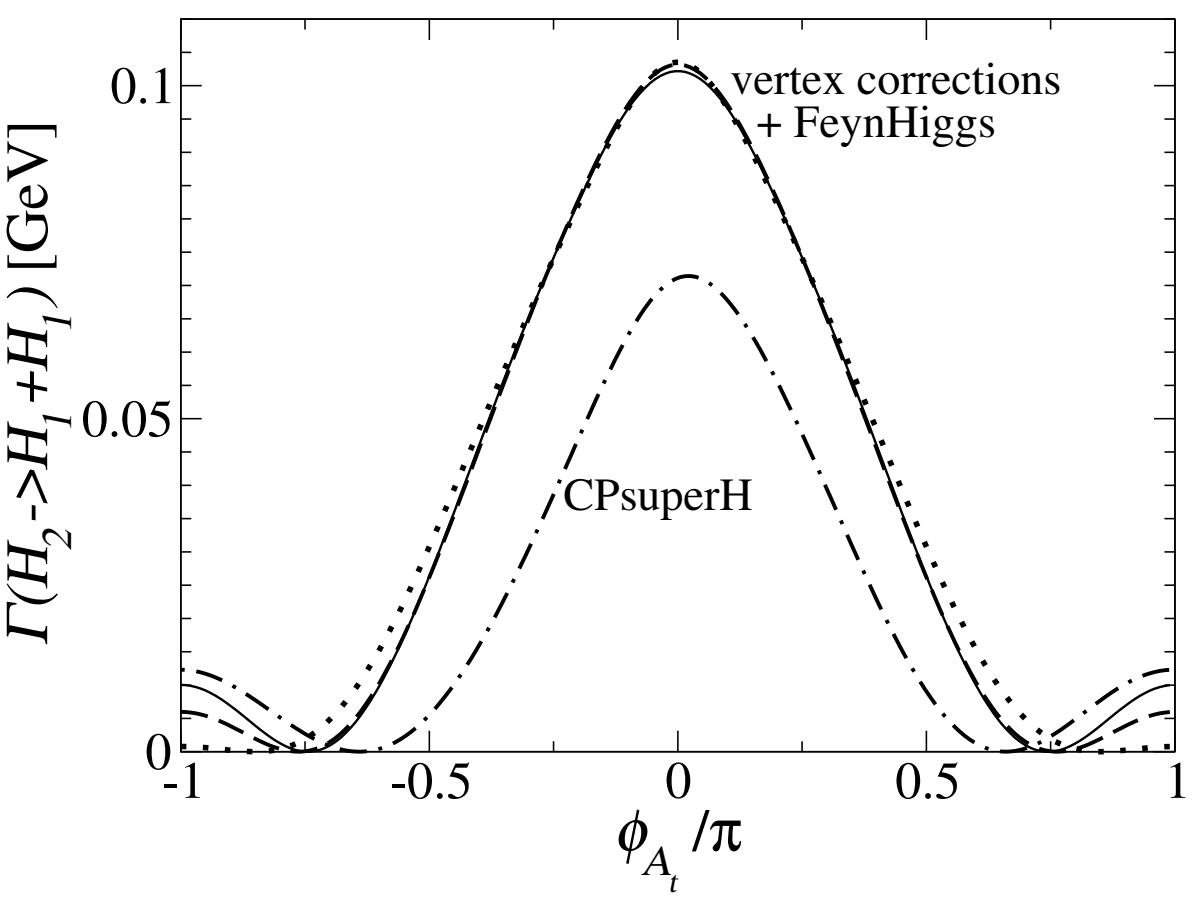

Fig. 3.23: Decay width $\Gamma(H_2 \to H_1 H_1)$ in the CPX scenario as function of the complex phase $\Phi_{A_t}$ for $M_{H^\pm} = 500$ GeV, $\tan\beta = 10$ (using the CPX value of $|A_t|$ as $\overline{\mathrm{DR}}$ parameter). Our results are shown including the genuine vertex contributions, supplemented with the external propagator corrections evaluated with `FeynHiggs` in different approximations (dotted, dashed and solid line, as in Fig. 3.22). They are compared with the result of the program `CPsuperH` (dot–dashed line).

In Fig. 3.23 our result for $\Gamma(H_2 \to H_1 H_1)$ is shown as a function of the complex phase $\Phi_{A_t}$ in the CPX scenario with $M_{\mathrm{SUSY}} = 500$ GeV in comparison with the result of the program `CPsuperH` [131]. While in `FeynHiggs`, based on the FD approach, the on-shell scheme is used (see Section 3.5.3 for details), the input parameters of the program `CPsuperH`, based on the EP/RG approach, correspond to the $\overline{\mathrm{DR}}$ scheme. In order to be able to compare the results of the two programs, the relevant input parameters have to be appropriately converted. We have used the CPX value of $|A_t|$ as $\overline{\mathrm{DR}}$ parameter at the scale $M_S \equiv (M_{\mathrm{SUSY}}^2 + m_t^2)^{1/2}$. This value has been taken as input for `CPsuperH`, while for our FD result we have used the corresponding on-shell parameter as obtained from the simple conversion relations given in Ref. [132].

Fig. 3.23 shows that the dependence of the decay width $\Gamma(H_2 \to H_1 H_1)$ on the complex phase $\Phi_{A_t}$ is very pronounced. The three implementations of the external propagator corrections evaluated with `FeynHiggs` yield very similar results. The qualitative behaviour of the result obtained from the program `CPsuperH` is similar to our FD result. A sizeable difference occurs, however, in particular in the region of small values of $\Phi_{A_t}$.

The implementation of the genuine vertex corrections for the decays of a Higgs boson into two other Higgses into the program `FeynHiggs` is in progress. While the numerical results shown above are based on the leading Yukawa corrections in the $t/\tilde{t}$ sector, the full one-loop vertex corrections in the CP-violating MSSM are currently being evaluated. Combining these with the most up-to-date propagator corrections should prove valuable in order to arrive at an accurate prediction for these decay processes.

### *3.5.3 The Code* FeynHiggs

FeynHiggs [60, 115, 130, 144, 147, 148, 235][6] is a Fortran code for the evaluation of the masses, decay properties and production processes of Higgs bosons in the MSSM with real or complex parameters. The calculation of the higher-order corrections is based on the FD approach [59]. The renormalisation has been performed in a hybrid $\overline{\rm DR}$/on-shell scheme. For the masses and mixings, the one-loop contributions incorporate the complete set of MSSM corrections, including the full momentum and phase dependence and the full $6 \times 6$ non-minimal flavor violation (NMFV) contributions [237, 238]. At the two-loop level all existing corrections from the real MSSM have been included (see Ref. [130] for a review). They are supplemented by the resummation of the leading effects from the (scalar) $b$ sector including the full complex phase dependence.

#### 3.5.3.1 Evaluation of observables

The evaluation of the Higgs-boson masses and mixing angles is supplemented with an estimate of the theory uncertainties from unknown higher-order corrections. The estimate for the total uncertainty is obtained as the sum of deviations from the central value[7], $\Delta X = \sum_{i=1}^{3} |X_i - X|$ with $X = \{M_{h_1,h_2,h_3,H^\pm}, \sin\alpha_{\rm eff}, U_{ij}\}$, where $\alpha_{\rm eff}$ is the loop corrected mixing angle in the CP-even Higgs sector (in the absence of CP-violating phases) and $U_{ij}$ is defined in Eq. (3.37). The $X_i$ are calculated as follows:

- $X_1$: varying the renormalization scale (entering via the $\overline{\rm DR}$ renormalization) within $1/2 m_t \leq \mu \leq 2 m_t$,
- $X_2$: using $m_t^{\rm pole}$ instead of the running $m_t$ in the two-loop corrections,
- $X_3$: using instead of a resummation in the (scalar) $b$ sector an unresummed bottom Yukawa coupling, $y_b$, i.e. an $y_b$ including the leading $\mathcal{O}(\alpha_s\alpha_b)$ corrections, but not resummed to all orders.

Besides predictions for the masses and mixing angles, FeynHiggs2.4 contains the evaluation of all relevant Higgs-boson decay widths and hadron collider production cross sections. These are in particular:

- the total width for the three neutral and the charged Higgs bosons,
- the couplings and branching ratios of the neutral Higgs bosons to
  - SM fermions (see also Ref. [239]), $h_i \to \bar{f}f$,
  - SM gauge bosons (possibly off-shell), $h_i \to \gamma\gamma, ZZ^*, WW^*, gg$,
  - gauge and Higgs bosons, $h_i \to Zh_j$, $h_i \to h_j h_k$,
  - scalar fermions, $h_i \to \tilde{f}^\dagger \tilde{f}$,
  - gauginos, $h_i \to \tilde{\chi}_k^\pm \tilde{\chi}_j^\mp$, $h_i \to \tilde{\chi}_l^0 \tilde{\chi}_m^0$,
- the couplings and branching ratios of the charged Higgs boson to
  - SM fermions, $H^- \to \bar{f}f'$,
  - a gauge and Higgs boson, $H^- \to h_i W^-$,
  - scalar fermions, $H^- \to \tilde{f}^\dagger \tilde{f}'$,
  - gauginos, $H^- \to \tilde{\chi}_k^- \tilde{\chi}_l^0$,
- the neutral Higgs boson production cross sections at the Tevatron and the LHC for all relevant channels (in an effective coupling approximation [240]).

For comparisons with the SM, the following quantities are also evaluated for SM Higgs bosons with the same mass as the three neutral MSSM Higgs bosons:

- the total decay width,
- the couplings and BRs of a SM Higgs boson to SM fermions,
- the couplings and BRs of a SM Higgs boson to SM gauge bosons (possibly off-shell),

[6] Current version: FeynHiggs2.4.0.

[7] Note that in FeynHiggs we use $h_i$ instead of $H_i$ as symbols for the Higgs boson states.

– the production cross sections at the Tevatron and the LHC for all relevant channels [240].

`FeynHiggs2.4` furthermore provides results for electroweak precision observables that give rise to constraints on the SUSY parameter space (see Ref. [47] and references therein)

– the leading corrections to the observables $M_W$ and $\sin^2\theta_{\text{eff}}$ entering via the quantity $\Delta\rho$, evaluated up to the two-loop level [241–246],
– an evaluation of $M_W$ and $\sin^2\theta_{\text{eff}}$ (via $\Delta\rho$) including at the one-loop level the dependence on complex phases from the scalar top/bottom sector [247] and NMFV effects [237],
– the anomalous magnetic moment of the muon, including a full one-loop calculation [248] as well as leading and subleading two-loop corrections [249–251],
– the evaluation of BR($b \to s\gamma$) including NMFV effects [238].

Some further features of `FeynHiggs2.4` are:

– Transformation of the input parameters from the $\overline{\text{DR}}$ to the on-shell scheme (for the scalar top and bottom parameters), including the full $\mathcal{O}(\alpha_s)$ and $\mathcal{O}(\alpha_{t,b})$ corrections.
– Processing of SUSY Les Houches Accord (SLHA 2) data [252–254]. `FeynHiggs2.4` reads the output of a spectrum generator file and evaluates the Higgs boson masses, branching ratios etc. The results are written in the SLHA format to a new output file.
– Predefined input files for the SPS benchmark scenarios [255] and the Les Houches benchmarks for Higgs boson searches at hadron colliders [234] are included.
– Detailed information about all the features of `FeynHiggs2.4` are provided in man pages.

#### 3.5.3.2 New features in `FeynHiggs2.4`

The main new features in `FeynHiggs2.4` as compared to older versions are summarized as follows:

– The imaginary parts of the Higgs-boson self-energies are taken into account in determining the poles of the propagators. The Higgs-boson pole masses are derived as the real parts of the complex poles of the complex propagator matrix.
– The mixing matrix (for internal Higgs bosons) is derived from the real part of the complex propagator matrix. This is also taken into account in the Higgs-boson couplings and decay widths.
– Neutral Higgs boson decays are evaluated with the full rotation to on-shell Higgs bosons. The corresponding rotation matrix for external Higgs bosons, derived from the complex propagator matrix, is provided.
– At the one-loop level the full $6 \times 6$ NMFV effects for the Higgs boson masses and mixings are included [237, 238].
– Negative entries are allowed for the squares of the soft SUSY-breaking parameters (for the diagonal entries for the sfermion mass matrices). The input is given as a negative mass, $-m$, that is then internally converted to $-(m^2)$.
– The two-loop corrections to $(g-2)_\mu$ have been extended, see Refs. [250, 251].
– The evaluation of BR($b \to s\gamma$) has been incorporated, including NMFV effects [238].

#### 3.5.3.3 Installation and use

The installation process is straightforward and should take no more than a few minutes:

– Download the latest version from `www.feynhiggs.de` and unpack the tar archive.
– The package is built with `./configure` and `make`. This creates the library `libFH.a` and the command-line frontend `FeynHiggs`.
– To build also the Mathematica frontend `MFeynHiggs`, invoke `make all`.
– `make install` installs the files into a platform-dependent directory tree,
for example `i586-linux/{bin,lib,include}`.

– Finally, remove the intermediate files with `make clean`.

`FeynHiggs2.4` has four modes of operation:

1. Library Mode: The core functionality of `FeynHiggs2.4` is implemented in a static Fortran 77 library `libFH.a`. All other interfaces are 'just' frontends to this library.
   The library provides the following functions:
   - `FHSetFlags` sets the flags for the calculation.
   - `FHSetPara` sets the input parameters directly, or
     `FHSetSLHA` sets the input parameters from SLHA data.
   - `FHSetCKM` sets the elements of the CKM matrix.
   - `FHSetNMFV` sets the off-diagonal soft SUSY-breaking parameters that induce NMFV effects.
   - `FHSetDebug` sets the debugging level.
   - `FHGetPara` retrieves (some of) the MSSM parameters calculated from the input parameters, e.g. the sfermion masses.
   - `FHHiggsCorr` computes the corrected Higgs masses and mixings.
   - `FHUncertainties` estimates the uncertainties of the Higgs masses and mixings.
   - `FHCouplings` computes the Higgs couplings and BRs.
   - `FHConstraints` evaluates further electroweak precision observables.

   These functions are described in detail on their respective man pages in the `FeynHiggs` package.
2. Command-line Mode: The `FeynHiggs` executable is a command-line frontend to the `libFH.a` library. It is invoked at the shell prompt as

   ```
   FeynHiggs inputfile [flags] [scalefactor]
   ```

   where
   - `inputfile` is the name of a parameter file (see below).
   - `flags` is an (optional) string of integers giving the flag values, e.g. `40030211`. If `flags` is not specified, `40020211` is used.
   - `scalefactor` is an optional factor multiplying the renormalization scale.

   `FeynHiggs` understands two kinds of parameter files:
   - Files in SUSY Les Houches Accord (SLHA) format [252] (using Ref. [253]). In this case `FeynHiggs` adds the Higgs masses and mixings to the SLHA data structure and writes the latter to a file *inputfile*`.fh`.
   - Files in its native format, for example

     ```
     MT          172.5
     MSusy       500
     MA0         200
     TB          5
     Abs(Xt)     1000
     ...
     ```

     Complex quantities can be given either in terms of absolute value `Abs(X)` and phase `Arg(X)`, or as real part `Re(X)` and imaginary part `Im(X)`. Abbreviations, summarizing several parameters (such as `MSusy`) can be used, or detailed information about the various soft SUSY-breaking parameters can be given. Furthermore, it is possible to define loops over parameters in order to scan parts of parameter space. The output is written in a human-readable form to the screen. The output can also be piped through the `table` filter to yield a machine-readable version appropriate for plotting etc.
3. WWW Mode: The `FeynHiggs` User Control Center (FHUCC) is a WWW interface to the command-line executable `FeynHiggs`. It provides a convenient way to play with parameters, but is of course not suited for large-scale parameter scans or extensive analyses. To use the FHUCC, point your favorite Web browser at `www.feynhiggs.de/fhucc`. adjust the parameters, and submit the form to see the results.

4. Mathematica Mode: The `MFeynHiggs` executable provides access to the `FeynHiggs` functions from Mathematica via the MathLink protocol. This is particularly convenient both because `FeynHiggs` can be used interactively this way and because Mathematica's sophisticated numerical and graphical tools, e.g. FindMinimum, are available. After starting Mathematica, install the package with

   `In[1]:= Install["MFeynHiggs"]`

   which makes all `FeynHiggs` subroutines available as Mathematica functions.

#### 3.5.4 Conclusions

We have presented new results on higher-order corrections in the MSSM with complex phases obtained in the Feynman diagrammatic approach. The Fortran code `FeynHiggs` provides the evaluation of masses, decay properties and production processes of Higgs bosons in the CP-violating MSSM. We have described the features of the program and its installation and use.

We have analysed the dependence of the two-loop corrections of $\mathcal{O}(\alpha_t \alpha_s)$ to the Higgs boson masses and mixings on the phases $\Phi_{A_t}$ and $\Phi_3$, i.e. the complex phase of the trilinear coupling in the stop sector and the gluino phase. The two-loop corrections significantly enhance the impact of $\Phi_{A_t}$ compared to the one-loop case. The gluino phase, which enters the Higgs-mass predictions only at the two-loop level, can give rise to a shift of the lightest Higgs-boson mass of several GeV.

A prediction for the decay $H_2 \to H_1 H_1$ has been obtained by combining genuine vertex contributions with external propagator corrections evaluated with `FeynHiggs`. The decay width depends sensitively on higher-order corrections. Varying the complex phase $\Phi_{A_t}$ has a very large effect on $\Gamma(H_2 \to H_1 H_1)$. The comparison with the program `CPsuperH` based on the EP/RG approach shows qualitative agreement in the phase dependence, while a sizeable difference occurs in the maximum value of $\Gamma(H_2 \to H_1 H_1)$.

### 3.6 Self-couplings of Higgs bosons in scenarios with mixing of CP-even/CP-odd states

*Elza Akhmetzyanova, Mikhail Dolgopolov and Mikhail Dubinin*

The effective two-doublet Higgs potential of the MSSM at the energy scale $m_{top}$ has the form of a general two-Higgs-doublet potential, see Eq. (2.1), with four real parameters $\lambda_1$-$\lambda_4$ and three complex-valued parameters $\lambda_5$, $\lambda_6$, $\lambda_7$ which explicitly violate CP invariance in the Higgs sector. The parameters $\lambda_1$-$\lambda_7$ can be calculated [50, 110, 256, 257] and expressed through the parameters of the MSSM in the sector of scalar quarks–Higgs bosons interaction. In this sense the MSSM Higgs sector as an effective field theory at the scale $m_{top}$ can be embedded in a general two-Higgs-doublet model, providing possibilities to interpret some special MSSM features in the language of the THDM parameter space.

In the following we are using the formalism described in [256, 258]. First the THDM mass eigenstates of CP conserving limit $\mathtt{Im}\lambda_{5,6,7}=0$ which are $h$, $H$ (CP-even scalars), $A$ (CP-odd scalar) and $H^\pm$ (charged scalar), see Eqs. (2.23), (2.25) and (2.29), are defined using the two mixing angles $\alpha$ and $\beta$. There is no CP violation at the scale $M_{SUSY}$, where $\lambda_i$ are real-valued, at the scale $m_{top}$ it is radiatively induced. The evaluation of $\lambda_{1-7}$ parameters is based on the effective field theory approach [257] using the MSSM potential of the Higgs bosons - scalar quarks interaction and including the contributions from the F-terms, leading and nonleading D-terms, the wave-function renormalization terms, and the leading two-loop Yukawa QCD-corrections.

In this section we calculate the trilinear and the quartic couplings of physical Higgs bosons in the CPX scenario [54] of the MSSM. Continious interest to the self-interactions of Higgs bosons both in the case of CP conservation [259–263] and the case of CP violation [131, 256, 258] is motivated by the experimental accessibility of the two and three Higgs bosons production signals [62, 64, 65, 78, 89, 91, 264, 265] providing possibilities to reconstruct experimentally the effective Higgs potential.

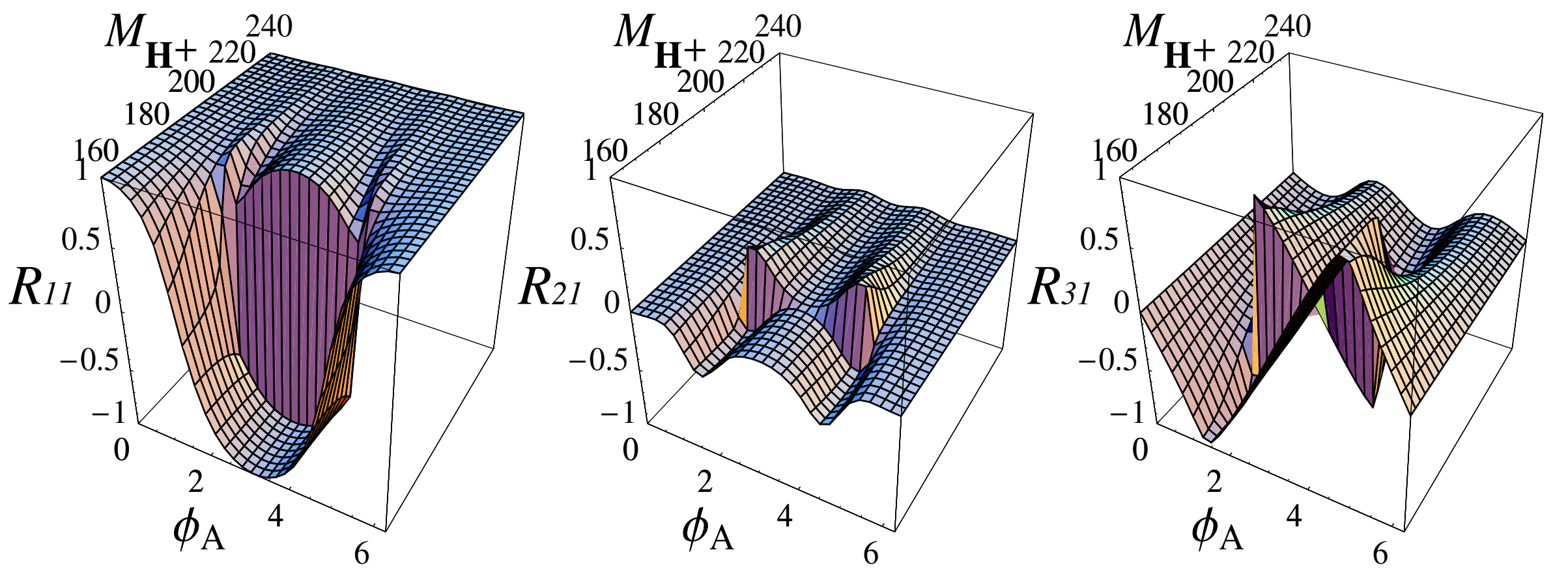

Fig. 3.24: The mixing matrix elements $R_{i1}$ ($i$ =1,2,3) as a two-dimensional functions of the mass $M_{H^\pm}$ (GeV) and the phase $\Phi_A$ (rad) calculated at the one-loop approximation for $\lambda_{1,...7}$ parameters of the MSSM two-doublet potential, CPX$_{500}$ scenario. With the discontinuities of $R_{i1}$ at $M_{H^\pm}$ =184 GeV and the discontinuites of $R_{i2}$, which are also introduced at the same charged scalar mass, the eigenvector basis is left-handed at any $\{\Phi_A, M_{H^\pm}\}$.

The effective trilinear and quartic couplings of physical Higgs bosons $H_1$, $H_2$ and $H_3$, Eq. (2.35) (i.e. their mass term $M^2_{ij}H_iH_j$ in the two-doublet potential is diagonal in the local minimum) can be written in a compact form, see [131]. For example, the trilinear couplings

$$\mathcal{L}_{3H} \;=\; v \sum_{i\geq j\geq k=1}^{3} g_{H_iH_jH_k} \frac{1}{N_S^{ijk}} H_iH_jH_k \;+\; v\sum_{i=1}^{3} g_{H_iH^+H^-}\, H_iH^+H^- \tag{3.38}$$

where $N_S$ are the combinatorial factors and

$$g_{H_iH_jH_k} \;=\; \sum_{\alpha\geq\beta\geq\gamma=1}^{3} \{R_{\alpha i}R_{\beta j}R_{\gamma k}\}\; g_{\alpha\beta\gamma}\,, \qquad g_{H_iH^+H^-} = \sum_{\alpha=1}^{3} R_{\alpha i}\, g_{\alpha H^+H^-}\,. \tag{3.39}$$

where curly brackets denote the symmetrization in the $i, j, k$ indices. Couplings $g_{\alpha\beta\gamma}$ and $g_{\alpha H^+H^-}$ are an intermediate expressions defined in the unphysical basis. Our mixing matrix $R = R_2^T\, R_3^T$ is specified by Eq. (2.35), $(h, H, A)^T = \|R_{ij}\|\,(H_1, H_2, H_3)^T$. The matrix elements $R_{ij}$ are defined in the orthonormal basis for eigenvectors, $R_{ij} = R'_{ij}/n_j$ :

$$R'_{11} = ((M_H^2 - M_{H_1}^2)(M_A^2 - M_{H_1}^2) - M'^4_{23}),\; R'_{21} = M'^2_{13}M'^2_{23},\; R'_{31} = -M'^2_{13}(M_H^2 - M_{H_1}^2), \tag{3.40}$$
$$R'_{12} = -M'^2_{13}M'^2_{23},\; R'_{22} = -((M_h^2 - M_{H_2}^2)(M_A^2 - M_{H_2}^2) - M'^4_{13}),\; R'_{32} = M'^2_{23}(M_h^2 - M_{H_2}^2), \tag{3.41}$$
$$R'_{13} = -M'^2_{13}(M_H^2 - M_{H_3}^2),\; R'_{23} = -M'^2_{23}(M_h^2 - M_{H_3}^2),\; R'_{33} = (M_h^2 - M_{H_3}^2)(M_H^2 - M_{H_3}^2), \tag{3.42}$$

where $n_i = k_i\sqrt{(R'^2_{1i} + R'^2_{2i} + R'^2_{3i})}$ and the sign factors $k_i$ are introduced to ensure definitely chosen (left-handed, $\mathtt{det}\|R_{ij}\|$ =1) orientation of the eigenvector basis at any phase $\Phi_A = \mathtt{arg}(\mu A_t) = \mathtt{arg}(\mu A_b)$ and charged scalar mass $M_{H^\pm}$, together with matching to the states $h$, $H$ and $A$ of the CP conserving limit.

For the two-Higgs doublet potential the off-diagonal mass matrix elements $M'^2_{13}$ and $M'^2_{23}$, Eq. (2.30), depend on the imaginary parts of $\lambda_5$, $\lambda_6$ and $\lambda_7$, see Eqs. (2.33–2.34) [256]. In the framework of MSSM the $\lambda_i$, i=1,...7 are calculated [50,110,256,257] by means of the effective potential method, taking into account the one-loop triangle and box squark insertions to the quartic vertices of the two-doublet potential. If the universal phase of complex parameters $\mathtt{arg}(\mu A_t) = \mathtt{arg}(\mu A_b)$ is introduced, the phases of $\lambda_{5,6,7}$

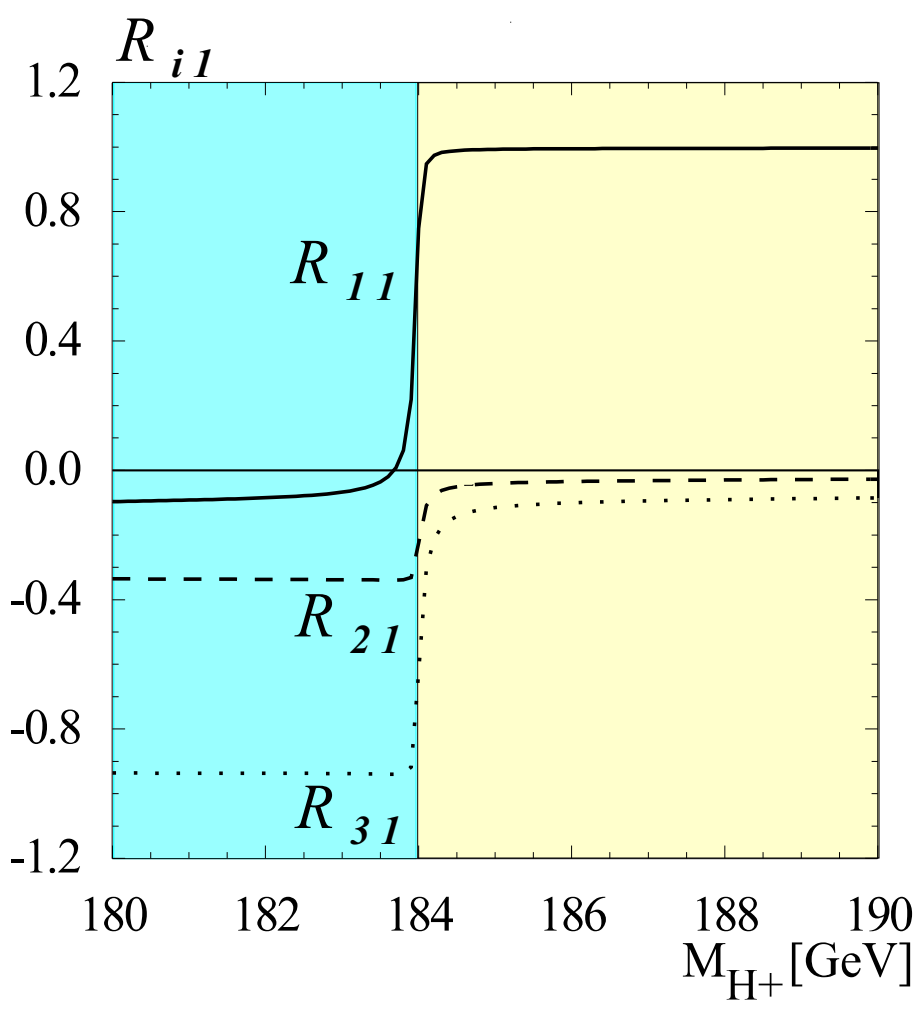

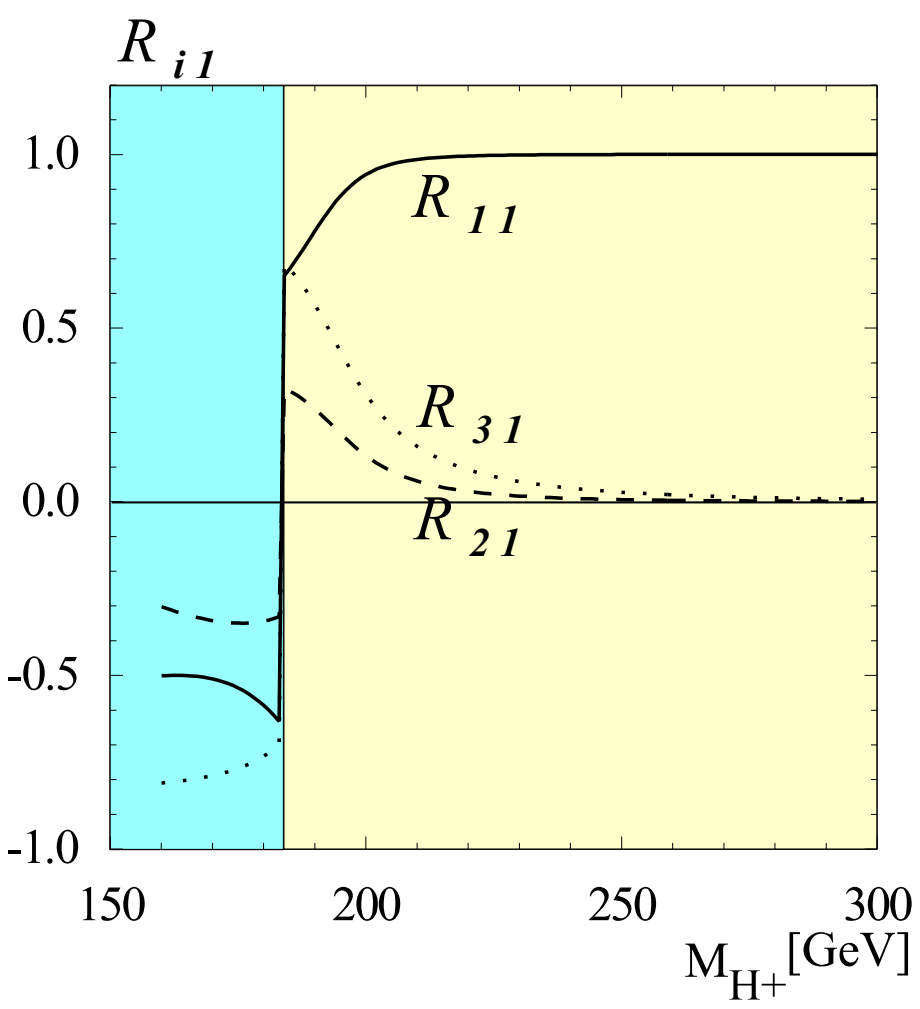

Fig. 3.25: The mixing matrix elements $R_{i1}$ as a function of the charged scalar mass $M_{H^\pm}$ (GeV) at the phases $\Phi_A = \pi/2$ (left plot) and $\Phi_A = 7\pi/12$ (right plot) calculated in the one-loop approximation for $\lambda_{1,...7}$ parameters of the MSSM two-doublet potential, CPX$_{500}$ scenario. Both pictures are the superimposed cross sections of the two-dimensional plots in Fig. 3.24 by a plane orthogonal to the $\Phi_A$ axis. The discontinuities of $R_{i1}$ at $M_{H^\pm}$ =184 GeV are introduced at the phase $\pi/2 < \Phi_A < 3\pi/2$. Then the eigenvector basis is left-handed at any $\{\Phi_A, M_{H^\pm}\}$

.

respect the constraint $\mathtt{arg}(\lambda_5) = 2\,\mathtt{arg}(\lambda_{6,7})$. In this way the off-diagonal mass matrix elements $M'^2_{13}$ and $M'^2_{23}$ can be fixed at a given point of the MSSM parameter space $\{\Phi_A, \mathtt{tan}\beta, M_{H^\pm}, \mu, A_{t,b}, M_{SUSY}\}$.

The squared masses of Higgs bosons are ($M^2_{H_1} \leq M^2_{H_2} \leq M^2_{H_3}$)

$$M^2_{H_1} = p\,\cos\left(\frac{\Theta + 2\pi}{3}\right) - \frac{a_2}{3}\,, M^2_{H_2} = p\,\cos\left(\frac{\Theta - 2\pi}{3}\right) - \frac{a_2}{3}\,, M^2_{H_3} = p\,\cos\left(\frac{\Theta}{3}\right) - \frac{a_2}{3}, \quad (3.43)$$

where $p = 2\,\sqrt{(-q)}$, $\cos\Theta = r/\sqrt{(-q^3)}$, $r = (9a_1a_2 - 27a_0 - 2a_2^3)/54$, $q = (3a_1 - a_2^2)/9$, $a_1 = M_h^2M_H^2 + M_h^2M_A^2 + M_H^2M_A^2 - M'^4_{13} - M'^4_{23}$, $a_2 = -M_h^2 - M_H^2 - M_A^2$, $a_0 = M'^4_{13}M_H^2 + M'^4_{23}M_h^2 - M_h^2M_H^2M_A^2$. Squared masses of Higgs bosons $M^2_{H_i}$ are the roots of the cubic equation $(M^2_{H_i})^3 + a_2(M^2_{H_i})^2 + a_1M^2_{H_i} + a_0 = 0$ which can be rewritten in the equivalent form

$$(M_h^2 - M^2_{H_i})[(M_H^2 - M^2_{H_i})(M_A^2 - M^2_{H_i}) - M'^4_{23}] - M'^4_{13}(M_H^2 - M^2_{H_i}) = 0. \quad (3.44)$$

If $i = 1$ then Eq. (3.44) takes the form $(M_h^2 - M^2_{H_1})R'_{11} + M'^2_{13}R'_{31} = 0$, so if $M'^2_{13} = 0$ then either $R'_{11} = 0$ or $M_{H_1} = M_h$ precisely. The special case of degenerate masses $M_{H_1} = M_{H_2}$ takes place when $\Theta$ =0, see Eq. (3.43). For CPX scenario at $M_{SUSY}$ =500 GeV, $\mathtt{tan}\beta$ =5 (denoted by CPX$_{500}$ everywhere in the following) $\Theta$ =0 at $M_{H^\pm}$ = 184 GeV. The case $M'^2_{13}$ =0, when the mixing matrix elements $R'_{31}$ and $R'_{21}$ change their sign crossing zero, is distinguished in mixing scenarios. Such property of the off-diagonal mass matrix is inherent to the MSSM. In other nonstandard models it may not take place. In the CPX$_{500}$ scenario of the MSSM $M'^2_{13}(\Phi_A)$ =0 at the phase very close to $\pi/2$. For example, the normalized matrix element $R_{11}(\Phi_A)$, see Fig. 3.24 and Fig. 3.25, decreases with increasing $\Phi_A$, but reaches zero $R_{11}(\Phi_A)$ =0 only if $M_{H_\pm}$ < 184 GeV. The cubic equation (3.44) for eigenvalues is respected in this case because $R'_{11} = 0$. The negative sign of $R_{11}(\Phi_A)$ (i.e. $k_1$ =-1 above) must be taken to keep proper orientation of the eigenvector basis (always left-handed). In the case $M_{H^\pm} > 184$ GeV $R'_{11}$ does not reach zero, remaining always positive. The cubic equation (3.44) for eigenvalues is then respected because $M_h = M_{H_1}$. No change of sign for $R_{11}$ is possible here. Different parametric behaviour of $R_{ij}(\Phi_A)$ at $M_{H^\pm}$ less or greater than 184 GeV leads to discontinuities of matrix elements $R_{i1}$ and $R_{i2}$ as a functions of $M_{H^\pm}$ in the vicinity of $M_{H^\pm}$=184 GeV (Fig. 3.24,3.25). With the two-loop calculation of $\lambda_i$ the situation remains qualitatively the same, but discontinuites of $R_{i1}$ and $R_{i2}$ are shifted to lower charged Higgs boson mass $M_{H^\pm}$ =162 GeV. Discontinuites are not a special feature of our approach. They take place in the `CPsuperH` [131] and `FeynHiggs` [60] packages, which are using different sign conventions for the basis, so other pattern of discontinuites exists there. Our convention is implemented in `CompHEP` [266].

The effective trilinear and quartic Higgs boson self-couplings of the general two-Higgs-doublet model can be written down in two equivalent representations. First one uses $\lambda_i$ parameters and the second representation expresses the effective couplings by means of Higgs boson masses in the CP conserving limit $\Phi_A$ =0. The effective charged Higgs boson triple couplings, see Eq. (3.38), in the $\lambda_i$ representation and the mass representation can be found in [256]. In the MSSM CPX$_{500}$ scenario the coupling $g_{H^+H^-H_1}$ goes through zero at $\Phi_A \sim \pi/2$, see Fig. 3.26, with the overall variation range approximately from -100 GeV to 100 GeV. The $\Theta$ parameter (3.43) is close to maximum in the vicinity of weak self-interaction, when $M_{H_1} \sim M_{H_2}$. Representations of quartic self-couplings in the $\lambda_i$ basis are more complicated. For example (using compact notation $\sin\alpha \equiv s_\alpha$ and so on)

$$g_{H_1H_1H_1H_1} = \sum_{i=1}^{4} l_i\,\lambda_i + \sum_{i=5}^{7} R_i\,\mathtt{Re}\lambda_i + \sum_{i=5}^{7} I_i\,\mathtt{Im}\lambda_i\,, \tag{3.45}$$

$$\begin{aligned}
l_1 &= -3(1+(-R_{11}^2+R_{21}^2)c_{2\alpha} - R_{31}^2 c_{2\beta} - 2R_{11}R_{21}s_{2\alpha})^2/4,\\
l_2 &= -3(1+(R_{11}^2-R_{21}^2)c_{2\alpha} + R_{31}^2 c_{2\beta} + 2R_{11}R_{21}s_{2\alpha})^2/4,\\
l_3 &= l_4 = 3(-(R_{11}^2+R_{21}^2)^2 - 4(R_{11}^2+R_{21}^2)R_{31}^2 - R_{31}^4 + (R_{11}^4 - 6R_{11}^2R_{21}^2 + R_{21}^4)c_{4\alpha} + R_{31}^4 c_{4\beta}\\
&\quad +4R_{31}^2 c_{2\beta}((R_{11}^2-R_{21}^2)c_{2\alpha} + 2R_{11}R_{21}s_{2\alpha}) + 4R_{11}(R_{11}^2-R_{21}^2)R_{21}s_{4\alpha})/4,
\end{aligned} \tag{3.46}$$

$$\begin{aligned}
R_5 &= 3(-(R_{11}^2+R_{21}^2)^2 + 4(R_{11}^2+R_{21}^2)R_{31}^2 - R_{31}^4 + (R_{11}^4 - 6R_{11}^2R_{21}^2 + R_{21}^4)c_{4\alpha}\\
&\quad +R_{31}^4 c_{4\beta} + 4R_{11}(R_{11}^2-R_{21}^2)R_{21}s_{4\alpha} + 4R_{31}^2 c_{2\alpha}((-R_{11}^2+R_{21}^2)c_{2\beta} + 4R_{11}R_{21}s_{2\beta})\\
&\quad -8R_{31}^2 s_{2\alpha}(R_{11}R_{21}c_{2\beta} + (R_{11}^2-R_{21}^2)s_{2\beta}))/4,\\
R_6 &= (-6R_{31}^2 c_{2\beta}(-2R_{11}R_{21}c_{2\alpha} + (R_{11}^2-R_{21}^2)s_{2\alpha}) - 6(-1+(R_{11}^2-R_{21}^2)c_{2\alpha}\\
&\quad +2R_{11}R_{21}s_{2\alpha})(-2R_{11}R_{21}c_{2\alpha} + (R_{11}^2-R_{21}^2)s_{2\alpha})\\
&\quad +6R_{31}^2(1+(-R_{11}^2+R_{21}^2)c_{2\alpha} - 2R_{11}R_{21}s_{2\alpha})s_{2\beta} - 3R_{31}^4 s_{4\beta})/2,\\
R_7 &= (6R_{31}^2 c_{2\beta}(-2R_{11}R_{21}c_{2\alpha} + (R_{11}^2-R_{21}^2)s_{2\alpha}) + 6(1+(R_{11}^2-R_{21}^2)c_{2\alpha}\\
&\quad +2R_{11}R_{21}s_{2\alpha})(-2R_{11}R_{21}c_{2\alpha} + (R_{11}^2-R_{21}^2)s_{2\alpha})\\
&\quad +6R_{31}^2(1+(R_{11}^2-R_{21}^2)c_{2\alpha} + 2R_{11}R_{21}s_{2\alpha})s_{2\beta} + 3R_{31}^4 s_{4\beta})/2,
\end{aligned} \tag{3.47}$$

$$\begin{aligned}
I_5 &= -6R_{31}(R_{21}c_{\alpha-\beta} - R_{11}s_{\alpha-\beta})(-2R_{11}R_{21}c_{2\alpha} + (R_{11}^2-R_{21}^2)s_{2\alpha} + R_{31}^2 s_{2\beta}),\\
I_6 &= 6R_{31}(1+(-R_{11}^2+R_{21}^2)c_{2\alpha} - R_{31}^2 c_{2\beta} - 2R_{11}R_{21}s_{2\alpha})(R_{21}c_{\alpha-\beta} - R_{11}s_{\alpha-\beta}),\\
I_7 &= 6R_{31}(1+(R_{11}^2-R_{21}^2)c_{2\alpha} + R_{31}^2 c_{2\beta} + 2R_{11}R_{21}s_{2\alpha})(R_{21}c_{\alpha-\beta} - R_{11}s_{\alpha-\beta}).
\end{aligned} \tag{3.48}$$

Various physical self-couplings in the CPX$_{500}$ scenario are shown in Fig. 3.26-3.28. Note that in the CPX$_{500}$ large contributions to them come from the terms with $\lambda_6$ ~0.5. The $\lambda_6$ parameter has the Yukawa coupling $h_{top}^4$ in front of the main power term $\mu^3 A_t/M_{SUSY}^4$.

Our calculations demonstrate that the structure of Higgs boson self-interactions for the two-doublet model with complex parameters in the CP violating potential is extremely strongly sensitive to radiative corrections and phases of effective parameters. Some detailed numerical evaluations illustrating this sensitivity were performed in the framework of the CPX scenario at the SUSY scale $M_{SUSY}$ =500

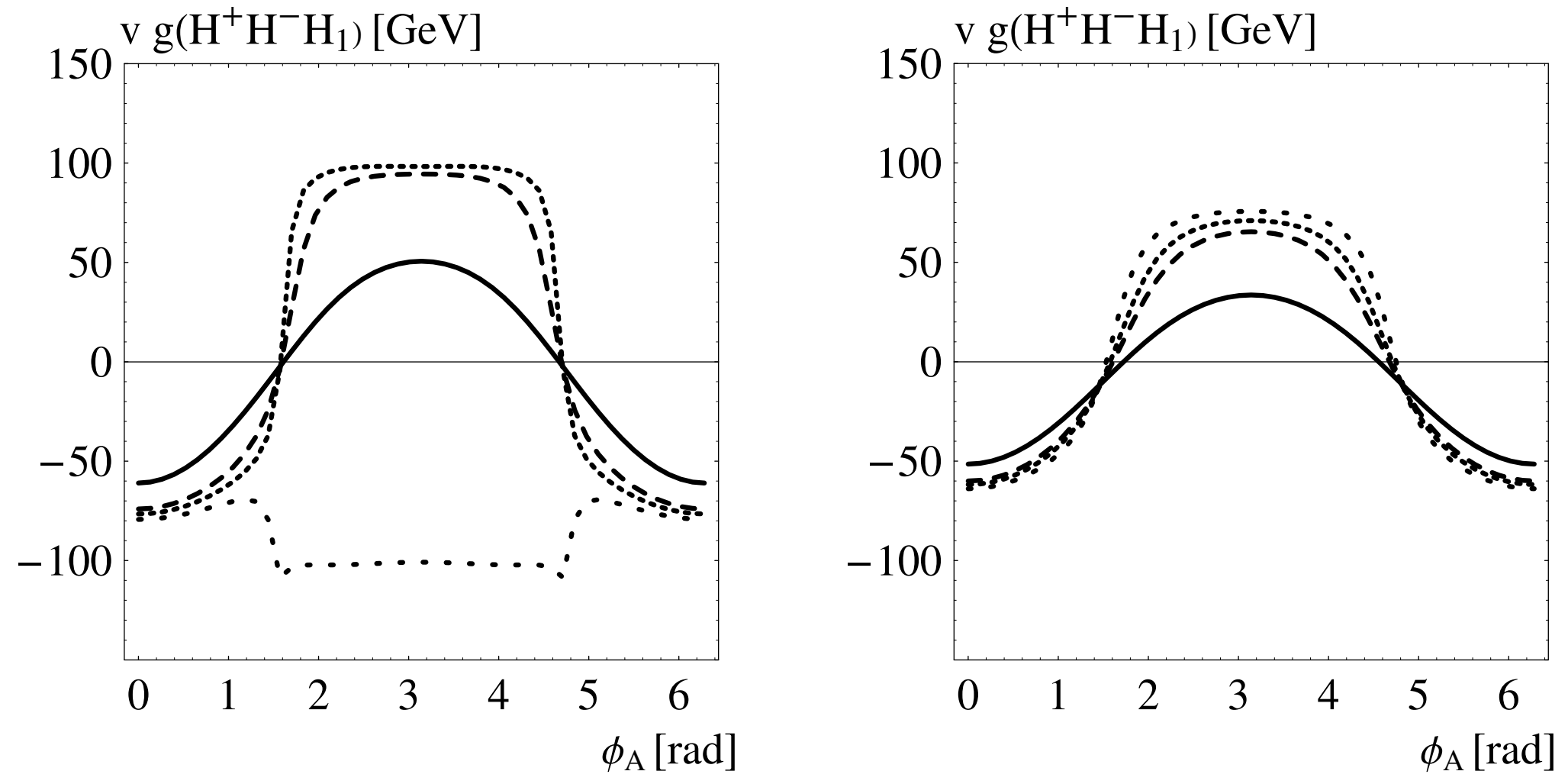

Fig. 3.26: The triple Higgs boson interaction vertex $v \cdot g_{H^+\, H^-\, H_1}$ (GeV) vs the phase $\texttt{arg}(\mu A_{t,b})$ (left figure for 1-loop approximation and the right figure with additional leading QCD Yukawa corrections to $\lambda_i$ included) at parameter values $M_{SUSY} = 500$ GeV, tg$\beta$ =5, $A_{t,b}$ =1000 GeV, $\mu = 2000$ GeV. Solid line – $M_{H^\pm} = 300$ GeV, long dashed line – $M_{H^\pm} = 200$ GeV, short dashed line – $M_{H^\pm} = 190$ GeV, rare dotted line – $M_{H^\pm} = 180$ GeV.

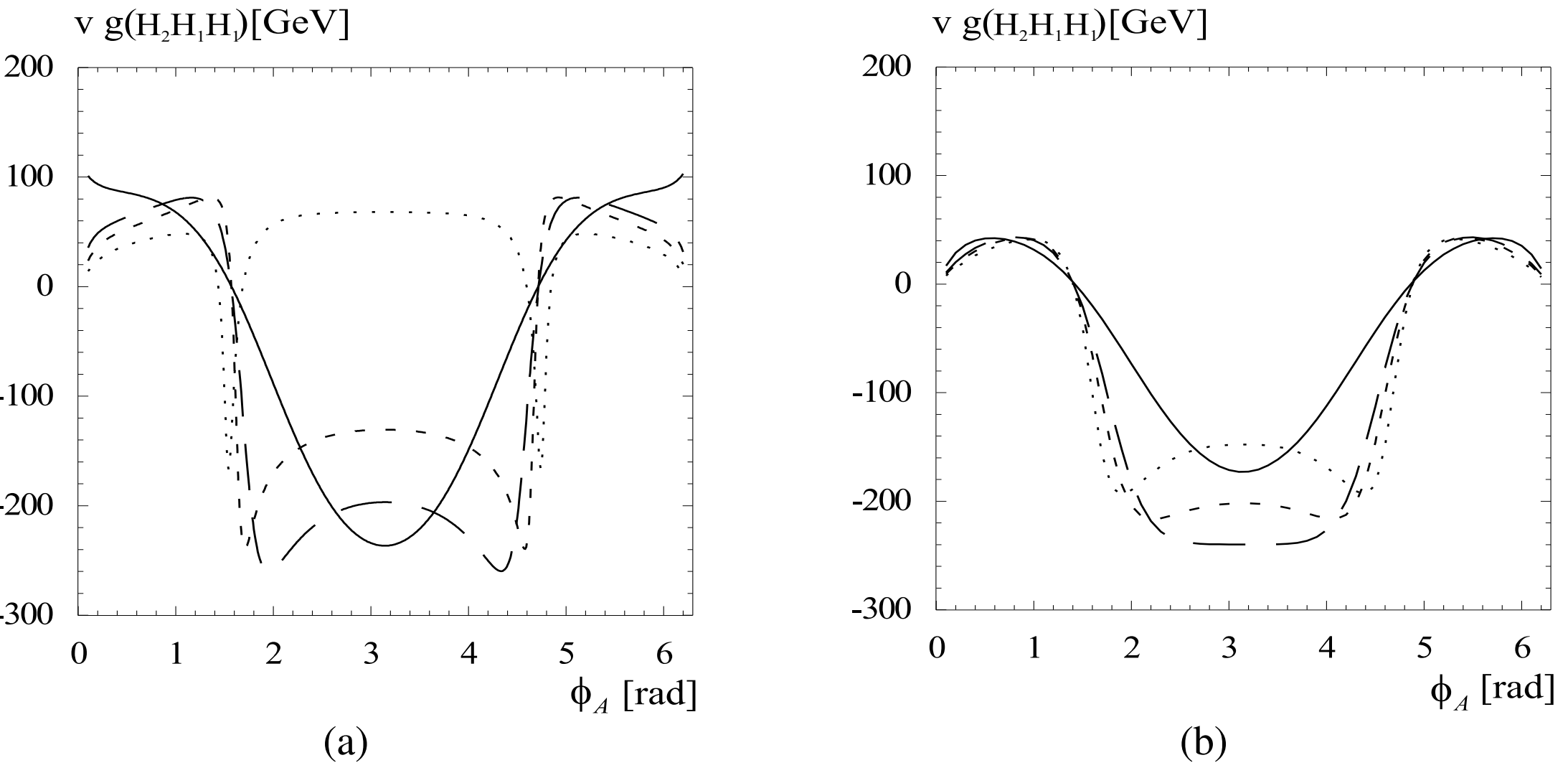

Fig. 3.27: The triple Higgs boson interaction vertex $v \cdot g_{H_1 H_1 H_2}$ (GeV) vs the phase $\texttt{arg}(\mu A_{t,b})$ at parameter values $M_{SUSY} = 500$ GeV, tg$\beta = 5$, $A_{t,b} = 1000$ GeV, $\mu = 2000$ GeV. Solid line – $M_{H^\pm} = 300$ GeV, long dashed line – $M_{H^\pm} = 200$ GeV, short dashed line – $M_{H^\pm} = 190$ GeV, dotted line – $M_{H^\pm} = 180$ GeV. (a) – the case of effective one-loop potential, (b) – leading two-loop corrections included.

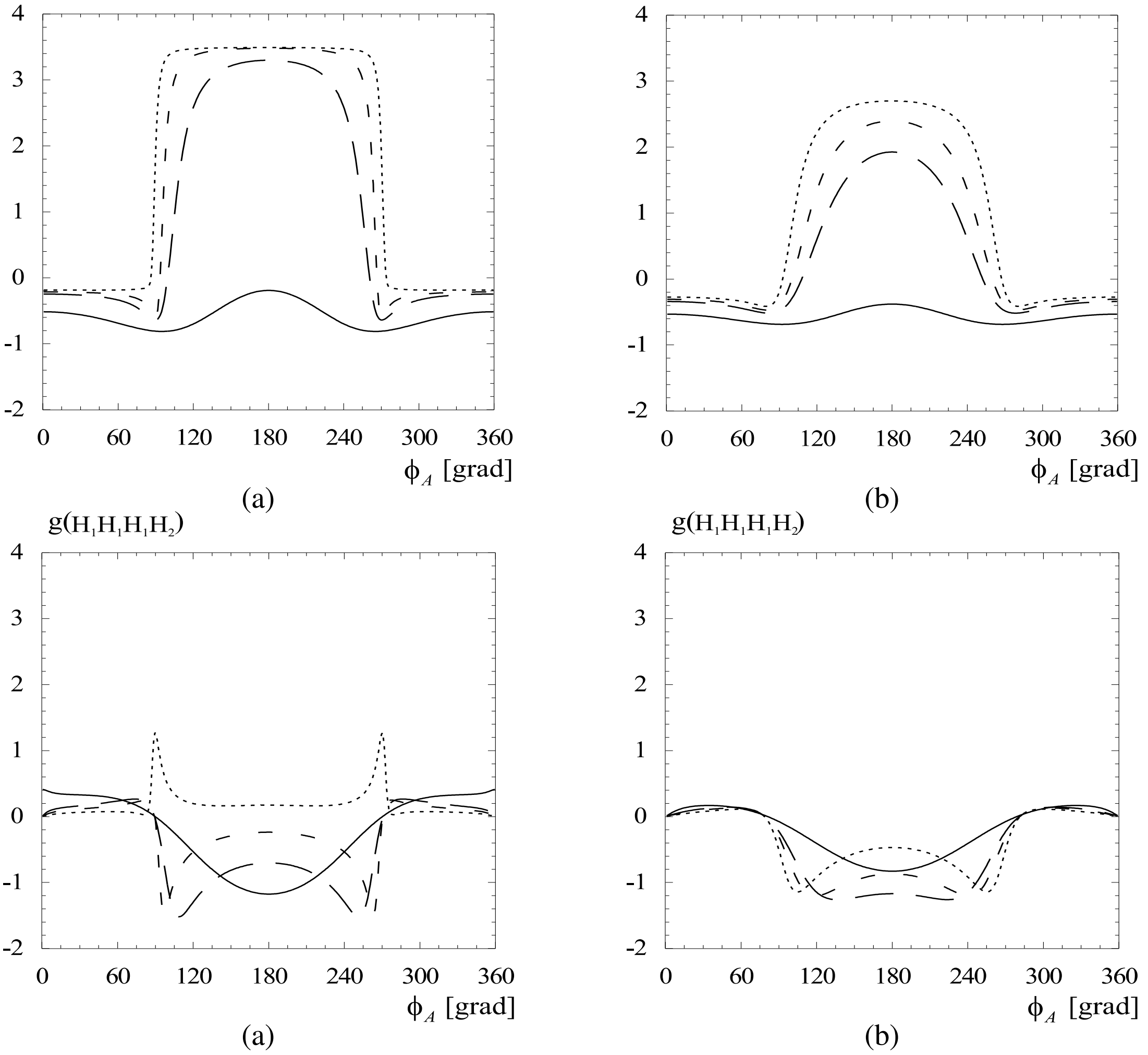

Fig. 3.28: The quartic interaction vertices $g_{H_1 H_1 H_1 H_i}$, $i$ =1,2, *vs* the phase $\mathtt{arg}(\mu A_{t,b})$ at parameter values $M_{SUSY}$ = 500 GeV, $\mathtt{tg}\beta$ = 5, $A_{t,b}$ = 1000 GeV, $\mu$ = 2000 GeV. Solid line – $M_{H^\pm}$ = 300 GeV, long dashed line – $M_{H^\pm}$ = 200 GeV, short dashed line – $M_{H^\pm}$ = 190 GeV, dotted line – $M_{H^\pm}$ = 180 GeV. (a) – effective one-loop approximation, (b) – with leading two-loop corrections included.

GeV and $\mathtt{tan}\beta$ =5 ($\text{CPX}_{500}$). A lot of self-couplings of physical scalars are very small at the phase $\Phi_A = \mathtt{arg}(\mu A_{t,b}) \sim \pi/2$, which is related to vanishing off-diagonal Higgs boson mass matrix element $M'^2_{13}(\Phi_A)$, and mass degeneracy of the states $h$,$H_1$ and $H_2$, that takes place in the vicinity of $\Theta$ =0, see Eq. (3.43). The availability of zero for the mass matrix element $M'^2_{13}(\Phi_A)$ is connected with the relation between phases $\mathtt{arg}(\lambda_5)$ =2 $\mathtt{arg}(\lambda_{6,7})$, inherent to the MSSM. In other representations of the THDM or other MSSM mixing scenarios the situation may be different. Mass degeneracy of the states $H_1$ and $H_2$ takes place in the MSSM, $\text{CPX}_{500}$ scenario, for the charged scalar mass $M_{H^\pm}$ =184 GeV at the one-loop approximation for $\lambda_i$. We point out an interesting property of the $\text{CPX}_{500}$ scenario, namely, definite (always left-handed, $\mathtt{det}\|R_{ij}\|$ =1) orientation of the eigenvector basis for scalars $H_{1,2,3}$ is respected in the $(\Phi_A, M_{H^\pm})$ parameter space (together with the mass ordering and matching to the ($h$,$H$,$A$) states in the CP-conserving limit), if the discontinuity of the matrix elements $R_{ij}(\Phi_A, M_{H^\pm})$ at $M_{H^\pm}$ =184 GeV is introduced (see Fig. 3.24 and Fig. 3.25). The two-dimensional functions $R_{ij}(\Phi_A, M_{H^\pm})$, if taken continious, define different orientations of the eigenvector basis for $H_{1,2,3}$ states inside the three intervals of phase variation. A discontinuity of the mixing matrix elements $R_{ij}$ leads to a discontinuity in the couplings, see for example Eq. (3.39), where terms linear in $R_{ij}$ appear. Such property could be relevant for systems that evolve in the phase and charged scalar mass, related to the phase transitions in cosmological models. Toy model with phase transitions can be found in [267]. Within the perturbation theory the discontinuites do not show up in the amplitudes (with the sign compensation in the product $R_{i1}\,R_{j1}$ for each $H_1$ propagator), however, this could be not the case for the nonperturbative insertions to diagrams. Specific features of self-interactions in the case of complex $\lambda_{5,6,7}$ in the THDM technically appear as a consequence of the eigenvalue and eigenvector problems for the $3{\times}3$ neutral Higgs bosons mixing matrix, dependent on several parameters. In such schemes very small radiative corrections to the input parameters may lead to large changes of a physical observables evaluated.

### 3.7 Production of neutral Higgs bosons through $b$-quark fusion in CP-violating SUSY scenarios

*Francesca Borzumati and Jae Sik Lee*

The $b\bar{b}$ fusion process can be one of the leading production channels of the two heaviest neutral Higgs bosons at the Tevatron and at the LHC for values of $\tan\beta$ ranging from intermediate up to large or very large. In scenarios with large CP-violating mixing among the neutral Higgs states [48–58], this channel can be relevant also for the lightest neutral Higgs boson. Moreover, the vertex and $m_b$ corrections induced by supersymmetric particles [124–128, 169–176] can affect substantially the size of the production cross sections of all three neutral Higgs bosons [77].

To illustrate these effects, we consider the CPX scenario defined in Eq. (3.13) of Section 3.1, that is in general used to highlight CP-violating effects in the Higgs sector. We choose the two free parameters $M_{\rm SUSY}$ and $\tan\beta$ to be: $M_{\rm SUSY} = 0.5\,$TeV and $\tan\beta = 10$. (We shall comment later on the rational for this choice of $\tan\beta$.) Moreover, after we fix the phases $\Phi_A \equiv {\rm Arg}(A_t\mu) = {\rm Arg}(A_b\mu)$ and $\Phi_3 \equiv {\rm Arg}(M_3\mu)$, the charged Higgs-boson mass is solved to give $M_{H_1} = 115\,$GeV. Our numerical analyses make use of the program `CPsuperH` [131].

We show in Fig. 3.29 masses and widths of the three neutral Higgs bosons obtained in such a scenario, as functions of $\Phi_A$ for three different values of $\Phi_3$: $0^\circ, 90^\circ, 180^\circ$. We observe that all three neutral Higgs bosons are relatively light and widths reaching few GeV for $\Phi_3 = 180^\circ$.

The effective Lagrangian for the interaction of these neutral Higgs bosons to $b$ quarks can be written as

$$\mathcal{L} \;=\; -\frac{m_b}{v}\,\bar{b}\left(g^S_{H_i\bar{b}b} + i g^P_{H_i\bar{b}b}\gamma_5\right) b\, H_i\,, \tag{3.49}$$

where the couplings $g^{S,P}_{H_i\bar{b}b}$ are $g^{S,P}_{H_i\bar{b}b} = \sum_\alpha O_{\alpha i}\, g^{S,P}_\alpha$ with $\alpha = (\phi_1, \phi_2, a)$ and the $3{\times}3$ matrix $O$ describing the CP-violating neutral Higgs-boson mixing. After including vertex and $m_b$ corrections induced by supersymmetric particles, the couplings $g^{S,P}_\alpha$ are given by [152]

$$\begin{aligned}
g^S_{\phi_1} &= \frac{1}{\cos\beta}\,{\rm Re\,e}\left(\frac{1}{R_b}\right), & g^P_{\phi_1} &= \frac{\tan\beta}{\cos\beta}\,{\rm Im\,m}\left(\frac{\kappa_b}{R_b}\right),\\
g^S_{\phi_2} &= \frac{1}{\cos\beta}\,{\rm Re\,e}\left(\frac{\kappa_b}{R_b}\right), & g^P_{\phi_2} &= -\frac{1}{\cos\beta}\,{\rm Im\,m}\left(\frac{\kappa_b}{R_b}\right),\\
g^S_{a} &= (\tan^2\beta + 1)\,{\rm Im\,m}\left(\frac{\kappa_b}{R_b}\right), & g^P_{a} &= -\,{\rm Re\,e}\left(\frac{\tan\beta - \kappa_b}{R_b}\right).
\end{aligned} \tag{3.50}$$

Here $R_b$ and $\kappa_b$ are defined through:

$$h_b = \frac{\sqrt{2}\,m_b}{v\cos\beta}\,\frac{1}{R_b} = \frac{\sqrt{2}\,m_b}{v\cos\beta}\,\frac{1}{1+\kappa_b\tan\beta}\,, \tag{3.51}$$

with $v \simeq 254\,$GeV. (Corrections not enhanced by $\tan\beta$ are also included in our numerical analysis.) In the above expression, the finite corrections to the $b$-quark mass are collected in $\kappa_b = \epsilon_g + \epsilon_H$ in which the contributions from the sbottom-gluino exchange diagram and those from the stop-Higgsino diagram,

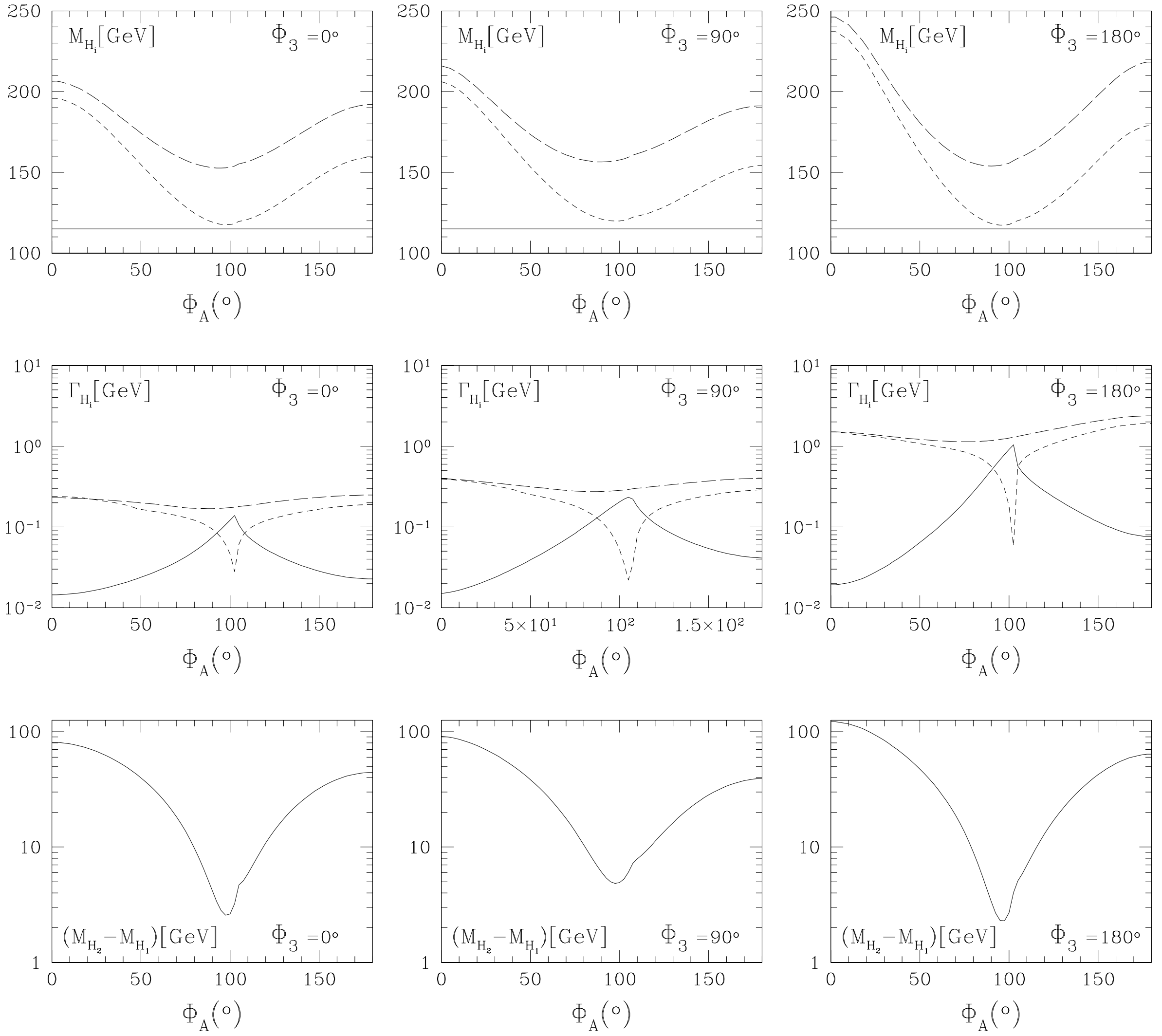

Fig. 3.29: The masses (top) and widths (middle) of the neutral Higgs bosons as functions of $\Phi_A$ for the spectrum in Eq. (3.13) with $M_{\rm SUSY} = 0.5$ TeV and $\tan\beta = 10$, for three different values of $\Phi_3$: 0° (left column), 90° (central column), and 180° (right column). The solid lines are for $H_1$, the dashed ones for $H_2$, and the long-dashed ones for $H_3$. The bottom frames are for the mass difference between $H_2$ and $H_1$.

respectively $\epsilon_g$ and $\epsilon_H$, are

$$\epsilon_g = \frac{2\alpha_s}{3\pi}\, M_3^*\mu^* I(m_{\tilde{b}_1}^2, m_{\tilde{b}_2}^2, |M_3|^2)\,, \qquad \epsilon_H = \frac{|h_t|^2}{16\pi^2}\, A_t^*\mu^* I(m_{\tilde{t}_1}^2, m_{\tilde{t}_2}^2, |\mu|^2)\,. \tag{3.52}$$

The one-loop function $I(a,b,c)$ is well known and can be found, for example, in Ref. [172].

Note that, in general, $\kappa_b$ is complex due to the CP phases of the combinations $M_3\mu$ and/or $A_t\mu$. In particular, the value $\Phi_3 \sim 180°$ plays an important role in the CPX scenario. While it does not affect considerably the masses of the three neutral Higgs bosons, and modifies their widths only by a factor 2-3, see Fig. 3.29, it can have consequences for the positivity of the lightest sbottom squark squared mass. Indeed, for $\Phi_3 \sim 180° \pm 30°$, the mass eigenstate $\tilde{b}_1$ is tachyonic at values of $\tan\beta$ ranging from intermediate to large, depending on the value of $\Phi_A$. See Ref. [77] for details. In this study, we choose $\tan\beta = 10$, at which all values of $\Phi_3$ and $\Phi_A$ are allowed.

For values of $\tan\beta$ such that $|\kappa_b|\tan\beta \sim 1$, with $\mathrm{Re\,e}(\kappa_b)\tan\beta \sim 1$ and/or $\mathrm{Im\,m}(\kappa_b)\tan\beta \sim 1$

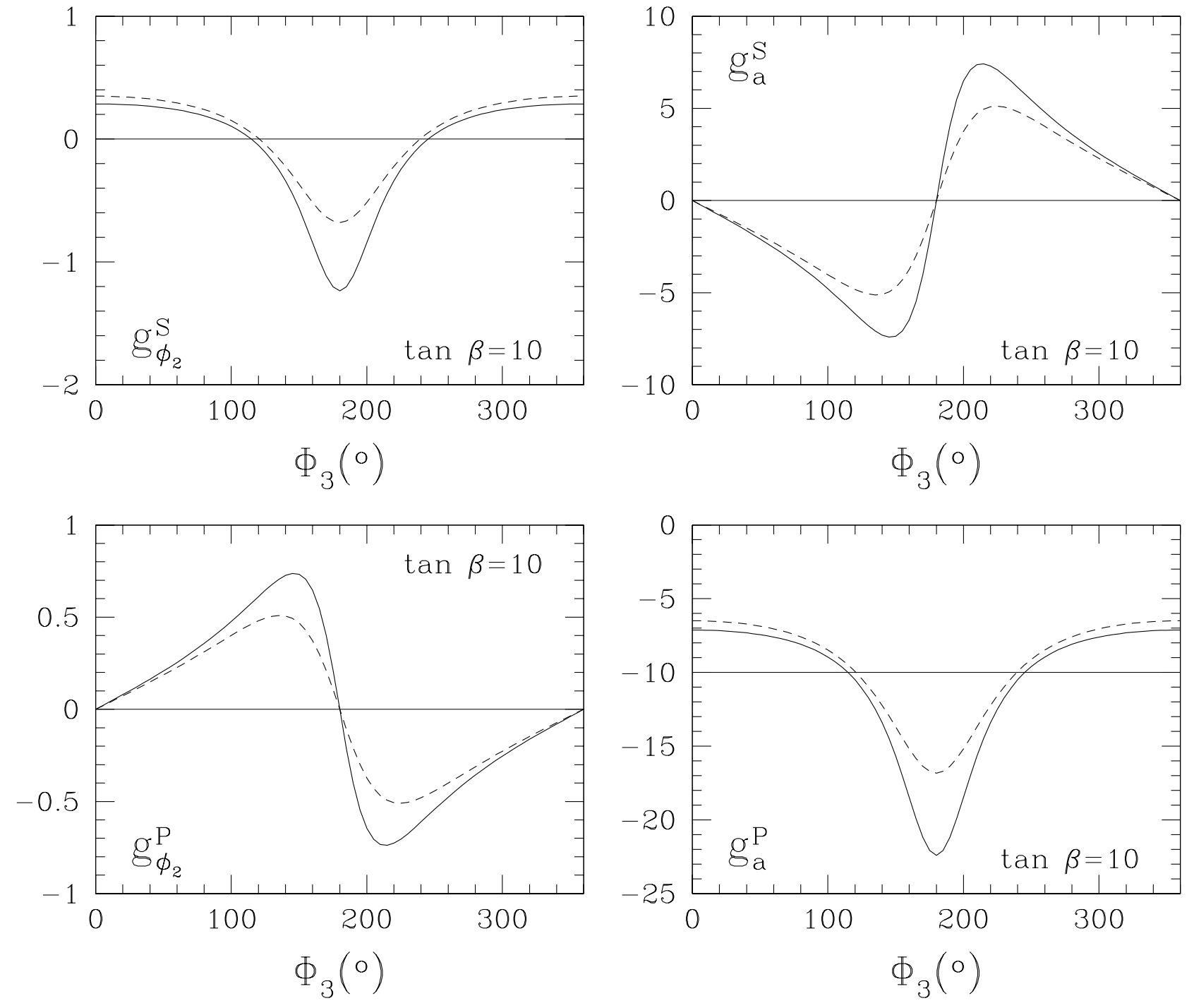

Fig. 3.30: Couplings $g_{\phi_2}^{S,P}$ and $g_a^{S,P}$ vs. $\Phi_3$, for the spectrum in Eq. (3.13) with $M_{\rm SUSY} = 0.5$ TeV and $\tan\beta = 10$. The dashed lines are for $\Phi_A = 0^{\rm o}$, the solid ones for $\Phi_A = 180^{\rm o}$. The horizontal lines indicate the values of the uncorrected couplings.

($\tan\beta = 10$ is one of those), the expressions for the couplings $g_\alpha^{S,P}$ reduce to

$$
\begin{aligned}
g_{\phi_1}^S &= \frac{\tan\beta}{|R_b|^2}\,[1 + {\rm Re\,e}(\kappa_b)\tan\beta]\,, & g_{\phi_1}^P &= \frac{\tan\beta}{|R_b|^2}\,[{\rm Im\,m}(\kappa_b)\tan\beta]\,,\\
g_{\phi_2}^S &= \frac{1}{|R_b|^2}\left[{\rm Re\,e}(\kappa_b)\tan\beta + |\kappa_b|^2\tan^2\beta\right], & g_{\phi_2}^P &= -\frac{1}{|R_b|^2}\,[{\rm Im\,m}(\kappa_b)\tan\beta]\,,\\
g_a^S &= \frac{\tan\beta}{|R_b|^2}\,[{\rm Im\,m}(\kappa_b)\tan\beta]\,, & g_a^P &= -\frac{\tan\beta}{|R_b|^2}\,[1 + {\rm Re\,e}(\kappa_b)\tan\beta]
\end{aligned}
\tag{3.53}
$$

If no threshold corrections are included, the only nonvanishing couplings are $g_{\phi_1}^S = 1/\cos\beta$ and $g_a^P = -\tan\beta$. The inclusion of these corrections affects these two couplings mainly through the factor ${\rm Re\,e}(1/R_b)$, which is a suppression or an enhancement factor, depending on the value of ${\rm Arg}(\kappa_b)$, and varies between $1/(1+|\kappa_b|\tan\beta)$ and $1/(1-|\kappa_b|\tan\beta)$. Note that the factor ${\rm Re\,e}(1/R_b)$ is larger than 1 for $\cos({\rm Arg}(\kappa_b)) \lesssim -|\kappa_b|\tan\beta$. The other four couplings are loop-induced. Among these, $g_{\phi_2}^S$ is the only one present if there are no CP-violating phases, and the couplings $g_a^S \approx g_{\phi_1}^P$ have an overall $\tan\beta$ enhancement factor compared to the couplings $g_{\phi_2}^{S,P}$. We show explicitly in Fig. 3.30 the couplings $g_{\phi_2}^{S,P}$ and $g_a^{S,P}$. The remaining two, in the same approximation of Eq. (3.53), are $g_{\phi_1}^S = -g_a^P$ and $g_{\phi_1}^P = g_a^S$, respectively. This figure shows clearly that the couplings $g_{\phi_2}^{S,P}$ and $g_a^{S,P}$ have a dependence on the CP phase $\Phi_A$ weaker than that on $\Phi_3$. This is because $|\epsilon_g|$ is about one order of magnitude larger than $|\epsilon_H|$ for the scenario under consideration.

We are now in position to discuss the production cross sections of the neutral Higgs bosons $H_i$ via $b$-quark fusion at hadron colliders. These cross sections can be expressed as:

$$
\sigma({\rm had_1 had_2} \to b\bar{b} \to H_i) = \sigma(b\bar{b} \to H_i)\int_{\tau_i}^1 {\rm d}x\,\left[\frac{\tau_i}{x}\,b_{\rm had_1}(x,Q)\,\bar{b}_{\rm had_2}\Big(\frac{\tau_i}{x},Q\Big) + (b \leftrightarrow \bar{b})\right]\,, \tag{3.54}
$$

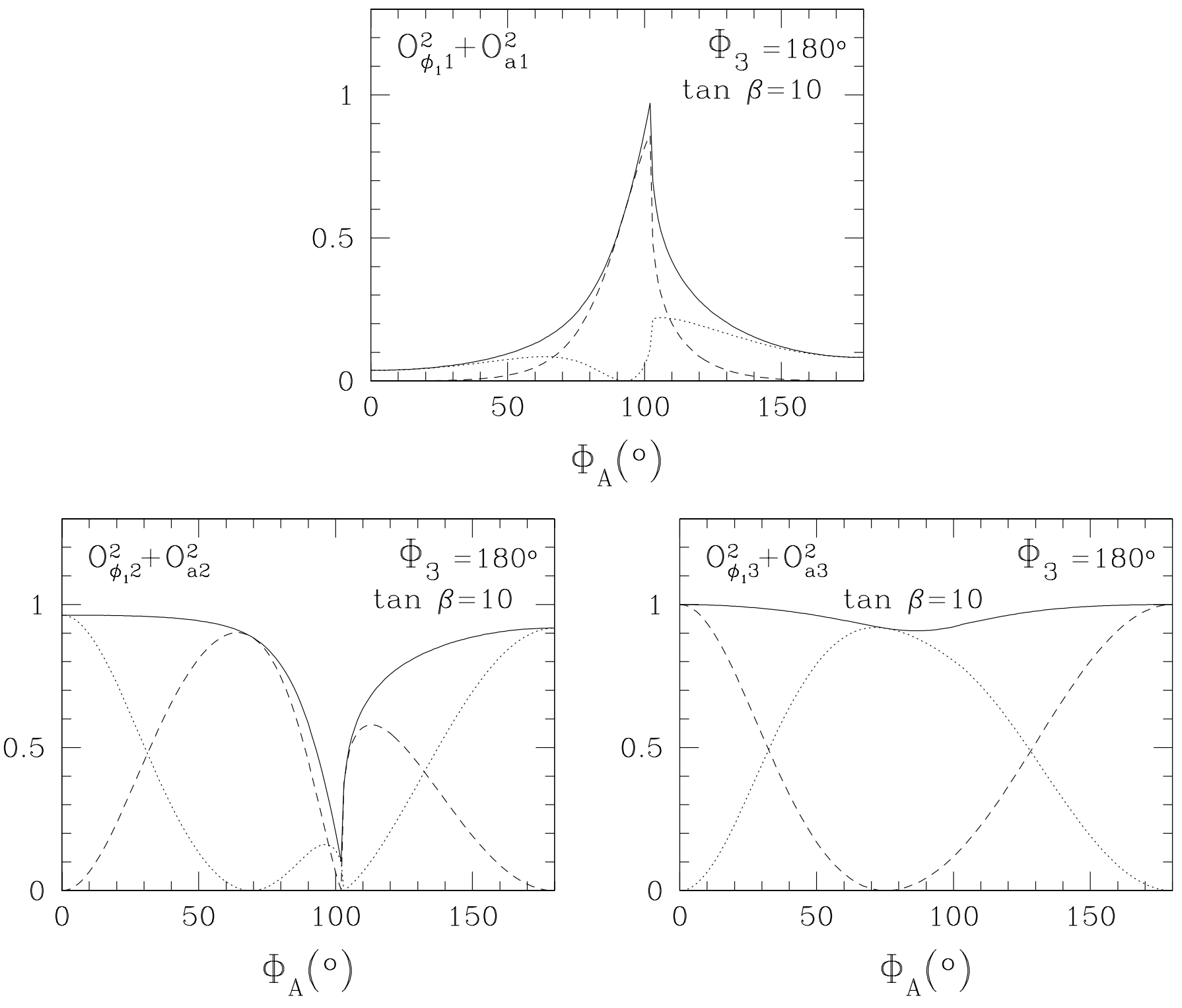

Fig. 3.31: The sums $O^2_{\phi_1 i} + O^2_{ai}$ vs. $\Phi_A$, for the spectrum of Eq. (3.13), with $M_{\rm SUSY} = 0.5$ TeV, $\tan\beta = 10$ and $\Phi_3 = 180^{\rm o}$. The dashed lines show $O^2_{ai}$, the dotted ones, $O^2_{\phi_1 i}$.

where $b_{{\rm had}_i}(x, Q)$ and $\bar{b}_{{\rm had}_i}(x, Q)$ are the $b$- and $\bar{b}$-quark distribution functions in the hadron ${\rm had}_i$, $Q$ is the factorization scale, and $\tau_i$ the Drell–Yan variable $\tau_i = M^2_{H_i}/s$, with $s$ the invariant hadron-collider energy squared. The partonic cross section is

$$\sigma(b\bar{b} \to H_i) = \frac{m_b^2}{v^2}\frac{\pi}{6M^2_{H_i}}\left[(g^S_{H_i\bar{b}b})^2 + (g^P_{H_i\bar{b}b})^2\right] \approx \frac{m_b^2}{v^2}\frac{\pi}{6M^2_{H_i}}\left[O^2_{\phi_1 i} + O^2_{ai}\right]\left[(g^S_a)^2 + (g^P_a)^2\right] . \qquad (3.55)$$

where the last approximation is valid when $g^{S,P}_{\phi_2}$ can be neglected. See discussion in Ref. [77]. The sums $O^2_{\phi_1 i} + O^2_{ai}$ are shown explicitly in Fig. 3.31, for the value $\Phi_3 = 180°$. This is sufficient since the dependence of these sums on $\Phi_3$ is rather weak, coming from the two-loop corrections to the Higgs potential. Notice that, for $\Phi_A \approx 100^{\rm o}$, $H_1$ is predominantly the CP-odd $a$ boson, whereas $H_2$ and $H_3$ are mainly $\phi_2$ and $\phi_1$, respectively.

The hadronic cross sections for the Tevatron ($\sqrt{s} = 1.96$ TeV) and the LHC ($\sqrt{s} = 14$ TeV) are obtained using the leading-order CTEQ6L [268] parton distribution functions, with the factorization scale fixed at $Q = M_{H_i}/4$. This has been suggested in most of the papers in Ref. [269–275] as the scale that minimizes the next-to-leading-order QCD corrections to these cross sections when no threshold corrections to $m_b$ are kept into account. Although this should be explicitly checked, we believe that the inclusion of these corrections should not affect substantially this result. We notice also that these supersymmetric threshold corrections capture the main part of all supersymmetric corrections to the production cross sections of neutral Higgs bosons through $b$-quark fusion. Other corrections, with a nontrivial dependence on the momenta of the $H_i$ bosons are of decoupling nature, and therefore subleading.

The cross sections are shown in Fig. 3.32 vs. $\Phi_A$, for two values of $\Phi_3$: $0^{\rm o}$ (dashed lines) and $180^{\rm o}$ (solid lines). These two curves delimit all cross sections obtained for all values of $\Phi_3$ between $0°$ and $180°$. We observe that these cross section can deviate substantially from those obtained in CP conserving scenarios, thanks to the nontrivial role played by the threshold corrections to $m_b$ and the CP-violating mixing effects. The largest deviations in the case of $H_1$ and $H_2$ are around $\Phi_A = 100^\circ$, with a large enhancement for the production cross section of $H_1$ and a large suppression for that of $H_2$. The former is due to the fact that the component of the field $a$ in $H_1$ around these values of $\Phi_A$ is large, where it is depleted by the same large amount in $H_2$. The cross section for $H_3$ can also deviate substantially from that in CP conserving scenarios, but this deviation is roughly independent of $\Phi_A$.

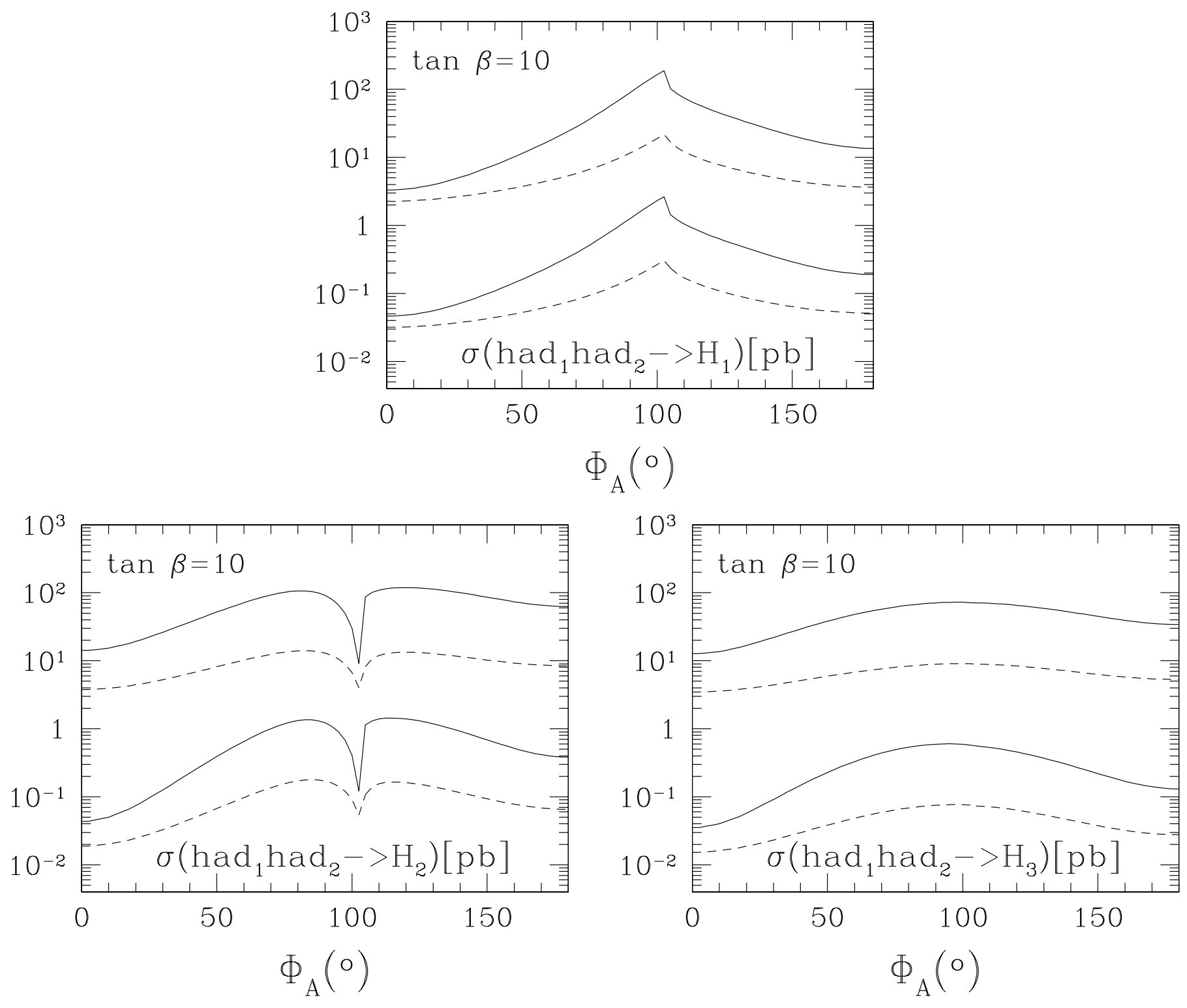

Fig. 3.32: Cross sections for the $b$-fusion production of $H_1$, $H_2$, and $H_3$ vs. $\Phi_A$, for $\Phi_3 = 180^{\rm o}$ (solid lines) and $0^{\rm o}$ (dashed lines), at the LHC with $\sqrt{s} = 14\,$TeV (two upper lines) and at the Tevatron with $\sqrt{s} = 1.96\,$TeV (two lower lines). The symbol $\mathrm{had}_1\mathrm{had}_2$ indicates $pp$ for the LHC, $p\bar{p}$ for the Tevatron.

The region of maximal enhancement or suppression of the production cross sections, around $\Phi_A = 100^\circ$, is also the region in which $H_1$ and $H_2$ are nearly degenerate. Compare with Fig. 3.29. Thus, we should worry about the fact that a further transition $H_1 \leftrightarrow H_2$ may occur during propagation (before decays) due to the off-diagonal absorbitive parts in the $3\times 3$ matrix for the neutral Higgs boson propagator considered in Ref. [158]. In the present case, we observe that $\sqrt{\Gamma_{H_1}\Gamma_{H_2}}$ is much smaller than twice the $H_1$–$H_2$ mass difference. This may imply that such a transition does not occur. We have numerically checked that this is the case [276].

The mass difference between $H_1$ and $H_2$ is, however, still small enough to question whether it is possible to disentangle the two corresponding peaks in the invariant mass distributions of the $H_1$- and $H_2$-decay products. There is no similar problem for the $H_3$ eigenstate, that has a mass always larger than $\sim 160\,$GeV, and therefore a splitting from $H_2$ always larger than $\sim 10\,$GeV. Having a width smaller than $\sim 2\,$GeV, $H_3$ can be easily separated from $H_2$, and therefore also from $H_1$. In the case of $H_1$ and $H_2$, on the contrary, the mass difference can be as small $2\,$GeV around $\Phi_A = 100^\circ$. It will therefore be very challenging to disentangle $H_2$ from $H_1$ experimentally. An analysis of the decay modes and their resolution can help in this sense. At the LHC, the best energy and momentum resolution is for the Higgs-boson decays into muon and photon pairs, with $\delta M_{\gamma\gamma} \sim 1\,$GeV and $\delta M_{\mu\mu} \sim 3\,$GeV, respectively [196]. Thus, it is presumably by combining the analyses of these two decay modes that $H_2$ can be disentagled from $H_1$ when the mass difference is as small as the resolution for the dimuon invariant mass [276].

### 3.8 CP-violating Higgs in diffraction at the LHC

*Valery A. Khoze, Alan D. Martin and Mikhail G. Ryskin*

Recently much attention has focussed on the use of forward proton tagging as a way to discover new physics at the LHC; see, for example, [277–281]. This method promises to provide an exceptionally clean environment to search for, and to identify the nature of, new objects at the LHC. A key motivation behind the recent proposal [282] to add forward proton taggers to the CMS and ATLAS detectors is the study of the central exclusive diffractive (CED) Higgs production process: $pp \to p + H + p$, where the $+$ signs denote the presence of large rapidity gaps.

In some MSSM Higgs scenarios, CED processes provide an opportunity for lineshape analyses [281, 283] and offer a way for direct observation of a CP-violating signal in the Higgs sector [279, 281]. Here, following Ref. [279] we illustrate the phenomenological consequences of CED Higgs production, using a benchmark scenario of maximal CP-violation (called CPX) which was introduced in Ref. [284]. The parameters are fixed according to Eq. (3.13) As shown in [284] the LEP2 data do not exclude the existence of a light Higgs boson with mass $M_H < 60$ GeV (40 GeV) in the model with $\tan\beta \sim$ 3–4 (2–3) and CP phase $\Phi_A = \Phi_3 = 90^\circ(60^\circ)$.

As discussed in [279, 281, 285], CED production (which we show in Fig. 3.33) has unique advantages in hunting for CP-violating Higgses as compared to the traditional non-diffractive processes. For numerical estimates, we use the formalism of [277, 286] to describe CED production, together with the Higgs parameters given by the code `CPsuperH` [287], where we choose $\Phi_A = \Phi_3 = 90^\circ$, $\tan\beta = 4$, $M_{\rm SUSY} = 0.5$ TeV, (that is $|A_f| = 1$ TeV, $|\mu| = 2$ TeV, $|M_3| = 1$ TeV) and $M_{H^\pm} = 135.72$ GeV, so that the mass of the lightest Higgs boson, $H_1$, is $M_{H_1} = 40$ GeV. The cross section is written [277, 286] as the product of the effective gluon–gluon luminosity $\mathcal{L}$, and the square of the matrix element of the subprocess $gg \to H$. Note that the hard subprocess is mediated by the quark/squark triangles in Fig. 3.33. For a CP-violating Higgs the $gg \to H$ matrix element contains two terms

$$\mathcal{M} = g_S \cdot (e_1^\perp \cdot e_2^\perp) - g_P \cdot \varepsilon^{\mu\nu\alpha\beta} e_{1\mu} e_{2\nu} p_{1\alpha} p_{2\beta} / (p_1 \cdot p_2) \tag{3.56}$$

where $e^\perp$ are the gluon polarisation vectors and $\varepsilon^{\mu\nu\alpha\beta}$ is the antisymmetric tensor. In forward CED production, the gluon polarisations are correlated, in such a way that the effective luminosity satisfies the P-even, $J_z = 0$ selection rule [277, 288] . Therefore only the first term contributes to the strictly forward cross section. However, at non-zero transverse momenta of the recoil protons, $p_{1,2}^\perp \neq 0$, there is an admixture of the P-odd $J_z = 0$ amplitude of order $p_1^\perp p_2^\perp / Q_\perp^2$, on account of the $g_P$ term becoming active. For non-zero recoil proton transverse momenta, the interference between the CP-even ($g_S$) and CP-odd ($g_P$) terms leads to left-right asymmetry in the azimuthal distribution of the outgoing protons.

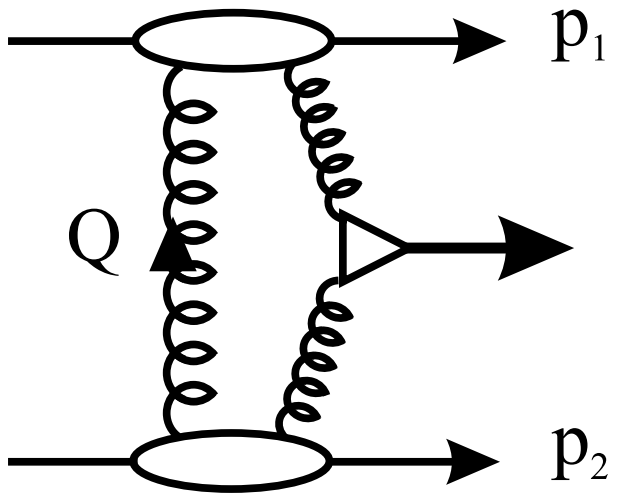

Fig. 3.33: Schematic diagram for the exclusive central diffractive (CED) production of a Higgs boson.

#### *3.8.1 Exclusive diffractive $H_1$ production followed by $b\bar{b}$ decay*

Let us consider the CED process

$$pp \to p + (H \to b\bar{b}) + p \, . \tag{3.57}$$

The signal-to-background ratio is given by the ratio of the cross sections for the hard subprocesses, since the effective gluon–gluon luminosity $\mathcal{L}$ cancels out. The cross section for the $gg \to H$ subprocess[8] [277]

$$\hat{\sigma}(gg \to H) \;=\; \frac{2\pi^2\Gamma(H \to gg)}{M_H^3}\delta\left(1 - \frac{M_{b\bar{b}}^2}{M_H^2}\right) \;\sim\; \text{constant} \times \delta\left(1 - \frac{M_{b\bar{b}}^2}{M_H^2}\right), \tag{3.58}$$

On the other hand, at leading order, the QCD background is given by the $gg \to b\bar{b}$ subprocess

$$\frac{d\hat{\sigma}_{\rm QCD}}{dE_T^2} \sim \frac{m_b^2}{E_T^2}\frac{\alpha_S^2}{M_{b\bar{b}}^2 E_T^2}, \tag{3.59}$$

where $E_T$ is the transverse energy of the $b$ and $\bar{b}$ jets. At leading order (LO), the cross section is suppressed by the $J_z = 0$ selection rule in comparison with the inclusive process. It was shown in [278] that it is possible to achieve a signal-to-background ratio of about 1 for the detection of a SM Higgs with $M_H \sim 120$ GeV, by selecting $b\bar{b}$ exclusive events where the polar angle $\theta$ between the outgoing jets lies in the interval $60^\circ < \theta < 120^\circ$ if the mass window $\Delta m_{\rm missing} = 3$ GeV. The situation is much worse for a light Higgs, since the signal-to-background ratio behaves as

$$\frac{\displaystyle\int \frac{d\mathcal{L}}{d\ln M_{b\bar{b}}^2}\,\hat{\sigma}(gg \to H)\, d\ln M_{b\bar{b}}^2}{\displaystyle\int \frac{d\mathcal{L}}{d\ln M_{b\bar{b}}^2}\,\hat{\sigma}_{\rm QCD}\, d\ln M_{b\bar{b}}^2} \;\sim\; M_{b\bar{b}}^5 \tag{3.60}$$

where we have used $\Delta \ln M_{b\bar{b}}^2 = 2\Delta M_{b\bar{b}}/M_{b\bar{b}}$. The $M^5$ behaviour comes just from dimensional counting. Thus, in going from $M_H \sim 120$ GeV to $M_H \sim 40$ GeV, the expected LO QCD $b\bar{b}$ background increases by a factor of 240 in comparison with that for $M_{b\bar{b}} = 120$ GeV.

Strictly speaking, there are other sources of background [278]. However, for $M_{H_1} \sim 40$ GeV, the LO $b\bar{b}$ contribution dominates. Finally, with the cuts of Ref. [278], we predict that the cross section of the $H_1$ signal is

$$\sigma^{\rm CED}(pp \to p + (H_1 \to b\bar{b}) + p) \simeq 14 \text{ fb}$$

as compared to the QCD background cross section, with the same cuts, of

$$\sigma^{\rm CED}(pp \to p + (b\bar{b}) + p) \simeq 1.4\frac{\Delta M}{1 \text{ GeV}} \text{ pb}.$$

That is the signal-to-background ratio is only $S/B \sim 1/300$, and so even for an integrated luminosity $\mathcal{L} = 300$ fb$^{-1}$ for $\Delta M = 3$ GeV the significance of the signal is too low. Therefore, to identify a light Higgs, it is desirable to study a decay mode other than $H_1 \to b\bar{b}$. The next largest mode is $H_1 \to \tau\tau$, with a branching fraction of about 0.07.

The dependence of the results on the mass of the $H_1$ Higgs boson is illustrated in Table 3.2. Clearly the cross section decreases with increasing mass. On the other hand the signal-to-background ratio increases.

### *3.8.2 The $\tau\tau$ decay mode*

At the LHC energy, the expected CED cross section for $H_1$ production, followed by $\tau\tau$ decay, is

$$\sigma\,(pp \to p + (H \to \tau\tau) + p) \sim 1.1 \text{ fb}, \tag{3.61}$$

[8]In [277] we denoted the initial state by $gg^{PP}$ to indicate that each of the incoming gluons belongs to colour-singlet Pomeron exchange. Here this notation is assumed to be implicit.

Table 3.2: The cross sections (in fb) of CED production of $H_i$ neutral Higgs bosons, together with those of the QCD($b\bar{b}$) and QED($\tau\tau$) backgrounds. The acceptance cuts applied are (a) the polar angle cut $60° < \theta(b \text{ or } \tau) < 120°$ in the Higgs rest frame, (b) $p_i^\perp > 300$ MeV for the forward outgoing protons and (c) the polar angle cut $45° < \theta(b) < 135°$. The azimuthal asymmetries $A_i$ are defined in Eq. (3.63).

| $M(H_1)$ GeV | cuts | 30 | 40 | 50 |
|---|---|---|---|---|
| $\sigma(H_1) \times \text{Br}(b\bar{b})$ | $a$ | 45 | 14 | 6 |
| $\sigma^{\text{QCD}}(b\bar{b})$ | $a$ | 48000 | 4200 | 600 |
| $A_{b\bar{b}}$ | | 0.14 | 0.07 | 0.04 |
| $\sigma(H_1) \times \text{Br}(\tau\tau)$ | $a, b$ | 1.9 | 0.6 | 0.3 |
| $\sigma^{\text{QED}}(\tau\tau)$ | $a, b$ | 0.6 | 0.3 | 0.12 |
| $A_{\tau\tau}$ | $b$ | 0.2 | 0.1 | 0.05 |
| $M(H_2)$ GeV | | 103.4 | 104.7 | 106.2 |
| $\sigma \times \text{Br}(H_2 \to 2H_1 \to 4b)$ | $c$ | 0.5 | 0.5 | 0.5 |
| $\sigma \times \text{Br}(H_2 \to 2b)$ | $a$ | 0.1 | 0.1 | 0.2 |
| $M(H_3)$ GeV | | 141.9 | 143.6 | 146.0 |
| $\sigma \times \text{Br}(H_3 \to 2H_1 \to 4b)$ | $c$ | 0.14 | 0.2 | 0.18 |
| $\sigma \times \text{Br}(H_3 \to 2b)$ | $a$ | 0.04 | 0.07 | 0.1 |

where the $60° < \theta < 120°$ polar angle cut has already been included. Despite the low Higgs mass, we note that the exclusive cross section is rather small. As we already saw in (3.58), the cross section of the hard subprocess $\hat{\sigma}(gg \to H)$ is approximately independent of $M_H$. Of course, we expect some enhancement from the larger effective gluon–gluon luminosity $\mathcal{L}$ for smaller $M_H$. This gives an enhancement of about 20 (for $M_H = 40$ GeV in comparison with that for $M_H = 120$ GeV).

On the other hand, in the appropriate region of SUSY parameter space, the CP-even $H \to gg$ vertex, $g_S$, is almost 2 times smaller [285, 287] than that of a SM Higgs, giving a suppression of 4. Also the ratio $B(H \to \tau\tau)/B(H \to b\bar{b})$ gives a further suppression of about 12. Although the $\tau\tau$ signal has the advantage that there is practically no QCD background, exclusive $\tau^+\tau^-$ events may be produced by $\gamma\gamma$ fusion. To suppress this QED background, one may select events with relatively large transverse momenta of the outgoing protons. For example, if $p_{1,2}^\perp > 300$ MeV, then the cross section for the QED background, for $M_{\tau\tau} = 40$ GeV, is about

$$\sigma_{\text{QED}}(pp \to p + \tau\tau + p) \simeq 0.1 \frac{\Delta M}{1 \text{ GeV}} \text{ fb}, \tag{3.62}$$

while the signal (3.61) contribution is diminished by the cuts, $p_{1,2}^\perp > 300$ MeV, down to 0.6 fb. Thus, assuming an experimental resolution of $\Delta M \sim 3$ GeV, we obtain a signal-to-background ratio of $S/B \sim 2$ for $M_{H_1} \sim 40$ GeV.

Note that in all the estimates given above, we include the appropriate soft survival factors $S^2$—that is the probabilities that the rapidity gaps are not populated by the secondaries produced in the soft rescattering [290]. Moreover, here we account for the fact that only events with proton transverse momenta $p_{1,2}^\perp > 300$ MeV were selected.

### 3.8.3 *Azimuthal asymmetry of the outgoing protons*

A specific prediction, in the case of a CP-violating Higgs boson, is the asymmetry in the azimuthal $\varphi$ distribution of the outgoing protons, caused by the interference between the two terms in (3.56). The polarisations of the active gluons are aligned along their respective transverse momenta, $Q_\perp - p_1^\perp$ and $Q_\perp + p_2^\perp$. Hence the contribution caused by the second term, $g_P$, is proportional to the vector product

$$\vec{n}_0 \cdot (\vec{p}_1^\perp \times \vec{p}_2^\perp) \sim \sin\varphi,$$

where $\vec{n}_0$ is a unit vector in the beam direction, $\vec{p}_1$. The sign of the angle $\varphi$ is fixed by the four-dimensional structure of the second term in (3.56). Of course, due to the selection rule, this contribution is suppressed in the amplitude by $p_1^\perp p_2^\perp / Q_\perp^2$, in comparison with that of the $g_S$ term. Note that there is a partial compensation of the suppression due to the ratio $g_P/g_S \sim 2$. Also the soft survival factors $S^2$ are higher for the pseudoscalar and interference terms, than for the scalar term.

An observation of the azimuthal asymmetry may therefore be a direct indication of the existence of CP-violation in the Higgs sector. Neglecting rescattering effects, we find, for example, an asymmetry

$$A = \frac{\sigma(\varphi < \pi) - \sigma(\varphi > \pi)}{\sigma(\varphi < \pi) + \sigma(\varphi > \pi)} = 2\mathrm{Re}(g_S g_P^*) r_{S/P} (2/\pi)/(|g_S|^2 + |r_{S/P} g_P|^2/2). \tag{3.63}$$

Here the parameter $r_{S/P}$ reflects the suppression of the P-odd contribution. At the LHC energy $A \simeq 0.09$ for $M_{H_1} = 40$ GeV. However we find soft rescattering tends to wash out the azimuthal distribution, and to weaken the asymmetry. Besides this the real part of the rescattering amplitude multiplied by the imaginary part of the pseudoscalar vertex $g_P$ (with respect to $g_S$) gives some negative contribution. So finally we predict[9] $A \simeq 0.07$.

The asymmetries expected at the LHC, with and without the cut $p_{1,2}^\perp > 300$ MeV, are shown for different $H_1$ masses in Table 1. The asymmetry decreases with increasing Higgs mass, first, due to the decrease of $|g_P|/|g_S|$ ratio in this mass range and, second, due to the extra suppression of the P-odd amplitude arising from the factor $p_1^\perp p_2^\perp / Q_\perp^2$ in which the typical value of $Q_\perp$ in the gluon loop increases with mass.

### 3.8.4 *Heavy $H_2$ and $H_3$ Higgs production with $H_1 H_1$ decay*

Another possibility to study the Higgs sector in the CPX scenario is to observe CED production of the heavy neutral $H_2$ and $H_3$ Higgs bosons, using the $H_2, H_3 \to H_1 + H_1$ decay modes. For the case we considered above ($\tan\beta = 4$, $\phi_{\mathrm{CPX}} = 90°$, $M_{H_1} = 40$ GeV), the masses of the heavy bosons bosons are $M_{H_2} = 104.7$ GeV and $M_{H_3} = 143.6$ GeV. At the LHC energy, the CED cross sections of the $H_2$ and $H_3$ bosons are not too small – $\sigma^{\mathrm{CED}} = 1.5$ and 0.9 fb respectively. When the branching fractions, $\mathrm{Br}(H_2 \to H_1 H_1) = 0.84$, $\mathrm{Br}(H_3 \to H_1 H_1) = 0.54$ and $\mathrm{Br}(H_1 \to b\bar{b}) = 0.92$, are included, we find

$$\sigma(pp \to p + (H \to b\bar{b}\, b\bar{b}) + p) = 1.1 \text{ and } 0.4 \text{ fb}$$

for $H_2$ and $H_3$ respectively. Thus there is a chance to observe, and to identify, the CED production of all three neutral Higgs bosons, $H_1, H_2$ and $H_3$, at the LHC. The QCD background for exclusive diffractive production of four $b$-jets is significantly less than the signal.

### 3.8.5 *Central Higgs production with double diffractive dissociation*

To enhance the Higgs signal we study a less exclusive reaction in which we allow both of the incoming protons to dissociate. In Ref. [277] it was called double diffractive *inclusive* production (here denoted CDD), and was written

$$pp \to X + H + Y. \tag{3.64}$$

Typical results, for the LHC energy, are shown in Table 3.3.

Of course, the missing mass method cannot be used to measure the mass of the Higgs for CDD production. Therefore the mass resolution will be not so good as for CED. Moreover, with the absence of the selection rule, the LO QCD $b\bar{b}$-background is not suppressed. Hence we study only the $\tau\tau$ decay mode for the light boson, $H_1$, and the four $b$-jet final state for the heavy $H_2$ and $H_3$ bosons.

[9] We expect a similar asymmetry in the tri-mixing scenario of Ref. [281].

Table 3.3: The cross sections (in fb) for the central production of $H_i$ neutral Higgs bosons by *inclusive* double diffractive dissociation (CDD), together with that of the QED($\tau\tau$) background. A polar angle acceptance cuts of $60° < \theta(b \text{ or } \tau) < 120°$ ($45° < \theta(b) < 145°$) in the Higgs rest frame is applied for the case of $H_1$ ($H_2, H_3$) bosons. The numbers in brackets correspond to the imposition of the additional cut of $E_i^\perp > 7$ GeV for the proton dissociated systems.

| $M(H_1)$ GeV | 30 | 40 | 50 |
|---|---|---|---|
| $\sigma(H_1) \times \mathrm{Br}(\tau\tau)$ | 19 (4) | 6 (2) | 2.6 (0.8) |
| $\sigma^{\mathrm{QED}}(\tau\tau)$ | 66 (2.2) | 30 (1.5) | 15 (0.9) |
| $M(H_2)$ GeV | 103.4 | 104.7 | 106.2 |
| $\sigma \times \mathrm{Br}(H_2 \to 2H_1 \to 4b)$ | 4 (2) | 4 (2) | 3.5 (2) |
| $M(H_3)$ GeV | 141.9 | 143.6 | 146.0 |
| $\sigma \times \mathrm{Br}(H_3 \to 2H_1 \to 4b)$ | 1.5 (0.8) | 2.2 (1.2) | 2 (1.1) |

The background to the $H_1 \to \tau\tau$ signal arises from the $\gamma\gamma \to \tau\tau$ QED process. It is evaluated in the equivalent photon approximation. From Table 2 we see that the $H_1$ signal for CDD production, (3.64), exceeds the exclusive signal by more than a factor of ten. On the other hand the signal-to-background ratio is worse; $S/B_{QED}$ is about 1/5. Moreover there could be a huge background due the misidentification of a gluon dijet as a $\tau\tau$-system.

For the four $b$-jet signals of the heavy $H_2$ and $H_3$ bosons, the QCD background can be suppressed by requiring each of the four $b$-jets to have polar angle in the interval $(45°, 135°)$, in the frame where the four $b$-jet system has zero rapidity. However in the absence of a good mass resolution, that is with only $\Delta M = 10$ GeV, we expect the four $b$-jet background to be 3-5 times the signal. Nevertheless these signals are still feasible, with cross sections of the order of a few fb. For example, with an integrated luminosity of $\mathcal{L} = 300$ fb$^{-1}$ and an efficiency of $4b$-tagging of $(0.6)^2$ [278], we predict about 400 $H_2$ events and 200 $H_3$ events.

The CDD kinematics allow a study of CP-violation, and the separation of the contributions coming from the scalar and pseudoscalar couplings, $g_S$ and $g_P$ of (3.56), respectively. Indeed, the polarizations of the incoming active gluons are aligned along their transverse momenta, $\vec{Q}_\perp - \vec{p}_1^{\,\perp}$ and $\vec{Q}_\perp + \vec{p}_2^{\,\perp}$. Hence the $gg \to H$ fusion vertices take the forms

$$V_S = (\vec{Q}_\perp - \vec{p}_1^{\,\perp}) \cdot (\vec{Q}_\perp + \vec{p}_2^{\,\perp}) g_S \tag{3.65}$$

$$V_P = \vec{n}_0 \cdot [(\vec{Q}_\perp - \vec{p}_1^{\,\perp}) \times (\vec{Q}_\perp + \vec{p}_2^{\,\perp})] g_P, \tag{3.66}$$

where $g_S$ and $g_P$ are defined in (3.56).

For the exclusive (CED) process the momenta $p_{1,2}^\perp$ were limited by the proton form factor, and typically $Q^2 \gg p_{1,2}^2$. Thus

$$V_S = g_S\, Q_\perp^2 \quad \text{while} \quad V_P = g_P\, (\vec{n}_0 \cdot [\vec{p}_2^{\,\perp} \times \vec{p}_1^{\,\perp}]). \tag{3.67}$$

On the contrary, for double diffractive dissociation production (CDD) $Q^2 < p_{1,2}^2$. In this case

$$V_S = g_S\, p_1^\perp p_2^\perp \cos\varphi \quad \text{and} \quad V_P = g_P\, p_1^\perp p_2^\perp \sin\varphi. \tag{3.68}$$

Moreover we can select events with large outgoing transverse momenta of the dissociating systems, say $p_{1,2}^\perp > 7$ GeV, in order to make reasonable measurements of the directions of the vectors $\vec{p}_1^{\,\perp} = \vec{E}_1^\perp$ and $\vec{p}_2^{\perp} = \vec{E}_2^{\perp}$. Here $E_{1,2}^{\perp}$ are the transverse energy flows of the dissociating systems of the incoming protons. At LO, this transverse energy is carried mainly by the jet with minimal rapidity in the overall centre-of-mass frame. The azimuthal angular distribution has the form

$$\frac{d\sigma}{d\varphi} = \sigma_0(1 + a\,\sin 2\varphi + b\,\cos 2\varphi), \tag{3.69}$$

where the coefficients are given by

$$a = \frac{2\mathrm{Re}(g_S g_P^*)}{|g_S|^2 + |g_P|^2} \quad \text{and} \quad b = \frac{|g_S|^2 - |g_P|^2}{|g_S|^2 + |g_P|^2}. \tag{3.70}$$

Table 3.4: The coefficients in the azimuthal distribution $d\sigma/d\varphi = \sigma_0(1 + a\sin 2\varphi + b\cos 2\varphi)$, where $\varphi$ is the azimuthal angle between the $E^{\perp}$ flows of the two proton dissociated systems. If there were no CP-violation, then the coefficients would be $a = 0$ and $|b| = 1$.

| $M(H_1)$ GeV | 30 | | 40 | | 50 | |
|---|---|---|---|---|---|---|
| | $a$ | $b$ | $a$ | $b$ | $a$ | $b$ |
| $H_1$ | −0.53 | −0.73 | −0.56 | −0.55 | −0.53 | −0.33 |
| $H_2$ | 0.44 | 0.90 | 0.41 | 0.91 | 0.37 | 0.92 |
| $H_3$ | −0.38 | 0.92 | −0.40 | 0.91 | −0.42 | 0.90 |

Note that the coefficient $a$ arises from scalar-pseudoscalar interference, and reflects the presence of a T-odd effect. Its observation would signal an explicit CP-violating mixing in the Higgs sector.

The predictions for the coefficients are given in Table 3.4 for different values of the Higgs mass, namely $M_{H_1}$ = 30, 40 and 50 GeV. The coefficients are of appreciable size and, given sufficient luminosity, may be measured at the LHC. Imposing the cuts $E_i^{\perp} > 7$ GeV reduces the cross sections by about a factor of two, but does not alter the signal-to-background ratio, $S/B_{QCD}$. However the cuts do give increased suppression of the QED $\tau\tau$ background and now, for the light $H_1$ boson, the ratio $S/B_{QED}$ exceeds one. We emphasize here that, since we have relatively large $E^{\perp}$, the angular dependences are quite insensitive to the soft rescattering corrections.

### *3.8.6 Conclusions*

We have evaluated the cross sections, and the corresponding backgrounds, for the central double-diffractive production of the (three neutral) CP-violating Higgs bosons at the LHC using, for illustration, the CPX scenario of Ref. [284]. We have studied the production of the three states, $H_1, H_2, H_3$, both with exclusive kinematics, $pp \to p + H + p$ which we denoted CED, and in double-diffractive reactions where both the incoming protons may be destroyed, $pp \to X + H + Y$ which we denoted CDD. Proton taggers are required in the former processes, but not in the latter. Typical results are summarised in Tables 1 and 2, respectively. The cross sections are not large, but should be accessible at the LHC. The azimuthal asymmetries of the outgoing protons, induced by CP-violation, are quite sizeable, of order $10\%$.

It would be very informative to measure the azimuthal angular dependence of the outgoing proton systems, for both the CED and CDD processes. Such measurements would reveal explicitly any CP-violating effect, via the interference of the scalar and pseudoscalar $gg \to H$ vertices.

### 3.9 CP violation in supersymmetric charged Higgs production at the LHC

*Jennifer Williams*

We investigate the possibility of observing CP violation in the production of MSSM charged Higgs bosons at the LHC. The CP violation arises from allowing the trilinear scalar couplings in the soft breaking Lagrangian to be complex, leading to complex phases. We have chosen to investigate the effect of a complex $A_t$, leaving the other phases zero. Initially, we set the phase of $A_t$, $\Phi_{A_t}$, to be maximal ($\frac{\pi}{2}$) and vary the magnitudes of $A_t$ and $A_b$, with $|A_t| = |A_b|$. In a study which is currently in preparation [291] we also consider fixing the magnitudes of the trilinear couplings and vary the phases of $A_t$ and $A_b$.

The main production modes for charged Higgs bosons at the LHC are through $b$ quark induced processes, due to the large coupling of the Higgs bosons to heavy quarks. The dominant production process is $b$ quark – gluon fusion, in which a charged Higgs boson is produced in association with a $t$ quark, this is shown in Fig. 3.34. The cross section for this process was found using HERWIG [292] to be $135 \times 10^{-3}$ pb. The cross section for the gluon – gluon fusion was found to be $8.24 \times 10^{-3}$ pb. For this reason this investigation only considers charged Higgs boson production by $b$ quark – gluon fusion.

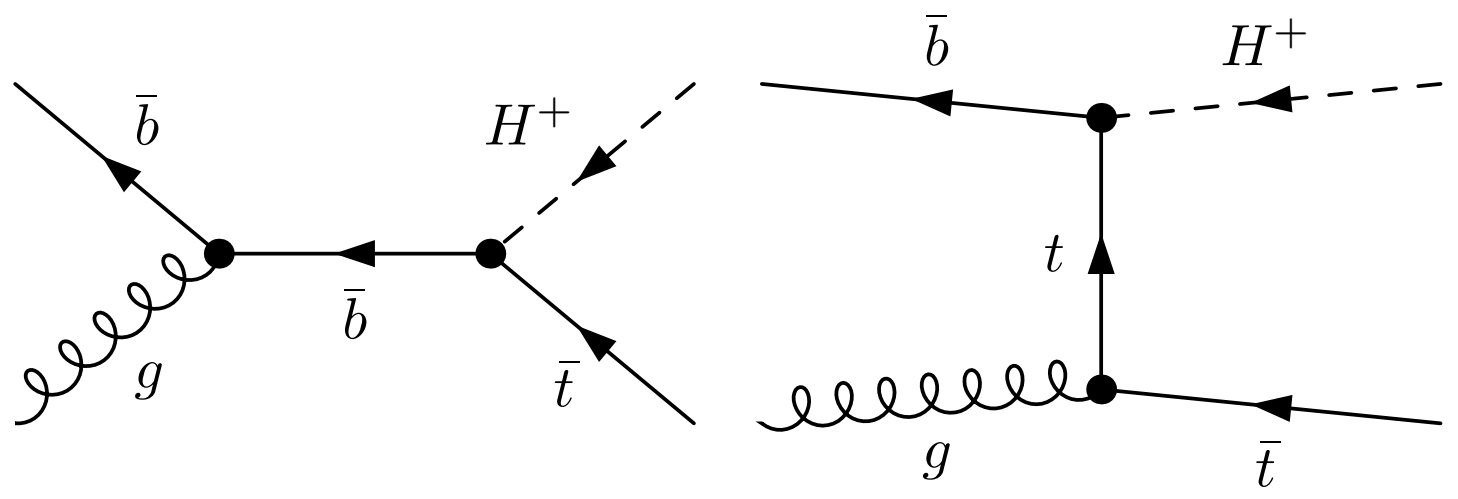

Fig. 3.34: Tree level production of charged Higgs bosons.

In order for the CP violation which is present in the complex phases to be manifest, it is necessary for there to be interference between the tree level and loop level processes. This is because we require the matrix element squared for the two CP conjugate processes (one producing $H^+$ and the other $H^-$ bosons) to be different. If there is no interference term, then $|\mathcal{M}|^2 = |\text{tree}|^2 + |\text{loop}|^2$ which must be the same for both CP conjugate processes.

Since there is no CP violation in the tree level process ($A_t$ does not enter at tree level), it is also necessary for the loop level matrix element to be complex in order for the tree – loop interference term to be different for the two CP conjugate processes. It is an intermediate result of the Optical Theorem that the matrix element at one loop level will have an imaginary part if the energy of the process is sufficient for the particles in the loop to be produced on mass shell.

At one loop level, the loops which contribute to the asymmetry are those involving stop and sbottom squarks. A selection of these loops is shown in Fig. 3.35. In later figures they are referred to in groups, with TB $\tilde{t}\tilde{b}\tilde{g}$ referring to the triangle and box diagrams containing stop and sbottom squarks and gluinos, SE $\tilde{t}\tilde{b}$ referring to the self energy diagrams containing stop and sbottom squarks and TB $\tilde{t}\tilde{b}\tilde{\chi}^0$ referring to the triangle and box diagrams containing stop and sbottom squarks and neutralinos.

The CP asymmetry in the production of $H^+$ and $H^-$ bosons at the parton level was calculated using FormCalc [228, 236] as

$$\mathcal{A}_{\text{parton}} = \frac{\sigma(H^+) - \sigma(H^-)}{\sigma(H^+) + \sigma(H^-)}. \tag{3.71}$$

The MSSM parameters which were used as input for FormCalc were chosen to give parameter space points based on SPS 1a [255], similar to that used in a study of the decay of charged Higgs bosons by Christova et al. [68]. The MSSM parameters used are given in Table 3.5. The Standard Model parameters used were the default parameters in `FormCalc`.

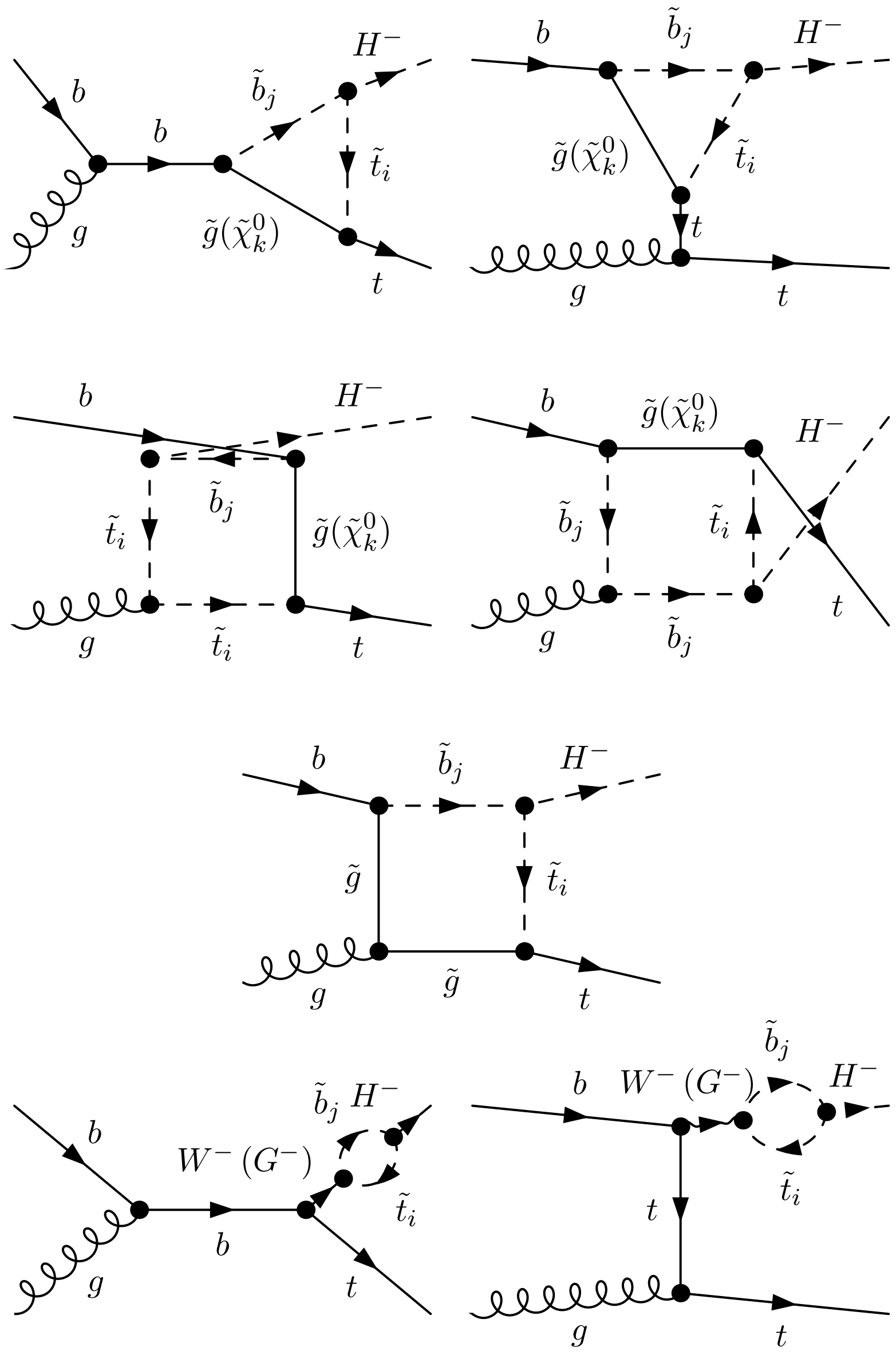

Fig. 3.35: A selection of diagrams contributing to the CP violating asymmetry in charged Higgs boson production at one loop level.

The CP asymmetry at parton level is shown in Fig. 3.36, both versus the partonic centre of mass energy, $\sqrt{\hat{s}}$ and versus changing Higgs mass, $M_{H^\pm}$. The very clear thresholds (marked with vertical lines) that can be seen occur at values of the partonic centre of mass energy or charged Higgs boson mass when the loop particles can be produced on mass shell, thus increasing the imaginary part of the loop matrix element. For example the thresholds in $\sqrt{\hat{s}}$ for the loop containing a gluino and stop and sbottom squarks occurs at values of $M_{\tilde{g}} + M_{\tilde{b}_j}$ and the thresholds in $M_{H^\pm}$ for the self energy loop containing stop and sbottom squarks occurs at values of $M_{\tilde{t}_i} + M_{\tilde{b}_j}$.

Up to this point we have considered the $b$ quark and the gluon as free on shell particles. In reality they are constituents of the protons and it is the protons that will be collided at the LHC. We now convolute the parton level results with the parton distribution functions; this effectively integrates over the partonic centre of mass energy. We used the 2004 MRST pdfs at next-to-leading order at a $q^2$ value of $(M_{H^\pm} + M_t)^2$ [293]. The CP violating asymmetry for the hadronic production of charged Higgs bosons is given by

$$\mathcal{A}_{\text{hadron}} = \frac{\sigma(pp \to \bar{b}g \to H^+\bar{t} + X) - \sigma(pp \to bg \to H^- t + X)}{\sigma(pp \to \bar{b}g \to H^+\bar{t} + X) + \sigma(pp \to bg \to H^- t + X)}. \tag{3.72}$$

Table 3.5: Constrained Minimal Supersymmetric Standard Model parameters used for this study.

| Parameter | Value | |
|---|---|---|
| | SPS 1a | SPS 1b |
| $\tan\beta$ | 10 | 30 |
| $\|\mu\|$ | 352.39 GeV | 501.08 GeV |
| $M_2$ | 192 GeV | 311.38 |
| $\|A_t\|$ | varied 250 $\rightarrow$ 1000 GeV usually 500 or 1000 GeV | |
| $\Phi_{A_t}$ | $\frac{\pi}{2}$ | |
| $\|A_b\|$ | $= \|A_t\|$ | |
| $\Phi_{A_b}$ | 0 | |
| $M_{\text{SUSY}}$ | 490 GeV | 707.06 GeV |
| $M_{\tilde{Q}_3}$ | 535 GeV | 767 GeV |
| $M_{\tilde{U}_3}$ | 340 GeV | 672 GeV |
| $M_{\tilde{D}_3}$ | 490 GeV | 788 GeV |

The results for the CP asymmetry at hadron level are shown in Fig. 3.37. The thresholds marked in Fig. 3.37a are thresholds in the charged Higgs boson mass at values of $M_{\tilde{t}_i} + M_{\tilde{b}_j}$.

We combined our results for the CP asymmetry in the production of charged Higgs bosons with those of Christova et al [68] for the decay after correcting a conjugation error in their decay results. The combined asymmetry for the production and the decay is given by

$$\mathcal{A}_{\text{total}} = \frac{\sigma\left(pp \to \bar{b}g \to H^+\bar{t} + X\right)\Gamma\left(H^+ \to t\bar{b}\right) - \sigma\left(pp \to bg \to H^- t + X\right)\Gamma\left(H^- \to \bar{t}b\right)}{\sigma\left(pp \to \bar{b}g \to H^+\bar{t} + X\right)\Gamma\left(H^+ \to t\bar{b}\right) + \sigma\left(pp \to bg \to H^- t + X\right)\Gamma\left(H^- \to \bar{t}b\right)}. \tag{3.73}$$

Because the loop contributions are small compared to the tree level this can be approximated as

$$\mathcal{A}_{\text{total}} = \mathcal{A}_{\text{hadron}} + \mathcal{A}_{\text{decay}}. \tag{3.74}$$

The results for combining the production and decay asymmetries are shown in Fig. 3.38.

Finally, we consider the possibility of observing this asymmetry at the LHC. The number of charged Higgs events which will be seen in the detector is given by

$$N = \sigma\left(pp \to bg \to H^\pm t\right) \text{BR}\left(H^\pm \to tb\right) \times \text{acceptance} \times \text{luminosity}. \tag{3.75}$$

We consider an optimistic acceptance of 0.05 and an integrated luminosity of 300 fb$^{-1}$. The acceptance is based on the acceptance given in the ATLAS TDR for $b$ quarks [196]. The significance of the signal over the background, measured in standard deviations is then, $f = \sqrt{N}\mathcal{A}$. This significance is shown in Fig. 3.39. The significance is reduced by the poor acceptance to an insignificant level, meaning that it will not be possible to observe this asymmetry at the LHC.

It should be born in mind however, that we have only considered one production method and one decay in this study. The inclusion of other processes could increase the CP asymmetry. It is also possible to investigate the variation of other phases in the soft supersymmetry breaking Lagrangian. This has been done in [291]. It would also be worthwhile to investigate the possibility of observing a CP violating asymmetry in a similar process at an $e^+e^-$ collider which is a cleaner environment.

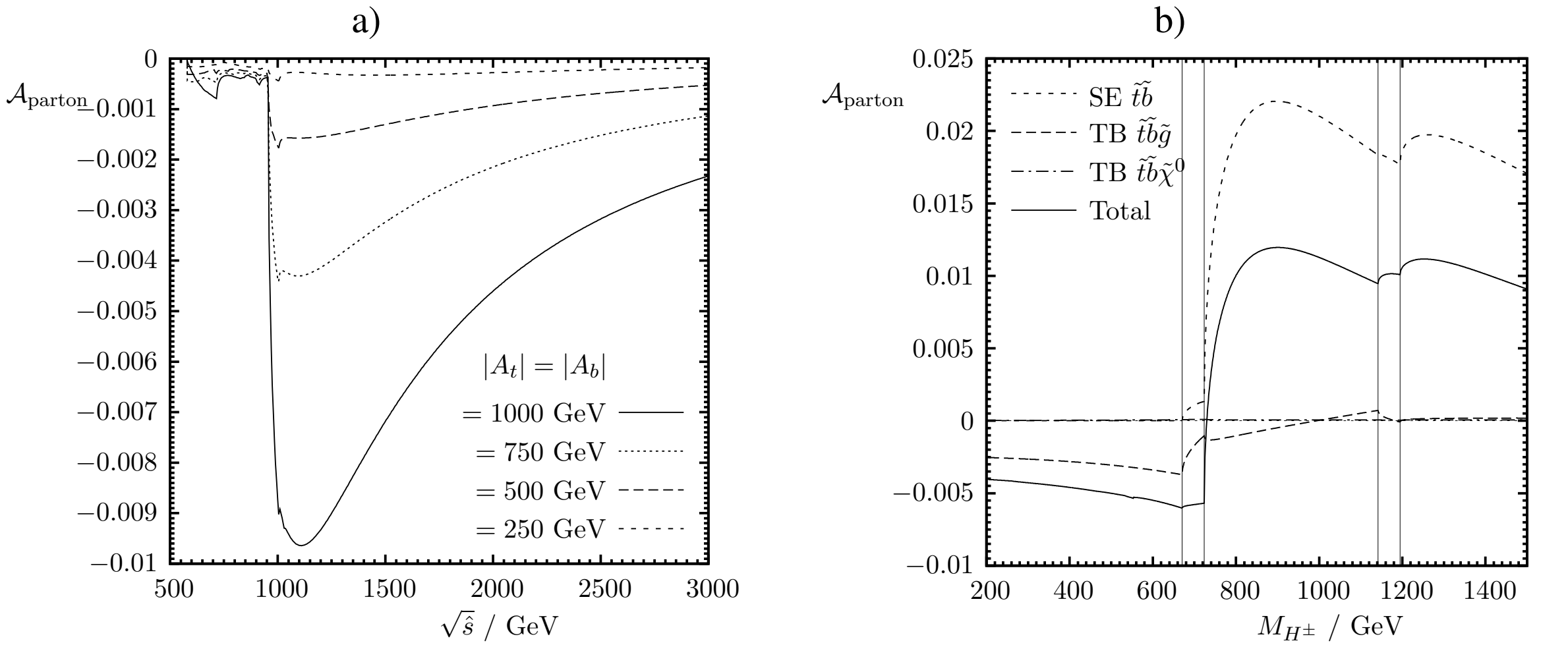

Fig. 3.36: CP asymmetry at parton level, for $\Phi_{A_t} = \frac{\pi}{2}$ and $\Phi_{A_b} = 0$. a) Plotted vs $\sqrt{\hat{s}}$, for $M_{H^\pm} = 402$ GeV and several values of $|A_t| = |A_b|$. b) Plotted vs $M_{H^\pm}$, for $\sqrt{\hat{s}} = 2000$ GeV, showing the contribution from different loops with $|A_t| = |A_b| = 1000$.

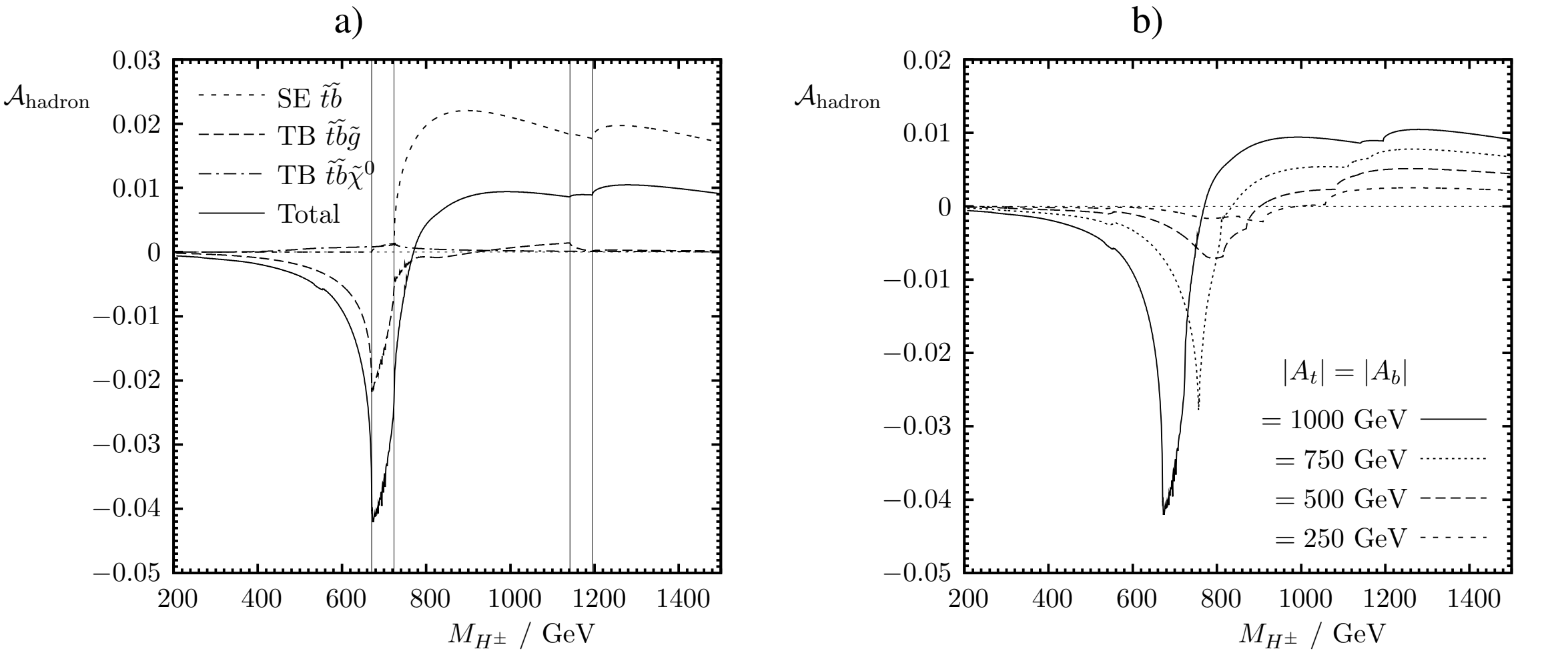

Fig. 3.37: The CP violating asymmetry at hadron level, plotted versus $M_{H^\pm}$, for $\Phi_{A_t} = \frac{\pi}{2}$ and $\Phi_{A_b} = 0$. a) contributions from different loops, with $|A_t| = |A_b| = 1000$. b) asymmetry for a range of trilinear couplings, $|A_t| = |A_b|$.

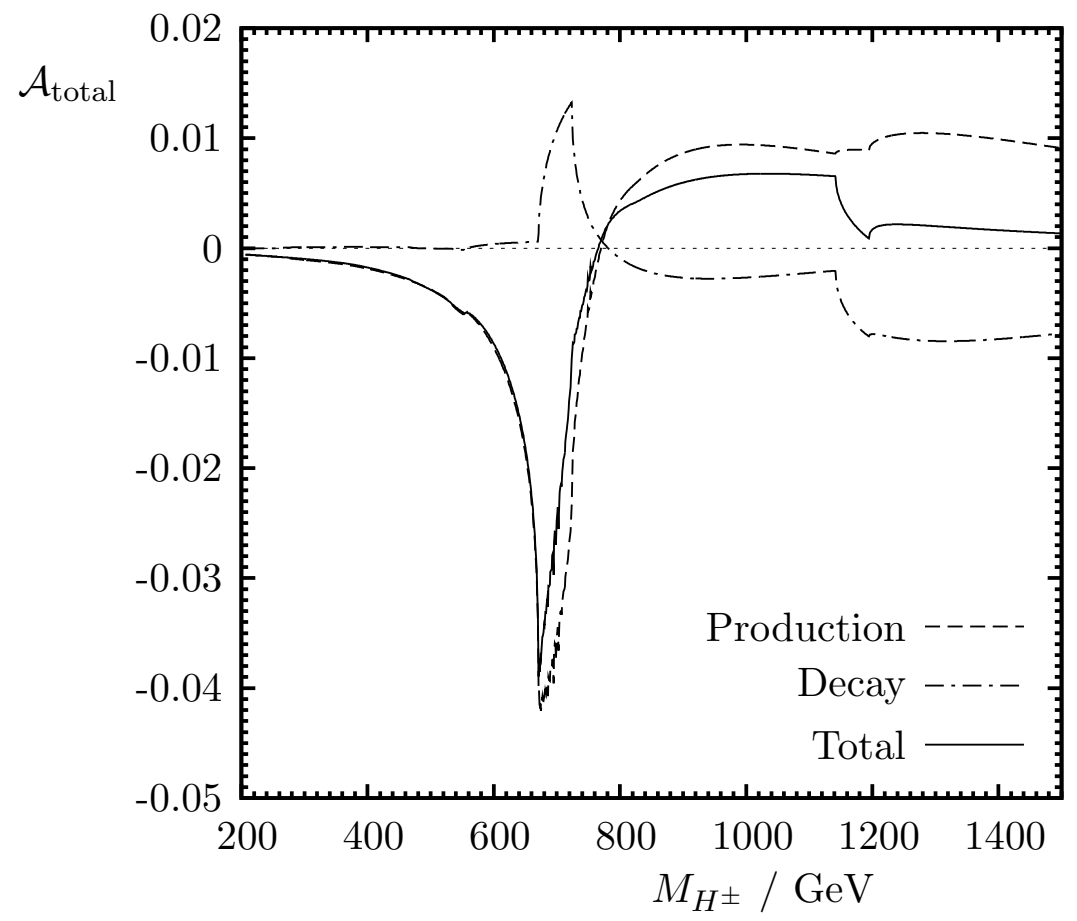

Fig. 3.38: Combined asymmetry for the production and decay, for $\Phi_{A_t} = \frac{\pi}{2}$ and $\Phi_{A_b} = 0$ and $|A_t| = |A_b| = 1000$.

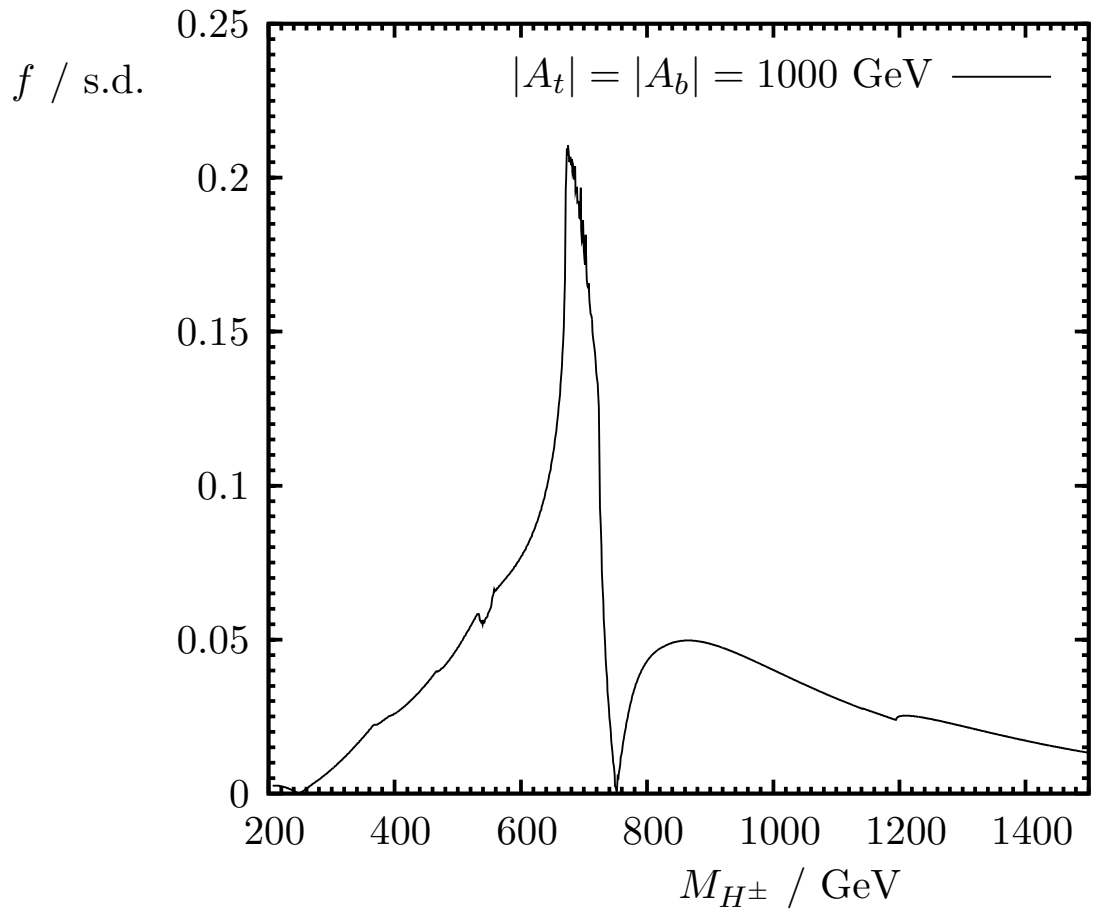

Fig. 3.39: The significance of the CP violating asymmetry expected to be seen in the ATLAS detector, for $\Phi_{A_t} = \frac{\pi}{2}$ and $\Phi_{A_b} = 0$ and $|A_t| = |A_b| = 1000$.

## 3.10 Exploring CP phases of the MSSM at future colliders

*Sven Heinemeyer and Mayda Velasco*

Measuring the phases of the CP-violating MSSM will be one of the important tasks of future high-energy colliders. We discuss the impact of complex phases within the MSSM on various Higgs boson production and decay channels (entering via loop corrections [48–53,55,56,59–61]). Results are compared for the LHC, the ILC, and a $\gamma\gamma$ collider ($\gamma$C). While the precision of the branching ratio measurement at the LHC is not accurate enough, both the ILC and the $\gamma$C could in principle be sensitive to the effects of complex phases (depending on the scenario). The precisions for the various Higgs boson decay channels at the three colliders are summarized in Table 3.6. The Higgs boson mass is set to "typical" values below the upper bound of $M_h \lesssim 135$ GeV [115, 130], which is valid in the real as well as in the complex MSSM.

### *3.10.1 Comparison of different colliders*

We compare the sensitivity of a future $\gamma$C with that of the LHC and the ILC. The comparison is based on two different physics scenarios:

<u>The CPX scenario:</u>
This scenario has been designed to give maximum effects of CP-violating phases [54]. The definition is given in Eq. (3.13). For the sake of comparison with the BGX scenario, the parameters are

$$\begin{aligned} &M_{\text{SUSY}} = 500 \text{ GeV}, |A_t| = 1000 \text{ GeV}, A_t = A_b = A_\tau \\ &M_2 = 500 \text{ GeV}, |M_3| = 1000 \text{ GeV}, \mu = 2000 \text{ GeV} \\ &\Phi = \Phi_{A_{t,b,\tau}} = \Phi_3 \ . \end{aligned} \tag{3.76}$$

$M_{\text{SUSY}}$ denotes a common soft SUSY-breaking mass in the sfermion mass matrices. $A_f$ is the trilinear Higgs-Sfermion coupling with the phase $\Phi_{A_f}$. $M_2$ is a gaugino mass parameter, $M_3$ denotes the gluino mass parameter, and $\mu$ is the Higgs mixing parameter.

<u>The BGX scenario:</u>
This scenario is motivated by baryogenesis. It has been shown in [154] that in this scenario (depending on the Higgs sector parameters) baryogenesis in the early universe could be possible. It is thus a physics motivated scenario, not emphazising possible effects of complex phases. The parameters are

$$\begin{aligned} &M_{\tilde{Q}_3} = 1.5 \text{ TeV}, M_{\tilde{U}_3} = 0, M_{\tilde{Q}_{1,2}} = 1.2 \text{ TeV}, M_{\tilde{L}_{1,2}} = 1.0 \text{ TeV} \\ &|X_t| = 0.7 \text{ TeV}, A_t = A_b = A_\tau \\ &M_2 = 220 \text{ GeV}, M_3 = 1 \text{ TeV}, \mu = 200 \text{ GeV} \\ &\Phi = \Phi_{A_{t,b,\tau}} = \Phi_3 \end{aligned} \tag{3.77}$$

Here $M_{\tilde{Q}_3,\tilde{U}_3}$ are the soft SUSY-breaking parameters in the scalar top mass matrix. $M_{\tilde{Q}_{1,2}}$ are the corresponding parameters for the squarks of the first two generations, while $M_{\tilde{L}_{1,2}}$ refer to the sleptons of the first two generations. $m_t X_t$ is the off-diagonal entry in the scalar top mass matrix with $X_t = A_t - \mu^*/\tan\beta$.

The results presented here have been obtained with the code `FeynHiggs2.2` [59,115,130,144,148, 235]. It should be noted that the higher-order uncertainties in these evaluations are somewhat less under control as compared to the real case, see e.g. Ref. [61]. The same applies to the parametric uncertainties due to the experimental errors of the input parameters [47,61,294,295]. Results for branching ratios obtained with an alternative code, `CPsuperH` [131], can differ quantitatively to some extent from the results shown here. A main difference between the two codes is the more complete inclusion of real two-loop corrections in `FeynHiggs2.2`, resulting in somewhat higher values for the lightest Higgs boson mass. While the complex phase dependence at the one-loop level is included completely in `FeynHiggs2.2`, at the two-loop level it is more complete in `CPsuperH`, which makes it difficult to disentangle the source of possible deviations. A more complete discussion can be found in [296].

Table 3.6: Expected experimental precision of the branching ratio measurement of $h \to X$ at the LHC, the ILC operating at $\sqrt{s} = 500, 1000$ GeV, and the $\gamma$C (based on the CLICHE design [297]). "—" means that no analysis exists or that the channel is not accessible.

| Study | $M_h$ | $b\bar{b}$ | $WW^*$ | $\tau^+\tau^-$ | $c\bar{c}$ | $gg$ | $\gamma\gamma$ |
|---|---|---|---|---|---|---|---|
| LHC [216,217] | 120 GeV | $\sim 20\%$ | $\sim 10\%$ | $\sim 15\%$ | — | — | — |
| ILC ($\sqrt{s} = 500$ GeV) [220,298] | 120 GeV | 1.5% | 3% | 4.5% | 6% | 4% | 19% |
| ILC ($\sqrt{s} = 1000$ GeV) [220,299] | 120 GeV | 1.5% | 2% | — | — | 2.3% | 5.4% |
| $\gamma$C [297,300] | 115 GeV | 2% | 5% | — | — | — | 22% |

#### 3.10.1.1 The CPX scenario

We start our analysis by the investigation of the CPX scenario, see Eq. (3.76). We first show the results for the $\gamma$C in Fig. 3.40 for the decay channel $h \to b\bar{b}$, which has the best sensitivity at this collider. The variation of $\Gamma_{\gamma\gamma} \times \text{BR}(h \to b\bar{b})$ is shown in the $\Phi_{A_{t,b}}$–$\tan\beta$ plane. The strips correspond to constant values of the lightest Higgs mass, while the color code shows the deviation from the corresponding SM value. It should be kept in mind that the Higgs boson mass will be measured to very high accuracy so that one will be confined to one of the strips. We are neglecting the parametric errors from the imperfect knowledge of the input parameters. In reality these parametric errors would widen the strips. The future intrinsic error of $\sim 0.5$ GeV [47], however, is included in the width of the strips. One can see that this channel can be strongly enhanced as compared to the SM. The variation along each strip is much larger than the anticipated precision of $\sim 2\%$ for this channel. This would allow to constrain the values of the complex phases. The picture becomes of course more complicated if the complex phases are varied independently. Various channels will have to be combined to disentangle the different effects.

Results for the LHC are shown in Fig. 3.41. The left plots gives the results for the channel $gg \to h \to \gamma\gamma$, while the right plots depicts $WW \to h \to \tau^+\tau^-$. The latter channel (like $h \to b\bar{b}$) is usually somewhat enhanced in the MSSM, the BR($h \to WW^*$) (not shown) and BR($h \to \gamma\gamma$) (see the left plot of Fig. 3.41) are normally suppressed in this scenario. The precision of the LHC will not be good enough to obtain information about complex phases in this way.

Finally in Fig. 3.42 shows the ILC results in the CPX scenario. The left plot shows the BR($h \to b\bar{b}$), while the right plot depicts BR($h \to \tau^+\tau^-$). Both channels are enhances as compared to the SM in this scenario. The high precision of the ILC (see Table 3.6) shows that this collider has a good potential to disentangle the complex phases.

Since in the examples shown here for the $\gamma$C and the ILC the largest deviations occur for different regions of the parameter space, the results from both colliders could be combined in order to extract the maximum information on $\Phi_{A_{t,b}}$.

#### 3.10.1.2 The BGX scenario

Now we turn to the investigation of the baryogenesis motivated BGX scenario, see Eq. (3.77). The effects in this scenario are expected to be smaller than in the CPX scenario that had been designed to give maximum effects of the complex phases.

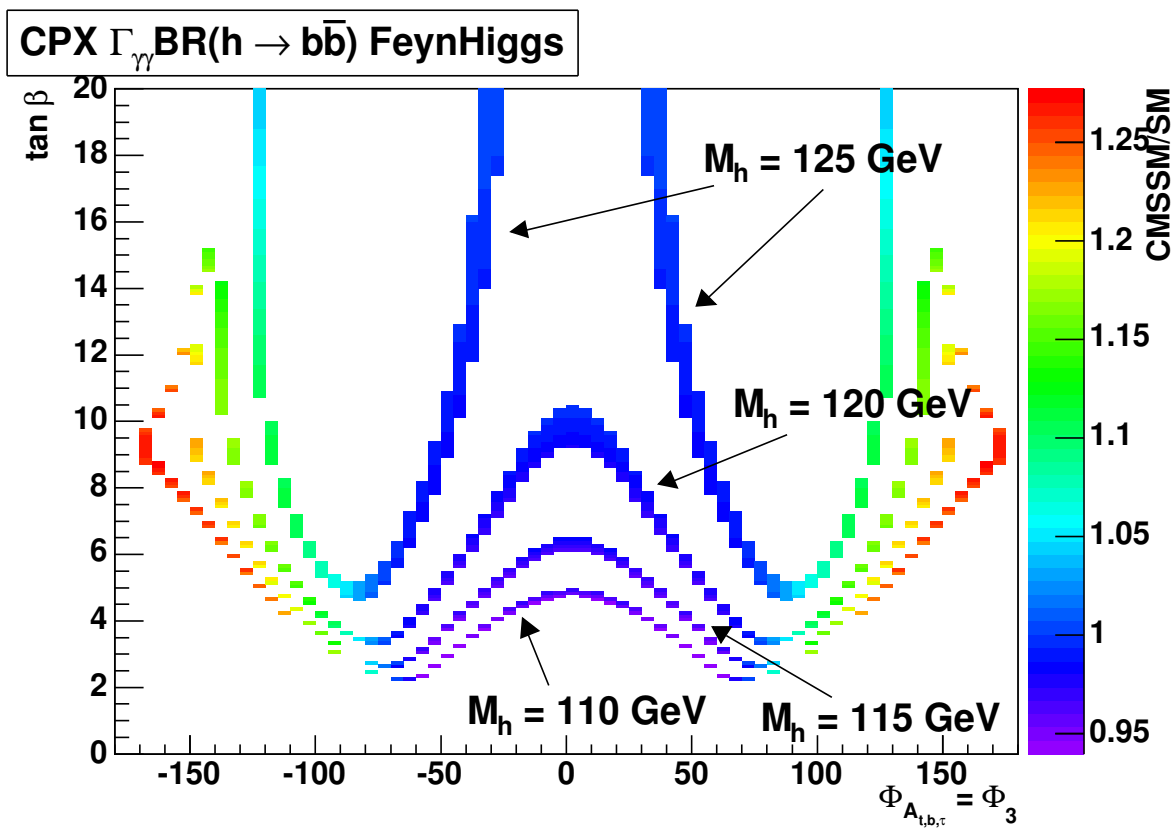

Fig. 3.40: The deviations of $\Gamma_{\gamma\gamma} \times \text{BR}(h \to b\bar{b})$ within the CPX scenario from the SM value is shown in the $\Phi_{A_{t,b}}$–$\tan\beta$ plane. The corresponding precision obtainable at a $\gamma$C is $\sim 2\%$.

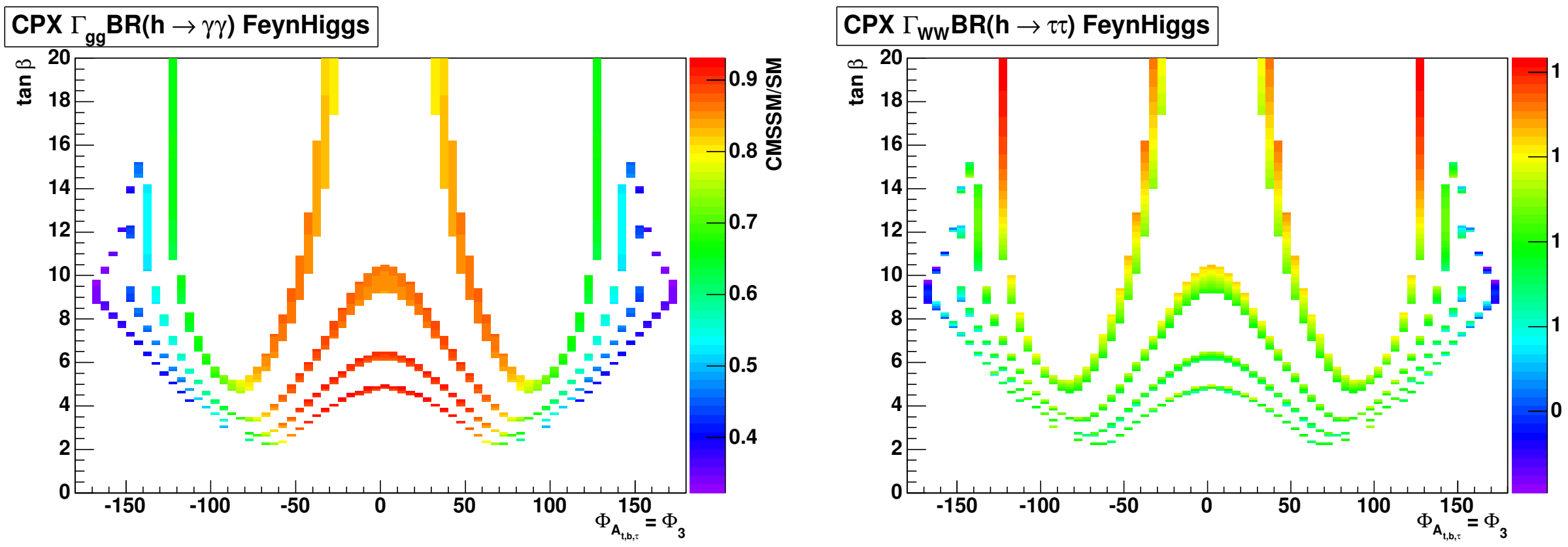

Fig. 3.41: The deviations of $\Gamma_{gg} \times \text{BR}(h \to \gamma\gamma)$ (left) and of $\Gamma_{WW} \times \text{BR}(h \to \tau^+\tau^-)$ (right) within the CPX scenario from the SM value is shown in the $\Phi_{A_{t,b}}$–$\tan\beta$ plane. The corresponding experimental precision can be found in Table 3.6.

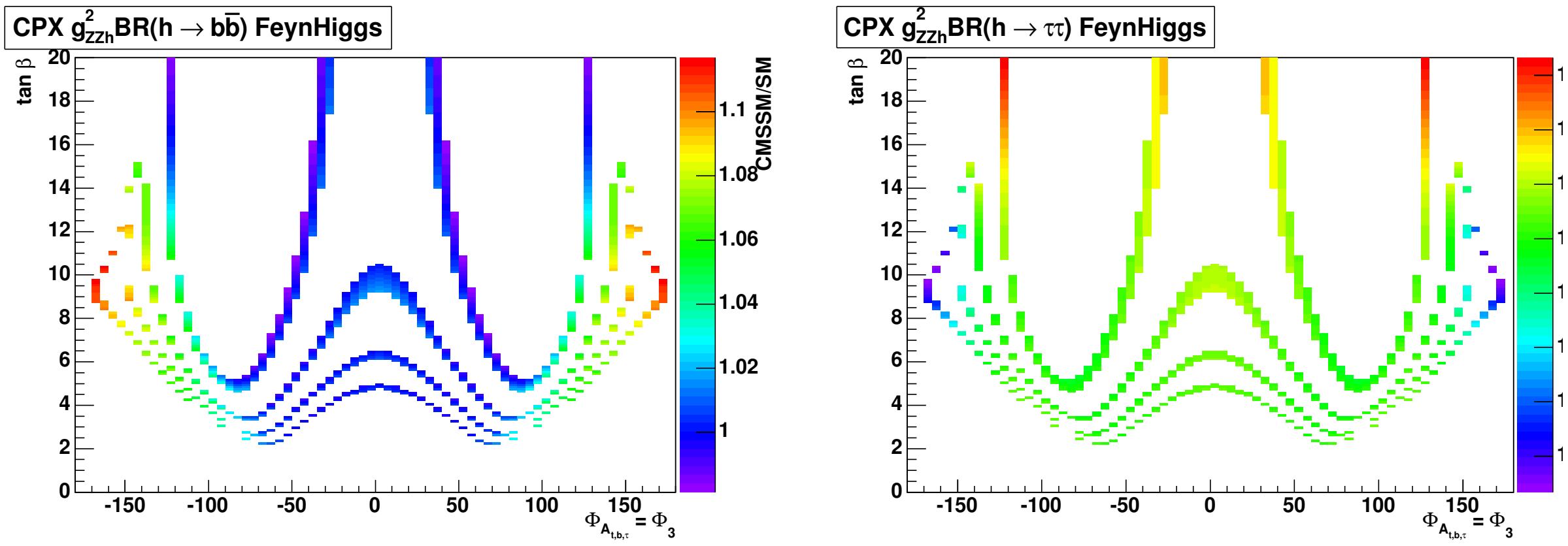

Fig. 3.42: The deviations of $g^2_{ZZh} \times \text{BR}(h \to b\bar{b})$ (left) and of $g^2_{ZZh} \times \text{BR}(h \to \tau^+\tau^-)$ (right) within the CPX scenario from the SM value is shown in the $\Phi_{A_{t,b}}$–$\tan\beta$ plane. The corresponding experimental precision can be found in Table 3.6.

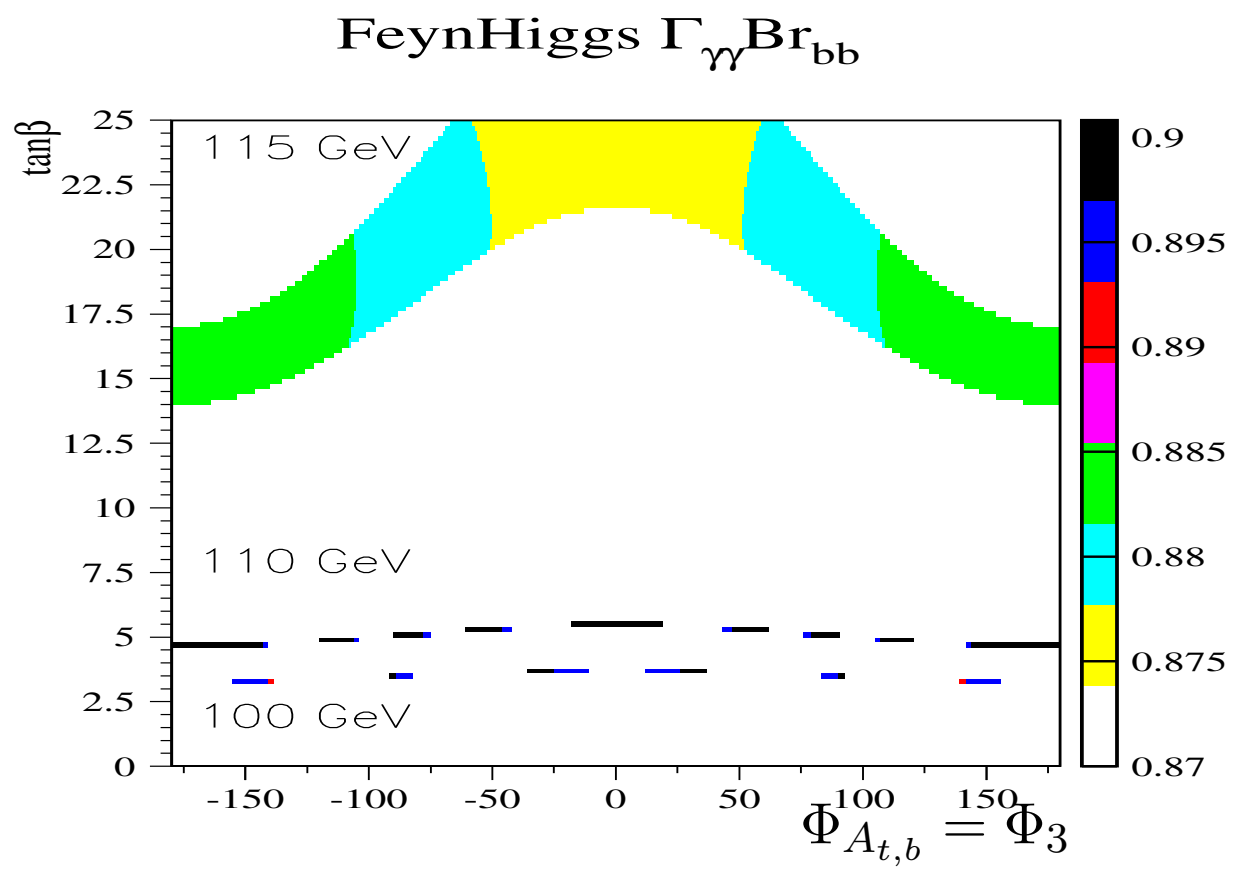

Fig. 3.43: The deviations of $\Gamma_{\gamma\gamma} \times$ BR$(h \to b\bar{b})$ within the BGX scenario from the SM value is shown in the $\Phi_{A_{t,b}}$–$\tan\beta$ plane. The corresponding precision obtainable at a $\gamma$C is $\sim 2\%$.

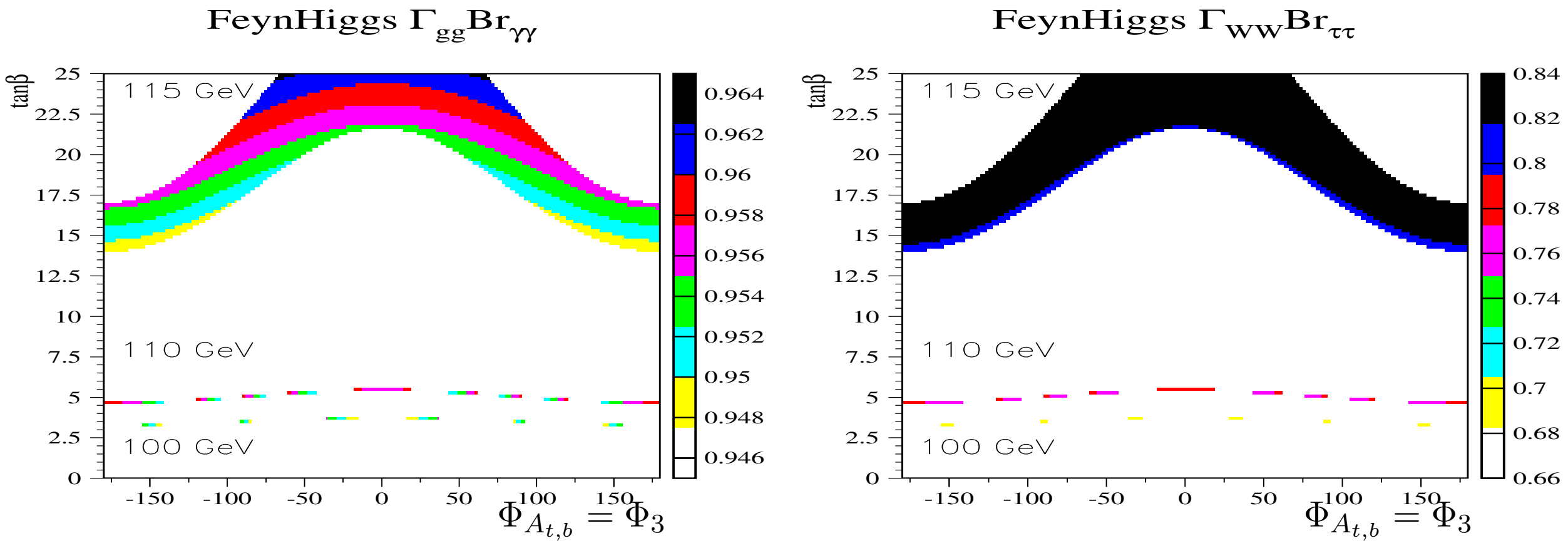

Fig. 3.44: The deviations of $\Gamma_{gg} \times$ BR$(h \to \gamma\gamma)$ (left) and of $\Gamma_{WW} \times$ BR$(h \to \tau^+\tau^-)$ (right) within the BGX scenario from the SM value is shown in the $\Phi_{A_{t,b}}$–$\tan\beta$ plane. The corresponding experimental precision can be found in Table 3.6.

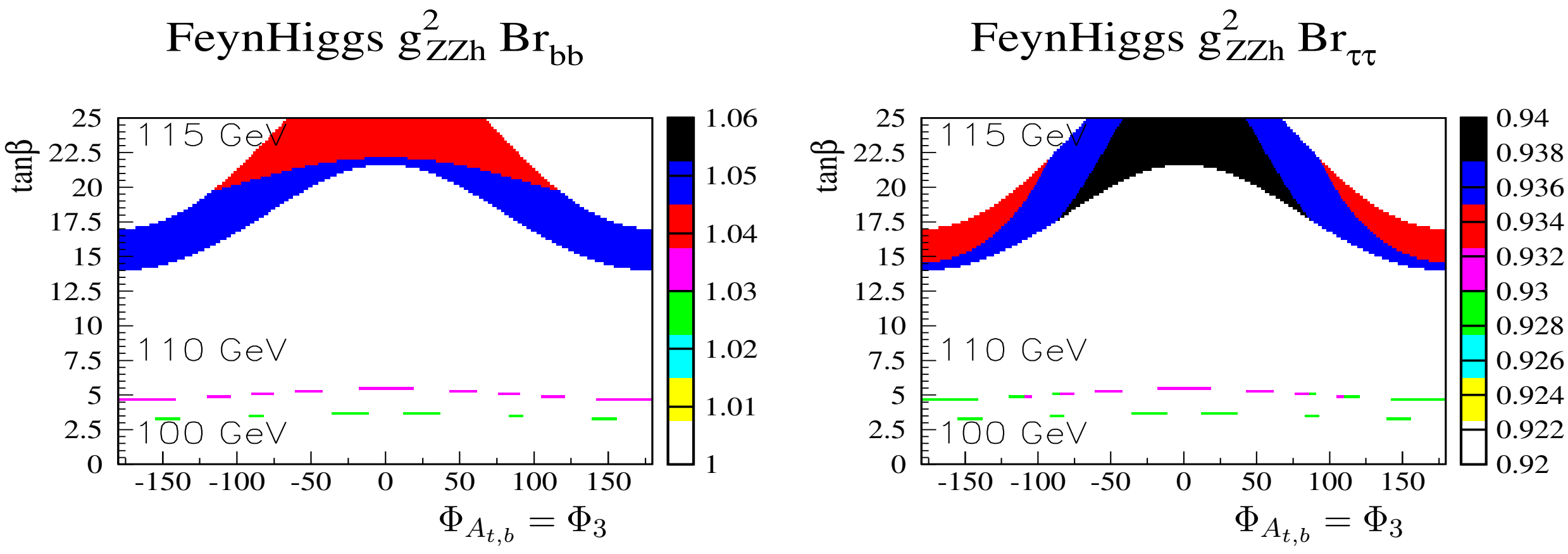

Fig. 3.45: The deviations of $g^2_{ZZh} \times$ BR$(h \to b\bar{b})$ (left) and of $g^2_{ZZh} \times$ BR$(h \to \tau^+\tau^-)$ (right) within the BGX scenario from the SM value is shown in the $\Phi_{A_{t,b}}$–$\tan\beta$ plane. The corresponding experimental precision can be found in Table 3.6.

In Fig. 3.43 we show the $h \to b\bar{b}$ channel at the $\gamma$C. A substantial suppression with respect to the SM can be observed. However, the variation of $\Gamma_{\gamma\gamma} \times \mathrm{BR}(h \to b\bar{b})$ for fixed Higgs boson mass (which will be known with high precision) with the complex phase $\Phi_{A_{t,b}}$ is very small. Thus a precise measurement of this channel at the $\gamma$C will not reveal any information about the complex phases entering the MSSM Higgs sector.

The two LHC channels in the BGX scenario are shown in Fig. 3.44, while the two ILC channels are given in Fig. 3.45. As for the CPX scenario no phase determination can be expected from he LHC measurements. The situation at the ILC in the BGX scenario is similar to the $\gamma$C. A deviation from the SM value can be measured, but the variation of $g^2_{ZZh} \times \mathrm{BR}(h \to b\bar{b}, \tau^+\tau^-)$ is too small to reveal any information on $\Phi_{A_{t,b}}$.

### *3.10.2 Conclusions*

We have compared the LHC, the ILC and the $\gamma$C in view of their power to determine the complex phases of the CP-violating MSSM. We have focused on the Higgs sector, where the complex phases enter via radiative corrections. Especially we have investigated the most promising combinations of Higgs production and decay ($\sigma \times$ BR) for each collider.

The analysis has been performed in two scenarios: The CPX scenario designed to maximize the effect of complex phases in the MSSM Higgs sector. The other scenario (BGX) is based on a part of the CP-violating MSSM that is motivated by baryogenesis.

The CPX scenario may offer good prospects for the $\gamma$C and the ILC to determine $\Phi_{A_{t,b}}$ via Higgs branching ratio measurements. On the other hand, the BGX scenario will only show a deviation from the SM. The variation of the analyzed channels is too small to give information on the complex phases.

It should be kept in mind that we have neglected the future parametric errors on the SUSY parameters (see e.g. Ref. [224] and references therein). These uncertainties will further widen the bands shown in Figs. 3.40–3.45.

## 3.11 Probing CP-violating Higgs contributions in $\gamma\gamma \to f\bar{f}$

*Rohini M. Godbole, Sabine Kraml, Saurabh D. Rindani and Ritesh K. Singh*

At a photon collider, fermion-pair production proceeds through the QED diagrams of Figs. 3.46(a,b) as well as through $s$-channel Higgs mediation, Fig. 3.46(c). Here $\phi$ denotes a generic Higgs boson which may or may not be a CP eigenstate; in the MSSM with CP phases we have of course $\phi = H_{1,2,3}$. The QED vertex $\gamma f\bar{f}$ conserves chirality, whereas the $\phi f\bar{f}$ coupling mixes different chiralities. In the absence of the Higgs contribution, the fermion mass $m_f$ is the only source of chirality-mixing. With unpolarized photons in the initial state, the QED contribution leads to unpolarized fermions in the final state. The same is true for the Higgs exchange should the Higgs boson(s) be a CP eigenstate; CP violation (CPV) in the Higgs sector leads to a net, though very small, polarization of the fermions. With polarized initial-state photons, the QED contribution alone gives rise to a finite polarization. The additional chirality-mixing contribution from the Higgs boson exchange diagram causes a change in this polarization in both the CP-conserving and the CP-violating case. It is thus possible to construct observables involving the polarizations of the initial-state photons and those of the final-state fermions ($\tau/t$), which can probe the Higgs boson couplings, including possible CP violation in the Higgs sector.

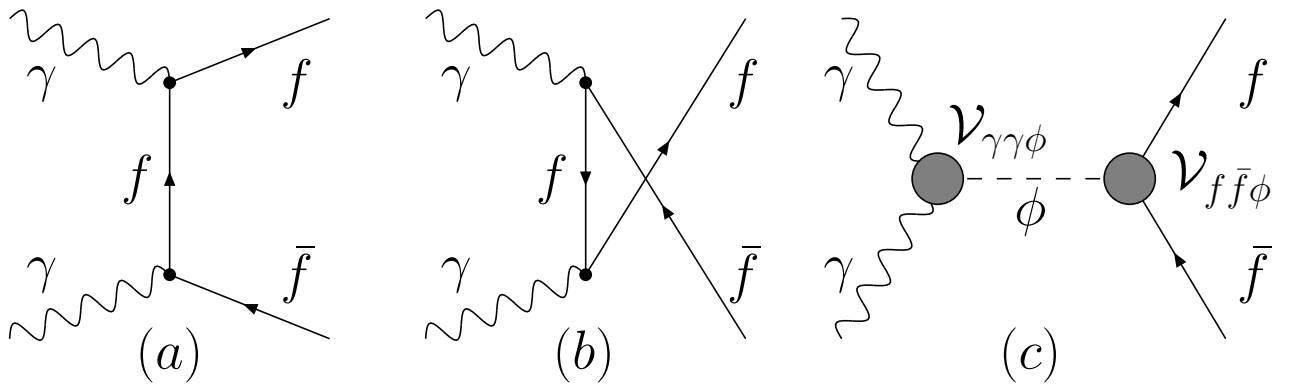

Fig. 3.46: Feynman diagrams contributing to $\gamma\gamma \to f\bar{f}$ production.

Table 3.7: Combinations of form factors that occur in the helicity amplitudes of Eqs. (3.78) and (3.79).

| Combination | Alias | CP | Combination | Alias | CP |
|---|---|---|---|---|---|
| $s_f\,\mathrm{Re}\,e(S_\gamma)$ | $x_1$ | even | $s_f\,\mathrm{Im}\,m(S_\gamma)$ | $x_2$ | even |
| $s_f\,\mathrm{Re}\,e(P_\gamma)$ | $y_1$ | odd | $s_f\,\mathrm{Im}\,m(P_\gamma)$ | $y_2$ | odd |
| $p_f\,\mathrm{Re}\,e(S_\gamma)$ | $y_3$ | odd | $p_f\,\mathrm{Im}\,m(S_\gamma)$ | $y_4$ | odd |
| $p_f\,\mathrm{Re}\,e(P_\gamma)$ | $x_3$ | even | $p_f\,\mathrm{Im}\,m(P_\gamma)$ | $x_4$ | even |

### 3.11.1 Helicity amplitudes

The helicity amplitudes for the fermion-pair production in the $t/u$- and $s$-channels are [90, 91]

$$M_{QED}(\lambda_1,\lambda_2;\lambda_f,\lambda_{\bar f}) = \frac{-i4\pi\alpha Q^2}{1-\beta^2\cos^2\theta_f}\Bigg[\frac{4m_f}{\sqrt{s}}\,(\lambda_1+\lambda_f\beta)\,\delta_{\lambda_1,\lambda_2}\delta_{\lambda_f,\lambda_{\bar f}}$$
$$-\frac{4m_f}{\sqrt{s}}\,\lambda_f\beta\sin^2\theta_f\,\delta_{\lambda_1,-\lambda_2}\delta_{\lambda_f,\lambda_{\bar f}} - 2\beta\,(\cos\theta_f+\lambda_1\lambda_f)\sin\theta_f\,\delta_{\lambda_1,-\lambda_2}\delta_{\lambda_f,-\lambda_{\bar f}}\Bigg], \quad (3.78)$$
$$M_\phi(\lambda_1,\lambda_2;\lambda_f,\lambda_{\bar f}) = \frac{-ie\alpha m_f}{4\pi M_W}\,\frac{s}{s-m_\phi^2+iM_\phi\Gamma_\phi}$$
$$\times\,[S_\gamma(s)+i\lambda_1P_\gamma(s)]\;[\lambda_f\beta s_f - ip_f]\;\delta_{\lambda_1,\lambda_2}\delta_{\lambda_f,\lambda_{\bar f}}\,, \quad (3.79)$$

respectively. Here, $\lambda_{1,2}$ are the helicities of the incoming photons and $\lambda_{f,\bar f}$ those of the produced fermions; $\beta = \sqrt{1-4m_f^2/s}$ and $\theta_f$ is the scattering angle. The form factors $S_\gamma$ and $P_\gamma$ are complex, whereas $s_f$ and $p_f$ may be taken to be real without loss of generality. In fact, only some specific combinations of these form factors occur as listed in Table 3.7. Only five of these eight combinations are independent, the other three can be obtained by inter-relations such as $x_1x_3 = y_1y_3$, etc. Simultaneous existence of $s_f$ and $p_f$, or $S_\gamma$ and $P_\gamma$, signifies CP violation and leads to non-vanishing values of $y_i$ $(i = 1, ..., 4)$. Even if no CPV is present, so that only the $x_i$'s are nonzero, the Higgs-boson contribution still alters the polarization of the fermions $f$ from that predicted by pure QED. CP violation, giving rise to nonzero $y_i$'s, gives an additional contribution to this fermion polarization.

### 3.11.2 Fermion polarization

The fermion polarization is defined as the fractional surplus of positive helicity fermions over negative helicity ones,

$$P_f^{ij} = \frac{N_+^{ij}-N_-^{ij}}{N_+^{ij}+N_-^{ij}}\,, \quad (3.80)$$

where the superscript $ij$ stands for the polarizations of the parent $e^+e^-$ beams $(P_e,\ P_{\bar e})$ of the ILC. $N_+$ and $N_-$ stand for the number of fermions with positive and negative helicities, respectively. The polarization of anti-fermions, $\bar{P}_f^{ij}$, is defined analogously; conservation of linear and angular momenta implies $P_f^{ij} = \bar{P}_f^{ij}$.

Table 3.8: Polarization observables, interactions and the form-factor combinations that they can probe.

| Observable | Type of interaction | Combinations probed |
|---|---|---|
| $P_f^U$ | P/CP violating | $y_i$'s |
| $\delta P_f^+ = P_f^{++} - (P_f^{++})^{QED}$ | chirality mixing | $x_i$'s, $y_i$'s |
| $\delta P_f^- = P_f^{--} - (P_f^{--})^{QED}$ | chirality mixing | $x_i$'s, $y_i$'s |
| $\delta P_f^{CP} = P_f^{++} + P_f^{--}$ | P/CP violating | $y_i$'s |

From Eqs. (3.79) and (3.78) we see that the chirality mixing amplitudes, containing $\delta_{\lambda_f,\lambda_{\bar{f}}}$, are proportional to the fermion mass $m_f$. Hence these are small for light fermions. Further, the Higgs-exchange diagram contributes only when the helicities of the colliding photons are equal; at the same time the QED contribution can be suppressed by choosing $\lambda_1 = \lambda_2$. For the case of the polarized, peaked photon spectrum [83] this amounts to choosing $(P_e,\ P_{\bar{e}}) = (\pm, \pm)$ for the parent $e^+/e^-$ beams.

The final state fermion polarization with unpolarized initial state, $P_f^U$, is zero should the Higgs boson have a definite CP quantum number. Nonzero values of $P_f^U$ only arise for $y_i \neq 0$, thus being a signal of CP violation in the Higgs sector. For polarized initial states, the final-state fermion polarization is always nonzero. Regardless whether CP is violated or not, any deviation of $P_f^{++}$ and $P_f^{--}$ from their QED predictions probes the Higgs boson contribution. Moreover, since P-invariance implies $P_f^{++} = -P_f^{--}$ for the QED contribution, $P_f^{++} + P_f^{--} \neq 0$ is a signal of CP violation. The polarization observables are summarized in Table 3.8. Here note that $\delta P_f^{CP} = \delta P_f^+ + \delta P_f^-$. In the following, we choose $\delta P_f^-$ and $\delta P_f^{CP}$ as the independent observables.

In order to test the relevance of our polarization observables, we perform a numerical analysis for $\gamma\gamma \to \tau^+\tau^-$ and $\gamma\gamma \to t\bar{t}$ in the CPX scenario, Eq. (3.13). We vary $M_{H^\pm} =$ 150–500 GeV and $\tan\beta =$ 3–40 and consider different phases $\Phi_A$, keeping $\Phi_3 = 90°$ fixed. The Higgs boson masses, couplings and widths are computed with both `CPsuperH` [131] and `FeynHiggs` [59]. For the polarized photon beams, we use the ideal Compton-backscattered photon spectrum of [83]. The beam energy $E_b$ for the parent $e^+e^-$ collider is chosen such that the peak in the spectrum of the $\gamma\gamma$ invariant mass corresponds to the relevant Higgs boson mass(es). This choice explores the ultimate potential of the polarization observables; we call it the "peak $E_B$" choice. In general, $P_f^U \neq 0$ would be a clear signal of CPV. However, $P_f^U$ is found to be very small, well below experimental sensitivity. So we have to work with polarized beams and consider $\delta P_f^-$ and $\delta P_f^{CP}$.

Let us start with $f = \tau$. Due to the small $\tau$ mass, the contribution to the $\tau$ polarization from the Higgs boson exchange diagram is very small unless one puts a cut on the $\tau^+\tau^-$ invariant mass. We use a cut $|m_{\tau\tau} - M_{H_1}| \leq \max(dE_m,\ 5\ \Gamma_{H_1})$ with $dE_m = 1$ GeV. In Fig. 3.47 we show $\delta P_\tau^-$ for both `CPsuperH` and `FeynHiggs` for zero and maximal phase $\Phi_A$. The $e^\pm$ beam energy is chosen such that $\sqrt{s}_{\gamma\gamma}$ at the peak of the photon spectrum is equal to the mass of the lightest Higgs boson $H_1$. The deviation in the polarization due to the $H_1$ exchange is large for both $\Phi_A = 0$ and $90°$. $\delta P_\tau^-$ increases with $\tan\beta$ because the $\tau$ Yukawa coupling increases. However, it turns out that $\delta P_\tau^+ \simeq -\delta P_\tau^-$, so that $\delta P_\tau^{CP} \simeq 0$ over all the $\tan\beta$–$M_{H^\pm}$ plane. The difference between `CPsuperH` and `FeynHiggs` in Fig. 3.47 can be traced to somewhat different predictions of the masses, couplings and decay widths as a result of the different approximations used in the two programs.

In the case of $t\bar{t}$ production, it is the heavier Higgs bosons $H_{2,3}$ which contribute. Since the masses of $H_{2,3}$ are in general close to each other, we choose the beam energy such that the mean value $(M_{H_2}+M_{H_3})/2$ matches with $\sqrt{s}_{\gamma\gamma}$ at the peak of the photon spectrum. We find that the top polarization is sensitive not only to the Higgs contribution in general ($\delta P_t^{\pm}$) but also to CP violation ($\delta P_t^{CP}$). In Fig. 3.48 we show contours of constant $\delta P_t^{CP}$ in the $\tan\beta$–$M_{H^\pm}$ plane for $\Phi_A = 90°$. Of course one needs $M_{H^\pm} \geq 2m_t$. Since the top Yukawa coupling decreases with $\tan\beta$, $\delta P_t^{CP}$ is only sizable for small values of $\tan\beta$. Note that due to the large top-quark mass, no cut on the $t\bar{t}$ invariant mass is needed to increase the sensitivity. The difference in the sign of $\delta P_t^{CP}$ in the two panels in Fig. 3.48 is due to different conventions in `CPsuperH` and `FeynHiggs` leading to the opposite signs of $y_i$, $i = 1...4$, for the same input MSSM parameters.

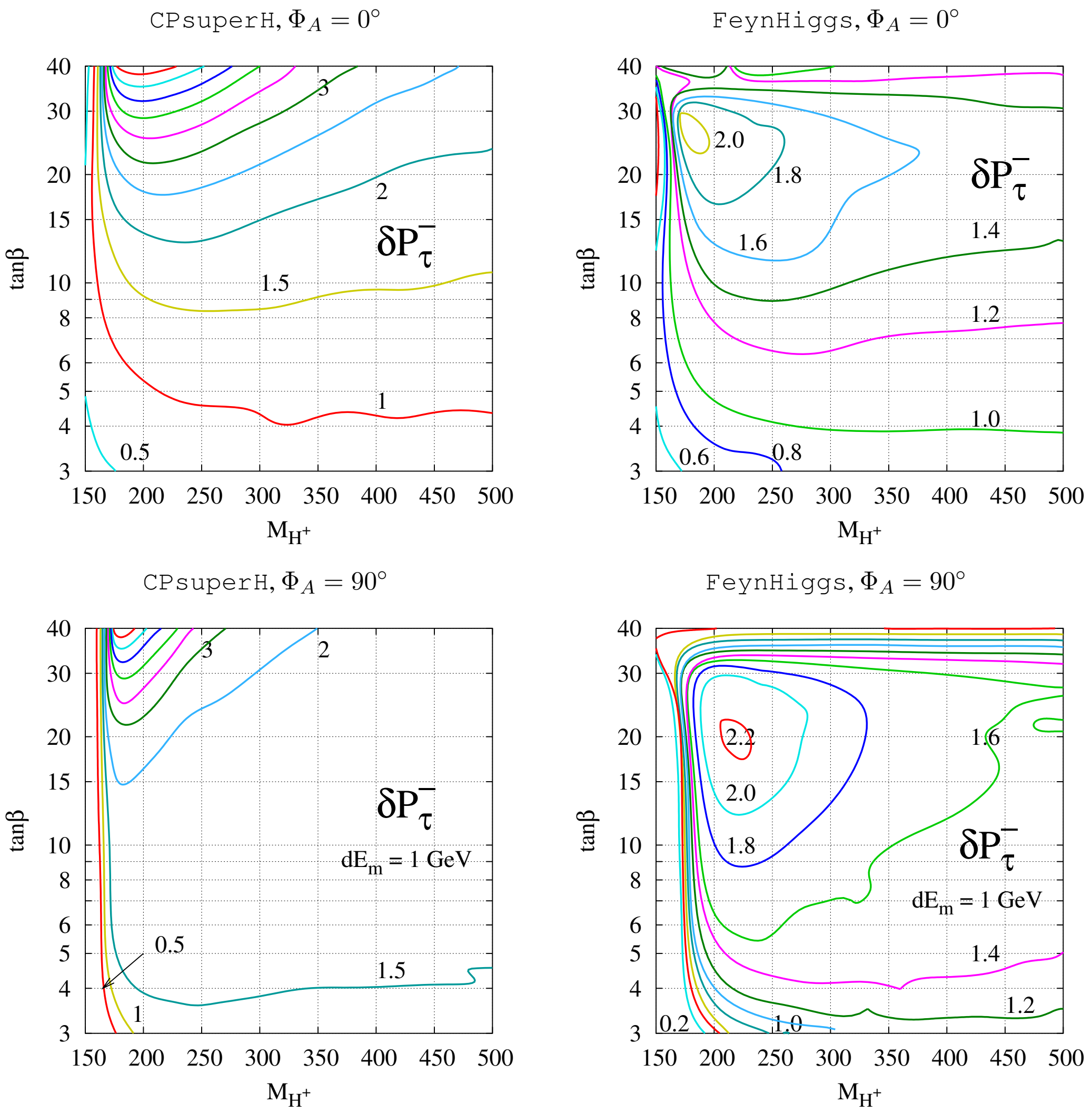

Fig. 3.47: Contours of constant $\delta P_\tau^-$ in units of $10^{-2}$ in the $(\tan\beta$–$M_{H^\pm})$ plane for the CPX scenario with $\Phi_A = 0°$ (top panel) and $\Phi_A = 90°$ (bottom panel) with $dE_m = 1$ GeV. The left panels show the results obtained with `CPsuperH`, the rights panels those obtained with `FeynHiggs`.

### *3.11.3 Leptonic asymmetries*

The polarization of $\tau$ leptons can be measured using the energy distribution of the decay pions [301–305]. The polarization of top quarks can be measured using energy distribution of $b$ quarks [306] or the angular distribution of decay leptons [307–309]. This kind of analysis requires the full reconstruction of the top momentum. Such a reconstruction may not always be possible for the semi-leptonic decay of the $t$ (or $\bar{t}$) quark. On the other hand, it is possible to construct simple asymmetries involving the polarization of the initial-state $e^\pm$ (and hence of the photons) and the charge of the final-state lepton, which are sensitive to CP violation. We denote the integrated cross-section for the process $\gamma\gamma \to t\bar{t} \to l^+\nu b\bar{t}$ $(tl^-\bar{\nu}\bar{b})$ by $\sigma(\lambda_{e^-}, Q_l)$, where $\lambda_{e^-}$ is the polarization of the electron beam in the parent collider and $Q_l$ the charge of the secondary lepton coming from the $t/(\bar{t})$ decay. The polarizations of all the other beams are adjusted to get a peaked spectrum and equal helicities for the incident photons. With this setup, we can define the following asymmetries [91]:

$$\mathcal{A}_1 = \frac{\sigma(+,+) - \sigma(-,-)}{\sigma(+,+) + \sigma(-,-)}, \qquad \mathcal{A}_2 = \frac{\sigma(+,-) - \sigma(-,+)}{\sigma(+,-) + \sigma(-,+)},$$
$$\mathcal{A}_3 = \frac{\sigma(+,+) - \sigma(-,+)}{\sigma(+,+) + \sigma(-,+)}, \qquad \mathcal{A}_4 = \frac{\sigma(+,-) - \sigma(-,-)}{\sigma(+,-) + \sigma(-,-)}. \tag{3.81}$$

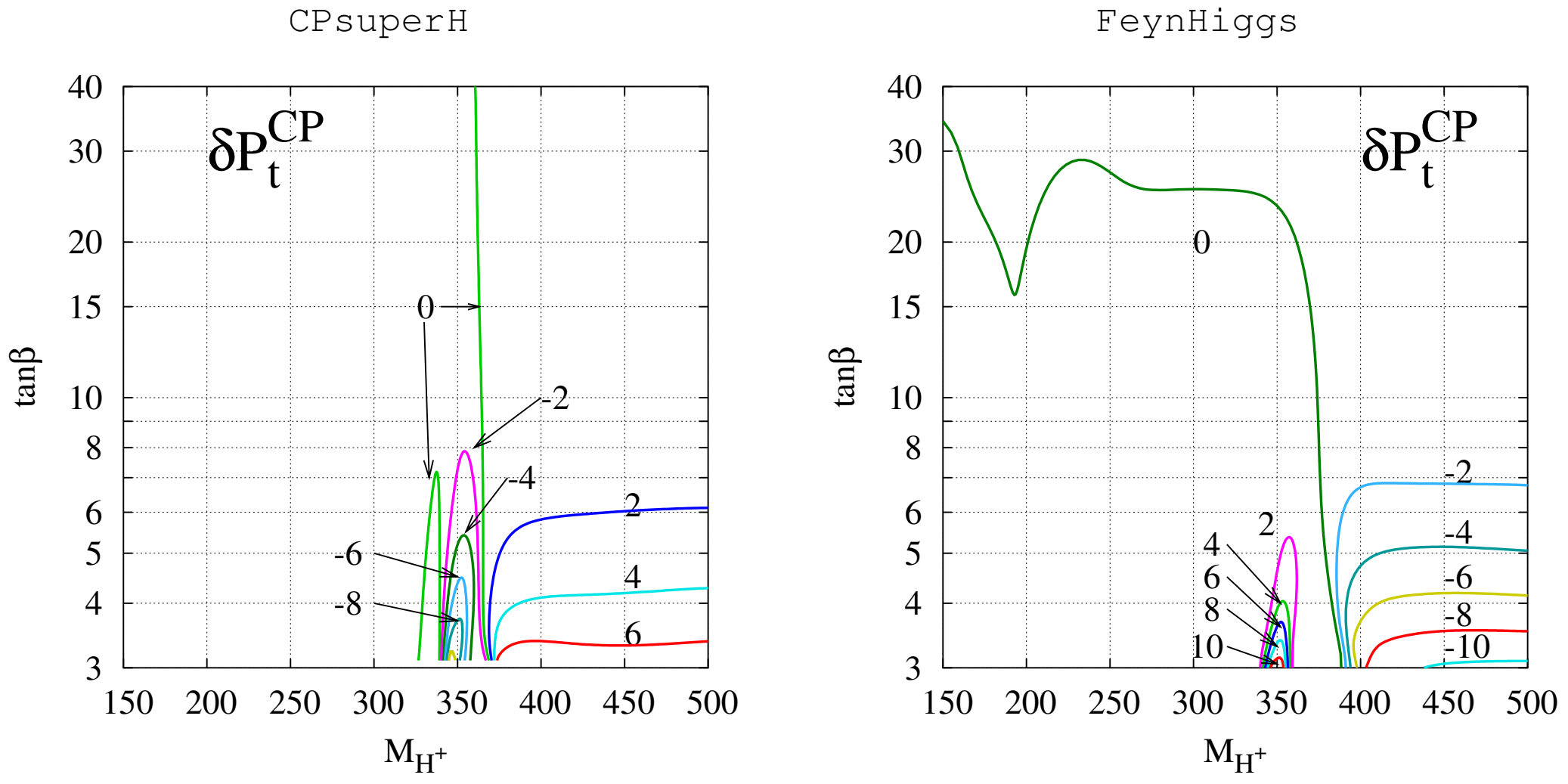

Fig. 3.48: Contours of constant $\delta P_t^{CP}$ in units of $10^{-2}$ in the $(\tan\beta$–$M_{H^\pm})$ plane for the CPX scenario with $\Phi_A = 90°$. The left panel shows the results obtained with `CPsuperH`, the right panel obtained with `FeynHiggs`.

Only one of the above asymmetries is independent [91] if no cut is put on the lepton's polar angle in the laboratory frame. Even with a finite cut on the polar angle, the $\mathcal{A}_{1\ldots4}$ have almost identical sensitivities to the Higgs couplings. We use a 20° beam-pipe cut on the lepton. Figure 3.49 shows contours of constant $\mathcal{A}_3$ for $\Phi_A = 30°$ and 90°, as obtained with `CPsuperH`. The asymmetry is large for large values of $\Phi_A$ and decreases rapidly as $\Phi_A$ decreases. Hence $\mathcal{A}_i$ can probe large regions in the $\tan\beta - M_{H^+}$ plane should $\Phi_A$ be large.

Last but not least we note that, as shown in [91], the lepton asymmetries of Eq. (3.81) are sensitive only to the CP-odd combinations of the form factors, i.e. to the $y_i$'s. This should be contrasted with the polarization observables $\delta P_f^\pm$, which are sensitive to both the CP-odd and CP-even combinations.

### *3.11.4 Summary*

In summary, the polarization of heavy fermions is a good probe of the coupling of the Higgs boson including CP-violation. We have analyzed this in the MSSM with CPV in the CPX scenario . We find that the polarization of $\tau$-leptons may be used to probe the couplings of the lightest Higgs boson, especially in the large $\tan\beta$ region. The $t$-quark polarization, which is sensitive to the contribution of the two heavier Higgs bosons, can be used in the low $\tan\beta$ and large $M_{H^\pm}$ region of the MSSM parameter space. The leptonic asymmetries constructed using the secondary $t/\bar{t}$ decay leptons, which involve only a simple number counting experiment, can probe CPV contributions in $\gamma\gamma \to t\bar{t}$.

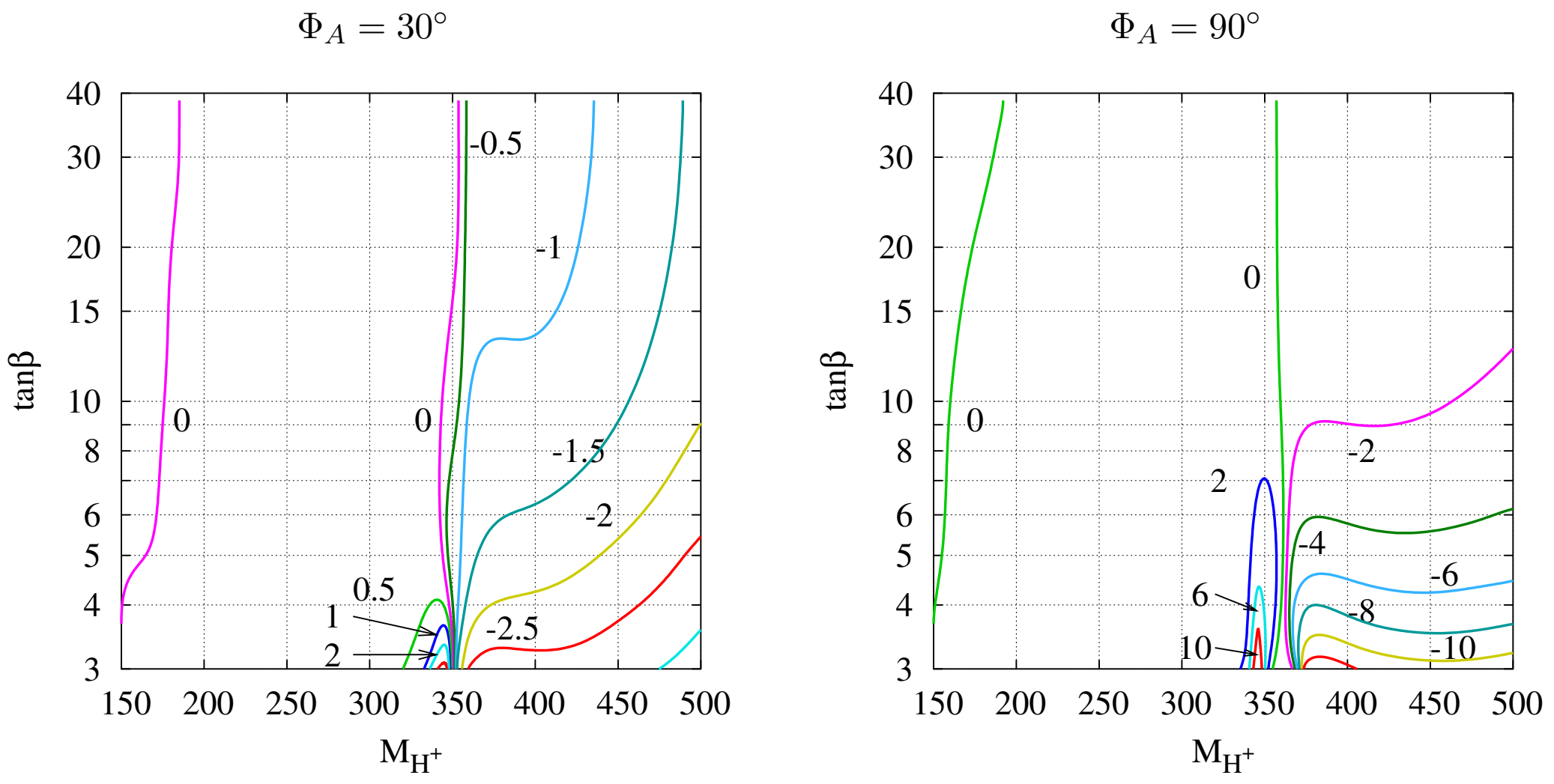

Fig. 3.49: Contours of constant $\mathcal{A}_3$ for $\Phi_A = 30°$ and $90°$; computed with CPsuperH.

### 3.12 Resonant $H$ and $A$ mixing in the CP-noninvariant MSSM

*Seong Youl Choi, Jan Kalinowski, Yi Liao and Peter M. Zerwas*

In CP-noninvariant extensions of the MSSM the three neutral Higgs boson states mix to form a triplet with both even and odd components in the wave–functions under CP transformations [48–53,78,159,181, 310]. The mixing can become very large if the states are nearly mass–degenerate [159]. This situation is naturally realized for supersymmetric theories in the decoupling limit [107] in which two of the neutral states are heavy.

In the present section we describe a simple quantum mechanical (QM) formalism for the CP-violating resonant $H/A$ mixing in the decoupling limit. Subsequently we discuss some experimental signatures of CP-violating mixing in Higgs production and decay processes at a photon collider with polarized photon beams [178].

#### *3.12.1 Mixing formalism*

The self–interaction of two Higgs doublets in a CP-noninvariant theory is described by the potential given in Eq. (2.1), where the coefficients are in general all non–zero and $m_{12}^2$, $\lambda_{5,6,7}$ can be complex. The complex $Y = 1$, SU(2)$_L$ iso–doublet fields can be rotated by an angle $\beta$ [where $\tan\beta = v_2/v_1$ is the ratio of the vev's of the original neutral fields] to a new basis in which only one doublet field acquires a non–zero vev. The real matrix $\mathcal{M}_0^2$ of the three neutral physical Higgs fields in this basis, $H_a$, $H_b$ and $A$, can be written in the form

$$\mathcal{M}_0^2 = v^2 \begin{pmatrix} \lambda & -\hat{\lambda} & -\hat{\lambda}_p \\ -\hat{\lambda} & \lambda - \lambda_A + M_A^2/v^2 & -\lambda_p \\ -\hat{\lambda}_p & -\lambda_p & M_A^2/v^2 \end{pmatrix} \tag{3.82}$$

The mass matrix is hermitian and symmetric by CPT invariance. The parameters $\lambda$, $\hat{\lambda}$ and $\lambda_A$ are combinations of the real parts of the coefficients of bilinear and quartic terms in the Higgs potential, while $\lambda_p$ and $\hat{\lambda}_p$ are given by the imaginary parts of the coefficients; their explicit form can be found in Ref. [159]. The *auxiliary* parameter $M_A^2$ is also a derivative of the real parts of bilinear and quartic coefficients in the Higgs potential; it plays a crucial rôle in characterizing the mass scale of the Higgs system.

In a CP-invariant theory all quartic couplings in the Higgs potential are real and the off–diagonal elements $\lambda_p$, $\hat{\lambda}_p$ vanish. Thus the neutral mass matrix breaks into the CP-even $2 \times 2$ part, and the [stand–alone] CP-odd part. The $2 \times 2$ part gives rise to two CP-even neutral mass eigenstates $h$ and $H$, while $M_A$ is identified as the mass of the CP-odd Higgs boson $A$. In the CP-violating case, however, all three states mix leading to $H_{1,2,3}$ mass eigenstates with no definite CP parities.

For small mass differences, the mixing of the states is strongly affected by their widths. This is a well–known phenomenon for resonant mixing [311] and has also been recognized for the Higgs sector [96, 157, 312–314]. The hermitian mass matrix (3.82) has therefore to be supplemented by the anti–hermitian part $-iM\Gamma$ incorporating the decay matrix [315]

$$\mathcal{M}^2 = \mathcal{M}_0^2 - iM\Gamma \tag{3.83}$$

This matrix includes the widths of the Higgs states in the diagonal elements as well as the transition elements within any combination of pairs. Following from the uncertainty principle, they are particularly important in the case of nearly mass–degenerate states. All these elements $M\Gamma$ are built by loops of virtual particles in the self–energy matrix of the Higgs fields.

In general, the light Higgs boson, the fermions and electroweak gauge bosons, and in supersymmetric theories, gauginos, higgsinos and scalar states may contribute to the loops in the propagator matrix. In the physically interesting case of decoupling, the mixing structure simplifies considerably, allowing a simple and transparent analysis [159]. Alternatively a full coupled–channel analysis may be applied [76, 93, 153, 158].

#### 3.12.1.1 Decoupling limit

The decoupling limit [107] is defined by the inequality $M_A^2 \gg |\lambda_i|\, v^2$ with the quartic couplings in the Higgs potential $|\lambda_i| \lesssim O(1)$. In this limit the $H_a$ state becomes the CP-even light Higgs boson $h$ and decouples from $H = H_b$ and $A$. The heavy states $H$ and $A$ are nearly mass degenerate, which turns out to be crucial for large mixing effects between $H$ and $A$. It is therefore sufficient to consider the lower–right $2 \times 2$ submatrix of the matrix (3.82) for the heavy $H/A$ states which we write as follows

$$\mathcal{M}_{HA}^2 = \begin{pmatrix} M_H^2 - iM_H\Gamma_H & \Delta_{HA}^2 \\ \Delta_{HA}^2 & M_A^2 - iM_A\Gamma_A \end{pmatrix} \tag{3.84}$$

The mixing element $\Delta_{HA}^2$ includes a real dissipative part and an imaginary absorptive part. The couplings of the heavy Higgs bosons to gauge bosons and their supersymmetric partners are suppressed. In the case in which all supersymmetric particle contributions are suppressed either by couplings or by phase space in $M\Gamma$, it is sufficient to consider only loops built by the light Higgs boson and the top quark; the explicit form of the light Higgs and top contributions to the matrix $M\Gamma$ is presented in Ref. [159]. The loops also contribute to the real part of the mass matrix, either renormalizing the $\lambda$ parameters of the Higgs potential or generating such parameters if not present yet at the tree level.

#### 3.12.1.2 Physical masses and states

The symmetric complex mass–squared matrix $\mathcal{M}_{HA}^2$ in Eq. (3.84) can be diagonalized [10] through a *complex rotation*

$$\mathcal{M}_{H_iH_j}^2 = \begin{pmatrix} M_{H_2}^2 - iM_{H_2}\Gamma_{H_2} & 0 \\ 0 & M_{H_3}^2 - iM_{H_3}\Gamma_{H_3} \end{pmatrix} = C\mathcal{M}_{HA}^2 C^{-1} \tag{3.85}$$

[10] The states $H_2$ and $H_3$ are in general not ordered in ascending mass values. Thus, if $M_{H_2} > M_{H_3}$ the indices may be interchanged *ad hoc* to comply with the convention in the Introduction.

where the mixing matrix and the mixing angle are given by

$$C = \begin{pmatrix} \cos\theta & \sin\theta \\ -\sin\theta & \cos\theta \end{pmatrix}, \qquad X = \frac{1}{2}\tan 2\theta = \frac{\Delta^2_{HA}}{M^2_H - M^2_A - i\,[M_H\Gamma_H - M_A\Gamma_A]} \tag{3.86}$$

A non–vanishing (complex) mixing parameter $X \neq 0$ requires CP-violating transitions between $H$ and $A$ either in the real mass matrix, $\lambda_p \neq 0$, or in the decay mass matrix, $(M\Gamma)_{HA} \neq 0$, [or both]. However, even for nearly degenerate masses, the mixing could be suppressed if the widths are significantly different. As a result, the mixing phenomena are strongly affected by the form of the decay matrix $M\Gamma$. Since the difference of the widths enters through the denominator in $X$, the modulus $|X|$ becomes large for small mass differences and small widths.[11]

The mixing shifts the Higgs masses and widths in a characteristic pattern [311]. The two complex mass values after and before diagonalization are related by the complex mixing angle $\theta$:

$$M^2_{H_3} - M^2_{H_2} - i\,(M_{H_3}\Gamma_{H_3} - M_{H_2}\Gamma_{H_2}) = \left[M^2_A - M^2_H - i(M_A\Gamma_A - M_H\Gamma_H)\right] \times \sqrt{1+4X^2} \tag{3.87}$$

Since the eigenstates of the complex, non–hermitian matrix $\mathcal{M}^2_{HA}$ are no longer orthogonal, the ket and bra mass eigenstates have to be defined separately: $|H_i\rangle = C_{i\alpha}|H_\alpha\rangle$ and $\langle\widetilde{H}_i| = C_{i\alpha}\langle H_\alpha|$ ($i = 2, 3$ and $H_\alpha = H, A$). The final state $F$ in heavy Higgs formation from the initial state $I$ is described by the amplitude

$$\langle F|H|I\rangle = \sum_{i=2,3} \langle F|H_i\rangle \frac{1}{s - M^2_{H_i} + iM_{H_i}\Gamma_{H_i}} \langle\widetilde{H}_i|I\rangle \tag{3.88}$$

where the sum runs only over diagonal transitions in the mass–eigenstate basis.

### 3.12.2 *Experimental signatures*

To illustrate the general QM results in a realistic example, we adopt a specific MSSM scenario with the source of CP-violation localized in the complex trilinear coupling $A_t$ of the soft supersymmetry breaking part involving the stop.[12] All other interactions are assumed to be CP-conserving. For complex $A_t$ the stop–loop corrections induce CP-violation in the effective Higgs potential. The effective $\lambda_i$ parameters have been calculated in Ref. [50] to two–loop accuracy; to illustrate the crucial points we focus on the dominant one–loop $t/\tilde{t}$ contributions.

More specifically, we take a typical set of parameters from Ref. [131],

$$M_{\rm SUSY} = 0.5 \text{ TeV}, \quad |A_t| = 1.0 \text{ TeV}, \quad \mu = 1.0 \text{ TeV}; \quad \tan\beta = 5 \tag{3.89}$$

and change the phase $\Phi_A$ of the trilinear parameter $A_t$. With $\Phi_A = 0$ we find the following values of the light and heavy Higgs masses and decay widths, and the stop masses:

$$M_h = 129.6\,\text{GeV},\; M_H = 500.3\,\text{GeV},\; M_A = 500.0\,\text{GeV}$$
$$\Gamma_H = 1.2\,\text{GeV},\; \Gamma_A = 1.5\,\text{GeV};\; m_{\tilde{t}_{1/2}} = 372/647\,\text{GeV}$$

Clearly, with the mass splitting of 0.3 GeV, the heavy Higgs states are not distinguishable. When the phase $\Phi_A$ is turned on,[13] the CP composition, the masses and the decay widths of heavy states are strongly affected, as shown in Figs. 3.50(a), (b) and (c), while the mass of the light Higgs boson $h$ is not. The heavy two–state system shows a very sharp resonant CP-violating mixing, purely imaginary a little above $\Phi_A = 3\pi/4$, Fig. 3.50(a). The mass shift $\Delta M = M_{H_2} - M_{H_3}$ is enhanced by more than an order of magnitude if the CP-violating phase rises to non–zero values[14], reaching a maximal value of $\sim 5.3$ GeV; the shift of the width $\Delta\Gamma = \Gamma_2 - \Gamma_3$ changes from $-0.3$ GeV to a range extending up to $+0.4$ GeV. As a result, the two mass–eigenstates should become clearly distinguishable at future colliders, in particular at a photon collider [179]. Moreover, both states have significant admixtures of CP-even and CP-odd components in the wave–functions. Since $\gamma\gamma$ colliders offer unique conditions for probing the CP-mixing [85, 90–92, 95, 159, 322–328], we discuss two experimental examples: (a) Higgs formation in polarized $\gamma\gamma$ collisions and (b) polarization of top quarks in Higgs decays, where spectacular signatures of resonant mixing can be expected.

[11] Though $H, A$ masses and widths are very close in the decoupling regime of supersymmetric models, they are not expected to be exactly identical if artificially large fine–tuning of unrelated parameters is disregarded; for comments see Ref. [157].

[12] This assignment is compatible with the bounds on CP-violating SUSY phases from experiments on electric dipole moments [232, 316–321].

[13] With one phase $\Phi_A$, the complex mixing parameter $X$ obeys the relation $X(2\pi - \Phi_A) = X^*(\Phi_A)$, implying all CP-even quantities to be symmetric and all CP-odd quantities to be anti–symmetric about $\pi$.

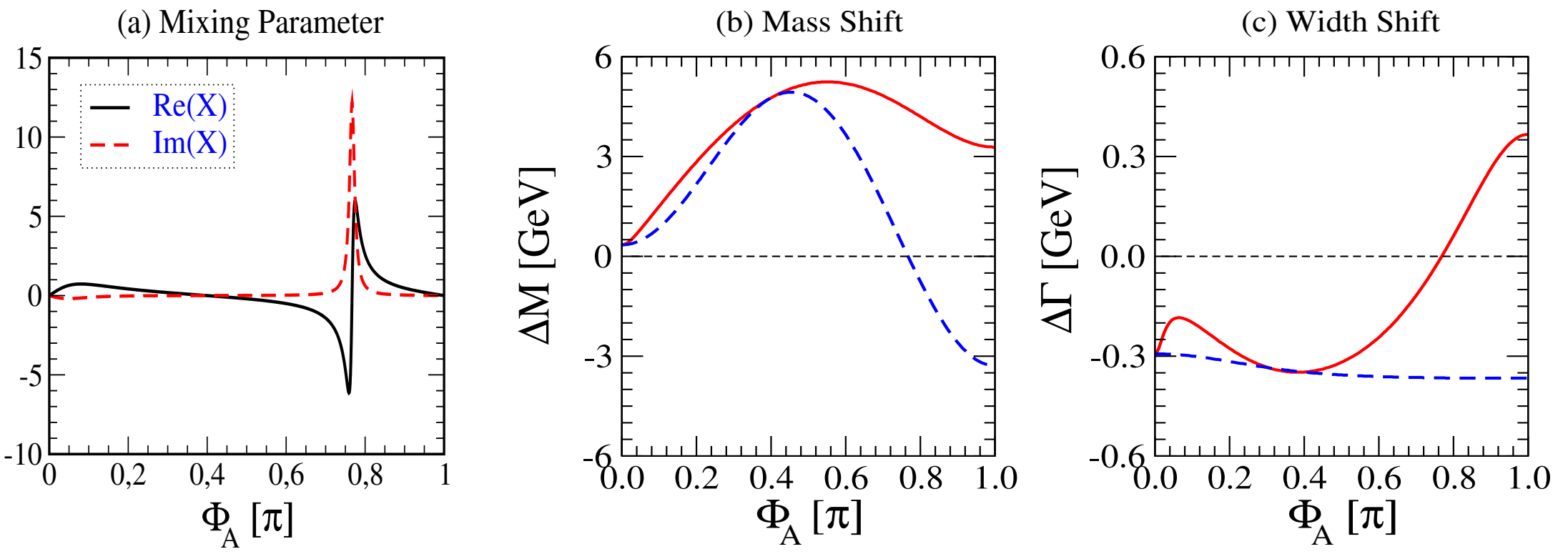

Fig. 3.50: The $\Phi_A$ dependence of (a) the mixing parameter $X$ and of the shifts of (b) masses and (c) widths with the phase $\Phi_A$ evolving from 0 to $\pi$ for $\tan\beta = 5$, $M_A = 0.5$ TeV and couplings as specified in the text; in (b,c) the mass and width differences without mixing are shown by the broken lines. $\mathrm{Re}\,e/\mathrm{Im}\,\mathrm{m}X(2\pi - \Phi_A) = +\mathrm{Re}\,e/- \mathrm{Im}\,\mathrm{m}X(\Phi_A)$ for angles above $\pi$.

**(a)** The amplitude of the reaction $\gamma\gamma \to H_i \to F$ is a superposition of $H_2$ and $H_3$ Higgs exchanges. For equal helicities $\lambda = \pm 1$ of the two photons, the amplitude reads

$$\mathcal{M}^F_\lambda = \sum_{i=2,3} \langle F|H_i\rangle \frac{1}{s - M^2_{H_i} + iM_{H_i}\Gamma_{H_i}} \left[S^\gamma_i(s) + i\lambda P^\gamma_i(s)\right] \tag{3.90}$$

where $\sqrt{s}$ is the $\gamma\gamma$ energy. The loop–induced $\gamma\gamma H_i$ scalar and pseudoscalar form factors, $S^\gamma_i(s)$ and $P^\gamma_i(s)$, are related to the well–known conventional $\gamma\gamma H/A$ form factors, $S^\gamma_{H,A}$ and $P^\gamma_{H,A}$ (explicit form in Refs. [159] and [131]). In our scenario the Higgs–$tt$ couplings are assumed to be CP-conserving, implying negligible top–loop contributions to $P^\gamma_H$ and $S^\gamma_A$ since the gluino mass is sufficiently heavy compared with the stop masses, while the $\tilde{t}_1$ loop generates a non–negligible CP-violating amplitude $S^\gamma_A$. In the region of strong mixing on which we focus, the CP-violating vertex corrections have only a small effect however on the experimental asymmetries compared with the large impact of CP-violating Higgs–boson mixing.

Polarized photons provide a very powerful tool to investigate the CP properties of Higgs bosons. With linearly polarized photons one can project out the CP-even and CP-odd components of the $H_i$ wave–functions by arranging the photon polarization vectors to be parallel or perpendicular. On the other hand, circular polarization provides us with direct insight into the CP-violating nature of Higgs

[14] Note that in this illustrative example $H_2$ is heavier than $H_3$ across the entire $\Phi_A$ range. To avoid confusion with the elaborate paper Ref. [159], we have chosen not to relabel the states in this report.

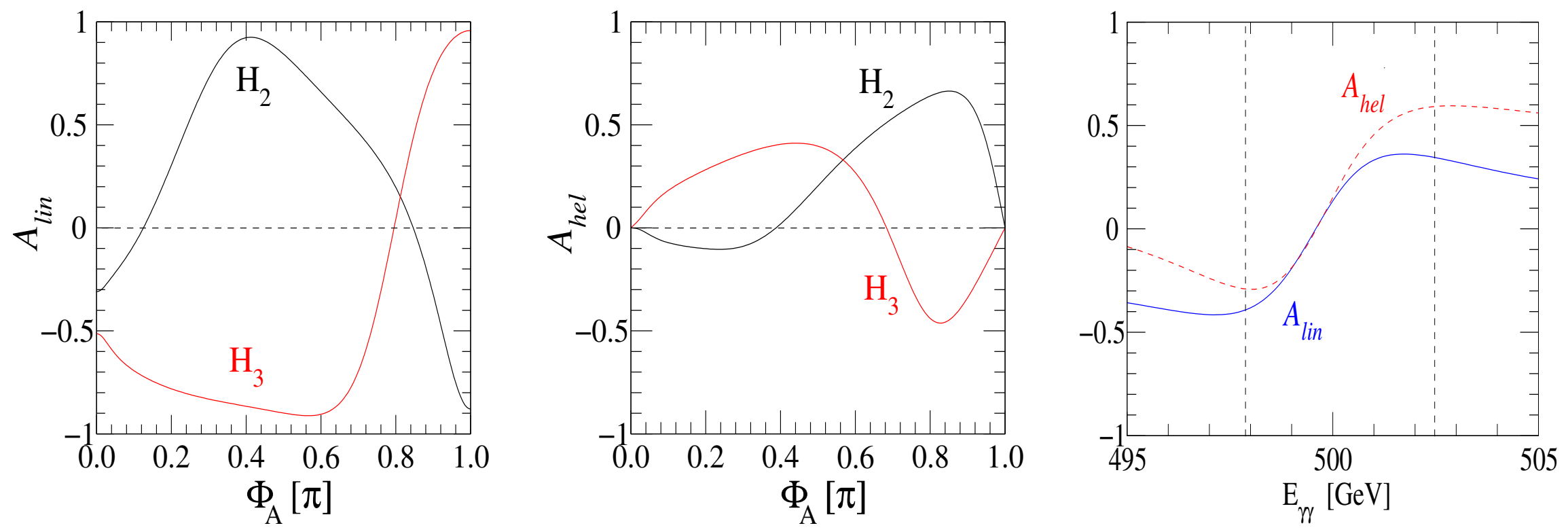

Fig. 3.51: The $\Phi_A$ dependence of the CP-even and CP-odd correlators, $\mathcal{A}_{lin}$ (leftmost panel) and $\mathcal{A}_{hel}$ (center panel), at the poles of the heavy Higgs bosons $H_2$ and $H_3$, respectively; and the $\gamma\gamma$ energy dependence (rightmost panel) of the correlators, $\mathcal{A}_{lin,hel}$, for $\Phi_A = 3\pi/4$ in the production process $\gamma\gamma \to H_i$ in the limit in which $H/A$ mixing is the dominant CP-violating effect. The same parameter set as in Fig. 3.50 is employed. The vertical lines on the right panel mark positions of the two mass eigenvalues, $M_{H_3}$ and $M_{H_2}$.

bosons. Two asymmetries are of interest

$$\mathcal{A}_{lin} = \frac{\sigma_{\parallel} - \sigma_{\perp}}{\sigma_{\parallel} + \sigma_{\perp}}, \qquad \mathcal{A}_{hel} = \frac{\sigma_{++} - \sigma_{--}}{\sigma_{++} + \sigma_{--}} \tag{3.91}$$

where $\sigma_{\parallel}$, $\sigma_{\perp}$ and $\sigma_{++}$, $\sigma_{--}$ are the corresponding total $\gamma\gamma$ fusion cross sections for linear and circular polarizations, respectively. Though CP-even, the asymmetry $\mathcal{A}_{lin}$ can serve as a powerful tool nevertheless to probe CP-violating admixtures to the Higgs states since $|\mathcal{A}_{lin}| < 1$ requires both $S_i^\gamma$ and $P_i^\gamma$ couplings to be non–zero. A more direct probe of CP-violation due to $H/A$ mixing is provided by the CP-odd asymmetry $\mathcal{A}_{hel}$.

In Fig. 3.51 the $\Phi_A$ dependence of the asymmetries $\mathcal{A}_{lin}$ and $\mathcal{A}_{hel}$ is shown at the poles of the heavy Higgs bosons $H_2$ and $H_3$ for the same parameter set as in Fig. 3.50 and with the common SUSY scale $M_{\tilde{Q}_3} = M_{\tilde{U}_3} = M_{\rm SUSY} = 0.5$ TeV for the soft SUSY breaking stop mass parameters. By varying the $\gamma\gamma$ energy from below $M_{H_3}$ to above $M_{H_2}$, the asymmetries, $\mathcal{A}_{lin}$ (blue solid line) and $\mathcal{A}_{hel}$ (red dashed line), move from $-0.39$ to $0.34$ and from $-0.29$ to $0.59$, respectively, as shown in the rightmost panel of Fig. 3.51 with $\Phi_A = 3\pi/4$, a phase value close to resonant CP-mixing.

**(b)** A second observable of interest is the polarization of the top quarks in $H_{2,3} \to t\bar{t}$ decays produced by $\gamma\gamma$ fusion or elsewhere in various production processes at an $e^+e^-$ linear collider and LHC. Even if the $H/Att$ couplings are [approximately] CP-conserving, the complex rotation matrix $C$ may mix the CP-even $H$ and the CP-odd $A$ states, leading to CP-violation. In the production–decay process $\gamma\gamma \to H_i \to t\bar{t}$, two CP-even and CP-odd correlators between the transverse $t$ and $\bar{t}$ polarization vectors $s_\perp$ and $\bar{s}_\perp$,

$$\mathcal{C}_{\parallel} = \langle s_\perp \cdot \bar{s}_\perp \rangle \qquad \text{and,} \qquad \mathcal{C}_{\perp} = \langle \hat{p}_t \cdot (s_\perp \times \bar{s}_\perp) \rangle \tag{3.92}$$

can be extracted from the azimuthal–angle correlation between the two decay planes $t \to bW^+$ and $\bar{t} \to \bar{b}W^-$ [322, 323].

Fig. 3.52 shows the $\Phi_A$ dependence of the CP-even and CP-odd asymmetries, $\mathcal{C}_{\parallel}$ and $\mathcal{C}_{\perp}$, at the poles of $H_2$ and of $H_3$ (leftmost and center panels, respectively). If the invariant $t\bar{t}$ energy is varied throughout the resonance region, the correlators $\mathcal{C}_{\parallel}$ (blue solid line) and $\mathcal{C}_{\perp}$ (red dashed line) vary characteristically from $-0.43$ to $-0.27$ and from $0.84$ to $-0.94$, respectively, as shown in the rightmost panel of Fig. 3.52.

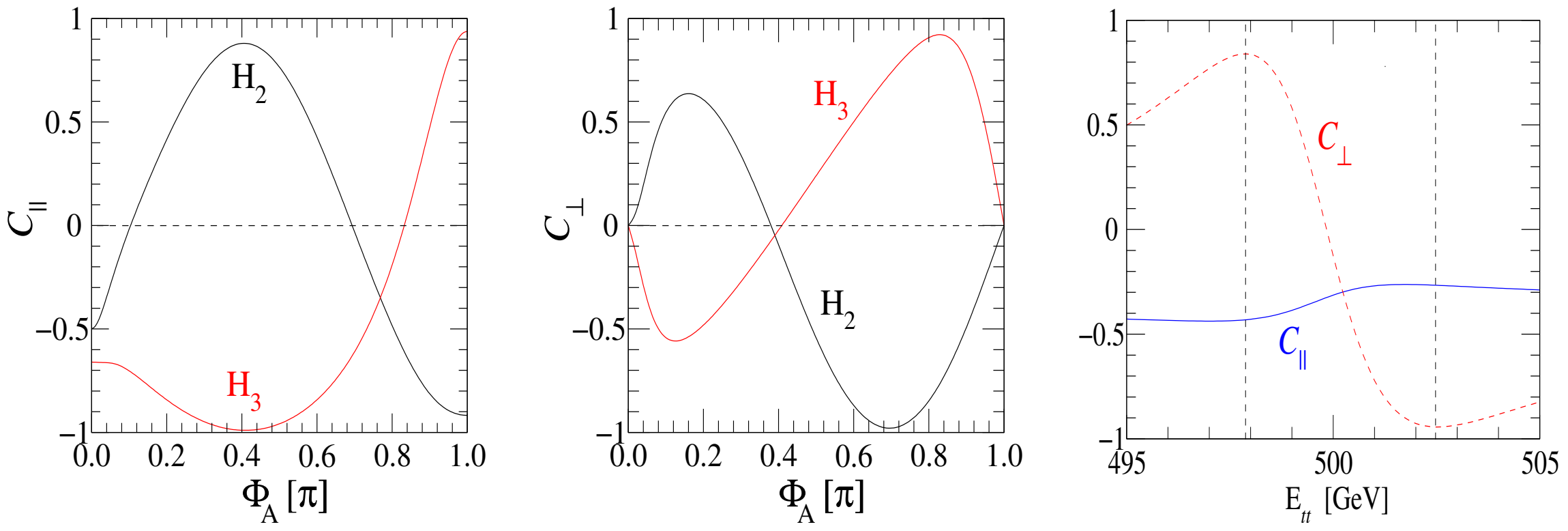

Fig. 3.52: The $\Phi_A$ dependence of the CP-even and CP-odd correlators, $\mathcal{C}_{\|}$ (leftmost panel) and $\mathcal{C}_{\perp}$ (center panel), at the pole of $H_2$ and $H_3$, and the invariant $t\bar{t}$ energy dependence (rightmost panel) of the correlators $\mathcal{C}_{\|,\perp}$ for $\Phi_A = 3\pi/4$ in the production–decay chain $\gamma\gamma \to H_i \to t\bar{t}$. [Same SUSY parameter set as in Fig.3.51.]

### *3.12.3 Conclusions*

Exciting mixing effects can occur in the supersymmetric Higgs sector if CP-noninvariant interactions are realized. In the decoupling regime these effects can become very large, leading to interesting experimental consequences. Higgs formation in $\gamma\gamma$ collisions with polarized beams proves particularly exciting for observing such effects. However, valuable experimental effects are also predicted in such scenarios for $t\bar{t}$ final–state analyses in decays of the heavy Higgs bosons at LHC and in the $e^+e^-$ mode of linear colliders.

Detailed experimental simulations would be needed to estimate the accuracy with which the asymmetries presented here can be measured. Though not easy to measure, the large magnitude and the rapid and significant variation of the CP-even and CP-odd asymmetries through the resonance region with respect to both the phase $\Phi_A$ and the $\gamma\gamma$ energy would be a very exciting effect to observe in any case.

## 3.13 Higgs boson interferences in chargino and neutralino production at a muon collider

*Hans Fraas, Olaf Kittel and Federico von der Pahlen*

A muon collider is an excellent tool to study the masses, widths and couplings of the heavy neutral MSSM Higgs bosons, since they are resonantly produced in s-channels [100,102,329–335]. In particular, interference effects of two nearly degenerate Higgs bosons can give valuable information on their CP properties [101, 336]. For the production of neutralinos [337] and charginos [338] with longitudinally polarized beams, it has been shown recently that energy distributions of their decay products can be used to analyze the CP-even and CP-odd Higgs boson couplings in the CP-conserving MSSM. We extend these studies [337,338] to the CP-violating MSSM with a nonvanishing physical phase $\Phi_A$ of the trilinear scalar coupling $A = A_t = A_b = A_\tau$, which induces CP violation in the Higgs sector at loop level. In the decoupling limit the heavier neutral Higgs bosons are nearly degenerate and CP-violating effects may be resonantly enhanced [93, 96, 157–159, 311–314]. In this report we define CP-sensitive polarization and charge asymmetries, which we analyze in CP-violating MSSM scenarios.

For chargino $\tilde{\chi}_i = \tilde{\chi}_i^{\pm}$ or neutralino $\tilde{\chi}_i = \tilde{\chi}_i^0$ production

$$\mu^+ + \mu^- \to \tilde{\chi}_i + \tilde{\chi}_j \qquad (3.93)$$

via Higgs exchange $H_k$ with $k = 1,2,3$, the effective CP-violating MSSM interaction Lagrangians are

$$\mathcal{L}_{\mu^+\mu^- H} = \bar{\mu}\,[g^S_{H_k\mu\mu} + i\gamma^5 g^P_{H_k\mu\mu}]\,\mu\, H_k, \tag{3.94}$$
$$\mathcal{L}_{\tilde{\chi}\tilde{\chi}H} = \kappa_{\tilde{\chi}}\bar{\tilde{\chi}}_i[g^S_{H_k\chi_i\chi_j} + i\gamma^5 g^P_{H_k\chi_i\chi_j}]\tilde{\chi}_j\, H_k, \tag{3.95}$$

with $\kappa_{\tilde{\chi}^\pm} = 1$ and $\kappa_{\tilde{\chi}^0} = 1/2$. The effective Higgs couplings to the initial muons and the final charginos/neutralinos are obtained from their tree level couplings $g^{S,P}_{\phi_\alpha\mu\mu}$ and $g^{S,P}_{\phi_\alpha\chi_i\chi_j}$, respectively [6],

$$g^{S,P}_{H_k\mu\mu} = C_{k\alpha} g^{S,P}_{\phi_\alpha\mu\mu}, \tag{3.96}$$
$$g^{S,P}_{H_k\chi_i\chi_j} = C_{k\alpha} g^{S,P}_{\phi_\alpha\chi_i\chi_j}, \quad \phi_\alpha = h, H, A, \tag{3.97}$$

where $C$ is a complex orthogonal matrix which diagonalizes the complex Weisskopf-Wigner mass matrix $\mathcal{M}^2$ in the $h, H, A$ basis, in analogy to Eq. (3.85) in Section 3.12. The leading radiative corrections are thus included into the Higgs couplings, as well as their masses and widths. In particular, observables can be defined which are sensitive to the off-diagonal absorptive parts of the Higgs-boson self energies [158], see Eq. (3.23) in Section 3.4. These CP$\tilde{\text{T}}$-odd observables, where $\tilde{\text{T}}$ denotes naive time reversal, depend strongly on the mass difference and widths of the overlapping Higgs bosons.

To analyze the longitudinal polarizations of the produced charginos or neutralinos, we consider their subsequent CP-conserving but P-violating leptonic two-body decays [337, 338]

$$\tilde{\chi}^\pm_j \to \ell^\pm + \tilde{\nu}^{(*)}_\ell, \qquad \tilde{\chi}^0_j \to \ell^\pm + \tilde{\ell}^\mp_a, \quad a = 1,2. \tag{3.98}$$

In the center-of-mass system the differential cross section in the energy $E_\ell$ of the decay lepton $\ell^\pm$ is

$$\frac{d\sigma_{\ell^\pm}}{dE_\ell} = \frac{\sigma(\mu^+\mu^- \to \tilde{\chi}_i\tilde{\chi}_j)\mathrm{BR}(\tilde{\chi}_j \to \ell^\pm \ldots)}{E^{\max}_\ell - E^{\min}_\ell}\left[1 + \eta_{\ell^\pm}\frac{\bar{\Sigma}^3}{\bar{P}}\frac{2(E_\ell - \hat{E}_\ell)}{E^{\max}_\ell - E^{\min}_\ell}\right], \tag{3.99}$$

with the mean lepton energy $\hat{E}_\ell = (E^{\max}_\ell + E^{\min}_\ell)/2$ and the kinematical end-points $E^{\max}_\ell$ and $E^{\min}_\ell$ [337, 338]. The coefficient $|\eta_{\ell^\pm}| \le 1$ is a measure of parity violation in the chargino/neutralino decay [337, 338]. The coefficients $\bar{P}$ and $\bar{\Sigma}^3$ of the production spin-density matrix are averaged over the chargino/neutralino production solid angle, indicated by a bar in our notation. The cross section $\sigma(\mu^+\mu^- \to \tilde{\chi}_i\tilde{\chi}_j)$ is proportional to $\bar{P}$, whereas the longitudinal chargino/neutralino polarization is proportional to $\bar{\Sigma}^3$. Both coefficients have resonant (res) contributions from Higgs $H_{1,2,3}$ exchange, and continuum (cont) contributions from gauge boson and slepton exchange

$$\bar{P} = P_{\text{res}} + \bar{P}_{\text{cont}}, \qquad \bar{\Sigma}^3 = \Sigma^3_{\text{res}} + \bar{\Sigma}^3_{\text{cont}}. \tag{3.100}$$

The dependence of the isotropic resonant contributions $P_{\text{res}}$ and $\Sigma^3_{\text{res}}$ on the longitudinal $\mu^+$ and $\mu^-$ beam polarizations $\mathcal{P}_+$ and $\mathcal{P}_-$, respectively, is given by

$$P_{\text{res}} = (1 + \mathcal{P}_+\mathcal{P}_-)a_0 + (\mathcal{P}_+ + \mathcal{P}_-)a_1, \tag{3.101}$$
$$\Sigma^3_{\text{res}} = (1 + \mathcal{P}_+\mathcal{P}_-)b_0 + (\mathcal{P}_+ + \mathcal{P}_-)b_1, \tag{3.102}$$

with

$$a_n = \sum_{H_k, H_l (k\le l)} (2 - \delta_{kl}) a^{kl}_n, \quad b_n = \sum_{H_k, H_l (k\le l)} (2 - \delta_{kl}) b^{kl}_n; \quad n = 0, 1; \quad k, l = 1,2,3 \tag{3.103}$$

and, suppressing the chargino/neutralino indices $i$ and $j$ of the couplings,

$$a^{kl}_0 = \frac{s}{2}|\Delta_{(kl)}|\Big[|c^+_\mu||c^+_\chi|f_{ij}\cos(\delta^+_\mu + \delta^+_\chi + \delta_\Delta) - |c^+_\mu||c^{RL}_\chi|m_i m_j \cos(\delta^+_\mu + \delta^{RL}_\chi + \delta_\Delta)\Big]_{(kl)} \tag{3.104}$$
$$a^{kl}_1 = \frac{s}{2}|\Delta_{(kl)}|\Big[|c^-_\mu||c^+_\chi|f_{ij}\cos(\delta^-_\mu + \delta^+_\chi + \delta_\Delta) - |c^-_\mu||c^{RL}_\chi|m_i m_j \cos(\delta^-_\mu + \delta^{RL}_\chi + \delta_\Delta)\Big]_{(kl)} \tag{3.105}$$
$$b^{kl}_0 = -\frac{s}{2}|\Delta_{(kl)}|\Big[|c^+_\mu||c^-_\chi|g_{ij}\cos(\delta^+_\mu + \delta^-_\chi + \delta_\Delta)\Big]_{(kl)} \tag{3.106}$$
$$b^{kl}_1 = -\frac{s}{2}|\Delta_{(kl)}|\Big[|c^-_\mu||c^-_\chi|g_{ij}\cos(\delta^-_\mu + \delta^-_\chi + \delta_\Delta)\Big]_{(kl)} \tag{3.107}$$

We have defined the products of couplings,

$$c^{+}_{\lambda(kl)} = g^{S}_{H_k\lambda\lambda}g^{S*}_{H_l\lambda\lambda} + g^{P}_{H_k\lambda\lambda}g^{P*}_{H_l\lambda\lambda} = \left[|c^{+}_{\lambda}|\exp(i\delta^{+}_{\lambda})\right]_{(kl)}, \quad \lambda = \mu, \chi, \tag{3.108}$$

$$c^{-}_{\lambda(kl)} = -i(g^{S}_{H_k\lambda\lambda}g^{P*}_{H_l\lambda\lambda} - g^{P}_{H_k\lambda\lambda}g^{S*}_{H_l\lambda\lambda}) = \left[|c^{-}_{\lambda}|\exp(i\delta^{-}_{\lambda})\right]_{(kl)}, \quad \lambda = \mu, \chi, \tag{3.109}$$

$$c^{RL}_{\chi(kl)} = g^{S}_{H_k\chi\chi}g^{S*}_{H_l\chi\chi} - g^{P}_{H_k\chi\chi}g^{P*}_{H_l\chi\chi} = \left[|c^{RL}_{\chi}|\exp(i\delta^{RL}_{\chi})\right]_{(kl)}, \tag{3.110}$$

the product of the Higgs boson propagators,

$$\Delta_{(kl)} = \Delta(H_k)\Delta(H_l)^* = \left[|\Delta|\exp(i\delta_{\Delta})\right]_{(kl)}, \tag{3.111}$$

and the kinematical functions $f_{ij} = (s - m_i^2 - m_j^2)/2$ and $g_{ij} = \sqrt{\lambda(s, m_i^2, m_j^2)}/2$ of the chargino/neutralino masses $m_i$, $m_j$, and the center of mass energy $s$. For longitudinally polarized muon beams, the two combinations $a_1$, Eq. (3.101), and $b_1$, Eq. (3.102), of products of Higgs boson couplings to the muons and charginos/neutralinos can be determined, e.g., by polarization asymmetries. A muon collider provides a good beam energy resolution and thus will be the ideal tool to analyze the strong $\sqrt{s}$ dependence of these observables.

### *3.13.1 Asymmetries of the chargino and neutralino production cross section*

For the cross section $\sigma_{ij}$ of chargino $\sigma(\mu^+\mu^- \to \tilde{\chi}_i^+\tilde{\chi}_j^-)$ or neutralino $\sigma(\mu^+\mu^- \to \tilde{\chi}_i^0\tilde{\chi}_j^0)$ pair production, Eq. (3.93), we define for equal beam polarizations $\mathcal{P}_+ = \mathcal{P}_- \equiv \mathcal{P}$ the asymmetries

$$\mathcal{A}^{\text{pol}\pm}_{\text{prod}} = \frac{[\sigma_{ij}(\mathcal{P}) - \sigma_{ij}(-\mathcal{P})] \pm [i \leftrightarrow j]}{[\sigma_{ij}(\mathcal{P}) + \sigma_{ij}(-\mathcal{P})] + [i \leftrightarrow j]}. \tag{3.112}$$

The asymmetry $\mathcal{A}^{\text{pol}+}_{\text{prod}}$ is CP-odd and CP$\tilde{\text{T}}$-odd, with $\tilde{\text{T}}$ naive time reversal $t \to -t$, and is thus non-zero only for complex transition amplitudes with absorptive phases. It is therefore sensitive to the CP phases of the Higgs boson couplings to the charginos/neutralinos and to the muons, and is largest if the mass difference of the two heavy Higgs bosons is of the order of their widths. In Fig. 3.53 we show, for $\tilde{\chi}_1^+\tilde{\chi}_1^-$ production, $\mathcal{A}^{\text{pol}+}_{\text{prod}}$ and $\sigma_{11}$ for $\mathcal{P} = 0.3$, $\mathcal{P} = -0.3$ and $\Phi_A = 0.2\pi$. We obtain an asymmetry of 30% which can be measured at a muon collider with longitudinally polarized beams. For neutralino production $\mu^+\mu^- \to \tilde{\chi}_1^0\tilde{\chi}_2^0$, the asymmetry $\mathcal{A}^{\text{pol}+}_{\text{prod}}$ reaches 16%, for the parameters as given in the caption of Fig. 3.53, and the neutralino production cross section $\sigma_{12}$ is of the order of 400 fb. The asymmetries are proportional to $a_1/a_0$, Eq. (3.101), if the continuum contributions $\bar{P}_{\text{cont}}$ are subtracted, e.g., through extrapolation of the cross section around the resonances [339], or by chargino/neutralino cross section measurements at the ILC [220]. The CP-even asymmetry $\mathcal{A}^{\text{pol}-}_{\text{prod}}$ vanishes for the production of neutralinos, due to their Majorana character, as well as for the production of equal charginos. For the production of unequal charginos $\tilde{\chi}_i^+\tilde{\chi}_j^-$, with $i \neq j$, measurements of $\mathcal{A}^{\text{pol}\pm}_{\text{prod}}$ allow to separate the CP-even and CP-odd parts of the coefficient $a_1$.

Similarly, for $\tilde{\chi}_i^+\tilde{\chi}_j^-$ production with $i \neq j$, the coefficients $a_0$, $b_0$ and $b_1$ can be separated into their symmetric and antisymmetric parts under exchange of $i$ and $j$, to obtain CP-even and CP-odd observables. We define the charge asymmetry

$$\mathcal{A}^{C}_{\text{prod}} = \frac{\sigma_{12}(\mathcal{P}) - \sigma_{21}(\mathcal{P}) + \sigma_{12}(-\mathcal{P}) - \sigma_{21}(-\mathcal{P})}{\sigma_{12}(\mathcal{P}) + \sigma_{21}(\mathcal{P}) + \sigma_{12}(-\mathcal{P}) + \sigma_{21}(-\mathcal{P})} \tag{3.113}$$

of the chargino production cross sections $\sigma_{ij}$. This asymmetry is CP-odd and CP$\tilde{\text{T}}$-even. In Fig. 3.54 we show $\mathcal{A}^{C}_{\text{prod}}$ for two scenarios with different scalar mass parameters $M_{\text{SUSY}}$ and trilinear coupling parameters $|A|$ for unpolarized beams $\mathcal{P} = 0$. The production of $\tilde{t}_1$ pair production strongly suppresses one chargino production amplitude of the Higgs boson, enhancing $\mathcal{A}^{C}_{\text{prod}}$.

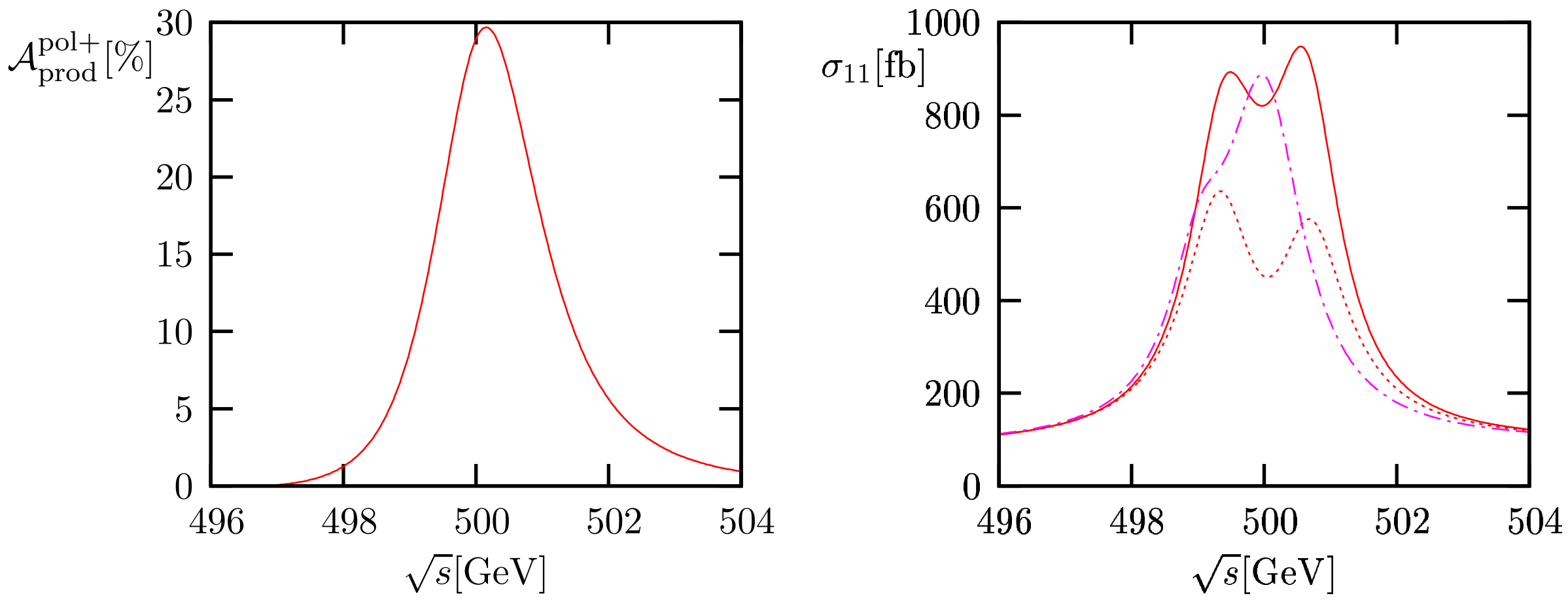

Fig. 3.53: Chargino production asymmetry $\mathcal{A}_{\rm prod}^{\rm pol+}$, Eq. (3.112), and cross section $\sigma_{11} = \sigma(\mu^+\mu^- \to \tilde{\chi}_1^+\tilde{\chi}_1^-)$ for $\Phi_A = 0.2\pi$, $\mathcal{P}_+ = \mathcal{P}_- = \mathcal{P} = 0.3$ (solid), $\mathcal{P} = -0.3$ (dotted), and for $\Phi_A = 0$, $\mathcal{P} = \pm 0.3$ (dash-dotted), with $M_A = 500$ GeV ($M_{H^\pm} = 506.9$ GeV), $\tan\beta = 10$, $\mu = 500$ GeV, $M_2 = 220$ GeV, $|A| = 2\,M_{\rm SUSY} = 1$ TeV, evaluated using `FeynHiggs2.3` [144]. The Higgs masses and widths for $\Phi_A = 0.2\pi$ are $M_{H_{2,3}} = 499.4$ GeV, 500.7 GeV and $\Gamma_{H_{2,3}} = 1.3$ GeV, respectively, and $m_{\chi_1^\pm} = 210$ GeV.

### *3.13.2 Asymmetries of the energy distributions for the chargino and neutralino decay products*

For the cross section $\sigma_{\ell^\pm}$ of chargino/neutralino production, Eq. (3.93), followed by their subsequent decay into a lepton $\ell^\pm$, Eq. (3.98), we define the asymmetry of the leptonic energy distribution, Eq. (3.99),

$$\mathcal{A}_{\ell^\pm} = \frac{\Delta\sigma_{\ell^\pm}}{\sigma_{\ell^\pm}} = \frac{1}{2}\eta_{\ell^\pm}\frac{\bar{\Sigma}^3}{\bar{P}}, \tag{3.114}$$

with $\Delta\sigma_{\ell^\pm} = \sigma_{\ell^\pm}(E_\ell > \hat{E}_\ell) - \sigma_{\ell^\pm}(E_\ell < \hat{E}_\ell)$. Since $\mathcal{A}_{\ell^\pm}$ is proportional to the averaged longitudinal chargino/neutralino polarization $\bar{\Sigma}^3/\bar{P}$ it allows to determine the coefficients $b_1$ and $b_0$, Eq. (3.102), with, respectively, the polarization asymmetries

$$\mathcal{A}_{\ell^\pm}^{\rm pol} = \frac{\Delta\sigma_{\ell^\pm}(\mathcal{P}) - \Delta\sigma_{\ell^\pm}(-\mathcal{P})}{\sigma_{\ell^\pm}(\mathcal{P}) + \sigma_{\ell^\pm}(-\mathcal{P})} = \eta_{\ell^\pm}\frac{\mathcal{P}b_1}{(1+\mathcal{P}^2)a_0 + \bar{P}_{\rm cont}}, \tag{3.115}$$

$$\mathcal{A}_{\ell^\pm}^{\prime\rm pol} = \frac{\Delta\sigma_{\ell^\pm}(\mathcal{P}) + \Delta\sigma_{\ell^\pm}(-\mathcal{P})}{\sigma_{\ell^\pm}(\mathcal{P}) + \sigma_{\ell^\pm}(-\mathcal{P})} = \frac{1}{2}\eta_{\ell^\pm}\frac{(1+\mathcal{P}^2)b_0 + \bar{\Sigma}^3_{\rm cont}}{(1+\mathcal{P}^2)a_0 + \bar{P}_{\rm cont}}, \tag{3.116}$$

for equal muon beam polarizations $\mathcal{P}_+ = \mathcal{P}_- \equiv \mathcal{P}$. The asymmetry $\mathcal{A}_{\ell^\pm}^{\rm pol}$ measures the correlation between initial and final longitudinal polarizations, and is CP-even for the production of neutralinos or equal charginos. Large values of $\mathcal{A}_{\ell^\pm}^{\rm pol}$ are obtained if both resonances are degenerate and their amplitudes are of the same magnitude. As in the CP-conserving MSSM [337, 338], the relative phase of the interfering resonances is approximately $\pi/2$ in the Higgs decoupling limit. However, resonantly enhanced CP violation tends to widen the mass difference of the heavy Higgs bosons [159] and thus suppress this asymmetry, as can be observed on the left hand side of Fig. 3.55, where we show the asymmetry $\mathcal{A}_{\ell^+}^{\rm pol}$ for light chargino pair production, both for $\Phi_A = 0.2\pi$ and for $\Phi_A = 0$. The corresponding asymmetry $\mathcal{A}_{\ell^+}^{\prime\rm pol}$, Eq. (3.116), depends on the continuum contributions of the chargino polarization $\bar{\Sigma}^3_{\rm cont}$, which can be eliminated in the charge asymmetry

$$\mathcal{A}_{\ell^+}^{\prime\rm C} = \frac{1}{2}(\mathcal{A}_{\ell^+}^{\prime\rm pol} - \mathcal{A}_{\ell^-}^{\prime\rm pol}) = \frac{1}{2}\eta_{\ell^\pm}\frac{(1+\mathcal{P}^2)b_0}{(1+\mathcal{P}^2)a_0 + \bar{P}_{\rm cont}}, \tag{3.117}$$

shown on the r.h.s. of Fig. 3.55. For neutralino production the continuum $\bar{\Sigma}^3_{\rm cont} = 0$ vanishes naturally due to their Majorana character [340], thus $\mathcal{A}_{\ell^+}^{\prime\rm C} = \mathcal{A}_{\ell^+}^{\prime\rm pol}$. For the production of neutralinos, $\mu^+\mu^- \to$ $\tilde{\chi}_1^0\tilde{\chi}_2^0$, with subsequent leptonic decay, Eq. (3.98), the asymmetry $\mathcal{A}'^{\rm pol}_{\ell+}$ reaches 8%, for the scenario as given in the caption of Fig. 3.55. The asymmetry $\mathcal{A}^{\rm pol}_{\ell+}$, Eq. (3.115), is similar in size and shape as the corresponding asymmetry for chargino production, see Fig. 3.55.

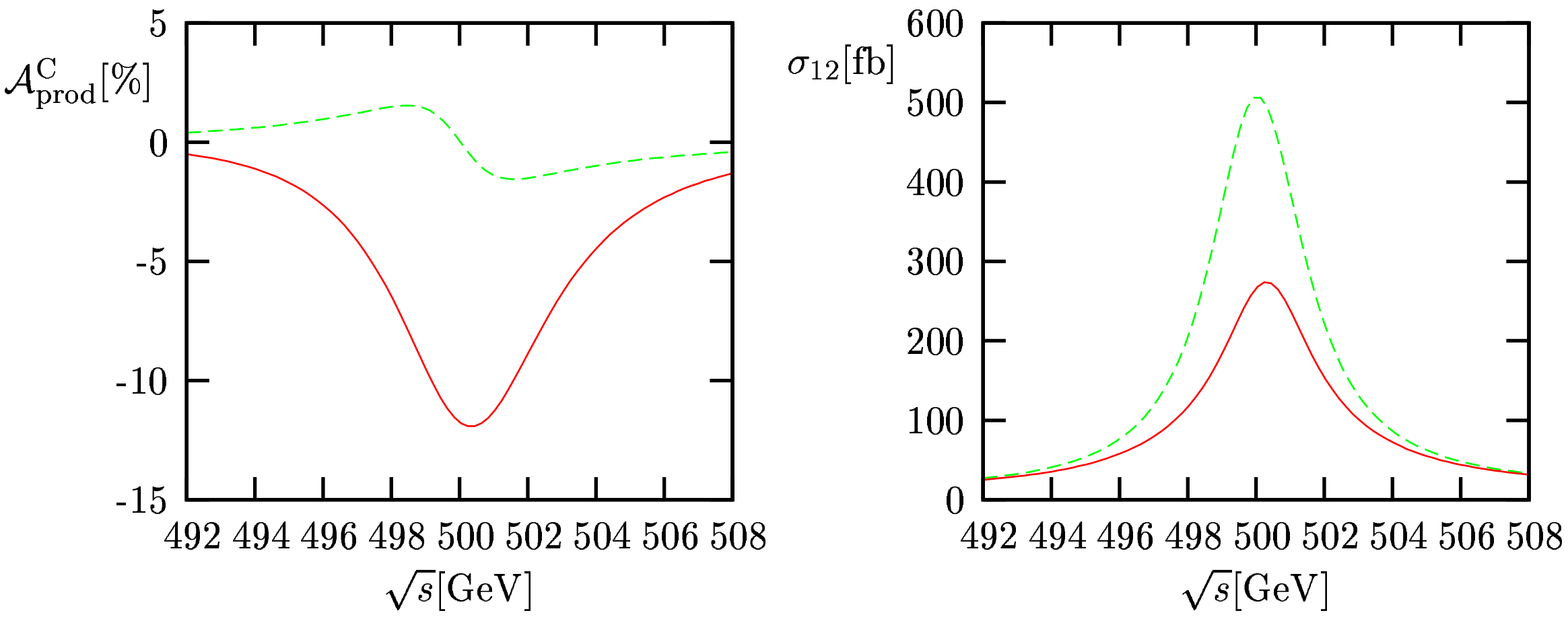

Fig. 3.54: Chargino production asymmetry $\mathcal{A}^{C}_{\rm prod}$, Eq. (3.113), and cross section $\sigma_{12} = \sigma(\mu^+\mu^- \to \tilde{\chi}_1^+\tilde{\chi}_2^-)$ for $|A| = 500$ GeV, $M_{\rm SUSY} = 300$ GeV (solid), and for $|A| = 2\,M_{\rm SUSY} = 1$ TeV (dashed), with $\mathcal{P}_+ = \mathcal{P}_- = 0$, $\Phi_A = 0.2\pi$, $M_A = 500$ GeV ($M_{H^\pm} = 505.7$ GeV), $\tan\beta = 10$, $\mu = 320$ GeV, $M_2 = 120$ GeV, evaluated using `FeynHiggs2.3` [144]. The Higgs masses and widths are $M_{H_{2,3}} = 500$ GeV, 500.3 GeV, $\Gamma_{H_{2,3}} = 7.5$ GeV, 3.2 GeV (solid), and $M_{H_{2,3}} = 499.8$ GeV, 500.3 GeV, $\Gamma_{H_{2,3}} = 3.2$ GeV (dashed), and $m_{\chi_1^\pm} = 108$ GeV, $m_{\chi_2^\pm} = 344$ GeV.

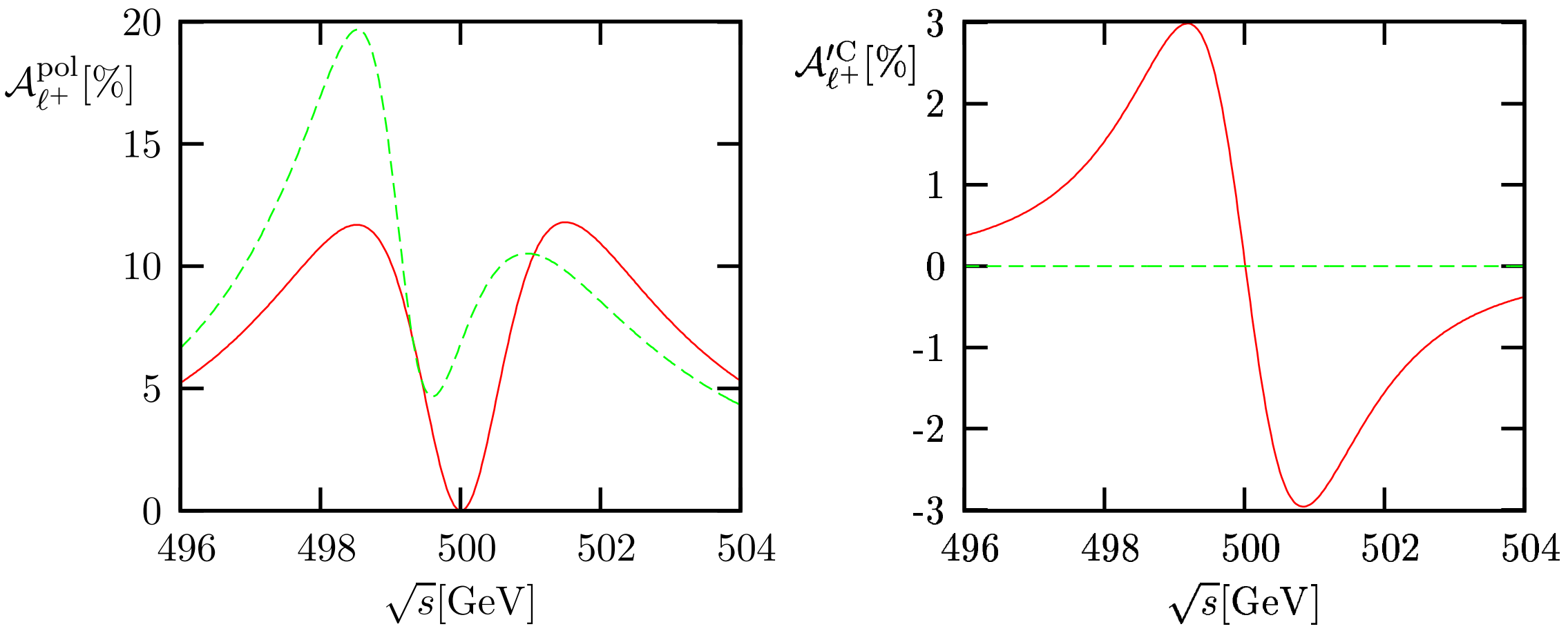

Fig. 3.55: Polarization asymmetry $\mathcal{A}^{\rm pol}_{\ell+}$, Eq. (3.115), and $\mathcal{A}'^{\rm C}_{\ell+}$, Eq. (3.117), for $\mu^+\mu^- \to \tilde{\chi}_1^+\tilde{\chi}_1^-$ with subsequent leptonic chargino decay, for $\Phi_A = 0.2\pi$ (solid), $\Phi_A = 0$ (dashed), with $\mathcal{P}_+ = \mathcal{P}_- = \mathcal{P} = -0.3$, evaluated using `FeynHiggs2.3` [144]. For cross sections and other parameters, see Fig. 3.53. The Higgs masses and widths for $\Phi_A = 0$ are $M_{H_{2,3}} = 499.0$ GeV, 500.0 GeV and $\Gamma_{H_{2,3}} = 1.1$ GeV, 1.4 GeV.

### *3.13.3 Summary and conclusions*

We have studied chargino and neutralino production and their leptonic decays at the muon collider with longitudinally polarized beams. We have defined polarization and charge asymmetries to study the interference of the heavy neutral CP-violating MSSM Higgs bosons with CP violation, radiatively induced by the common phase $\Phi_A$ of the trilinear scalar couplings. For nearly degenerate neutral Higgs bosons, with mass differences comparable to their decay widths, the asymmetries for chargino production can be as large as 30% with cross sections of the order of several hundred fb. In addition, we have defined and analyzed asymmetries of the energy distributions of the chargino and neutralino decay products which probe the longitudinal chargino/neutralino polarizations. Their dependence on the Higgs interference and mixing effects can be used, in addition to the polarization and charge asymmetries of the production cross sections, to study the CP-violating effects in the MSSM Higgs sector at the muon collider.

### 3.14 Impact of Higgs CP mixing on the neutralino relic density

*Geneviève Bélanger, Fawzi Boudjema, Sabine Kraml, Alexander Pukhov and Alexander Semenov*

In supersymmetric models with R-parity conservation the lightest supersymmetric particle (LSP), typically the lightest neutralino $\tilde{\chi}^0_1$, is an excellent cold dark matter candidate [30, 31] (see e.g. [341] for a recent review). With the precision measurements by WMAP [342, 343] the relic density of cold dark matter can be constrained to $0.0945 < \Omega h^2 < 0.1287$ at $2\sigma$. This in turn puts strong constraints on the neutralino LSP as a thermal relic from the Big Bang. In particular, some efficient mechanism for $\tilde{\chi}^0_1$ annihilation has to be at work to ensure $\Omega h^2 \sim 0.1$. One such mechanism is annihilation through $s$-channel Higgs exchange near resonance. In this contribution, we investigate this case in the context of the MSSM with CP violation. This topic has been studied recently in [344, 345], and we here discuss it in more detail. An extensive analysis of the neutralino relic density in the presence of CP phases for various scenarios of neutralino (co)annihilation is given in [346].

We consider the general MSSM with parameters defined at the weak scale. In general, one can have complex parameters in the neutralino/chargino sector with $M_i = |M_i|e^{i\Phi_i}$, $\mu = |\mu|e^{i\Phi_\mu}$ as well as for the trilinear couplings, $A_f = |A_f|e^{i\Phi_{A_f}}$. The phase of $M_2$ can be rotated away. Among the trilinear couplings, $A_t$ has the largest effect on the Higgs sector, with the loop-induced CP mixing proportional to $\mathrm{Im}\,m(A_t\mu)/(m^2_{\tilde{t}_2} - m^2_{\tilde{t}_1})$. Since the phase of $\mu$ is the most severely constrained by electric dipole moment (EDM) measurements, we set it to zero, hence being left with only two relevant phases, $\Phi_1$ and $\Phi_t \equiv \Phi_{A_t}$.

Owing to Fermi statistics, the $s$-wave state of two identical Majorana fermions has CP $= -1$. The $p$-wave state has CP $= +1$. In the CP-conserving MSSM, the annihilation of two LSP's through the scalar $h$ or $H$ is hence $p$-wave suppressed at small velocities, while annihilation through the pseudoscalar $A$ is preferred. For mass-degenerate $H$ and $A$, the scalar exchange therefore only amounts to $\mathcal{O}(10\%)$ of the pseudoscalar exchange at $2m_{\tilde{\chi}^0_1} \sim M_{A,H}$.

In the presence of CP-violating phases, the interaction of the lightest neutralino with a Higgs $H_i$, $i = 1...3$, which now does not have definite CP properties any more, is given by

$$\mathcal{L}_{H_i\tilde{\chi}^0_1\tilde{\chi}^0_1} = \frac{g}{2}\sum_{i=1}^{3}\overline{\tilde{\chi}^0_1}(g^S_{H_i\tilde{\chi}^0_1\tilde{\chi}^0_1} + i\gamma_5 g^P_{H_i\tilde{\chi}^0_1\tilde{\chi}^0_1})\tilde{\chi}^0_1 H_i \tag{3.118}$$

with the scalar part of the coupling

$$g^S_{H_i\tilde{\chi}^0_1\tilde{\chi}^0_1} = \mathrm{Re}\,e\,[(N^*_{12} - t_W N^*_{11})\,(O_{1i}N^*_{13} - O_{2i}N^*_{14} - iO_{3i}(s_\beta N^*_{13} - c_\beta N^*_{14}))]\,, \tag{3.119}$$

where $N$ is the neutralino mixing matrix in the SLHA notation [252] and $O$ is the Higgs mixing matrix defined in Eq. (3.5). The pseudoscalar coupling $g^P_{H_i\tilde{\chi}^0_1\tilde{\chi}^0_1}$ corresponds to the imaginary part of the same expression. From Eqs. (3.118) and (3.119) it is clear that the neutralino relic density, being inversely proportional to the thermally averaged annihilation cross section, $\Omega_\chi \sim 1/\langle\sigma v\rangle$, will be affected both by $\Phi_t$, which induces scalar-pseudoscalar mixing in the Higgs sector, as well as by $\Phi_1$, which modifies the neutralino mixing. Here note that not only the couplings but also the masses depend on the phases. In what follows it will therefore be important to disentangle effects due to CP violation in the couplings from purely kinematic effects. Note also that there is a kind of sum rule relating the couplings squared of the Higgses to neutralinos. Therefore, for the two heavy eigenstates which are in general close in mass, we do not expect a huge effect on the resulting relic density from Higgs mixing alone. A noteworthy exception occurs when, for kinematical reasons, one of the resonances completly dominates the neutralino annihilation. That is for instance the case for $M_{H_2} < 2m_{\tilde\chi^0_1} \simeq M_{H_3}$, or when the mass splitting between the heavy Higgs bosons becomes very large.

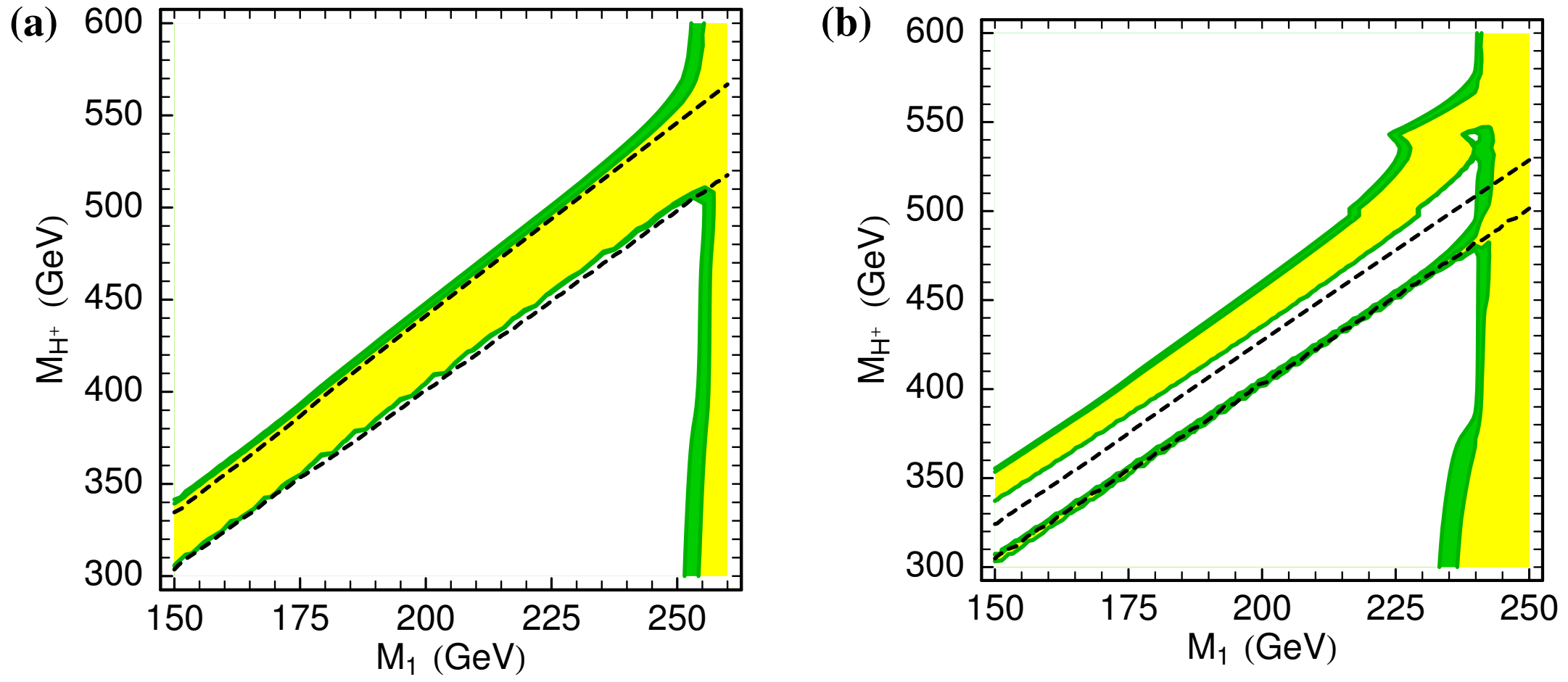

Fig. 3.56: WMAP-allowed bands in the $M_{H^+}$–$M_1$ plane for (a) $\mu = 1$ TeV and (b) $\mu = 2$ TeV with $\Phi_t = 90°$, $M_{\rm SUSY} = 0.5$ TeV, $|A_t| = 1.2$ TeV and $\tan\beta = 5$. In the narrow green (dark grey) bands $0.0945 < \Omega h^2 < 0.1287$, while in the yellow (light grey) regions $\Omega h^2 < 0.0945$. The positions of the WMAP-allowed bands for $\Phi_t = 0$ are shown as dashed lines.

For the numerical analysis, we are using an extension [347] of `micrOMEGAs` [348, 349] that allows for complex parameters in the MSSM. Using `LanHEP` [350], a new MSSM model file with complex parameters was rebuilt in the `CalcHEP` [351] notation, thus specifying all relevant Feynman rules. For the Higgs sector, an effective potential is written in order to include in a consistent way higher-order effects. To compute masses, mixing matrices and parameters of the effective potential the program is interfaced to `CPsuperH` [131]. All cross sections for annihilation and coannihilation processes are computed automatically with `CalcHEP`, and the standard `micrOMEGAs` routines are used to calculate the effective annihilation cross section and the relic density of dark matter. This CPV-MSSM version of `micrOMEGAs` has first been presented in [344].

Let us now turn to the numerical results. In order not to vary too many parameters, we choose $\tan\beta = 5$, $M_{\rm SUSY} \equiv M_{\tilde Q_3,\tilde U_3,\tilde D_3} = 0.5$ TeV and $|A_t| = 1.2$ TeV throughout this study. Moreover, we assume GUT relations for the gaugino masses, hence $M_2 \simeq 2M_1$. EDM constraints are avoided by setting $\Phi_\mu = 0$ and pushing the masses of the 1st and 2nd generation sfermions to 10 TeV. Last but not least in this contribution we are interested in the influence of CP violation in the Higgs sector. Therefore we also choose $\Phi_1 = 0$ and concentrate on the effect of $\Phi_t$. The effect of $\Phi_1 \neq 0$ is discussed in [344, 346].

Figure 3.56 shows the WMAP-allowed regions in the $M_{H^+}$–$M_1$ plane for this choice, maximal phase of $A_t$ ($\Phi_t = 90°$) and two values of $\mu$: $\mu = 1$ TeV and $\mu = 2$ TeV. The regions for which $0.094 < \Omega h^2 < 0.129$ are shown in green, and those for which $\Omega h^2 < 0.094$ in yellow. In addition, the positions of the WMAP-allowed strips for $\Phi_t = 0$ are shown as dashed lines. In the CP-conserving case, $H_3$ is a pure pseudoscalar and $H_2$ a pure scalar, while for $\Phi_t = 90°$ it is just the opposite and $H_2$ is dominantly pseudoscalar. The crossovers of 50% scalar-pseudoscalar mixing of $H_{2,3}$ occur at $\Phi_t \sim 15°$ and 145°. For $\mu = 1$ TeV, Fig. 3.56a, the mass splitting between $H_{2,3}$ is about 10 GeV for $\Phi_t = 90°$, as compared to about 2 GeV for $\Phi_t = 0$. Masses and the pseudoscalar content of $H_{2,3}$ are depicted in Fig. 3.57 as functions of $\Phi_t$. Here note that it is $H_2$, i.e. the state which changes from scalar to pseudoscalar with increasing $\Phi_t$, which shows the more pronounced change in mass. For $M_1$ values up to 250 GeV, we therefore find in both the CP-conserving and the CP-violating case two narrow bands where $0.094 < \Omega h^2 < 0.129$. For $\Phi_t = 0$ (and also for $\Phi_t = 180°$) both these bands are mainly due to pseudoscalar $H_3$ exchange, with one band just below and the other one above the pseudoscalar resonance. For $\Phi_t = 90°$ the situation is different: in the lower WMAP-allowed band the LSP annihilates through the scalar $H_3$, with the pseudoscalar $H_2$ not accessible because $M_{H_2} < 2m_{\tilde\chi^0_1} \simeq M_{H_3}$, while in the upper band both $H_2$ and $H_3$ contribute (with $H_2$ exchange of course dominating). In between the two WMAP-allowed green bands one is too close to the pseudoscalar resonance and $\Omega h^2$ falls below the WMAP bound; this holds for both $\Phi_t = 0$ and $\Phi_t = 90°$. The positions of the WMAP-allowed bands for $\Phi_t = 0$ and $\Phi_t = 90°$ are not very different from each other. Still the difference in the relic density between $\Phi_t = 0$ and $\Phi_t = 90°$ is typically a factor of a few in the WMAP-bands, and can reach orders of magnitudes at a pole. For $M_1 \gtrsim 250$ GeV and $\Phi_t = 90°$, one enters the region of coannihilation with stops, leading to a vertical WMAP-allowed band. For $\Phi_t = 0$, the $\tilde t_1$ is 55 GeV heavier, so the stop coannihilation occurs only at $M_1 \sim 305$ GeV (for $\Phi_t = 180°$ on the other hand, $m_{\tilde t_1} \simeq 230$ GeV and coannihilation already sets in at $M_1 \sim 200$ GeV).

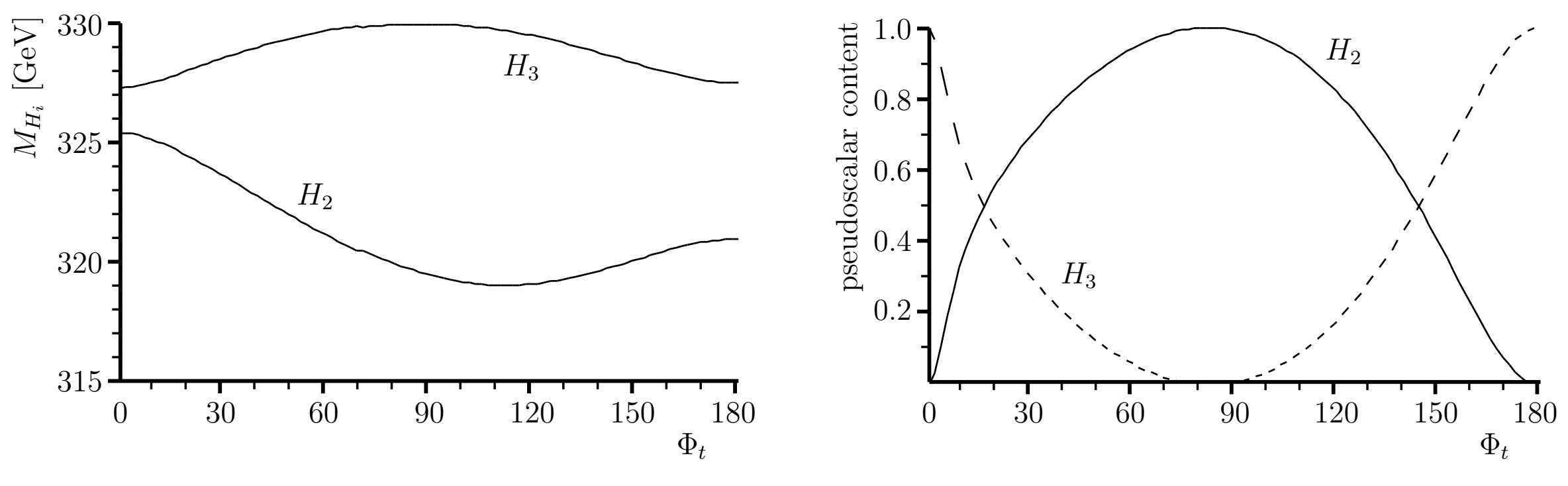

Fig. 3.57: Masses and pseudoscalar content of $H_2$ and $H_3$ as function of $\Phi_t$ for $M_{H^+} = 335$ GeV, $\mu = 1$ TeV, $|A_t| = 1.2$ TeV, $M_{\rm SUSY} = 0.5$ TeV and $\tan\beta = 5$. The light Higgs $H_1$ has a mass of $M_{H_1} \simeq 117$ GeV and a pseudoscalar content of $\leq 10^{-4}$.

For $\mu = 2$ TeV, Fig. 3.56b, there is an even stronger CP-mixing of $H_{2,3}$ and the mass splitting between the two states becomes $\sim$ 45 GeV for $\Phi_t = 90°$. The pseudoscalar contents are similar to those in Fig. 3.57 with the 50% cross-over at $\Phi_t \sim 20°$. Moreover, because the LSP has less higgsino admixture, one has to be closer to resonance to obtain the right relic density. As a result, the scalar and pseudoscalar funnels become separated by a region where $\Omega h^2$ is too large. In fact both the $H_2$ and $H_3$ exchange each lead to two WMAP-allowed bands, one above and one below the respective resonance. For the $H_3$ (scalar) exchange, however, these two regions are so close to each other that they appear as one line in Fig. 3.56b. This is in sharp contrast to the CP-conserving case, $\Phi_t = 0$, where the scalar and pseudoscalar states are close in mass, hence leading to only two WMAP-allowed bands. These are again shown as dashed lines in Fig. 3.56b and origin dominantly from the pseudoscalar resonance, the scalar resonance being 'hidden' within.

We next study the explicit dependence on $\Phi_t$, disentangling the effects due to scalar-pseudoscalar mixing from those due to changes in the Higgs masses. For this aim we fix $M_1 = 150$ GeV and $\mu = 1$ TeV. This gives $m_{\tilde\chi^0_1} = 149$ GeV with the LSP being 99.8% bino. Figure 3.58 shows the corresponding WMAP-allowed bands in the $M_{H^+}$–$\Phi_t$ plane. We observe a strong dependence on the phase of $A_t$, leading to huge shifts of up to two orders of magnitude in the relic density for constant $M_{H^+}$. To understand these huge effects, let us first discuss the upper WMAP-allowed band at $M_{H^+} \sim$ 335 GeV, shown in Fig. 3.58a, in more detail. As has been pointed out in [352, 353], the relic density is very sensitive to mass difference $\Delta M_{\tilde\chi^0_1 H_i} = M_{H_i} - 2m_{\tilde\chi^0_1}$, i.e. to the distance from the Higgs poles.

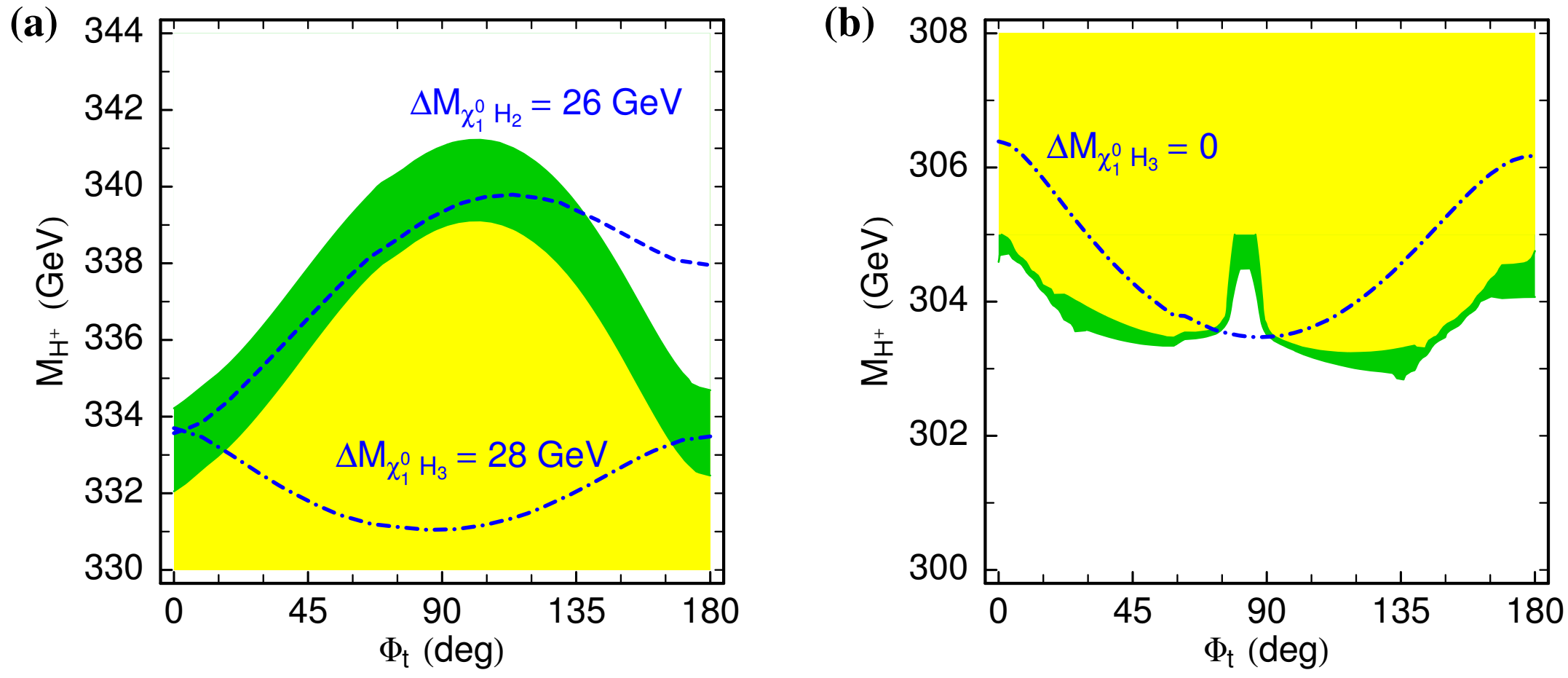

Fig. 3.58: The WMAP-allowed bands (green/dark grey) in the $M_{H^+}$–$\Phi_t$ plane for $M_1 = 150$ GeV, $\mu = 1$ TeV, $|A_t| = 1.2$ TeV, $M_{\rm SUSY} = 0.5$ TeV and $\tan\beta = 5$. Contours of constant mass differences $\Delta M_{\tilde\chi_1^0 H_i} \equiv M_{H_i} - 2m_{\tilde\chi_1^0}$ are also displayed. In the yellow (light grey) regions, $\Omega h^2$ is below the WMAP range.

At $M_{H^+} = 335$ GeV, the neutral Higgs masses range between $M_{H_2} \simeq 325$–319 GeV and $M_{H_3} \simeq 327$–330 GeV for $\Phi_t = 0$–$90°$. The LSP annihilates more efficiently through the Higgs which has the larger pseudoscalar content. For $\Phi_t \lesssim 15°$ ($\Phi_t \gtrsim 145°$) this is $H_3$, while for maximal phase it is $H_2$. Consequently in Fig. 3.58a agreement with WMAP is reached for $\Delta M_{\tilde\chi_1^0 H_i} \sim 26$–28 GeV with $H_i = H_3$ at $\Phi_t = 0$ and 180°, and $H_i = H_2$ at $\Phi_t = 90°$. When considering the Higgs–LSP couplings, we find $(g^S, g^P)_{H_2\tilde\chi_1^0\tilde\chi_1^0} \simeq (0.02, -10^{-5})$ and $(g^S, g^P)_{H_3\tilde\chi_1^0\tilde\chi_1^0} \simeq (-10^{-5}, -0.02)$ at $M_{H^+} = 335$ GeV and $\Phi_t = 0$, while at $\Phi_t = 90°$ $(g^S, g^P)_{H_2\tilde\chi_1^0\tilde\chi_1^0} \simeq (10^{-4}, 0.02)$ and $(g^S, g^P)_{H_3\tilde\chi_1^0\tilde\chi_1^0} \simeq (0.02, -10^{-4})$. We see that in the case where both $H_2$ and $H_3$ are accessible, the phase dependence of $\Omega h^2$ is directly linked to the position of the (dominantly) pseudoscalar resonance. For $\Phi_t = 0-90°$ and $\Phi_t \simeq 180°$, in the WMAP-allowed green band the dominant annihilation channels are about 75–80% into $b\bar b$ and about 10% into $\tau^+\tau^-$, corresponding to the pseudoscalar branching ratios. For $\Phi_t > 90°$, where the WMAP-allowed band deviates from the contour of constant $\Delta M_{\tilde\chi_1^0 H_i}$, there is also a sizeable, up to $\sim$ 25%, contribution from $\tilde\chi_1^0\tilde\chi_1^0 \to H_1 H_1$ with a constructive interference between $s$-channel $H_3$ and $t$-channel neutralino exchange. This is acompanied by roughly 10% annihilation into $WW$ and $ZZ$. For constant $\Delta M_{\tilde\chi_1^0 H_i}$, the variation in $\Omega h^2$ due to changes in the Higgs couplings alone can be $\mathcal{O}(100\%)$.

When the LSP mass is very near the heaviest Higgs resonance one finds another region where the relic density falls within the WMAP range. This is shown in Fig. 3.58b (corresponding to the phase dependence of the lower WMAP-allowed band in Fig. 3.56a). In the real case one needs $M_{H^+} = 305$ GeV, giving a mass difference $\Delta M_{\tilde\chi_1^0 H_3} = -1.5$ GeV. Note that annihilation is efficient enough even though one catches only the tail of the pseudoscalar resonance. For the same charged Higgs mass, the mass of $H_3$ increases when one increases $\Phi_t$, so that neutralino annihilation becomes more efficient despite the fact that $H_3$ becomes scalar-like and $g^P_{H_3\tilde\chi_1^0\tilde\chi_1^0}$ decreases. When $\Phi_t \sim 75°$–$90°$, the coupling $g^P_{H_3\tilde\chi_1^0\tilde\chi_1^0}$ becomes very small and one needs $\Delta M_{\tilde\chi_1^0 H_3} = 0$–1.5 GeV to achieve agreement with WMAP. Here we are in the special case where $M_{H_2} < 2m_{\tilde\chi_1^0} \simeq M_{H_3}$, so that only $H_3$ contributes significantly to the relic density. Figure 3.59 shows the $\tilde\chi_1^0\tilde\chi_1^0 \to b\bar b$ annihilation cross section as a function of $M_{H^+}$ and various values of $\Phi_t$. As can be seen, not only the position but also the hight of the peak changes with $\Phi_t$, corresponding to the change in the pseudoscalar content of $H_3$. In fact, at $M_{H^+} = 305$ GeV and $\Phi_t = 0$, the LSP annihilates to about 80% into $b\bar b$, 10% into $\tau\tau$ and 10% into $ZH_1$, while at $\Phi_t = 90°$, it annihilates to about 50% into $b\bar b$, 30% into $H_1 H_1$ and 10% into $WW/ZZ$. At $\Phi_t = 180°$, the rates are about 70% $b\bar b$, 10% $\tau\tau$ and 20% $ZH_1$,

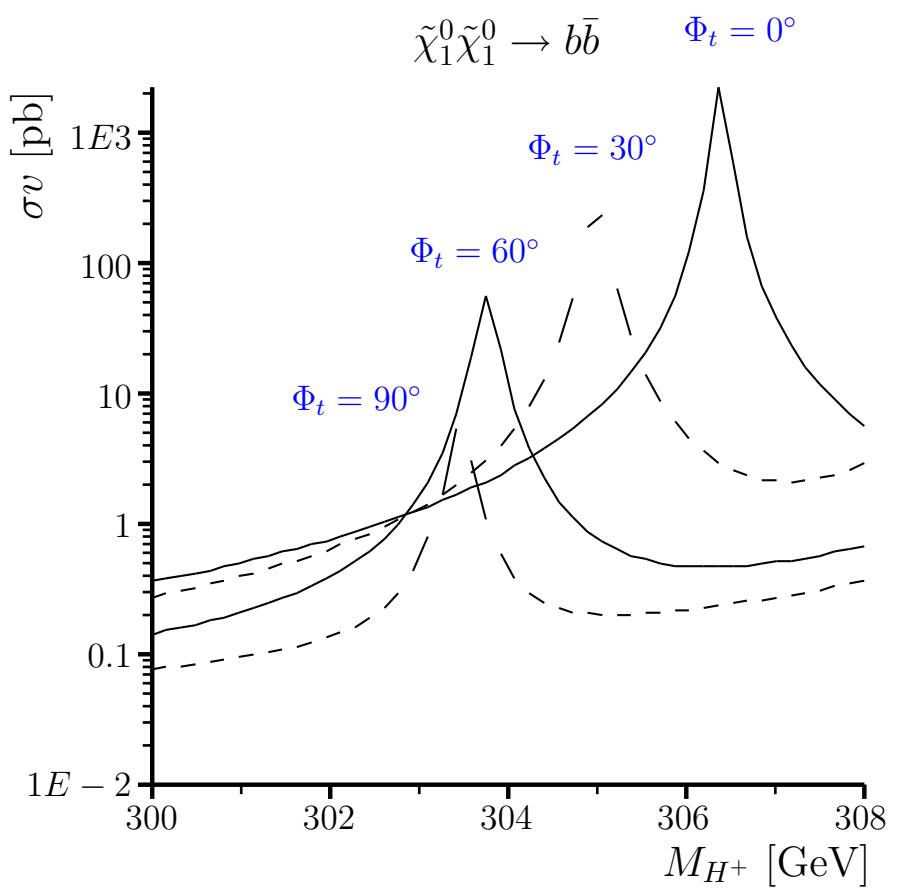

Fig. 3.59: $\sigma v(\tilde{\chi}^0_1\tilde{\chi}^0_1 \to b\bar{b})$ as a function of $M_{H^+}$ and various values of $\Phi_t$; other parameters as in Fig. 3.58.

In conclusion, CP-violating phases can lead to huge variations in the neutralino relic density. For fixed $M_{H^+}$, we find shifts in $\Omega h^2$ of up to two orders of magnitude, which agrees with the observations in [35] (although in that paper only $\sigma v$ and not $\Omega h^2$ was computed). From the discussion above it is clear that a large part of this can be attributed to changes in the Higgs masses. When disentangling the kinematic effects, we still find a significant dependence of $\Omega h^2$ on the CP-mixing in the Higgs sector. For example, for $\Delta M_{\tilde{\chi}^0_1 H_3} = -1.5$ GeV in Fig. 3.58b, we get an increase in $\Omega h^2$ relative to the $\Phi_t = 0$ case by almost an order of magnitude. Also the relative importance of different final states depends on the CP phases. To infer the relic density of the LSP it is therefore important to pin down the Higgs sector with good precision. This includes not only precise measurements of the Higgs masses and decay widths but also of a possible CP mixing. Last but not least note that loop corrections to neutralino annihilation processes will also be important for a precise prediction of $\Omega h^2$. For more details and other scenarios of neutralino annihilation and coannihilation, see [346].

## 3.15 Decays of third generation sfermions into Higgs bosons

*Alfred Bartl, Stefan Hesselbach, Keisyo Hidaka, Thomas Kernreiter and Werner Porod*

A precise knowledge of third generation sfermion parameters is important for Higgs physics as the dominant loop corrections to the masses of the MSSM Higgs bosons are due to stops and sbottoms [354]. Moreover, if the parameters $\mu$, $A_\tau$, $A_t$, $A_b$ are complex, they induce a mixing between neutral scalar Higgs bosons and the pseudoscalar Higgs boson which within the MSSM is impossible at tree-level. Furthermore, these parameters enter in the mixing matrices of the sfermions as well as in the couplings of sfermions to Higgs bosons. This leads to strong effects on sfermion decay widths and branching ratios which have been analyzed in [355, 356] for the stau sector and in [357–359] for the stop/sbottom sector. Thus sfermion production and subsequent decays into Higgs bosons are an additional source of Higgs bosons at future colliders with a potentially strong dependence on the SUSY CP phases.

### *3.15.1 SUSY CP phases in sfermion mixing and Higgs-sfermion couplings*

The left-right mixing of the stops and sbottoms is described by a hermitian $2 \times 2$ mass matrix which in the basis $(\tilde{q}_L, \tilde{q}_R)$ reads

$$\mathcal{L}^{\tilde{q}}_M = -(\tilde{q}^\dagger_L, \tilde{q}^\dagger_R) \begin{pmatrix} M^2_{\tilde{q}LL} & M^2_{\tilde{q}LR} \\ M^2_{\tilde{q}RL} & M^2_{\tilde{q}RR} \end{pmatrix} \begin{pmatrix} \tilde{q}_L \\ \tilde{q}_R \end{pmatrix}, \tag{3.120}$$

with

$$
\begin{aligned}
M^2_{\tilde{q}_{LL}} &= M^2_{\tilde{Q}_3} + (I^q_{3L} - e_q \sin^2\theta_W)\cos 2\beta\, m_Z^2 + m_q^2, &(3.121)\\
M^2_{\tilde{q}_{RR}} &= M^2_{\tilde{Q}'_3} + e_q \sin^2\theta_W \cos 2\beta\, m_Z^2 + m_q^2, &(3.122)\\
M^2_{\tilde{q}_{RL}} &= (M^2_{\tilde{q}_{LR}})^* = m_q\left(|A_q|e^{i\Phi_{A_q}} - |\mu|e^{-i\Phi_\mu}(\tan\beta)^{-2I^q_{3L}}\right), &(3.123)
\end{aligned}
$$

where $m_q$, $e_q$ and $I^q_{3L}$ are the mass, electric charge and weak isospin of the quark $q = b, t$. $\theta_W$ denotes the weak mixing angle, $\tan\beta = v_2/v_1$ with $v_1$ ($v_2$) being the vacuum expectation value of the Higgs field $H_1^0$ ($H_2^0$) and $M_{\tilde{Q}'_3} = M_{\tilde{D}_3}$ ($M_{\tilde{U}_3}$) for $q = b$ ($t$). $M_{\tilde{Q}_3}$, $M_{\tilde{D}_3}$, $M_{\tilde{U}_3}$, $A_b$ and $A_t$ are the soft SUSY-breaking parameters of the top squark and bottom squark system. In the case of complex parameters $\mu$ and $A_q$ the off-diagonal elements $M^2_{\tilde{q}_{RL}} = (M^2_{\tilde{q}_{LR}})^*$ are also complex with the phase

$$
\Phi_{\tilde{q}} = \arg\left[M^2_{\tilde{q}_{RL}}\right] = \arg\left[|A_q|e^{i\Phi_{A_q}} - |\mu|e^{-i\Phi_\mu}(\tan\beta)^{-2I^q_{3L}}\right]. \qquad (3.124)
$$

The mass eigenstates are

$$
\begin{pmatrix}\tilde{q}_1\\ \tilde{q}_2\end{pmatrix} = \mathcal{R}^{\tilde{q}}\begin{pmatrix}\tilde{q}_L\\ \tilde{q}_R\end{pmatrix} \qquad (3.125)
$$

with the $\tilde{q}$-mixing matrix

$$
\mathcal{R}^{\tilde{q}} = \begin{pmatrix} e^{i\Phi_{\tilde{q}}}\cos\theta_{\tilde{q}} & \sin\theta_{\tilde{q}} \\ -\sin\theta_{\tilde{q}} & e^{-i\Phi_{\tilde{q}}}\cos\theta_{\tilde{q}} \end{pmatrix}, \qquad (3.126)
$$

$$
\cos\theta_{\tilde{q}} = \frac{-|M^2_{\tilde{q}_{LR}}|}{\sqrt{|M^2_{\tilde{q}_{LR}}|^2 + (m^2_{\tilde{q}_1} - M^2_{\tilde{q}_{LL}})^2}}, \quad \sin\theta_{\tilde{q}} = \frac{M^2_{\tilde{q}_{LL}} - m^2_{\tilde{q}_1}}{\sqrt{|M^2_{\tilde{q}_{LR}}|^2 + (m^2_{\tilde{q}_1} - M^2_{\tilde{q}_{LL}})^2}} \qquad (3.127)
$$

and the mass eigenvalues

$$
m^2_{\tilde{q}_{1,2}} = \frac{1}{2}\left(M^2_{\tilde{q}_{LL}} + M^2_{\tilde{q}_{RR}} \mp \sqrt{(M^2_{\tilde{q}_{LL}} - M^2_{\tilde{q}_{RR}})^2 + 4|M^2_{\tilde{q}_{LR}}|^2}\right), \qquad m_{\tilde{q}_1} < m_{\tilde{q}_2}\,. \qquad (3.128)
$$

The respective mass and mixing matrices in the stau sector are obtained from those of the sbottoms by the replacement of the soft SUSY-breaking parameters $(M_{\tilde{Q}_3}, M_{\tilde{D}_3}, A_b) \to (M_{\tilde{L}_3}, M_{\tilde{E}_3}, A_\tau)$.

The $\tilde{q}_i - \tilde{q}'_j - H^\pm$ couplings are defined by

$$
\mathcal{L}_{\tilde{q}\tilde{q}H^\pm} = g\left(C^H_{\tilde{t}_j\tilde{b}_i} H^+ \tilde{t}^\dagger_j\,\tilde{b}_i + C^H_{\tilde{b}_j\tilde{t}_i} H^- \tilde{b}^\dagger_j\,\tilde{t}_i\right) \qquad (3.129)
$$

with

$$
C^H_{\tilde{t}_i\tilde{b}_j} = (C^H_{\tilde{b}_j\tilde{t}_i})^* = \frac{1}{\sqrt{2}\,m_W}(\mathcal{R}^{\tilde{t}} G \mathcal{R}^{\tilde{b}^\dagger})_{ij} \qquad (3.130)
$$

and

$$
G = \begin{pmatrix} m_b^2\tan\beta + m_t^2\cot\beta - m_W^2\sin 2\beta & m_b\,(|A_b|e^{-i\Phi_{A_b}}\tan\beta + |\mu|e^{i\Phi_\mu}) \\ m_t\,(|A_t|e^{i\Phi_{A_t}}\cot\beta + |\mu|e^{-i\Phi_\mu}) & \dfrac{2m_t m_b}{\sin 2\beta} \end{pmatrix}. \qquad (3.131)
$$

For the couplings of squarks to neutral Higgs bosons we have the Lagrangian

$$
\mathcal{L}_{\tilde{q}\tilde{q}H} = -g\,C(\tilde{q}^\dagger_k H_i \tilde{q}_j)\,\tilde{q}^\dagger_k H_i \tilde{q}_j \quad (k, j = 1, 2) \qquad (3.132)
$$

with

$$
C(\tilde{q}^\dagger_k H_i \tilde{q}_j) = \mathcal{R}_{\tilde{q}} \cdot \begin{pmatrix} C(\tilde{q}^\dagger_L H_i \tilde{q}_L) & C(\tilde{q}^\dagger_L H_i \tilde{q}_R) \\ C(\tilde{q}^\dagger_R H_i \tilde{q}_L) & C(\tilde{q}^\dagger_R H_i \tilde{q}_R) \end{pmatrix} \cdot \mathcal{R}_{\tilde{q}}{}^\dagger, \qquad (3.133)
$$

where for $\tilde{q} = \tilde{t}$

$$C(\tilde{t}_L^\dagger H_i \tilde{t}_L) = \frac{m_t^2}{m_W \sin\beta} O_{2i} + \frac{m_Z}{\cos\theta_W}\left(\frac{1}{2} - e_t \sin^2\theta_W\right)(\cos\beta O_{1i} - \sin\beta O_{2i}), \quad (3.134)$$

$$C(\tilde{t}_R^\dagger H_i \tilde{t}_R) = \frac{m_t^2}{m_W \sin\beta} O_{2i} + \frac{e_t m_Z}{\cos\theta_W} \sin^2\theta_W (\cos\beta O_{1i} - \sin\beta O_{2i}), \quad (3.135)$$

$$C(\tilde{t}_L^\dagger H_i \tilde{t}_R) = [C(\tilde{t}_R^\dagger H_i \tilde{t}_L)]^* = \frac{m_t}{2 m_W \sin\beta}\{-i\left(\cos\beta|A_t|e^{-i\Phi_{A_t}} + \sin\beta|\mu|e^{i\Phi_\mu}\right) O_{3i}$$
$$- \left(|\mu|e^{i\Phi_\mu} O_{1i} - |A_t|e^{-i\Phi_{A_t}} O_{2i}\right)\}, \quad (3.136)$$

while for $\tilde{q} = \tilde{b}$

$$C(\tilde{b}_L^\dagger H_i \tilde{b}_L) = \frac{m_b^2}{m_W \cos\beta} O_{1i} + \frac{m_Z}{\cos\theta_W}\left(-\frac{1}{2} - e_b \sin^2\theta_W\right)(\cos\beta O_{1i} - \sin\beta O_{2i}), \quad (3.137)$$

$$C(\tilde{b}_R^\dagger H_i \tilde{b}_R) = \frac{m_b^2}{m_W \cos\beta} O_{1i} + \frac{e_b m_Z}{\cos\theta_W} \sin^2\theta_W (\cos\beta O_{1i} - \sin\beta O_{2i}), \quad (3.138)$$

$$C(\tilde{b}_L^\dagger H_i \tilde{b}_R) = [C(\tilde{b}_R^\dagger H_i \tilde{b}_L)]^* = \frac{m_b}{2 m_W \cos\beta}\{-i\left(\sin\beta|A_b|e^{-i\Phi_{A_b}} + \cos\beta|\mu|e^{i\Phi_\mu}\right) O_{3i}$$
$$- \left(|\mu|e^{i\Phi_\mu} O_{2i} - |A_b|e^{-i\Phi_{A_b}} O_{1i}\right)\}. \quad (3.139)$$

The $3 \times 3$ matrix $O$ denotes the mixing matrix of the neutral Higgs bosons as defined in Eq. (3.5). The couplings of the staus to neutral Higgs bosons can be obtained from Eqs. (3.137)–(3.139) by replacing $b$ by $\tau$.

### *3.15.2 Numerical results*

In [357–359] the effects of the phases of the parameters $A_t$, $A_b$, $\mu$ and $M_1$ on the phenomenology of the third generation squarks, the stops $\tilde{t}_{1,2}$ and the sbottoms $\tilde{b}_{1,2}$ in the complex MSSM have been studied. The third generation squark sector is particularly interesting because of the effects of the large Yukawa couplings. The phases of $A_f$ and $\mu$ enter directly the squark mass matrices and the squark-Higgs couplings, which can cause a strong phase dependence of observables. The off-diagonal mass matrix element $M^2_{\tilde{q}RL}$, which describes the mixing between the left and right squark states, is given in Eq. (3.123). In the case of stops the $\mu$ term is suppressed by $1/\tan\beta$, hence the phase $\Phi_{\tilde{t}}$ of $M^2_{\tilde{t}RL}$ is dominated by $\Phi_{A_t}$. Therefore, the phase in the mixing matrix is in practice given by $\Phi_{A_t}$ and appears in several couplings due to the strong mixing in the stop sector. In the case of sbottoms the mixing is smaller because of the small bottom mass. It is mainly important for large $\tan\beta$, when the $\mu$ term is dominant in $M^2_{\tilde{b}RL}$. Hence the phase of $A_b$ has only minor impact on the sbottom mixing in a large part of the SUSY parameter space. However, in the squark-Higgs couplings, for example in the $H^\pm \tilde{t}_L \tilde{b}_R$ couplings, Eq. (3.131), the phase $\Phi_{A_b}$ appears independent of the sbottom mixing. This can lead to a strong $\Phi_{A_b}$ dependence of sbottom *and* stop partial decay widths into Higgs bosons. The stau sector behaves similar to the sbottom sector.

In the following we give examples where a strong dependence on phases occurs. We want to stress, that this is a general feature provided the decays into Higgs bosons are kinematically allowed. The masses and mixing matrix $O$ of the neutral Higgs bosons have been calculated with the program `FeynHiggs2.0.2` [59,60]. In Fig. 3.60 we show branching ratios of $\tilde{t}_2$ decays. As can be seen, the sum of the branching ratios into Higgs bosons is about 30% implying that stop decays serve as an additional source for Higgs bosons. As discussed in detail in [359], the partial widths for decays into fermions and the $Z$-boson have a $1 \pm \cos\Phi_{A_t}$ dependence. In the case of the Higgs bosons the dependence on the phases is much more involved as the parameters $A_f$ and $\mu$ appear directly in the couplings, see

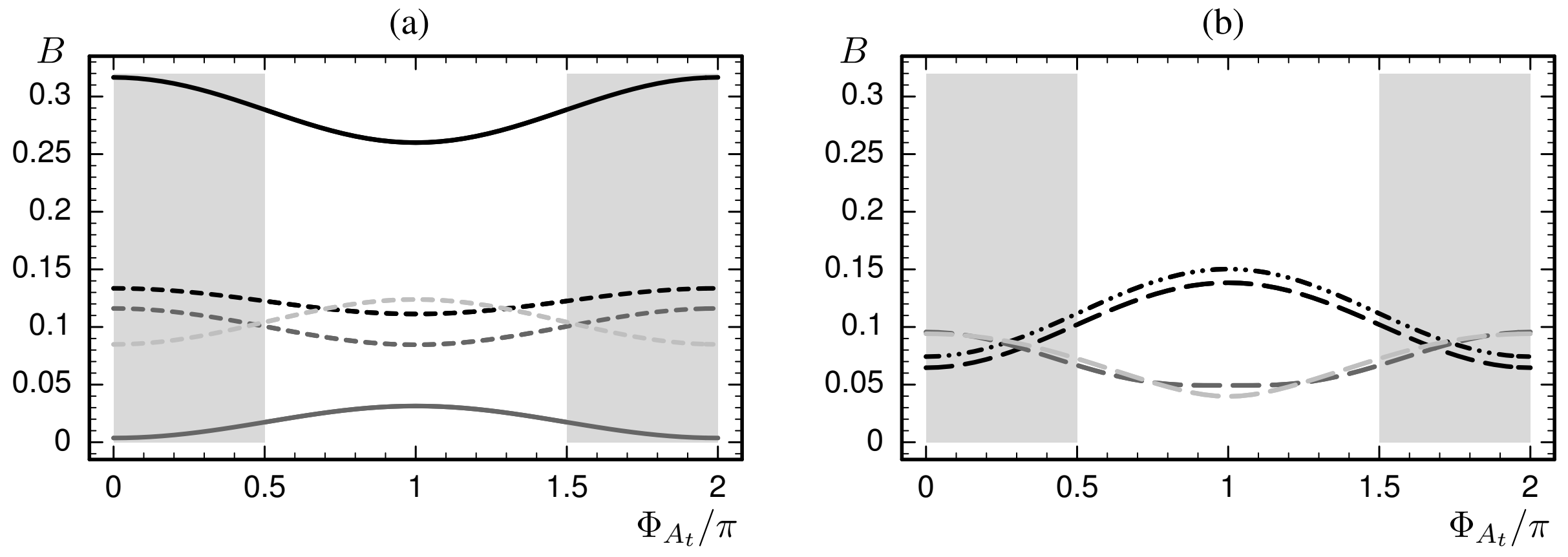

Fig. 3.60: $\Phi_{A_t}$ dependence of branching ratios of the decays (a) $\tilde{t}_2 \to \tilde{\chi}^+_{1/2} b$ (solid, black/gray), $\tilde{t}_2 \to \tilde{\chi}^0_{2/3/4} t$ (dashed, black/gray/light gray) and (b) $\tilde{t}_2 \to Z\tilde{t}_1$ (dashdotdotted), $\tilde{t}_2 \to H_{1/2/3}\tilde{t}_1$ (long dashed, black/gray/light gray) for $\tan\beta = 6$, $M_2 = 300$ GeV, $|\mu| = 500$ GeV, $|A_b| = |A_t| = 500$ GeV, $\Phi_\mu = \Phi_1 = \Phi_{A_b} = 0$, $m_{\tilde{t}_1} = 350$ GeV, $m_{\tilde{t}_2} = 800$ GeV, $m_{\tilde{b}_1} = 170$ GeV and $m_{H^\pm} = 350$ GeV, assuming $M_{\tilde{Q}_3} > M_{\tilde{U}_3}$. Only the decay modes with $B \gtrsim 1\,\%$ are shown. The shaded areas mark the region excluded by the experimental limit $B(b \to s\gamma) < 4.5 \times 10^{-4}$. From [359].

Eqs. (3.131), (3.136) and (3.139). Part of the phase dependence is due to the change of the Higgs masses as they depend on the phases. However, this effect is very small. As a test we have kept the Higgs masses constant and the lines in the plots are only shifted in the order of the line-thickness. In Figs. 3.61 and 3.62 it is demonstrated that (i) also the sbottom and stau decay branching ratios show a pronounced dependence on the phases and (ii) the branching ratios into Higgs bosons can be sizable and, thus, serve as an additional source for Higgs bosons. In contrast to the stop sector this dependence is mainly caused by the variations of the partial widths into Higgs bosons. For this reason it is also important if $\tan\beta$ is larger than $\sim 20$.

### *3.15.3 Parameter determination via global fit*

In order to estimate the precision, which can be expected in the determination of the underlying SUSY parameters, a global fit of many observables in the stop/sbottom sector has been made in [359]. In order to achieve this the following assumptions have been made: (i) At the ILC the masses of the charginos, neutralinos and the lightest Higgs boson can be measured with high precision. If the masses of the squarks and heavier Higgs bosons are below 500 GeV, they can be measured with an error of $1\,\%$ and 1.5 GeV, respectively. (ii) The masses of the squarks and heavier Higgs bosons, which are heavier than 500 GeV, can be measured at a 2 TeV $e^+e^-$ collider like CLIC with an error of $3\,\%$ and $1\,\%$, respectively. (iii) The gluino mass can be measured at the LHC with an error of $3\,\%$. (iv) For the production cross sections $\sigma(e^+e^- \to \tilde{t}_i\bar{\tilde{t}}_j)$ and $\sigma(e^+e^- \to \tilde{b}_i\bar{\tilde{b}}_j)$ and the branching ratios of the $\tilde{t}_i$ and $\tilde{b}_i$ decays we have taken the statistical errors, which we have doubled to be on the conservative side. We have analyzed two scenarios, one with small $\tan\beta = 6$ and one with large $\tan\beta = 30$. In both scenarios we have found that $\mathrm{Re}(A_t)$ and $|\mathrm{Im}(A_t)|$ can be determined with relative errors of 2–3%. For $A_b$ the situation is considerably worse because of the weaker dependence of the observables on this parameter. Here the corresponding errors are of the order of 50–100%. For the squark mass parameters $M_{\tilde{Q}_3}$, $M_{\tilde{U}_3}$, $M_{\tilde{D}_3}$ the relative errors are of order of 1%, for $\tan\beta$ of order of $3\,\%$ and for $\mu$ and the other fundamental SUSY parameters of order of 1–2%. In a similar analysis in the stau sector [356] it has been found that for $\tan\beta = 3$ (30) the relative errors of $\mathrm{Re}(A_\tau)$ and $|\mathrm{Im}(A_\tau)|$ are 22% and 7% (7% and 3%), respectively, whereas the errors of $M_{\tilde{L}_3}$, $M_{\tilde{E}_3}$ are of the order of 1%. In particular the expected precision in the stop sector will be necessary for the comparison of the theoretical calculations in the Higgs sector and experimental data.

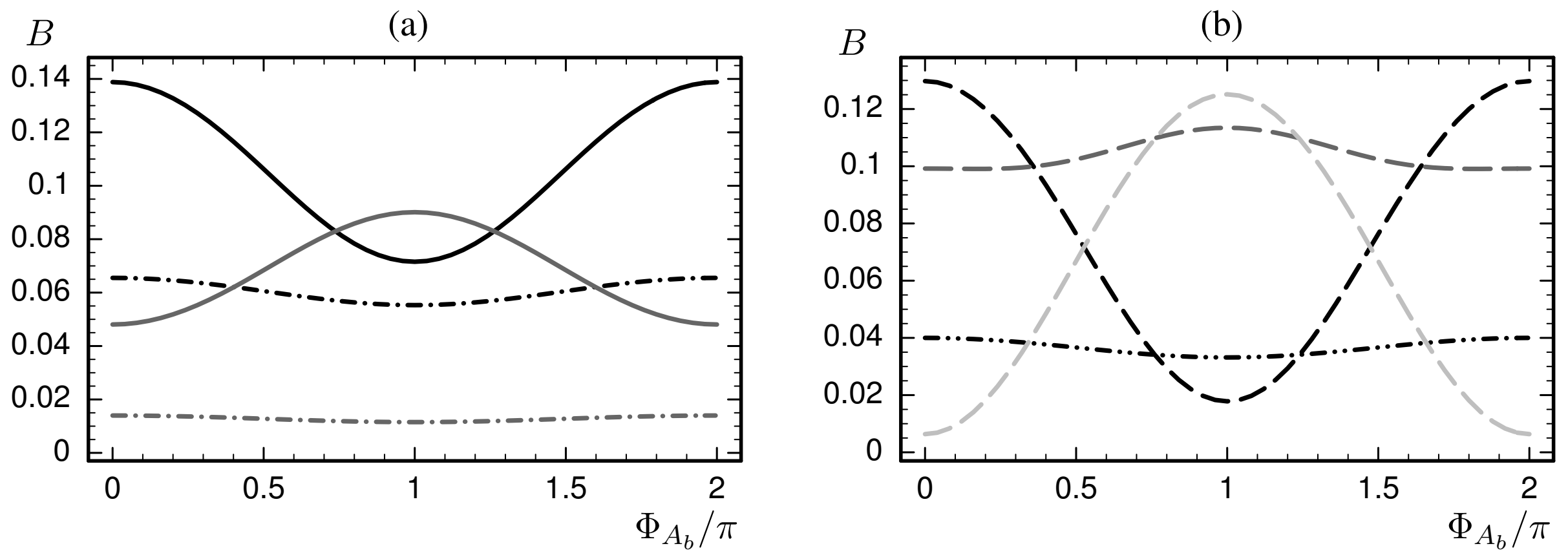

Fig. 3.61: $\Phi_{A_b}$ dependences of the branching ratios of the bosonic decays (a) $\tilde{b}_2 \to W^- \tilde{t}_{1/2}$ (dashdotted, black/gray), $\tilde{b}_2 \to H^- \tilde{t}_{1/2}$ (solid, black/gray) and (b) $\tilde{b}_2 \to Z\tilde{b}_1$ (dashdotdotted), $\tilde{b}_2 \to H_{1/2/3}\tilde{b}_1$ (long dashed, black/gray/light gray) for $\tan\beta = 30$, $M_2 = 200$ GeV, $|\mu| = 350$ GeV, $|A_b| = |A_t| = 600$ GeV, $\Phi_\mu = \Phi_{A_t} = \pi$, $\Phi_1 = 0$, $m_{\tilde{b}_1} = 350$ GeV, $m_{\tilde{b}_2} = 700$ GeV, $m_{\tilde{t}_1} = 170$ GeV and $m_{H^\pm} = 150$ GeV, assuming $M_{\tilde{Q}_3} < M_{\tilde{D}_3}$. From [359].

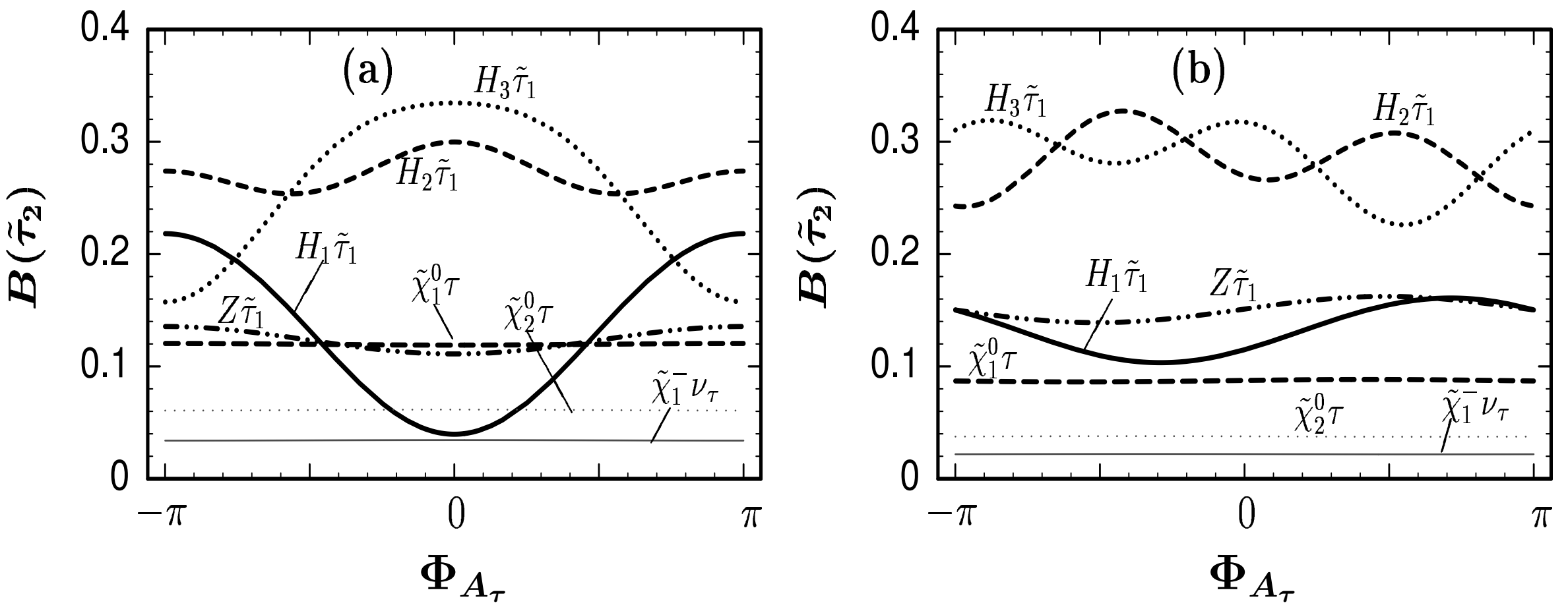

Fig. 3.62: Branching ratios of $\tilde{\tau}_2 \to H_{1,2,3}\tilde{\tau}_1$, $\tilde{\tau}_2 \to Z\tilde{\tau}_1$, $\tilde{\tau}_2 \to \tilde{\chi}^0_{1,2}\tau$ and $\tilde{\tau}_2 \to \tilde{\chi}^-_1\nu_\tau$ as a function of $\Phi_{A_\tau}$ for a) $\Phi_\mu = 0$ and b) $\Phi_\mu = \pi/2$, with the other parameters $m_{\tilde{\tau}_1} = 240$ GeV, $m_{\tilde{\tau}_2} = 500$ GeV, $m_{H^\pm} = 160$ GeV, $|\mu| = 600$ GeV, $M_2 = 450$ GeV, $\Phi_1 = 0$, $\tan\beta = 30$, and $|A_\tau| = 900$ GeV, assuming $M_{\tilde{L}_3} > M_{\tilde{E}_3}$. From [356].

## 4 SUPERSYMMETRIC MODELS WITH AN EXTRA SINGLET

### 4.1 Introduction

*Stephanie Baffioni, John F. Gunion, David J. Miller, Apostolos Pilaftsis and Dirk Zerwas*

The Minimal Supersymmetric Standard Model (MSSM) suffers from a serious theoretical flaw known as the "$\mu$-problem". The dimensionful parameter $\mu$ is required in the MSSM in order to have mixing between the two chiral Higgs doublet superfields. The difficulty is that $\mu$ has no *a priori* knowledge of electroweak symmetry breaking, but must, for phenomenological reasons, be around the electroweak scale. The very simplest means for resolving this problem is the introduction of a singlet Higgs chiral superfield. It is easy to arrange for the scalar component of the singlet superfield to acquire a vacuum expectation value of electroweak or SUSY breaking magnitude, and this automatically generates an effective $\mu$ with electroweak magnitude. The resulting models are very attractive and not only succeed in solving the $\mu$ problem but do not encounter the various problems of the MSSM associated with the LEP lower bounds on the masses of Higgs bosons. For instance, unlike the MSSM, the singlet-extended scenarios with low values of $\tan\beta \lesssim 3$ are viable due to additional tree-level contributions to the theoretical upper bound on the Higgs-boson mass. A SM-like Higgs boson with mass of order 100 GeV is also a possibility, evading conflict with LEP limits by virtue of Higgs to Higgs-pair decays. In addition, the parameters of the singlet models can be chosen so that the so-called "little hierarchy" problem can be significantly alleviated and electroweak baryogenesis is successful. In general, the Higgs sector phenomenology of the singlet models is much less constrained than that of the MSSM, leading to a far greater range of possible experimental signatures and needed search strategies.

In section 4.1.1 we begin by describing the $\mu$ problem in more detail and in section 4.1.2 show how it can be solved by the introduction of a new Higgs singlet chiral superfield. In section 4.1.3 we then outline the major variants of such models that emerge when cosmological issues associated with domain walls in the early universe are taken into consideration. In section 4.1.4, a summary of the main model and phenomenological features of the two simplest models, the NMSSM and the MNSSM, is provided. Experimentalists may wish to focus on this summary section.

#### *4.1.1 The $\mu$-problem*

Any renormalizable supersymmetric model can be completely specified by a choice of particle content, gauge symmetries and a *superpotential*. The gauge symmetries dictate the form of the gauge interactions in the Lagrangian, while the superpotential yields non-gauge interaction terms proportional to its second derivatives with respect to the scalar fields. A superpotential $W$ results in terms,

$$\mathcal{L} \supset -\frac{1}{2}\left(\frac{\partial^2 W}{\partial\phi_i\partial\phi_j}\psi_i\psi_j + \frac{\partial^2 W^*}{\partial\phi_i^*\partial\phi_j^*}\psi_i^\dagger\psi_j^\dagger\right) - \frac{\partial W}{\partial\phi_i}\frac{\partial W^*}{\partial\phi^{*\,i}}, \tag{4.1}$$

where $\phi_i$ are the scalar fields, $\psi_i$ are their fermion partners, and summation over repeated indices is understood. The scalar and fermion fields have dimension [mass] and [mass]$^{3/2}$ respectively, and the Lagrangian is dimension [mass]$^4$, so it is easy to see that the superpotential has dimension [mass]$^3$. Since supersymmetry (SUSY) is broken, one must also introduce soft SUSY breaking terms, which generally include soft mass terms in addition to terms of the same form as those in the superpotential multiplied by arbitrary (but $\mathcal{O}(M_{\rm SUSY})$) dimensionful coefficients.

The superpotential of the MSSM can be written as,[1]

$$W_{\rm MSSM} \;=\; \widehat{Q}\widehat{H}_u\mathbf{h_u}\widehat{U}^C \;+\; \widehat{H}_d\widehat{Q}\mathbf{h_d}\widehat{D}^C \;+\; \widehat{H}_d\widehat{L}\mathbf{h_e}\widehat{E}^C \;+\mu\widehat{H}_u\widehat{H}_d\,. \tag{4.2}$$

Here and in the following, fields with a hat denote superfields, whilst those without a hat stand for scalar superfield components. In addition to the Yukawa couplings of the two Higgs doublets to quarks and leptons, Eq. (4.2) contains a term involving a dimensionful parameter $\mu$. This performs two essential functions in the MSSM. Firstly, it provides a contribution to the masses of the Higgs bosons and their higgsino partners (for this reason $\mu$ is often called the "Higgs-higgsino mass parameter"). Without this contribution the lightest chargino, which is a mixture of a higgsino and a gaugino, would have a mass of order $M_W^2/M_{\rm SUSY}$ which is small enough to be excluded by experimental searches. Secondly, the accompanying soft supersymmetry breaking term $B\mu\, H_u H_d$ provides a mixing between the two Higgs doublets. Without this term, any electroweak symmetry breaking generated in the up-quark sector (caused by $M_{H_u}^2 < 0$) could not be communicated to the down-quark sector; the field $H_d$ would not gain a vacuum expectation value (vev) and the down-type quarks and leptons would remain massless. It is therefore essential that $\mu$ be non-zero and of order the electroweak or supersymmetry breaking scales.

[1] Our notation of the Higgs superfields with respect to the one given in Section 3 is: $\widehat{H}_u \equiv \widehat{H}_2$ and $\widehat{H}_d \equiv \widehat{H}_2$.

However, since the parameter $\mu$ appears in the superpotential it does not break supersymmetry and is present when supersymmetry is unbroken. Its value is therefore completely unrelated to the electroweak or supersymmetry breaking scales. In fact, within a supergravity (SUGRA) framework, the $\mu$-parameter is naturally expected to be of order $M_{\rm Planck}$. This huge disparity between the natural and phenomenologically needed scales of $\mu$ is known as the "$\mu$-problem".

### *4.1.2 Solving the $\mu$-problem with an extra singlet*

Many scenarios, all based on extensions of the MSSM, have been proposed in the existing literature [1–5] to provide a natural explanation of the $\mu$-term. The simplest approach, and that which concerns us here, is to introduce an extra Higgs iso-singlet superfield, $\widehat{S}$, into the model. If we replace the $\mu$-term of the MSSM with a term coupling this new superfield to the Higgs boson doublets, i.e.

$$W_\lambda \;=\; \widehat{Q}\widehat{H}_u \mathbf{h_u} \widehat{U}^C \;+\; \widehat{H}_d \widehat{Q} \mathbf{h_d} \widehat{D}^C \;+\; \widehat{H}_d \widehat{L} \mathbf{h_e} \widehat{E}^C \;+\; \lambda \widehat{S}\widehat{H}_u \widehat{H}_d \,, \tag{4.3}$$

where $\lambda$ is some dimensionless coupling, then an effective $\mu$-term will be generated if the real scalar component of $\widehat{S}$ develops a vacuum expectation value (vev). The *effective* $\mu$ parameter is given by

$$\mu \;=\; \mu_{\rm eff} \;=\; \lambda\langle S\rangle. \tag{4.4}$$

The constraints which arise when the resulting Higgs potential is minimized to find the vacuum state relate the vevs of the three neutral scalars, $H_u^0$, $H_d^0$ and $S$, to their soft supersymmetry breaking masses. Therefore, in the absence of fine tuning, one expects that these vevs should all be of order $M_{\rm SUSY}$, and the $\mu$ problem is solved. These three vevs are usually then replaced by the phenomenological parameters $\mu_{\rm eff}$, $M_Z$ and $\tan\beta$.

Additionally, one must introduce soft supersymmetry breaking terms into the Lagrangian. These must be of the same form as for the MSSM except that the term involving the supersymmetry breaking parameter $B$ must be removed (since it was associated with the $\mu$-term) and instead a soft mass for the new singlet should be added, together with soft supersymmetry breaking terms for the extra interactions. The soft supersymmetry breaking terms associated with the Higgs sector are then,

$$-\mathcal{L}_{\rm soft} \;\supset\; m_{H_u}^2|H_u|^2 + m_{H_d}^2|H_d|^2 + m_S^2|S|^2 + \Big(\lambda A_\lambda S H_u H_d + \text{h.c.}\Big) \,, \tag{4.5}$$

where $A_\lambda$ is a dimensionful parameter of order $\sim M_{\rm SUSY}$.

The new singlet superfield provides an additional scalar Higgs field, a pseudoscalar Higgs field, and an accompanying higgsino. The new Higgs fields will mix with the neutral Higgs fields from the usual Higgs doublets, and so the model will in total have five neutral Higgs bosons (three scalars and two pseudoscalars if CP is conserved). The extra higgsino will mix with the higgsinos from the doublets and the gauginos to provide an extra neutralino state, for a total of five. The charged Higgs and chargino mass spectrum remain unchanged.

However, the superpotential presented in Eq. (4.3), and its derived Lagrangian, contain an extra global U(1) symmetry, known as a Peccei-Quinn (PQ) Symmetry [6]. Assigning PQ charges, $Q^{PQ}$, according to

$$\widehat{Q}:-1,\qquad \widehat{U}^C:0,\qquad \widehat{D}^C:0,\qquad \widehat{L}:-1,\qquad \widehat{E}^C:0,\qquad \widehat{H}_u:1,\qquad \widehat{H}_d:1,\qquad \widehat{S}:-2, \tag{4.6}$$

the model is invariant under the global $U(1)$ transformation $\widehat{\Psi}_i \to e^{iQ_i^{PQ}\theta}\widehat{\Psi}_i$, where

$$\widehat{\Psi}_i \in \{\widehat{Q},\, \widehat{U}^C,\, \widehat{D}^C,\, \widehat{L},\, \widehat{E}^C,\, \widehat{H}_u,\, \widehat{H}_d,\, \widehat{S}\}\,.$$

The PQ symmetry will spontaneously break when the Higgs scalars gain vevs, and a pseudo[2]-Nambu-Goldstone boson, known as the PQ axion (it is actually one of the pseudoscalar Higgs bosons), will be generated. For values of $\lambda \sim \mathcal{O}(1)$, this axion would have been detected in experiment and this model ruled out. There are three ways that this model can be saved.

Firstly, one can simply decouple the axion by making $\lambda$ very small [7–12]. One finds that astrophysical constraints from the cooling of stars in globular clusters are most restrictive, requiring $\lambda \lesssim 10^{-6}$. Interestingly, since the singlet vev is always multiplied by $\lambda$, i.e. appears as $\lambda\langle S\rangle$, in the minimization equations which set the vevs, in the absence of fine tuning $\mu_{\text{eff}}$ will still naturally be of order the electroweak scale. Additionally, the presence of an axion automatically solves the strong CP problem via its effective coupling to the gluon. However, since there is no good explanation of why $\lambda$ should be so small, we are really just replacing one problem with another.

There is also an issue of how much dark matter is present in this model. Usually in R-parity conserving supersymmetry, the lightest supersymmetric partner (LSP) will, if neutral, provide a contribution to dark matter. In this case the LSP is the supersymmetric partner of the axion, often called the axino (it is actually a neutralino). It is very light, typically $\sim 10^{-6}$eV, and, like its partner state, is very decoupled. Therefore its annihilation rate in the early universe would be very small and it should naively provide a dark matter contribution so large that the model can be ruled out. However, the axino is so decoupled that it may never have come into equilibrium in the early universe. In this case, there would be no need to have a large annihilation cross-section to reduce its dark matter contribution; one could simply have very few axinos before annihilation starts.

A second possibility is to promote the PQ symmetry to a local symmetry. This requires the introduction of a new gauge boson, traditionally called $Z'$, mediating a new force, which will gain a mass when the PQ symmetry is broken. As usual, the Goldstone boson will be "eaten" by the gauge boson to provide the extra degree of freedom needed for its longitudinal polarization, and consequently there would be no axion to be found in low energy experiments.

The existence of additional U(1) gauge groups at TeV energies is well motivated by GUT and string models [13–16]. In particular, compactification of the extra dimensions in string theories often leads to large gauge groups such as $E_6$ or $E_8$. These gauge groups can then break down to the gauge groups of the SM with extra (local) U(1)'s. For example, one possible breaking would be $E_6 \to SO(10)\times U(1)_\phi$ followed by $SO(10) \to SU(5)\times U(1)_\chi$. In general, the gauge bosons of these two new $U(1)$ symmetries mix, and one can arrange the symmetry breaking such that one combination maintains a GUT scale mass, while the other is manifest at (just above) the electroweak scale and becomes the $Z'$.

The existence of an extra $Z'$ is already strongly constrained by experiment [17]. Direct searches at the Tevatron [18, 19] constrain the $Z'$ mass by looking for its decay to leptons or jets. These direct searches typically require a $Z'$ of the form described above to be heavier than a few hundred GeV. Indirect searches for virtual $Z'$ exchange and/or $Z$-$Z'$ mixing yield similar limits. Models with extra gauge groups are discussed in Section 6.

[2] The axion is only a "pseudo"-Nambu-Goldstone boson since the PQ symmetry is explicitly broken by the QCD triangle anomaly. The axion then acquires a small mass from its mixing with the pion.

### 4.1.3 Breaking the Peccei-Quinn symmetry

The last (but by no means least) way of avoiding the PQ axion constraints is to explicitly break the PQ symmetry. The new superfield $\widehat{S}$ has no gauge couplings but has a PQ charge, so one can naively introduce any term of the form $\widehat{S}^n$ with $n \in \mathbb{Z}$ into the superpotential in order to break the PQ symmetry. However, since the superpotential is of dimension 3, any power with $n \neq 3$ will require a dimensionful coefficient naturally of the GUT or Planck scale, naively making the term either negligible (for $n > 3$) or unacceptably large (for $n < 3$). For this reason, it is usual to postulate some extra discrete symmetry, e.g. $\mathbb{Z}_3$, in order to forbid terms with dimensionful coefficients. The superpotential of the model then becomes,

$$W_{\rm NMSSM} \;=\; W_\lambda \;+\; \frac{1}{3}\kappa\widehat{S}^3 \;, \tag{4.7}$$

where $\kappa$ is a dimensionless constant which measures the size of the PQ breaking.

Additionally, one must also introduce an extra soft supersymmetry breaking term to accompany the new trilinear self coupling. The complete soft SUSY-breaking Higgs sector becomes then,

$$-\mathcal{L}_{\rm soft} \;\supset\; m_{H_u}^2|H_u|^2 + m_{H_d}^2|H_d|^2 + m_S^2|S|^2 + \left(\lambda A_\lambda S H_u H_d + \frac{1}{3}\kappa A_\kappa S^3 + {\rm h.c.}\right) , \tag{4.8}$$

where, like $A_\lambda$, $A_\kappa$ is a dimensionful coefficient of order $\sim M_{\rm SUSY}$.

This model is known as the Next-to-Minimal Supersymmetric Standard Model (NMSSM) and has generated much interest in the literature [13, 15, 20–31]. Just as for the PQ symmetric model discussed above, the neutral Higgs sector will consist of three scalars and two pseudoscalars. The masses and singlet contents of the physical fields depend strongly on the parameters of the model, in particular how well the PQ symmetry is broken. Also, there will be five neutralinos instead of the usual four. The charged Higgs sector and the chargino sector remain unchanged. Some aspects of the phenomenology of the NMSSM will be summarized later and in separate contributions.

Phenomenologically, this model is rather interesting. Notice that we have introduced extra fields with no gauge couplings and mixed them with the usual fields of the MSSM. This will dilute the couplings of the Higgs bosons and neutralinos when compared to the MSSM. Furthermore, it is possible to have a rather light pseudoscalar Higgs boson, which is a bit more difficult to have in the MSSM. Potentially, heavier Higgs bosons may decay into this light pseudoscalar rather than via more conventional decays to, say, $b$ quarks. Therefore, one may find that the usual search channels at the LHC are not as successful as they are for the MSSM. Some of the related phenomenological issues will be summarized later and some will be discussed in separate contributions.

Here, we focus on the solution to a possible cosmological problem for the NMSSM. The $\mathbb{Z}_3$ symmetry, which we enforced on the model to ensure the absence of dimensionful couplings, cannot be completely unbroken. If it were, a "domain wall problem" would arise. In particular, if $\mathbb{Z}_3$ symmetry is exact, observables are unchanged when we (globally) transform all the fields according to $\Psi \to e^{i2\pi/3}\Psi$. Therefore the model will have three separate but degenerate vacua, and which one of these ends up being the "true" vacuum is a random decision taken at the time of electroweak symmetry breaking. However, one expects that causally disconnected regions of space would not necessarily choose the same vacuum, and our observable universe should consist of different domains with different ground states, separated by domain walls [32]. Such domain wall structures create unacceptably large anisotropies in the cosmic microwave background [33]. Historically, it was always assumed that the $\mathbb{Z}_3$ symmetry could be broken by an appropriate type of unification with gravity at the Planck scale. Non-renormalizable operators will generally be introduced into the superpotential and Kähler potential which break $\mathbb{Z}_3$ and lead to a preference for one particular vacuum, thereby solving the problem. However, the same operators may give rise at the loop level to quadratically divergent tadpole contributions in the Lagrangian, of the form [34–41]

$$\mathcal{L}_{\rm soft} \;\supset\; t_S\, S \;\sim\; \frac{1}{(16\pi^2)^n}\, M_{\rm P}\, M_{\rm SUSY}^2\, S \;, \tag{4.9}$$

where $n$ is the number of loops. Clearly, this tadpole breaks the $\mathbb{Z}_3$ symmetry as desired. But, if $n < 5$, $t_S$ is several orders of magnitude larger than the soft-SUSY breaking scale $M_{\rm SUSY}$. This leads to an unacceptably large would-be $\mu$-term since $t_S\, S$ combines with the $\sim M^2_{\rm SUSY} S^* S$ soft mass term to yield a shift in the vev of $S$ to a value of order $\langle S\rangle \sim \frac{t_S}{M^2_{\rm SUSY}} \sim \frac{1}{(16\pi^2)^n}\, M_{\rm P}$ and corresponding $\mu_{\rm eff} \sim \lambda\langle S\rangle$. For example, if the tadpole were generated at the one-loop level, the effective $\mu$-term would be huge, of order $10^{16}$–$10^{17}$ GeV i.e. close to the GUT scale, whereas $\mu$ should be of order of the electroweak scale to realize a natural Higgs mechanism. Hence, it was argued in [42] that the NMSSM is either ruled out cosmologically or suffers from a naturalness problem related to the destabilization of the gauge hierarchy. However, there are at least two simple escapes.

One obvious way out of this problem would be to gauge the $U(1)_{\rm PQ}$ symmetry [13–16]. In this case, the $\mathbb{Z}_3$ symmetry is embedded into the local $U(1)$ symmetry. The would-be PQ axion is then eaten by the longitudinal component of the extra gauge boson. However, from a low-energy perspective, the price one has to pay here is that the field content needs to be extended by adding new chiral quark and lepton states in order to ensure anomaly cancellation related to the gauged $U(1)_{\rm PQ}$ symmetry.

The second approach is to find symmetry scenarios [43] where all harmful destabilizing tadpoles are absent. This can be achieved by imposing a discrete $Z_2^R$ symmetry under which all superfields and the superpotential flip sign. To avoid destabilization while curing the domain wall problem, this symmetry has to be extended to the non-renormalizable part of the superpotential and to the Kähler potential. As happens to all $R$-symmetries, $Z_2^R$ symmetry is broken by the soft-SUSY breaking terms, giving rise to harmless tadpoles of order $\frac{1}{(16\pi^2)^n}\, M^3_{\rm SUSY}$, with $2 \leq n \leq 4$. Although these terms are phenomenologically irrelevant, they are entirely sufficient to break the global $Z_3$ symmetry and make the domain walls collapse.

Another potentially interesting alternative for breaking the PQ symmetry is the realization of the so-called Minimal Nonminimal Supersymmetric Standard Model (MNSSM) [44]. The basic idea is to find discrete $R$ symmetries, such that the destabilizing tadpoles do appear but are naturally suppressed because they arise at loops higher than 5 [45]. In particular, these symmetries may lead to a superpotential whose renormalizable part has exactly the form $W_\lambda$ in (4.3). Hence, the effective renormalizable superpotential of such a model reads [44]:

$$W^{\rm eff}_{\rm MNSSM} \;=\; W_\lambda \;+\; t_F \widehat{S}\,. \tag{4.10}$$

where $t_F$ is a radiatively induced tadpole of the electroweak scale. In addition, there will be a soft SUSY-breaking tadpole term $t_S S$ as given in (4.9). The key point in the construction of the renormalizable MNSSM superpotential is that the simple form (4.10) can be enforced by discrete $R$-symmetries [44–47]. These discrete $R$-symmetries govern the complete gravity-induced non-renormalizable superpotential and Kähler potential. Within the SUGRA framework of SUSY-breaking, it has then been possible to show [44] that the potentially dangerous tadpole $t_S$ will appear at a loop level $n$ higher than 5. From (4.9), the size of the tadpole parameter $t_S$ can be estimated to be in the right ballpark, i.e. $|t_S| \lesssim$ 1–10 TeV$^3$ for $n = 6, 7$, such that the gauge hierarchy does not get destabilized. To be specific, the tadpole $t_S\, S$ together with the soft SUSY-breaking mass term $m_S^2 S^* S \sim M^2_{\rm SUSY} S^* S$ lead to a vacuum expectation value (VEV) for $S$, $\big\langle S\big\rangle = \frac{1}{\sqrt{2}} v_S$, of order $M_{\rm SUSY} \sim 1$ TeV. The latter gives rise to a $\mu$-parameter at the required electroweak scale. Thus, another natural explanation for the origin of the $\mu$-parameter can be obtained.

#### 4.1.4 *Model features and phenomenological highlights of the NMSSM and the MNSSM*

In the following, we summarize the basic field-theoretic, phenomenological and cosmological features of the NMSSM and the MNSSM, and how these compare with the MSSM.

(i) Even though the mechanisms are different, both the NMSSM and the MNSSM can provide a *minimal* and an *elegant* solution to the $\mu$-problem of the MSSM. The $\mu$-parameter arises from the superpotential term $\lambda \widehat{S}\widehat{H}_u\widehat{H}_d$, through the vev of the scalar component $S$ of the singlet superfield $\widehat{S}$, i.e. $\lambda\langle S\rangle\widehat{H}_u\widehat{H}_d = \mu_{\rm eff}\widehat{H}_u\widehat{H}_d$. Such a term is essential for acceptable phenomenology.

(ii) The difference between the NMSSM and MNSSM arises from the symmetries imposed upon the superpotential and Kähler potential within a SUGRA framework. In the NMSSM case, they are such as to allow an additional superpotential component of form $\frac{1}{3}\kappa\widehat{S}^3$ and disallow all superpotential terms with dimensionful parameters, with the result that the scale of electroweak symmetry breaking is generated by the scale of SUSY breaking only. In the MNSSM case, symmetries are chosen so as to forbid the cubic singlet superpotential term but allow a linear tadpole term $t_S\, S$ ($t_F\widehat{S}$) where $t_S$ ($t_F$) is dimensionful. The symmetries in the two cases are set up so that the tadpole term proportional to $S$ coming from multi-loop-induced operators arising from physics above the unification scale is phenomenologically irrelevant in the NMSSM case, whereas it is crucial in the MNSSM case. In both cases, however, these symmetries play an important role in naturally solving the cosmological domain wall and visible axion problems.

(iii) After employing the minimization conditions for the Higgs potential and demanding the known value of the $Z$ mass, the parameters specifying the Higgs sectors in the two models are as follows. In the NMSSM a convenient set is

$$\lambda, \quad \kappa, \quad A_\lambda, \quad A_\kappa, \quad \mu = \mu_{\rm eff}, \quad \tan\beta\,. \tag{4.11}$$

In the case of the MNSSM, a convenient parameter set is

$$\lambda, \quad A_\lambda\ (\text{or } M_{H^+}), \quad \mu = \mu_{\rm eff}, \quad \tan\beta, \quad \lambda\, t_S/\mu\,, \tag{4.12}$$

whereas the extra parameter $t_F$ can be ignored as it appears usually suppressed in generic SUGRA-mediated SUSY-breaking scenarios. In addition, the stop squark masses strongly influence the Higgs boson masses and mixings through radiative corrections.

(iv) Since the NMSSM and the MNSSM introduce only a singlet superfield $\widehat{S}$ which is uncharged under the SM gauge group, the good property of gauge coupling unification in the MSSM is preserved. In the same context, the radiative electroweak symmetry breaking mechanism remains natural in both cases.

(v) The “little fine tuning problem”, which results in the MSSM due to the fact that LEP II failed to detect a CP-even Higgs boson, is less severe within the NMSSM and the MNSSM. In particular, the scenarios that arise from the requirement of minimal fine tuning point to certain phenomenologies [48, 49]. For example, one might have complete absence of the fine-tuning problem, if the lightest CP-odd Higgs particle $A_1$ is light enough to allow for the decay $H_1 \to A_1A_1$ [48, 49]. Indeed, the models with absolute minimum fine tuning and with $A_1$ mass below $2m_b$ are especially interesting since they predict a rate for $ZH_1$ with $H_1 \to b\bar{b}$, which is also consistent with a possible event excess in the LEP data for Higgs mases in the vicinity of 100 GeV. Such a possibility of a light ‘CP-odd’ Higgs may arise within the MSSM [50], which can also describe possible LEP excesses, but the associated scenarios are not related with reduced fine tuning in the Higgs sector.

(vi) In the NMSSM and the MNSSM, the upper bound on the mass of the lightest SM-like Higgs boson, e.g. $H_1$, increases by an amount of $\sim$ 30 GeV with respect to the MSSM, for small values of $\tan\beta$, i.e. $M_{H_1} \lesssim$ 145 GeV, for maximal stop mixing and $\tan\beta = 2$. Notice that MSSM scenarios using such low values of $\tan\beta$ have already been ruled out by LEP data or are at the verge of being ruled out at the Tevatron.

(vii) The NMSSM and the MNSSM both predict the existence of stable or quasi-stable light neutralinos that could be responsible for the Dark Matter (DM) of the universe [51–54]. There are many new possibilities as compared to the MSSM. In particular, in the NMSSM and MNSSM it is possible that the lightest neutralino is extremely light (100 MeV to 10 GeV) and can annihilate sufficiently through a light $A_1$ that the correct DM relic density is obtained. In general, the parameter regions for which correct DM relic abundance can be obtained in the NMSSM and MNSSM are far more extensive as compared to the MSSM and each such region will have interesting and significant consequences for the phenomenology to be expected at colliders.

(viii) Finally, an important cosmological feature of the MNSSM and the NMSSM is that they can comfortably explain the Baryon Asymmetry in the Universe by means of a strong first order electroweak phase transition [51, 55–57]. In contrast, baryogenesis considerations leave the MSSM in slight disfavor, requiring the right handed stop squark to be lighter than the top quark and the Higgs lighter than about 117 GeV [58, 59]. In these scenarios, the heavier stop quark is extremely heavy, leading to large fine-tuning.

Apart from the common features listed above, the two models, the NMSSM and the MNSSM, have some characteristic phenomenological differences, especially when compared with the MSSM. More explicitly, the following points can be made:

(a) Unlike the NMSSM, the MNSSM satisfies the tree-level mass sum rule [44]:

$$M_{H_1}^2 + M_{H_2}^2 + M_{H_3}^2 = M_Z^2 + M_{A_1}^2 + M_{A_2}^2 \,, \tag{4.13}$$

where $H_{1,2,3}$ and $A_{1,2}$ are the three CP-even and two CP-odd Higgs fields, respectively. The tree-level mass sum rule (4.13) is very analogous to the corresponding one of the MSSM [60,61], where the two heavier Higgs states $H_3$ and $A_2$ are absent in the latter. Radiative effects may violate (4.13) by an additional term of order $M_Z^2$. Hence, possible observation of a large violation of the sum rule (4.13) can rule out the MNSSM, pointing explicitly towards the NMSSM.

(b) The decoupling properties of a large tadpole in the MNSSM open up further possibilities in the Higgs-boson mass spectrum. In particular, the charged Higgs boson $H^+$ can be much lighter than the neutral Higgs boson with a SM-type coupling to the $Z$ boson. In fact, the $H^+$ boson could be as light as 80 GeV, so it could be the *lightest* particle in the *entire* Higgs spectrum. The planned colliders, i.e. the upgraded Tevatron collider and the LHC, are expected to be able to to test scenarios with a relatively light $H^+$, e.g. through the top-quark decay channel $t \to H^+ b$ [62]. It is crucial to notice that such light charged Higgs-boson scenarios, with $M_{H^+} \lesssim$ 100–120 GeV, are very difficult to obtain both in the MSSM and the NMSSM, for phenomenologically favoured values of the $\mu$-parameter, i.e. for $\mu > 100$ GeV.

(c) A clear phenomenological distinction of the NMSSM from the MNSSM and the MSSM as well can be obtained by means of the complementarity relations of the Higgs-boson couplings to the $Z$ boson:

$$g_{H_1 ZZ}^2 = g_{H_2 A_1 Z}^2 \,, \qquad g_{H_2 ZZ}^2 = g_{H_1 A_1 Z}^2 \,. \tag{4.14}$$

Specifically, the above relations (4.14) are not generically valid in the NMSSM [44]. A future $e^+e^-$ linear collider will have the capability to experimentally determine the $H_{1,2}ZZ$- and $H_{2,1}A_1Z$- couplings to a precision as high as 3% and so test, to a high degree of accuracy, the complementarity relations (4.14) which are an essential phenomenological feature of the MNSSM (with an unsuppressed $t_S$) and the MSSM.

The following sections will shed more light upon the field-theoretic, cosmological and phenomenological properties and implications of the NMSSM and the MNSSM.

### 4.2 The NMSSM Higgs mass spectrum

*David J. Miller*

We have seen in the introduction that the NMSSM provides an elegant solution to the $\mu$ problem of the MSSM by introducing an extra complex scalar Higgs superfield. After explicity breaking the Peccei-Quinn symmetry, the NMSSM results in the superpotential of Eq. (4.7) and the corresponding Higgs potential of Eq. (4.8). The Higgs sector consists of three scalars, $H_1$, $H_2$ and $H_3$, two pseudoscalars, $A_1$ and $A_2$, and two charged Higgs bosons, $H^{\pm}$, while the gaugino/higgsino sector consists of five neutralinos, $\tilde{\chi}_i^0$, $i = 1..5$, and four charginos $\tilde{\chi}_i^{\pm}$, $i = 1, 2$.

The first question to ask is, what is the mass spectrum of these particles, and how does it change as we alter the parameters of the model? In general, one would like to answer this question directly from the analytic formulae for the masses. However, even in the MSSM, these formulae are sufficiently complicated that it is difficult to untangle the effects of the different paramaters, and one must resort to numerical analysis. Principally, this is due to the large top quark Yukawa couplings which cause the Higgs masses to have large radiative corrections that must be taken into account.

This complicated nature largely carries over to the NMSSM, so that one must also include radiative corrections to get a complete picture of the mass hierarchy. However, the *extra* singlet fields introduced have no initial couplings to top quarks, and only gain a coupling through mixing with the other Higgs states. If the extra Higgs states are rather decoupled then their coupling to the top quark will be small and radiative corrections can be ignored for an approximate first look at their masses. Indeed, even if they are not decoupled, the top quark coupling will still be 'shared out' amongst the larger number of Higgs bosons and diluted. Since the radiative corrections are to the mass-squared rather than the mass, this will lead to a decrease in their importance. Therefore, neglecting radiative corrections in the NMSSM Higgs sector does a better job of approximating the masses than one would otherwise first suppose [30].

#### *4.2.1 The Higgs sector*

Since no new charged states have been added, the **charged Higgs boson** field content remains exactly the same as in the MSSM. Once the extra singlet field obtains a vev, the charged Higgs mass terms in the Higgs potential will be identical to the MSSM for both the D-terms and the soft-supersymmetry breaking terms. However, the term in the superpotential $\lambda S H_u^+ H_d^-$ provides an interaction between the charged Higgs bosons and the new singlet, so it will lead to slightly different F-terms. This causes a slightly altered parameterisation of the charged Higgs mass as compared to the MSSM,

$$M_{H^\pm}^2 = \frac{2\mu_{\text{eff}}}{\sin 2\beta}\left(A_\lambda + \frac{\kappa\mu_{\text{eff}}}{\lambda}\right) + M_W^2 - \frac{1}{2}\lambda^2 v^2. \tag{4.15}$$

Of course, since the overall scale is an input (via $A_\lambda$), one can regard the charged Higgs mass spectrum as being identical to that of the MSSM, and reparameterize the other Higgs masses accordingly. Indeed, this is what is normally done in the MNSSM and we will ocasionally follow the same approach here and regard $M_{H^\pm}$ as an input replacing $A_\lambda$.

The **pseudoscalar Higgs bosons** now have a $2 \times 2$ mass matrix, which rather easily lends itself to an analytic solution. However, the expression is still rather opaque, and it is useful to further approximate the masses by expanding in the (usually) small ratios $v/M_{H^\pm}$ and $1/\tan\beta$. This gives, for the two mass states,

$$M_{A_1}^2 \approx -3\frac{\kappa\mu_{\text{eff}}}{\lambda}A_\kappa, \tag{4.16}$$

$$M_{A_2}^2 \approx \frac{2\mu_{\text{eff}}}{\sin 2\beta}\left(A_\lambda + \frac{\kappa\mu_{\text{eff}}}{\lambda}\right)\left(1 + \lambda^2\frac{v^2}{8\mu_{\text{eff}}^2}\sin^2 2\beta\right). \tag{4.17}$$

The ordering of the solutions, normally in order of *ascending* mass, clearly depends on the choice of parameters, but one can see that one of the pseudoscalar Higgs bosons has a mass of order $M_{H^\pm}$, while the other depends on the root of the magnitude of the supersymmetry breaking term corresponding to the Peccei-Quinn breaking term in the supersopential, i.e. $\kappa A_\kappa$. Notice that vacuum stability requires that the quantity $A_\kappa$ should be (approximately) negative in order to keep $M^2_{A_1} > 0$.

The **scalar Higgs bosons** have a $3 \times 3$ mass matrix, so the exact tree-level masses are rather complicated, but again simplify considerably when we make our approximation. This gives,

$$M^2_{H_{1/2}} \approx \frac{1}{2}\left[M_Z^2\cos^2 2\beta + \frac{\kappa\mu_{\rm eff}}{\lambda}\left(A_\kappa + 4\frac{\kappa\mu_{\rm eff}}{\lambda}\right) \mp \left\{\left(M_Z^2\cos^2 2\beta - \frac{\kappa\mu_{\rm eff}}{\lambda}\left(A_\kappa + 4\frac{\kappa\mu_{\rm eff}}{\lambda}\right)\right)^2 \right.\right.$$
$$\left.\left. +2\lambda^2 v^2\left(2\mu_{\rm eff} - \left(A_\lambda + 2\frac{\kappa\mu_{\rm eff}}{\lambda}\right)\sin 2\beta\right)^2\right\}^{1/2}\right], \tag{4.18}$$

$$M^2_{H_3} \approx \frac{2\mu_{\rm eff}}{\sin 2\beta}\left(A_\lambda + \frac{\kappa\mu_{\rm eff}}{\lambda}\right)\left(1 + \lambda^2\frac{v^2}{8\mu^2_{\rm eff}}\sin^2 2\beta\right). \tag{4.19}$$

Again, the distinction of which scalar is $H_1$, $H_2$ or $H_3$ depends on the values of the parameters. We can see that $H_3$ is approximately degenerate with $A_2$ and $H^\pm$, which is directly analagous to the degeneracy of the heavy Higgs sector of the MSSM. All of these masses increase with increaing $A_\lambda$.

The dependence of the lighter scalars on $A_\lambda$ is entirely in the last term under the square root of Eq.(4.18). One finds that if this term becomes too large, i.e. if $A_\lambda$ deviates too far from $2\mu_{\rm eff}/\sin 2\beta - 2\kappa\mu_{\rm eff}/\lambda$, then the lightest Higgs boson will become tachyonic and the vacuum unstable. When this term is minimised, at $A_\lambda = 2\mu_{\rm eff}/\sin 2\beta - 2\kappa\mu_{\rm eff}/\lambda$, the two lightest scalars will take masses,

$$M^2_{H_1} \approx \frac{\kappa\mu_{\rm eff}}{\lambda}\left(A_\kappa + 4\frac{\kappa\mu_{\rm eff}}{\lambda}\right), \qquad M^2_{H_2} \approx M_Z^2\cos^2 2\beta. \tag{4.20}$$

Similarly to the lightest pseudoscalar, the lightest scalar mass depends on the square-root of $A_\kappa$. However, notice the opposite sign compared with Eq.(4.16). This effectively sets a constraint on the values which $A_\kappa$ for which the vacuum is stable,

$$-4\frac{\kappa\mu_{\rm eff}}{\lambda} \lesssim A_\kappa \lesssim 0. \tag{4.21}$$

A small value of $|A_\kappa|$ is phenomenologically interesting [48] since it leads to a very light pseudoscalar but a moderately heavy lightest scalar. Even if the scalar is still significantly below the current LEP Higgs boounds, it may have escaped detection by decaying into the light pseudoscalar. See Section 4.3 for further details.

The one-loop masses of $H_1$ and $A_1$ are shown in Fig.(4.1) as a function of $A_\kappa$ for a typical scenario. Increasing or decreasing $\kappa\mu_{\rm eff}/\lambda$ allows one to increase or decrease the masses of $H_1$ and $A_1$ simultaneously, while changing $A_\kappa$ allows one to shift mass from one state to the other. These effects are nicely summarized by the approximate sum rule (at $A_\lambda = 2\mu_{\rm eff}/\sin 2\beta - 2\kappa\mu_{\rm eff}/\lambda$),

$$M^2_{H_1} + \frac{1}{3}M^2_{A_1} \approx 4\left(\frac{\kappa\mu_{\rm eff}}{\lambda}\right)^2. \tag{4.22}$$

The approximate expressions are also plotted (dotted curves) and show at least a qualitative agreement with the one-loop results.

The approximate tree-level mass of the second lightest scalar (for this critical $A_\lambda$ value) should look familiar; it is the same as the tree-level approximation to the lightest scalar mass in the MSSM, and will similarly gain large radiative corrections.

Fig. 4.2 (*left*) shows the one-loop Higgs masses for the same scenario as Fig. 4.1 but now as a function of $M_{H^\pm}$, with $A_\kappa = 100$ GeV. As predicted, the heaviest scalar and pseudoscalar are approximately degenerate with the charged Higgs boson, and one scalar is roughly of the mass one would

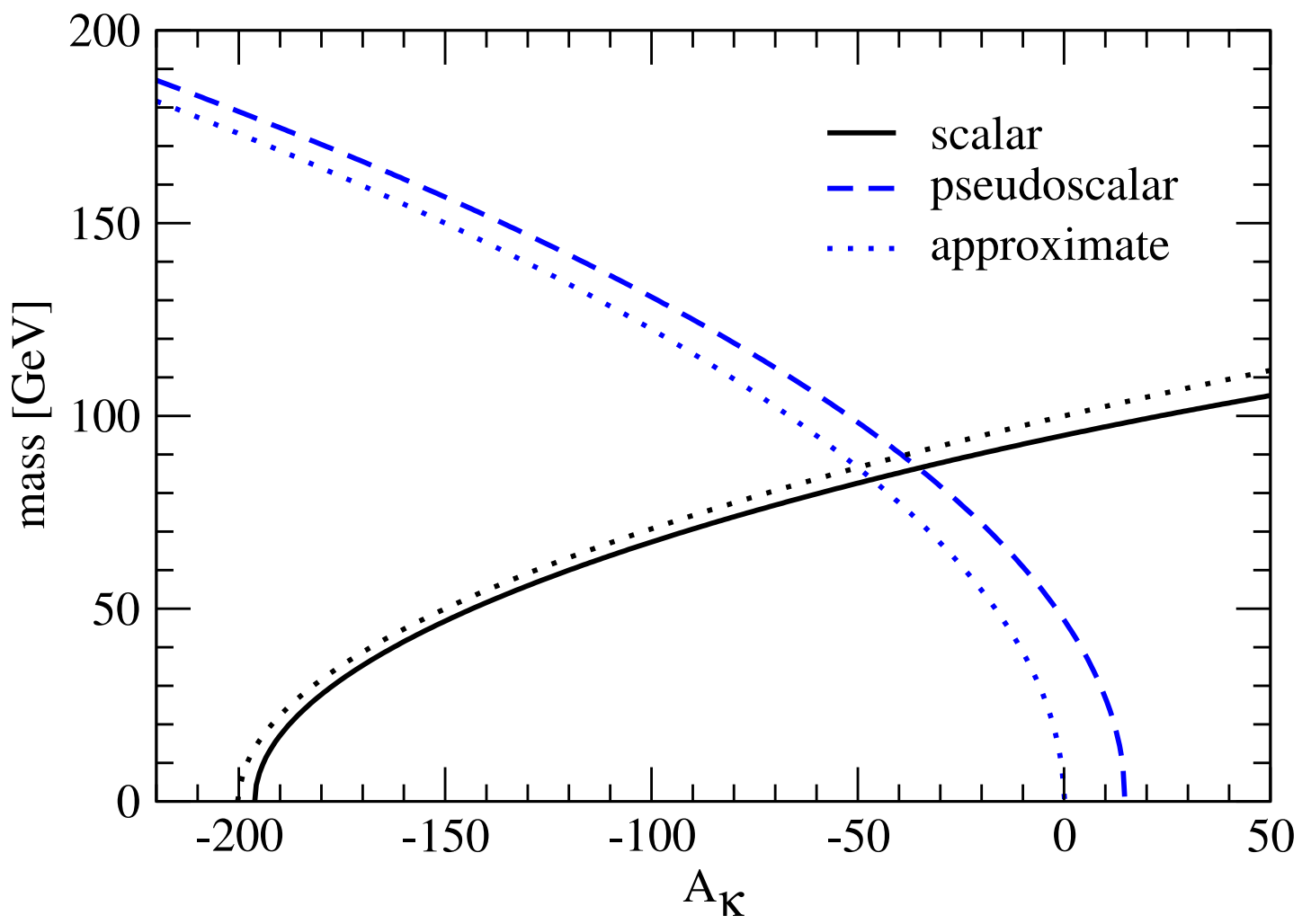

Fig. 4.1: The lightest scalar (solid) and pseudoscalar (dashed) Higgs masses at one-loop, as a function of $A_\kappa$, for $\lambda = 0.3$, $\kappa = 0.1$, $\tan\beta = 3$, $\mu_{\rm eff} = 150$ GeV, $A_\lambda = 450$ GeV and $M_{\rm SUSY} = 1$ TeV. Also shown by the dotted curves are the corresponding approximate expressions given in Eqs.(4.16) and (4.20).

expect of $h$ in the MSSM. The other two Higgs bosons are predominantly singlets, with masses set by the size of the Peccei-Quinn breaking terms, as previously discussed. Also shown (*dotted*) are the approximate masses of Eqs. (4.16-4.19). These approximate masses do very well indeed at predicting the pseudoscalar masses and the heaviest scalar mass. They do not do so well with the two lighter scalars, which is entirely due to the radiative corrections. If one were to plot the tree-level Higgs masses, one would see that the approximate solutions match the tree-level result almost perfectly. It is interesting to note that the lightest scalar mass is very well predicted at its maximal value (which is why the curves in Fig. 4.1 match so well). At this point, the lightest scalar is almost entirely the new singlet state, which has no coupling to top quarks and therefore no sizable radiative corrections. The second lightest scalar is, for this value of $M_{H^\pm}$, almost identical to the lightest scalar $h$ of the MSSM, and so will gain the same large radiative corrections from top/stop loops. As one moves away from this point, to the left or right, the Higgs bosons mix, the lightest scalar inherits a top quark coupling and the radiative corrections are shared out between them.

In Fig. 4.2 (*right*) we show the same masses for a larger value of $\kappa = 0.4$. As predicted, the increased size of the Peccei-Quinn breaking terms raises the masses of the singlet dominated fields. The singlet dominated scalar is now the heaviest scalar for low values of $M_{H^\pm}$ and the second lightest for high values of $M_{H^\pm}$. The increased mass contribution from the enhanced Peccei-Quinn breaking terms also reduces its mixing with the '$h$-like' scalar, and therefore its radiative corrections. In fact, now both scalar and pseudoscalar singlet dominated fields have masses which match very well with the approximate expressions (apart from where they become approximately degenerate with the other states). Once again, we can reduce the singlet dominated scalar mass while increasing the pseudoscalar mass, or vice versa, by altering the parmeter $A_\kappa$. Note that the charged Higgs mass is contrained by the requirement that the lightest scalar mass-squared be positive (i.e. vacuum stability).

### *4.2.2 The Neutralino Sector*

The Neutralinos do not suffer from the same large radiative corrections seen in the Higgs sector. However, the Neutralino mass matrix is now $5 \times 5$, so analytic expressions for the tree-level masses cannot be obtained in closed form. In order to get a handle on the behaviour of the masses with respect to variations in the parameters we must again resort to approximate expressions [31]. The Chargino masses and mixings are unaffected by the additional fields (although they may have extra decays if kinematically allowed) so we will not discuss them further here.

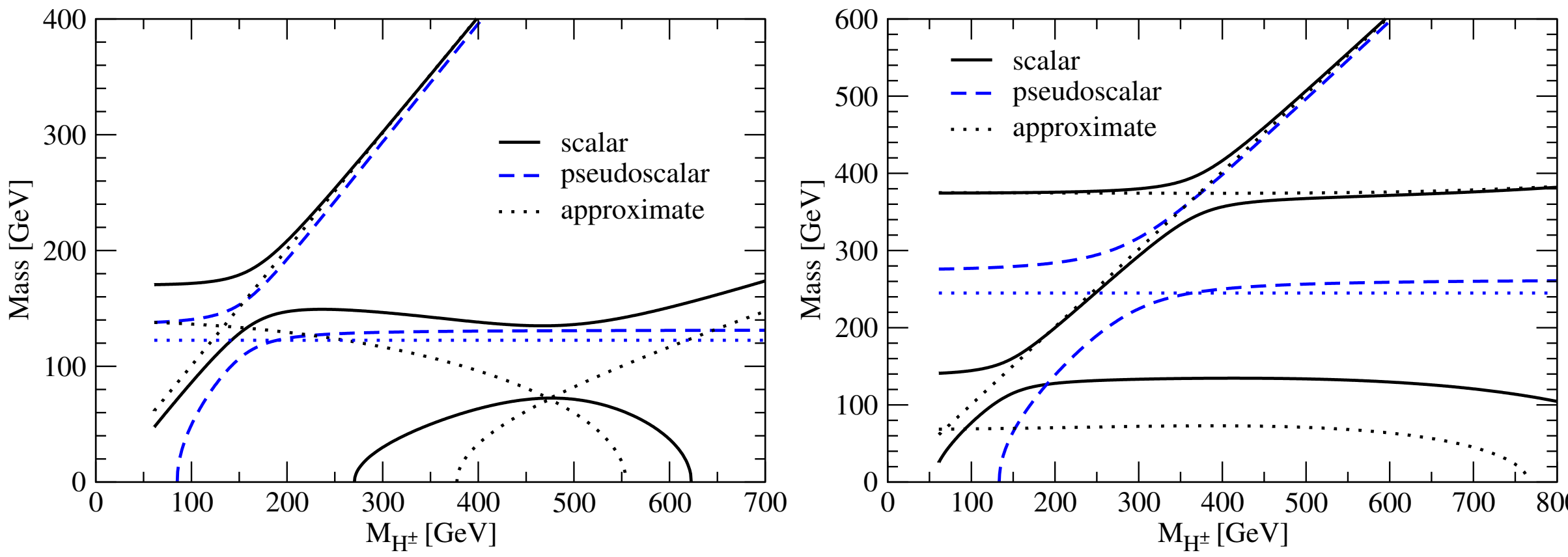

Fig. 4.2: The scalar (solid) and pseudoscalar (dashed) Higgs masses at one-loop, as a function of $M_{H^\pm}$, for $\lambda = 0.3$, $\tan\beta = 3$, $\mu_{\text{eff}} = 150$ GeV, $A_\kappa = 100$ GeV, $M_{\text{SUSY}} = 1$ TeV and $\kappa = 0.1$ (left) or 0.4 (right). Also shown by the dotted curves are the corresponding approximate expressions given in Eqs. (4.16-4.19).

When considering the neutralino sector we can forget about the soft supersymmetry breaking parameters $A_\lambda$ and $A_\kappa$, which have no effect, but must instead include the soft supersymmetry breaking gaugino masses $M_1$ and $M_2$. The relevant parameters are then $\lambda$, $\kappa$, $\mu_{\text{eff}}$, $\tan\beta$, $M_1$ and $M_2$. For illustrative purposes we will here consider the case where $M_{1,2} \gg |\mu_{\text{eff}}| \gg M_Z$, and $|\mu_{\text{eff}}| \gg \lambda v/\sqrt{2}$ (this last condition is saying that the vev of the new field should be substantially larger than that of the usual doublets, i.e. $\langle S\rangle \gg v/\sqrt{2}$). With this approximation, we find,

$$\begin{aligned}
m_1 &\approx M_1 + \frac{M_Z^2}{M_1} s_W^2, \\
m_2 &\approx M_2 + \frac{M_Z^2}{M_2} c_W^2, \\
m_3 &\approx -\mu_{\text{eff}} - \frac{M_{12}}{2M_1M_2} M_Z^2 (1 - \sin 2\beta) - \frac{\lambda^2 v^2}{4\mu_{\text{eff}}}(1 + \sin 2\beta), \\
m_4 &\approx \mu_{\text{eff}} - \frac{M_{12}}{2M_1M_2} M_Z^2 (1 + \sin 2\beta) + \frac{\lambda^2 v^2}{4\mu_{\text{eff}}}(1 - \sin 2\beta), \\
m_5 &\approx 2\frac{\kappa\mu_{\text{eff}}}{\lambda} + \frac{\lambda^2 v^2}{2\mu_{\text{eff}}} \sin 2\beta,
\end{aligned} \tag{4.23}$$

where $s_W$ and $c_W$ are the sine and cosine of the Weinberg angle and $M_{12} = M_1 c_W^2 + M_2 s_W^2$. As in the Higgs sector, one would normally reorder these states in order of ascending mass (and, unlike the Higgs sector, include phase rotations to render them positive), depending on the parameters. One finds that the singlet dominated neutralino is that labeled '5' above with a mass which grows as the Peccei-Quinn breaking (i.e. $\kappa$) is increased. For particularly low values of $\kappa$, the second term may dominate.

The tree-level neutralino masses are plotted in Fig. 4.3 as a function of $\kappa$, together with the approximate expressions. We see that two of the neutrinos match the input soft supersymmetry breaking gaugino masses very well, with two more slightly above and below $\mu_{\text{eff}}$. The last neutralino is singlet dominated and increases linearly with $\kappa$. The approximate forms match rather well with the exact (though tree-level) results, except when the states are nearly degenerate, and with the proviso that one must re-lable the approximate expressions depending on the hierarchy. Only the prediction for the lightest state is rather low. For other hierarchies of the paremeters e.g. $\mu_{\text{eff}} \gg M_{1,2} \gg M_z$, different approximations

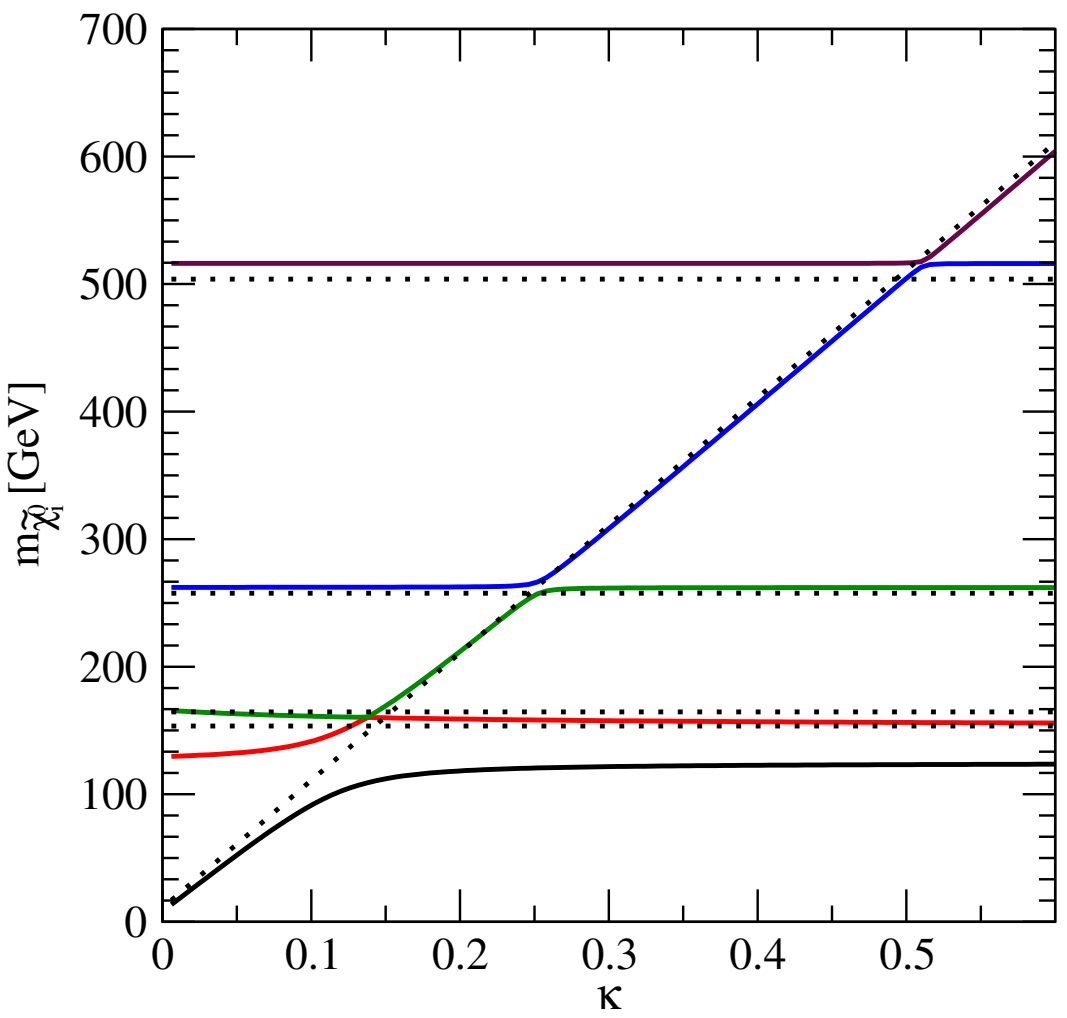

Fig. 4.3: The tree-level neutralino masses (*solid*) as a function of $\kappa$ with $M_1 = 250$ GeV, $M_2 = 500$ GeV, $\lambda = 0.3$, $\mu_{\rm eff} = 120$ GeV, $\tan\beta = 3$. Also shown (*dashed*) are the approximate forms of Eq. (4.23).

are more appropriate but give similar results [31].

### 4.3 Low fine-tuning scenarios in the NMSSM and LHC/ILC implications

*Radovan Dermisek and John F. Gunion*

In this contribution, we describe how the NMSSM can achieve extremely low fine-tuning and how it is that low fine-tuning scenarios can provide a very nice explanation of the LEP Higgs event excess in the vicinity of Higgs mass $\sim 100$ GeV. In these highly preferred scenarios, $e^+e^-$ collisions produce a CP-even Higgs boson (generically denoted by $H$) with SM-like $ZZ, WW$ couplings in association with the $Z$, but the Higgs decays predominantly to two light CP-odd Higgs bosons (generically denoted by $A$), where each CP-odd Higgs boson decays to $\tau^+\tau^-$ or jets (because its mass is below $2m_b$). This serves to suppress the decays of the CP-even Higgs to $b\overline{b}$ to more or less exactly the right level to describe the LEP event excess in the $Z + b\overline{b}$ channel. Implications of such scenarios for future colliders are reviewed.

Recall from the introduction that the Higgs sector of the NMSSM contains the CP-even Higgs bosons $H_1, H_2, H_3$, the CP-odd Higgs bosons $A_1, A_2$ and the usual charged Higgs pair $H^\pm$, and that the properties of the Higgs bosons are fixed by the six parameters

$$\lambda,\ \kappa,\ A_\lambda,\ A_\kappa,\ \tan\beta,\ \mu_{\rm eff}\ . \tag{4.24}$$

along with input values for the gaugino masses and for the soft terms related to the (third generation) squarks and sleptons that contribute to the radiative corrections in the Higgs sector and to the Higgs decay widths. Exploration of the NMSSM Higgs sector is greatly simplified by employing the `NMHDECAY` [63, 64] program. All available radiative corrections are implemented therein.

In general, an important issue for NMSSM Higgs phenomenology is the mass and nature of the lightest CP-even and CP-odd Higgs bosons. In particular, if the $A_1$ is very light or even just moderately light there are are dramatic modifications in the phenomenology of Higgs discovery at both the LHC and ILC [30, 63–72]. A light $A_1$ is natural in the context of the model. Indeed, the NMSSM can contain either an approximate global $U(1)$ R-symmetry in the limit that the Higgs-sector trilinear soft SUSY breaking terms are small ($\kappa A_\kappa, \lambda A_\lambda \to 0$), or a $U(1)$ Peccei-Quinn symmetry in the limit that the cubic singlet term in the superpotential and its soft partner vanish ($\kappa, \kappa A_\kappa \to 0$) [66, 67]. In either case, one ends up with the lightest CP-odd Higgs boson, $A_1$, as the pseudo-Nambu-goldstone boson of this broken symmetry, implying that it can naturally be light. If one of these symmetries were unbroken, it would lead to a massless CP-odd $A_1$ which is ruled out. However, a very light $A_1$ is not ruled out. The low fine-tuning scenarios are associated with a small breaking of the $U(1)$ R-symmetry that can arise from explicit non-zero values for $A_\lambda$ and $A_\kappa$ and/or radiative corrections to $A_\lambda$ and to the pseudoscalar Higgs mass-squared matrix that are present even when $A_\kappa$ and $A_\lambda$ are zero at tree-level.

#### *4.3.1 Fine-tuning in the MSSM and NMSSM*

We begin with a discussion of how it is that in the NMSSM, adding a Higgs singlet superfield allows one to reduce [48, 55] the fine tuning problem, which is present in the case of the MSSM due to the fact that LEP II excludes a SM-like CP-even Higgs boson with mass below 114 GeV that decays primarily to $b\bar{b}$.

One standard measure of fine-tuning is [73]

$$F = \mathrm{Max}_p F_p \equiv \mathrm{Max}_p \left| \frac{d \log M_Z}{d \log p} \right| , \tag{4.25}$$

where the parameters $p$ comprise all GUT-scale soft-SUSY-breaking parameters. We will show that $F$ can be much smaller in the NMSSM than in the MSSM [48,55]. In particular, in the NMSSM, fine-tuning can even be eliminated if the lightest CP-odd Higgs is light enough to allow $H_1 \to A_1 A_1$ decays [48] and the $A_1$ has mass below $2m_b$ so that it decays to $\tau^+\tau^-$, $q\bar{q}$ and/or $gg$.

In the MSSM model constraints are such that the lightest CP-even Higgs boson ($h$) is most naturally very SM-like, in which case $M_h \gtrsim 114$ GeV is required by LEP limits (except for a small window in parameter space where the CP-odd MSSM $A$ has mass between $\sim 90$ GeV and $\sim 114$ GeV and $\tan\beta$ is large). Such a large $M_h$ is not easily obtained without having a very substantial value for the root-mean stop mass, $\sqrt{m_{\tilde{t}_1} m_{\tilde{t}_2}}$ and/or large stop mixing (parameterized by the soft stop mixing parameter $A_t$), upon which the radiative corrections to $M_h$ in the MSSM primarily depend. As a result, the MSSM is very fine-tuned and the associated hierarchy problem is substantial (see, for example, [48]).

The NMSSM can be much less fine-tuned in several interesting ways. Let us recall the formula for the maximum tree-level mass-squared of a SM-like Higgs boson in the NMSSM:

$$[M_H{}^2]_{tree} \leq M_Z^2 \left( \cos^2 2\beta + \frac{2\lambda^2}{g^2 + g'^2} \sin^2 2\beta \right) , \tag{4.26}$$

where typically this applies to $H = H_1$ or $H = H_2$, depending upon which is SM-like. To this tree-level result one must add the radiative corrections from the stop squarks and top quark loops. For small $\lambda$ and/or large $\tan\beta$, Eq. (4.26) reduces to the MSSM result of $[M_H{}^2]_{tree} \leq M_Z^2 \cos^2 2\beta$. However, if $\lambda$ is taken large compared to $g, g'$, and $\tan\beta$ is not far from unity, the 2nd term can dominate and a value of $M_H{}^2 > (114 \text{ GeV})^2$ is possible without having to employ extreme $\sqrt{m_{\tilde{t}_1} m_{\tilde{t}_2}}$ or stop mixing parameter $A_t$ values. The result is a lower level of fine-tuning [29, 55] as compared to the MSSM. However, to get values substantially below those found in the MSSM requires $\lambda$ (at scale $M_Z$) to be $\mathcal{O}(1)$, above the limit $\lambda \leq 0.7$ for which $\lambda$ remains perturbative under evolution all the way up to the GUT scale $M_U$.

The alternative [48] is to choose parameters for which fine-tuning is "automatically" minimized. The lowest fine-tuning is achieved for scenarios in which the lightest Higgs boson of the NMSSM is SM-like in its normal couplings and has mass below 114 GeV and yet escapes LEP constraints by virtue of having unusual decay modes for which LEP limits are weaker. In particular, parameters for which $H_1 \to A_1 A_1$ decays are dominant are consistent with LEP constraints for $M_{H_1}$ as low as 90 GeV (110 GeV) if the dominant $A_1$ decay is to $\tau^+\tau^-$ ($b\bar{b}$). This immediately allows lower $\sqrt{m_{\tilde{t}_1} m_{\tilde{t}_2}}$ and $A_t$ and, therefore, smaller $F$ values. The very low values of $F$ that can be achieved are illustrated in Fig. 4.4. The points plotted are those from a large scan over NMSSM parameters at fixed $\tan\beta = 10$ and $M_{1,2,3}(M_Z) = 100, 200, 300$ GeV. All points plotted pass `NMHDECAY` constraints, which include in particular perturbativity for $\lambda$ up to $M_U$ and all single-channel (the meaning and importance of this restriction will be explained in Section 4.3.2) LEP constraints. The points with lowest $F$ ($F < 10$) correspond to $\sqrt{m_{\tilde{t}_1} m_{\tilde{t}_2}} \sim 250 - 400$ GeV and have 98 GeV $\leq M_{H_1} \leq 105$ GeV. For about 2/3 of these points $M_{A_1} < 2m_b$ and the main $A_1$ decay is to $\tau^+\tau^-$. For the remaining 1/3 of the $F < 10$ points, $M_{A_1} > 2m_b$ and $BR(A_1 \to b\bar{b}) \sim 0.92$ and $BR(A_1 \to \tau^+\tau^-) \sim 0.08$. As discussed in the next section, the $M_{A_1} > 2m_b$ scenarios are not consistent with the full LEP-Higgs Working Group (LHWG) LEP analysis, whereas the $M_{A_1} < 2m_b$ scenarios describe very nicely the $M_H \sim 100$ GeV excess in the $Z + b$'s final state. Implications for the LHC and the ILC are discussed in the final section of this report.

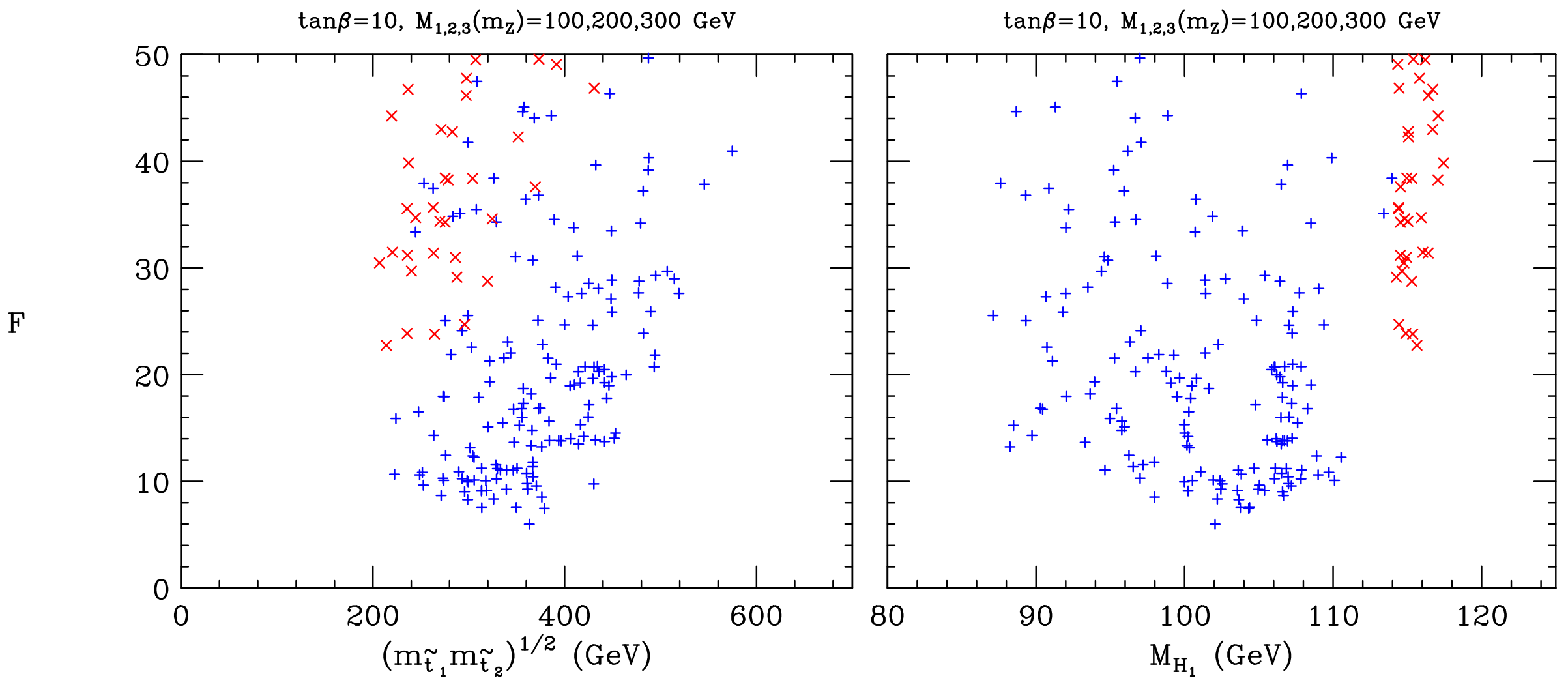

Fig. 4.4: For the NMSSM, we plot the  ne-tuning measure $F$ vs. $\sqrt{m_{\tilde{t}_1} m_{\tilde{t}_2}}$ (left) and vs. $M_{H_1}$ (right) for NMHDECAY-accepted scenarios with $\tan\beta = 10$ and $M_{1,2,3}(M_Z) = 100, 200, 300$ GeV. Points marked by '+' ('×') have $M_{H_1} < 114$ GeV ($M_{H_1} \geq 114$ GeV) and escape LEP *single-channel* limits primarily due to dominance of $H_1 \to A_1 A_1$ decays (due to $M_{H_1} > 114$ GeV).

Overall, the $M_{A_1} < 2m_b$ scenarios are very appealing since they are extremely consistent with the two primary features of the LEP data: i) the consistency of the LEP precision electroweak data with the presence of a CP-even Higgs boson with SM-like $ZZH$ coupling and $M_H \sim 100$ GeV and ii) a $M_H \sim 100$ GeV Higgs boson that has SM-like $WW, ZZ$ couplings but decays to $b\bar{b}$ with about 1/10 the branching ratio of a SM Higgs boson as a result of primary decays to state(s) that do not contain $b$-quarks.

### *4.3.2 NMSSM scenarios with low fine-tuning*

Let us now give more details regarding how the NMSSM can achieve much lower fine-tuning than the MSSM. In Ref. [48], two types of scenarios were examined for parameter choices such that $F < 10$. In both scenarios, $BR(H_1 \to b\bar{b}) \sim 0.07 - 0.2$ and $BR(H_1 \to A_1 A_1) \sim 0.88 - 0.75$. In scenarios of type I (II), $M_{A_1} > 2m_b$ ($M_{A_1} < 2m_b$) and $BR(A_1 \to b\bar{b}) \sim 0.92$ (0).

To relate this to LEP data, let us discuss the observed and expected 95% CL limits in the $Z4b$ channel from [74] and in the $Z2b$ channel from [75]. Both show event excesses. In particular, for $M_H$ in the vicinity of $105 - 110$ GeV and $M_A$ in the 30 GeV to 50 GeV zone there is a sharp deviation of the observed limit on $C_{\rm eff}^{4b} = [g_{ZZH}^2/g_{ZZH_{SM}}^2] BR(H \to ZAA)[BR[A \to b\bar{b})]^2$ to values above the expected limit, implying the presence of excess events. A similar deviation has been evident in the $Zb\bar{b}$ final state for a number of years [75]. One finds a higher observed 95% CL for $C_{\rm eff}^{2b} = [g_{ZZH}^2/g_{ZZH_{SM}}^2] BR(H \to b\bar{b})$ in the $Zb\bar{b}$ final state as a function of $M_H$ than expected in the $M_H \sim 100-110$ GeV vicinity. The statistical significance of this excess is in the $1\sigma-2\sigma$ range. It would seem that there is a significant possibility that both the $Z2b$ and $Z4b$ excesses arise from the decays of a single CP-even Higgs boson with SM-like coupling strength to gauge bosons and fermions, but with additional coupling to a light, perhaps CP-odd, Higgs boson. This is precisely the scenario that applies in the NMSSM (with $H = H_1$ and $A = A_1$) for those parameter choices that correspond to low values of $F$ and $M_{A_1} > 2m_b$. Typically, the low-$F$ scenarios with $M_{A_1} > 2m_b$ have $BR(H_1 \to b\bar{b}) \sim 0.05 - 0.2$ and $M_{A_1} \sim 30 - 45$ GeV with $BR(H_1 \to A_1A_1) \sim 0.85 - 0.75$ [with $BR(A_1 \to b\bar{b}) \sim 0.93$]. Note, however, that for simultaneous consistency with the separate $C_{\rm eff}^{2b}$ and $C_{\rm eff}^{4b}$ limits, $M_{H_1} \gtrsim 106$ GeV is required, a value which is not particularly ideal for the $M_H \sim 100$ GeV location of the largest $Z2b$ excess.

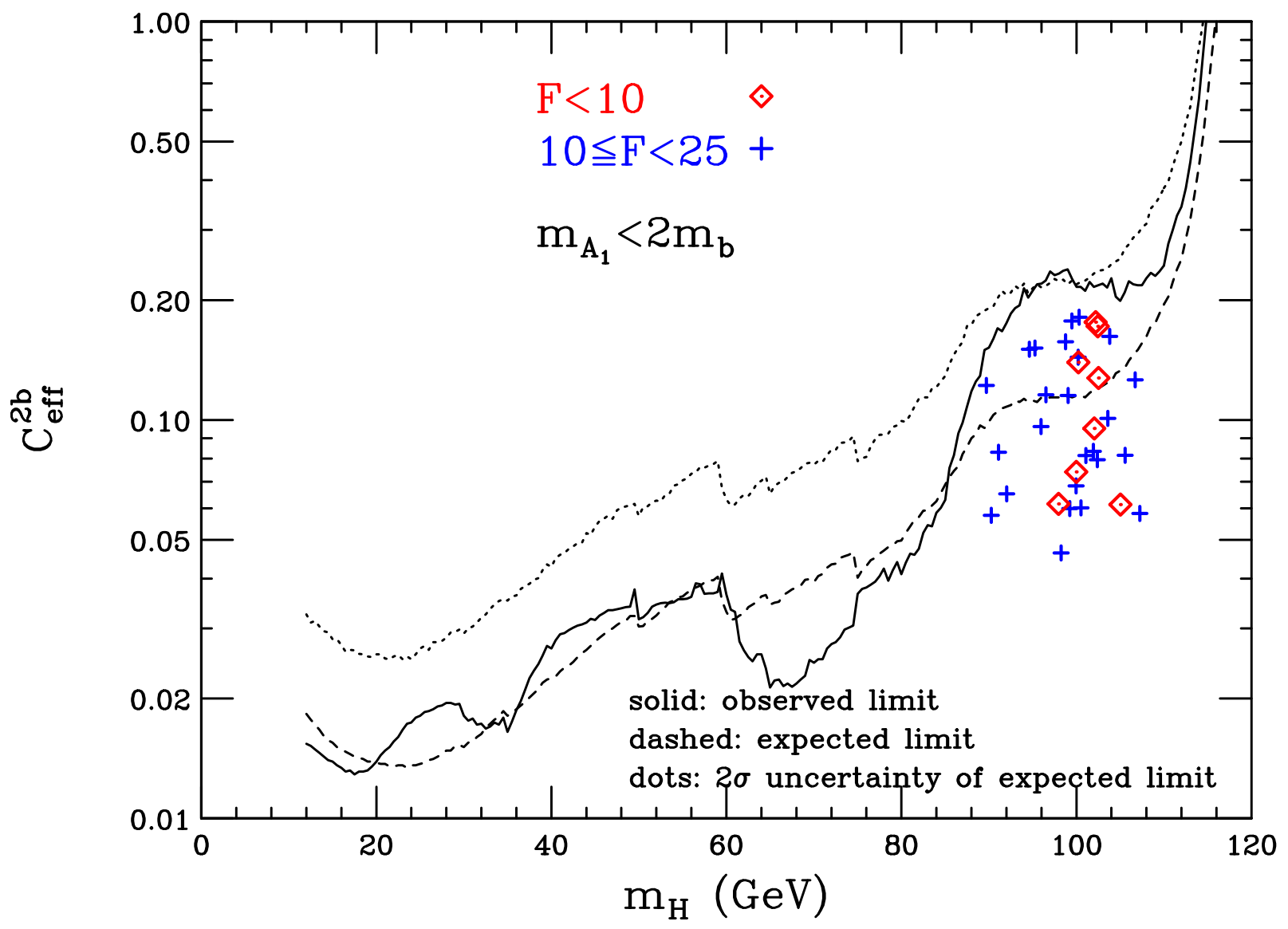

Fig. 4.5: Expected and observed 95% CL limits on $C_{\rm eff}^{2b} = [g^2_{ZZH}/g^2_{ZZH_{SM}}]BR(H \to b\bar{b})$ from Ref. [75] are shown vs. $M_H$. Also plotted are the predictions for NMSSM parameter choices in our scan that give ne-tuning measure $F < 25$ and $M_{A_1} < 2m_b$ with xed $\tan\beta = 10$ and gaugino masses of $M_{1,2,3}(M_Z) = 100, 200, 300$ GeV.

In fact, there is an even more severe problem. Although these type-I ($M_{A_1} > 2m_b$) scenarios satisfy the separate $Z2b$ and $Z4b$ LHWG limits, the overlap between these two analyzes is such that when both channels are present with the rates predicted, all type-I scenarios with $F < 10$ are excluded. This conclusion was reached only after processing the type-I scenario predictions through the full $1-CL_b$ LHWG analysis machinery. The result is that the only $F < 10$ scenarios consistent with the full LHWG analysis are of type-II ($M_{A_1} < 2m_b$). Type-I scenarios with $F < 10$ are typically excluded at better than the 99% CL after data from all experiments are combined. However, it should be remarked that the OPAL experiment, which has the best $Z2b$ vs. $Z4b$ discrimination, does not on its own exclude such a scenario and does see excesses in both the $Z2b$ and $Z4b$ channels. For $10 < F < 100$ the range of possibilities is expanded. In particular, there are $M_{H_1} \gtrsim 108$ GeV points that make a net contribution to the $Z2b$ and $Z4b$ channels that is reduced compared to the $F < 10$ cases and that probably would not be excluded by the combined analysis. However, again the $M_{H_1}$ mass is too large to explain the largest $Z2b$ excess in the $M_H \sim 100$ GeV region.

As a result of the problems with the $M_{A_1} > 2m_b$ scenarios as outlined above, it is clearly important to analyze the scenarios with $M_{A_1} < 2m_b$. As noted earlier, a light $A_1$ is natural in the NMSSM in the $\kappa A_\kappa, \lambda A_\lambda \to 0$ limit due to the presence of a global $U(1)_R$ symmetry of the scalar potential which is spontaneously broken by the vevs, resulting in a Nambu-Goldstone boson in the spectrum [66]. This symmetry is explicitly broken by the trilinear soft terms so that for small $\kappa A_\kappa, \lambda A_\lambda$ the lightest

CP odd Higgs boson is naturally much lighter than other Higgs bosons. For the $F < 10$ scenarios, $\lambda(M_Z) \sim 0.15 - 0.25$, $\kappa(M_Z) \sim 0.15 - 0.3$, $|A_\kappa(M_Z)| < 4$ GeV and $|A_\lambda(M_Z)| < 200$ GeV, implying small $\kappa A_\kappa$ and moderate $\lambda A_\lambda$. The effect of $\lambda A_\lambda$ on $M_{A_1}$ is further suppressed when the $A_1$ is largely singlet in nature, as is the case for low-$F$ scenarios. Therefore, we always obtain small $M_{A_1}$. We note that small soft SUSY-breaking trilinear couplings at the unification scale are generic in SUSY breaking scenarios where SUSY breaking is mediated by the gauge sector, as, for instance, in gauge or gaugino mediation. Although the value $A_\lambda(M_Z)$ might be sizable due to contributions from gaugino masses after renormalization group running between the unification scale and the weak scale, $A_\kappa$ receives only a small correction from the running (such corrections being one loop suppressed compared to those for $A_\lambda$). Altogether, a light, singlet $A_1$ is very natural in models with small soft SUSY-breaking trilinear couplings at the unification scale. Finally, we note that the above $\lambda(M_Z)$ values are such that $\lambda$ will remain perturbative when evolved up to the unification scale, implying that the resulting unification-scale $\lambda$ values are natural in the context of model structures that might yield the NMSSM as an effective theory below the unification scale.

A very important feature of the $M_{A_1} < 2m_b$, $F < 10$ scenarios is that a significant fraction of them can easily explain the well-known excess in the $ZH \to Zb\bar{b}$ final state in the vicinity of $M_H \sim 100$ GeV. This is illustrated in Fig. 4.5 from [49], where all $F < 10$ NMSSM parameter choices (from a lengthy scan) with $M_{A_1} < 2m_b$ are shown to predict $M_{H_1} \sim 98 - 105$ GeV with about half the points predicting $M_{H_1} \sim 100 - 102$ GeV along with a $C_{\rm eff}^{2b}$ that would explain the observed excess with respect to the expected limit. The other primary decay mode for all the plotted points is $H_1 \to A_1 A_1$ with $A_1 \to \tau^+\tau^-$ or light quarks and gluons (when $M_{A_1} < 2m_\tau$). Thus, unlike the type-I ($M_{A_1} > 2m_b$) scenarios, there is no additional contribution to the $Z + b$'s final state — the $C_{\rm eff}^{2b}$ limits are the most relevant single-channel limits. However, to really decide if a given scenario is consistent with the LEP data, and at what level, it must be processed through the complete LHWG confidence level/likelihood analysis. This processing was performed for us by P. Bechtle. In Table 4.1, we give the precise masses and branching ratios of the $H_1$ and $A_1$ for all the $F < 10$ points. We also give the number of standard deviations, $n_{\rm obs}$ ($n_{\rm exp}$), by which the observed rate (expected rate obtained for the predicted signal+background) exceeds the predicted background. These are derived from $(1 - CL_b)_{\rm observed}$ and $(1 - CL_b)_{\rm expected}$ using the usual tables: e.g. $(1 - CL_b) = 0.32$, 0.045, 0.0027 correspond to $1\sigma$, $2\sigma$, $3\sigma$ excesses, respectively. The quantity $s95$ is the factor by which the signal predicted in a given case would have to be multiplied in order to exceed the 95% CL. All these quantities are obtained by processing each scenario through the full preliminary LHWG confidence level/likelihood analysis. If $n_{\rm exp}$ is larger than $n_{\rm obs}$ then the excess predicted by the signal plus background Monte Carlo is larger than the excess actually observed and vice versa. The points with $M_{H_1} \lesssim 100$ GeV have the largest $n_{\rm obs}$. Point 2 gives the best consistency between $n_{\rm obs}$ and $n_{\rm exp}$, with a predicted excess only slightly smaller than that observed. Points 1 and 3 also show substantial consistency. For the 4th and 7th points, the predicted excess is only modestly larger (roughly within $1\sigma$) compared to that observed. The 5th and 6th points are very close to the 95% CL borderline and have a predicted signal that is significantly larger than the excess observed. LEP is not very sensitive to point 8. Thus, a significant fraction of the $F < 10$ points are very consistent with the observed event excess.

We wish to emphasize that in our scan there are many, many points that satisfy all constraints and have $M_{A_1} < 2m_b$. The remarkable result is that those with $F < 10$ have a substantial probability that they predict the Higgs boson properties that would imply a LEP $ZH_1 \to Z + b$'s excess of the sort seen. The smaller number of $F < 10$ points with $M_{A_1}$ substantially above $2m_b$ all predict a net $Z + b$'s signal that is ruled out at better than 99% CL by LEP data. Indeed, all $F < 25$ points have a net $H_1 \to b$'s branching ratio, $BR(H_1 \to b\bar{b}) + BR(H_1 \to A_1 A_1 \to b\bar{b}b\bar{b}) \gtrsim 0.85$, which is too large for LEP consistency if $M_{H_1}$ is near 100 GeV. (Analysis of points with $M_{A_1}$ very near $b\bar{b}$ decay threshold, but such that $A_1 \to b\bar{b}$ is dominant, is very subtle. Such points arise for $F < 10$ and require further analysis in cooperation with the LHWG.)

Table 4.1: Some properties of the $H_1$ and $A_1$ for the eight allowed points with $F < 10$ and $M_{A_1} < 2m_b$ from our $\tan\beta = 10$, $M_{1,2,3}(M_Z) = 100, 200, 300$ GeV NMSSM scan. The $n_{\rm obs}$, $n_{\rm exp}$ and $s95$ values are obtained after full processing of all $ZH$ nal states using the preliminary LHWG analysis code (thanks to P. Bechtle). See text for details. $N_{SD}^{LHC}$ is the statistical signi cance of the best standard LHC Higgs detection channel for integrated luminosity of $L = 300$ fb$^{-1}$.

| $M_{H_1}/M_{A_1}$ (GeV) | Branching Ratios $H_1 \to b\bar{b}$ | $H_1 \to A_1A_1$ | $A_1 \to \tau\bar{\tau}$ | $n_{\rm obs}/n_{\rm exp}$ units of $1\sigma$ | $s95$ | $N_{SD}^{LHC}$ |
|---|---|---|---|---|---|---|
| 98.0/2.6 | 0.062 | 0.926 | 0.000 | 2.25/1.72 | 2.79 | 1.2 |
| 100.0/9.3 | 0.075 | 0.910 | 0.852 | 1.98/1.88 | 2.40 | 1.5 |
| 100.2/3.1 | 0.141 | 0.832 | 0.000 | 2.26/2.78 | 1.31 | 2.5 |
| 102.0/7.3 | 0.095 | 0.887 | 0.923 | 1.44/2.08 | 1.58 | 1.6 |
| 102.2/3.6 | 0.177 | 0.789 | 0.814 | 1.80/3.12 | 1.03 | 3.3 |
| 102.4/9.0 | 0.173 | 0.793 | 0.875 | 1.79/3.03 | 1.07 | 3.6 |
| 102.5/5.4 | 0.128 | 0.848 | 0.938 | 1.64/2.46 | 1.24 | 2.4 |
| 105.0/5.3 | 0.062 | 0.926 | 0.938 | 1.11/1.52 | 2.74 | 1.2 |

As already noted, these low-$F$ NMSSM scenarios have an $A_1$ that is fairly singlet in nature. This means that $Z^* \to ZA_1A_1$ at LEP (and indeed all $A_1$ production mechanisms based on the $ZZA_1A_1$ and $WWA_1A_1$ quartic interactions) would have a very low rate. The $A_1WW$ and $A_1ZZ$ couplings arise first at one loop and the $A_1t\bar{t}$ coupling is also very suppressed. At $\tan\beta = 10$, the suppression from the $A_1$'s predominantly singlet composition is compensated by the $\tan\beta$ factor yielding $A_1b\bar{b}\,\gamma_5$ coupling strength that is of order the $H_{SM}b\bar{b}$ scalar coupling strength. A final feature of the low-$F$ points that should be noted is that all the other Higgs bosons are fairly heavy, typically above 400 GeV in mass.

### *4.3.3 Higgs detection in the low-$F$ scenarios*

An important question is the extent to which the type of $H \to AA$ Higgs scenario (whether NMSSM or other) described here can be explored at the Tevatron, the LHC and a future $e^+e^-$ linear collider. This has been examined in the case of the NMSSM in [65, 68, 72], with the conclusion that observation of any of the NMSSM Higgs bosons may be difficult at hadron colliders. At a naive level, the $H_1 \to A_1A_1$ decay mode renders inadequate the usual Higgs search modes that might allow $H_1$ discovery at the LHC. Since the other NMSSM Higgs bosons are rather heavy and have couplings to $b$ quarks that are not greatly enhanced, they too cannot be detected at the LHC. The last column of Table 4.1 shows the statistical significance of the most significant signal for *any* of the NMSSM Higgs bosons in the "standard" SM/MSSM search channels for the eight $F < 10$ NMSSM parameter choices. For the $H_1$ and $A_1$, the most important detection channels are $H_1 \to \gamma\gamma$, $WH_1 + t\bar{t}H_1 \to \gamma\gamma\ell^\pm X$, $t\bar{t}H_1/A_1 \to t\bar{t}b\bar{b}$, $b\bar{b}H_1/A_1 \to b\bar{b}\tau^+\tau^-$ and $WW \to H_1 \to \tau^+\tau^-$ – see [72]. Even after $L = 300$ fb$^{-1}$ of accumulated luminosity, the typical maximal signal strength is at best $3.5\sigma$. For the eight points of Table 4.1, this largest signal derives from the $WH_1 + t\bar{t}H_1 \to \gamma\gamma\ell^\pm X$ channel. There is a clear need to develop detection modes sensitive to the dominant $H_1 \to A_1A_1 \to \tau^+\tau^-\tau^+\tau^-$ decay channel.

Let us consider the possibilities. Two detection modes that can be considered are $WW \to H_1 \to A_1A_1 \to 4\tau$ and $gg \to t\bar{t}H_1 \to t\bar{t}A_1A_1 \to t\bar{t}4\tau$. Second, recall that the $\tilde{\chi}_2^0 \to H_1\tilde{\chi}_1^0$ channel provides a signal in the MSSM when $H_1 \to b\bar{b}$ decays are dominant. See, for example, [76]. It has not been studied for $H_1 \to A_1A_1 \to 4\tau$ decays. If a light $\tilde{\chi}_1^0$ provides the dark matter of the universe (as possible because of the $\tilde{\chi}_1^0\tilde{\chi}_1^0 \to A_1 \to X$ annihilation channels for a light $A_1$, see [53, 54] and references therein as well as the separate contribution on this subject, the $m_{\tilde{\chi}_2^0} - m_{\tilde{\chi}_1^0}$ mass difference might be large enough to allow such decays. Diffractive production [77–79], $pp \to ppH_1 \to ppX$, where the mass $M_X$ can be reconstructed with roughly a $1-2$ GeV resolution, can potentially reveal a Higgs peak, independent of the decay of the Higgs. A study [80] is underway to see if this discovery mode works for the $H_1 \to A_1A_1 \to 4\tau$ decay mode as well as it appears to work for the simpler SM $h_{SM} \to b\bar{b}$ case. The main issue may be whether events can be triggered despite the soft nature of the decay products of the $\tau$'s present in $X$ when $H_1 \to A_1A_1 \to 4\tau$ as compared to $h_{SM} \to b\bar{b}$.

At the Tevatron it is possible that $ZH_1$ and $WH_1$ production, with $H_1 \to A_1A_1 \to 4\tau$, will provide the most favorable channels. If backgrounds are small, one must simply accumulate enough events. However, efficiencies for triggering on and isolating the $4\tau$ final state will not be large. Perhaps one could also consider $gg \to H_1 \to A_1A_1 \to 4\tau$ which would have substantially larger rate. Studies are needed. If supersymmetry is detected at the Tevatron, but no Higgs is seen, and if LHC discovery of the $H_1$ remains uncertain, Tevatron studies of the $4\tau$ final state might be essential. However, rates imply that the $H_1$ signal could only be seen if Tevatron running is extended until $L > 10\ \mathrm{fb}^{-1}$ has been accumulated. Even if both the Tevatron and the LHC are unable to detect the $H_1$, the LHC *would* observe numerous supersymmetry signals and *would* confirm that $WW \to WW$ scattering is perturbative, implying that something like a light Higgs boson must be present.

Of course, discovery of the $H_1$ will be straightforward at an $e^+e^-$ linear collider via the inclusive $ZH_1 \to \ell^+\ell^-X$ reconstructed $M_X$ approach (which allows Higgs discovery independent of the Higgs decay mode). Direct detection in both the $ZH_1 \to \ell^+\ell^-b\bar{b}$ and $ZH_1 \to \ell^+\ell^-4\tau$ modes will also be possible. At a $\gamma\gamma$ collider, the $\gamma\gamma \to H_1 \to 4\tau$ signal will be easily seen [81].

In contrast, since (as already noted) the $A_1$ in these low-$F$ NMSSM scenarios is fairly singlet in nature, its *direct* (i.e. not in $H_1$ decays) detection will be very challenging even at the ILC. Further, the low-$F$ points are all such that the other Higgs bosons are fairly heavy, typically above 400 GeV in mass, and essentially inaccessible at both the LHC and all but a $\gtrsim 1$ TeV ILC.

We should note that much of the discussion above regarding Higgs discovery when $H \to AA$ decays are dominant is quite generic. Whether the $A$ is truly the NMSSM CP-odd $A_1$ or just a lighter Higgs boson into which the SM-like $H$ pair-decays, hadron collider detection of the $H$ in its $H \to AA$ decay mode will be very challenging — only an $e^+e^-$ linear collider can currently guarantee its discovery.

### 4.4 Di-photon Higgs signals at the LHC as a probe of an NMSSM Higgs sector

*Stefano Moretti and Shoaib Munir*

In view of the upcoming CERN LHC, quite some work has been dedicated to probing the NMSSM Higgs sector over recent years. Primarily, there have been attempts to extend the so-called 'No-lose theorem' of the MSSM [82] to the case of the NMSSM [68–71]. From this perspective, it was realised that at least one NMSSM Higgs boson should remain observable at the LHC over the NMSSM parameter space that does not allow any Higgs-to-Higgs decay. However, when the only light non-singlet (and, therefore, potentially visible) CP-even Higgs boson, $H_1$ or $H_2$, decays mainly to two very light CP-odd Higgs bosons, $A_1A_1$, one may not have a Higgs signal of statistical significance at the LHC [83]. While the jury is still out on whether a 'No-lose theorem' can be proved for the NMSSM, we are here concerned with an orthogonal approach. We asked ourselves if a, so to say, 'More-to-gain theorem' can be formulated in the NMSSM. That is, whether there exist regions of the NMSSM parameter space where more Higgs states of the NMSSM are visible at the LHC than those available within the MSSM. In our attempt to overview all such possibilities, we start by considering here the case of the di-photon decay channel of a neutral Higgs boson. This mode can be successfully probed in the MSSM, but limitedly to the case of one Higgs boson only, which is CP-even and rather light. We will argue that in the NMSSM one can instead potentially have up to three di-photon signals of Higgs bosons, involving not only CP-even but also CP-odd states, the latter with masses up to 600 GeV or so. In fact, even when only one di-photon signal can be extracted in the NMSSM, this may well be other than the $H_1$ state. When only the latter is visible, finally, it can happen that its mass is larger than the maximum value achievable within the MSSM. In all such cases then, the existence of a non-minimal SUSY Higgs sector would be manifest.

For a general study of the NMSSM Higgs sector (without any assumption on the underlying SUSY-breaking mechanism) we used here the `NMHDECAY` code (version 1.1) [63]. This program computes the masses, couplings and decay Branching Ratios (BRs) of all NMSSM Higgs bosons in terms of model parameters taken at the EW scale. For our purpose, instead of postulating unification, we fixed the soft SUSY breaking terms to a very high value, so that they have little or no contribution to the outputs of the parameter scans. Consequently, we are left with six free parameters: the Yukawa couplings $\lambda$ and $\kappa$, the soft trilinear terms $A_\lambda$ and $A_\kappa$, plus tan$\beta$ and $\mu_{\rm eff} = \lambda\langle S\rangle$. The computation of the spectrum includes leading two-loop terms, EW corrections and propagator corrections. `NMHDECAY` also takes into account theoretical as well as experimental constraints from negative Higgs searches at collider experiments.

We have used `NMHDECAY` to scan over the NMSSM parameter space defined in [84] (borrowed from [72]), where also the configuration of the remaining SUSY soft terms can be found. The allowed decay modes for neutral NMSSM Higgs bosons are into any SM particle, plus into any final state involving all possible combinations of two Higgs bosons (neutral and/or charged) or of one Higgs boson and a gauge vector as well as into all possible sparticles. We have performed our scan over several millions of randomly selected points in the specified parameter space. The data points surviving all constraints are then used to determine the cross-sections for NMSSM Higgs hadro-production. As the SUSY mass scales have been set well above the EW one, the production modes exploitable in simulations at the LHC are those involving couplings to heavy ordinary matter only, i.e., the so-called 'direct' Higgs production modes of [85]. Production and decay rates for NMSSM neutral Higgs bosons have then been multiplied together to yield inclusive event rates, assuming a LHC luminosity of 100 fb$^{-1}$ throughout.

As an initial step we computed the total cross-section times BR into $\gamma\gamma$ pairs against each of the six parameters of the NMSSM, for each neutral Higgs boson. We have assumed all production modes mentioned above and started by computing fully inclusive rates. We are focusing on the $\gamma\gamma$ decay mode since it is the most promising channel for the discovery of a (neutral) Higgs boson at the LHC in the moderate Higgs mass range (say, below 130 GeV). However, since the tail of the $\gamma\gamma$ background falls rapidly with increasing invariant mass of the di-photon pair, signal peaks for heavier Higgses could also be visible in addition to (or instead of) the lightest one [86]. As the starting point of our signal-to-background study, based on the ATLAS analysis of Ref. [87], we argue that cross-section times BR rates of 10 fb or so are interesting from a phenomenological point of view, in the sense that they may yield visible signal events, the more so the heavier the decaying Higgs state (also because the photon detection efficiency grows with the Higgs mass [87]). Hereafter, we will refer to such NMSSM parameter configurations as 'potentially visible'.

We have then plotted the NMSSM configurations with three potentially visible Higgses $H_1, H_2$ and $A_1$ (in selected combinations, as detailed in the captions) against the various model parameters in Figs. 4.6–4.8. Their spread is quite homogeneous over the NMSSM parameter space and not located in some specific parameter areas (i.e., in a sense, not 'fine-tuned'). The distribution of the same points in terms of cross-sections times BR as a function of the corresponding Higgs masses can be found in Fig. 4.9. Of particular relevance is the distribution of points in which only the NMSSM $H_1$ state is visible, when its mass is beyond the upper mass limit for the corresponding CP-even MSSM Higgs state, which is shown in Fig. 4.10[3]. This plot reveals that about 93% of the NMSSM $H_1$ masses visible alone are expected to be within 2–3 GeV beyond the MSSM bound, hence the two models would be indistinguishable[4]. Nonetheless, there is a fraction of a percent of such points with $M_{H_1}$ values even beyond 125 GeV or so (the higher the mass the smaller the density, though), which should indeed allow one to distinguish between the two models. Moreover, by studying the cross-section times BR of the Higgses when two of them are observable against their respective mass differences (see Figs. 9–11 of [84]) and widths, one sees that the former are larger than the typical mass resolution in the di-photon channel, so that the two decaying objects should indeed appear in the data as separate resonances.

[3]Notice that the value obtained for $M_{H_1}^{\rm max}$ from `NMHDECAY` version 1.1, of $\sim$ 130 GeV, based on the leading two-loop approximations described in [63], is a few GeV lower than the value declared in Sect. 4.1.4. Besides, for consistency, we use the value of 120 GeV (obtained at the same level of accuracy) as upper mass limit on the lightest CP-even Higgs boson of the MSSM. (Notice that a slightly modified $M_{H_1}^{\rm max}$ value is obtained for the NMSSM from `NMHDECAY` version 2.1 [64], because of the improved mass approximations with respect to the earlier version of the program adopted here.) Eventually, when the LHC is on line, the exercise that we are proposing can be performed with the then state-of-the-art calculations.

Table 4.2: Higgs events potentially visible at the LHC through the $\gamma\gamma$ decay mode. Percentage refers to the portion of NMSSM parameter space involved for each discovery scenario.

| Higgs Flavor | Points Visible | | Percentage |
|---|---|---|---|
| $H_1$ | Total: | 1345884 | 99.7468 |
| | Alone: | 1345199 | 99.6961 |
| | With $H_2$: | 528 | 0.0391 |
| | With $A_1$: | 152 | 0.0113 |
| | With $H_2$ and $A_1$: | 5 | 0.0004 |
| $H_2$ | Total: | 1253 | 0.0929 |
| | Alone: | 717 | 0.0531 |
| | With $H_1$: | 528 | 0.0391 |
| | With $A_1$: | 3 | 0.0002 |
| | With $A_1$ and $A_1$: | 5 | 0.0004 |
| $H_3$ | Total: | 0 | 0 |
| $A_1$ | Total: | 165 | 0.0122 |
| | Alone: | 5 | 0.0004 |
| | With $H_1$: | 152 | 0.0113 |
| | With $H_2$: | 3 | 0.0002 |
| | With $H_1$ and $H_2$: | 5 | 0.0004 |
| $A_2$ | Total: | 0 | 0 |

Table 4.2 recaps the potential observability of one or more NMSSM Higgs states in the di-photon mode at the LHC. It is obvious from the table that one light CP-even Higgs should be observable almost throughout the NMSSM parameter space. However, there is also a fair number of points where two Higgses may be visible simultaneously ($H_1$ and $H_2$ or – more rarely – $H_1$ and $A_1$), while production and decay of the three lightest Higgses ($H_1$, $H_2$ and $A_1$) at the same time, although possible, occurs for only a negligible number of points in the parameter space. Furthermore, the percentage of points for which only the second lightest Higgs state is visible is also non-negligible. These last two conditions are clearly specific to the NMSSM, as they are never realised in the MSSM. Finally, none of the two heaviest NMSSM neutral Higgs states ($H_3$ and $A_2$) will be visible in the di-photon channel at the LHC (given their large masses).

Next, we have proceeded to a dedicated parton level analysis of signal and background processes, the latter involving both tree-level $q\bar{q} \to \gamma\gamma$ and one-loop $gg \to \gamma\gamma$ contributions. We have adopted standard cuts on the two photons [87]: $p_T^{\gamma} > 25$ GeV and $|\eta^{\gamma}| < 2.4$ on transverse energy and pseudorapidity, respectively. As illustrative examples of a possible NMSSM Higgs phenomenology appearing at the LHC in the di-photon channel, we have picked up the following three configurations:

1. $\lambda = 0.6554$, $\kappa = 0.2672$, $\mu_{\rm eff} = -426.48$ GeV, $\tan\beta = 2.68$, $A_\lambda = -963.30$ GeV, $A_\kappa = 30.48$ GeV;
2. $\lambda = 0.6445$, $\kappa = 0.2714$, $\mu_{\rm eff} = -167.82$ GeV, $\tan\beta = 2.62$, $A_\lambda = -391.16$ GeV, $A_\kappa = 50.02$ GeV;
3. $\lambda = 0.4865$, $\kappa = 0.3516$, $\mu_{\rm eff} = 355.63$ GeV, $\tan\beta = 2.35$, $A_\lambda = 519.72$ GeV, $A_\kappa = -445.71$ GeV.

[4]Other than an experimental di-photon mass resolution of 2 GeV or so [87] one should also bear in mind here that the mass bounds in both models come at present with a theoretical error of comparable size.

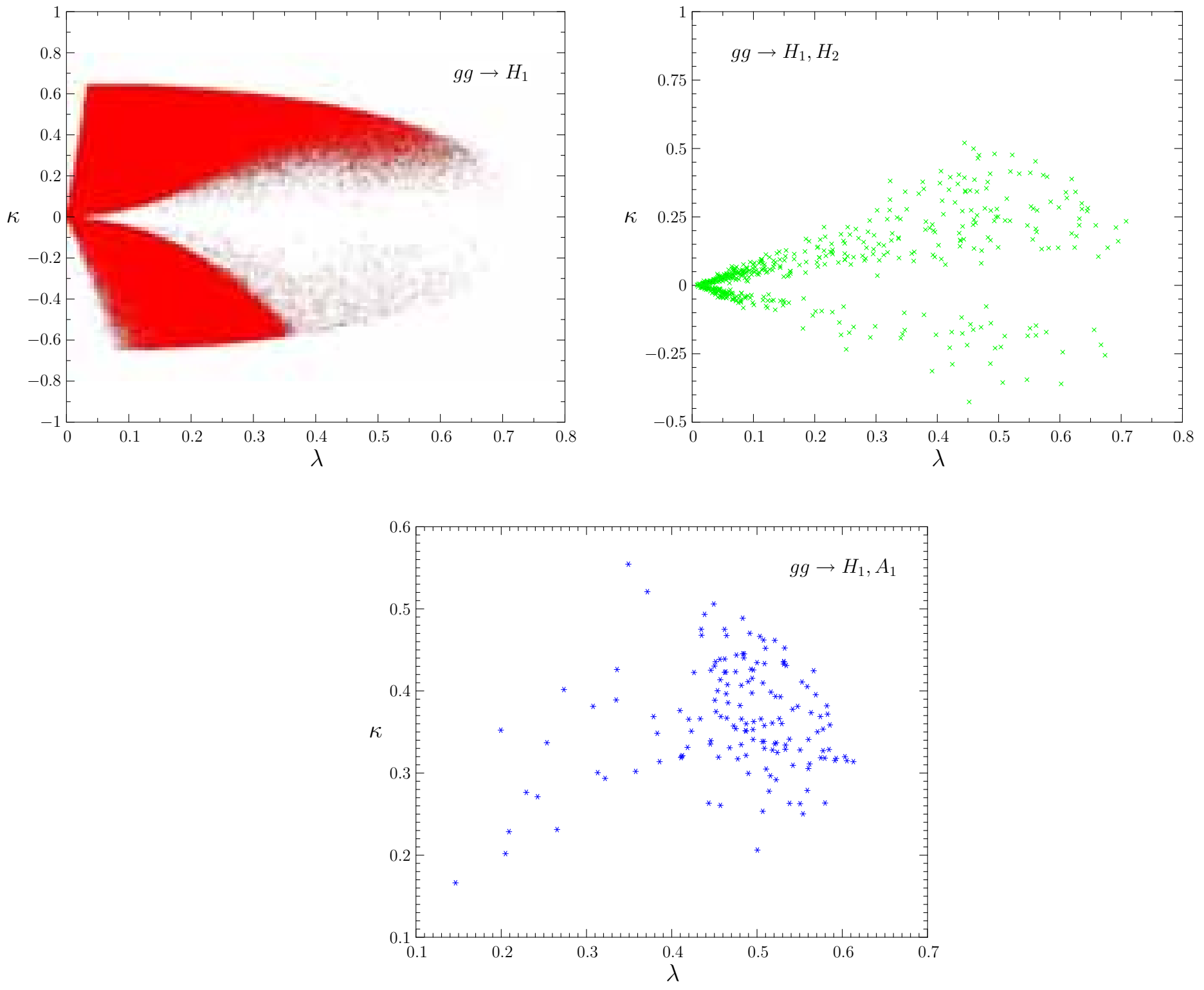

Fig. 4.6: The NMSSM parameter space when $H_1$ (red/dots), $H_1, H_2$ (green/crosses) and $H_1, A_1$ (blue/stars) are potentially visible (individually and simultaneously), plotted against $\lambda$ and $\kappa$.

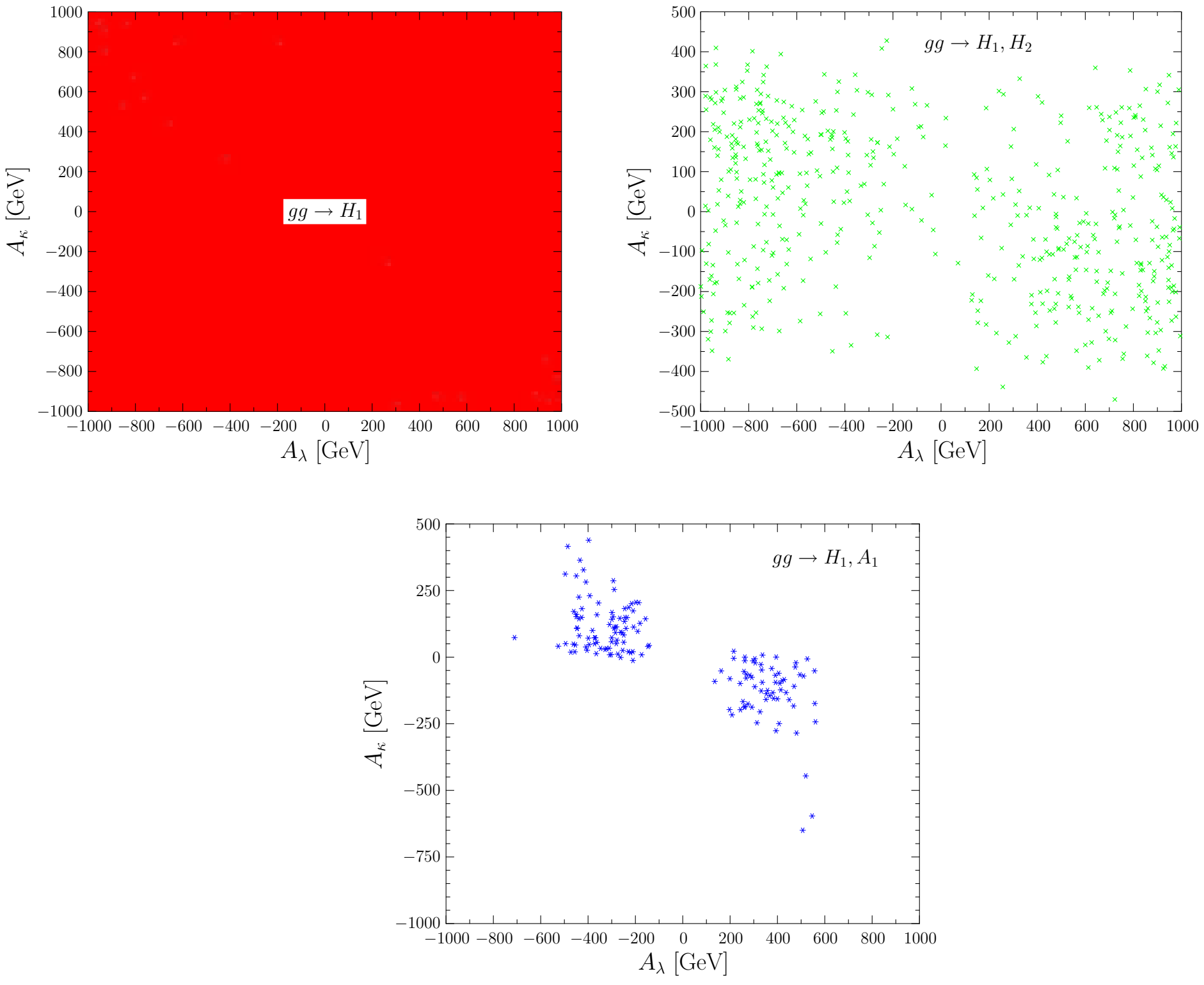

Fig. 4.7: As above, plotted against $A_\lambda$ and $A_\kappa$.

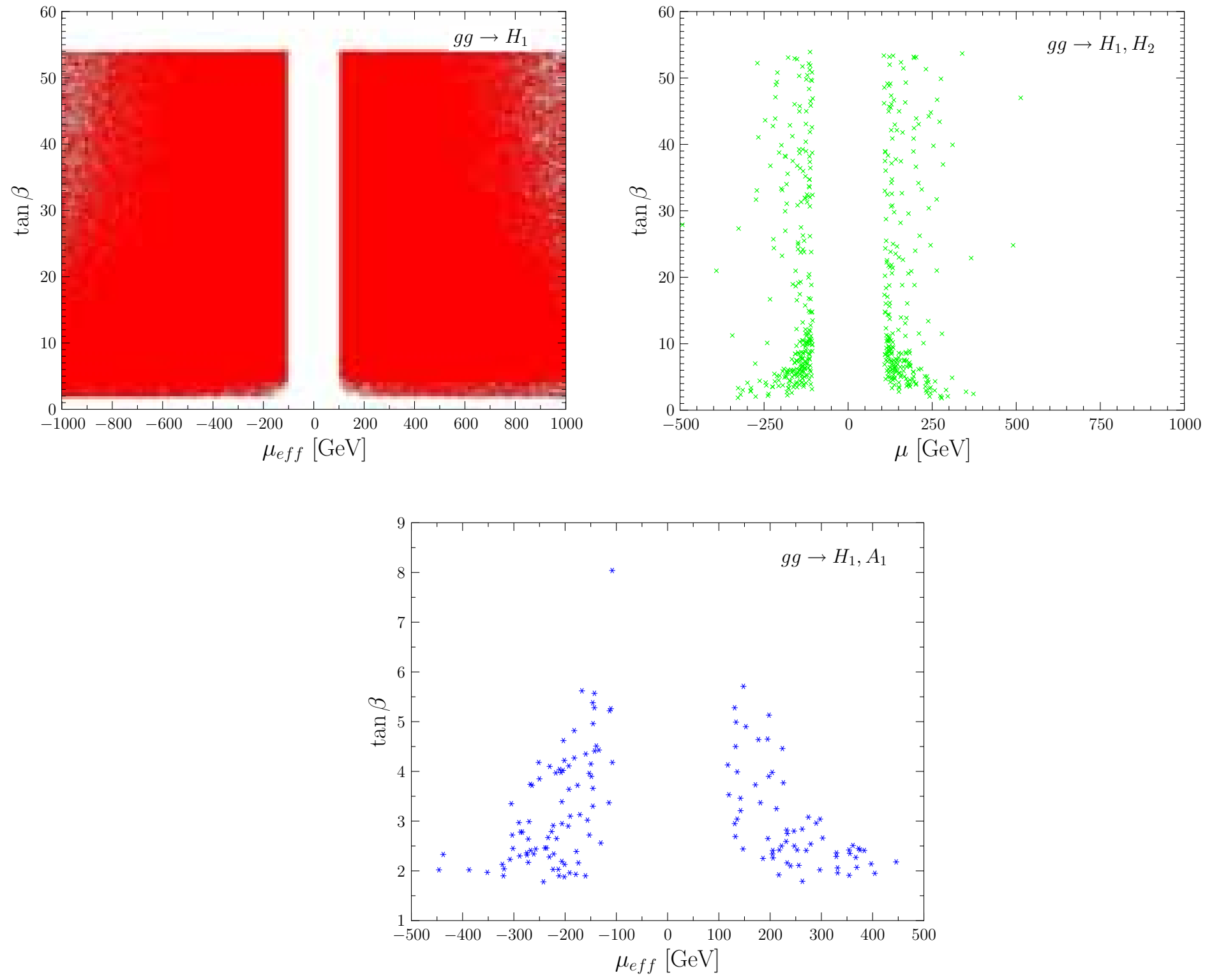

Fig. 4.8: As above, plotted against $\mu_{\rm eff}$ and $\tan\beta$.

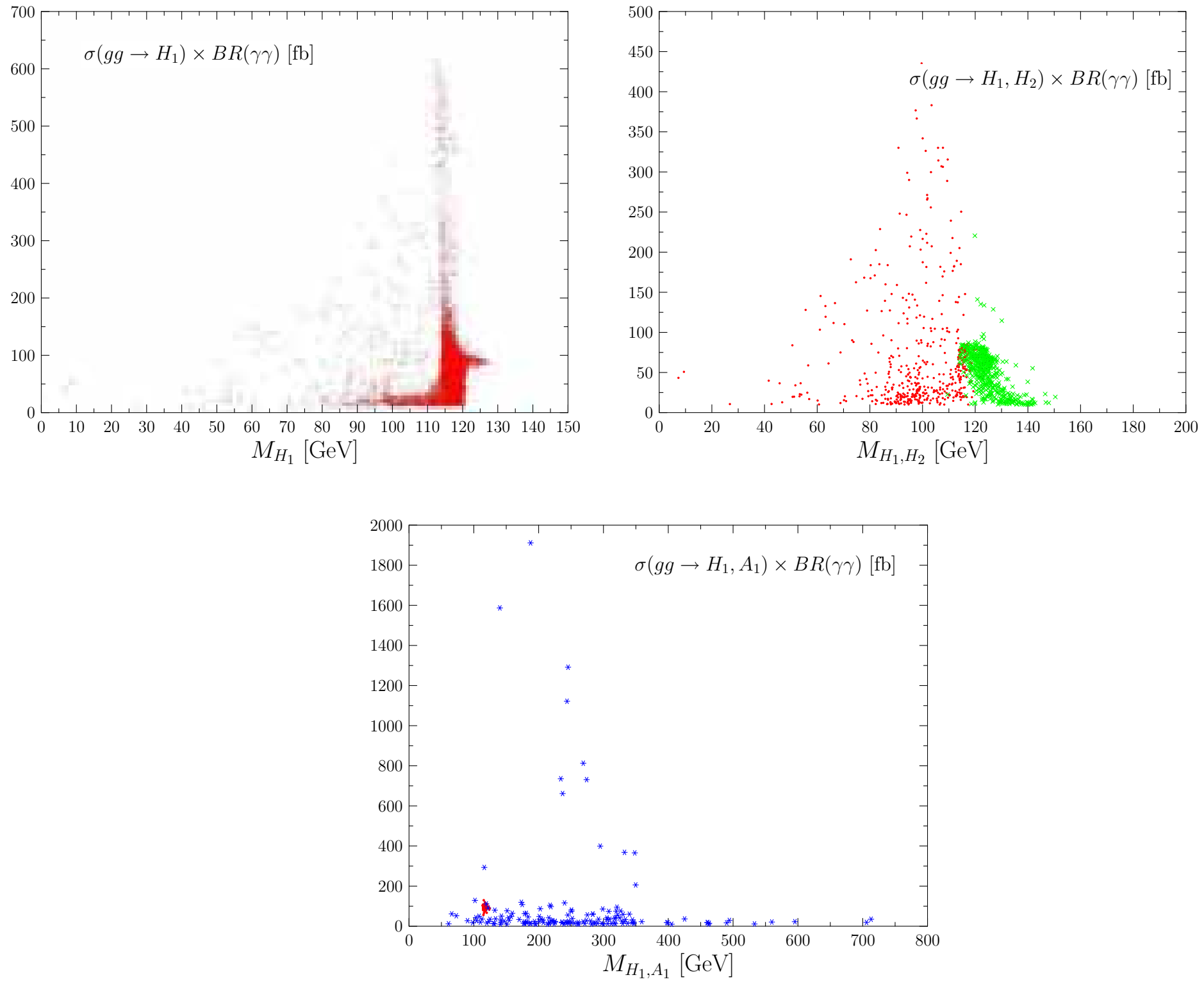

Fig. 4.9: Cross-section times BR for $H_1$ (red/dots), $H_2$ (green/crosses) and $A_1$ (blue/stars), when potentially visible (individually and simultaneously) plotted against their respective masses.

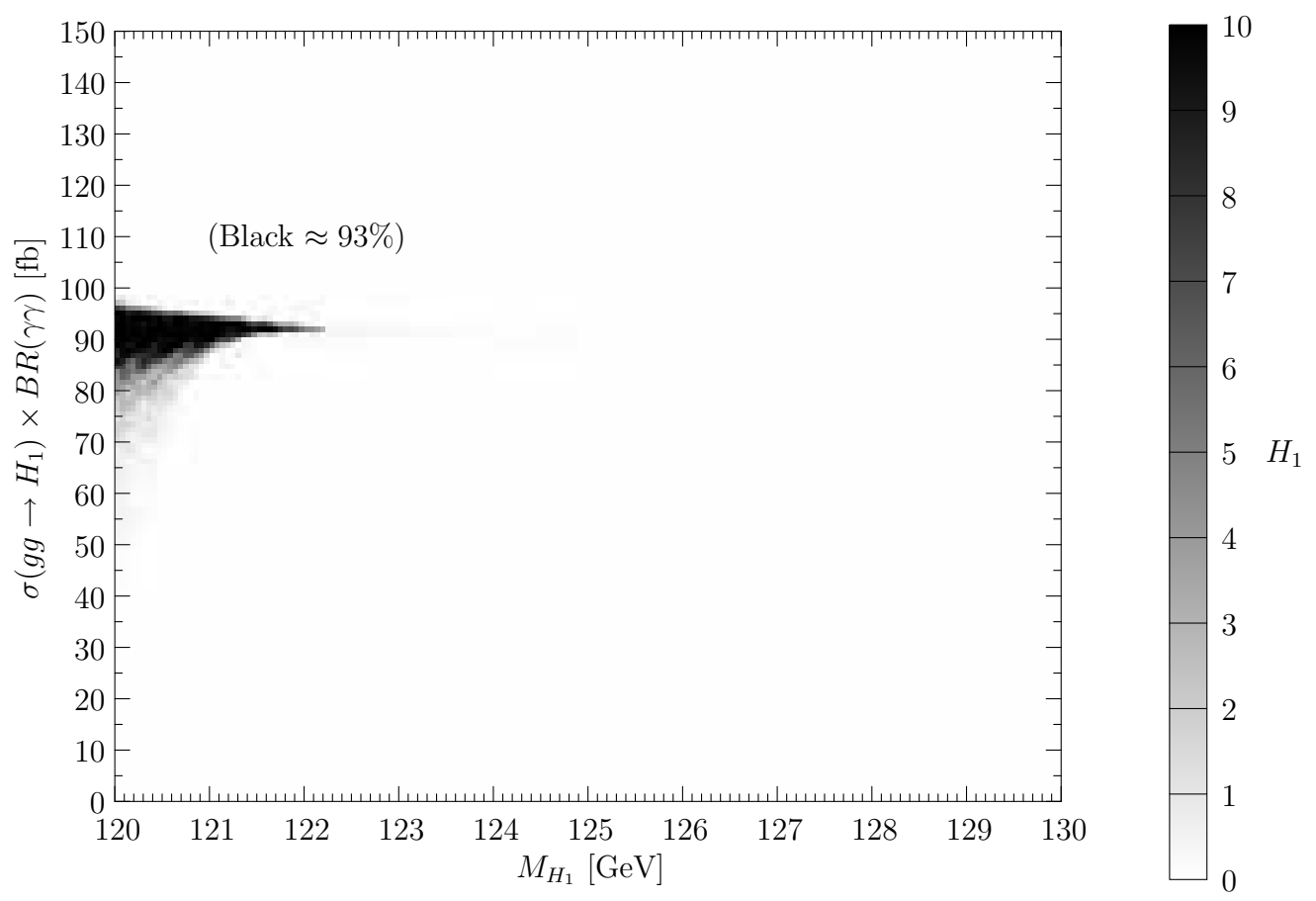

Fig. 4.10: The distribution of points with one potentially visible NMSSM $H_1$ state with mass beyond the MSSM upper mass limit on the corresponding Higgs state. The scale on the right represents a measure of point density.

The first is representative of the case in which only the NMSSM $H_1$ boson is visible, but with mass larger than the MSSM upper limit on the corresponding Higgs state. The second and third refer instead to the case when also the $H_2$ or $A_1$ state are visible, respectively. The final results are found in Fig. 4.11. The corresponding mass resonances are clearly visible above the continuum di-photon background and discoverable beyond the $5\sigma$ level. Indeed, similar situations can be found for each of the combinations listed in Tab. 4.2 and most of these correspond to phenomenological scenarios which are distinctive of the NMSSM and that cannot be reproduced in the MSSM.

In short, while the bulk of the NMSSM parameter space is in a configuration degenerate with the MSSM case (as far as di-photon Higgs signals at the LHC are concerned), non-negligible areas exist with the potential to unveil a non-minimal nature of the underlying SUSY model in this search channel alone.

## 4.5 Dark matter in the NMSSM and relations to the NMSSM Higgs sector

*John F. Gunion, Dan Hooper and Bob McElrath*

Since the NMSSM has five neutralinos and two CP-odd Higgs bosons, there are many new ways in which the relic density of the $\widetilde{\chi}_1^0$ could match the observed dark matter density. Dedicated work on NMSSM scenarios appears in [53, 54]. The latter group has made their code publicly available. Let us recall that in the MSSM there is a significant constraint on the mass $M_A$ of the single CP-odd state. This in turn constrains the values of $m_{\widetilde{\chi}_1^0}$ that would lie in the "funnel" region of $m_{\widetilde{\chi}_1^0} \sim 2M_A$ where $\widetilde{\chi}_1^0\widetilde{\chi}_1^0 \to A \to X$ can be sufficiently efficient to adequately reduce the $\widetilde{\chi}_1^0$ relic density to a level at or below that observed. In contrast, in the NMSSM there are two CP-odd states and their masses, $M_{A_1}$ and $M_{A_2}$, are quite unconstrained by LEP data and theoretical model structure, implying that $\widetilde{\chi}_1^0\widetilde{\chi}_1^0 \to A_{1,2} \to X$ could be the primary annihilation mechanism for large swaths of parameter space.

Let us first discuss the MSSM situation in a bit more detail. Neutralinos produced in the early Universe must annihilate into Standard Model particles at a sufficient rate to avoid overproducing the density of dark matter. Within the framework of the Minimal Supersymmetric Standard Model (MSSM), the lightest neutralino can annihilate through a variety of channels, exchanging other sparticles, $Z$ bosons, or Higgs bosons. The masses of sparticles such as sleptons or squarks, as well as the masses of Higgs bosons, are limited by collider constraints, with typical lower limits of around $\sim$100 GeV. For lighter neutralinos, it becomes increasingly difficult for these heavy propagators to generate neutralino annihilation cross sections that are large enough. The most efficient annihilation channel for very light neutralinos in the MSSM is the $s$-channel exchange of a pseudoscalar Higgs boson. It has been shown that this channel can, in principle, be sufficiently efficient to allow for neutralinos as light as 6 GeV [88, 89]. Such models require a careful matching of a number of independent parameters, however, making viable models with neutralinos lighter than ∼20 GeV rather unlikely [90]. Measurements of rare $B$-decays are also particularly constraining in this regime.

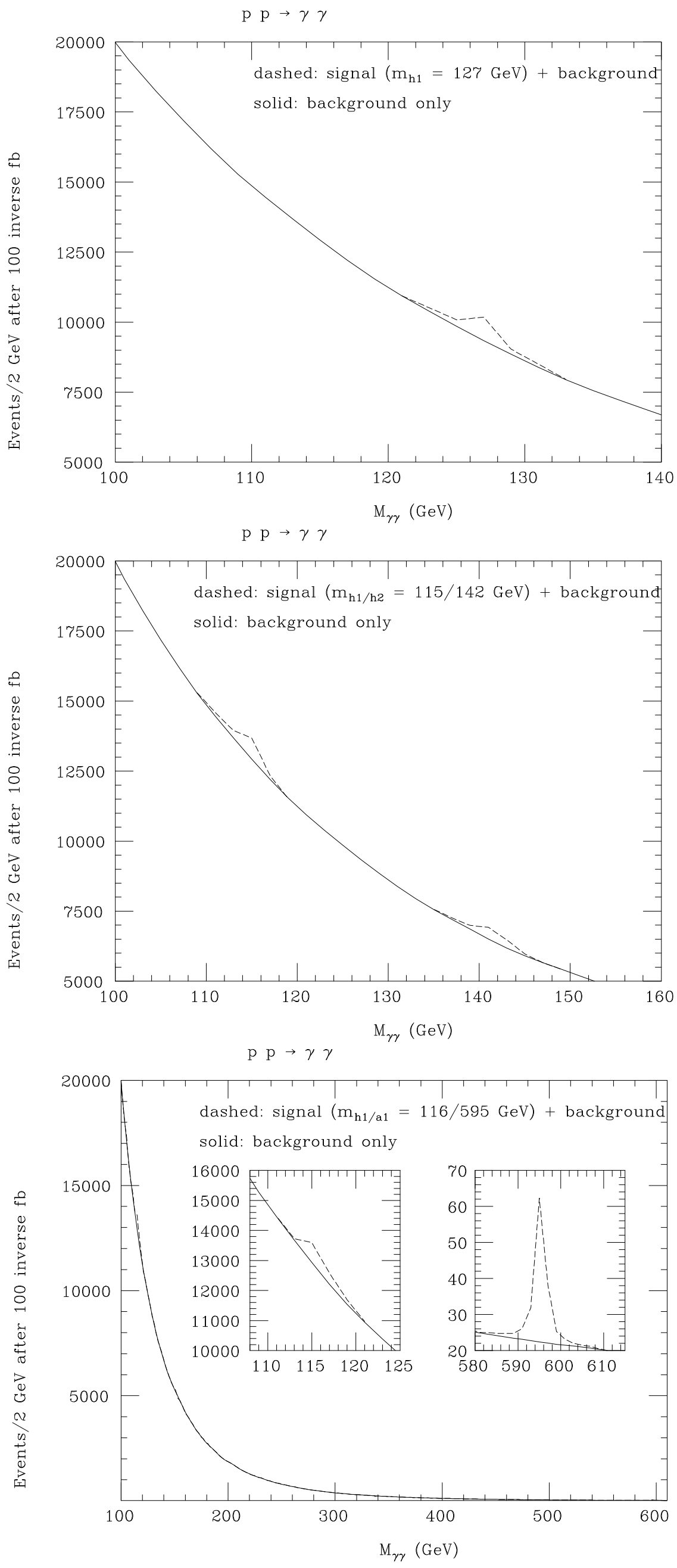

Fig. 4.11: The differential distribution in invariant mass of the di-photon pair after the cuts in $p_T^\gamma$ and $\eta^\gamma$ mentioned in the text, for 100 fb$^{-1}$ of luminosity, in the case of the background (solid) and the sum of signal and background (dashed), for the example points 1. 3. described in the text (from left to right, in correspondence).

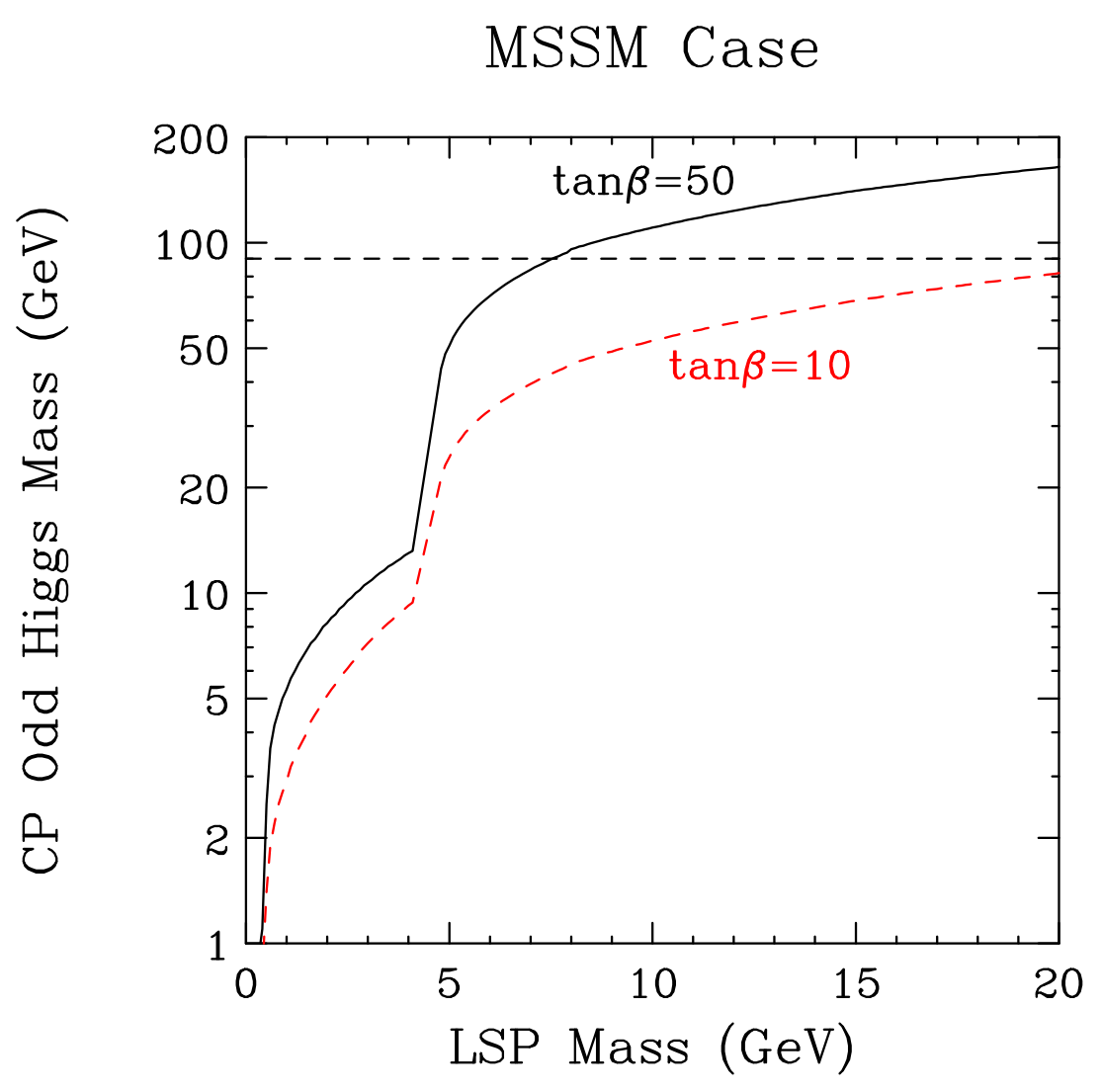

Fig. 4.12: The CP-odd Higgs mass required to obtain the measured relic density for a light neutralino in the MSSM. Models above the curves produce more dark matter than in observed. These results are for the case of a bino-like neutralino with a small higgsino admixture ($\epsilon_B^2 = 0.94$, $\epsilon_u^2 = 0.06$). Results for two values of $\tan\beta$ (10 and 50) are shown. The horizontal dashed line represents the lower limit on the CP-odd Higgs mass in the MSSM from collider constraints. To avoid overproducing dark matter, the neutralino must be heavier than about 8 (22) GeV for $\tan\beta = 50$ (10).

This result should be contrasted with that found for the NMSSM. In the NMSSM (and other supersymmetric models with an extended Higgs sector), a very light CP-odd Higgs boson can naturally arise making it possible for a very light neutralino to annihilate efficiently enough to avoid being overproduced in the early Universe. In fact, it is relatively easy to construct NMSSM models yielding the correct relic density even for a very light neutralino, 100 MeV $< m_{\widetilde{\chi}_1^0} <$ 20 GeV. Even after including constraints from Upsilon decays, $b \to s\gamma$, $B_s \to \mu^+\mu^-$ and the magnetic moment of the muon, a light bino or singlino neutralino is allowed that can generate the appropriate relic density.

### *4.5.1 Models with a light LSP*

Above we outlined the general possibilities for dark matter in the NMSSM context, focusing on the fact that a light or very light neutralino would not yield an over abundance of dark matter. In contrast, it was stated that in the MSSM context it is very difficult to have a light neutralino that is consistent with $\Omega h^2 < 0.1$. As a more specific benchmark for comparison, we consider a light bino in the MSSM which annihilates through the exchange of the CP-odd $A$. The results for this case are shown in Fig. 4.12. In this figure, the thermal relic density of LSP neutralinos exceeds the measured density for $M_A$ above the solid and dashed curves, for values of $\tan\beta$ of 50 and 10, respectively. Shown as a horizontal dashed line is the lower limit on $M_A$ from collider constraints. This figure demonstrates that even in the case of very large $\tan\beta$, the lightest neutralino must be heavier than about 7 GeV. For moderate values of $\tan\beta$, the neutralino must be heavier than about 20 GeV.

Turning to the NMSSM, as we have noted the physical LSP is a mixture of the bino, neutral wino, neutral higgsinos and singlino. The lightest neutralino therefore has, in addition to the four MSSM components, a singlino component which is the superpartner of the singlet Higgs. The eigenvector of the lightest neutralino, $\widetilde{\chi}_1^0$, in terms of gauge eigenstates can be written in the form

$$\widetilde{\chi}_1^0 = \epsilon_u \tilde{H}_u^0 + \epsilon_d \tilde{H}_d^0 + \epsilon_W \tilde{W}^0 + \epsilon_B \tilde{B} + \epsilon_s \tilde{S}, \tag{4.27}$$

where $\epsilon_u$, $\epsilon_d$ are the up-type and down-type higgsino components, $\epsilon_W$, $\epsilon_B$ are the wino and bino components and $\epsilon_s$ is the singlet component of the lightest neutralino. Similarly, the CP-even and CP-odd Higgs states are mixtures of MSSM-like Higgses and singlets. For the lightest CP-even Higgs state we can define:

$$H_1 = \sqrt{2}\left[\xi_u \mathrm{Re}(H_u^0 - v_u) + \xi_d \mathrm{Re}(H_d^0 - v_d) + \xi_s \mathrm{Re}(S - x)\right], \tag{4.28}$$

where $x \equiv \langle S \rangle$. Here, Re denotes the real component of the respective state. Lastly, the lightest CP-odd Higgs can be written as (a similar formula also applies for the heavier $A_2$)

$$A_1 = \cos\theta_{A_1} A_{\mathrm{MSSM}} + \sin\theta_{A_1} A_s, \tag{4.29}$$

where $A_s$ is the CP-odd [$\sqrt{2}\mathrm{Im}(S)$] component of the scalar singlet field and $A_{\mathrm{MSSM}}$ is the combination of the imaginary components of $H_u$ and $H_d$ that would be the MSSM pseudoscalar Higgs if the singlet were not present. Here, $\theta_{A_1}$ is the mixing angle between these two states. There is also a third linear combination of the imaginary components of $H_u^0$, $H_d^0$ and $S$ that we have removed by a rotation in $\beta$. This field becomes the longitudinal component of the $Z$ after electroweak symmetry is broken.

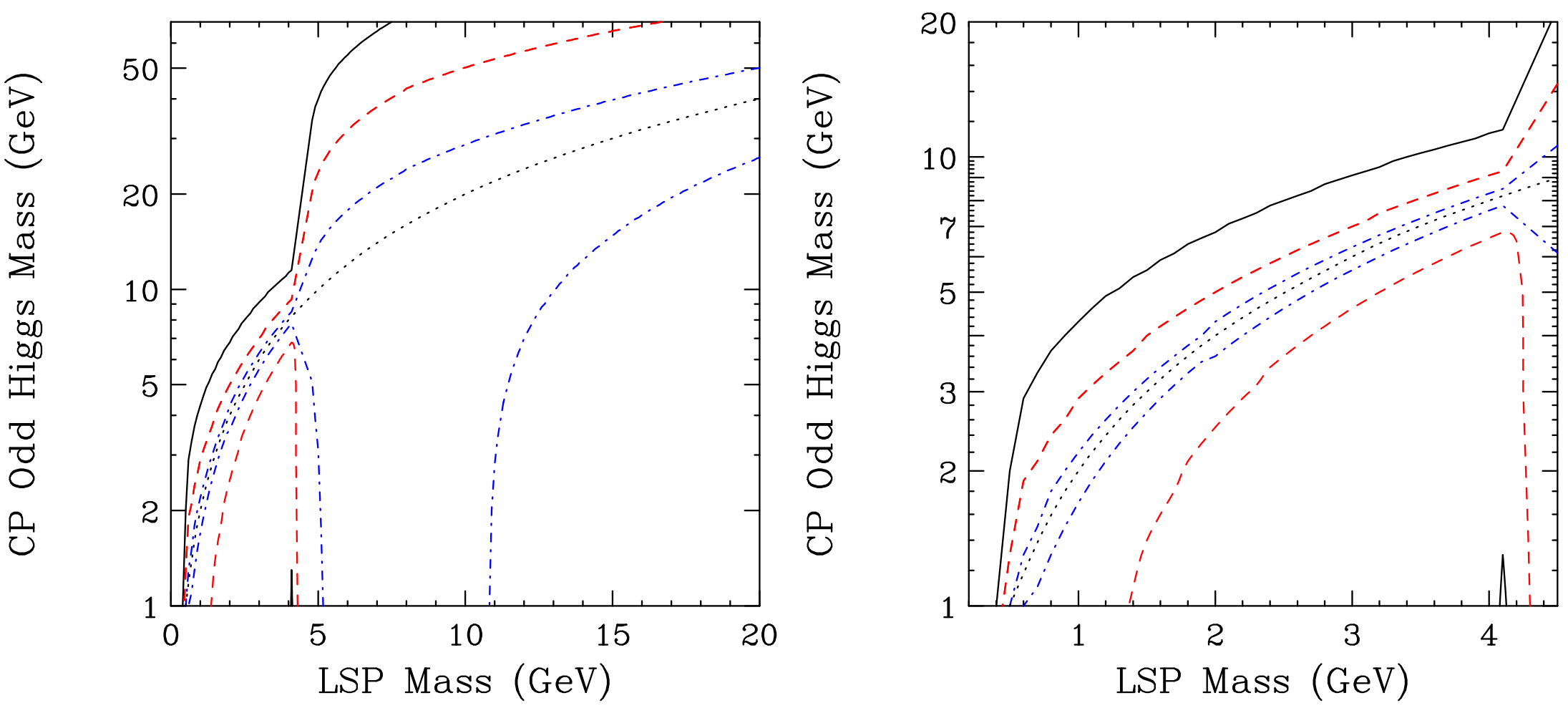

Fig. 4.13: We display contours in $M_{A_1}$ $m_{\widetilde{\chi}_1^0}$ parameter space for which $\Omega h^2 = 0.1$. Points above or below each pair of curves produce more dark matter than is observed; inside each set of curves less dark matter is produced than is observed. These results are for a bino-like neutralino with a small higgsino admixture ($\epsilon_B^2 = 0.94$, $\epsilon_u^2 = 0.06$). Three values of $\tan\beta$ (50, 15 and 3) have been used, shown as solid black, dashed red, and dot-dashed blue lines, respectively. The dotted line is the contour corresponding to $2m_{\widetilde{\chi}_1^0} = M_{A_1}$. For each set of lines, we have set $\cos^2\theta_{A_1} = 0.6$. The $\tan\beta = 50$ case is highly constrained for very light neutralinos, and is primarily shown for comparison with the MSSM case.

In the NMSSM context, when annihilation proceeds via one of the CP-odd Higgs bosons the calculation of the relic $\widetilde{\chi}_1^0$ density is much more flexible than in the MSSM. For annihilation via the $A_1$, the thermally averaged cross section takes the form [using the usual expansion in terms $t = T/m_{\widetilde{\chi}_1^0}$ and writing $\langle\sigma v\rangle = a + bt + \mathcal{O}(t^2)$]:

$$
\begin{aligned}
a_{\chi\chi\to A_1\to f\bar{f}} &= \frac{g_2^4 c_f m_f^2 \cos^4\theta_{A_1}\tan^2\beta}{8\pi m_W^2}\frac{m^2_{\tilde{\chi}^0_1}\sqrt{1-m_f^2/m^2_{\tilde{\chi}^0_1}}}{(4m^2_{\tilde{\chi}^0_1}-M^2_{A_1})^2+M^2_{A_1}\Gamma^2_{A_1}} \qquad (4.30)\\
\times\Bigg[-\epsilon_u(\epsilon_W &- \epsilon_B\tan\theta_W)\sin\beta+\epsilon_d(\epsilon_W-\epsilon_B\tan\theta_W)\cos\beta \\
&+ \sqrt{2}\frac{\lambda}{g_2}\epsilon_s(\epsilon_u\sin\beta+\epsilon_d\cos\beta)+\frac{\tan\theta_{A_1}}{g_2}\sqrt{2}(\lambda\epsilon_u\epsilon_d-\kappa\epsilon_s^2)\Bigg]^2, \\
b_{\chi\chi\to A_1\to f\bar{f}} &\simeq 0, \qquad (4.31)
\end{aligned}
$$

where $c_f$ is a color factor, equal to 3 for quarks and 1 otherwise. For this result, we have assumed that the final state fermions are down-type. If they are instead up-type fermions, the $\tan^2\beta$ factor should be replaced by $\cot^2\beta$. A similar formula holds for the $A_2$.

Fig. 4.13 shows how the MSSM results can be modified within the framework of the NMSSM. There, we give results for the case where the NMSSM CP-odd Higgs $A_1$ is taken to be a mixture of MSSM-like and singlet components specified by $\cos^2\theta_{A_1} = 0.6$ and the neutralino composition is taken to be specified by $\epsilon_B^2 = 0.94$ and $\epsilon_u^2 = 0.06$. These specific values are representative of those that can be achieved for various NMSSM parameter choices satisfying all constraints. For each pair of contours (solid black, dashed red, and dot-dashed blue), $\Omega h^2 = 0.1$ along the contours and the region between the lines is the space in which the neutralino's relic density obeys $\Omega h^2 < 0.1$. The solid black, dashed red, and dot-dashed blue lines correspond to $\tan\beta$=50, 15 and 3, respectively. Also shown as a dotted line is the contour corresponding to the resonance condition, $2m_{\tilde{\chi}^0_1} = M_{A_1}$.

For the $\tan\beta$=50 or 15 cases, neutralino dark matter can avoid being overproduced for any $A_1$ mass below $\sim 20-60$ GeV, as long as $m_{\tilde{\chi}^0_1} > m_b$. For smaller values of $\tan\beta$, a lower limit on $M_{A_1}$ can apply as well.

For neutralinos lighter than the mass of the $b$-quark, annihilation is generally less efficient. This region is shown in detail in the right frame of Fig. 4.13. In this funnel region, annihilations to $c\bar{c}$, $\tau^+\tau^-$ and $s\bar{s}$ all contribute significantly. Despite the much smaller mass of the strange quark, its couplings are enhanced by a factor proportional to $\tan\beta$ (as with bottom quarks) and thus can play an important role in this mass range. In this mass range, constraints from Upsilon and $J/\psi$ decays can be very important, often requiring fairly small values of $\cos\theta_{A_1}$.

For annihilations to light quarks, $c\bar{c}$, $s\bar{s}$, etc., the Higgs couplings to various meson final states should be considered, which include effective Higgs-gluon couplings induced through quark loops. The calculations shown employed a conservative approximation of keeping only the Higgs-quark-quark couplings alone, even for these light quarks, but with kinematic thresholds set by the mass of the lightest meson containing a given type of quark, rather than the quark mass itself. This corresponds to thresholds of 9.4 GeV, 1.87 GeV, 498 MeV and 135 MeV for bottom, charm, strange and down quarks, respectively. A more detailed treatment, which was not undertaken, would include the proper meson form factors as well as allowing for the possibility of virtual meson states.

The above discussion focused on the case of a mainly bino LSP. If the LSP is mostly singlino, it is also possible to generate the observed relic abundance in the NMSSM. A number of features differ for the singlino-like case in contrast to a bino-like LSP, however. Most importantly, an LSP mass that is chosen to be precisely at the Higgs resonance, $M_{A_1} \simeq 2m_{\tilde{\chi}^0_1}$, is not possible for this case: $M_{A_1}$ is always less than $2m_{\tilde{\chi}^0_1}$ by a significant amount. Second, in models with a singlino-like LSP, the $A_1$ is generally also singlet-like and the product of $\tan^2\beta$ and $\cos^4\theta_{A_1}$, to which annihilation rates are proportional, see Eq. (4.30), is typically very small. This limits the ability of a singlino-like LSP to generate the observed relic abundance. The result is that annihilation is too inefficient for an LSP that is more than 80% singlino. However, there is no problem having $m_{\widetilde{\chi}_1^0} \sim M_{A_1}/2$ so as to achieve the correct relic density when the $\widetilde{\chi}_1^0$ is mainly bino while the $A_1$ is mainly singlet.

Of course, we should also discuss the implications for direct dark matter detection in the NMSSM. As above, we focus on scenarios with a light $\widetilde{\chi}_1^0$ that are a somewhat unique feature of the NMSSM. The spin-independent elastic scattering cross section of a light neutralino with nuclei is generally dominated by the $t$-channel exchange of a CP-even Higgs boson. For a bino-like LSP and the $H_1$ with composition as in Eq. (4.28), the elastic cross section is approximated by

$$\sigma_{\rm elastic}^{\rm bino} \sim \frac{8G_F^2 M_Z^2}{\pi {M_{H_1}}^4}\left(\frac{m_p m_{\widetilde{\chi}_1^0}}{m_p + m_{\widetilde{\chi}_1^0}}\right)^2 \epsilon_B^2 \sin^2\theta_W \,(\epsilon_d\xi_u - \epsilon_u\xi_d)^2 \times \left(\sum_{q=d,s,b}\frac{m_q\xi_d}{\cos\beta} < N|q\bar{q}|N > + \sum_{q=u,c}\frac{m_q\xi_u}{\sin\beta} < N|q\bar{q}|N >\right)^2. \tag{4.32}$$

If the LSP is singlino-like, on the other hand, the appropriate approximation is

$$\sigma_{\rm elastic}^{\rm singlino} \sim \frac{8G_F^2 M_Z^2}{\pi {M_{H_1}}^4}\left(\frac{m_p m_{\widetilde{\chi}_1^0}}{m_p + m_{\widetilde{\chi}_1^0}}\right)^2 \frac{2\lambda^2\,\epsilon_s^2\cos^2\theta_W}{g_2^2} (\epsilon_d\xi_d + \epsilon_u\xi_u)^2 \times \left(\sum_{q=d,s,b}\frac{m_q\xi_d}{\cos\beta} < N|q\bar{q}|N > + \sum_{q=u,c}\frac{m_q\xi_u}{\sin\beta} < N|q\bar{q}|N >\right)^2. \tag{4.33}$$

In assessing the implications of the above, it is useful to note that LEP limits on the $H_1$ if it decays to $b\bar{b}$ (we return to the $H_1 \to A_1A_1$ type scenario later) with $M_{H_1} < 120$ GeV roughly imply

$$\xi_{u,d} \lesssim \left(\frac{M_{H_1}}{120\mathrm{GeV}}\right)^{3/2} + 0.1, \tag{4.34}$$

and for a light $\widetilde{\chi}_1^0$ LEP limits on invisible $Z$ decays roughly imply $\epsilon_{u,d} < 0.06$. If we assume that the $s$-quark contribution dominates and use $m_s < N|s\bar{s}|N > \approx 0.2$ GeV, the resulting cross section for a bino-like or singlino-like $\widetilde{\chi}_1^0$ is then roughly given by:

$$\sigma_{\rm elastic} \lesssim 1.4\times10^{-42} cm^2 \left(\frac{120\text{ GeV}}{M_{H_1}}\right)^4 \left(\left(\frac{M_{H_1}}{120\text{ GeV}}\right)^{3/2} + 0.1\right)^2 \left(\frac{\tan\beta}{50}\right)^2 F_\lambda \tag{4.35}$$

assuming $m_{\widetilde{\chi}_1^0} > m_p$ and $\tan\beta > 1$, using the $\xi_{u,d}$ limit of Eq. (4.34) and adopting $\epsilon_{u,d} \sim 0.06$. In the above, $F_\lambda = 1$ for the bino-like case and $F_\lambda = 2\lambda^2/(g_2^2\tan^2\theta_W) \approx 0.67\times(\lambda/0.2)^2$ for the singlino-like case. For $\tan\beta = 50$, $\lambda = 0.2$ and a Higgs mass of 120 GeV, we estimate a neutralino-proton elastic scattering cross section on the order of $4\times10^{-42}$ cm$^2$ ($4\times10^{-3}$ fb) for either a bino-like or a singlino-like LSP. This value may be of interest to direct detection searches such as CDMS, DAMA, Edelweiss, ZEPLIN and CRESST. To account for the DAMA data, the cross section would have to be enhanced by a local over-density of dark matter.

It is interesting to consider whether there are any special features related to the very attractive scenarios motivated by minimizing the fine-tuning measure $F$. In those scenarios, the $H_1$ can have mass below the LEP limit (e.g. of order 100 GeV) even though its $WW, ZZ$ couplings are very SM-like. This is possible provided $M_{A_1} < 2m_b$ so that the $H_1$ decays predominantly via $H_1 \to A_1A_1$ with $A_1 \to \tau^+\tau^-$ (or, if $M_{A_1} < 2m_\tau$, $A_1 \to gg, c\bar{c}, \ldots$) since the $H_1 \to A_1A_1 \to \tau^+\tau^-\tau^+\tau^-$ decay channel is not constrained by LEP data for $M_{H_1} \lesssim 90$ GeV. For our purposes, the important feature of such a scenario is that, the $A_1$ turns out to be very singlet-like, with $\cos^2\theta_{A_1} \lesssim 0.015$. In this case, adequate annihilation of a very light $\widetilde{\chi}_1^0$ via $\widetilde{\chi}_1^0\widetilde{\chi}_1^0 \to A_1 \to X$ occurs only if $m_{\widetilde{\chi}_1^0} \simeq M_{A_1}/2$. This requires a rather fine adjustment of the $M_1$ bino soft mass relative to $M_{A_1}$ that has no immediately obvious theoretical motivation. Because the $A_1$ is so light in the low fine-tuning scenarios, if $m_{\widetilde{\chi}_1^0}$ is significantly above $2m_b$ then consistency with relic abundance limits requires that $\widetilde{\chi}_1^0\widetilde{\chi}_1^0$ annihilation proceed via one of the more conventional co-annihilation channels or via $\widetilde{\chi}_1^0\widetilde{\chi}_1^0 \to A_2 \to X$. The latter case is only applicable if $m_{\widetilde{\chi}_1^0} \gtrsim 200$ GeV, since the $A_2$ is typically quite heavy in the low-$F$ scenarios, $M_{A_2} \gtrsim 400$ GeV.

Another issue is direct detection of dark matter. Since the $A_1$ is so singlet in nature, the only exchange of importance is $H_1$ exchange. In the low-$F$ scenarios, $H_1$ is almost entirely $H_u$. In particular, the $H_d$ composition component of the $H_1$ is $\xi_d \sim 0.1$, and correspondingly $\xi_u \sim 0.99$. For the $\widetilde{\chi}_1^0$, $\epsilon_B > 0.8$ and $\epsilon_u$ and $\epsilon_d$ can take a range of values from 0.1 up to 0.5. Referring to Eq. (4.32), again keeping only the $s$-quark contribution and keeping only the dominant $\epsilon_d\xi_u$ piece in the external factor, we obtain

$$\sigma_{\rm elastic} \sim 5\times 10^{-6}\epsilon_B^2 \left(\frac{\epsilon_d}{0.25}\right)^2 \left(\frac{100~{\rm GeV}}{M_{H_1}}\right)^4 \left(\frac{\tan\beta}{10}\right)^2 ~{\rm fb}\,. \tag{4.36}$$

If $m_{\widetilde{\chi}_1^0}$ is in the $15 - 100$ GeV mass range that is optimal for experiments like ZEPLIN and CRESST, direct dark matter detection would might be possible. If $m_{\widetilde{\chi}_1^0} < m_b$ (the $\widetilde{\chi}_1^0\widetilde{\chi}_1^0 \to A_1 \to X$ possibility), the sensitivity of ZEPLIN and CRESST is greatly reduced and dark matter detection would be very difficult.

So, where does all this leave us with respect to the ILC program. First consider the case where $m_{\widetilde{\chi}_1^0} \sim M_{A_1}/2 < m_b$, the best that a hadron collider can do will probably be to set an upper limit on $m_{\widetilde{\chi}_1^0}$. Determining its composition is almost certain to be very difficult. Note that the $m_{\widetilde{\chi}_2^0} - m_{\widetilde{\chi}_1^0}$ mass difference should be large, implying adequate room for $\widetilde{\chi}_2^0 \to Z\widetilde{\chi}_1^0$ and a search for lepton kinematic edges and the like. (Of course, $\widetilde{\chi}_2^0 \to H_1\widetilde{\chi}_1^0$ will also probably be an allowed channel, with associated implications for $H_1$ detection in SUSY cascade decays.) A light singlet-like $A_1$ is very hard to detect. At best, it might be possible to bound $\cos\theta_{A_1}$ by experimentally establishing an upper bound on the $WW \to A_1A_1$ rate (proportional to $\cos^4\theta_{A_1}$). Thus, the ILC would be absolutely essential. Precise measurement of the $\widetilde{\chi}_1^0$ mass and composition using the standard ILC techniques should be straightforward. A bigger question is how best to learn about the $A_1$ at the precision level. Of course, we will have lots of $A_1$'s to study from $ZH_1$ production followed by $H_1 \to A_1A_1$ decays. The events will give precise measurements of $g_{ZZH_1}^2 BR(H_1 \to A_1A_1)BR(A_1 \to X)BR(A_1 \to Y)$, where $X, Y = \tau^+\tau^-, gg, c\overline{c}, \ldots$. The problem will be to unfold the individual branching ratios so as to learn about the $A_1$ itself. Particularly crucial would be some sort of determination of $\cos\theta_{A_1}$ which enters so critically into the annihilation rate. (I assume that a $\tan\beta$ measurement could come from other supersymmetry particle production measurements and so take it as given.) There is some chance that $WW \to A_1A_1$ and $Z^* \to ZA_1A_1$, with rates proportional to $\cos^4\theta_{A_1}$, could be detected. The $\cos^2\theta_{A_1} = 1$ rates are very large, implying that observation might be possible despite the fact that the low-$F$ scenarios have $\cos^2\theta_{A_1} \lesssim 0.01$. One could also consider whether $\gamma\gamma \to A_1$ production would have an observable signal despite the suppression due to the singlet nature of the $A_1$. Hopefully, enough precision could be achieved for the $A_1$ measurements that they could be combined with the $\widetilde{\chi}_1^0$ precision measurements so as to allow a precision calculation of the expected $\widetilde{\chi}_1^0$ relic density. A study of the errors in the dark matter density computation using the above measurements as compared to the expected experimental error for the $\Omega h^2$ measurement is needed.

If the $\widetilde{\chi}_1^0$ is not light, but the low fine-tuning scenario applies with $A_1$ mass below $2m_b$, then early universe $\widetilde{\chi}_1^0\widetilde{\chi}_1^0$ annihilation cannot occur via the $A_1$ channel. In this case, proper relic density must be achieved using co-annihilation or $\widetilde{\chi}_1^0\widetilde{\chi}_1^0 \to A_2$ annihilation (where $A_2 \sim A_{MSSM}$ and $M_{A_2}$ is relatively large) — there is no point in repeating the relevant analyses here. We only note that the precision needed to compute the $\widetilde{\chi}_1^0\widetilde{\chi}_1^0$ annihilation rate and compare to the measured $\Omega h^2$ should be achievable at the ILC. As already noted, the $2m_{\widetilde{\chi}_1^0} \simeq M_{A_1}$ scenario seems relatively fine-tuned and we regard the large $m_{\widetilde{\chi}_1^0}$ scenarios as much more likely.

As an overall summary, we simply reiterate the fact that the NMSSM provides a huge increase in the possibilities for achieving the correct relic density for the $\tilde{\chi}_1^0$ and can drastically alter expectations for direct detection of dark matter.

### 4.6 Relic density of neutralino dark matter in the NMSSM

*Geneviève Bélanger, Fawzi Boudjema, Cyril Hugonie, Alexander Pukhov and Alexander Semenov*

In any supersymmetric (SUSY) extension of the Standard Model with conserved R-parity, the lightest SUSY particle (LSP) constitutes a good candidate for cold dark matter. Recent measurements from WMAP [91, 92] have constrained the value for the relic density of dark matter within 10% ($0.0945 < \Omega h^2 < 0.1287$ at $2\sigma$). The forthcoming PLANCK experiment should reduce this interval down to 2-3%. It is therefore important to calculate the relic density as accurately as possible in any given SUSY model, in order to match this experimental accuracy. Here we perform a precise calculation of the relic density of dark matter within the NMSSM using an extension of `micrOMEGAs` [93] and an interface with the program `NMHDECAY` [63] that calculates the spectrum of the model, in particular that of the Higgs sector. The NMSSM contains, in addition to the MSSM fields, an extra scalar and pseudo-scalar neutral Higgs bosons, as well as an additional neutralino. The phenomenology of the model can be markedly different from the MSSM [26, 27, 29]. In particular the possibility of light Higgs states [72] or light neutralinos that may have escaped LEP searches [94, 95] could impact significantly on the value of the relic density. We present a selection of scenarios that pedict a value of the relic density in agreement with WMAP [54].

#### 4.6.1 The model

We consider the general NMSSM with parameters defined at the weak scale. As free parameters, we take the parameters of the Higgs sector, Eq. (4.11), as well as the gaugino masses $M_1, M_2$ that enter the neutralino mass matrix. In the gaugino sector, we assume universality at the GUT scale, which at the EW scale corresponds to $M_2 = 2M_1$ and $M_3 = 3.3M_2$. The soft terms in the squark and slepton sector (which enter the radiative corrections in the Higgs sector) are also fixed at the EW scale. We assume very heavy sfermions $m_{\tilde{f}} = 1$ TeV and fix the trilinear mixing to $A_f = 1.5$ TeV. We thus consider as independent parameters the following set of variables

$$\lambda,\ \kappa,\ \tan\beta,\ \mu,\ A_\lambda,\ A_\kappa,\ M_1. \tag{4.37}$$

For the SM parameters, we assume $\alpha_s = 0.118$, $m_t^{\rm pole} = 175$ GeV and $m_b(m_b) = 4.24$ GeV.

We set all these parameters in the program `NMHDECAY` [63]. For each point in the parameter space, the program `NMHDECAY` first checks the absence of Landau singularities for $\lambda$, $\kappa$, $h_t$ and $h_b$ below the GUT scale. For $m_t^{\rm pole} = 175$GeV, this translates into $\lambda < .75$, $\kappa < .65$, and $3. < \tan\beta < 85$. `NMHDECAY` also checks the absence of an unphysical global minimum of the scalar potential with vanishing Higgs vevs. `NMHDECAY` then computes scalar, pseudo-scalar and charged Higgs masses and mixings, taking into account one and two loop radiative corrections, as well as chargino and neutralino masses and mixings. Finally, all available experimental constraints from LEP are checked.

The couplings of the LSP to the scalar and pseudo-scalar Higgs states will enter the computation of LSP annihilation through a Higgs resonance or $t$-channel annihilation into Higgs pairs. The Feynman rule for the LSP-scalar-scalar vertex reads

$$\begin{aligned} g_{\tilde{\chi}_1^0\tilde{\chi}_1^0 h_i} &= g(N_{12} - N_{11}\tan\theta_W)(S_{i1}N_{13} - S_{i2}N_{14}) \\ &\quad + \sqrt{2}\lambda N_{15}(S_{i1}N_{14} + S_{i2}N_{13}) + \sqrt{2}S_{i3}(\lambda N_{13}N_{14} - \kappa N_{15}^2)\,. \end{aligned} \tag{4.38}$$

Here $N$ describes the neutralino mixing and $S$ the scalar mixing [63]. The first term is equivalent to the $\tilde{\chi}_1^0\tilde{\chi}_1^0 h$ coupling in the MSSM by replacing $S_{11} = S_{22} = \cos\alpha$ and $S_{12} = -S_{21} = \sin\alpha$ while the last two terms are specific of the NMSSM. The second term is proportional to the singlino component of the LSP while the last one is proportional to the singlet component of the scalar Higgs. Similarly, the LSP coupling to a pseudo-scalar also contains terms proportional to the singlino component or to the singlet component of the Higgs.

### *4.6.2 Relic density*

In the context of the MSSM, the publicly available program `micrOMEGAs` [96, 97] computes the relic density of the lightest neutralino LSP by evaluating the thermally averaged cross section for its annihilation as well as, when necessary, for its coannihilation with other SUSY particles. It then solves the density evolution equation numerically, without using the freeze-out approximation. We have extended `micrOMEGAs` [93] to perform the relic density calculation within the NMSSM. An interface with `NMHDECAY` allows a precise calculation of the particle spectrum in the NMSSM, as well as a complete check of all the available experimental constraints from LEP [63].

In the MSSM with universal gaugino masses, one can classify the main scenarios for dark matter annihilation as follows: a bino scenario with light sfermions where neutralino annihilate into fermion pairs, a sfermion coannihilation scenario, a mixed bino/Higgsino scenario where neutralino annihilate dominantly into gauge boson pairs or into $t\bar{t}$ and finally a Higgs funnel scenario where neutralino annihilate in fermion pairs near a $s$-channel Higgs resonance. When sfermions are very heavy only the latter two scenarios predict $\Omega h^2 \approx 0.1$, in agreement with WMAP. To achieve this, the bino-Higgsino scenario requires $M_1 \approx \mu$, indeed higher higgsino content ($M_1 \gg \mu$) leads to very efficient annihilation and coannihilation while a smaller higgsino content ($M_1 \ll \mu$) to values of the relic density that are too high. The Higgs funnnel scenario requires that $M_H \approx 2m_{\tilde{\chi}_1^0}$, here $H$ corresponds to either the light scalar or the heavy scalar/pseudoscalar. The latter is enhances at large values of $\tan\beta$.

In the NMSSM, the same mechanisms as for the MSSM are at work for neutralino annihilation: into fermion pairs through $s$-channel exchange of a $Z$ or Higgs, into gauge boson pairs through either $Z$/H $s$-channel exchange or $t$-channel exchange of heavier neutralinos or charginos. The new features of the NMSSM are first a richer scalar/pseudoscalar Higgs sector that leads to more resonances but also to new decay modes for Higgses (into lighter Higgses) and second a neutralino LSP which because of its mixing with a singlino feature different couplings to gauge bosons, Higgs and sparticles. These new features imply new possibilities to either increase or decrease the relic density of neutralinos as compared to the MSSM. We next describe some typical scenarios that lead to a value for the relic density of dark matter in agreement with WMAP.

### *4.6.3 Results*

We concentrate on models which can differ markedly from the MSSM predictions, in particular models with $\tan\beta \leq 5$ for which annihilation through a Higgs resonance is marginal in the MSSM. We also consider models where the presence of light Higgs states opens up new channels for efficient neutralino annihilation as well as models where the LSP is dominantly singlino.

**Case 1: annihilation through Higgs resonances**

The presence of additional Higgs states in the NMSSM means additional regions of parameter space where rapid annihilation through a $s$-channel resonance can take place. In fact we found that such annihilation is dominant in large regions of the parameter space and this even at low to intermediate values of $\tan\beta$. For example, starting with a value of $\mu$ and $M_1$ for which one would expect $\Omega h^2 > .13$ in the MSSM, and varying the parameters $A_\lambda, A_\kappa$ one can tune the value of the scalars/pseudoscalars masses such that for at least one scalar/pseudoscalar satisfies $m_{H_i,A_i} \approx 2m_{\tilde{\chi}_1^0}$ . Note that the neutralino sector does not depend on $A_\lambda, A_\kappa$. We found scenarios consistent with the WMAP measurement where rapid neutralino annihilation proceeds through either the $H_2$ resonance (an example is given in Table 4.3, Case 1), the light pseudoscalar, $A_1$, or the lightest scalar $H_1$. The latter can also occur in the MSSM.

Table 4.3: Benchmark points satisfying both LEP and WMAP constraints

| Case | 1 | 2 | 3a | 3b | 3c |
|---|---|---|---|---|---|
| $\lambda$ | 0.1 | 0.35 | 0.6 | 0.23 | 0.035 |
| $\kappa$ | 0.11 | 0.2 | 0.12 | 0.003 | 0.0124 |
| $\tan\beta$ | 5 | 5 | 2 | 3.2 | 5 |
| $\mu$ [GeV] | 300 | 230 | 265 | 195 | 285 |
| $A_\lambda$ [GeV] | -100 | 400 | 450 | 590 | -28 |
| $A_\kappa$ [GeV] | -100 | 0 | -50 | -20 | -150 |
| $M_1$ [GeV] | 150 | 160 | 500 | 100 | 235 |
| $m_{\tilde{\chi}_1^0}$ [GeV] | 142 | 141 | 127 | 8 | 206 |
| $N_{13}^2 + N_{14}^2$ | 0.02 | 0.09 | 0.105 | 0.04 | 0.02 |
| $N_{15}^2$ | 0 | 0.02 | 0.90 | 0.95 | 0.94 |
| $m_{\tilde{\chi}_2^0}$ [GeV] | 250 | 209 | 270 | 85 | 215 |
| $m_{\tilde{\chi}_1^\pm}$ [GeV] | 246 | 218 | 269 | 138 | 273 |
| $m_{H_1}$ [GeV] | 118 | 113 | 102 | 18 | 115 |
| $m_{H_2}$ [GeV] | 561 | 258 | 130 | 115 | 158 |
| $m_{A_1}$ [GeV] | 297 | 54 | 122 | 14 | 211 |
| $\Omega h^2$ | 0.104 | 0.116 | 0.1155 | 0.124 | 0.111 |
| Channels | $qq$ (83%)<br>$ll$ (10%)<br>$VV$ (4%)<br>$HH$ (2%)<br>$ZH$ (1%) | $VV$ (51%)<br>$HA$ (31%)<br>$HH$ (15%)<br>$ZH$ (2%) | $HA$ (60%)<br>$VV$ (26%)<br>$ZH$ (10%)<br>$HH$ (3%)<br>$ff$(2%) | $qq$ (92%)<br>$ll$ (8%) | $\tilde{\chi}_2^0\tilde{\chi}_2^0 \to X$ (77%)<br>$\tilde{\chi}_1^0\tilde{\chi}_2^0 \to X$ (18%)<br>$\tilde{\chi}_1^0\tilde{\chi}_1^\pm \to X$ (1%)<br>$qq$ (2%) |

**Case 2: the mixed bino/higgsino: $\mu \approx M_1$**

We consider a scenario where $\mu = 230$, $M_1 = 160$ GeV, $\tan\beta = 5$, $A_\lambda = 500$ GeV and $A_\kappa = 0$. In the MSSM limit, that is when $\lambda \to 0$, $\Omega h^2 \approx 0.2$, a value slightly above WMAP. The LSP is a mixed bino/Higgsino and its main annihilation channel is into W pairs. For moderate values of $\kappa$, say $\kappa = 0.2$, increasing $\lambda$ affects the Higgs spectrum and increases the singlino component of the LSP. This leads either to a sharp drop in $\Omega h^2$ when one encounters the $H_2$ resonance or to a more moderate drop for large values of $\lambda$. For example, for $\lambda = 0.35$, we observe much enhanced cross sections for $\tilde{\chi}_1^0\tilde{\chi}_1^0 \to HH, HA$, this leads to $\Omega h^2 = 0.116$. Details of this scenario are presented in Table 4.3, Case 2. For even larger values of $\lambda$, the singlino component of the LSP becomes significant. Then the $\tilde{\chi}_1^0\tilde{\chi}_1^0 H_1$ and $\tilde{\chi}_1^0\tilde{\chi}_1^0 A_1$ couplings are large leading to an even larger contribution of the $\tilde{\chi}_1^0\tilde{\chi}_1^0 \to H_1 A_1$ annihilation through $t$-channel $\tilde{\chi}_1^0$ exchange. However, this area of the parameter space is excluded by Higgs searches at LEP.

**Case 3 : the singlino LSP**

We explore now scenarios satisfying both LEP and WMAP constraints with a predominantly singlino LSP. For this we scanned over the whole parameter space of the NMSSM in the range $\lambda < 0.75$, $\kappa < 0.65$, $2 < \tan\beta < 10$, $100 < \mu < 500$ GeV, $100 < M_2 < 1000$ GeV, $0 < A_\lambda < 1000$ GeV and $0 < -A_\kappa < 500$ GeV. We found three classes of models: a mixed singlino/higgsino LSP that annihilates mainly into $H_1 A_1$ and $VV$ ($V = W, Z$), an almost pure singlino that annihilates through a $Z$ or Higgs resonance and a singlino where dominant channels are coannihilation ones. In Table 4.3 we show a selection of benchmark points along these lines (Case 3a,3b,3c).

The first scenario, Case 3a in Table 4.3, is one for which $\mu \ll M_1$ and the LSP is a mixed higgsino/singlino. In this example, the LSP is 90% singlino and 10% higgsino, with a mass of 127 GeV. The main annihilation mode is $H_1 A_1$ through $t$-channel $\tilde{\chi}_1^0$ exchange, $H_1$ and $A_1$ being both mainly singlet (88% and 99% respectively). This is due to enhanced couplings $\tilde{\chi}_1^0\tilde{\chi}_1^0 A_1(H_1)$ which occur for large values of $\lambda$ (Eq. 4.38). Annihilation of the higgsino component into W/Z pairs accounts for the subdominant channel.

In Table 4.3 we also give an example, Case 3b, of a scenario with a light singlino LSP, here with a mass of 10 GeV. The only efficient mode for such a light singlino is via a Higgs resonance, here a light scalar dominantly singlet. This scalar decays into $b\bar{b}$, or when kinematically accessible into $A_1 A_1$, the $A_1$ being also mainly singlet. The Higgs sector of such models is of course severely constrained by LEP, in particular the limit on the SM-like scalar, here the second scalar, $H_2$. For this reason most scenarios with light singlinos have $\tan\beta \approx 3$ which is the value for which the lightest visible (i.e. non singlet) Higgs mass, $M_{H_2}$, is maximized [28, 98–100]. Note that a light singlino requires $\kappa \ll \lambda$ and not too large value for $\mu$.

For $\kappa \leq \lambda \ll 1$, the LSP is heavy with a large singlino component. No efficient annihilation mechanism is then available. However coannihilation with heavier neutralinos and charginos can be very efficient especially for a higgsino-like NLSP. Case 3c in Table 4.3 gives an example of such a scenario. The LSP is 96% singlino with a mass of 203 GeV. The mass difference with the NLSP $\widetilde{\chi}_2^0$ is 11 GeV. The coannihilation channels are overwhelmingly dominant. The $\widetilde{\chi}_2^0$ higgsino component is just enough (28%) for efficient annihilation. The main channels are $\widetilde{\chi}_2^0\widetilde{\chi}_2^0(\widetilde{\chi}_1^0) \to t\bar{t}, b\bar{b}$ and correspond to annihilation through $H_3$ and $A_2$ exchange. For this point, $H_3$ and $A_2$ belong to the heavy Higgs doublet with $M_A \approx 475$ GeV, so that we are close to a (double) resonance. Such a resonance is not necessary though, in order to have efficient $\widetilde{\chi}_2^0$ annihilation. We also found points in the parameter space with a heavy singlino where the dominant channel was $\widetilde{\chi}_2^0\widetilde{\chi}_2^0(\widetilde{\chi}_1^0) \to VV$ through $Z$ exchange.

### *4.6.4 Conclusion*

In the NMSSM, basically the same mechanisms as for the MSSM are at work for neutralino annihilation, nevertheless the presence of additional Higgs states provides additional possibilities for efficient neutralino annihilation. Specifically this means additional regions of parameter space where rapid annihilation through a $s$-channel resonance can take place, as well as new annihilation channels when light Higgs states are present. However, annihilation of neutralinos is not always favoured in the NMSSM. In general the singlino component of the LSP tends to reduce the annihilation cross-section. We found however regions of the parameter space where a singlino LSP gives the right amount of dark matter, either for large $\lambda$, $s$-channel resonances into a $Z$ or a Higgs, or coannihilation with $\widetilde{\chi}_2^0$, $\widetilde{\chi}_1^\pm$.

For scenarios for which the relic density is within the WMAP allowed region, one can ask whether it would be possible to see signatures of the model at colliders. In the case of a mixed bino/higgsino LSP, provided the singlino state is not heavy and decouples, i.e. $\lambda$ not too small and $\kappa$ not too large, the five neutralino states might be visible at the LHC/ILC. This would be a clear signature of the NMSSM. Finally, in the singlino LSP case, $\mu$ cannot be too large. One therefore would expect visible higgsinos at the LHC. The singlino LSP would however appear at the end of the decay chain in any sparticle pair production process, which might complicate the detection task as it was the case at LEP [94].

### 4.7 Comparison of Higgs bosons in the extended MSSM models

*Vernon Barger, Paul Langacker, Hye-Sung Lee and Gabe Shaughnessy*

When the $\mu$ parameter of the Minimal Supersymmetric Standard Model (MSSM) is promoted to a Standard Model (SM) singlet field, the fine-tuning problem of the MSSM [2] can be naturally solved with a dynamically generated $\mu_{\rm eff} \equiv \lambda \langle S \rangle = \lambda v_s/\sqrt{2}$. The new Higgs singlet is accompanied by a new symmetry that governs the interaction of the singlet. Depending on the symmetry, the MSSM can be extended to different models such as the Next-to-Minimal Supersymmetric SM (NMSSM) [26, 30, 55], the Minimal Nonminimal Supersymmetric SM (MNSSM) [44–46, 51], and the $U(1)'$-extended Minimal Supersymmetric SM (UMSSM) [13, 101]. Table 4.4 shows the symmetry, superpotential and the Higgs spectrum of several models. The Exceptional Supersymmetric SM (ESSM) [16] is, to a large extent, similar to the UMSSM. The secluded $U(1)'$ model (sMSSM) [102] has multiple Higgs singlets and, in a decoupling limit of the extra singlets, the low energy spectrum is similar to the MNSSM.

It is important to compare the implications of the MSSM and its various extensions. In this note, we treat the different models in a consistent way to compare and contrast their features. (For a full study by the authors, see [103].) The neutralino sectors are also extended by the singlino and $Z'$-ino in these models and were studied in [52].

#### *4.7.1 Models*

The tree-level Higgs mass-squared matrices are found from the potential, $V$, which is a sum of the $F$-term, $D$-term and soft-terms in the lagrangian, as follows.

$$V_F = |\lambda H_u \cdot H_d + t_F + \kappa S^2|^2 + |\lambda S|^2 \left(|H_d|^2 + |H_u|^2\right) \quad (4.39)$$

$$V_D = \frac{G^2}{8}\left(|H_d|^2 - |H_u|^2\right)^2 + \frac{g_2^2}{2}\left(|H_d|^2|H_u|^2 - |H_u \cdot H_d|^2\right) + \frac{{g_{1'}}^2}{2}\left(Q_{H_d}|H_d|^2 + Q_{H_u}|H_u|^2 + Q_S|S|^2\right)^2 \quad (4.40)$$

$$V_{\rm soft} = m_{H_d}^2|H_d|^2 + m_{H_u}^2|H_u|^2 + m_S^2|S|^2 + \left(A_\lambda \lambda S H_u \cdot H_d + \frac{\kappa}{3}A_\kappa S^3 + t_S S + h.c.\right) \quad (4.41)$$

This scalar potential is a collective form of all extended MSSM models considered here and, for a particular model, the parameters in $V$ are understood to be *turned-off* appropriately.

$$\begin{array}{rcl} \text{MNSSM/sMSSM} & : & g_{1'} = 0, \kappa = 0, A_\kappa = 0 \\ \text{NMSSM} & : & g_{1'} = 0, t_{F,S} = 0 \\ \text{UMSSM} & : & t_{F,S} = 0, \kappa = 0, A_\kappa = 0 \end{array} \quad (4.42)$$

The $F$-term and the soft terms contain the model dependence of the NMSSM and MNSSM/sMSSM, while the $D$-term contains that of the UMSSM. We ignore possible CP violation in the Higgs sector.

Table 4.4: Higgs bosons of the MSSM and several of its extensions

| Model | Symmetry | Superpotential | CP-even | CP-odd | Charged |
|---|---|---|---|---|---|
| MSSM | – | $\mu \hat{H}_u \cdot \hat{H}_d$ | $H_1, H_2$ | $A_1$ | $H^\pm$ |
| NMSSM | $\mathbf{Z_3}$ | $\lambda \hat{S}\hat{H}_u \cdot \hat{H}_d + \frac{\kappa}{3}\hat{S}^3$ | $H_1, H_2, H_3$ | $A_1, A_2$ | $H^\pm$ |
| MNSSM | $\mathbf{Z_5^R}, \mathbf{Z_7^R}$ | $\lambda \hat{S}\hat{H}_u \cdot \hat{H}_d + t_F \hat{S}$ | $H_1, H_2, H_3$ | $A_1, A_2$ | $H^\pm$ |
| UMSSM | $U(1)'$ | $\lambda \hat{S}\hat{H}_u \cdot \hat{H}_d$ | $H_1, H_2, H_3$ | $A_1$ | $H^\pm$ |
| sMSSM | $U(1)'$ | $\lambda \hat{S}\hat{H}_u \cdot \hat{H}_d + \lambda_s \hat{S}_1\hat{S}_2\hat{S}_3$ | $H_1, \cdots, H_6$ | $A_1, \cdots, A_4$ | $H^\pm$ |

### 4.7.2 Higgs mass matrices

The collective tree-level CP-even Higgs mass matrix elements in the $(H_d^0, H_u^0, S)$ basis are:

$$(\mathcal{M}_+^0)_{11} = \left[\frac{G^2}{4} + Q_{H_d}^2 g_{1'}^2\right] v_d^2 + \left(\frac{\lambda A_\lambda}{\sqrt{2}} + \frac{\lambda\kappa v_s}{2} + \frac{\lambda t_F}{v_s}\right)\frac{v_u v_s}{v_d} \quad (4.43)$$

$$(\mathcal{M}_+^0)_{12} = -\left[\frac{G^2}{4} - \lambda^2 - Q_{H_d} Q_{H_u} g_{1'}^2\right] v_d v_u - \left(\frac{\lambda A_\lambda}{\sqrt{2}} + \frac{\lambda\kappa v_s}{2} + \frac{\lambda t_F}{v_s}\right) v_s \quad (4.44)$$

$$(\mathcal{M}_+^0)_{13} = \left[\lambda^2 + Q_{H_d} Q_S g_{1'}^2\right] v_d v_s - \left(\frac{\lambda A_\lambda}{\sqrt{2}} + \lambda\kappa v_s\right) v_u \quad (4.45)$$

$$(\mathcal{M}_+^0)_{22} = \left[\frac{G^2}{4} + Q_{H_u}^2 g_{1'}^2\right] v_u^2 + \left(\frac{\lambda A_\lambda}{\sqrt{2}} + \frac{\lambda\kappa v_s}{2} + \frac{\lambda t_F}{v_s}\right)\frac{v_d v_s}{v_u} \quad (4.46)$$

$$(\mathcal{M}_+^0)_{23} = \left[\lambda^2 + Q_{H_u} Q_S g_{1'}^2\right] v_u v_s - \left(\frac{\lambda A_\lambda}{\sqrt{2}} + \lambda\kappa v_s\right) v_d \quad (4.47)$$

$$(\mathcal{M}_+^0)_{33} = \left[Q_S^2 g_{1'}^2 + 2\kappa^2\right] v_s^2 + \left(\frac{\lambda A_\lambda}{\sqrt{2}} - \frac{\sqrt{2} t_S}{v_d v_u}\right)\frac{v_d v_u}{v_s} + \frac{\kappa A_\kappa}{\sqrt{2}} v_s \quad (4.48)$$

where $v_{d,u} = \sqrt{2}\langle H_{d,u}^0\rangle$. The similarly modified matrix elements for the CP-odd and charged Higgs masses in the extended models can be found in [103]. We consider only the dominant 1-loop correction which comes from the common top/stop contributions to keep a consistent analysis.

The mass-squared sum rules are:

$$\begin{aligned}
\text{MSSM} &: \left(M_{H_1}^2 + M_{H_2}^2\right) - \left(M_{A_1}^2\right) = M_Z^2 + \delta M^2 \\
\text{MNSSM/sMSSM} &: \left(M_{H_1}^2 + M_{H_2}^2 + M_{H_3}^2\right) - \left(M_{A_1}^2 + M_{A_2}^2\right) = M_Z^2 + \delta M^2 \\
\text{NMSSM} &: \left(M_{H_1}^2 + M_{H_2}^2 + M_{H_3}^2\right) - \left(M_{A_1}^2 + M_{A_2}^2\right) \\
&\qquad = M_Z^2 + \kappa(-\lambda v_d v_u + v_s(\sqrt{2}A_\kappa + v_s\kappa)) + \delta M^2 \\
\text{UMSSM} &: \left(M_{H_1}^2 + M_{H_2}^2 + M_{H_3}^2\right) - \left(M_{A_1}^2\right) = M_Z^2 + M_{Z'}^2 + \delta M^2
\end{aligned} \quad (4.49)$$

With a one-loop radiative correction, the common loop effect, $\delta M^2$, could be as large as $\mathcal{O}((100 \text{ GeV})^2)$ unless $\tan\beta$ is very small.

The physical Higgs boson masses are found by diagonalizing the mass-squared matrices, $\mathcal{M}_D = R_+ \mathcal{M} R_+^{-1}$, where $\mathcal{M}$ also includes the radiative corrections. The rotation matrices, $R_+$, may then be used to construct the physical Higgs fields as

$$H_i = R_+^{i1}\phi_d + R_+^{i2}\phi_u + R_+^{i3}\phi_s \quad (4.50)$$

where the physical states are ordered by their mass as $M_{H_1} \le M_{H_2} \le M_{H_3}$, and similarly for the CP-odd Higgses.

### 4.7.3 Interesting limits

The extended MSSM models share the common characteristics of the near Peccei-Quinn symmetry limit [6] when the model-dependent terms (such as $\kappa$, $A_\kappa$, $t_{F,S}$, $g_{1'}$) are very small, and the lightest CP-odd Higgs boson ($Z'$ gauge boson for the UMSSM case) becomes nearly massless. The exact global Peccei-Quinn symmetry should be avoided though, to be compatible with the non-observation of the Weinberg-Wilczek axion [104, 105].

When the singlet VEV, $v_s$, is large while $\mu_{\text{eff}}$ is kept at the EW scale (i.e., $\lambda$ is small), they approach the MSSM limit when the model-dependent terms are not large. In the large $v_s$ limit (i.e., $M_c/v_s \sim \epsilon$ where $M_c$ is the common mass scale other than $v_s$, and $\epsilon \ll 1$), we show the explicit tree-level approximations of the CP-even Higgs masses [103].

NMSSM (with an additional assumption of $\kappa \sim \epsilon$) :

$$\begin{aligned}
M^2_{H_1} &\approx \frac{1}{2} v_s \kappa \left(4 v_s \kappa + \sqrt{2} A_\kappa\right) \\
M^2_{H_{2,3}} &\approx \frac{1}{2} M_Z^2 + \left(A_\lambda + \frac{v_s \kappa}{\sqrt{2}}\right) \mu_{\rm eff} \csc 2\beta \\
&\mp \sqrt{\left(\frac{1}{2} M_Z^2 - \left(A_\lambda + \frac{v_s \kappa}{\sqrt{2}}\right) \mu_{\rm eff} \csc 2\beta\right)^2 + 2 M_Z^2 \left(A_\lambda + \frac{v_s \kappa}{\sqrt{2}}\right) \mu_{\rm eff} \sin 2\beta}
\end{aligned} \tag{4.51}$$

MNSSM/sMSSM (with an additional assumption of $t_S/M_c^3 \sim \epsilon$) :

$$\begin{aligned}
M^2_{H_1} &\approx \frac{1}{v_s^2} \frac{\mu_{\rm eff} \sec^2 2\beta}{2 G^2 A_\lambda} \left(32 \mu_{\rm eff}^2 \sin 2\beta A_\lambda^2 - G^2 v^2 \sin^3 2\beta \left(4 \mu_{\rm eff}^2 + A_\lambda^2\right)\right. \\
&\left. + 2 \mu_{\rm eff} A_\lambda \left(G^2 v^2 - 16 \mu_{\rm eff}^2 - 2 A_\lambda^2 + \cos 4\beta \left(-G^2 v^2 + 2 A_\lambda^2\right)\right)\right) - \frac{\sqrt{2} t_S}{v_s} \\
M^2_{H_{2,3}} &\approx \frac{1}{2} M_Z^2 + A_\lambda \mu_{\rm eff} \csc 2\beta \mp \sqrt{\left(\frac{1}{2} M_Z^2 - A_\lambda \mu_{\rm eff} \csc 2\beta\right)^2 + 2 M_Z^2 A_\lambda \mu_{\rm eff} \sin 2\beta}
\end{aligned} \tag{4.52}$$

UMSSM :

$$\begin{aligned}
M^2_{H_{1,2}} &\approx \frac{1}{2} M_Z^2 + A_\lambda \mu_{\rm eff} \csc 2\beta \mp \sqrt{\left(\frac{1}{2} M_Z^2 - A_\lambda \mu_{\rm eff} \csc 2\beta\right)^2 + 2 M_Z^2 A_\lambda \mu_{\rm eff} \sin 2\beta} \\
M^2_{H_3} &\approx M^2_{Z'} \quad \text{(with } Z' \text{ mass given by } M^2_{Z'} = {g_{1'}}^2 (Q^2_{H_d} v_d^2 + Q^2_{H_u} v_u^2 + Q_S^2 v_s^2))
\end{aligned} \tag{4.53}$$

With large $v_s$, when $\kappa$ (and $\kappa A_\kappa$) $\to 0$, $g_{1'} \to 0$, $t_{F,S} \to 0$, all of the above Higgs masses reach the MSSM limits (with the identification of $A_\lambda = B$ and $\mu_{\rm eff} = \mu$) with an additional scalar decoupled with either negligible or very heavy mass. The first solution in the MNSSM/sMSSM is not valid when $\tan\beta$ is near 1 (or $\sec^2 2\beta \to \infty$), but an exact solution can be obtained in this limit.

### *4.7.4 Theoretical upper bounds on the lightest Higgs mass*

From the mass matrix of Eq. (4.43-4.48), the upper bounds on the lightest CP-even Higgs can be obtained.

$$\begin{aligned}
\text{MSSM} &: \quad M^2_{H_1} \le M_Z^2 \cos^2 2\beta + \tilde{\mathcal{M}}^{(1)} \\
\text{NMSSM/MNSSM/sMSSM} &: \quad M^2_{H_1} \le M_Z^2 \cos^2 2\beta + \frac{1}{2} \lambda^2 v^2 \sin^2 2\beta + \tilde{\mathcal{M}}^{(1)} \\
\text{UMSSM} &: \quad M^2_{H_1} \le M_Z^2 \cos^2 2\beta + \frac{1}{2} \lambda^2 v^2 \sin^2 2\beta \\
&\qquad + g_{1'}^2 v^2 (Q_{H_d} \cos^2\beta + Q_{H_u} \sin^2\beta)^2 + \tilde{\mathcal{M}}^{(1)}
\end{aligned} \tag{4.54}$$

where $\tilde{\mathcal{M}}^{(1)}$ is the common contribution from the 1-loop correction.

All extended models have larger upper bounds for the lightest CP-even Higgs than that of the MSSM due to the contribution of the singlet scalar. The UMSSM has an additional contribution in the quartic coupling from the gauge coupling constant, $g_{1'}$ of the $U(1)'$ symmetry. In the MSSM, large $\tan\beta$ values are suggested by the conflict between the experimental lower bound and the theoretical upper bound on $M_{H_1}$. Since the extended models contain additional terms which relax the theoretical bound, they allow smaller $\tan\beta$ values than the MSSM does (see Fig. 4.15b).

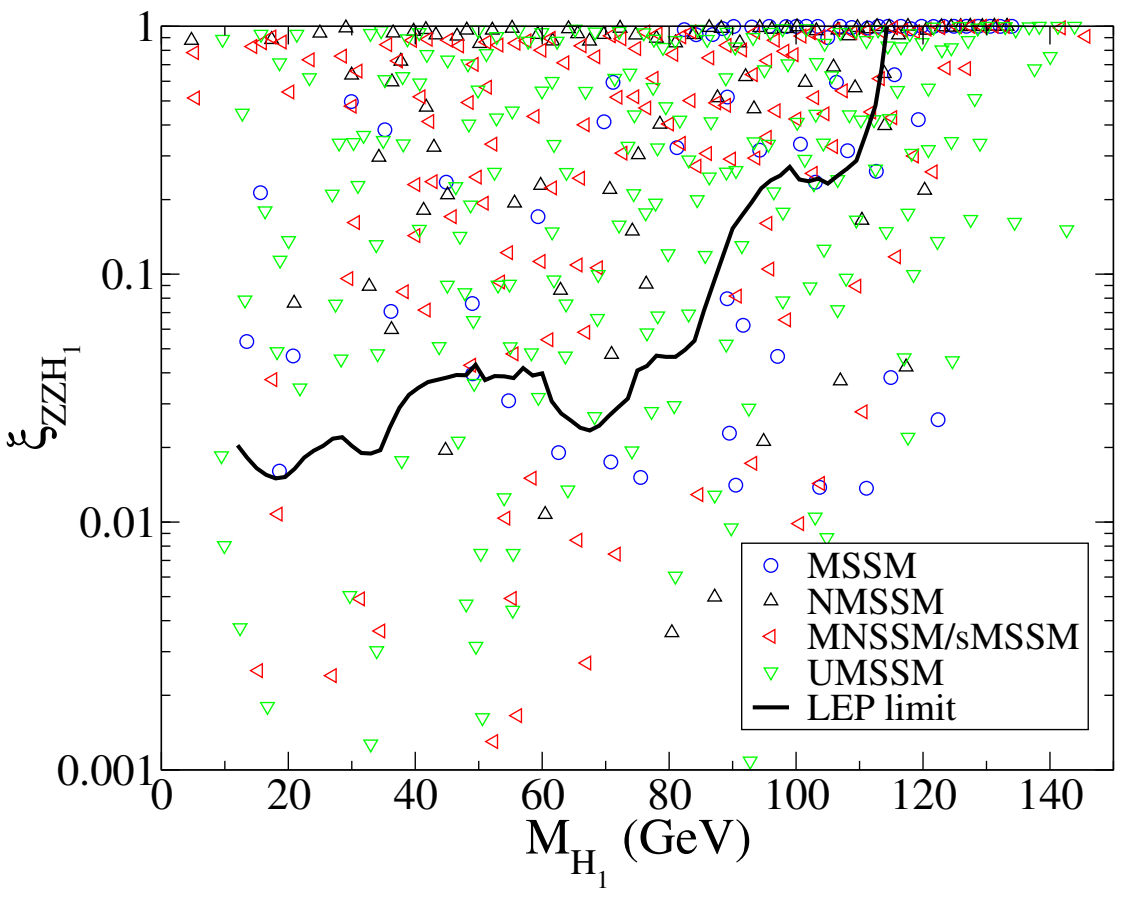

(a)

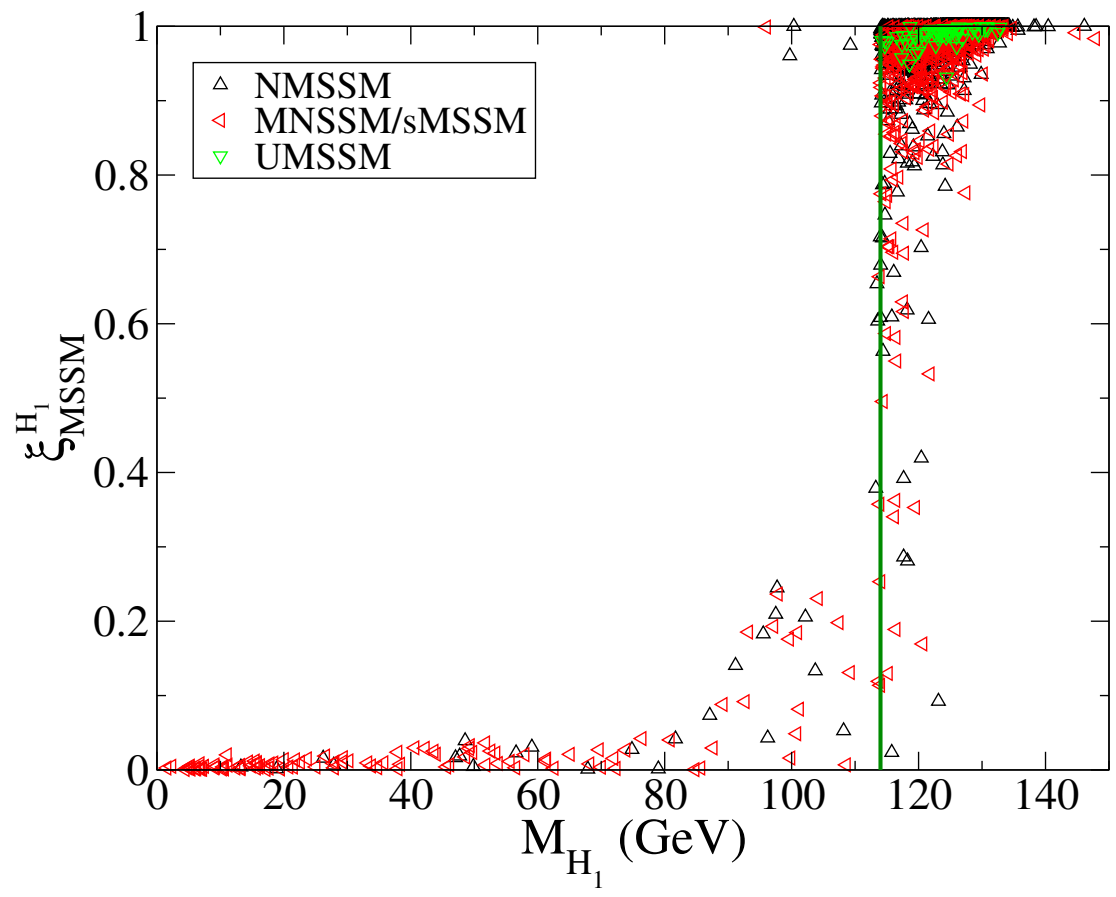

(b)

Fig. 4.14: (a) LEP limit [106] on $\xi_{ZZH_1} = \left(g_{ZZH_1}/g^{SM}_{ZZh}\right)^2 = \Gamma_{Z\to ZH_1}/\Gamma^{SM}_{Z\to Zh}$, the scaled $ZZH_1$ coupling in various Supersymmetric models, vs. the lightest CP-even Higgs mass. The other constraints are not applied. The solid black curve is the observed limit at 95% C.L. Points falling below this curve pass the $ZZH_1$ constraint. (b) The lightest CP-even Higgs masses vs. $\xi_{\rm MSSM}$ (MSSM fraction) after all constraints are applied. The vertical line is the LEP lower bound on the MSSM (SM-like) Higgs mass.

### 4.7.5 *Experimental constraints on the Higgs*

(i) LEP bound on $ZZh$ coupling: The $ZZh$ coupling limits from LEP [106] can be used to limit the mass of the lightest CP-even Higgs boson of the extended MSSM models. Fig. 4.14a shows the LEP limit (95% C.L.) on the $ZZH_1$ coupling relative to the SM coupling with the $H_1$ mass. The relative coupling is given by

$$\xi_{ZZH_i} = \left(g_{ZZH_i}/g^{SM}_{ZZh}\right)^2 = (R^{i1}_+ \cos\beta + R^{i2}_+ \sin\beta)^2. \tag{4.55}$$

As the scatter plot shows, when the Higgs coupling is diluted by the singlet component (i.e., $\xi_{ZZH_1} < 1$), it may have a mass smaller than the SM-like Higgs limit of 114.4 GeV.

(ii) LEP bound on $M_{H_1}$ and $M_{A_1}$: For the channel of $Z \to A_i H_j$ with $A_i \to b\bar{b}$ and $H_j \to b\bar{b}$, LEP gives bounds on the MSSM Higgs masses of $M_{H_1} \geq 92.9$ GeV and $M_{A_1} \geq 93.4$ GeV assuming maximal stop mixing, yielding the most conservative limit [74]. With the maximum LEP energy, $\sqrt{s} = 209$ GeV, mass limits on the $H_1$ and $A_1$ in the extended MSSM models can be obtained with the upper bound of the cross section for $e^+e^- \to A_i H_j$ at 40 fb. In practice, we find that the LEP $Z \to A_i H_j$ constraint eliminates a significant fraction of the points generated with a low CP-odd Higgs mass.

(iii) LEP bound on $M_{H^\pm}$: The Higgs singlet does not alter the charged Higgs part, and the LEP bound on the MSSM charged Higgs mass of $M_{H^\pm} \geq 78.6$ GeV is imposed [107].

(iv) LEP bound on $M_{\chi^\pm_1}$: The LEP bound on the chargino mass of $M_{\chi^\pm_1} > 104$ GeV is imposed [108].

(v) LEP invisible $Z$ decay width: The LEP bound on the invisible $Z$ decay width by new physics of $\Delta\Gamma_Z < 1.9$ MeV is imposed [109].

(vi) LEP $Z - Z'$ mixing angle: The LEP bound on the $Z - Z'$ mixing angle (for the UMSSM), $\alpha_{ZZ'} < 2 \times 10^{-3}$ is imposed [110–112]. The exact bound depends on the model.

### 4.7.6 *Numerical results*

The model-independent parameters are scanned over $\tan\beta = 1 \sim 50$, $v_s = 50 \sim 2000$ GeV, $\mu_{\rm eff} = 50 \sim 1000$ GeV, $A_\lambda = 0 \sim 1000$ GeV, $A_t = -1000 \sim 1000$ GeV, $M_2 = -500 \sim 500$ GeV.

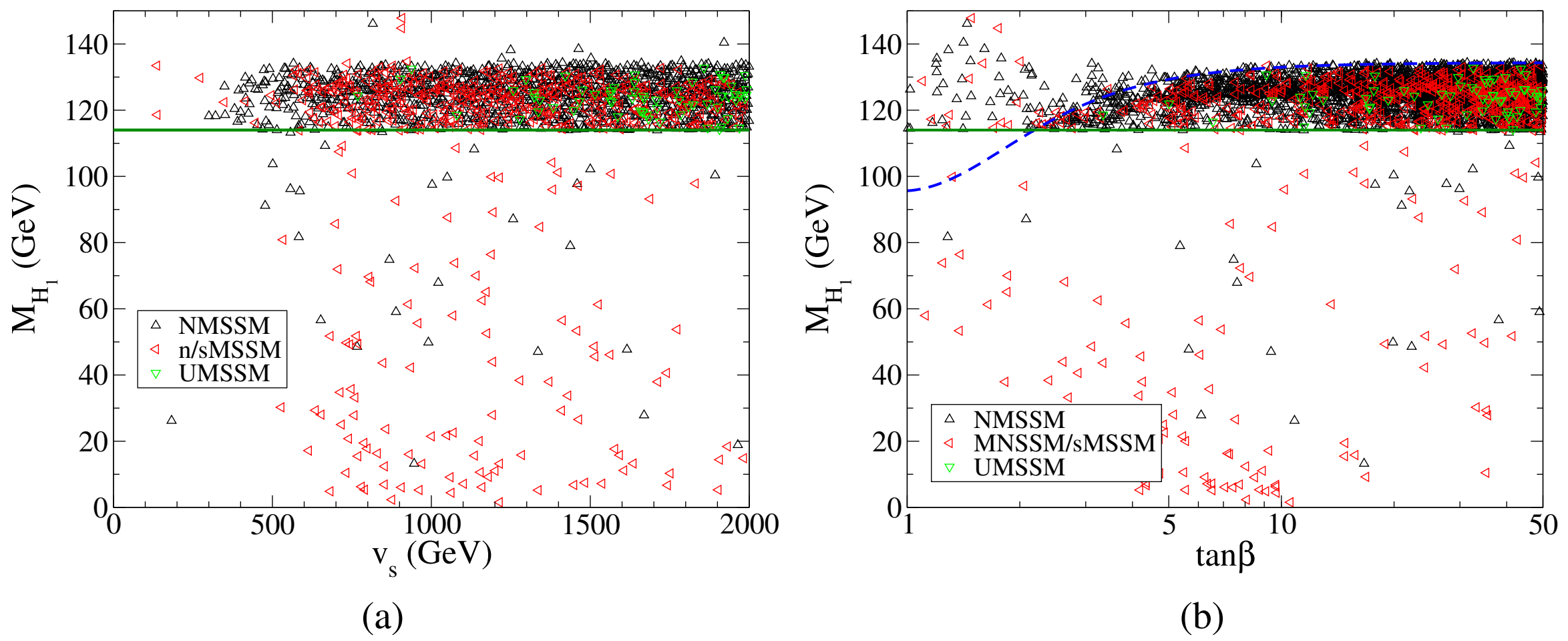

Fig. 4.15: The lightest CP-even Higgs masses vs. (a) $v_s$ and (b) $\tan\beta$. The horizontal line is the LEP lower bound on the SM-like Higgs. The dashed curve is the MSSM bound for a maximum stop mixing.

The model-dependent parameters are scanned over $\kappa = -0.75 \sim 0.75$, $A_\kappa = -1000 \sim 1000$ GeV, $t_F = -500^2 \sim 500^2$ GeV$^2$, $t_S = -500^3 \sim 500^3$ GeV$^3$, $\theta_{E_6} = 0 \sim \pi$. We assume gaugino mass unification $M_1 = M_{1'} = \frac{5g_1^2}{3g_2^2}M_2$ and fix the stop soft mass at $M_{\widetilde{Q}} = M_{\widetilde{U}} = 1000$ GeV and the renormalization scale for the loop correction at $Q = 300$ GeV. Additional constraints of $0.1 \le \lambda \le 0.75$ and $0.1 \le \sqrt{\kappa^2 + \lambda^2} \le 0.75$ for perturbativity and naturalness are also applied.

The relative coupling strength of a particular Higgs boson, $H_i$, to the MSSM fields may be quantified as the MSSM fraction

$$\xi_{\rm MSSM}^{H_i} = \sum_{j=1}^{2} (R_+^{ij})^2. \tag{4.56}$$

In Fig. 4.14b, we plot the MSSM fraction versus the lightest CP-even Higgs boson mass in extended MSSM models after all constraints are applied. When the singlet composition is large (i.e., $\xi_{\rm MSSM}^{H_1}$ is small), a lighter mass is allowed by the LEP constraint. The UMSSM has the additional constraint on the singlet VEV from the $Z - Z'$ mixing angle constraint, and it pushes the allowed points to more MSSM-like as shown in Fig. 4.15a. However, there are ways to allow lower $v_s$ values, such as leptophobic couplings [113, 114] or additional singlet contributions [102]. The lightest CP-even Higgs boson mass versus $\tan\beta$ in each model shown in Fig. 4.15b has a majority of generated points in the band $114.4$ GeV $< M_{H_1} < 135$ GeV and $\tan\beta > 2$. This is, as the dashed curve indicates, one of the salient features of the MSSM after the experimental constraints, which implies that most of those points are MSSM-like.

In Fig. 4.16a, we present the parameter scan results of the mass ranges for the lightest CP-even Higgs in the MSSM and its extensions after all constraints are applied. The NMSSM and the MNSSM/sMSSM have pretty similar mass ranges and they can be extremely light due to effect of the singlet. The additional constraint on $s$ makes the lower bound of the UMSSM to be more MSSM-like. The upper bounds in the extended MSSM models are about $30 \sim 40$ GeV larger than that of the MSSM in accordance with Eq. (4.54). For the CP-odd and charged Higgses (Fig. 4.16b,c) as well as the other aspects including the Higgs production and decays, see [103].

### *4.7.7 Conclusions*

Even though low energy Supersymmetry is well-motivated, the $\mu$-problem suggests the MSSM may not be the full Supersymmetric model that describes TeV scale physics. The introduction of a Higgs singlet can solve this problem naturally, but depending on the new symmetry that governs the interaction of the singlet, there are more than one direction to extend the MSSM. We presented a formalism that is convenient for comparing different models in a consistent way.

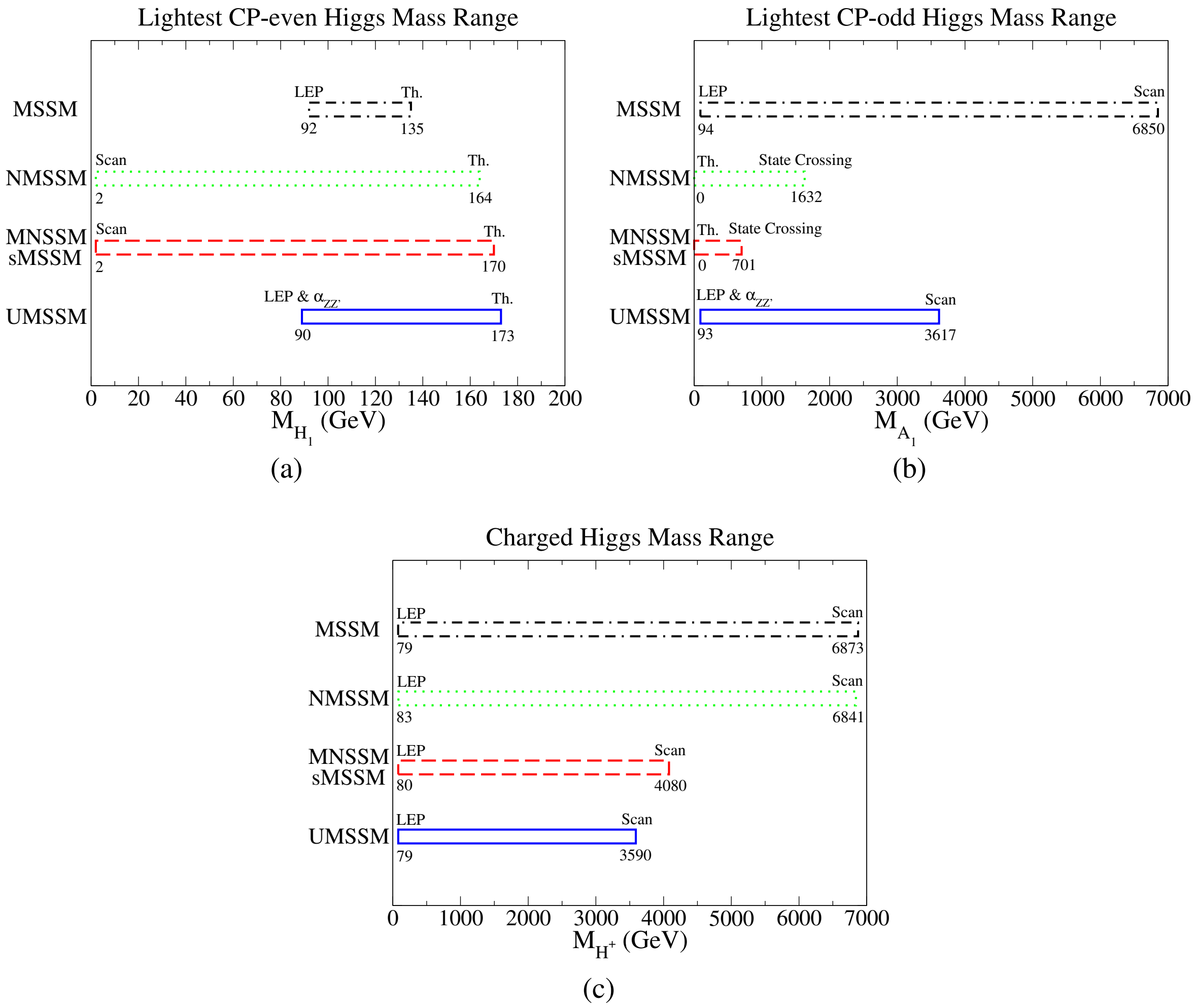

Fig. 4.16: Mass ranges of (a) the lightest CP-even, (b) the lightest CP-odd, and (c) charged Higgses

The extended MSSM have many similar features including:

- The $\mu$-problem is elegantly solved with a new Higgs singlet.
- The lightest CP-even Higgs can be considerably lighter than the SM or the MSSM bounds from LEP experiments.
- Low values of $\tan\beta$ are allowed unlike in the MSSM.
- The near Peccei-Quinn limit can be achieved when the model-dependent parameters are small, where the lightest CP-odd Higgs (or $Z'$ for the UMSSM) are expected to be very light.
- The MSSM limit can be achieved when the singlet VEV, $v_s$, is large compared to other mass parameters.

Due to the different governing symmetry, the models also have distinguishable characteristics, including:

- The UMSSM predicts an EW/TeV scale $Z'$ gauge boson (and $v_s$ receives additional experimental constraints).
- The NMSSM may have a domain wall problem related to the discrete symmetry [42].
- The Higgs spectra and mass sum rules are model-dependent (Table 4.4 and Eq. (4.49)).

– The neutralino properties are also distinguishable [52].

– In the large $v_s$ limit, the mass of the non-MSSM-like Higgs depends on the model parameters.

More studies are necessary to understand how to distinguish the extended MSSM models from the MSSM and from each other.

### 4.8 Distinction between NMSSM and MSSM in combined LHC and ILC analyses

*Hans Fraas, Fabian Franke, Stefan Hesselbach and Gudrid Moortgat-Pick*

In some parts of SUSY parameter space the experimentally accessible Higgs sector of the NMSSM is very similar to the MSSM Higgs sector and does not allow the identification of the underlying SUSY model. In such cases additional information from the neutralino sector can be crucial. In this contribution we analyze an NMSSM scenario for which only a combined analysis of LHC and ILC data will be sufficient to distinguish the models.

#### *4.8.1 Neutralino sector*

The NMSSM contains five neutralinos $\tilde{\chi}_i^0$, the mass eigenstates of the photino, zino and neutral higgsinos, and two charginos $\tilde{\chi}_i^\pm$, being mixtures of wino and charged higgsino. The neutralino/chargino sector depends at tree level on four parameters of Eq. (4.11), $\lambda$, $\kappa$, $\mu_{\rm eff}$, $\tan\beta$, and additionally on the U(1) and SU(2) gaugino masses $M_1$ and $M_2$ [31, 115–117]; see also Section 4.2. The additional fifth neutralino may significantly change the phenomenology of the neutralino sector. In scenarios where the lightest supersymmetric particle is a nearly pure singlino, the existence of displaced vertices may lead to a particularly interesting experimental signature [94, 95, 118–120] which allows the distinction between the models. Furthermore sfermion decays into fermions and singlino-dominated neutralinos can have large branching ratios resulting in modified signatures of the sfermions [121]. Especially the modified cascade decays of the squarks at the LHC can be important for the identification of the model. If however, only a part of the particle spectrum is kinematically accessible this distinction may become challenging. We start with a scenario with the parameters

$$M_1 = 360 \text{ GeV}, \quad M_2 = 147 \text{ GeV}, \quad \lambda = 0.5, \quad \kappa = 0.2, \quad \mu_{\rm eff} = 458 \text{ GeV}, \quad \tan\beta = 10, \tag{4.57}$$

and the following gaugino/higgsino masses and eigenstates:

$$m_{\tilde{\chi}_1^0} = 138 \text{ GeV}, \qquad \tilde{\chi}_1^0 = (-0.02, +0.97, -0.20, +0.09, -0.07), \tag{4.58}$$
$$m_{\tilde{\chi}_2^0} = 337 \text{ GeV}, \qquad \tilde{\chi}_2^0 = (+0.62, +0.14, +0.25, -0.31, +0.65), \tag{4.59}$$
$$m_{\tilde{\chi}_3^0} = 367 \text{ GeV}, \qquad \tilde{\chi}_3^0 = (-0.75, +0.04, +0.01, -0.12, +0.65), \tag{4.60}$$
$$m_{\tilde{\chi}_4^0} = 468 \text{ GeV}, \qquad \tilde{\chi}_4^0 = (-0.03, +0.08, +0.70, +0.70, +0.08), \tag{4.61}$$
$$m_{\tilde{\chi}_5^0} = 499 \text{ GeV}, \qquad \tilde{\chi}_5^0 = (+0.21, -0.16, -0.64, +0.62, +0.37), \tag{4.62}$$

where the neutralino eigenstates are given in the basis $(\tilde{B}^0, \tilde{W}^0, \tilde{H}_d^0, \tilde{H}_u^0, \tilde{S})$. As can be seen from Eqs. (4.59) and (4.60), the particles $\tilde{\chi}_2^0$ and $\tilde{\chi}_3^0$ have a rather strong singlino admixture.

This scenario translates at the $e^+e^-$ International Linear Collider (ILC) with $\sqrt{s} = 500$ GeV into the experimental observables of Table 4.5 for the measurement of the masses and production cross sections for several polarization configurations of the light neutralinos and charginos. We assume mass uncertainties of $\mathcal{O}(1-2\%)$ [122,123], a polarization uncertainty of $\Delta P_{e^\pm}/P_{e^\pm} = 0.5\%$ and one standard deviation statistical errors. The masses and cross sections in different beam polarization configurations provide the experimental input for deriving the supersymmetric parameters within the MSSM using standard methods [124–127]. Note that beam polarization may be crucial for distinguishing the two models [128–130].

Table 4.5: Masses with 1.5% ($\tilde{\chi}^0_{2,3}$, $\tilde{e}_{L,R}$, $\tilde{\nu}_e$) and 2% ($\tilde{\chi}^0_1$, $\tilde{\chi}^\pm_1$) uncertainty [122, 123] and the kinematically allowed cross sections with an error composed of the error due to the mass uncertainties, polarization uncertainty of $\Delta P_{e^\pm}/P_{e^\pm} = 0.5\%$ and one standard deviation statistical error based on $\int \mathcal{L} = 100$ fb$^{-1}$, for both unpolarized beams and polarized beams with $(P_{e^-}, P_{e^+}) = (\mp 90\%, \pm 60\%)$, in analogy to the study in [131].

| | | $\sigma(e^+e^- \to \tilde{\chi}^\pm_1 \tilde{\chi}^\mp_1)$/fb | | $\sigma(e^+e^- \to \tilde{\chi}^0_1 \tilde{\chi}^0_2)$/fb |
|---|---|---|---|---|
| $m_{\tilde{\chi}^0_1} = 138 \pm 2.8$ GeV | | | | |
| $m_{\tilde{\chi}^0_2} = 337 \pm 5.1$ GeV | $(P_{e^-}, P_{e^+})$ | $\sqrt{s} = 400$ GeV | $\sqrt{s} = 500$ GeV | $\sqrt{s} = 500$ GeV |
| $m_{\tilde{\chi}^\pm_1} = 139 \pm 2.8$ GeV | Unpolarized | $323.9 \pm 33.5$ | $287.5 \pm 16.5$ | $4.0 \pm 1.2$ |
| $m_{\tilde{e}_L} = 240 \pm 3.6$ GeV | $(-90\%, +60\%)$ | $984.0 \pm 101.6$ | $873.9 \pm 50.1$ | $12.1 \pm 3.8$ |
| $m_{\tilde{e}_R} = 220 \pm 3.3$ GeV | $(+90\%, -60\%)$ | $13.6 \pm 1.6$ | $11.7 \pm 1.2$ | $0.2 \pm 0.1$ |
| $m_{\tilde{\nu}_e} = 226 \pm 3.4$ GeV | | | | |

Table 4.6: Masses and mixing character in the basis $(H_u, H_d, S)$ of the NMSSM Higgs bosons for the parameters $\lambda = 0.5$, $\kappa = 0.2$, $\mu_{\text{eff}} = 458$ GeV, $\tan\beta = 10$, $A_\lambda = 4000$ GeV and $A_\kappa = -200$ GeV and the branching ratios of the lightest scalar Higgs $H_1$ calculated with `NMHDECAY` [63]. Only decay channels with $BR > 1\%$ are listed.

| | $m_H$/GeV | mixing | | $BR(H_1)$ |
|---|---|---|---|---|
| $H_1$ | 125 | $(-0.9949, -0.0992, 0.0165)$ | $H_1 \to gg$ | 5% |
| $H_2$ | 293 | $(-0.0145, -0.0211, -0.9997)$ | $H_1 \to \tau\tau$ | 7% |
| $H_3$ | 4415 | $(0.0995, -0.9948, 0.0196)$ | $H_1 \to cc$ | 3% |
| $A_1$ | 333 | $(0.0017, 0.0166, -0.9999)$ | $H_1 \to bb$ | 63% |
| $A_2$ | 4415 | $(0.0995, 0.9949, 0.0167)$ | $H_1 \to WW^*$ | 20% |
| $H^\pm$ | 4417 | | $H_1 \to ZZ^*$ | 2% |

### *4.8.2 Higgs sector*

The Higgs sector of the NMSSM [30, 99] depends on two additional parameters, the trilinear soft scalar mass parameters $A_\lambda$ and $A_\kappa$. The Higgs bosons with dominant singlet character may escape detection in large regions of these parameters, thus the Higgs sector does not allow the identification of the NMSSM. A scan with `NMHDECAY` [63] in our scenario, Eq. (4.57), over $A_\lambda$ and $A_\kappa$ results in parameter points which survive the theoretical and experimental constraints in the region 2614 GeV $< A_\lambda <$ 5583 GeV and $-564$ GeV $< A_\kappa <$ 5 GeV. For $-396$ GeV $< A_\kappa < -92$ GeV the second lightest scalar ($H_2$) and the lightest pseudoscalar ($A_1$) Higgs particle have very pure singlet character and are heavier than the mass difference $m_{\tilde{\chi}^0_3} - m_{\tilde{\chi}^0_1}$, hence the decays of the neutralinos $\tilde{\chi}^0_2$ and $\tilde{\chi}^0_3$, which will be discussed in the following, are not affected by $H_2$ and $A_1$. The dependence of the masses of $H_1$, $H_2$ and $A_1$ on $A_\kappa$ is illustrated in Fig. 4.17 (left panel). The mass of the lightest scalar Higgs $H_1$, which has MSSM-like character in this parameter range, depends only weakly on $A_\kappa$ and is about 125 GeV. The masses of $H_3$, $A_2$ and $H^\pm$ are of the order of $A_\lambda$. For $A_\kappa < -396$ GeV the smaller mass of the $H_2$ and a stronger mixing between the singlet and MSSM-like states in $H_1$ and $H_2$ might allow a discrimination in the Higgs sector while for $A_\kappa > -92$ GeV the existence of a light pseudoscalar $A_1$ may give first hints of the NMSSM [72]. For our specific case study we choose $A_\lambda = 4000$ GeV and $A_\kappa = -200$ GeV, which leads to to the Higgs masses and mixing characters as listed in Table 4.6. Here $H_3$ and $A_2$ are kinematically not accessible while $H_2$ and $A_1$ are not produced due to their nearly pure singlet character. Then only $H_1$ can be detected with the branching ratios given in Table 4.6, which are very similar to those of an SM Higgs boson of the same mass. Also the branching ratio of $\tilde{\chi}^0_2$ in the lightest Higgs particle differs only by a factor two in both scenarios. If a precise measurement of this branching ratio is possible first hints for the inconsistency of the model could be derived at the ILC.

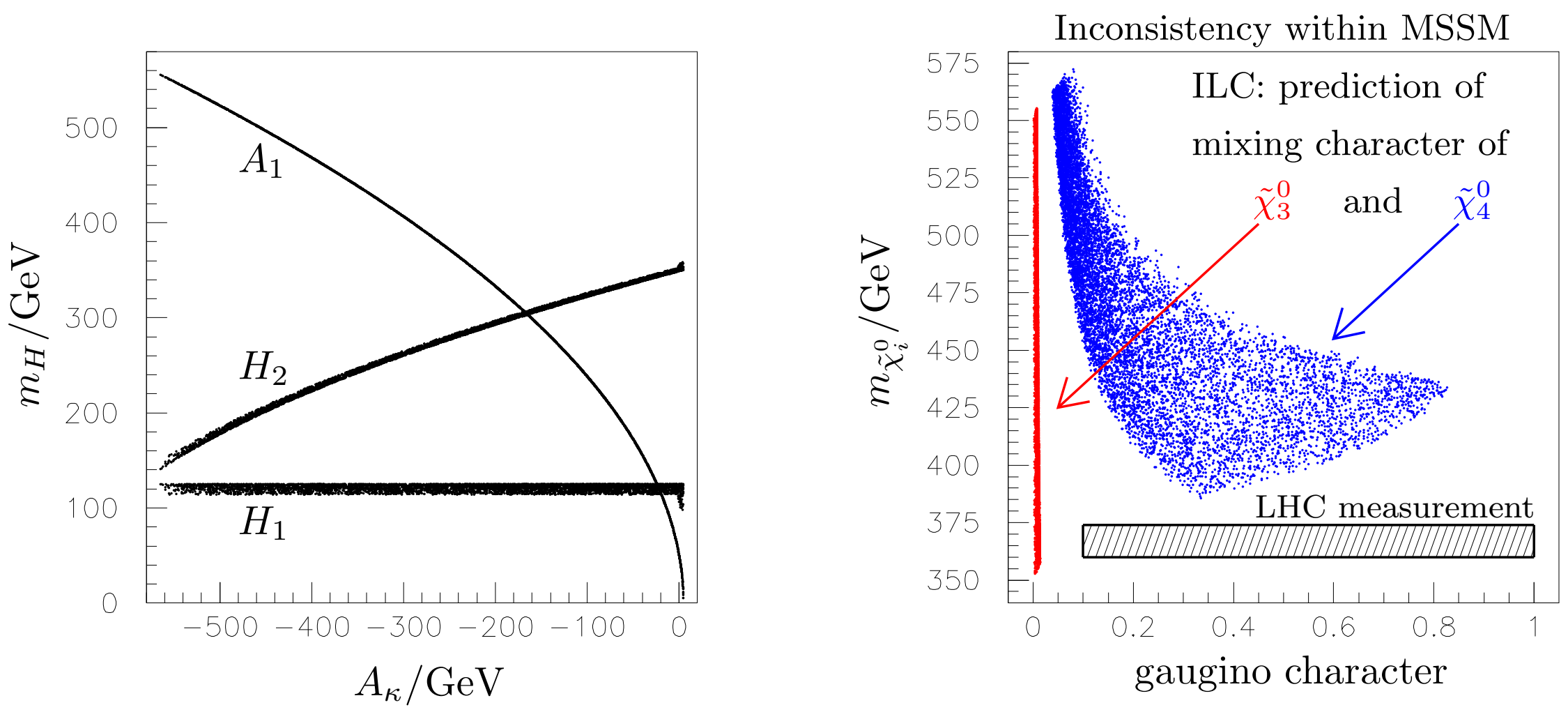

Fig. 4.17: Left: The possible masses of the two light scalar Higgs bosons, $m_{H_1}$, $m_{H_2}$, and of the lightest pseudoscalar Higgs boson $m_{A_1}$ as function of the trilinear Higgs parameter $A_\kappa$ in the NMSSM. In our chosen scenario, $H_1$ is MSSM-like and $H_2$ and $A_1$ are heavy singlet-dominated Higgs particles. Right: Predicted masses and gaugino admixture for the heavier neutralinos $\tilde{\chi}_3^0$ and $\tilde{\chi}_4^0$ within the consistent parameter ranges derived at the $\text{ILC}_{500}$ analysis in the MSSM and measured mass $m_{\tilde{\chi}_i^0} = 367 \pm 7$ GeV of a neutralino with suf ciently high gaugino admixture in cascade decays at the LHC. We require a gaugino admixture of $\gtrsim 10\%$ for the heavy neutralinos, cf. [137 139].

### *4.8.3 Gaugino/higgsino parameter determination at the ILC*

For the determination of the supersymmetric parameters in the MSSM straightforward strategies [124, 125, 132, 133] have been worked out even if only the light neutralinos and charginos $\tilde{\chi}_1^0$, $\tilde{\chi}_2^0$ and $\tilde{\chi}_1^\pm$ are kinematically accessible at the first stage of the ILC [126,127]. Using the methods described in [134,135] we derive constraints for the parameters $M_1$, $M_2$, $\mu$ and $\tan\beta$ in two steps. First, the measured masses and cross sections at two energies in the chargino sector constrain the chargino mixing matrix elements $U_{11}^2$ and $V_{11}^2$ [136]. Adding then mass and cross section measurements in the neutralino sector allows to constrain the parameters

$$M_1 = 377 \pm 42 \text{ GeV}, \quad (4.63)$$
$$M_2 = 150 \pm 20 \text{ GeV}, \quad (4.64)$$
$$\mu = 450 \pm 100 \text{ GeV}, \quad (4.65)$$
$$\tan\beta \leq 30. \quad (4.66)$$

Since the heavier neutralino and chargino states are not produced, the parameters $\mu$ and $\tan\beta$ can only be determined with a considerable uncertainty.

With help of the determined parameter ranges, Eqs. (4.63)–(4.66), the masses of heavier charginos and neutralinos can be calculated:

$$352 \text{ GeV} \leq m_{\tilde{\chi}_3^0} \leq 555 \text{ GeV}, \qquad 386 \text{ GeV} \leq m_{\tilde{\chi}_4^0} \leq 573 \text{ GeV}, \qquad 350 \text{ GeV} \leq m_{\tilde{\chi}_2^\pm} \leq 600 \text{ GeV}. \quad (4.67)$$

In Fig. 4.17 (right panel) the masses of $\tilde{\chi}_3^0$ and $\tilde{\chi}_4^0$ are shown as a function of its gaugino admixture for parameter points within the constraints of Eqs. (4.63)–(4.66). Obviously, the heavy neutralino $\tilde{\chi}_3^0$ should be almost a pure higgsino within the MSSM prediction. These predicted properties of the heavier particles can be compared with mass measurements of SUSY particles at the LHC within cascade decays [131].

We emphasize that although we started with an NMSSM scenario where $\tilde{\chi}_2^0$ and $\tilde{\chi}_3^0$ have large singlino admixtures, the MSSM parameter strategy does not fail and the experimental results from the $\mathrm{ILC}_{500}$ with $\sqrt{s} = 400$ GeV and 500 GeV lead to a consistent parameter determination in the MSSM. Hence in the considered scenario the analyses at the $\mathrm{ILC}_{500}$ or LHC alone do not allow a clear discrimination between MSSM and NMSSM. All predictions for the heavier gaugino/higgsino masses are consistent with both models. However, the $\mathrm{ILC}_{500}$ analysis predicts an almost pure higgsino-like state for $\tilde{\chi}^0_3$ and a mixed gaugino-higgsino-like $\tilde{\chi}^0_4$, see Fig. 4.17 (right panel). This allows the identification of the underlying supersymmetric model in combined analyses at the LHC and the $\mathrm{ILC}_{650}^{\mathcal{L}=1/3}$.

Table 4.7: Expected cross sections for the associated production of the heavier neutralinos and charginos in the NMSSM scenario for the $\mathrm{ILC}_{650}^{\mathcal{L}=1/3}$ option with one sigma statistical error based on $\int \mathcal{L} = 33\ \mathrm{fb}^{-1}$ for both unpolarized and polarized beams.

| | $\sigma(e^+e^- \to \tilde{\chi}^0_1\tilde{\chi}^0_j)$/fb at $\sqrt{s} = 650$ GeV | | | $\sigma(e^+e^- \to \tilde{\chi}^\pm_1\tilde{\chi}^\mp_2)$/fb at $\sqrt{s} = 650$ GeV |
|---|---|---|---|---|
| | $j = 3$ | $j = 4$ | $j = 5$ | |
| Unpolarized beams | $12.2 \pm 0.6$ | $5.5 \pm 0.4$ | $\leq 0.02$ | $2.4 \pm 0.3$ |
| $(P_{e^-}, P_{e^+}) = (-90\%, +60\%)$ | $36.9 \pm 1.1$ | $14.8 \pm 0.7$ | $\leq 0.07$ | $5.8 \pm 0.4$ |
| $(P_{e^-}, P_{e^+}) = (+90\%, -60\%)$ | $0.6 \pm 0.1$ | $2.2 \pm 0.3$ | $\leq 0.01$ | $1.6 \pm 0.2$ |

### *4.8.4 Combined LHC and ILC analysis*

In our original NMSSM scenario, Eq. (4.57), the neutralinos $\tilde{\chi}^0_2$ and $\tilde{\chi}^0_3$ have a large bino-admixture and therefore appear in the squark decay cascades. The dominant decay mode of $\tilde{\chi}^0_2$ has a branching ratio $BR(\tilde{\chi}^0_2 \to \tilde{\chi}^\pm_1 W^\mp) \sim 50\%$, while for the $\tilde{\chi}^0_3$ decays $BR(\tilde{\chi}^0_3 \to \tilde{\ell}^\pm_{L,R}\ell^\mp) \sim 45\%$ is largest. Since the heavier neutralinos, $\tilde{\chi}^0_4$, $\tilde{\chi}^0_5$, are mainly higgsino-like, no visible edges from these particles occur in the cascades. It is expected to see the edges for $\tilde{\chi}^0_2 \to \tilde{\ell}^\pm_R\ell^\mp$, $\tilde{\chi}^0_2 \to \tilde{\ell}^\pm_L\ell^\mp$, $\tilde{\chi}^0_3 \to \tilde{\ell}^\pm_R\ell^\mp$ and for $\tilde{\chi}^0_3 \to \tilde{\ell}^\pm_L\ell^\mp$ [140].

With a precise mass measurement of $\tilde{\chi}^0_1$,$\tilde{\chi}^0_2$, $\tilde{\ell}_{L,R}$ and $\tilde{\nu}$ from the $\mathrm{ILC}_{500}$ analysis, a clear identification and separation of the edges of the two gauginos at the LHC is possible without imposing specific model assumptions. We therefore assume a precision of about 2% for the measurement of $m_{\tilde{\chi}^0_3}$, in analogy to [137–139]:

$$m_{\tilde{\chi}^0_3} = 367 \pm 7\ \mathrm{GeV}. \tag{4.68}$$

The precise mass measurement of $\tilde{\chi}^0_3$ is compatible with the mass predictions of the $\mathrm{ILC}_{500}$ for the $\tilde{\chi}^0_3$ in the MSSM but not with the prediction of the small gaugino admixture, see Fig. 4.17 (right panel). The $\tilde{\chi}^0_3$ as predicted in the MSSM would not be visible in the decay cascades at the LHC. The other possible interpretation of the measured neutralino as the $\tilde{\chi}^0_4$ in the MSSM is incompatible with the cross section measurements at the ILC. We point out that a measurement of the neutralino masses $m_{\tilde{\chi}^0_1}$, $m_{\tilde{\chi}^0_2}$, $m_{\tilde{\chi}^0_3}$ which could take place at the LHC alone is not sufficient to distinguish the SUSY models since rather similar mass spectra could exist [134, 135]. Therefore the cross sections in different beam polarization configurations at the ILC have to be included in the analysis.

The obvious inconsistency of the combined results from the LHC and the $\mathrm{ILC}_{500}$ analyses and the predictions for the missing chargino/neutralino masses could motivate the immediate use of the low-luminosity but higher-energy option $\mathrm{ILC}_{650}^{\mathcal{L}=1/3}$ in order to resolve model ambiguities even at an early stage of the experiment and outline future search strategies at the upgraded ILC at 1 TeV. This would finally lead to the correct identification of the underlying model. The expected polarized and unpolarized cross sections, including the statistical error on the basis of one third of the luminosity of the $\mathrm{ILC}_{500}$, are given in Table 4.7. The neutralino $\tilde{\chi}^0_3$ as well as the higgsino-like heavy neutralino $\tilde{\chi}^0_4$ and the chargino $\tilde{\chi}^\pm_2$ are now accessible at the $\mathrm{ILC}_{650}^{\mathcal{L}=1/3}$.

The cross sections together with the precisely measured masses $m_{\tilde{\chi}^0_4}$ and $m_{\tilde{\chi}^\pm_2}$ constitute the observables for a fit of the NMSSM parameters. This will be achieved by extending the fit program `Fittino` [141] to include also the NMSSM [142], where the SUSY particle spectrum is calculated with `SPheno` [143] and the Higgs spectrum with `NMHDECAY` [63].

### *4.8.5 Concluding remarks*

We have presented an NMSSM scenario where the measurement of masses and cross sections in the neutralino and chargino sector as well as measurements in the Higgs sector do not allow a distinction from the MSSM at the LHC or at the ILC$_{500}$ with $\sqrt{s} = 500$ GeV alone. Precision measurements of the neutralino branching ratio into the lightest Higgs particle and of the mass difference between the lightest and next-to-lightest SUSY particle [122] may give first evidence for the SUSY model but are difficult to realize in the presented scenario. Therefore the identification of the underlying model requires precision measurements of the heavier neutralinos by combined analyses of LHC and ILC and the higher energy but lower luminosity option of the ILC at $\sqrt{s} = 650$ GeV. This gives access to the necessary observables for a fit of the underlying NMSSM parameters.

## 4.9 Moderately light charged Higgs bosons in the NMSSM and CPV-MSSM

*Rohini M. Godbole and Durga P. Roy*

We discuss some aspects of the phenomenology of a light charged Higgs ($M_{H^+} \lesssim 150$ GeV), allowed at low and moderate values of $\tan\beta$, in the NMSSM and CP-violating MSSM (CPV-MSSM), respecting all the LEP-II bounds. In the NMSSM with the $H^\pm$ near its lower mass limit ($M_{H^+} \simeq 120$ GeV), and a light pseudoscalar ($M_{A_1^0} \simeq 50$ GeV) with a very significant doublet component, the charged Higgs boson is expected to decay dominantly via the standard $H^+ \to \tau^+\nu$ mode. One can probe this mass range via the $t \to bH^+ \to b\tau^+\nu$ channel at Tevatron and especially at LHC. For somewhat heavier charged Higgs boson ($M_{H^\pm} > 130$ GeV) the dominant decay via the $H^+ \to W^+ A_1^0$ channel provides a probe for not only a light $H^+$ but also a light $A_1^0$ [144] in the moderate $\tan\beta$ region, where its dominant decay mode is into a $b\bar{b}$ final state. A similar situation also attains in the CP-violating MSSM as well. The CPV-MSSM allows the existence of a light neutral Higgs boson ($M_{H_1} \lesssim 50$ GeV) in the CPX scenario in the low $\tan\beta(\lesssim 5)$ region, which could have escaped the LEP searches due to a strongly suppressed $H_1ZZ$ coupling. The light charged $H^+$ decays dominantly into the $WH_1$ channel again giving rise to a striking $t\bar{t}$ signal at the LHC, where one of the top quarks decays into the $bb\bar{b}W$ channel, via $t \to bH^\pm, H^\pm \to WH_1$ and $H_1 \to b\bar{b}$. The characteristic correlation between the $b\bar{b}$, $b\bar{b}W$ and $bb\bar{b}W$ invariant mass peaks helps reduce the SM background, drastically.

### *4.9.1 Moderately light $H^\pm$ in the NMSSM*

As discussed in Section 4.1.2, the solution of the the so called $\mu$-problem of the MSSM was the original motivation for the NMSSM. The effect of the additional complex singlet scalar S in the NMSSM on the charged Higgs phenomenology mainly comes through a relaxation of the mass limits of the $A^0$ and the $H^+$ in the MSSM. This arises from the modification of the MSSM mass relations between the doublet scalars $H_{1,2}$ and pseudoscalar $A$ and the resulting modification of the $H_1$ mass bound. The masses of the $A_i^0, i = 1,2$ and $H_i, i = 1,3$ in terms of the various parameters of the NMSSM: the dimensionless parameters $\lambda, \kappa$ appearing in the superpotential of Eq. (4.7) as well as the corresponding soft trilinear terms $A_\lambda, A_\kappa$ and the vacuum expectation value of the singlet scalar field $\langle S \rangle = x = v_S/\sqrt{2}$, are given by Eqs. (4.16)–(4.19). In particular, the resulting upper bound of the lightest Higgs scalar mass including the radiative correction $\epsilon$, is [25,28,29,145–147]

$$M_{H_1}^2 \leq M_Z^2 \cos^2(2\beta) + \frac{2\lambda^2 M_W^2}{g^2} \sin^2(2\beta) + \epsilon, \tag{4.69}$$

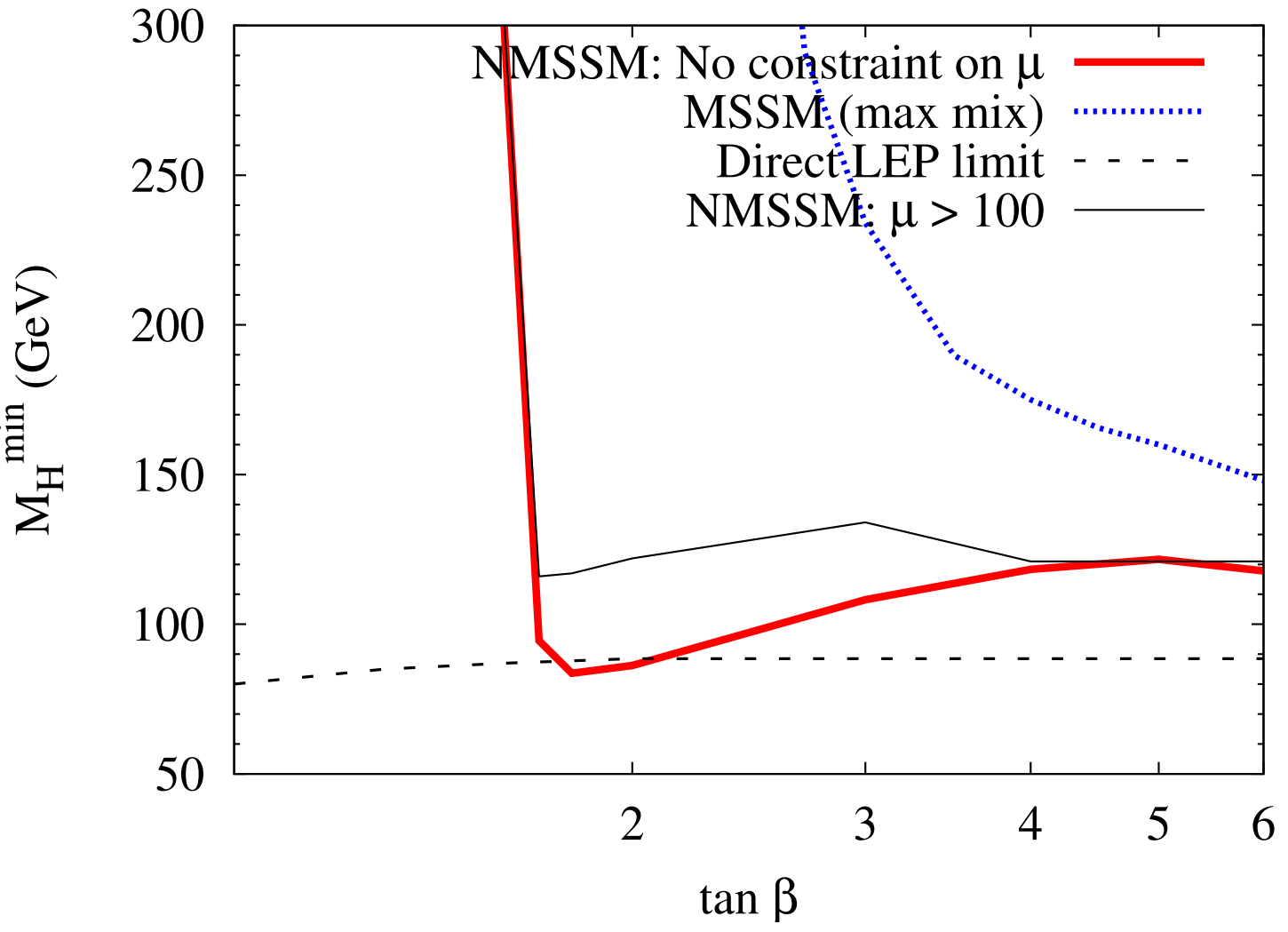

Fig. 4.18: The indirect lower bounds on the charged Higgs boson mass following from the LEP limits on the neutral Higgs bosons in the MSSM (Maximal Stop Mixing) and the NMSSM. The direct LEP limit on the charged Higgs boson mass is also shown for comparison .

where contribution specific to the NMSSM in addition to the terms in MSSM is given by the middle term. This is most pronounced in the low to moderate $\tan\beta$ region, where the MSSM mass bound coming from the first term of Eq. (4.69) is very small. Therefore it relaxes the MSSM bound on $M_{H_1}$ and the resulting lower limit of $M_{A_i}$, most significantly over this range of $\tan\beta$. This in turn relaxes the lower limit of the charged Higgs mass, which is related to the doublet pseudoscalar mass via

$$M_{H^+}^2 = M_A^2 + M_W^2 \left(1 - \frac{2\lambda^2}{g^2}\right) \tag{4.70}$$

along with a small radiative correction. This is helped further due to the additional (negative) contribution in Eq. (4.70). Note that the additional contributions of Eqs. (4.69) and (4.70) depend only on the $\hat{S}\hat{H}_u\hat{H}_d$ coupling $\lambda$, in the superpotential of Eq. (4.7). Therefore the Eqs. (4.69) and (4.70) hold also for the minimal nonminimal supersymmetric standard model (MNSSM), which assumes only this term in the superpotential [44, 45, 47]. Finally the upper bound of Eq. (4.69) will only be useful if one can find an upper limit on $\lambda$. Such a limit can be derived [25, 29, 145] from the requirement that all the couplings of the model remain perturbative upto some high energy scale, usually taken to be the GUT scale. Such an upper limit on $\lambda$ has been estimated in [148] as a function of $\tan\beta$ using two-loop renormalization group equations.

For quantitative evaluation of the NMSSM Higgs spectrum we consider the complete Higgs potential as given in terms of these parameters in [55]. The lower limit of the $H^\pm$ mass has been estimated as a function of $\tan\beta$ in [148] by varying all these five NMSSM parameters over the allowed ranges, which include the constraints from LEP-2. The resulting $H^\pm$ mass limit is shown in Fig. 4.18 along with the most conservative MSSM limit, corresponding to maximal stop mixing, which gives the largest radiative correction $\epsilon$. The NMSSM limit has practically no sensitivity to stop mixing. The LEP-2 mass limit from direct search of $H^+ \to \tau^+\nu$ events is also shown for comparison [109]. There is no limit from Tevatron in the moderate $\tan\beta$ region shown in Fig. 4.18.

One sees from Fig. 4.18 that even the most conservative MSSM limit implies $H^\pm$ mass $\geq$ 150 GeV (175 GeV) for $\tan\beta \leq 6$ (4). In contrast in the NMSSM one can have a $H^\pm$ mass $\lesssim$ 120 GeV over this moderate $\tan\beta$ region, going down to the direct LEP-2 limit of 86 GeV at $\tan\beta \simeq 2$. Note however that requiring that the effective $\mu$ parameter $\mu_{eff} = \langle S\rangle\lambda$ be greater than 100 GeV, as favored by the LEP chargino search, increases this mass limit to $\gtrsim$ 120 GeV [47]. The steep vertical rise at left reflects the well-known fixed-point solution at $\tan\beta = 1.55$, where the top Yukawa coupling blows up at the GUT scale. Thus allowing for possible intermediate scale physics one can evade the steep NMSSM mass limit at low $\tan\beta$ [149]. In contrast the MSSM limit holds independent of any intermediate scale physics ansatz.

Table 4.8: Examples of dominant $H^\pm \to WA_1^0$ decay in the NMSSM. These decay branching fractions are shown along with the Higgs boson masses and the other model parameters.

| $\tan\beta$ | $M_{H^+}$ (GeV) | $M_{A_1}$ (GeV) | $B_{A_1}$ (%) | $\lambda, \kappa$ | $x = v_s/\sqrt{2}, A_\lambda, A_\kappa$ (GeV) |
|---|---|---|---|---|---|
| 2 | 147 | 38 | 94 | 0.45, −0.69 | 224, −8, 2 |
| 3 | 159 | 65 | 83 | 0.33, −0.70 | 305, 40, 38 |
| 4 | 145 | 48 | 89 | 0.28, −0.70 | 563, 170, 85 |
| 5 | 150 | 10 | 91 | 0.26, −0.54 | 503, 109, 38 |

We have investigated the neutral scalar and pseudoscalar Higgs spectrum of the NMSSM, when the $H^\pm$ lies near its lower mass limit ($M_{H^+} \simeq 120$ GeV). The lightest scalar is dominantly singlet ($M_{H_1} \simeq 100$ GeV), while the doublet scalars are relatively heavy ($M_{H_{2,3}} > 120$ GeV). On the other hand there is often a light pseudoscalar ($M_{A_1^0} \simeq 50$ GeV) with a very significant doublet component. Consequently a light charged Higgs boson of mass $\simeq$ 120 GeV is expected to decay dominantly via the standard $H^+ \to \tau^+\nu$ mode. Thus one can probe this mass range via the $t \to bH^+ \to b\tau^+\nu$ channel at Tevatron and especially at LHC. On the other hand a somewhat heavier charged Higgs boson ($M_{H^\pm} > 130$ GeV) can dominantly decay via the $H^+ \to W^+ A_1^0$ channel [144]. In fact this seems to be a very favorable channel to probe for not only $H^+$ but also a light $A^0$ in the moderate $\tan\beta$ region, where the $A^0$ is expected to decay mainly in to the $b\bar{b}$ or $\tau^+\tau^-$ mode. Table 4.8 shows some illustrative samples of NMSSM Higgs spectra where $H^+$ decays dominantly into the $W^+A_1^0$ mode. These results are obtained by scanning the NMSSM parameter space. Note that in each case the effective $\mu$ parameter $\mu_{eff} = \lambda\langle S\rangle$ is greater than 100 GeV as favored by the LEP chargino limit. The decay branching fractions are shown along with the Higgs boson masses and the other model parameters.

### *4.9.2 Light $H^\pm$ in the CP-violating MSSM*

Interestingly one can have a similar signal in the CP violating MSSM due to large scalar-pseudoscalar mixing. The CP-violating MSSM allows existence of a light neutral Higgs boson ($M_{H_1} \lesssim 50$ GeV) in the CPX scenario in the low $\tan\beta(\lesssim 5)$ region, which could have escaped the LEP searches due to a strongly suppressed $H_1ZZ$ coupling. The light charged $H^+$ decays dominantly into the $WH_1$ channel giving rise to a striking $t\bar{t}$ signal at the LHC, where one of the top quarks decays into the $bb\bar{b}W$ channel, via $t \to bH^\pm, H^\pm \to WH_1$ and $H_1 \to b\bar{b}$. The characteristic correlation between the $b\bar{b}$, $b\bar{b}W$ and $bb\bar{b}W$ invariant mass peaks helps reduce the SM background, drastically [62]. Note that this signal is identical to the NMSSM case discussed above.

As already mentioned, a combined analysis of all the LEP results, shows that a light neutral Higgs is still allowed in the CPX [150] scenario in the CPV-MSSM. The experiments provide exclusion regions in the $M_{H_1} - \tan\beta$ plane for different values of the CP-violating phase, with the various parameters taking value as given in the CPX scenario in Section 3.1, Eq. (3.13). Combining the results of Higgs searches from ALEPH, DELPHI, L3 and OPAL, the authors in Ref. [50, 151] have provided exclusion regions in the $M_{H_1}$–$\tan\beta$ plane as well as in the $M_{H^+}$–$\tan\beta$ plane. A more recent analysis of the LEP exclusion limits is given in Section 3.2 of this report. While the exact exclusion regions differ somewhat in the three analysis they all show that for phases $\Phi_{\rm CP} = 90°$ and $60°$ LEP cannot exclude the presence of a light Higgs boson at low $\tan\beta$, mainly because of the suppressed $H_1ZZ$ coupling. The analysis of Ref. [50] further shows that in the same region the $H_1t\bar{t}$ coupling is suppressed as well. Thus this particular region in the parameter space can not be probed either at the Tevatron where the associated production $W/ZH_1$ mode is the most promising one; neither can this be probed at the LHC as the reduced $t\bar{t}H_1$ coupling suppresses the inclusive production mode and the associated production modes $W/ZH_1$ and $t\bar{t}H_1$, are suppressed as well. This region of Ref. [50] corresponds to $\tan\beta \sim 3.5 - 5, M_{H^+} \sim 125 - 140$ GeV, $M_{H_1} \lesssim 50$ GeV and $\tan\beta \sim 2 - 3, M_{H^+} \sim 105 - 130$ GeV$, M_{H_1} \lesssim 40$ GeV, for $\Phi_{CP} = 90°$ and $60°$ respectively. In the same region of the parameter space where $H_1ZZ$ coupling is suppressed, the $H^+W^-H_1$ coupling is enhanced because these two sets of couplings satisfy a sum-rule. Further, in the MSSM a light pseudo-scalar implies a light charged Higgs, lighter than the top quark.

Table 4.9: Range of values for BR ($H^+ \to H_1W^+$) and BR ($t \to bH^+$) for different values of $\tan\beta$ corresponding to the LEP allowed window in the CPX scenario, for the common phase $\Phi_{\rm CP} = 90°$, along with the corresponding range for the $H_1$ and $H^+$ masses. The quantities in the bracket in each column give the values at the edge of the kinematic region where the decay $H^+ \to H_1W^+$ is allowed.

| $\tan\beta$ | 3.6 | 4 | 4.6 | 5 |
|---|---|---|---|---|
| $\mathrm{Br}(H^+ \to H_1W^+)[\%]$ | > 90 (87.45) | > 90 (57.65) | > 90 (50.95) | > 90 (46.57) |
| $\mathrm{Br}(t \to bH^+)[\%]$ | ∼ 0.7 | 0.7 – 1.1 | 0.9 – 1.3 | 1.0 – 1.3 |
| $M_{H^+}$ [GeV] | < 148.5 (149.9) | < 139 (145.8) | < 130.1 (137.5) | < 126.2 (134) |
| $M_{H_1}$ [GeV] | < 60.62 (63.56) | < 49.51 (65.4) | < 36.62 (57.01) | < 29.78 (53.49) |

Table 4.9 shows the behaviour of the $M_{H^+}$, $M_{H_1}$ and the BR ($H^+ \to H_1W^+$), for values of $\tan\beta$ corresponding to the above mentioned window in the $\tan\beta$–$M_{H_1}$ plane, of Ref. [50]. It is to be noted here that indeed the $H^\pm$ is light (lighter than the top) over the entire range, making its production in $t$ decay possible. Further, the $H^\pm$ decays dominantly into $H_1W$, with a branching ratio larger than $47\%$ over the entire range where the decay is kinematically allowed, which covers practically the entire parameter range of interest; viz. $M_{H_1} < 50$ GeV for $\Phi_{\rm CP} = 90°$. It can be also seen from the table that the BR($H^\pm \to H_1W$) is larger than $90\%$ over most of the parameter space of interest. So not only that $H^+$ can be produced abundantly in the $t$ decay giving rise to a possible production channel of $H_1$ through the decay $H^\pm \to H_1W^\pm$, but this decay mode will be the only decay channel to see this light ($M_{H^\pm} < M_t$) $H^\pm$. The traditional decay mode of $H^\pm \to \tau\nu$ is suppressed by over an order of magnitude and thus will no longer be viable. Thus the process

$$
\begin{array}{llllll}
pp \to t & & + & \bar{t} & + & X \\
\;\;\hookrightarrow b\, H^+ & & & \;\;\hookrightarrow \bar{b}\, W & & \\
\qquad\;\hookrightarrow W & H_1 & & \qquad\;\hookrightarrow q\bar{q}(\ell\nu) & & \\
\qquad\qquad\hookrightarrow \ell\nu(q\bar{q}) & \hookrightarrow b\bar{b} & & & &
\end{array}
$$

allows a probe of both the light $H_1$ **and** a light $H^\pm$ in this parameter window in the CP-violating MSSM in the CPX scenario.

As can be seen from the Fig. 4.19 the largest signal cross-section case is $\sim 38$ fb and the signal cross-section is $\gtrsim 20$ fb for $M_{H_1} \gtrsim 15$ GeV. It is clear from the right panel of the Fig. 4.19, that there is simultaneous clustering in the $m_{b\bar{b}}$ distribution around $\simeq M_{H_1}$ and in the $m_{b\bar{b}W}$ distribution around $M_{H^\pm}$. This clustering feature can be used to distinguish the signal over the standard model background.

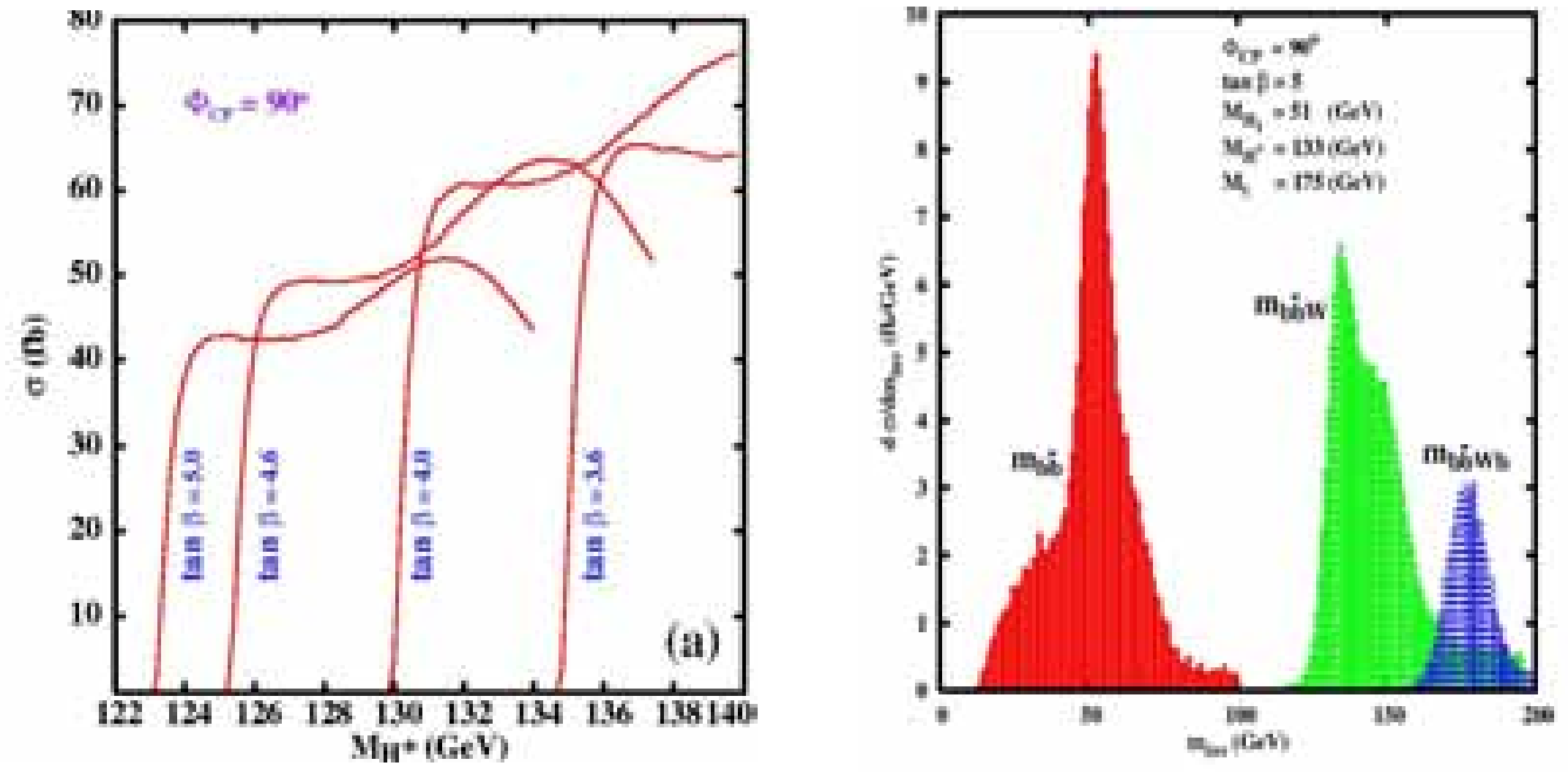

Fig. 4.19: Variation of the cross-section with $M_{H^+}$ for four values of $\tan\beta = 3.6, 4, 4.6$ and 5 is shown in the left panel, for the CP-violating phase $\Phi_{\rm CP} = 90°$. These numbers should be multiplied by $\sim 0.5$ to get the signal cross-section to take into account the b tagging ef ciency. The right panel shows the $m_{b\bar{b}}, m_{b\bar{b}W}$ and the $m_{b\bar{b}Wb} = M_t$ invariant mass distributions, for this choice of CP-violating phase and $\tan\beta = 5, M_{H^+} = 133$ GeV, corresponding to a light neutral Higgs $H_1$ with mass $M_{H_1} = 51$ GeV $M_t, M_W$ mass window cuts have been applied [62].

As a matter of fact the estimated background to the signal coming from the QCD production of $t\bar{t}b\bar{b}$ once all the cuts (including the mass window cuts) are applied, to the signal type events is less than 0.5 fb, in spite of a starting cross-section of 8.5 pb. The major reduction is brought about by requiring that the invariant mass of the $bbbW$ be within 25 GeV of $M_t$.

### *4.9.3 Summary*

Thus in conclusion, both in the NMSSM and in the CPV-MSSM the moderately light charged Higgs that is allowed at moderately low values of $\tan\beta$, provides interesting and novel phenomenology at the LHC.

## 5 THE MSSM WITH R-PARITY VIOLATION

### 5.1 Introduction

*Marc Besançon and Werner Porod*

#### 5.1.1 Explicit R-parity violation

The Standard Model conserves baryon number $B$ and lepton number $L$ separately at the perturbative level. On the contrast, its minimal supersymmetric extension does allow for the breaking of $B$ and $L$ if one requires 'only' gauge invariance and supersymmetry. The case of lepton number violation is easily seen by noting that the Higgs superfield $\hat{H}_d$ and the lepton superfields $\hat{L}_i$ have the same gauge quantum numbers and differ only by lepton number. The most general superpotential containing only the SM fields and being compatible with its gauge symmetry $G_{SM} = SU(3)_c \times SU(2)_L \times U(1)_Y$ is given as [1,2]:

$$W = W_{MSSM} + W_{\not{R}_p}\,, \tag{5.1}$$

$$W_{MSSM} = h^{ij}_E \hat{L}_i \hat{H}_d \hat{E}^c_j + h^{ij}_D \hat{Q}_i \hat{H}_d \hat{D}^c_j + h^{ij}_U \hat{Q}_i \hat{H}_u \hat{U}^c_j - \mu \hat{H}_d \hat{H}_u\,, \tag{5.2}$$

$$W_{\not{R}_p} = \frac{1}{2}\lambda_{ijk}\hat{L}_i\hat{L}_j\hat{E}^c_k + \lambda'_{ijk}\hat{L}_i\hat{Q}_j\hat{D}^c_k + \frac{1}{2}\lambda''_{ijk}\hat{U}^c_i\hat{D}^c_j\hat{D}^c_k + \epsilon_i\hat{L}_i\hat{H}_u\,, \tag{5.3}$$

$i, j, k = 1, 2, 3$ are generation indices. $\hat{L}_i$ ($\hat{Q}_i$) are the lepton (quark) $SU(2)_L$ doublet superfields. $\hat{E}^c_j$ ($\hat{D}^c_j, \hat{U}^c_j$) are the electron (down- and up-quark) $SU(2)_L$ singlet superfields. $\lambda_{ijk}$, $\lambda'_{ijk}$, and $\lambda''_{ijk}$ are dimensionless Yukawa couplings whereas the $\epsilon_i$ are dimensionful mass parameters. Gauge invariance implies that the first term in $W_{\not{R}_p}$ is anti-symmetric in $\{i, j\}$ and the third one is anti-symmetric in $\{j, k\}$. Equation (5.3) thus contains $9 + 27 + 9 + 3 = 48$ new terms beyond those of the MSSM. Once lepton number is broken, it is obvious from Eqs. (5.2) and (5.3) the MSSM seems to consist of three quark superfields, five $SU(2)$ doublet Higgs superfields and three charged $SU(2)$ singlet Higgs superfields as there are no means to distinguish between lepton and Higgs superfields. From this point of view, the known charged and neutral leptons are higgsinos.

The simultaneous appearance of lepton and baryon number breaking terms leads in general to a phenomenological catastrophe if all involved particles have masses of the order of the electroweak scale: rapid proton decay [1,2]. To avoid this problem a discrete multiplicative symmetry, called R-parity ($R_p$), had been invented [3] which can be written as

$$R_p = (-\mathbf{1})^{3B+L+2S}\,, \tag{5.4}$$

where $S$ is the spin of the corresponding particle. For all superfields of MSSM, the SM field has $R_p = +1$ and its superpartner has $R_p = -1$, e.g. the electron has $R_p = +1$ and the selectron has $R_p = -1$. In this way all terms in Eq. (5.3) are forbidden and one is left with the superpotential given in Eq. (5.2).

Recent neutrino experiments have shown that neutrinos are massive particles which mix among themselves (for a review see e.g. [4]). In contrast to leptons and quarks, neutrinos need not to be Dirac particles but can be Majorana particles. In the latter case the Lagrangian contains a mass term which violates explicitly lepton number by two units. This motivates one to allow the lepton number breaking terms in the superpotential in particular as they automatically imply the existence of massive neutrinos without the need of introducing right-handed neutrinos [5]. The $\lambda''$ terms can still be forbidden by a discrete symmetry which transforms ($\hat{U}^c, \hat{D}^c, \hat{Q}$) into ($-\hat{U}^c$, $-\hat{D}^c$, $-\hat{Q}$) while leaving the other fields unchanged. Breaking lepton number has two interesting consequences for the phenomenology of Higgs bosons in supersymmetric theories: (i) the Higgs bosons can mix with the sleptons and (ii) Higgs cascade decays into SUSY particles get altered.

Let us briefly comment on the number of free parameters before discussing the phenomenology in more detail. The last term in Eq. (5.3), $\hat{L}_i\hat{H}_u$, mixes the lepton and the Higgs superfields. In supersymmetry $\hat{L}_i$ and $\hat{H}_d$ have the same gauge and Lorentz quantum numbers and we can redefine them by a rotation in $(\hat{H}_d, \hat{L}_i)$. The terms $\epsilon_i \hat{L}_i \hat{H}_u$ can then be rotated to zero in the superpotential [5]. However, there are still the corresponding terms in the soft supersymmetry breaking Lagrangian

$$V_{\not R_p,soft} = B_i \epsilon_i \tilde{L}_i H_u \tag{5.5}$$

which can only be rotated away if $B_i = B$ and $M^2_{H_d} = M^2_{L,i}$ [5]. Such an alignment of the superpotential terms with the soft breaking terms is not stable under the renormalization group equations [6]. Assuming an alignment at the unification scale, the resulting effects are small [6] except for neutrino masses [6–10]. Models containing only bilinear terms do not introduce trilinear terms as can easily be seen from the fact that bilinear terms have dimension mass whereas the trilinear are dimensionless. For this reason we will keep in the following explicitly the bilinear terms in the superpotential.

The presence of the bilinear terms in the soft SUSY breaking potential, Eq. (5.5), implies that not only the usual Higgs bosons get vacuum expectations values (vevs) but also the sneutrinos (for details see [10–12]). As a consequence the neutral Higgs bosons mix with the sneutrinos resulting in five neutral scalar bosons and four neutral pseudoscalar bosons. In addition the charged Higgs boson mixes with the charged sleptons resulting in seven charged states, $S^\pm_i$ $(i = 1, ..7)$. In the following we will simplify the notation by denoting the particles with their MSSM notation indicating their main particle content. The complete set of the corresponding mass matrices is given in ref. [10]. The mixing between sleptons and Higgs bosons leads to additional decay modes for the Higgs bosons [13]:

$$\phi \;\rightarrow\; \nu\tilde{\chi}^0_i\,, \qquad l^\pm\tilde{\chi}^\mp_k\,, \qquad \bar{\nu}\nu \tag{5.6}$$

$$H^+ \;\rightarrow\; l^+\tilde{\chi}^0_i\,, \qquad \nu\tilde{\chi}^+_k\,, \tag{5.7}$$

where $\phi$ denotes $h^0$, $H^0$ and $A^0$. Moreover, there is the possibility of associate production of Higgs bosons together with sleptons or slepton-strahlung of t-quarks (in analogy to Higgs-strahlung) as discussed in Section 5.4. Also the sleptons have additional decay modes compared to the MSSM:

$$\tilde{\nu} \;\rightarrow\; q\bar{q}\,, \qquad l^+l^-\,, \qquad \nu\bar{\nu} \tag{5.8}$$

$$\tilde{l} \;\rightarrow\; l^+\nu\,, \qquad q\bar{q}'\,, \tag{5.9}$$

e.g. the sleptons have the same signatures apart from the $\nu\bar{\nu}$ channel as the usual Higgs bosons if the R-parity violating decay modes dominate. We want to stress here again, that although we use the MSSM symbols, Higgs bosons and sleptons mix and that the sleptons have to be considered as additional Higgs bosons once lepton number is broken.

How large can the branching ratio for those decay modes be? To answer this question one has to take into account existing constraints on R-parity violating parameters from low energy physics. As most of them are given in terms of trilinear couplings, we will work in the "$\epsilon$-less" basis, e.g. rotate away the bilinear terms in the superpotential Eq. (5.3). Therefore, the trilinear couplings get additional contributions. Assuming, without loss of generality, that the lepton and down type Yukawa couplings are diagonal they are given to leading order in $\epsilon_i/\mu$ as [14–16]:

$$\lambda'_{ijk} \rightarrow \lambda'_{ijk} + \delta_{jk} h_{d_k} \frac{\epsilon_i}{\mu} \tag{5.10}$$

and

$$\begin{aligned} &\lambda_{ijk} \rightarrow \lambda_{ijk} + \delta\lambda_{ijk}, \\ &\delta\lambda_{121} = h_e \frac{\epsilon_2}{\mu}, \quad \delta\lambda_{122} = h_\mu \frac{\epsilon_1}{\mu}, \quad \delta\lambda_{123} = 0 \\ &\delta\lambda_{131} = h_e \frac{\epsilon_3}{\mu}, \quad \delta\lambda_{132} = 0, \quad \delta\lambda_{133} = h_\tau \frac{\epsilon_1}{\mu} \\ &\delta\lambda_{231} = 0, \quad \delta\lambda_{232} = h_\mu \frac{\epsilon_3}{\mu}, \quad \delta\lambda_{233} = h_\tau \frac{\epsilon_2}{\mu} \end{aligned} \tag{5.11}$$

Table 5.1: R-parity violating decays of sfermions via trilinear $\not{R}_p$ operators $\lambda L_i L_j E^c_k$, $\lambda' L_i Q_j D^c_k$ and $\lambda'' U^c_i D^c_j D^c_k$.

| Supersymmetric particles | Couplings $\lambda$ | $\lambda'$ | $\lambda''$ |
|---|---|---|---|
| $\tilde{\nu}_{i,L}$ | $\ell^+_{j,L}\ell^-_{k,R}$ | $\bar{d}_{j,L}d_{k,R}$ | |
| $\tilde{l}^-_{i,L}$ | $\bar{\nu}_{j,L}\ell^-_{k,R}$ | $\bar{u}_{j,L}d_{k,R}$ | |
| $\tilde{l}^-_{k,R}$ | $\nu_{i,L}\ell^-_{j,L}$ , $\ell^-_{i,L}\nu_{j,L}$ | | |
| $\tilde{u}_{i,R}$ | | | $\bar{d}_{j,R}\bar{d}_{k,R}$ |
| $\tilde{u}_{j,L}$ | | $\ell^+_{i,L}d_{k,R}$ | |
| $\tilde{d}_{j,L}$ | | $\bar{\nu}_{i,L}d_{k,R}$ | |
| $\tilde{d}_{k,R}$ | | $\nu_{i,L}d_{j,L}$, $\ell^-_{i,L}u_{j,L}$ | $\bar{u}_{i,R}\bar{d}_{j,R}$ |

where we have used the fact that neutrino physics requires $|\epsilon_i/\mu| \ll 1$ [10]. An essential point to notice is that the additional contributions in Eqs. (5.10) and (5.11) follow the hierarchy dictated by the down quark and charged lepton masses of the standard model.

A comprehensive list of bounds on various R-parity violating parameters can be found in [17]. However, there the recent data from neutrino experiments like Super-Kamiokande [18], SNO [19] and KamLAND [20] are not taken into account. These experiments yield strong bounds on trilinear couplings involving the third generation [21,22]. In addition also the sneutrino vevs are constrained by neutrino data [10,21]. Most of the trilinear couplings have a bound of the order $(10^{-2} - 10^{-1}) * m_{\tilde{f}}/(100GeV)$ where $m_{\tilde{f}}$ is the mass of the sfermion in the process under considerations. The cases with stronger limits are: $|\lambda'_{111}| \lesssim O(10^{-4})$ due to neutrinoless double beta decay and $|\lambda_{i33}| \simeq 5|\lambda'_{i33}| \simeq O(10^{-4})$ due to neutrino oscillation data. Moreover, neutrino oscillation data imply $|\mu^2(v_1^2+v_2^2+v_3^2)/\det(\mathcal{M}_{\chi^0})| \lesssim 10^{-12}$ where $v_i$ are the sneutrino vevs and $\det(\mathcal{M}_{\chi^0})$ is the determinant of the MSSM neutralino mass matrix.

In particular the last constraint implies that in general there is only a small mixing between sleptons and Higgs bosons and usually the R-parity violating decay modes of both, Higgs bosons and sleptons, have only tiny branching ratios of the order $10^{-6}$ and below. One exception is if by chance a Higgs boson is nearly mass degenerate with one of the sleptons which requires quite some fine-tuning. The other exception is if all R-parity conserving decay modes are kinematically forbidden. This can occur if either the sneutrinos or the right sleptons are the lightest supersymmetric particles (LSPs), which will be discussed in detail in Section 5.2. The case of left-slepton LSPs is practically excluded as the sneutrinos are always lighter provided $\tan\beta \geq 1$. In the case of sneutrino LSPs one finds the usual MSSM but misses the ordinary sneutrinos and finds instead additional neutral states behaving nearly like neutral doublet Higgs bosons [16,23]. The two main differences are: (i) The existence of lepton flavour violating decays modes such as $\tilde{\nu}_\tau \to e^\pm\tau^\mp$ which are sizable. (ii) The invisible decay mode into $\bar{\nu}\nu$, which turns out to be small with branching ratios in the order of $10^{-4}$. In the case of charged slepton LSPs the situation is reverse: one finds the MSSM sneutrinos but misses the right sleptons and finds instead three additional electrically charged but $SU(2)$ singlet Higgs bosons [24,25]. In both cases one finds that either sneutrinos or charged sleptons have in general couplings to quarks and leptons proportional to the usual Yukawa couplings. The main effect of additional trilinear couplings is to change the SM hierarchy of the couplings enhancing in particular those couplings containing only first and second generation indices.

A further aspect of R-parity violation is that the LSP becomes unstable[1]. This is important for the Higgs sector if the Higgs bosons have sizable decay modes into SUSY particles. For R-parity violating couplings larger than $\mathcal{O}(10^{-8} - 10^{-6})$ these decays can be observed in a typical $\mathcal{O}(10)$m diameter collider experiment. In the range up to $\mathcal{O}(10^{-5} - 10^{-4})$ for those couplings displaced vertices can be observed. The LSP decays are important in those cases where the usual MSSM Higgs bosons have sizable branching ratios in SUSY particles, e.g. decays like $A^0 \to \tilde{\chi}^0_i \tilde{\chi}^0_k$. In models with conserved R-parity such decays contain large missing momenta as part of their signatures as the LSP, usually the lightest neutralino, escapes detection. In the case of R-parity violation several things change, e.g. all particles can be the LSPs. Tables 5.1 and 5.2 list the R-parity violating final states induced by trilinear couplings of all particles which have tree-level couplings to Higgs bosons. All lepton number violating final states are also induced by sneutrino vevs. The sneutrino vevs induce additional decay modes: $\tilde{\chi}^0_1 \to W^\pm l^\mp$, $\tilde{\chi}^0_1 \to Z\nu$, $\tilde{\chi}^0_1 \to h^0\nu$, $\tilde{\nu} \to \nu\nu$, $\tilde{\nu} \to t\bar{t}$, and $\tilde{\chi}^+_1 \to W^+\nu$. Several of the R-parity violating decay channels do not have the usual missing energy signal. In other cases it is considerably reduced as the neutrinos carry in average less missing energy compared to neutralinos. R-parity violation implies an enhancement of jet and lepton multiplicities in the final states. For all these reasons decays of Higgs bosons into SUSY particle will look completely different if R-parity is broken compared to the case where it is conserved. Further detailed discussions of R-parity violating decays of SUSY particles can be found in [26,29] for the case of trilinear R-parity violation and in [23,24,28,30,31] for bilinear R-parity violation.

[1]As a side remark we note that it has been shown that a LSP cannot be considered as a cold dark matter candidate in the presence of a single $\not{R}_p$ coupling with value even as small as $\mathcal{O}(10^{-20})$. The only exception is the case of a light gravitino LSP with a mass in the order of 100 eV with a life-time in the order $10^{75}$ Hubble times [27,28].

Table 5.2: R-parity violating decays of neutralinos and charginos with trilinear $\not{R}_p$ operators $\lambda L_i L_j E^c_k$, $\lambda' L_i Q_j D^c_k$ and $\lambda'' U^c_i D^c_j D^c_k$ (from [26]).

| Supersymmetric particles | Couplings $\lambda_{ijk}$ | $\lambda'_{ijk}$ | $\lambda''_{ijk}$ |
|---|---|---|---|
| $\tilde{\chi}^o$ | $\ell^+_i \bar{\nu}_j \ell^-_k$ , $\ell^-_i \nu_j \ell^+_k$, $\bar{\nu}_i \ell^+_j \ell^-_k$ , $\nu_i \ell^-_j \ell^+_k$ | $\ell^+_i \bar{u}_j d_k$ , $\ell^-_i u_j \bar{d}_k$, $\bar{\nu}_i \bar{d}_j d_k$ , $\nu_i d_j \bar{d}_k$ | $\bar{u}_i \bar{d}_j \bar{d}_k$ , $u_i d_j d_k$ |
| $\tilde{\chi}^+$ | $\ell^+_i \ell^+_j \ell^-_k$ , $\ell^+_i \bar{\nu}_j \nu_k$ $\bar{\nu}_i \ell^+_j \nu_k$ , $\nu_i \nu_j \ell^+_k$ | $\ell^+_i \bar{d}_j d_k$ , $\ell^+_i \bar{u}_j u_k$ $\bar{\nu}_i \bar{d}_j u_k$ , $\nu_i u_j \bar{d}_k$ | $u_i d_j u_k$ , $u_i u_j d_k$ $\bar{d}_i \bar{d}_j \bar{d}_k$ |

### 5.1.2 *Spontaneous R-parity violation*

Up to now we have only considered explicit R-parity violation keeping the particle content of the MSSM. In the case that one enlarges the spectrum by gauge singlets one can obtain models where lepton number and, thus, R-parity is broken spontaneously together with $SU(2) \otimes U(1)$ [32–36]. A second possibility to break R-parity spontaneously is to enlarge the gauge symmetry [37].

The most general superpotential terms involving the Minimal Supersymmetric Standard Model (MSSM) superfields in the presence of the $SU(2) \otimes U(1)$ singlet superfields $(\widehat{\nu}^c_i, \widehat{S}_i, \widehat{\Phi})$ carrying a conserved lepton number assigned as $(-1, 1, 0)$, respectively, is given as [38]

$$\begin{aligned} \mathcal{W} &= \varepsilon_{ab}\Big(h^{ij}_U \widehat{Q}^a_i \widehat{U}_j \widehat{H}^b_u + h^{ij}_D \widehat{Q}^b_i \widehat{D}_j \widehat{H}^a_d + h^{ij}_E \widehat{L}^b_i \widehat{E}_j \widehat{H}^a_d + h^{ij}_\nu \widehat{L}^a_i \widehat{\nu}^c_j \widehat{H}^b_u - \hat{\mu} \widehat{H}^a_d \widehat{H}^b_u - h_0 \widehat{H}^a_d \widehat{H}^b_u \widehat{\Phi}\Big) \\ &+ \quad h^{ij} \widehat{\Phi} \widehat{\nu}^c_i \widehat{S}_j + M^{ij}_R \widehat{\nu}^c_i \widehat{S}_j + \frac{1}{2} M_\Phi \widehat{\Phi}^2 + \frac{\lambda}{3!} \widehat{\Phi}^3 \end{aligned} \tag{5.12}$$

The first three terms together with the $\hat{\mu}$ term define the R-parity conserving MSSM, the terms in the second line only involve the $SU(2) \otimes U(1)$ singlet superfields $(\widehat{\nu}^c_i, \widehat{S}_i, \widehat{\Phi})$, while the remaining terms couple the singlets to the MSSM fields. For completeness we note, that lepton number is fixed via the Dirac-Yukawa $h_\nu$ connecting the right-handed neutrino superfields to the lepton doublet superfields. For simplicity we assume in the discussion below that only one generation of $(\widehat{\nu}^c_i, \widehat{S}_i)$ is present.

The presence of singlets in the model is essential in order to drive the spontaneous violation of R-parity and electroweak symmetries in a phenomenologically consistent way. As in the case of explicit

R-parity violation all sneutrinos obtain a vev beside the Higgs bosons as well as the $\tilde{S}$ field and the singlet field $\Phi$. For completeness we want to note that in the limit, that all sneutrino vevs vanish and all singlets carrying lepton number are very heavy one obtains the NMSSM as an effective theory. The spontaneous breaking of R-parity also entails the spontaneous violation of total lepton number. This implies that one of the neutral CP–odd scalars, which we call majoron $J$ and which is approximately given by the imaginary part of

$$\frac{\sum_i v_i^2}{V v^2}(v_u H_u^0 - v_d H_d^0) + \sum_i \frac{v_i}{V}\tilde{\nu}_i + \frac{v_S}{V}S - \frac{v_R}{V}\tilde{\nu}^c \tag{5.13}$$

remains massless, as it is the Nambu-Goldstone boson associated to the breaking of lepton number. $v_R$ and $v_S$ are the vevs of $\tilde{\nu}^c$ and $\tilde{S}$, respectively and $V = \sqrt{v_R^2 + v_S^2}$. Clearly, the presence of these additional singlets enhances further the number of neutral scalar and pseudoscalar bosons. Explicit formulas for the mass matrices of scalar and pseudoscalar bosons can be found e.g. in [39].

The presence of the singlet fields implies in many respects similar features to the addition of the singlet Higgs in the NMSSM, see Section 4, e.g. the Higgs bosons have reduced couplings to the $Z$-boson:

$$\mathcal{L}_{ZZH} = \sum_{i=1}^{8} (\sqrt{2}G_F)^{1/2} M_Z^2 Z_\mu Z^\mu \, \eta_i H_i^0 \,. \tag{5.14}$$

In the basis (Re($H_1^0$), Re($H_2^0$), Re($\tilde{\nu}_i$), Re($\tilde{\nu}^c$), Re($S$),Re($\Phi$)), the $\eta_i$ read as:

$$\eta_i = \frac{v_d}{v} R^{S^0}_{i1} + \frac{v_u}{v} R^{S^0}_{i2} + \sum_{j=1}^{3} \frac{v_j}{v} R^{S^0}_{ij+2} \tag{5.15}$$

where $R^{S^0}$ is the $8 \times 8$ mixing matrix of the neutral scalars. As a consequence the production cross section $e^+e^- \to H_i^0 Z$ can be reduced compared to the MSSM implying that one gets weaker bounds from the LEP data. Another feature similar to the NMSSM is that there is an upper bound on the mainly doublet Higgs boson of 150 GeV. Further details are discussed in Section 5.3.

The case of an enlarged gauge symmetry can be obtained for example in left-right symmetric models, e.g. with the gauge group $SU(2)_L \times SU(2)_R \times U(1)_{B-L}$ [37]. Additional details on extra gauge groups can be found in Section 6. The corresponding superpotential is given by:

$$\begin{aligned} W \;=\; & h_{\phi Q}\widehat{Q}_L^T i\tau_2 \widehat{\phi}\widehat{Q}_R^c + h_{\chi Q}\widehat{Q}_L^T i\tau_2 \widehat{\chi}\widehat{Q}_R^c \\ & + h_{\phi L}\widehat{L}_L^T i\tau_2 \widehat{\phi}\widehat{L}_R^c + h_{\chi L}\widehat{L}_L^T i\tau_2 \widehat{\chi}\widehat{L}_R^c + h_\Delta \widehat{L}_R^{cT} i\tau_2 \widehat{\Delta}\widehat{L}_R^c \\ & + \mu_1 \mathrm{Tr}(i\tau_2 \widehat{\phi}^T i\tau_2 \widehat{\chi}) + \mu_2 \mathrm{Tr}(\widehat{\Delta}\widehat{\delta}), \end{aligned} \tag{5.16}$$

where the Higgs sector consists of two triplet and two bi-doublet Higgs superfields with the following $SU(2)_L \times SU(2)_R \times U(1)_{B-L}$ quantum numbers:

$$\begin{aligned} &\widehat{\Delta} = \begin{pmatrix} \widehat{\Delta}^-/\sqrt{2} & \widehat{\Delta}^0 \\ \widehat{\Delta}^{--} & -\widehat{\Delta}^-/\sqrt{2} \end{pmatrix} \sim (\mathbf{1}, \mathbf{3}, -2), \\ &\widehat{\delta} = \begin{pmatrix} \widehat{\delta}^+/\sqrt{2} & \widehat{\delta}^{++} \\ \widehat{\delta}^0 & -\widehat{\delta}^+/\sqrt{2} \end{pmatrix} \sim (\mathbf{1}, \mathbf{3}, 2), \\ &\widehat{\phi} = \begin{pmatrix} \widehat{\phi}_1^0 & \widehat{\phi}_1^+ \\ \widehat{\phi}_2^- & \widehat{\phi}_2^0 \end{pmatrix} \sim (\mathbf{2}, \mathbf{2}, 0), \qquad \widehat{\chi} = \begin{pmatrix} \widehat{\chi}_1^0 & \widehat{\chi}_1^+ \\ \widehat{\chi}_2^- & \widehat{\chi}_2^0 \end{pmatrix} \sim (\mathbf{2}, \mathbf{2}, 0). \end{aligned} \tag{5.17}$$

In the fermion sector the 'right-handed' matter superfields are combined to $SU(2)_R$ doublets which requires the existence of right-handed neutrinos. The corresponding superfields are denoted by $\widehat{Q}_R^c$ and $\widehat{L}^c_R$ for quark and lepton superfield respectively. Also in this case all neutral components of the Higgs fields and all sneutrinos get vevs. However, the majoron now becomes the longitudinal component of the extra $Z'$ gauge boson.

Looking at the decays of the Higgs bosons, one has to distinguish two scenarios: (i) Lepton number is gauged and, thus, the majoron becomes the longitudinal part of an additional neutral gauge boson. (ii) The majoron remains a physical particle in the spectrum. In the first case one has a situation similar to the case of explicit R-parity violation augmented with the possibilities of the NMSSM. There are for example regions in the parameter space where the scalar Higgs, which is mainly a doublet, decays into two pseudoscalar singlet Higgs bosons yielding e.g.

$$H^0_1 \to A^0_1 A^0_1 \to b\bar{b}b\bar{b}\,. \tag{5.18}$$

In the case of the enlarged gauge group there are additional doubly charged Higgs bosons $H^{--}_i$ which have lepton number violating couplings. In $e^-e^-$ collisions they can be produced according to

$$e^-e^- \to H^{--}_i \tag{5.19}$$

and have decays of the type

$$H^{--}_i \;\to\; H^-_j\, H^-_k \tag{5.20}$$
$$H^{--}_i \;\to\; l^-_j\, l^-_k \tag{5.21}$$

where $l$ denotes $e$, $\mu$ and $\tau$. Further details on the phenomenology of doubly charged Higgs bosons can be found in Section 13.

The second case, where the majoron is part of the spectrum, leads to additional decay modes of the Higgs bosons. For example, the scalar Higgs bosons can decay according to

$$H^0_i \;\to\; A^0_j\, J \tag{5.22}$$
$$H^0_i \;\to\; J\, J \tag{5.23}$$

Note, that the later one is completely invisible. It has been shown that there is a sizable region in parameter space with a light scalar Higgs boson which is mainly a doublet and which decays mainly into the invisible mode above [38, 39]. The existence of the majoron leads also to new decay modes of the pseudoscalar Higgs bosons:

$$A^0_i \;\to\; H^0_j\, J \tag{5.24}$$
$$A^0_i \;\to\; J\, J\, J\,. \tag{5.25}$$

The later one is also completely invisible. However, either the production of the decaying pseudoscalar boson or the branching ratio into the invisible state are quite suppressed as discussed in Section 5.3. Therefore, this mode is phenomenologically less important than the decay $H^0_i \to J\,J$.

For the decays of supersymmetric particles the same general statements hold as for the Higgs bosons. In case (i) from above, the phenomenology is similar to the case of explicit R-parity breaking. The main difference is the existence of additional singlet neutralinos and/or gauginos which can be produced in the various cascade decays. In the case that all these singlet states turn out to be much heavier than the MSSM states one ends up with the bilinear model of R-parity breaking. In the case that the majoron is present, charginos and neutralinos have additional decay modes:

$$\tilde{\chi}^+_i \to J\, l^+ \tag{5.26}$$
$$\tilde{\chi}^0_j \to J\, \nu\,. \tag{5.27}$$

The latter one is completely invisible. In the case that it dominates one recovers the usual missing energy of the MSSM although R-parity is broken. Further details on the phenomenology of SUSY particles in models with spontaneously broken R-parity can be found e.g. in [40, 41].

### 5.1.3 Constraints from colliders

A brief summary of the constraints on R-parity violation couplings from low energy effects has been given in Section 5.1.1 and we refer the reader to [26] for a more detailed review. In the following we focus on direct searches at colliders in models with broken R-parity, which have been carried by HERA, LEP and Tevatron collaborations over the past decade. The pair production of supersymmetric particles with the usual R-parity conserving supersymmetric couplings followed by direct or indirect decays involving R-parity violating couplings as well as singly produced supersymmetric particles involving directly the R-parity violating couplings (followed again by direct or indirect decays) have been extensively searched for. No evidence for supersymmetry with R-parity violation have been found at those colliders.

Constraints have been set on the masses of supersymmetric particles produced in pair where it has been assumed that the effect of the R-parity violating couplings is only important in the decays. An example for these constraints is shown in Table 5.3 for pair produced sfermions at LEP from [42–45]. In the case of the lower limits on the mass of $\tilde{e}_R$ and $\tilde{\nu}_e$ the Aleph collaboration assumes $\mu = -200$ GeV and $\tan\beta = 2$. For the lower limits for indirect decays $m_{\tilde{l},\tilde{\nu}} - m_{\tilde{\chi}} > 10$ GeV is assumed for the $\lambda''_{ijk}$ couplings. The lower limit on the $\tilde{t}_L$ mass (for direct decay) is obtained assuming BR$(\tilde{t}_L \to q\tau) = 1$. The Delphi collaboration takes $\mu = -200$ GeV, $\tan\beta = 1.5$ and $m_{\tilde{f}} - m_{\tilde{\chi}_1^0} \geq 5$ GeV. They also assume the lower mass limit on the lightest neutralino. The $\tilde{t}_1$ and $\tilde{b}_1$ limits from L3 are derived for a squark mixing angle minimizing the cross-section. The Opal collaboration take $m_{\tilde{\chi}_1^0} = 10$ GeV to derive the lower limits on charged sleptons and refer to $\tilde{l}_R$ ($\tilde{l}_L$) for the indirect (direct) decays. In case of the sneutrinos (indirect decays) $m_{\tilde{\chi}_1^0}$ is set to 60 GeV. The limit on $\tilde{t}_L$ assumes $i = 3$ (for $i = 1, 2$ this limit rises to 100 GeV). Moreover, for large sfermion masses an absolute limit of 103 GeV has been set on the chargino mass by the Aleph collaboration [42], irrespective of the R-parity violating coupling. The Delphi collaboration [43] has set a lower limit of 39.5 GeV (103 GeV) for $m_{\tilde{\chi}_1^0}$ ($m_{\tilde{\chi}_1^+}$) for $\lambda_{ijk}$ couplings. For $\lambda''_{ijk}$ these lower limits are 38.0 GeV and 102.5 GeV, respectively

Squark pair production and gluino pair production have been considered by CDF and D0. For example, a lower limit on the stop mass of 122 GeV has been set assuming BR$(\tilde{t}_1 \to b\tau) = 1$ [46]. The CDF collaboration [47] considered the processes $p\bar{p} \to \tilde{g}\tilde{g} \to c\tilde{c}c\tilde{c} \to c(e^{\pm}d)c(e^{\pm}d)$ and $p\bar{p} \to \tilde{d}\bar{\tilde{q}} \to q\tilde{\chi}_1^0\bar{q}\tilde{\chi}_1^0 \to q(ce^{\pm}d)\bar{q}(ce^{\pm}d)$ taking only $\lambda'_{121}$ to be non-zero. This resulted into the constraint $\sigma \times B > 0.18$ pb for the $\tilde{c}$ search and into lower limits on squark masses i.e. 260 GeV for mass degenerate squarks and 135 GeV for $\tilde{t}_1$ assuming $m_{\tilde{g}} = 200$ GeV and a heavy $\tilde{\chi}_1^0$. The D0 collaboration [48] considered gluino and squark cascade decay till the lightest neutralino and then lightest neutralino decay via $\lambda'_{2jk}$ with $j = 1, 2$ and $k = 1, 2, 3$. They obtained a lower limit on squark masses of 240 GeV independent of $m_{\tilde{g}}$ and a lower limit on $m_{\tilde{g}}$ of 224 GeV for all squark masses. In the case of $m_{\tilde{g}} = m_{tildeq}$ a lower limit of 265 GeV has been obtained. In all cases $\tan\beta = 2$, $A_0 = 0$ and $\mu < 0$ has been assumed. The D0 collaboration [49] has also searched for gauginos pair production followed by decays mediated by the $\lambda_{121}$ and $\lambda_{122}$ couplings which allows one to exclude a large region of the parameter space for coupling values of the order of $10^{-4}$.

R-parity violation allows for the possibility of singly produced supersymmetric particles. For example the Delphi collaboration [50] has searched for resonant sneutrino production and decay involving the $\lambda_{121}$ and $\lambda_{131}$ couplings. The obtained constrained on these couplings are in the order of 2-3 $\cdot 10^{-3}$ for 180 GeV $\leq m_{\tilde{\nu}} \leq$ 208 GeV. The $e^{\pm}p$ HERA collider is ideally suited to the search for single squark production involving $\lambda'$ couplings. For example the H1 collaboration [51,52] (see also [53]) has excluded large regions of the planes $(\lambda'_{1j1}, m_{\tilde{q}})$ for j=1,2,3, e.g. $\lambda'_{1j1} \lesssim 10^{-2}$ for $m_{\tilde{q}} \lesssim 200$ GeV, $|\mu| \leq 300$ GeV, 70 GeV $\leq M_2 \leq$ 350 and $\tan\beta = 6$. The CDF collaboration [54] has searched for single sneutrino production and direct decays via $\lambda'$ leading to $e\mu$ (respectively $e\tau$ and $\tau\mu$) final states which resulted into the lower limits $\sigma \times B > 0.14$ pb (respectively 1.2 pb and 1.9 pb). The D0 collaboration [55] has searched for resonant smuon and muon sneutrino single production via $\lambda'_{211}$ and has put lower limits of 280 GeV on the corresponding masses.

Table 5.3: Lower limits at the 95% confidence level on the masses of sleptons (unit GeV) from LEP assuming pair production followed by direct or indirect decay involving the R-parity violating couplings $\lambda_{ijk}$, $\lambda'_{ijk}$ and $\lambda''_{ijk}$ from [42–45]. The acronyms A, D, L and O indicate respectively the Aleph, Delphi, L3 and Opal collaborations. Details on the various assumptions are given in the text.

| | | | $\lambda_{ijk}$ | | $\lambda'_{ijk}$ | | $\lambda''_{ijk}$ | |
|---|---|---|---|---|---|---|---|---|
| | | | direct | indirect | direct | indirect | direct | indirect |
| $\tilde{e}$ | A | $(\tilde{e}_R)$ | 96.0 | 96.0 | | 93.0 | | 94.0 |
| | D | $(\tilde{e}_R)$ | | 95.0 | | | | 92.0 |
| | L | $(\tilde{e}_R)$ | 69.0 | 79.0 | | | | 96.0 |
| | O | | 89.0 | 99.0 | 89.0 | 92.0 | | |
| $\tilde{\mu}$ | A | $(\tilde{\mu}_R)$ | 87.0 | 96.0 | 81.0 $(\tilde{\mu}_L)$ | 90.0 | | 85.0 |
| | D | $(\tilde{\mu}_R)$ | | 90.0 | | | | 87.0 |
| | L | $(\tilde{\mu}_R)$ | 61.0 | 87.0 | | | | 86.0 |
| | O | | 74.0 | 94.0 | 75.0 | 87.0 | | |
| $\tilde{\tau}$ | A | $(\tilde{\tau}_R)$ | 87.0 | 95.0 | | 76.0 | | 70.0 |
| | D | $(\tilde{\tau}_R)$ | | 90.0 | | | | |
| | L | $(\tilde{\tau}_R)$ | 61.0 | 86.0 | | | | 75.0 |
| | O | | 74.0 | 92.0 | 75.0 | | | |
| $\tilde{\nu}_e$ | A | | 100.0 | 98.0 | | 91.0 | | 88.0 |
| | D | | 96.0 | 98.0 | | | | |
| | L | | 95.0 | 99.0 | | | | 99 |
| | O | | 89.0 | 95.0 | 89.0 | 88.0 | | |
| $\tilde{\nu}_\mu$ | A | | 90.0 | 89.0 | 79.0 | 78.0 | | 65.0 |
| | D | | 83.0 | 85.0 | | | | |
| | L | | 65.0 | 78.0 | | | | 70.0 |
| | O | | 79.0 | 81.0 | 74.0 | | | |
| $\tilde{\nu}_\tau$ | A | | | 89.0 | | 78.0 | | 65.0 |
| | D | | 91.0 | 85.0 | | | | |
| | L | | 65.0 | 78.0 | | | | 70.0 |
| | O | | 79.0 | 81.0 | 74.0 | | | |
| $\tilde{u}_L$ (u-type) | A | | | | | | 82.5 | |
| | L | | | | | | 87.0 | 87.0 |
| $\tilde{d}_L$ (d-type) | A | | | | | | 77.0 | |
| | L | | | | | | 86.0 | 86.0 |
| $\tilde{t}$ | A | $(\tilde{t}_L)$ | | 91.0 | 97.0 | 85.0 | | 71.5 |
| | D | $(\tilde{t}_L)$ | | 92.0 | | | | 87.0 |
| | L | $(\tilde{t}_1)$ | | | | | 77.0 | 77.0 |
| | O | $(\tilde{t}_L)$ | | | 98.0 | | 88.0 | |
| $\tilde{b}$ | A | $(\tilde{b}_L)$ | | 90.0 | | 80.0 | | 71.5 |
| | D | $(\tilde{b}_L)$ | | | | | | 78.0 |
| | L | $(\tilde{b}_1)$ | | | | | 55.0 | 48.0 |

## 5.2 The Higgs sector in models with explicitly broken R-parity

*Martin Hirsch and Werner Porod*

The down-type Higgs field $\hat{H}_d$ and the left slepton $\hat{L}_i$ fields of the MSSM carry the same $SU(3) \times SU_L(2) \times U_Y(1)$ quantum numbers. In models with conserved R-parity they can only be distinguished by lepton number. Therefore, in models where lepton number and, thus, R-parity is broken these fields are not distinguishable and the MSSM appears to consist of five Higgs doublets and three electrically charged but $SU_L(2)$ singlet Higgs fields.

The breaking of R-parity can be realized by introducing explicit R-parity breaking terms [5] or by a spontaneous break-down of lepton number [35]. The first class of models can be obtained in mSUGRA scenarios where depending on the choice of discrete symmetries various combinations of R-parity violating parameters are present at the GUT or Planck scale [14]. The latter class of models leads after electroweak symmetry breaking to effective terms, the so-called bilinear terms, which are a sub-class of the terms present in the models with explicit R-parity breaking. These bilinear terms have an interesting feature: They do not introduce trilinear terms when evolved from one scale to another with renormalization group equations (RGEs). In contrast, trilinear terms do generate bilinear terms when evolved from one scale to another.

From the point of view of Higgs physics these bilinear terms have the interesting feature that they lead to a mixing between the usual Higgs fields and sleptons, more precisely the charged Higgs boson mixes with the charged sleptons, the real part of the sneutrinos with the neutral scalar Higgs bosons and the imaginary part of the sneutrinos with the pseudoscalar Higgs boson. Therefore we will concentrate on the effect of bilinear R-parity breaking terms and comment on the case of additional tri-linear couplings at the end of this contribution. The model is specified by the following superpotential $W$ and soft SUSY breaking Lagrangian $V_{soft}$:

$$W = W_{MSSM} + \epsilon_i \hat{L}_i \hat{H}_u \tag{5.28}$$
$$V_{soft} = V_{soft,MSSM} + B_i \epsilon_i \tilde{L}_i H_u \tag{5.29}$$

where $W_{MSSM}$ and $V_{soft,MSSM}$ contain the usual MSSM terms. The $B_i$ induce vevs $v_i$ for the sneutrinos which, however, are not independent quantities. In the following we will trade them against the $B_i$ and take the $v_i$ as free parameters. In the discussion below we will work in the '$\epsilon$-less' basis presented in Section 5.1.1. In this basis effective trilinear couplings of the form $\lambda_{ijk} = (\epsilon_i/\mu) \cdot h_E^j \delta^{jk}$ and $\lambda'_{ijk} = (\epsilon_i/\mu) \cdot h_D^j \delta^{jk}$ are present, see Eqs. (5.10) and (5.11).

The R-parity violating parameters are constrained due to data from rare decays of leptons and mesons and other low-energy data. As discussed in Section 5.1.1 the most important ones for this model arise from the observed neutrino data implying that $|\epsilon_i/\mu| \simeq O(10^{-3}) - O(10^{-2})$ and $v/M_2 \simeq 10^{-6}$ where $v = \sqrt{v_1^2 + v_2^2 + v_3^2}$. The smallness of these couplings imply that the mixing between sleptons and Higgs fields is in general small. The largest effects are observable if either the right-sleptons or the sneutrinos are the lightest supersymmetric particles (the left-sleptons are in general heavier than the sneutrinos). In the following we will discuss these two cases. Although lepton number is not defined anymore in this class of models we will nevertheless use the MSSM notation for simplicity.

### 5.2.1 *The charged scalars*

In this section we will discuss the case that the right-sleptons are the LSPs. From the experimental point of view they appear to be charged Higgs bosons decaying mainly into leptons:

$$\tilde{l}_R \to e\nu,\ \mu\nu,\ \tau\nu \qquad (l = e, \mu, \tau) \tag{5.30}$$

where $\nu$ denotes the sum over all neutrinos. Decays into quarks are suppressed by the corresponding left-right mixings and only in case of the right stau one can expect a sizable branching ratios for quark

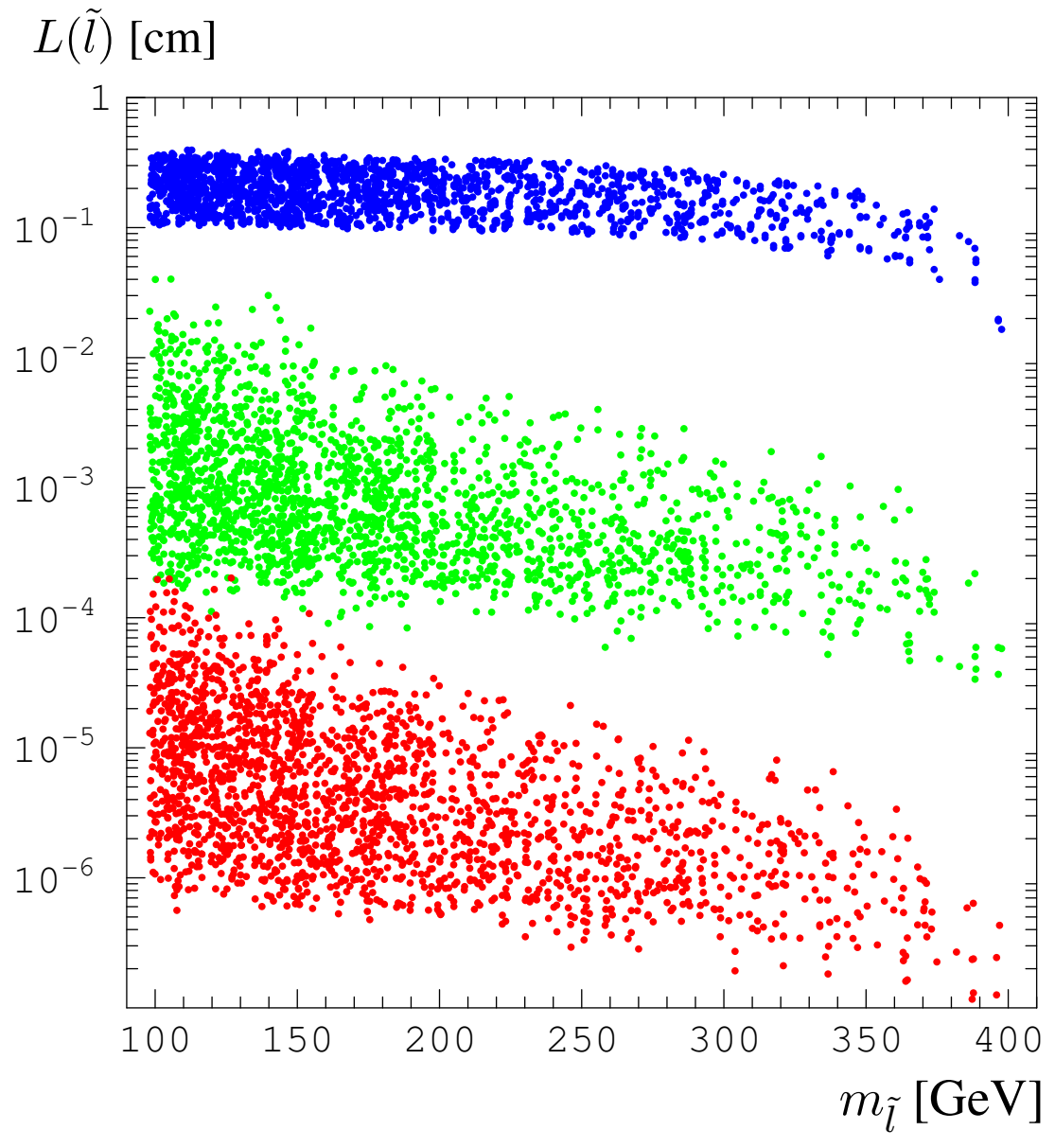

Fig. 5.1: Charged slepton decay length as a function of $m_{\tilde{l}}$ at a linear collider with 0.8 TeV c.m.s. energy. From top to bottom: $\tilde{e}$ (dark, blue), $\tilde{\mu}$ (light shaded, green) and $\tilde{\tau}$ (medium gray, red).

final states if $\tan\beta$ is sufficiently large.

Before discussing the decays in some detail, let us briefly comment on slepton production at future collider experiments. Due to the smallness of the R-parity violating couplings the production of supersymmetric particles is MSSM like. Therefore, at the LHC the direct production of right–sleptons is small, e.g. about 110 (20) fb if $m_{\tilde{l}} \simeq 100(200)$ GeV [56]. As a result, they will be produced mainly in cascade decays. The relative $\tilde{e}_R$, $\tilde{\mu}_R$ and $\tilde{\tau}_R$ yields will depend on the details of the cascade decays involved. In the cascade decays of squarks and the gluino several neutralinos and charginos will be produced. The gaugino like states will decay into an equal number of $\tilde{e}_R$, $\tilde{\mu}_R$ and $\tilde{\tau}_R$ except for kinematics. In particular the bino-like neutralino is expected to have a large branching ratio into $\tilde{l}_R$ as these are the particles with the biggest hypercharge. In the case of higgsino like states the corresponding branching ratios are proportional to the corresponding Yukawa coupling squared. At a future international linear collider the sleptons can be directly produced in $e^+e^-$ annihilation: $e^+e^- \to \tilde{l}^-\tilde{l}^+$. Typical cross sections are of the order of a 100 fb (10 fb) for $\tilde{e}$ ($\tilde{\mu}$ and $\tilde{\tau}$).

All three sleptons can decay into all charged leptons as can be seen from Eq. (5.30) and, thus, the question arises if there are any means to distinguish them. It turns out that different generations of sleptons have quite different life times as discussed in detail in Ref. [24]. In Fig. 5.1 we show the charged slepton decay lengths ($\tilde{e}$, $\tilde{\mu}$ and $\tilde{\tau}$, from top to bottom) as a function of the scalar lepton masses performing a scan of the parameter space: $0 \leq M_2 \leq 1.2$ TeV, $0 \leq |\mu| \leq 2.5$ TeV, $0 \leq m_0 \leq 0.5$ TeV, $-3 \leq A_0/m_0, B_0/m_0 \leq 3$ and $2.5 \leq \tan\beta \leq 10$. The R-parity violating parameters are chosen in such a way [10] that the neutrino masses and mixing angles are approximately consistent with the experimental data as described in detail in [24]. As can be seen, all decay lengths are small compared to typical detector sizes, despite the smallness of the neutrino masses. The three generations of sleptons decay with quite different decay lengths and thus it should be possible to separate the different generations experimentally at a future linear collider. Note that the ratio of the decay lengths $L(\tilde{\tau})/L(\tilde{\mu})$ is approximately given by $(h_\mu/h_\tau)^2$ which can easily be understood from Eq. (5.11).

From Eq. (5.11) one expects that ratios of the branching ratios of the 'flavour' violating decay

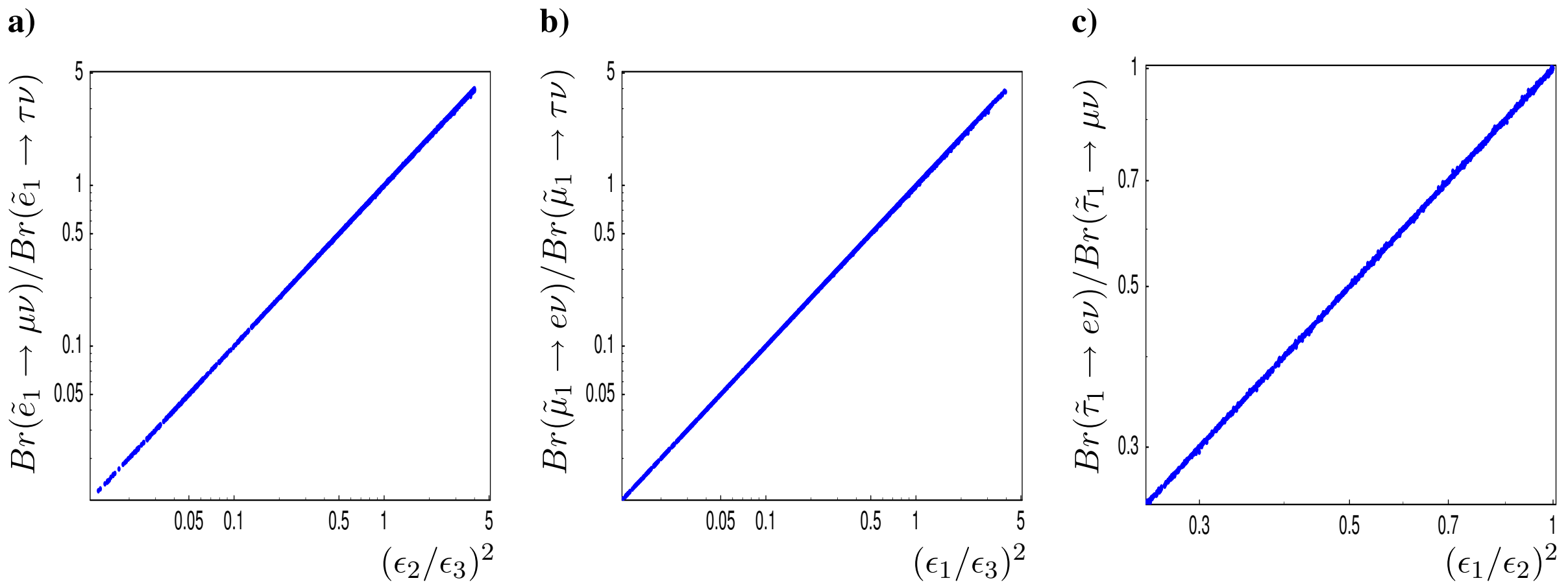

Fig. 5.2: Ratios of branching ratios for (a) selectrons decays versus $(\epsilon_2/\epsilon_3)^2$, (b) smuon decays versus $(\epsilon_1/\epsilon_3)^2$ and (c) stau decays versus $(\epsilon_1/\epsilon_2)^2$ scanning over the SUSY parameter space. Here $\nu$ is the sum over all neutrinos. From [24].

modes should be proportional to ratios of $\epsilon_i$ squared:

$$\frac{BR(\tilde{l}_i \to l_j \sum_r \nu_r)}{BR(\tilde{l}_i \to l_k \sum_r \nu_r)} \simeq \frac{\epsilon_j^2}{\epsilon_k^2} \tag{5.31}$$

with $l_1 = e$, $l_2 = \mu$, $l_3 = \tau$ and $i \neq j \neq k$. Moreover, this feature should remain valid after taking into account all the mixing effects between SM particles and supersymmetric particles as has been shown in [24] semi-analytically. That this is indeed the case is shown in Fig. 5.2. As can be seen from these figures, the ratio of charged slepton branching ratios are correlated with the ratios of the corresponding BRpV parameters $\epsilon_i$, following very closely the expectation from Eq. (5.31), nearly insensitive to variation of the other parameters. Recall, that all the points were generated through a rather generous scan over the MSSM parameters. Ratios of $\epsilon_i$'s should therefore be very precisely measurable. Moreover, since only two of the three ratios of $\epsilon_i$'s are independent it is possible to derive the following *prediction*:

$$\frac{BR(\tilde{\tau}_1 \to e \sum \nu_i)/BR(\tilde{\tau}_1 \to \mu \sum \nu_i)}{BR(\tilde{\mu}_1 \to e \sum \nu_i)/BR(\tilde{\mu}_1 \to \tau \sum \nu_i)} \simeq \frac{BR(\tilde{e}_1 \to \mu \sum \nu_i)}{BR(\tilde{e}_1 \to \tau \sum \nu_i)} \tag{5.32}$$

which provides an important cross check of the validity of the bilinear R-parity model. Any significant departure from this equality would be a clear sign that the bilinear model is incomplete. In the parameter ranges compatible with neutrino data, it turns out that the branching ratios of 'flavour diagonal' decay modes hardly vary with the underlying parameters: $BR(\tilde{\tau} \to \tau \sum_i \nu_i) \simeq BR(\tilde{\mu} \to \mu \sum_i \nu_i) \simeq 0.5$. The variations are of the order of $1\%$. The selectrons behave differently due to the smallness of lepton Yukawa coupling yielding that $0.96 \lesssim BR(\tilde{e}_R \to e \sum_i \nu_i) \lesssim 0.999$.

### 5.2.2 *The neutral scalars*

In this section we will discuss the scenario where the sneutrinos are the LSPs. The occurrence of sneutrino vevs implies in principle a splitting between real and imaginary parts of the sneutrino in analogy to the neutral Higgs sector (see e.g. [10] for the corresponding mass formulas). However, as a consequence of the smallness of the R-parity violating parameters this splitting is well below the expected accuracy of future collider experiments. Moreover, for the same reason R-parity violating production processes like $e^+e^- \to Z\mathrm{Re}(\tilde{\nu})$ have tiny cross sections in the order of $10^{-5}$ fb and the branching ratios of charginos and neutralinos into the real and imaginary parts of the sneutrinos occur with practically the same probability, e.g. $1 - BR(\tilde{\chi} \to \mathrm{Re}(\tilde{\nu}))/BR(\tilde{\chi} \to \mathrm{Im}(\tilde{\nu})) \simeq 10^{-5}$. For these reasons we will speak of 'the' sneutrino instead of making the distinction between scalar and pseudoscalar particles.

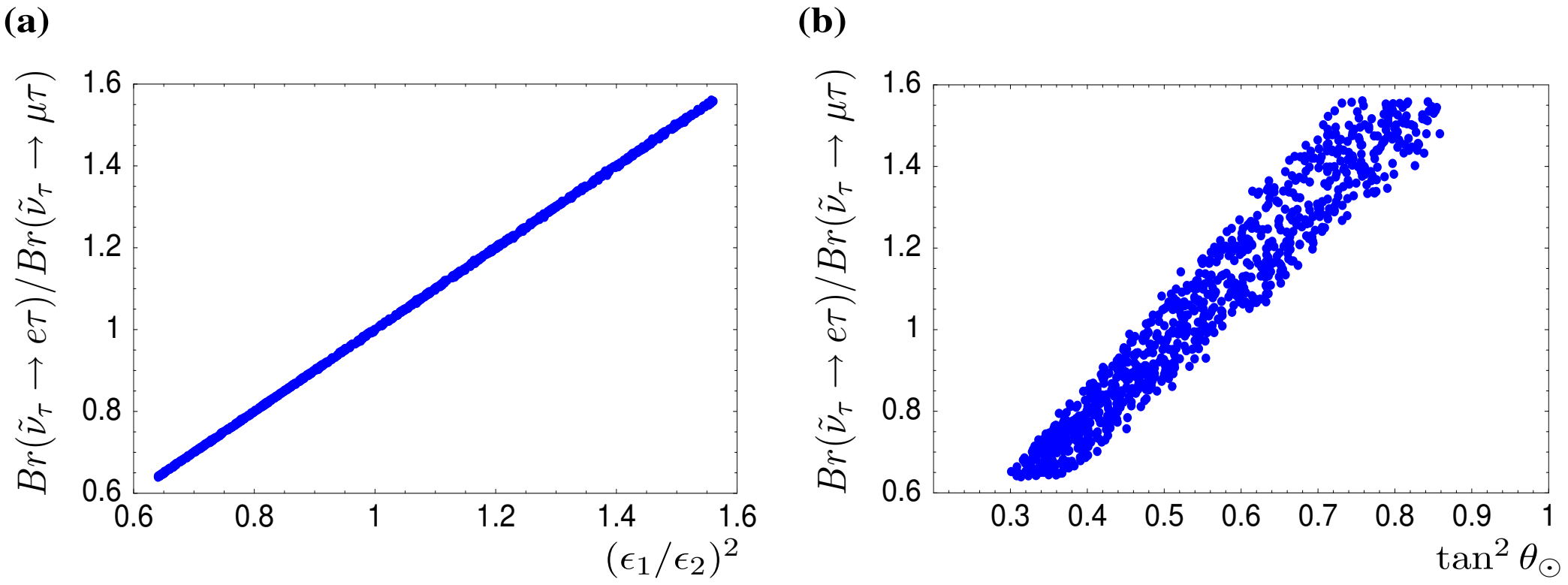

Fig. 5.3: Ratio of branching ratios $BR(\tilde{\nu}_\tau \to e\tau)/BR(\tilde{\nu}_\tau \to \mu\tau)$ versus a) $(\epsilon_1/\epsilon_2)^2$ and b) $\tan^2\theta_\odot$. From Ref. [16].

In this scenario the trilinear couplings of the sneutrinos to down quarks and charged leptons follow a hierarchy dictated by the standard model quark and charged lepton masses, see Eqs. (5.10) and (5.11). One expects therefore that the most important final state for sneutrinos is $b\bar{b}$, independent of the sneutrino generation. Electron and muon sneutrinos will decay also to $\tau\bar{\tau}$ final states with a relative ratio of

$$\frac{BR(\tilde{\nu}_{e,\mu} \to \tau\bar{\tau})}{BR(\tilde{\nu}_{e,\mu} \to b\bar{b})} \simeq \frac{h_\tau^2}{3h_b^2(1+\Delta_{QCD})} \tag{5.33}$$

independent of all other parameters. Here $\Delta_{QCD}$ are the QCD radiative corrections. Decays to $\mu\bar{\mu}$ (and non-$b$ jets) final states are suppressed by the corresponding Yukawa couplings squared. From this point of view they behave as a pure down-type Higgs boson of the MSSM. The two main differences are the occurrence of the invisible decay mode $\nu\nu$ and a small decay width implying a finite decay length as discussed below.

Tau sneutrinos, on the other hand, will decay to final states $e\tau$ and $\mu\tau$ with sizable branching ratios

$$\frac{BR(\tilde{\nu}_\tau \to e\tau)[BR(\tilde{\nu}_\tau \to \mu\tau)]}{BR(\tilde{\nu}_\tau \to b\bar{b})} \simeq \frac{h_\tau^2}{3h_b^2(1+\Delta_{QCD})}\frac{\epsilon_1^2[\epsilon_2^2]}{\epsilon_3^2} \tag{5.34}$$

The above relation allows one to cross check the consistency of the bilinear scenario with neutrino data, as demonstrated in Fig. 5.3. The current $3\sigma$ allowed range for the solar neutrino mixing angle $\theta_\odot$ of $0.30 \le \tan^2\theta_\odot \le 0.59$ fixes $BR(\tilde{\nu}_\tau \to e\tau)/BR(\tilde{\nu}_\tau \to \mu\tau)$ to be in the range from about 0.55 to about 1.25, as can be seen in Fig. 5.3.

Non-zero sneutrino vevs induce the decay $\tilde{\nu} \to \nu\nu$, i.e. by measuring non-zero branching ratios for invisible decays one could establish that sneutrino vevs exist. From the estimate on $\frac{v}{M_2}$ and $\frac{\epsilon}{\mu}$ discussed above one can estimate that branching ratios of sneutrino decays to invisible states should be of the order $\mathcal{O}(10^{-4})$. Figure 5.4a shows the calculated branching ratios for invisible final states, $BR(\tilde{\nu}_i \to \sum \nu_j\nu_k)$, as a function of the sneutrino mass. The figure shows that the estimate discussed above is correct within an order of magnitude. It also demonstrates that for sneutrinos below $m_{\tilde{\nu}} \le 500$ GeV one expects $BR(\tilde{\nu}_i \to \sum \nu_j\nu_k) \ge 10^{-5}$ a few events of the form $e^+e^- \to \tilde{\nu}\tilde{\nu} \to b\bar{b}\nu\nu$ are expected per year at a future international linear collider with a center of mass energy of 1 TeV.

To measure absolute values of R-parity violating parameters it would be necessary to measure the decay widths of the sneutrinos. Given the current neutrino data, however, such a measurement seems to be very difficult for the next generation of colliders. Figure 5.4b shows calculated decay lengths, assuming a center of mass energy of $\sqrt{s} = 1$ TeV, versus sneutrino mass. The decay lengths are short compared to sensitivities expected at a future linear collider which are of order 10 $\mu$m [57]. One can turn this argument around to conclude that observing decay lengths much larger than those shown in Fig. 5.4 would rule out explicit R-parity violation as the dominant source of neutrino mass.

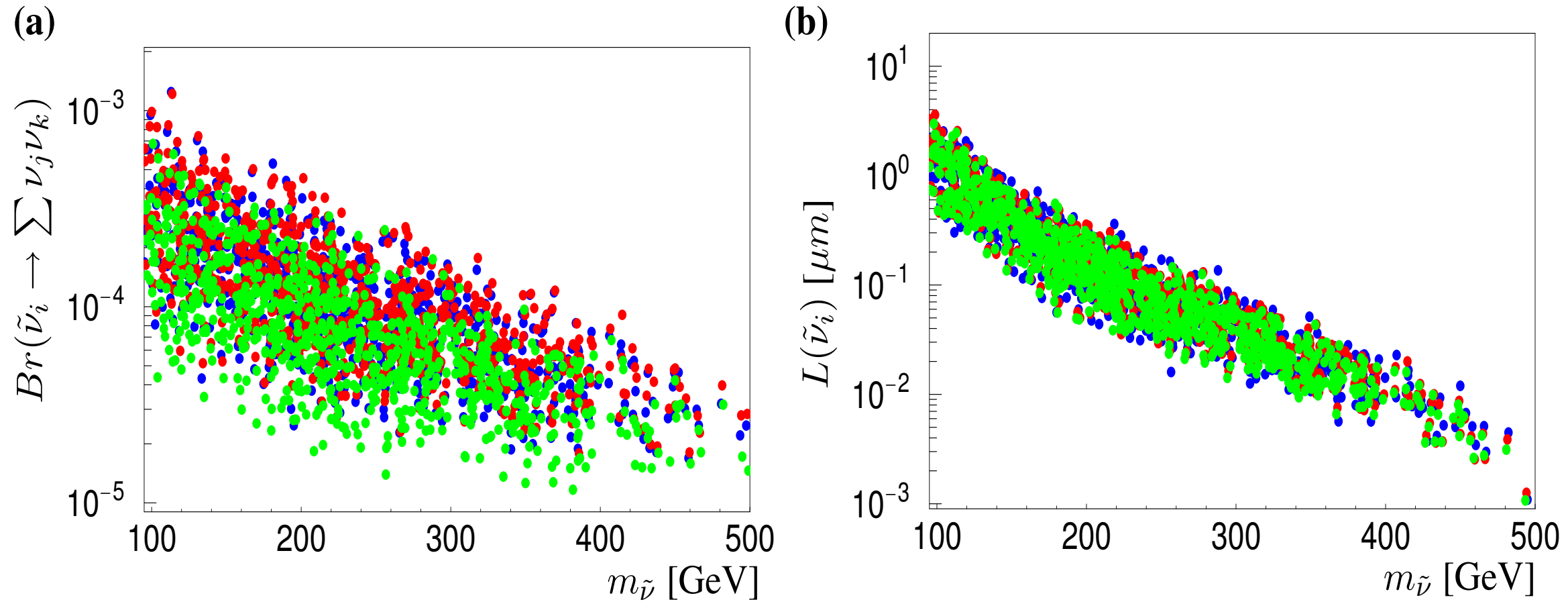

Fig. 5.4: (a) Invisible sneutrino decay branching ratio versus sneutrino mass and (b) sneutrino decay length versus sneutrino mass. Light (medium, dark) points (green, red, blue) are for $\tilde{\nu}_e$ ($\tilde{\nu}_\mu$, $\tilde{\nu}_\tau$).

### *5.2.3 Summary and comments*

We have discussed in some detail the Higgs sector in models where R-parity is broken by bilinear terms. In this class of models the Higgs bosons mix with the slepton fields. However, the bounds on the R-parity violating parameters due to requirement of correctly explaining the observed neutrino data imply that large effects occur mainly if one of the following requirements on the SUSY spectrum is fulfilled.

The first possibility is that the right sleptons are the LSPs. In this case their signature is that of three electrically charged but $SU_L(2)$ singlet Higgs bosons decaying into all generations of charged leptons. Decays into quarks are in general suppressed. An important property of these charged Higgs bosons is that their life time is quite different and that at least two of them should have a visible decay length at future collider experiments, see e.g. Fig. 5.1.

The second possibility is that the sneutrinos are the LSPs. In this case their signatures are close to those expected for the down-type Higgs boson of the MSSM. The main differences are: (i) the occurrence of the invisible mode $\nu\nu$, (ii) small decay widths resulting in decay lengths of the order $\mu m$ and (iii) that one of them has sizable lepton flavour violating decay modes into $e\tau$ and $\mu\tau$.

Let us finally comment on the occurrence of additional trilinear couplings $\lambda_{ijk}$ and $\lambda'_{ijk}$. Their main effects are: (i) The hierarchy of the branching ratios discussed above will be distorted in general. (ii) They can give rise to significantly larger decay widths, in particular if their structure is anti-hierarchical compared to the usual lepton Yukawa couplings. (iii) The invisible decay mode gets tiny. Corresponding scenarios are discussed e.g. in Refs. [16, 25].

## 5.3 Phenomenology of the neutral Higgs sector in a model with spontaneously broken R-parity

*Albert Villanova del Moral*

Current neutrino data can be explained in the framework of the Spontaneously Broken R-Parity Model (SBRPM). This model contains a massless Nambu-Goldstone boson associated to spontaneous lepton number violation called majoron which opens additional invisible decay channels for Higgs bosons. We analyze the full neutral Higgs boson sector of the model and demonstrate that there is always a neutral CP-even Higgs boson, whose mass is bounded from above as in the Minimal Supersymmetric Standard Model, which is copiously produced. Moreover we show that its invisible decay mode to two majorons can be dominant in some regions of parameter space. We also study the associated channel where a neutral CP-odd Higgs boson is produced together with a neutral CP-even one. We show how the lightest CP-odd Higgs boson can have a sizable production cross section and decay to a neutral CP-even Higgs boson and a majoron.

### 5.3.1 The model

The superpotential of this model is given in Eq. (5.12). Note, that if only trilinear terms are non-zero in the superpotential, this specific realization of the SBRPM would solve the $\mu$-problem in the same way as the Next to Minimal Supersymmetric Standard Model (NMSSM) [58, 59].

Considering only one generation of the superfields $(\widehat{\nu}^c_i, \widehat{S}_i)$ for simplicity, the scalar potential for the electrically neutral fields reads

$$\begin{aligned} V_{\rm total} &= |h\Phi\tilde{S} + h^i_\nu\tilde{\nu}_i H^0_u + M_R\tilde{S}|^2 + |h_0\Phi H^0_u + \hat{\mu}H^0_u|^2 + |h\Phi\tilde{\nu^c} + M_R\tilde{\nu^c}|^2 \\ &+ |-h_0\Phi H^0_d - \hat{\mu}H^0_d + h^i_\nu\tilde{\nu}_i\tilde{\nu^c}|^2 + |-h_0 H^0_u H^0_d + h\tilde{\nu^c}\tilde{S} - \delta^2 + M_\Phi\Phi + \frac{\lambda}{2}\Phi^2|^2 \\ &+ \sum_{i=1}^{3}|h^i_\nu\tilde{\nu^c}H^0_u|^2 + \Big[A_h h\Phi\tilde{\nu^c}\tilde{S} - A_{h_0}h_0\Phi H^0_u H^0_d + A_{h_\nu}h^i_\nu\tilde{\nu}_i H^0_u\tilde{\nu^c} - B\hat{\mu}H^0_u H^0_d \\ &- C_\delta\delta^2\Phi + B_{M_R}M_R\tilde{\nu^c}\tilde{S} + \frac{1}{2}B_{M_\Phi}M_\Phi\Phi^2 + \frac{1}{3!}A_\lambda\lambda\Phi^3 + h.c.\Big] \\ &+ \sum_\alpha \tilde{m}^2_\alpha|z_\alpha|^2 + \frac{1}{8}(g^2 + g'^2)\Big(|H^0_u|^2 - |H^0_d|^2 - \sum_{i=1}^{3}|\tilde{\nu}_i|^2\Big)^2, \end{aligned} \tag{5.35}$$

where $z_\alpha$ denotes any neutral scalar of the model.

As usual, electroweak symmetry is broken by the isodoublet nonzero vacuum expectation values (vevs)

$$\langle H^0_u\rangle = \frac{v_u}{\sqrt{2}}, \qquad \langle H^0_d\rangle = \frac{v_d}{\sqrt{2}}. \tag{5.36}$$

R-parity is broken by the lepton-number-carrying isosinglet vevs

$$\langle\tilde{S}\rangle = \frac{v_S}{\sqrt{2}}, \qquad \langle\tilde{\nu^c}\rangle = \frac{v_R}{\sqrt{2}}. \tag{5.37}$$

as well as by the (tiny) left sneutrino vevs

$$\langle\tilde{\nu}_{Li}\rangle = \frac{v_i}{\sqrt{2}}. \tag{5.38}$$

Last but not least, another important vev which is the key ingredient to generate an effective $\mu$-term (and so, solving the $\mu$-problem), is

$$\langle\Phi\rangle = \frac{v_\Phi}{\sqrt{2}}. \tag{5.39}$$

### 5.3.2 Neutral Higgs boson masses

The neutral Higgs boson sector of the SBRPM consists of eight CP-even states $H^0_i$, six CP-odd states $A^0_i$ and one massless majoron $J$ (as the electroweak Goldstone boson $G^0$ is eaten by the $Z$). The $8\times 8$ mass matrices for the neutral CP-even and CP-odd Higgs bosons (given in ref. [39]) can be analytically understood in certain limits [60]. First of all, we note that doublet sneutrinos practically do not mix with the rest of the Higgs bosons, as the corresponding entries in the mass matrices are proportional to $h^i_\nu$ and these parameters are small because of neutrino physics phenomenology. If there were also no mixing between the $(H_d^0,\, H_u^0)$ doublet sector, the $\tilde{\Phi}$ singlet sector and the $(\tilde{S},\, \tilde{\nu}^c)$ singlet sector[2], we would have as CP-odd mass eigenstates

$$m^2_{G^0} = 0\,, \qquad\qquad m^2_{A^0_D} = \Omega\left(\frac{v_u}{v_d}+\frac{v_d}{v_u}\right) \tag{5.40}$$

in the $(H_d^0,\, H_u^0)$ doublet sector and

$$m^2_{J} = 0\,, \qquad\qquad m^2_{A^0_S} = -\Gamma\left(\frac{v_R}{v_S}+\frac{v_S}{v_R}\right) \tag{5.41}$$

in the $(\tilde{S},\, \tilde{\nu}^c)$ singlet sector, with

$$\Omega = B\hat{\mu} - \delta^2 h_0 + \frac{\lambda}{4}h_0 v_\Phi^2 + \frac{1}{2}hh_0 v_R v_S + \frac{\sqrt{2}}{2}A_{h_0}h_0 v_\Phi + \frac{\sqrt{2}}{2}h_0 M_\Phi v_\Phi \tag{5.42}$$

$$\Gamma = B_{M_R}M_R - \delta^2 h + \frac{1}{4}h\lambda v_\Phi^2 - \frac{1}{2}hh_0 v_u v_d + \frac{\sqrt{2}}{2}h\left(A_h + M_\Phi\right)v_\Phi\,. \tag{5.43}$$

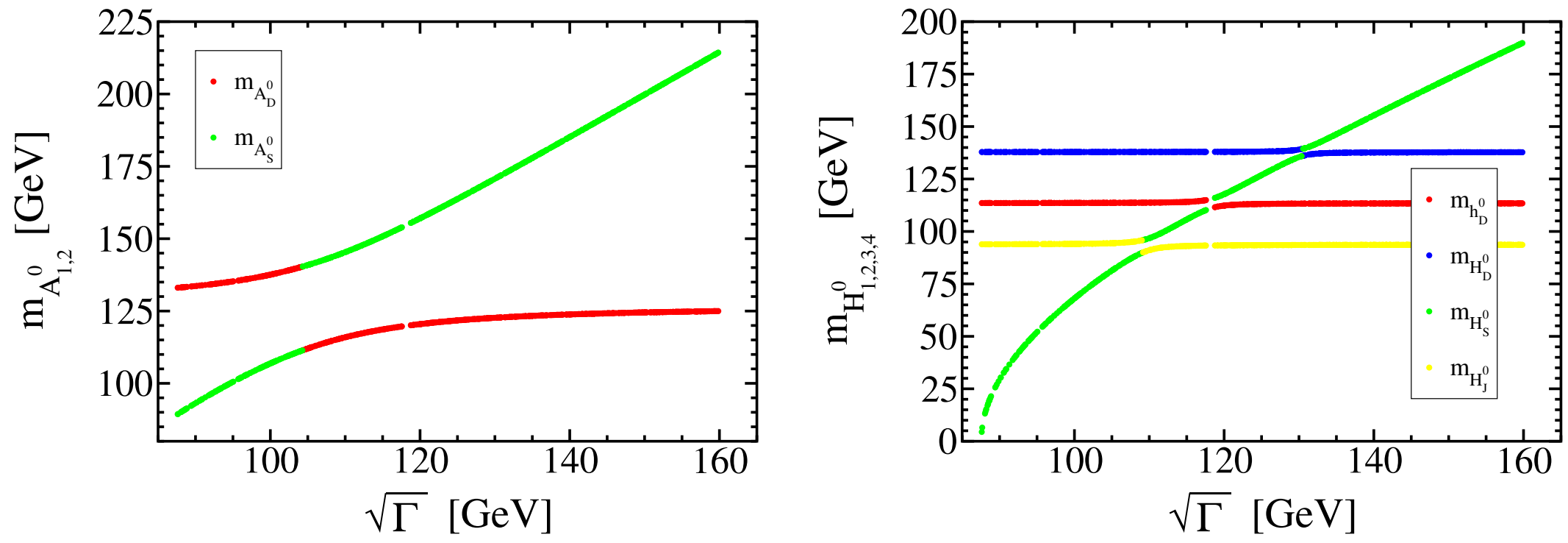

Fig. 5.5: Some of the neutral CP-odd (on the left) and CP-even (on the right) Higgs masses scanned versus the parameter $\sqrt{\Gamma}$. The main flavour component of each mass eigenstate is identified by means of each colour, as described in the text. From ref. [60].

In addition we would have as CP-even mass eigenstates

$$m^2_{H^0_J} = 2h^2\frac{v_R^2 v_S^2}{v_R^2+v_S^2}\,, \qquad\qquad m^2_{H^0_S} = -\Gamma\left(\frac{v_R}{v_S}+\frac{v_S}{v_R}\right) - 2h^2\frac{v_R^2 v_S^2}{v_R^2+v_S^2} \tag{5.44}$$

in the $(\tilde{S},\, \tilde{\nu}^c)$ singlet sector, besides the states $h_D^0$ and $H_D^0$ in the $(H_d^0,\, H_u^0)$ doublet sector, which are analogous to their MSSM counterparts. In Fig. 5.5 a typical scanned Higgs mass spectrum is plotted as a function of the parameter $\sqrt{\Gamma}$ and we can identify the $A_S^0$ and $H_S^0$ states as those which depend on $\sqrt{\Gamma}$. The various gray-shadings (colors) indicate that the asymptotic states given in Eqs. (5.40), (5.41) and (5.44) constitute more the 50% of the corresponding particle.

Taking into account the phenomenological relation $v_i \ll v_R, v_S$ we find the following approximation for the majoron

$$J \simeq \frac{v_S}{V}\,\mathrm{Im}(\tilde{S}) - \frac{v_R}{V}\,\mathrm{Im}(\tilde{\nu}^c) \tag{5.45}$$

where

$$V^2 = v_S^2 + v_R^2\,. \tag{5.46}$$

[2]This assumption is not strictly valid, as some of their mixings are not negligible, but it is useful to gain some insight on the parameter dependence of the eigenvalues.

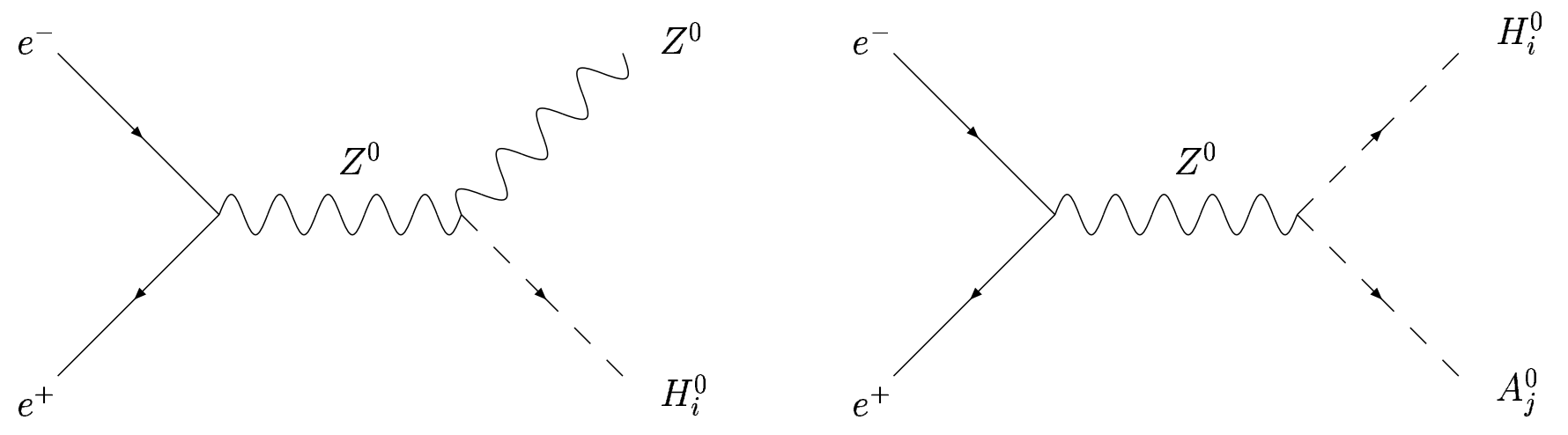

Fig. 5.6: Neutral Higgs boson production processes in an $e^+e^-$ collider. On the left, we can see the direct production or Bjorken process. On the right, we can see the associated production process.

### *5.3.3 Higgs boson production*

Neutral Higgs bosons can be produced at an $e^+e^-$ collider via the Bjorken process (or direct production), as shown in Fig. 5.6, on the left. The relevant Lagrangian terms for this production mode are

$$\mathcal{L}_{ZZH} = \sum_{i=1}^{8} (\sqrt{2}G_F)^{1/2} M_Z^2 Z_\mu Z^\mu \, \eta_i H_i^0 \tag{5.47}$$

where $\eta_i$ is the direct production parameter given by

$$\eta_i = \frac{v_d}{v} R_{i1}^{S^0} + \frac{v_u}{v} R_{i2}^{S^0} + \sum_{j=1}^{3} \frac{v_j}{v} R_{ij+2}^{S^0} \tag{5.48}$$

We note that if $\eta_i^2$ is nearly one (zero), then $H_i^0$ is mainly a doublet (singlet).

From Fig. 5.7, on the left, we can see that when the direct production parameter $\eta_1^2$ of the lightest CP-even Higgs boson $H_1^0$ is nearly zero, then the direct production parameter $\eta_2^2$ of the next-to-lightest Higgs boson $H_2^0$ approaches one, i.e. $H_2^0$ is largely produced. From Fig. 5.7, on the right, we can see that there is always a state with a mass smaller than about 150 GeV. Combining both plots, we conclude that there is always a light Higgs boson with a large production cross section [60].

Another way of producing neutral Higgs bosons is via the associated production process shown in Fig. 5.6 on the right. The relevant Lagrangian terms for this production mechanism are

$$\mathcal{L}_{ZHA} = \sum_{i,j=1}^{8} (\sqrt{2}G_F)^{1/2} M_Z \, \zeta_{ij} \left( Z^\mu H_i^0 \overleftrightarrow{\partial_\mu} P_j^0 \right) \tag{5.49}$$

where $\zeta_{ij}$ is the associated production parameter, which is given by

$$\zeta_{ij} = R_{i1}^{S^0} R_{j1}^{P^0} - R_{i2}^{S^0} R_{j2}^{P^0} + \sum_{k=1}^{3} R_{ik+2}^{S^0} R_{jk+2}^{P^0} \tag{5.50}$$

In the MSSM exists a sum rule which relates both the direct and the associated production parameters. In the SBRPM we can construct an analogous but more complicated sum rule taking into account all possible final states [60]. The conclusion is that always at least one state will be produced.

### *5.3.4 Higgs boson decays*

#### 5.3.4.1 CP-even Higgs boson decays

The main decay channels for the lightest (or next-to-lightest) neutral CP-even Higgs bosons $H_{1,2}^0$ are

$$H_{1,2}^0 \to f_i \bar{f}_i \qquad \text{if} \quad m_{H_{1,2}^0} > 2m_{f_i} \tag{5.51a}$$

$$H_{1,2}^0 \to JJ \tag{5.51b}$$

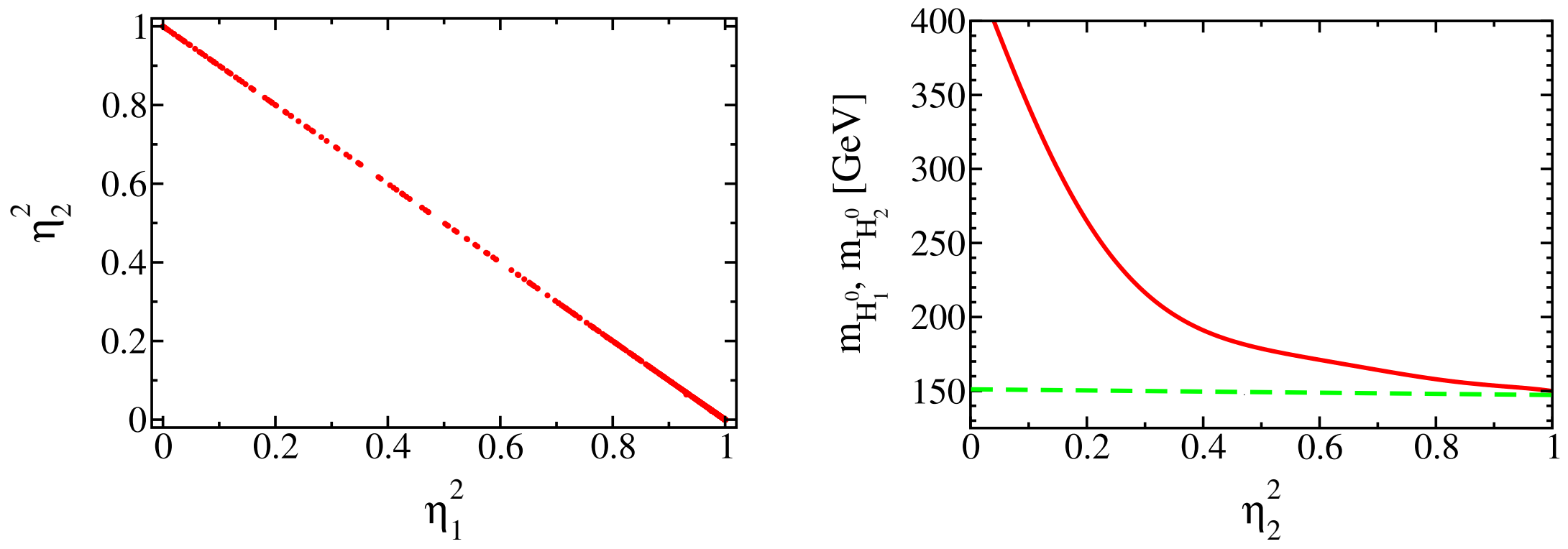

Fig. 5.7: On the left, direct production parameter for the second lightest neutral CP-even Higgs boson $\eta_2^2$ as function of the direct production parameter for the lightest neutral CP-even Higgs boson $\eta_1^2$. On the right, upper bounds on the masses of the first (dashed green) and second (red solid) lightest CP-even Higgs boson, $m_{H_1^0}$ and $m_{H_2^0}$, as a function of the direct production parameter for the second lightest CP-even Higgs boson, $\eta_2^2$.

where the invisible decay width to majorons is

$$\Gamma(H_{1,2}^0 \to JJ) = \frac{g^2_{H^0_{1,2}JJ}}{32\pi m_{H^0_{1,2}}} \tag{5.52}$$

and for the fermionic decay widths all possible final states have been considered. We have taken into account the most important QCD corrections for the quark final states as given in [61].

We define the ratio between the invisible decay width and the visible one as

$$R_{1,2} = \frac{\Gamma(H_{1,2}^0 \to JJ)}{\sum_j \Gamma(H_{1,2}^0 \to f_j \bar{f}_j)} \,. \tag{5.53}$$

These ratios depend on the couplings $g_{H_i^0 JJ}$ which are in general complicated functions of the underlying parameters. However, using Eq. (5.45) one obtains the following couplings to the *unrotated* doublet Higgs fields ($H_1^{\prime 0} = \Re(H_d^0)$ and $H_2^{\prime 0} = \Re(H_u^0)$):

$$g_1' \simeq h h_0 v_u \frac{v_S v_R}{V^2} \tag{5.54a}$$

$$g_2' \simeq h h_0 v_d \frac{v_S v_R}{V^2} - \frac{2 v_u}{V^2} \sum_{j=1}^{3} \epsilon_j^2 \tag{5.54b}$$

Equations (5.54a) and (5.54b) imply that the doublet Higgs bosons can have large branching ratios for the invisible decay mode if the product of the couplings $h$ and $h_0$ is large. Therefore, scenarios exists where neutral Higgs bosons have at the same time a large cross section and a large invisible decay branching ratio [38,39]. This is shown explicitly in Fig. 5.8 where the ratios $R_i$ for the two lightest Higgs bosons are shown as a function of their direct production parameter $\eta_i$ for different values of the parameter $h$ [60].

#### 5.3.4.2 CP-odd Higgs boson decays

The main decay channels for the lightest neutral CP-odd Higgs boson $A_1^0$ are:

$$A_1^0 \to f_i \bar{f}_i \qquad \text{if} \quad m_{A_1^0} > 2 m_{f_i} \tag{5.55a}$$

$$A_1^0 \to H_j^0 J \qquad \text{if} \quad m_{A_1^0} > m_{H_j^0} \tag{5.55b}$$

$$A_1^0 \to JJJ \tag{5.55c}$$

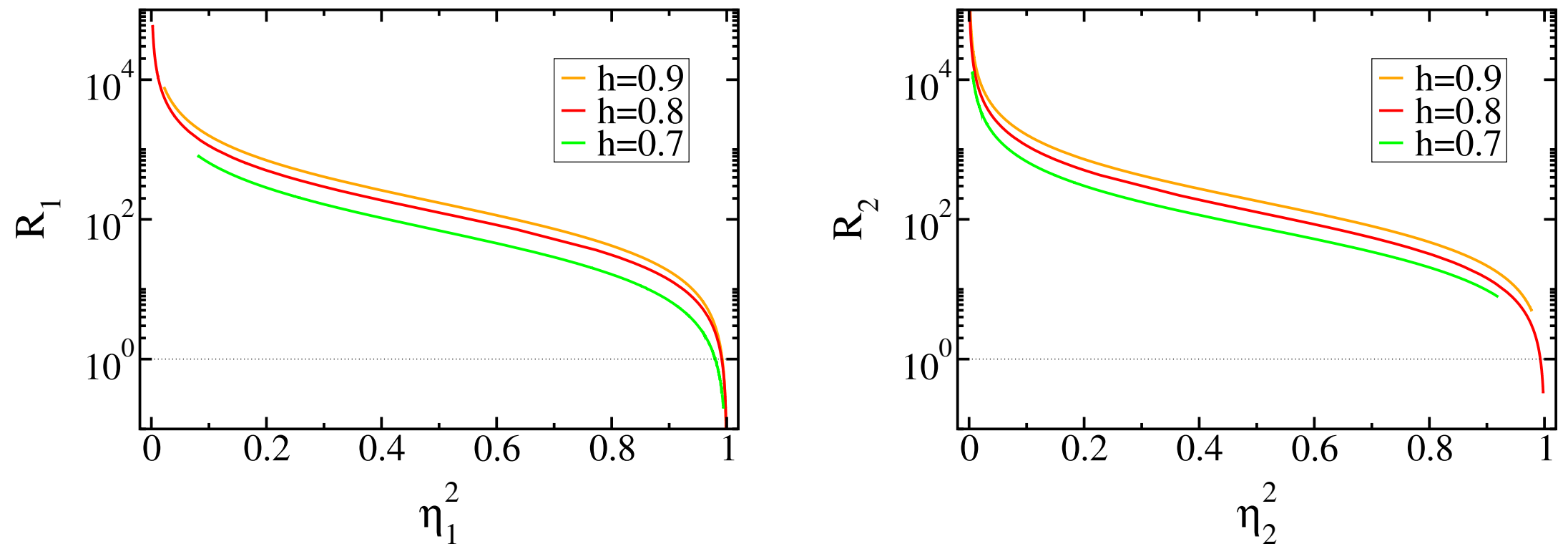

Fig. 5.8: On the left, ratio between the invisible and the visible decay widths $R_1$ of the lightest neutral CP-even Higgs boson as function of its direct production parameter $\eta_1^2$ for different values of the parameter $h$. On the right, ratio between the invisible and the visible decay widths $R_2$ of the second lightest neutral CP-even Higgs boson as function of its direct production parameter $\eta_2^2$ for different values of the parameter $h$.

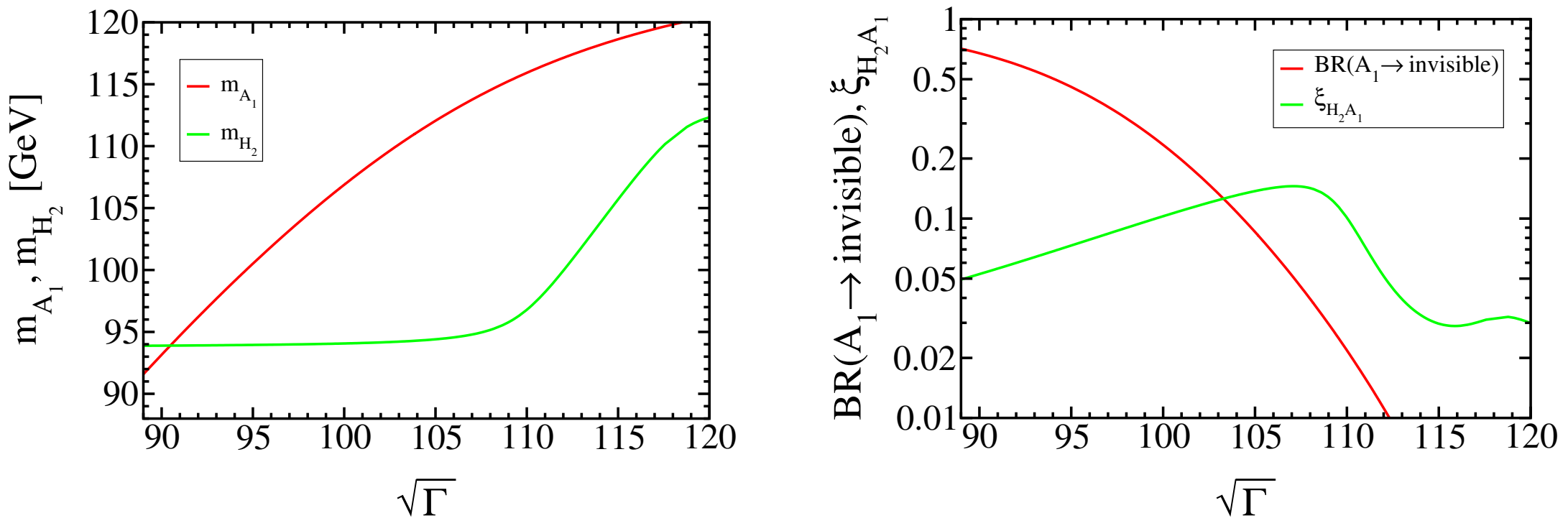

Fig. 5.9: On the left, masses of both the lightest CP-odd Higgs boson and the second lightest CP-even Higgs boson, as a function of the parameter $\sqrt{\Gamma}$. On the right, invisible branching ratio through majorons of the lightest CP-odd Higgs boson, as well as associated production parameter of this and the second lightest CP-even Higgs boson, as a function of the parameter $\sqrt{\Gamma}$.

The width of the CP-odd Higgs boson to a CP-even Higgs boson and a majoron reads

$$\Gamma(A_1^0 \to H_i^0 J) = \frac{g^2_{H_i^0 A_1^0 J}}{16\pi m^3_{A_1^0}} \left( m^2_{A_1^0} - m^2_{H_i^0} \right) , \tag{5.56}$$

and to three majorons

$$\Gamma(A_1^0 \to JJJ) = \frac{m_{A_1^0}\, g^2_{A_1^0 JJJ}}{3072\pi^3} . \tag{5.57}$$

Contrary to the neutral CP-even case, the corresponding couplings of the majorons to the *unrotated* doublet Higgs boson ($\Im(H_d^0)$ and $\Im(H_u^{0I})$) appearing in Eqs. (5.56) and (5.57) are zero in first order approximation, using Eq. (5.45) (detailed expressions are given in Ref. [60]). Therefore, the pseudoscalar Higgs boson has to have sizable admixtures of both, doublet and singlet components, if it should be produced at a sizable rate while having at the same time a significant invisible branching ratio. As an example we show in Fig. 5.9 the masses of $A_1^0$ and $H_2^0$, the associated production parameter and the invisible branching ratio of $A_1^0$ as a function of the parameter $\sqrt{\Gamma}$. One sees that in the region where the production cross section is at least 1% of the corresponding MSSM cross section, the branching ratio for the invisible mode varies between 5 and 10% [60].

### 5.3.5 Conclusions

We have shown that the model with spontaneously broken R-parity contains a light CP-even Higgs boson which is mainly doublet and which has a mass below about 150 GeV (like in the NMSSM). As a new feature we have demonstrated that its invisible decay mode into two majorons can be dominant.

In the case of the CP-even Higgs bosons we have seen that they decay mainly their MSSM counterpart if the doublet component dominates. However, in certain regions of parameter space the singlet component can be large enough to obtain a branching ratio of the invisible mode up to 10%.

## 5.4 Charged-Higgs-boson and charged-slepton radiation off a top quark at hadron colliders

*Francesca Borzumati and Jean-Loïc Kneur*

In this contribution we study the production of charged scalar particles radiated off a top quark at hadron colliders. The charged particles we consider are charged Higgs bosons and charged sleptons in R-parity-violating models. The remnant of the radiating top quark is the bottom quark in the charged-Higgs-boson case; in the charged-slepton case, it is most likely to be the down quark, but possibly also the strange and the bottom quark. Hereafter we shall refer to this production mechanism as strahlung production.

### 5.4.1 The charged-Higgs-boson case

Charged Higgs bosons, if detected, would be a clear signal of an extended Higgs sector. They are present in supersymmetric models, which, as is well known, require at least two Higgs doublets of opposite hypercharge. At hadron colliders, charged Higgs bosons can be pair produced in quark-initiated processes, such as the Drell–Yan process mediated by an off-shell photon or $Z$ boson, the $b$-quark fusion and the $W$-bosons fusion. All these are tree-level processes. Alternatively, they can be pair produced through gluon-fusion processes, which however proceed at the one-loop level. In both cases, the cross sections are not very large, reaching at the LHC $\sim$ 2–3 fb for $m_{H^\pm} = 400\,\text{GeV}$ and $\tan\beta = 30$ [62]. They can also be singly produced in association with other bosons, such as neutral Higgs bosons, or the $W$ boson. These processes are quark-initiated at the tree level, but they can be gluon-initiated at the quantum level. Their cross sections may reach up to $\sim$ 100 fb for the same values of $m_{H^\pm}$ and $\tan\beta$ [62].

Strahlung off a third-generation quark, which can be gluon-initiated also at the tree level, can give similarly large or slightly larger cross sections [62–69]. Such a production mechanism can proceed through the $2 \to 2$ elementary process $gb \to tH^-$ and the $2 \to 3$ processes $gg, q\bar{q} \to tH^-\bar{b}$, which formally give rise to the hadronic processes $p\bar{p}, pp \to tH^-X$ and $p\bar{p}, pp \to tH^-\bar{b}X$, respectively.

Leading-order predictions for the two production cross sections are given in the two top frames of Fig. 5.10 for the Tevatron and LHC energies, as functions of the charged-Higgs-boson mass, for three different values of $\tan\beta$: $\tan\beta = 2$, 10, and 50. (For discussions about limits on the $H^\pm$ mass in Two-Higgs-Doublet Models and supersymmetric models, see Refs. [70–75].) The hadronic cross sections $\sigma(p\bar{p}, pp \to tH^-\bar{b}X)$ are shown by solid lines, $\sigma(p\bar{p}, pp \to tH^-X)$ by dashed lines. The integrations needed to obtain these cross sections are performed by the Monte Carlo integration routine `VEGAS` [76]. Moreover, the leading-order parton distribution functions `CTEQ4L` [77] were used, and the renormalization ($\mu_R$) and factorization ($\mu_f$) scales were fixed to the threshold value $m_t + m_{H^\pm}$. A variation of these scales in the interval between $(m_t + m_{H^\pm})/2$ and $2(m_t + m_{H^\pm})$ results in changes up to $\pm 30\%$ in both cross sections. QCD corrections, therefore, may be important, but they have been completed only for the $2 \to 2$ processes [78–81]. Part of these corrections are captured by the QCD correction to the $b$-quark mass, on which these cross sections depend quite sensitively. A study of their variation for different values of the $b$-quark mass can be found in Ref. [82]. (Supersymmetric corrections to both decays have also been calculated [83, 84].)

In the kinematical region $m_t > m_{H^\pm} + m_b$ the $2 \to 3$ elementary processes give the largest hadronic cross section. This can be well approximated by the much simpler resonant production cross section, given by the on-shell $t\bar{t}$ production cross section times the branching fraction for the decay $\bar{t} \to H^-\bar{b}$. In the region $m_t \sim m_{H^\pm} + m_b$, however, this approximation fails to account for the correct mechanism of production and decay of the charged Higgs boson [85, 86]. When $m_t < m_{H^\pm} + m_b$, the relative size of the two classes of cross sections depends on $\sqrt{s}$ and $m_{H^\pm}$. At the Tevatron centre-of-mass energy, the quark-initiated $2 \to 3$ processes still have the dominant role up to intermediate values of $m_{H^\pm}$, i.e. up to $m_{H^\pm} \sim 265\,\text{GeV}$. Both classes of cross sections show the typical behaviour as a function of $\tan\beta$, with a minimum at around $(m_t/m_b)^{1/2}$.

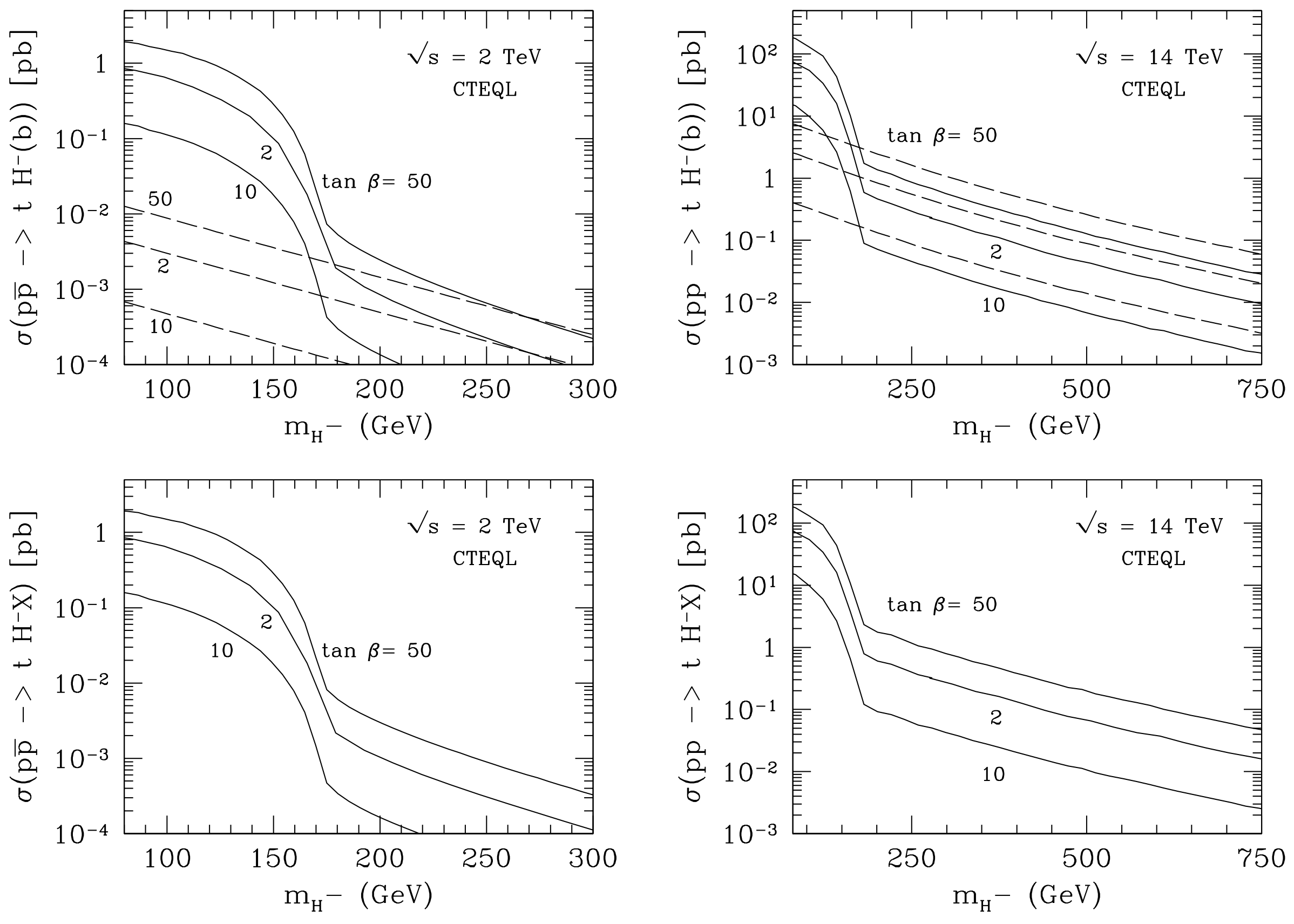

Fig. 5.10: Cross-sections $\sigma(p\bar{p}\ (pp) \to t(\bar{b})H^-X)$ versus $m_{H^-}$, at the Tevatron and the LHC, for $\tan\beta = 2, 10, 50$, with $m_t = 175\,\text{GeV}$, $m_b = 3\,\text{GeV}$. Renormalization and factorization scales are fixed as $\mu_R = \mu_f = m_t + m_{H^-}$. In the two upper frames the solid lines correspond to the $2 \to 3$ processes, the dashed lines to the $2 \to 2$ process. In the two lower frames are shown the cross sections obtained by adding the contributions from the $2 \to 2$ and $2 \to 3$ processes, and subtracting overlapping terms.

When the charged Higgs boson decays leptonically, $H^- \to \tau^-\nu_\tau$, the two production mechanisms, which lead to two and one $b$ quark in the final state are independent. (We assume here that it is possible to detect two $b$'s and one $\tau$.) This decay channel is suitable for the discovery of the charged Higgs boson in the region of large and possibly intermediate values of $\tan\beta$ [87], since it is not plagued by QCD background as the $H^- \to b\bar{t}$ mode [69]. The two production mechanisms can be experimentally distinguished, and studied separately.

When $H^-$ decays hadronically, typically into $\bar{t}b$, the final state to be identified contains at least three $b$'s for the $2 \to 2$ production mechanism, and at least four for the $2 \to 3$ one. Since tagging so many $b$'s seems very difficult, even at the LHC, the two production mechanisms result into final states that are indistinguishable. In this case, a sum of the two cross sections is necessary. Care must be taken, however, not to double count the overlapping part, obtained when one of the two initial gluons in the $2 \to 3$ processes produces a $b\bar{b}$ pair collinear to the initial $p$ or $\bar{p}$ [65]. Predictions for the appropriately summed inclusive cross section are shown in lower frames of Fig. 5.10, for both the Tevatron and the LHC. At the LHC, the cross section for $m_{H^\pm} = 400\,\mathrm{GeV}$ and $\tan\beta = 30$ is about 140 fb. These cross sections have the same theoretical uncertainty as the individual ones, as well as the same $\tan\beta$ dependence.

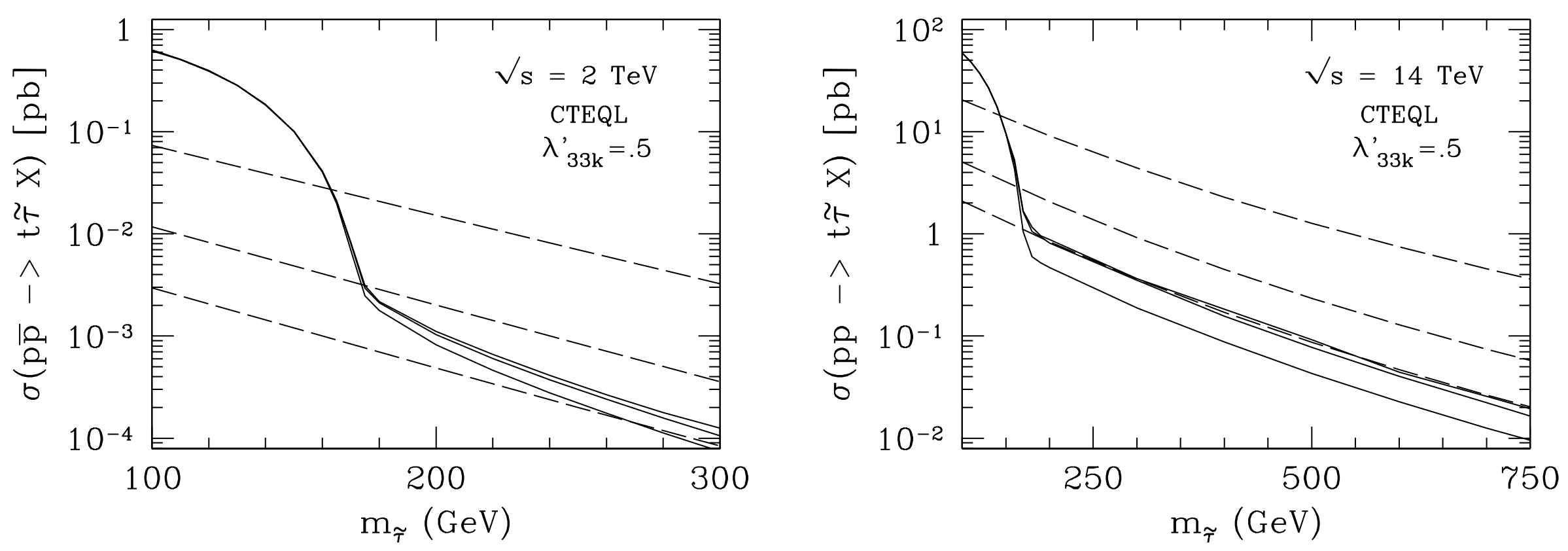

Fig. 5.11: Cross sections $\sigma(p\bar{p}\ (pp) \to t(d_k)\tilde{\tau}X)$ versus $m_{\tilde{\tau}}$, at the Tevatron and LHC, for $\lambda'_{33k} = 0.5$. The solid lines correspond to the $2 \to 3$ processes, the dashed lines to the $2 \to 2$ process. The three curves in each of the two sets correspond, from top to bottom to $d_k = d, s, b$. Renormalization and factorization scales are fixed as $\mu_R = \mu_f = m_t + m_{\tilde{\tau}}$.

### *5.4.2 The charged-slepton case*

As is well known, the component of the charged Higgs boson with hypercharge $-1/2$ has the same quantum numbers as the three superpartners of the charged leptons, $\tilde{l}$, except for the lepton number $L$. In $R_p$-violating models, in which $L$ is violated (by operators with $\Delta L = 1$), these fields cannot be distinguished. Thus, some of the knowledge acquired by studying the strahlung of $H^\pm$ off a top-quark line can be applied to investigate a similar production mechanism for charged sleptons.

The relevant operators for this discussion are the superpotential trilinear term $-\lambda'_{ijk} L_i Q_j D^c_k$, with $i, j, k = 1, 2, 3$. In a basis in which all right-handed quarks and the left-handed down ones are diagonal, the trilinear superpotential operator gives rise to the lagrangian interaction terms:

$$\mathcal{L}_{\not{R}} \supset \lambda'_{imk} V_{mj}\, \overline{u}_{Lj}\, d_{Rk} \tilde{l}^*_{Li} \;-\; \lambda'^*_{imk}\, \overline{d}_{Lm}\, d_{Rk} \tilde{\nu}^*_{Li} + \mathrm{H.c.}\,, \qquad (5.58)$$

where $V_{mj}$ are elements of the CKM matrix. The second operators in this equation give rise to contributions to neutrino masses. Among the first ones, those with couplings such as $\lambda'_{i3k}$, induce the production of single charged sleptons in association with the top quark, in complete analogy with the strahlung production of the charged Higgs boson described before. Couplings like these, with at least one third-generation index, are only very weakly constrained by present experiments, except for those giving indications on the values of the neutrino masses. If we postpone for a moment the discussion of the impact of neutrino physics experiments, values of $\mathcal{O}(1)$ for these couplings, for squark masses of 300 GeV, are still not ruled out by other indirect processes [26].

Also in this case, two classes of elementary processes $q\bar{q}, gg \to t\, \bar{d}_k\, \tilde{l}_{Li}$ and $gd_k \to t\tilde{l}_{Li}$, with $d_k = d, s$ or $b$ for $k = 1, 2, 3$ respectively, are induced by the couplings $\lambda'_{i3k}$. Strictly speaking, strahlung off a top-quark line is obtained, more generally, also from couplings $\lambda'_{imk}$ with $m \neq 3$. The cross sections from these couplings, however, are suppressed by the factor $|V_{m3}|^2$ ($m = 1, 2$), which is smaller than $10^{-4}$. We therefore neglect this possibility and consider only the contribution from $\lambda'_{i3k}$.

In Fig. 5.11, we show the cross sections for strahlung production of $\tilde{\tau}_L$ (i.e. $i = 3$) at Tevatron and LHC energies, for the reference value $\lambda'_{33k} = 0.5$: solid lines denote $\sigma(p\bar{p}\ (pp) \to t\bar{d}_k\tilde{\tau}_L X)$, dashed lines denote $\sigma(p\bar{p}\ (pp) \to t\tilde{\tau}_L X)$. We assume here that the left–right mixing terms in the slepton mass matrix are small enough to render $\tilde{\tau}_L$ nearly a mass eigenstate, which we indicate simply by $\tilde{\tau}$ in the following. Obviously, the same cross sections are also obtained for the strahlung production of $\tilde{\mu}_L$ and $\tilde{e}_L$, or simply $\tilde{\mu}$ and $\tilde{e}$, when $\lambda'_{23k}$ and $\lambda'_{13k}$ are equal to 0.5. For each of the two sets of cross sections, induced by the $2 \to 2$ and the $2 \to 3$ processes, the three lines correspond, from top to bottom, to $k = 1, 2, 3$ in the coupling for $\lambda'_{33k}$, and therefore to $d_k = d, s, b$. We observe that among the hadronic cross sections induced by the $2 \to 2$ elementary processes, those indicated by the top lines in the two frames of Fig. 5.11 are initiated by a gluon and mainly a valence $d$ quark; those denoted by the central and bottom lines are initiated by a gluon and a sea quark, respectively $s$ and $b$. This is the reason for the larger values of the cross sections induced by $\lambda'_{331}$. Their enhancement with respect to those induced by $\lambda'_{333}$ is $\gtrsim 10$, the enhancement with respect to the cross sections induced by $\lambda'_{332}$ is roughly a factor of 5 at both colliders. Being mainly light-quark- or gluon-initiated, the cross sections induced by the $2 \to 3$ elementary processes, on the contrary, have all similar sizes for any value of $k$ in the couplings $\lambda'_{33k}$. All three curves are practically indistinguishable in the region $m_t > m_{\tilde{\tau}}$. The two curves corresponding to $\lambda'_{331}$ and $\lambda'_{332}$ are still very similar when $m_{\tilde{\tau}} \gtrsim m_t$. In this same region, they deviate from the curve corresponding to $\lambda'_{333}$, only slightly at the Tevatron, but by a factor of 2 at the LHC. This is essentially due to the large logarithms $\alpha_s(\mu_f)\ln(\mu_f/m_{d_k})$, originating from the $gg \to t\bar{d}_k\tilde{\tau}$ diagrams containing a virtual $d_k$ propagator [65], thus enhancing the cross-section for $d_k = s, d$ with respect to that for $d_k = b$. This effect is particularly evident at the LHC, where the $gg$-initiated processes largely dominate over the $q\bar{q}$ ones.

At both colliders, the overall situation for the production cross sections obtained for $k = 3$ is similar to that for the production of the charged Higgs boson: the resonant production, described by the cross section induced by the $2 \to 3$ processes, well exceeds the production induced by the $2 \to 2$ process in the region $m_t \gtrsim m_{\tilde{\tau}} + m_b$. The two production mechanisms are of similar size outside this region. The situation is different in the case in which $k = 2$, and much more so when $k = 1$: the cross section induced by the $2 \to 3$ processes can be neglected with respect to that due to the $2 \to 2$ process for $m_{\tilde{\tau}} \gtrsim m_t$ (at the precision of our calculation), and starts exceeding that induced by the $2 \to 2$ process only when $m_{\tilde{\tau}} < m_t$. There is indeed a region at $m_{\tilde{\tau}} \lesssim m_t$ in which the $2 \to 2$ process gives rise to a cross section still larger than that due to the decay of one of the two pair-produced top quarks, which, as already mentioned, is well described by the cross section induced by the $2 \to 3$ processes. For $k = 1$, this region is $150\,\mathrm{GeV} \lesssim m_{\tilde{\tau}} \lesssim m_t$ at the LHC, and $160\,\mathrm{GeV} \lesssim m_{\tilde{\tau}} \lesssim m_t$ at the Tevatron.

To obtain these cross sections, we have assumed that only one of the couplings $\lambda'_{33k}$ is present at a time. There is however no reason why this should be the case. Since $d$ and $s$ jets cannot be distinguished, at least the two sets of cross sections obtained for $k = 1$ and $k = 2$ give rise to the same final states. If the value of 0.5 is allowed for both couplings $\lambda'_{331}$ and $\lambda'_{332}$, for each of the two sets of cross sections, those obtained with these two couplings should be added. (Notice that, in this case, the width of the top quark in the $2 \to 3$ processes, in the kinematical region $m_t \lesssim m_{\tilde{\tau}} + m_b$, should be calculated accordingly, i.e. by considering the contribution from both couplings.) The case of $\lambda'_{333}$ is a little more complex and whether the corresponding cross sections can or cannot be distinguished from those induced by the couplings $\lambda'_{33k}$ with $k = 1, 2$ depends on the decay modes of $\tilde{\tau}$ and on how many $b$ quarks can be tagged. If it cannot be distinguished, and the same value for the three couplings $\lambda'_{33k}$ is allowed, the overall production cross sections for $\tilde{\tau}$ in the region $m_{\tilde{\tau}} < m_t$ can be considerably larger: three times the values indicated by solid lines in the two frames of Fig. 5.11. In the region $m_{\tilde{\tau}} > m_t$ the overall production cross section, however, still remains that induced by the coupling $\lambda'_{331}$.

This observation brings us back to the aforementioned issue of possible constraints induced on these couplings by neutrino physics. Neutrino masses get contributions induced by these couplings at the one- and two-loop-level. The expressions for these contributions are [22,26,67,88]

$$m_{\nu,ii'} \sim \frac{3}{8\pi^2}\,\lambda'_{ikj}\lambda'_{i'jk}\,m_{d_k}\,m_{d_j}\,\frac{(\mathcal{M}^k_{\tilde{d}\mathrm{LR}})_{kk}}{m^2_{\tilde{d}_k}}\,, \tag{5.59}$$

$$m_{\nu,ii'} \sim \frac{3g_j^2}{2(16\pi^2)^2}\,\lambda'_{ikk}\lambda'_{i'kk}\,m^2_{d_k}\,\frac{m_{\tilde{\chi}^0_j}}{m^2_{\tilde{\nu}}}\,\ln\!\left(\frac{m^2_{\tilde{d}_{k1}}\,m^2_{\tilde{d}_{k2}}}{m^4_{d_k}}\right), \tag{5.60}$$

where $j = 1, 2$, $g_1$ and $g_2$ are the $U(1)$ and $SU(2)$ gauge couplings, $m_{\tilde{\nu}}$ is the sneutrino mass, $m_{\tilde{\chi}^0_j}$ the $j$-th neutralino eigenvalue, $m_{d_k}$ the mass of the $k$-th down quark, $m_{d_k}(\mathcal{M}^k_{\tilde{d}\mathrm{LR}})_{kk}$ the LR mixing term of the $k$-th down-squark mass matrix, $m^2_{\tilde{d}_{ki}}$ are the two $k$-th down-squark eigenvalues, and $m^2_{\tilde{d}_k}$ is an average value between these two. Clearly, we have already assumed that intergenerational mixing terms in the squark mass matrix are negligible. Under this assumption, second and third generation indices in the $R_p$-violating couplings in the two-loop contribution must be equal, while this is not the case in the one-loop contribution. The importance of the two-loop contribution stem from the fact that the parameter $\mathcal{M}^k_{\tilde{d}\mathrm{LR}}$ can be very small. Even if this is not the case, if we take $m_{\tilde{d}_k} \sim 3m_{\tilde{\nu}}$, $m_{\tilde{\nu}} \sim m_{\tilde{\chi}^0_j}$, $\mathcal{M}^k_{\tilde{d}\mathrm{LR}} \sim m_{\tilde{d}_k}$, and squark masses at 300 GeV, it is evident that the one- and two-loop contributions to neutrino masses differ by only one order of magnitude, when $k = j$, therefore giving one- and two-loop constraints on the $\lambda'_{ikk}$ couplings that are numerically very similar. By imposing that the contributions to the neutrino mass in Eqs. (5.59) and (5.60), does not exceed the value of 1 eV, we obtain, up to coefficients of $\mathcal{O}(1)$:

$$\left|(\lambda'_{ikj}\lambda'_{i'jk})\right|_{1-\mathrm{loop}} \lesssim 10^{-6}\left(\frac{3\,\mathrm{GeV}}{m_{d_k}}\right)\left(\frac{3\,\mathrm{GeV}}{m_{d_j}}\right)\left(\frac{m_{\tilde{d}_k}}{300\,\mathrm{GeV}}\right)\left(\frac{300\,\mathrm{GeV}}{|\mathcal{M}^k_{\tilde{d}\mathrm{LR}}|}\right), \tag{5.61}$$

$$\left|(\lambda'_{ikk}\lambda'_{i'kk})\right|_{2-\mathrm{loop}} \lesssim 10^{-5}\left(\frac{3\,\mathrm{GeV}}{m_{d_k}}\right)^2\left(\frac{m_{\tilde{\nu}}}{100\,\mathrm{GeV}}\right)^2\left(\frac{100\,\mathrm{GeV}}{|m_{\tilde{\chi}^0}|}\right)\left(\frac{\ln(100)}{\ln(m_{\tilde{d}_k}/m_{d_k})}\right) \tag{5.62}$$

The constraints imposed on $\lambda'_{i33}$ ($j = k = 3$) by the two previous equations are rather severe. Even in the case in which $|\mathcal{M}^k_{\tilde{d}LR}|$ is practically vanishing, the two-loop consraints say that $|\lambda'_{i33}|$ cannot be larger than $3 \times 10^{-3}$. When $k = 3$ and $j \neq 3$, a constraint on $\lambda'_{i3j}$, the coupling responsible for the production of a charged slepton in association with a top quark, can only come from Eq. (5.59). Strictly speaking, however, this is a constraint on the product of two different couplings $\lambda'_{i3j}$ and $\lambda'_{ij3}$, and there are no a priori reasons why the suppression on the right-hand side has to be inherited only by $\lambda'_{i3j}$ and does not have to be shared equally between the two couplings, or even be completely borne out by $\lambda'_{ij3}$. (For $|\lambda'_{i3j}| \sim |\lambda'_{ij3}|$, when $j = 2$ it should be $|\lambda'_{i32}| \lesssim 5 \times 10^{-3}$, for $j = 1$, $|\lambda'_{i31}| \lesssim 3 \times 10^{-2}$. A suppression of $|\mathcal{M}^k_{\tilde{d}\mathrm{LR}}|$ could, however, ease out these upper bounds.) Clearly all these constraints can still be evaded if the contributions to neutrino masses in Eqs. (5.59) and (5.60) are cancelled by other one- and two-loop contributions induced by other $R_p$-violating couplings [22,67]. If this were the case and the value of 0.5 were allowed for both couplings $\lambda'_{i32}$ and $\lambda'_{i31}$, then the two sets of cross sections in Fig. 5.11 would have to be summed, as mentioned a little earlier. This would have practically no consequences for the cross sections induced by the $2 \to 2$ processes, but it would double the charged-slepton rate of production at the Tevatron (LHC) in the region $m_{\tilde{l}} \lesssim 160$ (150) GeV, where the $2 \to 3$ processes give the dominant contribution. If no more than $2b$'s can be detected in the final state obtained after the decay of the charged slepton, and if it could be $\lambda'_{i33} = 0.5$, also the cross sections generated by this coupling should be combined to the other two.

We focus on the couplings $\lambda'_{i3j}$ ($j = 1, 2$) to discuss the signature of the final state obtained after the slepton decay. Its main decay mode is $\tilde{l}_i \to l_i\chi^0$. The neutralino, in turn, can decay through the same coupling $\lambda'_{i31}$ into [89]: $b\bar{d}\nu_i$, $d\bar{b}\nu_i$, $t\bar{d}l_i$, $d\bar{t}l_i$. Particularly interesting is the mode $t\bar{d}l_i$; for a mass of $\tilde{l}_i$ larger than 160 GeV at the Tevatron or 150 GeV at the LHC, this gives rise to

$$t\,\tilde{l}_i \;\to\; t\,l_i\chi^0 \;\to\; 2t + 2l_i + \mathrm{jet} \;\to\; 2b + 2W + 2l_i + \mathrm{jet}\,, \tag{5.63}$$

with two equal-sign leptons. Notice that these two leptons do not have to be of the same type since two $\lambda'_{i31}$ couplings, with two different $i$ indices, may intervene at the level of production and of decay of the charged slepton. If one of the two $W$'s decays leptonically, the final state with three leptons, two $b$'s jets and missing energy cannot be overlooked. The obvious background for such a final state would be given by the decay products of a $t\bar{t}$ pair with $t\bar{t} \to b\bar{b}W^+W^-$, with both $W$'s decaying leptonically and one $b$ quark semileptonically. The identification of two $b$'s would then allow us to distinguish the signal from the background without too much loss in the signal, at least in the case in which there are no $\tau$'s among the leptons.

# 6 Extra Gauge Groups

## 6.1 Introduction

*Paul Langacker, Alexei Raspereza and Sabine Riemann*

### 6.1.1 Classes of models

Extended gauge symmetries and/or extra gauge bosons appear in many extensions of the standard model, such as left-right symmetric models [1], superstring motivated models [2], GUT (grand unification theory) [3], little Higgs models [4], large extra dimensions [5], and dynamical symmetry breaking [6]. In many cases, the extra symmetry is broken at the TeV scale, leading not only to additional gauge bosons, but also to an extended Higgs sector (needed to break the gauge symmetry), extended neutralino/chargino sectors [7, 8] (with implications for dark matter), new sources of CP violation at tree level in the Higgs sector [9, 10] (important for collider physics and baryogenesis), new fermions [11] (for anomaly cancellation), new sources of Higgs [12] or $Z'$ [13]-mediated flavor-changing neutral currents, and new constraints or parameter ranges for the Standard Model (SM) or MSSM. Here, we focus on the additional Higgs bosons, which may dramatically affect the Higgs collider signatures, taking the cases of $SU(2)_L \times U(1)_Y \times U(1)'$ [3] and left-right symmetric $SU(2)_L \times SU(2)_R \times U(1)_{B-L}$ [1] for definiteness.

Especially common are new $U(1)'$ gauge symmetries, broken by the expectation values of standard model singlets $S$. In most supersymmetric examples, the $U(1)'$ symmetry forbids elementary $\mu$ terms, but may allow trilinear superpotential couplings

$$W = h_s S H_u H_d. \tag{6.1}$$

If $S$ acquires a vacuum expectation value $\langle S\rangle$, an effective $\mu$ parameter $\mu_{eff} = h_s\langle S\rangle$ is generated. This is in the needed range for $h_s < O(0.8)$ (needed if $h_s$ is to remain perturbative up to the Planck scale) and $\langle S\rangle$ is in the 100 GeV-1 TeV range (expected if the $U(1)'$ and electroweak scales are both set by the scale of soft supersymmetry breaking). In this respect, the $U(1)'$ models are similar to the NMSSM and related models. However, there are no discrete symmetries and therefore no danger of cosmological domain walls.

The simplest class of models involve a single $S$ field. Then, the potential for $S$ and the neutral components of $H_{u,d}$ is given by

$$V = V_F + V_D + V_{soft}, \tag{6.2}$$

where

$$V_F = h_s^2\left(|H_d|^2|H_u|^2 + |S|^2|H_d|^2 + |S|^2|H_u|^2\right), \tag{6.3}$$

$$V_D = \frac{g_1^2+g_2^2}{8}\left(|H_u|^2 - |H_d|^2\right)^2 + \frac{1}{2}g_{Z'}^2\left(Q_S|S|^2 + Q_{H_d}|H_d|^2 + Q_{H_u}|H_u|^2\right)^2, \tag{6.4}$$

and

$$V_{soft} = m_{H_d}^2|H_d|^2 + m_{H_u}^2|H_u|^2 + m_S^2|S|^2 - (A_h h_s S H_d H_u + \text{H.C.}), \tag{6.5}$$

where $g_1,\ g_2$, and $g_{Z'}$ are respectively the $U(1)$, $SU(2)$, and $U(1)'$ gauge couplings, and $Q_i$ is the $U(1)'$ charge of particle $i$. Of course, the coupling (6.1) requires that $Q_S + Q_{H_d} + Q_{H_u} = 0$. The last $(A_h)$ term in $V_{soft}$ is the analog of the $B\mu$ term of the MSSM. (The MSSM limit of the model is obtained for $h_s \to 0$ with $\mu_{eff}$ held fixed.)

The spectrum of physical Higgses after symmetry breaking [14–26] consists of three neutral CP even scalars ($h_i^0$, $i = 1, 2, 3$), one CP odd pseudoscalar ($A^0$) and a pair of charged Higgses ($H^\pm$), i.e., it has one scalar more than in the MSSM. Masses for the three neutral scalars are obtained by diagonalizing the corresponding $3\times 3$ mass matrix. The tree level mass of the lightest scalar $h_1^0$ satisfies the bound

$$m_{h_1^0}^2 \le M_Z^2\cos^2 2\beta + \frac{1}{2}h_s^2 v^2\sin^2 2\beta + g'^2\overline{Q}_H^2 v^2, \tag{6.6}$$

where $\overline{Q}_H = \cos^2\beta Q_1 + \sin^2\beta Q_2$; $v^2 \equiv v_1^2 + v_2^2 \sim (246\text{ GeV})^2$, where $v_i \equiv \sqrt{2}\langle H_i^0\rangle$; and $\tan\beta \equiv v_2/v_1$. In contrast to the MSSM, $h_1^0$ can be heavier than $M_Z$ at tree level, both due to the F-term contributions (similar to the NMSSM) and the D-terms. Including the radiative corrections (which are similar to the MSSM), the MSSM upper bound of $\sim$ 130 GeV can be relaxed to $O(170)$ GeV.

The Higgs spectrum is particularly simple in the large $s \equiv \sqrt{2}\langle S\rangle$ case [22]. The mass of the lightest Higgs boson $h_1^0$ remains below the bound (6.6) and approaches

$$m^2_{h_1^0} \leq M_Z^2 \cos^2 2\beta + h_s^2 v^2 \left[\frac{1}{2}\sin^2 2\beta - \frac{h_s^2}{g'^2 Q_S^2} - 2\frac{\overline{Q}_H}{Q_S}\right]. \tag{6.7}$$

The limiting value (6.7) for $m_{h_1^0}$ can be larger or smaller than the MSSM upper bound $M_Z^2\cos^2 2\beta$, depending on couplings and charge assignments.

The pseudoscalar $A^0$ mass $m^2_{A^0} \simeq \sqrt{2}Ah_s s/\sin 2\beta$ is expected to be large (unless $Ah_s$ is very small), and one of the neutral scalars and the charged Higgs are then approximately degenerate with $A^0$, completing a full $SU(2)_L$ doublet $(H^0, A^0, H^\pm)$ not involved in $SU(2)_L$ breaking. The lightest neutral scalar is basically the (real part of the) neutral component of the Higgs doublet involved in $SU(2)_L$ breaking and has then a very small singlet component. The third neutral scalar has mass controlled by $M_{Z'}$ and is basically the singlet.

Most of these $U(1)'$ models require the existence of new heavy fermions carrying standard model charges to cancel anomalies [11], such as a heavy, $SU(2)$-singlet, quark $D_L + D_R$ with electric charge $-1/3$. These can be consistent with gauge unification if they fall into complete $SU(5)$ representations, as in the $E_6$ model [27]. The physics of a particular $E_6$ model is discussed in detail in [28, 29] and in Section 6.3. The Higgs sectors in $U(1)'$ models with a single Higgs field are compared and contrasted with other models involving a dynamical $\mu$ parameter in [30] and in Section 4 in this report.

In the single $S$ model, $\langle S\rangle$ is responsible both for $\mu_{eff}$ and the $Z'$ mass. The experimental lower limits on $M_{Z'}$ are model dependent, but are typically of order 600-900 GeV unless the$Z'$ has very weak couplings to ordinary quarks and leptons. In the former case, there is a tension between obtaining a large enough $Z'$ mass while generating the much lower electroweak scale [22, 27], requiring at least a small amount of tuning. This difficulty is resolved in the secluded sector models involving several $S$ fields. In particular, one $S$, whose expectation value is comparable to that of the doublet Higgs fields, generates $\mu_{eff}$, while all of the fields, some of which can have much larger expectation values, contribute to $M_{Z'}$. An explicit model in which this occurs naturally was constructed in [31]. In addition to the $S$ field related to $\mu_{eff}$ there are three addition complex scalar fields $S_i, i = 1, 2, 3$, with superpotential

$$W = h_s S H_u H_d + \lambda S_1 S_2 S_3. \tag{6.8}$$

In the limit $\lambda \to 0$ there would be an F and D flat direction involving the $S_i$ fields only, which would therefore acquire very large expectation values for appropriate soft breaking terms. For $\lambda$ small but nonzero (e.g., 0.05), one finds $\langle S_i\rangle \sim m_{S_i}/\lambda$, where $m_{S_i}$ is a soft mass, and large $M_{Z'}$. The secluded sector model has a very rich Higgs sector, involving 6 scalars and 4 pseudoscalars. (There can also be tree-level CP violation in the Higgs sector, which would lead to scalar-pseudoscalar mixing and with implications for baryogenesis [10].) The upper limit on the lightest scalar is relaxed, as in (6.6). However, the experimental *lower* limit of 114.4 GeV from LEP is also relaxed, because there can be considerable mixing between Higgs singlets and doublets (reducing the production rates), or the lightest scalar can be mainly singlet. There is often a light pseudoscalar, and low $\tan\beta$ values (e.g., $\sim 1-3$) are favored (these values are disfavored in the MSSM because of the Higgs mass limit). The Higgs sector was analyzed in detail in [32, 33] and in section 6.2. It was found in a parameter scan that the lightest Higgs could be as heavy as 168 GeV consistent with perturbatively to the Planck scale. A wide range of possibilities were found, depending on the parameters. These included both small and large values for the masses of the lightest scalar and pseudoscalar, and typically a fairly light neutralino. Many possibilities for Higgs decays were found, including MSSM-like decays into $b\bar{b}$, etc., Higgs decays into lighter Higgs (e.g., scalar into two pseudoscalars), invisible decays into the lightest neutralino, and cascade decays involving a heavier neutralino. A particular limit of the model, in which three of the singlets essentially decouple, is discussed in [30]. There are three Higgs scalars, consisting mainly of $S, H_d$, and $H_u$, one of which can be light, and three additional singlets, one of which is very heavy (around $M_{Z'}$).

Many authors have considered the gauge group $SU(2)_L \times SU(2)_R \times U(1)_{B-L}$ [1], which may emerge as a subgroup of the grand-unified $SO(10)$ group. A principal motivation is that it allows a left-right interchange symmetry $\psi_L \leftrightarrow \psi_R$ between left and right-handed fermions, so that parity is broken spontaneously, with $SU(2)_R \times U(1)_{B-L}$ breaking to $U(1)_Y$ (weak hypercharge). Original versions assumed that both $SU(2)_R$ and the LR symmetry are broken at low energies (e.g., the TeV scale). However, there are variants (e.g., motivated by gauge unification) involving low scale $SU(2)_R$ breaking and high scale (e.g. $10^{10}$ GeV) LR breaking [34, 35]. It is also possible for both to be broken at a high scale, as may be necessary in some supersymmetric versions to avoid the spontaneous breaking of electric charge [36].

The simplest non-supersymmetric version involves a Higgs bi-doublet

$$\phi = \begin{pmatrix} \phi_1^0 & \phi_1^+ \\ \phi_2^- & \phi_2^0 \end{pmatrix}, \tag{6.9}$$

which transforms as $(2, 2)$ under $SU(2)_L \times SU(2)_R$ with $B - L = 0$, and $\phi \leftrightarrow \phi^\dagger$ under the LR symmetry. The most popular version (which can also lead to a neutrino seesaw [1]) also introduces Higgs multiplets $\Delta_L$, and $\Delta_R$, which are respectively triplets under $SU(2)_L$ and $SU(2)_R$, with $B - L = 2$ and $\Delta_L \leftrightarrow \Delta_R$ under LR. They can be represented by the matrices

$$\Delta_{L,R} = \begin{pmatrix} \delta^+_{L,R}/\sqrt{2} & \delta^{++}_{L,R} \\ \delta^0_{L,R} & -\delta^+_{L,R}/\sqrt{2} \end{pmatrix}. \tag{6.10}$$

The expectation values of the neutral components are

$$\begin{aligned} \langle\phi\rangle &= \begin{pmatrix} v_1/\sqrt{2} & 0 \\ 0 & v_2/\sqrt{2} \end{pmatrix} \\ \langle\Delta_{L,R}\rangle &= \begin{pmatrix} 0 & 0 \\ V_{L,R}/\sqrt{2} & 0 \end{pmatrix}, \end{aligned} \tag{6.11}$$

which may be complex. The $SU(2)_R$ breaking is due to $V_R^0 \gg v_{1,2}^0 \gg V_L^0$, while the normal electroweak breaking is from $v_{1,2}^0$. Variant forms of the model replace $\Delta_{L,R}$ by $\kappa_{L,R}$, which transform respectively as $(2, 1)$ and $(1, 2)$ under $SU(2)_L \times SU(2)_R$.

The Yukawa couplings are

$$-\mathcal{L}_{\text{Yukawa}} = \bar{\Psi}_L \left(\gamma\phi + \tilde{\gamma}\tilde{\phi}\right) \Psi_R + L_L^T C i\tau_2 \Gamma_L \Delta_L L_L + L_R^T C i\tau_2 \Gamma_R \Delta_R L_R + \text{h.c.} \tag{6.12}$$

where $\psi_{L,R}$ can be left (right) handed quark or lepton doublets, $L_{L,R}$ are lepton doublets, $\tilde{\phi} = \tau_2\phi^*\tau_2$ is the charge-conjugated bi-doublet, and $C$ is the charge conjugation matrix. $\gamma$ and $\tilde{\gamma}$ are $3 \times 3$ Yukawa matrices (Hermitian by LR symmetry), while $\Gamma_{L,R}$ (equal by LR) are the $3 \times 3$ lepton-triplet couplings.

The minimal (non-supersymmetric) model in (6.12) therefore involves two Higgs doublets $\phi_{1,2}$ and two triplets $\Delta_{L,R}$. (More information on Higgs triplets may be found in Section 13.) After removing the Goldstone bosons eaten by the Higgs mechanism there are 14 physical Higgs degrees of freedom: 4 scalars, 2 pseudoscalars (which can mix with the scalars if CP is broken), two charged bosons and their charge conjugates, and two doubly charged bosons and their charge conjugates. Because the $\gamma$ and $\tilde{\gamma}$ couplings will in general not be diagonalized by the same unitary transformations, the fermion matrices will not be proportional to the physical Higgs-scalar Yukawa matrices (as in the standard model or MSSM). Therefore, the associated physical Higgs scalars will in general mediate flavor changing neutral currents. Explicit or spontaneous CP violation in the Higgs sector is also possible. Detailed studies of the Higgs sector include [37–44].

Supersymmetric generalizations of (6.12) cannot have the $\tilde{\gamma}$ term. (6.12) alone would then imply that the mass matrices for the charge $2/3$ and charge $-1/3$ quarks are proportional, leading to incorrect mass ratios and no CKM mixing. Such models would require either a second Higgs bi-doublet or other effects, such as significant soft supersymmetry breaking $A$ terms associated with the Yukawa matrix, which however are not aligned with $\gamma$. There are also doubly charged Higgs triplets associated with $\Delta_{L,R}$. These could be light [36], even if $SU(2)_R$ is broken at a high scale, as is assumed in most supersymmetric studies. Implications of these doubly charged states for leptonic flavor changing processes and for collider physics have been studied, e.g., in [45] and [46] and in section 6.4. The Higgs structure of $SU(2)_L \times SU(2)_R \times U(1)_{B-L}$ is very rich. More systematic studies, especially of possible collider signatures, would be very useful.

### *6.1.2 Experimental signatures of Higgs bosons*

Higgs boson phenomenology at future collider experiments can be illustrated using as an example the secluded $SU(2)_L \times U(1)_Y \times U(1)'$ model. The Higgs sector of this model consists of two MSSM like $SU(2)$ Higgs doublets and four additional Higgs singlets which are charged under an extra $U(1)'$ gauge symmetry. The spectrum of physical states comprises 6 CP-even scalars and 4 CP-odd states, denoted as $H_1$... $H_6$ and $A_1$...$A_4$, respectively, in order of increasing mass. One of the most striking features of this model is that $A_1$ is allowed to be very light, a feature shared with the NMSSM. To compare signatures of this model with the MSSM, it is convenient to introduce an "MSSM fraction" for a given state $H_i$ ($A_i$)

$$\epsilon^i_{MSSM} = \sum_{j=1}^{2} (R^{ji})^2, \tag{6.13}$$

where $R$ is the matrix relating interaction eigenstates to the mass eigenstates and $j$ runs over MSSM states. When $\epsilon^i_{MSSM}$ = 1, the mass eigenstate contains no admixture of singlet Higgs bosons and approaches the properties of the MSSM Higgs boson. With increasing singlet fraction in the mass state, which corresponds to decreasing $\epsilon^i_{MSSM}$, the Higgs state deviates in its properties from the MSSM Higgs boson. A large admixture of the singlet Higgs results in a reduced $ZZH$ coupling,

$$g^2_{ZZH_i} = (R^{i1}_H \sin\beta - R^{i2}_H \cos\beta)^2 g^2_{ZZH,SM}, \tag{6.14}$$

relative to the standard model, where $R_H$ is the matrix rotating CP-even interaction eigenstates to the mass basis. As a consequence, the Higgs-strahlung cross section is reduced with respect to the SM expectation, allowing a relaxation of the lower limit of 114 GeV from LEP.

Due to rather distinctive features of the Higgs sector from the SM and MSSM, it is important to study decay properties of the lightest CP-even and CP-odd Higgs bosons in order to explore their possible observation at future collider experiments. For the lightest CP-even Higgs boson with a mass below approximately 100 GeV, the LEP constraints require the Higgs to be mostly singlet. Thus, the decay modes to $A_1A_1$ and $\tilde{\chi}^0_1\tilde{\chi}^0_1$ are dominant when they are kinematically allowed, due to the presence of the extra $U(1)'$ gauge coupling and trilinear superpotential terms proportional to $h_s$ and $\lambda$ (Eq. 6.8). If these channels are kinematically disallowed, the properties of the lightest CP-even state become similar to those of MSSM Higgs bosons and decays to the heaviest SM fermions, $b\bar{b}$, $c\bar{c}$ and $\tau^+\tau^-$ become dominant. When the lightest scalar is heavier than the LEP2 bound, it may have a substantial "MSSM fraction" and can decay to $A_1A_1$ and $\tilde{\chi}^0_1\tilde{\chi}^0_1$ and SM particles. The light $A_1$ will decay dominantly to neutralinos when it is kinematically possible. Otherwise the $A_1$ decays into the heaviest accessible fermions, which are usually $b$ quarks, unless $A_1$ is lighter than the $b\bar{b}$ pair mass. Charm and tau decays can also be significant, depending on $\tan\beta$. For heavy $A_1 \geq 200$ GeV, decays to neutralinos and charginos universally dominate due to their gauge strength, suppressing $b\bar{b}$ below 10%. The $A_1$ and $H_1$ bosons can be lighter than $\tilde{\chi}_1^0$. However, in models with R-parity conservation, decays of $\tilde{\chi}_1^0$ to Higgs bosons are not allowed and the lightest neutralino is considered to be the lightest stable supersymmetric particle. Hence, the decay of lightest Higgs states into neutralinos are assumed to be invisible.

If Higgs bosons are discovered at the LHC, the linear collider will be an ideal machine to probe this class of models and disentangle them from the SM or MSSM. If due to specific decay modes the observation of Higgs bosons will be difficult at the LHC, the linear collider will serve as a discovery machine.

In electron-positron collisions the dominant production mechanisms for CP-even Higgs bosons are Higgs-strahlung, $e^+e^- \to ZH_i$, and $W$-fusion, $e^+e^- \to \nu\bar{\nu}H_i$ processes. The cross sections of these processes are reduced compared to the SM cross sections by a factor $(R_H^{i1}\sin\beta - R_H^{i2}\cos\beta)^2$, as in (6.14). As can be seen, the SM Higgs-strahlung cross section gets suppressed for a small amount of mixing into SM-like or MSSM-like Higgs for a given Higgs state and vanishes when the Higgs boson is mostly singlet. The CP-odd states are produced mainly through the Higgs pair production mechanism, $e^+e^- \to H_iA_j$, with the cross section given by

$$\sigma(e^+e^- \to H_iA_j) = (R_H^{i1}R_A^{j1} - R_H^{i2}R_A^{j2})^2 K\sigma_{SM}(e^+e^- \to Zh), \tag{6.15}$$

where $R_A$ is the matrix that diagonalizes the CP-odd Higgs mass matrix, $\sigma_{SM}(e^+e^- \to Zh)$ is the SM cross section for Higgs-strahlung, and $K$ is a kinematic factor given in [32].

The Higgs phenomenology at future colliders depends on the specific model point, defining Higgs boson mass spectrum, their production cross section and decay branching ratios. In the following we consider a set of representative scenarios, reflecting various experimental signatures in this class of models.

**MSSM like scenario**

When MSSM fractions are close to one, the model is MSSM like, with the production rates and decay modes similar to those expected in the MSSM. The observation of Higgs states is possible in the standard discovery channels anticipated for the SM and MSSM, even if production rates are reduced due to the admixture of the singlet component for a given Higgs state.

**Multijet final states**

For some parameter choices, the decay modes $H_i \to H_jH_j$ or $H_i \to A_jA_j$ become dominant, with subsequent decays of $H_j$ and $A_j$ into hadrons or tau-leptons. This scenario provides a challenge for the LHC because of large QCD backgrounds. At the linear $e^+e^-$ collider, the rich spectrum of signal topologies will be available for the detection of Higgs bosons.

The most promising channel is the Higgs-strahlung followed by decays of $Z$ into $e^+e^-$ and $\mu^+\mu^-$. The signal can be identified as the peak in the mass distribution of the system recoiling against the dilepton pair. The multi-jet channels, such as

- $H_1A_1 \to 4\,jets$,
- $ZH_2 \to q\bar{q}H_1H_1, q\bar{q}A_1A_1 \to 6\,jets$,
- $H_2A_1, H_1A_2 \to 6\,jets$,
- $H_2A_2 \to 8\,jets$

can also be exploited to detect Higgs bosons and measure their properties. Excellent performance of the vertex detector is very crucial for identification of these final states as they will include $b$ or $c$ quarks, stemming from the light CP-even and CP-odd Higgs decays. Good capability of identifying vertex charges would also be desirable as it will allow a reduction of the combinatorial background in multi-jet final states.

**Invisible decays**

Scenarios are possible when MSSM or SM decay modes of $H_i$ states are suppressed to the benefit of the $H_i \to \tilde{\chi}_1^0\tilde{\chi}_1^0, A_jA_j, H_jH_j$ channels with $A_j$ and $H_j$ decaying into the lightest neutralino. This scenario will present a challenge at the LHC. At the LC, however, the reconstruction of the mass peak is possible exploiting the Higgs-strahlung process followed by a visible decay of the $Z$, $Z \to q\bar{q}, e^+e^-, \mu^+\mu^-$. Dedicated analyses showed [47] that the invisible branching ratio of the Higgs can be measured with a relative precision of few % down to $Br(H \to inv.)$ = 0.1 for Higgs masses up to 160 GeV. This result is obtained for a Higgs-strahlung cross section close to the value expected in the SM.[1]

**Cascade decays involving supersymmetric particles**

For certain model parameter points one or more Higgs bosons can decay into heavy neutralinos. In this case, cascade decays of $\tilde{\chi}^0{}_{i>1}$ to the LSP will produce multi-fermion final states which may include both jets and leptons accompanied by large missing energy.

The class of models with extra gauge groups significantly enriches the Higgs boson phenomenology compared to the SM or MSSM. At this point it should emphasized that efficient identification of exotic channels, involving multi-jet final states, invisible decays and cascade decays of Higgs bosons to the LSP will be of crucial importance for disentangling these models from the SM and MSSM.

**Constraints from $Z'$ Searches**

Results from searches for extra gauge bosons constrain $U(1)'$ models. Dilepton searches at the Tevatron require $M_{Z'} > 600 - 900$ GeV, depending on the model [48], while results from weak neutral currents and LEP2 [49] restrict the $Z'$ mass and mixing. The strongest restrictions arise from the mixing between $Z$ and $Z'$ induced by electroweak symmetry breaking:

$$\alpha_{Z-Z'} = \frac{1}{2}\arctan\left(\frac{2M_{ZZ'}^2}{M_{Z'}^2 - M_Z^2}\right), \tag{6.16}$$

with the off-diagonal entry $M_{ZZ'}^2$ and the diagonal elements $M_{Z'}^2$, $M_Z^2$ in the mass-squared matrix. For typical models the $Z - Z'$ mixing is restricted by measurements at the $Z$ resonance to be less than a few $10^{-3}$ [49–51]. The small $Z - Z'$ mixing angle requires $M_{Z'} \gg M_Z$ or, with respect to existing results, $M_{Z'} > 500$ GeV.

If the fermions receive mass through the usual Higgs mechanism some of them must be charged under $U(1)'$ to keep the superpotential Yukawa terms gauge invariant (assuming the Higgs fields are charged). The exact $Z'$ production cross section depends on the fermion $U(1)'$ charges, but bosons with $M_{Z'} > 500$ GeV would be produced at tree level at the Tevatron and future colliders.

$Z'$ models as constructed in [31,32] can easily be satisfied for $M_{Z'}$ in the TeV range. Even higher $M_{Z'}$ values are allowed for large vacuum expectation values. Accordingly, the $Z'$ bosons can also be light if the singlets have smaller vacuum expectation values. This is possible since the singlets do not couple directly to the Standard Model.

In general, establishing the existence of new $U(1)'$ gauge symmetries necessitates experimental evidence also for $Z'$ bosons. Distinguishing different $U(1)'$ models would require certain signatures based on the nature of $Z'$ couplings which are not considered in this context of Higgs studies here.

[1] Invisible Higgs decays occur in a variety of models and are discussed in detail in the context of large extra dimensions in Section 8.

## 6.2 The Higgs sector in a secluded sector $U(1)'$ model

*Tao Han, Paul Langacker and Bob McElrath*

### 6.2.1 The model

#### 6.2.1.1 General structure

The model we [32] consider, first introduced in [31], has the superpotential:

$$W = hSH_uH_d + \lambda S_1S_2S_3 + W_{\rm MSSM}|_{\mu=0} \tag{6.17}$$

$S$, $S_1$, $S_2$, and $S_3$ are standard model singlets, but are charged under an extra $U(1)'$ gauge symmetry. The off-diagonal nature of the second term is inspired by string constructions, and the model is such that the potential has an $F$ and $D$-flat direction in the limit $\lambda \to 0$, allowing a large (TeV scale ) $Z'$ mass for small $\lambda$. The use of an $S$ field different from the $S_i$ in the first term allows a decoupling of $M_{Z'}$ from the effective $\mu$. $W$ leads to the $F$-term scalar potential:

$$\begin{aligned} V_F = & \; h^2\left(|H_u|^2|H_d|^2 + |S|^2|H_u|^2 + |S|^2|H_d|^2\right) \\ + & \; \lambda^2\left(|S_1|^2|S_2|^2 + |S_2|^2|S_3|^2 + |S_3|^2|S_1|^2\right) \end{aligned} \tag{6.18}$$

The $D$-term potential is:

$$\begin{aligned} V_D &= \frac{G^2}{8}\left(|H_u|^2 - |H_d|^2\right)^2 \\ &+ \frac{1}{2}g_{Z'}^2\left(Q_S|S|^2 + Q_{H_d}|H_1|^2 + Q_{H_u}|H_u|^2 + \sum_{i=1}^{3} Q_{S_i}|S_i|^2\right)^2 , \end{aligned} \tag{6.19}$$

where $G^2 = g_1^2 + g_2^2 = g_2^2/\cos^2\theta_W$. $g_1, g_2$, and $g_{Z'}$ are the coupling constants for $U(1), SU(2)$ and $U(1)'$, respectively, and $\theta_W$ is the weak angle. $Q_\phi$ is the $U(1)'$ charge of the field $\phi$. We will take $g_{Z'} \sim \sqrt{5/3}g_1$ (motivated by gauge unification) for definiteness.

We do not specify a SUSY breaking mechanism but rather parameterize the breaking with the soft terms

$$\begin{aligned} V_{\rm soft} &= m_{H_d}^2|H_d|^2 + m_{H_u}^2|H_u|^2 + m_S^2|S|^2 + \sum_{i=1}^{3} m_{S_i}^2|S_i|^2 \\ &- (A_h hSH_uH_d + A_\lambda \lambda S_1S_2S_3 + \text{H.C.}) \\ &+ (m_{SS_1}^2 SS_1 + m_{SS_2}^2 SS_2 + \text{H.C.}) \end{aligned} \tag{6.20}$$

The last two terms are necessary to break two unwanted global $U(1)$ symmetries, and require $Q_{S_1} = Q_{S_2} = -Q_S$. The potential $V = V_F + V_D + V_{soft}$ was studied in [31], where it was shown that for appropriate parameter ranges it is free of unwanted runaway directions and has an appropriate minimum. We denote the vacuum expectation values of $H_i$, $S$, and $S_i$ by $v_i, v_s$, and $v_{si}$, respectively, i.e., without a factor of $1/\sqrt{2}$. Without loss of generality we can choose $A_h h > 0$, $A_\lambda \lambda > 0$ and $m_{SS_i}^2 < 0$ in which case the minimum occurs for the expectation values all real and positive.

So far we have only specified the Higgs sector, which is the focus of this study. Fermions must also be charged under the $U(1)'$ symmetry in order for the fermion superpotential Yukawa terms $W_{fermion} = \bar{u}\mathbf{y_u}QH_u - \bar{d}\mathbf{y_d}QH_d - \bar{e}\mathbf{y_e}LH_d$ to be gauge invariant. The $U(1)'$ charges for fermions do not contribute significantly to Higgs production or decay, if sfermions and the $Z'$ superpartner are heavy. We therefore ignore them in this study.

Anomaly cancellation in $U(1)'$ models generally requires the introduction of additional chiral supermultiplets with exotic SM quantum numbers [2, 10, 11, 22, 27, 52]. These can be consistent with gauge unification, but do introduce additional model-dependence. The exotics can be given masses by the same scalars that give rise to the heavy $Z'$ mass. The exotic sector is not the focus of this study. We therefore consider the scenario in which the $Z'$ and other matter necessary to cancel anomalies is too heavy to significantly affect the production and decays of the lighter Higgs particles.

6.2.1.2 Higgs sector and electroweak symmetry breaking

The Higgs sector for this model contains 6 CP-even scalars and 4 physical CP-odd scalars, which we label $H_1 ... H_6$ and $A_1 ... A_4$, respectively, in order of increasing mass.

We find viable electroweak symmetry breaking minima by scanning over the vacuum expectation values of the six CP-even scalar fields. We require that the CP-even mass matrix be positive definite numerically, which guarantees a local minima, while simultaneously eliminating the soft mass squared for each field. The masses reported are evaluated including the dominant 1-loop correction coming from the top and stop loops. The CP-odd mass matrix is guaranteed to be positive semi-definite at tree level (and thus, all VEV's are real) by appropriate redefinitions of the fields and choices of parameters.

We scan over vacuum expectation values such that the three singlets $S_1$, $S_2$, and $S_3$ typically have larger VEV's than the other three fields. We allow points in our Monte Carlo scan that fluctuate from all VEV's equal up to $\langle S \rangle$ approximately 1 TeV and $\langle S_i \rangle$ approximately 10 TeV. This generically results in a spectrum with 1-5 relatively light CP-even states, often with one of them lighter than the LEP2 mass bound, but having a relatively small mixing with the MSSM $H_u$ and $H_d$. It is necessary that at least one of the singlets have an $\mathcal{O}(\mathrm{TeV})$ vacuum expectation value, so that the mass of the $Z'$ gauge boson is sufficently heavy that it evades current experimental bounds, and any extra matter needed to cancel anomalies is heavy enough to not significantly affect light Higgs production or decay.

A bound exists on the mass of the lightest Higgs particle in any perturbatively valid supersymmetric theory [18, 53]. The limit on the lightest MSSM-like CP-even Higgs mass in this model is:

$$\begin{aligned} M_h^2 \leq h^2 v^2 \quad &+ \quad (M_Z^2 - h^2 v^2) \cos^2 2\beta + 2 g_{Z'}^2 v^2 (Q_{H_u} \cos^2 \beta + \sin^2 \beta Q_{H_d})^2 \\ &+ \quad \frac{3}{4} \frac{m_t^4}{\pi^2 v^2} \ln \frac{m_{\tilde{t}_1} m_{\tilde{t}_2}}{m_t^2}. \end{aligned} \tag{6.21}$$

This is obtained by taking the limit as the equivalent of the $B$-term in the MSSM goes to infinity, $B = A_h h v_s \to \infty$, in the $2 \times 2$ submatrix containing $H_u$ and $H_d$. In the MSSM this is equivalent to taking $M_A \to \infty$, the decoupling limit. This expression is the same as in the NMSSM, except for the $g_{Z'}$ ($D$-term) contribution. Perturbativity to a GUT or Planck scale places an upper limit $\mathcal{O}(0.8)$ on $h$ [31], which is less stringent than the corresponding limit in the NMSSM [54–56] due to the $U(1)'$ contributions to its renormalization group equations. Larger values would be allowed if another scale entered before the Planck scale. We will allow $h$ as large as 1 in the interest of exploring the low energy effective potential. The second term of Eq. (6.21) vanishes for $\tan\beta = 1$. Since $\tan\beta \simeq 1$ generically in these models, the lightest Higgs mass is determined mostly by the new $F$ and $D$-term contributions proportional to $h^2$ and $g_{Z'}^2$. In this model, as with any model with many Higgs particles, a situation can arise in which the MSSM-like couplings are shared among many states, allowing unusually heavy states or unusually light states that evade current experimental bounds.

The four CP-odd masses can in principle be found algebraically but the results are complicated and not very illuminating. Perhaps the most striking feature of the mass spectrum is that the $A_1$ is allowed to be very light, a feature shared with the NMSSM [57–63]. This can lead to a much lighter CP-even higgs due to $H_1 \to A_1 A_1$ decays [64, 65] and very light dark matter due to the new s-channel annihilation through the $A_1$ [66]. This light $A_1$ is caused by a combination of small $m_{SS_1}^2$ or $m_{SS_2}^2$ and a small value of $v_s$ compared to the $v_{si}$. In the limit that $v_{si}$ ($i = 1$ or 2) is the largest scale in the problem, the lightest $A$ mass is

$$m_{A_1}^2 = -m_{SS_i}^2 \frac{v_s v_{si}}{v_{si}^2 + v_{s3}^2} + \mathcal{O}\left(\frac{1}{v_{si}^4}\right). \tag{6.22}$$

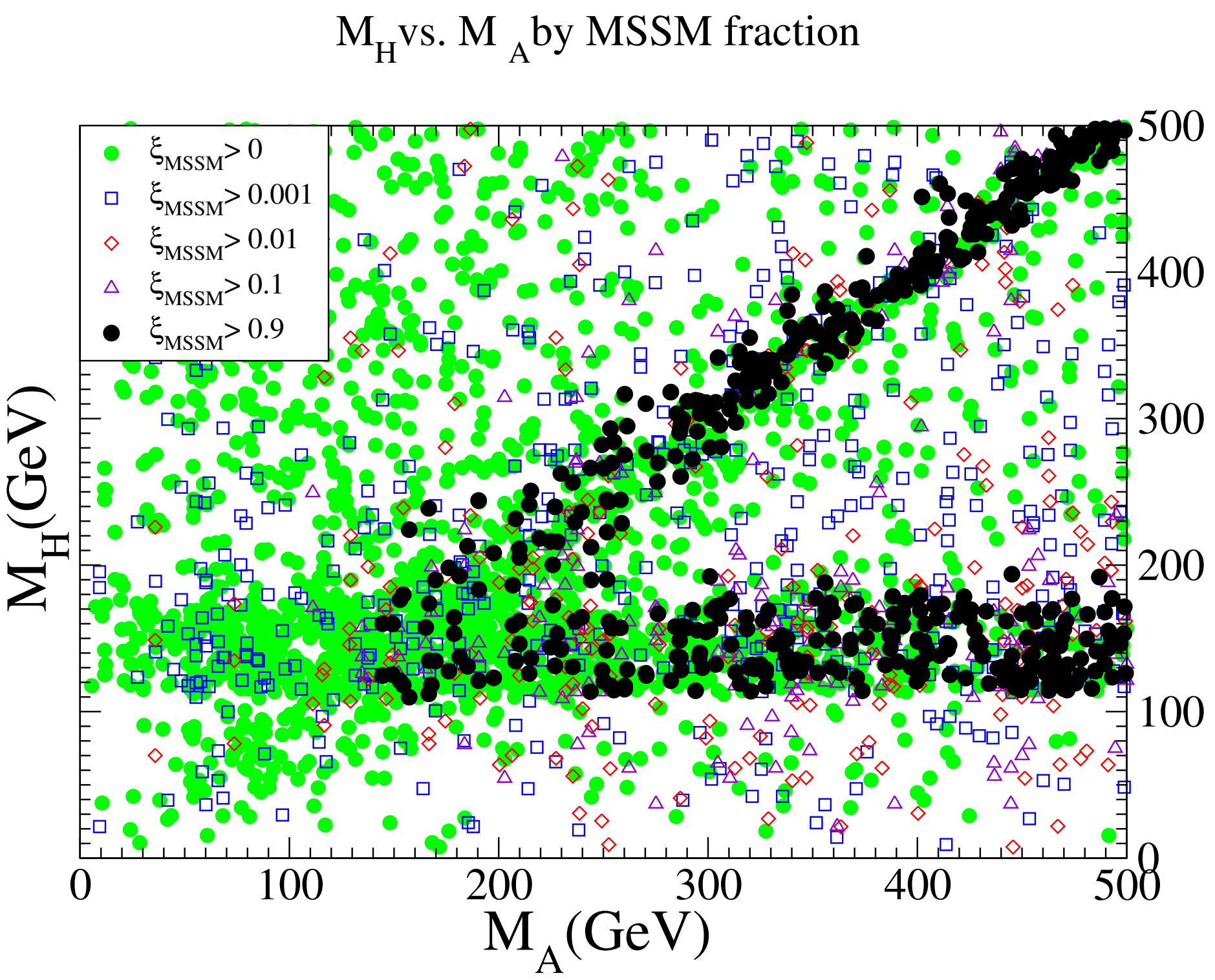

Fig. 6.1: $M_H - M_A$ mass plane, labeled according to MSSM fraction $\xi_{\rm MSSM}$. For each point both $H_i$ and $A_i$ satisfy the condition $\xi_{\rm MSSM} > 0, 0.001, 0.01, 0.1$, or $0.9$. All pairs $(M_{H_i}, M_{A_j})$ are plotted.

In the limit that $s_3$ is large we obtain

$$m^2_{A_1} = -4m^2_{SS_i}\frac{v_s v_{si}}{v_s^2 + v_{s1}^2 + v_{s2}^2} + \mathcal{O}\left(\frac{1}{v_{s3}}\right). \tag{6.23}$$

In our scans, $-m^2_{SS_i}$ is approximately in the range $(0 - 1000$ GeV$)^2$. However, this requires a hierarchy between the off-diagonal soft masses $m_{SS_i}$ and the other soft masses $m_S$ and $m_{S_i}$. This might be difficult to achieve depending on the SUSY breaking mechanism. A similar analysis holds for $H_1$, but an algebraic expression cannot be derived since the eigenvalues of a $6 \times 6$ matrix cannot be expressed algebraically.

### *6.2.2 Phenomenological constraints*

Due to the introduction of the Higgs singlets, there are several more parameters than in the MSSM Higgs sector. We follow the global symmetry breaking structure of Model I of Ref. [31]. Existing experimental measurements already constrain any new model. In our parameter space scans, we apply the constraints as outlined in Ref. [32]. All the model points shown on our figures are consistent with all the important constraints from LEP2.

### *6.2.3 Mass spectrum and couplings for Higgs bosons*

We first point out the relaxed upper bound on the mass of the lightest CP-even Higgs boson. As given in Eq. (6.21), the lightest CP-even Higgs boson mass at tree level would vanish in the limit $h \to 0$, $g_{Z'} \to 0$ and $\tan\beta \to 1$. Using the parameters discussed in Ref. [32], the upper limit on the lightest Higgs boson mass at tree level as given by the first two terms in Eq. (6.21) is $142\,\text{GeV}$. Including the effects of Higgs mixing and the one-loop top correction, we find masses up to $\sim 168\,\text{GeV}$. The mass could be made even larger if we allowed $h > 1$, although the perturbativity requirement up to the GUT scale at 1-loop level would imply that $h \leq 0.8$. We know that new heavy exotic matter must enter this model to cancel anomalies, so it is not necessarily justified to require $h$ to be perturbative to the Planck scale by calculating its 1-loop running using only low energy fields.

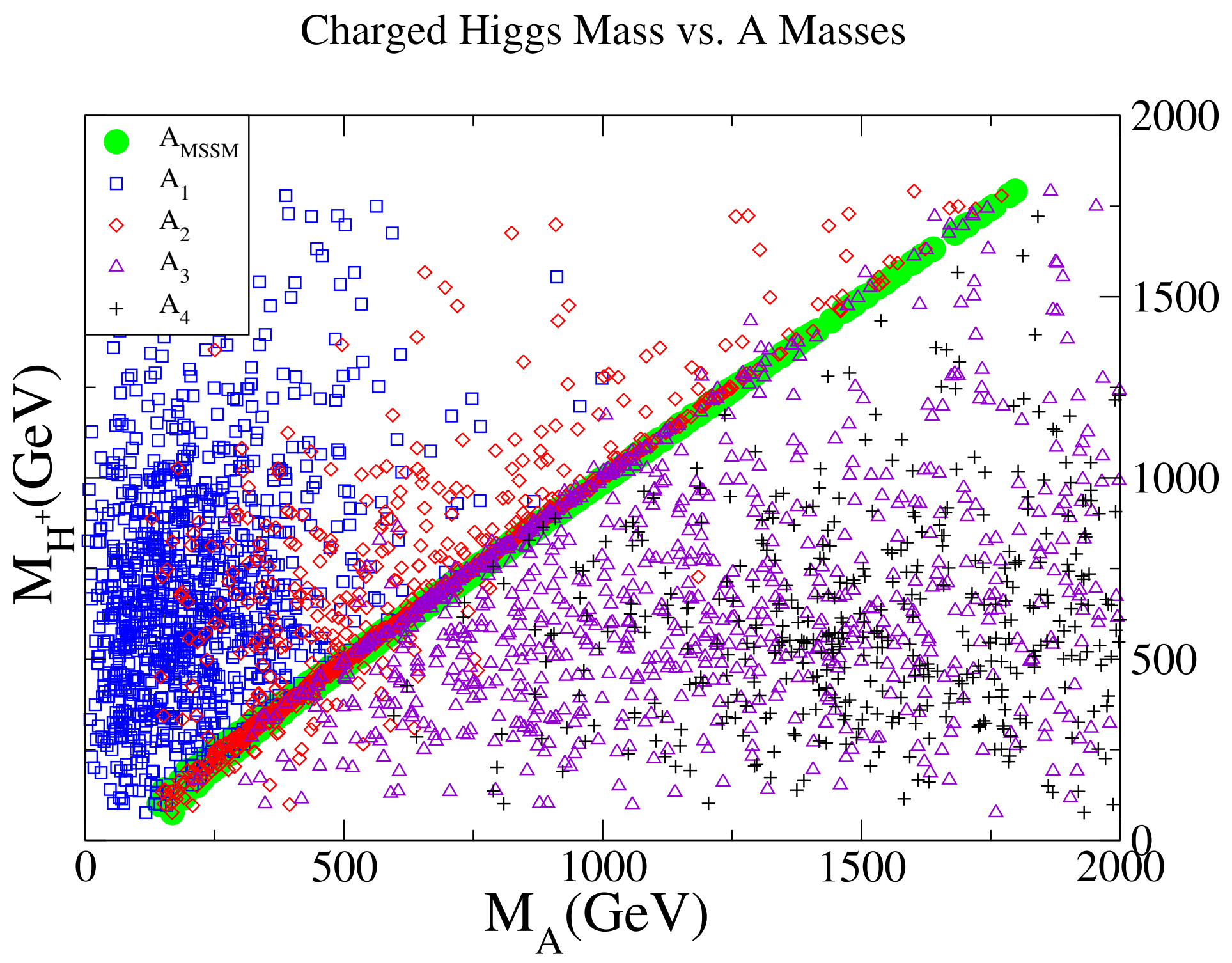

Fig. 6.2: $M_{H^+} - M_A$ mass plane with the MSSM $A^{\rm MSSM}$ mass $M_A^{\rm MSSM} = 2A_h h v_s / \sin 2\beta$ included for comparison. All pairs $(M_{H_i}, M_{A_j})$ are plotted.

The masses of the various Higgs particles are a function of the mixing parameters, and most of the simple MSSM relations among masses are broken. It is quite common to have a light singlet with sizable MSSM fraction that can still evade the LEP2 bounds. Typical allowed light CP-even and odd masses are shown in Fig. 6.1 for various ranges of MSSM fractions. We see that it is possible to have light MSSM Higgs bosons below about 100 GeV without conflicting the LEP2 searches. This is because of the reduced couplings to the $Z$ when the MSSM fraction becomes small. One can clearly make out the usual MSSM structure when $\xi_{\rm MSSM}$ is large, with the diagonal band for $\xi_{\rm MSSM} > 0.9$ being $M_H^{\rm MSSM} \simeq M_A^{\rm MSSM}$, and the horizontal band being the saturation of $M_h^{\rm MSSM}$ at its upper bound in the decoupling limit. As $\xi_{MSSM}$ decreases, we can see points in the lower left that are able to evade the LEP2 bounds on $M_{h,H}$ and $M_A$.

The mass range for the charged Higgs boson is demonstrated in Fig. 6.2. There is still a linear relationship between the charged Higgs mass and the MSSM $A$ mass since the singlets do not affect the $H^+$ mass. However, after mixing there is not necessarily a state with that mass, or the identity of the state is obscured. Most of the parameter space has a single state that can be identified as MSSM-like, with $\xi_{\rm MSSM} \sim 1$; in such circumstances there is also generally an $H$ very close in mass to both the $A$ and $H^+$. However, the difference between $M_{H^+}$ and the $M_{A_i}$ can be 50 GeV or more due to mixing, especially when the MSSM-like state is not clearly identifiable.

One of the most important parameters in the SUSY Higgs sector is $\tan\beta$. In the model under consideration, $\tan\beta \approx 1$ is favored (because $A_h$ must be large enough to ensure $SU(2)$ breaking).

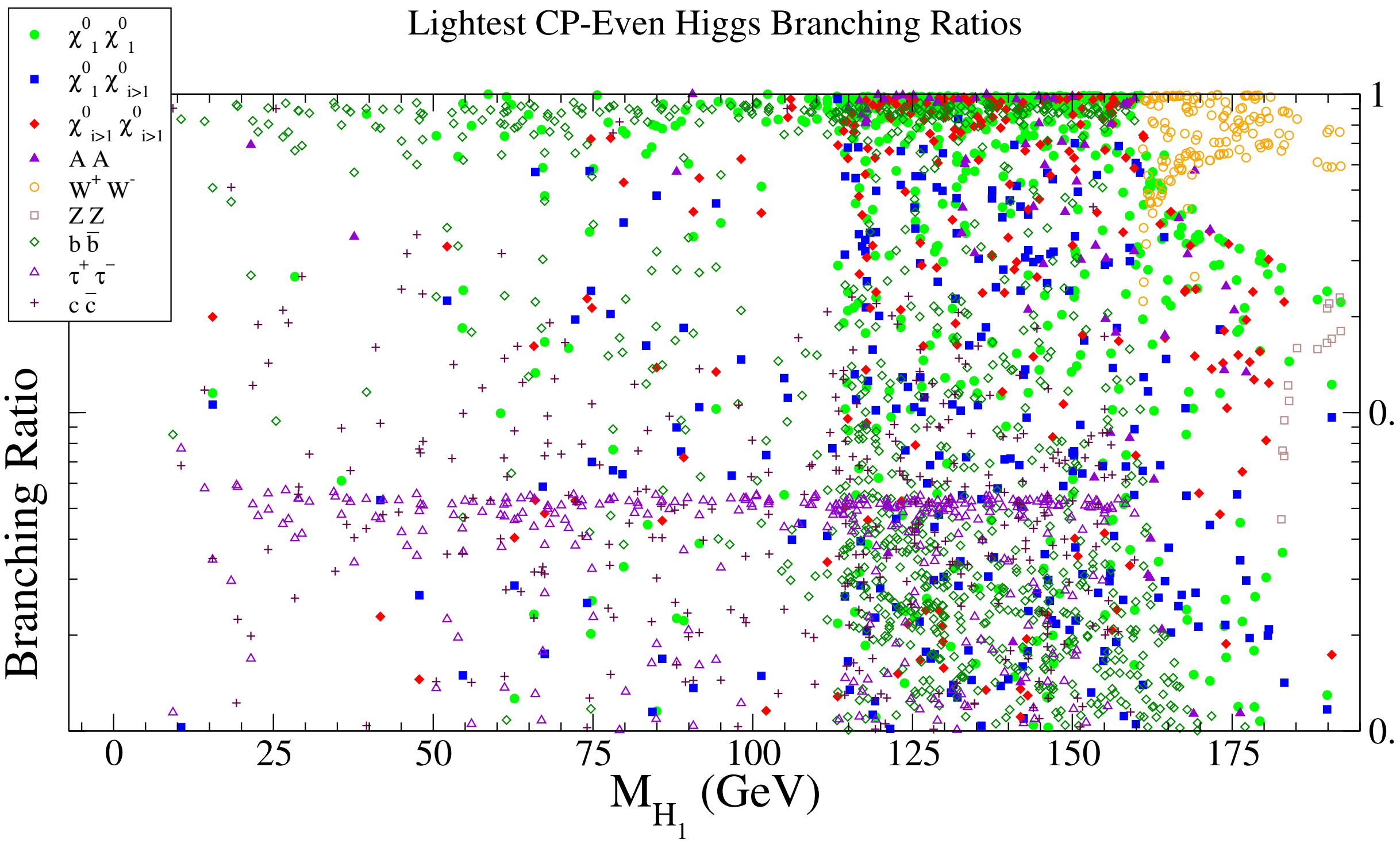

Fig. 6.3: Branching ratios of the lightest CP-even Higgs.

### 6.2.4 *Higgs boson decay and production in $e^+e^-$ collisions*

Due to the rather distinctive features of the Higgs sector different from the SM and MSSM, it is important to study how the lightest Higgs bosons decay in order to explore their possible observation at future collider experiments. The lightest Higgs bosons can decay to quite non-standard channels, leading to distinctive, yet sometimes difficult experimental signatures. For the Higgs boson production and signal observation, we concentrate on an $e^+e^-$ linear collider. It is known that a linear collider can provide a clean experimental environment to sensitively search for and accurately study new physics signatures. If the Higgs bosons are discovered at the LHC, a linear collider would be needed to disentangle the complicated signals in this class of models. If, on the other hand, a Higgs boson is not observed at the LHC due to the decay modes difficult to observe at the hadron collider environment, a linear collider will serve as a discovery machine.

#### 6.2.4.1 Lightest CP-even state $H_1$

The main decay modes and corresponding branching fractions for the lightest CP-even Higgs $H_1$ are presented in Fig. 6.3(left). For lightest Higgs masses below approximately 100 GeV, the LEP2 constraint is very tight, and the lightest Higgs must be mostly singlet. Thus, the decay modes to $A_1A_1$ and $\chi_1^0\chi_1^0$ are dominant when they are kinematically allowed, due to the presence of the extra $U(1)'$ gauge coupling and trilinear superpotential terms proportional to $h$ and $\lambda$. When those modes are not kinematically accessible, the decays are very similar to the MSSM modulo an eigenvector factor that is essentially how much of $H_u$ and $H_d$ are in the lightest state. Therefore $b\bar{b}$, $c\bar{c}$ and $\tau^+\tau^-$ decays dominate, with $c\bar{c}$ and $\tau^+\tau^-$ approximately an order of magnitude smaller than $b\bar{b}$, due to the difference in their Yukawa couplings. Since $\tan\beta \approx 1$, the $c\bar{c}$ mode can be competitive with both $\tau^+\tau^-$ and $b\bar{b}$ since their masses are similar. In the MSSM the $c\bar{c}$ mode is suppressed because $\tan\beta$ is expected to be larger.

When the lightest Higgs is heavier than the LEP2 bound, it does not need to be mostly singlet, and there can be a continuum of branching ratios to $A_1A_1$, $\chi_1^0\chi_1^0$ or SM particles, depending on how much singlet is in the lightest state. This is indeed seen in Fig. 6.3(right) for a heavier $H_1$ where the modes $H_1 \to W^+W^-$, $ZZ$ become substantial.

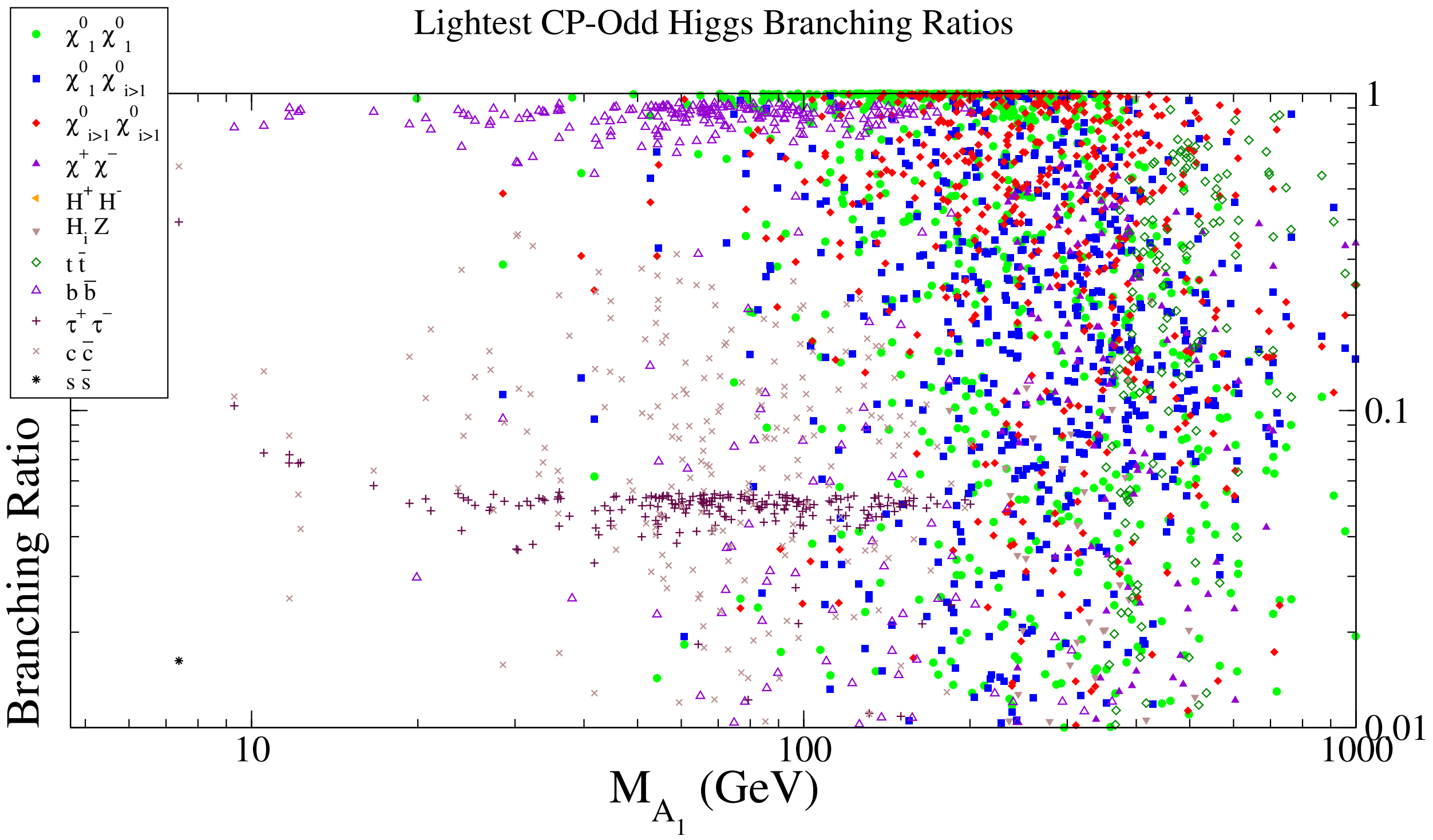

Fig. 6.4: Branching ratios for the lightest CP-odd Higgs.

A striking feature of this graph is that the usual "discovery" modes for $M_{H_1} < 140$, $H_1 \to b\bar{b}$, $\tau^+\tau^-$ are often strongly suppressed by decays to $A_1$ and $\chi^0_1$. Only $H_1 \to W^+W^-$, $ZZ$ decays are able to compete with the new $A_1$ and $\chi^0_1$ decays, which are all of gauge strength. One can see that the traditional shape of the $W^+W^-$ and $ZZ$ threshold is obscured by the presence of $\chi^0_1$ and $A$ decays, depending on what is kinematically accessible. For a $H_1$ heavy enough for these decay modes to be open, however, the coupling $h$ is typically greater than 0.8, large enough that it will become non-perturbative before the Planck scale unless new thresholds enter at a lower scale to modify its running.

The $A_1$ or $H_1$ can be lighter than the $\chi^0_1$. However, we assume R-parity is conserved. Therefore, decays of $\chi^0_1$ to $A_1$ or $H_1$ are not allowed and the lightest neutralino is assumed to be the (stable) LSP. We do not analyze the sfermion sector, which can produce a sfermion LSP in some regions of parameter space, but these scenarios are phenomenologically disfavored. We therefore assume $H$ and $A$ decays to $\chi^0_1$ are invisible at a collider. We separate the heavier neutralinos $\chi^0_{i>1}$ which may decay visibly [67].

### 6.2.4.2 CP-odd

The decays of the CP-odd Higgs bosons are presented in Fig. 6.4. The light $A_1$ will decay dominantly to neutralinos when it is kinematically possible. When it is not, it decays dominantly into the nearest mass SM fermion, which is usually $b$ unless the $A_1$ is lighter than the $b\bar{b}$ pair mass. Charm and tau decays can also be significant, depending on the value of $\tan\beta$. The $c\bar{c}$ decays are about 3 times more likely than the $\tau^+\tau^-$ due to the color factor. However, for larger $\tan\beta$ the $\tau^+\tau^-$ dominates.

For heavy $A_1 \gtrsim 200$ GeV, decays to neutralinos and charginos universally dominate due to their gauge strength, suppressing the $b\bar{b}$ mode below 10%.

The lightest $A$ can decay only into light SM fermions, the photon, and neutralinos. Hadronic bottom and charm decays are difficult to separate from background, and $\tau$'s are obscured by missing

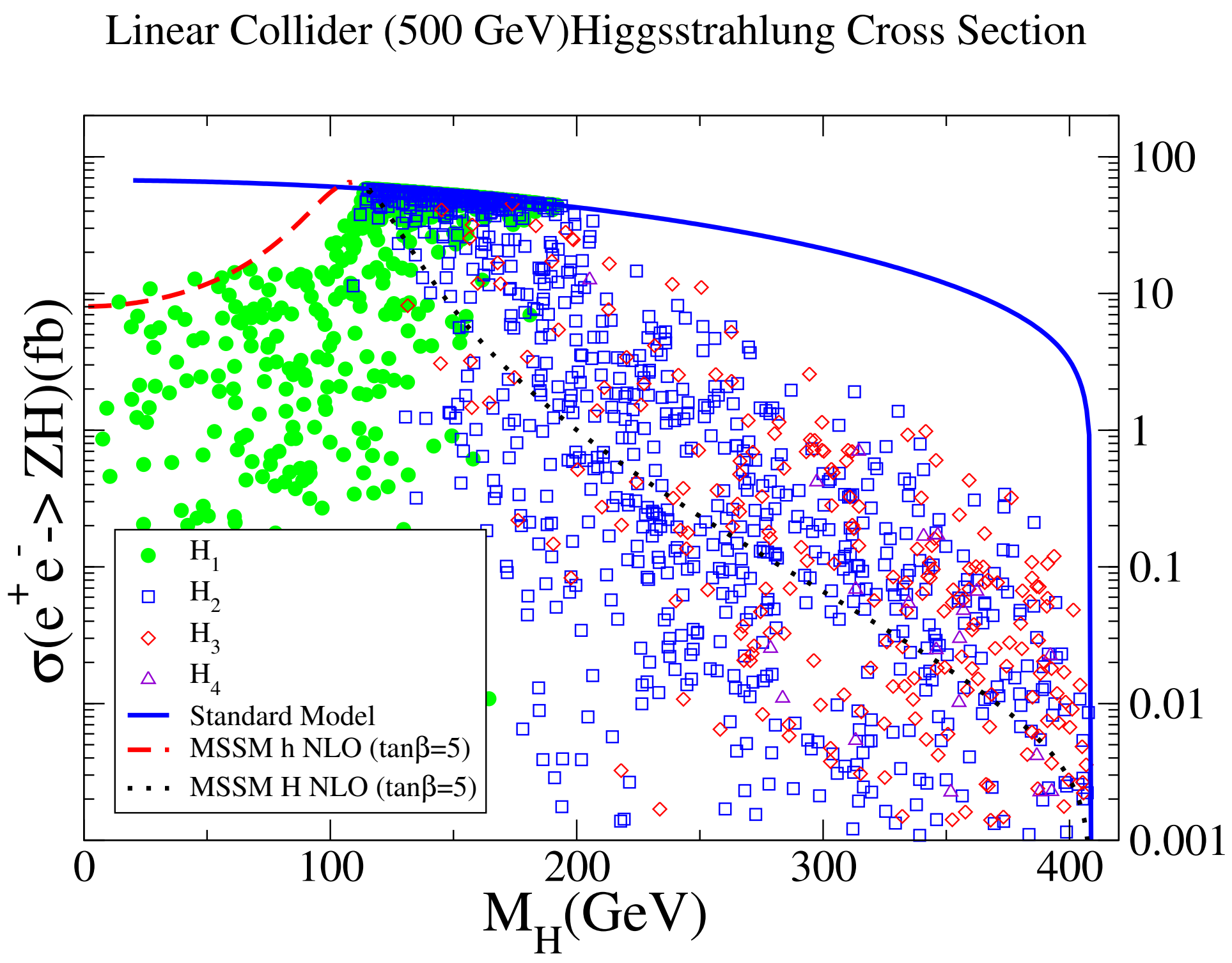

Fig. 6.5: Cross section at a 500 GeV linear collider for $ZH_i$ production. The solid curve is the SM production, and the dashed and dot-dashed for MSSM $h$ and $H$ production with $\tan\beta = 5$.

energy and hadronic background.

#### 6.2.4.3 The Higgs signatures at a linear collider

The production via radiation of a Higgs from a virtual $Z$ boson is the dominant mechanism for CP-even Higgs production at a linear collider. We show this cross section in Fig. 6.5, where each point is a viable model solution satisfying all the constraints. The curves present the SM and MSSM cross sections for comparison. Model points with $M_H < 114.1$ are only those with suppressed coupling to the $Z$, and those with large MSSM fraction are removed by the LEP2 bounds. The ratio between the Standard Model cross section and that for any model point simply reflects the amount of mixing into the SM-like or MSSM-like Higgs for a given Higgs state.

The production cross sections for the heavier Higgs particles are very small. For heavy states (that correspond to the $H$ in MSSM), $\cos(\alpha - \beta) \to 0$ as the $H$ gets heavier. In this decoupling limit of the MSSM the heavy $H$ has no coupling to the $Z$.

At 500 GeV the weak boson fusion production modes $e^+e^- \to \nu\bar{\nu}H$, $e^+e^-H$ are comparable in size to the Higgsstrahlung mode. At higher energies, the weak boson fusion becomes larger than Higgsstrahlung and is the most important production mode. These curves are similar to Fig. 6.4(b), reflecting that all of these single Higgs production modes are simply a mixing factor times the Standard Model curve. It is particularly interesting to note that the $ZZ$ fusion channel $e^+e^- \to e^+e^-H$ can serve as a model-independent process to measure the $ZZH$ coupling regardless the decay of $H$, even if $H$ is invisible [68].

As anticipated for the next generation linear collider with $\sqrt{s} = 500$ GeV and an integrated luminosity of the order of $500 - 1000$ fb$^{-1}$, one should be able to cover a substantial region of the parameter space. For instance, with a cross section of the order of 0.1 fb, this may lead to about 50−100 events. As for further exploration of signal searches, it depends on specific model parameters. We have provided a comprehensive list of representative models in the Appendices of Ref. [32].

It is clear that the model studied in this paper presents very rich physics in the Higgs sector. An $e^+e^-$ linear collider will be ideally suited for the detailed exploration of the non-standard Higgs physics. Analyses for the LHC should also be performed, particularly for the non-MSSM modes [69, 70].

### 6.3 Higgs spectrum in the exceptional supersymmetric standard model

*Steve F. King, Stefano Moretti and Roman Nevzorov*

#### 6.3.1 The model

A solution to the $\mu$-problem discussed in the Introduction of this section naturally arises within superstring inspired models based on the $E_6$ gauge group. At the string scale, $E_6$ can be broken directly to the rank-6 subgroup $SU(3)_C \times SU(2)_L \times U(1)_Y \times U(1)_\psi \times U(1)_\chi$ via the Hosotani mechanism [71]. Two anomaly-free $U(1)_\psi$ and $U(1)_\chi$ symmetries of the rank-6 model are defined by: $E_6 \to SO(10) \times U(1)_\psi$, $SO(10) \to SU(5) \times U(1)_\chi$. Near the string scale the rank-6 model can be reduced further to an effective rank-5 model with only one extra $U(1)'$ gauge symmetry. Thus in general the extra $U(1)'$ that appears at low energies in superstring inspired models is a linear combination of $U(1)_\chi$ and $U(1)_\psi$, i.e. $U(1)' = U(1)_\chi \cos\theta + U(1)_\psi \sin\theta$. If $\theta \neq 0$ or $\pi$ the extra $U(1)'$ gauge symmetry forbids an elementary $\mu$-term but allows the interaction $h_s S H_d H_u$ in the superpotential. After electroweak symmetry breaking (EWSB) the scalar component of the standard model (SM) singlet superfield $S$ acquires a non-zero vacuum expectation value (VEV) breaking $U(1)'$ and giving rise to an effective $\mu$ term.

Here we explore the Higgs sector of a particular $E_6$ inspired supersymmetric model with extra $U(1)_N$ gauge symmetry in which right handed neutrinos do not participate in the gauge interactions ($\theta = \arctan\sqrt{15}$). Only in this exceptional supersymmetric standard model ($E_6$SSM) right-handed neutrinos may be superheavy, shedding light on the origin of the mass hierarchy in the lepton sector and providing a mechanism for the generation of lepton and baryon asymmetry of the Universe [28]- [29]. Recently the implications of SUSY models with an additional $U(1)_N$ gauge symmetry have been studied for the neutrino physics [72]- [73], leptogenesis [74] and electroweak baryogenesis [75]. Previously supersymmetric models with an extra $U(1)_N$ factor have been also considered in [76]- [77] in the context of $Z - Z'$ mixing and a discussion of the neutralino sector and in [24] where the one-loop upper bound on the lightest Higgs was examined.

To ensure anomaly cancellation the particle content of the $E_6$SSM should include complete fundamental 27 representations of $E_6$. These multiplets decompose under the $SU(5) \times U(1)_N$ subgroup of $E_6$ [78] as follows:

$$27_i \to \left(10, \frac{1}{\sqrt{40}}\right)_i + \left(5^*, \frac{2}{\sqrt{40}}\right)_i + \left(5^*, -\frac{3}{\sqrt{40}}\right)_i + \left(5, -\frac{2}{\sqrt{40}}\right)_i + \left(1, \frac{5}{\sqrt{40}}\right)_i + (1, 0)_i \,. \tag{6.24}$$

The first and second quantities in the brackets are the $SU(5)$ representation and extra $U(1)_N$ charge while $i$ is a family index that runs from 1 to 3. An ordinary SM family which contains the doublets of left-handed quarks $Q_i$ and leptons $L_i$, right-handed up- and down-quarks ($u_i^c$ and $d_i^c$) as well as right-handed charged leptons, is assigned to $\left(10, \frac{1}{\sqrt{40}}\right)_i + \left(5^*, \frac{2}{\sqrt{40}}\right)_i$. Right-handed neutrinos $N_i^c$ should be associated with the last term in Eq. (6.24), $(1, 0)_i$. The next-to-last term in Eq. (6.24), $\left(1, \frac{5}{\sqrt{40}}\right)_i$, represents SM-type singlet fields $S_i$ which carry non-zero $U(1)_N$ charges and therefore survive down to the EW scale. The pair of $SU(2)$-doublets ($H_{1i}$ and $H_{2i}$) that are contained in $\left(5^*, -\frac{3}{\sqrt{40}}\right)_i$ and $\left(5, -\frac{2}{\sqrt{40}}\right)_i$ have the quantum numbers of Higgs doublets. Other components of these $SU(5)$ multiplets form color triplets of exotic quarks $\overline{D}_i$ and $D_i$ with electric charges $-1/3$ and $+1/3$ respectively. However these exotic quark states carry a $B-L$ charge $\left(\pm\frac{2}{3}\right)$ twice larger than that of ordinary ones. Therefore in the phenomenologically viable $E_6$ inspired models they can be either diquarks or leptoquarks.

In addition to the complete $27_i$ multiplets some components of the extra $27'$ and $\overline{27'}$ representations must survive to low energies in order to preserve gauge coupling unification. We assume that an additional $SU(2)$ doublet components $H'$ of $\left(5^*, \frac{2}{\sqrt{40}}\right)$ from a $27'$ and corresponding anti-doublet $\overline{H}'$ from $\overline{27'}$ survive to low energies. Thus in addition to a $Z'$ the E$_6$SSM involves extra matter beyond the MSSM that forms three $5+5^*$ representations of $SU(5)$ plus three $SU(5)$ singlets with $U(1)_N$ charges.

The superpotential in $E_6$ inspired models involves a lot of new Yukawa couplings in comparison to the SM. In general these new interactions induce non-diagonal flavour transitions. To suppress flavour changing processes one can postulate a $Z_2^H$ symmetry under which all superfields except one pair of $H_{1i}$ and $H_{2i}$ (say $H_d \equiv H_{13}$ and $H_u \equiv H_{23}$) and one SM-type singlet field ($S \equiv S_3$) are odd. The $Z_2^H$ symmetry reduces the structure of the Yukawa interactions to:

$$
\begin{aligned}
W_{\text{ESSM}} \;\simeq\;& \lambda_i S(H_{1i}H_{2i}) + \kappa_i S(D_i\overline{D}_i) + f_{\alpha\beta}S_\alpha(H_dH_{2\beta}) + \tilde{f}_{\alpha\beta}S_\alpha(H_{1\beta}H_u)+ \\
&+h_t(H_uQ)t^c + h_b(H_dQ)b^c + h_\tau(H_dL)\tau^c\,,
\end{aligned}
\tag{6.25}
$$

where $\alpha,\beta = 1,2$ and $i = 1,2,3$. In Eq. (6.25) we keep only Yukawa interactions whose couplings are allowed to be of order unity and ignore $H'$ and $\overline{H}'$ for simplicity. Here we define $h_s \equiv \lambda_3$. The $SU(2)$ doublets $H_u$ and $H_d$ play the role of Higgs fields generating the masses of quarks and leptons after EWSB. Therefore it is natural to assume that only $S$, $H_u$ and $H_d$ acquire non-zero VEVs. If $h_s$ or $\kappa_i$ are large at the grand unification (GUT) scale $M_X$ they affect the evolution of the soft scalar mass $m_S^2$ of the singlet field $S$ rather strongly resulting in negative values of $m_S^2$ at low energies that trigger the breakdown of the $U(1)_N$ symmetry. To guarantee that only $H_u$, $H_d$ and $S$ acquire a VEV we impose a certain hierarchy between the couplings $H_{1i}$ and $H_{2i}$ to the SM-type singlet superfields $S_i$: $h_s \gg \lambda_{1,2},\ f_{\alpha\beta}$ and $\tilde{f}_{\alpha\beta}$.

Although $Z_2^H$ eliminates any problem related with non-diagonal flavour transitions it also forbids all Yukawa interactions that would allow the exotic quarks to decay. Since models with stable charged exotic particles are ruled out by different experiments [79] the $Z_2^H$ symmetry must be broken. However even a small violation of this discrete symmetry permits to get a phenomenologically acceptable model. Because the Yukawa interactions of exotic particles to quarks and leptons of the first two generations give an appreciable contribution to the amplitude of $K^0-\overline{K}^0$ oscillations and give rise to new muon decay channels like $\mu \to e^-e^+e^-$ we assume that the violation of the $Z_2^H$ symmetry in the E$_6$SSM is mainly caused by the Yukawa couplings of the exotic particles to the quarks and leptons of the third generations.

### *6.3.2 Higgs and collider phenomenology*

The potential of the E$_6$SSM Higgs sector that involves two $SU(2)$ doublets $H_u$ and $H_d$ as well as the SM-type singlet field $S$ is given by Eqs. (6.2)-(6.5) in the Introduction of this section. The value of the extra $U(1)_N$ gauge coupling $g_{Z'}$ appearing in the Higgs scalar potential can be determined assuming gauge coupling unification. It turns out that for any renormalisation scale $Q$ below the unification scale ($Q < M_X$) $g_{Z'}(Q) \simeq \sqrt{\frac{5}{3}}g_1(Q)$, where $g_1(Q)$ is the $U(1)_Y$ gauge coupling ($g_1(M_Z) \simeq 0.36$) [28]. The only new coupling in the Higgs sector is then $h_sSH_dH_u$ which shows that the Higgs sector of the E$_6$SSM contains only one additional singlet field and one extra parameter compared to the MSSM. Therefore it can be regarded as the simplest extension of the Higgs sector of the MSSM.

At the physical vacuum Higgs fields develop the VEVs $\langle H_d\rangle = \frac{v_d}{\sqrt{2}}$, $\langle H_u\rangle = \frac{v_u}{\sqrt{2}}$ and $\langle S\rangle = \frac{s}{\sqrt{2}}$, thus breaking the $SU(2)_L\times U(1)_Y\times U(1)_N$ symmetry to $U(1)_{\text{EM}}$, associated with electromagnetism. Instead of $v_d$ and $v_u$ it is more convenient to use $\tan\beta = \frac{v_u}{v_d}$ and $v = \sqrt{v_d^2+v_u^2}$, where $v = 246\,\text{GeV}$. After the breakdown of the gauge symmetry two CP-odd and two charged Goldstone modes in the Higgs sector are absorbed by the $Z$, $Z'$ and $W^{\pm}$ gauge bosons so that only six physical degrees of freedom are left. They represent three CP-even (as in the NMSSM), one CP-odd and two charged Higgs states (as in the MSSM).

When the mass of the CP-odd Higgs boson $m_{A^0}$ is considerably larger than $M_Z$ the tree-level masses of the Higgs particles can be written as [28]

$$m^2_{A^0} \simeq m^2_{H^\pm} \simeq m^2_{h^0_3} \simeq \frac{2h_s^2 s^2 x}{\sin^2 2\beta}\,, \qquad\qquad m^2_{h^0_2} \simeq g^2_{Z'} Q_S^2 s^2\,, \tag{6.26}$$

$$\begin{aligned} m^2_{h^0_1} \;\simeq\; & \frac{h_s^2}{2} v^2 \sin^2 2\beta + \frac{\bar{g}^2}{4} v^2 \cos^2 2\beta + g^2_{Z'} v^2 \biggl( Q_{H_1} \cos^2\beta + Q_{H_2} \sin^2\beta \biggr)^2 - \\ & - \frac{h_s^4 v^2}{g^2_{Z'} Q_S^2} \biggl( 1 - x + \frac{g^2_{Z'}}{h_s^2} \biggl( Q_{H_1} \cos^2\beta + Q_{H_2} \sin^2\beta \biggr) Q_S \biggr)^2 \,, \end{aligned} \tag{6.27}$$

where $m_{H^\pm}$ and $m_{h^0_i}$ are the masses of charged and CP-even states respectively while $x = \dfrac{A_h}{\sqrt{2} h_s s} \sin 2\beta$. From Eqs. (6.26)-(6.27) it follows that at tree level the Higgs spectrum can be parametrised in terms of four variables only: $h_s$, $s$, $\tan\beta$, $m_{A^0}$ (or $x$). As one can see at least one CP-even Higgs boson is always heavy preventing the distinction between the E$_6$SSM and MSSM Higgs sectors. Indeed the mass of the singlet dominated Higgs scalar particle $m_{h^0_2}$ is always close to the mass of the $Z'$ boson $M_{Z'} \simeq g_{Z'} Q_S s \sim g_{Z'} s$ that has to be heavier than $600 - 800\,$GeV. The masses of the charged, CP-odd and one CP-even Higgs states are governed by $m_{A^0}$. The mass of the SM-like Higgs boson given by Eq. (6.27) is set by $M_Z$. The last term in Eq. (6.27) must not be allowed to dominate since it is negative. This constrains $x$ around unity for $h_s > g_{Z'}$. As a consequence $m_{A^0}$ is confined in the vicinity of $\dfrac{h_s s}{\sqrt{2}} \tan\beta$ and is much larger than the masses of the $Z'$ and $Z$ bosons. At so large values of $m_{A^0}$ the masses of the heaviest CP-even, CP-odd and charged states are almost degenerate around $m_{A^0}$.

The qualitative pattern of the Higgs spectrum obtained for $h_s > g_{Z'}$ is shown in Fig. 6.6 where we plot the masses of the Higgs bosons as a function of $m_{A^0}$. As a representative example we fix $\tan\beta = 2$ and the VEV of the singlet field $s = 1.9\,$TeV, corresponding to $M_{Z'} \simeq 700\,$GeV, which is quite close to the current limit on the $Z'$ boson mass. For our numerical study we also choose the maximum possible value of $h_s(M_t) \simeq 0.794$ which does not spoil the validity of perturbation theory up to the GUT scale for $\tan\beta = 2$. In order to obtain a realistic spectrum, we include the leading one-loop corrections from the top and stop loops that depend rather strongly on the soft masses of the superpartners of the top-quark ($m_Q^2$ and $m_U^2$) and on the stop mixing parameter $X_t$. Here and in the following we set $m_Q = m_U = M_S = 700\,$GeV while $X_t$ is taken to be $\sqrt{6}\,M_S$ in order to enhance stop-radiative effects.

The numerical analysis confirms the analytic tree-level results discussed above. From Fig. 6.6 it becomes clear that for $m_{A^0}$ below $2\,$TeV or above $3\,$TeV the mass squared of the lightest Higgs boson tends to be negative. Negative value of $m^2_{h^0_1}$ implies that the considered vacuum configuration is unstable, i.e. there is a direction in field space along which the energy density decreases. The requirement of stability of the physical vacuum therefore limits the range of variations of $m_{A^0}$ from below and above. Together with the experimental lower limit on the mass of the $Z'$ boson it maintains the mass hierarchy in the spectrum of the Higgs particles seen in Fig. 6.6. The numerical analysis also reveals that the heaviest CP-even, CP-odd and charged Higgs states lie beyond the TeV range when $h_s > g_{Z'}$. The second lightest CP-even Higgs boson is predominantly singlet so that it will be quite difficult to observe at colliders.

When $h_s < g_{Z'}$ the allowed range of $m_{A^0}$ enlarges. Although the requirement of vacuum stability still prevents having very high values of $m_{A^0}$ (or $x$) the mass squared of the lightest Higgs boson remains positive even if charged, CP-odd and second lightest CP-even Higgs states lie in the $200-300\,$GeV range. But for $m_{A^0} < 500\,$GeV and $h_s < g_{Z'}$ we get an MSSM-type Higgs spectrum with the lightest SM-like Higgs boson below $130\,$GeV and with the heaviest scalar above $600 - 800\,$GeV being singlet dominated and phenomenologically irrelevant. The non-observation of Higgs particles at LEP rules out most parts of the $E_6$SSM parameter space in this case.

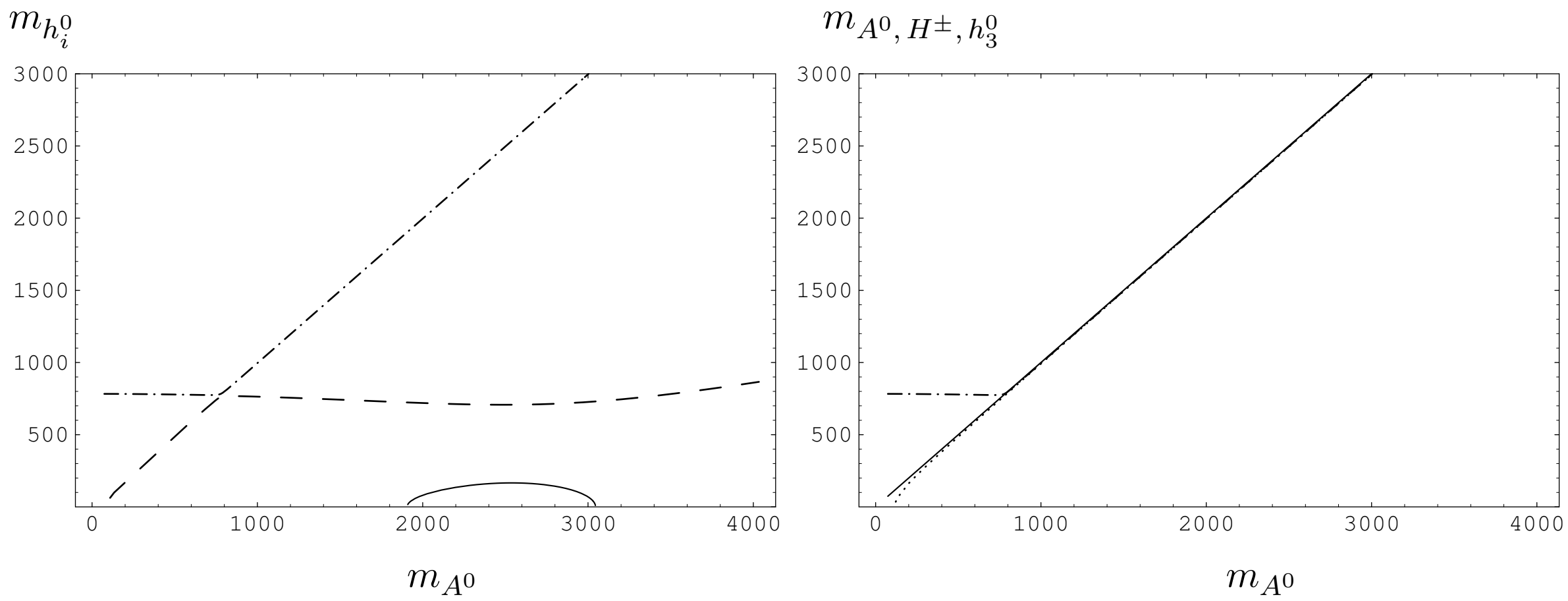

Fig. 6.6: Higgs masses for $h_s(M_t) = 0.794$, $\tan\beta = 2$, $M_{Z'} = M_S = 700\,\mathrm{GeV}$ and $X_t = \sqrt{6}M_S$. Left: One-loop masses of the CP-even Higgs bosons versus $m_{A^0}$. Solid, dashed and dashed-dotted lines correspond to the masses of the lightest, second lightest and heaviest Higgs scalars respectively. Right: One-loop masses of the CP-odd, heaviest CP-even and charged Higgs bosons versus $m_{A^0}$. Dotted, dashed-dotted and solid lines correspond to the masses of the charged, heaviest scalar and pseudoscalar states.

From Fig. 6.6 and Eq. (6.27) it becomes clear that at some value of $m_{A^0}$ (or $x$) the lightest CP-even Higgs boson mass $m_{h_1^0}$ attains its maximum value. At tree level the upper bound on $m_{h_1^0}$ is given by the sum of the first three terms in Eq. (6.27). The inclusion of loop corrections increases the bound on the lightest CP-even Higgs boson mass in models of supersymmetry (SUSY) substantially. When the soft masses of the superpartners of the top-quark are equal to $M_S^2$, the upper limit on $m_{h_1^0}$ in the $E_6$SSM in the leading two-loop approximation can be written in the following form [28]- [29]:

$$\begin{aligned} m_{h_1^0}^2 \;\le\;& \left[\frac{h_s^2}{2}v^2\sin^2 2\beta + M_Z^2\cos^2 2\beta + g_{Z'}^2 v^2\left(\tilde{Q}_1\cos^2\beta + \tilde{Q}_2\sin^2\beta\right)^2\right]\left(1 - \frac{3h_t^2}{8\pi^2}l\right) \\ &+ \frac{3h_t^4 v^2\sin^4\beta}{8\pi^2}\left\{\frac{1}{2}U_t + l + \frac{1}{16\pi^2}\left(\frac{3}{2}h_t^2 - 8g_3^2\right)(U_t + l)l\right\}, \end{aligned} \tag{6.28}$$

where $U_t = 2\dfrac{X_t^2}{M_S^2}\left(1 - \dfrac{1}{12}\dfrac{X_t^2}{M_S^2}\right)$, $l = \ln\left[\dfrac{M_S^2}{m_t^2}\right]$. Eq. (6.28) is a simple generalisation of the approximate expressions for the two-loop theoretical restriction on the mass of the lightest Higgs particle obtained in the MSSM [80] and NMSSM [56]. If as before we assume that $M_S = 700\,\mathrm{GeV}$ and $X_t = \sqrt{6}\,M_S$ then the theoretical restriction on the lightest Higgs mass given by Eq. (6.28) depends on $h_s$ and $\tan\beta$ only. The requirement of validity of perturbation theory up to the GUT scale constrains the parameter space further setting a limit on the Yukawa coupling $h_s(M_t)$ for each value of $\tan\beta$. Relying on the results of the analysis of the renormalisation group flow in the $E_6$SSM presented in [28] one can obtain the maximum possible value of the lightest Higgs scalar for each particular choice of $\tan\beta$.

The dependence of the tree-level and two-loop upper bounds on the mass of the lightest Higgs particle is examined in Fig. 6.7 where it is compared with the corresponding limits in the MSSM and NMSSM. At moderate values of $\tan\beta$ $(1.6 - 3.5)$ the upper limit on the lightest Higgs boson mass in the $E_6$SSM is considerably higher than in the MSSM and NMSSM. In the leading two-loop approximation it reaches the maximum value $150 - 155\,\mathrm{GeV}$ at $\tan\beta = 1.5 - 2$ [28]- [29]. Remarkably, we find that in the interval of $\tan\beta$ from 1.2 to 3.4 the absolute maximum value of the mass of the lightest Higgs scalar in the $E_6$SSM is larger than the experimental lower limit on the SM-like Higgs boson even at tree level. Therefore the non-observation of Higgs bosons at LEP does not cause any trouble for the $E_6$SSM.

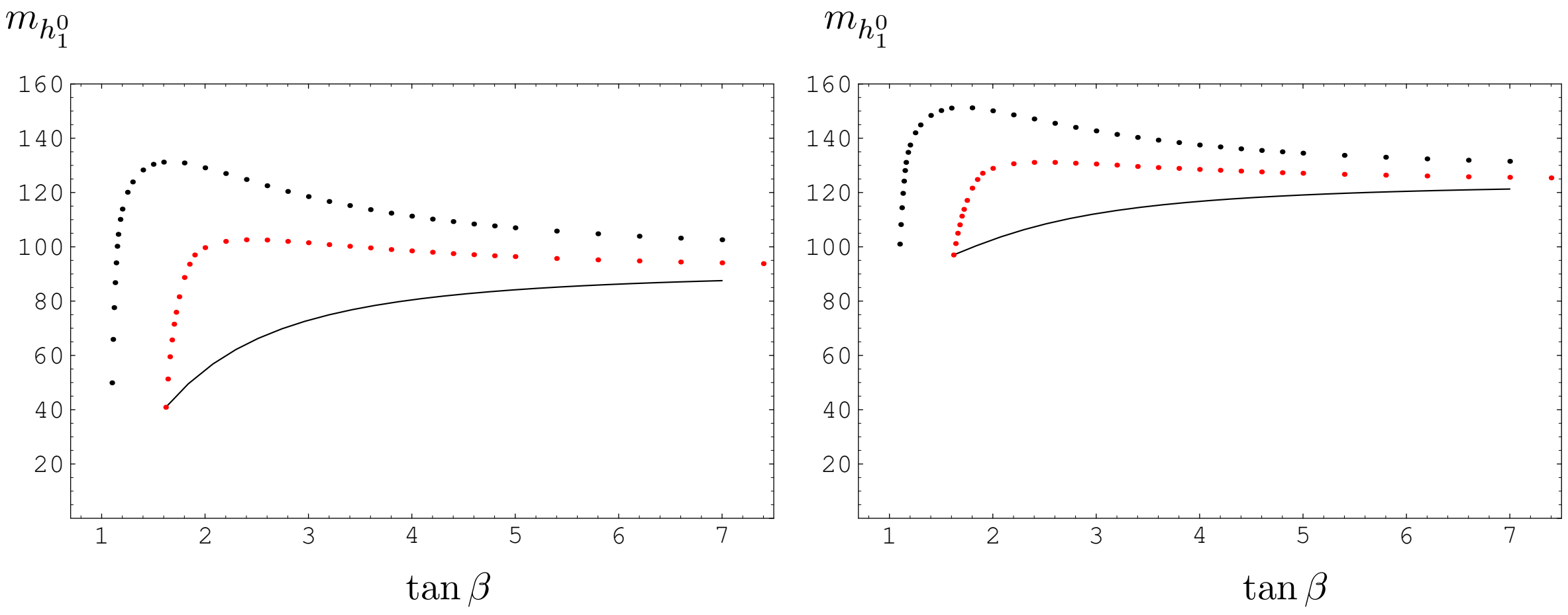

Fig. 6.7: Upper bound on the mass of the lightest Higgs boson. The solid, lower and upper dotted lines correspond to the theoretical restrictions on the lightest Higgs mass in the MSSM, NMSSM and $E_6$SSM respectively. Left: tree-level upper limit on the mass of the lightest Higgs particle as a function of $\tan\beta$. Right: Two-loop upper bound on the lightest Higgs mass versus $\tan\beta$.

In the considered part of the parameter space the theoretical restriction on the mass of the lightest CP-even Higgs boson in the NMSSM exceeds the corresponding limit in the MSSM because of the extra contribution to $m^2_{h_1^0}$ induced by the additional $F$-term in the Higgs scalar potential of the NMSSM. The size of this contribution, which is described by the first term in Eq. (6.27), is determined by the Yukawa coupling $h_s$. The upper limit on $h_s$ caused by the validity of perturbation theory in the NMSSM is more stringent than in the $E_6$SSM. Indeed new exotic $5 + \overline{5}$-plets of matter in the particle spectrum of the $E_6$SSM change the running of the gauge couplings so that their values at the intermediate scale rise preventing the appearance of the Landau pole in the evolution of the Yukawa couplings. It means that for each value of $\tan\beta$ the maximum allowed value of $h_s(M_t)$ in the $E_6$SSM is larger than in the NMSSM. The increase of $h_s(M_t)$ is accompanied by the growth of the theoretical restriction on the mass of the lightest CP-even Higgs particle. This is the main reason why the upper bound on $m_{h_1^0}$ in the $E_6$SSM exceeds that in the NMSSM.

At large $\tan\beta > 10$ the contribution of the $F$-term of the SM-type singlet field to $m^2_{h_1^0}$ vanishes. Therefore with increasing $\tan\beta$ the upper bound on the lightest Higgs boson mass in the NMSSM approaches the corresponding limit in the MSSM. In the $E_6$SSM the theoretical restriction on the mass of the lightest Higgs scalar also diminishes when $\tan\beta$ rises. But even at very large values of $\tan\beta$ the two-loop upper limit on $m_{h_1^0}$ in the $E_6$SSM is still $4 - 5\,\text{GeV}$ larger than the ones in the MSSM and NMSSM because of the $U(1)_N$ $D$-term contribution to $m_{h_1^0}$ (the third term in Eq. (6.27)). This contribution is especially important in the case of minimal mixing between the superpartners of the top quark. In the considered case the two-loop theoretical restriction on $m_{h_1^0}$ in the MSSM and NMSSM is less than the experimental limit on the SM-like Higgs boson mass set by LEP. As a result the scenario with $X_t = 0$ is ruled out in the MSSM. The contribution of an extra $U(1)_N$ $D$-term to $m^2_{h_1^0}$ raises the upper bound given by Eq. (6.28) at large $\tan\beta \geq 10$ slightly above the existing LEP limit thus relaxing the constraints on the $E_6$SSM parameter space. The discovery at future colliders of a relatively heavy SM-like Higgs boson with mass $140 - 155\,\text{GeV}$, corresponding to $h_s > g_{Z'}$ in the $E_6$SSM, will permit to distinguish the $E_6$SSM from the MSSM and NMSSM.

Other possible manifestations of our exceptional SUSY model at the LHC are related to the presence of a $Z'$ and of exotic multiplets of matter. For instance, a relatively light $Z'$ will lead to enhanced production of $l^+l^-$ pairs ($l = e, \mu$). The analysis performed in [81] revealed that a $Z'$ boson in $E_6$ inspired models can be discovered at the LHC if its mass is less than $4 - 4.5\,\text{TeV}$. At the same time the determination of its couplings should be possible up to $M_{Z'} \sim 2-2.5\,\text{TeV}$ [82]. Moreover in the E$_6$SSM the exotic fermions can be relatively light since their masses are set by the Yukawa couplings $\kappa_i$ and $\lambda_i$ that may be small. This happens, for example, when the Yukawa couplings of the exotic particles have a hierarchical structure which is similar to the one observed in the ordinary quark and lepton sectors. Then the production cross section of exotic quark pairs at the LHC can be comparable with the cross section of $t\bar{t}$ production. The lifetime of new exotic particles is defined by the extent to which the $Z_2^H$ symmetry is broken. Since we have assumed that $Z_2^H$ is mainly broken by operators involving quarks and leptons of the third generation the lightest exotic quarks decay into either two heavy quarks $Q\bar{Q}$ or a heavy quark and a lepton $Q\tau(\nu_\tau)$, where $Q$ is either a $b$- or $t$-quark. In the case when $Z_2^H$ is broken significantly this results in the growth of the cross section of either $pp \to Q\bar{Q}Q^{(\prime)}\bar{Q}^{(\prime)} + X$ or $pp \to Q\bar{Q}l^+l^- + X$. If the violation of the $Z_2^H$ invariance is extremely small then a set of new composite scalar leptons or baryons containing quasi-stable exotic quarks could be discovered at the LHC. The discovery of the $Z'$ and exotic quarks predicted by the E$_6$SSM would represent a possible indirect signature of an underlying $E_6$ gauge structure at high energies and provide a window into string theory.

## 6.4 Doubly charged Higgs bosons from the left-right symmetric model at the LHC

*Georges Azuelos, Kamal Benslama and Jonathan Ferland*

The Left-Right Symmetric Model (LRSM) [83–85], based on the group $SU(2)_L \otimes SU(2)_R \otimes U(1)_{B-L}$, is a natural extension of the Standard Model, deriving from Grand Unified Theories. The breaking of $SU(2)_R \otimes U(1)_{B-L} \to U(1)_Y$ occurs at a high energy scale due to a triplet of complex Higgs fields[2] with physical states consisting of $\Delta_R^0$, $\Delta_R^+$ and $\Delta_R^{++}$, when the neutral component acquires a non-vanishing vacuum expectation value (An overview of triplet models is given in Section 13.1). The Higgs sector of the model therefore contains a doubly charged Higgs boson, which could provide a clean signature at the LHC since charge conservation prevents it from decaying to a pair of quarks. Doubly charged scalars are also predicted in Little Higgs models [89, 90], see Section 7, and in 3-3-1 models [91–94], where doubly charged vector bilepton states are also predicted. Very light, $\mathcal{O}(\sim 100)$ GeV, doubly-charged Higgs particles can also be expected in supersymmetric left-right models [36, 95]. Here, we summarize the results [46] of an analysis, performed for ATLAS, which expands on previous phenomenological studies [39, 96–98] by including the effects of backgrounds as well as detector acceptance and resolution.

Other signatures involving the decay of the new heavy gauge bosons of the LRSM have been studied in ATLAS [99–102]. As a complement to these searches, observation of a doubly charged Higgs would clearly provide an important confirmation of the nature of the new physics. In fact, heavy gauge bosons could be out of kinematical reach, and the Majorana neutrinos could be extremely heavy ($\sim 10^{11}$ GeV) if the see-saw mechanism explains the mass of the light neutrinos. Thus the observation of a doubly-charged Higgs boson could serve as the discovery channel for the LRSM.

The Higgs sector [39] of the LRSM consists of (i) the right-handed complex triplet $\Delta_R$ mentioned above, with weights (0,1,2), meaning singlet in $SU(2)_L$, triplet in $SU(2)_R$ and $B - L = 2$, (ii) a left-handed triplet $\Delta_L$ (1,0,2) (if the Lagrangian is to be symmetric under $L \leftrightarrow R$ transformation); and a bidoublet $\phi$ (1/2,1/2,0). The vacuum expectation values (vev) of the neutral members of the scalar triplets, $v_L$ and $v_R$, break the symmetry $SU(2)_L \times SU(2)_R \times U(1)_{B-L} \to SU(2)_L \times U(1)_Y$. The non-vanishing vev of the bidoublet breaks the SM $SU(2)_L \times U(1)_Y$ symmetry. It is characterized by two parameters $\kappa_1$ and $\kappa_2$, with $\kappa = \sqrt{\kappa_1^2 + \kappa_2^2} = 246$ GeV. To prevent flavour changing neutral currents (FCNC), one must have $\kappa_2 \ll \kappa_1$, implying minimal mixing between $W_L$ and $W_R$ [103]. The mass eigenstate of the singly charged Higgs is a mixed state of the charged components of the bidoublet and of the triplet. Bounds on the parameters are given in [39, 98, 104]: custodial symmetry constrains $v_L \lesssim 9$ GeV and present Tevatron lower bounds on $M_{W_R}$ impose a limit $v_R > 1.4$ TeV, or

[2] Alternative minimal Left-Right symmetric models exist with only doublets of scalar fields [86–88]. They do not lead to Majorana couplings of the right-handed neutrinos.

$m_{W_R} > 650$ GeV, assuming equal gauge couplings $g_L = g_R$. Direct limits from the Tevatron on the mass of the doubly charged Higgs from di-leptonic decays have recently been reported in [105, 106]. Indirect limits on the mass and couplings of the triplet Higgs bosons, obtained from various processes, are given in [39, 104, 107] (see also Section 13).

We will assume a truly symmetric Left-Right model, with equal gauge couplings $g_L = g_R = e/\sin(\theta_W) = 0.64$. We relate the mass of $W_R$ to $v_R$ by: $m^2_{W_R} = g^2_R v^2_R/2$, which is a valid approximation in the limit where $v_L = 0$ and $\kappa_1 \ll v_R$.

### 6.4.1 *Phenomenology of the doubly-charged Higgs boson*

Single production of $\Delta^{++}_R$ production is dominated by the vector fusion process $W^{\pm}_R W^{\pm}_R$, as long as the mass of the $W_R$ is of the TeV scale [98]. For the process $W^+W^+ \to \Delta^{++}_L$, the suppression due to the small value of the $v_L$ is somewhat compensated by the fact that the incoming quarks radiate a lower mass vector gauge boson.

Double production of the doubly charged Higgs is also possible via a Drell-Yan process, with $\gamma$, $Z$ or $Z_R$ exchanged in the $s$-channel, but at a high kinematic price since enough energy is required to produce two heavy particles. In the case of $\Delta^{++}_L$, double production may nevertheless be the only possibility if $v_L$ is very small or vanishing.

The decay of a doubly charged Higgs can proceed by several channels. Present bounds [98, 107] on the diagonal couplings $h_{ee,\mu\mu,\tau\tau}$ to charged leptons are consistent with values $\sim O(1)$ if the mass scale of the triplet is larger than a few hundred GeV. For the $\Delta^{++}_L$, this may be the dominant coupling if $v_L$ is very small. For very low Yukawa couplings ($h_{\ell\ell} \lesssim 10^{-8}$), the doubly charged Higgs boson could be quasi-stable [108], leaving a characteristic dE/dx signature in the detector, but this case is not considered here. The decay $\Delta^{++}_{R,L} \to W^{+}_{R,L} W^{+}_{R,L}$ can also be significant. However, it is kinematically suppressed in the case of $\Delta^{++}_R$, and suppressed by the small coupling $v_L$ in the case of $\Delta^{++}_L$.

Here, we discuss only dilepton ($ee$ or $\mu\mu$) decay, which provides a clean signature, kinematically enhanced, although the branching ratios will depend on the unknown Yukawa couplings. A complete description of this analysis as well as of other channels, including $\tau\tau$ and $WW$ can be found in [46].

### 6.4.2 *Simulation of the signal and backgrounds*

The processes of single and double production of doubly charged Higgs are implemented in the PYTHIA generator [109]. Events were generated using the CTEQ5L parton distribution functions, taking account of initial and final state interactions as well as hadronization.

Detector effects and acceptance were simulated using ATLFAST [110], a fast simulation program for the ATLAS detector, where efficiencies and resolutions are parametrized according to the expected detector performance, as evaluated in [99].

PYTHIA was used to generate the $t\bar{t}$ background, which has a very large cross section of $\sim 500$ pb. Other backgrounds were simulated using the CompHep generator [111]: (i) The Standard Model processes $qq \to W^+W^+qq$ and (ii) $qq \to W^+Zqq$ and (iii) $pp \to Wt\bar{t}$.

A number of systematic uncertainties, some of which are difficult to evaluate reliably before experimental data are available, will apply. No k-factors have been used here, although next-to-leading-order corrections can be substantial for these high mass resonance states. Experimental systematic uncertainties involve: the luminosity measurement ($\sim 5 - 10\%$), the efficiency of lepton reconstruction in ATLAS, here taken to be 90%, the uncertainty in the energy resolution, especially for high energy electrons, charge misidentification, estimated to be small, and misidentification of jets as electrons, for which preliminary estimates suggest that it will have a small effect.

Table 6.1: Number of events of signal and backgrounds after successive application of cuts, for the case $\Delta_R^{++} \to \ell^+ \ell^+$, around $m_{\ell\ell}$ = 300 or 800 GeV (shown as $n_{300}/n_{800}$ for the backgrounds), for $m_{W_R}$ = 650 GeV and for 100 fb$^{-1}$. Mass windows $\pm 2\sigma$ around the resonances have been chosen. In parentheses is shown the number of events without the mass window cut.

| | $\Delta^{++}$ 300 GeV | $\Delta^{++}$ 800 GeV | $W^+W^+ qq$ | $W\, t\bar{t}$ | $WZqq$ | $t\bar{t}$ | total backg |
|---|---|---|---|---|---|---|---|
| Isolated leptons | 278 (327) | 63 (95) | 109/12 | 7.6/0.6 | 0/0.8 | 17/0 | 133/13 |
| Lepton $P_T > 50$ GeV | 256 (301) | 63 (94) | 63/11 | 5.9/0.5 | 0/0.8 | 1.1/0 | 70/12 |
| $2.4(P_T^{l_1} + P_T^{l_2}) - M_{ll} > 480$ | 191(227) | 59(85) | 10/2.1 | 1.3/0.3 | 0 | 0 | 12/2.4 |
| Fwd Jet tagging | 156(186) | 56(74) | 6.0/1.3 | 0.1/0 | 0 | 0 | 6/1.3 |
| ptmiss $<$ 100 GeV | 154(181) | 56(68) | 3.0/0.3 | 0/0 | 0 | 0 | 3.1/0.3 |

### 6.4.3 *Search for* $\boldsymbol{\Delta_R^{++}}$

The cross section for single production of $\Delta_R^{++}$ is of the order of $\sim$ fb: for example, it is 0.9 fb for the case $m(\Delta_R^{++}) = 800$ GeV, $m(W_R^{++}) = 850$ GeV. We consider signals for doubly positively charged Higgs bosons, as they are about 1.6 times more abundant than the negatively charged ones, at the LHC. The same ratio of positively charged to negatively charged leptons can be expected from the backgrounds, to the extent that $qqWW$ dominates, and hence the improvement in the significances obtained below can be estimated at 22%.

The selection criteria for this channel are summarized in Table 6.1, which also shows the number of events of signal for typical cases of signal where $m_{\Delta_R^{++}}$ = 300 or 800 GeV and $m_{W_R^+} = 650$ GeV, and of the various backgrounds after successive application of cuts. A clean signal is found for an integrated luminosity of 100 fb$^{-1}$. A window of $\pm 2\times$ the width of the reconstructed mass of the $\Delta_R^{++}$ has been selected. The intrinsic width depends on the assumed $\Delta_R^{++} - \ell\ell$ couplings, but is expected, in any case to be very narrow [98]. The width is therefore dominated by the detector resolution, which is measured to be $\sigma_R = 20$, 55 and 123 GeV for the cases of $\Delta_R^{++}$= 300, 800, 1500 GeV respectively. The cuts involve forward jet tagging (for details, see [46]), since the primary partons from which the $W_R$ are radiated will tend to continue in the forward and background directions, and hadronize as jets.

Since the background is negligible, discovery can be claimed if the number of signal events is 10 or higher. With this definition, the contour of discovery, in the plane $m_{W_R^+}$ versus $m_{\Delta_R^{++}}$ (or $v_R$) has been estimated from a sample of test cases. The discovery reach at the LHC is shown in Fig. 6.8, for integrated luminosities of 100 fb$^{-1}$ and 300 fb$^{-1}$ and assuming 100% BR to lepton pairs.

Pair production of $\Delta_R^{++}\Delta_R^{--}$ is suppressed by the expected high mass of the $\Delta_R^{++}$ but can have a dominant cross section in some region of phase space (see [98]). The diagrams with $s$-channel $Z$ and $Z'$ exchange have been added to the $\gamma$ exchange diagram in the implementation of the Drell-Yan process in the `PYTHIAg` generator, taking the coupling of $Z,\ Z'$ to fermions and to $\Delta_L^{++}$ from references [96, 112]. In principle, the branching ratio depends on the assumed mass of $\Delta_L^{++}$, as well as that of $\Delta_R^{++}$, but since the $Z'$ has a large partial width to fermions, such that $BR(Z' \to \Delta^{++}\Delta^{--})$ is of the order of 1%, the contribution of these decay channels to the total width of the $Z'$ was neglected. For the case of leptonic decays of the doubly-charged Higgs bosons, the process constitutes a golden channel and the background will be negligible.

Fig. 6.9 shows the contours of discovery, defined as observation of 10 events, if all four leptons are detected or if any 3 of the leptons are observed. As $m(Z_R)$ increases, the mass reach for $m(\Delta_R^{++})$ increases at first, as the $s$-channel diagram with $Z_R$ produced on mass shell becomes the dominant contribution. However, for very large masses of $Z_R$, the contribution of this diagram is kinematically suppressed. Being an $s$-channel process not involving the $W_R$, this channel is not sensitive to the mass of this heavy gauge boson.

Table 6.2: Number of events of signal and total background after successive application of cuts, for the case $\Delta^{++} \to \ell^+ \ell^+$, for $m_{\Delta_L^{++}}$ = 300 or 800 GeV (shown as $n_{300}/n_{800}$ for the background) and $v_L = 9$ GeV, for 100 fb$^{-1}$. Mass windows $\pm 2\sigma$ around the resonances have been chosen. In parentheses is shown the number of events without the mass window cut.

| | $\Delta^{++}$ 300 GeV | $\Delta^{++}$ 800 GeV | total backg |
|---|---|---|---|
| Isolated leptons | 330 (384) | 59 (69) | 133/13 |
| $\lvert\Delta\phi_{\ell\ell} > 2.5\rvert$ | 253 (289) | 56 (65) | 75/8.3 |
| $\Delta_{P_T^{ll}} > (\frac{M_{ll}}{2}$+50) | 220 (260) | 50 (59) | 37/2.5 |
| Fwd Jet tagging | 144(170) | 38 (45) | 11/0.6 |
| ptmiss | 140(165) | 33 (38) | 2.0/0.07 |

### 6.4.4 *Search for* $\Delta_L^{++}$

The search for $\Delta_L^{++}$ follows closely the strategy used for $\Delta_R^{++}$. However, some major differences in the kinematics of the events force the use of different selection criteria. In particular, $\Delta_L^{++}$ single production occurs via fusion of a pair of $W_L$'s, which are much lighter than the $W_R$'s in the case of single production of $\Delta_R^{++}$. The distribution of forward jets is strongly affected, as well as the final transverse momentum of the $\Delta^{++}$. For that reason, an independent analysis has been performed, using cuts similar to the case of $\Delta_R^{++}$ to the extent possible (for details, see [46]).

As for the case of the $\Delta_R^{++}$ the dilepton channel provides a clean signature. Although the Yukawa coupling of $\Delta_L^{++}$ to leptons remains a parameter of the theory, this channel can, in fact, be dominant since the alternative decay to gauge bosons is possibly negligible, being proportional to the very small value of the vev $v_L$. In the limit where $v_L = 0$, it will be the only open channel, but production of $\Delta_L^{++}$ will only occur in pairs, through $s$-channel $\gamma/Z/Z'$ exchange. As before, we will assume below 100% branching ratio to leptons, but results can be reinterpreted in a straightforward way for different values of this branching ratio.

Table 6.2 gives the number of expected signal and background events for the cases $m_{\Delta_L^{++}} = 300$ GeV and $m_{\Delta_L^{++}} = 800$ GeV respectively. A mass window of $\pm 2\times$ the width of the resonance was selected. The discovery reach in the plane $v_L$ vs $m_{\Delta_L^{++}}$ is shown in Fig. 6.10.

As for the case of the right-handed sector, pair production of $\Delta_L$ is a possible discovery channel. The diagram with $s$-channel $Z'$ exchange has been added to the implementation of this Drell-Yan process in the `PYTHIA` generator, taking the coupling of $Z'$ to fermions and to $\Delta_L^{++}$ from references [96, 112]. Assuming leptonic decays, the background will be negligible. Fig. 6.11 shows the contours of discovery, defined as observation of 10 events, if all four leptons are detected or if at least any 3 of the leptons are observed. The reach has the same qualitative dependence on the mass of $Z_R$ as for the case of $\Delta_R^{++}$ pair production.

### 6.4.5 *Summary and Conclusion*

Left-Right symmetric models predict the existence of doubly-charged Higgs bosons which should yield a striking signature at the LHC. The principal production and decay modes, including $\Delta \to \tau\tau$ and $\Delta \to WW$ have been investigated in [46] but only the dilepton channel is reported here. It must be emphasized that these results have assumed that the decay to two leptons ($e$ or $\mu$) dominate and that they should be rescaled if there is a substantial branching ratio to $\tau\tau$. It is found that the LHC will be able to probe a large region of unexplored parameter space in the triplet Higgs sector. This analysis complements previous ATLAS studies searching for signals of the Left-Right symmetric model.

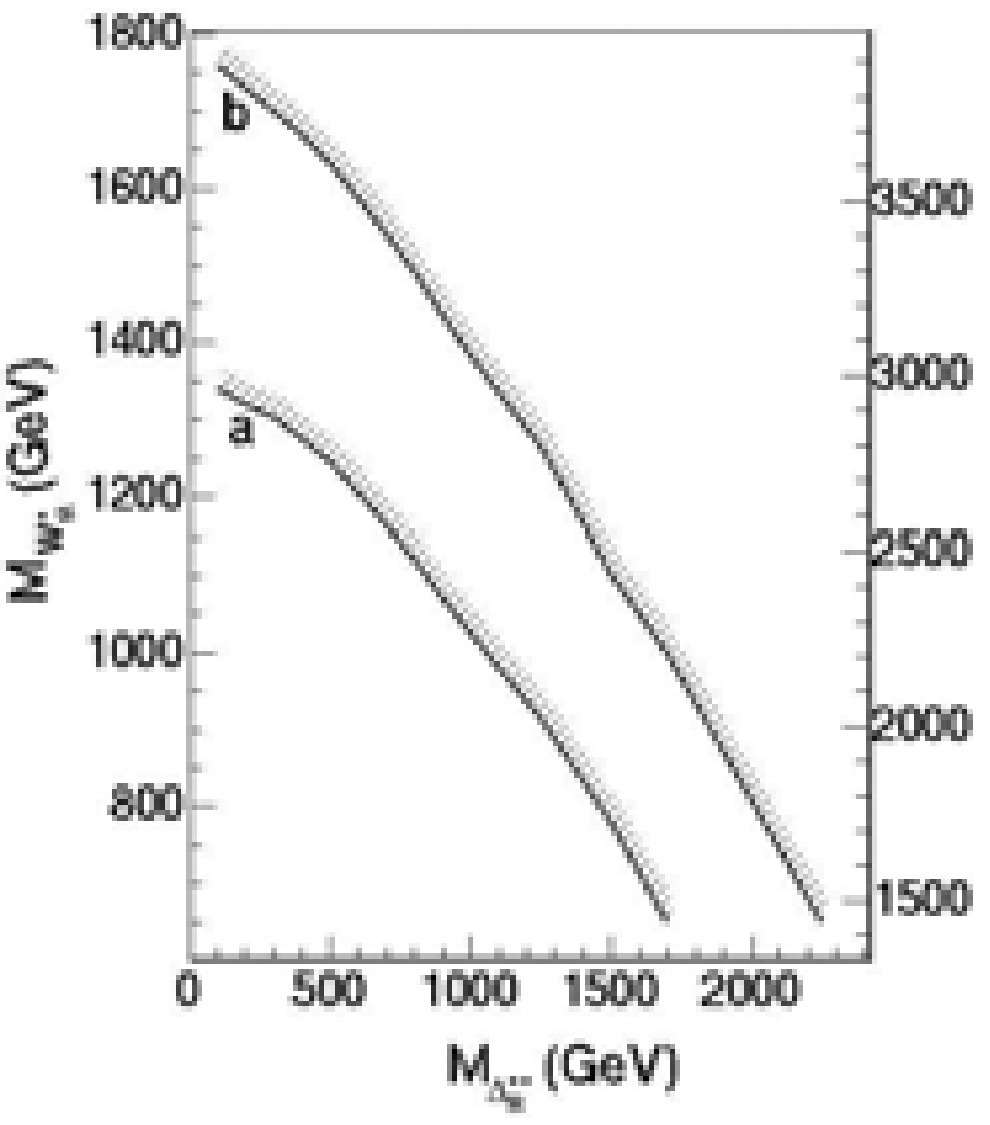

Fig. 6.8: Discovery reach for $\Delta_R^{++} \to l^+l^+$ in the plane $m_{W_R^+}$ versus $m_{\Delta_R^{++}}$ (or $v_R$) for integrated luminosities of 100 fb$^{-1}$(a) and 300 fb$^{-1}$(b), and assuming 100% BR to dileptons. The region where discovery is not possible is on the hatched side of the line.

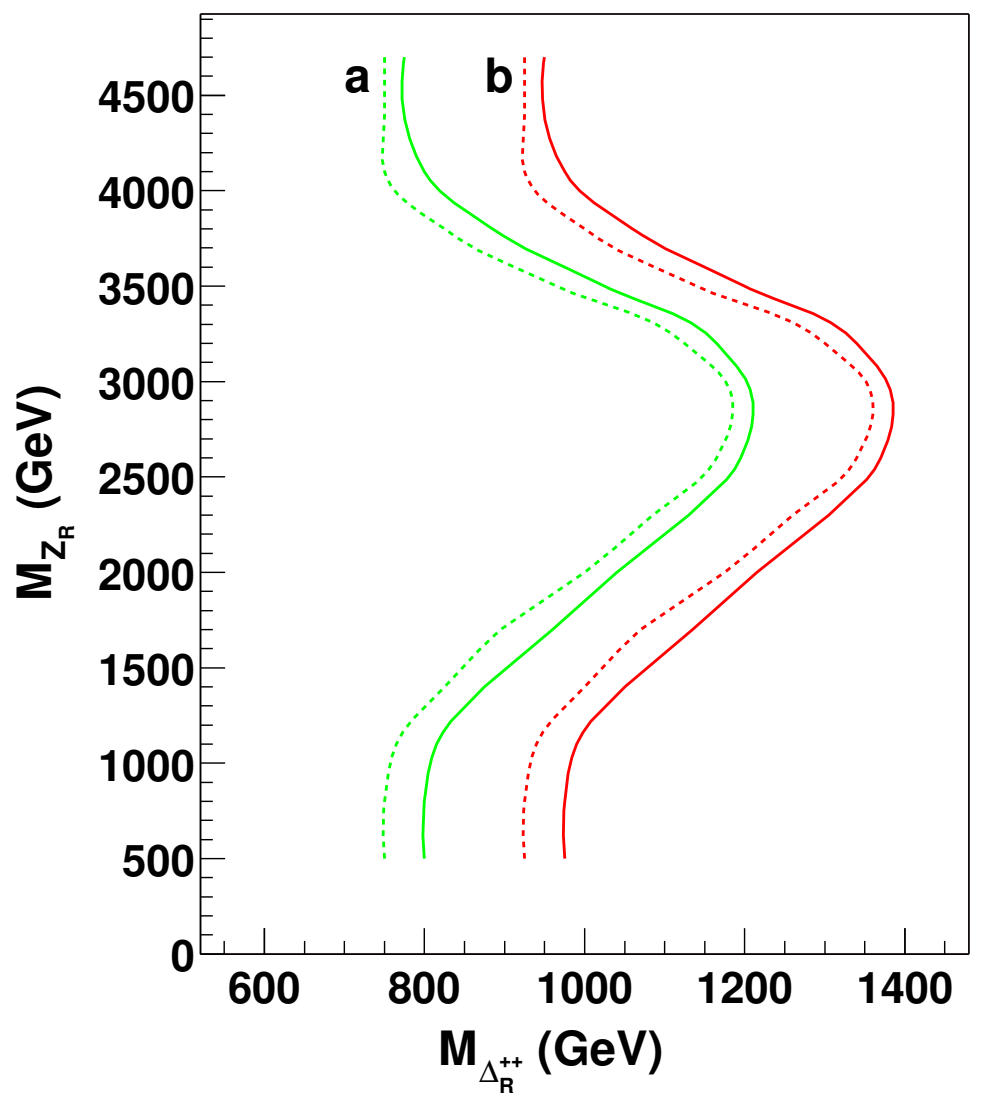

Fig. 6.9: Contours of discovery for 100 fb$^{-1}$(a) and for 300 fb$^{-1}$(b) in the plane $m_{Z'}$ vs $m_{\Delta_R^{++}}$. The dashed curves are for the case where all four leptons are observed, and the full curves are when only three leptons are detected.

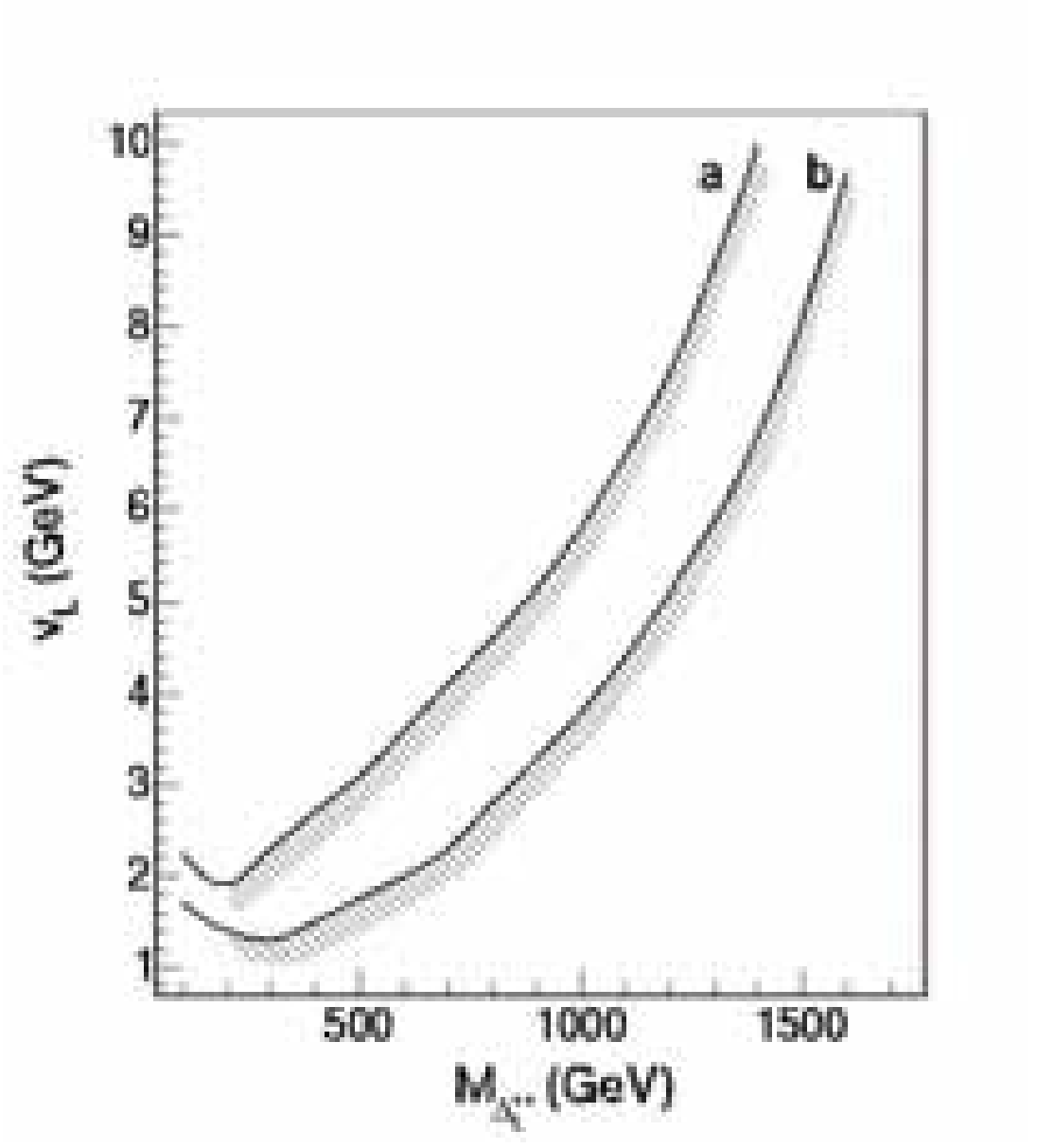

Fig. 6.10: Discovery reach for $\Delta_L^{++} \to \ell^+\ell^+$ in the plane $v_L$ versus $m_{\Delta_L^{++}}$ for integrated luminosities of 100 fb$^{-1}$(a) and 300 fb$^{-1}$(b) and assuming 100% BR to dileptons.

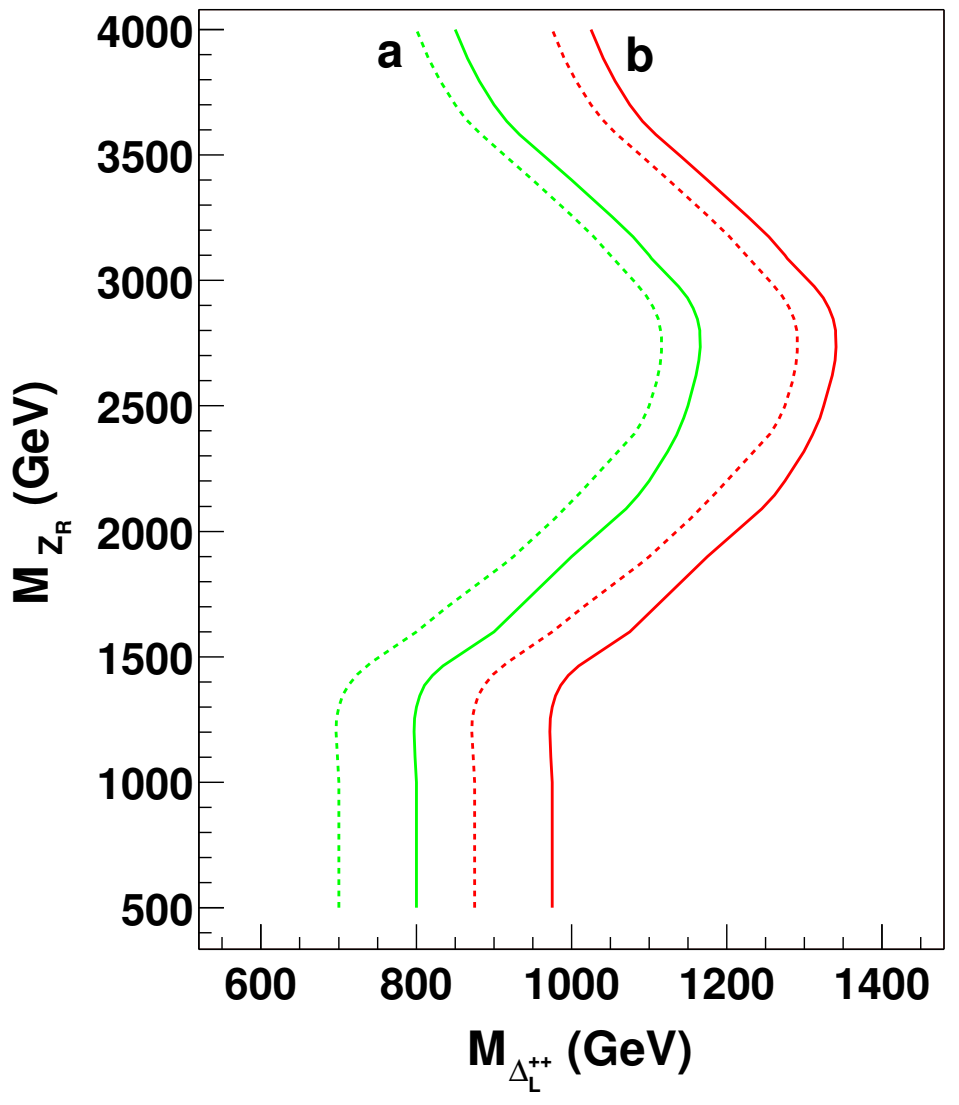

Fig. 6.11: Contours of discovery in the plane $m_{Z'}$ vs $m_{\Delta_L^{++}}$ for 100 fb$^{-1}$(a) and 300 fb$^{-1}$(b). The dashed curves are for the case where all four leptons are observed, and the full curves are when only three leptons are detected.

# 7 LITTLE HIGGS MODELS

## 7.1 Introduction

*Thomas Grégoire, Heather E. Logan, and Bob McElrath*

### 7.1.1 The little Higgs mechanism and collective symmetry breaking

We should learn soon from the LHC how electroweak symmetry is broken. Electroweak precision tests suggest that the physics responsible for this phenomenon is weakly coupled, or in other words, it is expected that a Higgs particle will be discovered. To have a *natural* theory of electroweak symmetry breaking, the Higgs mass needs to be protected from radiative corrections that would drive it toward the ultraviolet (UV) cutoff of the theory, presumably with the help of a symmetry. Two possible symmetries exist. The first one is supersymmetry, and the well studied MSSM relies on this symmetry to protect the Higgs mass. The other possible symmetry is a shift symmetry, and in fact the only light scalar particles that we know in nature, the pions, are light thanks to this kind of symmetry: they are pseudo-Goldstone bosons.

Goldstone bosons arise whenever a global symmetry is spontaneously broken. Due to the shift symmetry they have derivative couplings, but no potential: they are massless. The strength of their derivative interactions is set by an energy scale $f$, the decay constant. At energies larger than $\Lambda \sim 4\pi f$, the Goldstone bosons become strongly coupled and some new physics is needed to regulate this behavior. The regulating physics can be strongly coupled at scale $\Lambda$, like in QCD, or it can be weakly coupled if the global symmetry is spontaneously broken by an elementary scalar (for example in the SM, the Higgs field regulates $WW$ scattering). Small explicit breaking of the global symmetry can generate a potential for the pseudo-Goldstone bosons. For example, in QCD, the quark masses explicitly break the flavor symmetry and as a result, the pions are not exactly massless. The gauging of electromagnetism also breaks the global symmetry and a quadratically divergent photon loop is responsible for the $\pi^+ - \pi^0$ mass difference:

$$m_{\pi^+}^2 - m_{\pi^0}^2 \sim \frac{\alpha_{\rm em}}{4\pi}\Lambda_{\rm QCD}^2, \tag{7.1}$$

which is parametrically of order $gf_\pi$ ($f_\pi$ is the pion decay constant). Early attempts to write down a theory of a Higgs as a pseudo-Goldstone boson by Georgi and Kaplan [1–4] faced the following problem. The typical potential generated for a pseudo-Goldstone boson $\phi$ is of the form

$$V(\phi) = f^4(c_1\phi^2/f^2 + c_2\phi^4/f^4)\,, \tag{7.2}$$

with $c_1, c_2$ of order one, which leads to a minimum (if $c_1 < 0$) at $\phi \sim f$. If $\phi$ is the Higgs, $f \sim 100$ GeV, and strong physics at 1 TeV (or a linear sigma model field at $\sim 100$ GeV, which is not any better than having a fundamental Higgs in the first place) is needed. The philosophy of little Higgs models is to avoid having to deal with potentially dangerous contributions to electroweak precision observables coming from strongly-coupled physics by pushing the strong coupling scale up to 10 TeV.

The idea of little Higgs models [5, 6] is to break the global symmetry in such a way that the mass of the Higgs is parametrically *two* loop factors smaller than $\Lambda$ instead of one. We could then have $f \sim 1$ TeV and $\Lambda \sim 10$ TeV. This is achieved through collective breaking of the symmetry. The idea is that any one global symmetry breaking coupling by itself leaves enough of the global symmetry intact so that the Higgs is still an exact Goldstone. However, once all couplings are turned on, the Higgs gets a mass parametrically of order:

$$m_h^2 \sim \left(\frac{g^2}{16\pi^2}\right)^2 \Lambda^2 \tag{7.3}$$

so that the strong coupling scale $\Lambda$ could be as high as 10 TeV, a scale that is out of reach of forthcoming experiments and safe with respect to electroweak precision measurements. In order for this mechanism to work, the global symmetry group needs to be quite large, which implies the presence of extra particles typically at scale $f \sim 1$ TeV. Those particles are responsible for canceling the one loop quadratic divergences to the Higgs mass. The following ingredients are needed to build a little Higgs model:

– The spontaneous breaking of a global symmetry. The mechanism by which the symmetry breaking happens is not specified. It could be strongly coupled physics at 10 TeV, or weakly coupled physics at 1 TeV. The breaking produces a set of Goldstone bosons, among which is the Higgs, and at low energies these 'pions' are described by a non-linear sigma model field which is written as an exponential of the broken generators $T^a$ of the global symmetry: $\Sigma(x) = \exp\left(i\pi^a(x)T^a\right)$.
– Gauge couplings for the Higgs that implement the collective symmetry breaking principle. To achieve this, one needs to gauge a group larger than the Standard Model gauge group, which breaks to the Standard Model at the scale $f$. There will then be extra gauge bosons at the scale $f$ that cancel the quadratically divergent contributions of the Standard Model gauge bosons to the Higgs mass.
– Yukawa couplings that implement the collective symmetry breaking principle. This leads to extra heavy fermions that cancel the quadratically divergent contribution of the Standard Model top quark loop to the Higgs mass. Note that the size of the quadratic divergence from light fermions is small up to the cutoff at 10 TeV; the contributions of the light fermions to the Higgs mass quadratic divergence need not be canceled and we can couple the light fermions to the Higgs in the usual way.
– Higgs quartic couplings that implement the collective symmetry breaking principle. Once again, this leads to additional scalars and higher-dimensional Higgs self-interactions that cancel the SM Higgs self coupling quadratic divergence at $f \sim 1$ TeV. Some little Higgs models however do not have these features and need some fine tuning to get the light Higgs vacuum expectation value (vev).

As an example of a coupling that respects the collective breaking principle, we write down a typical top Yukawa coupling of a little Higgs model. Consider a 3×3 nonlinear sigmal model field describing the breaking of a SU(3)$_L$×SU(3)$_R$ global symmetry to the diagonal SU(3)$_D$:

$$\Sigma(x) = \exp(i\pi^a(x)T^a) \qquad \Sigma \to \exp(i\alpha^a_L T^a)\Sigma(x)\exp(-i\alpha^a_R T^a) \tag{7.4}$$

$$\Pi = \pi^a T^a = \begin{pmatrix} \phi+\eta & h \\ h^\dagger & -2\eta \end{pmatrix} \qquad \pi^a \to \pi^a + \alpha^a_L - \alpha^a_R + \cdots \tag{7.5}$$

where $T^a$ are the SU(3) generators; $\alpha^a_L$ and $\alpha^a_R$ are infinitesimal vectors in $SU(3)_L$ and $SU(3)_R$ respectively; $\phi$ is an SU(2)$_D$ triplet; $h$ is an SU(2)$_D$ doublet; and $\eta$ is an SU(2)$_D$ singlet. This symmetry shifts the $\pi^a$ by a constant. We consider the coupling of this nonlinear sigma model to a fundamental of SU(3)$_L$, $Q = \begin{pmatrix} u & d & T \end{pmatrix}$, and to one singlet fermion $t^c$. We add a second singlet fermion $T^c$ to give mass to the extra fermion:

$$\lambda Q \Sigma \begin{pmatrix} 0 \\ 0 \\ t^c \end{pmatrix} + M T T^c. \tag{7.6}$$

If $M = 0$, the coupling $\lambda$ respects the SU(3)$_L$ global symmetry and the Higgs is still protected, while with $\lambda = 0$ and $M \neq 0$ the SU(3)$_R$ symmetry under which the $\pi^a$ also shift remains. Therefore both couplings are needed to give masses to the $\pi^a$. The cancellation of one loop quadratic divergences is discussed in more detail in Sections 7.1.7 and 7.1.8.

In the next subsections we will examine some specific models. They all have sligthly different properties, but they all have a set of TeV-scale gauge bosons, colored fermions and scalars that cancel the quadratic divergences due to the ususal SM gauge bosons, Yukawa and Higgs quartic couplings, respectively. The models are summarized in Table 7.1.

Table 7.1: Various little Higgs models classified by their type: theory space (t.s.), product gauge group (p.g.g.), or simple gauge group (s.g.g).

| Model | Global group | Gauge group | Type | Comments |
|---|---|---|---|---|
| Minimal moose [7] | $SU(3)^8/SU(3)^4$ | $SU(3)\times SU(2)\times U(1)$ | t.s. | can contain extra light triplet and singlet scalars |
| Minimal moose with $SU(2)_C$ [8] | $SO(5)^8/SO(5)^4$ | $SO(5)\times SU(2)\times U(1)$ | t.s. | less constrained from electroweak precision tests (EWPT) |
| Moose with T-parity [9] | $SO(5)^{10}/SO(5)^5$ | $(SU(2)\times U(1))^3$ | t.s. | very few constraints from EWPT, larg spectrum, complicated plaquettes |
| Littlest Higgs [6] | SU(5)/SO(5) | $(SU(2)\times U(1))^2$ | p.g.g | Minimal field content |
| SU(6)/Sp(6) model [10] | SU(6)/Sp(6) | $(SU(2)\times U(1))^2$ | p.g.g | Small field content, contains a heavy vector-like quark doublet |
| Littlest Higgs with $SU(2)_C$ [11] | SO(9)/ $(SO(5)\times SO(4))$ | $SU(2)^3\times U(1)$ | p.g.g. | less constraints from EWPT |
| Littlest Higgs with T-parity [12] | SU(5)/SO(5) | $(SU(2)\times U(1))^2$ | p.g.g | Minimal field content, very few constraints from EWPT |
| SU(3) simple group [13, 14] | $(SU(3)\times U(1))^2/$ $(SU(2)\times U(1))^2$ | $SU(3)\times U(1)$ | s.g.g. | no large quartic |
| SU(4) simple group [13] | $(SU(4)\times U(1))^4/$ $(SU(3)\times U(1))^4$ | $SU(4)\times U(1)$ | s.g.g. | Two Higgs doublets, large quartic |
| SU(9)/SU(8) simple group [15] | SU(9)/SU(8) | $SU(3)\times U(1)$ | s.g.g. | Two Higgs doublets, large quartic |

### 7.1.2 *Theory space models*

Theory space models [5, 7–9, 16] were the first little Higgs models and were inspired by the deconstruction [17–19] of extra dimensional models where the Higgs is the fifth component of a gauge field. Theory spaces are sets of sites and links, also called moose diagrams. Sites represent gauge groups, and links are $N \times N$ nonlinear sigma model fields transforming as bifundamentals under the gauge groups associated with the sites they touch (see Fig. 7.1). Each link breaks a global $SU(N)^2$ symmetry to the diagonal $SU(N)$. This results in the presence of Goldstone bosons. The gauge symmetry explicitly breaks the large global symmetry group. However, no single gauge coupling alone breaks enough symmetry to give the Goldstone bosons a mass.

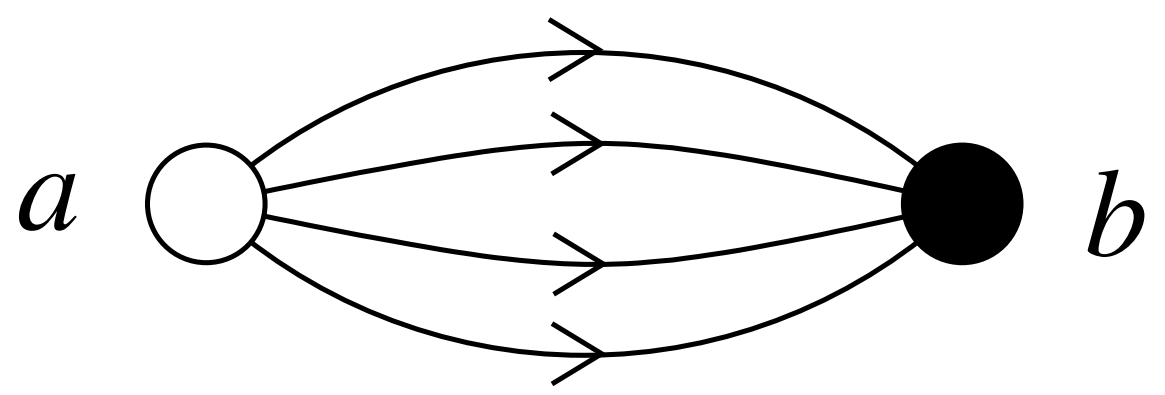

Fig. 7.1: Moose diagram for the minimal moose from Ref. [20]. The open site corresponds to an SU(3) gauge group, while the filled site corresponds to an $SU(2)\times U(1)$ gauge group.

In Fig. 7.1 we show the theory space of the 'minimal moose' [7], the most simple little Higgs of this type. The kinetic term for the link fields is given by:

$$\sum_{i=1}^{4} |D_\mu \Sigma_i|^2 \,, \tag{7.7}$$

with

$$D_\mu \Sigma_i = \partial_\mu \Sigma_i + iA_1\Sigma_i - i\Sigma_i A_2 \quad \text{and} \quad \Sigma_i = \exp(i\pi_i^a T^a)\,, \tag{7.8}$$

where $T^a$ are the generators of SU(3). The global symmetry group is $\mathrm{SU}(3)_L^4 \times \mathrm{SU}(3)_R^4$ (i.e., one copy of $\mathrm{SU}(3)_L \times \mathrm{SU}(3)_R$ for each link) broken down to the diagonal $\mathrm{SU}(3)_D^4$, resulting in $4 \times 8 = 32$ Goldstone bosons. The spontaneous breaking of the global group also breaks the $\mathrm{SU}(2)_L \times \mathrm{U}(1)_L \times \mathrm{SU}(3)_R$ gauge symmetry down to the diagonal $\mathrm{SU}(2)_L \times \mathrm{U}(1)_Y$ subgroup, which eats 8 Goldstone bosons leaving 24 pseudo-Goldstone bosons. The gauge group also explicitly breaks the global symmetry and the 24 pseudo-Goldstone bosons will get a potential generated by gauge interactions. Note that if only the SU(3) gauge coupling is nonzero, there is an exact $\mathrm{SU}(3)_L^4 \times \mathrm{SU}(3)_R$ global symmetry broken to the diagonal SU(3). In this case there would be 32 Goldsone bosons, 8 of which would be eaten, leaving 24 exact Goldstone bosons. This tells us that we need the gauge couplings of both sites to generate any potential for the 24 pseudo-Goldsone bosons. The potential is in fact parametrically of the form:

$$\frac{g_1^2 g_2^2}{16\pi^2}(c_1 f^2 \phi^2 + c_2 \phi^4 + \cdots), \tag{7.9}$$

where $c_1$ and $c_2$ are coefficients of order one and $\phi$ represents the various pseudo-Goldstone bosons. As in Eq. (7.2), the minimum is either at zero or parametrically at $\phi \sim f$. To correct this situation, we need to generate a large quartic coupling for $\phi$. This can be achieved with 'plaquette' interactions of the form

$$\lambda \mathrm{Tr} \Sigma_1 \Sigma_2^\dagger \Sigma_3 \Sigma_4^\dagger. \tag{7.10}$$

These interactions also break some of the global symmetries. They do not respect the collective breaking principle for all the pseudo-Goldsones, and they therefore give mass of order $f$ to 8 of them. But still, each of these interactions respects enough global symmetries to protect the mass of the remaining 16 pseudo-Goldstones, which consist of two Higgs doublets, two triplets and two singlets. The important feature of the plaquette terms is that, at tree level, they give no mass to the Higgs doublets, but they do give them an order one quartic coupling. In the extra-dimensional picture, the plaquette term corresponds to the $F_{56}$ part of the gauge kinetic term. These plaquettes are also the reason why we need four link fields. With fewer link fields, the global symmetry structure is not large enough to allow for the desired plaquette terms.

Finally, a top Yukawa coupling can be introduced in a way very similar to Eq. (7.6):

$$\lambda Q \Sigma_1 \Sigma_2^\dagger \begin{pmatrix} 0 \\ 0 \\ t^c \end{pmatrix} + M T T^c. \tag{7.11}$$

The variety of little Higgs models that can be built from theory space is infinite provided one follows a simple set of rules [20]. Along with the minimal moose model, other interesting 'mooses' include a model with a custodial SU(2) symmetry built in [8], and a model with T-parity [9], both constructed to relax constraints from electroweak precision measurements.

Typically, the little Higgs models based on theory spaces are slightly more involved than the 'Littlest Higgs' type model that we will present in the next subsection. However, as already mentioned, they can have interesting extra dimensional interpretations, and many models of a Higgs as the extra-dimensional component of gauge fields [21–27] can be reinterpreted as theory space little Higgs models once the extra dimension is deconstructed.

### *7.1.3 Product gauge group models*

Product gauge group models [6, 10–12, 28] do not have any extra dimensional interpretations. In these models, the gauge groups are subgroups of a single global symmetry. The typical example, and the most studied little Higgs model, is the Littlest Higgs [6]. The global group structure is SU(5)/SO(5). This generates $24 - 10 = 14$ Goldstone bosons that can be parametrized by the following nonlinear sigma

model:

$$\Sigma(x) = \exp(i\Pi)\Sigma_0 \exp(i\Pi^T) \qquad \Sigma_0 = \begin{pmatrix} & & \mathbf{1} \\ & 1 & \\ \mathbf{1} & & \end{pmatrix} \tag{7.12}$$

$$\Pi = \begin{pmatrix} & h & \phi \\ h^\dagger & & h^T \\ \phi^\dagger & h^* & \end{pmatrix}, \tag{7.13}$$

where $\mathbf{1}$ is the $2 \times 2$ identity matrix, $\phi$ is an SU(2) triplet, and $h$ is an SU(2) doublet.

Two SU(2)×U(1) subgroups of SU(5) are gauged with the following generators:

$$Q_{1a} = \begin{pmatrix} \sigma_a & & \\ & & \\ & & \end{pmatrix} \qquad Q_{2a} = \begin{pmatrix} & & \\ & & \\ & & \sigma_a^* \end{pmatrix} \tag{7.14}$$

$$Y_1 = \frac{1}{10}\mathrm{diag}(2,2,-3,-3,-3) \qquad Y_2 = \frac{1}{10}\mathrm{diag}(3,3,3,-2,-2). \tag{7.15}$$

The diagonal subgroup belongs to SO(5) and is unbroken by the $\Sigma$ vev. The gauging explicitly breaks the SU(5) and generates a potential for the Goldstone bosons. Out of the 14 original Goldstones, 4 are eaten by the gauge bosons that become massive. There are 10 left. If only one SU(2)×U(1) gauge coupling constant is turned on, the global symmetry breaking pattern is (SU(3)×SU(2))/(SO(3)×U(1)). This leaves 7 exact Goldstones, three of which are eaten, and four of which remain massless. These massless Goldstone bosons are the Higgs bosons, whose mass is protected by collective symmetry breaking. To summarize, out of the 10 uneaten pseudo-Goldstone bosons, 6, forming an electroweak triplet, do not have their mass protected by collective symmetry breaking and get a mass of order $f$ (and possibly a vev $v'$ of order $v^2/f$), while 4, corresponding to an electroweak doublet, get a lower mass of order the electroweak scale.

An interesting feature of the littlest Higgs model is that gauge boson loops generate the following operators:

$$f^4 \left( c_1 g_1^2 \mathrm{Tr}\Sigma Q_{1a} \Sigma^* Q_{1a}^* + c_2 g_2^2 \mathrm{Tr}\Sigma Q_{2a} \Sigma^* Q_{2a}^* \right) . \tag{7.16}$$

These operators give a mass of order $f$ for the electroweak triplet $\phi$, and generate a quartic coupling of order one for the Higgs. Therefore we do not need to add a plaquette term 'by hand'. It is naturally there, generated by gauge interactions.

Similarly, a top quark Yukawa coupling can be written down in a way analogous to Eq. (7.6) (see Section 7.1.8 for details):

$$\mathcal{L}_Y = \frac{i}{2}\lambda_1 f \epsilon_{ijk}\epsilon_{xy}\chi_i \Sigma_{jx}\Sigma_{ky} u_3'^c + \lambda_2 f \tilde{t}\tilde{t}'^c + \mathrm{h.c.} \tag{7.17}$$

From these couplings, one finds that the heavy vector-like SU(2)-singlet quark $T$ is heavier than $\sqrt{2}f$.

There are many possible variations on the Littlest Higgs theme. The simplest one is the SU(6)/Sp(6) model [10]. This model trades the electroweak triplet scalar of the Littlest Higgs model for an electroweak singlet and an extra light Higgs doublet. To relax constrains from electroweak precision measurements it is also possible to build in a custodial SU(2) symmetry in the gauge sector of the Littlest Higgs model by having an SO(9)/(SO(5)×SO(4)) coset space [11]. Finally one can build a Littlest Higgs model with T-parity [12, 28], which we will discuss further in Section 7.1.5; the phenomenology of models with T-parity will be reviewed in more detail in Section 7.5.

### 7.1.4 Simple gauge group models

In the previous models we could obtain the desired low energy gauge couplings in a way that respects the collective symmetry breaking by using a product of gauge groups. When one gauge coupling of the product was set to zero, the Higgs was exactly massless and that is how collective symmetry breaking was achieved. One can also use a simple gauge group and get the collective symmetry breaking by having two nonlinear sigma fields that get a vacuum expectation value [13–15]. Each field alone 'thinks' that it is the one breaking the symmetry and getting absorbed by the massive gauge bosons, and the couplings of both fields are needed to generate a potential for the uneaten pseudo-Goldstone bosons. The simplest model [14] of this type is an SU(3)×U(1) gauge theory broken down to SU(2)×U(1) by the vev of two different SU(3) fundamentals:

$$\langle \phi_1 \rangle = \begin{pmatrix} 0 \\ 0 \\ f_1 \end{pmatrix}, \qquad \langle \phi_2 \rangle = \begin{pmatrix} 0 \\ 0 \\ f_2 \end{pmatrix}. \tag{7.18}$$

The global symmetry in this case is an $(\mathrm{SU(3)}\times\mathrm{U(1)})^2/(\mathrm{SU(2)}\times\mathrm{U(1)})^2$ that rotates the two fields independently. The pseudo-Goldstone bosons can be parameterized by fluctuations about the vacuum:

$$\phi_1(x) = \exp(iT^a\pi_1^a/f_1)\begin{pmatrix} 0 \\ 0 \\ f_1 \end{pmatrix}, \qquad \phi_2(x) = \exp(iT^a\pi_2^a/f_2)\begin{pmatrix} 0 \\ 0 \\ f_2 \end{pmatrix}. \tag{7.19}$$

Once again, the gauge couplings explicitly break the global symmetry, but couplings to both $\phi_1$ and $\phi_2$ are needed to generate a potential for the pseudo-Goldstone bosons. The Higgs mass is then suppressed relative to the $f$ scale; however, the Higgs quartic coupling is also small. An extra 'plaquette' operator that breaks the $(\mathrm{SU(3)}\times\mathrm{U(1)})^2$ global symmetry must be added to give a large enough quartic coupling [14]. Alternately, a large quartic can be produced if the theory is enlarged to an SU(4) gauge theory with four fundamentals breaking it to SU(2) [13]. Another model, consisting of a SU(9)/SU(8) global symmetry with SU(3)×U(1) gauged, contains two light Higgs doublets and also generates a large enough quartic [15].

The field content of simple group models is slightly different than in the other models. Instead of an electroweak triplet of vector bosons at the scale $f$, there are the broken SU(3) or SU(4) generators: one or two $Z'$ bosons and an extra electroweak doublet of vector bosons. Also, the spectrum of fermions is enlarged. Before we only needed an extra fermion in the top sector to cancel the one loop quadratic divergence of the Standard Model top quark. Here, because of the extended gauge group, extra fermions for all generations of the Standard Model are needed. Finally, because of the two vevs $f_1, f_2$, there is no simple relationship between the mass of the heavy vector bosons and the mass of the heavy top. As we will see next, this helps in avoiding constraints from electroweak precision measurements.

### 7.1.5 Constraints from electroweak precision measurements

Even if the extra states of little Higgs models are predicted to be out of reach of LEP II and the Tevatron, precision electroweak tests provide stringent constraints on the properties of these particles. This is sometimes referred to as the LEP paradox [29, 30]: we need new states at about 1 TeV to stabilize the Higgs mass, however LEP precision data have probed physics at the TeV scale via its influence on radiative corrections and do not see anything new. Little Higgs models suffer from this paradox; the new TeV scale states that are responsible for the cancellation of the Higgs quadratic divergences can be exchanged at tree level, and this can result in a significant departure from the LEP I and LEP II data. There are many studies on the subject [31–39], and a more detailed review will be presented in Section 7.2. In the Littlest Higgs model for example, exchange of the $B'$ and $W'$, $Z'$ gauge bosons, as well as a vev for the heavy triplet, can all cause trouble. In product group models and theory space models, the couplings of the new gauge bosons can be written solely in terms of Standard Model currents [40], and the deviation from the Standard Model can be parametrized by the oblique parameters $S, T, Y$ and $W$ [37, 40]. In general, none of the dangerous couplings that give large contributions to these parameters are tied to the couplings that ensure the cancellation of the Higgs mass quadratic divergences. Therefore it is in general possible to find regions of parameter space where the constraints are satisfied with reasonable fine tuning in the Higgs mass ($\sim 10\%$). However, the allowed region is in general quite small. One reason is that, in both the product group models and the theory space models, to avoid contrains from exchange of heavy gauge bosons one needs a largish $f$. However, in these models the mass of the heavy top partner, responsible for the cancellation of the top quark quadratic divergence, is tied to the scale $f$. Since the top loop quadratic divergence is the largest, the heavy top quark partner cannot be too heavy without reintroducing fine-tuning, and this tends to push the models into a small corner of parameter space. In simple gauge group models, the relationship between the heavy gauge boson and the heavy top partner masses is not as direct. Therefore, one gains a little bit. In particular, in simple gauge group models the electroweak precision measurements typically give strong constraints on $\sqrt{f_1^2 + f_2^2}$, while the heavy top quark partner mass is not directly tied to this combination, and can be made relatively light.

Several models have been built with the specific intention of reducing the constraints of electroweak precision measurements. One straightforward option to improve the Littlest Higgs model is to gauge only the diagonal U(1)$_Y$ instead of a U(1)$^2$ [32, 34, 41]. This eliminates the constraints coming from the exchange of the $B'$, which is removed from the spectrum, at the expense of not cancelling the quadratic divergence due to the hypercharge gauge boson. This is not a serious problem since with a cutoff of 10 TeV, the quadratic divergence due to the hypercharge gauge boson is not very big. There are also models with a custodial SU(2) symmetry built in to eliminate the worst contributions to the $T$ parameter; these models are typically more complicated but slightly less constrained [8, 11].

The most interesting direction in trying to avoid electroweak precision measurements is probably the idea of T-parity [9, 12, 28, 42, 43]. Just as in the MSSM where R-parity forbids the coupling of one superpartner with two Standard Model particles, T-parity tries to avoid tree-level exchange of the heavy states by making them odd under a new parity, while all the Standard Model particles are even. This has the additional advantage of ensuring the presence of a stable heavy particle which could play the role of dark matter [44]. The main drawback of this approach is that it requires the addition of one new TeV scale fermion for each of the fermions of the Standard Model [12]. This in turns raises flavor questions similar to those in the MSSM. T-parity has been introduced in theory space models, where the parity has a nice geometric interpretation, and in product group models, but not in simple group models. The phenomenology of models with T-parity will be discussed in more detail in Section 7.5.

#### *7.1.6 Theoretical constraints*

In addition to the electroweak precision constraints, there are additional constraints on little Higgs models from unitarity and from considering the log-divergent terms in the Higgs potential. We also discuss here the prospects for little Higgs models to incorporate dark matter, neutrino masses, and the baryon asymmetry of the universe.

One can analyze the scattering of all possible pairs of Goldstone bosons in little Higgs models to find where unitarity is violated. The violation of unitarity at some scale indicates that the theory is not valid above that scale, or that perturbation theory has broken down. Due to the large number of Goldstones in little Higgs models, this unitarity analysis generically predicts an upper cutoff $\Lambda \simeq (3-4)f$ depending on the model, which is somewhat less than the $4\pi f \sim$ 10–30 TeV usually quoted using Naive Dimensional Analysis [45].

There are also constraints on the scale $f$. The Naive Dimensional Analysis used to predict that $f \simeq 1$ TeV neglects the contributions to the potential that go like

$$\mathrm{Tr} M^4(\Sigma) \log \frac{M^2(\Sigma)}{\Lambda^2}. \tag{7.20}$$

Including these terms, it is found that a light Higgs can only be achieved with $f$ somewhat smaller than 1 TeV. For large $f \gg 1$ TeV, the Higgs mass is pulled up toward the scale $f$, destroying the desired hierarchy [46]. Therefore the desired hierarchy $v \ll f \ll \Lambda$ can be preserved, but the separation between each of these scales may only be a factor of 3–5 instead of $4\pi$. While worsening the electroweak precision constraints, these observations significantly improve the possibility of finding not only the $f$-scale particles at the LHC, but also the $\Lambda$-scale particles as well.

Because little Higgs models have a cutoff at a relatively low scale $\Lambda \sim 10$ TeV, the issues of dark matter, neutrino masses, and the baryon asymmetry of the universe can be deferred to energy scales above the cutoff. However, there have been some attempts to incorporate this physics within little Higgs models themselves.

Dark matter appears naturally as the lightest T-odd particle in little Higgs models with T-parity [44]. Even without T-parity, theory space models often contain discrete symmetries, some part of which can remain unbroken even after electroweak symmetry breaking; the dark matter could then consist of a nonlinear sigma model field made stable by this accidental exact global symmetry [47].

There have been two main approaches to neutrino mass generation in little Higgs models. First, some models (such as the Littlest Higgs) contain a scalar triplet with a nonzero vev. This triplet can be used to generate neutrino Majorana masses through a lepton number violating coupling to two left-handed SM neutrinos [35, 48–50]. Second, simple group models naturally contain a pair of extra SM gauge singlets $N, N^c$ at the $f$ scale due to the expansion of the lepton doublets into fundamentals of the enlarged gauge group. If lepton number is broken at a small scale $M \sim$ keV, generating a small Majorana mass for $N^c$, then the SM neutrinos can get a radiatively generated Majorana mass [51] of the correct size through their mixing with $N$, without requiring extremely tiny Yukawa couplings.

Electroweak baryogenesis relies on the restoration of electroweak symmetry at high temperature. This happens as a result of an effective positive mass squared term $m^2_{\rm eff} \sim T^2$ acquired by the Higgs from interactions with the ambient thermal plasma. However, this effective mass is generated precisely by the Higgs self-energy diagrams that are quadratically divergent at $T = 0$ [52], i.e., those that are canceled by the little Higgs mechanism. A similar cancellation happens in the MSSM, in which the quadratically divergent contributions cancel between bosonic and fermionic degrees of freedom; at finite temperature, these contributions enter the thermal mass with different coefficients due to the different statistics of the relevant particles in the thermal bath, and thus no longer cancel. In little Higgs models, however, the quadratic divergences cancel between particles *of the same statistics*, so that the thermal mass is also canceled [52]. A detailed study [52] of the Littlest Higgs model with SU(2)$^2\times$U(1) gauged shows an initial symmetry restoration as in the Standard Model as $T$ is increased, followed by a rebreaking at $T \sim f$ to a new global minimum.

The baryon asymmetry of the universe could also arise through leptogenesis, with an initial lepton asymmetry transmitted to the baryon sector through electroweak sphalerons. Leptogenesis generates the initial CP asymmetry through out-of-equilibrium decay of heavy right-handed neutrinos. To generate a large enough asymmetry, the right-handed neutrinos must have large enough CP-violating couplings to the light neutrinos and the SM Higgs. Normally this forces the right-handed neutrino scale to be near the GUT scale ($\sim 10^{16}$ GeV) so that the SM neutrinos will be kept light enough by the see-saw mechanism. Such a scenario cannot be fit into a little Higgs model because the cutoff is much lower, $\Lambda \sim 10$ TeV.

However, recently it was shown [53] how to implement TeV-scale leptogenesis in little Higgs models, both in simple group models and in Littlest Higgs-type models. In simple group models the SM neutrino masses can be radiatively generated as discussed above [51], so that the CP-violating couplings relevant for leptogenesis can still be large without generating too large a neutrino mass. In Littlest Higgs-type models with the SM neutrino masses generated through couplings to a scalar triplet, leptogenesis can be implemented by adding a moderately heavy fourth neutrino family which carries the large CP-violating coupling.

### 7.1.7 New gauge bosons

Little Higgs models extend the electroweak gauge group at the TeV scale. The structure of the extended electroweak gauge group determines crucial properties of the model, which can be revealed by studying the new gauge bosons at the TeV scale. Experimental studies for the LHC will be presented in Sections 7.6 and 7.7; prospects for ILC measurements will be discussed in Section 7.8. In the Littlest Higgs model [6], the heavy gauge bosons consist of an SU(2)$_L$ triplet $Z_H, W_H^{\pm}$ from the breaking of SU(2)×SU(2) down to the electroweak SU(2)$_L$. A similar structure arises in many of the product group and theory space models. In the SU(3)×U(1) simple group model [13, 14], the heavy gauge bosons consist of an SU(2)$_L$ doublet $(Y^0, X^-)$ corresponding to the broken off-diagonal generators of SU(3), and a $Z'$ gauge boson corresponding to the broken linear combination of the $T^8$ generator of SU(3) and the U(1). Again, a similar pattern arises in other simple group models.

The extra gauge bosons get their masses from the $f$ condensate, which breaks the extended gauge symmetry. For example, in the Littlest Higgs and the SU(3) simple group models, the gauge boson masses are given in terms of the model parameters by

$$\left.\begin{array}{l} M_{W_H} = M_{Z_H} = gf/2sc = 0.65f/\sin 2\theta \\ M_{A_H} = g s_W f/2\sqrt{5} c_W s' c' = 0.16 f/\sin 2\theta' \end{array}\right\} \text{in the Littlest Higgs model,}$$
$$\left.\begin{array}{l} M_{Z'} = \sqrt{2} g f/\sqrt{3 - t_W^2} = 0.56 f \\ M_X = M_Y = gf/\sqrt{2} = 0.46 f = 0.82 M_{Z'} \end{array}\right\} \text{in the SU(3) simple group model.} \quad (7.21)$$

In the SU(3) simple group model the heavy gauge boson masses are determined by only one free parameter, the scale $f = \sqrt{f_1^2 + f_2^2}$. The Littlest Higgs model has two additional gauge sector parameters, $\tan\theta = s/c = g_2/g_1$ [in the SU(2)$^2$ →SU(2) breaking sector] and $\tan\theta' = s'/c' = g'_2/g'_1$ [in the U(1)$^2$ →U(1) breaking sector]. If only one copy of U(1) is gauged [32], the $A_H$ state is not present and the gauge sector of the Littlest Higgs model is controlled by only two free parameters, $f$ and $\tan\theta$.

The gauge couplings of the Higgs doublet take the general form [54]

$$\mathcal{L} = \left\{ \begin{array}{l} \left[G_{HHVV} VV + G_{HHV'V'} V'V' + G_{HHVV'} VV'\right] H^2 \\ \left[G_{HHV^+V^-} V^+V^- + G_{HHV'^+V'^-} V'^+V'^- + G_{HHV^+V'^-}(V^+V'^- + V^-V'^+)\right] H^2, \end{array}\right. \quad (7.22)$$

where the top line is for $V$ neutral and the bottom line is for $V$ charged. Here $V$ and $V'$ stand for the SM and heavy gauge bosons, respectively. This Lagrangian leads to two quadratically divergent diagrams contributing to the Higgs mass: one involving a loop of $V$, proportional to $G_{HHVV}$, and the other involving a loop of $V'$, proportional to $G_{HHV'V'}$. The divergence cancellation in the gauge sector can thus be written as

$$\sum_i G_{HHV_iV_i} = 0, \quad (7.23)$$

where the sum runs over all gauge bosons in the model. The couplings in the Littlest Higgs and SU(3) simple group models are given, e.g., in Table 3 of Ref. [54]. In the SU(3) simple group model, the quadratic divergence cancels between the $Z$ and $Z'$ loops and between the $W$ and $X$ loops. In the Littlest Higgs model, the quadratic divergence cancels between the $W$ and $W_H$ loops and there is a partial cancellation between the $Z$ and $Z_H$ loops. Including the $A_H$ loop leads to a complete cancellation of the quadratic divergence from the $Z$ loop. The key test of the little Higgs mechanism in the gauge sector is the experimental verification of Eq. (7.23).

After EWSB, the couplings of $H^2$ to one heavy and one SM gauge boson induce mixing between the heavy and SM gauge bosons:

$$V' = V'_0 - \delta_V V_0, \qquad \delta_V = -v^2 G_{HHVV'}/M_{V'}^2, \quad (7.24)$$

where $V'_0$, $V_0$ stand for the states before EWSB. This mixing gives rise to triple gauge couplings between one heavy and two SM gauge bosons.

In the Littlest Higgs model, the couplings of the heavy gauge bosons to the SU(2)$_L$ fermion currents take the form

$$Z_H^\mu \overline{f} f: \quad ig \cot\theta \, T_f^3 \gamma^\mu P_L, \qquad\qquad W_H^{+\mu} \overline{u} d: \quad -\frac{ig}{\sqrt{2}} \cot\theta \, \gamma^\mu P_L, \tag{7.25}$$

where $T_f^3 = 1/2$ $(-1/2)$ for up (down) type fermions. Below the TeV scale, exchange of $W_H$ and $Z_H$ gives rise to four-fermi operators, which are constrained by the electroweak precision data. The experimental constraints are loosened by going to small values of $\cot\theta$, for which the couplings of the heavy gauge bosons are suppressed. In the SU(3) simple group model, the $Z'$ couples to SM fermions with gauge strength, while the $X, Y$ gauge bosons couple only via the mixing between SM fermions and their TeV-scale partners. The $Z'$ couplings are fixed by the charges of the SM fermions under SU(3)×U(1)$_X$, and cannot be written in terms of the usual SM currents. The electroweak precision constraints in this case cannot be parameterized solely in terms of the oblique parameters [37, 40].

### 7.1.8 *New fermions and the top partner*

The new heavy quark sector in the Littlest Higgs model [6] consists of a pair of vectorlike SU(2)-singlet quarks that couple to the top sector. The Lagrangian is [54]

$$\mathcal{L}_Y = \frac{i}{2} \lambda_1 f \epsilon_{ijk} \epsilon_{xy} \chi_i \Sigma_{jx} \Sigma_{ky} u_3^{\prime c} + \lambda_2 f \tilde{t} \tilde{t}^{\prime c} + \text{h.c.}, \tag{7.26}$$

where $\chi_i = (b_3, t_3, i\tilde{t})$ and the factors of $i$ in Eq. (7.26) and $\chi_i$ are inserted to make the masses and mixing angles real. The summation indices are $i, j, k = 1, 2, 3$ and $x, y = 4, 5$, and $\epsilon_{ijk}$, $\epsilon_{xy}$ are antisymmetric tensors. The vacuum expectation value $\langle \Sigma \rangle \equiv \Sigma_0$ marries $\tilde{t}$ to a linear combination of $u_3^{\prime c}$ and $\tilde{t}^{\prime c}$, giving it a mass of order $f \sim$ TeV. The resulting new charge 2/3 quark $T$ is an isospin singlet up to its small mixing with the SM top quark (generated after EWSB). The orthogonal linear combination of $u_3^{\prime c}$ and $\tilde{t}^{\prime c}$ becomes the right-handed top quark and marries $t_3$.

In the SU(3)×U(1) simple group model [13, 14], the top quark mass is generated by the Lagrangian [54]

$$\mathcal{L}_Y = i\lambda_1^t u_1^c \Phi_1^\dagger Q_3 + i\lambda_2^t u_2^c \Phi_2^\dagger Q_3, \tag{7.27}$$

where $Q_3^T = (t, b, iT)$ and the factors of $i$ in Eq. (7.27) and $Q_3$ are again inserted to make the masses and mixing angles real. The $\Phi$ vevs marry $T$ to a linear combination of $u_1^c$ and $u_2^c$, giving it a mass of order $f \sim$ TeV. The new charge 2/3 quark $T$ is a singlet under SU(2)$_L$ up to its small mixing with the SM top quark (generated after EWSB). The orthogonal linear combination of $u_1^c$ and $u_2^c$ becomes the right-handed top quark. For the rest of the quarks, the scalar interactions depend on the choice of their embedding into SU(3). The most straightforward choice is to embed all three generations in a universal way, $Q_m^T = (u, d, iU)_m$, so that each quark generation contains a new heavy charge 2/3 quark. This embedding leaves the SU(3) and U(1)$_X$ gauge groups anomalous; the anomalies can be canceled by adding new spectator fermions at the cutoff scale $\Lambda \sim 4\pi f$. An alternate, anomaly-free embedding [55, 56] puts the quarks of the first two generations into antifundamentals of SU(3), $Q_m^T = (d, -u, iD)_m$, with $m = 1, 2$, so that the first two quark generations each contain a new heavy charge $-1/3$ quark. Interestingly, an anomaly-free embedding of the SM fermions into SU(3)$_c$×SU(3)×U(1)$_X$ is only possible if the number of generations is a multiple of three [55–58].[1]

The masses of the top quark $t$ and its heavy partner $T$ are given in terms of the model parameters

[1] This rule can be violated in models containing fermion generations with non-SM quantum numbers, e.g., mirror families [59].

by

$$m_t = \lambda_t v = \begin{cases} \dfrac{\lambda_1\lambda_2}{\sqrt{\lambda_1^2+\lambda_2^2}}v & \text{in the Littlest Higgs model,} \\ \dfrac{\lambda_1\lambda_2}{\sqrt{2}\sqrt{\lambda_1^2 c_\beta^2+\lambda_2^2 s_\beta^2}}v & \text{in the SU(3) simple group model;} \end{cases}$$

$$M_T = \begin{cases} \sqrt{\lambda_1^2+\lambda_2^2}\, f = (x_\lambda + x_\lambda^{-1})\dfrac{m_t}{v}f & \text{in the Littlest Higgs model,} \\ \sqrt{\lambda_1^2 c_\beta^2+\lambda_2^2 s_\beta^2} f = \sqrt{2}\,\dfrac{t_\beta^2+x_\lambda^2}{(1+t_\beta^2)x_\lambda}\dfrac{m_t}{v}\, f & \text{in the SU(3) simple group model.} \end{cases}$$

Fixing the top quark mass $m_t$ leaves two free parameters in the Littlest Higgs model, which can be chosen to be $f$ and $x_\lambda \equiv \lambda_1/\lambda_2$. We see that the SU(3) simple group model contains one additional parameter, $t_\beta \equiv \tan\beta = f_2/f_1$. In the SU(3) simple group model, we define $f \equiv \sqrt{f_1^2+f_2^2}$.

To reduce fine-tuning in the Higgs mass, the top-partner $T$ should be as light as possible. The lower bound on $M_T$ is obtained for certain parameter choices:

$$M_T \geq \begin{cases} 2\dfrac{m_t}{v}f \approx \sqrt{2}f & \text{for } x_\lambda = 1 \text{ in the Littlest Higgs model,} \\ 2\sqrt{2}s_\beta c_\beta\dfrac{m_t}{v}f \approx f\sin 2\beta & \text{for } x_\lambda = t_\beta \text{ in the SU(3) simple group model,} \end{cases}$$

where in the last step we used $m_t/v \approx 1/\sqrt{2}$. The $T$ mass can be lowered in the SU(3) model for fixed $f$ by choosing $t_\beta \neq 1$, thereby introducing a mild hierarchy between $f_1$ and $f_2$.

The couplings of the Higgs doublet to the $t$ and $T$ mass eigenstates can be written in terms of an effective Lagrangian [54],

$$\mathcal{L}_Y \supset \lambda_t H t^c t + \lambda_T H T^c t + \frac{\lambda_T'}{2M_T} HHT^cT + \text{h.c.}, \tag{7.28}$$

where the four-point coupling arises from the expansion of the nonlinear sigma model field. This effective Lagrangian leads to three diagrams contributing to the Higgs mass corrections at one-loop level, shown in Fig. 7.2: (a) the SM top quark diagram, which depends on the well-known SM top Yukawa coupling $\lambda_t$; (b) the diagram involving a top quark and a top-partner $T$, which depends on the $HTt$ coupling $\lambda_T$; and (c) the diagram involving a $T$ loop coupled to the Higgs doublet via the dimension-five $HHTT$ coupling. The couplings in the three diagrams of Fig. 7.2 must satisfy the following relation [60] in order for the quadratic divergences to cancel:

$$\lambda_T' = \lambda_t^2 + \lambda_T^2. \tag{7.29}$$

This equation embodies the cancellation of the Higgs mass quadratic divergence in any little Higgs theory. The couplings in the Littlest Higgs and SU(3) simple group models are given, e.g., in Table 1 of Ref. [54]. If the little Higgs mechanism is realized in nature, it will be of fundamental importance to establish the relation in Eq. (7.29) experimentally.

After EWSB, the coupling $\lambda_T$ induces a small mixing of electroweak doublet into $T$,

$$T = T_0 - \delta_T t_0, \quad \delta_T = \lambda_T \frac{v}{M_T}, \tag{7.30}$$

where $T_0, t_0$ stand for the electroweak eigenstates before the mass diagonalization at the order of $v/f$. This mixing gives rise to the couplings of $T$ to the SM states $bW$ and $tZ$ with the same form as the corresponding SM couplings of the top quark except suppressed by the mixing factor $\delta_T$.

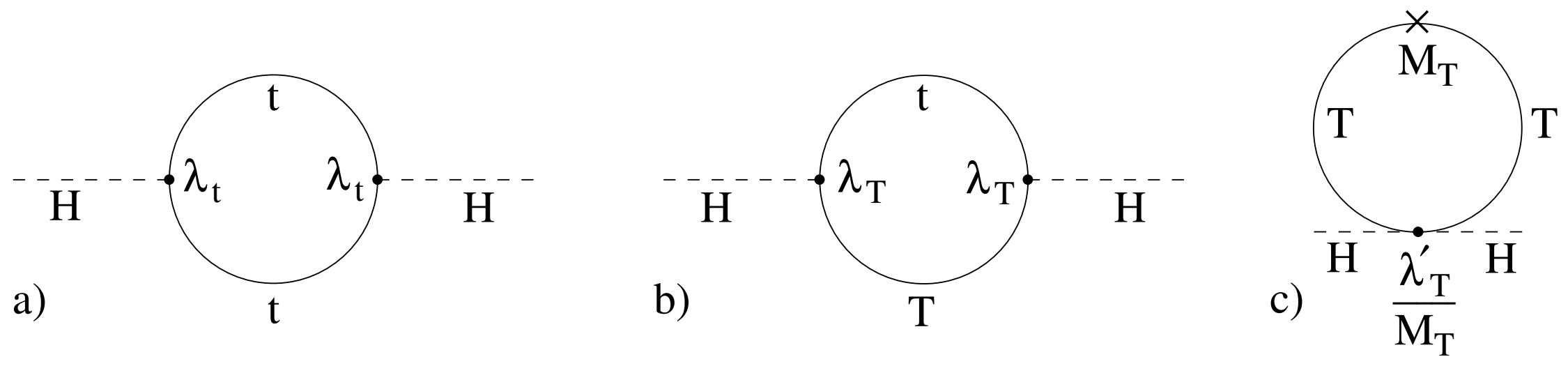

Fig. 7.2: Quadratically divergent one-loop contributions to the Higgs boson mass-squared from the top sector in little Higgs models. From Ref. [54].

Table 7.2: Particle content of the scalar sectors of little Higgs models. SU(2) doublets, triplets, complex singlets, and singlet pseudoscalars are denoted by $h$, $\phi$, $\sigma$, and $\eta$, respectively. SU(2) multiplets are complex unless specified otherwise; the real triplet and singlet are denoted by $\phi^r$ and $\sigma^r$, respectively. In the minimal moose with SU(2)$_C$, the $\sigma^\pm$, $\sigma^r$ fields form a triplet under the custodial SU(2) symmetry but are SU(2)$_L$ singlets.

| Model | EW-scale scalars | TeV-scale scalars |
|---|---|---|
| Minimal moose [7] | $h_1$, $h_2$, $\phi$, $\sigma$ | (none) |
| Minimal moose with SU(2)$_C$ [8] | $h_1$, $h_2$ | $\phi^r$, $\sigma^\pm$, $\sigma^r$ |
| Moose with T-parity [9] | $h_1$, $h_2$ | $h_{3,4,5}$, $\phi^r_{1,2,3}$, $\sigma_{1,2,3,4,5}$, $\eta_{1,2,3}$ |
| Littlest Higgs [6] | $h$ | $\phi$ |
| SU(6)/Sp(6) model [10] | $h_1$, $h_2$ | $\sigma$ |
| Littlest Higgs with SU(2)$_C$ [11] | $h$ | $\phi$, $\phi^r$, $\eta$ |
| Littlest Higgs with T-parity [12] | $h$ | $\phi$ |
| SU(3) simple group [13, 14] | $h$, $\eta$ | (none) |
| SU(4) simple group [13] | $h_1$, $h_2$, $\eta_1$, $\eta_2$ | $\sigma_1$, $\sigma_2$, $\sigma_3$ |
| SU(9)/SU(8) simple group [15] | $h_1$, $h_2$ | $\sigma_1$, $\sigma_2$ |

### *7.1.9 The scalar sector*

The scalar sectors of little Higgs models are very model dependent, because they correspond to the coset space of the broken global symmetries minus those exact Goldstone bosons that are eaten by the broken gauge generators. The phenomenology of the scalar sector thus provides a very important experimental handle on the global symmetry of the model and the symmetry breaking pattern.

The states in the scalar sector are characterized by their SU(2)$_L\times$U(1) and CP quantum numbers. The scalar sector must contain at least one SU(2)-doublet Higgs field with mass near the electroweak scale to reproduce SM electroweak symmetry breaking. The scalar content of various little Higgs models is summarized in Table 7.2. We denote SU(2) doublets as $h$, SU(2) triplets as $\phi$, complex SU(2) singlets as $\sigma$, and SU(2) singlet pseudoscalars as $\eta$.

Some models, including the Littlest Higgs [6], its extensions with custodial SU(2)$_C$ symmetry [11] and with T-parity [12], and the SU(3) simple group model [13, 14], contain a single SU(2) doublet Higgs field at the electroweak scale. The physical Higgs boson in these models has couplings that are identical to those of the SM Higgs up to corrections suppressed by the ratio of scales $v/f$; the corrections to the Higgs production cross sections and decay partial widths are then proportional to $(v/f)^2 \sim$ few percent [61–63]. These corrections come from the mixing between SM and TeV-scale states, from the higher-order terms in the expansion of the nonlinear sigma model, and from corrections to the SM input parameters such as $G_F$. In such models, high-precision Higgs coupling measurements will be a useful test of the model structure, and could shed light on strongly-coupled new physics at the UV-completion scale around 10 TeV. This will be reviewed in more detail in Section 7.3.

Other models give rise to two Higgs doublets at the electroweak scale. The Higgs phenomenology below the TeV scale is then that of a two Higgs doublet model, typically with a constrained form of the scalar potential that can give rise to characteristic relations between the Higgs masses and mixing angles.

Little Higgs models often contain at least one additional U(1) global symmetry that is broken by the $f$ vev. This gives rise to an additional physical pseudoscalar mode, $\eta$, typically with mass near the electroweak scale, which can have significant effects on the phenomenology [64]. This occurs in, e.g., the SU(3) simple group model [13, 14]. A pseudoscalar also arises in the Littlest Higgs model [6] when only $SU(2)^2 \times U(1)$ is gauged, instead of the usual $[SU(2)\times U(1)]^2$. The origin and phenomenology of these pseudoscalars will be reviewed in more detail in Section 7.4.

Finally, some models contain Higgs triplets at the TeV scale, or even at the electroweak scale. These triplets can give rise to potentially dangerous contributions to electroweak precision observables through their nonzero vevs $v'$. They can also yield interesting phenomenology such as decays of the doubly-charged member of the triplet into pairs of like-sign $W$ bosons or, in versions of the models with lepton number violation [35, 48–50], into like-sign dileptons.

## 7.2 Impact of electroweak precision data on the little Higgs models

*Aldo Deandrea*

The electroweak sector of the SM has been tested to a very high accuracy and an important test of the validity of little Higgs models is therefore through comparison with precision data (for reviews treating this subject see [65, 66]). The strategy to compute limits from the electroweak precision data is not unique and indeed different methods are discussed in the literature. It is possible to compute directly quantities which are constrained by the experimental data and fit the whole set in order to get constraints on the model. One can also rely on the computation of a restricted set of relevant quantities. Finally one can integrate out the heavy fields and study the effective low energy lagrangian. The originally proposed models are tightly constrained while more recent ones, such as the Littlest Higgs model with T-parity, satisfy the electroweak constraints in a larger region of the parameter space.

A special feature of the SM with one Higgs doublet is the validity of the tree level relation

$$\rho = 1 = \frac{M_W^2}{M_Z^2 c_\theta^2} \tag{7.31}$$

due to the tree level custodial symmetry. In many little Higgs models the custodial symmetry is no longer a good symmetry of the model, *i.e.* $\rho \neq 1$ already at the tree level. Another source of constraints from the electroweak precision data are $SU(2)_L$ triplet Higgs, as a trilinear coupling between the doublet and the triplet Higgs, $H^T \Phi^\dagger H$, is allowed by the gauge symmetry $\mathrm{SU}(2)_L \times \mathrm{U}(1)_Y$. Unless a discrete symmetry is imposed to forbid such a trilinear interaction, the vev of the triplet is non-zero and leads to a new input parameter in the gauge sector, and many predictions of the Standard Model are changed by the presence of such a term.

### 7.2.1 Littlest Higgs

Many studies in the literature concern the little Higgs model and its extensions [31–33, 35, 37–39, 60, 67, 68]. As an example we show in Fig. 7.3 the limits obtained in the $\mathrm{SU}(5)/\mathrm{SO}(5)$ Littlest Higgs in [32]. The leading corrections are given by the tree-level exchanges of heavy gauge bosons and the effects of the non-zero triplet scalar vev. Weak isospin violating contributions arise at tree level due to the absence of a custodial SU(2) symmetry. The main component of the corrections come from heavy gauge boson exchanges, while a smaller contribution is due to the triplet vev $v'$.

The input parameters in the analysis of the electroweak data can be chosen to be the Fermi constant $G_F$, the mass of the $Z$ vector boson $m_Z$ and the fine–structure coupling $\alpha(m_Z)$. One can first look at the modification to $G_F$. We have two types of modifications: one directly from the mixing of the heavy $W_H$ bosons to the coupling of the charged current and the second one from the contribution of the charged current to the equations of motion of the heavy gauge bosons. In terms of the model parameters:

$$\frac{G_F}{\sqrt{2}} = \frac{\alpha\pi(g^2+g'^2)}{2g^2g'^2m_Z^2}\left(1 - c^2(c^2-s^2)\frac{v^2}{f^2} + 2c^4\frac{v^2}{f^2} - \frac{5}{4}(c'^2-s'^2)^2\frac{v^2}{f^2}\right) . \tag{7.32}$$

where $s, c, s'$, and $c'$ denote the sines and cosines of two mixing angles, respectively. They can be expressed with the help of the coupling constants:

$$\begin{aligned} c' &= g'/g_2' \qquad s' = g'/g_1' \\ c &= g/g_2 \qquad s = g/g_1 \ , \end{aligned} \tag{7.33}$$

with the usual SM couplings $g, g'$, related to $g_1$, $g_2$, $g_1'$ and $g_2'$ by

$$\frac{1}{g^2} = \frac{1}{g_1^2} + \frac{1}{g_2^2}, \qquad \frac{1}{g'^2} = \frac{1}{g_1'^2} + \frac{1}{g_2'^2} \ . \tag{7.34}$$

The Weinberg angle is defined through

$$\frac{G_F}{\sqrt{2}} = \frac{\alpha\pi}{2s_\theta^2c_\theta^2m_Z^2} \ . \tag{7.35}$$

In terms of the model parameters the mass of the $Z$-boson is given by

$$m_Z^2 = (g^2+g'^2)\frac{v^2}{4}\left[1 - \frac{v^2}{f^2}\left(\frac{1}{6} + \frac{(c^2-s^2)^2}{4} + \frac{5}{4}(c'^2-s'^2)\right) + 8\frac{v'^2}{v^2}\right] , \tag{7.36}$$

whereas the $W$-mass is

$$m_W^2 = \frac{g^2v^2}{4}\left[1 - \frac{v^2}{f^2}\left(\frac{1}{6} + \frac{(c^2-s^2)^2}{4}\right) + 4\frac{v'^2}{v^2}\right] . \tag{7.37}$$

The expression for the $Z$-mass can be used to determine the value of $v$ for a given ratio $v/f$. One can also compute from the previous results the correction to the $\rho$ parameter.

By doing a complete analysis one can establish that the symmetry breaking scale is generically bounded by $f > 4$ TeV at 95% C.L. with more stringent bounds for particular choices of the couplings. Modifying the way in which the gauged U(1) generators are embedded, the fermion $U(1)$ charges, or gauging a single U(1), gives the possibility of relaxing the constraints on the scale $f$ [32]. For example one can modify the U(1) charges of the first two generations in the form $RY_F$ under the first U(1) and $(1-R)Y_F$ under the second U(1), where $Y_F$ is the SM hypercharge of the fermion. By requiring the invariance of the Yukawa couplings under the U(1)'s (for details see [32]) one can show that $R$ can take only values which are integer multiples of $1/5$. Results depend also on an additional parameter $a$ expected to be $\mathcal{O}(1)$ in the Coleman-Weinberg potential of the Littlest Higgs. This parameter affects the size of the triplet vev $v'$.

### 7.2.2 *Other little Higgs models*

Little Higgs models implementing custodial symmetry were discussed in [8, 11]. Precision electroweak constraints were considered in the previous papers and also in [38, 69]. The breaking scale $f$ can be as low as 700 GeV without contradiction with the precision electroweak data.

A different approach to reduce the impact of electroweak constraints is based on a discrete parity called T-parity [9, 12, 28] in analogy with R-parity in supersymmetric models. T-parity forbids tree-level contributions from the heavy gauge bosons to observables involving only standard model particles as external states. In the case of the Littlest Higgs model, it also forbids the interactions that induce the triplet vev. Corrections to precision electroweak observables are therefore loop level effects. Analysis of the electroweak precision data allows the scale $f$ to be as low as 500 GeV in these models [43]. However one should keep in mind that T-parity introduces new mirror fermions and their presence leads to tree level flavour changing currents which must be kept under control by an appropriate choice of the mirror fermions mass spectrum and mixing parameters [70].

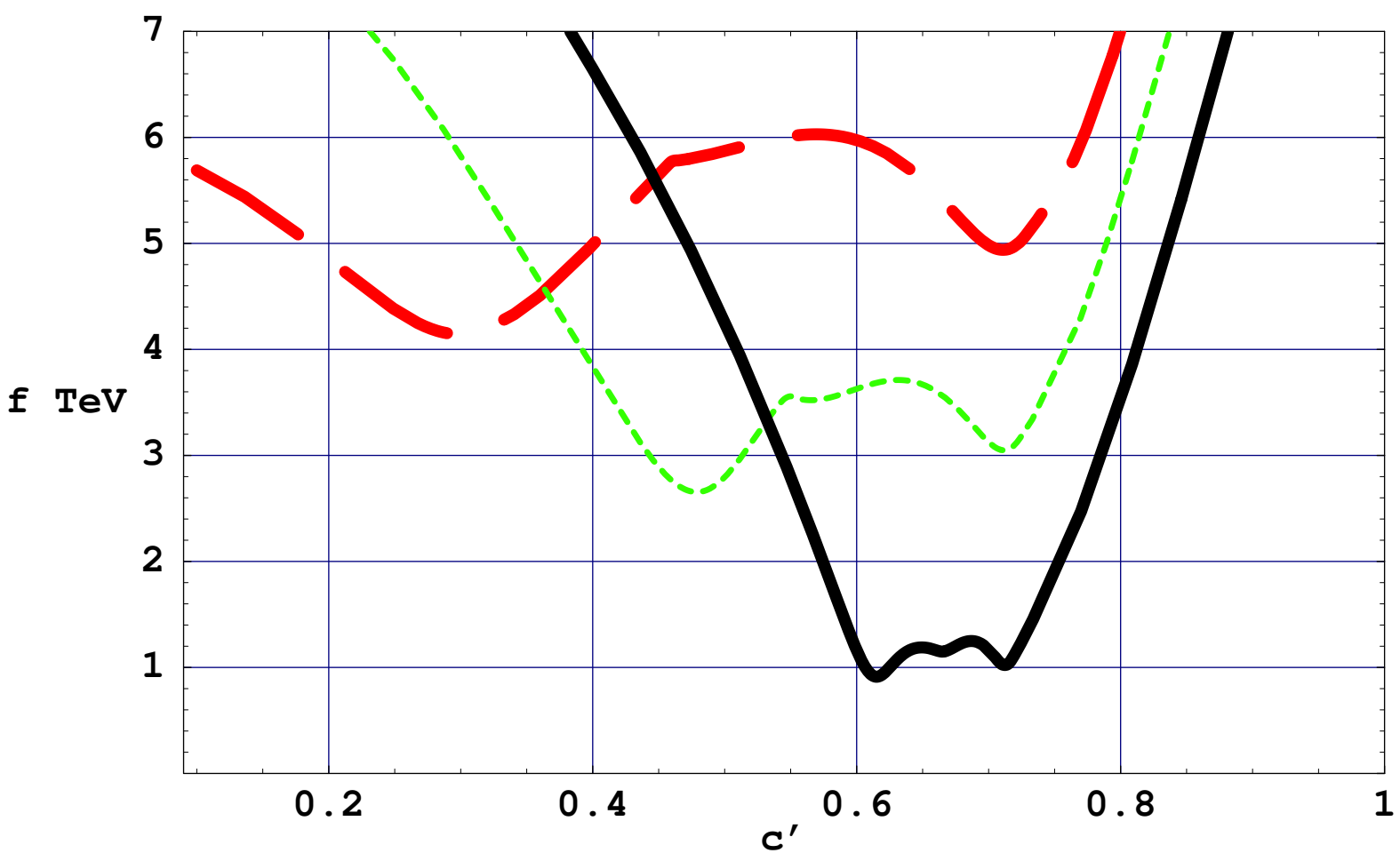

Fig. 7.3: 95% confidence level bound on $f$ for $a = 1$ and $R = 1$ (dashed), $R = 4/5$ (dotted), and $R = 3/5$ (solid). The bounds for $R = 2/5, 1/5$ and 0 can be obtained by reflection around $c'^2 = 1/2$ due to the $R \to 1 - R$, $c'^2 \to 1 - c'^2$ symmetry of the expressions for the electroweak corrections. From [32].

### *7.2.3 Low energy precision data*

Precision experiments at low energy allow a determination of the $g-2$ of the muon and of the weak charge of cesium atoms. These data can be used to put constraints on little Higgs models [38]. Concerning the $g-2$ of the muon, the contributions of the additional heavy particles are completely negligible and the dominant contributions arise from the corrections to the light $Z$ and $W$ couplings. On the contrary the measure of the weak charge of cesium atoms, gives constraints on the little Higgs models, even if weaker that those at LEP energies. Parity violation in atoms is due to the electron-quark effective Lagrangian

$$\mathcal{L}_{\mathit{eff}} = \frac{G_F}{\sqrt{2}}(\bar{e}\gamma_\mu\gamma_5 e)(C_{1u}\bar{u}\gamma^\mu u + C_{1d}\bar{d}\gamma^\mu d)\ . \tag{7.38}$$

The experimentally measured quantity is the so-called “weak charge” defined as

$$Q_W = -2\left(C_{1u}(2Z+N) + C_{1d}(Z+2N)\right)\ , \tag{7.39}$$

where $Z$ and $N$ are the number of protons and neutrons of the atom, respectively. The value of the scale $f$ should be in the range of few TeV in order to obtain the measured deviation. The allowed scale is slightly lower in the custodial model with respect to the non-custodial one as the custodial model is closer to the standard model in its predictions. When the scale $f$ is too large the new physics effects become negligible. The scale $f$ in the few TeV range is consistent with what is expected on the model-building side and from the LEP data for little Higgs model. Obviously this result should be taken only as a first indication as the error on $\delta Q_W(Cs)$ is large.

## 7.3 Couplings of the Littlest Higgs boson

*Heather E. Logan*

A light Higgs boson is the central feature of the little Higgs models. In general, the couplings of the light Higgs boson to Standard Model particles receive corrections due to the structure of the Higgs sector and the presence of new TeV-scale particles [35, 38, 54, 61–63, 67]. In models containing only one light Higgs doublet, such as the Littlest Higgs model [6], the SO(9)/SO(5)×SO(4) model of Ref. [11], the SU(3) simple group model [13, 14], and the "Littlest Higgs" model with T-parity [12], these corrections are suppressed by the square of the ratio of the electroweak scale to the TeV scale, $v^2/f^2$, and are thus parametrically at the level of a few percent. In this contribution we give a convenient parameterization for these corrections, discuss their sources, and summarize their generic features, focusing on the Littlest Higgs model. We also discuss the outstanding issue of tree-level heavy particle exchange in Higgs production and decay, and sketch the generalization to other little Higgs models.

### *7.3.1 Higgs couplings*

In general, all masses that originate with the Standard Model Higgs mechanism and all couplings of SM particles to the Higgs boson $H$ are modified in the Littlest Higgs model at order $v^2/f^2$. We parameterize the modifications by the factors $y_i$, which are the couplings of $H$ to SM particle $i$ normalized according to [71, 72]

$$\mathcal{L} = -\frac{m_t}{v} y_t \bar{t} t H - \frac{m_f}{v} y_f \bar{f} f H + 2\frac{M_W^2}{v} y_W W^+ W^- H + \frac{M_Z^2}{v} y_Z Z Z H. \tag{7.40}$$

The coupling factors $y_i$ in Eq. (7.40) are of order $1 + \mathcal{O}(v^2/f^2)$ and are given for the Littlest Higgs model in Table 7.3. The Higgs coupling to the top quark gets a different correction than the couplings to the light fermions due to the mixing between $t$ and $T$ in the Littlest Higgs model, which can be parameterized by $c_t \equiv \lambda_1/\sqrt{\lambda_1^2 + \lambda_2^2}$ (with $0 < c_t < 1$). The remaining corrections arise from (i) the mixing between $H$ and the neutral CP-even component of the scalar triplet, controlled by $x \equiv 4fv'/v^2$ (with $0 \leq x < 1$), where $v'$ is the triplet vev; (ii) mixing between $W^\pm$ and $W_H^\pm$ (parameterized by $c$) and between $Z$ and $Z_H, A_H$ (parameterized by $c, c'$), which affects both the Higgs couplings and the physical $W$ and $Z$ masses; and (iii) the difference in the contribution to masses and couplings of genuine dimension-six terms arising from the expansion of the nonlinear sigma model in powers of the Higgs field.

In any theory of new physics, corrections to observables must be calculated relative to the SM predictions for a given set of electroweak inputs. These electroweak inputs are usually taken to be the Fermi constant $G_F$ defined in muon decay, the $Z$ mass $M_Z$, and the electromagnetic fine structure constant $\alpha$. Thus, a calculation of corrections to the Higgs couplings due to new physics must necessarily involve a calculation of the corrections to the SM electroweak input parameters due to the same new physics. In the Littlest Higgs model, it is most straightforward to calculate corrections to the Higgs couplings with the SM Higgs vev $v \simeq 246$ GeV as an input, as in Eq. (7.40). To obtain useful predictions for the couplings, however, this must be related to the Fermi constant in the Littlest Higgs model according to $v^{-2} = \sqrt{2} G_F y_{G_F}^2$, where $y_{G_F}^2 = 1 + \mathcal{O}(v^2/f^2)$ (given in Table 7.3) is the correction factor to the relation between the Higgs vev $v$ and $G_F$ as measured in muon decay.

The partial widths of the Higgs boson into $Z$ boson pairs ($\Gamma_Z$), top quark pairs ($\Gamma_t$), and pairs of other fermions ($\Gamma_f$) normalized to their SM values are given by [35, 62]

$$\Gamma_Z/\Gamma_Z^{\rm SM} = y_{G_F}^2 y_Z^2, \qquad \Gamma_t/\Gamma_t^{\rm SM} = y_{G_F}^2 y_t^2, \qquad \Gamma_f/\Gamma_f^{\rm SM} = y_{G_F}^2 y_f^2. \tag{7.41}$$

The calculation of the partial width for the Higgs decay to $W$ bosons is a little subtle when $G_F$, $M_Z$ and $\alpha$ are used as inputs because the relation between these inputs and the physical $W$ boson mass receives corrections from the Littlest Higgs model. The partial width of $H \to WW^{(*)}$ depends on the $W$ mass in the kinematics, especially in the intermediate Higgs mass range, 115 GeV $\lesssim M_H \lesssim 2M_W$. To deal with this, one can follow the approach taken by the program HDECAY [73] for the Minimal Supersymmetric Standard Model (MSSM), which is to define the $H \to WW^{(*)}$ partial width in the MSSM in terms of the SM partial width simply by scaling by the ratio of the $WWH$ couplings-squared in the two models, ignoring the shift in the kinematic $W$ mass. Calculating only the correction to the coupling-squared in the Littlest Higgs model and ignoring the shift due to the $W$ mass correction in the kinematics, one finds [35, 62]

$$\Gamma_W/\Gamma_W^{\rm SM} = y^2_{G_F} y^2_W \frac{y^4_{M_W}}{y^4_{M_Z}} y^4_{c_W}. \tag{7.42}$$

Table 7.3: Coupling factors $y_i$ in the Littlest Higgs model, in terms of the inputs $f, c_t, x, c$, and $c'$.

| | |
|---|---|
| $y_t$ | $1 + \frac{v^2}{f^2}\left[-\frac{2}{3} + \frac{1}{2}x - \frac{1}{4}x^2 + c_t^2 s_t^2\right]$ |
| $y_f$ | $1 + \frac{v^2}{f^2}\left[-\frac{2}{3} + \frac{1}{2}x - \frac{1}{4}x^2\right]$ |
| $y_W$ | $1 + \frac{v^2}{f^2}\left[-\frac{1}{6} - \frac{1}{4}(c^2 - s^2)^2\right]$ |
| $y_Z$ | $1 + \frac{v^2}{f^2}\left[-\frac{1}{6} - \frac{1}{4}(c^2 - s^2)^2 - \frac{5}{4}(c'^2 - s'^2)^2 + \frac{1}{4}x^2\right]$ |
| $y^2_{G_F}$ | $1 + \frac{v^2}{f^2}\left[-\frac{5}{12} + \frac{1}{4}x^2\right]$ |
| $y^2_{M_Z}$ | $1 + \frac{v^2}{f^2}\left[-\frac{1}{6} - \frac{1}{4}(c^2 - s^2)^2 - \frac{5}{4}(c'^2 - s'^2)^2 + \frac{1}{2}x^2\right]$ |
| $y^2_{M_W}$ | $1 + \frac{v^2}{f^2}\left[-\frac{1}{6} - \frac{1}{4}(c^2 - s^2)^2 + \frac{1}{4}x^2\right]$ |
| $y^2_{c_W}$ | $1 + \frac{v^2}{f^2}\frac{s_W^2}{c_W^2 - s_W^2}\left[-\frac{1}{4} + \frac{1}{4}(c^2 - s^2)^2 + \frac{5}{4}(c'^2 - s'^2)^2 - \frac{1}{4}x^2\right]$ |
| $y_T$ | $-c_t^2 s_t^2 \frac{v^2}{f^2}$ |
| $y_{W_H}$ | $-s^2 c^2 \frac{v^2}{f^2}$ |
| $y_{\Phi^+}$ | $\frac{v^2}{f^2}\left[-\frac{1}{3} + \frac{1}{4}x^2\right]$ |
| $y_{\Phi^{++}}$ | $\mathcal{O}(v^4/f^4)$ |

The additional correction factors $y^2_{M_W}$, $y^2_{M_Z}$, and $y^2_{c_W}$ are of order $1 + \mathcal{O}(v^2/f^2)$ and are listed in Table 7.3. An alternative approach [63] uses the $W$ mass directly as an input; in this case one has

$$\Gamma_W/\Gamma_W^{\rm SM} = y^2_{G_F} y^2_W. \tag{7.43}$$

In order to calculate the contributions to the loop induced Higgs couplings to $gg$, $\gamma\gamma$, and $\gamma Z$, the couplings of the Higgs to the colored and/or charged TeV-scale particles are also needed. In the Littlest Higgs model, these are,

$$\mathcal{L} = -\frac{M_T}{v} y_T \bar{T} T H + 2\frac{M^2_{W_H}}{v} y_{W_H} W^+_H W^-_H H - 2\frac{M^2_\Phi}{v} y_{\Phi^+} \Phi^+ \Phi^- H - 2\frac{M^2_\Phi}{v} y_{\Phi^{++}} \Phi^{++}\Phi^{--} H. \tag{7.44}$$

(For this calculation it is sufficient to use a common mass $M_\Phi$ for the components $\Phi^+$, $\Phi^{++}$ of the TeV-scale scalar triplet.) Because the masses of the TeV-scale particles do not arise from their couplings to the Higgs boson, the coupling factors $y_i$ for these particles are generically of order $v^2/f^2$. They are listed in Table 7.3.

The partial width of the Higgs boson into two photons, normalized to its SM value, is given in the Littlest Higgs model by [61]

$$\Gamma_\gamma/\Gamma_\gamma^{\rm SM} = y^2_{G_F} \frac{\left|\sum_{i,{\rm LH}} y_i N_{ci} Q_i^2 F_i(\tau_i)\right|^2}{\left|\sum_{i,{\rm SM}} N_{ci} Q_i^2 F_i(\tau_i)\right|^2}, \tag{7.45}$$

where $N_{ci}$ is the color factor (= 1 or 3), $Q_i$ is the electric charge, $\tau_i = 4m_i^2/m_H^2$, and $m_i$ is the mass, respectively, for each particle $i$ running in the loop: $t$, $T$, $W$, $W_H$, and $\Phi^+$ in the Littlest Higgs (LH) case (the $\Phi^{++}$ loop can be neglected at order $v^2/f^2$ [61]; see Table 7.3); and $t$ and $W$ in the SM case. The standard dimensionless loop factors $F_i$ for particles of spin 1, 1/2, and 0 can be found in Ref. [71]. Likewise, the partial width of the Higgs boson into two gluons, normalized to its SM value, is given in the Littlest Higgs model by [61]

$$\Gamma_g/\Gamma_g^{\rm SM} = y_{G_F}^2 \frac{\left|\sum_{i,{\rm LH}} y_i F_{1/2}(\tau_i)\right|^2}{\left|\sum_{i,{\rm SM}} F_{1/2}(\tau_i)\right|^2}, \tag{7.46}$$

where $i$ runs over the fermions in the loop: $t$ and $T$ in the Littlest Higgs case, and $t$ in the SM case.

The partial width of the Higgs boson into $\gamma Z$, normalized to its SM value, is given in the Littlest Higgs model by [63]

$$\Gamma_{\gamma Z}/\Gamma_{\gamma Z}^{\rm SM} = y_{G_F}^2 \frac{\left|\sum_{i,{\rm LH}} A_i^{\rm LH}\right|^2}{\left|\sum_{i,{\rm SM}} A_i^{\rm SM}\right|^2}, \tag{7.47}$$

where the amplitude factors $A_i^{\rm LH}$ are given in Ref. [63] and contain the appropriate scaling factors $y_i$. In this process the corrections to the $Z$ couplings to the particles in the loop must also be taken into account [63].

### *7.3.2 Generic features*

The Higgs decay branching ratio to a final state $X$, ${\rm BR}(H \to X) = \Gamma_X/\Gamma_{\rm tot}$, is computed in terms of the SM branching ratio as

$$\frac{{\rm BR}(H \to X)}{{\rm BR}(H \to X)^{\rm SM}} = \frac{\Gamma_X/\Gamma_X^{\rm SM}}{\Gamma_{\rm tot}/\Gamma_{\rm tot}^{\rm SM}}. \tag{7.48}$$

The numerator can be read off from the partial width ratios given above. The denominator requires a calculation of the Higgs total width in both the SM and the little Higgs model. This can be computed using, e.g., HDECAY [73] to calculate the SM Higgs partial width into each final state for a given Higgs mass; the SM total width $\Gamma_{\rm tot}^{\rm SM}$ is of course the sum of these partial widths, while the total width in the Littlest Higgs model is found by scaling each partial width in the sum by the appropriate ratio.

A quick examination of the corrections to the Higgs partial widths given above reveals that the corrections are all parametrically of order $v^2/f^2$. In particular, no coupling receives especially large corrections. This is in contrast to the MSSM, in which the corrections to the couplings of the light SM-like Higgs boson to fermions are parametrically larger than those to $W$ and $Z$ bosons (the deviations in the down-type fermion sector are also enhanced by $\tan\beta$); this coupling structure is due to the two-Higgs-doublet nature of the MSSM Higgs sector [74]. Thus in the Littlest Higgs model there is no “golden channel” in which one expects to see especially large deviations from the SM Higgs couplings. We therefore expect the experimentally best-measured channel to give the highest sensitivity to TeV-scale effects. For example, with $f = 1$ TeV and $M_H = 115$ GeV, the rate for $\gamma\gamma \to H \to b\bar{b}$ in the Littlest Higgs model is reduced by about 6–7% compared to that in the SM [62]. The shifts in the other Higgs branching fractions are of a comparable magnitude.

### *7.3.3 Heavy particle exchange in Higgs production and decay*

The partial width ratios given above can immediately be used to find the corrections to the Higgs boson production cross sections in gluon fusion and in two-photon fusion, since the production cross section is simply proportional to the corresponding Higgs partial width (detailed results were given in Ref. [61]). For other Higgs boson production channels, the cross section corrections are more complicated because in addition to the corrections to the Higgs couplings to SM particles, tree-level exchange of the TeV-scale particles in the production diagrams must also be taken into account. This has been studied in the Littlest Higgs model for Higgs production at an $e^+e^-$ linear collider via $ZH$ associated production [75, 76], $W$ boson fusion [77], and associated $t\bar{t}H$ production [78].

The process $e^+e^- \to ZH$ receives a correction in little Higgs models from s-channel exchange of the neutral TeV-scale gauge bosons [75, 76]. If the $e^+e^-$ center-of-mass energy $\sqrt{s}$ is well below the mass scale of the heavy gauge bosons, their exchange is propagator-suppressed and the dominant effect comes from the interference term between the SM process and the new diagrams. This correction is parametrically of order $v^2/f^2$, i.e., the same size as the corrections to the Higgs couplings. In this case the corrections to the SM $e^+e^- \to ZH$ amplitude due to the modifications of the $ZZH$ and $e^+e^-Z$ couplings at order $v^2/f^2$ must also be taken into account. In general, however, the effect of the TeV-scale gauge boson exchange varies with $\sqrt{s}$, providing a valuable additional handle on the model parameters, and dramatic resonance effects appear when $\sqrt{s}$ is close to the mass of one of the heavy gauge bosons. This process can be used to probe the crucial $ZZ_HH$ coupling with high precision [76].

Similarly, the $WW$ fusion process $e^+e^- \to \nu\bar{\nu}H$ receives corrections in little Higgs models from substitution of one or both of the t-channel $W$ bosons with their TeV-scale counterparts [77]. Again, for relatively low $\sqrt{s}$, propagator suppression ensures that the dominant effect comes from the interference term between the SM process and diagrams in which one of the two $W$ bosons is replaced with a $W_H$. This correction is parametrically of order $v^2/f^2$, again the same size as the corrections to the Higgs couplings. For this reason, the corrections to the SM $WW$ fusion amplitude due to modifications of the $WWH$ and $We\nu$ couplings at order $v^2/f^2$ must also be taken into account. As in $ZH$ production, the correction due to tree-level $W_H$ exchange will depend on $\sqrt{s}$. The new diagrams can also potentially modify the final-state kinematic distributions, leading to additional observables; these have not yet been studied.

The process $e^+e^- \to t\bar{t}H$ receives corrections in little Higgs models from the substitution of the internal top quark line with the heavy top-partner and from the substitution of the s-channel $Z$ or $\gamma$ with a TeV-scale gauge boson [78]. Diagrams in which the Higgs is radiated off the s-channel gauge boson also contribute. As before, for relatively low $\sqrt{s}$, propagator suppression ensures that the dominant effect comes from the interference term between the SM process and diagrams containing one TeV-scale particle, leading to corrections of order $v^2/f^2$. Again, corrections due to order $v^2/f^2$ modifications of the SM couplings must be included. The corrections will depend on $\sqrt{s}$, with resonances appearing when $\sqrt{s}$ is close to the mass of one of the heavy neutral gauge bosons. Final-state kinematic distributions can provide additional observables; these have likewise not yet been studied.

Tree-level exchange of the TeV-scale particles can also appear in off-shell contributions to Higgs decays. We expect their largest effect to appear in decays in which the SM exchange is also off shell, e.g., $H \to W^{(*)}W^*, Z^{(*)}Z^*$ for $M_H$ below the $WW$ or $ZZ$ threshold, respectively. In this case the decay products of the off-shell $W$ or $Z$ boson(s) have a broad invariant mass distribution, allowing a potentially non-negligible correction from interference of $H \to W^{(*)}W_H^*, Z^{(*)}Z_H^*$ with the SM amplitude. Propagator suppression ensures that the corrections are again of order $v^2/f^2$. These effects can modify the invariant mass distribution of the relevant final-state fermion pair, leading to an additional observable and introducing a potential dependence of the measured decay branching fraction on details of the experimental selection. These contributions to $H \to WW^*, ZZ^*$ have not been studied at all.

Finally, we note that in little Higgs models with T-parity, the TeV-scale gauge bosons are T-parity odd and therefore cannot contribute at tree-level to Higgs production or decay. However, models with T-parity typically contain a T-parity even top-partner which can contribute at tree level to $t\bar{t}H$ production.

### *7.3.4 Generalization to other models*

In the preceding we have discussed the modifications of the light Higgs boson couplings in the Littlest Higgs model [6]. The other little Higgs models that contain only one light Higgs doublet [11, 13, 14] can be fit into the same structure and exhibit the same generic corrections of order $v^2/f^2$. A partial list of coupling correction factors as in Table 7.3 has been worked out [54] only for the SU(3) simple group model [13, 14]. The Higgs coupling corrections in this model differ in their details from those in the Littlest Higgs model. In particular, (i) the model contains no scalar triplet, so there is no correction from Higgs mixing with the triplet; (ii) there is no mixing between $W^{\pm}$ and the charged TeV-scale gauge bosons; and (iii) corrections to $G_F$ come from the mixing of neutrinos with their TeV-scale partners while the contribution from tree-level exchange of the TeV-scale charged gauge boson is negligible.

Higgs production and decay in the “Littlest Higgs” model with T-parity was studied in detail in Ref. [79]. Because almost all of the new TeV-scale particles in this model are odd under T-parity (the exception being a single T-even top-partner), there are no corrections to tree-level Higgs couplings (aside from $Ht\bar{t}$) due to mixing or exchange of the TeV-scale particles; the only corrections to these couplings come at order $v^2/f^2$ from the genuine dimension-six terms arising from the expansion of the nonlinear sigma model in powers of the Higgs field. The $Ht\bar{t}$ coupling receives additional corrections from the mixing of the top quark with its heavy T-even partner. Finally, the loop-induced Higgs couplings to photon or gluon pairs receive corrections from loops of the new TeV-scale particles, including the T-odd states. All the corrections to Higgs couplings are parametrically of order $v^2/f^2$. However, because electroweak precision constraints on the mass scale $f$ of the heavy particles are much weaker in models with T-parity [9, 12, 28, 43], the spectrum of new particles can be significantly lighter resulting in much larger modifications of Higgs couplings than in models without T-parity.

Many little Higgs models contain two light Higgs doublets [7, 8, 10, 13, 15]. In such models, the dominant corrections to the couplings of the lightest CP-even Higgs boson typically come from the two-doublet mixing effects, rather than the genuine $v^2/f^2$-suppressed effects of the TeV-scale states. In this case the Higgs phenomenology will look more like that of the MSSM Higgs sector. If the dimensionless couplings in the Higgs potential are not large, the Higgs sector exhibits a *decoupling limit* (for a review see Ref. [80]) in which one doublet becomes heavy, leaving a single light SM-like Higgs boson. The corrections to the couplings will then follow a pattern similar to a general two Higgs doublet model, with the parametrically largest deviations expected in the couplings of the Higgs to fermions. Beyond the couplings of the lightest Higgs, the Higgs potential in these two-doublet models exhibits interesting restrictions due to the global symmetry structure; further study in this direction would be interesting.

Finally, we note here that corrections to the Higgs couplings can also be induced by the UV completion at $\sim 10$ TeV. For example, the loop-induced Higgs coupling to photon pairs receives corrections from the dimension-six operator $(c/\Lambda^2)h^\dagger h F^{\mu\nu}F_{\mu\nu}$ suppressed by the UV completion scale $\Lambda$ [62, 81]; this can be thought of as arising from the particles of the UV completion running in the loop. If the UV completion is weakly coupled, these corrections should naively be suppressed by the square of the ratio of the electroweak scale to the 10 TeV scale, $v^2/\Lambda^2$, and thus be too small to detect; in particular, the corrections to the Higgs couplings will then be accurately predicted by the TeV-scale theory alone. However, if the UV completion is strongly coupled, the strong-coupling enhancement can counteract the suppression from the high mass scale, leading to corrections naively of the same order as those from the TeV scale physics.

### 7.4 Pseudo-axions in Little Higgs models

*Wolfgang Kilian, David Rainwater and Jürgen Reuter*

Little Higgs models have an extended structure of global symmetries, broken both spontaneously and explicitly. Among these global symmetries there can appear U(1) factors, which lead to the presence of light pseudoscalar particles, Goldstone bosons associated with this U(1) group, in the spectrum of little Higgs models [64]. These global U(1) factors arise from two different mechanisms.

In the first case, in the Littlest Higgs type models, there is a $[\mathrm{SU}(2)\times\mathrm{U}(1)]^2$ product structure, where U(1) groups happen to be quasi arbitrary additional factors (we do not discuss possible embeddings into larger symmetry groups here). The doubled gauge group $[\mathrm{SU}(2)]^2$ of weak interactions is broken to the diagonal SU(2) of the Standard Model. Analogously, both U(1) factors can be gauged and be broken down to the diagonal U(1) of hypercharge. In many models, especially those without a built-in custodial symmetry or T-parity, the second U(1) gauge boson is tightly constrained from direct searches (Tevatron) and electroweak precision observables. So considering the second U(1) factor ungauged as a removal of the constraints, leaves one with a global U(1) symmetry.

Secondly, in simple group models the breaking of the global symmetries can be understood as a breaking from e.g. in the SU(3) simple group model [14] U(3) to U(2) instead of SU(3) to SU(2) with an additional U(1) symmetry left unbroken at this stage. If this symmetry were exact, the corresponding Goldstone boson, parameterized by

$$\xi = \exp[i\eta/f], \qquad (7.49)$$

would be exactly massless and would have only derivative interactions. But in these models, there is always an explicit breaking of the global symmetries, necessary to give a large enough quartic coupling to the Higgs boson. This explicit breaking generates a mass for the pseudoscalar particle. The U(1) quantum numbers of the pseudoscalar in this second case are predicted by the model because the U(1) symmetry is embedded in a larger symmetry whose structure is known. In contrast, in the Littlest Higgs type models, where the global U(1) is an additional factor not connected to non-Abelian group structure there is no prediction for the quantum numbers. Note that anomaly cancellation is not an issue, since the global U(1) symmetry may well be anomalous. This second type of pseudoscalars resembles the breaking of chiral symmetries in QCD, with the $\eta$ meson playing the role of the pseudoscalar here.

Since these U(1)-Pseudo-Goldstone bosons are pseudoscalars and electroweak singlets, they do not have couplings to the Standard Model gauge bosons. All their couplings to Standard Model fermions are suppressed by the ratio $v/f$ of the electroweak scale over the TeV scale. In Littlest Higgs type models, the mass of the pseudoaxions is not predicted , while in simple group models it is connected to the masses of the Higgs bosons by the Coleman-Weinberg potential. In order not to reintroduce fine tuning and to destabilize the so called little hierarchy between the electroweak and the TeV scale, the pseudoaxion mass should be approximately bounded by $v, m_\eta \lesssim$ 250 GeV. In principle, the pseudoaxions can become quite light, of the order of a few GeV or less, but for masses in that range there exist constraints from rare $\Upsilon$ decays and other flavour processes [82, 83]. Even in simple group models, the pseudoscalar mass – like the Higgs mass – cannot be fixed, because there are too many free parameters, e.g. the mass parameter $\mu$ and a mixing angle $t_\beta \equiv \tan\beta = f_2/f_1$ [54, 64] similar to the Minimal Supersymmetric Standard Model. As an example we show the connection of the pseudoscalar and Higgs masses as functions of the $\mu$ parameter in the SU(3) simple group model on the left of Fig. 7.4.

From the structure of the couplings above a quite generic pattern of branching ratios can be deduced. All decays to electroweak vector bosons (even off-shell) are absent. The dominant decay modes are into the heaviest available Standard Model fermions, i.e. $b\bar{b}$, $c\bar{c}$ and $\tau^+\tau^-$. Compared to the Higgs branching ratios, the decays to two gluons or two photons are more important for the pseudoaxions due to two reasons: Firstly, the absence of the $WW$ and $ZZ$ decays and the $v^2/f^2$ suppression for the fermionic decays enhance these final states. Secondly, the triangle anomaly graphs responsible for the $gg$ and $\gamma\gamma$ decays are enhanced by the additional heavy top state and possibly further heavy fermionic states present in the simple group Models (this second point partly also applies to the Higgs boson). A typical set of branching ratios as a function of the pseudoscalar's mass is shown on the right in Fig. 7.4. The decay to $Zh$ is special for simple group Models, and does not appear in Littlest Higgs type models.

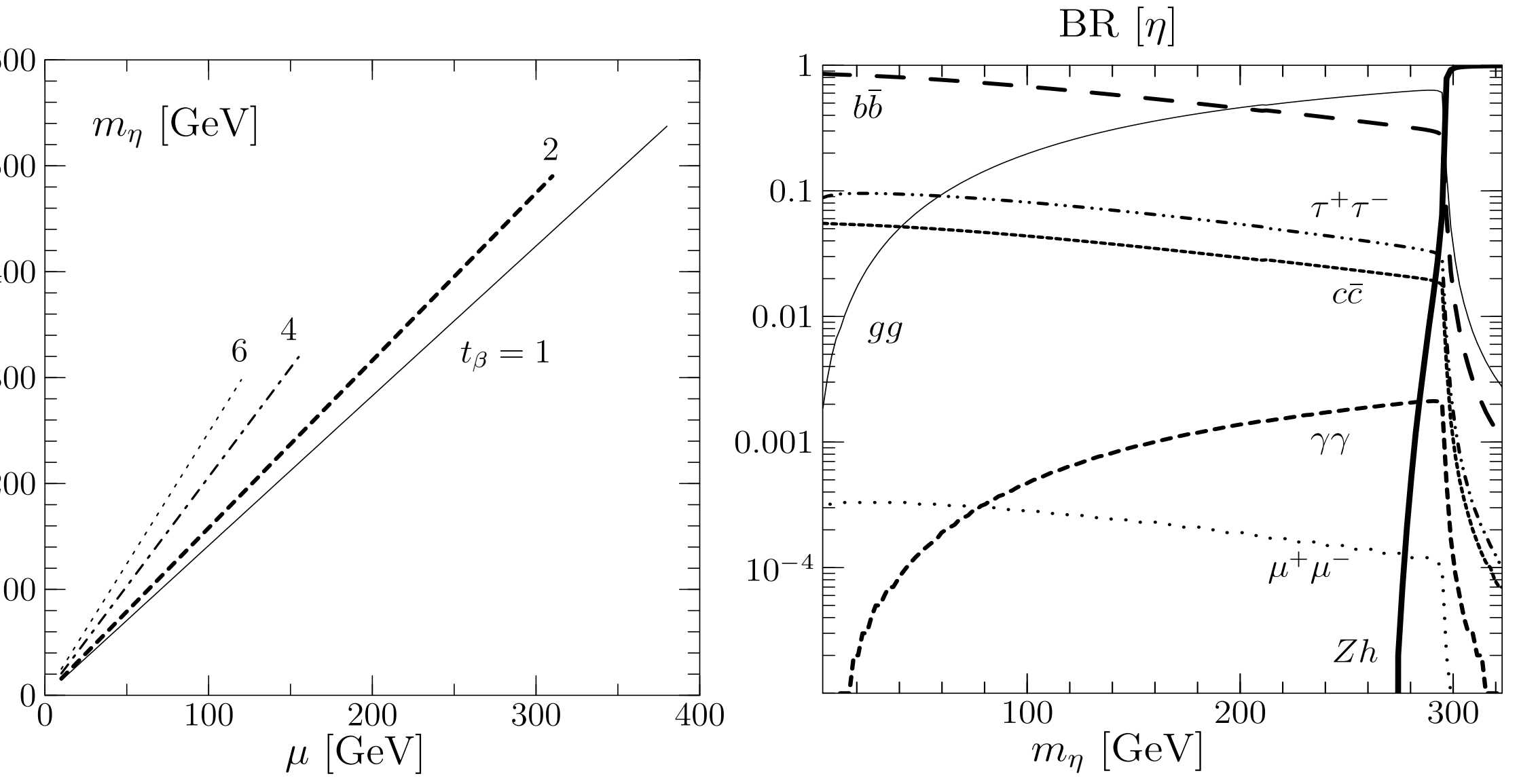

Fig. 7.4: On the left: Typical masses of the pseudoaxion in the SU(3) simple group model as a function of the $\mu$ mass parameter. The four lines correspond to different values of $\tan\beta$. On the right: Branching ratios for the pseudoaxion in the SU(3) simple group model as a function of its mass.

The couplings of the pseudoaxion to two gluons or two photons generated by the triangle anomalies offer the best search possibility for these particles at the LHC. There the pseudoaxions can be produced in analogy to the Higgs boson in gluon fusion, while the decay to two photons gives the cleanest decay signature as a peak in the diphoton spectrum. Fig. 7.5 shows the cross section for this process as a function of the invariant diphoton mass and hence the pseudoaxion mass, for the Littlest Higgs and the SU(3) simple group model. In principle, the associated production $t\bar{t}\eta$ can also be used, but is plagued by huge backgrounds at the LHC (note again the $v/f$ suppression of this coupling). At a future ILC, this would be the dominant search mode, by looking for narrow peaks in the invariant mass of pairs of $b$ jets in the final state $b\bar{b}b\bar{b}$ and missing energy. As was shown in [64], this is in fact the only viable search possibility for 50 GeV $< m_\eta <$ 85 GeV. A search for an $s$-channel resonance at the photon collider is the best search option for masses of the pseudoaxion well above the $Z$ threshold and allows for precision measurements of such a state; however, the photon spectrum deteriorates dramatically for energies as low as the $Z$ mass.

Introducing T-parity into little Higgs models, one finds that generically the pseudoaxion becomes T-odd. This means, if it is the lightest T-odd particle (LTP), it can be a Cold Dark Matter candidate. As was discussed in [44], in the Littlest Higgs model with T-parity the heavy partner of the hypercharge boson, called $A_H$, becomes the LTP. Note that an ungauging of the additional U(1) is unnecessary from the point of view of the electroweak precision observables, since T-parity already forbids tree-level contributions to gauge-boson self energies. Nevertheless, if this second U(1) *is* ungauged, its gauge boson $A_H$ is traded for the pseudo-Goldstone boson $\eta$, which takes over the role of the LTP. The consequences for the dark matter content in [44] remains the same, since it is mainly the Goldstone boson couplings in the $A_H$ which are responsible for the dominant annihilation channel to Higgs or longitudinal electroweak gauge bosons.

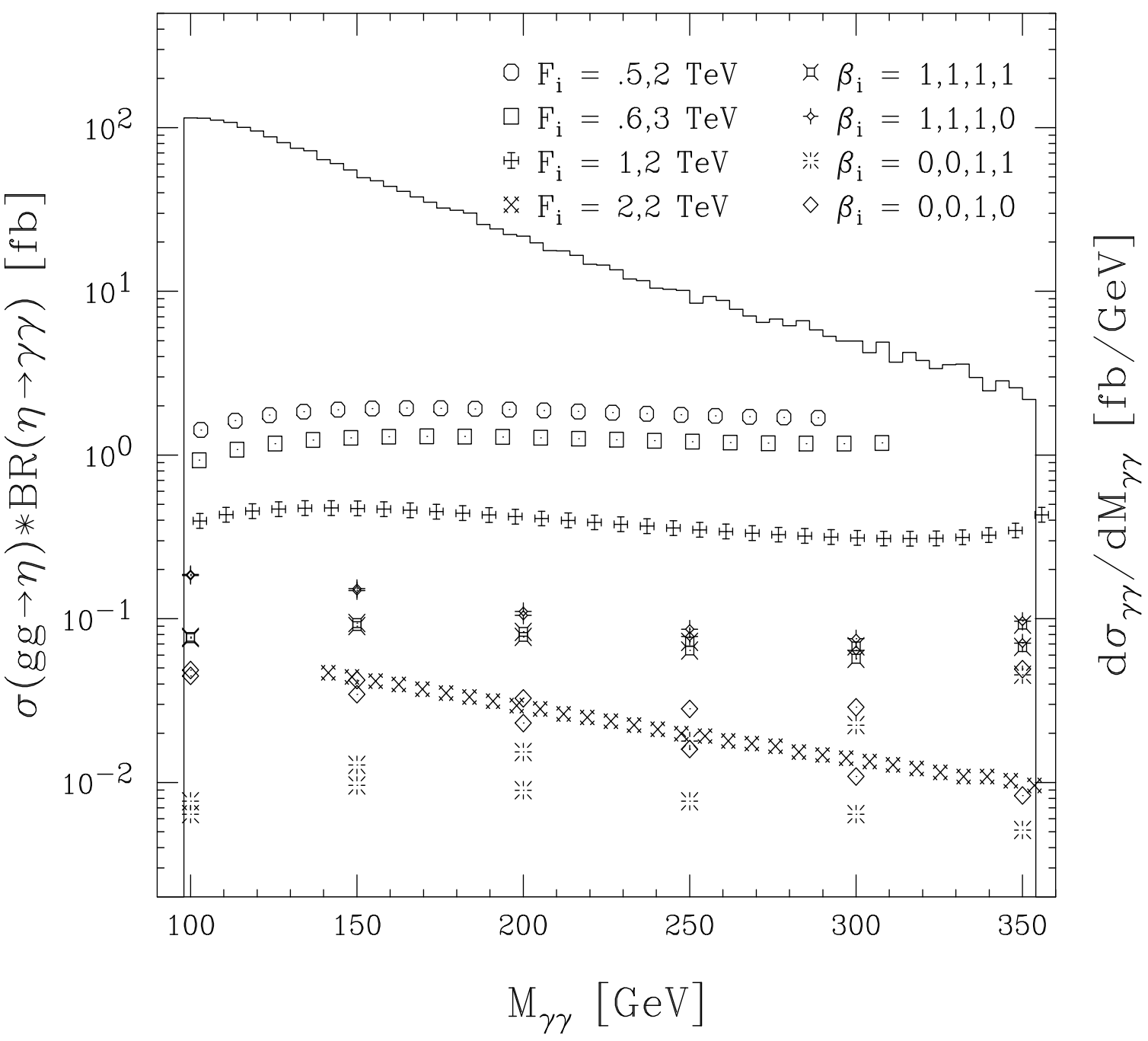

Fig. 7.5: Cross section times branching ratio for the gluon fusion production of the pseudoaxion in the Littlest and the SU(3) simple group model and subsequent decay into two photons. The symbols on the left correspond to the SU(3) simple group model, showing different scales (which can also be expressed as $f$ and $\tan\beta$). On the right, the Littlest Higgs, where the $\beta$'s are different assignments of $U(1)_\eta$ quantum numbers. For more details see [64].

### 7.5 Little Higgs with T-parity

*Jay Hubisz*

The earliest implementations of the little Higgs structure suffered from electroweak precision (EWP) constraints [31–33]. After electroweak symmetry breaking, mixing is generally induced between the standard model gauge bosons and their TeV scale partners. This mixing can lead to, for example, violations of custodial SU(2). This leads to a tree level shift in the $\rho$ parameter, a relation between the $W$ and $Z$ mass which is tightly constrained. In addition, SM fermions couple to the heavy gauge bosons, leading to large four-fermion operators which must be suppressed.

T-parity is a postulated discrete symmetry under which the SM particles are neutral, while most new heavy states are odd [9, 12, 28]. This is in analogy with R-parity (or matter parity) where the supersymmetric partners of the SM fields are odd. This discrete symmetry, if unbroken, then forbids mixing of the SM particles with the new states. Contributions to EWP observables and four-fermion operators are then not generated at tree level, but at the one-loop level.

T-parity is a symmetry that is inherited from an automorphism of the gauge group algebra of little Higgs models. In the Littlest Higgs model T-parity exchanges the two copies of $\mathrm{SU}(2) \times \mathrm{U}(1)$ gauge bosons. In the moose models with T-parity, this symmetry has a geometrical interpretation as a parity symmetry of the moose diagram. Implementing T-parity as a symmetry of the theory requires that the gauge couplings for the two $\mathrm{SU}(2) \times \mathrm{U}(1)$ gauge groups be equal. In this way, the diagonal subgroup (the standard model gauge group) is even under T-parity, while the other combinations of gauge bosons, which receive $f$ scale masses, are odd. In addition, if one wishes to implement this symmetry consistently throughout the entire model, the matter sector of the model must also be symmetric under this interchange. For every multiplet that transforms under $[SU(2) \times U(1)]_1$, there must be a partner multiplet that transforms under $[SU(2) \times U(1)]_2$ [12]. This opens up a new flavor structure in the low energy effective theory, which is constrained by studies of neutral meson mixing and rare decays [70]. This discrete symmetry, while it eliminates the tree level shifts in standard model observables, drastically changes the phenomenology of little Higgs models [44].

If the discrete symmetry is made exact, the lightest T-odd particle is stabilized, and is a potential dark matter candidate. In collider phenomenology, this lightest particle becomes a missing energy signal, making observation of this new physics more complicated. In particular, it is likely that this type of model will look very much like supersymmetry. This is similar to studies of universal extra dimensions, where the signals are also similar to those of supersymmetry [84, 85]. In the Littlest Higgs model with T-parity, the heavy partner of the hypercharge gauge boson, the $A_H$, is the dark matter candidate, and can account for the WMAP observed relic density [44, 86].

### *7.5.1 New features in models with T-parity*

In the Littlest Higgs model, the action of T-parity on the gauge bosons and scalars is as follows:

$$\begin{aligned} T &: A_1 \to A_2 \\ T &: \Pi \to -\Omega\Pi\Omega, \end{aligned} \tag{7.50}$$

where $\Omega = \mathrm{diag}(1, 1, -1, 1, 1)$. It is easily verified that the Higgs doublet is neutral under this transformation, whereas the scalar triplet is odd under T-parity. This assignment forbids a vev for the triplet which would break custodial SU(2).

Implementing T-parity fixes the gauge couplings such that the angles defined in the introduction are set equal: $s = s' = 1/\sqrt{2}$. Thus T-parity imposes restrictions on the mass spectrum that are not present in models without T-parity. To match onto the standard model, the gauge couplings are given by

$$\begin{aligned} g_1 &= g_2 = \sqrt{2}g \\ g_1' &= g_2' = \sqrt{2}g', \end{aligned} \tag{7.51}$$

where $g$ and $g'$ are the SU(2)$_L$ and hypercharge gauge couplings, respectively.

In the Littlest Higgs with T-parity, this implies that the masses of the new T-odd gauge bosons with respect to the overall breaking scale $f$ are

$$M_{W_H^\pm} = M_{Z_H} = gf, \qquad M_{A_H} = \frac{g' f}{\sqrt{5}}. \tag{7.52}$$

In models with T-parity, the standard model fermion doublet spectrum needs to be doubled. This is to ensure that there is equal matter content charged under each copy of SU(2).[2] For each lepton/quark doublet, two fermion doublets $\psi_1 \in (\mathbf{2}, \mathbf{1})$ and $\psi_2 \in (\mathbf{1}, \mathbf{2})$ are introduced. (The quantum numbers refer to representations under the $\mathrm{SU(2)}_1 \times \mathrm{SU(2)}_2$ gauge symmetry.) These can be embedded in incomplete representations $\Psi_1, \Psi_2$ of the global SU(5) symmetry. An additional set of fermions forming an SO(5) multiplet $\Psi^c$, which transforms nonlinearly under the full SU(5), is introduced to give mass to the extra fermions; the field content can be expressed as follows:

$$\Psi_1 = \begin{pmatrix} \psi_1 \\ 0 \\ 0 \end{pmatrix}, \quad \Psi_2 = \begin{pmatrix} 0 \\ 0 \\ \psi_2 \end{pmatrix}, \quad \Psi^c = \begin{pmatrix} \tilde{\psi}^c \\ \chi^c \\ \psi^c \end{pmatrix}. \tag{7.53}$$

[2]In principle, the standard model fermions could transform non-linearly under the full SU(5), and thus only under the SU(2)$_L$ unbroken gauge symmetry [9]. In this case, the T-odd fermions are not present. However, this leads to large contributions to four fermion operators which are constrained primarily by studies at LEP, CDF, and D0. These constraints are referred to as compositeness bounds on quarks and leptons.

These fields transform under the SU(5) global symmetry as follows:

$$\Psi_1 \to V^* \Psi_1\,, \quad \Psi_2 \to V\Psi_2\,, \quad \Psi^c \to U\Psi^c, \tag{7.54}$$

where $U$ is the nonlinear transformation matrix defined in Refs. [12, 28, 44]. The action of T-parity on the multiplets takes

$$\Psi_1 \leftrightarrow \Sigma_0\Psi_2, \quad \Psi^c \to -\Psi^c. \tag{7.55}$$

These assignments allow a term in the Lagrangian of the form

$$\kappa f(\bar{\Psi}_2\xi\Psi^c - \bar{\Psi}_1\Sigma_0\Omega\xi^\dagger\Omega\Psi^c), \tag{7.56}$$

where $\xi = \exp(i\Pi/f)$. $\xi$ transforms linearly on the left, and non-linearly on the right, rendering Eq. (7.56) invariant under SU(5) transformations. Eq. (7.56) gives a Dirac mass $M_- = \sqrt{2}\kappa f$ to the T-odd linear combination of $\psi_1$ and $\psi_2$, $\psi_- = (\psi_1 - \psi_2)/\sqrt{2}$, together with $\tilde{\psi}^c$; the T-even linear combination, $\psi_+ = (\psi_1 + \psi_2)/\sqrt{2}$, remains massless and is identified with the standard model lepton or quark doublet. To give Dirac masses to the remaining T-odd states $\chi^c$ and $\psi^c$, a spinor multiplet of $SO(5)$ can be introduced, along with an additional singlet.[3]

To complete the discussion of the fermion sector, we introduce the usual SM set of the SU(2)$_L$-singlet leptons and quarks, which are T-even and can participate in the SM Yukawa interactions with $\psi_+$. The Yukawa interactions induce a one-loop quadratic divergence in the Higgs mass; however, the effect is numerically small except for the third generation of quarks. The Yukawa couplings of the third generation must be modified to incorporate the collective symmetry breaking pattern.

In order to avoid large one-loop quadratic divergences from the top sector, the $\Psi_1$ and $\Psi_2$ multiplets for the third generation must be completed to representations of the SU(3)$_1$ and SU(3)$_2$ subgroups of SU(5). We write these as

$$\chi_1 = \begin{pmatrix} q_1 \\ U_{L1} \\ 0 \end{pmatrix}, \quad \chi_2 = \begin{pmatrix} 0 \\ U_{L2} \\ q_2 \end{pmatrix}. \tag{7.57}$$

These obey the same transformation laws under T-parity and the SU(5) symmetry as do $\Psi_1$ and $\Psi_2$. The quark doublets are embedded such that

$$q_i = -\sigma_2 \begin{pmatrix} u_{Li} \\ b_{Li} \end{pmatrix}. \tag{7.58}$$

In addition to the SM right-handed top quark field $u_R$, which is assumed to be T-even, the model contains two SU(2)$_L$-singlet fermions $U_{R1}$ and $U_{R2}$ of hypercharge 2/3, which transform under T-parity as

$$U_{R1} \leftrightarrow U_{R2}. \tag{7.59}$$

The top Yukawa couplings arise from the Lagrangian of the form

$$\begin{aligned}\mathcal{L}_t &= \frac{1}{2\sqrt{2}}\lambda_1 f\epsilon_{ijk}\epsilon_{xy}\big[(\bar{\chi}_1)_i\Sigma_{jx}\Sigma_{ky} + (\bar{\chi}_2\Sigma_0)_i\tilde{\Sigma}_{jx}\tilde{\Sigma}_{ky}\big]u_R \\ &\quad +\lambda_2 f(\bar{U}_{L1}U_{R1} + \bar{U}_{L2}U_{R2}) + \text{h.c.}\end{aligned} \tag{7.60}$$

where $\tilde{\Sigma} = \Sigma_0\Omega\Sigma^\dagger\Omega\Sigma_0$ is the image of the $\Sigma$ field under T-parity. The indices $i, j, k$ run from 1 to 3 whereas $x, y = 4, 5$. The T-parity eigenstates are given by

$$q_\pm = \frac{1}{\sqrt{2}}(q_1 \pm q_2), \quad U_{L\pm} = \frac{1}{\sqrt{2}}(U_{L1} \pm U_{L2}), \quad U_{R\pm} = \frac{1}{\sqrt{2}}(U_{R1} \pm U_{R2}). \tag{7.61}$$

[3] In other extensions of the Littlest Higgs model with T-parity which contain more sigma model fields and an enlarged gauge symmetry (such as the $\mathrm{SU}(5)^2/\mathrm{SO}(5)$ model of [12]), the non-linearly transforming multiplet can be avoided altogether. We choose here to focus on the most compact phenomenologically consistent model with T-parity.

The T-odd states $U_{L-}$ and $U_{R-}$ combine to form a Dirac fermion $T_-$, with mass $m_{T_-} = \lambda_2 f$. The remaining T-odd states $q_-$ receive a Dirac mass from the interaction in Eq. (7.56).

To leading order in $v/f$, after diagonalizing to the mass eigenbasis, the T-even states have masses given by

$$m_t = \frac{\lambda_1 \lambda_2 v}{\sqrt{\lambda_1^2 + \lambda_2^2}}, \quad m_T = \sqrt{\lambda_1^2 + \lambda_2^2}\, f, \tag{7.62}$$

identical to the Littlest Higgs without T-parity. It is interesting to note that in this model, the T-odd states do not participate in the cancellation of quadratic divergences in the top sector: the cancellation only involves loops of $t$ and $T_+$.[4]

### *7.5.2 Electroweak precision constraints*

As mentioned above, there are no tree level contributions to electroweak precision observables in the Littlest Higgs model with T-parity. Contributions enter at the one loop level, however, and these restrict the available parameter space in the model. A global fit analysis has been performed in [43]. The one loop constraints arising from the SU(2)$_L$ triplet, the T-odd gauge bosons, the T-even partner of the top-quark, and the T-odd fermion doublets are all taken into account.

An interesting feature of this fit is that the contributions to $\Delta\rho$ from the T-even singlet partner of the top quark come in with the correct sign to allow for a larger Higgs mass. It is shown in [43] that the Higgs mass can be increased up to the unitarity bound for certain choices of the free parameters of the theory. As we will discuss below, larger Higgs masses are preferred for dark matter as well. In Fig. 7.6, the 95, 99, and 99.9 confidence level contours are shown for $x_\lambda = 2$, and $\Lambda = 4\pi f$.

There are certain contributions to EWP that are in fact log divergent [28,43], and thus sensitive to the UV completion of the model. In particular, SM gauge boson self energy diagrams receive divergent contributions from loops where the T-odd gauge bosons run in the loop. This is not a sign of a sickness of the theory, but is rather a consequence of working in the context of a non-linear sigma model. Just as the Higgs mass enters into electroweak precision constraints through loop diagrams, the physics of the UV completion will provide the scale which cuts off these logarithmic divergences.

The dominant contribution to EWP comes from the $T$ parameter, which is directly related to $\Delta\rho$. We summarize here the contributions to the $T$-parameter; expressions for the remaining oblique and non-oblique corrections needed for a full global fit appear in [43]. The expressions for the one loop little Higgs contributions to the T-parameter are

$$\begin{aligned}
T_{T_+} &= \frac{3}{8\pi}\frac{1}{s_w^2 c_w^2} x_\lambda^2 \frac{m_t^4}{m_{T_+}^2 m_Z^2}\left[\log\frac{m_{T_+}^2}{m_t^2} - 1 + \frac{1}{2}x_\lambda^2\right] \\
T_{\rm gauge} &= -\frac{9g^2}{128\pi c_w^2 s_w^2 m_Z^2}\frac{v^4}{f^2}\log\frac{\Lambda^2}{m_{W_H}^2} \\
T_{\Psi_-} &= -\sum_i \frac{\kappa_i^2}{192\pi^2\alpha}\frac{v^2}{f^2},
\end{aligned} \tag{7.63}$$

where the $\kappa_i$ are the T-odd fermion Yukawa couplings. The presence of the logarithmic divergence in $T_{\rm gauge}$ signifies that a counterterm is necessary [28]. In Fig. 7.6, it is assumed that this counterterm is zero, and $\Lambda = 4\pi f$. The contribution from the T-odd fermions includes a sum over all T-odd fermion doublets. This does not take into account color factors, and so each T-odd quark doublet gets an additional factor of 3. Of interest is the fact that the contributions of the T-odd fermion doublets do not decouple with increasing $\kappa$. This reason for this is similar to non-decoupling of the top quark in the SM. The coupling constants $\kappa$ are proportional to the masses, and so the propagator suppression is compensated for by the coupling constant. The deeper reason is that these fermions are tied into gauge invariance of the low energy effective theory, and that there are scattering amplitudes that become non-unitary as these fermion masses approach the cutoff, $\Lambda = 4\pi f$.

[4] It has been recently discovered that it is possible to realize the cancellation of the top quark divergence with a T-odd partner of the top quark [42]. In these models, there are additional T-even fields, however these are allowed to be rather heavy, as they do not participate in the quadratic divergence cancellation.

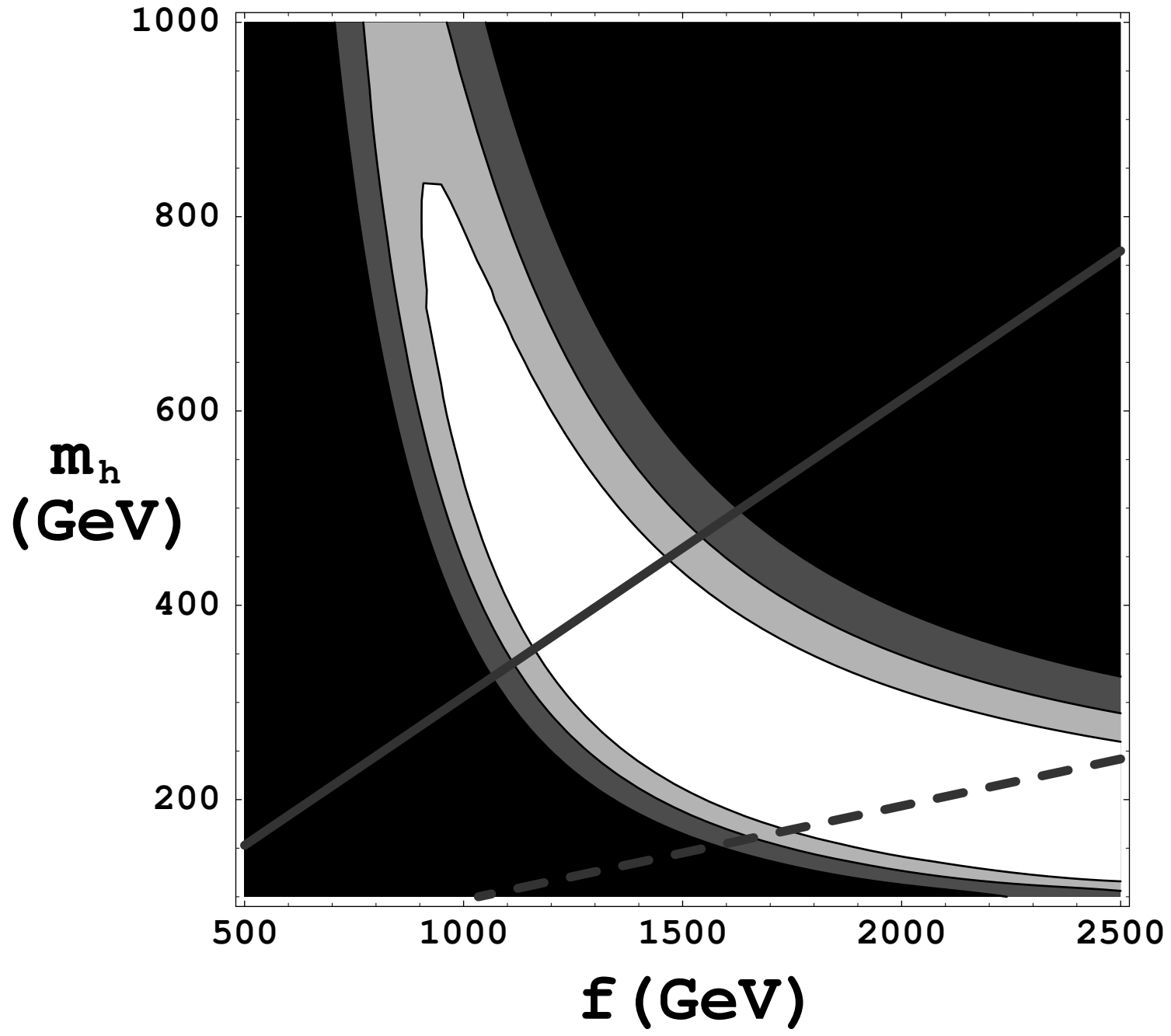

Fig. 7.6: Exclusion contours in terms of the Higgs mass $m_h$ and the symmetry breaking scale $f$. From lightest to darkest, the contours correspond to the 95, 99, and 99.9 confidence level exclusion. The white region is consistent with EWP measurements. Contours of constant values of an estimate of fine-tuning are also shown; the solid and dashed lines correspond to $10\%$ and $1\%$ fine tuning, respectively.

### *7.5.3 Flavor constraints*

In addition to electroweak precision, there are also constraints from neutral meson mixing and rare decays. In the Littlest Higgs without T-parity, these are due to the mixing of the $T_3 = 1/2$ eigenstate, and the singlet which is responsible for canceling the top quark quadratic divergence.

With the addition of T-parity, and the consequential necessity of introducing the mirror fermions, there are new and potentially very large contributions to flavor observables. These arise from one loop box diagrams where the T-odd mirror fermions and T-odd gauge bosons run in the loop.

The origin of these interactions can be understood as follows. In terms of T-parity eigenstates, the fermion kinetic terms can schematically be expanded in the following way:

$$\bar{\Psi}_1 i\not{D}_1 \Psi_1 + \bar{\Psi}_2 i\not{D}_1 \Psi_2 = \bar{\Psi}_{\rm SM} i\not{D}_{\rm SM} \Psi_{\rm SM} + \bar{\Psi}_- i\not{D}_{\rm SM} \Psi_- + ig\bar{\Psi}_- \not{A}_- \Psi_{\rm SM} + ig\bar{\Psi}_{\rm SM} \not{A}_- \Psi_- \quad (7.64)$$

The T-odd fermion mass term in Eq. (7.56) can be extended to include generational mixing. After rewriting Eq. (7.64) in the mass eigenbasis, the last two terms generically involve flavor changing T-odd neutral and charged currents between a standard model fermion and T-odd fermion.

In [70], the contributions to neutral meson mixing observables are computed for arbitrary values of the free parameters associated with the mirror fermions. Is is found that in some regions of parameter space, the T-odd fermion spectrum must be degenerate to within a few percent to satisfy these flavor constraints. This degeneracy can be relaxed with particular choices of mass textures. However, either the degeneracy, or these tuned values of the mixing matrices must be explained by any UV completion of this low energy effective theory.

In addition, there are in principle contributions from physics above the cutoff scale, $\Lambda = 4\pi f$, however these are sensitive to the UV completion, and thus model dependent. Flavor analyses of little Higgs models usually take these contributions to be zero, assuming that the UV completion gives no new contributions to flavor physics.

As in the flavor problem associated with supersymmetry, it is the the $\epsilon_K$ observable (associated with CP violation in the K-meson system) that gives the strongest bounds. Constraints on the fermion mass spectrum are greatly reduced if the CP violating phase which gives contributions to $\epsilon_K$ is set to zero in the new mixing matrices.

In addition, it is found that there are allowed regions of parameter space where one finds enhancements in the $B_s$ mass splitting relative to the standard model prediction. The mass splitting can be as much as a factor of 10 or more larger than in the standard model.

### *7.5.4 The dark matter candidate*

We calculate the relic density of the lightest T-odd particle assuming that T-parity is an exact symmetry, and that the T-odd fermions are heavy. The mass spectrum is sufficiently non-degenerate that coannihilation effects are unimportant, and only direct annihilation channels need be considered. The dominant channels are those involving s-channel Higgs exchange with $W^{\pm}$, $Z$, Higgs, or top quarks in the final state. As a result, the annihilation cross section is primarily a function of the Higgs mass, and the mass of the dark matter candidate. Imposing the constraints given by the WMAP collaboration [87] leads to Fig. 7.7. We see that there is a strong correlation between the scale $f$ and the Higgs mass if the dark matter is to come purely from little Higgs physics. This is due to the s-channel pole present when $m_{A_H} = m_H/2$. Notably, for larger values of $f$, larger Higgs masses than the standard model best fit value are preferred.

We consider regions as ruled out where the relic density exceeds the 95% confidence limits imposed by the WMAP bound. In Fig. 7.7, these regions are shown in black. In regions where the relic density of the $A_H$ is below the WMAP 95% confidence band, there is the possibility that there is another form of dark matter, such as axions, which could make up the difference. These are the lighter contours in Fig. 7.7. Finally, the second darkest region is where the relic density of $A_H$ lies within the 95% confidence bounds given by WMAP.

The narrow region where $f$ is below 600 GeV is where $M_{A_H}$ drops below the $W$ boson mass, and can only annihilate to SM fermions. Because the s-channel Higgs exchange is the only contribution, and the coupling of the Higgs to the accessible fermions is small, it is required that the annihilation happen very close to the Higgs resonance to enhance the cross section enough to get the correct relic density. For values of $f$ below 600 GeV then, the Higgs mass must be very close to $M_{A_H}/2$ in order to get the right abundance of dark matter.

A study of the one loop electroweak precision corrections in this model reveals that certain contributions to $\Delta\rho$ from one-loop diagrams arise with the opposite sign as the terms which are logarithmic in the Higgs mass [43]. This effect is due to the contributions from singlet-doublet quark mass mixing in the third generation Yukawa. Consequentially, the Higgs mass can be raised far above its standard electroweak precision bound while remaining consistent with LEP. Thus, for certain ranges of the parameters in the top-quark Yukawa sector, both dark matter and EWP bounds may be satisfied simultaneously.

We note that the T-odd fermion doublets may in principle be quite light, such that they play a significant role in the relic abundance calculation through coannihilation channels. This has been considered in detail in [86]. In addition, this paper also discusses the potential for direct and indirect detection of the relic $A_H$. Currently, the best way to search for this type of dark matter is with the upcoming GLAST gamma ray telescope. The nucleon scattering cross section turns out to be quite small, as the amplitude is dominated by T-channel Higgs exchange, which couples to the nucleon through the $hgg$ vertex. The ultimate projected sensitivity of CDMS will begin probing the parameter space relevant to dark matter in the Littlest Higgs model, however the current experimental precision is orders of magnitude away from being able to discover the $A_H$.

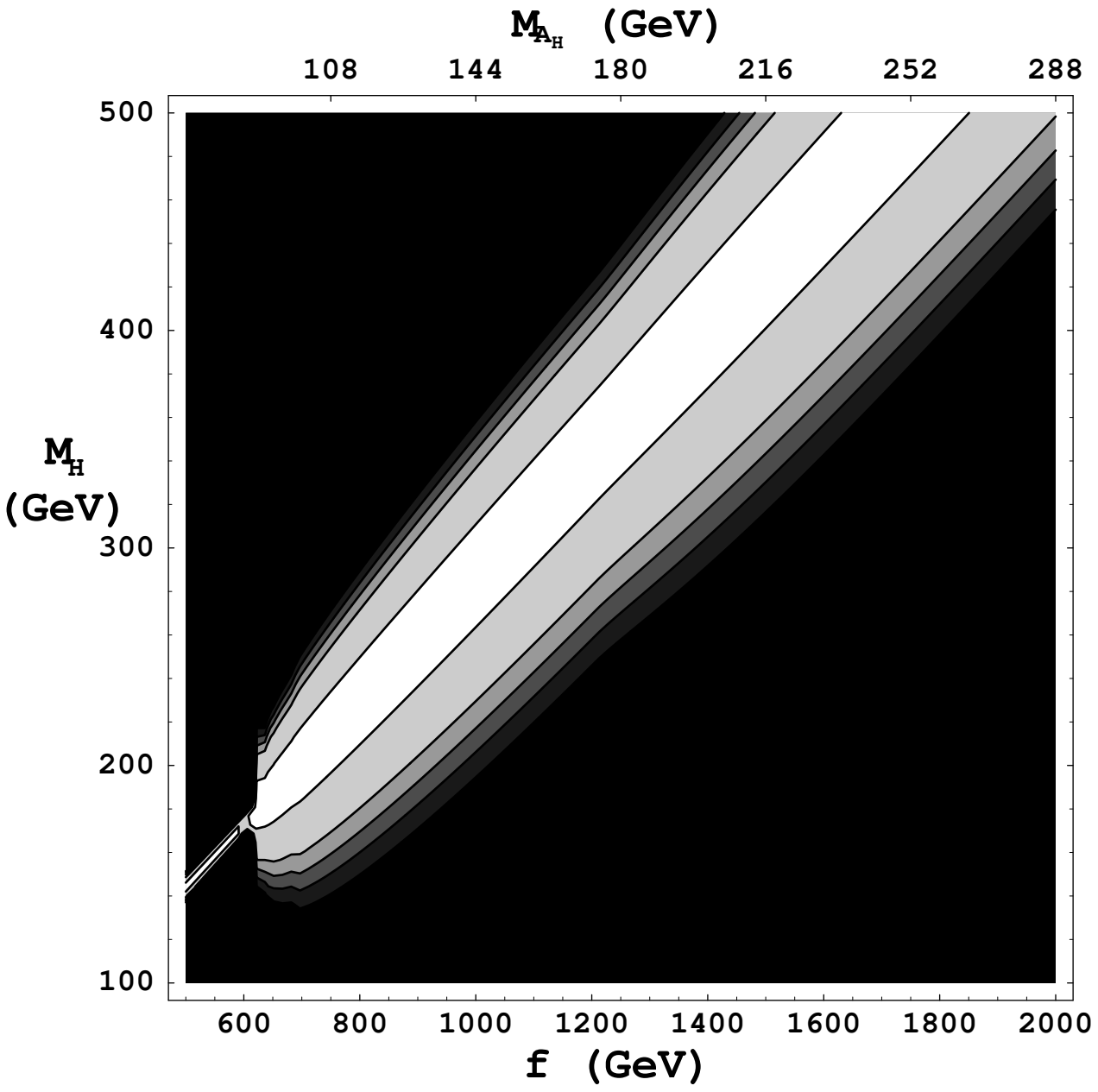

Fig. 7.7: This plot depicts the variation of the relic density with respect to the Higgs mass and the symmetry breaking scale, $f$. In order from lightest to darkest regions, the $A_H$ makes up $(0 - 10\%, 10 - 50\%, 50 - 70\%, 70 - 100\%, 100\%, > 100\%)$ of the observed relic abundance of dark matter. The second darkest region is the preferred region, where the $A_H$ dark matter candidate relic density lies within the 95% confidence level bounds determined by the WMAP collaboration. The black region is excluded at the 95% confidence level.

### *7.5.5 Collider phenomenology*

Nearly all of the phenomenology of T-parity models is distinct from little Higgs models without T-parity. Because almost all of the new particles are odd under T-parity, they must be pair produced, which reduces the production cross sections due to the additional energy cost. Also, the new T-odd particles will not appear as resonances in detectable particles, as all T-odd particles cascade decay to the lightest T-odd particle, the $A_H$. This makes identification of the little Higgs mechanism nearly impossible at the LHC. The exception is the $T_+$, which is a new T-even state. The production mechanisms and cross section for the $T_+$ are precisely the same as in the original Littlest Higgs. This has been well studied in [67]. The potential for discovery and parameter extraction in the Littlest Higgs without T-parity at Atlas and CMS respectively are discussed in Sections 7.6 and 7.7. However, the decay modes are modified as the $T_+$ now has the channel $T_+ \to T_- A_H \to t A_H A_H$ open to it. This means that the $T_+$ has a sizable portion of its width which cannot be reconstructed. This further complicates attempts to identify the little Higgs mechanism at the LHC.

The pattern of cascade decays in models with T-parity resembles the decay chains of supersymmetry, meaning that it could potentially be quite easy to mistake one for the other at the LHC. There is a phenomenology dictionary between the Littlest Higgs model with T-parity and supersymmetry:

$$\begin{aligned} \text{electroweak gauginos} &\leftrightarrow \mathrm{T-odd\ gauge\ bosons} \\ \text{sfermions} &\leftrightarrow \mathrm{T-odd\ fermion\ doublets} \\ \text{second Higgs doublet} &\leftrightarrow \text{scalar triplet} \\ \text{higgsinos} &\leftrightarrow \mathrm{NONE} \\ \text{gluinos} &\leftrightarrow \mathrm{NONE} \\ \mathrm{NONE} &\leftrightarrow \mathrm{T-even\ partner\ of\ top} \end{aligned} \quad (7.65)$$

For a certain choice of spectra, a cascade decay in the Littlest Higgs with T-parity can be duplicated in supersymmetry using this dictionary. Clearly there are distinguishing features of the two, such that not all regions of MSSM parameter space could be confused with a corresponding region of parameter space in this particular little Higgs model. For example, there is no analog of the gluino in little Higgs models. Similarly, there is no translation for the $T_+$ in supersymmetry. However, modifications and extensions can be made on both sides. The Littlest Higgs model is only the most compact way to extend the SM to include collective symmetry breaking (just as the MSSM is the most compact way to extend the SM to include supersymmetry).

To date, only the T-odd gauge bosons, scalars, and singlet fermions ($T_+$ and $T_-$) have been studied in detail, in the limit that the T-odd fermion doublets are taken to be heavy. The phenomenology of the T-odd fermion doublets is potentially quite rich, especially as EWP, flavor physics, and compositeness bounds all favor them being light with respect to the breaking scale, $f$. There are numerous studies currently underway which will study the phenomenology of these states.

## 7.6 Little Higgs studies with ATLAS

*Eduardo Ros and David Rousseau*

Observability of new particles predicted by little Higgs models at the LHC has been studied using a simulation of the ATLAS detector. We discuss first the channels available for the discovery of the new heavy quark $T$, then for new gauge bosons $A_H$, $Z_H$ and $W_H$, and finally for the doubly charged Higgs boson $\phi^{++}$. Most of the results presented here are extracted from [88], with more recent studies in [90–92], where further details can be found. The Monte Carlo program PYTHIA 6.203 [93] with suitably normalised rates was used to generate signal events. The Higgs boson branching ratios were taken to be as in the standard model. These events were passed through the ATLAS fast simulation which provides a parametrised response of the ATLAS detector to jets, electrons, muons, isolated photons and missing transverse energy. This fast simulation has been validated using a large number of studies [94] where it was adjusted to agree with the results of a full, GEANT based, simulation. Jets are reconstructed using a cone algorithm with a size of $\Delta R = 0.4$. Performance for the high luminosity ($10^{34}$ cm$^{-2}$ sec$^{-1}$) is assumed. Results will be in general quoted for 300 fb$^{-1}$, which correspond approximately to the amount of data collected during three years running at high luminosity. It is assumed that the Higgs boson will have been found and its mass measured. The event selections are based on the characteristics of the signal being searched for, and are such that they will pass the ATLAS trigger criteria. The most important triggers arise from the isolated leptons, jets or photons present in the signal. PYTHIA was also used for simulation of the backgrounds. Other event generators were used if backgrounds were needed in regions of phase space where PYTHIA is not reliable.

### *7.6.1 Search for the heavy quark $T$*

The $T$ quark can be produced at the LHC via two mechanisms: QCD production via the processes $gg \to T\overline{T}$ and $q\overline{q} \to T\overline{T}$ which depend only on the mass of $T$; and production via $W$ exchange $qb \to q'T$ which leads to a single $T$ in the final state and therefore falls off much more slowly as $M_T$ increases. This latter process depends on the model parameters and, in particular, upon the mixing of the $T$ with the conventional top quark. The Yukawa couplings of the new $T$ are given by two constants $\lambda_1$ and $\lambda_2$ (following the notation from [67]). The physical top quark mass eigenstate is a mixture of $t$ and $T$, and the various couplings contain three parameters $\lambda_1$, $\lambda_2$ and $f$ that determine the masses of $T$ and the top quark as well as their mixings. Two of the parameters can be reinterpreted as the top mass and the $T$ mass. The third can then be taken to be $\lambda_1/\lambda_2$. This determines the mixings and hence the coupling strength $TbW$ which controls the production rate via the $qb \to q'T$ process. The production rates have been calculated in [67]. It is found that single production dominates for masses above 700 GeV. As we expect that we are sensitive to masses larger than this, we consider only the single production process in what follows. We assume a cross-section of $\sigma = 200$ fb for $M_T = 1.0$ TeV and $\lambda_1/\lambda_2 = 1$. Events generated using `PYTHIA` were normalised to these values. The decay rates of $T$ are as follows

$$\Gamma(T \to tZ) = \Gamma(T \to tH) = \frac{1}{2}\Gamma(T \to bW) = \frac{\kappa^2}{32\pi} M_T \tag{7.66}$$

with $\kappa = \lambda_1^2/\sqrt{\lambda_1^2 + \lambda_2^2}$ implying that $T$ is a narrow resonance. The last of these decays would be expected for a charged 2/3 $4^{th}$ generation quark; the first two are special to the "little Higgs Model". We now discuss the reconstruction of these channels.

#### 7.6.1.1 Study of the decay $T \to Zt$

This channel can be observed via the final state $Zt \to \ell^+\ell^-\ell\nu b$, which implies that the events contain three isolated leptons, a pair of which reconstructs to the $Z$ mass, one $b-$jet and missing transverse energy. The background is dominated by $WZ$, $ZZ$ and $tbZ$. Events were selected as follows.

- Three isolated leptons (either $e$ or $\mu$) with $p_T > 40$ GeV and $|\eta| < 2.5$. One of these is required to have $p_T > 100$ GeV.
- No other leptons with $p_T > 15$ GeV.
- $E_T^{miss} > 100$ GeV.
- At least one tagged $b-$jet with $p_T > 30$ GeV.

The presence of the leptons ensures that the events are triggered. A pair of leptons of same flavour and opposite sign is required to have an invariant mass within 10 GeV of $Z$ mass. The third lepton is then assumed to arise from a $W$ and the $W$'s momentum reconstructed using it and the measured $E_T^{miss}$. The selection efficiency is 3.3% for $M_T = 1$ TeV. The invariant mass of the $Zt$ system can then be reconstructed by including the $b-$jet. This is shown in Fig. 7.8 for $M_T = 1$ TeV where a clear peak is visible above the background. Following the cuts, the background is dominated by $tbZ$ which is more than 10 times greater than all the others combined. Using this analysis, the discovery potential in this channel can be estimated. The signal to background ratio is excellent as can be seen from Fig. 7.8. Requiring a peak of at least $5\sigma$ significance containing at least 10 reconstructed events implies that for $\lambda_1/\lambda_2 = 1(2)$ and 300 fb$^{-1}$ the quark of mass $M_T < 1050(1400)$ GeV is observable. At these values, the single $T$ production process dominates, justifying *a posteriori* the neglect of $T\overline{T}$ production in this simulation.

#### 7.6.1.2 Study of the decay $T \to Wb$

This channel can be reconstructed via the final state $\ell\nu b$. The following event selection was applied.

- At least one charged lepton with $p_T >$100 GeV.
- One $b$-jet with $p_T > 200$ GeV.
- No more than 2 jets with $p_T > 30$ GeV.
- Mass of the pair of jets with the highest $p_T$ is greater than 200 GeV.

– $E_T^{miss}$ >100 GeV.

The lepton provides a trigger. The backgrounds arise from $t\bar{t}$, single top production and QCD production of $Wb\bar{b}$. The requirement of only one tagged $b-$jet and the high $p_T$ lepton are effective against all of these backgrounds. The requirement of only two energetic jets is powerful against the dangerous $t\bar{t}$ source where the candidate $b-$jet arises from the $t$ and the lepton from the $\bar{t}$. The selection efficiency is 14% for $M_T = 1$ TeV. The signal to background ratio in the case of $T$ with 1 TeV mass is somewhat worse than in the previous case primarily due to the $t\bar{t}$ contribution. From this analysis, the discovery potential in this channel can be estimated. For $\lambda_1/\lambda_2 = 1(2)$ and 300 fb$^{-1}$, $M_T <$ 2000(2500) GeV has at least a $5\sigma$ significance.

#### 7.6.1.3 Study of the decay $T \to Ht$

In this final state, the event topology depends on the Higgs mass. For a Higgs mass of 120 GeV the decay to $b\bar{b}$ dominates. The semileptonic top decay $t \to Wb \to \ell\nu b$ produces a lepton that can provide a trigger. The final state containing an isolated lepton and several jets then needs to be identified. The initial event selection is as follows.

- One isolated $e$ or $\mu$ with $p_T >$ 100 GeV and $|\eta| < 2.5$.
- Three jets with $p_T >$ 130 GeV.
- At least one jet tagged as a $b-$jet.

Events were further selected by requiring that at least one di-jet combination have a mass in the range 110 to 130 GeV. If there is a pair of jets with invariant mass in the range 70 to 90 GeV, the event is rejected in order to reduce the $t\bar{t}$ background. The measured missing transverse energy and the lepton are then combined using the assumption that they arise from a $W \to \ell\nu$ decay. Events that are consistent with this hypothesis are retained and the $W$ momentum inferred. The selection efficiency is 2.3% for $M_T = 1$ TeV. The invariant mass of the reconstructed $W$, $H$ and one more jet is formed and the result is shown in Fig. 7.8. The width of the reconstructed $T$ resonance is dominated by experimental resolution. This analysis assumes that $\lambda_1/\lambda_2 = 1$. The background is dominated by $t\bar{t}$ events. The significance is lower than the previous channels, about $4\sigma$ for $M_T = 1$ TeV, down to $3\sigma$ for $M_T =$ 700 GeV, thus only providing a confirmation if the signal is seen in the previous channel.

### *7.6.2 Search for new gauge bosons*

The model predicts the existence of one charged $W_H$ and two neutral ($Z_H$ and $A_H$) heavy gauge bosons. $W_H$ and $Z_H$ are almost degenerate in mass and are typically heavier than $A_H$. From fine tuning arguments [95], an upper bound can be set: $M_{W_H,Z_H} <$ 6 TeV$(m_H/200 \text{ GeV})^2$, i.e. 2 TeV for $m_H =$ 120 GeV and 6 TeV for $m_H =$ 200 GeV. All these bosons are likely to be discovered via their decays to leptons. However, in order to distinguish these gauge bosons from those that can arise in other models, the characteristic decays $Z_H \to ZH$ and $W_H \to WH$ must be observed [96]. Two new couplings are present, in addition to those of the Standard Model. These additional parameters can be taken to be two angles $\theta$ and $\theta'$. Once the masses of the new bosons are specified, $\theta$ determines the couplings of $Z_H$ and $\theta'$ those of $A_H$. In the case of $Z_H$, the branching ratio into $e^+e^-$ and $\mu^+\mu^-$ rises with $\cot\theta$ to an asymptotic value of 4%.

#### 7.6.2.1 Discovery of $Z_H$, $A_H$ and $W_H^{\pm}$

A search for a peak in the invariant mass distribution of either $e^+e^-$ or $\mu^+\mu^-$ is sensitive to the presence of $A_H$ or $Z_H$. As an example, Fig. 7.9 shows the $e^+e^-$ mass distribution arising from a $Z_H$ of mass of 2 TeV for $\cot\theta = 1$ and $\cot\theta = 0.2$. The production cross-section for the former (latter) case is 1.2 (0.05) pb [67]. Events were required to have an isolated $e^+$ and $e^-$ of $p_T >$ 20 GeV and $|\eta| < 2.5$ which provides a trigger. The Standard Model background shown on the plot arises from the Drell-Yan process. In order to establish a signal we require at least 10 events in the peak of at least $5\sigma$ significance.

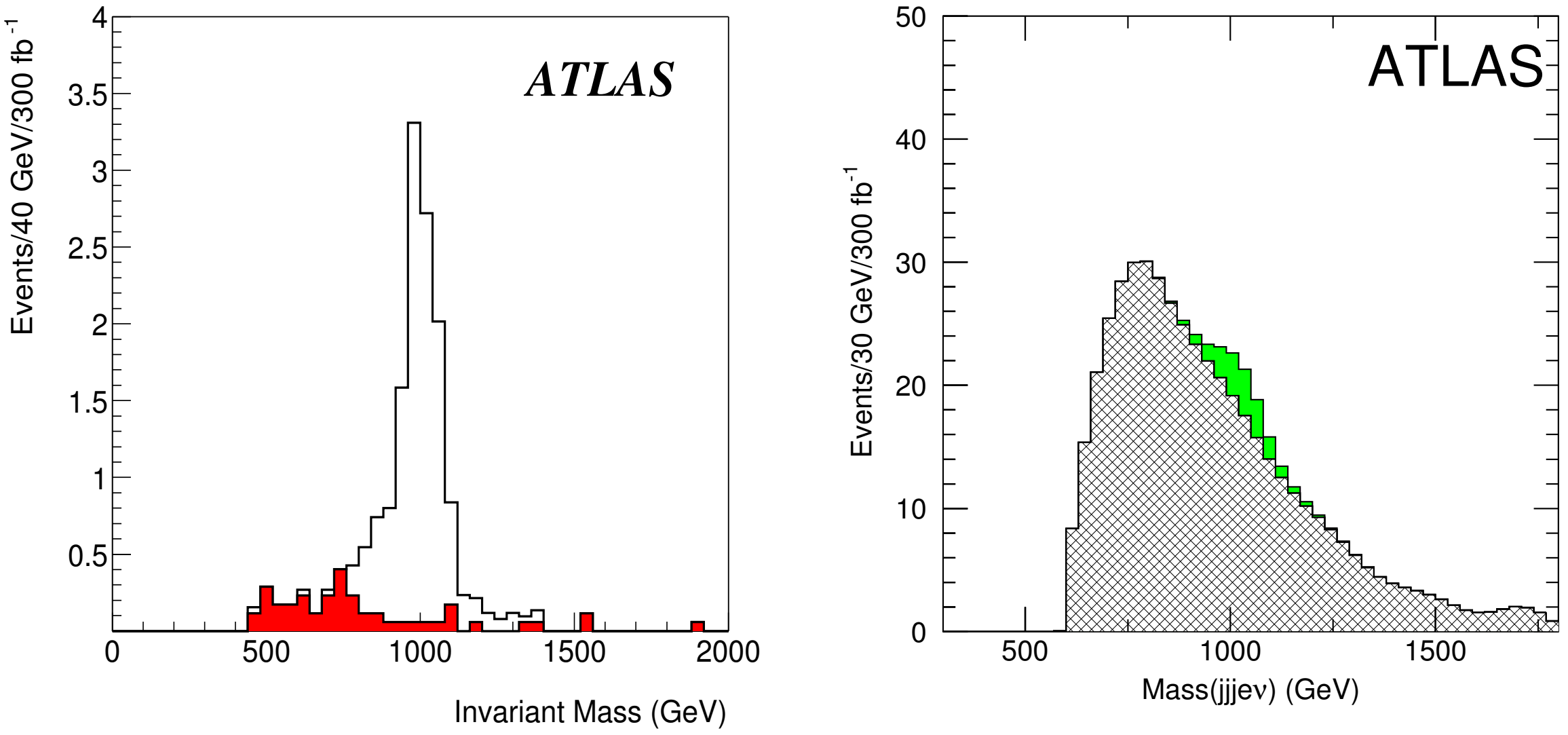

Fig. 7.8: The signal $T \rightarrow Zt$ (left) and $T \rightarrow Ht$ (right) is shown for a mass of 1 TeV. The background is dominated by $WZ$ and $tbZ$ production (left) and $t\bar{t}$ production (right).

Including the $\mu + \mu^-$ channel improves the reach slightly, given the poorer mass resolution. Fig. 7.14 top left shows the accessible region as a function of $\cot\theta$ and $M_{Z_H}$. A similar search for $A_H$ can be carried out and the accessible region as a function of $\tan\theta'$ and $M_{A_H}$ is shown in Fig. 7.14 top right. Masses greater than 3 TeV are not shown as these are not allowed in the model. There is a small region around $\tan\theta' \sim 1.3$ where the branching ratio to $\mu^+\mu^-$ and $e^+e^-$ is very small and the channel is insensitive. The decay $W_H^{\pm} \rightarrow \ell\nu$ manifests itself via events that contain an isolated charged lepton and missing transverse energy. Events were selected by requiring an isolated electron with $e^-$ or $e^+$ of $p_T > 200$ GeV, $|\eta| < 2.5$ and $E_T^{miss} > 200$ GeV. The transverse mass from $E_T^{miss}$ and the observed lepton is formed and the signal appears as a peak in this distribution. The main background arises from $\ell\nu$ production via a virtual $W$. In order to establish a signal we require at least 10 events in the signal region of at least $5\sigma$ significance. Fig. 7.14 top left shows the accessible region as a function of $\cot\theta$ and $M_{W_H}$.

#### 7.6.2.2 Observation of $Z_H \rightarrow ZH$, $A_H \rightarrow ZH$ and $W_H \rightarrow WH$ for $m_H = 120$ GeV

Observation of the cascade decays $Z_H \rightarrow ZH$, $A_H \rightarrow ZH$, and $W_H \rightarrow WH$ provides crucial evidence that an observed new gauge boson is of the type predicted in the little Higgs Models. For a Higgs mass of 120 GeV, two signatures have been searched for : the more abundant $H \rightarrow b\bar{b}$ (with a branching ratio of 68%) [90] , and the much rarer $H \rightarrow \gamma\gamma$ (with a branching ratio of 0.2%) compensated by a clearer signature [91] .

The decay $ZH \rightarrow \ell^+\ell^- b\bar{b}$ results in a final state with two $b-$jets that reconstruct to the Higgs mass and a $\ell^+\ell^-$ pair that reconstructs to the $Z$ mass. The coupling $Z_H ZH$ is proportional to $\cot 2\theta$. When combined with the coupling of $Z_H$ to quarks that controls the production cross-section, the $\cot\theta$ dependence of the rate in this channel is shown in Fig. 7.10, which shows that this decay vanishes for $\cot\theta \sim 1$. A typical value of $\cot\theta \sim 0.5$ is chosen, in the following. The signal is extracted from the $Z_H \rightarrow ZH$ state using the following event selection:

- Two leptons of opposite charge and same flavour with $p_T > 6(5)$ GeV for muons (electrons) and $|\eta| < 2.5$. One of them is required to satisfy $p_T > 25$ GeV in order to provide a trigger.
- The lepton pair has a mass between 76 and 106 GeV

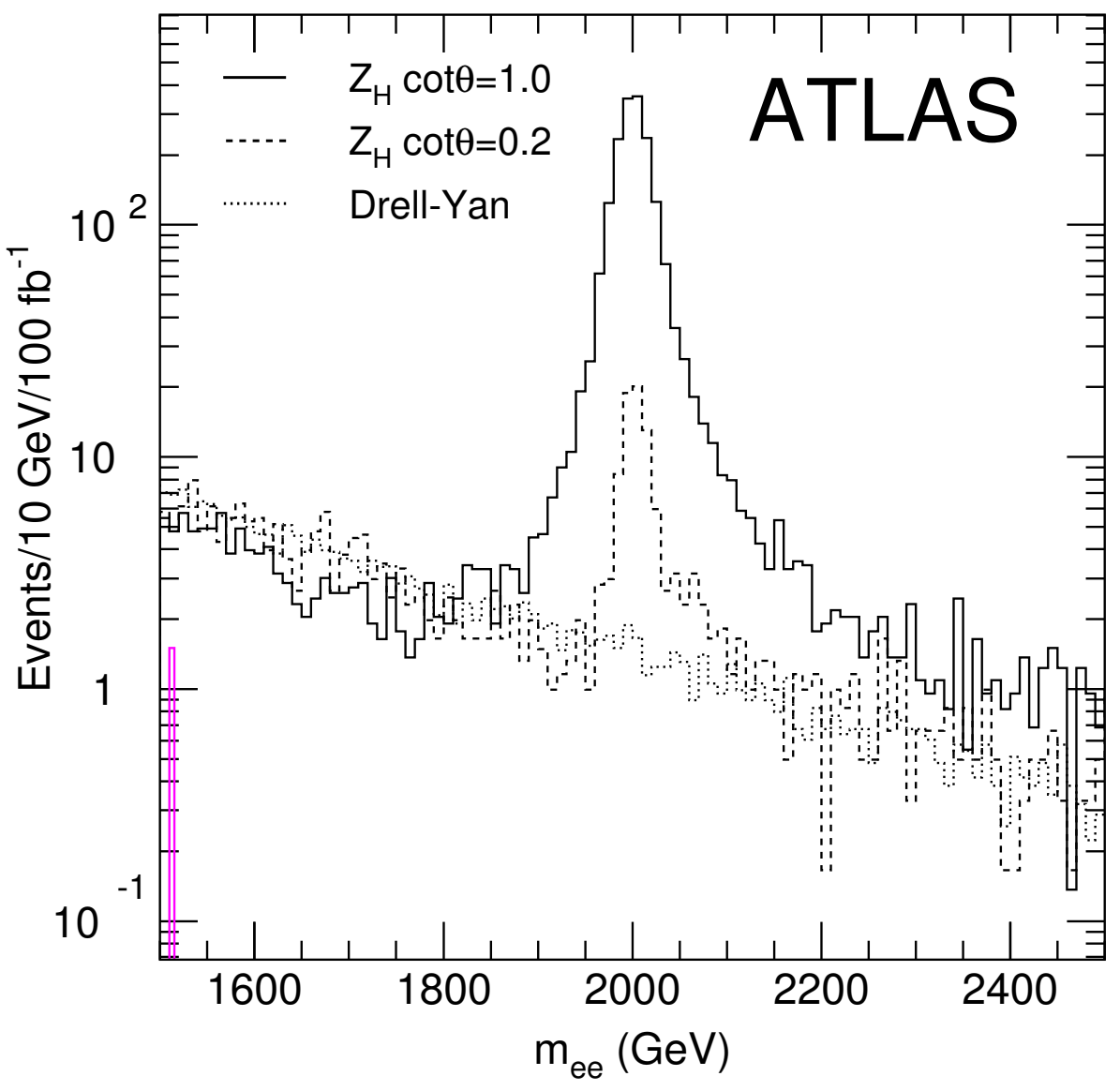

Fig. 7.9: The $e^+e^-$ mass distribution arising from a $Z_H$ of mass of 2 TeV for $\cot\theta = 1$ (upper, solid, histogram) and $\cot\theta = 0.2$ (middle, dashed, histogram). The lowest, dotted histogram shows the distribution from background only.

– Two reconstructed $b-$jets with $p_T > 25$ GeV and $|\eta| < 2.5$, which are within $\Delta R = \sqrt{(\Delta\eta)^2 + (\Delta\phi)^2} < 1.5$.
– The $b-$jet pair should have a mass between 60 and 180 GeV.

The efficiency for $M_{Z_H} = 1$ TeV is 35 %.The mass of the reconstructed $ZH$ system is shown in Fig. 7.11 for a $Z_H$ mass of 1 TeV and $\cot\theta = 0.5$. The presence of a leptonic $Z$ decay in the signal ensures that the background arises primarily from $Z + jet$ final states.

A similar method can be used to reconstruct the $W_H \to WH \to \ell\nu b\bar{b}$ decay. The $b-$jet selections were the same as above while the lepton selection is now as follows:

– One isolated $e$ or $\mu$ with $p_T > 25$ GeV and $|\eta| < 2.5$.
– $E_T^{miss} > 25$ GeV.

The missing transverse energy is assumed to arise only from the neutrino in the leptonic $W$ decay, and the $W$ momentum is then reconstructed. The efficiency for $M_{W_H} = 1$ TeV is 38 %. The background which is dominated by $W + jets$ and $t\bar{t}$ events is larger than in the previous case, nevertheless a signal can be extracted.

The decay $H \to \gamma\gamma$ provides a very characteristic signal. A preliminary event selection requiring two isolated photons, one having $p_T > 25$ GeV and the other $p_T > 40$ GeV and both with $|\eta| < 2.5$ was made. This requirement ensures that the events are triggered. The invariant mass of the two photon system is required to be within $2\sigma$ of the Higgs mass, $\sigma$ being the measured mass resolution of the diphoton system. The reconstructed jets in the event are then combined in pairs and the pair with invariant mass closest to $M_W$ was selected. If this pair has a combined $p_T > 200$ GeV, its mass was corrected to the $W$ mass and then combined with the $\gamma\gamma$ system. The efficiency for $M_{W_H,Z_H} = 1$ TeV is 50 %. The mass distribution of the resulting system is shown in Fig. 7.11. The contributions from $W_H$ and $Z_H$ are shown separately, the former dominates due to its larger production rate. The presence of the two photons with a mass comparable to the Higgs mass ensures that the background is small. This background arises from either direct Higgs production or the QCD production of di-photons.

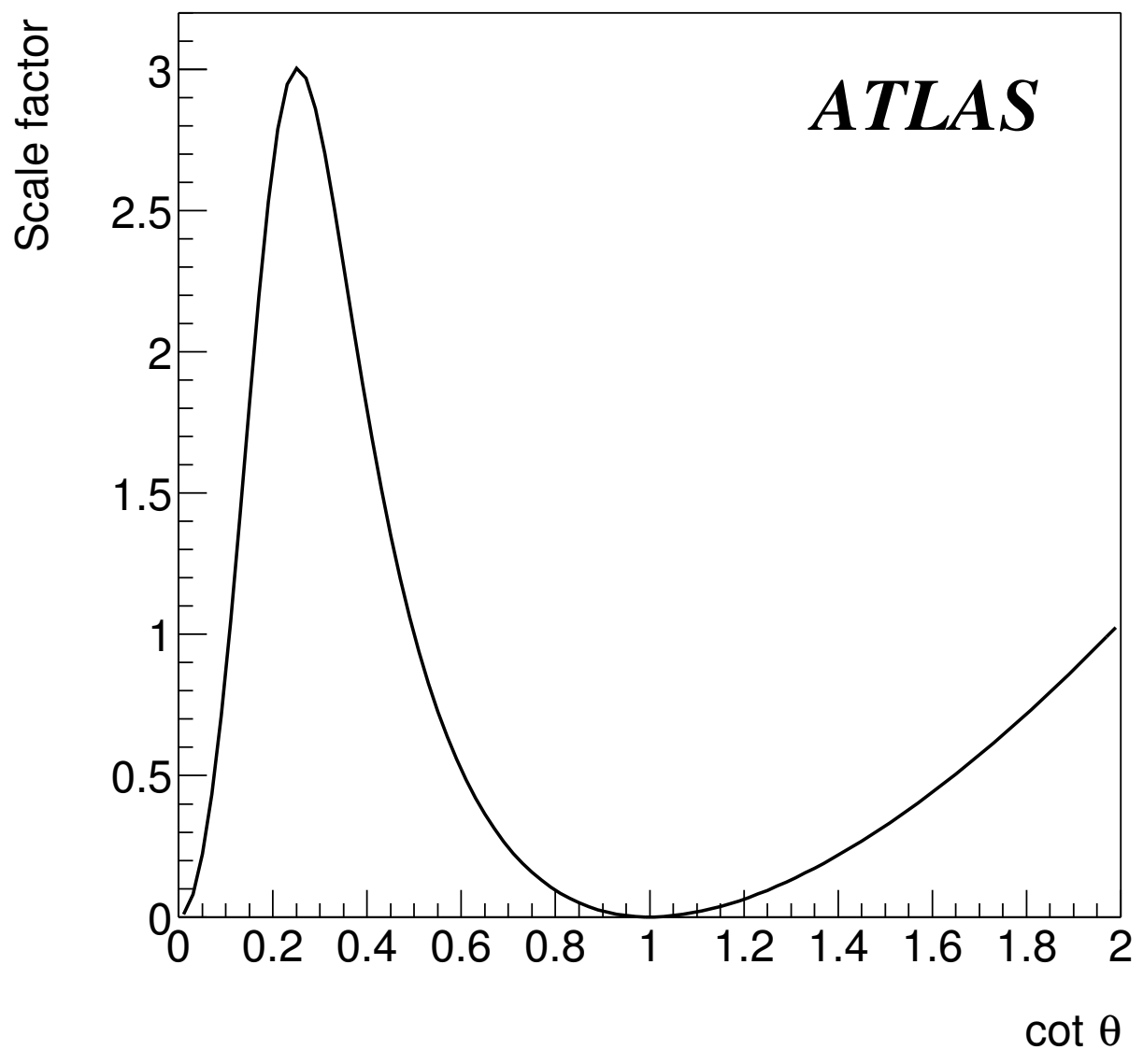

Fig. 7.10: The $\cot\theta$ dependence of the production rate times branching ratio $Z_H \to ZH$.

The analyses were redone for $M_{W_H,Z_H}$= 1, 1.5, and 2 TeV. The reach is shown in Fig. 7.14 bottom left. If $m_H = 120$ GeV, the mass of the heavy bosons is bound to be less than 2 TeV for fine-tuning considerations. A large fraction of the parameter space is hence covered, except for the region around $\cot\theta = 1$.

The search for $A_H \to ZH$ is identical to the search for $Z_H \to ZH$. However, the $A_H$ production and decay to $ZH$ depend on the mixings and so we present the sensitivity in terms of a cross-section that allows reinterpretation of these results to other models. Using the method described above, and assuming only that the $Z_H$ signal does not mask the $A_H$ signal, Fig. 7.12 shows the value of the production cross-section times branching ratio needed to obtain discovery in the channels $A_H \to ZH \to \ell\ell b\bar{b}$ and $A_H \to ZH \to$ jets $\gamma\gamma$.

### 7.6.2.3 Observation of $Z_H \to ZH$ and $W_H \to WH$ for $m_H = 200$ GeV

For a Higgs mass of 200 GeV, the main Higgs decays are $H \to W^+W^-$ (73 %) and $H \to ZZ$ (26 %). Different $ZH$ and $WH$ final states have been selected, resulting from a compromise between cross-section and signature, as listed in Table 7.4. For the A modes [90] , all leptons are isolated, and the Higgs boson final state is purely leptonic. For the B modes [91] , the Higgs boson final state contain one hadronic $W$ or $Z$.

For the sake of brevity, only the salient points of the analyses are reported here. In all the modes, the main background is inclusive top production, $t\bar{t} \to WbWb \to \ell^-\nu\ell^+\nu bb$ where a third lepton can arise from a $b$ jet. In the A1 and A2 modes, the missing transverse momentum is used to reconstruct the Higgs momentum, with the additional hypothesis that the neutrino is collinear to the leptons, a valid approximation given the high momentum. In addition, the $W$ mass constraint is applied in the B1 and B3 modes. The A3 and A4 modes have indistinguishable final states. For all B modes, some leptons may overlap with the hadronic $W$ or $Z$ decay, given the very high momentum of the Higgs boson (above 500 GeV). Hence a special tuning of the lepton isolation was applied. The hadronic decay of the high $p_T$ $W$ or $Z$ are reconstructed by looking for two high $p_T$ jets with mass close to the $W$ or $Z$ mass, or, if it fails, by taking the jet with largest $p_T$ (assuming that in this case the $W$ or $Z$ is reconstructed as a single jet). The efficiencies for the different modes for $M_{W_H,Z_H} = 1$ TeV are as follows : A1 34%, A2 12%, A3/A4 26%, B1 22%, B2 17%, B3 15%. For $M_{W_H,Z_H} = 2$ TeV, the efficiencies decrease by at most a factor of two, due to a more severe overlap of the Higgs boson decay products. An example of the expected reconstructed mass for the B1 modes is shown in Fig. 7.13.

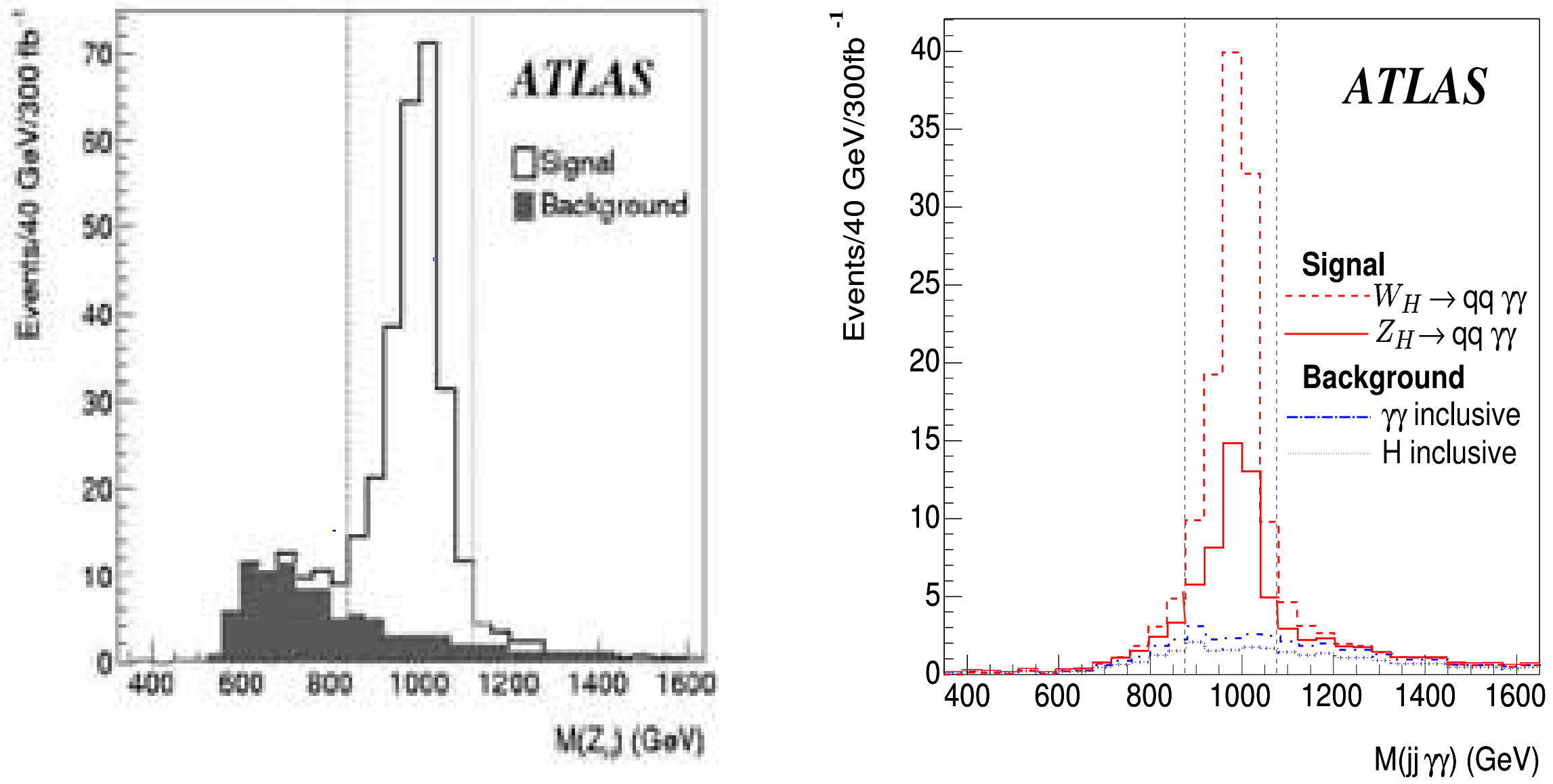

Fig. 7.11: Left : Invariant mass of the $ZH$ system reconstructed from the $\ell^+\ell^- b\bar{b}$ final state. Right : Invariant mass of the $ZH$ or $WH$ system reconstructed from the $jj\gamma\gamma$ final state. The following hypotheses are made: $M_{Z_H/W_H} = 1$ TeV, $m_H = 120$ GeV and $\cot\theta = 0.5$ .

Table 7.4: $W_H$ and $Z_H$ final states being studied. The branching ratios are computed assuming $\cot\theta = 0.5$

| Mode | BR ($10^{-4}$) | decay | signature |
|---|---|---|---|
| A1: | 1.0 | $Z_H \to ZH \to \ell^+\ell^- W^+W^- \to \ell^+\ell^-\ \ \ell^+\nu\ell^-\nu$ | 4 leptons + $E_T^{miss}$ |
| A2: | 3.0 | $W_H \to WH \to \ell\nu W^+W^- \to \ell\nu\ \ \ell^+\nu\ell^-\nu$ | 3 leptons + $E_T^{miss}$ |
| A3: | 0.4 | $Z_H \to ZH \to jjZZ \to jj\ \ \ell^+\ell^-\ell^+\ell^-$ | 4 leptons + jets |
| A4: | 0.4 | $W_H \to WH \to jjZZ \to jj\ \ \ell^+\ell^-\ell^+\ell^-$ | 4 leptons + jets |
| B1: | 6.8 | $Z_H \to ZH \to \ell^+\ell^- W^+W^- \to \ell^+\ell^-\ \ jj\ell\nu$ | 3 leptons + jets + $E_T^{miss}$ |
| B2: | 0.8 | $Z_H \to ZH \to \ell^+\ell^- ZZ \to \ell^+\ell^-\ \ jj\ell^+\ell^-$ | 4 leptons + jets |
| B3: | 2.4 | $W_H \to WH \to \ell\nu ZZ \to \ell\nu\ \ jj\ell^+\ell^-$ | 4 leptons + jets |

The reach of the analyses are combined separately for A modes and B modes and are summarised in Fig. 7.14 bottom right. The reach is very similar to the $m_H = 120$ GeV case, except that now the mass of the heavy bosons is only bound to be less than 6 TeV, hence a much smaller fraction of the parameter space is covered.

### 7.6.2.4 Search for hadronic $Z_H$ and $W_H$ decay

While the leptonic decays of $Z_H$ and $W_H$ allow the quicker discovery of the heavy bosons, a test of the little Higgs model necessitates the measurements of other decay modes, like $WH$ or $ZH$ as described in the previous sections but also the hadronic decay modes [92]. In particular, for $\cot\theta \sim 1$, $BR(W_H \to WH)$ and $BR(Z_H \to ZH)$ vanish, and the branching ratios to heavy quarks are [67] :

$$BR(Z_H \to b\bar{b}) = BR(Z_H \to t\bar{t}) = 1/8 = 12.5\% \tag{7.67}$$

$$BR(W_H \to tb) = 1/4 = 25\% \tag{7.68}$$

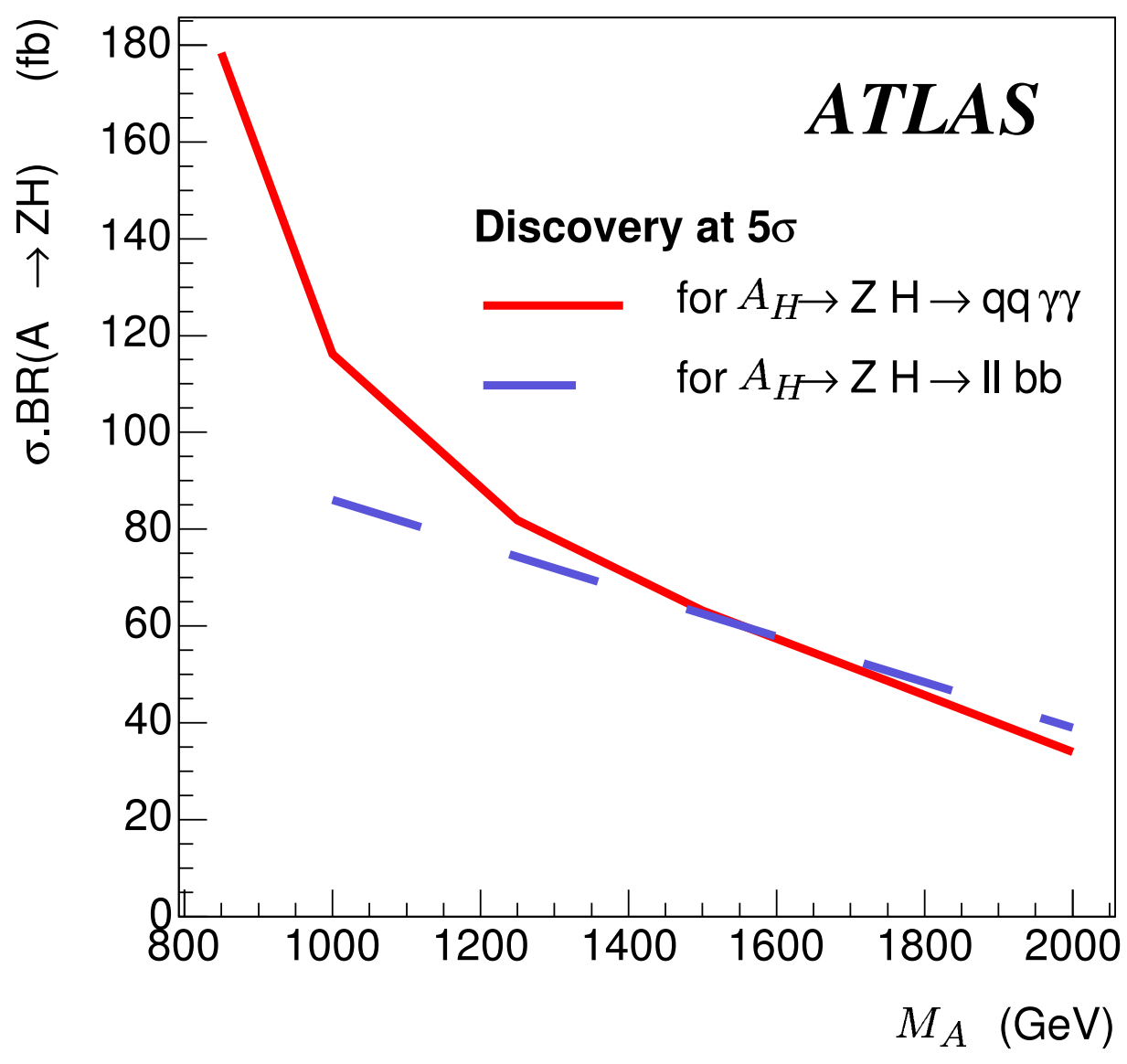

Fig. 7.12: Minimum value of the production cross-section times branching ratio needed to obtain discovery in the channels $A_H \to ZH \to \ell\ell b\bar{b}$ and $A_H \to ZH \to jets\gamma\gamma$ as a function of the $A_H$ mass, for a luminosity of 300 fb$^{-1}$.

The observability of these three final states has been assessed using fast simulation, with parameters tuned on full simulation, and with special care for $b$-tagging at very high jet $p_T$ (up to 1 TeV). While no convincing signal can be seen in the $Z_H$ case, the $W_H \to tb$ appears indeed to be visible and is now described in a few lines. The top is reconstructed in the $W(\ell\nu)b$ final state. One isolated high $p_T$ lepton is searched for, and two $b$-jets tagged, one close to the lepton, one recoiling against the lepton. The neutrino 3-momentum is estimated from the reconstructed missing transverse momentum and assuming it is parallel to the lepton momentum. The final state can be reconstructed with typical efficiency of 25% and mass resolution 110 GeV for $M_{W_H} = 1$ TeV. The background is mainly inclusive top production (irreducible) as well as $W + jets$ (reducible).

The reconstructed mass plot is shown in Fig. 7.13: the signal is clearly visible. The reach shown in Fig. 7.14 top left demonstrates that the $\cot\theta = 1$ region which was missing in the $W_H(Z_H) \to W(Z)H$ analyses is well covered up to $M_{W_H} = 2.5$ TeV.

### *7.6.3 Search for $\phi^{++}$*

The doubly-charged Higgs boson could be produced in pairs and decay into leptonic final states via $q\bar{q} \to \phi^{++}\phi^{--} \to 4\ell$. While this would provide a very clean signature, it will not be considered here since the mass reach in this channel is poor due to the small cross-section. The coupling of $\phi^{++}$ to $W^+W^+$ allows it to be produced singly via $WW$ fusion processes of the type $dd \to uu\phi^{++} \to uuW^+W^+$. This can lead to events containing two leptons of the same charge, and missing energy from the decays of the $W$'s. The $\phi WW$ coupling is determined by $v'$, the vacuum expectation value (vev) of the neutral member of the triplet. This cannot be too large as its presence causes a violation of custodial SU(2) which is constrained by measurements of the $W$ and $Z$ masses. We have examined the sensitivity of searches at the LHC in terms of $v'$ and the mass of $\phi^{++}$. For $v' = 25$ GeV and a mass of 1 TeV, the rate for production of $\phi^{++}$ followed by the decay to $WW$ is 4.9 fb if the $W$'s have $|\eta| < 3$ and $p_T > 200$ GeV [67]. As in the case of Standard Model Higgs searches using the $WW$ fusion process [94], the presence of jets at large rapidity must be used to suppress backgrounds. The event selection closely follows that used in searches for a heavy Standard Model Higgs via the $WW$ fusion process and is as follows [88, 89].

– Two reconstructed positively charged isolated leptons (electrons or muons) with $|\eta| < 2.5$.

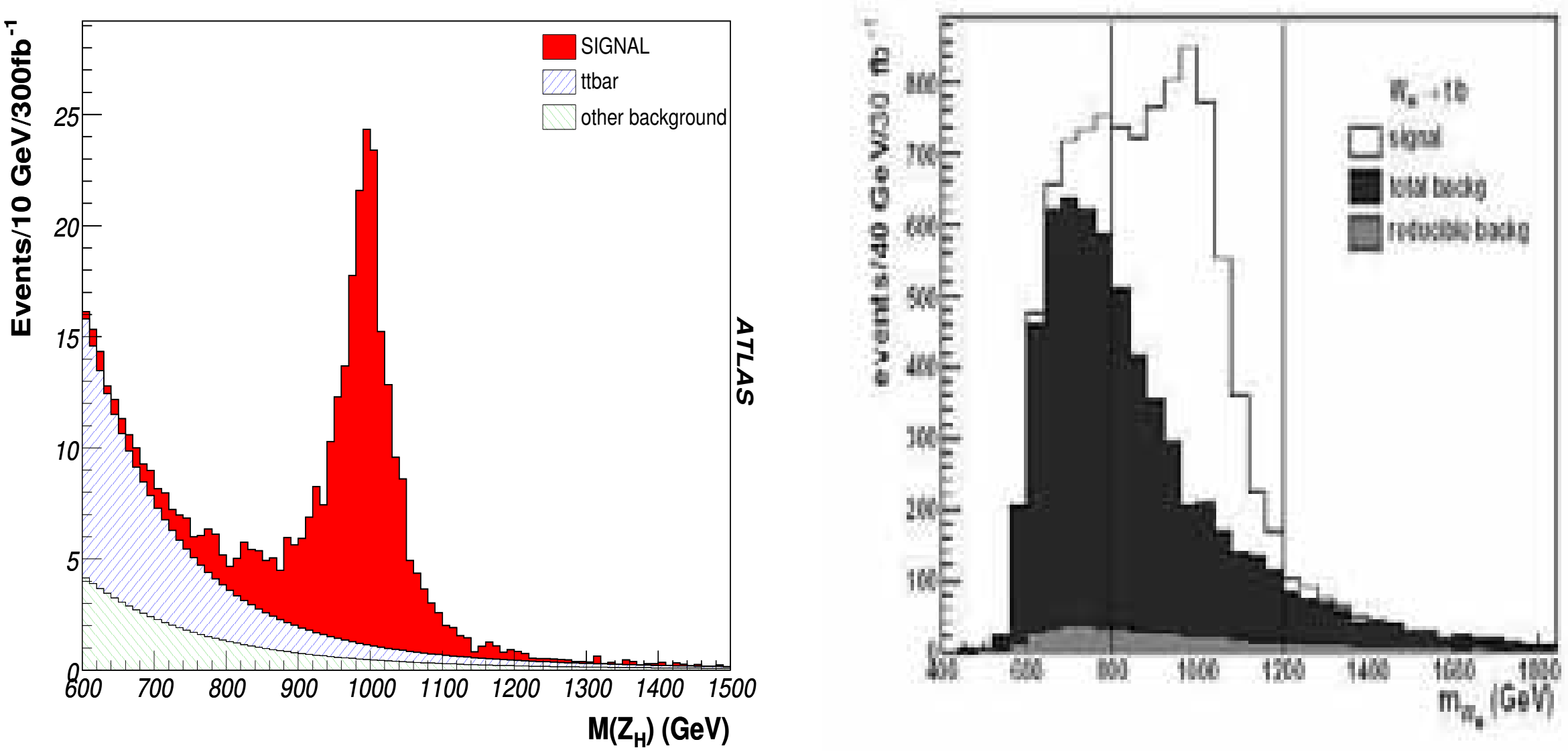

Fig. 7.13: Left : reconstructed mass of $ZH \to \ell^+\ell^- jj\ell\nu$ (B1 mode) system, for $M_{Z_H} = 1$ TeV, $m_H = 120$ GeV and $\cot\theta = 0.5$. Right: reconstructed $tb$ mass for $M_{W_H} = 1$ TeV and $\cot\theta = 1$.

- One of the leptons was required to have $p_T > 150$ GeV and the other $p_T > 20$ GeV.
- The leptons are not balanced in transverse momentum: $|p_{T1} - p_{T2}| > 200$ GeV.
- The difference in pseudorapidity of the two leptons should be $|\eta_1 - \eta_2| < 2$.
- $E_T^{miss} > 50$ GeV.
- Two jets each with $p_T > 15$ GeV, with rapidities of opposite sign, separated in rapidity $|\eta_1 - \eta_2| > 5$; one jet has $E > 200$ GeV and the other $E > 100$ GeV.

The presence of the leptons ensures that the events are triggered. The invariant mass of the $WW$ system cannot be reconstructed, but the signal can be observed using a mass variable $m_{trans}$ made from the observed leptons momenta ($\mathbf{p_1}$ and $\mathbf{p_2}$) and the missing transverse momentum $\mathbf{p_T^{miss}}$ as follows:

$$m_{trans}^2 = (E_1 + E_2 + \left|E_T^{miss}\right|)^2 - (\mathbf{p_1} + \mathbf{p_2} + \mathbf{p_T^{miss}})^2 \tag{7.69}$$

The reconstructed mass distribution is shown in Fig. 7.15 for a mass of 1 TeV. Standard Model backgrounds are shown separately on the figure. Note that the rate shown in this figure is small and the signal does not appear as a clear peak. The process is very demanding of luminosity, the ability to detect forward jets at relatively small $p_T$, and the ability to control backgrounds. These issues cannot be fully addressed until actual data is available. At this stage, we can only estimate our sensitivity using our current, best estimates, of these issues. Since the cross-section for a $\phi^{++}$ of a fixed mass is proportional to $(v')^2$, the simulation can be used to determine the sensitivity. Requiring at least 10 events with $m_{trans} > 700(1000)$ GeV for $M_\phi = 1000(1500)$ GeV and a value of $S/\sqrt{B} > 5$ implies that discovery is possible if $v' > 29(54)$ GeV. Such values are larger than the constraint of $v' < 25$ GeV from electro-weak fits [67].

### *7.6.4 Model constraints and conclusions*

We have shown, using a series of examples, how measurements using the ATLAS detector at the LHC can be used to reveal various particles predicted by little Higgs models. The $T$ quark is observable up to masses of approximately 2.5 TeV via its decay to $Wb$. Sensitivity in $Zt$ or $Ht$ is lower but it still extends over the range expected in the model provided that the Higgs mass is not too large. In the case of $Ht$ the sensitivity will depend on the Higgs mass. The $H \to b\bar{b}$ channel is effective until the Higgs mass exceeds 150 GeV. In this case ATLAS will be able to detect $T$ in its three decay channels and provide a test of the model.

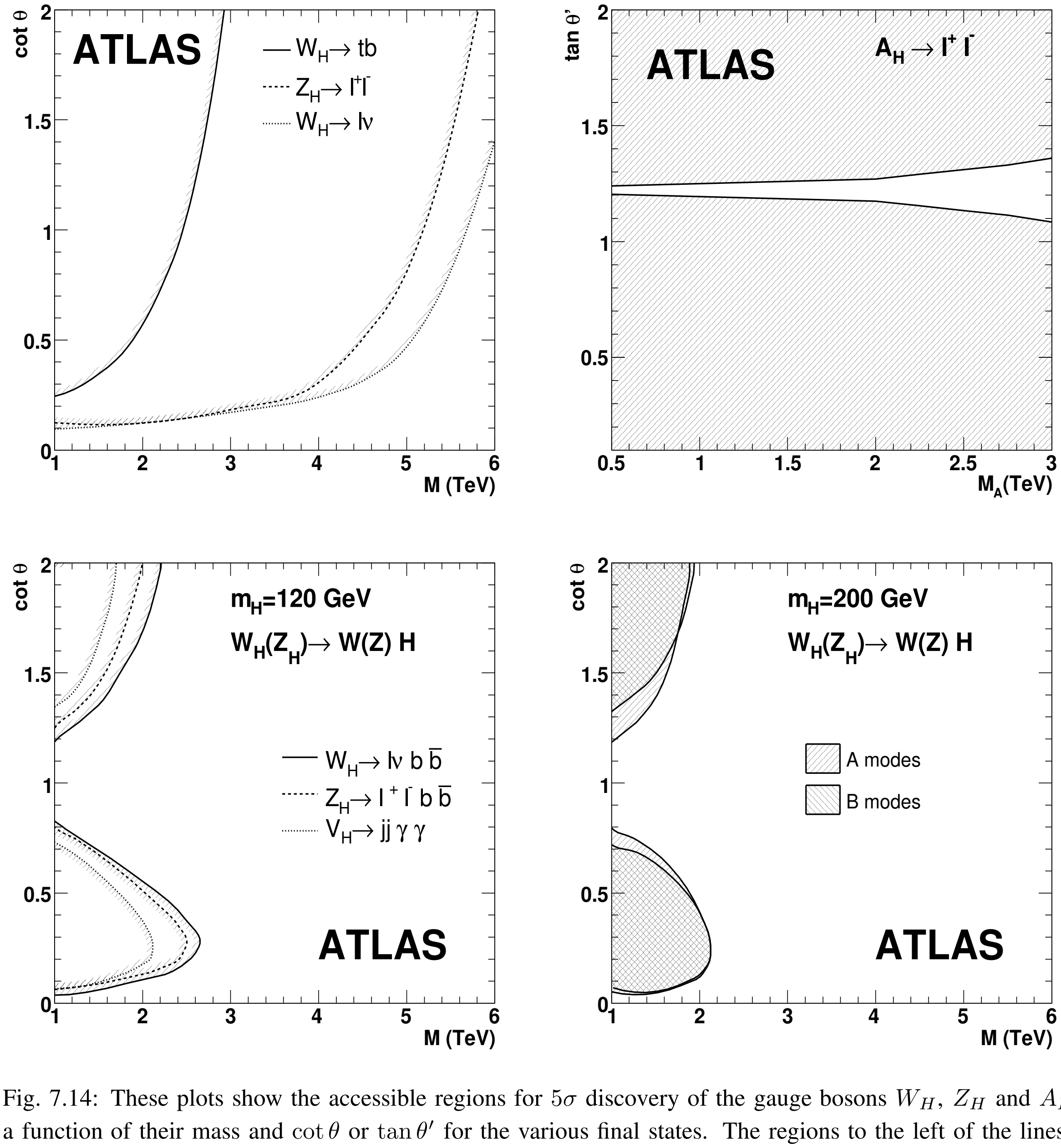

Fig. 7.14: These plots show the accessible regions for $5\sigma$ discovery of the gauge bosons $W_H$, $Z_H$ and $A_H$ as a function of their mass and $\cot\theta$ or $\tan\theta'$ for the various final states. The regions to the left of the lines are accessible with 300 fb$^{-1}$: top right for $A_H \to e^+e^-$, top left for $W_H$ or $Z_H$ leptonic and hadronic decays, bottom left for decays with a Higgs in the final state with $m_H = 120$ GeV, bottom right for decays with a Higgs in the final state with $m_H = 200$ GeV (see text for details).

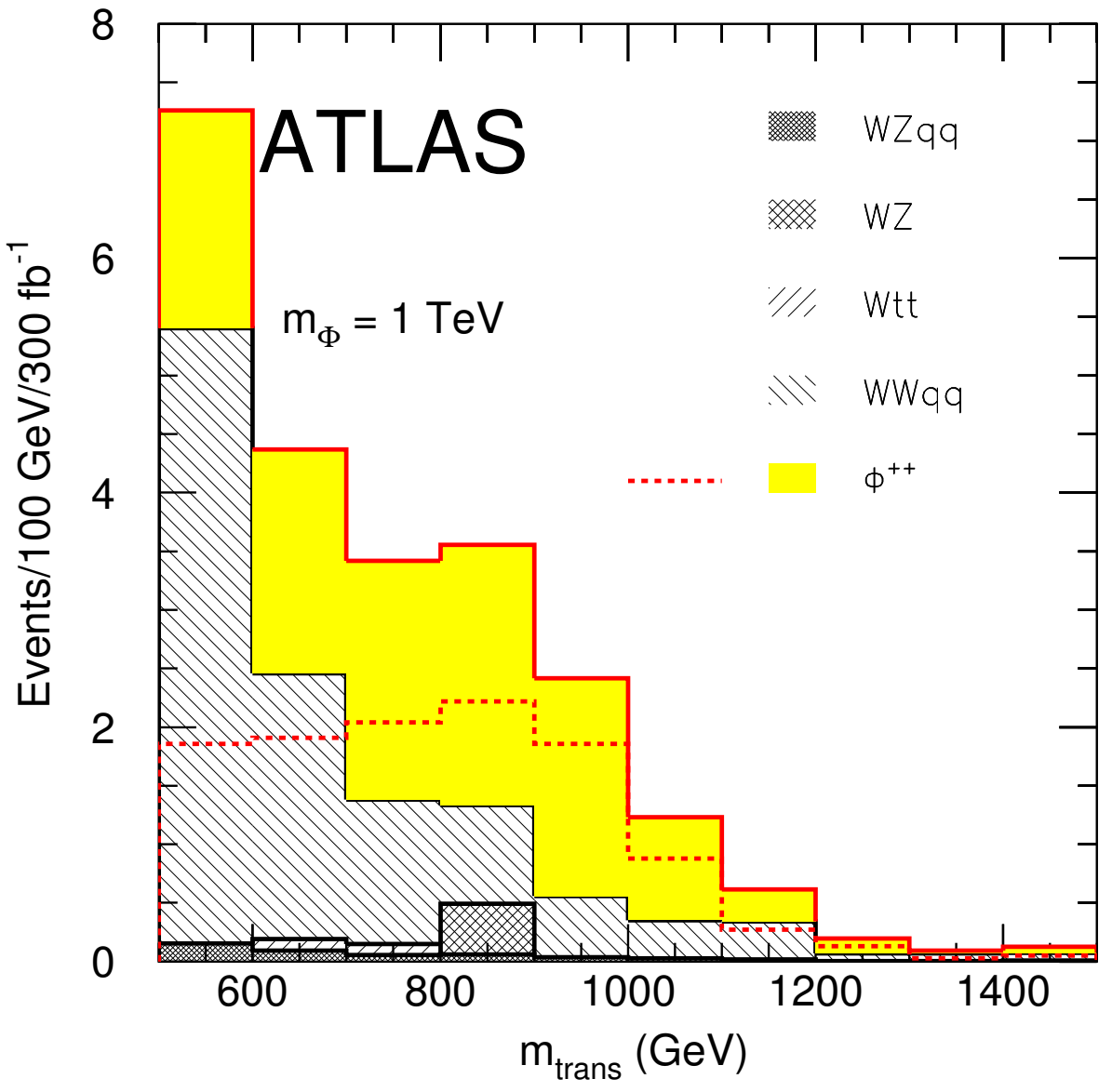

Fig. 7.15: The mass distribution $m_{trans}$, see text, for a $\phi^{++}$ of mass 1 TeV and $v' = 25$ GeV. The dashed histogram shows the signal alone and the solid shows the sum of signal and backgrounds. The components of the background are also shown separately.

In the case of the new gauge bosons, the situation is summarised in Fig. 7.14, that shows the accessible regions via leptonic final states of $Z_H$ and $W_H$ as a function of the mixing angle. However observation of such a gauge boson will not prove that it is of the type predicted in the little Higgs Models. In order to do this, the decays to the Standard Model bosons must be observed. Fig. 7.14 also shows the sensitive regions for decays of $Z_H$ and $W_H$ into various final states as a function of $\cot\theta$ and the masses. It can be seen that several decay modes are only observable for smaller masses over a restricted range of $\cot\theta$ where the characteristic decays $Z_H \to ZH$ and $W_H \to WH$ can be detected. The region of $\cot\theta \sim 1$ is covered searching for $W_H \to tb$ There is a small region at very small values of $\cot\theta$ where the leptonic decays are too small, and only the decays to $W$ or $Z$ can be seen.

In the case of $\phi^{++}$ the situation is not so promising. The Higgs sector is the least constrained by fine tuning arguments and this particle's mass can extend up to 10 TeV. We are only sensitive to masses up to 2 TeV or so provided that $v'$ is large enough. Other "little Higgs" models have a different Higgs structure that is similar to models with more than one Higgs doublet. Work is needed to evaluate the sensitivity of the LHC to these models.

## 7.7 Search for new heavy quark $T$ in CMS

*Aristotelis Kyriakis and Kajari Mazumdar*

Most extensions of the Standard Model contain an extended gauge sector and/or an extended Higgs sector but they are severely constrained by precision electroweak data. The little Higgs models [5,7,97] give an alternative solution to the fine-tuning problem present in the SM and consequently invoke a new set of particles. Since the mass upper limits depend on the relative importance of the contribution to the Higgs boson mass, we have a new singlet heavy quark, $T$ which is the least massive ($<$ 2 TeV) among all the new particles predicted and hence likely to be more easily produced at the LHC.

Table 7.5: Major background processes with their cross-sections folded with leptonic branching ratios, the expected number of events at integrated luminosity $\mathcal{L}$ = 30 fb$^{-1}$ and the number of events analyzed.

| Background | $\sigma\times$ BR (pb) | $N_{\rm expected}$ ($\mathcal{L}$=30 fb$^{-1}$) | $N_{\rm analyzed}$ |
|---|---|---|---|
| $t\bar{t}\rightarrow$ leptons | 85 | 2550K | 908K |
| inclusive $ZW\rightarrow$ leptons | 2.6 | 78K | 49K |
| inclusive $ZZ\rightarrow$ leptons | 0.16 | 4.8K | 93K |
| inclusive $WW\rightarrow$ leptons | 19.8 | 549K | 93K |
| $Zb\bar{b}$ | 116 | 3480 K | 220K |
| $Z(\rightarrow$ leptons)+jets | 161.7 | 4851K | 142K |

In this contribution, the potential of the CMS experiment at the LHC to discover $T$ is investigated for the production channel $q\ b \rightarrow q'\ T$ where the heavy quark is produced singly in the $t$-channel fusion process $Wb \rightarrow T$. This process is model dependent, being governed by the ratio of the Yukawa coupling constants involved in the model. The pair production of $T\bar{T}$ via gluon-gluon fusion is model independent and falls off more rapidly at higher values of $T$-mass [67]. The details of the CMS study can be found in [98].

The study is performed for the decay channel $T \rightarrow t\ Z$, which has a branching fraction of 25%. The cleanest signal is expected for the leptonic ($e,\ \mu$) decay modes of $Z$ and $W$ (from top decay), though the event rate is low. We have not considered their tau-decay modes. The complete process with the final state considered is $q\ b \rightarrow q'\ T,\ T \rightarrow Zt,\ Z \rightarrow \ell^+\ell^-,\ t \rightarrow bW,\ \ W \rightarrow \ell\nu.$ Hence there are three isolated, charged leptons, one $b$ jet, and genuine missing transverse energy in the central part and one forward going, light-quark jet in the event.

### *7.7.1 Event simulation and reconstruction*

The major background types with their cross-section folded with the leptonic branching ratios, the expected number of events for an integrated luminosity of 30 fb$^{-1}$ and the number of events analyzed for the present study are shown in Table 7.5.

We have used the `PYTHIA` package [93] for signal and background event generation. For signal we used the subprocess corresponding to 4th-generation heavy quark production and treating it as a resonance with mass 1 TeV. We have used CTEQ5L structure function for the event generation of the channel. For $t\bar{t}$ and double vector boson productions (*i.e.,* $WW$, $WZ$, $ZZ$) the accompanying jet is not very hard in `PYTHIA`. We plan to use dedicated event generators, based on matrix element calculation, in future where the accompanying jets in inclusive processes are much harder. The events for the process $Zb\bar{b}$ are produced with`ALPGEN` package [99]. We have also considered inclusive $Z$ production events, since, the production rate is very high (Z+jets, Drell-Yan $\sim$ 10 nb). The third lepton may be either from the jet or due to the initial state gluon radiation in DY events. In CMS detector the jet misidentification probability is very low ($10^{-4}$ for electron and $10^{-5}$ for muon). It is impossible to simulate the background channels for full statistics. To save on computing resources we have considered for $Z$ + jets background a specific kinematic region of ($\hat{p}_t$ =75-500 GeV) which overlaps with typical transverse momentum of $Z$ in the subprocess. We note here that the $Q^2$ scale for the signal channel is much higher than that in most of the simulated events for SM background processes. We are in the process of studying the SM events specially produced at higher $Q^2$ values.

Generated events are processed through `GEANT`-based CMS detector simulation package (`OSCAR` [1C and reconstructed subsequently using CMS-specific software (`ORCA` [101]). We have taken into account event pileup situation for low luminosity running phase of the LHC for an instantaneous luminosity of

Table 7.6: Efficiency of the selection criteria for the signal and various backgrounds analyzed.

| Selection | $T \to Zt(\%)$ | $t\bar{t} \to (\%)$ | $ZZ(\%)$ | $ZW(\%)$ | $WW(\%)$ | $Z$+jets (%) | $Zbb(\%)$ |
|---|---|---|---|---|---|---|---|
| Trigger | 95 | 43 | 59 | 16 | 25 | 43 | 92 |
| $Z$ | 63 | 0.240 | 4.160 | 1.130 | 0.14 | 11 | 7.4 |
| $W$ | 39 | 0.014 | 1.120 | 0.500 | 0. | 0.036 | 0.39 |
| $W + b$-jet | 13 | 0.005 | 0.020 | 0.002 | 0. | 0. | 0.09 |
| $SM$ top | 11 | 0.001 | 0.006 | 0.002 | 0. | 0. | 0.02 |
| $T$ | 9.7 | 0. | 0.001 | 0. | 0. | 0. | 0. |

$2 \times 10^{33}$ cm$^{-2}$ s$^{-1}$.

We have used standard reconstruction softwares of CMS. For jet reconstruction we used the iterative cone algorithm with cone radius of 0.5. A cut on jets with the minimum transverse energy 10 GeV is applied during jet reconstruction. The missing transverse energy in the event, $E_T^{\rm miss}$, is estimated from the balance of calorimeter tower energies used in jet reconstruction.

Lepton isolation is defined by choosing a cone of radius $\Delta R = \sqrt{\Delta\eta^2 + \Delta\phi^2} < 0.1$ around the candidate (electron or muon) track and searching for other tracks within the cone having $p_T > 0.9$ GeV. The sum $\Sigma\, p_T$ is required to be $< 4\%$ of the candidate track momentum for muon or transverse energy deposited in electromagnetic calorimeter for electron.

### *7.7.2 Event trigger and selection*

The reconstructed events are first checked if they pass standard CMS-trigger criteria. For 'double electron' and 'double muon' topology, the thesholds for lepton transverse momentum at higher level trigger are 17 GeV and 7 GeV respectively [102]. The combined trigger effeciency was evaluated to be 95%.

Our main selection conditions for off-line analysis are summarized below:

- The 'same flavour opposite sign dileptons:' $e^+e^-$ and $\mu^+\mu^-$ combinations should have a $p_T >$ 100 GeV (Fig. 7.16) and a mass of $\pm 10$ GeV around the nominal $Z$ mass (Fig. 7.17). This is referred to later as *Z criteria.*
- We further require a third lepton compatible with the leptonic decay of $W$. Hence the combination of lepton momentum and the missing transverse energy (nominally $E_t^{\rm miss} > 20$ GeV), should have a transverse momentum greater than 60 GeV and a transverse mass less than 120 GeV. This is referred to as *W criteria.*
- We allow only one jet with transverse momentum greater than 30 GeV within the tracker acceptance ($|\eta| < 2.5$) satisfying the conditions of a $b$-jet. The combination of $W$ and the $b$-jet should have a transverse momentum greater than 150 GeV , the condition referred to as *W + b-jet criteria.*
- The $(W, b)$ combination is required to have a mass in the range 110–220 GeV, referred to as *SM top criteria.*
- Finally we apply the *Heavy Top* characteristics: the combination $(Z,\ W,\ b)$ should have a mass in the range $850-1150$ GeV (Fig. 7.18).

In Table 7.6 we have summarized the efficiency of our selection cuts to signal and background events. The hard cuts applied during selection are quite effective in removing the backgrounds in almost all cases.

### *7.7.3 Preliminary Results*

The only SM background which survives all selections is $ZZ \to$ leptonic. The total efficiency for the signal selection is 9.7%. Taking into account the single heavy $T$ production cross-section (192 fb for equal Yukawa couplings $\lambda_1 = \lambda_2$) for heavy $T$ mass of 1 TeV and the various branching ratios we can calculate that a signal sample of only $N_S = 2.1$ events are expected with an integrated luminosity of 30 fb$^{-1}$. The significance could be calculated from $S_{stat} = 2(\sqrt{N_S + N_B} - \sqrt{N_B})$, that gives: $S_{stat} = 2.5$ with a signal-to-background ratio of 41.

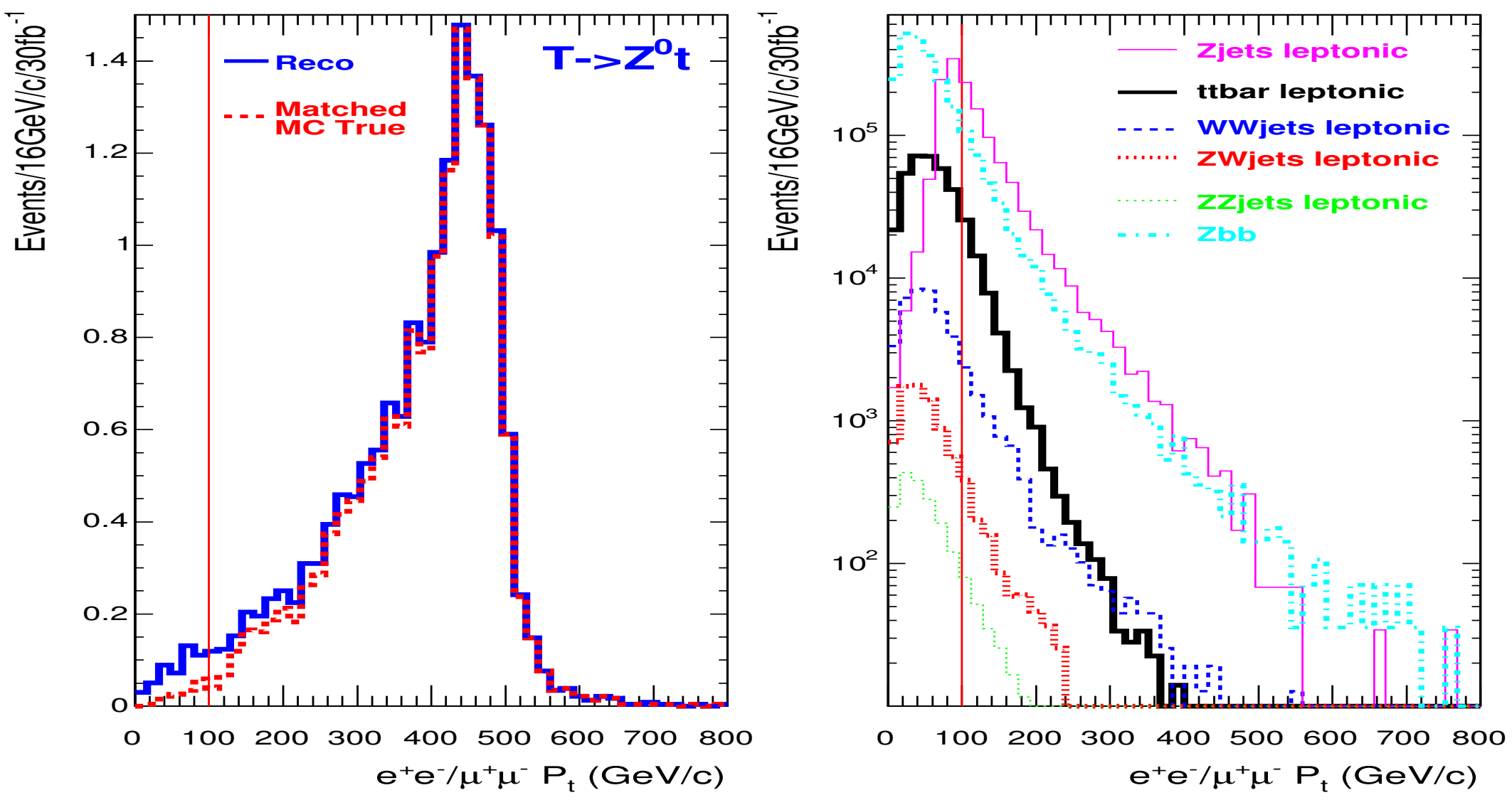

Fig. 7.16: The transverse momentum of the $e^+e^-$ and $\mu^+\mu^-$ combinations for signal (left) and background (right) events. The vertical lines show the threshold value.

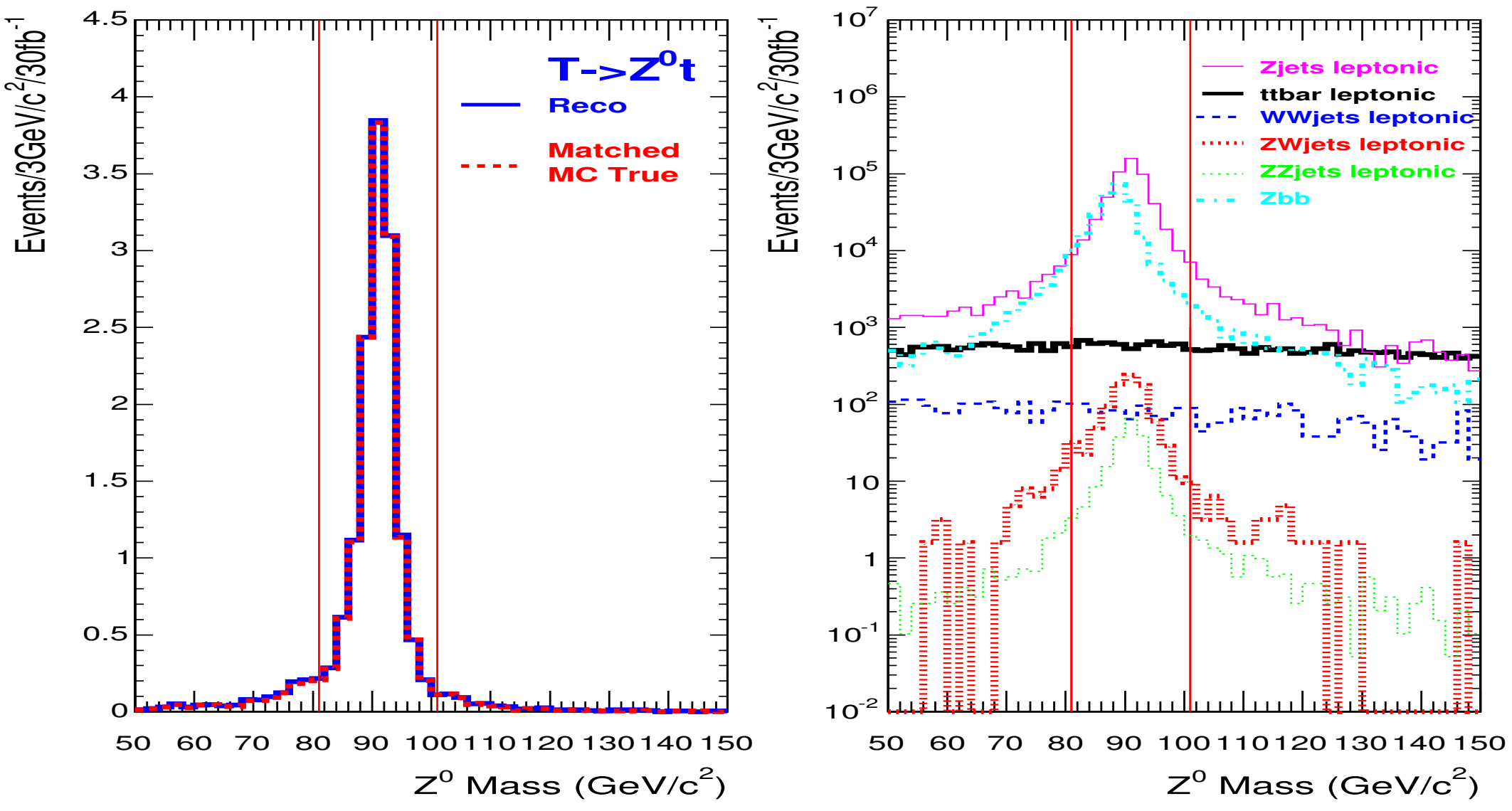

Fig. 7.17: Invariant mass of the $e^+e^-$ and $\mu^+\mu^-$ pairs for signal (left) and background (right) events (with and without $Z$ in the final state); the events within the vertical lines are accepeted combinations.

#### 7.7.3.1 Systematics

To study the systematic effect on the result we considered various experimental sources affecting the reconstruction of the observed leptons and jets, and estimated their impact on the selection efficiency.

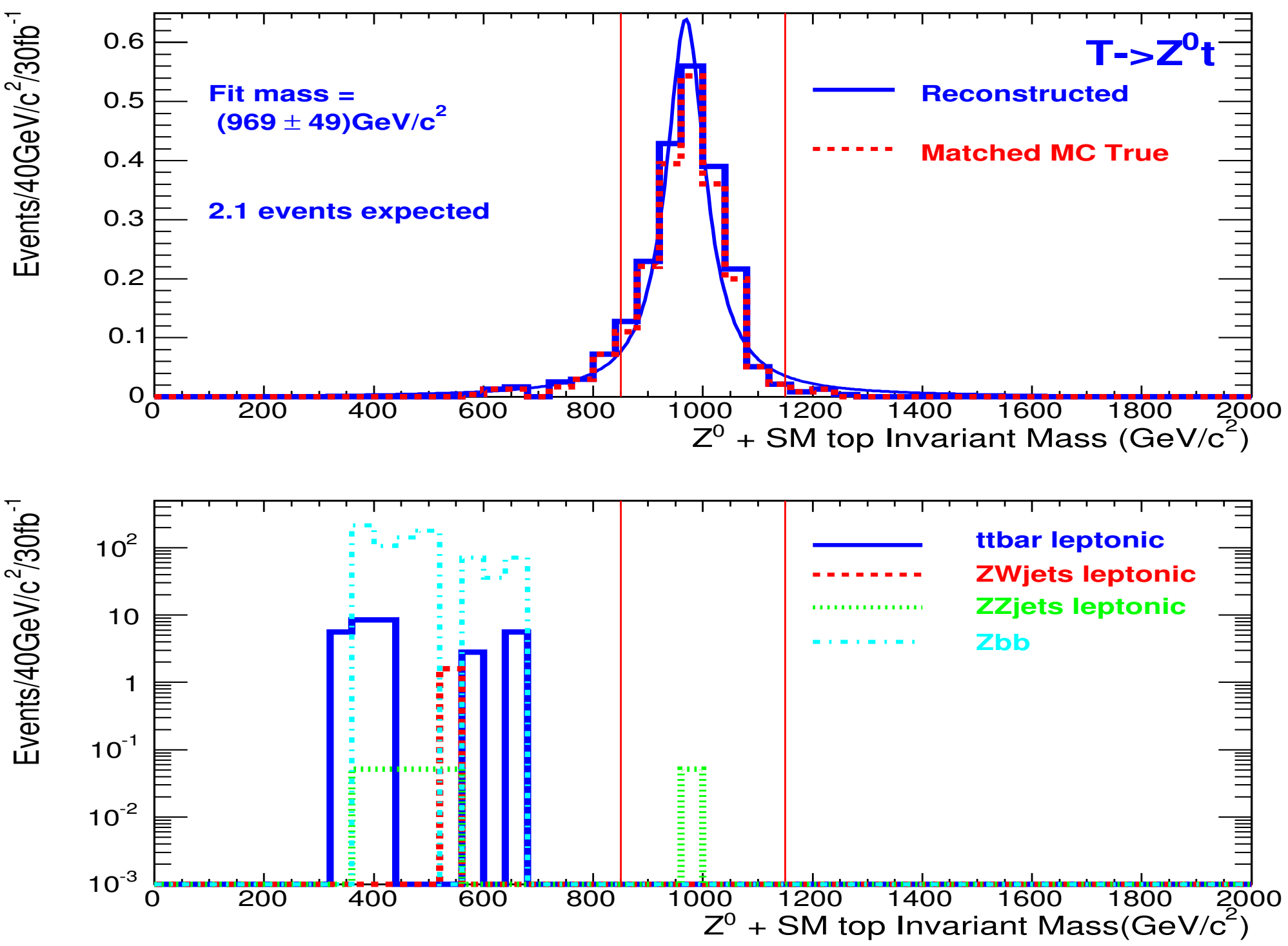

Fig. 7.18: Signal (left) and background (right) distributions for the invariant mass of the $Z$ and SM top combination for candidate heavy top with generated mass of 1 TeV.

Only the uncorrelated sources (or with negligible correlation) are included. We determined the surviving number of background events and fluctations are considered as the maximum shift for the central value without the systematic bias.

- Lepton energy scale: Due to imperfect knowledge of the detector material, the exact value of magnetic field at a given point of the detector or initial misalignments of detector units, estimates of 4-momenta of leptons have an uncertainty. This effect is accounted for by rescaling all reconstructed leptons' energy and momenta by a factor $\pm 0.005$. The error in efficiency is found to be 0.4%, whilst the background is not significantly affected.
- Jet and missing energy scale: The jet energy scale uncertainty (after 10 fb$^{-1}$ integrated luminosity) is expected to be about 5% for jets with $p_T$ =20 GeV and continuously decreasing to about 2.5% for jets with $p_T > 50$ GeV. With a $E_T^{\rm miss}$ estimated from jet energies, missing energy scale is totally correlated to the Jet Energy Scale and we considered a variation of 5%. The error in the efficiency is found to be 1% whilst the background is not significantly affected.
- $b$-tag uncertainty: The $b$-tagging of jet is important in this study and the experimental method is effective up to $|\eta| \leq 2.5$ with an efficiency of about 60% [103]. The $b$-tag uncertainty is assumed to be 4% after 10 fb$^{-1}$ integrated luminosity and it has large effect both on efficiency (5%) and on the background events ($\pm 0.15$ events).

So the significance after taking into account the systematcs is given as [104]:

$$S_{stat+syst} = S_{stat}\sqrt{\frac{N_B}{N_B + (\Delta N_B)^2}} = 2.0 \qquad (7.70)$$

Thus, the significance of the channel worsens after systematic effects are taken into account. The situation can improve significantly when the signal cross-section is higher as for the choice $\lambda_1/\lambda_2 = 2$.

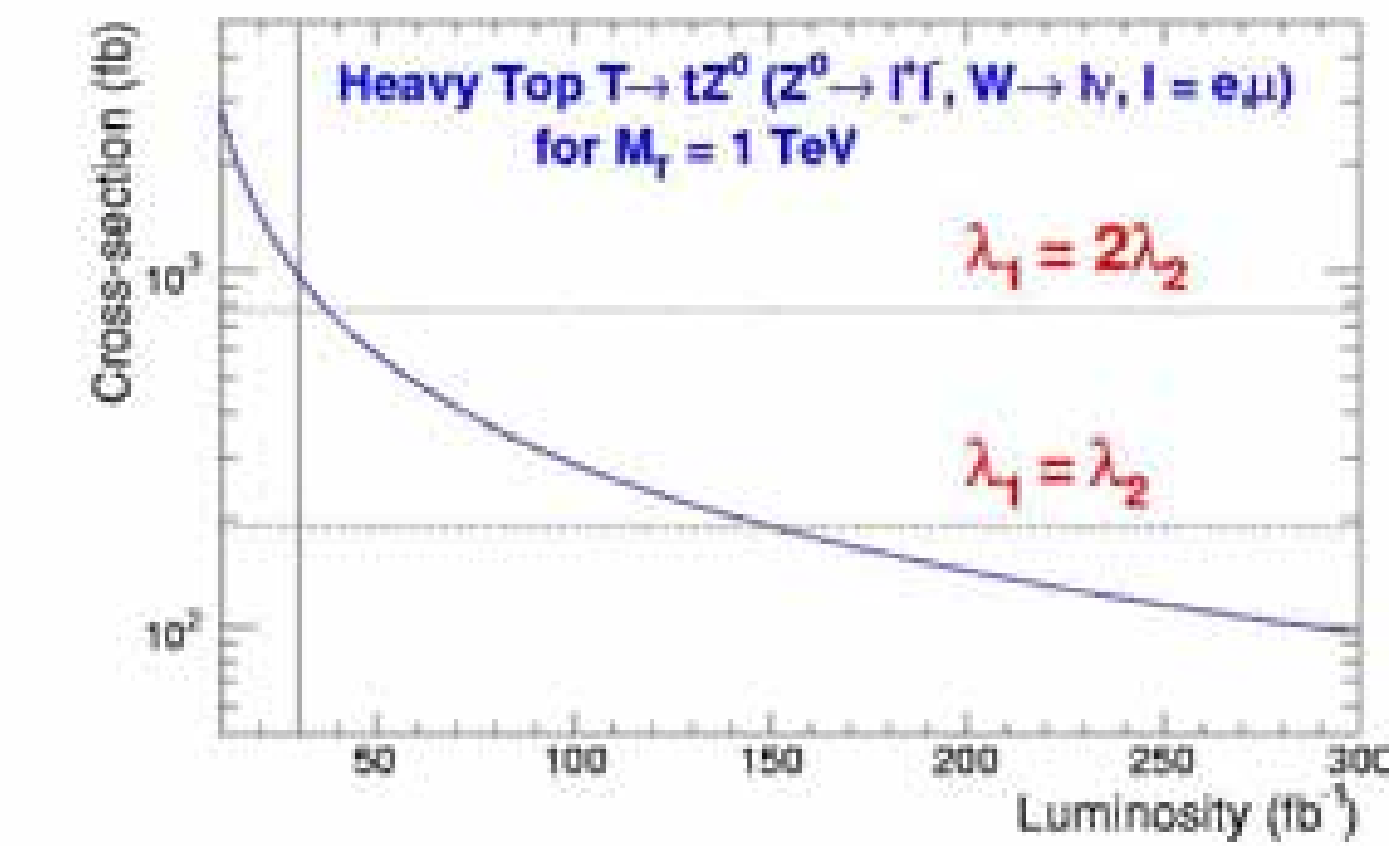

Fig. 7.19: The discovery plot. The curve represents the signal cross-section required as a function of integrated luminosity at LHC, for establishing single production of a heavy quark of mass = 1 TeV at $5\sigma$ level. The horizontal lines correspond to various choices of $\lambda_1/\lambda_2$. The vertical line corresponds to the luminosity used for this analysis *i.e.,* 30 fb $^{-1}$.

### 7.7.4 Conclusion

The experimental signature of single **T** production with subsequent decays in $T \to Zt,\ Z \to \ell^+\ell^-$ where $W$ from top-quark decays leptonically is investigated in the context of CMS experiment. The significance of the search is determined after taking into account various systematic effects. The study demonstrates that with an integrated luminosity of 30 fb$^{-1}$, the discovery potential of the channel $T \to tZ$, with leptonic decays of $Z$ and $W$, is rather limited. Fig. 7.19 shows signal cross-section required as a function of integrated luminosity, for establishing at $5\sigma$ level, single production of a heavy quark of mass = 1 TeV. The luminosity needed for $5\sigma$ evidence is estimated to be around 150 fb$^{-1}$ and 40 fb$^{-1}$ respectively for choice of parameters $\lambda_1 = \lambda_2$ and $\lambda_1 = 2\lambda_2$.

## 7.8 Determination of Littlest Higgs model parameters at the ILC

*J.A. Conley, J.L. Hewett, and M.P. Le*

The most economical little Higgs model is the so-called "Littlest Higgs" (LH) [6]. This scenario is based on a non-linear sigma model with an SU(5) global symmetry, which is broken to the subgroup SO(5) by a vev $f$. The natural scale for $f$ is around a TeV; if $f$ is much larger, the Higgs mass must again be finely tuned and this model no longer addresses the hierarchy problem. The SU(5) contains a gauged subgroup $[\mathrm{SU}(2) \times \mathrm{U}(1)]^2$ which is broken by the vev to the SM electroweak group $[\mathrm{SU}(2)_L \times \mathrm{U}(1)_Y]$. The global SU(5) breaking leaves 14 massless Goldstone bosons, four of which are eaten by the gauge bosons of the broken gauge groups, giving these gauge bosons a mass of order $f$. These new bosons correspond to two a heavy neutral bosons, $Z_H$ and $A_H$, and two heavy charged bosons $W_H^{\pm}$.

Here, we are mainly concerned with the extended neutral gauge sector, which contains 3 new parameters: $f$ and two mixing angles. Although we focus on the Littlest Higgs model, we note that an enlarged gauge sector with generic features is present in all little Higgs scenarios. After EWSB, the mass eigenstates are obtained via mixing

$$\begin{aligned} M^2_{A_L} &= 0, \quad M^2_{Z_L} = m^2_Z \left[1 - \frac{v^2}{f^2}\left(\frac{1}{6} + \frac{1}{4}(c^2 - s^2)^2 + \frac{5}{4}(c'^2 - s'^2)^2\right) + 8\frac{v'^2}{v^2}\right], \\ M^2_{A_H} &= m^2_Z s^2_w \left[\frac{f^2}{5s'^2c'^2v^2} - 1 + \frac{v^2}{2f^2}\left(\frac{5(c'^2 - s'^2)^2}{2s^2_w} - x_H \frac{g}{g'}\frac{c'^2s^2 + c^2s'^2}{cc'ss'}\right)\right], \\ M^2_{Z_H} &= m^2_W \left[\frac{f^2}{s^2c^2v^2} - 1 + \frac{v^2}{2f^2}\left(\frac{(c^2 - s^2)^2}{2c^2_w} + x_H \frac{g'}{g}\frac{c'^2s^2 + c^2s'^2}{cc'ss'}\right)\right], \end{aligned} \tag{7.71}$$

with $x_H$ being given in [67]. The mixing angles

$$s = \frac{g_2}{\sqrt{g_1^2 + g_2^2}} \quad \text{and} \quad s' = \frac{g'_2}{\sqrt{g_1'^2 + g_2'^2}} \tag{7.72}$$

relate the coupling strengths of the two copies of [SU(2) $\times$ U(1)]. The couplings of the neutral gauge bosons $Z_L$, $A_H$, and $Z_H$ to fermions and the light Higgs similarly depend on $s$, $s'$ and $f$:

$$\begin{aligned} g(A_L f\bar{f}) &= g_{SM}(Af\bar{f}), \quad g(Z_L f\bar{f}) = g_{SM}(Zf\bar{f})\left(1 + \frac{v^2}{f^2}a_i(s,s')\right), \\ g(A_H f\bar{f}) &= b_i \frac{g'}{2s'c'}\left(\frac{1}{5} - \frac{1}{2}c'^2\right), \\ g(Z_H f\bar{f}) &= \pm\frac{gc}{4s}, \quad g(Z_{L\mu}Z_{L\nu}H) = g_{SM}(Z_\mu Z_\nu H)\left(1 + \frac{v^2}{f^2}a(s,s')\right), \\ g(Z_{L\mu}Z_{H\nu}H) &= \frac{-i}{2}\frac{g^2}{c_W}v\frac{c^2 - s^2}{2sc}g_{\mu\nu}, \quad g(Z_{L\mu}A_{H\nu}H) = \frac{-i}{2}\frac{gg'}{c_W}v\frac{c'^2 - s'^2}{2s'c'}g_{\mu\nu}, \end{aligned} \tag{7.73}$$

where $g_{SM}$ represents the relevant coupling in the SM, and $a(b)_i$ are $\mathcal{O}(1)$ where $i$ labels the fermion species.

Equation (7.72) shows that for generic choices of $s$ and $s'$, $M_{A_H}/M_{Z_H} \simeq s_w m_Z/\sqrt{5}m_W \simeq 1/4$. This light $A_H$ is responsible for the most stringent experimental constraints on the model [31, 33]. As a result, phenomenologically viable variations of the Littlest Higgs models typically decouple the $A_H$ by modifying the gauge structure of the theory. To gain some understanding of models in which the $A_H$ decouples we take two approaches in our analysis: one is to choose a parameter value ($s' = \sqrt{3/5}$) for which the coupling of $A_H$ to fermions vanishes. Another is to artificially take $M_{A_H} \to \infty$ while letting all other quantities in the theory take on their usual, parameter-dependent values. While not theoretically consistent, this approach gives us a more general picture of the behavior of models in which the $A_H$ decouples.

We first examine the process $e^+e^- \to f\bar{f}$, where all of the LH neutral gauge bosons participate via s-channel exchange. We first study the constraints on the model from LEP II, taking as our observables the normalized, binned angular distribution and total cross section for $e^+e^- \to b\bar{b}$, $c\bar{c}$, and $\ell\bar{\ell}$, with $l = e$, $\mu$, or $\tau$. We use $\sqrt{s} = 200$ GeV and an integrated luminosity of 627 pb$^{-1}$. For the detection efficiencies, we take $\epsilon_e = 97\%$, $\epsilon_\mu = 88\%$, $\epsilon_\tau = 49\%$, $\epsilon_b = 40\%$, and $\epsilon_c = 10\%$ [105]. For the ILC, in addition to the above mentioned observables, we also include the angular binned left-right asymmetry $A_{LR}$ for each fermion pair. We use the energy $\sqrt{s} = 500$ GeV, an integrated luminosity of 500 fb$^{-1}$, and detection efficiencies of $\epsilon_e = 97\%$, $\epsilon_{\mu,\tau} = 95\%$, $\epsilon_b = 60\%$, and $\epsilon_c = 35\%$ [106].

The exclusion region at LEP II (taking $s' = s/2$) and the $5\sigma$ search reach at the ILC for various values of $s'$ are shown in Fig. 7.20. The $5\sigma$ discovery contour for the $Z_H$ at the LHC, as computed by an ATLAS based analysis [88], is included in the figure for comparison. We find that the search region at $\sqrt{s} = 1$ TeV reaches to somewhat higher values of the parameter $s$, but has essentially the same reach for $f$ as the 500 GeV results.

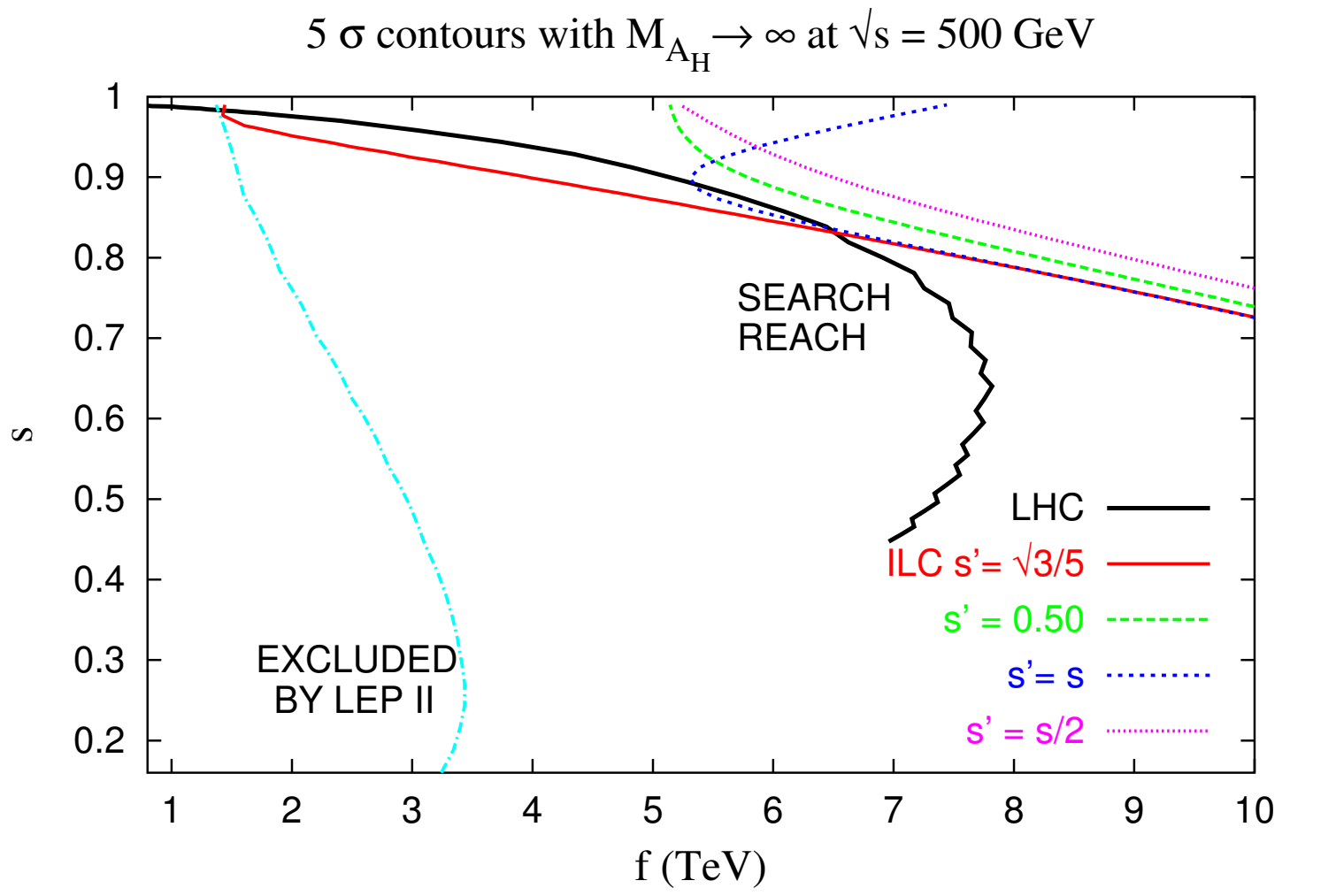

Fig. 7.20: LEP II exclusion region and ILC $5\sigma$ search reach in the $s - f$ parameter plane for various values of $s'$. The LHC result [88] is included for comparison.

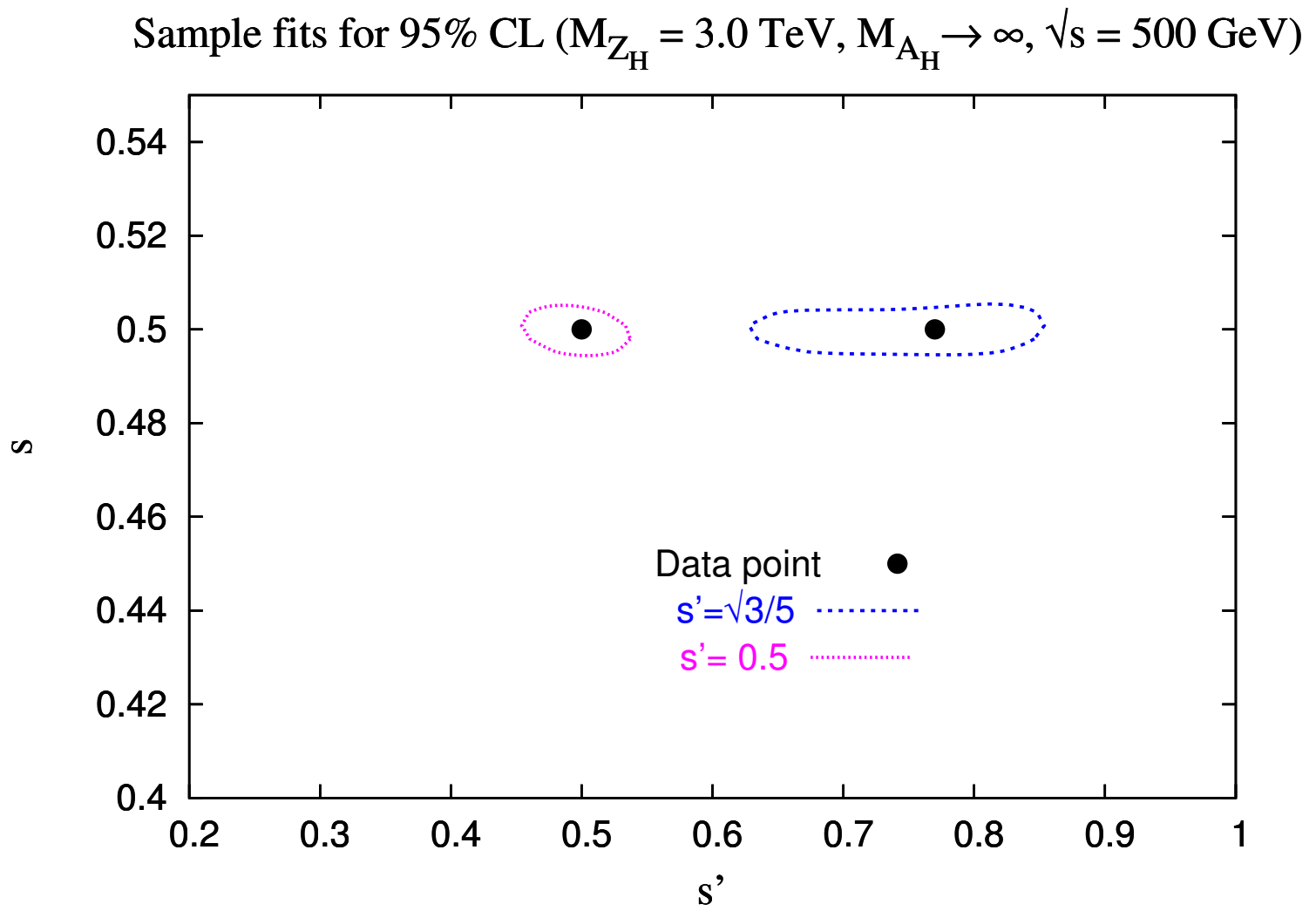

Fig. 7.21: 95% CL sample fits to the data points ($s = 0.5$, $s' = 0.5$) and ($s = 0.5$, $s' = \sqrt{3/5}$), at a 500 GeV ILC, taking $M_{Z_H} = 3.0$ TeV.

We have now determined the available parameter space accessible to the ILC and not already excluded by LEP II. It remains to ask, given the existence of an LH model with parameters in this accessible range, how accurately would the ILC be able to measure them? To answer this we perform some sample fits employing a $\chi$-square analysis. We use the same set of observables as before, and now take $M_{Z_H}$, $s$, and $s'$ as our free parameters. We choose a generic data point $(s, s', M_{Z_H})$ and use it to calculate the observables, which we then fluctuate according to statistical error. We assume that the Large Hadron Collider would have determined $M_{Z_H}$ relatively well, to the order of a few percent for $M_{Z_H} < 5 - 6$ TeV; we thus fix $M_{Z_H}$ and perform a 2-variable fit to $s$ and $s'$. Figure 7.21 shows the results of this fit for two sample data points. For both cases, the determination of $s$ is very accurate, due to the strong dependence of the $Z_H f\bar{f}$ couplings on this parameter.

In order to confirm that the LH model is the correct description of TeV-scale physics, it is important to measure the new particle couplings to the Higgs. Here we are concerned with the coupling of the $Z_H$ to the Higgs boson, which can be tested via the process $e^+e^- \to Z_L H$. In the LH model, deviations in this process from SM expectations arise from three sources: $Z_H$ and $A_H$ exchange in the s-channel and the deviation of the $Z_L Z_L H$ coupling from its SM value.

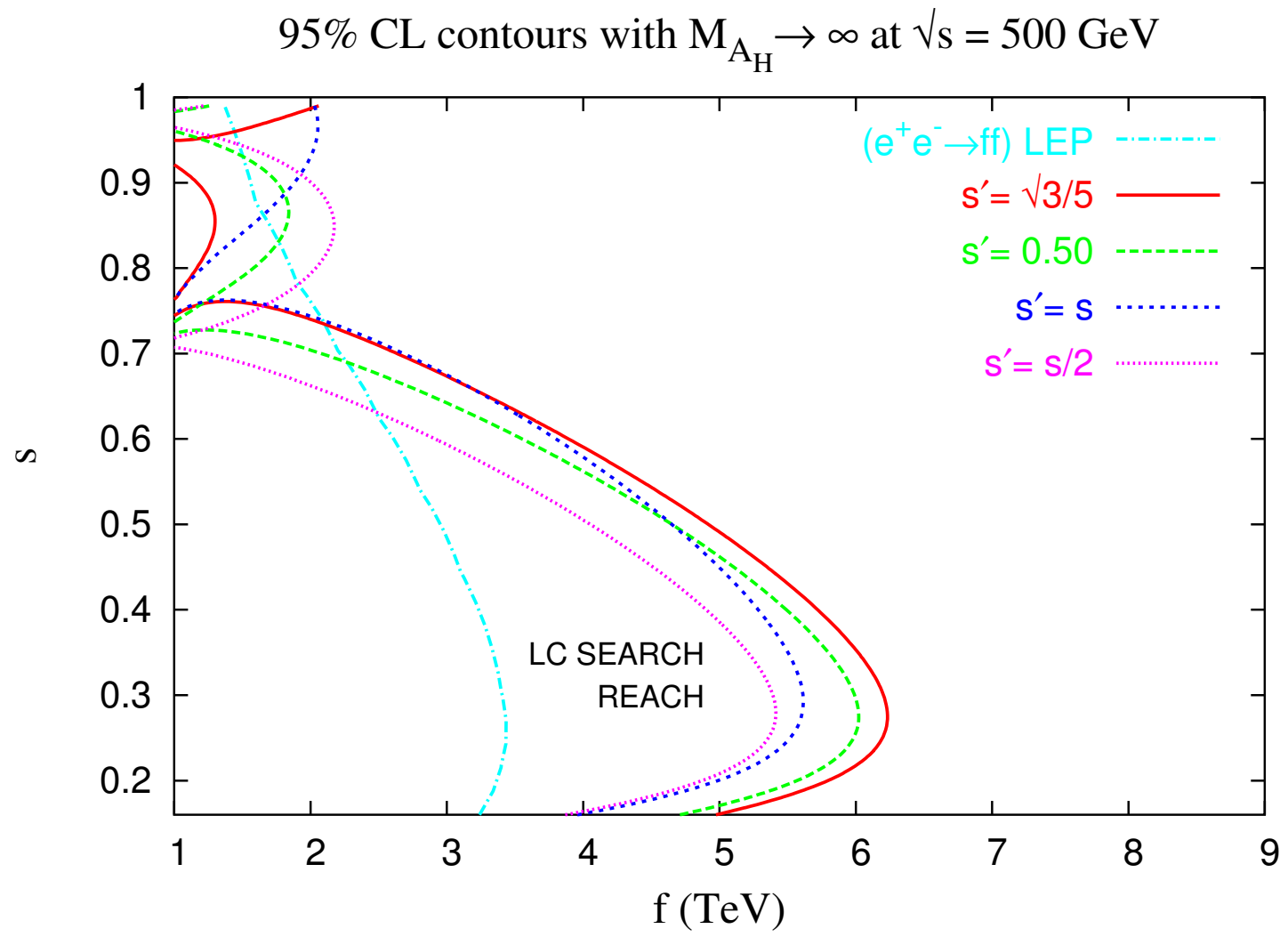

Fig. 7.22: The ILC 95% CL search reach in the $s - f$ parameter plane from the process $e^+e^- \to Z_L H$ for various values of $s'$ and $\sqrt{s} = 500$ GeV. The LEP II exclusion region from $e^+e^- \to f\bar{f}$ is shown for comparison.

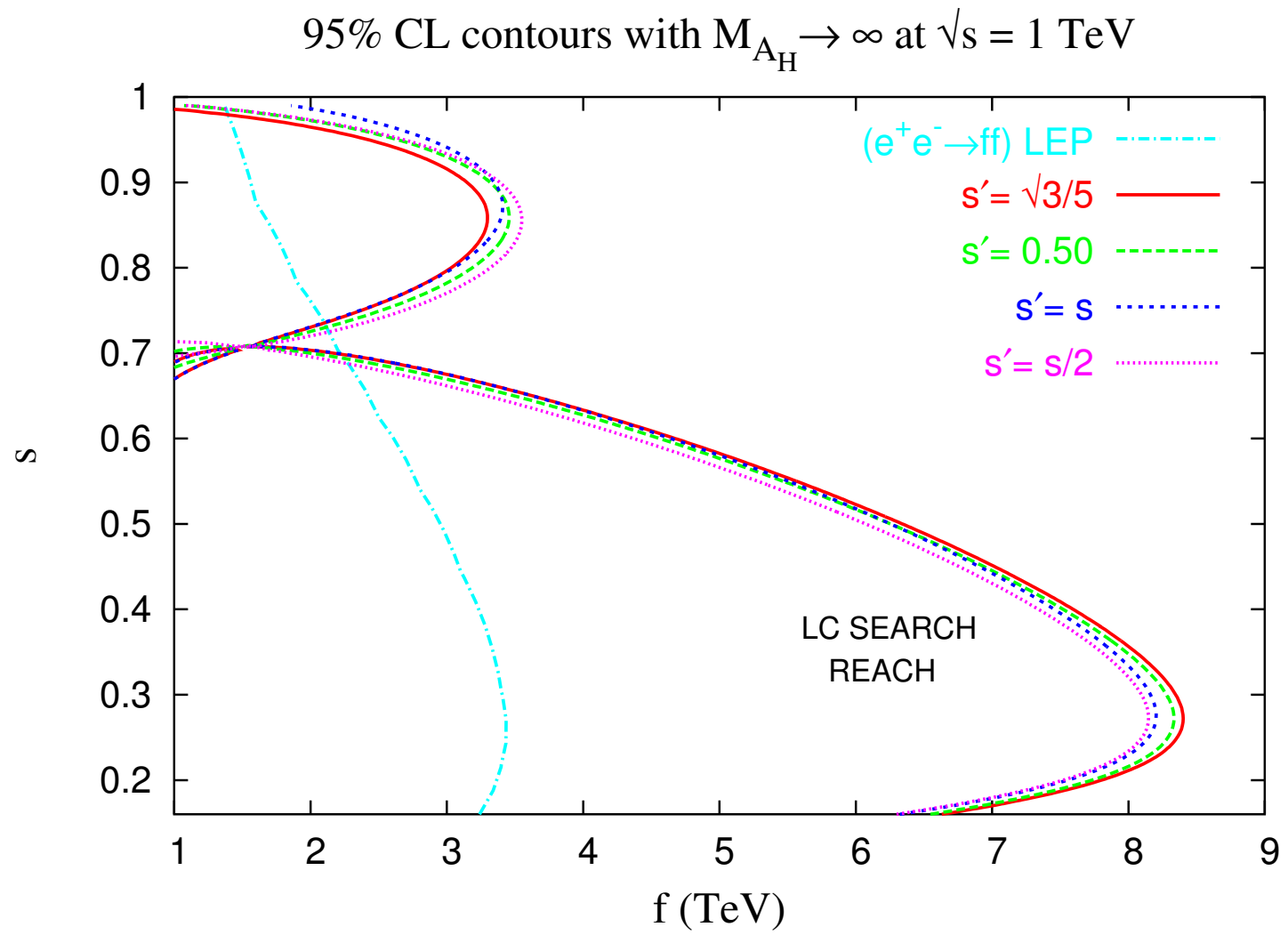

Fig. 7.23: Same as Fig. 7.22 but for $\sqrt{s} = 1$ TeV.

We then repeat our analysis using the process $e^+e^- \to Z_L H$ and taking the total cross section as our observable with $m_H = 120$ GeV. We assume that at a $\sqrt{s} = 500$ GeV ILC this cross section will be measured to an accuracy of 1.5% [106]. A $\chi$-squared analysis is carried out as before and our results for the ILC search reach in the LH parameter space are displayed in Figs. 7.22 and 7.23 for $\sqrt{s} = 0.5$ and 1 TeV, respectively.

In summary, we find that the reaction $e^+e^- \to f\bar{f}$ at a $\sqrt{s} = 500\ GeV$ ILC is sensitive to essentially the entire parameter region where the Littlest Higgs model is relevant to the gauge hierarchy problem. It also provides an accurate determination of the fundamental model parameters, to the precision of a few percent, provided that the LHC measures the mass of the heavy neutral gauge field. Additionally, we verified that the couplings of the extra gauge bosons to the light Higgs can be observed from the process $e^+e^- \to ZH$ for a significant region of the parameter space. Further details of our analysis can be found in [76].

# 8 LARGE EXTRA DIMENSIONS

## 8.1 Introduction

*Daniele Dominici and Samir Ferrag*

Recent theories with extra dimensions are an attempt to understand the large difference between the Planck mass $M_{Pl}$ and the Fermi scale $G_F^{-1/2}$ through a geometrical reformulation of the gauge hierarchy problem. In general these theories are formulated in a $D$-dimensional space time which has the geometry of a direct product $\mathcal{M}^4 \times X^\delta$ where $\mathcal{M}^4$ denotes the Minkowski space and $X^\delta$ the internal (compact) space [1, 2] but also non factorizable metrics and non compact spaces have been considered [3, 4]. In the large extra dimension models [1, 2] non gravitational interactions are confined to a 4-dimensional space time (the brane) while the gravitational interactions propagate in $D = 4 + \delta$ dimensions (the bulk). The standard Planck mass becomes an effective parameter and it is replaced by a fundamental parameter $\overline{M}_D$ which is of the order of a TeV. The big hierarchy between $\overline{M}_D$ and the Planck mass is explained by the large compactification volume $V_\delta$ appearing in the formula:

$$\overline{M}_{Pl}^{\,2} = V_\delta \overline{M}_D^{\,2+\delta} \tag{8.1}$$

where $\overline{M}_{Pl} = 1/\sqrt{8\pi G_N} = 2.4 \times 10^{18}$GeV is the reduced Planck mass. Assuming as internal space a $\delta$-torus with a common radius $R$, the compactification volume is given by $V_\delta = (2\pi R)^\delta$. The case $\delta = 1$ is excluded because it would give modifications to the Newton law at solar system distances; however for $\delta = 2$, assuming $\overline{M}_D = 1$ TeV, we can explain the large separation between the Planck and the electroweak scale with a radius $R \lesssim 10^{-2}$ mm, which is below the present limits on large extra dimensions from Newton law deviation experiments (200 $\mu$m) [5].

If the scale of the additional dimensions is small enough ($\leq$ TeV$^{-1}$) then also electroweak and strong interactions can propagate in the bulk. The existence of TeV sized compact extra dimensions has been also suggested for different reasons, related to the possibility of breaking supersymmetry [6] or a gauge symmetry by boundary conditions [7, 8] or explaining fermion mass spectrum delocalizing quarks and leptons in different regions of the extra dimensions [9]. This has motivated the construction of extensions of the Standard Model (SM), without gravity, in five or more dimensions, with matter and scalar fields living on branes or delocalized in the bulk. In the simplest version, known as Universal Extra Dimensions (UED) [10], all fields live in the bulk. These models are of interest because for the presence of a conserved $Z_2$ symmetry, which is a subgroup of the translation invariance in the fifth dimension, they are only weakly constrained by the electroweak precision measurements (for a recent analysis, see [11]). Furthermore for the same $Z_2$ symmetry the lightest Kaluza-Klein (KK) mode is stable and provides a candidate for dark matter [12, 13]. Models where matter and/or Higgses live on branes are more constrained by the electroweak precision measurements: 95% CL lower bound on the compactification scale are $M = 1/R \sim 5 - 6$ TeV [14–21].

All these higher dimensional theories are non renormalizable and therefore they should be interpreted as low energy effective theories valid up to some cut-off scale where some ultraviolet completion is necessary.

### *8.1.1 Review of the Arkani-Hamed-Dimopoulos-Dvali (ADD) model*

The action of the theory is given by the $D = 4+\delta$ dimensional Einstein term and a brane term containing the SM gauge interactions [22, 23]:

$$S = \frac{\overline{M}_D^{\,2+\delta}}{2} \int d^D x \sqrt{|g|} R + \int d^4 x \sqrt{-g_{\rm ind}} \mathcal{L}_{\rm SM} \tag{8.2}$$

where $R = g^{AB} R_{AB}$, $(A, B = 0, \cdots, 3+\delta)$ is the Ricci scalar curvature in $D$ dimensions and $(g_{\rm ind})_{\mu\nu}$ is the metric induced on the brane. If the internal space is a $\delta$-torus, $\int d^D x = \int d^4 x \int_0^{2\pi R} dy_1 \cdots \int_0^{2\pi R} dy_\delta$.

The action is computed by performing a linear expansion for weak gravity:

$$g_{AB} = \eta_{AB} + \frac{2}{\overline{M}_D^{\,1+\delta/2}} h_{AB} \tag{8.3}$$

up to second order in the fields $h_{AB}$ and expanding $h_{AB}$ in Fourier series:

$$h_{AB} = \sum_{n_1=-\infty}^{+\infty} \cdots \sum_{n_\delta=-\infty}^{+\infty} \frac{1}{\sqrt{V_\delta}} h_{AB}^{(n)}(x) e^{-i\sum_{j=1}^{\delta} n_j y_j} \tag{8.4}$$

where the KK modes $h_{AB}^{(n)}$ appear:

$$h_{AB}^{(n)} = \begin{pmatrix} h_{\mu\nu}^{(n)} & h_{\mu j}^{(n)} \\ h_{i\nu}^{(n)} & h_{ij}^{(n)} \end{pmatrix} \tag{8.5}$$

where $\mu, \nu = 0, 1, 2, 3$ and $i, j = 1, \cdots, \delta$.

The quadratic Lagrangian can be re-expressed in the unitary gauge as

$$\begin{aligned} \mathcal{L} \;=\; & \sum_{\vec{n}} \Big[ -\frac{1}{2} G^{(-\vec{n})\mu\nu}(\Box + m_n^2) G_{\mu\nu}^{(\vec{n})} + \frac{1}{2} G_\mu^{(-\vec{n})\mu}(\Box + m_n^2) G_\mu^{(\vec{n})\mu} \\ & - G^{(-\vec{n})\mu\nu}\partial_\mu\partial_\nu G_\lambda^{(\vec{n})\lambda} + G^{(-\vec{n})\mu\nu}\partial_\mu\partial_\lambda G_\nu^{(\vec{n})\lambda} \\ & -\frac{1}{4}|\partial_\mu V_{\nu j}^{(\vec{n})} - \partial_\nu V_{\mu j}^{(\vec{n})}|^2 + \frac{1}{2} m_n^2 V_{\mu j}^{(-\vec{n})} V^{(\vec{n})\mu j} - \frac{1}{2} S^{(-\vec{n})jk}(\Box + m_n^2) S_{jk}^{(\vec{n})} \\ & -\frac{1}{2} H^{(-\vec{n})}(\Box + m_n^2) H^{(\vec{n})} \Big] \end{aligned} \tag{8.6}$$

where $G_{\mu\nu}^{(\vec{n})}$, $V_{\mu j}^{(\vec{n})}$, $S_{jk}^{(\vec{n})}$ and $H^{(\vec{n})}$ are suitable linear combinations of the fields appearing in Eq. (8.5) [22, 23] and where the mass of the Kaluza-Klein excitations is given by

$$m_n = \frac{|\vec{n}|}{R} \tag{8.7}$$

Assuming again $\delta = 2$, and $\overline{M}_D = 1$ TeV, the range of the masses is of order $10^2$ mm$^{-1}$ $\sim 10^{-1}$ eV. Therefore in large extra dimension models KK gravitons tend to be very light and densely spaced ($\Delta m_n \sim 1/R$). Finally the brane action, when computed using the Fourier expansion, gives the following interaction Lagrangian

$$-\frac{1}{\overline{M}_{Pl}} \sum_{\vec{n}} \Big[ G^{(\vec{n})\mu\nu} - \frac{\kappa}{3} \eta^{\mu\nu} H^{(\vec{n})} \Big] T_{\mu\nu} \tag{8.8}$$

where $\kappa = \sqrt{\frac{3(\delta-1)}{\delta+2}}$ and $T_{\mu\nu}$ is the energy momentum tensor built from the SM Lagrangian, $\mathcal{L}_{SM}$. Notice that the vectors $V_{\mu j}^{(\vec{n})}$ and the scalars $S_{jk}^{(\vec{n})}$ do not couple to the ordinary matter. A mechanism for generating a small mass ($\geq 1\ \mathrm{mm}^{-1} \sim 10^{-3}$ eV) for the zero mode of the scalar fields $H^{(\vec{n})}$ is necessary in order to avoid deviations to the Newton law at the corresponding scale of 1 mm [24].

Feynman rules for the massive gravitons $G_{\mu\nu}^{(\vec{n})}$ and for the graviscalars $H^{(\vec{n})}$ can be derived from Eqs. (8.6)–(8.8) and are contained in [22, 23]. Since the couplings of these particles to ordinary matter are $O(1/\overline{M}_{Pl})$ their life-times tend to be very long. On the other side, even if their couplings are weak, the inclusive KK production cross section can be large at energies close to $\overline{M}_D$ because of the large multiplicity of the final state. In order to quantify this argument, let us consider the density of states: the number of modes $dN(|\vec{n}|)$ with the modulus $|\vec{n}| \equiv |n|$ being in the interval $(|n|, |n| + d|n|)$ is given by

$$dN(|\vec{n}|) = S_{\delta-1}|n|^{\delta-1}d|n| = S_{\delta-1}R^\delta m^{\delta-1}dm \tag{8.9}$$

Table 8.1: 95% CL bounds on $M_D$ (TeV)

| $\delta$ | 2 | 3 | 4 | 5 | 6 |
|---|---|---|---|---|---|
| 95% CL collider bounds on $M_D$ (TeV) | | | | | |
| LEP Exotica WG [25] | 1.60 | 1.20 | 0.94 | 0.77 | 0.66 |
| D0 mono-jets Run I data (K=1) [26] | 0.89 | 0.73 | 0.68 | 0.64 | 0.63 |
| CDF mono-jets Run I data (K=1) [27] | 1.00 | | 0.77 | | 0.71 |
| Astrophysical bounds on $M_D$ (TeV) | | | | | |
| SN1987A [28,29] | 22 | 2 | | | |
| Diffuse $\gamma$ rays from SN/NS [28,29] | 97 | 8 | 1.5 | | |
| Excess heat from $\gamma$ hitting the NS [28,29] | 1800 | 77 | 9.4 | 2.1 | |

with $m \equiv m_n = |n|/R$ and where $S_{\delta-1} = 2\pi^{\delta/2}/\Gamma(\delta/2)$ is the surface of the unit sphere in $\delta$ dimensions. We evaluate the sum over the KK excitations by converting it in the continuum notation

$$\begin{aligned} \sum_{\vec{n}} &\rightarrow \int dm^2 \rho_\delta(m) \\ &= \frac{1}{2} \frac{\overline{M}_{Pl}^2}{M_D^{2+\delta}} S_{\delta-1} \int dm^2 m^{\delta-2} \end{aligned} \tag{8.10}$$

It turns out that the density of states is

$$\rho_\delta(m) = \frac{L^\delta m^{\delta-2}}{(4\pi)^{\delta/2}\Gamma(\delta/2)} = \frac{\overline{M}_{Pl}^2}{2M_D^{2+\delta}} S_{\delta-1} m^{\delta-2} \tag{8.11}$$

where we have used $L = 2\pi R$ and we have defined a new effective $D$-dimensional Planck mass

$$M_D = (2\pi)^{\frac{\delta}{\delta+2}} \overline{M}_D \tag{8.12}$$

such that $1/(8\pi G_N) = R^\delta M_D^{\delta+2}$.

LEP has searched for graviton production in the process $e^+e^- \rightarrow \gamma(Z) + \text{missing energy}$. The corresponding 95% CL lower limits on $M_D$, using the best channel ($\gamma$ plus missing energy), are shown in Table 8.1 [25]. The Tevatron 95% CL limits on $M_D$ using the missing energy plus a single jet process are shown in Table 8.1 [26, 27] . LEP results are more sensitive for small extra dimensions. Some preliminary results of Run II have been presented for D0 [30].

As shown in Table 8.1, strong astrophysical bounds from supernova and neutron star processes involving emission of KK gravitons are present for $\delta \leq 3$ [28, 29]. However, as suggested [31], while astrophysics probes only the infrared part of the Kaluza-Klein spectrum of gravitons, high-energy experiments are mainly sensitive to the ultraviolet part. These limits can be evaded by assuming a small distortion of the $D$-dimensional space so that the mass of the lightest KK excitation is not given by the inverse radius $1/R$, but by a new intrinsic mass $\mu$. If $\mu \simeq 50$ MeV gravitons cannot be produced in astrophysical systems and therefore no bounds on the scale $M_D$ are derived. This idea has been pursuit also to study the case corresponding to $\delta = 1$ [31]. The model is built using the Randall Sundrum metric [3] (see also Section 9) but assuming the visible brane at $y = 0$ and the Planck brane at $y = \pi R$, in the limit in which $\mu$ is larger than $R^{-1}$ and both are much smaller than $\bar{M}_5$. In this model the hierarchy between the Fermi and the Planck scale is explained both by the warp factor and by the large extra dimension.

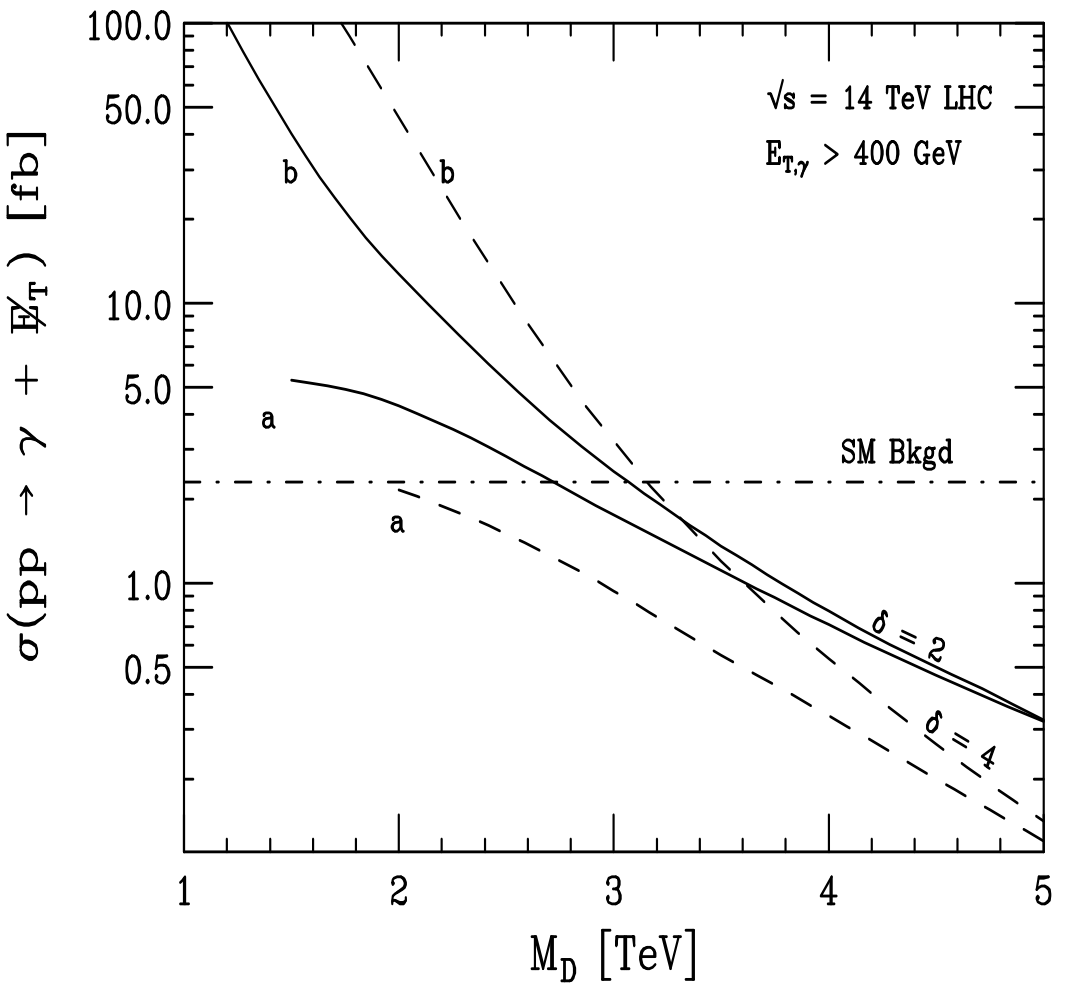

Fig. 8.1: $\gamma + \not{E}_T$ cross sections after integrating over a) $\hat{s} < M_D^2$ or b) all $\hat{s}$, where $\hat{s}$ is subprocess $s$. The SM background is the dot-dashed line. The signal is plotted as a solid (dashed) line for $\delta = 2(4)$. Taken from [32].

### *8.1.2 Direct graviton production*

The process which are relevant for LHC are the production of jets plus missing energy and photon plus missing energy. They are associated with the amplitudes $gg \to gG^{(\vec{n})}$, $qG \to qG^{(\vec{n})}$, $q\bar{q} \to gG^{(\vec{n})}$ and $q\bar{q} \to \gamma G^{(\vec{n})}$ respectively.

The relevant partonic differential cross section for single graviton emission $d^2\sigma/dtdm$ is proportional to the coupling $\sim 1/\overline{M}_{Pl}^2$ (as seen from Eq. (8.8)), however, summing over all the KK gravitons and taking into account the final state density, Eq. (8.11), the final form for the differential cross section is proportional to $1/M_D^{\delta+2}$. Figure 8.1 shows the total cross section for the final state photon plus missing energy for the signal for $\delta = 2, 4$ and for the SM background. This channel is less sensitive than the jet plus missing energy one, because of the smallness of the electromagnetic coupling and the lower luminosity of $q\bar{q}$ with respect to $qg$. Figure 8.2 shows the missing transverse energy distribution of the backgrounds and of the signals for several choices of $\delta$ and $M_D$ for the channel jet plus missing energy. The signals have been generated using the ISAJET implementation of the extra dimension model and the fast simulation of the ATLAS detector (ATLFAST) has been used. At the large values of missing transverse energy the dominant backgrounds arise from processes that can give rise to neutrinos in the final state, jet $+Z(\to \nu\nu)$, jet $+W(\to l\nu)$. The sensitivity for $\delta = 2, 3$ and 4 is respectively $M_D = 9.1, 7.0$ and 6.0 TeV for 100 fb$^{-1}$ [33]. For comparison we report here also the sensitivity of ILC with $\sqrt{s} = 800$ GeV, integrated luminosity $L = 1$ ab$^{-1}$, from the channel $e^+e^- \to \gamma G^{(\vec{n})}$: the 95% CL limits are $M_D$=5.9, 3.5 and 2.5 TeV for $\delta = 2$, 4 and 6 TeV respectively for unpolarized beams and 10.4, 5.1 and 3.3 TeV for $P_- = 0.8$, $P_+ = 0.6$ [34]. As shown in Fig. 8.3, the evolution of the $e^+e^- \to \gamma\not{E}_T$ cross section with the center of mass energy of the linear collider depends strongly on the number of extra dimensions [35]. Measurements of cross sections at different energies, as shown in [36], can determine the values of $M_D$ and $\delta$. Drell–Yan lepton pairs plus missing energy [37] have been also considered however the corresponding reach is lower than the single jet process.

The model corresponding to $\delta = 1$ has been recently studied in [31] by considering a warping of the 5-dimensional metric. The warping avoids standard conflicts with observations by introducing a mass gap in the KK graviton spectrum. LHC can be sensitive up to $M_D = 17$ TeV for 100 fb$^{-1}$.

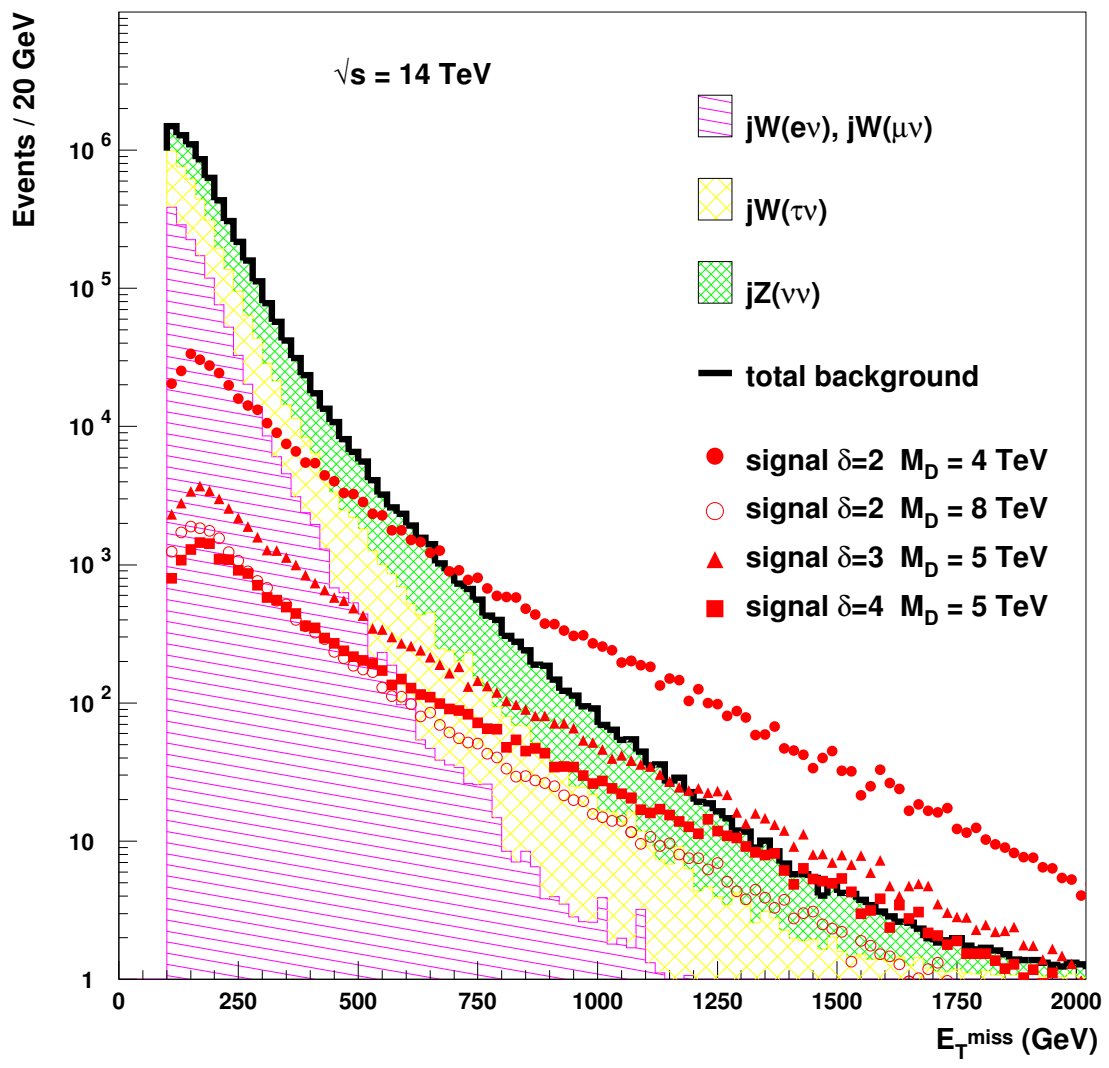

Fig. 8.2: Missing energy distribution (dots), shown here for various choices of the number of extra dimensions ($\delta$) and of the mass scale ($M_D$) and for SM backgrounds (histograms) for the channel jet $+\not{E}_T$. Taken from [33].

### 8.1.3 *Virtual graviton exchange*

The graviton, or any of its KK modes, can be exchanged in the $s$-channel. The multiplicity of KK states can give a large contribution to the production cross sections of any final state and the cross sections are divergent for $\delta \geq 2$. The expression of the cross section have been regularized in [23] by cutting off all the KK contributions above $M_S$ where $M_S$ presumably is of order $M_D$. The first ATLAS study [38] focused on the channels $pp \to l^+l^- + X$ and $pp \to \gamma\gamma + X$. Fig. 8.4 shows the signal shape for the two channels as a function of the two final state particle invariant mass. For a luminosity of 100 fb$^{-1}$ (one year of LHC at high luminosity) and combining the two channels, a $5\sigma$ sensitivity to an energy scale $M_S$ of 7 to 8 TeV is reached for a number of extra dimensions varying between 2 and 5.

In the simulations for the ILC the exchange of KK gravitons has been approximated by the following dimension eight operator:

$$\mathcal{L} = i\frac{4\lambda}{\Lambda^4}T^{\mu\nu}T_{\mu\nu} \tag{8.13}$$

where $|\lambda| = 1$ and $\Lambda$ is a cut-off related to $M_S$. Deviations with respect to the SM in fermion and $\gamma\gamma$ channels, left-right and center-edge asymmetries have been investigated [22, 39–43]. $5\sigma$ sensitivity for $\sqrt{s} = 500(1000)$ GeV and integrated luminosity of 500 fb$^{-1}$ turns out to be $\Lambda = 3.5$ (5.8) TeV [44]. The corresponding analysis for CLIC [45] gives a limit $M_S \sim 6\sqrt{s}$ for an integrated luminosity of 1 ab$^{-1}$.

### 8.1.4 *Invisible Higgs decay*

An entirely different class of signals is associated with the mixing between the Higgs boson and the very dense (continuum-like) graviscalar states. Instead of a single Higgs boson, one must consider the production of the full set of densely spaced mass eigenstates all of which are mixing with one another. The new signature that arises as a result of this mixing is that the Higgs boson will effectively acquire a possibly very large branching ratio to invisible final states composed primarily of graviscalars [32,46,47]. If the Higgs-graviscalar mixing parameter is of order one, then the Higgs decays to invisible final states will provide invaluable probes of the ADD model, often allowing detection of the extra dimensions in portions of parameter space for which the $jets/\gamma + \not{E}_T$ signal is not observable. If both types of signal are observable, a complete determination of the model parameters is generally possible [36]. The interaction between the Higgs complex doublet field $H$ and the Ricci scalar curvature $R$ of the induced 4-dimensional metric $g_{\rm ind}$ is derived from the following action

$$S = -\xi \int d^4x \sqrt{-g_{\rm ind}} R(g_{\rm ind}) \Phi^\dagger \Phi \,. \tag{8.14}$$

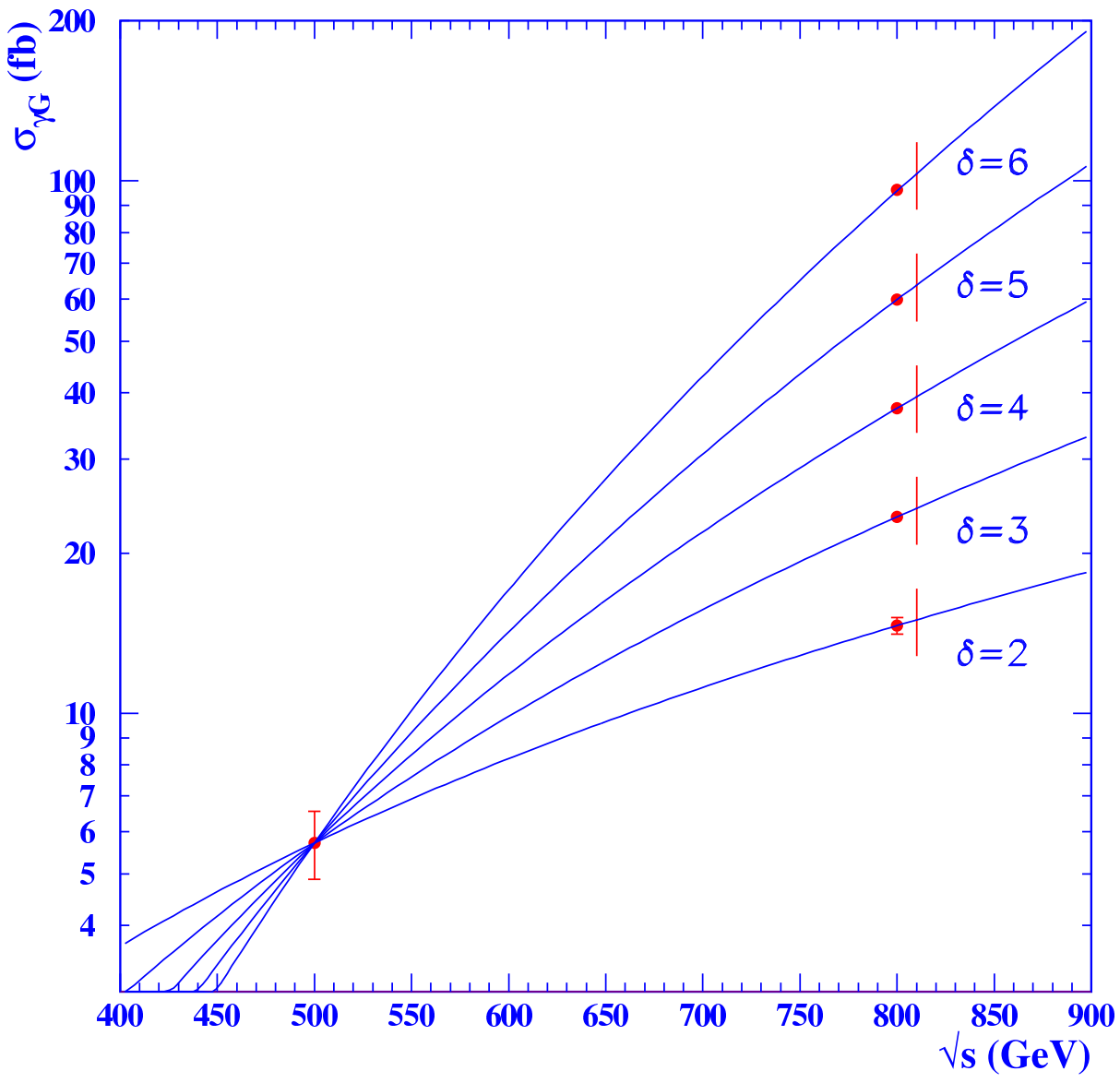

Fig. 8.3: $\gamma + \not{E}_T$ cross sections vs. $\sqrt{s}$, normalized to a common value at $\sqrt{s} = 500$ GeV. Thus, energy dependence gives $\delta$ via ratio of cross sections. Absolute normalization then gives $M_D$. Taken from [35].

After the usual shift $\Phi = (v + \frac{h}{\sqrt{2}}, 0)$, this interaction leads to the mixing term [32], (we have rewritten the graviscalars $H^{(\vec{n})} = 1/\sqrt{2}(s_{\vec{n}} + i a_{\vec{n}})$)

$$\mathcal{L}_{\rm mix} = \epsilon h \sum_{\vec{n}>0} s_{\vec{n}} \tag{8.15}$$

with

$$\epsilon = -\frac{2\sqrt{2}}{\overline{M}_{Pl}} \xi v m_h^2 \sqrt{\frac{3(\delta-1)}{\delta+2}} \,. \tag{8.16}$$

This mixing generates an oscillation of the Higgs itself into the closest KK graviscalar levels which decay invisibly. The invisible width $\Gamma_{h\to invisible} \equiv \Gamma_{inv}$ can be calculated by extracting the imaginary part of the mixing contribution to the Higgs self energy [32]. In an equivalent way the mixing requires diagonalization to the physical eigenstates $h'$ and $s'_{\vec{n}}$: the $s'_{\vec{n}}$ are nearly continuous and so those near in mass to the $h'$ act coherently together with the $h'$. Then when computing a process such as $WW \to h' + \sum_{\vec{m}>0} s'_{\vec{m}} \to F$, the full coherent sum over physical states must be performed. The result at the amplitude level is

$$\mathcal{A}(WW \to F)(p^2) \sim \frac{g_{WWh} g_{hF}}{p^2 - m_h^2 + i m_h \Gamma_h + i G(p^2) + F(p^2) + i\bar{\epsilon}} \tag{8.17}$$

where $\bar{\epsilon}$ provides the standard Feynman prescription and

$$F(p^2) \equiv -\epsilon^2 {\rm Re}\left[\sum_{\vec{m}>0} \frac{1}{p^2 - m_{\vec{m}}^2 + i\bar{\epsilon}}\right], \quad G(p^2) \equiv -\epsilon^2 {\rm Im}\left[\sum_{\vec{m}>0} \frac{1}{p^2 - m_{\vec{m}}^2 + i\bar{\epsilon}}\right] \tag{8.18}$$

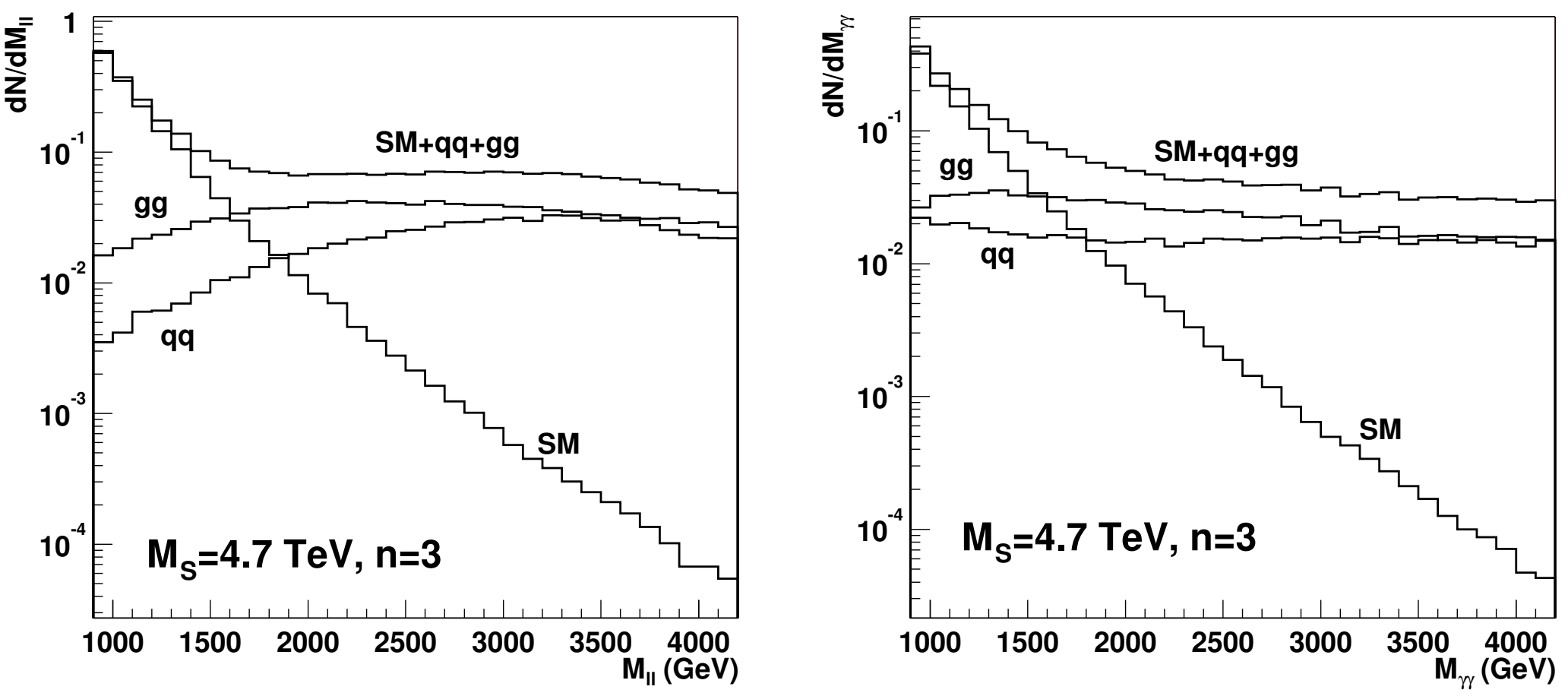

Fig. 8.4: Left: $pp \to l^+l^-$ cross section versus di-lepton invariant mass for SM and for 3 extra-dimensions with $M_S = 4.7$ TeV. Labeled contributions $q\bar{q}$ or $gg$ correspond to a graviton exchange with $q\bar{q}$ or $gg$ initial state. Right: same for $pp \to \gamma\gamma$. Taken from [38].

Writing $F(p^2) = F(m_{h_{eff}}^2) + (p^2 - m_{h_{eff}}^2)F'(m_{h_{eff}}^2) + \ldots$, where $m_{h_{eff}}^2 - m_h^2 + F(m_{h_{eff}}^2) = 0$, we obtain the structure

$$\mathcal{A}(WW \to F)(p^2) \sim \frac{g_{WWh}g_{hF}}{(p^2 - m_{h_{eff}}^2)[1 + F'(m_{h_{eff}}^2)] + im_h(\Gamma_h + \Gamma_{inv})} \tag{8.19}$$

with

$$m_h\Gamma_{inv} = G(p^2)|_{m_{h_{eff}}^2} = Im\Sigma(p^2)|_{m_{h_{eff}}^2} \tag{8.20}$$

A simple estimate of the mass renormalization is that $F(m_{h_{eff}}^2)$ should be of order $\xi^2 m_h^6/\Lambda^4$, where $\Lambda$ is an unknown ultraviolet cutoff energy presumably of order $\Lambda \sim M_D$ [22]. In this case, the contribution from $F(m_{h_{eff}}^2)$ is small for $m_h \ll M_D$. A simple estimate of the quantity $F'(m_{h_{eff}}^2)$, associated with wave function renormalization, suggests that it is of order $\xi^2 \frac{m_h^4}{\Lambda^4}$. In this case, $F'$ will provide a correction to coherently computed LHC production cross sections that is very probably quite small for $m_h \ll M_D$. Neglecting the terms $F, F'$, then $m_{h_{eff}} \sim m_h$. Taking the amplitude squared and integrating over $dp^2$ in the narrow width approximation we get

$$\sigma(WW \to h' + \sum_{\vec{n}>0} s_{\vec{n}} \to F) = \sigma_{SM}(WW \to h \to F) \times \left[\frac{\Gamma_{h\to F}^{SM}}{\Gamma_h^{SM} + \Gamma_{inv}}\right] \tag{8.21}$$

with

$$\begin{aligned} m_h\Gamma_{inv} = G({m_h}^2) \qquad & \to -\epsilon^2 Im\frac{1}{2}\int dm^2 \rho_\delta(m)\frac{1}{{m_h}^2 - m^2 + i\bar{\epsilon}} \\ & = \; -\epsilon^2\frac{1}{4}\frac{\overline{M}_P^2}{M_D^{2+\delta}}S_{\delta-1}(-\pi)({m_h}^2)^{(\delta-2)/2} \end{aligned} \tag{8.22}$$

So finally

$$\begin{aligned} \Gamma_{inv} = 2\pi\xi^2 v^2 \frac{3(\delta-1)}{\delta+2}\frac{m_h^{1+\delta}}{M_D^{2+\delta}}S_{\delta-1} \quad & \sim \quad (16\,MeV)20^{2-\delta}\xi^2 S_{\delta-1}\frac{3(\delta-1)}{\delta+2} \\ & \times \left(\frac{m_h}{150\,\mathrm{GeV}}\right)^{1+\delta}\left(\frac{3\,\mathrm{TeV}}{M_D}\right)^{2+\delta} \end{aligned} \tag{8.23}$$

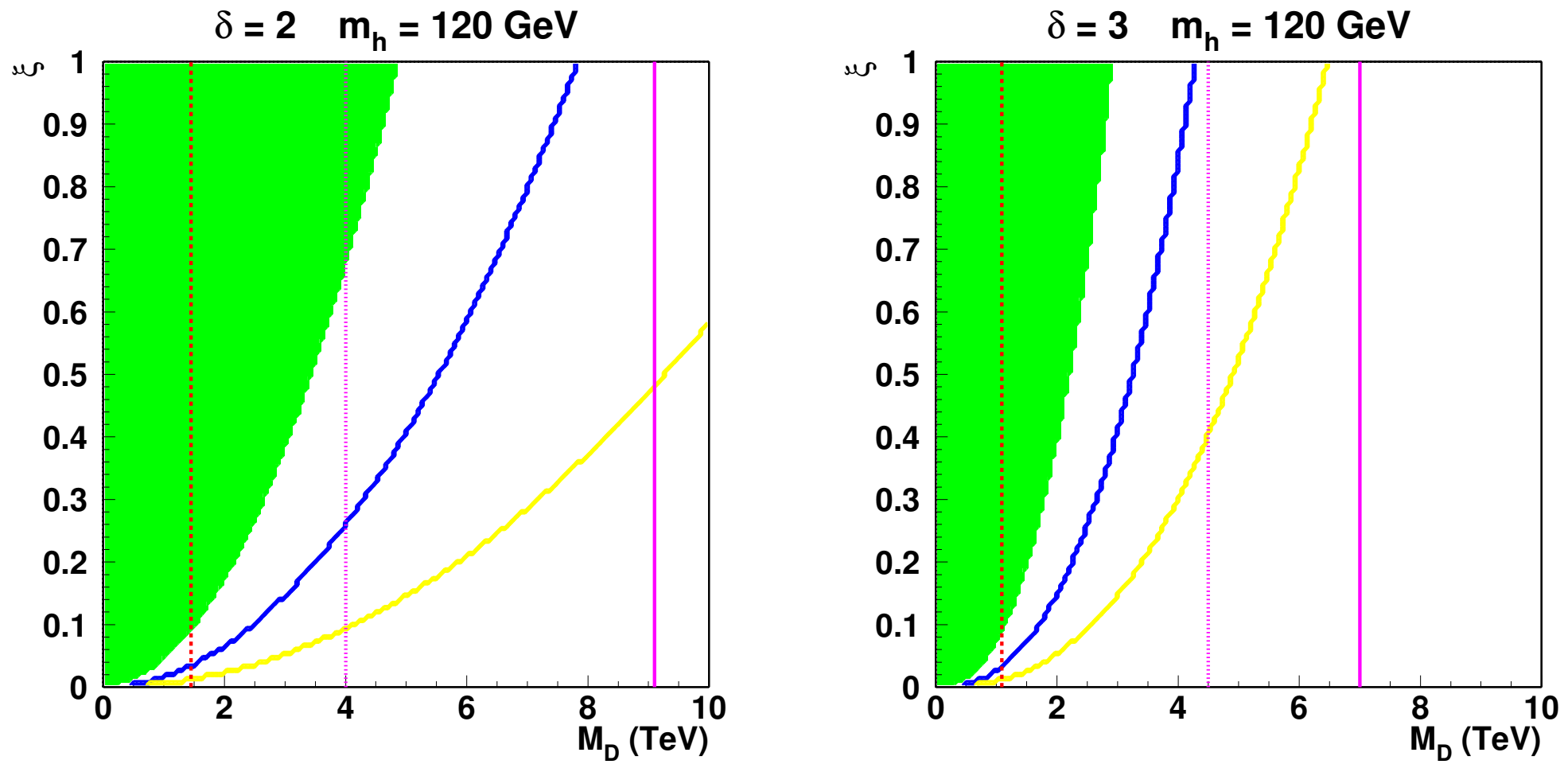

Fig. 8.5: Invisible decay width effects in the $\xi$ - $M_D$ plane for $m_h$ = 120 GeV. The plots are for $\delta = 2$ (left), 3 (right). The green (grey) regions indicate where the Higgs signal at the LHC drops below the $5\sigma$ threshold for 100 fb$^{-1}$ of data. The regions above the blue (bold) line are where the LHC invisible Higgs signal in the $WW$-fusion channel exceeds $5\sigma$ significance. The solid vertical line at the largest $M_D$ value in each figure shows the upper limit on $M_D$ at the $5\sigma$ level by the analysis of jets/$\gamma$ with missing energy at the LHC. The middle dotted vertical line shows the value of $M_D$ below which the theoretical computation at the LHC is ambiguous. The dashed vertical line at the lowest $M_D$ value is the 95% CL lower limit coming from combined Tevatron and LEP/LEP2 limits. The regions above the yellow (light grey) line are the parts of the parameter space where the ILC invisible Higgs signal will exceed $5\sigma$ assuming $\sqrt{s} = 350$ GeV and $L = 500$ fb$^{-1}$. Taken from [36].

In the ADD model, the statistical significance for detecting a SM Higgs in the standard visible channels is suppressed by the appearance of this invisible decay width. There are regions at high $\xi$ where the significance of the Higgs boson signal in the canonical channels drops below the $5\sigma$ threshold. Fortunately, the LHC experiments will also be sensitive to an invisibly decaying Higgs boson produced via $WW$-fusion, with tagged forward jets. In Ref. [48] the results of a detailed CMS study for this mode are given in Fig. 25, (see also Fig. 8.10). With only 10 fb$^{-1}$, a Higgs boson produced with the SM $WW \rightarrow$Higgs rate and decaying to an invisible final state with $BR(h \rightarrow invisible) = 0.12 - 0.28$ exceeds the 95% CL for 120 GeV $< m_h <$ 400 GeV. Fig. 8.5 summarizes the results for different values of $\delta$ when $m_h = 120$ GeV. In the green (light grey) region, the Higgs signal in standard channels drops below the $5\sigma$ threshold with 100 fb$^{-1}$ of LHC data. But in the area above the bold blue line the LHC search for invisible decays in the fusion channel yields a signal with an estimated significance exceeding $5\sigma$. It is important to observe that, whenever the Higgs boson sensitivity is lost due to the suppression of the canonical decay modes, the invisible rate is large enough to still ensure detection through a dedicated analysis. For increasing $m_h$ the invisible decay mode is important in a more limited range because the SM Higgs decay width is much larger in this latter case, $m_h$ being above the $WW, ZZ$ pair decay thresholds.

The $5\sigma$ upper reach in $M_D$ at each $\delta$ from the analysis of jets/$\gamma+$ missing energy [33] is shown in the figures by the solid medium gray (purple) line. The reliability of the theory prediction at the LHC fails for $M_D$ below the medium gray (purple) dotted line. Also shown in each figure is the 95% CL lower limit on $M_D$ coming from the combination of LEP, LEP2 and Tevatron data, as summarized in [49]. A TeV-class $e^+e^-$ linear collider will be able to see the Higgs signal regardless of the magnitude of the invisible branching ratio simply by looking for a peak in the $M_X$ mass spectrum in $e^+e^- \rightarrow ZX$ events. We should note that the $e^+e^- \rightarrow Z\not{E}_T$ events from direct graviton emission do not result in substantial background [50] to the Higgs signal when $\sqrt{s} \lesssim 500$ GeV. The region above the light grey (yellow) curves in Fig. 8.5 corresponds to the portion of $(M_D, \xi)$ parameter space for which the invisible Higgs signal will be observable at the ILC at the $5\sigma$ or better level. Not surprisingly, the ILC will be able to detect this signal over an even larger part of the parameter space than can the LHC.

The parameters of the model can be determined by combining several measurements that can be performed at LHC and at ILC [36]. In general the ability of the LHC to determine the model parameters is limited; however by including the ILC data, associated to the Higgs signals in both visible and invisible final states and also to the $\gamma + \not{E}_T$ signal, a good determination of $\delta$ and $M_D$ is possible so long as $M_D$ is not too big [36].

Invisible Higgs decays are discussed further in two contributions in this report: Section 8.2 discusses the weak boson fusion, $Z + h_{inv}$, $t\bar{t} + h_{inv}$ channels at the LHC and describes a new method to extract the Higgs mass from production cross sections. In Section 8.3, the CMS strategy for discovering an invisible Higgs at LHC is presented.

### *8.1.5 Universal extra dimension and TeV $^{-1}$ models*

More general constructions have been proposed, where gravity propagates in the entire 4+$\delta$ dimensional space-time, while the SM lives in a subspace with $p \geq 3$ space dimensions. The scale associated to these extra $p - 3$ compact dimensions is assumed of the TeV$^{-1}$ size.

The Universal Extra Dimensions model [10] is an extension of the ADD model [1,51] in which all the SM fields, fermions as well as bosons, propagate in the bulk, so that each SM particle has an infinite tower of KK partners. The spin of the KK particles is the same as their SM partners, as well as the strength of the couplings (up to a normalization factor such as $\sqrt{2}$). The minimal UED (mUED) [52,53] scenario is based on the following hypotheses: the fields of the theory propagate in a single extra dimension; the extra dimension is compactified on the orbifold $S^1/Z^2$ of size $R$ (compactification radius). The choice of the topology is very important since different topologies give different realizations of the low energy theory even when one starts with the same five-dimensional Lagrangian. Compactification on the orbifold has two advantages: only four of the five components of the vector fields and chiral fermions are present in the low energy spectrum. Therefore the SM could be the low energy regime of a UED theory. The tree level Lagrangian has a local five-dimensional Lorentz symmetry responsible for the conservation of the momentum in the extra dimension. The quantum number associated to the symmetry is the KK number. The symmetry, however, is broken due to the presence of additional interactions at the boundaries of the orbifold. But the KK-parity is still conserved with important phenomenological consequences: the lightest massive KK particle (LKP), the KK photon, is stable and can be a candidate for dark matter; the level 1 KK states must be pair produced. 95% CL lower bound on the compactification scale is $M = 1/R \sim 800$ GeV (for a recent analysis, see [11]). Dark matter constraints imply that $\mathrm{R}^{-1} \in [600, 1050]$ GeV [13]. The UED Lagrangian in $4 + \delta$ dimensions contains two parts, the bulk Lagrangian (SM like) and the boundary interaction terms. If only the bulk Lagrangian is taken into account, the mass of the $n$-th KK mode is

$$m_n = \sqrt{n^2/R^2 + m_0^2} \tag{8.24}$$

where $m_0$ is the zero mode mass which could be identified as the SM particles. Therefore the model has a highly degenerate spectrum at each KK level except for large $m_0$ like $t$, $W$, $Z$, etc. The boundary terms coefficients constitute new free parameters of the model renormalized by the bulk interactions, thus scale dependent. The mUED model assumes they are negligible at the scale $\Lambda > R^{-1}$. In conclusion the mUED has only three parameters $R, \Lambda, m_h$. The new terms of this Lagrangian, besides being responsible for breaking the KK number conservation down to the KK parity, split the near degeneracy of each KK level. The corrections to the masses are such that $m_{g_n} > m_{Q_n} > m_{q_n} > m_{W_n} \sim m_{Z_n} > m_{L_n} > m_{l_n} > m_{\gamma_n}$. Therefore the heaviest first level KK modes are pair produced and then cascade decay until the LKP. The experimental signature for KK modes production at hadron colliders will be the missing energy carried away by the LKPs in addition to soft SM leptons and/or jets radiated in the cascade decay process.

TeV$^{-1}$ models are extra dimension models where electroweak interactions propagate in the bulk, while matter and/or Higgses live on the branes [15, 54, 55]. These models are more constrained than the UED ones, since, due to fact that there is no KK number conservation, KK excitations of gauge bosons can mix with the standard electroweak gauge bosons. Using the electroweak precision measurements, 95% CL lower bound on the compactification scale is $R^{-1} \sim 5 - 6$ TeV [14–21, 56]. A second peculiar consequence is that these KK excitations of $W$, $Z$ and $\gamma$ can be singly produced at LHC [57, 58]. The discovery potential of LHC for the heavy neutral resonances using the $e^+e^-$ decay channel will be discussed in detail in 8.4.

## 8.2 Invisibly decaying Higgs at the LHC

*Heather E. Logan*

In models of large extra dimensions, the Higgs boson often acquires an invisible decay width. This invisible width, $\Gamma_{inv}$, can be due to the mixing of the Higgs with graviscalars which escape the detector [32, 46, 47]. It can also arise from Higgs decays to Kaluza-Klein neutrinos if neutrinos are allowed to propagate in the bulk [59]. In particular, if $m_h < 160$ GeV $\simeq 2\,m_W$ so that the Higgs partial width into SM particles is very small, the invisible width $\Gamma_{inv}$ can dominate the Higgs width, so that the Higgs decays predominantly into invisible modes. An invisibly decaying Higgs can also arise in supersymmetric models, with Higgs decays to pairs of lightest neutralinos or to a neutralino plus neutrino in models with Higgs-sneutrino mixing due to R-parity violation [60]; in Majoron models [61, 62]; and in generic models of dark matter containing a stable singlet scalar [63–65]. The combined LEP experimental bound on the mass of an invisibly-decaying Higgs boson is 114.4 GeV at 95% confidence level [66], assuming the Higgs is produced with Standard Model rate.

In this contribution we review existing studies [67–75] of detection of an invisibly decaying Higgs boson at the LHC. If the invisible branching fraction is large enough that the usual visible Higgs signals drop below the $5\sigma$ threshold, the Higgs mass will be difficult to measure at the LHC; we describe a new method [75] to extract the Higgs mass from production cross sections in a fairly model independent way.

### *8.2.1 Invisible Higgs detection at LHC*

Studies of an invisibly decaying Higgs $h_{inv}$ typically assume Standard Model Higgs production cross sections and a 100% invisible branching fraction. Results can easily be rescaled for non-SM Higgs production rates and partly-visible decay branching fractions. The signal rate is simply scaled by the production rate and invisible branching fraction:

$$S = S_0 \frac{\sigma}{\sigma_{SM}} \frac{\mathrm{BR}_{inv}}{1}, \tag{8.25}$$

where $S_0$ is the signal rate from the studies, $\sigma/\sigma_{SM}$ is the ratio of the nonstandard production cross section to that of the SM Higgs, and BR$_{inv}$ is the invisible branching fraction. Ignoring systematic uncertainties and assuming that the SM is the only source of background, the luminosity required for a given signal significance then scales like

$$\mathcal{L} = \mathcal{L}_0 \left[ \frac{\sigma}{\sigma_{SM}} \frac{\mathrm{BR}_{inv}}{1} \right]^{-2}, \tag{8.26}$$

where $\mathcal{L}_0$ is the luminosity required for a given significance found in the studies. Certainly, many models of new physics that can give rise to an invisibly decaying Higgs boson can also give rise to non-SM backgrounds with large missing energy, which must then be dealt with in order to isolate the invisibly-decaying Higgs signal. The studies discussed below assume only SM backgrounds.

*Weak boson fusion*

Production of an invisible Higgs via weak boson fusion (WBF) was studied for the LHC in Refs. [70,72] and in Section 8.3 of this report. These studies showed that WBF could provide significant signals for invisible Higgs discovery, even at low luminosity. The most important backgrounds are $Zjj$ (with $Z \to \nu\bar{\nu}$) and $Wjj$ (with $W \to \ell\nu$ and the lepton missed), with the jets produced by either QCD or electroweak (EW) processes. The QCD backgrounds are reduced by taking advantage of the colour-flow structure of these backgrounds versus the signal: the two jets in the QCD backgrounds are colour-connected while those in the signal process (and the EW backgrounds) are not. This leads to a depletion of gluon emission in the region between the two jets in the signal process; the QCD background can then be suppressed by vetoing additional soft jet activity in the central region [76]. Applying this central jet veto and characteristic "WBF cuts", the parton-level study in Ref. [70] found a significance of $S/(\sqrt{B}+\Delta B) \simeq 15$ (6.4) for $m_h = 120$ (400) GeV and 10 fb$^{-1}$. Note that this study takes into account a "systematic" uncertainty $\Delta B$ on the background normalization, arising from a separate analysis of the uncertainty on a direct background measurement from data on $Zjj$ (with $Z \to \ell\ell$) and $Wjj$ (with $W \to \ell\nu$ and the lepton detected) events. A $5\sigma$ detection of $h_{inv}$ is possible for Higgs masses up to 480 (770) GeV with 10 (100) fb$^{-1}$ [70]. This large Higgs mass reach is characteristic of WBF processes, which proceed through t-channel weak boson exchange and thus fall slowly with increasing Higgs mass.

The analysis was extended with a more realistic experimental simulation in Ref. [72], which considered the importance of triggering on high rapidity jets, as well as the impact of showering and detector effects on the central jet veto, which is not yet well understood. Signal and background cross sections in [72] are generally in good agreement with the parton-level results [70], with the exception of the central jet veto. Analytic calculations [77] of soft central jet production used in [70] lead to a factor 2 smaller QCD $Wjj$, $Zjj$ backgrounds compared to Pythia generation; taking the more pessimistic Pythia backgrounds leads to a significance of $S/(\sqrt{B}+\Delta B) \simeq 5.6$ (4.7) for $m_h = 120$ (250) GeV and 10 fb$^{-1}$ [72]. For the relatively soft central jets that dominate in the QCD background processes, it is believed that a resummation is needed and the perturbative showering in Pythia is unreliable; this was the purpose of the analytic calculations [77].

*$Z + h_{inv}$*

Discovery of the Higgs in the $Z + h_{inv}$ channel was studied for the LHC in Refs. [68, 69, 73, 75]. (This channel was also analyzed for the Tevatron in Ref. [78].) The signal is $Z(\to \ell\ell) + h_{inv}$, where $\ell = e, \mu$. The most important backgrounds are $Z(\to \ell\ell)Z(\to \nu\bar{\nu})$, $W(\to \ell\nu)W(\to \ell\nu)$, $Z(\to \ell\ell)W(\to \ell\nu)$ where the lepton from the $W$ decay is missed, $t\bar{t}$ with each top decay yielding a lepton, and $Z(\to \ell\ell) +$ jets with fake $\not{p}_T$ from jet energy mismeasurements or jets escaping down the beam hole.

The $WW$ background can be largely eliminated by requiring that the $\ell^+\ell^-$ invariant mass is close to the $Z$ mass. This requirement introduces a dependence on the electron and muon energy resolution of the LHC detectors. A cut on the azimuthal angle of the lepton pair eliminating back-to-back leptons also reduces the $WW$ background and eliminates Drell-Yan backgrounds with fake $\not{p}_T$ caused by mismeasurement of the lepton energies. The $WZ$ background is reduced by vetoing events with a third isolated lepton. The $Z +$ jets background with fake $\not{p}_T$ can be largely eliminated by vetoing events with hard jets; for this the large rapidity coverage of the LHC calorimetry is vital [69]. The jet veto and the cut on the $\ell^+\ell^-$ invariant mass also largely eliminate the $t\bar{t}$ background.

The $ZZ$ background is largely irreducible, but can be controlled to some extent with a cut on $\not{p}_T$. The number of $\ell^+\ell^-\not{p}_T$ signal events typically falls more slowly with $\not{p}_T$ than those of the $ZZ$ or $WW$ backgrounds. The $\not{p}_T$ of the $WW$ background is typically quite low because the $\not{p}_T$ comes from the two neutrinos emitted independently in the two $W$ decays. Although the $\not{p}_T$ of the $ZZ$ background is also typically not quite as large as that of the signal, due to the t-channel nature of the $ZZ$ background in which the $Z$ decaying to neutrinos itself tends to carry less $p_T$ than the $h_{inv}$ produced via s-channel Higgsstrahlung, this background still dominates after cuts. The $\not{p}_T$ distribution of the signal is somewhat sensitive to the Higgs mass; it falls off more slowly with increasing $\not{p}_T$ as $m_h$ gets larger. Thus a fit to the $\not{p}_T$ distribution can in principle give some (very) limited sensitivity to the Higgs mass.

The parton-level study in Ref. [75] found a significance of $S/\sqrt{B} \simeq 5.3$ (2.9) for $m_h = 120$ (160) GeV and 10 fb$^{-1}$. This is in good agreement with the results of the more realistic experimental simulation in Ref. [73], which included hadronization of the $Z + h_{inv}$ signal and backgrounds using Pythia/Herwig. For comparable cuts, Ref. [73] found a signal cross section smaller by about 30% and a total background cross section (dominated by $ZZ$ production) smaller by about 20% compared to Ref. [75]; this reduction in both signal and background cross sections is due to events being rejected by the jet veto imposed in Ref. [73] after including QCD initial-state radiation. However, the 30% reduction in signal cross section is compensated [78] by the known NLO QCD K-factor for $Z + h$ at LHC of about 1.3 [79, 80], and the reduction in the dominant $ZZ$ background is compensated by the known NLO QCD K-factor for $ZZ$ at LHC of about 1.2 [81, 82], yielding cross sections consistent with the leading order partonic results [75].

The channel $W + h_{inv}$ was also studied in Refs. [68, 73]; however, in the leptonic $W$ decay channels the signal is $\ell + \not{p}_T$ and is swamped by overwhelming backgrounds.

*$t\bar{t} + h_{inv}$*

Detection of $h_{inv}$ produced by Yukawa radiation off of a top quark pair was studied for the LHC in Refs. [67, 71]. The most important background is $t\bar{t}$ production, with $t\bar{t}Z$, $t\bar{t}W$, $b\bar{b}Z$, and $b\bar{b}W$ also contributing. The analysis in Ref. [71] reconstructs one top quark in its hadronic decay mode and requires an isolated lepton (electron or muon) from the decay of the second top quark along with large missing transverse energy. Both $b$ quarks from the two top decays are required to be tagged.

Ref. [71] found a background after cuts dominated by $t\bar{t}$ production with one top decaying leptonically and the other decaying to a tau. Vetoing the taus would significantly improve the results but was beyond the scope of the study in [71]. From the results of Ref. [71] we calculate a significance of $S/\sqrt{B} \simeq 2.0$ (0.7) for $m_h = 120$ (200) GeV and 10 fb$^{-1}$. If the background from top decays to taus could be eliminated, this would improve to $S/\sqrt{B} \simeq 3.2$ (1.1) for $m_h = 120$ (200) GeV. The signal observability should not degrade significantly for the high-luminosity LHC running; thus the $S/\sqrt{B}$ numbers quoted can be scaled up by $\sqrt{30}$ to estimate the ultimate LHC sensitivity with 300 fb$^{-1}$: i.e., $S/\sqrt{B} \simeq 11.1$ (3.8) [17.7 (6.0)] for $m_h = 120$ (200) GeV including [excluding] the $t \to \tau$ background. More experimental work is needed to understand the systematic uncertainties in both physics and detector simulation.

Although the discovery potential of this mode is much less than that of WBF or $Z + h_{inv}$, its study is well motivated because it offers access to the coupling of $h_{inv}$ to top quarks. This may well be the only Higgs coupling to SM fermions measurable at the LHC in the case that BR($h \to$ invisible) $\sim 100\%$, and thus provides a valuable input to the study of electroweak symmetry breaking.

*Diffraction*

Detection of $h_{inv}$ produced by central exclusive diffraction at the LHC was studied in Ref. [74]. The signal process is $pp \to p h_{inv} p$, with the two final-state protons very forward. In such a process, the mass of the Higgs boson can be very accurately measured using the missing-mass method; the sharp peak in the missing mass spectrum dramatically reduces background contributions. In fact, one can imagine using an ILC-style missing-mass analysis to measure the $pp \to php$ cross section and $h$ branching fractions in a model-independent way. Further, such a "Pomeron-Pomeron fusion" process can only produce neutral, colourless, flavourless particles of parity $P = (-1)^J$, enabling these quantum numbers of the invisibly decaying state to be pinned down.

The difficulty arises in detecting and triggering on $p h_{inv} p$ events. The final-state protons must be detected by far-forward proton detectors (roman pots or microstations) installed up to 400 m from the interaction point. These detectors would register protons that have lost a small fraction of their incoming energy through the diffractive process; the trigger would have to be on the far-forward protons, since $h_{inv}$ leaves "nothing" (except noise) in the central detector. The main backgrounds consist of soft inelastic Pomeron-Pomeron fusion yielding hadrons in the central detector and events in which the final-state protons lose energy through QED radiation. In order for the Higgs events to be separated, these backgrounds must be suppressed by forward calorimeters able to reject events with additional high-energy photons and charged pions with very high efficiency [74].

Table 8.2: Higgs mass determination from $Z + h_{inv}$ with 10 (100) fb$^{-1}$, assuming Standard Model production cross section and 100% invisible decays. The signal and background cross sections were taken from Table I of Ref. [75] for the cut $\not{p}_T > 75$ GeV. The total uncertainty includes a theoretical uncertainty on the signal cross section from QCD and PDF uncertainties of 7% [83] and an estimated lepton reconstruction efficiency uncertainty of 4% (2% per lepton) and luminosity normalization uncertainty of 5% [84]. From Ref. [75].

| $m_h$ (GeV) | 120 | 140 | 160 |
|---|---|---|---|
| $\rho = (d\sigma_S/dm_h)/\sigma_S$ (1/GeV) | $-0.013$ | $-0.015$ | $-0.017$ |
| Statistical uncert. | 21% (6.6%) | 28% (8.8%) | 37% (12%) |
| Background normalization uncert. | 33% (10%) | 45% (14%) | 60% (19%) |
| Total uncert. | 40% (16%) | 53% (19%) | 71% (24%) |
| $\Delta m_h$ (GeV) | 30 (12) | 35 (12) | 41 (14) |

### 8.2.2 *Higgs boson mass measurement*

The mass of an invisibly-decaying Higgs boson obviously cannot be reconstructed from the Higgs decay products. Unless the Higgs is also observed in a visible channel, our only chance of determining the Higgs mass at the LHC comes from the $m_h$ dependence of the production process. Here we describe the method of Ref. [75] to extract the Higgs boson mass from cross sections in a fairly model-independent way.

Extracting $m_h$ from the cross section of a single production channel requires the assumption that the production couplings are the same as in the SM. Non-observation of the Higgs in any visible final state implies that the invisible branching fraction is close to 100%. The Higgs mass extraction from LHC measurements of the production cross sections in $Z + h_{inv}$ and WBF under these assumptions are shown in Tables 8.2 and 8.3, respectively. There are two main sources of uncertainty in the signal: statistical and from background normalization. The statistical uncertainty is $\Delta\sigma_S/\sigma_S = \sqrt{\mathrm{S+B}}/\mathrm{S}$. Ref. [75] estimated the total background normalization uncertainty for $Z + h_{inv}$ to be the same size as that of the dominant process involving $Z \to \nu\bar{\nu}$: $\Delta\mathrm{B}/\mathrm{B} = \Delta\mathrm{B}(ZZ)/\mathrm{B}(ZZ)$. They assumed that this background can be measured via the corresponding channels in which $Z \to \ell^+\ell^-$ and took the uncertainty to be the statistical uncertainty on the $Z \to \ell^+\ell^-$ rate: $\Delta\mathrm{B}(ZZ)/\mathrm{B}(ZZ) \simeq 7.1\%$ (2.2%), for an integrated luminosity of 10 (100) fb$^{-1}$. Tables 8.2 and 8.3 quote the resulting uncertainty on the signal cross section, given by $\Delta\sigma_S/\sigma_S = (\mathrm{B/S}) \times \Delta\mathrm{B/B}$. The total uncertainty $[\Delta\sigma_S/\sigma_S]_{tot}$ presented in Tables 8.2 and 8.3 is then the sum, in quadrature, of the statistical and background uncertainties, as well as the other uncertainties described in the table captions. We then have $\Delta m_h = (1/\rho)[\Delta\sigma_S/\sigma_S]_{tot}$, where $\rho \equiv (d\sigma_S/dm_h)/\sigma_S$ is the "slope" of the cross section.

The cross section for $Z + h_{inv}$ production falls quickly with increasing $m_h$ due to the s-channel propagator suppression. This is in contrast to the WBF production, which provides a $> 5\sigma$ signal up to $m_h \simeq 480$ GeV with 10 fb$^{-1}$ if the Higgs decays completely invisibly [70]. Thus, while the statistics are much better on the WBF measurement than on $Z + h_{inv}$, the systematic uncertainties hurt WBF more because $(d\sigma_S/dm_h)/\sigma_S$ is much smaller for WBF than for $Z + h_{inv}$. The $Z + h_{inv}$ cross section is therefore more sensitive to the Higgs mass than the WBF cross section.

More importantly, however, taking the ratio of the $Z + h_{inv}$ and WBF cross sections allows for a more model-independent determination of the Higgs mass. This is due to the fact that the production couplings in $Z + h_{inv}$ ($hZZ$) and in WBF (contributions from $hWW$ and $hZZ$) are related by custodial

Table 8.3: Higgs mass determination from WBF $\to h_{inv}$ with 10 (100) fb$^{-1}$, assuming Standard Model production cross section and 100% invisible decays. The background and signal cross sections were taken from Tables II and III, respectively, of Ref. [70], and include a central jet veto. The total uncertainty includes a theoretical uncertainty from QCD and PDF uncertainties of 4% [85, 86], and an estimated uncertainty on the efficiency of the WBF jet tag and central jet veto of 5% and luminosity normalization uncertainty of 5% [84]. From Ref. [75].

| $m_h$ (GeV) | 120 | 130 | 150 | 200 |
|---|---|---|---|---|
| $\rho = (d\sigma_S/dm_h)/\sigma_S$ (1/GeV) | $-0.0026$ | $-0.0026$ | $-0.0028$ | $-0.0029$ |
| Statistical uncert. | 5.3% (1.7%) | 5.4% (1.7%) | 5.7% (1.8%) | 6.4% (2.0%) |
| Background normalization uncert. | 5.2% (2.1%) | 5.3% (2.1%) | 5.6% (2.2%) | 6.5% (2.6%) |
| Total uncert. | 11% (8.6%) | 11% (8.6%) | 11% (8.6%) | 12% (8.8%) |
| $\Delta m_h$ (GeV) | 42 (32) | 42 (33) | 41 (31) | 42 (30) |

Table 8.4: Higgs mass determination from the ratio method discussed in the text, with 10 (100) fb$^{-1}$. The event rates for WBF were interpolated linearly for Higgs masses of 140 and 160 GeV, which were not given explicitly in Ref. [70]. Statistical uncertainties were obtained assuming SM signal rates. The total uncertainty includes theoretical uncertainties from QCD and PDF uncertainties of 7% for $Z + h_{inv}$ [83] and 4% for WBF [85, 86], and estimated uncertainties on the lepton reconstruction efficiency in $Z + h_{inv}$ of 4% (2% per lepton) and on the efficiency of the WBF jet tag and central jet veto of 5% [84]. The luminosity normalization uncertainty cancels out in the ratio of cross sections and is therefore not included. From Ref. [75].

| $m_h$ (GeV) | 120 | 140 | 160 |
|---|---|---|---|
| $r = \sigma_S(Zh)/\sigma_S(\mathrm{WBF})$ | 0.132 | 0.102 | 0.0807 |
| $(dr/dm_h)/r$ (1/GeV) | $-0.011$ | $-0.013$ | $-0.013$ |
| Total uncert., $\Delta r/r$ | 41% (16%) | 54% (20%) | 72% (25%) |
| $\Delta m_h$ (GeV) | 36 (14) | 43 (16) | 53 (18) |

SU(2) symmetry in any model containing only Higgs doublets and/or singlets. The production couplings thus drop out of the ratio of rates in this wide class of models (which includes the MSSM, multi-Higgs-doublet models, and models of singlet scalar dark matter), leaving dependence only on the Higgs mass. (The dependence on the invisible branching fraction of the Higgs also cancels in the ratio.) The resulting Higgs mass extraction is illustrated in Table 8.4. Assuming SM event rates for the statistical uncertainties, the Higgs mass can be extracted with an uncertainty of $\pm$35–50 GeV ($\pm$15–20 GeV) with 10 (100) fb$^{-1}$ of integrated luminosity. The ratio method also allows a test of the SM cross section assumption by checking the consistency of the separate $m_h$ determination from the $Z + h_{inv}$ or WBF cross section alone with the $m_h$ value extracted from the ratio method. Furthermore, observation of the invisibly-decaying Higgs in WBF but not in $Z + h_{inv}$ allows one to set a lower limit on $m_h$ in this class of models.

### 8.3 Search for invisible Higgs decays in CMS

*Kajari Mazumdar and Alexandre Nikitenko*

There are several scenarios beyond Standard Model where the Higgs boson can decay invisibly [87]. These mechanisms also modify the production and decay rates of the Higgs in visible modes at the LHC and hence model-independent, experimental investigation for invisible decay of Higgs boson constitutes an essential aspect of the Higgs search program at the collider experiments..

The discovery potential of an invisible Higgs boson in the context of the LHC has been discussed in various production modes. The sensitivities of the CMS detector for the invisibly decaying Higgs boson, when it is produced via vector boson fusion (VBF) process [70], have been evaluated in [88] with CMS-specific detailed detector simulation and event reconstruction softwares. We discuss here the salient features.

### *8.3.1 Invisible Higgs boson signal and the background*

We note that in the invisible decay channel no mass can be reconstructed and hence the discovery is established by observing an excess of events compared to predicted backgrounds. Therefore sufficient signal cross-section and a good signal-to-background ratio are the requirements for the experimental search. In gluon-gluon fusion process the final state is *nothing* and hence can't even be identified. The VBF channel offers the highest cross-section among all the processes where Higgs is not produced alone in the final state and thus provides a handle to tag the event via accompanying particle(s).

The dynamics of the signal channel in the VBF process leads to energetic jets in the forward and backward directions due to the continuation of the interacting quarks in original direction, after the simultaneous emission of $W/Z$ bosons. The absence of colour exchange between the scattered quarks and the colourless Higgs boson leads to low hadronic activity in the central region. Consequently, this process has special signature of two jets with distinct topology in the final state: (i) the rapidity gap between the jets is large, (ii) the jets are in opposite hemispheres, (iii) the jets carry large energy and so invariant mass of jets is large.

There are several SM processes which can have similar final state as this signal. For the final state under study, *ie,* 2 jets plus missing energy, QCD di-jet production may mimic the signal characteristics, the rate being very high. The other SM processes, QCD and electroweak production of $W$ + 2jets and $Z$ + 2jets events with leptonic decays of the $W, Z$ bosons which, in particular, are potential backgrounds. For $W \to \ell\nu$ and $Z \to \ell\ell$ events with $(\ell = e, \mu, \tau)$ and $e, \mu$ decays of $\tau$, if the charged lepton is not identified within the detector acceptance, the final state effectively consists of 2 jets and missing energy. The genuine background is due to the process $Z \to \nu\bar{\nu}$ and the event is almost similar to the signal when $Z$ + 2 jets events are produced via weak interaction (t-channel exchange of $W$). As will be discussed later, fortunately the jets in the signal channel seem to balance the Higgs and hence the angle between the jets is smaller which is not the case for these background events when the satisfy VBF like criteria.

The search strategy heavily relies on the optimal performance of the calorimeters of the LHC detectors for jets and missing transverse energy ($E_T^{\rm miss}$) reconstruction as well as on a dedicated calorimeter trigger.

### *8.3.2 Event generation, simulation and reconstruction*

The signal events, $qq \to qqH,\ H \to invisible$ were generated with PYTHIA event generator, [89], and using the CTEQ5L structure function parameterization. Since the process $H \to invisible$ is not actually available in PYTHIA, we generated $H \to ZZ(^*) \to \nu\bar{\nu}\nu\bar{\nu}$. We produced signal samples for different Higgs masses ranging from $110 - 400$ GeV. The SM production cross section considered is calculated by the VV2H program [90]. The PYTHIA package was also used to simulate QCD di-jet production (MSEL=1) processes, in various $\hat{p_T}$ bins from $10 - 15$ GeV ($\sigma = 8.868 \times 10^{12}$ fb) up to 2600-3000 GeV ($\sigma = 11.25$ fb) for a total of about $10^6$ events.

For $W$ + 2jets and $Z$ + 2jets processes (2 $\to$ 3) parton level events were produced, according to Leading Order matrix element calculations, with dedicated event generators combined with forced leptonic decays. The QCD subprocess events were produced with MadCUP [91] based on the work of [92] using CTEQ4L structure function. The electroweak subprocess events were generated with the COMPHEP generator [93], with the CTEQ5L structure function. The parton level events were subsequently hadronized through PYTHIA. The production rates were found to be in good agreement with the values

Table 8.5: Cross sections (in pb) for backgrounds as given by LO matrix element calculations with preselection cuts described in the text. $\mathrm{BR}(Z \to \nu\nu)$ and $\mathrm{BR}(W \to \ell\nu)$ are included.

| QCD Wjj | QCD Zjj | EW Wjj | EW Zjj |
|---|---|---|---|
| 76.0 | 15.7 | 4.7 | 0.644 |

Table 8.6: Survival probabilities for signal and background for a veto on central jets with $E_T > 20$ GeV [77].

| Signal | QCD $W$ + 2jets & QCD $Z$ + 2jets | EW $W$ + 2jets & EW $Z$ + 2jets |
|---|---|---|
| 0.87 | 0.28 | 0.82 |

from other packages, Madgraph [94] and ALPGEN [95].

Loose selection criteria were used to generate $W$ + 2jets and $Z$ + 2jets events.

$$p_T^{1,2} > 20 \text{ GeV}, \;\; |\eta_{1,2}| < 5.0, \;\; |\eta_1 - \eta_2| > 4.2, \;\; \eta_1 \times \eta_2 < 0, \;\; M_{1,2} > 900 \text{ GeV}$$

where 1 and 2 refer to the partons (gluons and quarks) accompanying Z and W production. The cross sections (in pb) given by the matrix element calculations with these cuts are presented in Table 8.5. Cross sections include the leptonic branching ratios, $\mathrm{BR}(Z \to \nu\bar{\nu})$ and $\mathrm{BR}(W \to \ell\nu)$, for three lepton generations.

All the signal samples and QCD multijet events after generation are fully simulated in CMS detector, using GEANT-based detector simulation package CMSIM [96]. Subsequently the events were digitized and reconstructed with CMS-specific reconstruction software ORCA [97]. Event pile up, corresponding to instantaneous luminosity of $2{\times}10^{33}$ cm$^{-2}$ s$^{-1}$, was taken into account. The $W$ + 2jets and $Z$ + 2jets events were processed using the fast simulation package CMSJET [98].

A key point of the search is the use of a mini-jet veto, namely the rejection of events with additional soft ($E_T > 20$ GeV) jet(s) inside the rapidity gap between the two forward tagging jets as discussed before. The efficiency of the mini-jet veto is expected to be sensitive to detector effects like calibration, electronic noise and readout thresholds, interaction of soft particles in the tracker in front of the calorimeter, magnetic field, or pile up activity. Since the fast CMSJET simulation is not expected to properly reproduce these effects, the mini-jet veto efficiency was not evaluated for the background events with CMSJET. Instead the background efficiency was multiplied by $P_{\mathrm{surv}}$, defined as the probability for a jet (parton) to be radiated in the rapidity gap between the two tagging jets and with the $\mathrm{E_T} > 20$ GeV cut. The values of $P_{\mathrm{surv}}$ calculated in [77] and used in the analysis [70] are listed in Table 8.6. In the parton level study it has been assumed that such jets will be reconstructed with 100 % efficiency. The effect of the central jet veto is realistically evaluated for the Higgs boson signal with detailed simulations and it causes a 24 % reduction of the signal which is worse than that quoted in [70].

#### 8.3.2.1 Event Analysis

For the present analysis the signal-to-background ratio is effectively enhanced by identifying the forward jets obeying topological features and applying further requirements on the effective mass of the tagging jets ($M_{\mathrm{jj}}$), the missing transverse energy ($E_T^{\mathrm{miss}}$) and the azimuthal angle between the two jets in the transverse plane ($\phi_{\mathrm{jj}}$). These are

$$E_T^{\mathrm{j1,\,j2}} > \; 40 \text{ GeV}, \;\; |\eta_{\mathrm{j1,\,j2}}| < 5.0, \;\; |\eta_{\mathrm{j1}} - \eta_{\mathrm{j2}}| \; > \; 4.4, \;\; \eta_{\mathrm{j1}} \times \eta_{\mathrm{j2}} \; < \; 0, \tag{8.27}$$

$$E_T^{\mathrm{miss}} \; > \; 100 \text{ GeV} \tag{8.28}$$

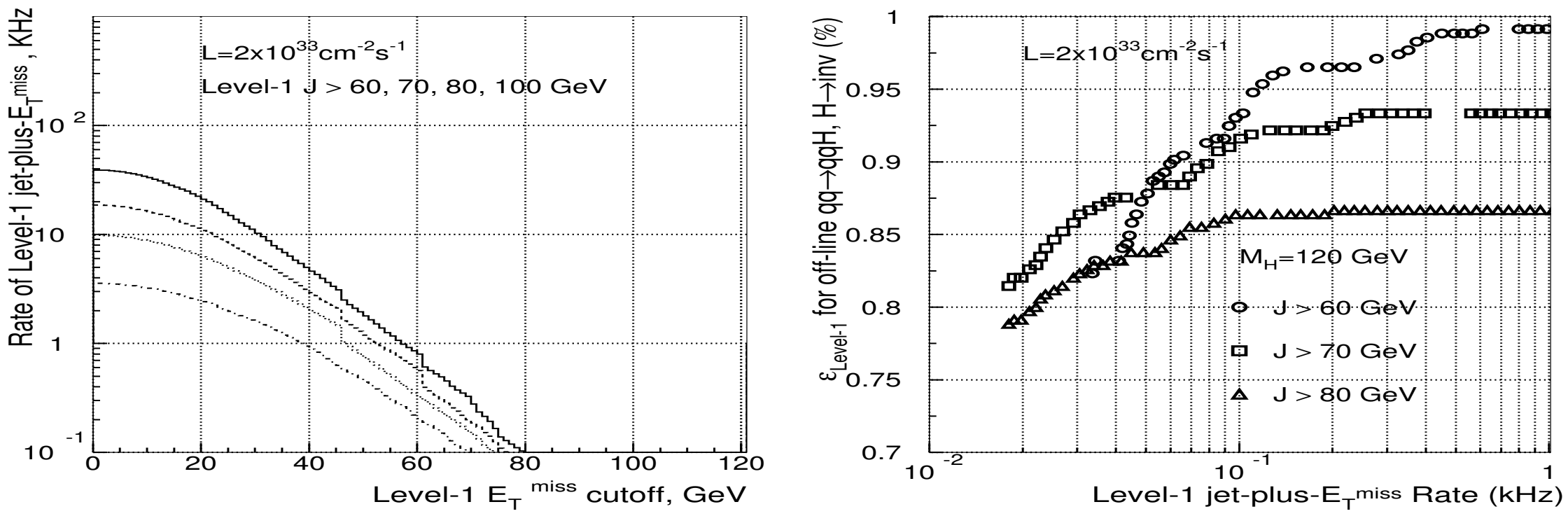

Fig. 8.6: Rate of L1 jet−plus−$E_T^{\rm miss}$ trigger as a function of $E_T^{\rm miss}$ threshold for given values of jet threshold. Trigger rate vs efficiency for $qq \to qqH,\ H \to invisible$ events which passed VBF cuts.

$$M_{\rm jj} > 1200\,{\rm GeV}, \qquad (8.29)$$

$$\phi_{\rm jj} < 1\,{\rm rad} \qquad (8.30)$$

In addition a mini-jet veto in the central region, and a lepton veto, i.e. the requirement that no lepton be reconstructed with transverse momentum $p_T > p_T^{\rm cut}$, $p_T^{\rm cut}$ = 10 (5) GeV for electron (muon) have also been used. In the present study the veto on taus is separated into a lepton veto and a jet veto, depending on whether the tau decays leptonically or hadronically. The full set of these cuts we shall refer to, hereafter, as VBF cuts.

### 8.3.2.2 Triggers

At LHC the collision and the overall data rate being much higher than the data archival storage capability the required rejection power, $\mathcal{O}(10^5)$, is achieved in two steps for CMS experiment. At lower level (L1) the trigger conditions are implemented through hardware and in the second/higher level (HLT) selection algorithms are executed in processor farm [99]. The invisible Higgs decay channel requires dedicated calorimeter trigger both at L1 and at HLT. The Hadron Forward calorimeter (HF) of the CMS detector plays a crucial role in the on-line and off-line selections. The combined jet−plus−$E_T^{\rm miss}$ trigger condition is the most effective for the invisible Higgs boson selection. The trigger bandwidth is dominated by QCD jet events which has huge cross-section. At low luminosity, the trigger threshold optimization was performed by studying the trigger rate for jet−plus−$E_T^{\rm miss}$ trigger vs. the signal efficiency by varying the $E_T^{\rm miss}$ threshold for a fixed set of single jet threshold values as illustrated in Fig.8.6. The optimum values were found to be 60 and 64 GeV, corresponding to a signal efficiency of 98%. In the off-line reconstruction both jet $E_T$ and $E_T^{\rm miss}$ are corrected for the effects of calorimeter non-linearity. Jet energy corrections are also applied at L1, while it is not foreseen to correct $E_T^{\rm miss}$ at L1.

At HLT, the off-line requirement (1) on the pseudorapidity gap between the two highest $E_T$ jets can be exploited concurrently with the minimum threshold requirements on $E_T^{\rm miss}$ and $M_{\rm jj}$. Full-granularity calorimeter information is available at HLT and hence jet and $E_T^{\rm miss}$ will be reconstructed like in the off-line analysis. The left plot in Figure 8.7 shows the rate of QCD multi-jet events after cut (1) as a function of the cutoff on $E_T^{\rm miss}$.

### 8.3.2.3 Event selection

One of the crucial problems of this study is a proper simulation of the tails in the $E_T^{\rm miss}$ distribution of the QCD multi-jet background events which could be due to real $E_T^{\rm miss}$ from heavy quarks decays, but also due to a number of detector effects. To make reliable estimates, a total of about one million of QCD multi-jet events was used. However, this statistics was still not enough to directly prove that the QCD background could be suppressed. The right plot of Figure 8.7 shows the $E_T^{\rm miss}$ distribution of the QCD jet background (open histogram) and of the Higgs boson signal (dark shaded area) after cuts (1) and (3). With an additional cut (4), the $E_T^{\rm miss}$ distribution for the signal events is superposed as the light shaded area. It can be observed that the tail in the background distribution goes well beyond 100 GeV. The QCD multi-jet events in the tail come from $\hat{p_T}$ bins between 300 and 600 GeV. Once the cut (4) on $\phi_{\rm jj}$ is applied, no background event with $E_T^{\rm miss} > 100$ GeV is left. With the statistics used in the analysis, this leads to an upper limit of $\simeq 1$ pb on the QCD background contribution, i.e., about of 10 times higher than the signal expected after the same selections (1-4). In the final analysis for signal observability in CMS, it is assumed that QCD multi-jet events can be entirely suppressed with a cut on the minimal angle in the transverse plane between missing $p_T$ and a jet as implemented in the ATLAS fast simulation study [100]. The $E_T^{\rm miss}$ requirement reduces the $W/Z$ + 2jets backgrounds as well. Figure 8.8 shows the $M_{jj}$ distributions for the signal and background events after cuts (1) and (2) and the same electron, muon veto. In signal events the transverse momenta of the tagging jets balance the $E_T^{\rm miss}$ due to the invisible Higgs boson. An upper threshold on the azimuthal angle between the two jets reduce further the background of W/Z+2 jets types of events as shown in Figure 8.9.

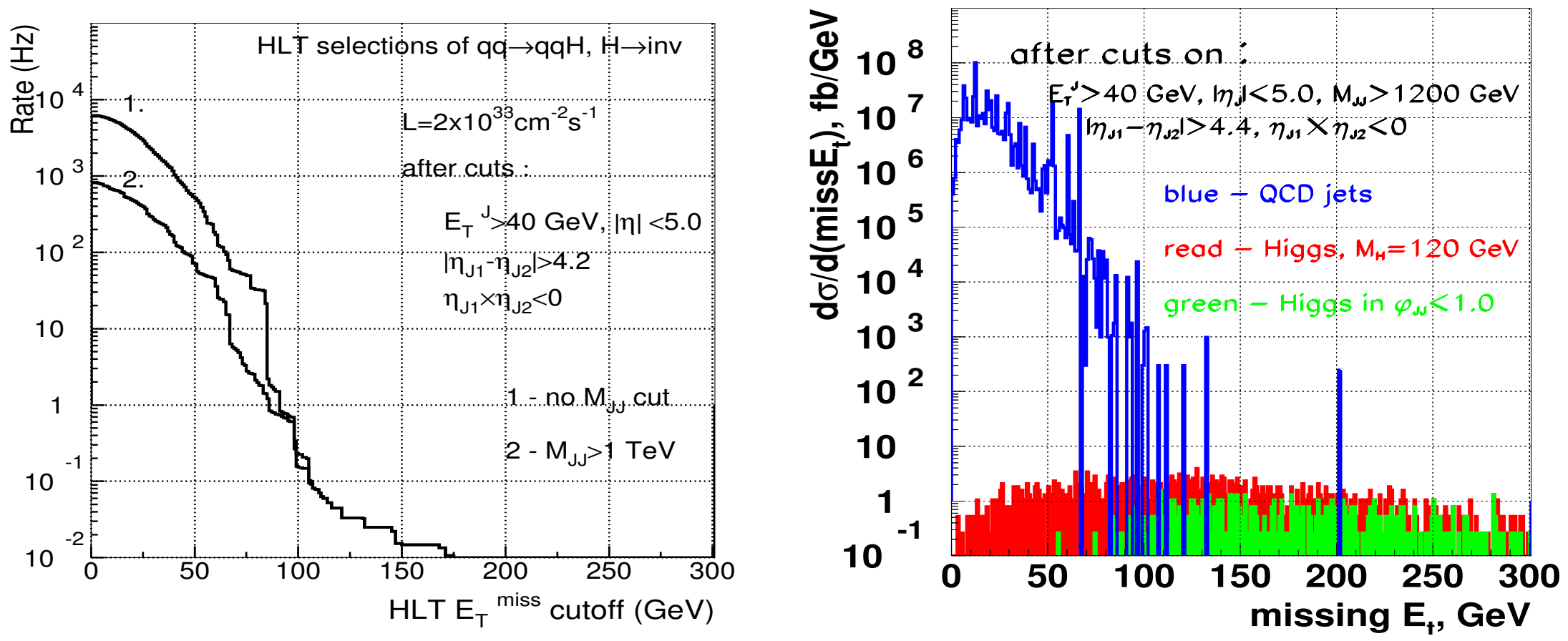

Fig. 8.7: Left:QCD di-jet background rate after jet topological selection (1) as a function of the threshold on $E_T^{\rm miss}$. Right: $E_T^{\rm miss}$ distribution for 120 GeV Higgs boson events (dark shaded area) and for the QCD multi-jet background (open histogram) after selections (1) and (3). $E_T^{\rm miss}$ distribution in signal events after cut (4) is shown as the light shaded area.

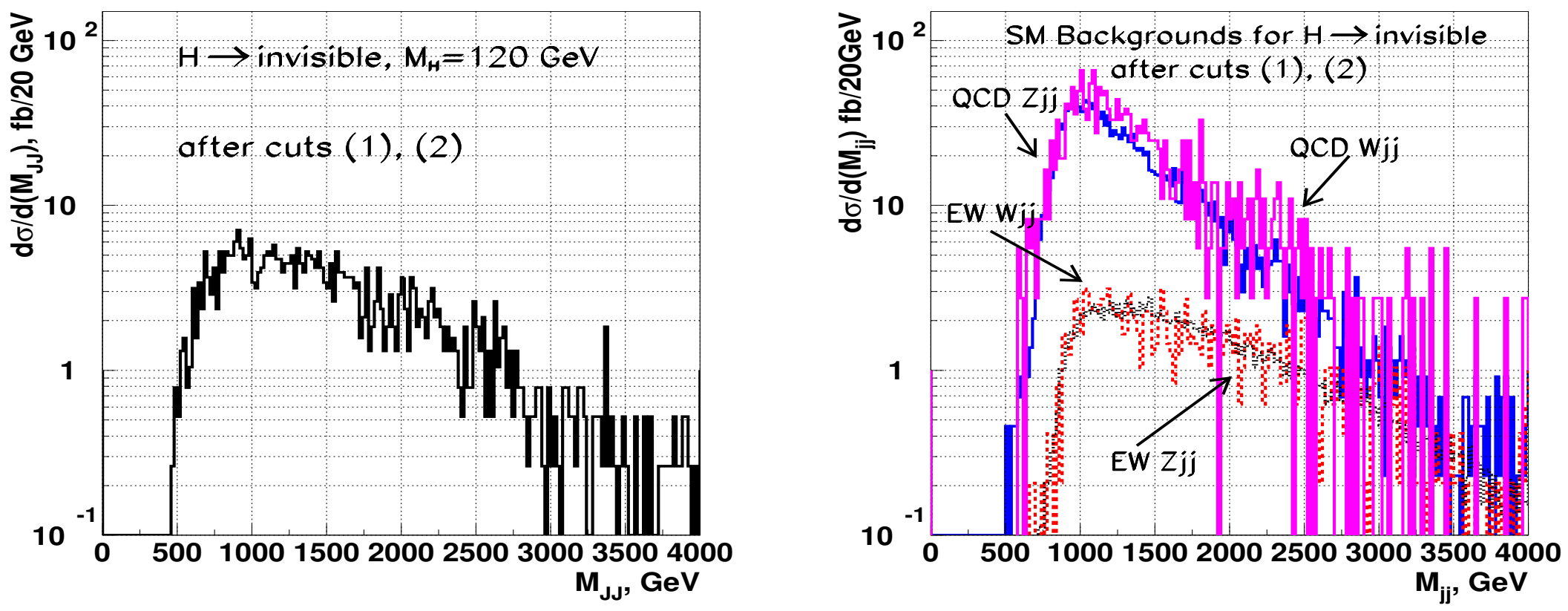

Fig. 8.8: The $M_{\rm jj}$ distribution for 120 GeV Higgs boson and background events after cuts (1-2).

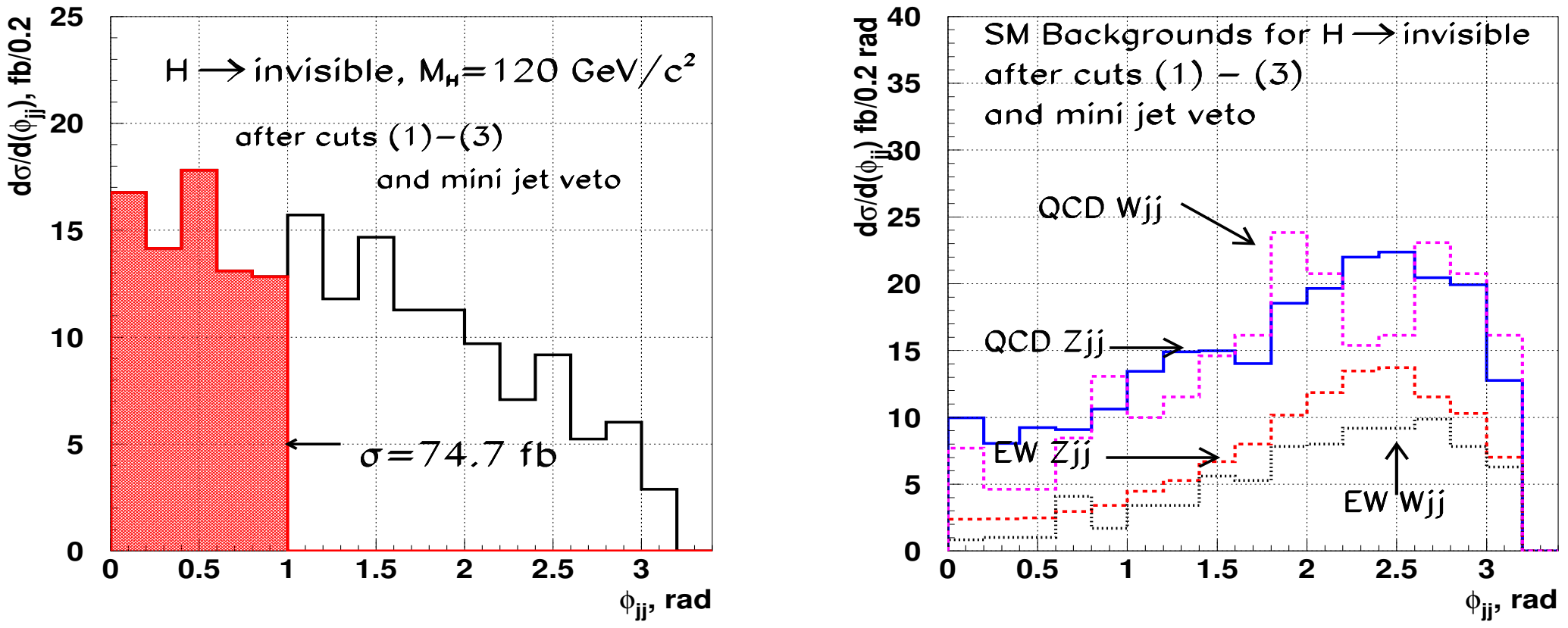

Fig. 8.9: The $\phi_{\rm jj}$ distribution for 120 GeV Higgs boson and background events after cuts (1-3) and mini-jet veto.

Table 8.7: Cross sections in fb for the background and a 120 GeV Higgs boson assuming $\mathrm{BR}(H \to inv.) = 100\%$ and Standard Model production cross section for the Higgs boson.

| cross section, fb | Higgs | QCD $Z+2$ jets | QCD $W+2$ jets | EW $Z+2$ jets | EW $W+2$ jets |
|---|---|---|---|---|---|
| after cuts (1-3), $e$, $\mu$ veto | 238 | 857 | 1165 | 141.5 | 145.1 |
| + mini-jet veto | 180 | 240 | 237 | 116 | 84.5 |
| + $\phi_{\rm jj} \leq 1$ rad. | 74.7 | 48.0 | 40.0 | 12.8 | 8.7 |

### *8.3.3 Results*

Estimated cross sections (in fb) for the Higgs boson and various types of backgrounds at different steps of the event selection are shown in Table 8.7. SM production cross sections and $\mathrm{BR}(H \to inv.)$ = 100 % are assumed. The first row of Table 8.7 presents the cross sections after cuts (1-3) and a veto on identified leptons. The second row presents the cross sections after the mini-jet veto where the survival efficiencies are obtained from Table 8.6. The last row of the Table 8.7 shows the cross sections after all selection cuts.

### *8.3.4 Limit on branching ratio for invisible decay*

It is evident that background estimation is a crucial aspect of this study. At present we have considered the rates of the background events as calculated at the Leading Order. It is expected that at the LHC we can estimate the cross-sections directly from data utilizing $W$ + 2jets and $Z$ + 2jets events for leptonic decays of $W$, $Z$ where the lepton is identified (and isolated). As estimated in [70], this method leads to a systematic uncertainty on the background evaluation to be about of $\sim 3\%$. It is taken into account along with the statistical uncertainty. The total error on the background cross section is evaluated to be 4.7 fb.

The sensitivity to invisible decay mode of the Higgs boson is defined as 1.96 standard deviation (95 % CL) from the expected background. The absence of a signal can be interpreted as an upper limit on branching ratio for invisible decay of the Higgs boson which can be probed at a given luminosity, assuming SM-like production rate. We define the parameter,

$$\xi = \mathrm{BR}(H \to invisible) \frac{\sigma(qq \to qqH)}{\sigma(qq \to qqH)_{SM}} \tag{8.31}$$

which can be probed at 95% CL as a function of Higgs mass with an integrated luminosity of 10 fb$^{-1}$ as displayed in Figure 8.10.

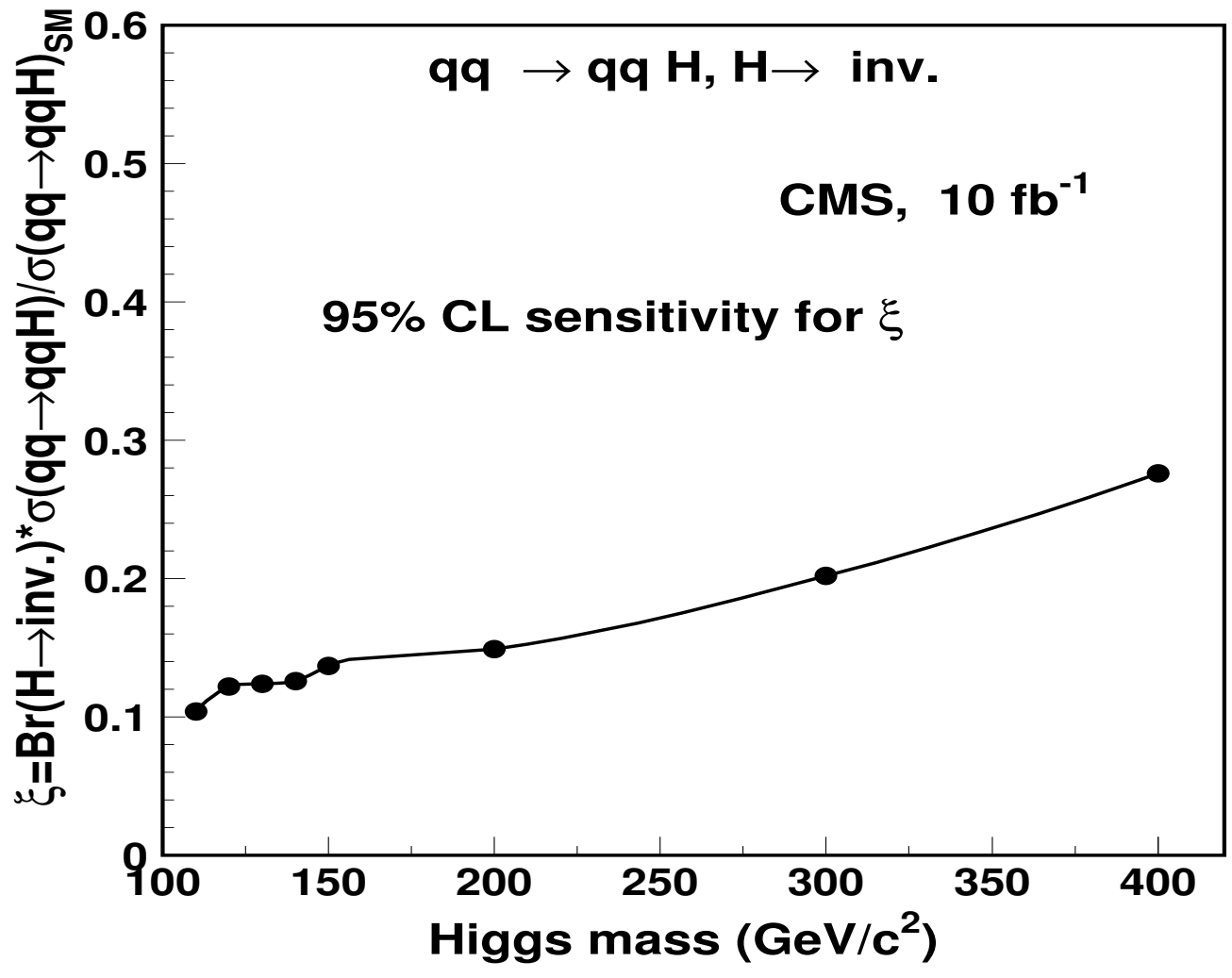

Fig. 8.10: Upper limit of parameter $\xi$ as a function of Higgs boson mass.

#### *8.3.5 Summary*

The potential of the CMS experiment for the observability of the Higgs boson in Vector Boson Fusion production channel with an invisible decay has been studied. The viability of the basic selection criteria is checked with realistic detector simulation. The signal channel has very high efficiency for the CMS trigger condition. A reasonably low value for the lower limit on the branching fraction for the invisible mode of the Higgs boson can be obtained with a limited luminosity for a light Higgs boson, for example about 12% for a Higgs mass of 120 GeV.

### 8.4 Search for heavy resonances

*Barbara Clerbaux, Tariq Mahmoud, Caroline Collard, Philippe Miné*

As explained in Section 8.1, in addition to large extra dimensions where only gravity may propagate, the electroweak interactions could possibly also propagate in TeV$^{-1}$-sized extra dimensions. This possibility allows for new model building, which address gauge coupling unification [101] or fermion mass hierarchy [9]. The phenomenological consequence of this scenario is the appearance of a KK tower of states for gauge boson fields. The masses of the gauge boson modes are given by: $M_n^2 = M_0^2 + n^2 M^2$, where $M_0$ is the mass of the zero$^{th}$ mode, corresponding to the SM fields, $n$ is the mode number and $M$ is the compactification scale, $M = 1/R$ ($R$ being the compactification radius). In this approach, the existence of only one extra dimension, of radius $R \simeq \mathrm{TeV}^{-1} \simeq 10^{-17}\mathrm{cm}$, compactified on a circle with an orbifold condition, (compactification on $S^1/Z^2$), is assumed. In the model considered here, all the SM fermions are localized at the same orbifold point. The couplings of fermions to KK gauge bosons are the same as in the SM, but scaled by a factor $\sqrt{2}$. The model has only one free parameter, the compactification scale $M$.

In fact, heavy resonances with mass above 1 TeV are predicted by several models beyond the Standard Model. Superstring-inspired $\mathrm{E}_6$ models [102] or left-right symmetry-breaking models [103, 104], predict the existence of an extra heavy neutral gauge boson, generically called $Z'$ (c.f. Section 6). Currents lower limits on the $Z'$ mass are (depending on the model) of the order of 600-900 GeV [105]. In addition to the KK $Z$ resonances, we consider six cases of $Z'$ bosons which are frequently discussed and whose properties are thought to be representative of a broad class of extra gauge bosons:

– $Z_{SSM}$ within the Sequential Standard Model (SSM), which has the same couplings as the Standard

Model $Z$ and is often used as a benchmark by experimentalists.

- $Z_\psi$, $Z_\eta$, and $Z_\chi$, arising in $E_6$ and SO(10) GUT groups.
- $Z_{LRM}$ and $Z_{ALRM}$, arising in the framework of the so-called "left-right" and "alternative left-right" models. Their couplings were calculated according to the formalism in Ref. [103, 104] with $g_R = g_L$.

In the framework of the Randall and Sundrum (RS) model, see Section 9 of this report, gravitons appear as massive resonances, with masses of order of TeV. Two parameters control the properties of the RS model: the mass of the first KK graviton excitation, and the coupling constant $c$ determining graviton couplings and widths.

In this contribution we present the CMS experiment discovery potential for new heavy resonances, decaying into an electron pair. The $e^+e^-$ decay channel provides a clean signature in the CMS detector. The presence of a heavy particle would be detected in CMS by the observation of a resonance peak in the dielectron mass spectrum over the Drell-Yan process ($pp \to \gamma/Z \to e^+e^-$) which constitutes the main Standard Model background. We also show how to contrast Standard Model Drell-Yan process and $Z'$ production (spin 1) from graviton production (spin 2). Details of the analyses presented in this Section can be found in [106] and [107].

### *8.4.1 Event selection and correction*

Two electrons are requested for this analysis. They are reconstructed as super-clusters (SC) in the CMS electromagnetic calorimeter ECAL in the barrel and the endcap regions [108]. For endcap SC, the energy loss in the preshower detector is taken into account. The two SC with highest energies are selected as the electron candidates.

An important characteristic of the signal events is that the final state electrons are very energetic and may have a significant energy leakage beyond the ECAL clusters. An improvement in the energy determination is achieved by including the hadronic calorimeter HCAL measurement, event by event.

Reducible backgrounds (like QCD jets and $\gamma$-jets) are suppressed by applying the following requirements:

- The ratio of the HCAL to ECAL energy deposits is required to be $H/E < 10\%$.
- The two SC must be isolated: the total additional transverse energy in a cone of radius $0.1 < \Delta R < 0.5$ is required to be below 2% of the SC transverse energy (where $\Delta R = \sqrt{\Delta\eta^2 + \Delta\phi^2}$).
- To identify electrons and reject neutral particles, a track is requested to be associated for each electron candidate. If a track is associated with only one of these SC, the event is however kept if it contains a third SC with $E > 300$ GeV with an associated track and satisfying the $H/E$ and isolation cuts described above.

The selected events are then corrected for the following effects:

- Saturation correction: for very energetic electrons and photons, saturation occurs in the ECAL electronics because of the limited dynamical range of the Multi-Gain-Pre-Amplifier. The saturation threshold has been established to be at 1.7 TeV in crystals of the barrel and 3.0 TeV in the endcaps. A correction method (for barrel only) has been developed using the energy deposit in crystals surrounding the saturated crystal. The correction leads to the correct estimate of energy deposit with a resolution of around $7\%$ [109].
- Energy correction: the ECAL measured electron energy after preshower, HCAL and saturation corrections, is smaller than the generated energy. Dedicated energy correction factors for very energetic electrons have been determined using calibration files. These factors depend on both energy, $\eta$ and whether saturation occurs or not. The resolution on the corrected SC energy is 0.6% at $E = 1000$ GeV.
- z-vertex distribution: the measurement in $\eta$ takes into account the knowledge of the $z$-vertex position.

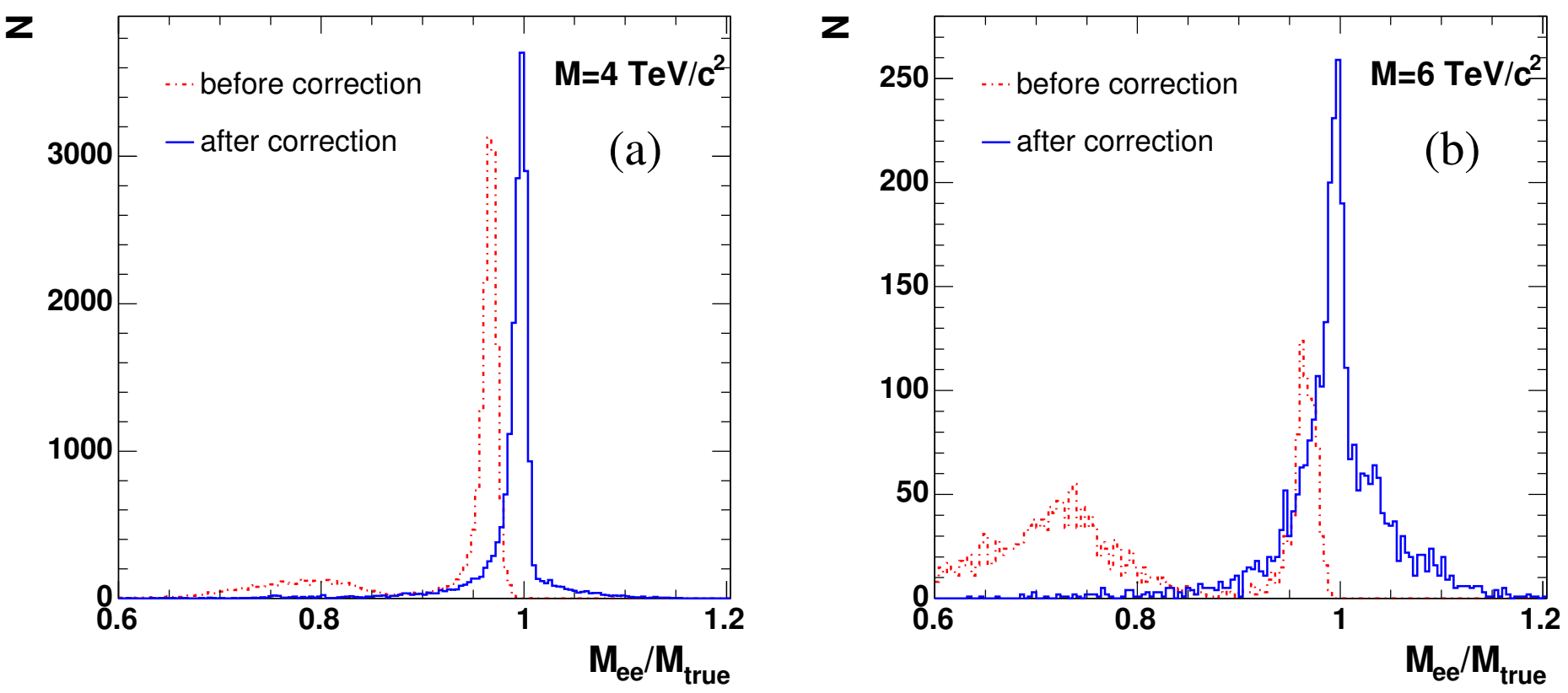

Fig. 8.11: Ratio $M_{\rm ee}/M_{\rm true}$ before and after corrections for KK $Z$ boson production, for $M = 4$ TeV (a) and $M = 6$ TeV (b).

– FSR recovery: hard photon emission from Final State Radiation can induce the detection in the event of a third energetic SC. If a SC with $E > 300$ GeV satisfying the $H/E$ and isolation cuts is observed very close to the SC of the electron candidates ($\Delta R < 0.1$ ), this additional SC is associated to the corresponding electron.

The signal efficiency for the three heavy resonance production models is typically of the order of 80%.

### 8.4.2 *Mass peak distributions*

The resonance mass is reconstructed from the energies and angles of the 2 electron candidates, after the selection cuts and energy corrections mentioned above. Figures 8.11(a) and (b) show the ratio of the reconstructed and the true masses, $M_{\rm ee}/M_{\rm true}$, before and after energy corrections for KK $Z$ production with $M$ =4 and 6 TeV, respectively. The peaks at low values of $M_{\rm ee}/M_{\rm true}$ correspond to events with saturated ECAL electronics. The final resolution on the resonance mass is around 0.6% for events with no saturation, and 7% in case of saturation.

Figure 8.12(a) presents the signal and the Drell-Yan background for KK $Z$ boson production with $M = 4$ TeV; Figure 8.12(b) for $Z'$ boson production with $M = 3.0$ TeV; Figure 8.12(c) for graviton production with $M = 1.5$ TeV and coupling parameter $c = 0.01$.

### 8.4.3 *Discovery potential of CMS*

The discovery potential of a new physics resonance is determined using the likelihood estimator $S$ based on event counting [110], suited for small event samples:

$$S = \sqrt{2[(N_{\rm s} + N_{\rm b}) \log(1 + \frac{N_{\rm s}}{N_{\rm b}}) - N_{\rm s}]}, \tag{8.32}$$

where $N_{\rm s}$ (resp. $N_{\rm b}$) is the number of signal (resp. background) events. The discovery limit is defined by $S > 5$.

The number of signal and background events, $N_{\rm s}$ and $N_{\rm b}$, computed for a given mass window around the peak, are presented in Table 8.8 for the three models, together with the corresponding significance, for an integrated luminosity of 30 fb$^{-1}$. The $5\sigma$ discovery limits as a function of mass are given in

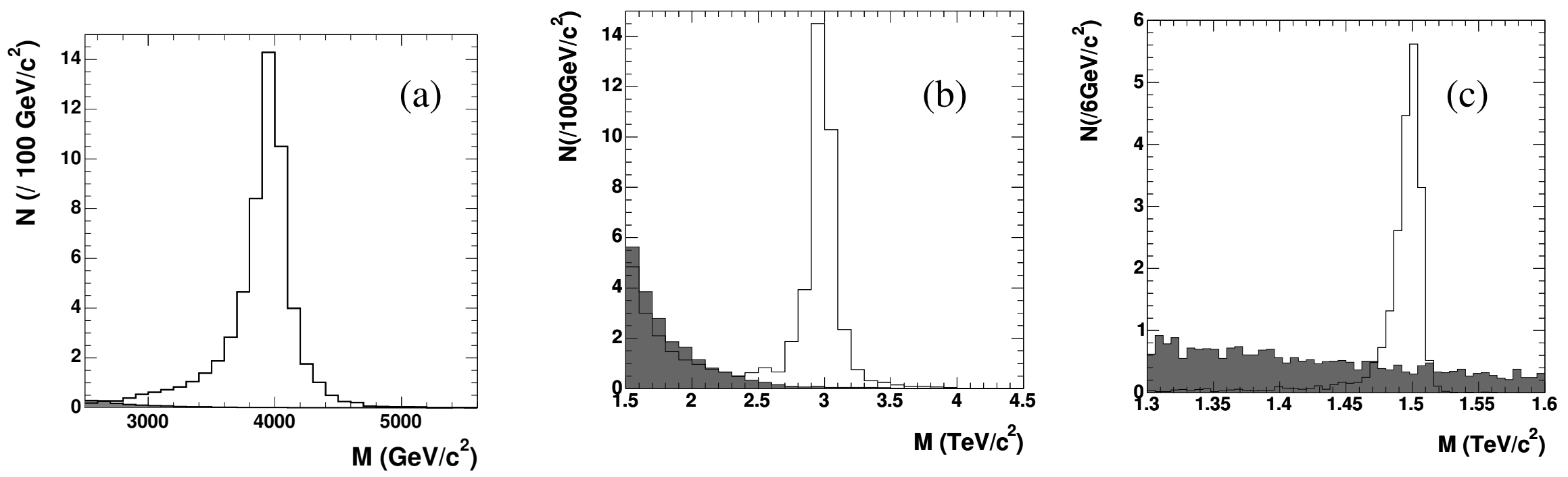

Fig. 8.12: Resonance signal (white histograms) and Drell-Yan background (shaded histograms) for KK $Z$ boson production with $M = 4.0$ TeV (a), SSM $Z'$ boson production with $M = 3.0$ TeV (b), and graviton production with $M = 1.5$ TeV, $c = 0.01$ (c), for an integrated luminosity of 30 fb$^{-1}$.

Fig. 8.13(a) and Fig. 8.13(b), for KK $Z$ boson production and $Z'$ production (for the 6 considered models), respectively. In the graviton case, the $5\sigma$ discovery plane as a function of the coupling parameter $c$ and the resonance mass is given in Fig. 8.13(c).

For KK $Z$ bosons, a $5\sigma$ discovery can be achieved for a resonance mass up to $M = 4.97$ TeV for an integrated luminosity of 10 fb$^{-1}$, $M = 5.53$ TeV for 30 fb$^{-1}$ and $M = 5.88$ TeV for 60 fb$^{-1}$.

For $Z'$ boson production, with an integrated luminosity of 30 fb$^{-1}$, a $5\sigma$ discovery can be extracted for masses up to 3.31 TeV for model $\psi$ and up to 4.27 TeV for model ARLM.

For gravitons, with an integrated luminosity of 30 fb$^{-1}$, a $5\sigma$ discovery can be extracted for masses up to 1.64 TeV for $c = 0.01$ and up to 3.81 TeV for $c = 0.1$. Similar discovery limits are obtained in the graviton di-photon decay channel (see Section 9.5).

The $5\sigma$ discovery limits on the resonance masses for 10, 30 and 60 fb$^{-1}$ are summarized in Table 8.9.

For KK $Z$ boson production, the luminosities needed for a $5\sigma$ discovery are 1.5, 4.0, 10.8, 29.4, and 81.4 fb$^{-1}$ for $M = 4.0$, 4.5, 5.0, 5.5 and 6.0 TeV, respectively; for SSM $Z'$ boson production, they are 0.015, 3.0 and 260 fb$^{-1}$ for $M = 1$, 3 and 5 TeV; for graviton production, most of the interesting region of the (mass, coupling) plane is already covered with 10 fb$^{-1}$.

For KK $Z$ and $Z'$ production, a K factor of 1 was taken for both the signal and the Drell-Yan background, since heavy $Z$ production interferes with $Z/\gamma$ Drell-Yan production. For the graviton analysis, as little interference is present with the Standard Model processes, a K factor of 1.0 is used for the signal and of 1.3 for the Drell-Yan background, in order to take into account the higher order terms in the cross section. The latter number comes from the CDF analysis [111] and is compatible with the K factor obtained from theoretical computations [112, 113].

### *8.4.4 Systematic uncertainties*

The uncertainty coming from the choice of the parton distribution function (PDF) was investigated using the set of 20 positive and 20 negative errors, of the CETQ6.1M "best fit" parameterization [114–116]. For each event, a weight factor is computed according to the $x_1$, $x_2$, and $Q^2$ variables, for each of the 40 PDF errors, in the case of graviton production with $M = 1.5$ TeV ($c = 0.01$) and $M = 3.5$ TeV ($c = 0.1$). The uncertainties on the PDF modify the number of signal events by a factor 1.20 (positive deviations) and 0.86 (negative deviations) for $M = 1.5$ TeV ($c = 0.01$). The corresponding numbers for $M = 3.5$ TeV ($c = 0.1$) are 1.47 and 0.78. For the Drell-Yan background, the re-weighting effects on the numbers of events are 1.065 and 0.941 for masses around 1.5 TeV , and 1.19 and 0.88 for masses around 3.5 TeV. For an integrated luminosity of 30 $\text{fb}^{-1}$, the significances with the ”best fit” and with the positive/negative deviations are equal respectively to 6.40 and 7.25/5.78 for $M = 1.5$ TeV ($c = 0.01$), and to 6.83 and 8.54/5.93 for $M = 3.5$ TeV ($c = 0.1$). A lower dependence is observed for the KK $Z$ and $Z'$ channels, which are produced by quark-antiquark annihilation. For KK $Z$ boson production at $M = 4$ TeV with an integrated luminosity of 30 $\text{fb}^{-1}$, the significances with the ”best fit” and with the positive/negative errors are equal respectively to 22.5 and 23.3/21.9.

Table 8.8: Number of events for resonant signal, $N_\mathrm{s}$, and for Drell-Yan background, $N_\mathrm{b}$, and corresponding significances as defined by Eq. 8.32, for an integrated luminosity of 30 fb$^{-1}$. The masses $M$ and the mass windows $M_\mathrm{w}$ are in TeV.

| | KK $Z$ | | SSM $Z'$ | | G, $c = 0.01$ | G, $c = 0.1$ |
|---|---|---|---|---|---|---|
| $M$ | 4.0 | 6.0 | 1.0 | 5.0 | 1.5 | 3.5 |
| $M_\mathrm{w}$ | 3.5-4.5 | 5.0-6.7 | 0.92-1.07 | 4.18-5.81 | 1.47-1.52 | 3.30-3.65 |
| $N_\mathrm{s}$ | 50.6 | 1.05 | 72020 | 0.58 | 18.8 | 7.30 |
| $N_\mathrm{b}$ | 0.13 | 0.005 | 85.5 | 0.025 | 4.16 | 0.121 |
| $S$ | 22.5 | 3.0 | 225 | 1.63 | 6.39 | 6.83 |

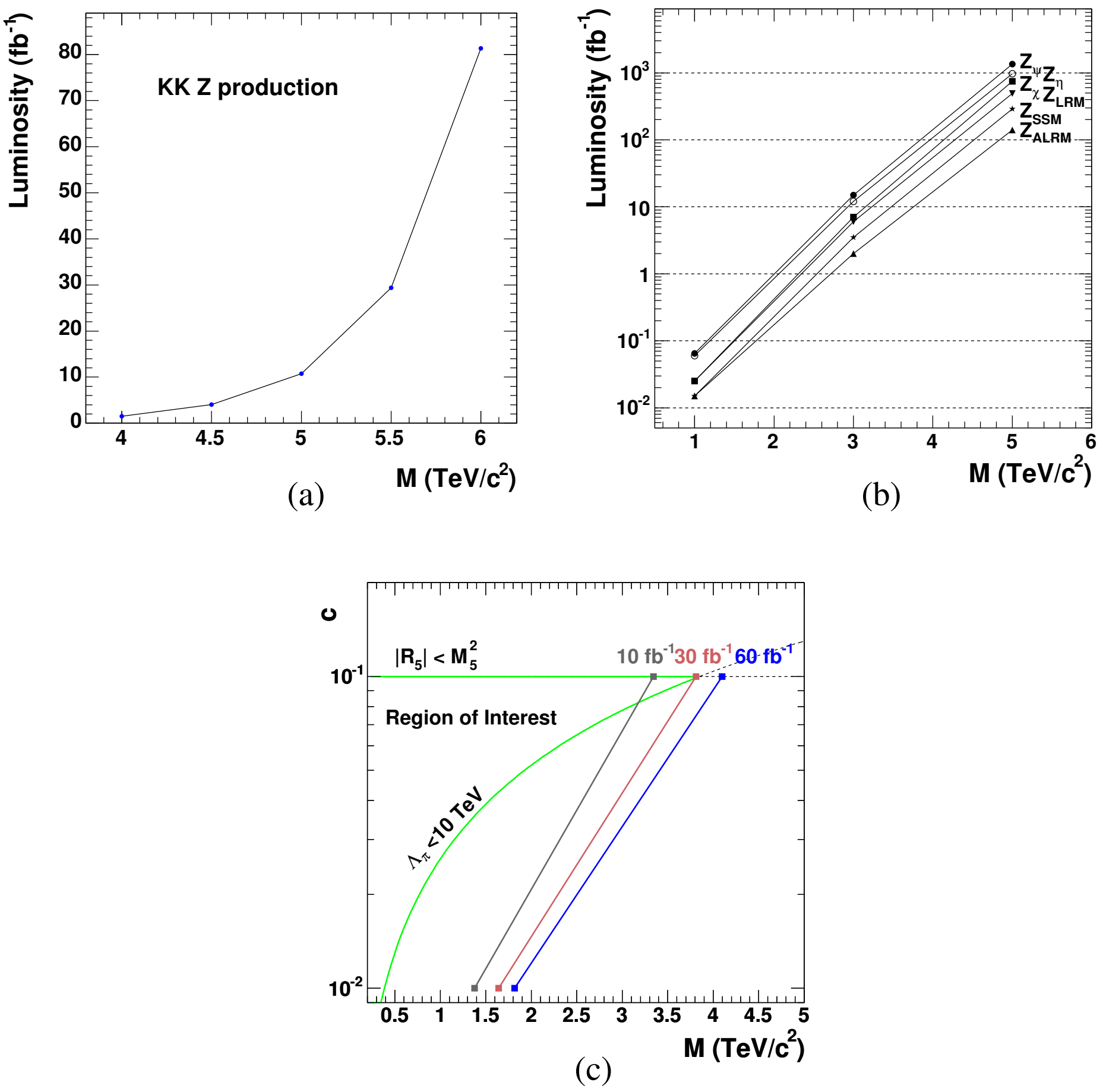

Fig. 8.13: $5\sigma$ discovery limit as a function of the resonance mass for KK $Z$ boson production (a), for the 6 $Z'$ models (b); $5\sigma$ discovery plane for graviton production as a function of the coupling parameter $c$ and the graviton mass (c).

Table 8.9: The $5\sigma$ discovery limit on the resonance mass (given in TeV), for an integrated luminosity of 10, 30 and 60 $\text{fb}^{-1}$.

| Model | Luminosity ($\text{fb}^{-1}$) | | |
|---|---|---|---|
| | 10 | 30 | 60 |
| KK $Z$ | 4.97 | 5.53 | 5.88 |
| $Z'$ ($\psi$) | 2.85 | 3.31 | 3.62 |
| $Z'$ (ALRM) | 3.76 | 4.27 | 4.60 |
| G ($c = 0.01$) | 1.38 | 1.64 | 1.82 |
| G ($c = 0.1$) | 3.34 | 3.81 | 4.10 |

Changing to 1 the value of the K factor of the DY background for RS graviton production increases the significance from 6.39 to 6.87 ($M = 1.5$ TeV , $c = 0.01$) and from 6.83 to 7.09 ( $M = 3.5$ TeV , $c = 0.1$). The discovery limits increase respectively from 1.64 to 1.68 TeV and from 3.81 to 3.84 TeV.

### *8.4.5 Identification of new particles*

Once a resonance is found, information will be gained on its characterization from the study of other decay channels, like $\gamma\gamma$ (see Section 9.5), of angular distributions and of asymmetries, in view of the spin determination (see also [117]).

As an example, RS gravitons with spin 2 can be distinguished from the Standard Model background and $Z'$ bosons with spin 1 using the distribution of the $\cos\theta^*$ variable, computed as the cosine of the polar angle between the electron and the boost direction of the heavy particle in the latter rest frame. In addition to the cuts defined above, the electron and positron candidates are requested to have opposite charges, in order to identify the electron, from which the $\cos\theta^*$ variable is computed.

The $\cos\theta^*$ distributions for graviton production with $M$ = 1.25 TeV, $c = 0.01$, and $M = 2.5$ TeV, $c = 0.1$, are presented in Fig. 8.14, for an integrated luminosity of 100 $\text{fb}^{-1}$. The error bars represent the corresponding statistical uncertainties, applied to the signal distribution obtained from a large statistics simulation. The spin-2 hypothesis is compared to the spin-1 hypothesis (red curve in the figures), formed by the Drell-Yan production (Figs. 8.14(a) and (b)) or the ALRM $Z'$ production (Figs. 8.14(c) and (d)). For graviton production, the expected background is included in the $\cos\theta^*$ distributions.

The spin 2 of RS gravitons can be determined by contrast with the Drell-Yan production or the $Z'$ boson production for an integrated luminosity of 100 $\text{fb}^{-1}$ up to 1.25 TeV for $c = 0.01$ and 2.5 TeV for $c = 0.1$.

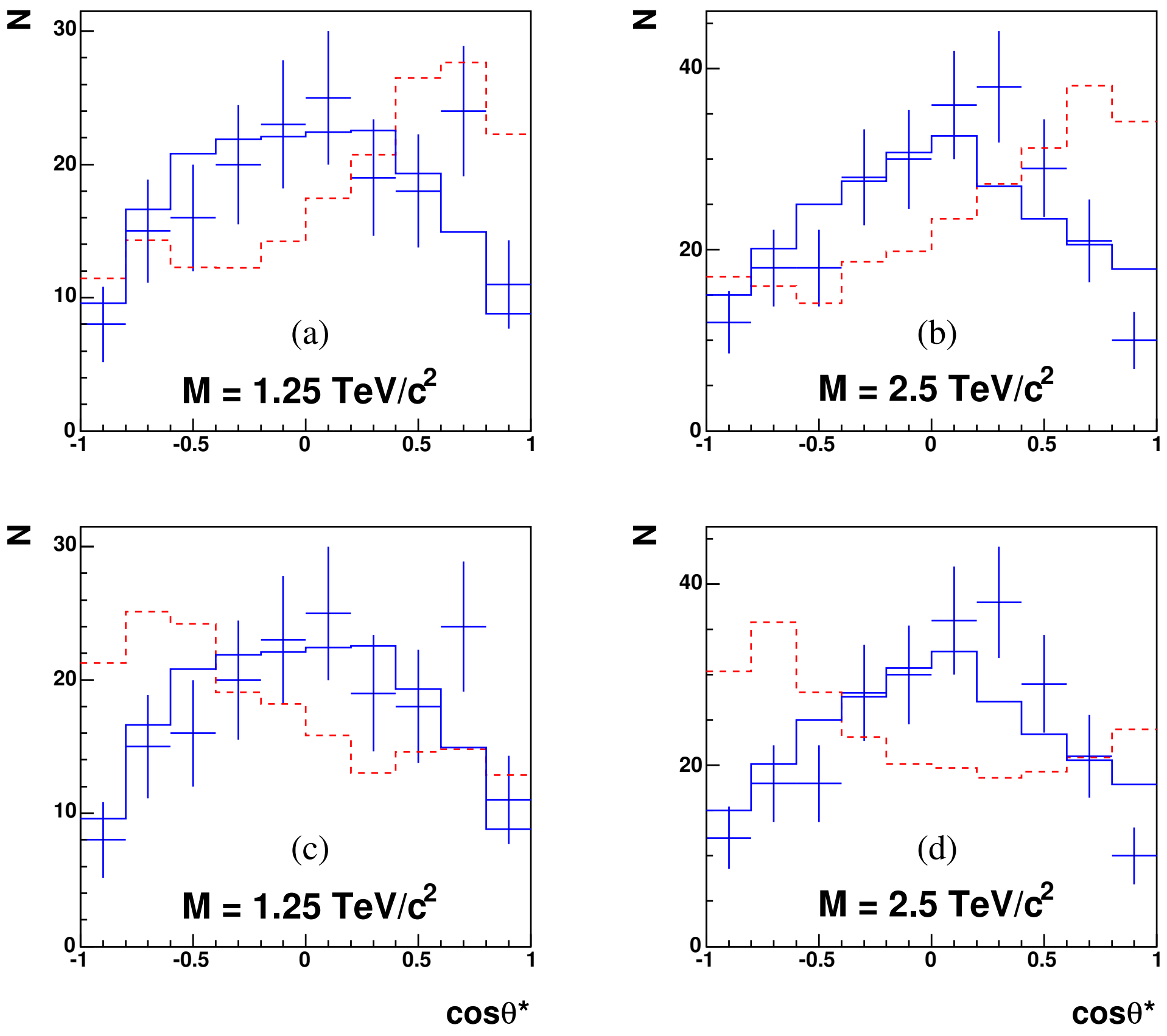

Fig. 8.14: Distributions of $\cos\theta^*$ for graviton production (blue curves) and for Drell-Yan production (red curves) normalized to the signal, for $M$ = 1.25 TeV (a) and 2.5 TeV (b), and for $Z'$ boson (ALRM model) (red curves), normalized to the signal, for $M$ = 1.25 TeV (c) and 2.5 TeV (d), with an integrated luminosity of 100 $\mathrm{fb}^{-1}$. The error bars represent the "1-experiment" distribution for the graviton production. For graviton production, the expected background is included in the $\cos\theta^*$ distributions.

# 9 WARPED EXTRA DIMENSIONS AND THE RANDALL-SUNDRUM MODEL

## 9.1 Introduction

*JoAnne L. Hewett and Thomas G. Rizzo*

The Randall-Sundrum (RS) model of localized gravity [1] offers a potential solution to the hierarchy problem that can be tested at present and future accelerators [2, 3]. In the original (and most simple) version of this model, all of the Standard Model (SM) fields are confined to one of two branes that are sited at the $S_1/Z_2$ orbifold fixed points embedded in a 5-dimensional anti-de Sitter space (AdS$_5$). The theory is described by the metric $ds^2 = e^{-2k|y|}\eta_{\mu\nu}dx^\mu dx^\nu - dy^2$, with $y$ being the extra dimension and where $r_c$ is the compactification radius; thus the two branes are separated by a distance $\pi r_c$. The parameter $k$ characterizes the curvature of the 5-dimensional space and is naturally of order the Planck scale. The two branes form the boundaries of the AdS$_5$ slice and gravity is localized on the Planck brane located at $y = 0$. Mass parameters on the TeV brane, located at $y = r_c\pi$, are red-shifted compared to those on the $y = 0$ brane and are typically given by $\Lambda_\pi = \overline{M}_{Pl}e^{-kr_c\pi}$, where $\overline{M}_{Pl}$ is the reduced Planck scale. In order to address the hierarchy problem, $\Lambda_\pi \sim$ TeV and hence the separation between the two branes, $r_c$, must have a value of $kr_c \sim 11 - 12$. It has been [4] demonstrated that this quantity can be naturally stabilized to this range of values.

There are very many variations on this basic model mostly having to do with placing at least some of the SM fields into the bulk for model building purposes [5–8]. In many cases it is also useful to include brane kinetic terms for these fields [9–15] to increase model flexibility. In almost all cases where a fundamental Higgs field is present it remains on the TeV brane (without Kaluza-Klein (KK) excitations) even when fermion and gauge fields are in the bulk. Up until recently this was thought to be necessary to avoid fine-tuning and phenomenological requirements. It has recently been shown that the fundamental Higgs can also be a bulk field [16, 17] which can lead to significant changes in Higgs phenomenology.

### *9.1.1 Graviton phenomenology*

Since in all cases the graviton is a bulk field in 5-d we not only have the familiar zero mode massless graviton but also the massive tower of Kaluza-Klein excitations. The wavefunctions of these states in the extra dimension (in this simple version of the model) are given by

$$\chi_n = \frac{e^{2\sigma}}{N_n} J_2\Big(x_n \epsilon e^{\sigma}\Big)\,, \tag{9.1}$$

with $J_2$ the usual Bessel function, $\sigma = k|y|$ and $N_n$ a normalization factor. The KK graviton masses are given by $m_n = x_n k\epsilon$ with $\epsilon = e^{-\pi k r_c} \simeq 10^{-16}$, while the $x_n$ roots can be obtained from the equation

$$J_1(x_n) = 0\,. \tag{9.2}$$

The wavefunction of the ordinary massless graviton is flat. Note that since $k\epsilon \sim$ TeV, the graviton KK excitations are TeV-scale.

In this simplest scenario, the graviton KK phenomenology is governed by 2 parameters, $k/\overline{M}_{Pl}$ and $m_1$ (or $\Lambda_\pi$). The action is computed by performing a linear expansion of the flat metric $g_{AB} = e^{-2ky}(\eta_{AB} + 2h_{AB}/M_5^{3/2})$ , which for this scenario includes the warp factor. The interactions of the graviton KK tower with the SM fields on the TeV-brane are given by

$$\mathcal{L} = -\frac{1}{\overline{M}_{Pl}} T^{\mu\nu}(x) h^0_{\mu\nu}(x) - \frac{1}{\Lambda_\pi} T^{\mu\nu}(x) \sum_{n=1}^{\infty} h^{(n)}_{\mu\nu}(x)\,, \tag{9.3}$$

where $T^{\mu\nu}$ is the conserved stress-energy tensor. The zero-mode decouples and the couplings of the excitation states are inverse-TeV strength. The hallmark signature for this scenario [2] is the presence of TeV-scale spin-2 graviton resonances at colliders; the KK spectrum in $e^+e^- \to \mu^+\mu^-$, taking $m_1 = 500$ GeV, is shown in Fig. 9.1. Note that the curvature parameter controls the width of the resonance. The LHC can discover these resonances in the Drell-Yan channel if $\Lambda_\pi < 10$ TeV [3], provided that the resonance width is not too narrow, and determine their spin-2 nature via the angular distributions of the final-state lepton pairs [18] if enough statistics are available. This is illustrated in Fig. 9.2 which displays the LHC search reach and the present experimental and theoretical constraints on the RS parameter space.

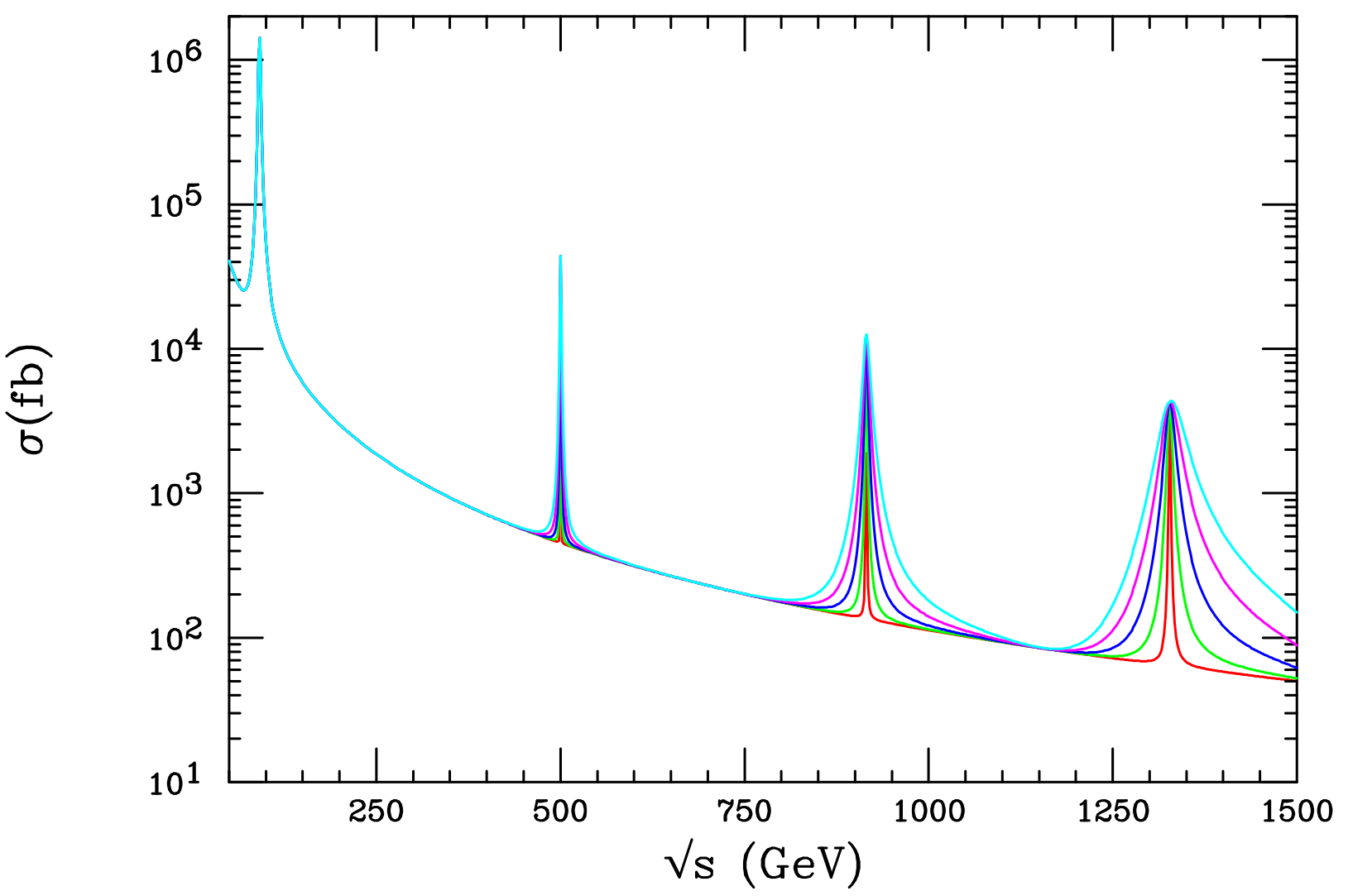

Fig. 9.1: The cross section for $e^+e^- \to \mu^+\mu^-$ including the exchange of a KK tower of gravitons in the RS model with $m_1 = 500$ GeV. The various curves correspond to $k/\overline{M}_{Pl}$ in the range $0.01 - 0.1$. From [2].

If the KK gravitons are too massive to be produced directly, their contributions to fermion pair production may still be felt via virtual exchange. In this case, the uncertainties associated with a cut-off (as present in the large extra dimensions scenario) are avoided, since there is only one additional dimension and thus the KK states may be neatly summed. The resulting sensitivity to the scale $\Lambda_\pi$ at the LHC is $\Lambda_\pi = 3.0 - 20.0$ TeV as $k$ varies in the range $k/\overline{M}_{Pl} = 0.01 - 1.0$. The 1 TeV International Linear Collider extends this reach by a factor of $1.5 - 2$.

Extensions of this simplest scenario modify the graviton KK spectrum and couplings. If the SM fields are allowed to propagate in the bulk, then each SM state also expands into a KK tower. The couplings of the bulk SM fields to the graviton KK states are cataloged in [3], where it is shown that the zero-mode fermion and gauge couplings to the graviton KK excitations are substantially reduced compared to the case where the SM fields are constrained to the TeV-brane. Graviton searches can then become more difficult in this scenario and are highly dependent on the explicit localization of the SM fields in the $5^{th}$ dimension.

The inclusion of brane kinetic terms, which arise from higher order effects, also alter the graviton KK phenomenology [10]. The graviton KK masses are again given by $m_n = x_n \Lambda_\pi k/\overline{M}_{Pl}$, where the $x_n$ are now roots of the equation $J_1(x_n) - \gamma_\pi x_n J_2(x_n) = 0$. Here, $\gamma_\pi$ represents the coefficient of the boundary term for the TeV-brane and is naturally of order unity. The couplings are modified to be

$$\mathcal{L} = -\frac{1}{\overline{M}_{Pl}} T^{\mu\nu}(x) h^0_{\mu\nu}(x) - \frac{1}{\Lambda_\pi} T^{\mu\nu}(x) \sum_{n=1}^{\infty} \lambda_n h^{(n)}_{\mu\nu}(x) \,, \tag{9.4}$$

where $\lambda_n$ depends on the coefficient of the boundary terms on both branes. This yields a dramatic reduction of the graviton KK couplings to SM fields on the TeV-brane, even for small values of the brane kinetic term coefficients. The resulting degradation in the graviton search reach at the LHC is displayed in Fig 9.3 for 100 fb$^{-1}$ of integrated luminosity. From this figure, it is clear that the LHC can no longer cover all of the interesting parameter space for this model. For example, a first graviton KK excitation of mass 600 GeV with $k/\overline{M}_{Pl} = 0.01$ may still miss detection.

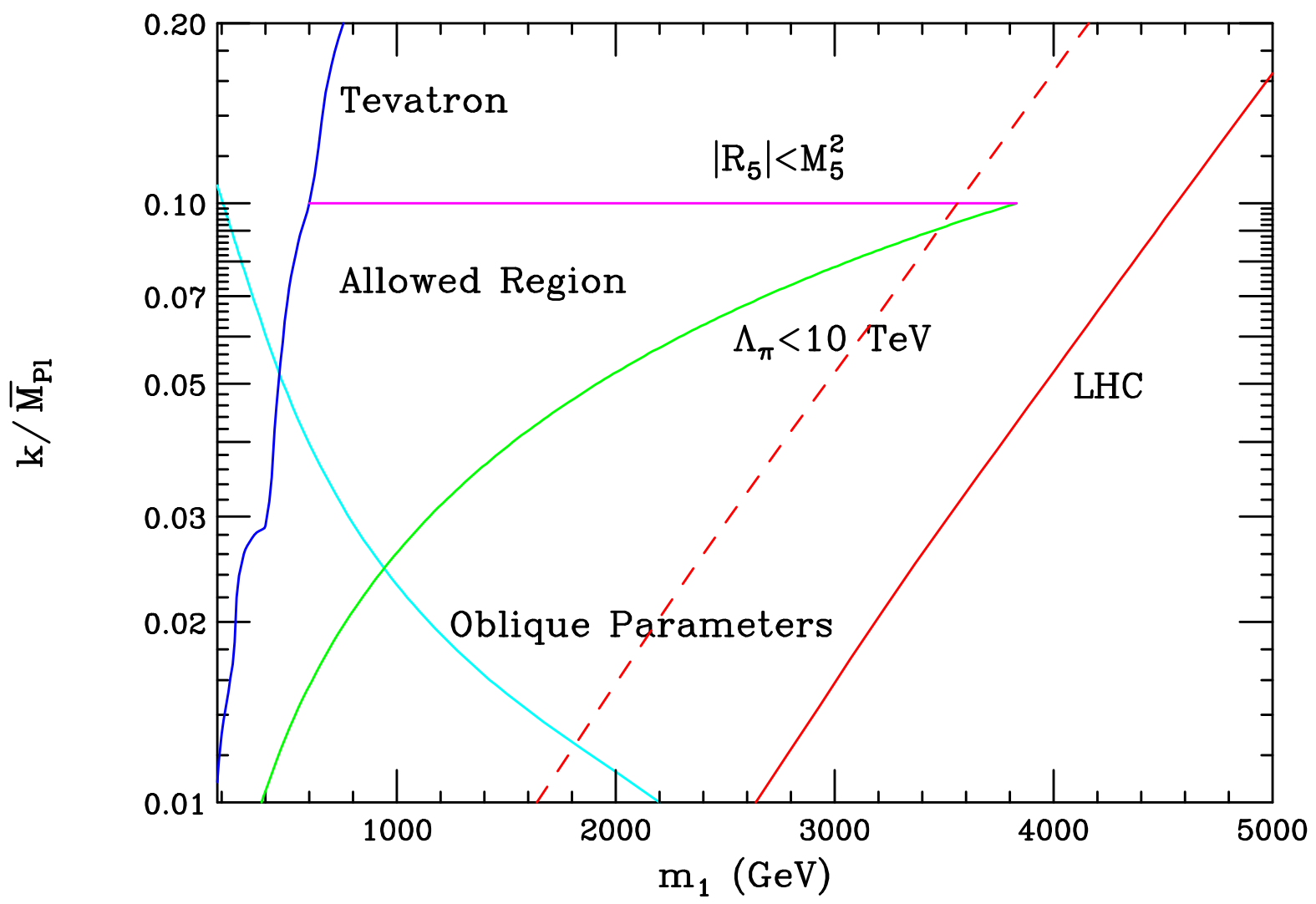

Fig. 9.2: Summary of experimental and theoretical constraints on the RS model in the two-parameter plane $k/\overline{M}_{Pl} - m_1$. The allowed region lies in the center. From [3].

From a theoretical perspective, the RS model may be viewed as an effective theory whose low energy features originate from a full theory of quantum gravity, such as string theory. One may thus expect that a more complete version of this scenario admits the presence of additional dimensions compactified on a manifold $\mathcal{M}^\delta$ of dimension $\delta$. The existence of an extra manifold also modifies the conventional RS phenomenology and collider signatures [19]. For the simplest scenario of an additional $S^1$ manifold, the RS metric is expanded to $ds^2 = e^{-2kr_c\phi}\eta_{\mu\nu}dx^\mu dx^\nu + r_c^2 d\phi^2 + R^2 d\theta^2$ , where $\theta$ parameterizes the $S^1$, and $R$ represents its radius. The masses of the KK states are now given by $m_{n\ell} = x_{n\ell}\Lambda_\pi k/\overline{M}_{Pl}$, where the $x_{n\ell}$ are solutions of the equation $2J_\nu(x_{n\ell}) + x_{n\ell}J'_\nu(x_{n\ell}) = 0$ , with $\nu \equiv \sqrt{4 + (\ell/kR)^2}$. The KK mode number $\ell$ corresponds to the orbital excitations, while $n$ denotes the usual RS $AdS_5$ mode levels. The couplings of the $m_{n\ell}$ graviton KK states are then given by

$$\mathcal{L} = -\frac{1}{\overline{M}_{Pl}} T^{\mu\nu}(x) h^0_{\mu\nu}(x) - \frac{1}{\Lambda_\pi} T^{\mu\nu}(x) \sum_{n=1}^{\infty} \xi(n\ell) h^{(n,\ell)}_{\mu\nu}(x) \, , \tag{9.5}$$

where $\xi(n\ell)$ depends on $k$ , $R$, and $x_{n\ell}$ [19]. In particular, the addition of the $S^\delta$ background to the RS setup results in the emergence of a forest of graviton KK resonances. These originate from the orbital excitations on the $S^\delta$ and occur in between the original RS resonances. A representative KK spectrum is depicted in Fig. 9.3 for the additional $S^1$ manifold.

Finally, we note that the graviton KK spectrum and couplings to matter fields on the TeV brane will be modified if higher curvature terms are present in the action [20].

### *9.1.2 Radion basics*

Fluctuations about the stabilized RS configuration allow for two massless excitations described by the metric tensor. The first of these corresponds to the usual graviton as discussed above while the second is a new scalar field essentially arising from the $\sim g_{55}$ component of the 5-d metric and is known as the radion ($\phi_0$). Recall that when the 5-d graviton field is decomposed into 4-d fields it consists of a tower of tensor fields $\sim g_{\mu\nu}$ in addition to a tower of vectors $\sim g_{5\mu}$ and a tower of scalars $\sim g_{55}$. When the graviton KK fields acquire mass by eating the corresponding vector and scalar fields. With the orbifold symmetry imposed only a massless graviton and radion remain. This scalar radion field corresponds to a quantum excitation of the separation between the two branes. The mass of the radion is proportional to the backreaction of the bulk scalar vacuum expectation value (vev) on the metric and is correlated with the stabilization mechanism.

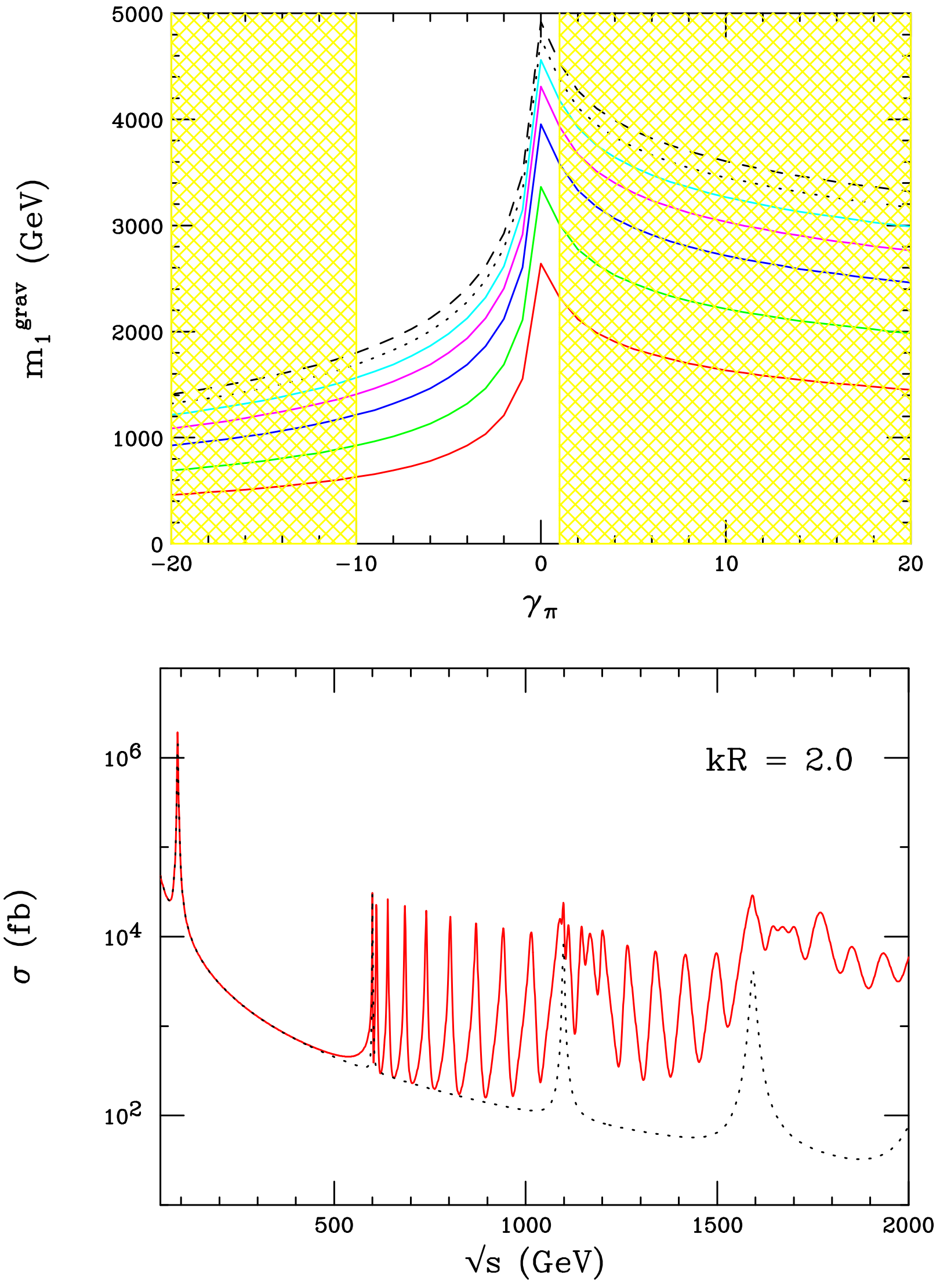

Fig. 9.3: Top: Search reach for the first graviton KK resonance employing the Drell-Yan channel at the LHC with an integrated luminosity of 100 $fb^{-1}$ as a function of the boundary term coefficient $\gamma_\pi$ assuming $\gamma_0 = 0$. From bottom to top on the RHS of the plot, the curves correspond to $k/\overline{M}_{Pl} = 0.01, 0.025, 0.05, 0.075, 0.10, 0.125$ and $0.15$, respectively. The unshaded region is that allowed by naturalness considerations and the requirement of a ghost-free radion sector. From [10]. Bottom: The solid (red) curve corresponds to the cross section for $e^+e^- \rightarrow \mu^+\mu^-$ when the additional dimension is orbifolded, i.e., for $S^1/Z_2$, with $m_{10} = 600$ GeV, $k/\overline{M}_{Pl} = 0.03$ and $kR = 2.0$. The result for the conventional RS model is also displayed, corresponding to the dotted curve. From [19].

Generally, one expects that the radion mass should be in the range of a few $\times 10$ GeV $\leq m_{\phi_0} \leq \Lambda_\pi$, where the lower limit arises from radiative corrections and the upper bound is the cutoff of the effective field theory. The radion mass $m_{\phi_0}$ is then expected to be below the scale $\Lambda_\pi$ implying that the radion may be the lightest new field present in the RS model. A basic introduction to the radion and its origins can be found in [21, 22] while some basic phenomenology can be found in [23–25].

Unlike the graviton which couples directly to the stress-energy tensor, $T^\mu_\nu$, as seen above, the radion's couplings to SM matter fields on the TeV brane are proportional to the trace of the stress tensor, $T^\mu_\mu$:

$$\mathcal{L}_{eff} = -\frac{\phi_0(x)\, T^\mu_\mu}{\sqrt{6}\Lambda_\pi} \,. \tag{9.6}$$

These simplified couplings can be modified by the existence of brane kinetic terms for the graviton [10, 16] as well as when SM fields are placed in the bulk. The explicit couplings of the (unmixed) radion to SM fields are qualitatively similar to that of the Higgs, e.g., for fermions and massive gauge bosons we have

$$\mathcal{L} = \frac{1}{v}\gamma(m_f \bar{f} f - m_V^2 V_\mu V^\mu)\phi_0 \,, \tag{9.7}$$

where $v$ is the SM Higgs vev and $\gamma$ is the ratio

$$\gamma = \frac{v}{\sqrt{6}\Lambda_\pi} \,, \tag{9.8}$$

which we might expect to be of order a few percent. The corresponding coupling to gluon pairs occurs through the trace anomaly and can be written as

$$\mathcal{L} = c_g^0 \frac{\alpha_s}{8\pi} G_{\mu\nu} G^{\mu\nu} \phi_0 \,, \tag{9.9}$$

with

$$c_g^0 = \frac{1}{2v}\gamma[F_g - 2b_3] \,. \tag{9.10}$$

Here $b_3 = 7$ is the $SU(3)$ $\beta$-function and $F_g$ is a well-known kinematic loop-function [26] of the ratio of masses of the top-quark to the radion while the second term originates from the anomaly. Similarly the radion coupling to two photon pairs is now given by

$$\mathcal{L} = c_\gamma^0 \frac{\alpha_{em}}{8\pi} F_{\mu\nu} F^{\mu\nu} \phi_0 \,, \tag{9.11}$$

where

$$c_\gamma^0 = \frac{1}{v}\gamma[F_\gamma - (b_2 + b_Y)] \,. \tag{9.12}$$

Here, $b_2 = 19/6$ and $b_Y = -41/6$ are the $SU(2) \times U(1)$ $\beta$-functions and $F_\gamma$ is another well-known kinematic function [26] of the ratios of the $W$ boson and top-quark masses to the radion and the second term again originates from the trace anomaly.

### *9.1.3 Radion-Higgs mixing*

Since the radion and Higgs are both real scalar fields they might mix. In fact such a mixing can proceed through a dimension-4 brane term when there is a single Higgs doublet on the TeV brane, through an interaction term of the form

$$S_{rH} = -\xi \int d^4x \sqrt{-g_{ind}} R^{(4)}[g_{ind}] H^\dagger H \,. \tag{9.13}$$

Here $H$ is the Higgs doublet field, $R^{(4)}[g_{ind}]$ is the 4-d Ricci scalar constructed out of the induced metric $g_{ind}$ on the SM brane, and $\xi$ is a dimensionless mixing parameter assumed to be of order unity (since there is no reason why such an operator should be suppressed) and with unknown sign. We note that the structure of this mixing is quite different when the Higgs field is in the bulk [16]. In any case, such an interaction term leads to both gauge kinetic and mass mixing between the unmixed Higgs, $h_0$, and the radion.

The resulting Lagrangian can be diagonalized by a set of field redefinitions and rotations [22]

$$\begin{aligned} h_0 &= Ah + B\phi\,, \\ \phi_0 &= Ch + D\phi\,, \end{aligned} \tag{9.14}$$

with

$$\begin{aligned} A &= \cos\theta - 6\xi\gamma/Z\sin\theta\,, \\ B &= \sin\theta + 6\xi\gamma/Z\cos\theta\,, \\ C &= -\sin\theta/Z\,, \\ D &= \cos\theta/Z\,, \end{aligned} \tag{9.15}$$

where $h\,,\phi$ represent the physical, mass-eigenstate fields, and

$$Z^2 = 1 + 6\xi(1-6\xi)\gamma^2\,, \tag{9.16}$$

The factor $Z$ serves to bring the physical radion kinetic term to canonical form and as such it must satisfy $Z > 0$ to avoid ghosts. For a fixed value of $\gamma$ this implies that the range of $\xi$ is bounded, i.e., $\xi_- \le \xi \le \xi_+$, where

$$\xi_\pm = \frac{1}{12}[1 \pm (1 + 4/\gamma^2)^{1/2}]\,, \tag{9.17}$$

For example, if $\gamma$ takes on the natural values, e.g., $\gamma = 0.1$ then $\xi$ must lie in the approximate range $-1.585 \le \xi \le 1.752$. The masses of the physical states, $\phi, h$, are then given by

$$m_\pm^2 = \frac{1}{2}\left[T \pm \sqrt{T^2 - 4F}\right]\,, \tag{9.18}$$

where $m_+(m_-)$ is the larger(smaller) of the two masses and

$$\begin{aligned} T &= (1+t^2)m_{h_0}^2 + m_{\phi_0}^2/Z\,, \\ F &= m_{h_0}^2 m_{\phi_0}^2/Z^2\,, \end{aligned} \tag{9.19}$$

with $m_{h_0,\phi_0}$ being the weak interaction eigenstate masses and $t = 6\xi\gamma/Z$. This mixing will clearly affect the phenomenology of both the radion and Higgs fields, for example, the bounds on the Higgs mass from LEP searches. Here, we note the modifications to the properties of the Higgs boson, in particular its decay widths and branching fractions, induced by this mixing and find that substantial differences from the SM expectations for the Higgs can be obtained. In the case where the Higgs is in the bulk a general procedure similar to the above can be followed to get to the mass eigenstate basis [16]. The main difference, in addition to the existence of the Higgs KK tower (all of whose members now mix with the radion), is that the Higgs-radion mixing term now occurs both in the bulk as well as on both branes.

### 9.2 Higgs–radion phenomenology

*Daniele Dominici and John F. Gunion*

The main consequence of Higgs-radion mixing is a modification of the prospects for discovering a light Higgs boson at the LHC [23,25,27]. In particular, this mixing could suppress or enhance the signal rate in the channel $gg \to H \to \gamma\gamma$. The effect of this mixing has been studied [25] by implementing the new Higgs and radion couplings in the HDECAY program [28]. Let us first recall the parameters of the model: when the mixing parameter $\xi \neq 0$, there are four independent parameters that must be specified to fix the mixing coefficients $A, B, C, D$ ($d, c, a, b$ in the notation of [25]) defining the mass eigenstates, $h$ and $\phi$, in terms of the unmixed Higgs and radion states, $h^0$ and $\phi^0$. These are

$$\xi\,, \quad \Lambda_\phi = \sqrt{6}\Lambda_\pi\,, \quad m_h\,, \quad m_\phi\,, \tag{9.20}$$

where $m_h$ and $m_\phi$ denote the eigenstate masses of the Higgs and of the radion, defined so that $h \to h^0$ and $\phi \to \phi^0$ in the $\xi \to 0$ limit. An additional parameter is required to determine the phenomenology of the scalar sector, including all possible decays: the mass, $m_1$, of the first KK graviton excitation.

Let us first review how the most relevant Higgs decays are modified. In Fig. 9.4, we plot the branching ratios for $h \to b\bar{b}$, $gg$, $WW^*$, $ZZ^*$ and $\gamma\gamma$ as a function of the mixing parameter $\xi$, taking $m_h = 120$ GeV and $\Lambda_\phi = 5$ TeV. Results are shown for three different $m_\phi$ values: 20, 55 and 200 GeV. These plots are limited to $\xi$ values allowed by the theoretical and experimental constraints [25]. Large values for the $gg$ branching ratio, due to the anomalous contribution to the $hgg$ coupling, suppress the other branching ratios to some extent. The anomalous contribution to the $h\gamma\gamma$ coupling is less important due to presence of the large $W$ loop contribution in this case. Second, for $m_\phi = 55$ GeV, $BR(h \to \phi\phi)$ is large at large $|\xi|$ and suppresses the conventional branching ratios. For larger $|\xi|$, the changes in the branching ratio of the $h$ with respect to the SM are at a level that would be observable, often at the LHC but with greatest certainty at the ILC. In addition, the anomalous contribution to the $ggh$ coupling implies substantial changes in the $gg \to h$ production rate at the LHC.

The above results imply that detection of the $h$ at the LHC could be significantly modified if $|\xi|$ is large. To illustrate this, we plot in Fig. 9.5 the ratio of the rates for $gg \to h \to \gamma\gamma$, $WW \to h \to \tau^+\tau^-$ and $gg \to t\bar{t}h \to t\bar{t}b\bar{b}$ (the latter two ratios being equal) to the corresponding rates for the SM Higgs boson. For this figure, we take $m_h = 120$ GeV and $\Lambda_\phi = 5$ TeV and show results for $m_\phi = 20$, 55 and 200 GeV. The result is that prospects for $h$ discovery in the $gg \to h \to \gamma\gamma$ and $WW \to h \to \tau^+\tau^-$ modes could be either substantially poorer or substantially better than for a SM Higgs boson of the same mass, depending on $\xi$ and $m_\phi$.

For $m_\phi > m_h$, the suppression is very substantial for large, negative values of $\xi$. This region of significant suppression becomes wider at large values of $m_\phi$ and $\Lambda_\phi$. In contrast, for $m_\phi < m_h$, the $gg \to h \to \gamma\gamma$ rate is generally only suppressed when $\xi > 0$. All this is shown, in a quantitative way, by the contours in Fig. 9.6 [29]. The outermost, hourglass shaped contours define the theoretically allowed region. Three main regions of non-detectability may appear. Two are located at large values of $m_\phi$ and $|\xi|$. A third region appears at low $m_\phi$ and positive $\xi$, where the above-noted $gg \to h \to \gamma\gamma$ suppression sets in. It becomes further expanded when $2m_\phi < m_h$ and the decay channel $h \to \phi\phi$ opens up, thus reducing the $h \to \gamma\gamma$ branching ratio. These regions shrink as $m_h$ increases, since additional channels, in particular $gg \to h \to ZZ^* \to 4\ \ell$, become available for Higgs discovery. At $m_h = 120$ GeV, these regions are reduced by considering either a larger data set or $qqh$ Higgs production, in association with forward jets. An integrated luminosity of 100 fb$^{-1}$ would remove the regions of non-detection shown in Fig. 9.6 at large positive $\xi$ in the case of $\Lambda_\phi = 5$ TeV. Similarly, including the $qqh$, $h \to WW^* \to \ell\ell\nu\bar{\nu}$ channel in the list of the discovery modes removes the same two regions and reduces the large region of $h$ non-observability at negative $\xi$ values. In all these regions, a complementarity is potentially offered by the process $gg \to \phi \to ZZ^* \to 4\ \ell$, which becomes important for $m_\phi > 140$ GeV. At the LHC, this process would have the same event structure as the golden SM

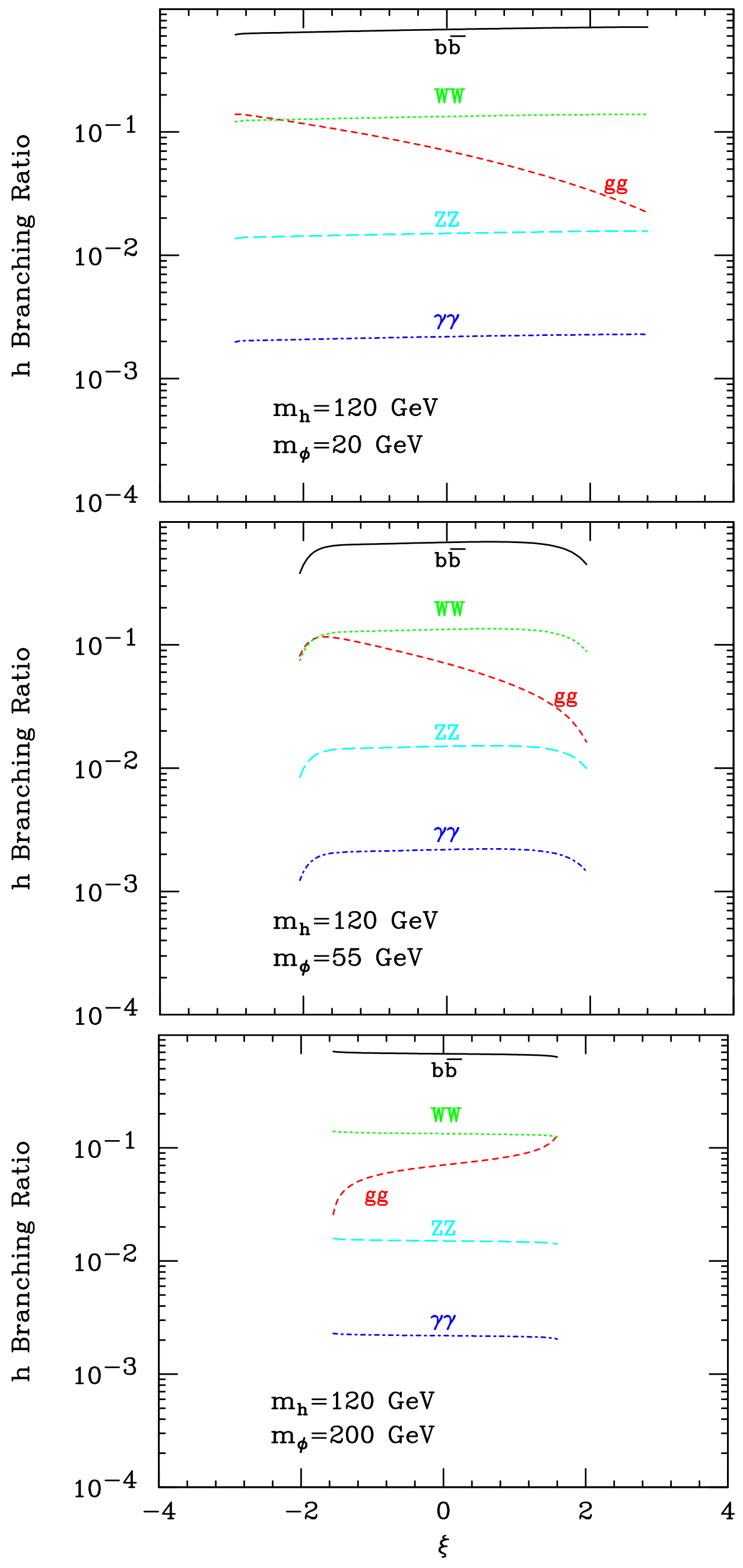

Fig. 9.4: The branching ratios for $h$ decays to $b\bar{b}$, $gg$, $WW^*$, $ZZ^*$ and $\gamma\gamma$ for $m_h = 120$ GeV and $\Lambda_\phi = 5$ TeV as functions of $\xi$ for $m_\phi = 20$, 55 and 200 GeV.

Higgs mode, $H \to ZZ^* \to 4\,\ell$, which has been thoroughly studied for an intermediate mass Higgs boson. By computing the $gg \to \phi \to ZZ^* \to 4\,\ell$ rate relative to that for the corresponding SM $H$ process and employing the LHC sensitivity curve for $H \to ZZ^*$ of [30], the significance for the $\phi$ signal in the 4 $\ell$ final state at the LHC can be extracted. Results are overlayed on Fig. 9.6 assuming 30 fb$^{-1}$ of data.

The observability of $\phi$ production in the four lepton channel fills most of the gaps in $(m_\phi, \xi)$ parameter space in which $h$ detection is not possible (mostly due to the suppression of the loop-induced $gg \to h \to \gamma\gamma$ process). For example, for $\Lambda_\phi = 5$ TeV and $L = 30$ fb$^{-1}$, the observation of at least one scalar is guaranteed over almost the full parameter phase space, with the exception of: (a) the region of large positive $\xi$ with $m_\phi < 120$ GeV, where the $\phi \to ZZ^*$ decay is phase-space-suppressed; and (b) a narrow region with $\xi < 0$ and $m_\phi \simeq 170$ GeV. The latter region arises due to the ramp-up of the $\phi \to W^+W^-$ channel; in this region a luminosity of order 100 fb$^{-1}$ is required to reach a $\geq 5\sigma$ signal for $\phi \to ZZ^*$. We should also note that the $\phi \to ZZ$ decay is reduced for $m_\phi > 2m_h$ by the onset of the $\phi \to hh$ decay, which can become the main decay mode. The resulting $hh \to b\bar{b}b\bar{b}$ topology, with di-jet mass constraints, may represent a viable signal for the LHC in its own right [31, 32].

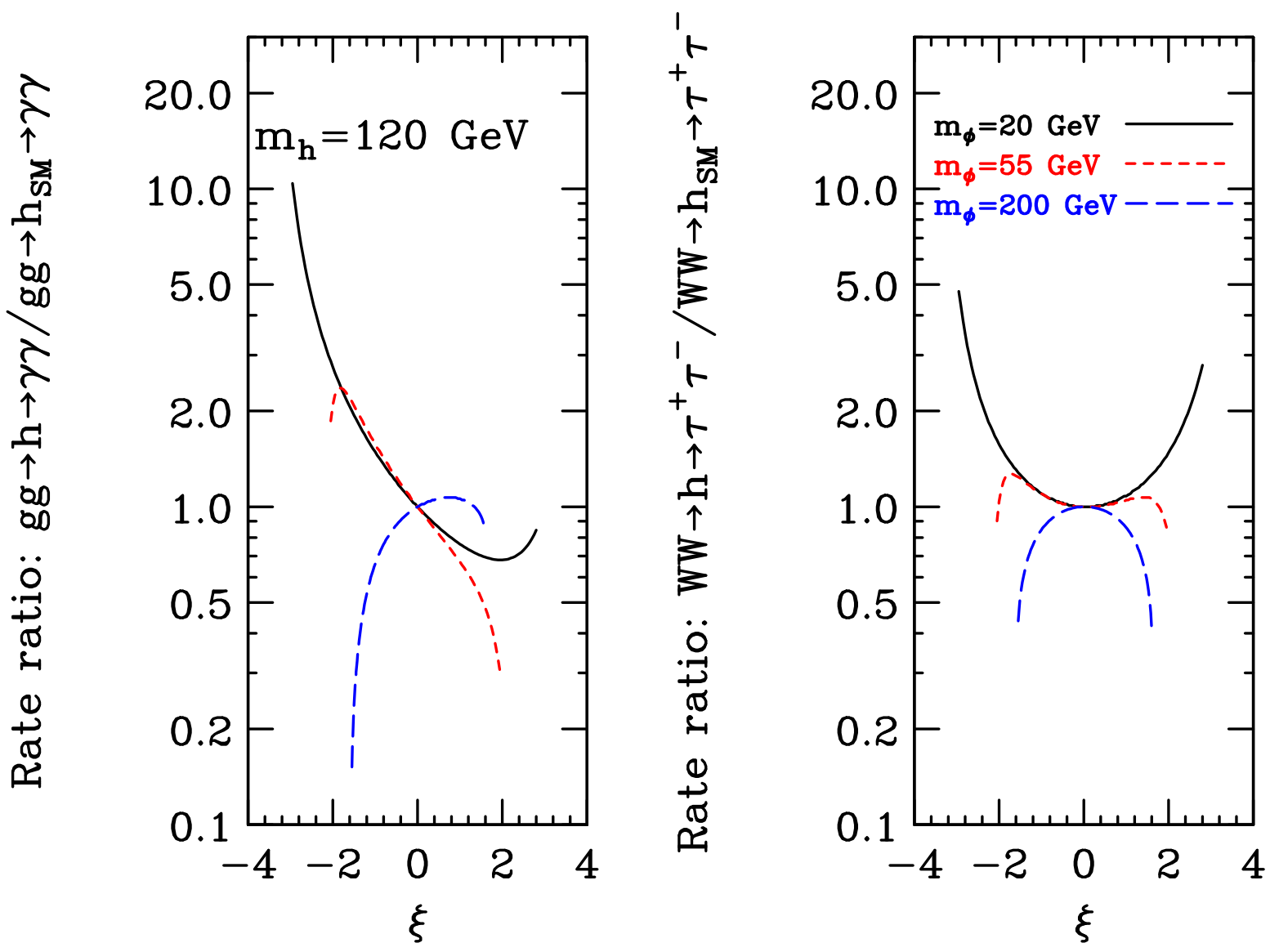

Fig. 9.5: The ratio of the rates for $gg \to h \to \gamma\gamma$ and $WW \to h \to \tau^+\tau^-$ (the latter is the same as that for $gg \to t\bar{t}h \to t\bar{t}b\bar{b}$) to the corresponding rates for the SM Higgs boson. Results are shown for $m_h = 120$ GeV and $\Lambda_\phi = 5$ TeV as functions of $\xi$ for $m_\phi = 20$, 55 and 200 GeV.

As seen in Fig. 9.6, there are regions of $(m_\phi, \xi)$ parameter space in which *both* the $h$ and $\phi$ mass eigenstates will be detectable. In these regions, the LHC will observe two scalar bosons somewhat separated in mass with the lighter (heavier) having a non-SM-like rate for the $gg$-induced $\gamma\gamma$ ($ZZ$) final state. Despite the ability to see both eigenstates, still more information will be required to ascertain whether these two Higgs bosons derive from a multi-doublet or other type of extended Higgs sector or from the present type of model with Higgs-radion mixing [29].

The ILC should guarantee observation of both the $h$ and the $\phi$ even in most of the regions within which detection of either at the LHC might be difficult. In particular, in the region with $m_\phi > m_h$ the $hZZ$ coupling is enhanced relative to the SM $HZZ$ coupling and $h$ detection in $e^+e^-$ collisions would be even easier than SM $H$ detection. Further, assuming that $e^+e^-$ collisions could also probe down to $\phi ZZ$ couplings of order $g^2_{\phi ZZ}/g^2_{HZZ} \simeq 0.01$, the $\phi$ would be seen in almost the entirety of the region for which $\phi$ detection at the LHC would not be possible. In this case, the measurements of the $ZZ$ boson couplings of both the Higgs and the radion particles would significantly constrain the values of the $\xi$ and $\Lambda_\phi$ parameters of the model. Furthermore, the ILC has the capability of extending the coupling measurements to all fermions separately with accuracies of order 1%-5% and achieves a determination of the total width to 4% - 6% accuracy [33]. This is shown in Fig. 9.7 by the additional contours, which indicate the regions where the discrepancy with the SM predictions for the Higgs couplings to pairs of $b$ quarks and $W$ bosons exceeds $2.5\sigma$.

As already noticed, the presence in the Higgs radion sector of trilinear terms opens up the important possibility of $\phi \to hh$ decay and $h \to \phi\phi$. For example, for $m_h = 120$ GeV, $\Lambda_\phi = 5$ TeV and $m_\phi \sim 250 - 350$ GeV one finds $BR(\phi \to hh) \sim 0.2 - 0.3$. In [34], the CMS discovery potential for the radion in its two Higgs decay mode ($\phi \to hh$) with $\gamma\gamma$+$b\bar{b}$, $\tau\tau$+$b\bar{b}$ and $b\bar{b}$+$b\bar{b}$ final states was esti-mated, assuming $m_h = 125$ GeV and $m_\phi = 300$ GeV. The $\gamma\gamma$+$b\bar{b}$ topology provides the best discovery potential. Details of this analysis and the corresponding analysis by ATLAS can be found respectively in Section 9.4 and Section 9.3.

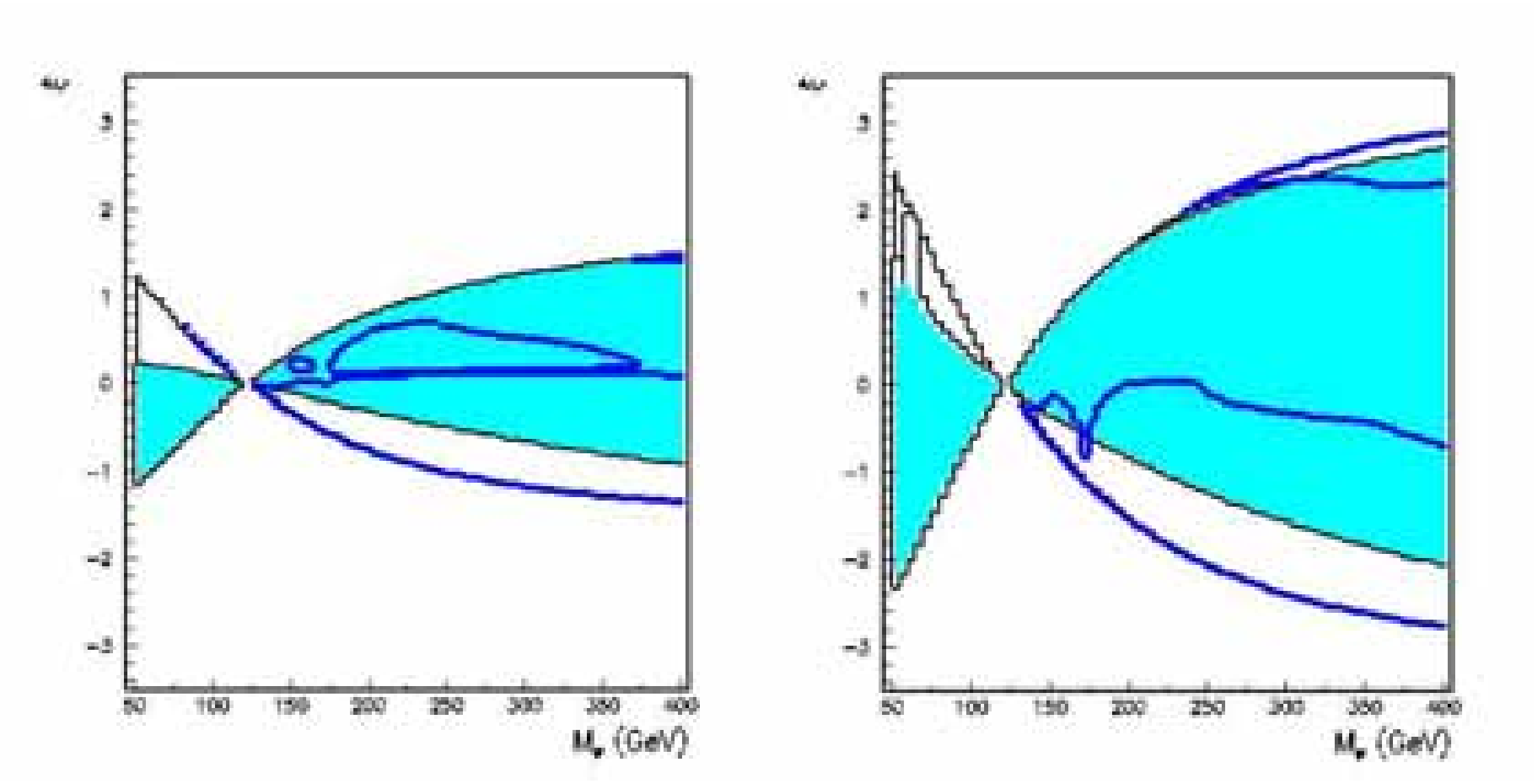

Fig. 9.6: Regions in $(m_\phi, \xi)$ parameter space of $h$ detectability (including $gg \to h \to \gamma\gamma$ and other modes) and of $gg \to \phi \to ZZ^* \to 4\,\ell$ detectability at the LHC for one experiment and 30 fb$^{-1}$. The outermost, hourglass shaped contours define the theoretically allowed region. The light gray (cyan) regions show the part of the parameter space where the net $h$ signal significance remains above $5\sigma$. In the empty regions between the shading and the outermost curves, the net $h$ signal drops below the $5\sigma$ level. The thick gray (blue) curves indicate the regions where the significance of the $gg \to \phi \to ZZ^* \to 4\,\ell$ signal exceeds $5\sigma$. Results are presented for $m_h$=120 GeV and $\Lambda_\phi$= 2.5 TeV (left), 5.0 TeV (right).

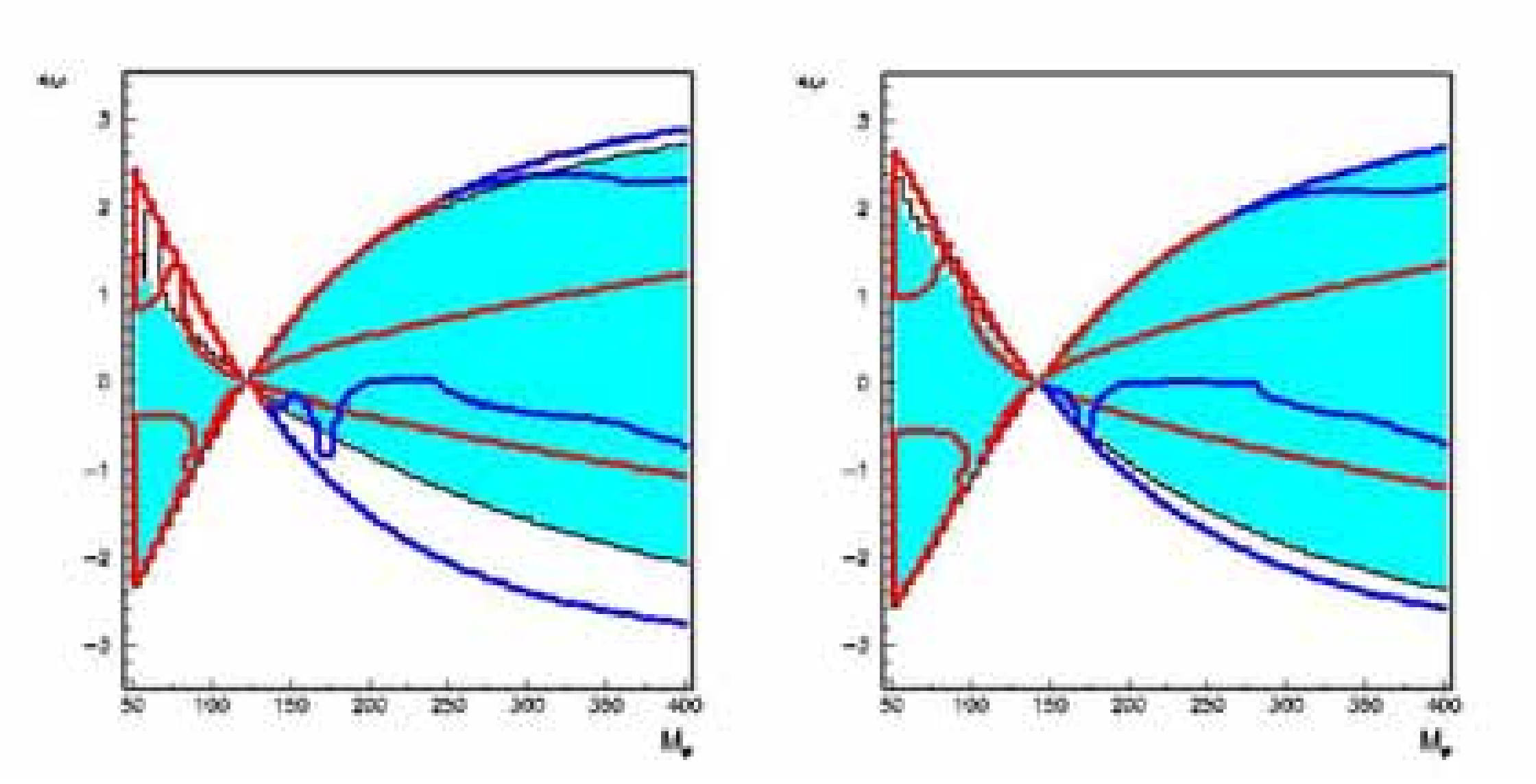

Fig. 9.7: Same as Fig. 9.6 for $m_h$ = 120 GeV (left), 140 GeV (right) and $\Lambda_\phi$ = 5 TeV with added contours, indicated by the medium gray (red) curves, showing the regions where the ILC measurements of the $h$ couplings to $b\bar{b}$ and $W^+W^-$ would provide a $> 2.5\sigma$ evidence for the radion mixing effect. (Note: the gray (red) lines are always present along the outer edge of the hourglass in the $m_\phi > m_h$ region, but are sometimes buried under the darker (blue) curves. In this region, the $> 2.5\sigma$ regions lie between the outer hourglass edges and the inner gray (red) curves.)

### 9.3 Radion search in ATLAS

*Georges Azuelos, Donatella Cavalli, Helenka Przysieniak and Laurent Vacavant*

As explained in the introduction, the radion is a physical scalar that remains in 4-d from the 5-d metric tensor and from fluctuations about a stabilized Randall-Sundrum (RS) configuration. The presence of the radion is one of the important phenomenological consequences of theories of warped extra dimensions and observation of this scalar therefore constitutes a crucial probe of the model. This section summarizes the main results of a study performed for ATLAS [35].

The radion, being a scalar, has a Higgs-like phenomenology [21, 22, 36] but has large effective coupling to gluons. The decay branching ratios and widths depend on three parameters: the physical mass of the radion $m_\phi$, the vacuum expectation value of the radion field, $\Lambda_\phi = \sqrt{6}\Lambda_\pi$ where $\Lambda_\pi$ is the mass scale at the TeV brane in the RS model, and the radion-SM Higgs mixing parameter $\xi$.

Here, we have investigated the possibility of observing a RS radion with the ATLAS detector through the following decays: $\phi \to \gamma\gamma$, $\phi \to ZZ^{(*)} \to 4\ell$, $\phi \to hh \to b\bar{b}\gamma\gamma$ and $\phi \to hh \to b\bar{b}\tau^+\tau^-$. Only the direct production of the radion $gg \to \phi$ has been considered since it is the main process at LHC and it benefits from the enhancement of the coupling $\phi gg$.

#### 9.3.1 $\phi \to \gamma\gamma$ and $ZZ^{(*)} \to 4\ell$

For the $\gamma\gamma$ ($m_\phi < 160$ GeV) and ZZ$^{(*)}$ ($m_\phi > 100$ GeV) decay channels, the radion signal significance is determined by reinterpreting the results from the SM Higgs analyses obtained with ATLAS [37], for 100 fb$^{-1}$. The ratio of the radion $\mathrm{S}/\sqrt{\mathrm{B}}$ over that of the SM Higgs is given by [21]:

$$\frac{\mathrm{S}/\sqrt{\mathrm{B}}\,(\phi)}{\mathrm{S}/\sqrt{\mathrm{B}}\,(h)} = \frac{\Gamma(\phi \to gg)\,\mathrm{BR}(\phi \to \gamma\gamma,\, ZZ)}{\Gamma(h \to gg)\,\mathrm{BR}(h \to \gamma\gamma,\, ZZ)}\sqrt{\frac{\max(\Gamma^h_{\mathrm{tot}}, \sigma_m)}{\max(\Gamma^\phi_{\mathrm{tot}}, \sigma_m)}} \tag{9.21}$$

where the $\sigma_m$ are the experimentally reconstructed resonance widths for each of the two decay processes. The radion signal significance thus determined is shown as a function of the mass of the radion, in Fig. 9.8, for $\Lambda_\phi = 1, 10$ TeV, $\xi = 0$, 1/6, and for an integrated luminosity of 100 fb$^{-1}$.

#### 9.3.2 $\phi \to hh \to \gamma\gamma b\bar{b}$

For the purpose of estimating the limits of observation of radion decay to a pair of SM Higgs bosons, two reference values are taken for the mass of the radion: 300 GeV and 600 GeV and a Higgs mass of 125 GeV (assumed to be known). The production cross sections in these cases are 58 pb and 8 pb respectively.

The specific decay channel $\phi \to hh \to \gamma\gamma b\bar{b}$ offers an interesting signature, with two high-$p_T$ isolated photons and two b-jets. The background rate is expected to be very low for the relevant mass region $m_h > 115$ GeV and $m_\phi > 2m_h$. In addition, triggering on such events is easy and the diphoton mass provides very good kinematical constraints for the reconstruction of $m_\phi$.

The backgrounds for this channel are $\gamma\gamma b\bar{b}$ (irreducible), $\gamma\gamma c\bar{c}$, $\gamma\gamma bj$, $\gamma\gamma cj$ and $\gamma\gamma jj$ (reducible with b-tagging). In the region of mass considered for the Higgs, the main production processes are the Born diagram $qq \to \gamma\gamma$ and the box diagram $gg \to \gamma\gamma$. The events were generated with PYTHIA [38] and some cuts had to be applied at the event generation: the sample was generated in seven different bins of $\hat{p}_\perp$, the transverse momentum defined in the rest frame of the hard interaction. Detector effects on the signal and background events were taken into account by the fast simulation code ATLFAST [39].

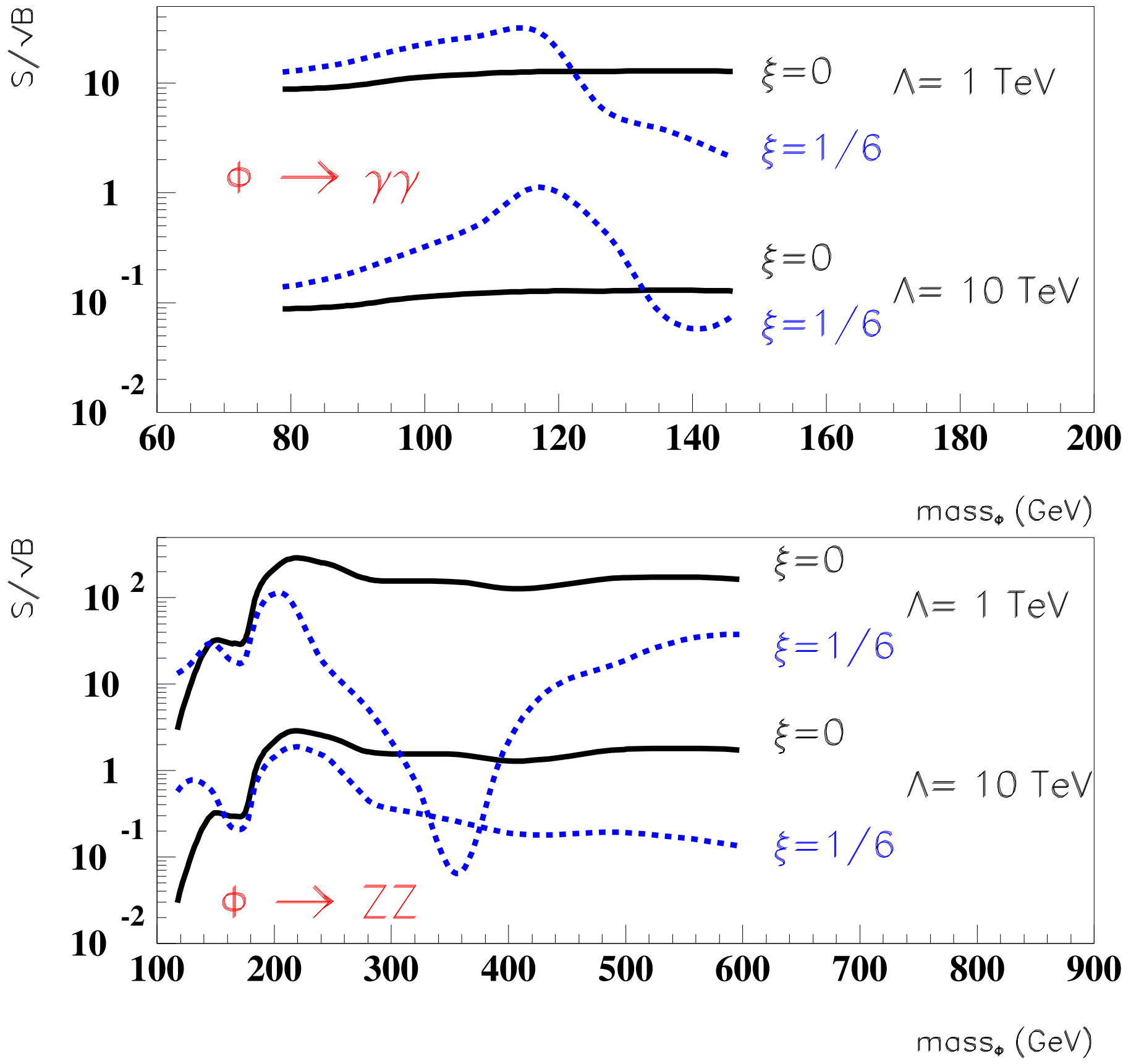

Fig. 9.8: Signal significance versus the mass of the radion, for the $\gamma\gamma$ channel (top) and for the ZZ$^{(*)}$ channel (bottom). In both plots, the values for $\Lambda_\phi = 1, 10$ TeV and $\xi = 0$, 1/6 are shown, for an integrated luminosity of 100 fb$^{-1}$.

To extract the signal, two isolated photons with $p_T > 20$ GeV and $|\eta| < 2.5$, and two jets with $p_T > 15$ GeV, $|\eta| < 2.5$ are required. At least one of the jets has to be tagged as a $b$. Fig. 9.9 shows the reconstructed radion mass for the case of $m_\phi = 300$ GeV and Table 9.1 shows the expected number of events for different cases of $\xi$ and $\Lambda_\phi$. The background is negligible. A few fb$^{-1}$ would be sufficient to observe the radion if $\Lambda_\phi = 1$ TeV, and it is estimated that with 30 fb$^{-1}$, the reach in $\Lambda_\phi$ would be $\sim 2.2$ TeV for $m_\phi = 300$ GeV.

### *9.3.3* $\phi \to hh \to b\bar{b}\tau^+\tau^-$

The channel $\phi \to hh \to b\bar{b}\tau^+\tau^-$ provides another potentially interesting signal for radion discovery, although the background is higher and the reconstructed mass resolutions are poorer than in the $\phi \to hh \to \gamma\gamma b\bar{b}$ channel.

The background here are: $hh \to b\bar{b}\ \tau^+\tau^-$, $t\bar{t} \to bW^+\ \bar{b}W^-$, $Z$ +jets, followed by $Z \to \tau^+\tau^-$ and $W$ + jets. In order to provide a trigger, a leptonic decay of one of the two $\tau$'s is required. Although the signal efficiency is low, the background rejection is high, after application of simple cuts (for details, see [35]). Figure 9.10 shows the reconstructed masses for signal when $m_\phi$=300 and 600 GeV respectively, for 30 fb$^{-1}$, $\Lambda_\phi = 1$ TeV and $\xi = 0$. The shape for a 300 GeV radion resonance is not distinguishable from the background (mostly $t\bar{t}$). Therefore systematic errors will most probably be dominated by the understanding of the level of this background.

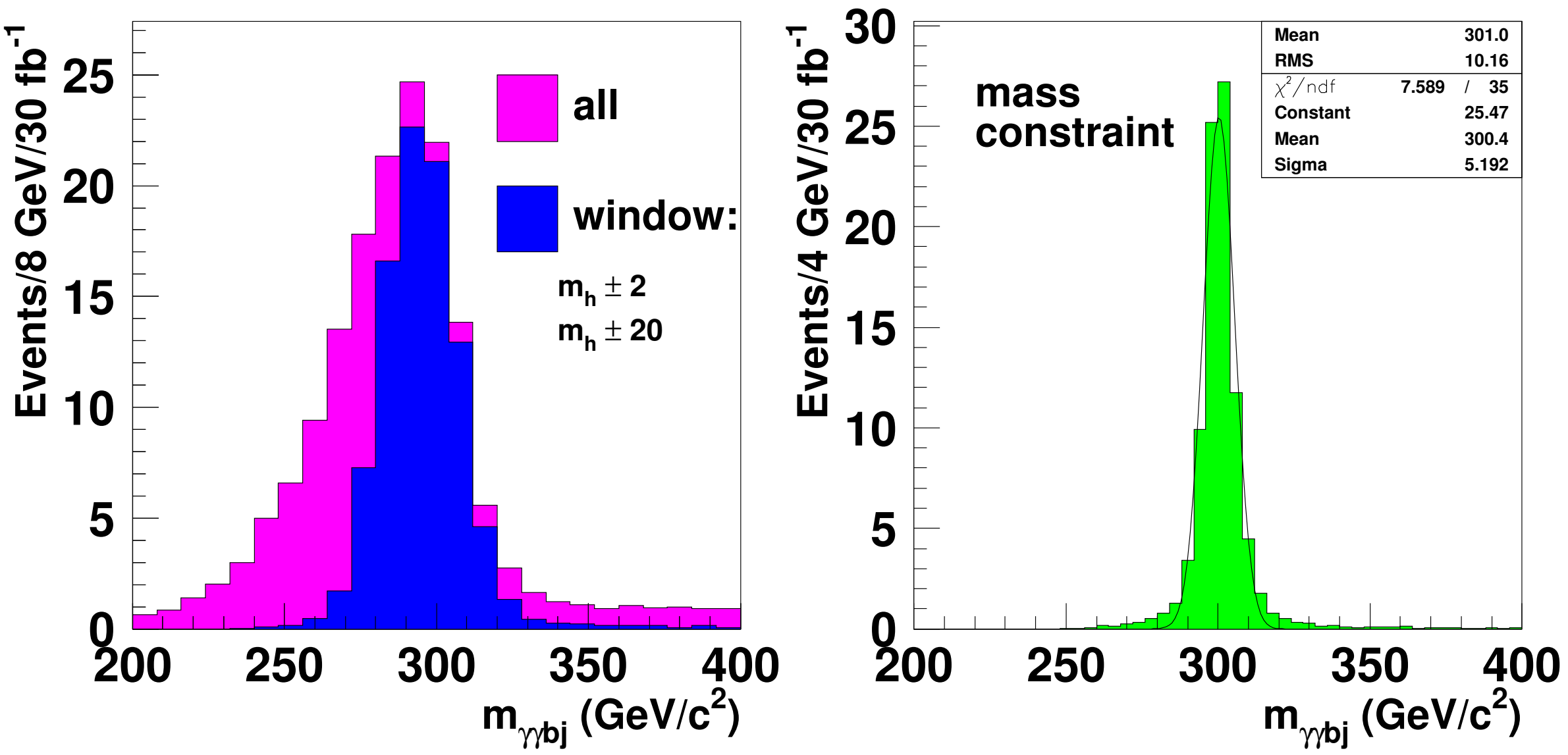

Fig. 9.9: Reconstructed $\gamma\gamma bj$ invariant mass distribution, for $m_\phi = 300$ GeV, $\xi = 0$, $\Lambda_\phi = 1$ TeV and for 30 fb$^{-1}$. The plots on the left show all combinations and the ones fulfilling the mass window cuts shown on the two Higgs resonances: $m_{\gamma\gamma} = m_h \pm 2$ GeV and $m_{bj} = m_h \pm 20$ GeV . The distribution on the right is obtained by constraining the reconstructed masses $m_{bj}$ and $m_{\gamma\gamma}$ to the light Higgs mass $m_h$, after the mass window cuts.

| | $m_\phi = 300$ GeV | $m_\phi = 600$ GeV |
|---|---|---|
| $\xi = 0, \Lambda_\phi = 1$ TeV | 84.5 | 7.0 |
| $\xi = 0, \Lambda_\phi = 10$ TeV | 0.9 | 0.1 |
| $\xi = 1/6, \Lambda_\phi = 1$ TeV | 150.9 | 5.3 |
| $\xi = 1/6, \Lambda_\phi = 10$ TeV | 1.2 | 0.1 |

Table 9.1: Number of events selected for signal, $\phi \to hh \to \gamma\gamma b\bar{b}$ for $m_\phi$ =300 and 600 GeV, for 30 fb$^{-1}$ and for $m_{\rm h} = 125$ GeV.

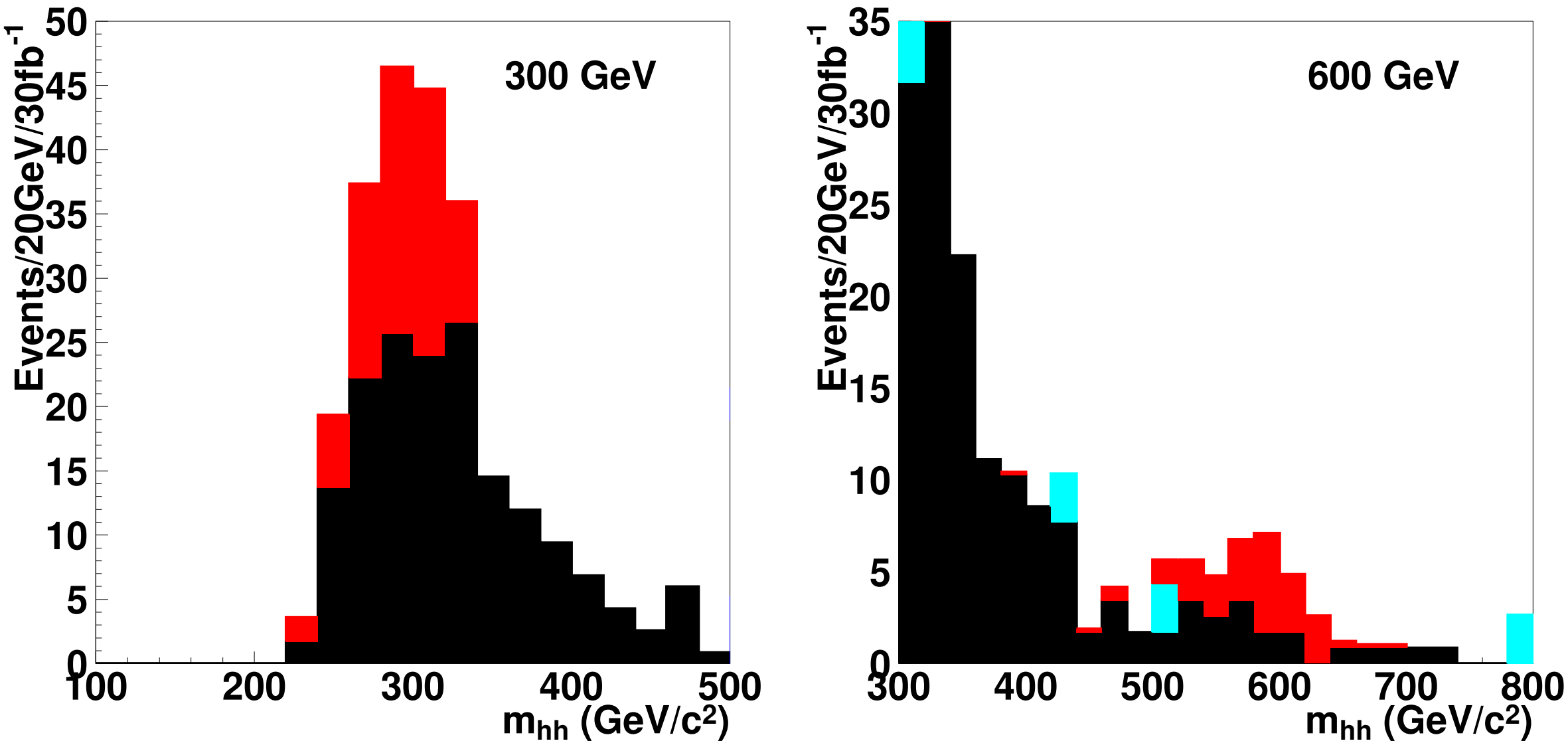

Fig. 9.10: Reconstructed mass of the radion, for 30 fb$^{-1}$ and $\Lambda_\phi = 1$ TeV, $\xi = 0$. Left plot: expected signal (light) and $t\bar{t}$ background (dark). Right plot: $t\bar{t}$ background (dark), expected signal (light) and $Z$+jets background (very light).

Requiring a minimum of 10 events and $S/\sqrt{B} \geq 5$, the maximum reach in $\Lambda_\phi$ is 1.04 TeV for both $m_\phi$=300 GeV and $m_\phi$=600 GeV, but the uncertainties in background subtraction may affect considerably the observability of this channel in the first case.

### *9.3.4 Conclusions*

The search for the radion at the LHC is similar to the case of the SM Higgs boson, and indeed the analyses already performed for the latter can be re-interpreted in terms of a radion search, after rescaling the branching ratios and widths. For an integrated luminosity of 100 fb$^{-1}$, the values $S/\sqrt{B}$ ~10 (0.1) are obtained for the $\gamma\gamma$ channel, with a mixing parameter $\xi$=0 and a scale $\Lambda_\phi$=1 (10) TeV, in the range $80 \text{ GeV} < m_\phi < 160 \text{ GeV}$. For the $ZZ^{(*)}$ channel, $S/\sqrt{B}$ ~100 (1) for $200 \text{ GeV} < m_\phi < 600 \text{ GeV}$ for the same conditions. Because the couplings are similar to those of the SM Higgs, a good measurement of the production cross section and branching ratios will be necessary to discriminate between the two scalars.

The radion can also decay into a pair of Higgs scalars, if the masses permit. Two cases were examined: $\phi \rightarrow hh \rightarrow \gamma\gamma b\bar{b}$ and $\phi \rightarrow hh \rightarrow \tau\tau b\bar{b}$, for radion masses of 300 and 600 GeV, for $m_h = 125$ GeV and for an integrated luminosity of 30 fb$^{-1}$. The maximal reach in $\Lambda_\phi$ is $\sim 1-2$ TeV. It must be remarked that a value of $\Lambda_\phi = 1$ TeV could be overly optimistic since, with the corresponding value of $\Lambda_\pi$ and with a value of $k/M_{Pl}$ even as large as 0.1, the mass of the first KK graviton state ($m_1 = 3.83(k/M_{Pl})\Lambda_\pi$) would be very low [40].

## 9.4 Radion search in CMS

*Albert De Roeck, Guy Dewhirst, Daniele Dominici, Livio Fano, Simone Gennai and Alexander Nikitenko*

The CMS discovery potential is estimated for the decay of the radion in a pair of Higgs bosons, with $\gamma\gamma b\bar{b}$, $\tau\tau b\bar{b}$ and $b\bar{b}b\bar{b}$ final states and for an integrated luminosity of 30 fb$^{-1}$. The study has been carried out for the radion mass of 300 GeV and the Higgs boson mass of 125 GeV . The sensitivity was evaluated in the ($\xi$,$\Lambda_\phi$) plane, with systematics uncertainties included. A detailed description of the analysis can be found in [34]. A brief summary of the analysis and the results is presented below.

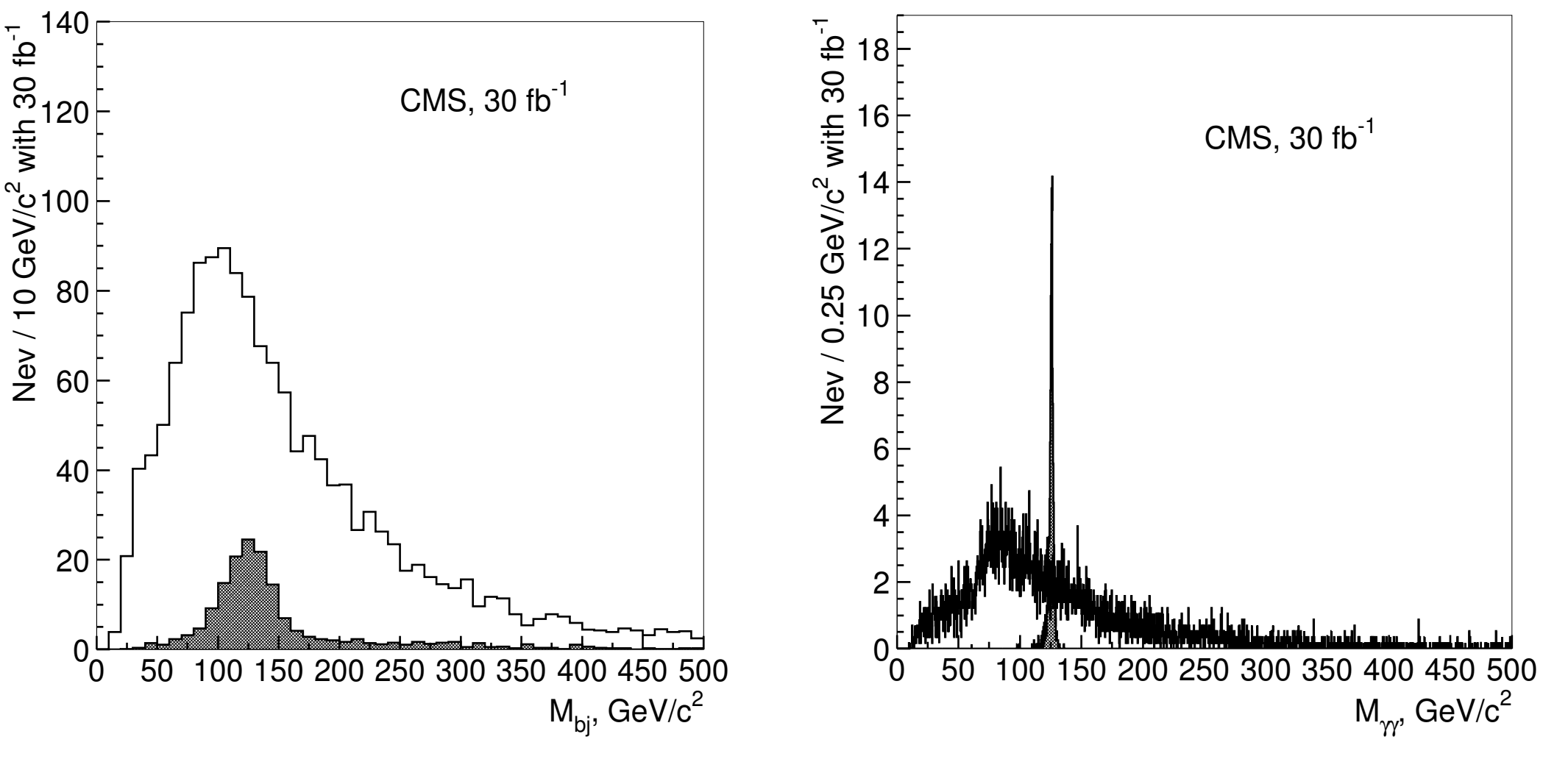

Fig. 9.11: The di-jet (left plot) and the di-photon (right plot) mass distribution for the background (open histogram) and the signal of $\phi \to hh \to \gamma\gamma b\bar{b}$ (full black histogram) after all selections except the mass window cuts with 30 fb$^{-1}$. The signal is shown for the maximal cross section times branching ratios point in ($\xi$-$\Lambda_\phi$) plane.

Signal events $gg \to \phi \to hh$ were generated with PYTHIA. The cross sections and branching ratios were evaluated using rescaled NLO cross sections for the SM Higgs boson and a modified HDECAY program. For the radion and Higgs boson mass points considered ($m_h$ = 125 GeV, $m_\phi$=300 GeV) and for $\Lambda_\phi$= 1 TeV the maximal cross section times branching ratio is 71 fb for $\gamma\gamma b\bar{b}$ final state. For the $\tau\tau b\bar{b}$ final state with the topology considered in the analysis, one $\tau$ lepton decaying leptonically and the other $\tau$ lepton decaying hadronically (producing a $\tau$ jet), the maximal cross section times branching ratio is 960 fb. This maximal cross section is reached for the radion mixing parameter $\xi = -0.35$.

For the $\gamma\gamma b\bar{b}$ final state the irreducible backgrounds $\gamma\gamma jj (j = u, d, s, g)$ (generated with CompHEP) and the $\gamma\gamma c\bar{c}$ and $\gamma\gamma b\bar{b}$ (generated with MadGraph) were studied. The reducible background from $\gamma$+three jets and four-jet processes was not evaluated directly, but assumed to be the same as in for the inclusive $h \to \gamma\gamma$ analysis, namely 40% of the total background after all selection. For the $\tau\tau b\bar{b}$ final state, the $t\bar{t}$, $Z$+jets, $W$+jets backgrounds (generated with PYTHIA) and the $b\bar{b}Z$ background (generated with CompHEP) were studied.

### *9.4.1 The $\phi \to hh \to \gamma\gamma b\bar{b}$ channel*

The $\gamma\gamma b\bar{b}$ events were required to pass the Level-1 and HLT di-photon trigger. In the off-line analysis two photon candidates with $E_T^{\gamma 1,\gamma 2} > 40$, 25 GeV were required to pass tracker cuts and calorimeter isolation cuts. Events with only two calorimeter jets of $E_T$ >30 GeV and within $|\eta|$ <2.4 were selected. At least one of these jets must be tagged as a b-jet. Finally, the di-photon mass, $M_{\gamma\gamma}$, was required to be in a window of ± 2 GeV, the di-jet mass, $M_{j\bar{b}}$, in a window of ± 30 GeV and the di-photon-di-jet mass, $M_{\gamma\gamma b\bar{b}}$, in a window ± 50 GeV around the Higgs and radion mass. Figure 9.11 shows the di-jet (left plot) and the di-photon (right plot) mass distribution for the background (open histogram) and the signal of $\phi \to hh \to \gamma\gamma b\bar{b}$ (full, black histogram) after all selections except the mass window cuts, and for 30 fb$^{-1}$. The signal is shown for the maximal cross section times branching ratios point in ($\xi$-$\Lambda_\phi$) plane. Figure 9.12 (left plot) shows the $M_{\gamma\gamma bj}$ distribution for the background (dashed histogram) and for the signal of $\phi \to hh \to \gamma\gamma b\bar{b}$ plus background (solid histogram) after all selections, and for 30 fb$^{-1}$.

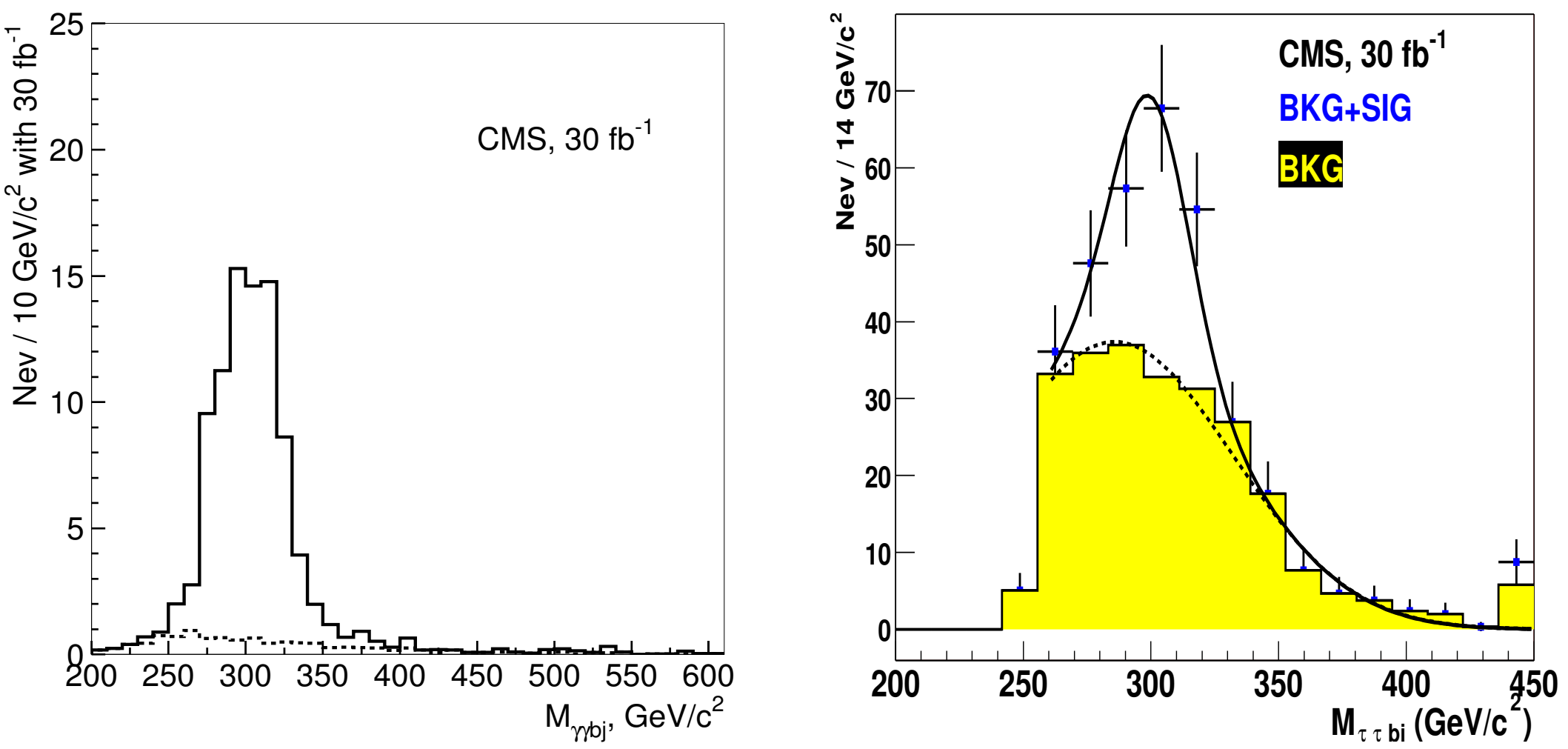

Fig. 9.12: Left plot: the $\mathrm{M}_{\gamma\gamma bj}$ distribution for the background (dashed histogram) and for the signal of $\phi \to hh \to \gamma\gamma b\bar{b}$ plus background (solid histogram) after all selections for 30 fb$^{-1}$. Right plot: the $\mathrm{M}_{\tau\tau bj}$ distribution for the background (full gray histogram) and for the signal of $\phi \to hh \to \tau\tau b\bar{b}$ plus background (black points with the error bars) after all selections for 30 fb$^{-1}$. The fitted curves for the background and signal plus background are superimposed. On both plots the signal is shown for the maximal cross section times branching ratios point in ($\xi$-$\Lambda_\phi$)

### *9.4.2 The $\phi \to hh \to \tau\tau b\bar{b}$ channel*

The $\tau\tau b\bar{b}$ events were selected by the single electron and muon triggers and by the combined $e$-plus-$\tau$-jet and the $\mu$-plus-$\tau$-jet triggers. In the off-line analysis a lepton and $\tau$-jet identification was performed. The requirements on the jets were similar to the ones used in the $\gamma\gamma b\bar{b}$ analysis. In addition a cut of the transverse mass of the lepton and missing transverse momentum, $M_T^{\ell\nu}$ <35 GeV was applied to suppress the $t\bar{t}$ and $W$+jets backgrounds. The di-$\tau$-lepton mass was reconstructed using the missing transverse energy. The significance of the discovery was calculated using expected number of the signal and background events after the mass window selections: 100< $\mathrm{M_{bj}}$ <150 GeV, 100< $\mathrm{M}_{\tau\tau}$ <160 GeV and 280< $\mathrm{M}_{\tau\tau}$ <330 GeV. Figure 9.12 (right plot) shows the $\mathrm{M}_{\tau\tau\mathrm{bj}}$ distribution for the background (full, gray histogram) and for the signal of $\phi \to hh \to \tau\tau b\bar{b}$ plus background (black points with the error bars) after all selections, for 30 fb$^{-1}$. Fitted curves for the background and the signal plus background are superimposed.

### *9.4.3 The $\phi \to hh \to b\bar{b}b\bar{b}$ channel*

The four b-jet final state yields the highest rate for the signal. The maximal cross section times branching ratio at $\Lambda_\phi$ = 1 TeV is 10.3 pb, which results in about 3.1 $\times$ 10$^5$ signal events for 30 fb$^{-1}$. The effective triggering and selection in the off-line analysis of the events is, however a big challenge due to the huge multi-jet background rate. In fact the remaining background is a few orders of magnitude larger that the signal in the relevant mass range. Techniques can be envisaged to normalize the background directly from a signal free region and predict the number of background events in the signal region. In order to make a 3$\sigma$ discovery, such extrapolation needs to be performed with a precision of about 0.1%, making this channel essentially hopeless.

### *9.4.4 Results*

The background contribution to the $\gamma\gamma b\bar{b}$ final state can be determined directly from the $\gamma\gamma$-plus-two-jets data obtained after all selections, except the final mass window cuts on the $\mathrm{M}_{\gamma\gamma}$, $\mathrm{M}_{j\bar{b}}$ and $\mathrm{M}_{\gamma\gamma b\bar{b}}$. The signal-to-background ratio is always less than 10% before the mass cuts are applied. The final cuts on the $\mathrm{M}_{\gamma\gamma}$, $\mathrm{M}_{j\bar{b}}$ and $\mathrm{M}_{\gamma\gamma b\bar{b}}$ introduce a systematic uncertainty on the number of the background events expected after these cuts. This uncertainty is determined by the following factors: the energy scale uncertainty for the photons and jets, and the theoretical uncertainty of the shape of the mass distributions due to the scale and PDF uncertainties. Figure 9.13 (left plot) shows the $5\sigma$ discovery contours for the $\phi \to hh \to \gamma\gamma b\bar{b}$ channel for 30 fb$^{-1}$. The solid (dashed) contour shows the discovery region without (with) the effects of the systematic uncertainties.

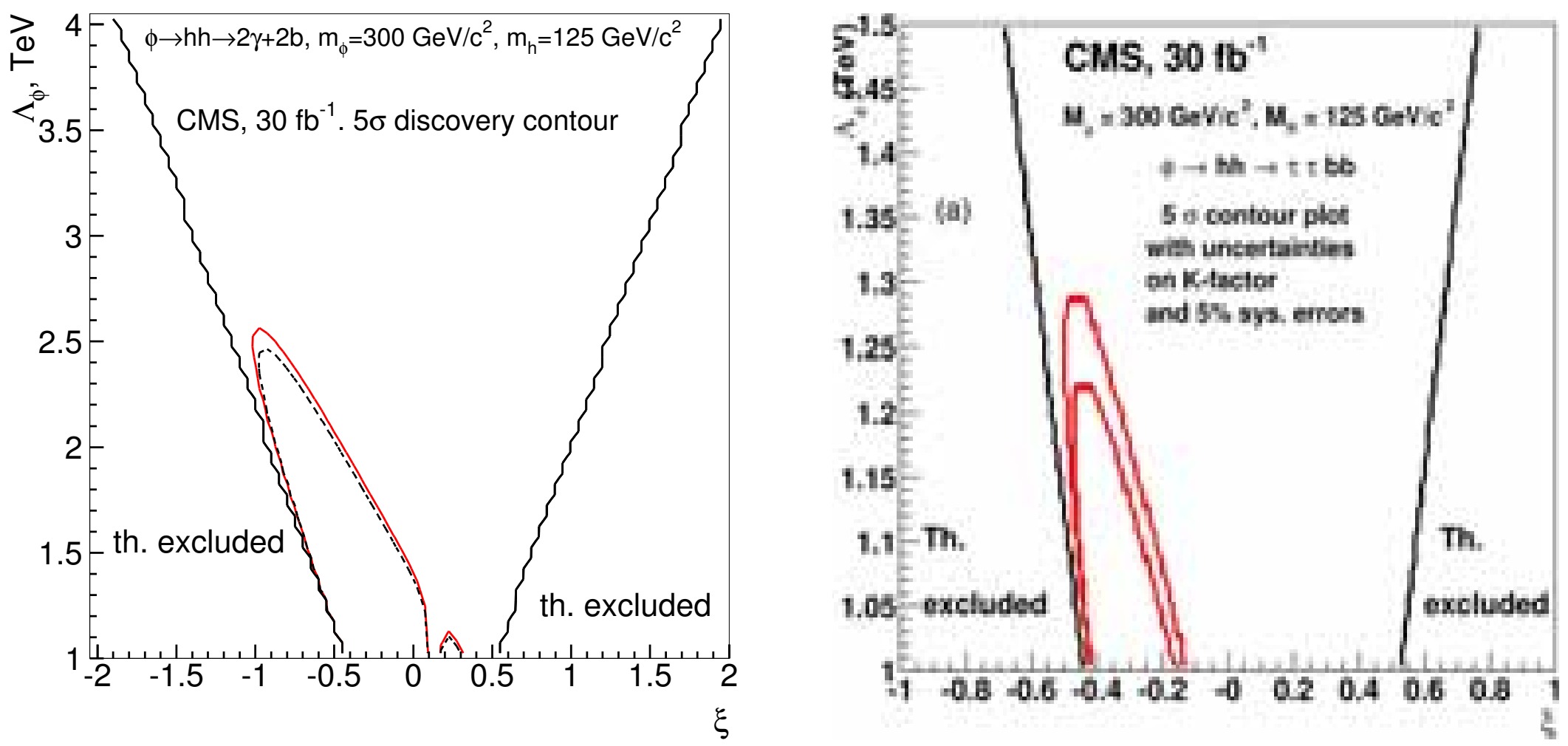

Fig. 9.13: Left plot: the $5\sigma$ discovery contours for the $\phi \to hh \to \gamma\gamma b\bar{b}$ channel for 30 fb$^{-1}$. The solid (dashed) contour shows the discovery region without (with) the effects of the systematic uncertainties (find more explanations in the text). Right plot: the $5\sigma$ discovery contours for the $\phi \to hh \to \tau\tau b\bar{b}$ channel for 30 fb$^{-1}$. The two contours corresponds to the variation of the background NLO cross sections due to the scale uncertainty. The 5% experimental systematics on the background is taken into account (see text).

For the $\tau\tau b\bar{b}$ final state the background uncertainty due to the experimental selections was estimated to be between 5% and 10% [34]. Figure 9.13 (right plot) shows the $5\sigma$ discovery contours for the $\phi \to hh \to \tau\tau b\bar{b}$ channel for 30 fb$^{-1}$. The two contours correspond to the variation of the background NLO cross sections due to the scale uncertainty. The 5% experimental systematics on the background is taken into account.

## 9.5 Search for Randall-Sundrum excitations of gravitons decaying into two photons in CMS

*Marie-Claude Lemaire, Vladimir Litvin and Harvey Newman*

To test the Randall-Sundrum model, not only Higgs/radion but also graviton production needs to be studied. This contribution discusses the CMS discovery potential; full simulation and reconstruction are used to study diphoton decay of Randall-Sundrum gravitons. For an integrated luminosity of 30 fb$^{-1}$ diphoton decays of Randall-Sundrum gravitons can be discovered at $5\sigma$ level for masses up to 1.61 TeV in case of weak coupling between graviton excitations and Standard Model particles $k/\bar{M}_{Pl} \equiv c\,(c = 0.01)$. Heavier resonances can be detected for larger coupling constant ($c = 0.1$), with mass reach of 3.95 TeV.

The search for the $G \to \gamma\gamma$ signal at LHC is affected by four types of backgrounds:

- The prompt di-photon production from the quark annihilation ("Born") and gluon fusion diagrams ("Box"), which provides an intrinsic or 'irreducible' background.
- The $\gamma$ + jets ("Brem") production consisting of two parts: i) prompt photon from hard interaction + the second photon coming from the outgoing quark due to final state radiation and ii) prompt photon from hard interaction + the decay of a neutral hadron (mostly isolated $\pi^0$) in a jet, which could fake a real photon.

- The background from QCD hadronic jets ("QCD"), where electromagnetic energy deposits result from the decay of neutral hadrons (especially isolated $\pi^0$s) in both jets.
- Drell Yan process with $e^+e^-$ in a final state ("DY") which could mimic photons when correspondent electron tracks will not be assigned to the superclusters during the reconstruction.

The analysis was done as follows:

1 To find two super-clusters (SCs) with $E_T > 150$ GeV and two HLT trigger bits triggered at the same time: 1) 2p (two isolated photons with trasnverse energies larger than 40 GeV and 25 GeV respectively, without any other energy deposited in electromagnetic calorimeter within distance R ($R = \sqrt{\delta\eta^2 + \delta\phi^2}$) less than 0.5 from the photon); and 2) r2p (two photons with trasnverse energies larger than 40 GeV and 25 GeV respectively, without any isolation).

2 Calorimeter isolation criteria: for each SC the energy in a cone of $\Delta R = 0.5$ (excluding SC itself) should be $< 0.02E_T(SC)$

3 $E(\mathrm{HCAL})/E(\mathrm{ECAL}) < 0.05$

4 Tracker isolation: the sum of the energy of all tracks in a cone $\Delta R = 0.5$ around the SC should be $< 0.01E_T(SC)$

5 Photon energy corrections are done in a simple way so far:
   - If the crystal in SC with largest deposited energy had less than 1.7 TeV, only simple energy dependent part of correction is applied (just a shift of the peak)
   - If the crystal in SC with largest deposited energy had more than 1.7 TeV, the Multi-Gain-Pre-Amplifier (MGPA) saturation correction (1d) was applied

To produce the final results and to calculate the expected statistical significance for RS-1 graviton search recently calculated next-to-leading order corrections (K factors) to the cross sections of different types of background are used: K = 1.5 for quark annihilation [41], K = 1.2 for gluon fusion [42], K = 1 for the $\gamma$ + hadronic jets [42] and K = 1 for QCD jets. For signal, a conservative K= 1 value is taken.

The graviton invariant mass is reconstructed from the two Super-Clusters. For each value of the generated graviton mass, the corresponding peak is fitted to a Gaussian distribution. The $\sigma$ of the fit is $\simeq$ 10 GeV for $M_G$ = 1.5 TeV and c=0.01, reflecting the detector energy resolution, which is slightly below 0.5% constant term, as obtained from 2004 test beam data; and an additional contribution of 0.16% which is due to the reconstruction. For $M_G$ = 3.5 TeV and c=0.1 it increases up to $\simeq$ 35 GeV, due to the natural width of the resonance.

A $\pm3\sigma$ window is defined around the fitted peak to compute the numbers of signal and background events, $N_s$ and $N_{bkg}$. The corresponding number of events, obtained through the successive analysis cuts described above are given for an integrated luminosity 30 fb$^{-1}$ in Table 9.2 for ($M_G = 1.5$ TeV, $c = 0.01$) and in Table 9.3 for ($M_G = 3.5$ TeV, $c = 0.1$).

Signals over backgrounds with all events satisfying all the selection cuts are displayed in Fig. 9.14 for ($M_G = 1.5$ TeV, $c = 0.01$), ($M_G = 3.0$ TeV, $c = 0.1$) and for an integrated luminosity of 30 fb$^{-1}$. In Fig. 9.15, signal over backgrounds are given for ($M_G = 1.0$ TeV, $c = 0.01$), ($M_G = 2.5$ TeV, $c = 0.1$) and for an integrated luminosity of 10 fb$^{-1}$.

Taking into account the K-factors described above, we have got next number of events for signal and background and calculated significance for $c = 0.01$ and $c = 0.1$ , for the $L = 30$ fb$^{-1}$. $S = \sqrt{2lnQ}$, with $Q = (1 + n_s/n_b)^{n_s+n_b}exp(-n_s)$ in Tables 9.4 and 9.5.
Expected statistical significance $S_{L2}$ is plotted for $(M_G, c)$ space for 10, 30 and 60 fb$^{-1}$ (Fig. 9.16). Uncertainties were not taken into account in Fig. 9.16.

The discovery region in the plane of the coupling parameter $c$ and the graviton mass is given in Fig. 9.17.

Table 9.2: Number of events passed through the analysis cuts defined above for $M_G = 1.5$ TeV, $c = 0.01$ and $L = 30$ fb$^{-1}$. Leading column is non-saturated events, all saturated events, passed through the analysis, were added in brackets, where applied.

| | Signal | Born (k=1.5) | Box (k=1.2) | Brem (k=1) | QCD (k=1) | DY (k=1) |
|---|---|---|---|---|---|---|
| trigger + 2SC | 28.9 | 8.6 | 0.10 | 29.2 | 798.7 | 4.3 |
| + EM isolation | 24.5 | 5.5 | 0.08 | 20.3 | 361.8 | 3.5 |
| + HCAL/ECAL | 24.3 | 5.4 | 0.08 | 4.4 | 12.8 | 3.5 |
| + tracker isolation | 17.6 | 4.2(+0.2) | 0.05 | 0.17 | 0.0 | 0.0 |

Table 9.3: Number of events passed through the analysis cuts defined above for $M_G = 3.5$ TeV, $c = 0.1$ and $L = 30$ fb$^{-1}$. Leading column is non-saturated events, all saturated events, passed through the analysis, were added in brackets, where applied.

| | Signal | Born (k=1.5) | Box (k=1.2) | Brem (k=1) | QCD (k=1) | DY (k=1) |
|---|---|---|---|---|---|---|
| trigger + 2SC | 11.6 | 0.20 | $4.4 * 10^{-4}$ | 0.78 | 821.9 | 0.10 |
| + EM isolation | 10.8 | 0.14 | $3.6 * 10^{-4}$ | 0.32 | 164.4 | 0.095 |
| + HCAL/ECAL | 10.6 | 0.13 | $3.4 * 10^{-4}$ | 0.016 | 0.0 | 0.095 |
| + tracker isolation | 8.9(+1.0) | 0.10(+0.02) | $2.7(+0.24) * 10^{-4}$ | $1.7 * 10^{-3}$ | 0.0 | $7.2 * 10^{-4}$ |

Table 9.4: Significance for $c = 0.01$ and $L = 30$ fb$^{-1}$

| | $M_G = 1.0$ TeV | $M_G = 1.25$ TeV | $M_G = 1.5$ TeV | $M_G = 1.75$ TeV | $M_G = 2.0$ TeV |
|---|---|---|---|---|---|
| $N_s$ | 135.8 | 44.0 | 17.6 | 7.3 | 3.9 |
| $N_{bkg}$ | 15.0 | 8.8 | 4.6 | 1.8 | 1.2 |
| Significance | 20.6 | 10.1 | 5.9 | 3.9 | 2.6 |

Table 9.5: Significance for $c = 0.1$ and $L = 30$ fb$^{-1}$

| | $M_G = 2.5$ TeV | $M_G = 3.0$ TeV | $M_G = 3.5$ TeV | $M_G = 4.0$ TeV | $M_G = 4.5$ TeV |
|---|---|---|---|---|---|
| $N_s$ | 103.8 | 31.6 | 9.9 | 3.44 | 1.11 |
| $N_{bkg}$ | 1.11 | 0.35 | 0.13 | 0.06 | 0.02 |
| Significance | 27.3 | 15.0 | 8.2 | 4.6 | 2.6 |

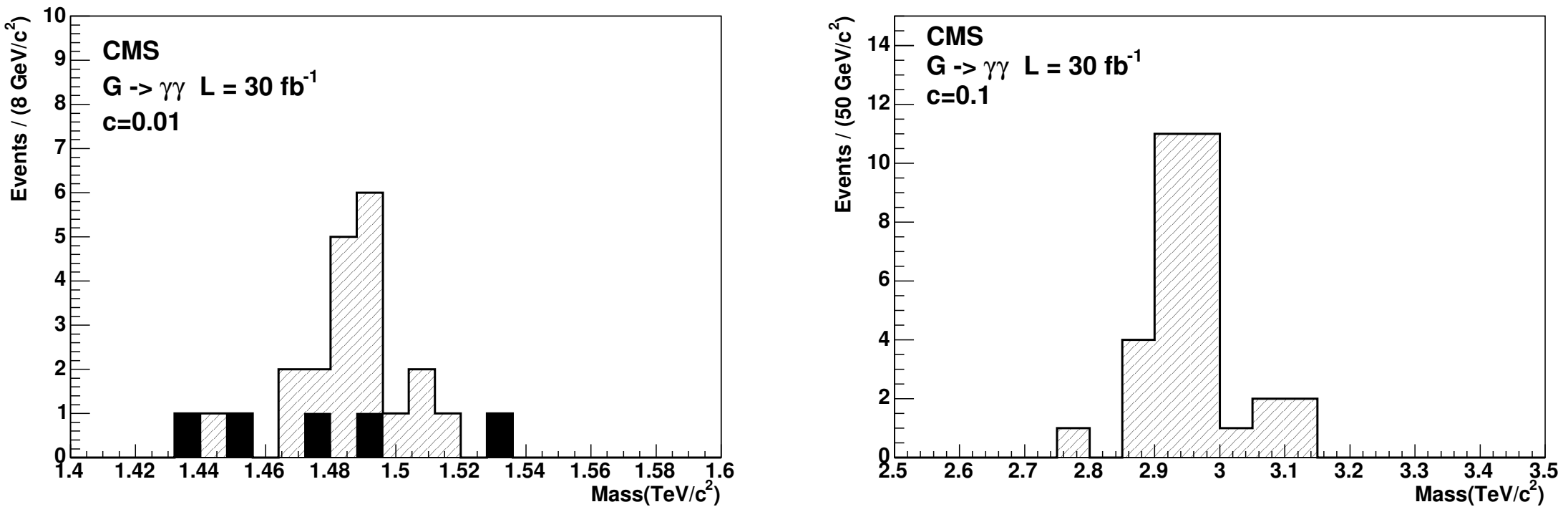

Fig. 9.14: Number of events passed through all cuts for (1.5 TeV, 0.01) (left) and (3.0 TeV, 0.1) (right) RS-1 gravitons for 30 fb$^{-1}$ integrated luminosity.

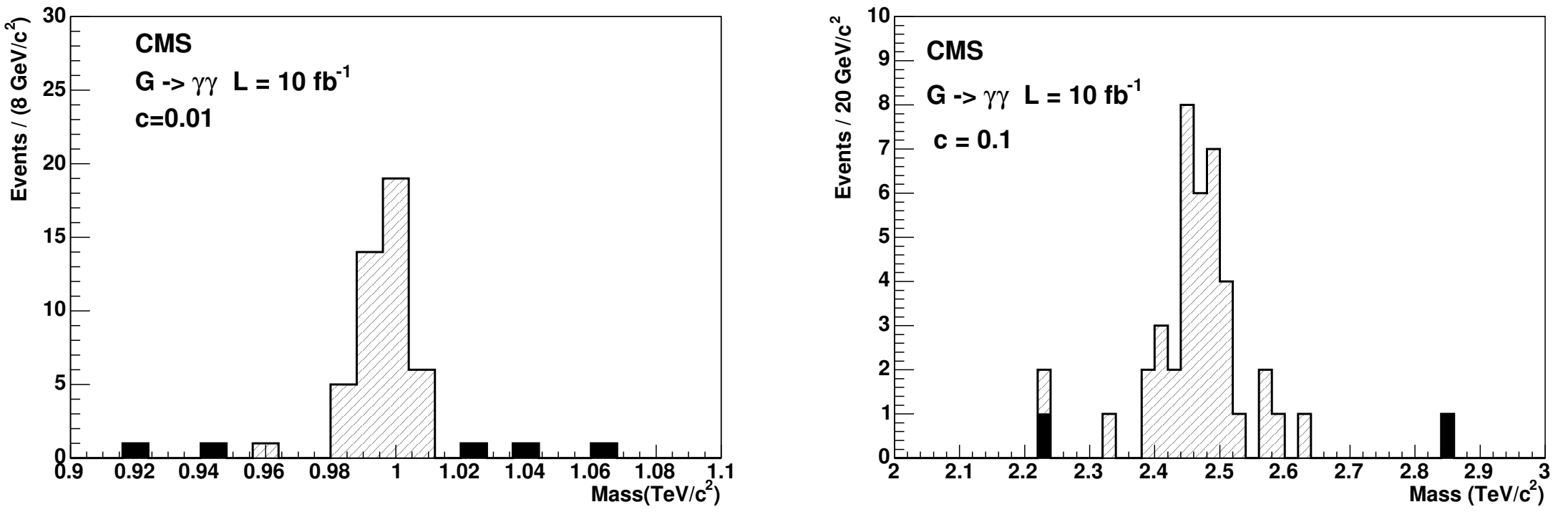

Fig. 9.15: Number of events passed through all cuts for (1.0 TeV, 0.01) (left) and (2.5 TeV,0.1) (right) RS-1 gravitons for 10 fb$^{-1}$ integrated luminosity.

Recent results for $5\sigma$ discovery limits for 10, 30 and 60 fb$^{-1}$ are presented in Table 9.6. Confidence limits uncertainties were calculated for 30 fb$^{-1}$ and also presented in Table 9.6. The first uncertainty corresponds to hard scale uncertainties (it was varied from $0.25\hat{s}$ to $4\hat{s}$, default value was $\hat{s}$ of the subprocess). The second uncertainty corresponds to cross section uncertainties because of PDF uncertainties. The third uncertainty corresponds to the fact, that Tevatron most recent measures pointed out that Born K-factor might be closer to 2 [43] instead of 1.5 as was used in this study.

Table 9.6: Results for $5\sigma$ discovery limits for 10, 30 and 60 fb$^{-1}$ and confidence limits uncertainties for 30 fb$^{-1}$

| | for 60 fb$^{-1}$, GeV | for 30 fb$^{-1}$, GeV | for 10 fb$^{-1}$, GeV |
|---|---|---|---|
| $c = 0.01$ | 1820 | $1610 + (+56 - 62) \pm 55 \pm 50$ | 1310 |
| $c = 0.1$ | 4270 | $3950 + (+42 - 47) \pm 152 \pm 30$ | 3470 |

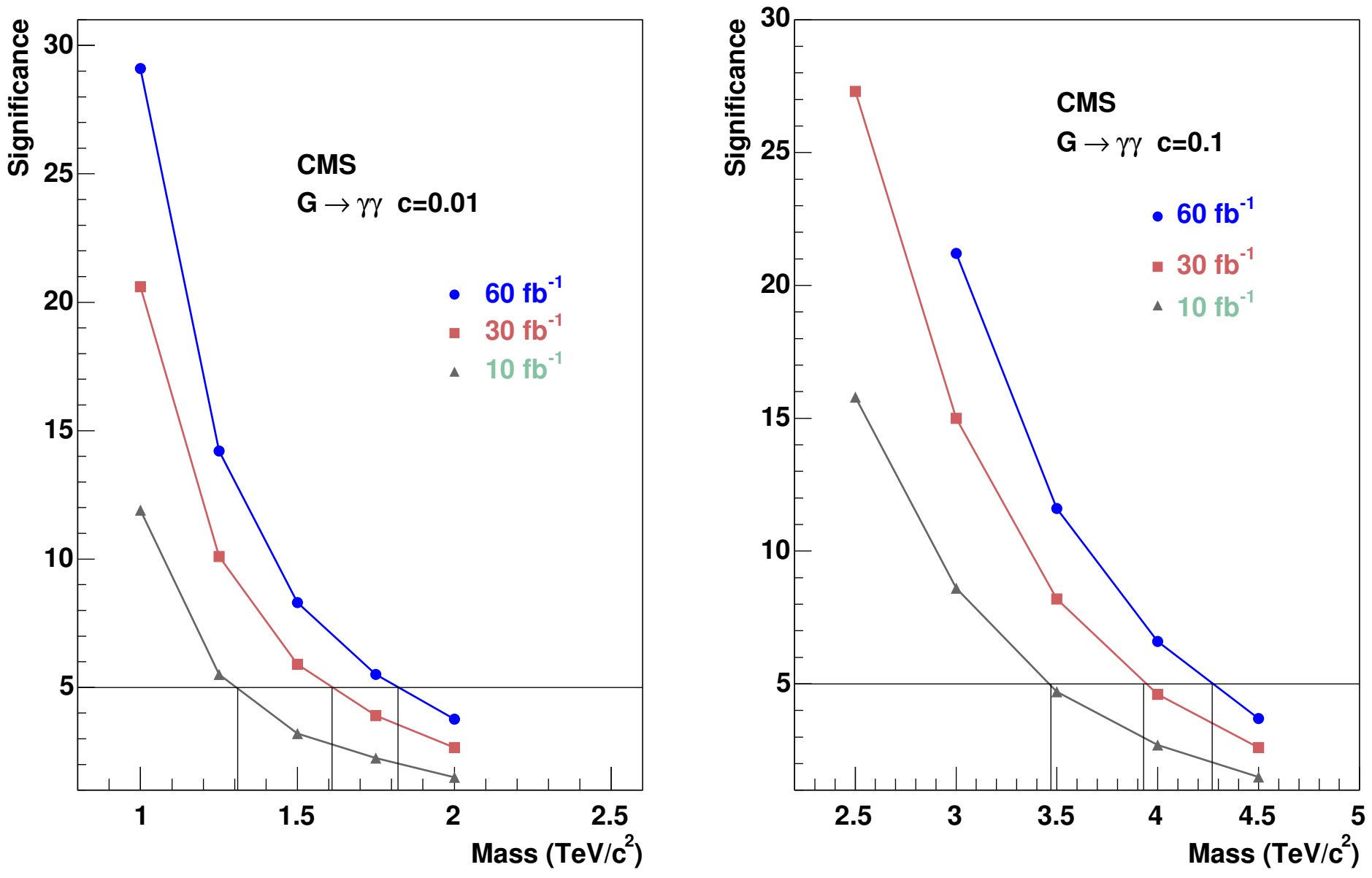

Fig. 9.16: Significance as a function of the graviton mass for 10 fb$^{-1}$ and 30 fb$^{-1}$ integrated luminositiies, c=0.01 (left) and c=0.1 (right)

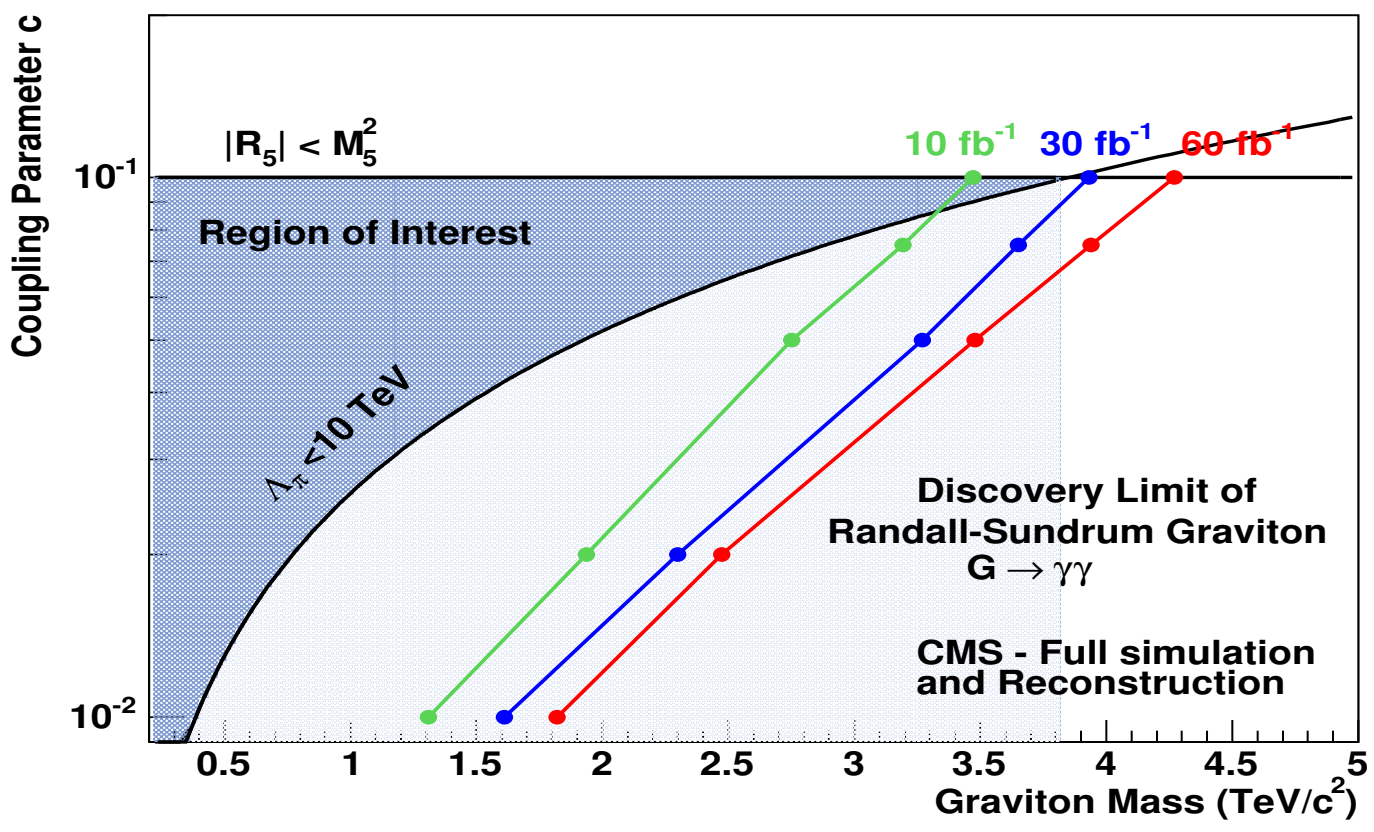

Fig. 9.17: Reach of the CMS experiment in the search for the Randall-Sundrum graviton decaying into diphoton channel as a function of the coupling parameter $c$ and the graviton mass for 10 fb$^{-1}$, 30 fb$^{-1}$ and 60 fb$^{-1}$. The left part of each curve is the region where the significance exceeds $5\sigma$.

## 9.6 SUSY Higgs production from 5-D warped supergravity

*Steve Fitzgerald*

Scenarios which combine the five-dimensional warped spacetime of the Randall-Sundrum model [1,44] with supersymmetry are interesting for a number of reasons. Firstly, they address the issue of radion stabilization – the energy scale at which SUSY is broken corresponds to the scale at which the scale-invariance of the Randall-Sundrum solution is broken, and hence sets the scale of the extra dimension; see for example [45–49]. Secondly, since they are derived from supergravity, the Giudice-Masiero mechanism [50] to resolve the $\mu$-problem can be implemented. Thirdly, it seems possible to make contact with Type IIB string theory via a supersymmetric warped intermediate sector [51–54]. In this work, Higgs production in vector boson fusion is investigated, in the context of a minimal supersymmetric extension of the Standard Model realized on $\mathcal{M}_4 \times S_1/Z_2$. The model, [55], is inspired by 5-D $\mathcal{N} = 2$ supergravity. The $VV - H_i^0$ channel is sensitive to these scenarios, as it receives contributions both from corrections to the $VV - H_i^0$ vertex itself, and from new 4-point vertices appearing in the 4-D effective Lagrangian. Production cross-sections, $k_T$ and rapidity distributions are calculated for a high-energy $e^+e^-$ collider, and compared with those of the MSSM.

### *9.6.1 The model*

The model under consideration first appeared in [55]. Starting from 5-D, $\mathcal{N} = 2$ supergravity realized on $\mathcal{M}_4 \otimes S_1/Z_2$, constant brane energy densities are added at the $Z_2$ fixed points. The 5-D Einstein equations then admit the Randall-Sundrum solution

$$ds^2 = e^{-2kr_c\phi}\eta_{\mu\nu}dx^\mu dx^\nu - r_c^2 d\phi^2 \tag{9.22}$$

and the 5-D, $\mathcal{N} = 2$ SUSY in bulk is broken to 4-D, $\mathcal{N} = 1$ on the brane by the localized energy densities and orbifold projection. If we give the modulus field for the fifth dimension ($T$) a VEV:

$$\langle T + T^\dagger \rangle = \log\left[\frac{3M_p^2}{\Lambda^2}\right]; \quad T = \langle T \rangle + \frac{t}{\Lambda} \tag{9.23}$$

we can extract the 4-D effective theory on the brane fixed at $\phi = \pi$, and arrive at a theory which looks like the MSSM with some important differences. In the equation above, $M_P$ is the unwarped 4-D Planck mass, related to the 5-D Planck mass via $kM_P^2 = M_5^3$; $\Lambda$ and $t$ are defined below.

We use the standard MSSM superpotential with $\mu$-term modified (see later):

$$\begin{aligned} W &= h_U^{ij} Q_{Li} \cdot H_2 U_{Rj} + h_D^{ij} H_1 \cdot Q_{Li} D_{Rj} \\ &+ h_E^{ij} H_1 \cdot L_{Li} E_{Rj} + \mu H_1 \cdot H_2. \end{aligned} \tag{9.24}$$

A non-minimal effective Kähler potential $K_{\text{eff}}$ is obtained by integrating the action over the compactified dimension and comparing the curvature term with 4-D SUGRA. This is due to the fact that we are trying to match an effective theory defined on a flat 4-D spacetime with a higher-dimensional bulk which is curved. This necessitates some important field redefinitions which result in an additional level of mixing (see later).

$$K_{\text{eff}} = \Lambda^2 \exp\left\{ -\frac{t+t^*}{\Lambda} + \frac{1}{\Lambda^2}\sum_i |\phi_i|^2 + (\lambda H_1 \cdot H_2 + \text{h.c.}) \right\} \tag{9.25}$$

In the above expression, $\Lambda$ represents the cutoff for the effective theory, $\phi_i$ represents all matter + Higgs scalars, and $\lambda$ is a parameter allowing "chiral" Higgs terms in $K$ ("$\cdot$" = $SU(2)$ product $H_1 \cdot H_2 = H_1^0 H_2^0 - H_1^- H_2^+$). There also arises an extra neutral scalar $t$ – the *radion*, which can mix with the neutral

Higgses, and its fermionic partner $\chi_t$, the *radino*, which can mix with the neutral gauginos and Higgsinos, resulting in five neutralinos, compared with four in the MSSM. $\lambda$ contributes to the $\mu$-parameter in the superpotential:

$$\mu = \mu_0 + \lambda m_{3/2}, \tag{9.26}$$

via the Giudice-Masiero mechanism. One can remove $\mu_0$ entirely, so $\mu \sim m_{3/2}$, or leave it in at $\mathcal{O}(M_P)$, naturally warped down to $\mathcal{O}(\Lambda)$. The 'free' parameters $\Lambda$, $\lambda$, $\mu_0$, and $m_{3/2}$ are constrained by electroweak symmetry breaking (EWSB) as follows [55]:

$$\lambda \sim \frac{-1}{3+\sqrt{3}} \left\{ 1 - (1+\sqrt{3})\frac{\mu}{m_{3/2}} + \frac{\mu^2}{m_{3/2}^2} \right\}. \tag{9.27}$$

The MSSM parameter $\tan\beta = 1$ after EWSB.

The scalar kinetic terms in the effective Lagrangian

$$\sim \frac{\partial^2 K}{\partial\phi_I \partial\phi_J^*} D_\mu \phi_I D^\mu \phi_J \equiv K_{IJ} D_\mu \phi_I D^\mu \phi_J, \tag{9.28}$$

where $K_{IJ}$ is the *Kähler metric*. For a minimal $K = \sum_i |\phi_i|^2$, $K_{IJ} = \delta_{IJ}$, and the usual structure is recovered. However, a non-minimal $K$ leads to a non-diagonal $K_{IJ}$ and what is known as *kinetic mixing* of the scalar fields, when one performs the field redefinition to return the kinetic terms to canonical form. The effect is least suppressed in the neutral scalar and neutralino sector.

These (holomorphic) transformations take the form (to $\mathcal{O}(1/\Lambda)$):

$$\begin{pmatrix} H_1^0 \\ H_2^0 \\ t \end{pmatrix} \longrightarrow \begin{pmatrix} 1 & 0 & \frac{1+\lambda}{4}\frac{v}{\Lambda} \\ 0 & 1 & \frac{1+\lambda}{4}\frac{v}{\Lambda} \\ \frac{1+\lambda}{4}\frac{v}{\Lambda} & \frac{1+\lambda}{4}\frac{v}{\Lambda} & 1 \end{pmatrix} \begin{pmatrix} H_1^0 \\ H_2^0 \\ t \end{pmatrix} \tag{9.29}$$

diagonalizing kinetic terms and rescaling fields. There are two levels of mixing: the above plus the normal weak mixing of neutral degrees of freedom. This leads to (CP-even scalar) mass matrix

$$M^2 = \begin{pmatrix} A + B/\Lambda^2 & 0 & 0 \\ 0 & C/\Lambda^2 & D/\Lambda \\ 0 & D/\Lambda & E + F/\Lambda^2 \end{pmatrix} \tag{9.30}$$

with $A$ to $F$ depending on other scales in the theory, $v$, $m_{3/2}$, $\mu$ (Appendix C of [55]). The first scalar is still an exact eigenstate – there is no 1–2, 1–3 mixing. As $\Lambda \to \infty$, $\tan\beta \to 1$, i.e. a D-flat direction in this limit. The finite-$\Lambda$ corrections break the D-flatness. Also, the lightest Higgs can be up to $\sim$ 700 GeV without violating the LEP bounds. Figure 9.18 shows the mass of the lightest scalar vs. $\mu$ for various values of $\Lambda$. One finds an an *upper* bound of $\Lambda \sim$ 4.4–9.5 TeV from $m_H > 114$ GeV (from breaking of D-flatness by $\Lambda$-suppressed terms).

### 9.6.2 SUSY Higgs production in $W^+W^-$ fusion

We now consider the process $e^+e^- \longrightarrow \nu\bar{\nu}H_0$. Below are the SM/MSSM tree-level diagrams:

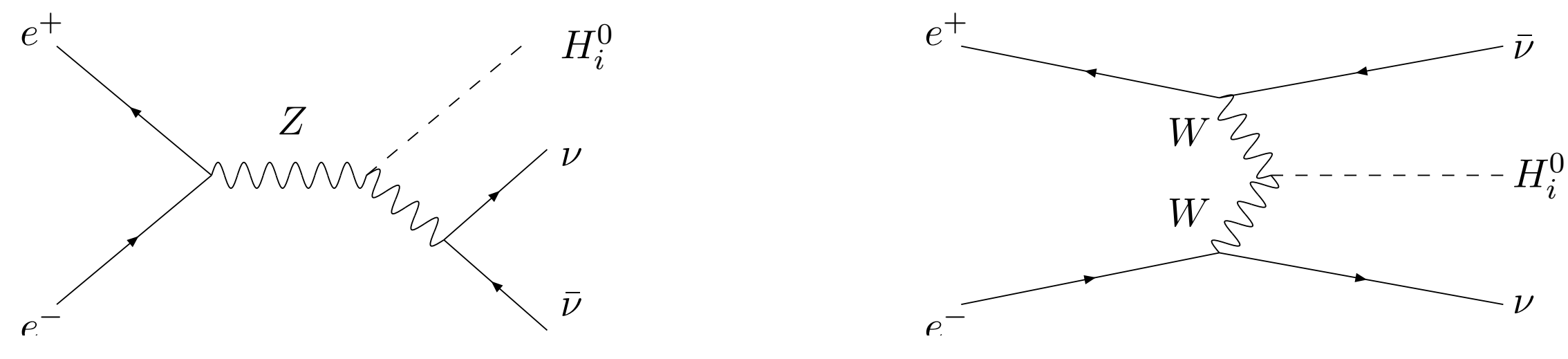

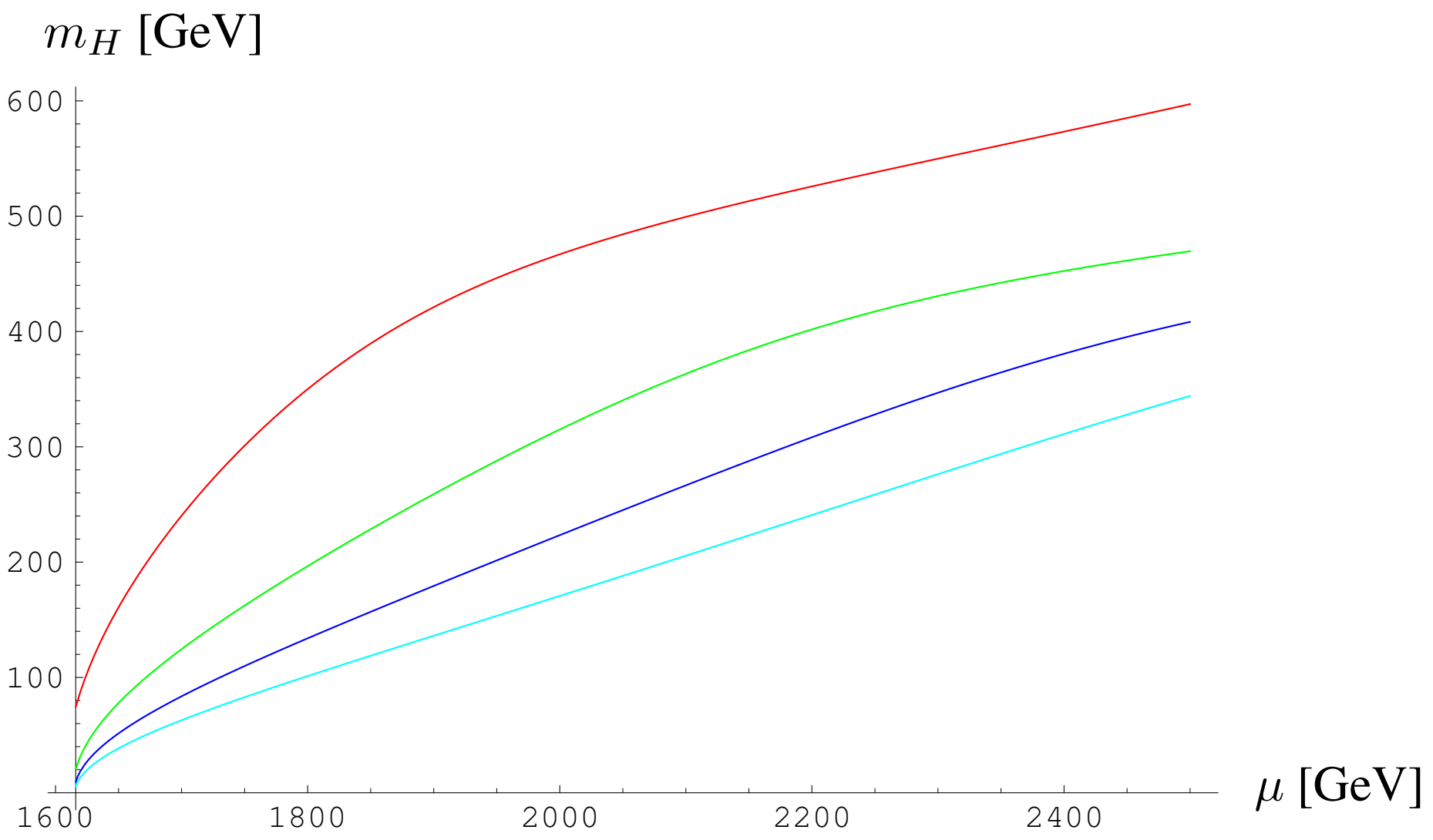

Fig. 9.18: The mass of the lightest scalar as a function of $\mu$ for (top to bottom) $\Lambda = 2, 4, 6, 8$ TeV.

We can safely neglect the Higgsstrahlung diagram at high $\sqrt{s}$, as $t$-channel $\ln s$ growth dominates, and we also neglect diagrams with Higgs radiation off $e^+, e^-$, due to the small Yukawas. New and modified vertices appear in the 4-D effective interaction Lagrangian: the $WWH$ vertex is modified and a new adjacency 4 vertex arises (with associated Feynman rules in Eq. (9.31)):

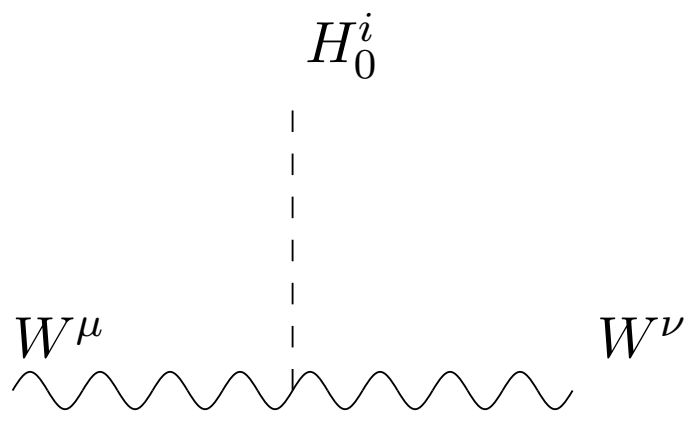

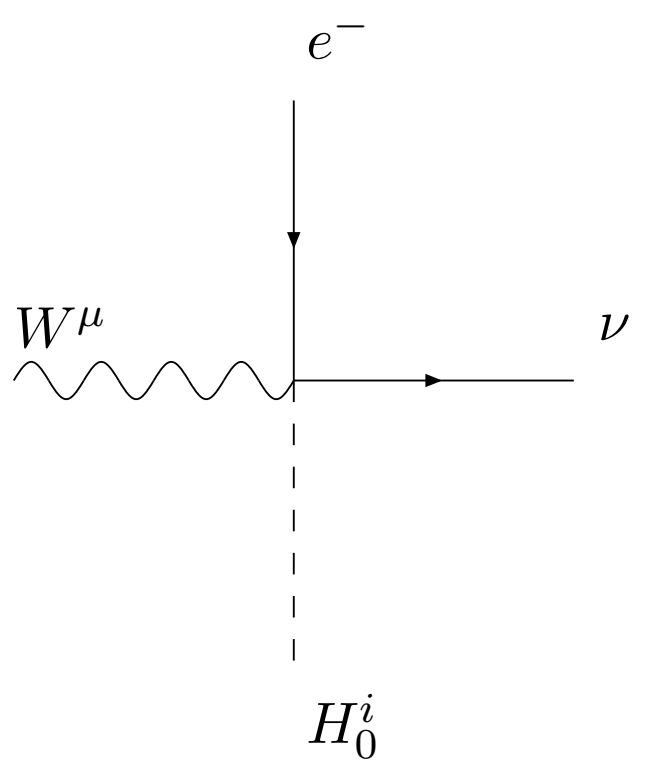

$$\frac{ie^2}{2\sin^2\theta_W}\left\{v(Z_R^{1i}+Z_R^{2i})+\frac{1-2\sqrt{2}+\lambda}{2}\frac{v^2}{\Lambda}Z_R^{3i}\right\}g^{\mu\nu};\quad \frac{ie}{\Lambda\sin\theta_W}\gamma^\mu P_L Z_R^{3i}. \tag{9.31}$$

The following new diagrams appear:

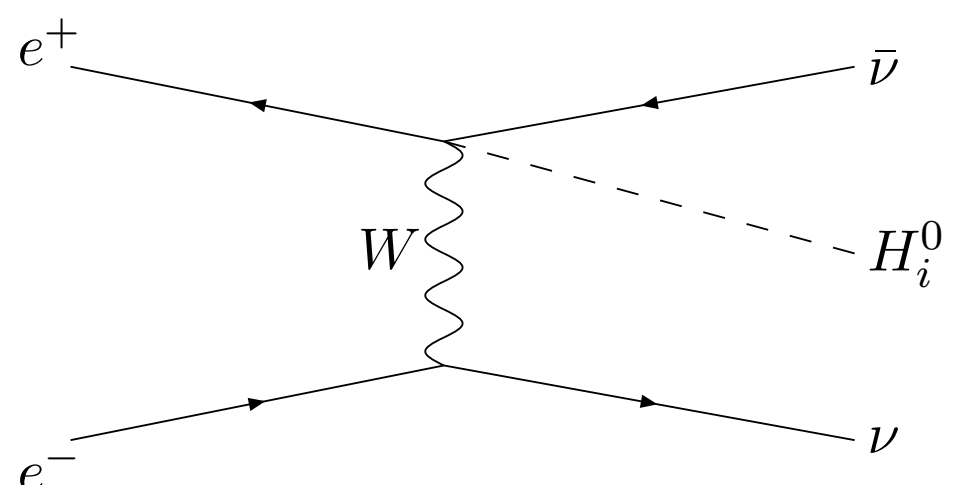

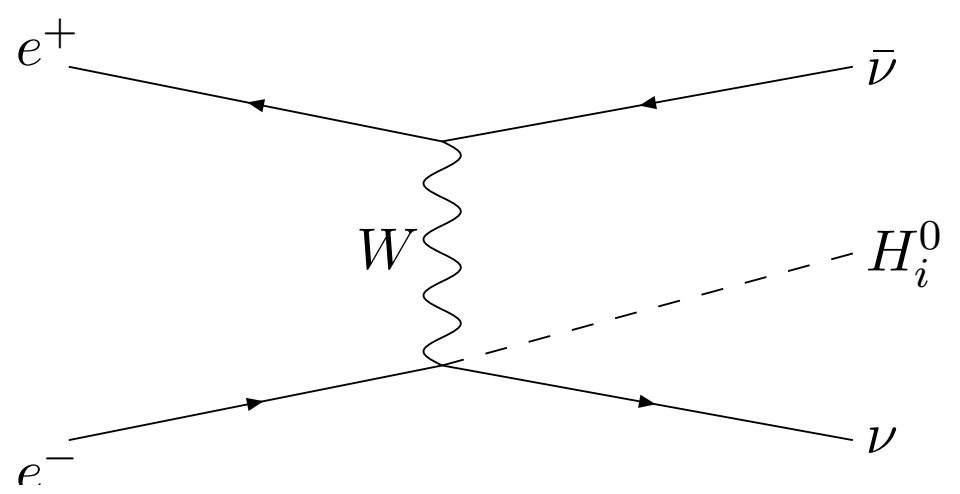

Table 9.7: Parameters of points used

| | Point 1 | Point 2 | Point 3 | MSSM | SM |
|---|---|---|---|---|---|
| $\Lambda$ (GeV) | 4000 | 4000 | 8000 | N/A | N/A |
| $m_{3/2}$ (GeV) | 500 | 350 | 500 | N/A | N/A |
| $R_1 + R_2$ | 0.97 | 0.96 | 0.99 | 0.90 | 0 |
| $R_3$ | $-8D-7$ | $-1.9D-6$ | $-5.5D-7$ | 0 | 0 |
| $\mu$ (GeV) | 1695 | 1234 | 1853 | 1850 | N/A |

The extra ($\frac{1}{\Lambda}$-suppressed) diagrams come from the following structure in the effective Lagrangian:

$$\mathcal{L} \sim \left(1 - \frac{t + t^*}{\Lambda}\right) \times \mathcal{L}^{\rm kinetic}_{\rm MSSM} \tag{9.32}$$

Also, the contribution from term $(1/\Lambda)\bar{\chi}_t \sigma^\mu \chi_t \partial_\mu t$ gives a correction to Yukawas, and hence the following diagrams appear:

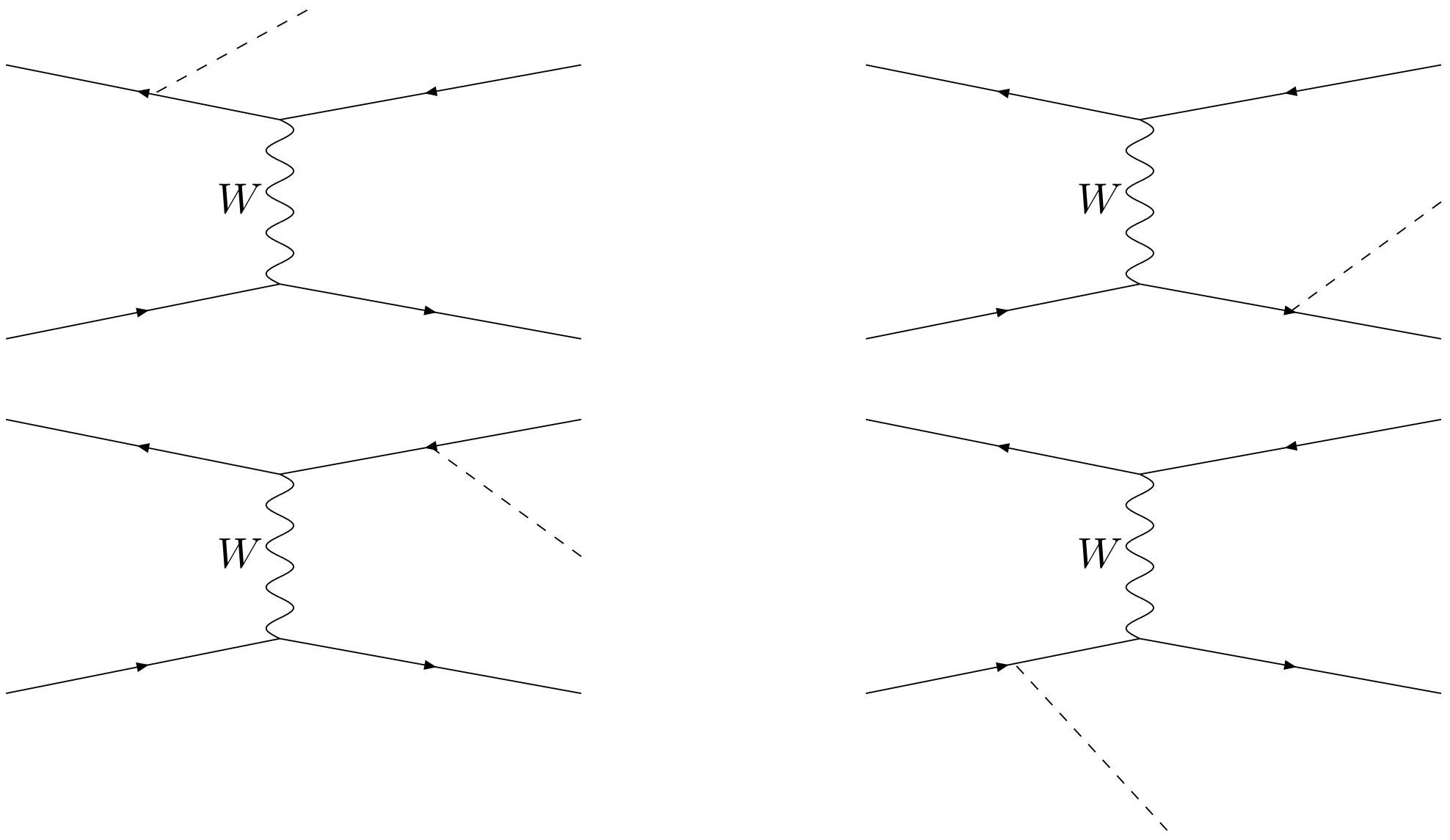

However, their contributions cancel exactly in the matrix element. Any potentially large effects on the decays, say, $\sqrt{s}/\Lambda$, say, cancel, leaving observables unaffected to $\mathcal{O}(m_f/\Lambda)$. We parameterize the matrix element as

$$|\mathcal{M}|^2 = |\mathcal{M}_{SM}|^2 \{R_1 + R_2 + R_3(p_1 \cdot p_1' + p_2 \cdot p_2' + m_W^2)\}, \tag{9.33}$$

where:

- $R_1$ : like $\sin^2(\beta - \alpha)$ in MSSM.
- $R_2$ : Correction to WWH vertex.
- $R_3$ : Contribution from new diagrams.

The dominant effect arises from the WWH vertex modification (see Table 9.7). Can it be distinguished from normal mSUGRA? We choose a Higgs mass of 120 GeV (we assume it is known from LHC). The mixing matrices, and hence the normalization of the cross-section, will differ from that of mSUGRA. Table 9.7 gives the parameters of the points chosen for comparison. The mSUGRA point used has lightest scalar at 120 GeV and $\tan\beta = 3$, (from using the low energy mass matrix, $\beta, m_A, m_Z$ in the standard way).

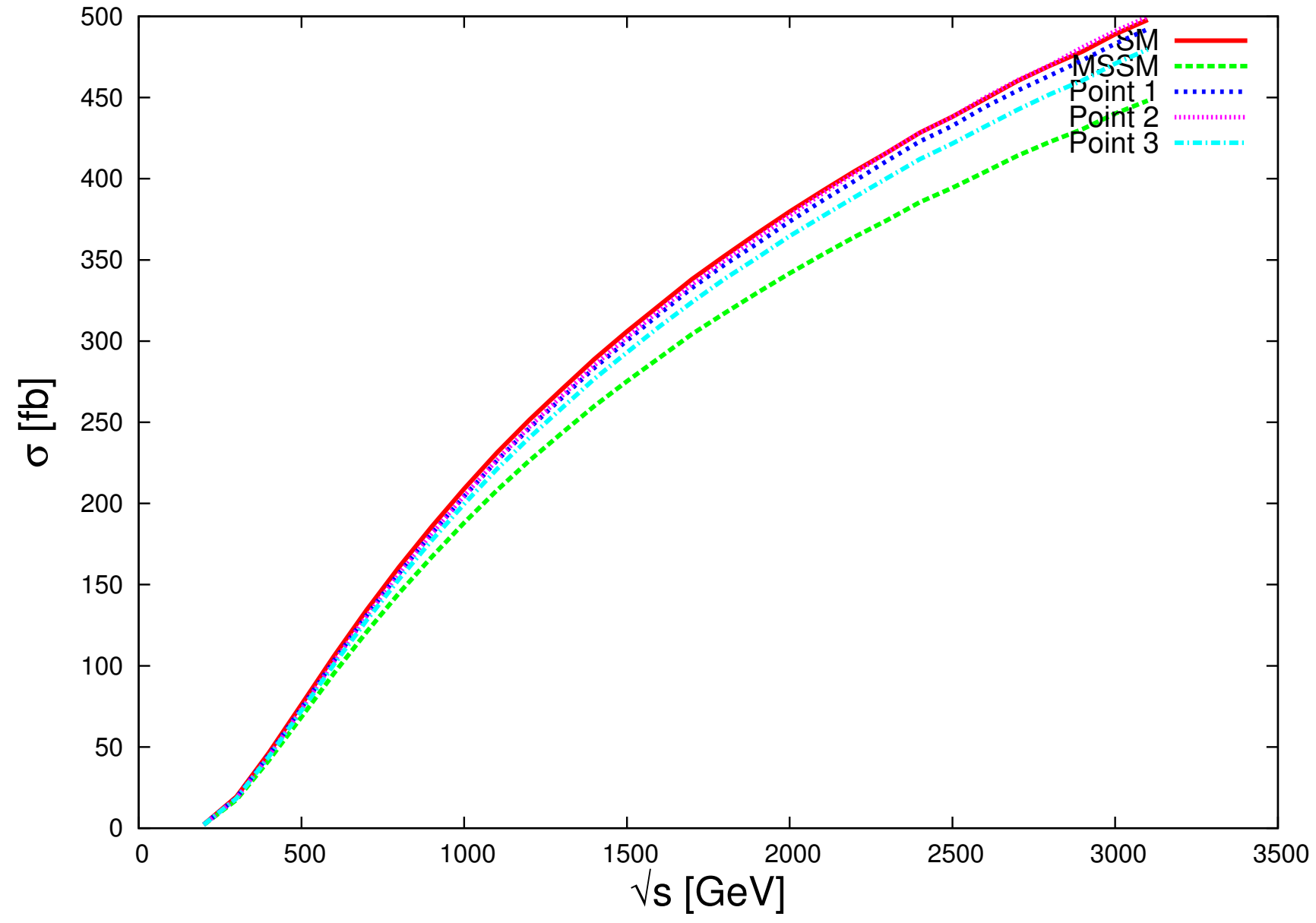

Fig. 9.19: Total cross sections for the points of Table 9.7.

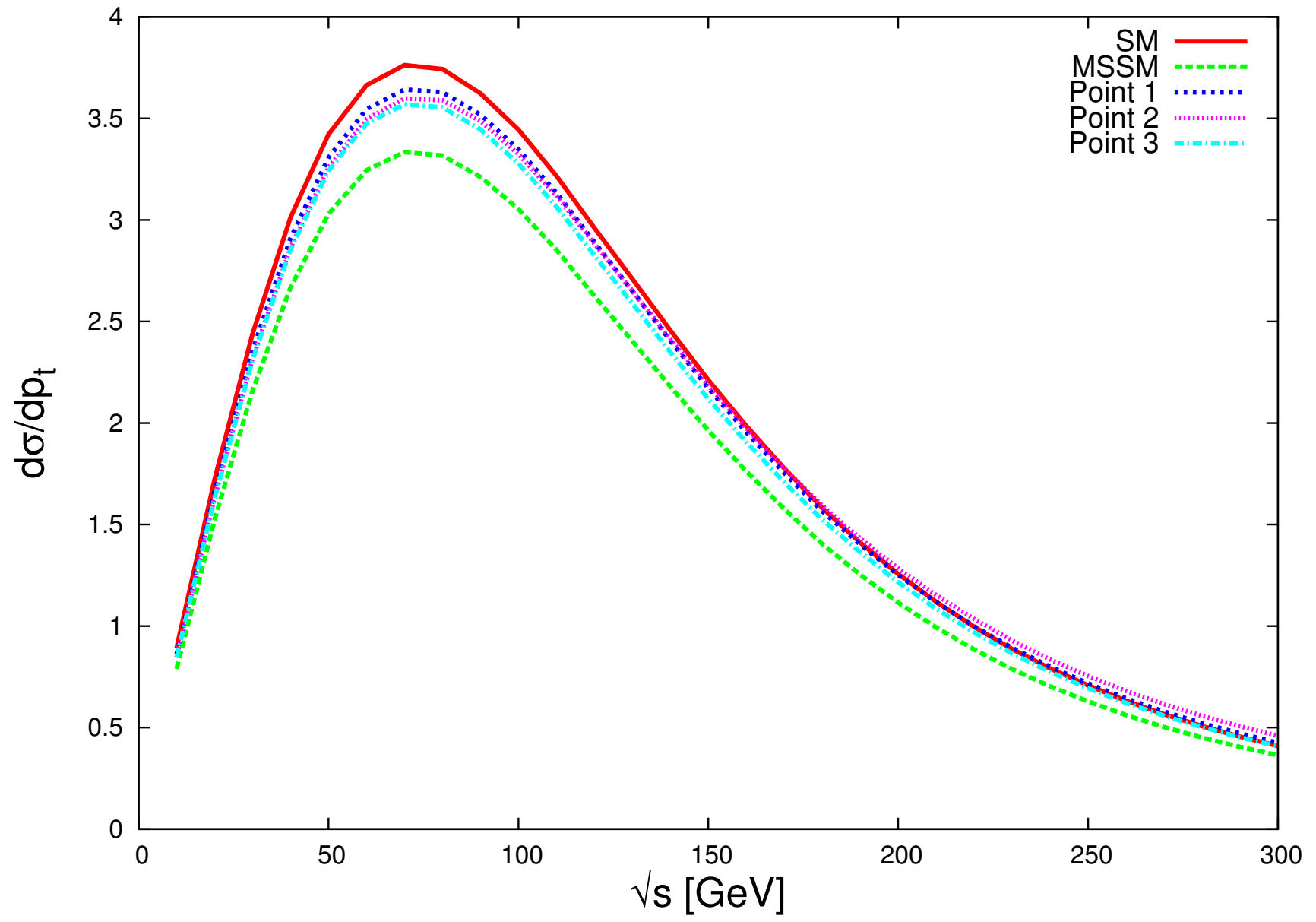

Fig. 9.20: $p_T$ distributions for the points of Table 9.7.

It should be noted that these example points above represent a 'worst case scenario,' where the model would be most difficult to distinguish from the closest mSUGRA point, and indeed in this scenario it would be difficult, as the differences are only about $10\%$. This is illustrated in Figs. 9.19 and 9.20, where the cross sections and $p_T$ distributions for the example points are plotted against $\sqrt{s}$. A more likely scenario would be an unusually heavy lightest scalar, which could be satisfactorily explained by a scenario like the one under consideration. Having a cutoff for an effective theory in the TeV region necessarily makes phenomenological effects small, and a linear collider would be indispensable in distinguishing this model from other SUSY scenarios.

## 10 HIGGSLESS MODELS

### 10.1 Introduction

*Ben Lillie and John Terning*

Recent developments in string theory have definitely had an impact of phenomenological model building. The possible existence of branes in large extra dimensions has opened up new classes of theories, especially in the area of electroweak symmetry breaking. The anti-de Sitter/conformal field theory (AdS/CFT) correspondence led to the Randall-Sundrum (RS) model [1], see Section 9, which allows for a new approach to the hierarchy problem. Thus discovering inverse TeV sized extra dimensions at the LHC has become a tantalizing possibility.

The existence of inverse TeV sized dimensions themselves allow for a completely new way to break electroweak symmetry: boundary conditions in the extra dimension [2]. Since this mechanism is intrinsically extra dimensional it leads to a very different phenomenology from the standard Higgs mechanism. In fact, the Dirichlet boundary condition required to break the gauge symmetry can be thought of as arising through the limit of a Higgs with an infinite VEV. Since the Higgs mass is of the order of its VEV, we see that boundary condition breaking is effectively a class of Higgsless models for electroweak symmetry breaking.

At asymptotically high energies it can be shown that the terms in the $WW$ scattering amplitude that grow with energy are cancelled by the exchange of $W$ Kaluza-Klein (KK) modes [2, 3]. Quark and lepton masses can also arise through boundary conditions [4]. In a warped AdS background (like the RS model) a custodial symmetry can ensure the correct ratio for the $W$ and $Z$ masses [5]. Most corrections to precision electroweak measurements and $Z'$ couplings can suppressed if the quarks and leptons are approximately uniformly spread out in the extra dimension.

However not everything is rosy in Higgsless models. If the $W'$ and $Z'$ resonances are too heavy (roughly $> 1$ TeV) then $WW$ scattering becomes strongly coupled [6–8]. Also implementing a mechanism to produce the top quark mass without messing up the $Zb\bar{b}$ coupling is quite difficult.

The most obvious implication of this model for colliders is that no physical Higgs state will be found. However, there are many new positive signals that can be searched for. In particular the Kaluza-Klein states of the gauge bosons will be easily visible in the Drell-Yan and dijet channels [8], and analysis of longitudinal gauge boson scattering can directly probe the Higgsless mechanism of electroweak symmetry breaking [9].

In the following we will review the requirements for maintaining perturbative unitarity, the constraints imposed by precision electroweak measurements, and the collider signatures.

#### *10.1.1 KK mode couplings*

We will mainly focus on the most interesting case of a warped extra dimension. We will use the conformally flat metric

$$ds^2 = h(z)^2 \left(\eta_{\mu\nu} dx^\mu dx^\nu - dz^2\right) \tag{10.1}$$

where the extra spatial dimension $z$ is on the interval $[R, R']$. A flat extra dimension can be recovered by taking $h(z)$ = constant, while AdS is obtained by taking $h(z) = R/z$. If one is uncomfortable with a non-renormalizable 5D theory, one can always deconstruct the theory [10, 11] which provides a renormalizable 4D ultraviolet completion of the theory with the same low-energy (TeV) predictions.

The 5D gauge boson decomposes into a 4D gauge boson $A^a_\mu$ and a 4D scalar $A^a_5$ in the adjoint representation. Since there is a quadratic term mixing $A_\mu$ and $A_5$ we need to add a gauge fixing term that eliminates this cross term. Thus (using $\sqrt{-g} = h^5(z)$) we write the action after gauge fixing in R$_\xi$

gauge as

$$\mathcal{S} = \int d^4x \int_R^{R'} dz\, h(z) \left( -\frac{1}{4} F^a_{\mu\nu} F^{a\mu\nu} - \frac{1}{2} F^a_{5\nu} F^{a5\nu} - \frac{1}{2\xi} (\partial_\mu A^{a\mu} - \xi \partial_5 A^a_5)^2 \right), \tag{10.2}$$

where $F^a_{MN} = \partial_M A^a_N - \partial_N A^a_M + g_5 f^{abc} A^b_M A^c_N$, and the $f^{abc}$'s are the structure constants of the gauge group. The gauge fixing term is chosen such that (as usual) the cross terms between the 4D gauge fields $A^a_\mu$ and the 4D scalars $A^a_5$ cancel (see also [12]). Taking $\xi \to \infty$ will result in the unitary gauge, where all the KK modes of the scalars fields $A^a_5$ are unphysical (they become the longitudinal modes of the 4D gauge bosons), except if there is a zero mode for the $A_5$'s. We will assume that every $A^a_5$ mode is massive, and thus that all the $A_5$'s are eliminated in unitary gauge.

The variation of the action (10.2) leads, as usual after integration by parts, to the bulk equations of motion as well as to boundary terms (we denote by $[F]$ the boundary quantity $F(R') - F(R)$):

$$\begin{aligned} \delta\mathcal{S} &= \int d^4x\, dz \left( \partial_M h F^{aM\nu} - g_5 f^{abc} h\, F^{bM\nu} A^c_M + \frac{h}{\xi} \partial^\nu \partial^\sigma A^a_\sigma - h \partial^\nu \partial_5 A^a_5 \right) \delta A^a_\nu \\ &\quad - \int d^4x\, dz \left( h \partial^\sigma F^a_{\sigma 5} - g_5 f^{abc} h\, F^b_{\sigma 5} A^{c\sigma} + \partial_5 h \partial_\sigma A^{a\sigma} - \xi \partial_5 h \partial_5 A^a_5 \right) \delta A^a_5 \\ &\quad + \int d^4x \left( [h F^a_{5\nu}\, \delta A^{a\nu}] + [h(\partial_\sigma A^{a\sigma} - \xi \partial_5 A^a_5) \delta A^a_5] \right). \end{aligned} \tag{10.3}$$

The bulk terms will give rise to the usual bulk equations of motion:

$$\begin{aligned} &\partial_M h F^{aM\nu} - g_5 f^{abc} h\, F^{bM\nu} A^c_M + \frac{h}{\xi} \partial^\nu \partial^\sigma A^a_\sigma - h \partial^\nu \partial_5 A^a_5 = 0, \\ &h \partial^\sigma F^a_{\sigma 5} - g_5 f^{abc} h\, F^b_{\sigma 5} A^{c\sigma} + \partial_5 h \partial_\sigma A^{a\sigma} - \xi \partial_5 h \partial_5 A^a_5 = 0. \end{aligned} \tag{10.4}$$

However, one has to ensure that the variation of the boundary pieces vanish as well. This will lead to the requirements

$$h F^a_{\nu 5}\, \delta A^{a\nu}|_{R,R'} = 0, \tag{10.5}$$

$$h(\partial_\sigma A^{a\sigma} - \xi \partial_5 A^a_5) \delta A^a_5|_{R,R'} = 0. \tag{10.6}$$

The boundary conditions (BCs) have to be such that the above equations be satisfied.

For the case of a scalar (Higgs) with a VEV localized at the endpoint the generic form of the BC for the gauge fields (in unitary gauge) will be of the form

$$\partial_5 A^a_\mu|_{R,R'} = V|^{ab}_{R,R'} A^b_\mu h|_{R,R'}, \tag{10.7}$$

where $V|^{ab}_R$and $V|^{ab}_{R'}$ are proportional to the VEV's squared at $R$ and $R'$. The BCs in (10.7) are mixed BCs that still ensure the hermiticity (self-adjointness) of the Hamiltonian. In the limit $V^{ab} \to 0$ the mixed BC reduces to a Neumann BC, while the limit $V^{ab} \to \infty$ the mixed BC reduces to a Dirichlet BC.

Finding the KK decomposition of the gauge field reduces to solving a Sturm–Liouville problem with Neumann or Dirichlet BCs, or in the case of boundary scalars with mixed BCs. Those general BCs lead to a Kaluza–Klein expansion of the gauge fields of the form

$$A^a_\mu(x, z) = \sum_n \epsilon_\mu\, \psi^a_n(z) e^{i p_n x}, \tag{10.8}$$

where $p_n^2 = M_n^2$ and $\epsilon_\mu$ is a polarization vector. These wavefunctions then satisfy the equation:

$$h(z) \psi^{a\prime\prime}_n(z) + h'(z) \psi^{a\prime}_n(z) + M^{a\,2}_n h(z) \psi^a_n(z) = 0, \;\; \psi^{a\prime}_n|_{R,R'} = h V|^{ab}_{R,R'}\, \psi^b_n|_{R,R'}. \tag{10.9}$$

The KK mode wavefunctions can be normalized by requiring

$$\int dz\, h(z)(\psi_n^a(z))^2 = 1\,. \tag{10.10}$$

The couplings between the different KK modes can then be obtained by substituting this expression into the Lagrangian (10.2) and integrating over the extra dimension. The resulting couplings are then the usual 4D Yang-Mills couplings, with the gauge coupling $g_4$ in the cubic and gauge coupling square in the quartic vertices replaced by the effective couplings involving the integrals of the wave functions of the KK modes over the extra dimension:

$$g_{cubic} \quad \to \quad g_{mnk}^{abc} = g_5 \int dz\, h(z)\psi_m^a(z)\psi_n^b(z)\psi_k^c(z), \tag{10.11}$$

$$g_{quartic}^2 \quad \to \quad g_{mnkl}^{2\,abcd} = g_5^2 \int dz\, h(z)\psi_m^a(z)\psi_n^b(z)\psi_k^c(z)\psi_l^d(z). \tag{10.12}$$

Here $a, b, c, d$ refer to the gauge index of the gauge bosons and $m, n, k, l$ to the KK number.

### *10.1.2 The elastic scattering amplitude*

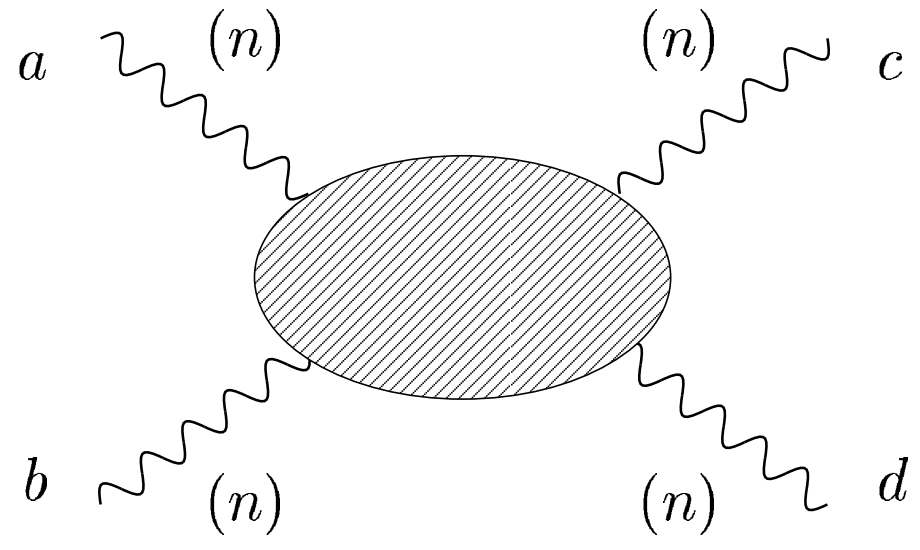

Fig. 10.1: Elastic scattering of longitudinal modes of KK gauge bosons, $n + n \to n + n$, with the gauge index structure $a + b \to c + d$.

We want to calculate the energy dependence of the amplitude of the elastic scattering of the longitudinal modes of the KK gauge bosons $n + n \to n + n$ with gauge index structure $a + b \to c + d$ (see Fig. 10.1), where this process involves both exchange of the $k$'th KK mode from the cubic vertex, and the direct contribution from the quartic vertex. There are four diagrams as shown in Fig. 10.2: the s, t and u-channel exchange of the KK modes, and the contribution of the quartic vertex. The kinematics assumed for this elastic scattering is in the center of mass frame, where the incoming momentum vectors are $p_\mu = (E, 0, 0, \pm\sqrt{E^2 - M_n^2})$, while the outgoing momenta are $(E, \pm\sqrt{E^2 - M_n^2}\sin\theta, 0, \pm\sqrt{E^2 - M_n^2}\cos\theta)$. $E$ is the incoming energy, and $\theta$ the scattering angle with forward scattering for $\theta = 0$. The longitudinal polarization vectors are as usual $\epsilon_\mu = (\frac{|\vec{p}|}{M}, \frac{E}{M}\frac{\vec{p}}{|\vec{p}|})$ and accordingly the contribution of each diagram can be as bad as $E^4/M_n^4$. It is straightforward to evaluate the full scattering amplitude, and extract the leading behavior for large values energies of this amplitude. The general structure of the expansion in energy contains three terms:

$$\mathcal{A} = A^{(4)}\frac{E^4}{M_n^4} + A^{(2)}\frac{E^2}{M_n^2} + A^{(0)}. \tag{10.13}$$

We would like to understand under what circumstances will $A^{(4)}$ and $A^{(2)}$ vanish, this does not imply that the conventional unitarity bounds on the finite amplitudes have to be satisfied for all processes.

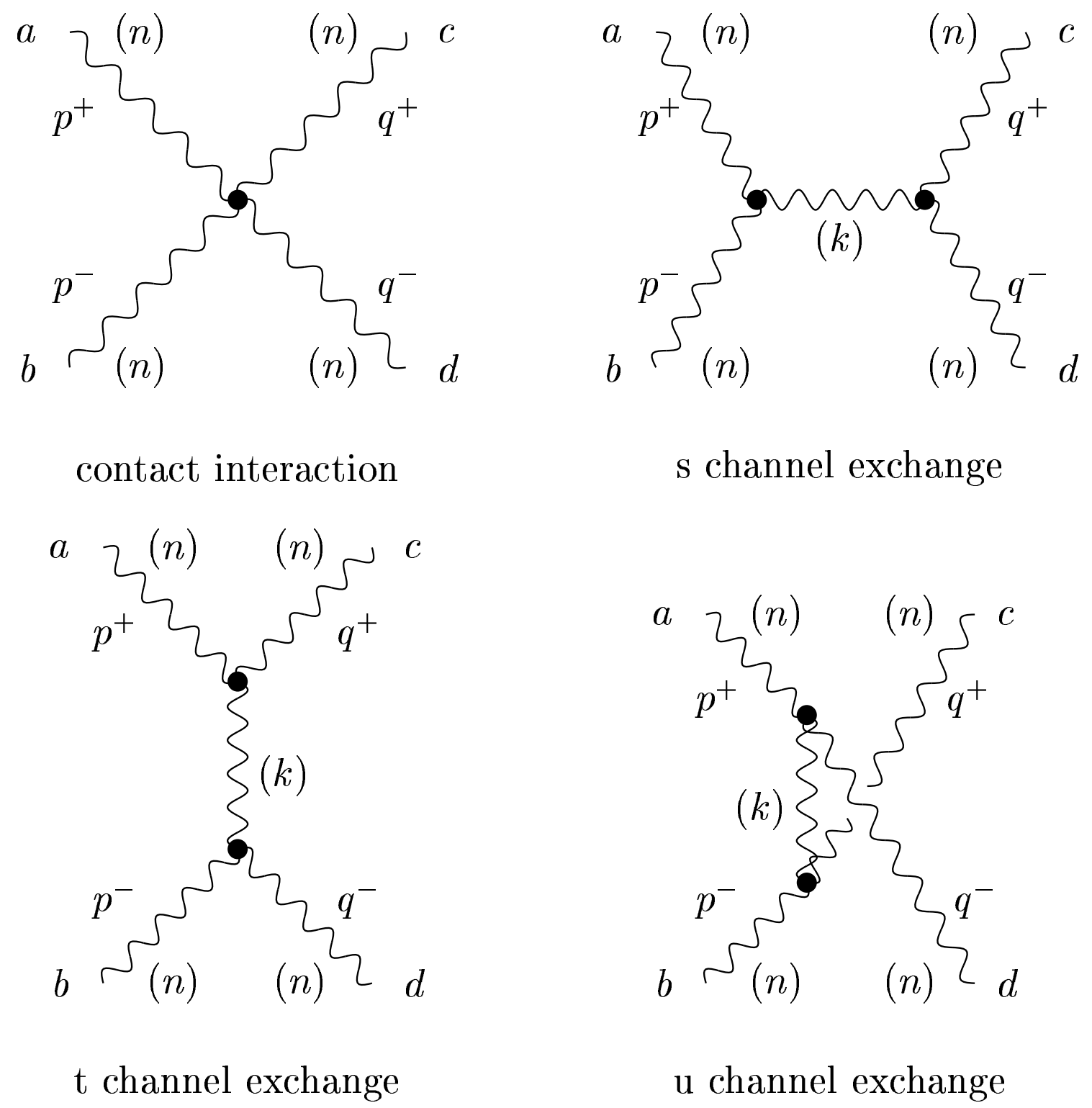

Fig. 10.2: The four gauge diagrams contributing at tree level to the gauge boson elastic scattering amplitude.

The term growing with $E^4$ depends only on the effective couplings and not on the mass spectrum. The condition for cancelling the coefficient of this term is:

$$g_{nnnn}^2 = \sum_k g_{nnk}^2. \tag{10.14}$$

Using this relation the condition for the $E^2$ terms to cancel can be simplified to:

$$4g_{nnnn}^2 M_n^2 \quad = \quad 3\sum_k g_{nnk}^2 M_k^2\ . \tag{10.15}$$

The goal of the remainder of this section is to examine under what circumstances the terms that grow with energy actually cancel. Consider first the $E^4$ term. According to (10.14) the requirement for cancellation is

$$\int_R^{R'} dz\, h(z)\psi_n^4(z) = \sum_k \int_R^{R'} dy \int_R^{R'} dz\; h(y)h(z)\psi_n^2(y)\psi_n^2(z)\psi_k(y)\psi_k(z). \tag{10.16}$$

One can easily see that this equation is in fact satisfied no matter what BC one is imposing, as long as that BC still maintains hermiticity of the kinetic operator

$$h\,\partial_z^2 + (\partial_z h)\partial_z\ . \tag{10.17}$$

In this case one can explicitly check that

$$\int_R^{R'} h\psi_n^* {\psi_m'}' + h'\psi_n^*\psi_m' = \int_R^{R'} h{\psi_n^*}''\psi_m + h'{\psi_n^*}'\psi_m\ . \tag{10.18}$$

For such hermitian operators one is guaranteed to get an orthonormal complete set of solutions $\psi_k(y)$, thus from the completeness it follows that

$$\sum_k \psi_k(y)\psi_k(z) = \frac{1}{h(z)}\delta(y-z), \tag{10.19}$$

which immediately implies (10.16).

The condition for the cancellation of the $E^2$ terms is as in (10.15)

$$3\sum_k M_k^2 \int_R^{R'} dy \int_R^{R'} dz\, h(y)h(z)\psi_n^2(y)\psi_n^2(z)\psi_k(y)\psi_k(z) = 4\,M_n^2 \int_R^{R'} dz\, h(z)\psi_n^4(z). \tag{10.20}$$

By integration by parts we find (we denote again by $[F]$ the boundary quantity $F(R') - F(R)$)

$$\begin{aligned}\sum_k M_k^2 \left(\int dz h(z)\psi_n^2(z)\psi_k(z)\right)^2 &= \tfrac{4}{3}M_n^2 \int dz h(z)\psi_n^4(z) - \tfrac{2}{3}[h\psi_n^3\psi_n'] \\ -\sum_k [h\psi_n^2\psi_k'] \int dz h(z)\psi_n^2(z)\psi_k(z) &+ 2\sum_k [h\psi_n\psi_n'\psi_k] \int dz h(z)\psi_n^2(z)\psi_k(z).\end{aligned} \tag{10.21}$$

Thus one can see that for arbitrary BCs the $E^2$ terms do not cancel. However, if one has pure Dirichlet or Neumann BCs for all modes then all the extra boundary terms will vanish, and thus the cancellation of the $E^2$ terms goes through [2]. The fact that in the absence of a Higgs VEV, or any other source of gauge symmetry breaking, (i.e. the Neumann BC) there is no problem with unitarity is not really surprising. It is somewhat surprising that with an infinite Higgs VEV (the Dirichlet BC) there is also no problem. To understand what is happening it is useful to recall what actually happens in the general mixed case. Even with mixed BCs there is no problem with unitary once one includes the diagrams corresponding to the exchange of the Higgs on the boundary. These diagrams cancel the $E^2$ terms just as they do in 4D. However in the limit that Higgs VEV is large the gauge boson wavefunctions are repelled from the boundary since it costs a lot of energy to reside there. A simple calculation shows that the product of the Higgs VEV times the gauge boson wavefunction squared evaluated on the boundary goes to zero in the large VEV limit. Thus the Higgs decouples in the infinite VEV limit and unitarity is preserved without any need for a physical Higgs boson.

### *10.1.3 Electroweak gauge bosons*

We will now apply the Higgsless idea to electroweak symmetry breaking in a mostly realistic RS type model. We denote by $A_M^{R\,a}$, $A_M^{L\,a}$ and $B_M$ the gauge bosons of $SU(2)_R$, $SU(2)_L$ and $U(1)_{B-L}$ respectively; $g_5$ is the gauge coupling of the two $SU(2)$'s and $\tilde{g}_5$, the gauge coupling of $U(1)_{B-L}$. We impose the following BCs:

$$\text{at } z = R': \quad \begin{cases} \partial_z(A_\mu^{L\,a} + A_\mu^{R\,a}) = 0,\ A_\mu^{L\,a} - A_\mu^{R\,a} = 0,\ \partial_z B_\mu = 0, \\ (A_5^{L\,a} + A_5^{R\,a}) = 0,\ \partial_z(A_5^{L\,a} - A_5^{R\,a}) = 0,\ B_5 = 0. \end{cases} \tag{10.22}$$

$$\text{at } z = R: \quad \begin{cases} \partial_5 A_\mu^{L\,a} = 0,\ A_\mu^{R\,1,2} = 0, \\ \partial_z(g_5 B_\mu + \tilde{g}_5 A_\mu^{R\,3}) = 0,\ \tilde{g}_5 B_\mu - g_5 A_\mu^{R\,3} = 0, \\ A_5^{L\,a} = 0,\ A_5^{R\,a} = 0,\ B_5 = 0. \end{cases} \tag{10.23}$$

The BCs break $SU(2)_R \times U(1)_{B-L}$ down to $U(1)_Y$ on the Planck brane ($z = R$) and break $SU(2)_L \times SU(2)_R$ down to a diagonal $SU(2)$ on the TeV brane ($z = R'$).

The Euclidean bulk equation of motion satisfied by spin-1 fields in AdS space is

$$(\partial_z^2 - \frac{1}{z}\partial_z + q^2)\psi(z) = 0, \tag{10.24}$$

where the solutions in the bulk are assumed to be of the form $A_\mu(q)e^{-iqx}\psi(z)$. The KK mode expansion is given by the solutions to this equation which are of the form

$$\psi_k^{(A)}(z) = z\left(a_k^{(A)} J_1(q_k z) + b_k^{(A)} Y_1(q_k z)\right) , \tag{10.25}$$

where $A$ labels the corresponding gauge boson.

To leading order in $1/R$ and for $\log(R'/R) \gg 1$, the lightest solution for eigenvalue equation for the mass of the $W^\pm$'s is

$$M_W^2 = \frac{1}{R'^2 \log\left(\frac{R'}{R}\right)} , \tag{10.26}$$

while the lowest mass in the $Z$ tower is approximately given by

$$M_Z^2 = \frac{g_5^2 + 2\tilde{g}_5^2}{g_5^2 + \tilde{g}_5^2} \frac{1}{R'^2 \log\left(\frac{R'}{R}\right)} . \tag{10.27}$$

The correct mass ratios (a small $T$ parameter) are guaranteed by the unbroken diagonal $SU(2)$ symmetry on the TeV brane which acts as a custodial symmetry [5].

From the expansion for small arguments of the Bessel functions appearing in (10.25), the wavefunction of a mode with mass $M \ll 1/R'$ can be written as [13]:

$$\psi^{(A)}(z) \approx c_0^{(A)} + M_A^2 z^2 \left(c_1^{(A)} - \frac{c_0^{(A)}}{2}\log(z/R)\right) + \mathcal{O}(M_A^4 z^4), \tag{10.28}$$

with $c_0^{(A)}$ at most of order one, $c_1^{(A)} \sim \mathcal{O}(\log(R'/R))$, and $M^2 \sim \mathcal{O}(1/\log(R'/R))$.

The boundary conditions on the bulk gauge fields give the following results for the leading and next-to-leading log terms in the wavefunction for the lightest charged gauge bosons

$$c_0^{(L\pm)} = c_\pm\,,\ c_0^{(R\pm)} \approx 0\,, \tag{10.29}$$

$$c_1^{(L\pm)} \approx 0\,,\ c_1^{(R\pm)} \approx \frac{c_\pm}{2}\log\left(\frac{R'}{R}\right) , \tag{10.30}$$

while for the neutral gauge bosons we find in the same approximation

$$c_0^{(L3)} \approx c\,,\ c_0^{(R3)} \approx -c\,\frac{\tilde{g}_5^2}{g_5^2+\tilde{g}_5^2}\,,\ c_0^{(B)} \approx -c\,\frac{g_5\,\tilde{g}_5}{g_5^2+\tilde{g}_5^2}\,. \tag{10.31}$$

To leading log order we also have:

$$c_1^{(L3)} \approx 0\,,\ c_1^{(R3)} \approx c\,\frac{g_5^2}{2\left(g_5^2+\tilde{g}_5^2\right)}\log\left(\frac{R'}{R}\right)\,,\ c_1^{(B)} \approx -c\,\frac{g_5\,\tilde{g}_5}{2\left(g_5^2+\tilde{g}_5^2\right)}\log\left(\frac{R'}{R}\right)\,. \tag{10.32}$$

### *10.1.4 Precision electroweak measurements*

The simplest way to calculate the contributions to precision electroweak measurements is to use "equivalent vacuum polarizations" that can be extracted from the gauge boson wavefunction renormalizations:

$$Z_\gamma = 1\,, \qquad Z_W = 1 - g^2\Pi'_{11}\,, \qquad Z_Z = 1 - (g^2 + g'^2)\Pi'_{33}\,, \tag{10.33}$$

Since the photon is massless it's wavefunction is exactly flat, so requiring that the quarks and leptons have the correct charge fixes $Z_\gamma = 1$. For the $W$ and $Z$ however the wavefunctions have some nontrivial shape so canonically normalizing the light quark and lepton modes and requiring that their overlaps with the $W$ and $Z$ reproduce the standard model couplings fixes $Z_W$, $Z_Z \neq 1$. Given these "equivalent vacuum polarizations" it is straightforward to compute the $S$, $T$, and $U$ parameters.

$$\begin{aligned} S &\equiv 16\pi(\Pi'_{33} - \Pi'_{3Q}), \\ T &\equiv \frac{4\pi}{s^2c^2M_Z^2}(\Pi_{11}(0) - \Pi_{33}(0)), \\ U &\equiv 16\pi(\Pi'_{11} - \Pi'_{33}), \end{aligned} \tag{10.34}$$

where $\Pi_{ii}(0)$ is simply extracted from the gauge boson mass terms. For fermions localized on the Planck brane the calculation can be perfomed analytically and we find there is a problem with $S$:

$$S \approx \frac{8\pi R'^2 M_z^2}{g^2 + g'^2} \approx 1.5 \tag{10.35}$$

Such a large value is reminiscent of technicolor models.

However if we allow the quarks and leptons can have an arbitrary bulk mass then the lightest modes can be localized with a wavefunction that is arbitrary power of $z$. Then since it is known that in RS models with quarks and leptons on the TeV brane (at $z = R'$) $S$ is large and negative [13], it is clear that $S$ will vanish for some intermediate localization. Indeed when the fermions are roughly uniformly distributed through the bulk $S$ goes through zero [10, 14–17]. The fact that $S$ vanishes has a deeper significance, since it follows from orthogonalilty of wavefunctions. If the currents that couple to the $W$ and $Z$ have the same profile as the gauge bosons then the overlap of the current with all higher gauge boson KK modes will be exactly zero. If the current and the gauge bosons have very similar profiles in the extra dimension then the gauge boson coupling is still relatively enhanced while the coupling to higher KK modes is suppressed. Thus in the region where the precision electroweak constraints are satisfied ($S \approx 0$) because the coupling to higher KK is suppressed, the bounds from the Tevatron and LEP on such KK modes are relaxed down to a bound around $500 - 600$ GeV.

The remaining problem with precision electroweak measurements is the compatibility of a large top quark mass with the observed $Zb\bar{b}$ coupling. The large top mass requires the left-handed and right-handed top quarks to have a profile which is localized toward the TeV brane. However if the left-handed top (and thus the left-handed bottom) are too close to the TeV brane then the gauge couplings of the bottom will be too different from the down and strange quarks. This problem may be avoided by separating the physics which generates the top quark mass [17, 18] or by allowing the Higgs VEV to extend slightly into the bulk [19].

### *10.1.5 Collider phenomenology*

The most distinctive feature of the Higgsless models is, of course, the absence of a physical scalar state in the spectrum. However, other models exist in which the Higgs is unobservable at the LHC. For this reason, identification of this as the mechanism of electroweak symmetry breaking will require examination of other sectors. There are two potential types of signals that will help to identify the model: those that are related to the RS physics required to realize the Higgsless mechanism, and those that directly probe this as the mechanism of symmetry breaking and unitarity restoration. In the first class are observations of Kaluza-Klein of electroweak gauge bosons and gluons [8]. In the second class are observations of resonances in the scattering of longitudinal electroweak bosons [20].

The easiest signal to see is the first gluon Kaluza-Klein (KK) resonance. This will show up as a resonance in the dijet spectrum, as seen in Fig. 10.3. Like all low-lying KK resonances, this is localized near the IR brane. In most RS models with fermions in the bulk the need to produce a large top mass forces the right-handed top and third generation quark doublet to also be localized near the IR brane. This means that the gluon resonances will generically couple more strongly to tops and bottoms than to light quarks, and observation of this will be a clue that the model may be RS.

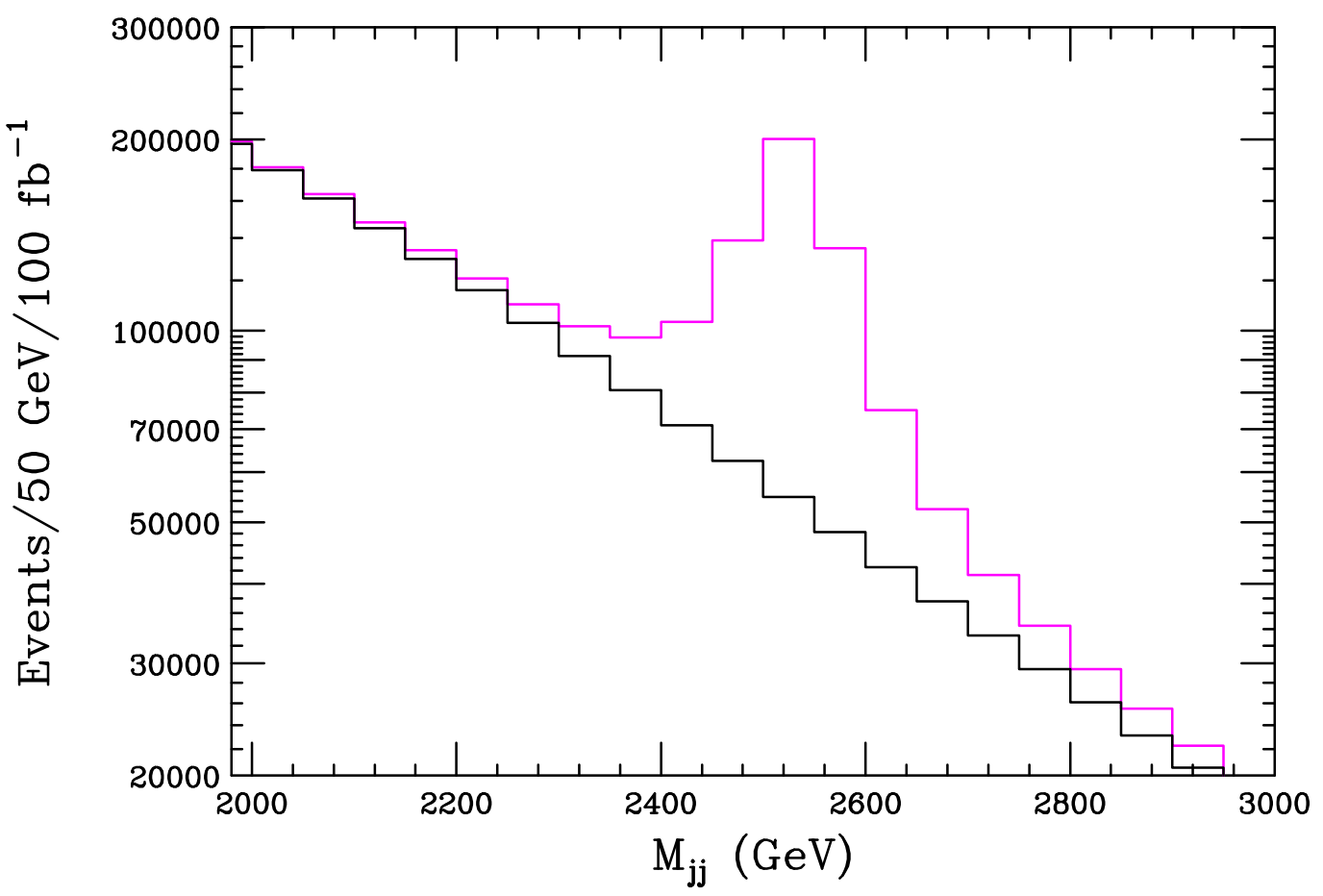

Fig. 10.3: Dijet invariant mass spectrum at the LHC showing a prominent resonance due to the first gluon Kaluza-Klein state.

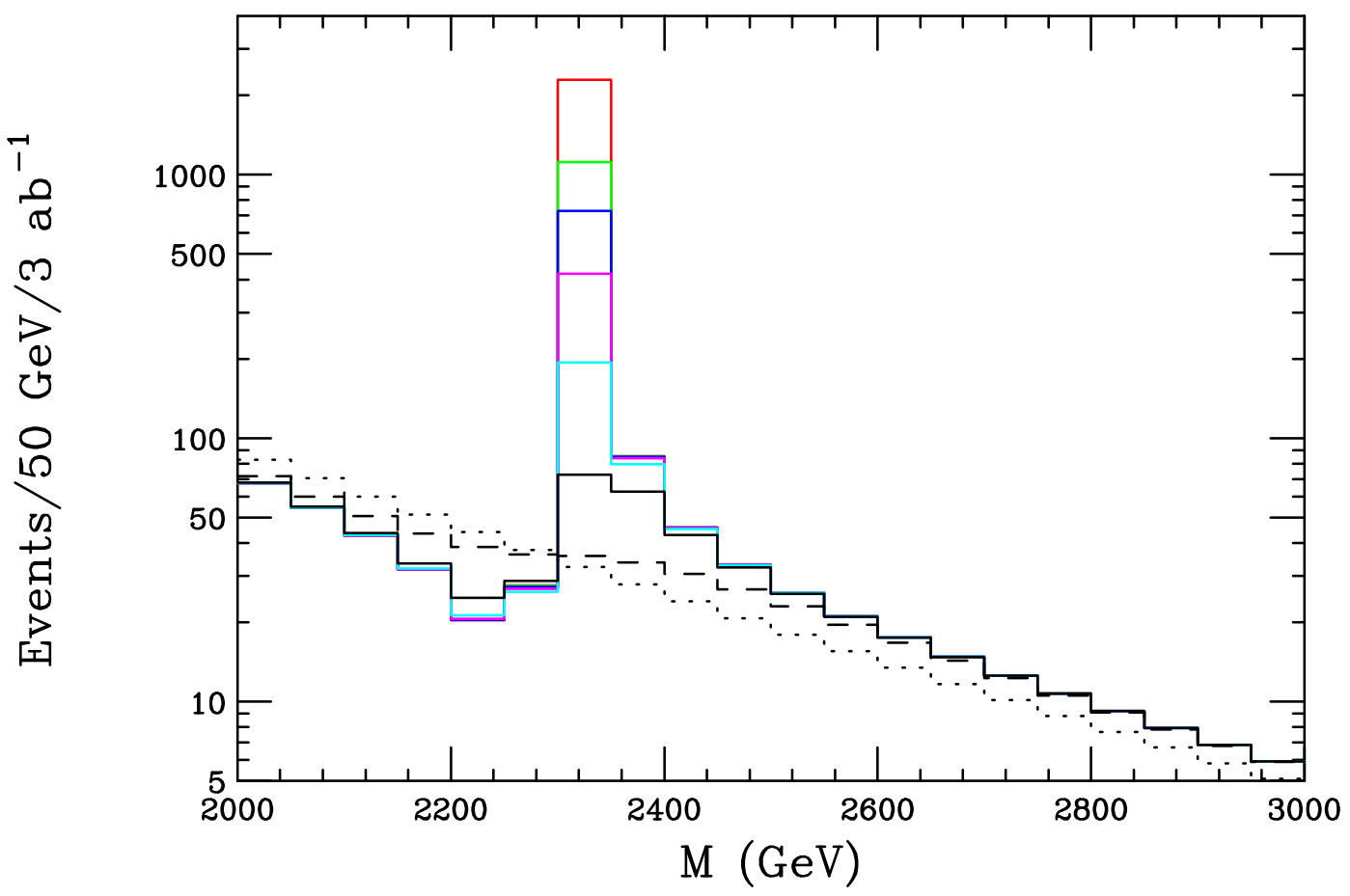

Fig. 10.4: Event rate for Drell-Yan production of the first neutral KK gauge boson, as a function of the invariant mass of the lepton pair. The dotted line is the SM background. The other histograms, from top to bottom, include the resonance with width parameter $c = (1, 2, 3, 5, 10, 25, 100)$.

More important for studies of electroweak symmetry breaking, of course, is the observation of gauge boson KK resonances. The simplest place to look is in the Drell-Yan spectrum. The couplings of these states to fermions depend on the fermion localization parameters, in particular the top localization.

Fig. 10.4 shows the Drell-Yan spectrum from a neutral KK at about 2.3 TeV. We can write the width as $\Gamma = c\Gamma_0$, where $\Gamma_0$ is what the width of the state would be if all fermions were localized to the Planck brane. The different curves in Fig. 10.4 correspond to different values of $c$, ranging from $c = 1$ to $c = 100$. Note that the state presented is at a high mass. In general, a successful Higgsless model is expected to have lighter states, and hence they should be more easily discoverable at the LHC.

As shown in [20] searches for the process $WZ \to WZ$ can directly probe the Higgsless mechanism for electroweak symmetry breaking. In particular, the sum rules that ensure unitarity can be directly probed by measurements of the couplings of the gauge KKs to longitudinal gauge bosons. Fig. 10.5 shows the production of the first charged KK resonance in this channel. For comparison, two resonances appearing in different technicolor-type models are also shown. As can be seen, the most striking feature of the Higgsless model is the narrow width of the resonance. Note that these searches have the additional advantage of being largely independent of the parameters in the fermion sector.

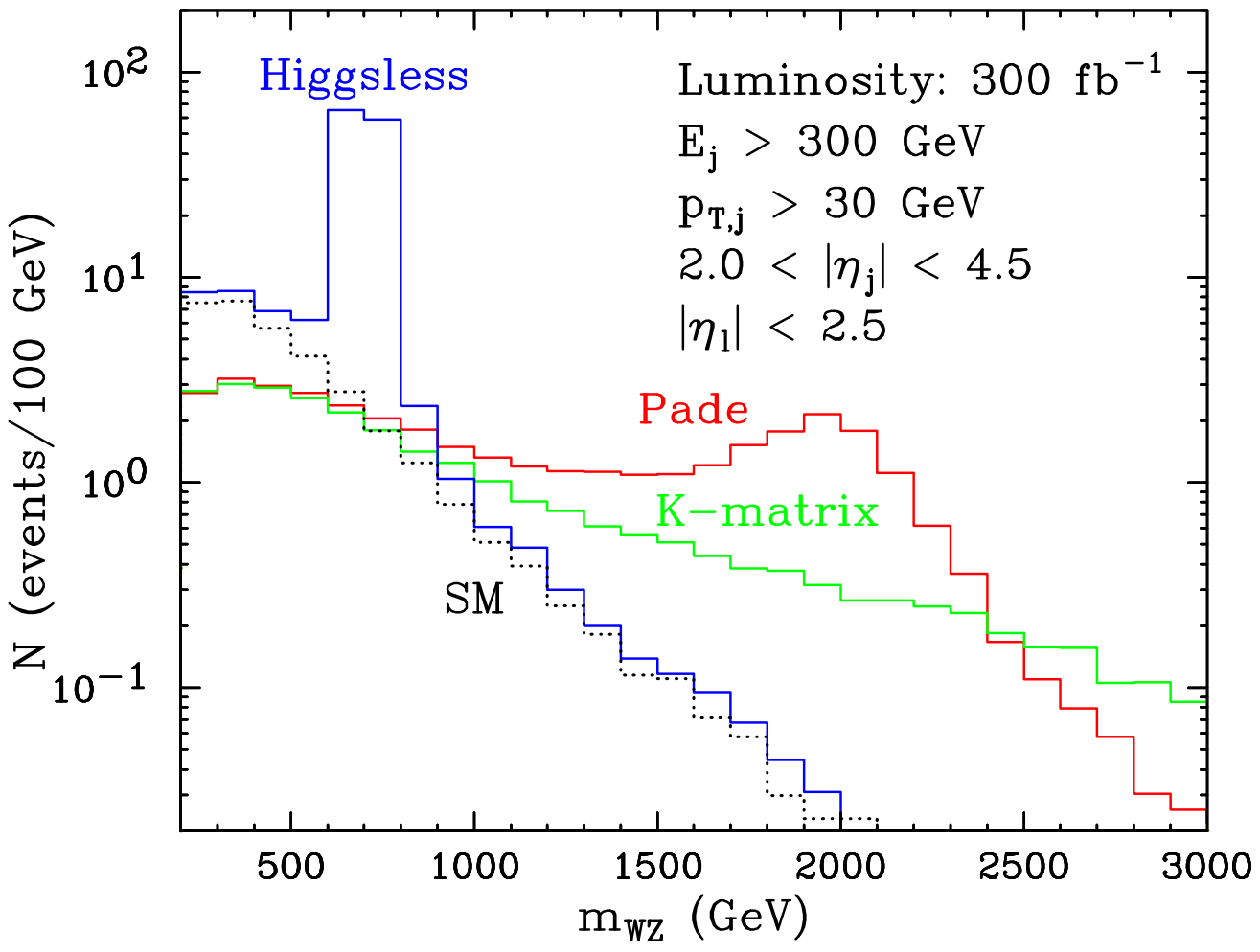

Fig. 10.5: Event rate for production of $2j + 3\ell + \nu$, corresponding to $WZ$ production with fully leptonic decays. From ref. [9].

## 10.2 Quark and lepton masses

*Christophe Grojean*

### *10.2.1 Chiral fermions from a 5D theory on an interval*

In the SM, quarks and leptons acquire a mass, after EWSB, through their Yukawa couplings to the Higgs. In absence of a Higgs, one cannot write any Yukawa coupling and one should expect the fermions to remain massless. However, as for the gauge fields, appropriate boundary conditions will force the fermions to acquire a momentum along the extra dimension and this is how they will become massive from the 4D point of view. We are now going to review this construction [4].

The SM fermions cannot be completely localized on the UV boundary: since the unbroken gauge group on that boundary coincides with the SM $SU(2)_L \times U(1)_Y$ symmetry, the theory on that brane would be chiral and there is no way for the chiral zero mode fermions to acquire a mass. The SM fermions cannot live on the IR brane either since the unbroken $SU(2)_D$ gauge symmetry will impose an isospin invariant spectrum and the up-type and down-type quarks will be degenerate. The only possibility is thus to embed the SM fermions into 5D fields living in the bulk and feeling the gauge symmetry breakings on both boundaries. Since the irreducible spin-1/2 representations of the 5D Lorentz group correspond to 4-component Dirac spinor, extra fermionic degrees of freedom are needed to complete the SM chiral spinors to 5D Dirac spinors and we are back to a vector-like spectrum. However, as it is well known, orbifold like projections (or equivalently appropriate boundary conditions) can get rid of half of the spectrum at the lowest KK level to actually provide a 4D effective chiral theory. This way we can embed the SM quarks and leptons into 5D Dirac spinors following Table 10.1.

### *10.2.2 Fermions in AdS background*

In principle when one is dealing with fermions in a non-trivial background, one needs to work with the "square-root" of the metric also known as *vielbeins* and to introduce the spin connection to covariantize derivatives. Fortunately, in an $AdS$ background, the spin connection drops out from the spin-1/2 action that simply reads

$$\mathcal{S} = \int d^5x \frac{R^4}{z^4} \left( -i\bar{\chi}\bar{\sigma}^\mu \partial_\mu \chi - i\psi\sigma^\mu\partial_\mu\bar{\psi} + (\psi \overleftrightarrow{\partial_5} \chi - \bar{\chi} \overleftrightarrow{\partial_5} \bar{\psi}) + \frac{c}{z}\left(\psi\chi + \bar{\chi}\bar{\psi}\right)\right) \qquad (10.36)$$

Table 10.1: Embedding of the SM fermions into 5D Dirac spinors. We have indicated the quantum numbers of the different components under the bulk $SU(2)_L \times SU(2)_R \times U(1)_{B-L}$ symmetry, the subgroup $SU(2)_L \times U(1)_Y$ that remains unbroken on the UV boundary, the subgroup $SU(2)_D \times U(1)_{B-L}$ unbroken on the IR brane and finally the electric charge. The shaded spinors are the fields with the right quantum numbers to be identified as the massless SM fermions while the other spinors correspond to partners needed to complete 5D Dirac spinors. The latter become massive by the orbifold projection/boundary conditions. Through the Dirac mass added on the IR boundary, there will be a mixing between the would be zero modes and some partners and at the end the guy that would be identified as the SM $u_L$ is a mix of $\chi_{u_L}$ and a small amount of $\chi_{u_R}$. Since this last field has wrong SM quantum numbers, we would end up with deviations in the couplings of the fermions to the gauge bosons. These deviations will be particularly sizable for the third generation due to the heaviness of the top.

| Particle | bulk<br>$L \times R \times (B-L)$ | UV<br>$L \times Y$ | IR<br>$D \times (B-L)$ | $Q_{em}$ |
|---|---|---|---|---|
| $\begin{pmatrix}\chi_u\\ \chi_d\end{pmatrix}_L$ | $(\square, 1, 1/6)$ | $(\square, 1/6)$ | $(\square, 1/6)$ | $2/3$<br>$-1/3$ |
| $\begin{pmatrix}\psi_u\\ \psi_d\end{pmatrix}_L$ | $(\bar{\square}, 1, -1/6)$ | $(\bar{\square}, -1/6)$ | $(\bar{\square}, -1/6)$ | $-2/3$<br>$1/3$ |
| $\begin{pmatrix}\chi_u\\ \chi_d\end{pmatrix}_R$ | $(1, \square, 1/6)$ | $(1, 2/3)$<br>$(1, -1/3)$ | $(\square, 1/6)$ | $2/3$<br>$-1/3$ |
| $\begin{pmatrix}\psi_u\\ \psi_d\end{pmatrix}_R$ | $(1, \bar{\square}, -1/6)$ | $(1, -2/3)$<br>$(1, 1/3)$ | $(\bar{\square}, -1/6)$ | $-2/3$<br>$1/3$ |

$Q_{em} = Y + T_{3L}$ $\qquad$ $Y = (B-L) + T_{3R}$

where the coefficient $c = mR$ is the bulk Dirac mass in units of the $AdS$ curvature (and $\overleftrightarrow{\partial_5} = (\overrightarrow{\partial_5} - \overleftarrow{\partial_5})/2$). The bulk equations of motion are:

$$-i\bar{\sigma}^\mu \partial_\mu \chi - \partial_5 \bar{\psi} + \frac{c+2}{z}\bar{\psi} = 0 \qquad -i\sigma^\mu \partial_\mu \bar{\psi} + \partial_5 \chi + \frac{c-2}{z}\chi = 0. \tag{10.37}$$

The KK decomposition is of the form

$$\chi = \sum_n g_n(z)\, \chi_n(x) \quad \text{and} \quad \bar{\psi} = \sum_n f_n(z)\, \bar{\psi}_n(x). \tag{10.38}$$

and the 5D Dirac equation is equivalent to the coupled first order differential equations

$$f_n' + m_n g_n - \frac{c+2}{z} f_n = 0, \quad g_n' - m_n f_n + \frac{c-2}{z} g_n = 0, \tag{10.39}$$

which can be combined into uncoupled second order differential equations

$$f_n'' - \tfrac{4}{z} f_n' + (m_n^2 - \tfrac{c^2-c-6}{z^2}) f_n = 0, \qquad g_n'' - \tfrac{4}{z} g_n' + (m_n^2 - \tfrac{c^2+c-6}{z^2}) g_n = 0. \tag{10.40}$$

The solutions are now linear combinations of Bessel functions, as opposed to $\sin$ and $\cos$ functions for the flat case:

$$g_n(z) = z^{\frac{5}{2}} \left( A_n J_{c+\frac{1}{2}}(m_n z) + B_n Y_{c+\frac{1}{2}}(m_n z) \right) \tag{10.41}$$

$$f_n(z) = z^{\frac{5}{2}} \left( C_n J_{c-\frac{1}{2}}(m_n z) + D_n Y_{c-\frac{1}{2}}(m_n z) \right). \tag{10.42}$$

The first order bulk equations of motion (10.39) further impose that

$$A_n = C_n \text{ and } B_n = D_n. \tag{10.43}$$

The remaining undetermined coefficients are determined by the boundary conditions, and the wave function normalization.

Finally, when the boundary conditions permit, there can also be a zero mode. For instance, if $\psi_{|R,R'} = 0$, the zero mode is given by

$$g_0(z) = A_0 \left(\frac{z}{R}\right)^{2-c}, \; f = 0. \tag{10.44}$$

The coefficient $A_0$ is determined by the normalization condition

$$\int_R^{R'} dz \left(\frac{R}{z}\right)^5 \frac{z}{R} A_0^2 \left(\frac{z}{R}\right)^{4-2c} = A_0^2 \int_R^{R'} \left(\frac{z}{R}\right)^{-2c} dz = 1. \tag{10.45}$$

To understand from these equations where the fermions are localized, we study the behavior of this integral as we vary the limits of integration. If we send $R'$ to infinity, we see that the integral remains convergent if $c > 1/2$, and the fermion is then localized on the UV brane. If we send $R$ to zero, the integral is convergent if $c < 1/2$, and the fermion is localized on the IR brane. The value of the Dirac mass determines whether the fermion is localized toward the UV or IR branes. We note that the opposite choice of boundary conditions that yields a zero mode ($\chi|_{R,R'} = 0$) results in a zero mode solution for $\psi$ with localization at the UV brane when $c < -1/2$, and at the IR brane when $c > -1/2$. The interesting feature in the warped case is that the localization transition occurs not when the bulk mass passes through zero, but at points where $|c| = 1/2$. This is due to the curvature effects of the extra dimension. The $CFT$ interpretation of the $c$ parameter is an anomalous dimension that controls the amount of compositeness of the fermion [21].

### *10.2.3 Higgsless fermions masses*

We have already explained how to embed SM fermions into 5D Dirac spinors. To get the zero modes we desire, the following boundary conditions have to be imposed

$$\begin{pmatrix} \chi_{u_L} \\ \bar{\psi}_{u_L} \\ \chi_{d_L} \\ \bar{\psi}_{d_L} \end{pmatrix} \begin{matrix} + & + \\ - & - \\ + & + \\ - & - \end{matrix} \qquad \begin{pmatrix} \chi_{u_R} \\ \bar{\psi}_{u_R} \\ \chi_{d_R} \\ \bar{\psi}_{d_R} \end{pmatrix} \begin{matrix} - & - \\ + & + \\ - & - \\ + & + \end{matrix} \tag{10.46}$$

Where the $+$ and $-$ refer to Neumann and Dirichlet boundary conditions, the first/second sign denoting the BC on the UV/IR brane respectively. These boundary conditions give massless chiral modes that match the fermion content of the standard model. However, the $u_L$, $d_L$, $u_R$, and $d_R$ are all massless at this stage, and we need to lift the zero modes to achieve the standard model mass spectrum. While simply giving certain boundary conditions for the fermions will enable us to lift these zero modes, in the following discussion, we talk about boundary operators, and the boundary conditions that these operators induce. There are some subtleties in dealing with boundary operators for fermions. These arise from the fact that the fields themselves are not always continuous in the presence of a boundary operator. This is due to the fact that the equations of motion for fermions are first order. The most straightforward approach is to enforce the boundary conditions that give the zero modes as shown in Eq. (10.46) on the real boundary at $z = R, R'$ while the boundary operators are added on a fictitious brane a distance $\epsilon$ away from it. The new boundary condition is then obtained by taking the distance $\epsilon$ to be small. This physical picture is quite helpful in understanding what the different boundary conditions will do. The details can be found in [4].

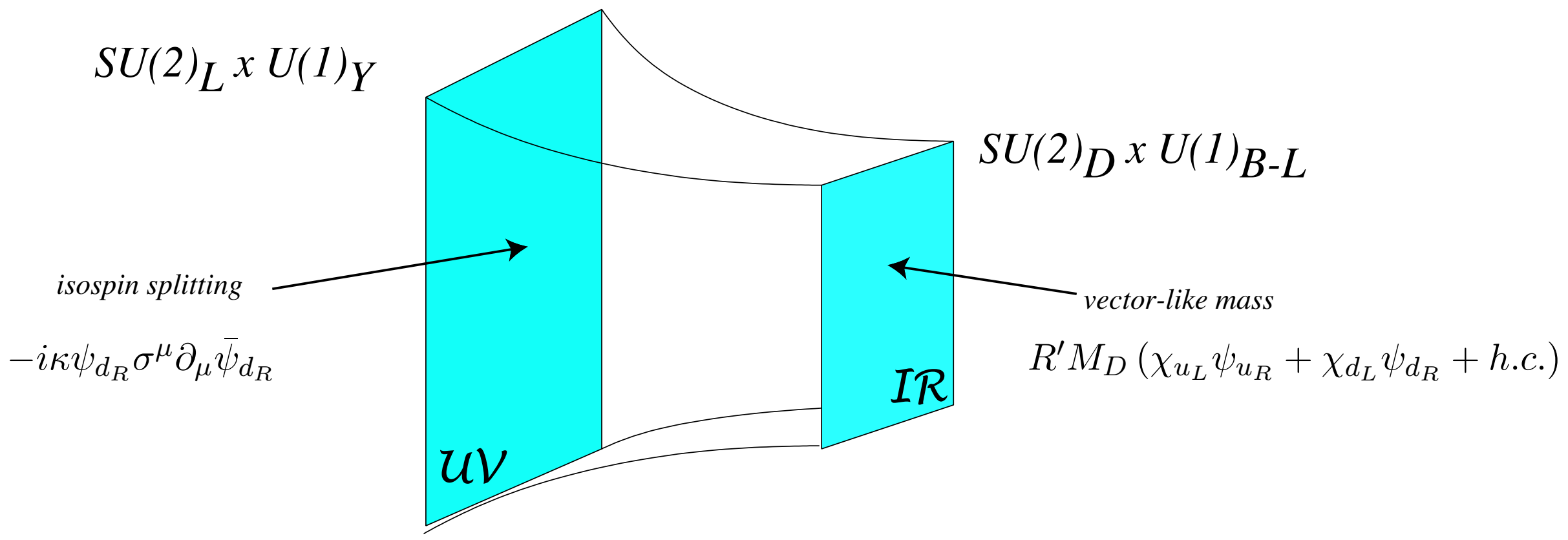

Fig. 10.6: Brane localized operators needed to lift up the masses of the SM fermions.

The IR brane being vector-like, we can now form an $SU(2)_D$ mass term that will mix the $L$ and $R$ SM helicities. However, this Dirac mass term has to be the same for the up and the down quarks (the mass term is isospin invariant). Fortunately, the $SU(2)_R$ invariance is broken on the UV brane and there we can introduce operators that will distinguish between $u_R$ and $d_R$. Technically, the effects of the brane localized operators is to modified the BCs. Explicitly, the IR Dirac mass affect the BCs as follows

$$\begin{array}{ll} \chi_L & + \\ \psi_L & - \\ \chi_R & - \\ \psi_R & + \end{array} \quad \begin{array}{c} \psi_{L|\text{IR}} = 0 \\ \chi_{R|\text{IR}} = 0 \end{array} \quad \begin{array}{c} M_D \\ \Longrightarrow \end{array} \quad \begin{array}{c} \text{discontinuities} \\ \text{in} \\ \chi_L \,\&\, \psi_R \end{array} \quad \begin{array}{c} \psi_{L|\text{IR}} = -M_D R'\,\psi_{R|\text{IR}} \\ \chi_{R|\text{IR}} = M_D R'\,\chi_{L|\text{IR}} \end{array}$$

In the same way, the UV brane operator will modify the BCs as follows

$$\begin{array}{ll} \chi_{u_R} & - \\ \psi_{u_R} & + \end{array} \quad \chi_{u_R|\text{UV}} = 0 \quad \begin{array}{c} \kappa \\ \Longrightarrow \end{array} \quad \begin{array}{c} \text{discontinuity} \\ \text{in} \\ \psi_{u_R} \end{array} \quad \chi_{u_R|\text{UV}} = \kappa m\,\psi_{u_R|\text{UV}}$$

It is now easy to enforce these modified boundary conditions using the general form of the wavefunctions (10.41)–(10.42) that satisfy the bulk equations of motion. For fermions localized toward the UV brane ($c_L > 1/2$ and $c_R < -1/2$), we obtain the approximate expression

$$m \approx \frac{\sqrt{2c_L - 1}}{\sqrt{\kappa^2 - 1/(2c_R + 1)}}\, M_D \left(\frac{R_{\text{UV}}}{R_{\text{IR}}}\right)^{c_L - c_R - 1}. \tag{10.47}$$

### *10.2.4 Top mass and $Zb_L\bar{b}_L$ deviation*

The spectrum of the light generations of quarks can be easily reproduced along these lines. The top mass poses a difficulty, however. Indeed, increasing $M_D$ won't arbitrarily increase the fermion mass which will saturate: the situation is similar to what happens with a large Higgs vev localized on the boundary, the gauge boson masses remain finite even when the vev is sent to infinity. The maximum value of the fermion mass can be inferred by noticing that in the infinite $M_D$ limit, there is a *chirality flip* in the BCs that become

$$\begin{array}{ll} \chi_L & + \\ \psi_L & - \\ \chi_R & - \\ \psi_R & + \end{array} \quad \begin{array}{c} \psi_{L|\text{IR}} = -M_D R'\,\psi_{R|\text{IR}} \\ \chi_{R|\text{IR}} = M_D R'\,\chi_{L|\text{IR}} \end{array} \quad \begin{array}{c} M_D \to \infty \\ \Longrightarrow \end{array} \quad \begin{array}{c} \psi_{R|\text{IR}} = 0 \\ \chi_{L|\text{IR}} = 0 \end{array} \quad \begin{array}{ll} \chi_L & - \\ \psi_L & + \\ \chi_R & + \\ \psi_R & - \end{array}$$

and the corresponding mass is

$$m^2 = \frac{2}{R'^2 \log R'/R} = 2M_W^2. \tag{10.48}$$

where in the last equality, we used the expression of the $W$ mass in terms of $R$ and $R'$ and we have assumed $g_{5R} = g_{5L}$. If we want to go above this saturated mass, one needs to localize the fermions toward the IR brane. However, even in this case a sizable Dirac mass term on the TeV brane is needed to obtain a heavy enough top quark. The consequence of this mass term is the boundary condition for the bottom quarks

$$\chi_{bR} = M_D R' \chi_{bL}. \tag{10.49}$$

This implies that if $M_D R' \sim 1$ then the left handed bottom quark has a sizable component also living in an $SU(2)_R$ multiplet, which however has a coupling to the $Z$ that is different from the SM value. Thus there will be a large deviation in the $Zb_L\bar{b}_L$ coupling. Note, that the same deviation will not appear in the $Zb_R\bar{b}_R$ coupling, since the extra kinetic term introduced on the Planck brane to split top and bottom will imply that the right handed $b$ lives mostly in the induced fermion on the Planck brane which has the correct coupling to the $Z$.

The only way of getting around this problem would be to raise the value of $1/R'$, and thus lower the necessary mixing on the TeV brane needed to obtain a heavy top quark. One way of raising the value of $1/R'$ is by increasing the ratio $g_{5R}/g_{5L}$ (at the price of also making the gauge KK modes heavier and thus the theory more strongly coupled). Another possibility for raising the value of $1/R'$ is to separate the physics responsible for electroweak symmetry breaking from that responsible for the generation of the top mass. In technicolor models this is usually achieved by introducing a new strong interaction called topcolor. In the extra dimensional setup this would correspond to adding two separate $AdS_5$ bulks, which meet at the Planck brane [18]. One bulk would then be mostly responsible for electroweak symmetry breaking, the other for generating the top mass. The details of such models have been worked out in [18] (see also [22]). The main consequences of such models would be the necessary appearance of an isotriplet pseudo-Goldstone boson called the top-pion, and depending on the detailed implementation of the model there could also be a scalar particle (called the top-Higgs) appearing. This top-Higgs would however not be playing a major role in the unitarization of the gauge boson scattering amplitudes, but rather serve as the source for the top-mass only.

### *10.2.5 Fermion delocalization and EW precision tests*

As already mentioned in earlier, the delocalization of SM fermions in the bulk is helpful in keeping the oblique corrections under control. In order to quantify this statement, it is sufficient to consider a toy model where all the three families of fermions are massless and have a universal delocalized profile in the bulk. When the profile of the fermion wavefunction is almost flat, $c_L \approx 1/2$, the leading contributions to $S$ are:

$$S = \frac{2\pi}{g^2 \log \frac{R'}{R}} \left( 1 + (2c_L - 1) \log \frac{R'}{R} + \mathcal{O}\left((2c_L-1)^2\right) \right) . \tag{10.50}$$

In the flat limit $c_L = 1/2$, $S$ is already suppressed by a factor of 3 with respect to the Planck brane localization case. Moreover, the leading terms cancel out for:

$$c_L = \frac{1}{2} - \frac{1}{2 \log \frac{R'}{R}} \approx 0.487 . \tag{10.51}$$

In Fig. 10.7 we have plotted the value of the NDA cut-off scale as well as the mass of the first resonance in the $(c_L - R)$ plane. Increasing $R$ also affects the oblique corrections. However, while it is always possible to reduce $S$ by delocalizing the fermions, $T$ increases and puts a limit on how far $R$ can be raised. One can also see from Fig. 10.8 that in the region where $|S| < 0.25$, the coupling of the first resonance with the light fermions is generically suppressed to less than $10\%$ of the SM value.

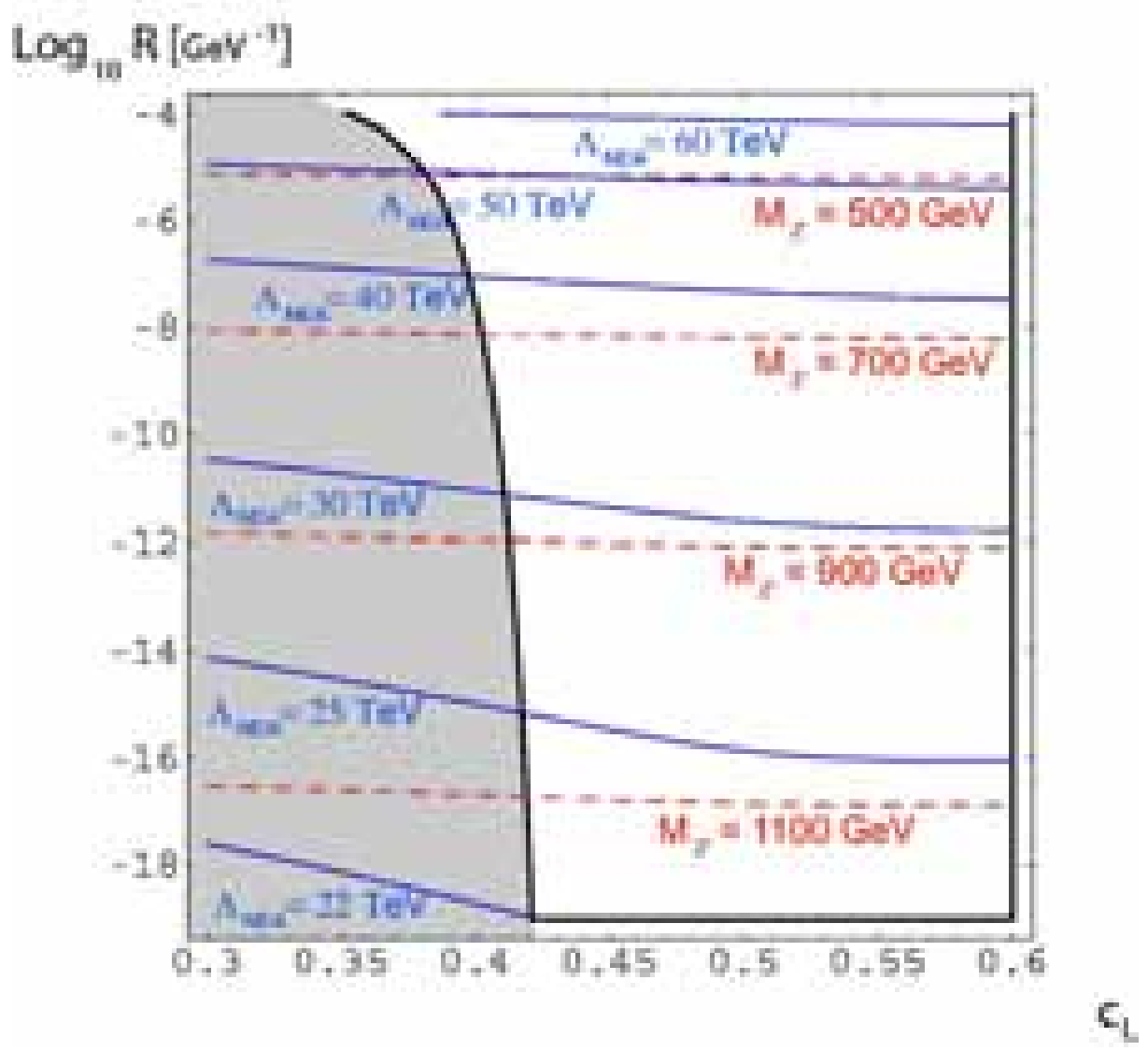

Fig. 10.7: Contour plots of $\Lambda_{\rm NDA}$ (solid blue lines) and $M_{Z^{(1)}}$ (dashed red lines) in the parameter space $c_L$–$R$. The shaded region is excluded by direct searches of light $Z'$ at LEP.

This means that the LEP bound of 2 TeV for SM–like $Z'$ is also decreased by a factor of 10 at least (the correction to the differential cross section is roughly proportional to $g^2/M^2_{Z'}$). In the end, values of $R$ as large as $10^{-7}$ GeV$^{-1}$ are allowed, where the resonance masses are around 600 GeV. So, even if, following the analysis of [23], we take into account a factor of roughly $1/4$ in the NDA scale, we see that the appearance of strong coupling regime can be delayed up to 10 TeV.

It is fair to say that, to date, the major challenge facing Higgsless models is really the incorporation of the third family of quarks while the oblique corrections can be kept under control, at a price of some conspiracy in the localization of the SM quarks and leptons along the extra dimension.

### 10.3 Higgsless electroweak symmetry breaking from moose models

*Stefania De Curtis and Daniele Dominici*

Higgsless models, in their "modern" version, are formulated as gauge theories in a five dimensional space-time and symmetry breaking is realized by means of field boundary conditions in the fifth dimension [2]. One of the interesting features of these schemes is the possibility to delay the unitarity violation scale via the exchange of massive (Kaluza Klein) KK modes [2, 3]. However, it is generally difficult to reconcile a delayed unitarity with the electroweak (EW) constraints. For instance in the framework of models with only ordinary fermions, it is possible to get a small or zero $S$ parameter [24], at the expenses of having a unitarity bound as in the Standard Model (SM) without the Higgs, that is of the order of 1 TeV. A recent solution to the problem, which does not spoil the unitarity requirement at low scales, has been found by delocalizing the fermions in five dimensional theories [14, 25]. We will investigate this possibility in the context of deconstructed gauge theories which come out when the extra dimension is discretized [26]. Through discretization of the fifth dimension we get a finite set of four-dimensional gauge theories each of them acting at a particular lattice site. In this construction, any connection field along the fifth dimension, $A_5$, goes naturally into the link variables $\Sigma_i = e^{-iaA_5^{i-1}}$ realizing the parallel transport between two lattice sites (here $a$ is the lattice spacing). The link variables satisfy the condition $\Sigma\Sigma^\dagger = 1$ and can be identified with chiral fields. In this way the discretized version of the original 5-dimensional gauge theory is substituted by a collection of four-dimensional gauge theories with gauge interacting chiral fields $\Sigma_i$, synthetically described by a moose diagram (an example is given in Fig. 10.9). Here we consider the simplest linear moose model for the Higgsless breaking of the EW symmetry and we delocalize fermions by introducing direct couplings between ordinary left-handed fermions and the gauge vector bosons along the moose string [27].

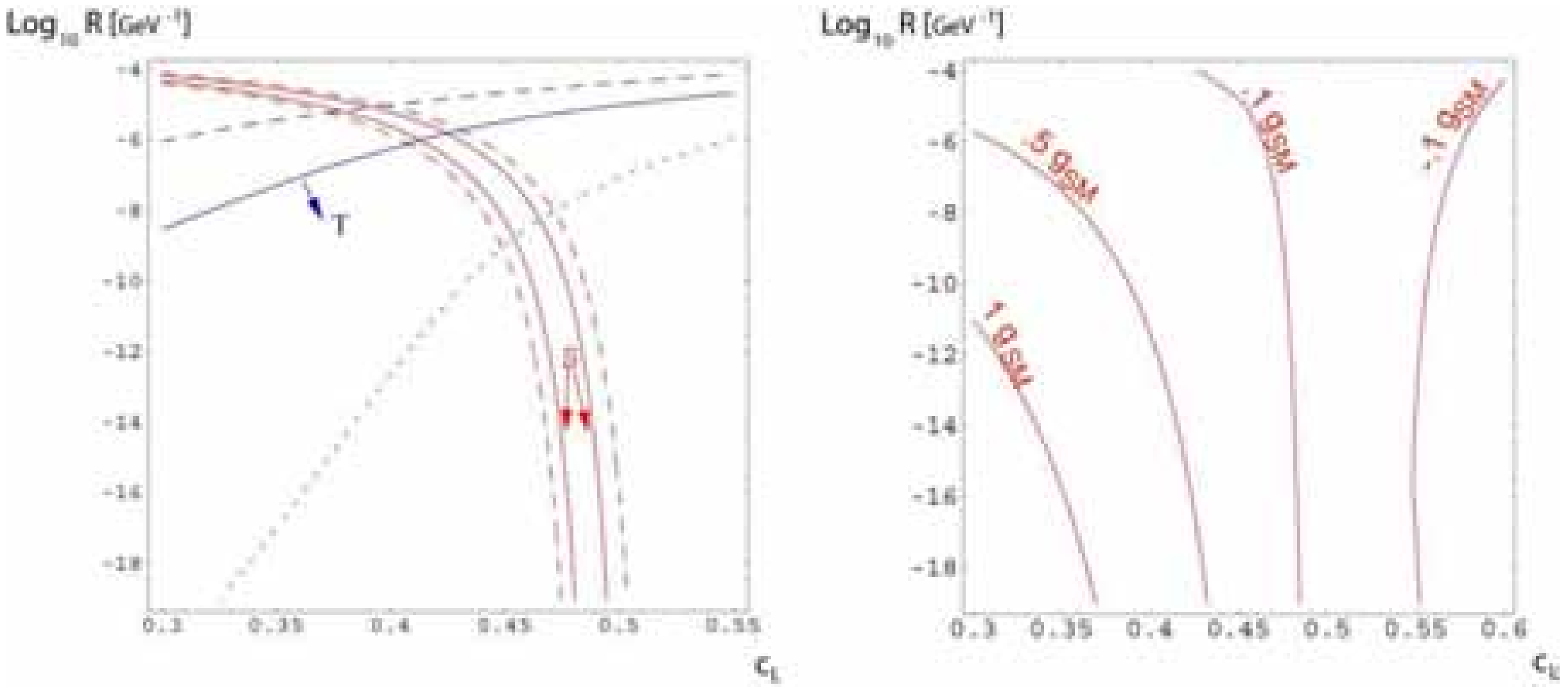

Fig. 10.8: In the left, the contours of $S$ (red), for $|S| = 0.25$ (solid) and $0.5$ (dashed) and $T$ (blue), for $|T| = 0.1$ (dotted), $0.3$ (solid) and $0.5$ (dashed), as function of $c_L$ and $R$ are shown. On the right, contours for the generic suppression of fermion couplings to the first resonance with respect to the SM value can be seen. The region for $c_L$, allowed by $S$, is between $0.43 \pm 0.5$, where the couplings are suppressed at least by a factor of 10.

Let us briefly review the linear moose model based on the $SU(2)$ symmetry [24,27]. We consider $K+1$ non linear $\sigma$-model scalar fields $\Sigma_i$, $i = 1, \cdots, K+1$, $K$ gauge groups $G_i$, $i = 1, \cdots, K$ and a global symmetry $G_L \otimes G_R$ as shown in Fig. 10.9. A minimal model of EW symmetry breaking is obtained by choosing $G_i = SU(2)$, $G_L \otimes G_R = SU(2)_L \otimes SU(2)_R$. The SM gauge group $SU(2)_L \times U(1)_Y$ is obtained by gauging a subgroup of $G_L \otimes G_R$. The $\Sigma_i$ fields can be parameterized as $\Sigma_i = \exp[i/(2f_i)\vec{\pi}_i \cdot \vec{\tau}]$ where $\vec{\tau}$ are the Pauli matrices and $f_i$ are $K+1$ constants that we will call link couplings. The Lagrangian of the linear moose model is given by

$$\mathcal{L} = \sum_{i=1}^{K+1} f_i^2 \mathrm{Tr}[D_\mu \Sigma_i^\dagger D^\mu \Sigma_i] - \frac{1}{2}\sum_{i=1}^{K} \mathrm{Tr}[(F_{\mu\nu}^i)^2] - \frac{1}{2}\mathrm{Tr}[(F_{\mu\nu}(\tilde{W}))^2 - \frac{1}{2}\mathrm{Tr}[(F_{\mu\nu}(\tilde{Y}))^2], \quad (10.52)$$

with the covariant derivatives defined as follows:

$$D_\mu \Sigma_1 = \partial_\mu \Sigma_1 - i\tilde{g}\tilde{W}_\mu \Sigma_1 + i\Sigma_1 g_1 V_\mu^1 \quad (10.53)$$

$$D_\mu \Sigma_i = \partial_\mu \Sigma_i - i g_{i-1} V_\mu^{i-1} \Sigma_i + i\Sigma_i g_i V_\mu^i \;\; (i = 2, \cdots, K) \quad (10.54)$$

$$D_\mu \Sigma_{K+1} = \partial_\mu \Sigma_{K+1} - i g_K V_\mu^K \Sigma_{K+1} + i\tilde{g}' \Sigma_{K+1} \tilde{Y}_\mu \quad (10.55)$$

where $V^i_\mu = V^{ia}_\mu \tau^a/2$, $g_i$ are the gauge fields and gauge coupling constants associated to the groups $G_i$, $i = 1, \cdots, K$, and $\tilde{W}_\mu = \tilde{W}^a_\mu \tau^a/2$, $\tilde{Y}_\mu = \tilde{\mathcal{Y}}_\mu \tau^3/2$ are the gauge fields associated to $SU(2)_L$ and $U(1)_Y$ respectively. Notice that, in the unitary gauge, all the $\Sigma_i$ fields are eaten up by the gauge bosons which acquire mass, except for the photon corresponding to the unbroken $U(1)_{em}$. By identifying the lowest mass eigenvalue in the charged sector with $M_W$, we get at $O(\tilde{g}^2/g_i^2)$ a relation between the EW scale $v$ ($\approx 250\ GeV$) and the link couplings of the chain:

$$\frac{4}{v^2} \equiv \frac{1}{f^2} = \sum_{i=1}^{K+1} \frac{1}{f_i^2}. \tag{10.56}$$

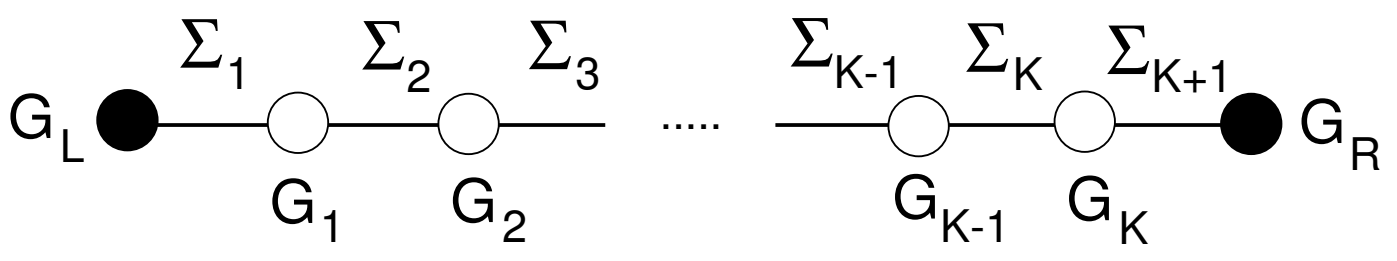

Fig. 10.9: The simplest moose diagram for the Higgsless breaking of the EW symmetry.

Concerning fermions, we will consider only the standard model ones, that is: left-handed fermions $\psi_L$ as $SU(2)_L$ doublets and singlet right-handed fermions $\psi_R$ coupled to the SM gauge fields through the groups $SU(2)_L$ and $U(1)_Y$ at the ends of the chain.

### *10.3.1 Constraints from perturbative unitarity and EW tests*

The worst high-energy behavior of the moose models arises from the scattering of longitudinal vector bosons whose calculation is simplified by using the equivalence theorem. This allows to evaluate these amplitudes in terms of the corresponding Goldstone boson ones. However this theorem holds in the approximation where the energy of the process is much higher than the mass of the vector bosons. Let us evaluate the amplitude for the SM $W$ scattering at energies $M_W \ll E \ll M_{V_i}$. The unitary gauge for the $V_i$ bosons is given by the choice $\Sigma_i = \exp[if\vec{\pi}\cdot\vec{\tau}/(2f_i^2)]$ with $f$ given in Eq. (10.56) and $\vec{\pi}$ the GB's giving mass to $W$ and $Z$. The resulting four-pion amplitude is

$$A_{\pi^+\pi^-\to\pi^+\pi^-} = -\frac{f^4 u}{4}\sum_{i=1}^{K+1}\frac{1}{f_i^6} + \frac{f^4}{4}\sum_{i,j=1}^{K} L_{ij}\left((u-t)(s-M_2)^{-1}_{ij} + (u-s)(t-M_2)^{-1}_{ij}\right), \tag{10.57}$$

with $(M_2)_{ij}$ the square mass matrix for the gauge fields, and

$$L_{ij} = g_i g_j (f_i^{-2} + f_{i+1}^{-2})(f_j^{-2} + f_{j+1}^{-2})\,. \tag{10.58}$$

Note that this amplitude grows linearly with the squared energy, for every choice of $f_i$. This reflects the fact that in the continuum limit the theory corresponds to a 5D gauge theory with boundary conditions which are not simply Neumann or Dirichlet, see 10.3.3 and [2]. In the high-energy limit, where we can neglect the second term in Eq. (10.57), the amplitude has a minimum for all the $f_i$'s being equal to a common value $f_c$. As a consequence, the scale at which unitarity is violated by this single channel contribution is delayed by a factor $(K+1)$ with respect to the one in the SM without the Higgs: $\Lambda_{moose} = (K+1)\Lambda_{HSM}$.

However the moose model has many other longitudinal vector bosons with bad behaving scattering amplitudes. For energies much higher than all the masses of the vector bosons, we can determine the unitarity bounds by considering the eigenchannel amplitudes corresponding to all the possible four-longitudinal vector bosons. Since in the unitary gauge for all the vector bosons $\Sigma_i$ are given by $\Sigma_i = \exp[i\vec{\pi}\cdot\vec{\tau}/(2f_i)]$, the amplitudes are diagonal, and the high-energy result is simply

$$A_{\pi_i\pi_i\to\pi_i\pi_i} \to -\frac{u}{4f_i^2}\,. \tag{10.59}$$

We see that, also in this case, the best unitarity limit is for all the link couplings being equal: $f_i = f_c$. Then: $\Lambda^{TOT}_{moose} = \sqrt{K+1}\ \Lambda_{HSM}$ (for similar results see ref. [23] in [27]). However, in order our approximation to be correct, we have to require $M^{max}_{V_i} \ll \Lambda^{TOT}_{moose}$. By using the explicit expression for the highest mass eigenvalue, in the case of equal couplings $g_i = g_c$, we get an upper bound $g_c \sim 5$. As we will see, this choice gives unacceptable large EW correction.

In this class of models all the corrections from new physics are "oblique" since they arise from the mixing of the SM vector bosons with the moose vector fields (we are assuming the standard couplings for the fermions to $SU(2)_L \otimes U(1)$). As well known, the oblique corrections are completely captured by the parameters $S$, $T$ and $U$ or, equivalently by the parameters $\epsilon_i$, $i = 1, 2, 3$. For the linear moose, the existence of the custodial symmetry $SU(2)_V$ ensures that $\epsilon_1 \approx \epsilon_2 \approx 0$. On the contrary, the new physics contribution to the EW parameter $\epsilon_3$ is sizeable and positive [24]:

$$\epsilon_3 = (\tilde{g}^2/g_i^2) \sum_{i=1}^{K} (1 - y_i) y_i \tag{10.60}$$

where $y_i = \sum_{j=1}^{i} f^2/f_j^2$. Since $0 \le y_i \le 1$ it follows $\epsilon_3 \ge 0$ (see also [28–30]). As an example, let us take equal couplings along the chain: $f_i = f_c$, $g_i = g_c$. Then $\epsilon_3 = \tilde{g}^2 \; K(K+2)/(6 \; g_c^2(K+1))$, which grows with the number of sites of the moose. The requirement of satisfying the experimental constraints ( $\epsilon_3 \approx 10^{-3}$), already for $K = 1$ would imply $g_c \ge 15.8\tilde{g}$, leading to a strong interacting gauge theory in the moose sector and unitarity violation. Notice also that, insisting on a weak gauge theory would imply $g_c$ of the order of $\tilde{g}$, then the natural value of $\epsilon_3$ would be of the order $10^{-1} - 10^{-2}$, incompatible with the experimental data.

### *10.3.2 Effects of fermion delocalization*

A way to reconcile perturbative unitarity requirements with the EW bounds is to allow for delocalized couplings of the SM fermions to the moose gauge fields and some amount of fine tuning [27]. In fact, by generalizing the procedure in [31, 32], the SM fermions can be coupled to any of the gauge fields at the lattice sites by means of Wilson lines.

Define $\chi_L^i = \Sigma_i^\dagger \Sigma_{i-1}^\dagger \cdots \Sigma_1^\dagger \psi_L$, for $i = 1, \cdots, K$. Since under a gauge transformation, $\chi_L^i \to U_i \chi_L^i$, with $U_i \in G_i$, at each site we can introduce a gauge invariant coupling given by

$$b_i \bar{\chi}_L^i \gamma^\mu \left( \partial_\mu + i g_i V_\mu^i + \frac{i}{2} \tilde{g}'(B - L) \tilde{Y}_\mu \right) \chi_L^i, \tag{10.61}$$

where $B(L)$ is the barion(lepton) number and $b_i$ are dimensionless parameters. The new fermion interactions give extra non-oblique contributions to the EW parameters. These are calculated in [27] by decoupling the $V_\mu^i$ fields and evaluating the corrections to the relevant physical quantities. To the first order in $b_i$ and to $O(\tilde{g}^2/g_i^2)$, the $\epsilon_i$ parameters are modified as follows:

$$\epsilon_1 \approx 0, \quad \epsilon_2 \approx 0, \quad \epsilon_3 \approx \sum_{i=1}^{K} y_i \left( \frac{\tilde{g}^2}{g_i^2}(1 - y_i) - b_i \right). \tag{10.62}$$

This final expression suggests that the introduction of the $b_i$ direct fermion couplings to $V_i$ can compensate for the contribution of the tower of gauge vectors to $\epsilon_3$. This would reconcile the Higgsless model with the EW precision measurements by fine-tuning the direct fermion couplings.

As shown in the left panel of Fig. 10.10, in the simplest model with all $f_i = const = f_c$, $g_i = const = g_c$ and $b_i = const = b_c$, the experimental bounds from the $\epsilon_3$ parameter can be satisfied by fine-tuning the direct fermion coupling $b_c$ along a strip in the plane $(Kb_c, \sqrt{K}/g_c)$ (we have chosen these parameters due to the scaling properties of $g_c$ and $b_c$ with $K$, see ref. [27] for details).

The expression for $\epsilon_3$ given in Eq. (10.62) suggests also the possibility of a site-by-site cancellation, provided by:

$$b_i = \delta(\tilde{g}^2/g_i^2)(1 - y_i). \tag{10.63}$$

This choice, for small $b_i$, gives $\epsilon_3 \approx 0$ for $\delta = 1$. Assuming again $f_i = f_c$, $g_i = g_c$, the allowed region in the space $(\delta, \sqrt{K}/g_c)$ is shown on the right panel of Fig. 10.10.

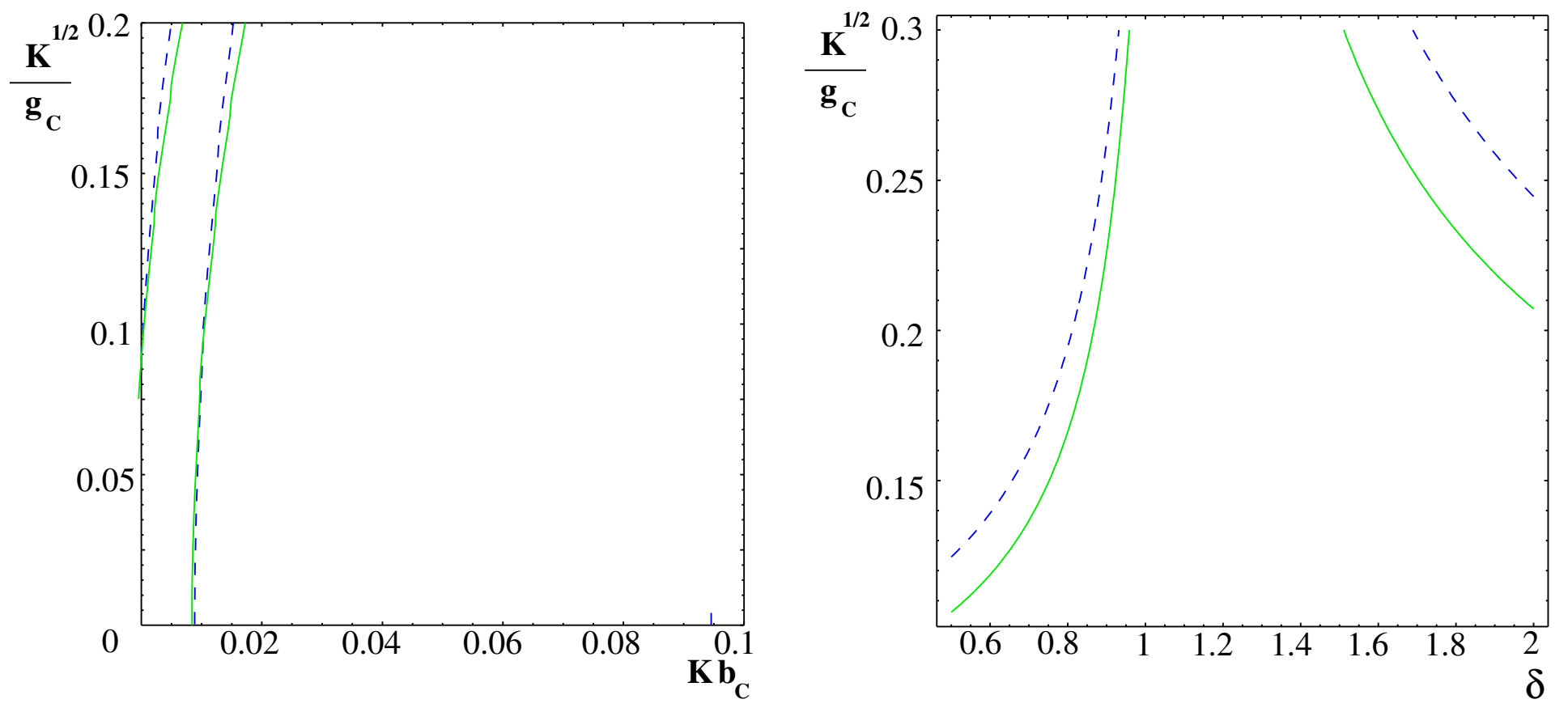

Fig. 10.10: The 95% CL bounds on the plane $(Kb_c, \sqrt{K}/g_c)$-left panel, $(\delta, \sqrt{K}/g_c)$-right panel, from the experimental value of $\epsilon_3$ for $K = 1$ (solid green lines), $K = 10$ (dash blue lines). The allowed regions are between the corresponding lines.

In conclusion, by fine tuning every direct fermion coupling at each site to compensate the corresponding contribution to $\epsilon_3$ from the moose gauge bosons (see also [15, 16]), it is possible to satisfy the EW constraints and improve the unitarity bound of the Higgsless SM at the same time.

### *10.3.3 Continuum limit*

We would like to discuss the continuum limit of the previously discussed moose model by taking $K \to \infty$ with the condition $Ka = \pi R$, where $\pi R$ is the length of the segment in the fifth dimension and $a \to 0$ is the lattice spacing. By defining

$$\lim_{a\to 0} a f_i^2 = f^2(y), \quad \lim_{a\to 0} a g_i^2 = g_5^2(y) \tag{10.64}$$

the action obtained as the continuum limit of the Lagrangian (10.52), for flat metric $g_5(y) = g_5$ and $f(y) = \bar{f}$, can be written as [11]

$$S = -\frac{1}{4}\int d^4x \int_0^{\pi R} dy \big[\frac{1}{g_5^2}(F^a_{MN})^2 + \frac{1}{\tilde{g}^2}(F^a_{\mu\nu})^2\delta(y) + \frac{1}{\tilde{g}'^2}(F^3_{\mu\nu})^2\delta(y-\pi R)\big] + S_{ferm}$$

where $F_{MN}$ is the tensor associated to the 5D field $A_N$ and the brane kinetic terms, arising from the left and right ends of the moose chain, are the terms which modify the otherwise linear mass spectrum of KK excitations of gauge bosons. These are necessary in order to avoid light KK excitations of the standard gauge bosons.

In the continuum limit, left- and right-handed fermions live at the opposite ends of the extra-dimension. However, in the discrete, we have introduced an interaction term invariant under all the symmetries of the model which delocalizes the left-handed fermions in the continuum limit. In fact, we have seen that the fermionic fields along the string are defined in terms of the operator

$$\Sigma_1\Sigma_2\cdots\Sigma_i\,. \tag{10.65}$$

In five-dimensions the fields $\Sigma$'s can be interpreted as the link variables along the fifth dimension. As such they can be written in terms the fifth component of the gauge fields $A_N$. As a consequence the operator given in Eq. (10.65) becomes a Wilson line

$$\Sigma_1\Sigma_2\cdots\Sigma_i \to P\left[\exp\left(-i\int_0^y dz A_5(z)\right)\right]. \tag{10.66}$$

In this way the original fermionic fields acquire a non-local interaction induced by Wilson lines. The fermion action in Eq. (10.65) is therefore given by

$$S_{ferm} = \int d^4x \int_0^{\pi R} dy \left[\delta(y) i\bar{\psi}_L \not{D}\psi_L + \delta(\pi R - y) i\bar{\psi}_R \not{D}\psi_R + b(y) i\bar{\chi}_L \not{D}\chi_L\right] \quad (10.67)$$

where

$$b(y) = \lim_{a\to 0}\frac{b_i}{a}, \quad \chi_L(y) = P\left[\exp\left(-i\int_0^y dz A_5(z)\right)\right]\psi_L \quad (10.68)$$

and

$$\not{D}\psi_L = \left(\not{\partial} + i\frac{\tau^a}{2}\not{A}^a(y) + iY_L\not{A}^3(\pi R)\right)\psi_L, \quad \not{D}\psi_R = \left(\not{\partial} + iY_R\not{A}^3(y)\right)\psi_R \quad (10.69)$$

$$\not{D}\chi_L = \left(\not{\partial} + i\frac{\tau^a}{2}\not{A}^a(y) + iY_L\not{A}^3(\pi R)\right)\chi_L \quad (10.70)$$

with $Y_{L,R}$ the left and right hypercharges. Mass terms for the fermions can be generated by

$$\lambda^{ij}\bar{\psi}^i_L P\Big[\exp(-i\int_0^{\pi R} dz A_5(z))\Big]\psi^j_R.$$

The breaking of $SU(2)$ to $U(1)_{em}$ is obtained by the following boundary conditions:

$$\partial_y A^{1,2}_\mu - \frac{g_5^2}{\tilde{g}^2}\Box A^{1,2}_\mu|_{y=0} = 0, \quad A^{1,2}_\mu|_{y=\pi R} = 0, \quad (10.71)$$

$$\partial_y A^3_\mu - \frac{g_5^2}{\tilde{g}^2}\Box A^3_\mu|_{y=0} = 0, \quad \partial_y A^3_\mu + \frac{g_5^2}{\tilde{g}'^2}\Box A^3_\mu|_{y=\pi R} = 0 \quad (10.72)$$

We would like to discuss what is the continuum limit for the direct fermionic couplings when we choose the $b_i$'s according to the Eq. (10.63) with $\delta = 1$ which corresponds to a site-by-site cancellation. By assuming $g_5(y) = g_5$, with $g_5$ constant, we get

$$b(y) = \frac{\tilde{g}^2}{g_5^2}\int_y^{\pi R} dt \frac{f^2}{f^2(t)}, \text{ with } \frac{1}{f^2} = \int_0^{\pi R}\frac{dy}{f^2(y)}. \quad (10.73)$$

From Eq. (10.73) we see that $b(0) = \tilde{g}^2/g_5^2,\ b(\pi R) = 0$. Therefore the direct fermionic coupling decreases along the fifth dimension going from the brane located at $y = 0$ to the brane at $y = \pi R$. For the case of constant $f(y) = \bar{f}$ we find

$$b(y) = \frac{\tilde{g}^2}{g_5^2}\left(1 - \frac{y}{\pi R}\right). \quad (10.74)$$

With this choice the contribution from the new delocalized fermion interactions to $\epsilon_3$ is given by

$$\epsilon_3|_{ferm} = -\frac{1}{\pi R}\int_0^{\pi R} dy\, y b(y) = -\frac{\tilde{g}^2}{g_5^2}\frac{\pi R}{6} \quad (10.75)$$

which is just the opposite of the contribution to $\epsilon_3$ in the linear moose [11, 24].

Another interesting case corresponds to a Randall-Sundrum metric along the fifth dimension [1, 33]. It corresponds to

$$f(y) = \bar{f}e^{ky} \quad (10.76)$$

and we find

$$b(y) = \frac{\tilde{g}^2}{g_5^2} \frac{e^{-2\pi kR} - e^{-2ky}}{e^{-2\pi kR} - 1}\,. \tag{10.77}$$

In this case we get

$$\epsilon_3|_{ferm} = -\int_0^{\pi R} dy \frac{e^{-2ky} - 1}{e^{-2k\pi R} - 1} b(y) = -\frac{\tilde{g}^2}{g_5^2} \frac{1}{4k} \frac{e^{4k\pi R} - 4k\pi R e^{2k\pi R} - 1}{(1 - e^{2k\pi R})^2} \tag{10.78}$$

which is the opposite of the contribution from the gauge bosons derived in [24].

Therefore by allowing for a fine tuning obtained with a convenient delocalization of the fermion couplings, the contribution to the $S$ parameter coming from the gauge and fermion sectors vanishes. We are currently investigating whether a geometrical mechanism can be found in order to guarantee such a cancellation. Notice that in the previous formulation we do not assume the existence of bulk fermions; under certain hypotheses these can generate the delocalization of the standard model fermionic couplings, parameterized by $b$.

## 11 STRONGLY INTERACTING HIGGS SECTOR AND ANOMALOUS COUPLINGS

### 11.1 Introduction

*Georges Azuelos, Tao Han and Wolfgang Kilian*

There is no fundamental principle that requires the physics responsible for electroweak symmetry breaking to be weakly interacting. In fact, there are many well-understood cases of spontaneous symmetry breaking in quantum physics, among them the breaking of electromagnetic gauge symmetry in superconductors and chiral symmetry breaking in QCD. In all these cases, the symmetry breaking is due to some interaction effectively becoming strong in a certain low-energy regime. In superconductors, this is the exchange of phonons near the Fermi surface. In QCD, the gluon coupling is effectively strong at energies below a GeV. These non-perturbative effects lead to a condensation of a field bilinear $\langle\psi\phi\rangle \neq 0$ which breaks the symmetry of the basic Lagrangian.

Mathematically, in such models of strong (dynamic) symmetry breaking the corresponding local operator $\psi\phi$ is a Higgs field. In general, the characteristics of the resulting Higgs boson will depend on the model. In QCD, for instance, the composite "Higgs" of chiral symmetry breaking is a heavy ($\sim$ GeV), poorly defined state. The lightest CP-even scalar, the broad $\sigma(600)$ meson resonance, may not even be thought of as a $q\bar{q}$ state [1] [cf. Section 12] and experimentally, its importance for low-energy hadronic interactions is minor compared to other states such as the $\rho$ vector resonance. In QCD-like theories of electroweak symmetry breaking, the Higgs scalar is therefore expected to be very heavy and broad ($\sim$ TeV). In other dynamical theories, however, it can be associated with a low-lying excitation, such as in models with specific symmetry structure [cf. Section 7], or in technicolor theories with fermions in a higher dimensional represenetions of the gauge group [cf. Section 12]. Other conceivable mechanisms of electroweak symmetry breaking extend beyond four-dimensional quantum field theory. For instance, spontaneous symmetry breaking due to string interactions, bulk-brane interplay, or explicit symmetry-breaking boundary interactions in extra dimensions also could lead to a strongly-interacting effective field theory at the electroweak scale that need not involve Higgs-like states [cf. Section 10].

In this section, we consider first the scenario of electroweak symmetry breaking where the Higgs boson is either absent from the spectrum, or it is merely one among several heavy resonances, similar to the QCD case. In that case, strong interactions of some of the known particles are guaranteed in the TeV energy range. From the experience with QCD we can deduce that direct detection of the underlying new physics, analogous to quark-induced jets, may then need multi-TeV energies. However, even without a-priori knowledge of the underlying model we can nevertheless investigate the low-energy effective theory and identify observables that carry nontrivial information.

This could be the first information on the Higgs sector that becomes available through collider experiments at the LHC and the ILC.

We then consider the scenario where a light Higgs is present. The new physics, described by dimension-six operators of an effective Chiral Lagrangian, is then manifest, at high energy colliders, by anomalous couplings of the Higgs and Standard Model gauge bosons.

#### *11.1.1 Weak interactions without a Higgs*

In principle, it is trivial to write down the generic low-energy effective theory for a strongly interacting Higgs sector: this is simply the Standard Model (SM) with the physical Higgs field omitted. More precisely, we can formally remove the Higgs in the tree-level action by pushing its self-coupling in the potential to infinity, keeping the vacuum expectation value and all gauge and Yukawa couplings constant.

However, there is a tricky point with this approach. Since, in a generic gauge, the Higgs boson is part of a complex $SU(2)_L$ doublet together with the Goldstone bosons that provide the longitudinal components of the massive $W$ and $Z$ bosons, removing the Higgs also removes the gauge symmetry. The natural choice of removing the Higgs in unitarity gauge where Goldstone bosons do not appear, results in an effective theory that does not exhibit any electroweak symmetry at all [2, 3]. This is a valid approach, but such a model does not naturally implement any of the well-established facts about electroweak symmetry, e.g., the left-handedness of charged currents, CKM unitarity for flavor mixing, or the $W$-$Z$ mass relation.

Therefore, it is customary to construct the low-energy effective theory for electroweak interactions *bottom-up* by explicitly implementing gauge invariance [4–12]. In the absence of a Higgs boson, the $SU(2)_L \times U(1)_Y$ electroweak symmetry has then to be nonlinearly realized on the Goldstone-boson fields.

For the construction of the effective Lagrangian, let us introduce some notation. As building blocks we need the left-handed quark and lepton doublet fields ($Q_L, L_L$) and corresponding right-handed fields. For notational convenience, we also can write them as doublets ($Q_R, L_R$). Next, we introduce the Goldstone bosons of $SU(2)_L \times U(1)_Y \to U(1)_{\mathrm{QED}}$ spontaneous symmetry breaking, $w^a$ ($a = 1, 2, 3$) and their contraction with the Pauli matrices $\tau^a$ to a matrix-valued scalar field $\mathbf{w} = w^a \tau^a$. A possible nonlinear representation is realized by a derived matrix-valued scalar field of the form

$$\Sigma(x) = \exp\left(-\frac{i}{v}\mathbf{w}(x)\right), \tag{11.1}$$

where $v$ is the electroweak constant $v = (\sqrt{2}\,G_F)^{-1/2} = 246\,\mathrm{GeV}$. This field is required to transform as a $SU(2)_L$ doublet, transformations applied to the left. Local $U(1)_Y$ (hypercharge) transformations apply as $\Sigma \to \Sigma\exp(i\beta(x)\,\tau^3/2)$. Note that $\Sigma$ is a unitary matrix ($\Sigma^\dagger\Sigma = 1$), and electroweak transformations act linearly on $\Sigma(x)$, but nonlinearly on $\mathbf{w}(x)$.

The electroweak gauge boson fields are $W^a_\mu$ ($a = 1, 2, 3$) and $B_\mu$, which we contract with the Pauli matrices $\tau^a$ to matrix-valued vector fields $\mathbf{W}_\mu = W^a_\mu \tau^a/2$ and $\mathbf{B}_\mu = \Sigma B_\mu(\tau^3/2)\Sigma^\dagger$. In accordance with the gauge transformation properties, the covariant derivative of $\Sigma$ is

$$\mathbf{D}_\mu\Sigma = (i\partial_\mu + g\mathbf{W}_\mu - g'\mathbf{B}_\mu)\Sigma. \tag{11.2}$$

The covariant derivatives of fermion fields involve the corresponding representation matrices of the $SU(2)_L \times U(1)_Y$ symmetry, which differ between left- and right-handed fermions. We also need field-strength tensors

$$\mathbf{W}_{\mu\nu} = \partial_\mu\mathbf{W}_\nu - \partial_\nu\mathbf{W}_\mu + ig[\mathbf{W}_\mu, \mathbf{W}_\nu] \qquad \text{and} \qquad \mathbf{B}_{\mu\nu} = \partial_\mu\mathbf{B}_\nu - \partial_\nu\mathbf{B}_\mu. \tag{11.3}$$

Finally, it is convenient to introduce two covariant fields derived from $\Sigma$, a vector and a scalar:

$$\mathbf{V}_\mu = \Sigma(\mathbf{D}_\mu\Sigma^\dagger), \qquad \mathbf{T} = \Sigma\tau^3\Sigma^\dagger. \tag{11.4}$$

With these definitions, the generic low-energy effective Lagrangian for the electroweak interactions of leptons, quarks, and gauge bosons is

$$\begin{aligned}\mathcal{L} = &-\frac{1}{2}\,\mathrm{tr}[\mathbf{W}_{\mu\nu}\mathbf{W}^{\mu\nu}] - \frac{1}{2}\,\mathrm{tr}[\mathbf{B}_{\mu\nu}\mathbf{B}^{\mu\nu}] + \frac{v^2}{4}\,\mathrm{tr}[(\mathbf{D}_\mu\Sigma)(\mathbf{D}^\mu\Sigma)^\dagger] - \beta'\frac{v^2}{8}\,\mathrm{tr}[\mathbf{T}\mathbf{V}_\mu]\,\mathrm{tr}[\mathbf{T}\mathbf{V}^\mu] \\ &+ \bar{Q}_L i\not{D} Q_L + \bar{Q}_R i\not{D} Q_R + \bar{L}_L i\not{D} L_L + \bar{L}_R i\not{D} L_R \\ &- (\bar{Q}_L\Sigma M_Q Q_R + \bar{L}_L\Sigma M_L L_R + \text{h.c.}) - \bar{L}^c_L\Sigma^* M_{N_L}\frac{1+\tau^3}{2}\Sigma L_L - \bar{L}^c_R M_{N_R}\frac{1+\tau^3}{2}L_R \\ &+ \cdots,\end{aligned} \tag{11.5}$$

where we wrote only the operators with the lowest dimension. In this Lagrangian, QED (and, implicitly, QCD) gauge invariance is still linearly realized.

Before continuing, let us add a few remarks on this Lagrangian: (i) the particular representation of $\Sigma$ in terms of $\mathbf{w}$ is irrelevant, only the symmetry properties matter; (ii) the unitarity gauge is recovered by $\mathbf{w} \equiv 0$, i.e., $\Sigma \equiv 1$, and yields a Lagrangian identical to the Standard Model with Higgs removed; (iii) we could write Majorana and Dirac masses for both left- and right-handed neutrinos.

The Lagrangian (11.5) uniquely determines the leading term in a low-energy expansion of all scattering amplitudes. The $W$ and $Z$ masses result from the first term in the $\Sigma$ expansion,

$$M_W = gv/2 \qquad \text{and} \qquad M_Z = \sqrt{g^2 + g'^2}(1 + \beta'/2)\, v/2 \tag{11.6}$$

The experimental results imply that the $\rho$ parameter, related to the coefficient $\beta'$, vanishes at tree level [1], hence $\beta' = 0$. This is understood as an extra $SU(2)_R$ symmetry of the dimension-2 part of the (bosonic) Lagrangian, called *custodial* symmetry [13–16]. It is often assumed to be an approximate symmetry of the fundamental physics responsible for electroweak symmetry breaking, even though there are further parameters in the tree-level Lagrangian that break the symmetry explicitly, most notably $g'$ and $m_t$.

This effective theory is formally non-renormalizable. While the Lagrangian already contains an infinite number of terms with prefactor $1/v^n$ due to the infinite series expansion of $\Sigma$, computing radiative corrections requires another infinite series of higher-dimensional counterterms, indicated by dots in (11.5). These extra counterterms carry prefactors $1/(4\pi v)^n$ and are therefore suppressed. Below energies of about $4\pi v = 3\,\text{TeV}$ the effective theory has some region of applicability. Above that scale, it has no physical content anymore.

In addition to counterterms, we can add extra contributions to higher-dimensional operators with arbitrary coefficients. Within a specific underlying strong-interaction model we expect the actual coefficients, computed in a given renormalization scheme, to be well-defined. A similar program, set up for low-energy hadronic interactions, has proven very successful [4, 17, 18], and is currently used as a convenient gauge for lattice QCD calculations. In the electroweak case, the list of CP-even dimension-4 bosonic operators reads [7, 8]

$$\mathcal{L}_1 = \alpha_1 g g' \,\text{tr}[\mathbf{B}_{\mu\nu}\mathbf{W}^{\mu\nu}], \qquad \mathcal{L}_6 = \alpha_6 \,\text{tr}[V_\mu V_\nu]\,\text{tr}[\mathbf{T}V^\mu]\,\text{tr}[\mathbf{T}V^\nu], \tag{11.7}$$

$$\mathcal{L}_2 = i\alpha_2 g' \,\text{tr}[\mathbf{B}_{\mu\nu}[V^\mu, V^\nu]], \qquad \mathcal{L}_7 = \alpha_7 \,\text{tr}[V_\mu V^\mu]\,\text{tr}[\mathbf{T}V_\nu]\,\text{tr}[\mathbf{T}V^\nu], \tag{11.8}$$

$$\mathcal{L}_3 = i\alpha_3 g \,\text{tr}[\mathbf{W}_{\mu\nu}[V^\mu, V^\nu]], \qquad \mathcal{L}_8 = \tfrac{1}{4}\alpha_8 g^2 (\text{tr}[\mathbf{T}\mathbf{W}_{\mu\nu}])^2, \tag{11.9}$$

$$\mathcal{L}_4 = \alpha_4 (\text{tr}[V_\mu V_\nu])^2, \qquad \mathcal{L}_9 = \tfrac{i}{2}\alpha_9 g \,\text{tr}[\mathbf{T}\mathbf{W}_{\mu\nu}]\,\text{tr}[\mathbf{T}[V^\mu, V^\nu]], \tag{11.10}$$

$$\mathcal{L}_5 = \alpha_5 (\text{tr}[V_\mu V^\mu])^2, \qquad \mathcal{L}_{10} = \tfrac{1}{2}\alpha_{10} (\text{tr}[\mathbf{T}V_\mu]\,\text{tr}[\mathbf{T}V_\nu])^2, \tag{11.11}$$

$$\mathcal{L}_{11} = \alpha_{11} g \epsilon^{\mu\nu\rho\lambda}\,\text{tr}[\mathbf{T}V_\mu]\,\text{tr}[V_\nu \mathbf{W}_{\rho\lambda}]. \tag{11.12}$$

CP-odd operators, operators involving fermions, and higher-dimensional terms may be added to this list, but are not considered in most studies.

In the above list, the coefficients $\alpha_i$ are dimensionless. As long as the operators are generated only as counterterms, their coefficients are naturally of order $1/16\pi^2$; for this reason, a different normalization that multiplies the coefficients by $16\pi^2$ is used frequently in the literature. Additional contributions due to new physics can in principle be of arbitrary magnitude, but in a strong-interaction scenario they tend to be somewhat larger than the loop contributions. Higher-dimensional terms, which we do not consider at this point, would be suppressed by additional factors of $1/(4\pi v)^2$.

Due to the fact that $\beta' = 0$ at tree level, bosonic loops generate those operators that involve $\mathbf{T}$ factors only with coefficients suppressed by $g'^2$, the small hypercharge coupling squared. Nevertheless, these terms may be present in the tree-level Lagrangian with unsuppressed coefficients. Loops involving top quarks also break the custodial symmetry.

### *11.1.2 Vector-Boson Scattering*

Despite the fact that the use of the effective Chiral Lagrangian is limited to the range up to a few TeV, it is nevertheless a valuable tool since this is exactly the energy range that will be accessed by the upcoming

LHC and ILC colliders. The bilinear couplings $\alpha_1$ and $\alpha_8$, related to the $S$ and $T$ parameters [19, 20], have already been measured at LEP1 in 2-fermion production [1, 21]. For the trilinear couplings $\alpha_2, \alpha_3, \alpha_9, \alpha_{11}$, 4-fermion processes ($W$ pair and single-$W$ or single-$Z$ production) are needed [21–23] There was some sensitivity at LEP2, but in order to reach the order of magnitude $\alpha_i \sim 1/16\pi^2$ implied by the perturbative expansion, LHC and ILC data will be necessary [24–27]. Finally, the remaining couplings are accessed at the LHC [28–33] and the ILC [27, 34–38] by 6-fermion processes (Fig. 11.1). Since the 4-fermion processes also depend on the anomalous two-boson couplings, and the 6-fermion processes also depend on all lower-order couplings, in practice a simultaneous fit of all couplings is necessary to extract their physical values.

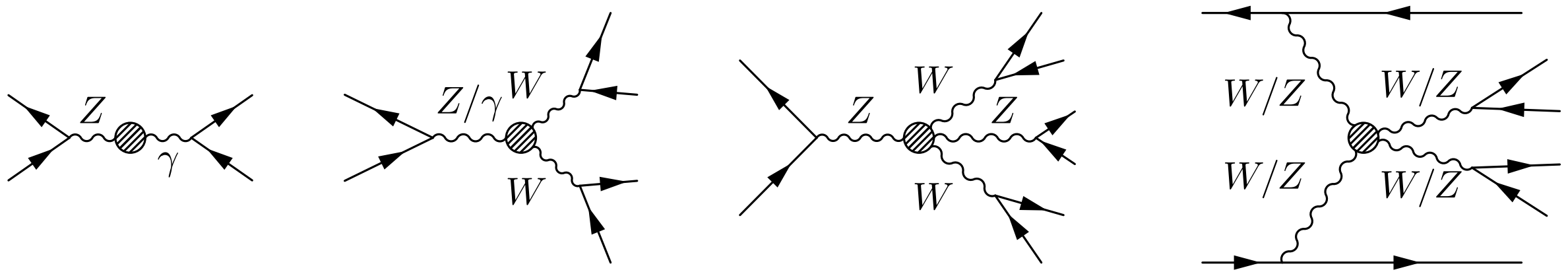

(a) (b) (c) (d)

Fig. 11.1: Processes that involve the anomalous couplings (a) $\alpha_{1,8}$; (b) $\alpha_{2,3,9,11}$; (c,d) $\alpha_{4,5,6,7,10}$, respectively. The fermions may be either quarks or leptons.

The 6-fermion processes of type Fig. 11.1d are of particular interest, since in the limit that the intermediate vector bosons get on-shell, they include $2 \to 2$ quasi-elastic vector boson scattering as subprocesses. (Processes of type Fig. 11.1c probe the same interactions in a far off-shell regime.) These are the only accessible $2 \to 2$ processes[1] that contain the Higgs, if it exists, at tree-level. Consequently, the absence of a Higgs boson has a strong effect: quasi-elastic vector-boson scattering amplitudes, as calculated at lowest order, rise without bound and surpass their unitarity limit in the TeV range [42, 43]. Higher-order corrections would remedy this, but depend on the unknown infinite series of higher-order counterterms and thus leave the actual result undetermined. Observing the presence or absence of strong interactions in 6-fermion production is thus the ultimate experimental test of the Higgs mechanism, independent of its particular realization: strong interactions are absent if light scalar state(s) are present in the model and couple exactly as required by the constraint that the symmetry is broken exclusively by their vacuum expectation value(s).

For quantitative estimates, it is convenient to start with the limit $g, g', m_f \to 0$ (but $v$ constant), where the gauge symmetry is formally broken, but the gauge-boson and fermion masses are arbitrarily small. In this limit, several simplifications apply to the processes of type Fig. 11.1d: (i) the final-state gauge bosons can be treated in the narrow-width approximation; (ii) the initial-state gauge bosons are approximately on-shell and can, at small angles, be treated as partons within the incoming fermions: the Effective $W$ Approximation [44–46]; (iii) the scattering amplitudes of longitudinally polarized vector bosons become equal to corresponding scattering amplitudes of Goldstone bosons, while the transversal degrees of freedom decouple: the Goldstone-boson Equivalence Theorem [47–49]; (iv) the resulting Lagrangian is exactly identical to the effective Lagrangian of low-energy pion scattering in the $m_\pi \to 0$ limit: Chiral Perturbation Theory [4, 17, 18].

The last analogy allows us to transfer QCD knowledge to the present case. Noting that the scattering amplitudes of transversal gauge bosons do not violate unitarity limits, we conclude that the dominant contributions to quasi-elastic vector boson scattering amplitudes fulfil, like pion scattering amplitudes, $SU(2)$ symmetry relations. We can diagonalize the $2 \to 2$ scattering matrix and derive the strongest bound on the low-energy effective theory: The amplitude projected onto $J = 0$ (angular momentum) and $I = 0$ (weak isospin) saturates the unitarity limit at $E = \sqrt{8\pi}\, v = 1.2\,\text{TeV}$ [42, 43]. Due to the

[1] with the exception of vector-boson scattering into heavy-quark pairs [39–41], which however is an experimental challenge.

rapidly falling $W/Z$ structure functions, the c.m. energy of the incoming fermions has to be considerably above that limit if this unitarity saturation is to have an observable effect.

In the actual environment of the LHC and ILC colliders with their limited energy reach, cuts and backgrounds, it turns out that these estimates in the gaugeless limit are useful for qualitative considerations, but fail if reasonably precise values for cross sections and simulated event samples are desired. Dropping the approximations (i) to (iv) above altogether, the physical picture becomes much less clear; in particular, the implications of unitarity saturation for the complete off-shell process are not obvious. While at the ILC it is unlikely that the unitarity limit can be reached, the kinematic capability of the LHC does extend into that range. However, the high background and the rapidly falling structure functions make it a challenge also at the LHC to detect observable consequences of unitarity violation in a naive tree-level extrapolation. Therefore, the Chiral Lagrangian (11.5), optionally augmented by resonances coupled with free coefficients, is a theoretical framework sufficient for all practical purposes. As a consequence, while we are not allowed to use simplifying approximations in calculating the 6-fermion processes of interest at the LHC or the ILC, with appropriate calculational tools it is possible to make reliable predictions and to compare them with data.

### *11.1.3 Low-Energy Parameters*

Without any knowledge about the underlying mechanism that triggers electroweak symmetry breaking, we have no idea about how the scattering amplitudes will evolve beyond the limit where the effective theory fails. At least, the $\alpha$ parameters allow us to parameterize scattering amplitudes, and once data are available, values for the parameters can be obtained. For the ILC studies, we essentially have a complete picture about the possible sensitivity on the $\alpha$ couplings [27,50,51] which involves a complete simulation of the 6-fermion process and does not rely on further simplifying assumptions at the theoretical level. For isolating the strong scattering signal, one looks at four-jet events in association with missing energy due to the forward-scattered neutrinos or electrons in the final state. The main uncertainties originate from the ambiguity in identifying $W$ and $Z$ bosons in their hadronic decay modes, which for the current detector designs appears to be manageable. These results currently include all known backgrounds and detector effects based on fast simulation.

### *11.1.4 Resonances*

It is likely that $WW$ scattering amplitudes do not just saturate unitary and remain featureless at higher energies. Just like in the analogous situation of $\pi\pi$ scattering, there may be strong resonances on top of that which could be observable at the LHC at energies above the $1.2\,\mathrm{TeV}$ cutoff. The best-studied cases are (i) a heavy scalar, or (ii) a heavy vector. The first case is just the heavy-mass extrapolation of the Standard Model, while the second one is modelled after the QCD case with its strong $\rho$ resonances [52–54].

It should be kept in mind, however, that the actual situation may be very different. For instance, in the context of models that entangle electroweak symmetry with extra dimensions and gravitation, tensor resonances could play an important role. So far, a few studies have considered the prospects for distinguishing qualitatively different scenarios at hadron and lepton colliders [28–30], and for the case of vector resonances, the possibility of resonance parameter measurements have been discussed. An experimental analysis of the general case that would allow for quantitative conclusions on all possible resonance couplings is desirable, but so far has not been attempted. In the ILC context, a detailed study that relates the estimated uncertainties of anomalous coupling measurements to the coupling parameters of heavy resonances can be found in [51] (see also Section 11.2).

Without a preferred underlying model, extrapolating the scattering amplitudes of vector bosons beyond the range where the lowest-order prediction saturates unitarity remains speculative. There exists an infinite set of extrapolation prescriptions that, at least, maintain elastic unitarity. Particularly popular are the $K$-matrix model [55, 56] that parameterizes featureless amplitudes (resonant at infinity), and the Padé or inverse-amplitude method [57, 58] that parameterizes in each channel, up to the order where it is typically evaluated, a single resonance with a specified mass and coupling strength. The latter method has proven successful in the description of low-energy QCD amplitudes [59]; this success is due to the particular property of QCD to contain dominant resonances in its hadron form factors and scattering amplitudes. Adopting this unitarization prescription and making further assumptions such as custodial symmetry, with which vector boson scattering is determined by the only coefficients $\alpha_4$ and $\alpha_5$, resonances can be mapped in two-dimensional parameter space [58]. Unfortunately, despite the good description of QCD amplitudes by such a model, we have no clue whether a strong-interaction theory of electroweak interactions would actually exhibit such a behavior.

LHC studies [30, 32, 33, 58, 60–62] on prospects for observing vector boson scattering remain at the parton or fast detector simulation level, but full detector simulation analyses are in progress. They generally rely on the abovementioned simplifying assumptions, and should therefore not be taken at face value. The results are valid, however, for a generic resonance and can be re-interpreted in the context of a more general Chiral Lagrangian model. They could serve, in principle, to derive limits on the $\alpha$ couplings. The $WZ$ scattering signal, with one or two leptons in the final state, can be observed up to a mass of $\sim 1500$ GeV [60] within a few years of LHC running at nominal luminosity (see Fig. 11.2). The major irreducible background is $WZjj$ from SM processes involving gluon or $\gamma/Z$ exchange diagrams, with transversely polarized vector bosons radiated from incoming or outgoing quarks. It has been shown [63, 64] that a complete description of the 6-fermion process is necessary for a correct evaluation of vector boson scattering. The reducible backgrounds are $t\bar{t}$ production and $W + j$ or $Z + j$, which have all very high cross sections. Using either the Padé or N/D unitarization protocol, it has also been shown [33] that scalar and vector resonances may occur in the channel $WW \to jj + \ell\nu$ in the TeV region and if so, should be observable at the LHC. In all cases, the requirement of forward jets, expected to result from the outgoing primary quarks in Fig. 11.1d, and of a veto on an excess of central jets is essential for an effective reduction of these backgrounds.

The BESS model [52, 53] (which stands for *Breaking Electoweak Symmetry Strongly*) is another general framework for describing resonances in gauge boson scattering. The model is without a Higgs, but with a triplet of new massive QCD-like vector bosons $V$. These bosons are originally auxiliary fields of a hidden SU(2) symmetry, but it is explicitly assumed that they become physical and a kinetic term is added for them. The new $V^{\pm,0}$ bosons mix with the $W$ and $Z$ of the SM through a mixing angle proportional to $x = g/g''$, where $g''$ is the self-coupling of the $V$. Another parameter, $b$, governs the coupling to fermions. A variant of this model, called the degenerate BESS model [65, 66], predicts the existence of two triplets of new gauge bosons $L^{\pm}, L_3, R^{\pm}, R_3$ which are almost degenerate in mass, the mass splitting depending on the above parameter $x$, and with the neutral bosons mixing with the SM electroweak boson. The BESS models could be an effective theory for which walking type technicolor theories (see Section 12) are possible prototypes of an underlying gauge theory [67,68]. Expected bounds on the parameters of the BESS model can be found in [27]. The degenerate model is best studied in the $f\bar{f}$ decay channels of the new resonances. Expected 95% limits in the parameter space $\{x, M\}$ in present and future colliders are given in [69, 70], where it is shown that with 100 fb$^{-1}$, the LHC can discover resonances up to $\sim 2$ TeV, for $x = 0.1$, and that CLIC could measure the width, mass, double peak structure and forward-backward asymmetries around the $L_3$ and $R_3$ resonances. A CMS study [71] has investigated in detail the discovery reach for the channel decaying to muons.

### *11.1.5 Weak interactions with a Higgs*

While the effective Lagrangian (11.5) may be obtained as the heavy-Higgs limit of the Standard Model, the reverse construction can also be made: we may introduce a CP-even neutral scalar resonance $h$ and couple it to any of the terms in (11.5) with a-priori arbitrary strength. This case is covered by the resonance studies mentioned above. If we make the particular choice of replacing $\Sigma$ by $(1 + h/v)\Sigma$ in all of its explicit occurences, and add some specific scalar self-couplings, we can apply a nonlinear field redefinition and transform the Chiral Lagrangian coupled to the resonance $h$ into the usual Standard Model. Since nonlinear field redefinitions, with proper renormalizations of higher-dimensional counterterms, do not affect the $S$-matrix, one thus proves that the Chiral Lagrangian coupled to a scalar in a specific way is exactly equivalent to the Standard Model. This also means that a fictitious heavy Higgs boson is equivalent to a formal cutoff for the extra loop divergences and, conversely, a cutoff can be interpreted as an effective Higgs mass.

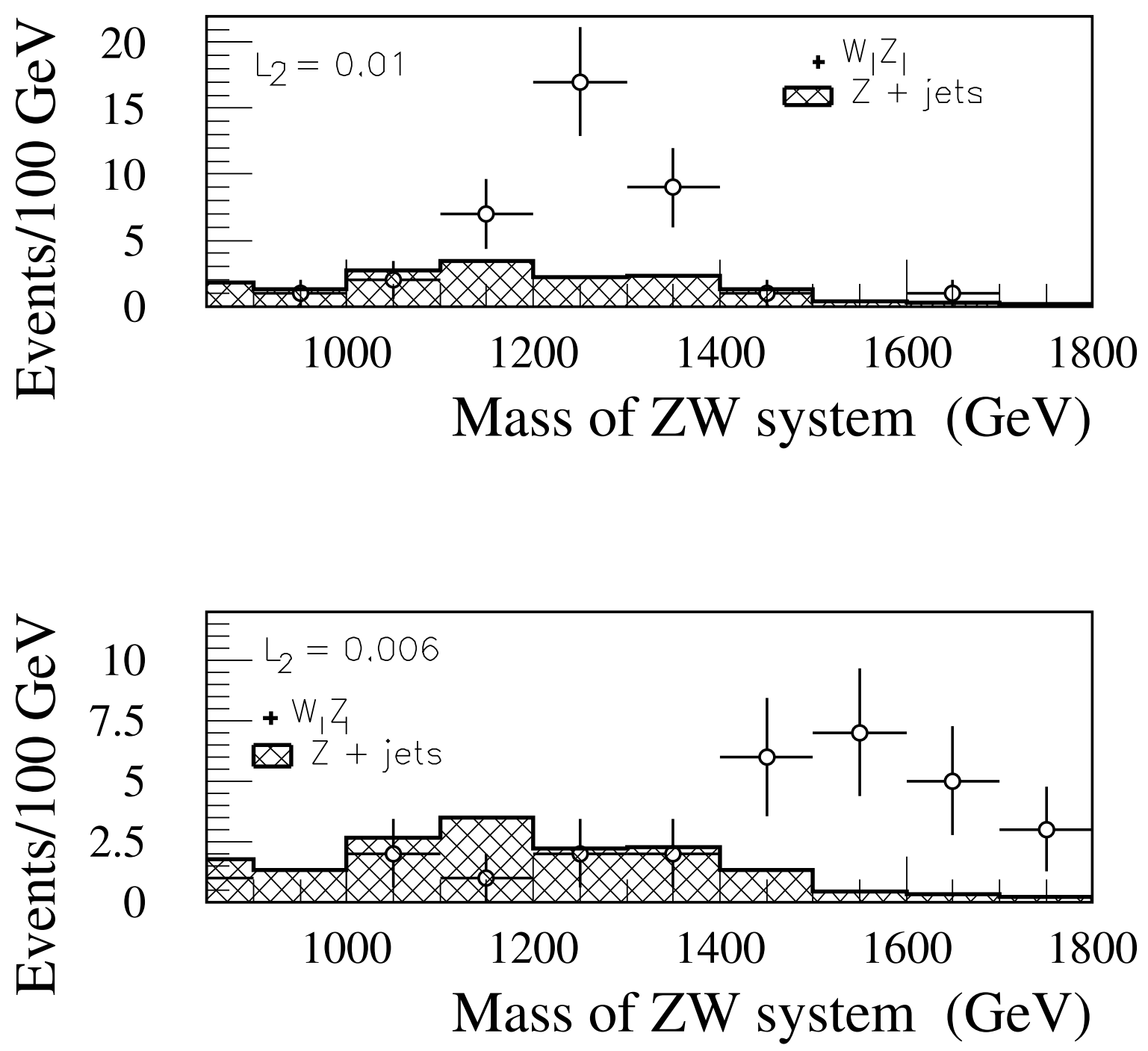

Fig. 11.2: Reconstructed mass distribution the $WZ \to jj\ \ell\ell$ system for different assumed resonance masses, and for an integrated luminosity of 300 fb$^{-1}$. The background shown is $Z$ + jets. Fast detector simulation was used. (from ref [60])

If the Higgs state is sufficiently light such that the linear realization of the Chiral Lagrangian appears appropriate as a theoretical framework, we should revise the power-counting implicit in our previous discussion. After the appropriate nonlinear redefinition of $h$ and $\mathbf{w}$ that turns a generic parameterization into the linear representation, we may consider the resulting matrix field

$$H = v(1 + h/v)\Sigma = (v + h')\mathbf{1} - i\mathbf{w}' \tag{11.13}$$

as a canonical scalar multiplet with mass dimension 1 and vacuum expectation value $\langle H \rangle = v\mathbf{1}$ that can alternatively be decomposed in terms of column vectors $\tilde{\Phi}, \Phi$ with $\tilde{\Phi} = i\tau^2\Phi^*$:

$$H = \begin{pmatrix} v + h - iz & -i\sqrt{2}w^+ \\ -i\sqrt{2}w^- & v + h + iz \end{pmatrix} = \sqrt{2}\begin{pmatrix}\tilde{\Phi} & \Phi\end{pmatrix}. \tag{11.14}$$

Rewriting everything in terms of the complex doublet $\Phi$, the lowest-order terms of the Chiral Lagrangian turn into the textbook expression for the SM Lagrangian.

There is no reason to discard the possibility of anomalous couplings in this linear realization, where the dimensionality of the operators is modified. Eliminating factors $\Sigma^\dagger\Sigma = 1$ as far as possible before the transformation to the linear representation is made, in (11.5) we thus assign dimension 6 to the operators multiplying $\beta'$ and $M_{N_L}$, dimension 4 to the kinetic energy of $\Sigma$, replaced by $H$, while only the operator multiplying $M_{N_R}$ retains dimension 2. In the list (11.7–11.12), the operators $\mathcal{L}_{1,2,3}$ are assigned dimension-6, $\mathcal{L}_{4,5,9,11}$ become dimension-8, and the others acquire even higher canonical dimension. Furthermore, at the level of dimension-6 operators we have to include four-fermion operators and additional terms that involve vector bosons and the Higgs field; various operator bases and detailed discussions can be found in Refs. [72–79].

Conventionally, the effective interactions are parameterized as

$$\mathcal{L}_{\text{eff}} = \sum_n \frac{f_n}{\Lambda^2} \mathcal{O}_n , \qquad (11.15)$$

where $f_n$'s are dimensionless "anomalous couplings", and $\mathcal{O}_n$ the gauge-invariant dimension-6 operators, constructed from the SM fields. If $\Lambda$ appropriately parameterizes the new physics scale (such as the mass of the next resonance), then one would expect $f_n$'s to be of the order of unity. The anomalous couplings of the Higgs boson and gauge bosons are of special interest since they may be directly related to the mechanism of EWSB. A possible parameterization of the $C$ and $P$ conserving dimension-6 effective operators of our current interests involving the $SU(2)$ gauge field $W^i_\mu$, the $U(1)$ gauge field $B_\mu$ as well as a Higgs doublet $h$ is given by

$$\mathcal{O}_{BW} = \Phi^\dagger \hat{B}_{\mu\nu} \hat{W}^{\mu\nu} \Phi, \qquad \mathcal{O}_{\Phi,1} = (D_\mu \Phi)^\dagger \, \Phi^\dagger \Phi \, (D^\mu \Phi) , \qquad (11.16)$$

$$\mathcal{O}_{\Phi,2} = \frac{1}{2} \partial^\mu \left(\Phi^\dagger \Phi\right) \partial_\mu \left(\Phi^\dagger \Phi\right), \qquad \mathcal{O}_{\Phi,3} = \frac{1}{3} \left(\Phi^\dagger \Phi\right)^3 , \qquad (11.17)$$

$$\mathcal{O}_W = (D_\mu \Phi)^\dagger \hat{W}^{\mu\nu} (D_\nu \Phi), \qquad \mathcal{O}_B = (D_\mu \Phi)^\dagger \hat{B}^{\mu\nu} (D_\nu \Phi),$$

$$\mathcal{O}_{WW} = \Phi^\dagger \hat{W}_{\mu\nu} \hat{W}^{\mu\nu} \Phi, \qquad \mathcal{O}_{BB} = \Phi^\dagger \hat{B}_{\mu\nu} \hat{B}^{\mu\nu} \Phi, \qquad (11.18)$$

where $\hat{B}_{\mu\nu}$ and $\hat{W}_{\mu\nu}$ stand for

$$\hat{B}_{\mu\nu} = i \frac{g'}{2} B_{\mu\nu}, \qquad \hat{W}_{\mu\nu} = i \frac{g}{2} \tau^a W^a_{\mu\nu},$$

in which $g$ and $g'$ are the $SU(2)$ and $U(1)$ gauge couplings, respectively.

Precision electroweak data and the measurements of the triple-gauge-boson couplings give considerable constraints on some of the anomalous couplings $f_n/\Lambda^2$ [78–80]. For instance, the oblique correction parameters $S$ and $T$ [19] give rise to constraints on the anomalous coupling constants $f_{BW}$ and $f_{\Phi,1}$ in Eq. (11.16). Assuming either $f_{BW}$ or $f_{\Phi,1}$ dominance, the $2\sigma$ constraints obtained are quite strong [80],

$$-0.07 < \frac{f_{BW}}{(\Lambda/\text{TeV})^2} < 0.04, \quad -0.02 < \frac{f_{\Phi,1}}{(\Lambda/\text{TeV})^2} < 0.02.$$

At the LHC, considerably improved bounds on anomalous triple gauge boson couplings can be expected [24], with 30 fb$^{-1}$. Systematic errors arise from higher order QCD contributions to vector boson pair production [81], and from uncertainties in parton distibution functions. These couplings, expressed as energy dependent form factors in order to safe-guard unitarity, are related [76] to $f_W, f_B$ given above, as well as $f_{WWW}$. A study of the LHC sensitiviy to anomalous quartic couplings is in progress [82]

The next two operators in Eq. (11.17) are purely Higgs boson self-interactions, and lead to corrections to the Higgs triple and quartic vertices. They have been dedicatedly studied in [79] at linear colliders and we will not pursue them further. However, the present experimental observables are not sensitive to the four anomalous coupling operators $O_W$, $O_{WW}$, $O_B$ and $O_{BB}$ (with anomalous coupling constants $f_W/\Lambda^2$, $f_{WW}/\Lambda^2$, $f_B/\Lambda^2$ and $f_{BB}/\Lambda^2$) in Eq. (11.18). The constraints from the existing experiments and the requirement of unitarity of the $S$ matrix element on these four anomalous coupling constants are rather weak. We summarize the above constraints on those four anomalous couplings in Table 11.1. The results are obtained by assuming that only one anomalous coupling is nonzero at a time.

Table 11.1: Current $2\sigma$ constraints on $f_n/\Lambda^2$ from existing studies. The results are obtained by assuming that only one anomalous coupling is nonzero at a time.

| Constraints from | $f_n/\Lambda^2$ in TeV$^{-2}$ |
|---|---|
| Precision EW fit [80]: | $-6 \leq \frac{f_W}{\Lambda^2} \leq 5$ |
| | $4.2 \leq \frac{f_B}{\Lambda^2} \leq 2.0$ |
| | $-5.0 \leq \frac{f_{WW}}{\Lambda^2} \leq 5.6$ |
| | $17 \leq \frac{f_{BB}}{\Lambda^2} \leq 20$ |
| Triple gauge couplings [78] | $-31 \leq \frac{f_W+f_B}{\Lambda^2} \leq 68$ |
| LEP2 Higgs searches [83]: | $-7.5 \leq \frac{f_{WW(BB)}}{\Lambda^2} \leq 18$ |
| Unitarity (at $\sqrt{s}$=2 TeV) [84, 85]: | $\lvert\frac{f_B}{\Lambda^2}\rvert \leq 24.5; \quad \lvert\frac{f_W}{\Lambda^2}\rvert \leq 7.8$ |
| | $-160 \leq \lvert\frac{f_{BB}}{\Lambda^2}\rvert \leq 197; \quad \lvert\frac{f_{WW}}{\Lambda^2}\rvert \leq 39.2$ |

It is perhaps more intuitive to express the new operators in terms of couplings of the explicit physical component fields. Taking into account of the mixing between $W^3_\mu$ and $B_\mu$, the effective couplings of the Higgs boson $H$ and the electroweak gauge bosons $V$ ($V = \gamma,\ W^\pm,\ Z$) in Eq. (11.18) can be cast into [78]

$$\begin{aligned}\mathcal{L}^H_{\text{eff}} &= g_{H\gamma\gamma} H A_{\mu\nu} A^{\mu\nu} + g^{(1)}_{HZ\gamma} A_{\mu\nu} Z^\mu \partial^\nu H + g^{(2)}_{HZ\gamma} H A_{\mu\nu} Z^{\mu\nu} + g^{(1)}_{HZZ} Z_{\mu\nu} Z^\mu \partial^\nu H \\ &+ g^{(2)}_{HZZ} H Z_{\mu\nu} Z^{\mu\nu} + g^{(1)}_{HWW} (W^+_{\mu\nu} W^{-\mu} \partial^\nu H + \text{h.c.}) + g^{(2)}_{HWW} H W^+_{\mu\nu} W^{-\mu\nu}, \quad (11.19)\end{aligned}$$

where the anomalous couplings $g_{HVV}$'s (of dimension $-1$) are related to those Lagrangian parameters $f_n$'s by

$$\begin{aligned} g_{H\gamma\gamma} &= -\alpha \frac{s^2(f_{BB}+f_{WW})}{2}, \\ g^{(1)}_{HZ\gamma} &= \alpha \frac{s(f_W - f_B)}{2c}, \qquad g^{(2)}_{HZ\gamma} = \alpha \frac{s[s^2 f_{BB} - c^2 f_{WW}]}{c}, \\ g^{(1)}_{HZZ} &= \alpha \frac{c^2 f_W + s^2 f_B}{2c^2}, \qquad g^{(2)}_{HZZ} = -\alpha \frac{s^4 f_{BB} + c^4 f_{WW}}{2c^2}, \\ g^{(1)}_{HWW} &= \alpha \frac{f_W}{2}, \qquad g^{(2)}_{HWW} = -\alpha f_{WW}, \end{aligned} \quad (11.20)$$

with the weak mixing $s \equiv \sin\theta_W,\ c \equiv \cos\theta_W$ and $\alpha = gM_W/\Lambda^2 \approx 0.053$ TeV$^{-1} \approx 1/(19$ TeV$)$ for $\Lambda = 1$ TeV . Roughly speaking, an order unity coupling of $f_n$ translates to $g^{(i)}_{HVV} \sim 1/(20$ TeV$)$=0.05 TeV$^{-1}$.

Since new physics responsible for the mechanism of the EWSB is more likely to show up with the Higgs couplings to gauge bosons, these couplings should be tested as thoroughly as possible at future high energy colliders. At the LHC the most sensitive constraints on $f_W/\Lambda^2$ and $f_{WW}/\Lambda^2$ will be from the measurement of the gauge-boson scattering $W^+W^+ \to W^+W^+$ [80]. The obtained $2\sigma$ level constraints on these two anomalous couplings are

$$-1.4 \text{ TeV}^{-2} < f_W/\Lambda^2 < 1.2 \text{ TeV}^{-2}, \quad \text{and} \quad 2.2 \text{ TeV}^{-2} \leq f_{WW}/\Lambda^2 < 2.2 \text{ TeV}^{-2}, \quad (11.21)$$

which may reach the parameter regime sensitive to TeV-scale new physics. Those processes are insensitive to $f_B/\Lambda^2$ and $f_{BB}/\Lambda^2$ however. At the $e^+e^-$ linear colliders on the other hand, the anomalous couplings $g^{(1)}_{HZZ}$ and $g^{(2)}_{HZZ}$ can be constrained at the $2\sigma$ sensitivity to $(10^{-3} - 10^{-2})$ TeV$^{-1}$ from the Higgs-strahlung process $e^+e^- \to Z^* \to ZH$ [86–91].

Due to the renormalizability of the dimension-4 part of the SM effective Lagrangian, there is no a-priori limit on the prefactors multiplying the higher-dimensional operators $1/\Lambda^n$: the cutoff $\Lambda$ may be arbitrarily high. However, the naturalness problem of the Higgs self-energy (an uncancelled quadratic divergence that would imply strong fine-tuning for high cutoff $\Lambda$) lets us argue that $\Lambda$ is nevertheless in the TeV range or below, and new particles or interactions should be expected there that induce anomalous couplings at the same level as in the nonlinear Higgs-less representation. From the power-counting in the linear realization we may draw the qualitative conclusions that custodial symmetry is approximately conserved (since $\beta'$ now multiplies a dimension-6 operator), that the Majorana mass parameters of right-handed neutrinos are large ($M_{N_R}$ multiplying a dimension-2 operator), effectively removing right-handed neutrinos from the low-energy spectrum, and that the Majorana mass parameters $M_{N_L}$ of left-handed neutrinos are small (dimension-6). These properties are apparently realized in Nature and therefore provide some support for the light-Higgs hypothesis.

In weakly-interacting extensions of the SM (e.g., Little-Higgs models), some combinations of the anomalous coefficients can be induced at tree-level and are thus suppressed by factors of $v^2/\Lambda^2$, while the rest requires loops and is thus suppressed by additional factors $1/16\pi^2$. In particular, in models where all extra low-lying particles carry a conserved parity quantum number (e.g., the MSSM, Little-Higgs models with T-parity, universal extra dimensions), anomalous couplings have a common suppression of at least $v^2/(16\pi^2\Lambda^2)$ and are very difficult to detect. This could be an explanation of the absence of deviations from the SM in the precision data obtained so far. However, the list of anomalous couplings has not been exhausted by previous measurements, and precise data from LHC and ILC will be necessary to complete the picture.

## 11.2 Anomalous quartic gauge couplings at the ILC

*Michael Beyer, Wolfgang Kilian, Predrag Krstonošić, Klaus Mönig, Jürgen Reuter, Erik Schmidt and Henning Schröder*

A measurement of quasi-elastic vector boson scattering, i.e., of quartic gauge couplings, is clearly the most direct probe of the Higgs mechanism. In this subsection we present a new determination of the sensitivity the ILC can provide for the couplings $\alpha_{4,5,6,7,10}$ that modify the quartic gauge couplings. Furthermore, we translate this sensitivity into the physics reach for new resonances that could be responsible for such anomalous couplings. Details of the study can be found in Ref. [51].

### *11.2.1 Analysis of triple weak-boson production*

We first consider the triple gauge-boson production processes $e^+e^- \to W^+W^-Z$ and $e^+e^- \to ZZZ$. In these processes not all anomalous couplings can be disentangled individually. The process $e^+e^- \to W^+W^-Z$ depends on the $\alpha$ parameters in the two linear combinations $\alpha_4 + \alpha_6$ and $\alpha_5 + \alpha_7$, while the process $e^+e^- \to ZZZ$ depends on the single combination $\alpha_4 + \alpha_5 + 2(\alpha_6 + \alpha_7 + \alpha_{10})$. For the study of triple gauge-boson production we concentrate on $\alpha_4$ and $\alpha_5$ as independent couplings.

The total cross section at $\sqrt{s} = 1000$ GeV as calculated with the event generator `WHIZARD` [92] is given in Table 11.2. For the analysis presented here, we produce SM events corresponding to a luminosity of 1000 fb$^{-1}$. Three-boson events are reconstructed via six (hadronic) jets utilizing the YCLUS jet-finding algorithm with the Durham recombination scheme. The dominant background is due to $t\bar{t} \to b\bar{b}WW \to 6$ jets. We select events with kinematical conditions for a combination of missing energy and transverse momentum and combine jets to form a $W$ or a $Z$ by requiring a window in invariant mass around the nominal mass. Finally, we take the combination that minimizes the deviation from the PDG values and do a kinematical fit of the bosonic momenta to the total energy and momentum. For a binned likelihood fit, we do not consider the bosonic spins, and choose as observables two invariant masses, $M^2_{WZ} = (p_W + p_Z)^2$, $M^2_{WW} = (p_{W^+} + p_{W^-})^2$, and the angle $\theta$ between the $e^-$ beam axis and the direction of the $Z$-boson.

Table 11.2: Cross section for triple boson production at $\sqrt{s} = 1000$ GeV for different initial state polarization. (A) unpolarized, (B) 80%R electrons, and (C) 80%R electrons with 60%L positrons.

| WWZ | | | ZZZ |
|---|---|---|---|
| no pol. | $e^-$pol. | both pol. | no pol. |
| 59.1 fb | 12.3 fb | 5.57 fb | 0.79 fb |

Table 11.3: Sensitivity of $\alpha_4$ and $\alpha_5$ expressed as $1\sigma$ errors. WWZ: two-parameter fit; ZZZ: one-parameter fit; best: best combination of both. Positive ($\sigma^+$) and negative ($\sigma^-$) errors are given separately.

| | | WWZ | | | ZZZ | best |
|---|---|---|---|---|---|---|
| | | no pol. (case A) | $e^-$ pol. (case B) | both pol. (case C) | no pol. | |
| $16\pi^2\Delta\alpha_4$ | $\sigma^+$ | 9.79 | 4.21 | 1.90 | 3.94 | 1.78 |
| | $\sigma^-$ | −4.40 | −3.34 | −1.71 | −3.53 | −1.48 |
| $16\pi^2\Delta\alpha_5$ | $\sigma^+$ | 3.05 | 2.69 | 1.17 | 3.94 | 1.14 |
| | $\sigma^-$ | −7.10 | −6.40 | −2.19 | −3.53 | −1.64 |

Results are shown in Fig. 11.3 and Tab. 11.3. For $WWZ$ we give in Fig. 11.3A the 90% contours for the different polarization cases A, B, and C, and for both beams polarized also the 68% contour. The respective $\Delta\alpha_i$ are given in Tab. 11.3. We find that the sensitivity strongly increases with polarization, cf. the different cases A, B, and C. A best combined fit for triple boson production is given in Fig. 11.3B.

### *11.2.2 Analysis of weak-boson scattering*

In this section we consider those six-fermion processes in $e^+e^-$ and $e^-e^-$ collisions that depend on quartic gauge couplings via quasi-elastic weak-boson scattering subprocesses, i.e., $VV \to VV$, where $V = W^\pm, Z$. We use full six-fermion matrix elements and thus do not rely on simplifications such as the effective $W$ approximation, the Goldstone-boson equivalence theorem, or the narrow-width approximation for vector bosons.

For the simulation we assume a c.m. energy of 1 TeV and a total luminosity of 1000 fb$^{-1}$ in the $e^+e^-$ mode. Beam polarization of 80% for electrons and 40% for positrons is also assumed. Since the six-fermion processes under consideration contain contributions from triple weak-boson production processes ($ZZ$ or $W^+W^-$ with neutrinos of second and third generation as well as a part of $\nu_e\bar{\nu}_e WW(ZZ)$, $e\nu_e WZ$ and $e^+e^-W^+W^-$ final states), there is no distinct separation of signal and background.

The present study extends the previous study [50] which considered a restricted set of channels and parameters. In addition to the backgrounds considered there, we include single weak-boson production in the background simulation for completeness. We take initial-state radiation into account when generating events. For the generation of $t\bar{t}$ events we use `PYTHIA` [93]. The event samples are generated by the multi-purpose event generator `O'Mega/WHIZARD` [92, 94–96], using exact six-fermion tree-level matrix elements. Hadronization is done with `PYTHIA`. We use the `SIMDET` [97] program to produce the detector response of a possible ILC detector.

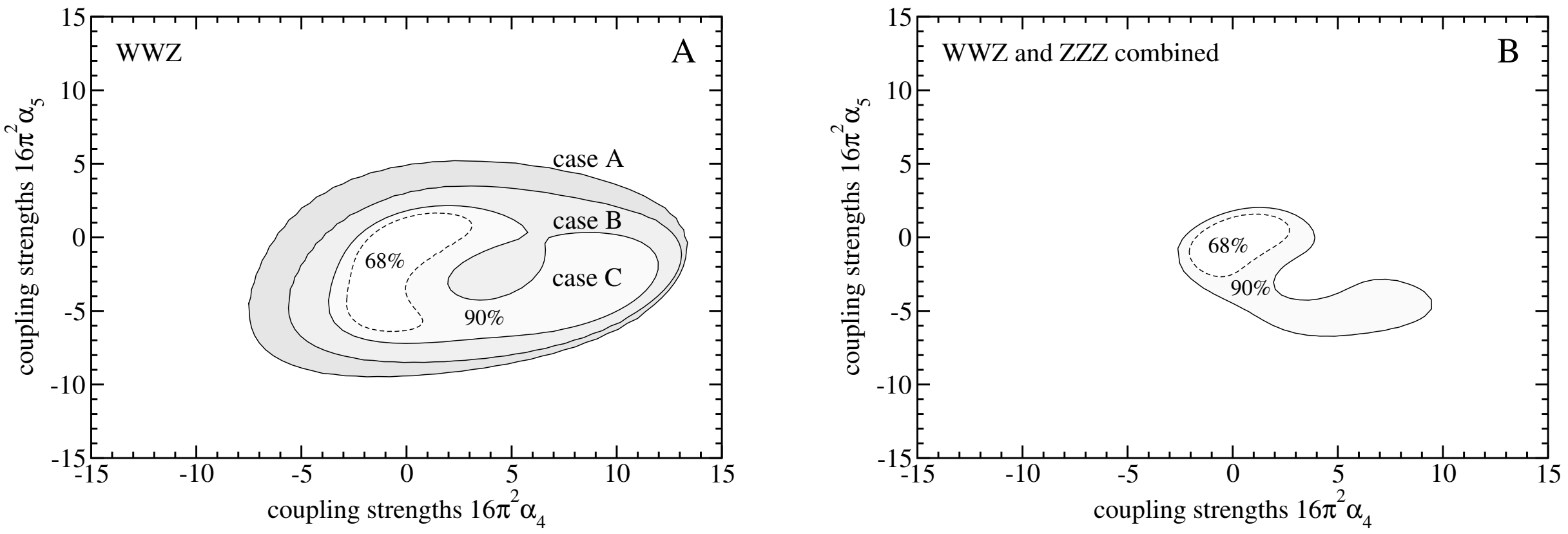

Fig. 11.3: Expected sensitivity for $\alpha_4/\alpha_6$ and $\alpha_5/\alpha_7$ at $\sqrt{s} = 1000\,\text{GeV}$. Luminosity assumption $1000$ fb$^{-1}$. A) $WWZ$-channel only, for an unpolarized beam (A) and the different polarizations cases, $e^-$ only polarized (B) and both beams polarized (C) as explained in the text. Solid lines represent $90\%$ confidence level, the dashed line is for $68\%$, i.e. $\Delta\chi^2 = 2.3$. B) Combined fit using $WWZ$ of case C and $ZZZ$ production. Lines represent $90\%$ (outer line), $68\%$ (inner line) confidence level.

Table 11.4 contains a summary of all generated processes used for analysis and their corresponding cross sections. For pure background processes a full $1\ \text{ab}^{-1}$ sample is generated. All signal processes are generated with higher statistics. Single weak-boson processes and $q\bar{q}$ events are generated with an additional cut on $M(q\bar{q}) > 130$ GeV to reduce the number of generated events.

The event selection is done by a cut-based approach, using again hadronic $W/Z$ decays. The observables sensitive to the quartic couplings are the total cross section (either reduction or increase depending on the interference term in the amplitude and the point in parameter space), and modification of the differential distributions in vector-boson production angle and decay angle. The extraction of quartic gauge couplings from reconstructed kinematic variables is done by a binned likelihood fit. To determine the dependence of the cross section on the parameters within each bin, starting from an unweighted event sample as generated by `WHIZARD`, we use the complete matrix elements encoded in the event generator itself to reweight each event as a function of the quartic gauge couplings. The obtained four-dimensional event distributions are fitted with `MINUIT` [98], maximizing the likelihood as a function of $\alpha_i$,$\alpha_j$.

### *11.2.3 Combined results and resonance interpretation*

In Tables 11.5 and 11.6 we combine our results for the measurement of anomalous electroweak couplings for an integrated luminosity of $1000$ fb$^{-1}$ in the $e^+e^-$ mode, assuming conservation of the custodial symmetry (i.e., $\alpha_{6,7,10} = 0$) and non-conservation, respectively. In Fig. 11.4, the results are displayed in graphical form, projecting the multi-dimensional exclusion region in $\alpha$ space around the reference point $\alpha_i \equiv 0$ onto the two-dimensional subspaces $(\alpha_4, \alpha_5)$ and $(\alpha_6, \alpha_7)$.

In order to get a more intuitive physical interpretation in terms of a new-physics scale, we can transform anomalous couplings into resonance parameters. In each spin/isospin channel we may place a single resonance, one at a time. For each measured value of some $\alpha$ parameter, we may deduce the properties of the resonance that would result in this particular value, assuming that no other contributions to the anomalous couplings are present. Inserting the values that correspond to the sensitivity bound obtained by the experimental analysis, we get a clear picture on the possible sensitivity to resonance-like new physics in the high-energy region.

A resonance in a given scattering channel has two parameters, the mass $M$ and the coupling to this channel. If we are just interested in the sensitivity reach, we have to get rid of the arbitrariness in the coupling. To this end, we first note that the total resonance width does not exceed the mass — otherwise the notion of a resonance is meaningless. To be more specific, we can introduce the ratio of width and mass as a parameter $f \equiv \Gamma/M$. Since the low-energy effect of tree-level resonance exchange is proportional to $f^2$, the ultimate sensitivity of a low-energy measurement can be associated with the possible maximum $f \approx 1$, i.e., a resonance that is as wide as heavy.

Table 11.4: Generated processes and cross sections for signal and background for $\sqrt{s} = 1$ TeV, polarization 80% left for electron and 40% right for positron beam. For each process, those final-state flavor combinations are included that correspond to the indicated signal or background subprocess.

| Process | Subprocess | $\sigma$ [fb] |
|---|---|---|
| $e^+e^- \to \nu_e\bar{\nu}_e q\bar{q}q\bar{q}$ | $W^+W^- \to W^+W^-$ | 23.19 |
| $e^+e^- \to \nu_e\bar{\nu}_e q\bar{q}q\bar{q}$ | $W^+W^- \to ZZ$ | 7.624 |
| $e^+e^- \to \nu\bar{\nu}q\bar{q}q\bar{q}$ | $V \to VVV$ | 9.344 |
| $e^+e^- \to \nu e q\bar{q}q\bar{q}$ | $WZ \to WZ$ | 132.3 |
| $e^+e^- \to e^+e^- q\bar{q}q\bar{q}$ | $ZZ \to ZZ$ | 2.09 |
| $e^+e^- \to e^+e^- q\bar{q}q\bar{q}$ | $ZZ \to W^+W^-$ | 414. |
| $e^+e^- \to b\bar{b}X$ | $e^+e^- \to t\bar{t}$ | 331.768 |
| $e^+e^- \to q\bar{q}q\bar{q}$ | $e^+e^- \to W^+W^-$ | 3560.108 |
| $e^+e^- \to q\bar{q}q\bar{q}$ | $e^+e^- \to ZZ$ | 173.221 |
| $e^+e^- \to e\nu q\bar{q}$ | $e^+e^- \to e\nu W$ | 279.588 |
| $e^+e^- \to e^+e^- q\bar{q}$ | $e^+e^- \to e^+e^- Z$ | 134.935 |
| $e^+e^- \to X$ | $e^+e^- \to q\bar{q}$ | 1637.405 |

Table 11.5: The expected sensitivity from 1000 fb$^{-1}e^+e^-$ sample at 1 TeV in the $SU(2)_c$ conserving case, positive and negative one sigma errors given separately.

| **coupling** | $\sigma-$ | $\sigma+$ |
|---|---|---|
| $16\pi^2\alpha_4$ | -1.41 | 1.38 |
| $16\pi^2\alpha_5$ | -1.16 | 1.09 |

Table 11.6: The expected sensitivity from 1000 fb$^{-1}e^+e^-$ sample at 1 TeV in the broken $SU(2)_c$ case, positive and negative 1 sigma errors given separately.

| **coupling** | $\sigma-$ | $\sigma+$ |
|---|---|---|
| $16\pi^2\alpha_4$ | -2.72 | 2.37 |
| $16\pi^2\alpha_5$ | -2.46 | 2.35 |
| $16\pi^2\alpha_6$ | -3.93 | 5.53 |
| $16\pi^2\alpha_7$ | -3.22 | 3.31 |
| $16\pi^2\alpha_{10}$ | -5.55 | 4.55 |

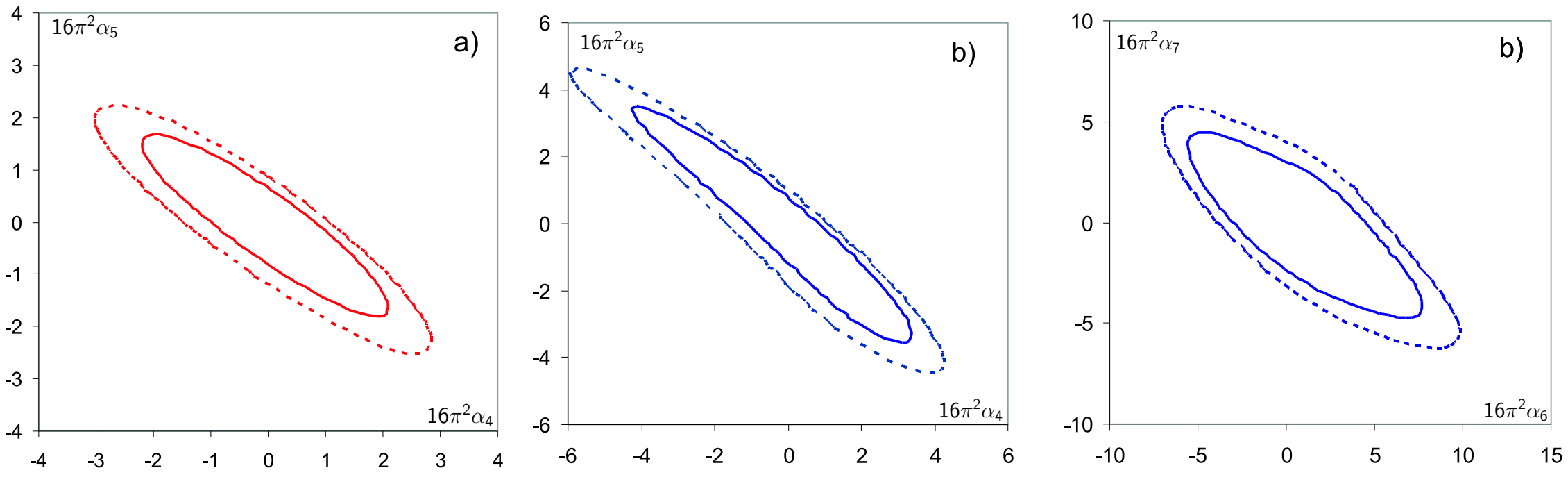

Fig. 11.4: Expected sensitivity (combined fit for all sensitive processes) to quartic anomalous couplings for a 1000 fb$^{-1}$ $e^+e^-$ sample. The full line (inner one) represents 68%, the dotted (outer) one 90% confidence level. a) conserved $SU(2)_c$ case b) broken $SU(2)_c$ case.

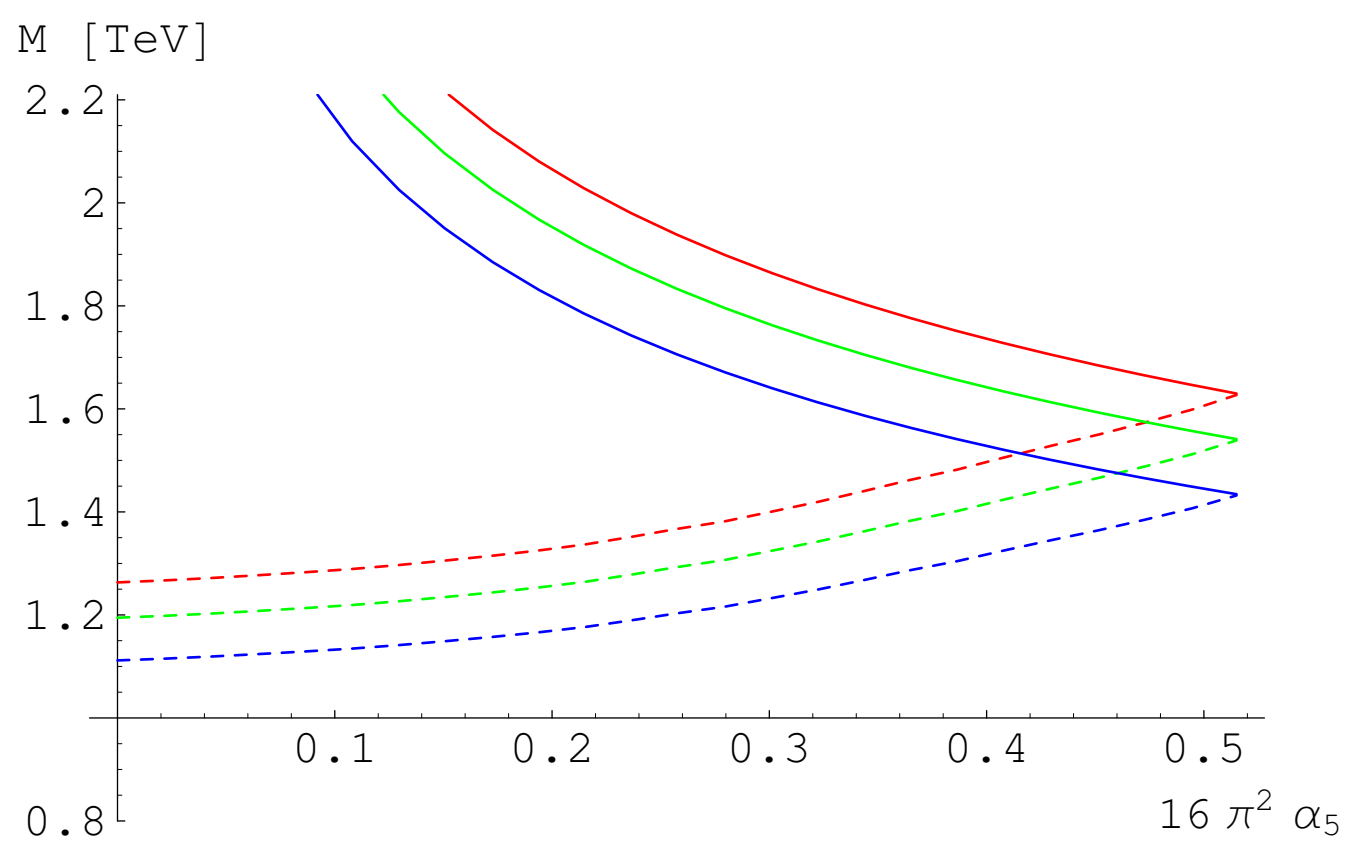

Fig. 11.5: Allowed region for a scalar singlet resonance with isospin breaking as a function of $\alpha_5$ between the upper (full) and lower bound (dashed). Ratio of width to mass of the resonance equal to $1.0$ (top curve, red), $0.8$ (middle curve, green), and $0.6$ (lower curve, blue), respectively.

Table 11.7: Mass reach for the scalar singlet resonance in the case of isospin breaking.

| $f_\sigma = \frac{\Gamma_\sigma}{M_\sigma}$ | 1.0 | 0.8 | 0.6 |
|---|---|---|---|
| $M_\sigma$ [TeV] | 1.39 | 1.32 | 1.23 |

As an example, in Fig. 11.5 we display the allowed mass range for an isosinglet scalar resonance as a function of the measured anomalous coupling $\alpha_5$, assuming no other new physics to be present. Inserting the sensitivity on $\alpha_5$ as obtained from the ILC analysis above, we end up with a mass reach, depending on the resonance width ratio, as listed in Table 11.7.

Similar analyses can be carried out for all possible spin/isospin channels, where for the particular case of vector resonances the results from oblique corrections and triple gauge couplings have to be included in the fit. Detailed results can be found in Ref. [51]. Here, we just quote the final results in Table 11.8.

To conclude, from purely bosonic interactions we find limits for the sensitivity of the ILC in the 1 to 3 TeV range, where the best reach corresponds to the highest-spin channel. These limits are not as striking as possible limits from contact interactions, but agree well with the expected direct-search limits for resonances at the LHC. Note that the selection of purely bosonic interaction depends on the choice of operator basis. In concrete models such as the Technicolor, in a generic basis fermionic couplings of new resonances have to be accounted for (cf. the discussion in [51]), and by including those, better limits can be obtained. Our results thus correspond to the 'worst-case' parameter set where new-physics contributions are minimized.

Performing global fits of all electroweak parameters, analogous to LEP analyses, and combining data from both colliders will be important for disentangling the contributions. Significant knowledge about the mechanism of electroweak symmetry breaking can thus be gained even in scenarios that do not lead to striking new-physics signatures at all.

Table 11.8: Accessible resonance mass in TeV for all possible spin/isospin channels. The results are derived from the analysis of vector-boson scattering processes at the ILC, assuming a single resonance with optimal properties. Left: custodial $SU(2)$ symmetry is assumed to hold; right: no constraints beyond the SM symmetries are assumed.

| Spin | $I = 0$ | $I = 1$ | $I = 2$ |
|---|---|---|---|
| 0 | 1.55 | – | 1.95 |
| 1 | – | 2.49 | – |
| 2 | 3.29 | – | 4.30 |

| Spin | $I = 0$ | $I = 1$ | $I = 2$ |
|---|---|---|---|
| 0 | 1.39 | 1.55 | 1.95 |
| 1 | 1.74 | 2.67 | – |
| 2 | 3.00 | 3.01 | 5.84 |

## 11.3 WW scattering at high WW centre-of-mass energy

*Jonathan M. Butterworth*

Given the issues already discussed in the introduction, it is clear that whatever scenario for electroweak symmetry breaking is realised in nature, the measurement of the vector-boson scattering process $qq \to qqWW \to qqWW$, where "W" implies both charged and neutral vector bosons, is a priority for the LHC experiments. Without a Higgs, the standard model makes no prediction for this cross section above 1.2 TeV; put another way - this cross section is almost entirely determined by the electroweak symmetry-breaking mechanism, and thus is the most model-independent probe of this mechanism.

These processes have been widely studied (see references in the previous section). However, many of these processes focus on Higgs searching. In this case, the mass range of interest is well below 1 TeV, and in addition, since one is searching for a resonance, it is acceptable to look for threshold enhancements in a lepton transverse momentum spectrum where both $W$'s decay leptonically. However, since the cross section is dominated by charged vector bosons, this implies the presence of two neutrinos in the final state, and prohibits an accurate measurement of the $WW$ centre-of-mass energy.

Here we briefly summarise a study [33] of the charged-$WW$ scattering process at $WW$ centre-of-mass energies of 0.6 TeV and above. This study was motivated by the desire to measure the $WW$ cross section as a function of $WW$ centre-of-mass energy as accurately as possible, regardless of the (unknown) structure of the cross section. Therefore the requirement of sufficient statistics at the extreme kinematic limit, as well as the requirement that a maximum of one neutrino be in the final state, and the control of QCD backgrounds, drives the study toward measuring the cross section where one $W$ decays leptonically and the other decays hadronically.

The high energy implies that the decaying $W$'s are very highly boosted. This aids reconstruction of the neutrino kinematics, but implies that the hadronically decaying $W$ is generally reconstructed as a single jet. The principal backgrounds are $W$+jet production where the jet fakes a hadronic $W$, and $t\bar{t}$ production.

One interesting feature of the study was the use of the $k_T$ jet algorithm [99] to reconstruct the jets. This is theoretically preferred to simple cone algorithms, meaning that comparisons to predictions can eventually be made with greater confidence. This is in part because it mirrors the structure of the QCD cascade itself. This property can be exploited in identifying the hadronically-decaying $W$, since in this case the highest $k_T$ "splitting" of the jet is expected to be the $W$ decay, i.e. at a characteristic scale of $M_W$, whereas for the QCD jet in the $W$+jet process, it is a gluon radiation which will typically be at a scale much lower than the $p_T$ of the jet. By decomposing the $W$-candidate jet into subjets in the $k_T$ algorithm, this splitting scale may be evaluated and used to suppress the $W$+jet background. This is similar in principle to rerunning the cone algorithm with a smaller cone, as was done in previous studies; however, the $k_T$ approach is better theoretically controlled, less ambiguous and is invariant under boosts along the $W$ direction. The latter property is particularly important when the desire is to study the shape of the cross section as a function of the $WW$ centre-of-mass energy. This technique is in fact generally applicable to any highly boosted massive particle decaying to hadrons. We note that the $k_T$ algorithm is available in a standard implementation, used by the experiments and theorists alike [100], and a recent reimplementation improves the speed dramatically [101], a factor which was previously a limitation on its use.

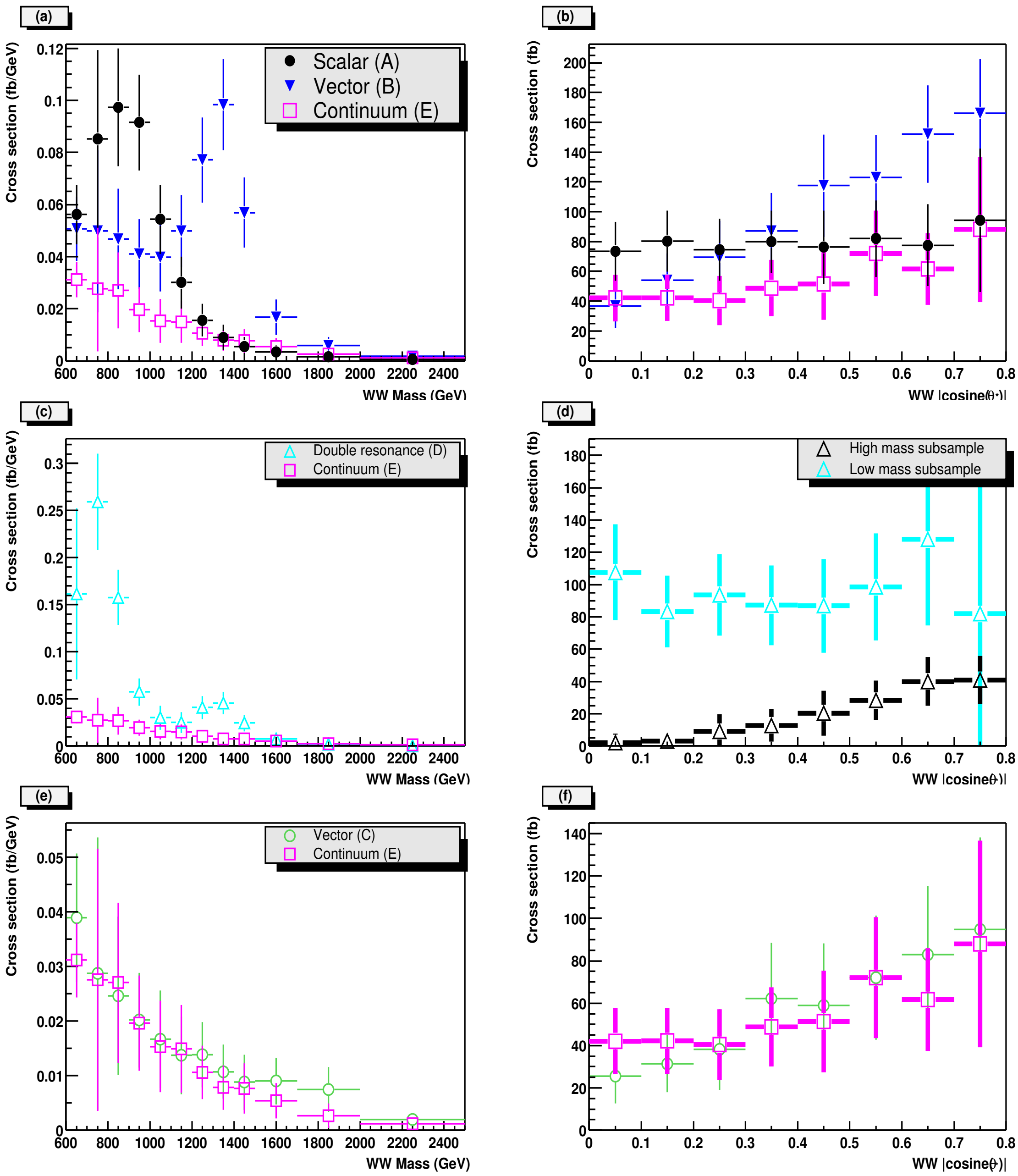

Fig. 11.6: Measurement expectation after 100 fb$^{-1}$ of LHC luminosity at 14 TeV cm energy. (a,c,e) $d\sigma/dM_{WW}$ and (b,d,f) $d\sigma/d|\cos\theta^{*}|$. (d) shows $d\sigma/d|\cos\theta^{*}|$ for the high and low mass subsamples for the double resonance model, separated by a cut at 1200 GeV. (Figure taken from [33]).

Other new features of the analysis included a top quark veto and a cut on the transverse momentum of the hard subsystem. In addition, the established tag jet and minijet veto cuts were applied.

Five different physics scenarios, representative of the different types of physics which might reasonably be expected at the LHC, were studied. The effect of uncertainties in the underlying event leads to a model dependent systematic error of 40-50%. The results compare very well with previous Higgs search studies in the semi-leptonic channel. Over a wide range of parameter space signal/background ratios of greater than unity can be obtained, and the cross-section can be measured differentially in the $WW$ centre-of-mass energy within one year of high luminosity LHC running (100 fb$^{-1}$). Vector and scalar resonances up to around 1.5 TeV may well be observable, and their spins measureable. Figure 11.6 shows the simulated measurements, after background subtraction, estimated for 100 fb$^{-1}$ of LHC luminosity at 14 TeV cm energy.

These studies are currently being repeated and updated using a simulation of the ATLAS detector [102–104]. Early indications are that the conclusions are robust against detector effects, but clearly much more detailed work is needed to realise these measurements in LHC data.

## 11.4 VV-fusion in CMS: a model-independent way to investigate EWSB

*Elena Accomando, Nicola Amapane, Alessandro Ballestrero, Aissa Belhouari, Riccardo Bellan, Giuseppe Bevilacqua, Sara Bolognesi, Gianluca Cerminara, Vladimir Kashkan, Ezio Maina and Chiara Mariotti*

Vector boson scattering is the reaction of choice to probe the nature of Electroweak Symmetry Breaking (EWSB), for which the Standard Model (SM) provides the simplest and most economical explanation. In the SM the Higgs particle is essential to the renormalizability of the theory and is also crucial to ensure that perturbative unitarity bounds are not violated in high energy reactions. Scattering processes between longitudinally polarized on shell vector bosons ($V_L$) are particularly sensitive in this regard. Without a Higgs the $V_L$'s interact strongly at high energy, violating perturbative unitarity at about one TeV. If, on the contrary, a relatively light Higgs exists then they are weakly coupled [105]. In the strong scattering case one is led to expect the presence of resonances in $V_L V_L$ interactions. Unfortunately the mass, spin and even number of these resonances are not uniquely determined [33]. If a Higgs particle is discovered it will nonetheless be necessary to verify that indeed longitudinally polarized vector bosons are weakly coupled at high energy by studying boson boson scattering in full detail. Studying the large mass region of boson-boson scattering could provide an alternative method to determine the Higgs mass range. This could be very useful in case of a light Higgs which will require several years of data taking for a reliable discovery.

In the absence of firm predictions in the strong scattering regime, trying to gauge the possibilities of discovering signals of new physics at the LHC requires the somewhat arbitrary definition of a model of $V_L V_L$ scattering beyond the boundaries of the SM. The simplest approach is to consider the SM in the presence of a very heavy Higgs. While this entails the violation of perturbative unitarity, the linear rise of the cross section with the invariant mass squared in the hard $VV$ scattering will be swamped by the decrease of the parton luminosities at large momentum fractions and, as a consequence, will be particularly challenging to detect. At the LHC, the offshellness of the incoming vector bosons will further increase the difference between the expectations based on the behaviour of on shell $VV$ scattering and the actual results. For $M_H$ >10 TeV, all Born diagrams with Higgs propagators become completely negligible in the Unitary gauge, and all expectations reduce to those in the $M_H \to \infty$ limit. Since this limit is gauge invariant, the results for the no Higgs case presented in the following do not depend on the gauge choice.

At the LHC no beam of on shell EW bosons will be available. Incoming quarks will emit spacelike virtual bosons which will then scatter among themselves and finally decay. All previous studies of boson boson scattering at high energy hadron colliders, with the exception of Refs. [64, 106], have resorted to some approximation, either the Equivalent Vector Boson Approximation (EVBA), or a production times decay approach. There are however issues that cannot be tackled without a full six fermion calculation like exact spin correlations between the decays of different heavy particles, the effect of the non resonant background, the relevance of the offshellness of boson decays, the question of interferences between different subamplitudes.

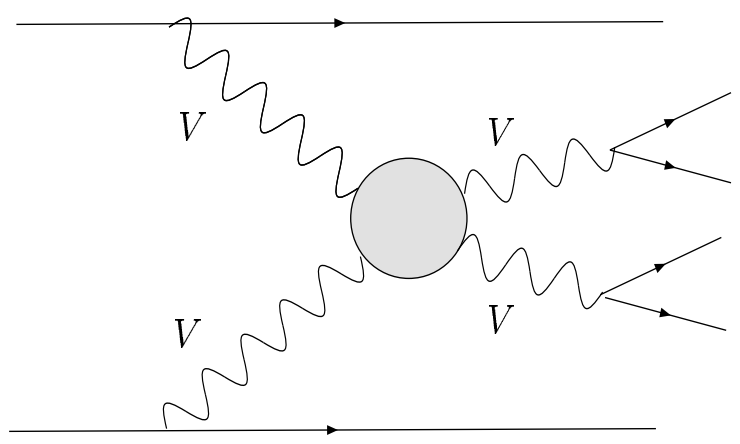

Fig. 11.7: The general diagram of vector boson fusion processes.

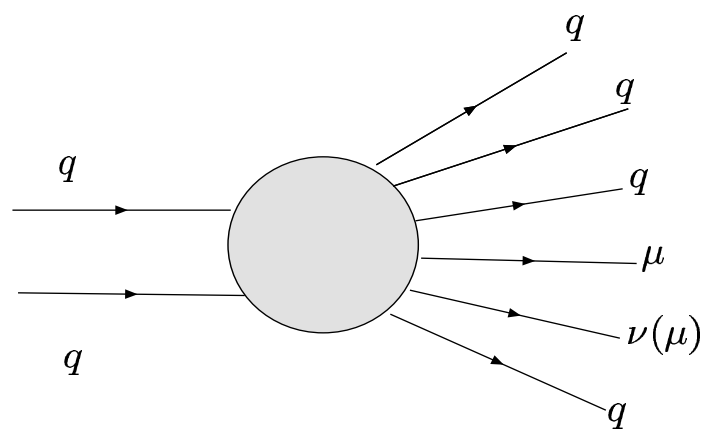

Fig. 11.8: The $qq \to qqqq\mu\nu/\mu$ process generic diagram.

### *11.4.1 Semileptonic final states*

At the LHC the large $VV$ invariant mass region can be explored studying the following processes

$$qq \to qqVW \to qqqq\mu\nu$$
$$qq \to qqVZ \to qqqq\mu\mu$$

They offer a clear experimental signature, because of the presence of high $p_T$ muons from the $W$ or $Z$ decay, together with the highest branching ratio among the final states which can be reconstructed in an hadronic environment. In fact boson boson scattering with a totally hadronic final state cannot be isolated from the non resonant QCD background whose cross section is much larger while final states where both bosons decay leptonically have a smaller rate. Moreover in $qqWW \to qq\mu\nu\mu\nu$ processes the $VV$ invariant mass cannot be reconstructed. Up to now, only final states with muons have been considered, but processes with electrons can be used as well. In the future four lepton final state channels will also be studied.

### *11.4.2 Signal definition and simulation*

In order to explore the EWSB mechanism through the analysis of $VV$ scattering processes, a precise knowledge of the cross section $\sigma(qq \to qqVV)$ over the whole $VV$ invariant mass spectrum is essential. The choice of the MC generator is therefore a key aspect of this study. The $qq \to qqqq\mu\nu$ processes have been simulated with PHASE [107, 108] and the $qq \to qqqq\mu\mu$ with PHANTOM which are exact leading order matrix element MC's at order $\alpha_{EW}^6$.

The general diagram of the vector boson fusion process is shown in Fig. 11.7. Once vector bosons are decayed we have a six fermion final state. If the virtuality of the incoming vector bosons is properly taken into account and the outgoing vector bosons are allowed to be off mass shell then the full set of diagrams describing $qq \to qqqqll'$ has to be considered in order to preserve gauge invariance (see Fig. 11.8). This process includes not only boson boson scattering but also all the irreducible backgrounds that interfere with the signal and cannot be computed and simulated separately. They include the production of a $VV$ pair produced without undergoing scattering as well as diagrams describing $t\bar{t}$ and single top electroweak production. Furthermore there can be subprocesses with three outgoing vector bosons from Triple or Quartic Gauge Couplings or from Higgs production via Higgstrahlung. Finally "non resonant" diagrams are considered where only one pair of fermions in the final state comes from a vector boson decay. For a detailed description of all these contributions see [64].

Table 11.9: Standard acceptance cuts applied during event simulation. Here lepton refers only to $l^{\pm}$.

| | |
|---|---|
| E(quarks)$>$ 20 GeV | E(lepton)$>$ 20 GeV |
| $p_T$(quarks)$>$ 10 GeV | $p_T$(lepton)$>$ 10 GeV |
| $\lvert\eta$(quarks)$\rvert < 6.5$ | $\lvert\eta$(lepton)$\rvert < 3$ |
| M(qq)$>$ 20 GeV | M(ll)$>$ 20 GeV |

Table 11.10: Cross sections and percentages of $qq \to 4q\mu\nu$ and $qq \to 4q\mu\mu$ events generated for the signal and the irreducible background.

| | $qq \to qqqq\mu\nu$ | | | | $qq \to qqqq\mu\mu$ | | | |
|---|---|---|---|---|---|---|---|---|
| | no Higgs | | 500 GeV | | no Higgs | | 500 GeV | |
| | $\sigma$ (pb) | perc. | $\sigma$ (pb) | perc. | $\sigma$ (pb) | perc. | $\sigma$ (pb) | perc. |
| total | 0.689 | 100% | 0.718 | 100% | 0.0430 | 100% | 0.0482 | 100% |
| signal | 0.158 | 23% | 0.184 | 26% | 0.0170 | 39% | 0.0213 | 44% |
| top | 0.495 | 72% | 0.494 | 69% | 0.0206 | 48% | 0.0206 | 43% |
| non resonant | 0.020 | 3% | 0.023 | 3% | 0.0039 | 9% | 0.0046 | 10% |
| three bosons | 0.016 | 2% | 0.017 | 2% | 0.0015 | 4% | 0.0017 | 3% |

#### 11.4.2.1 $qqVV$ signal selection at partonic level

In order to comply with the acceptance of the CMS detector and with the CMS trigger requirements, the cuts shown in Table 11.9 have been applied to all events.

With the aim to enhance the contribution of boson boson scattering with respect to the irreducible background and investigate EWSB, additional kinematical cuts have been applied at parton level. Single top and $t\bar{t}$ production are the main backgrounds. They represent about 70% (45%) of the total cross section in the $4q\mu\nu$ ($4q\mu^+\mu^-$) channel. To suppress them, events with a b-quark and two other quarks (with flavour compatible with W decay) are rejected if the invariant mass of these three particles is between 160 and 190 GeV. Analogously, events in which the muon, the neutrino and a b-quark have an invariant mass between 160 and 190 GeV are rejected.

The two leptons have to reconstruct the mass of a $W$ or a $Z$, so their invariant mass is required to be in the range $M_V \pm 10$ GeV. In $VV$ fusion an additional $W$ or a $Z$ decaying hadronically is expected to be present. Therefore events are required to contain two quarks with the correct flavours to be produced in $W$ or $Z$ decay and with an invariant mass of $\pm$ 10 GeV around the central value of the corresponding gauge boson. If more than one combination of two quarks satisfy these requirements, the one closest to the corresponding central mass value is selected. This combination will in the following be assumed to originate from the decay of an EW vector boson. The requirement of at least two reconstructed vector bosons in the final state eliminates about 3% (10%) of the total cross section. Finally one has to reject events with the production of three outgoing vector bosons: if the two remaining quarks have the right flavour to reconstruct a $W$ or a $Z$ boson and if their invariant mass is compatible within 10 GeV with the corresponding boson mass then the event is rejected. This happens in about 2% (3%) of the cases. As shown in [106] the requirement that quark pairs which have masses in the neighborhood of the EW vector boson masses also have the correct flavour content has a very modest impact on our results.

In the following we will refer as ”signal” to the events which pass the selection cuts in Table 11.9 and the additional kinematical cuts mentioned above: top veto and presence of two and only two particle pairs ($ll$ and $jj$) with masses close to the masses of the EW vector bosons. In Table 11.10 we list the cross sections for the signal and the irreducible backgrounds corresponding to the described cuts. In Fig. 11.9, the M($VW$) spectrum is shown.

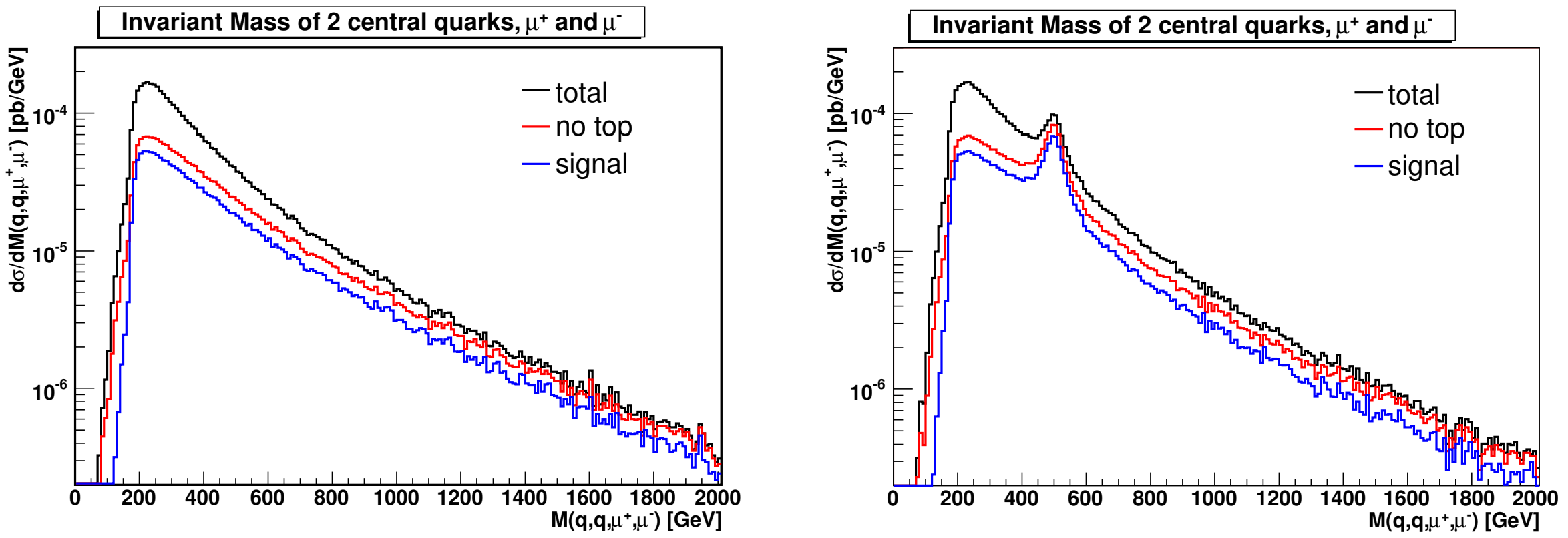

Fig. 11.9: Distributions of the invariant mass of the two central quarks, the muon and the neutrino for the no-Higgs case (left). Distributions of the invariant mass of the two central quarks and the two muons for M(H)=500 GeV (right). The top curve refers to the full sample. The intermediate one shows the effects of antitagging on the top. The lowest line corresponds to imposing all the mentioned cuts and antitagging on the presence of three vector bosons.

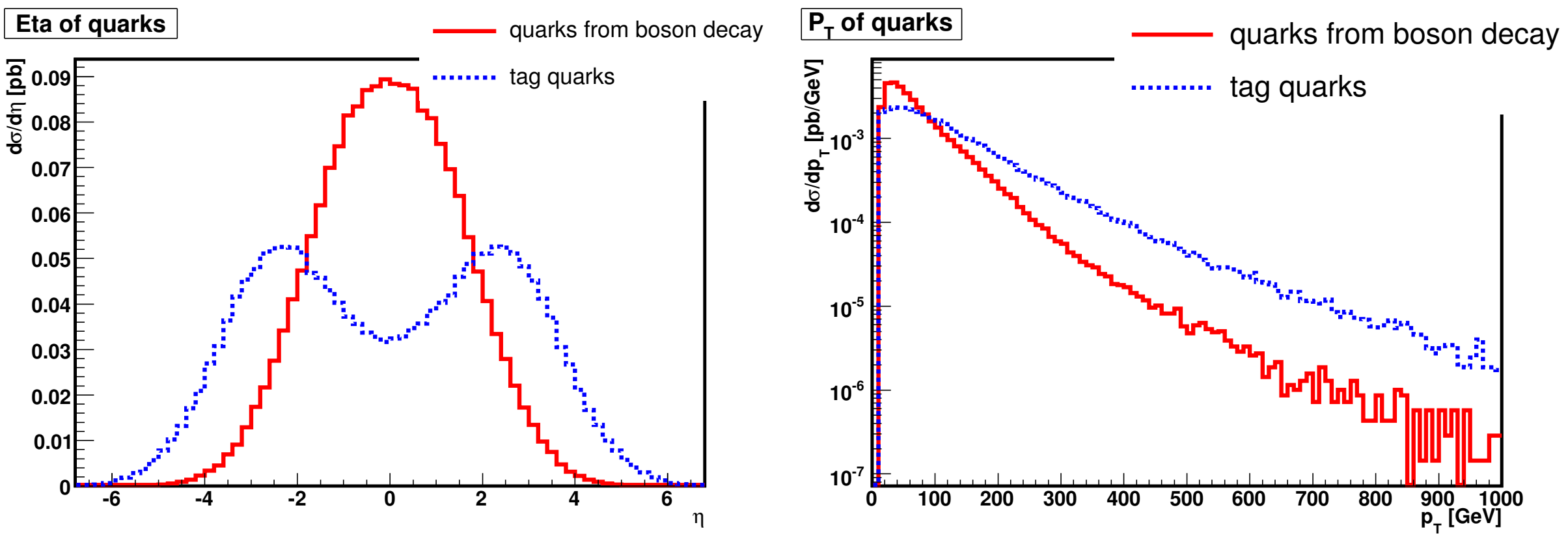

Fig. 11.10: Pseudorapidity and transverse momentum of the two quarks from the boson decay and of the two tag quarks in the no-Higgs scenario for the $4q\mu\nu$ channel.

#### 11.4.2.2 Signal topology

The lepton and the neutrino ($l^+l^-$ pair) in the final state come from the decay of a $W$ ($Z$). They are expected to have a quite high transverse momentum ($p_T$) and to be mostly produced centrally in the detector, i.e. at low absolute value of the pseudo-rapidity ($\eta$). Similarly the two quarks from the decay of the second vector boson. On the contrary, the two quarks which emit the two incoming vector bosons tend to go in the forward/backward region (high $|\eta|$) with very large energy and $p_T$. Thanks to their peculiar kinematical pattern these two spectator quarks are essential to tag the $VV$ fusion events among all six fermions final states, therefore they will also be called "tag quarks". In Fig. 11.10 the kinematics of the two quarks from the boson decay and of the two tag quarks are compared.

It is interesting to look at possible differences in the kinematics of the VV-fusion signal with respect to the irreducible background. In Fig. 11.11 the distributions of some kinematical variables are presented for the signal and for the full sample in case of the $qqqq\mu\nu$ final state. We expect similar results for the $qqq\mu^+\mu^-$ final state as well. In the figure the no-Higgs case is chosen as an example, but there are no significant differences with the case of a visible Higgs. Only some variables, which are connected to the mass of the Higgs boson, show the presence of the resonance. The total invariant mass of the six fermions in the final state is presented for the signal and for the full sample: the signal tends to have a very large final six fermion mass. The muon from the signal has a larger $p_T$ than the one from the background, and the same applies to the spectator quarks. The difference of the $\eta$'s of the tag quarks is also shown: the signal tends to have a larger $\Delta\eta$ with respect to the background.

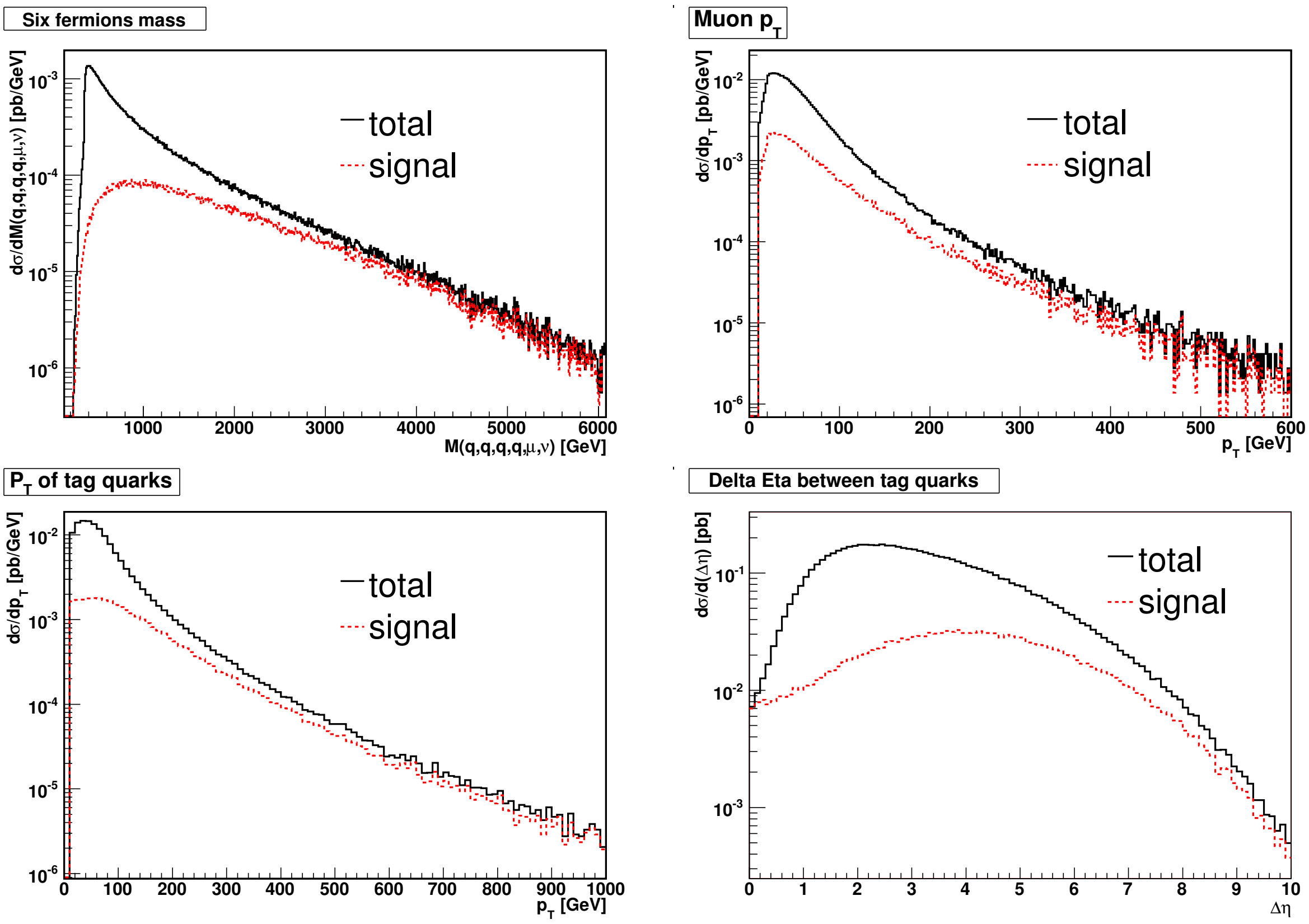

Fig. 11.11: Differential cross section in the $4q\mu\nu$ case as a function of the invariant mass of the six fermions in the final state, the transverse momentum of the muon, the transverse momentum of the two tag quarks, and the difference in pseudorapidity between the two tag quarks for all the events (black) and for the signal events (red). All plots refer to the no-Higgs case.

### 11.4.2.3 Main backgrounds

The most problematic background for the vector boson fusion signal is the production of a single $W$ (or $Z$) in association with n jets (n=1,2,3,4). In this case the outgoing jets come from gluons or quarks produced via QCD processes, so we expect a larger occupancy of the central pseudorapidity region with respect to the pure EW signal process.

$t\bar{t}$ and single top production via QCD processes, e.g. $q\bar{q} \rightarrow t\bar{t}$ and $gg \rightarrow t\bar{t}$, are other backgrounds with very large cross sections. Top quarks decay into a $W$ boson and a b-quark with a branching ratio of almost 100% giving a final state similar to that produced in $VV$ fusion events. However the outgoing b-quarks can be recognized through a b-tagging algorithm and they have a kinematical behaviour quite different from the signal tag quarks: the former are in fact more central and have much lower energy.

Exactly the same final state of the signal $jjVW \rightarrow jjqq\mu\nu$ (or $jjVZ \rightarrow jjqq\mu\mu$) can be pro-duced also at a different perturbative order: $\alpha_S^2 \alpha_{EW}^4$. In this case the two jets not coming from a boson decay are generated from QCD processes so they can come from a quark or a gluon. These background processes can be distinguished from the $VV$ fusion signal thanks to the fact that in the first case the outgoing bosons have very high energy and quite high pseudorapidity so they have lower transverse momentum with respect to the signal bosons that are produced in the central region. Moreover the two outgoing QCD quarks (or gluons) have very high energy, higher than the energy of the signal tag quarks, and their pseudorapidity distribution and their transverse momentum spectrum are intermediate between that of the signal quarks from the boson decay and that of the signal tag quarks.

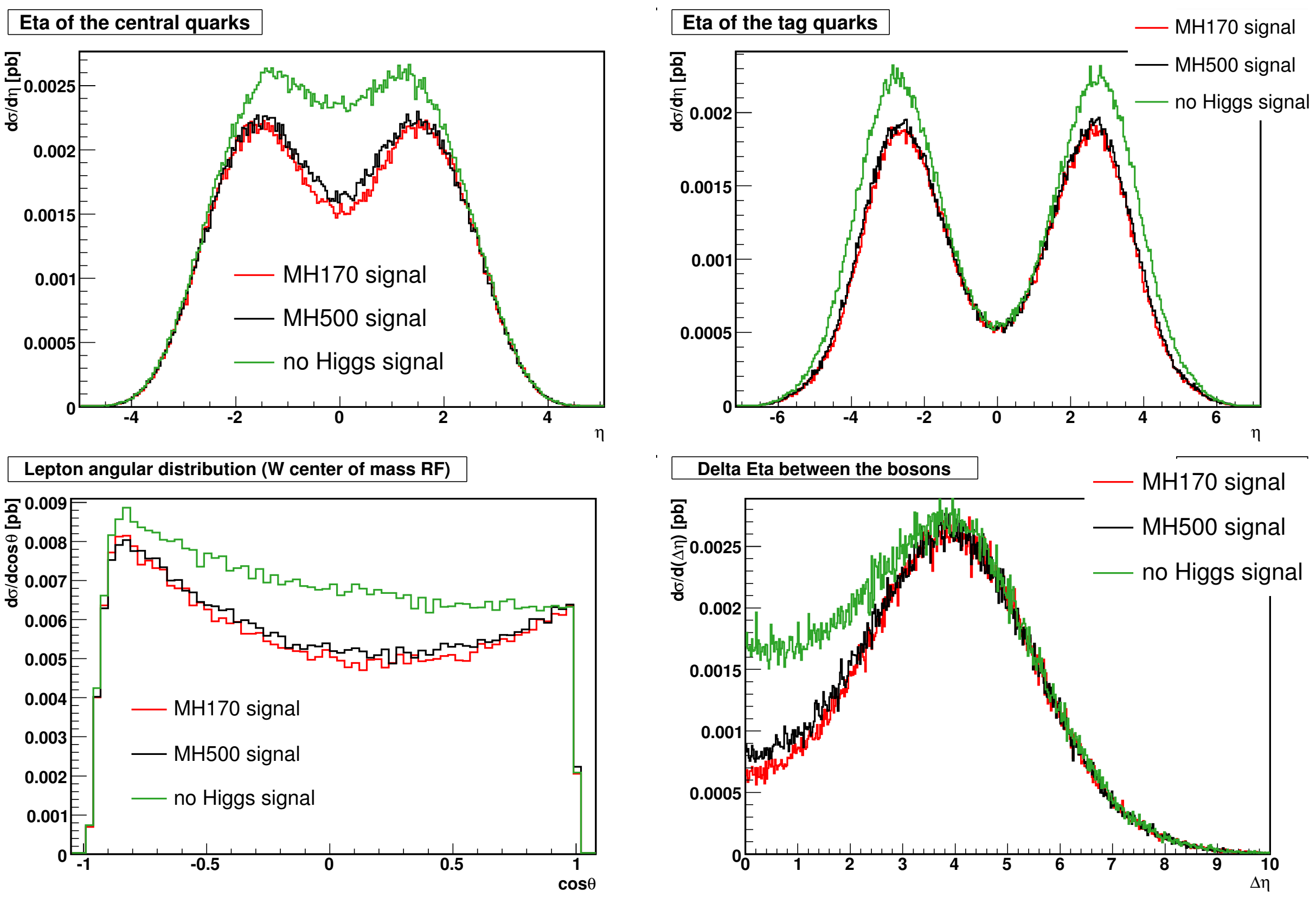

Fig. 11.12: The pseudorapidity $\eta$ of the two central jets, the $\eta$ of the forward quarks, the cosine of the angle between the lepton and the W boson in the W boson rest frame and the $\Delta\eta$ of the two vector bosons for $M(VW) > 800$ GeV in the $4q\mu\nu$ channel. In green for the no-Higgs case, in black for M(H)=500 GeV and in red for M(H)=170 GeV.

Some preliminary studies have been done with the CMS detector fast simulation (refs. [109], [110], [111], [112]). In these references you can find a detailed description of all mentioned backgrounds together with some preliminary strategies to eliminate them and to reconstruct the signal in the CMS detector.

### *11.4.3 The high mass region*

An interesting physics possibility is to investigate whether there exist or not an elementary Higgs boson by measuring the $VW$ cross section at large M($VW$). The rise of the cross section related to unitarity violation in the no-Higgs case is difficult to detect at the LHC, since the center-of-mass energy is still rather low and the decrease of the proton distribution functions at large $x$ has the dominant effect.

As discussed in [64], the unitarity violation is related only to the scattering of longitudinally polar-ized vector bosons. In fact, in presence of a light Higgs, the cross section at high $VW$ invariant masses is due essentially to transversely polarized bosons while in the no Higgs case the cross section the contributions of longitudinally and transversely polarized bosons are comparable. Thus, if it was possible to distinguish $W_LW_L$ from $W_TW_T$ the difference in cross section at high masses between the no Higgs and the light Higgs scenarios would be sizeable. In order to distinguish $W_LW_L$ from $W_TW_T$ we must exploit the different behaviour of the final state in the two cases. Preliminary studies in this direction have been presented in [64].

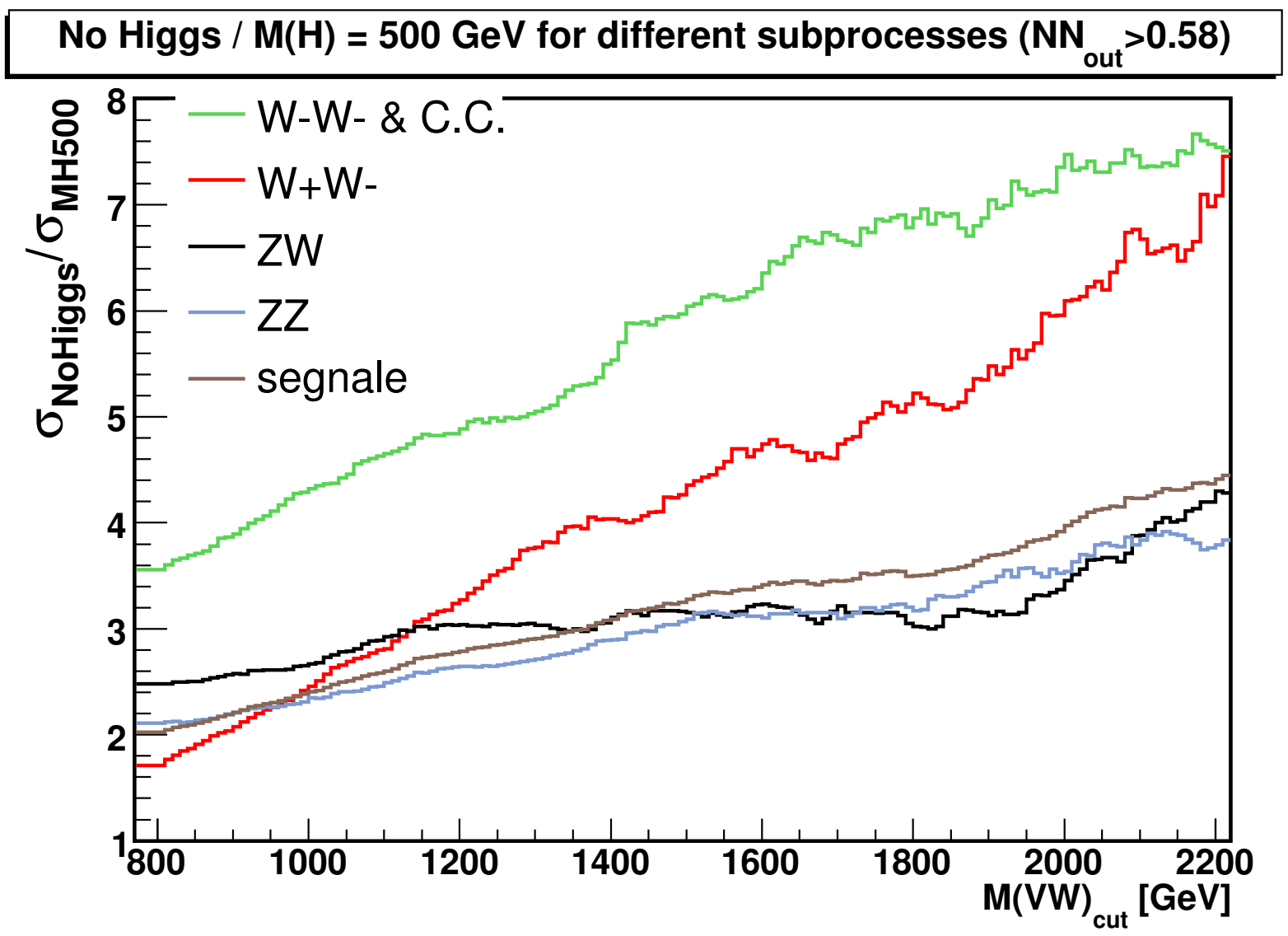

Fig. 11.13: The ratio in Eq.(11.22) for different groups of processes of $4q\mu\nu$ type, for a given cut on NN, the Neural Network output variable.

In Fig. 11.12 several kinematical distributions for M(H)=170 GeV, M(H)=500 GeV and the no-Higgs case have been compared for M($VW$)>800 GeV in the $qqqq\mu\nu$ final state[2]. The variables most sensitive to the different behaviour of the two cases (Higgs and no Higgs) have been selected and then used to train a Neural Network focusing on events in the high mass region (M($VV$)>800 GeV).

Thus the Neural Network becomes able to distinguish the light Higgs from the no Higgs scenario and a significative difference in the integrated cross section in these two cases can be achieved imposing a cut on the Neural Network output. It is also interesting to study how the difference at high invariant masses changes if we consider different processes. In Fig. 11.13 we show the ratio

$$\frac{\displaystyle\int_{M_{cut}}^{\infty} dM_{VW} \frac{d\sigma_{noHiggs}}{dM_{VW}}}{\displaystyle\int_{M_{cut}}^{\infty} dM_{VW} \frac{d\sigma_{M_H=500}}{dM_{VW}}} \tag{11.22}$$

for different groups of processes for a given cut on the Neural Network output (for details see [64]). The set which includes $W^\pm W^\pm \to W^\pm W^\pm$ is the one with the largest separation, while the sets including $ZZ \to W^\pm W^\mp$ and $ZW \to ZW$ scattering show a smaller difference between the Higgs/no–Higgs scenarios.

For the final states $4q\mu^+\mu^-$ simple kinematical cuts have been applied to enhance the difference between the case of Higgs and no Higgs [106]. With the requirement of $|\eta(V)| < 2$ we obtain a difference between the Higgs/no–Higgs case of a factor 2 to 3 with a cross section of 0.4 to 0.04 in the first case and 0.7 to 0.1 fb in the no Higgs scenario (see Fig. 11.14).

[2] Recall that only purely EW processes has been included in this study.

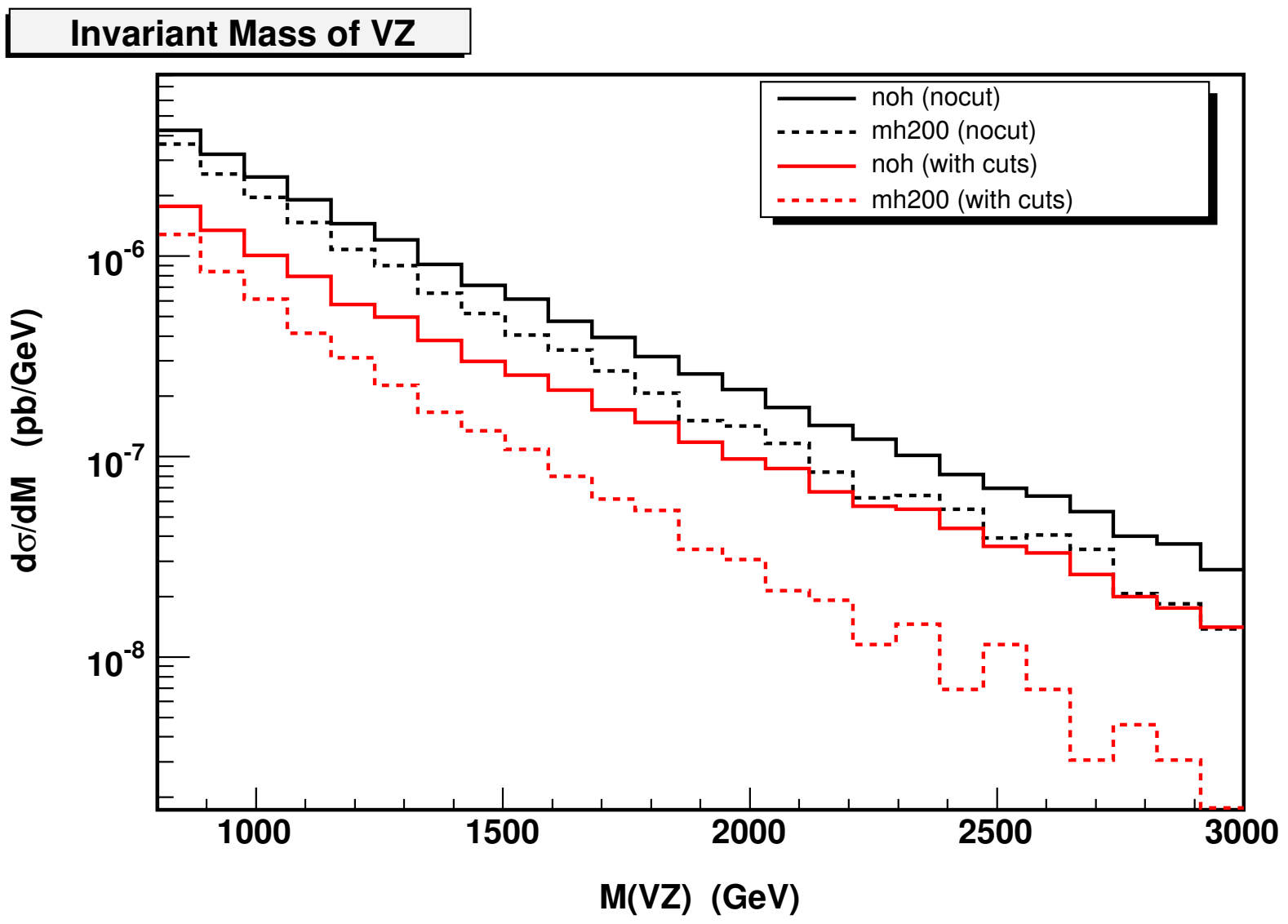

Fig. 11.14: Invariant mass distribution for $M(VZ) > 800$ GeV in the $4q\mu^+\mu^-$ case. The full line refers to the no-Higgs case, the dashed one to M(H)=200 GeV. The upper two curves present the results for our signal definition. For the two lower ones we have further required $|\eta(Z_{ll})| < 2$ and $|\eta(q_V)| < 2$.

### *11.4.4 Summary and future*

We have studied at parton level the processes $qq \to qqqq\mu\nu$ and $qq \to qqqq\mu\mu$ at the LHC using two new MonteCarlo generators `PHASE` and `PHANTOM`. A strategy to enhance boson boson scattering with respect to the irreducible backgrounds has been developed.

The SM predictions in the absence of the Higgs have been chosen as a benchmark scenario for possible signals of new physics in the EWSB sector: a comparison with the standard case of a visible Higgs has been performed focusing on the high M($VV$) region.

The effect of the CMS detector is under study. Preliminary analyses [112], using different generators and an old version of the detector simulation, showed that at CMS a good resolution on the M($VV$) variable can be achieved along with a resonable signal over background ratio. Since the cross section of the process is very small a few years of data taking will be necessary to be able to study these channels.

Channels with four leptons in the final state will be investigated in the near future.

# 12 TECHNICOLOR

## 12.1 Introduction

*Georges Azuelos and Francesco Sannino*

A number of possible generalizations of the standard model have been conceived. Such extensions are introduced on the basis of one or more guiding principles or prejudices.

By invoking the absence of fundamental scalars in Nature one is led to construct theories in which the electroweak symmetry breaks via a fermion bilinear condensate. The Higgs sector of the Standard Model becomes an effective description of a more fundamental fermionic theory. This is similar to the Ginzburg-Landau theory of superconductivity. If the force underlying the fermion condensate driving electroweak symmetry breaking is due to a strongly interacting gauge theory these models are termed technicolor. Here we will discuss the basic models and summarize experimental searches.

Technicolor, in brief, is an additional non-abelian and strongly interacting gauge theory augmented with (techni)fermions transforming under a given representation of the gauge group. The Higgs Lagrangian is replaced by a suitable new fermion sector interacting strongly via a fifth force (technicolor). Schematically:

$$L_{Higgs} \to -\frac{1}{4}F_{\mu\nu}F^{\mu\nu} + i\bar{Q}\gamma_{\mu}D^{\mu}Q \ , \qquad (12.1)$$

where, to be as general as possible, we have left unspecified the underlying nonabelian gauge group and the associated technifermion representation. The characteristic scale of the new theory is expected to be less than or of the order of one TeV. The chiral-flavor symmetries of this theory, as for ordinary QCD, break spontaneously when the technifermion condensate $\bar{Q}Q$ forms. It is possible to choose the fermion charges in such a way that there is, at least, a weak left-handed doublet of technifermions and the associated right-handed one which is a weak singlet. The covariant derivative contains the new gauge field as well as the electroweak ones. The condensate spontaneously breaks the electroweak symmetry down to the electromagnetic and weak interactions. The Higgs is now interpreted as the lightest scalar field with the same quantum numbers of the fermion-antifermion composite field. The Lagrangian part responsible for the mass-generation of the ordinary fermions will also be modified since the Higgs particle is no longer an elementary object.

Models of electroweak symmetry breaking via new strongly interacting theories of technicolor type [1,2] are a mature subject (for recent reviews see [3,4]). One of the main difficulties in constructing such extensions of the standard model is the very limited knowledge about generic strongly interacting theories. This has led theorists to consider specific models of technicolor which resemble ordinary quantum chromodynamics and for which the large body of experimental data at low energies can be directly exported to make predictions at high energies.

To be able to make contact with experiments we need to introduce some of the phenomenological key players of technicolor theories. These are some of the hadronic states of the theory.

As we have already explained above, the techniflavor global symmetries, as for ordinary QCD, break spontaneously and technipions $\pi_T$ will emerge as light states of the theory. Three of them become the longitudinal components of the $W$ and $Z$ gauge bosons. Since the quantum numbers of the remaining technipions are model dependent, it can happen that some carry weak charges and/or ordinary color charges. New sources of techniflavor symmetry breaking are needed to be able to provide large masses to the technipions. Vector mesons are relevant in QCD and their technivector cousins may equally play an important role for the technicolor theories. According to the underlying model, as for technipions, these technihadrons can also experience the color and electroweak force. A mass in the several hundred GeV range is expected.

The situation for the composite Higgs boson is more delicate. According to the common lore generic theories of composite Higgs contain large corrections with respect to the minimal standard model, similar to those of a heavy elementary Higgs boson [5]. However, a heavy composite Higgs is *not* necessarily an outcome of strong dynamics [6–10]. Here we are not referring to models in which the Higgs is a quasi Goldstone boson [11], investigated recently in [12] (see Section 7).

In the analysis of QCD-like technicolor models information on the non-perturbative dynamics at the electroweak scale is obtained by simply scaling up QCD phenomenology to the electroweak energy scale. The Higgs particle is then mapped into the $q\bar{q}$ scalar partner of the pions in QCD. However, the scalars are a complicated sector of QCD (see the PDG - review dedicated to this sector of the theory). There is a growing consensus that the low lying scalar object, i.e. $f_0(600)$, needed to provide a good description of low energy pion pion scattering [13, 14] is not the chiral partner of the pions but is of four quark nature à la Jaffe [15, 16]. Recent arguments, based on taking the 't Hooft limit of a large number of colors $N$, also demonstrate that the low energy scalar is not of $q\bar{q}$ nature [17]. The natural candidate for the scalar partner of the ordinary pions is very heavy, i.e. it has a mass larger than one GeV. When transposed to the electroweak theory by simply taking the scaled pion decay constant $F_\pi$ as the electroweak scale, one concludes that in technicolor theories with QCD-type dynamics the composite Higgs is very heavy, $m_H \sim 4\pi F_\pi$, of the order of the TeV scale. This implies that corrections are needed to compensate the effects of such a heavy Higgs in order not to be at odds with the electroweak precision measurements data. The presence of a heavy Higgs, however, does not exclude the possibility to observe a lighter and very broad techniscalar below the TeV region constructed with more than two technifermions in analogy with the $f_0(600)$ of QCD. For strongly interacting theories with non-QCD-like dynamics we are no longer guaranteed that the associated composite Higgs particle is heavy.

To generate standard model fermion masses in technicolor theories additional interactions are needed. Extended technicolor (ETC) models (Section 12.2), which couple technifermions to ordinary fermions [18], are an example of such interactions. Typically one imagines a very large gauge group in which color, flavor and technicolor are embedded simultaneously. Such a gauge group must then break to ordinary color and technicolor. The breaking of the ETC gauge group provides also masses to the ETC gauge bosons which are not part of the technicolor and color interactions. ETC massive gauge bosons and associated dynamics can lead to:

- a mass term for the standard model fermions,
- provide sufficiently large masses for some of the dangerously light technipions,
- technipions decaying into ordinary fermion pairs,
- couplings between fermions of different generations and hence to flavor-changing neutral currents.

Experimental constraints on interactions mediating flavor-changing neutral currents are obtained via the $K^0\ \bar{K}^0$ system [18]. These constraints for technicolor theories have been re-analyzed in [19] and found to be less restrictive. It turns out that the scale of ETC breaking must be of the order of hundreds to thousands of GeV. Such a high scale for the ETC interactions leads to very light technipions and quarks. This problem can be alleviated if the technifermion bilinear, whose condensate breaks the electroweak symmetry, can be dynamically enhanced. This is possible in theories in which the technicolor gauge coupling as function of the renormalization scale *walks* rather than run [20–24]. In Fig. 12.1 we provide a sketch of a typical walking coupling constant versus a standard running one. The enhancement of the condensate allows reasonable masses for light quarks and leptons, even for large ETC scales necessary to suppress flavor-changing neutral currents sufficiently well. However, to obtain the observed top mass, one can use the ETC type models presented in [19], or one must rely on additional dynamics, as in so-called non-commuting ETC models, where the ETC interaction does not commute with the electroweak interaction [25]. This last mechanism is similar in spirit to the topcolor assisted technicolor models [26, 27] which will be introduced later in the text.

Most of the models used in the literature have considered the technifermions in the fundamental representation of the gauge group. In this case one needs a very large number of technifermions, roughly of the order of $4N_{TC}$ with $N_{TC}$ the number of technicolors, to achieve walking [28–31]. It is, in general, hard to reliably compute physical quantities in walking theories [4]. However attempts have been made in the literature to provide an estimate of the S parameter [32]. One expects a reduced S parameter with respect to the naive one computed in perturbation theory [32–34].

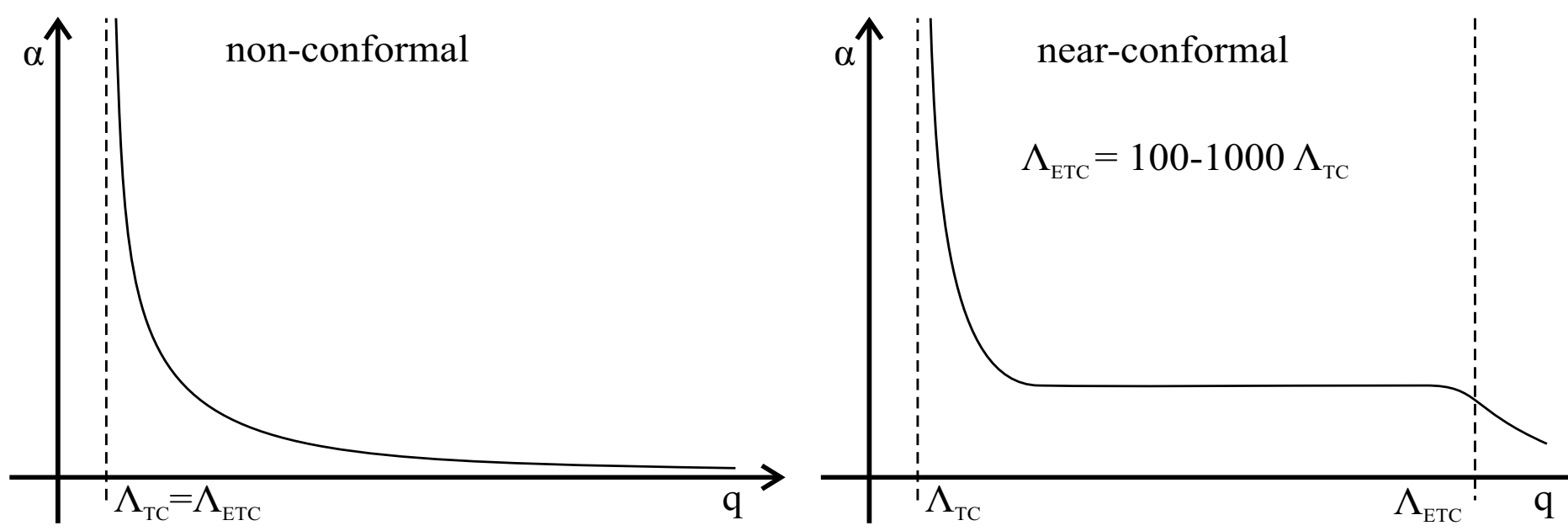

Fig. 12.1: Left Panel: A standard running behavior of a coupling constant in a generic asymptotically free theory. Right Panel: The walking behavior of the coupling constant when the number of flavors is near a conformal fixed point.

Some of the previous problems can be ameliorated [8–10, 35] if one considers technifermions in higher dimensional representations of the technicolor gauge group. Here one achieves walking for a very small number of technifermions. Technicolor-like theories with fermions in higher dimensional representations of the gauge group have also been considered in the past [36–38]. For the walking type theories introduced in [8] it is argued that: i) The S-parameter is naturally small [9, 10]; ii) One has zero or a very small number of technipions [8]. A possible feature of these theories is that the resulting composite Higgs can be light with a mass of the order of 150 GeV. The phenomenology of these theories leads to interesting signatures as shown in Section 12.5.

It is instructive to examine the constraints from new precision measurements on the model presented in [10]. The model consists of two techniflavors in the adjoint representation of a two technicolor theory. A new lepton family is needed to have a consistent theory while the hypercharge needs not be the one of the standard model. In Fig. 12.2 the ellipse corresponds to the one sigma contour in the T$-$S plane. The central values for S$= +0.07 \pm 0.10$ and $T = +0.13 \pm 0.10$ have been taken from reference [39]. The black area bounded by parabolas corresponds to the region in the T$-$S plane obtained when varying the relative Dirac masses of the two new leptons. The point at T$= 0$ where the inner parabola meets the S axis corresponds to the contribution due solely to the technicolor theory. The electroweak parameters are computed perturbatively. Fortunately for walking technicolor theories the nonperturbative corrections further reduce the S parameter contribution [32, 33] and hence our estimates are expected to be rather conservative.

The figure clearly shows that the walking technicolor type theories are still viable models for dynamical breaking of the electroweak symmetry [40].

We stress that the S parameter problem, per se, can be alleviated in different ways [3, 41], also for walking technicolor theories with technifermions in the fundamental representation of the gauge group [42].

Although the ETC problem it not yet solved, some progress in building ETC models has been made in [19, 35] also with respect to the problem of generating neutrino masses [43]. The improvement with respect to earlier models is due to a better assignment of left and right-handed fermions to specific representations of a given ETC gauge group as well as of a better control over the walking dynamics [35]. Also the intragenerational mass splitting problem, especially the top-bottom mass one, has been recently re-investigated [19,35,44] with promising results. Particular care was payed in avoiding generating large corrections to the electroweak precision parameters.

There is the possibility that a top quark condensation may be responsible for part or all of the electroweak symmetry breaking [45–48]. This may happen due to the fact that the top quark is very heavy and hence strongly coupled to the electroweak symmetry breaking sector. Unfortunately the top-quark condensation mechanism per se seems to yield a too large top mass [49]. This problem can be addressed by re-introducing a technicolor theory [27]. One has also to invent a new strongly interacting theory coupling to the third generation of quarks and an additional strongly coupled $U(1)$ forbidding the formation of the bottom condensate. In this model one predicts the existence of topgluons, i.e. a massive color octet of vectors coupling mostly to the third generation. Due to the presence of the $U(1)$ interaction one predicts also the presence of a topcolor $Z'$ particle.

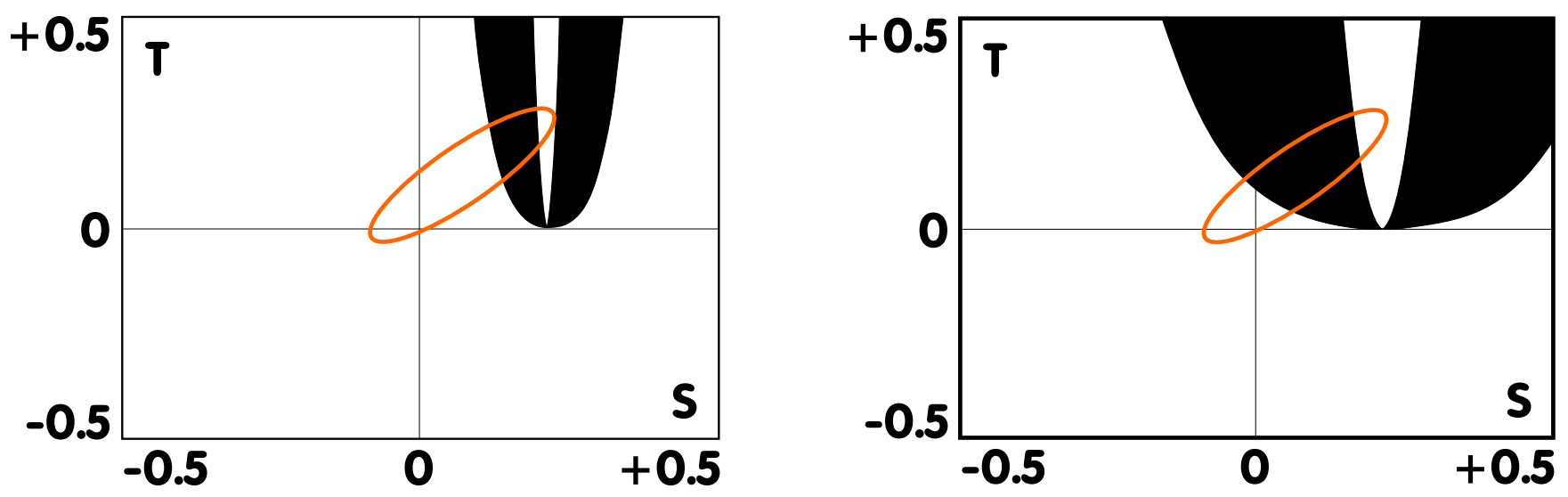

Fig. 12.2: Left Panel: The black shaded parabolic area corresponds to the accessible range of S and T for the extra neutrino and extra electron for masses from $m_Z$ to $10m_Z$ for the model of ref [10]. The perturbative estimate for the contribution to S from techniquarks equals $1/2\pi$. The ellipse is the $90\%$ confidence level contour for the global fit to the electroweak precision data with U kept at $0$. The contour is for the reference Higgs mass of $m_H = 150$ GeV. Right Panel: Here the plot is obtained with a larger value of the hypercharge choice, according to which one of the two fermions is doubly charged and the other is singly charged under the electromagnetic interactions.

Another promising idea is the top-seesaw model [6,50] in which the electroweak symmetry is broken thanks to the topcolor dynamics augmented with a seesaw mechanism involving an extra vectorlike quark, $\chi$. The Higgs boson is composite, resulting from a $I = 1/2$ condensate of a left-handed top quark and a right-handed state of the new isosinglet quark. With the condensate mass scale at $\sim$ 600 GeV, the vev of the Higgs field is at the right scale for electroweak symmetry breaking (EWSB) and the correct physical top mass will derive from the diagonalization of the mass matrix.

A summary of direct experimental limits on the existence of technicolor particles, as well as other resonances predicted in dynamical electroweak breaking scenarios can be found in [3,51].

Most of the searches for technicolor resonances, have been performed in the context of a "multi-scale" technicolor model [36,37] in which "walking" of $\alpha_{TC}$ is achieved by the presence of a large number of technifermions, which are copies of the fundamental representation of the technicolor gauge group, or which belong to a few higher representations, or both. It is then expected, in a "technicolor straw man model"(TCSM) [52–54], that the low energy phenomenology will be determined by the lowest-lying bound states associated to the lightest technifermion family doublet. The lowest technicolor scale could be of a few hundred GeV's, and therefore these bound states, the isovector technipions $\pi_T^{\pm,0}$ and technirho $\rho^{\pm,0}$ and the isoscalar $\pi'_T$ and techniomega $\omega_T$, would have a good chance of being seen at the Tevatron and should certainly be accessible at the LHC. A limited number of parameters is assumed in the TCSM model: (i) $N_{TC}$, the number of technicolors of the $SU(N_{TC})$ group, (ii) $N_D$, the number of technifermion families (iii) $\chi$, the mixing angle between the longitudinal vector bosons and the physical technipions, (iv) $Q = Q_U + Q_D$, the sum of the electric charges of the technifermions, (v) $m_V \sim m_A$, the mass parameters that control the strength of the technivector decay to a technipion and a transversely polarized electroweak boson (e.g., $\omega_T \to \pi_T^0 + \gamma$), (vi) $|\epsilon_{\rho\omega}|$, a mixing amplitude between $\rho_T^0$ and $\omega_T$, and (vii) $m_{\rho_T}, m_{\omega_T}, m_{\pi_T}$, the masses of the vector resonances and of the technipions.

At LEP, in technicolor searches based on the TCSM [55–58] the processes considered were: $e^+e^- \to \rho_T^0, \omega_T \to \pi_T^+\pi_T^- \to b\bar{q}\bar{b}q'$, as well as final states $\pi_T^0\gamma \to b\bar{b}\gamma$ and $W\pi_T$. As a result of these searches, an excluded region was obtained in the $(M_{\pi_T}, m_{\rho_T/\omega_T})$ plane. An upper bound of 87 GeV is obtained on the mass of $\pi_T$, at 95% CL, in the TCSM with $N_D = 9$, $(\chi = 1/3)$, independent of the mass of the vector states $m_{\rho_T}, m_{\omega_T}$. This limit is somewhat reduced for larger values of the angle $\chi$.

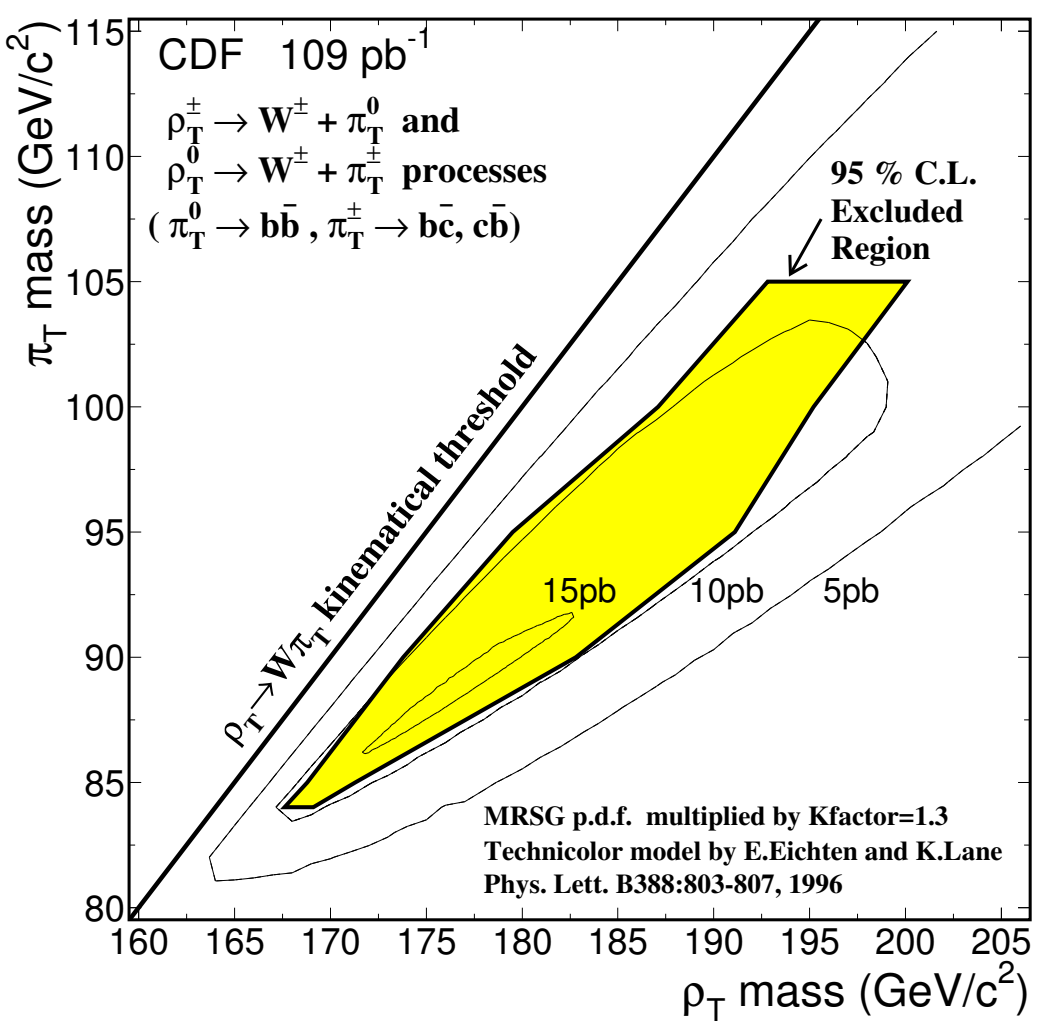

Fig. 12.3: Cross section contours, and 95% exclusion region in the $m(\pi_T), m(\rho_T)$ plane, from the lepton + 2 jets signal (from ref. [66]).

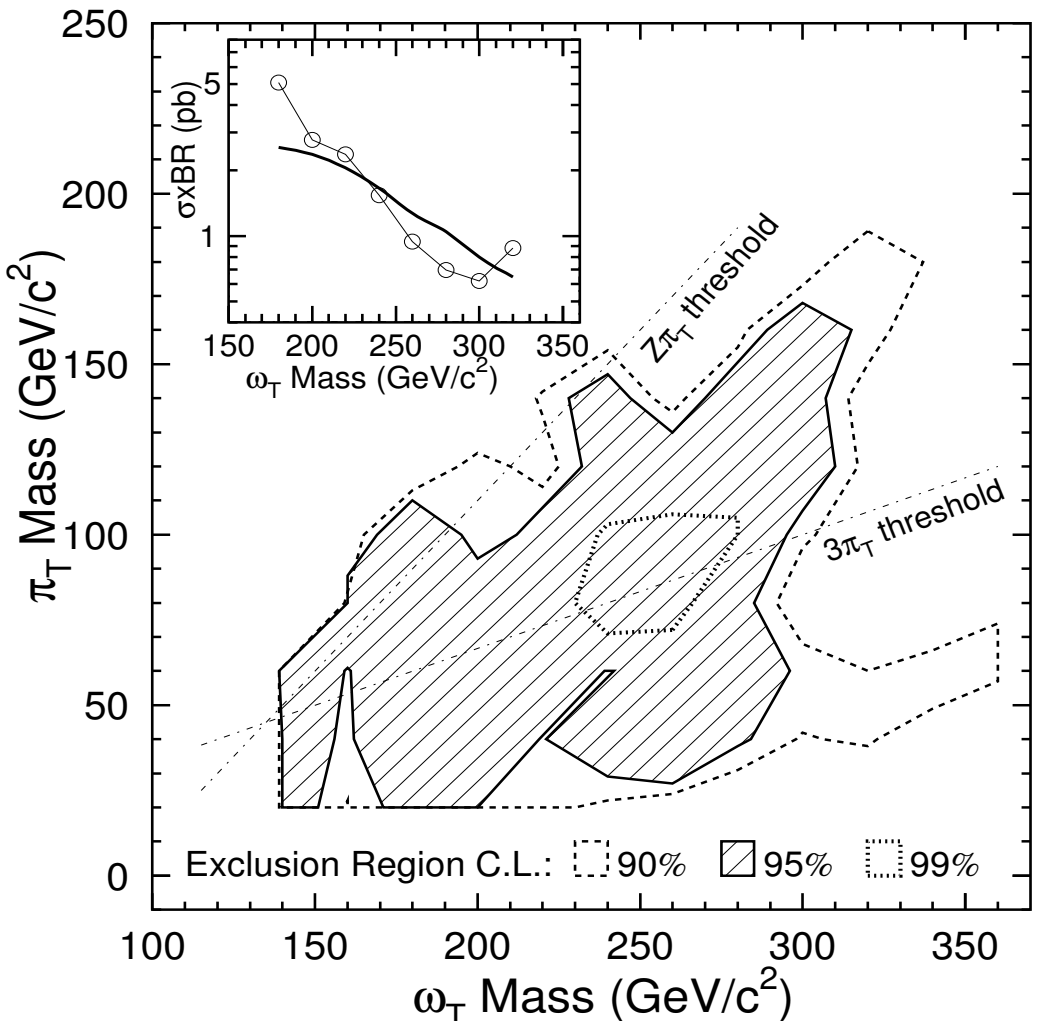

Fig. 12.4: Mass region excluded by the search for a light techniomega decaying to a technipion and a photon, with $\pi_T$ decaying to two jets, with one tagged b -jet. (Inset: cross section limit for $m_{\pi_T}$ = 120 GeV) [67]

Presently, the most stringent experimental constraints on masses of technicolor particles are derived from Tevatron searches. Since technipions, similarly to the Higgs, are expected to couple preferentially to the heavier fermions, one possible signal of technicolor would be the detection of leptoquarks (color-triplet technipion) of the second or third generation, decaying to $b\tau, c\nu$ or $b\nu$. Pair production of these leptoquarks should be enhanced by the s-channel exchange of the color octet technirho resonance, $\rho_8 \to 2\pi_{LQ}$, coupling in a vector dominance model (VDM) to the gluon propagator. Searches for these processes [59, 60] have excluded regions of the $m_{\pi_{LQ}} - m_{\rho_T}$ plane, in the kinematically allowed areas of phase space. Typically, the limit on $m_{\rho_T}$ is about 600 GeV (for $m_{\pi_T} < m_t$ in the $c\bar{c}\nu\bar{\nu}$ channel), but depends on the assumed value of the mass difference $\Delta M = m_{\pi_8} - m_{\pi_{LQ}}$. For high masses of technipions, a color-octet technirho could decay to di-jets. The mass range $350 < m_{\rho_8} < 440$ GeV, has been excluded [61, 62] in the $\rho_8 \to b\bar{b}$ channel. The absence of such $b\bar{b}$ mass peak also serves to set limits, depending on the assumed width of the resonance, on the masses of topgluon states. The above limits on the color octet vector resonance carry high uncertainty as it has been shown that the coupling of the state to two gluons is forbidden by gauge invariance [63] in a VDM, although higher order operators will lead to such couplings [64]. Furthermore, it has been argued [65] that in a model deviating even slightly from VDM, where some direct coupling exists between a quark and the interaction eigenstate of a technihadron, the coupling of the quark to the physical technirho would be highly suppressed.

Searches at the Tevatron have also been performed in the context of the TCSM model described above. The principal decay channels of $\rho_T$ are $\pi_T\pi_T$, $V\pi_T$ and $VV$, $(V = W$ or $Z)$, but depending on the mass relations, branching ratios vary or some decay channels could be closed. Searches for event topologies containing leptons + jets [66, 68], or 4-jets were therefore performed, with selection of b-jets in the final state to reduce the backgrounds. The resulting contours of exclusion are shown in Fig. 12.3. Even with b-tagging of the jets, large backgrounds from $W$ + heavy flavor, mistags, $t\bar{t}$ and single top events remain. Similarly, in the absence of evidence for peaks in the $\{m(\gamma bj) - m(bj)\}$ distribution, searches [67, 68] for $\omega_T \to \gamma\pi_T$ followed by $\pi_T \to b\bar{b}$ have also led to cross section limits, or to exclusion regions of approximately $150 < m_{\omega_T} < 300$ GeV when $m_{\omega_T} > m_Z + m_{\pi_T}$ (see Fig. 12.4). Another relatively clean possible signature, which will be enhanced if decays of the vector resonances to technipions are kinematically closed, is a $Z'$-like peak in the di-lepton invariant mass: $\rho_T, \omega_T \rightarrow e^+e^-$. Fig. 12.5 shows preliminary bounds from D0 [69, 70] on the cross section times branching ratio of this process, obtained in RunII.

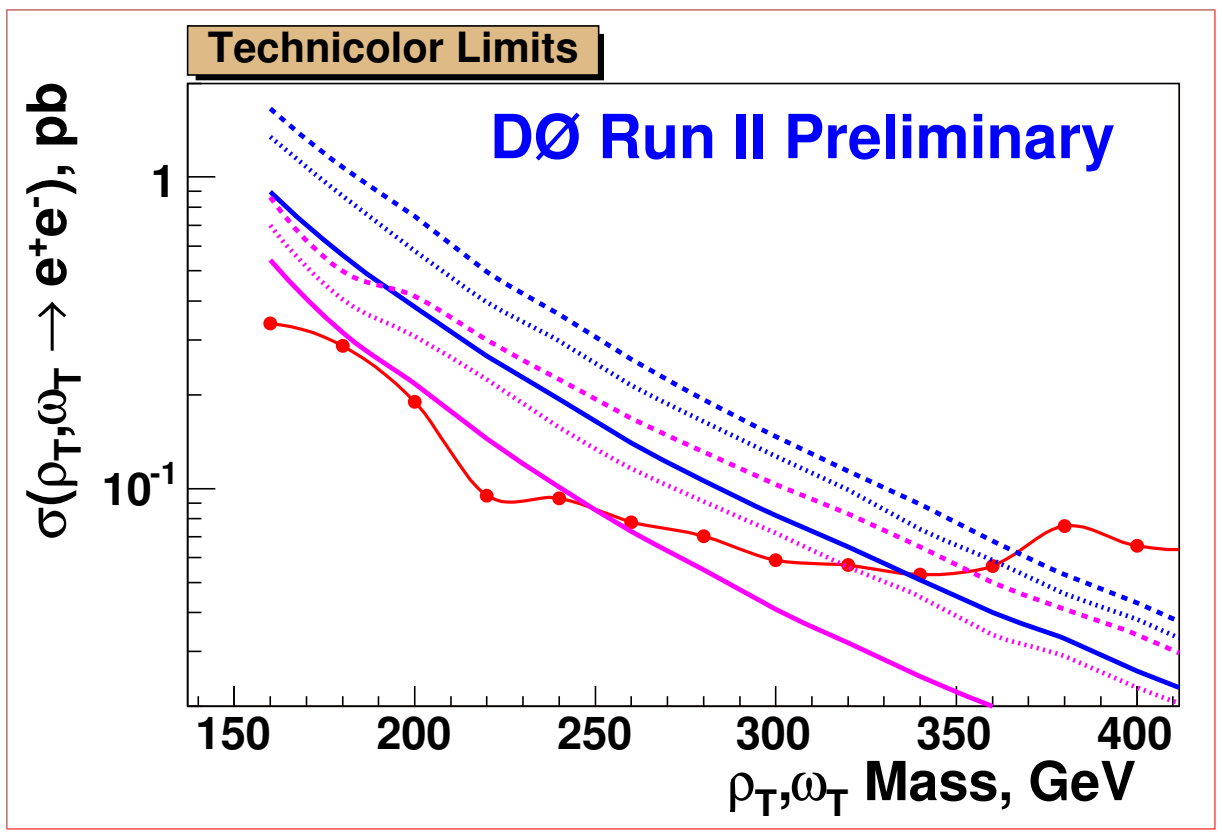

Fig. 12.5: Solid line with dots: 95% cross section upper bounds on $\rho_T, \omega_T$ cross section times branching ratio into $ee$. The various curves are predictions under certain assumptions of mass difference ($m(\rho_T) - m(\pi_T)$ = 60, 100 GeV) or of the parameter $m_V$ = 500, 200 and 100 GeV (see ref. [70]).

At the LHC, fast simulation studies [71], performed in the context of an earlier version of the TCSM model, implemented in PYTHIA, suggest that the technirho, techniomega and technipions could be detected up to masses of around 1 TeV. The LHC reach depends strongly on the assumed parameters, as the various branching ratios and decay widths are sensitive to the relative masses, as well as the mixing angles. Nevertheless, depending on parameters, a variety of signals can be investigated. For example, with an integrated luminosity of 30 $fb^{-1}$, the technirho can be observed up to masses of $\sim$ 800 GeV and the technipion up to $\sim$ 500 GeV, in channels: $\rho_T^{\pm} \rightarrow W^{\pm}Z$, $\rho_T^{\pm} \rightarrow \pi_T^{\pm}Z$ and $\rho_T^{\pm} \rightarrow W^{\pm}\pi_T^0$, when the vector bosons decay leptonically and the technipions decay to heavy flavor quarks. Fig. 12.6 shows an example of some specific cases of $\rho_T$ resonances. It has been assumed that the $\pi_T$ coupling to the top is small, as one would expect in topcolor models. As emphasized above, such plots are meant to be only indicative of the expected signals as the results are parameter dependent and as the branching ratios do not account for certain possible decays [53], implemented in later version of PYTHIA. The production of $\rho_T^{\pm}$ by a vector boson fusion process is another possibility [72], which could complement the $q\bar{q}$ fusion process. In all cases, it will be important to have efficient $b-$tagging of jets, and good lepton-jet discrimination.

At the ILC, a big advantage is the clean final state which allows reconstruction of hadronic decay modes of the W's, although di-jets will tend to have small opening angle because of the kinematic boost. Depending on the center of mass energy, resonances could be observed up to $\sim$ 2 TeV in channels of resonant vector boson scattering or of vector boson fusion producing fermion pairs, including top pairs. A summary of the potential for observing scalar and vector resonances at the ILC can be found, for example in [73–75]. In [74], expected bounds on the BESS model [76, 77] parameters are shown. This model assumes a triplet of vector resonances $V^{\pm,0}$ with parameters describing the mixing to the electroweak gauge bosons and the coupling to the fermions (see Section 11).

The topcolor models mentioned above have a rich phenomenology (for a review, see [3]). A triplet of top pions of mass $\sim$ 200 GeV could look like $W'$ or $Z'$ decaying to the third generation, and a scalar top Higgs could decay by the flavor changing process $\pi_t^{0'} \rightarrow \bar{t}c$. Color octet topgluons will produce resonances in the $t\bar{t}$ system. In the top condensation see-saw model, in general several composite scalars are predicted [6, 7]. From a phenomenological point of view, the mass of the isosinglet quark may be too high ($\sim$ 4 TeV) to be easily observed at the LHC, but a heavy Higgs ($m_H \sim$ TeV) is predicted and the mixing between the interaction states of $t$ and of $\chi$ of same chirality affects the interaction of the physical top with the gauge bosons [41].

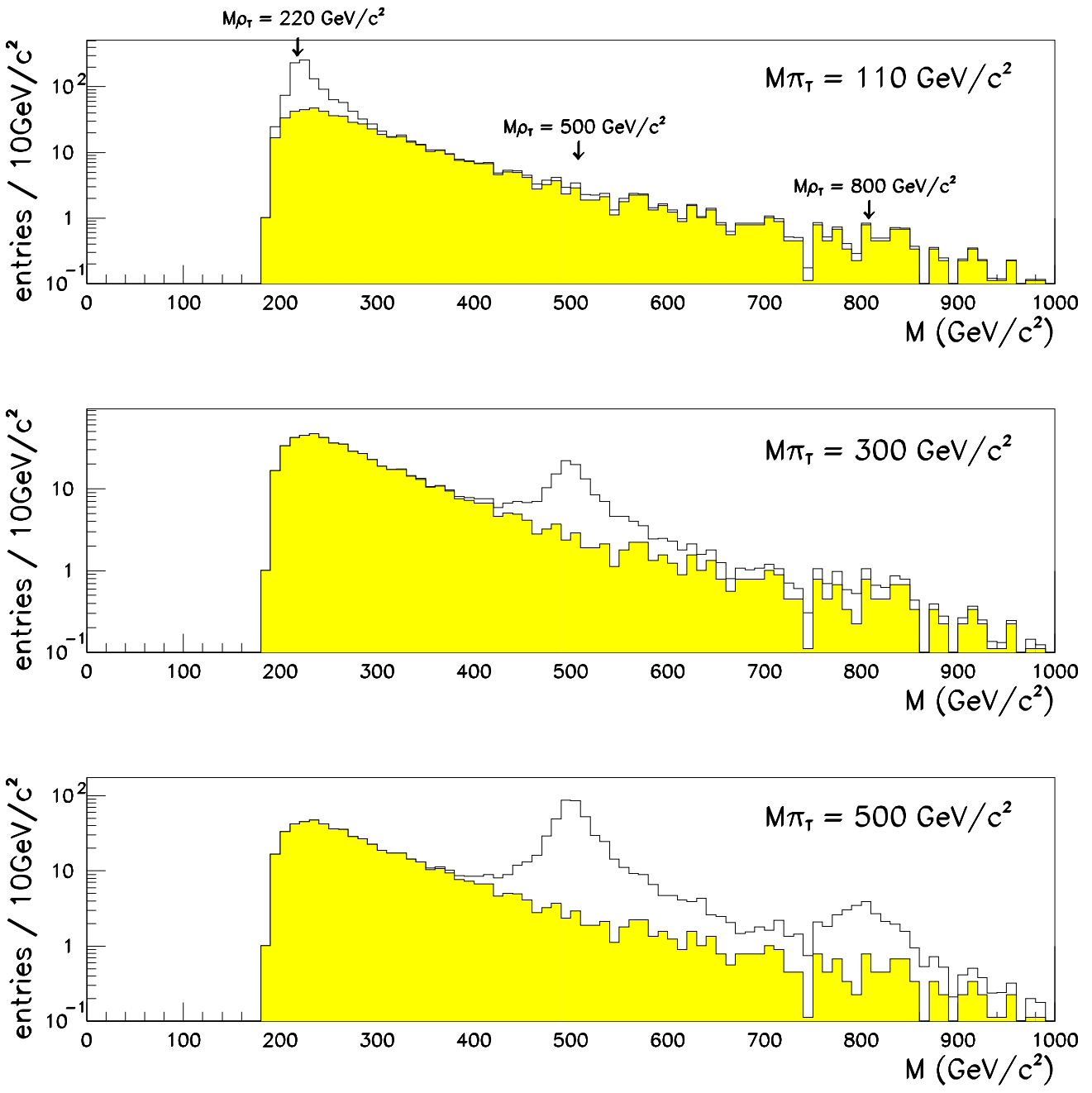

Fig. 12.6: $\rho_T^\pm \to W^\pm Z$: expected signals and background at the LHC, for 30 fb$^{-1}$ and for different masses of the techni-resonances (from ref. [71]).

Other promising signatures for technicolor have been recently suggested. One very interesting possibility is the discovery of a fourth family of leptons, which could serve to confirm the model, discussed above, of electroweak symmetry breaking from technifermions in higher dimensional representations of the technicolor gauge group [10]. In this theory one would also expect a light composite Higgs, and the associated production of this Higgs boson with vector gauge bosons could be enhanced with respect to the Standard Model [78] (see Section 12.5). Another striking signal [79] (see also Section 12.6) would be a strong narrow resonance in the $\tau\tau$ and $\gamma\gamma$ channels. Indeed, a light technipion could be abundantly produced by $gg$ fusion via techniquark loops, or by $b\bar{b}$ annihilation. The large enhancement factors predicted in models of dynamical symmetry breaking for resonances in these channels will allow to distinguish these technipions from the light scalars of the Standard Model or of Supersymmetry.

Many other phenomena possibly observed at the LHC could be interpreted in the context of a technicolor model, and indeed, it is by combining different signatures that confusion with other models can be cleared and that the nature of these resonances could be understood. For example, pair production of leptoquarks could be enhanced by a technirho resonance, as discussed above; or a $Z'$-like signature could signal a technivector resonance decay into leptons; a narrow $t\bar{t}$ or $b\bar{b}$ resonance which is non-flavour-universal can also signal topgluons (or leptophobic $Z'$). It is intriguing that an excess (not statistically significant) of $t\bar{t}$ events with an invariant mass around $\sim$ 500 GeV at the Tevatron [80, 81], could be a hint for the existence of such a resonance.

### 12.2 Extended technicolor

*Nick Evans*

The idea of breaking electroweak symmetry by a dynamically generated fermion condensate is both elegant and phenomenologically achievable. The harder challenge for such technicolor theories is to feed the electroweak symmetry breaking order parameter down to generate the diverse masses of the standard model fermions. The top mass presents a particular challenge because its large mass cannot be considered a small perturbation on the electroweak scale.

The most appealing mechanism for generating the standard model fermion, $f$, masses is extended technicolor [18, 82]. The technicolor gauge group is unified at high energies with a gauged flavour symmetry of the standard model fermions. This symmetry is then broken above the technicolor scale leaving massive gauge bosons that link the standard model fermions to the technicolour sector. The basic fermion mass generation mechanism is depicted in figure (12.7). The resulting mass generated is given in terms of the techniquark condensate $\langle \bar{T}T \rangle$ by

$$m_f \simeq \frac{g_{ETC}^2}{M_{ETC}^2} \langle \bar{T}T \rangle \tag{12.2}$$

where $g_{ETC}$ and $M_{ETC}$ are the coupling and mass of the ETC gauge boson.

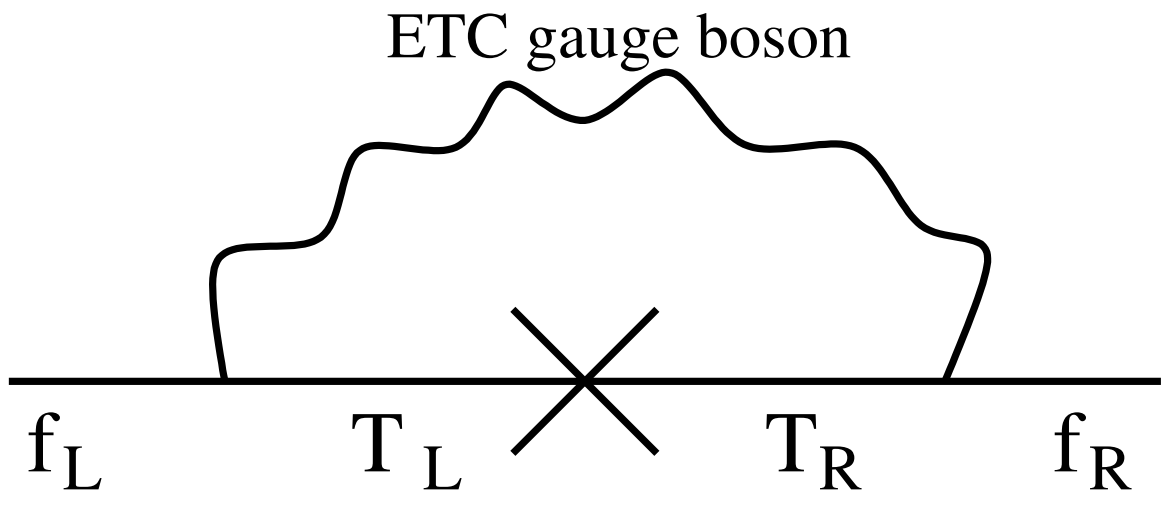

Fig. 12.7: ETC mass generation mechanism for a standard model fermion $f$. The ETC gauge boson converts $f$ to a techni-fermion, $T$. The chirality mixing of the interactions is explicitly shown.

#### *12.2.1 Model building*

Many ETC models exist in the literature including [18, 19, 35, 82–91]. There are two common patterns chosen for the ETC gauge symmetry (see Fig 12.8). A natural choice is to gauge the family symmetry of the standard model fermions and then unify it with technicolor to give an SU(N+3) ETC group. Such a model will have a partner techni-fermion for each member of a standard model family - so called one family technicolor models. The ETC group is envisaged to be broken in the cascading pattern

$$SU(N+3) \to SU(N+2) \to SU(N+1) \to SU(N) \tag{12.3}$$

If the breakings occur at the scales of a few 100 TeV, a few 10TeV and of order a few TeV then the rough structure of the three family mass hierarchy is reproduced.

Another commonly used pattern of ETC leaves a one doublet technicolor model and appeals to the ideas of Pati Salam unification of quarks and leptons [92]. For example for the third family one might gauge the SU(4) flavour symmetry on the three colors of quarks and the tau lepton doublet. This could then be unified at high scales with the technicolor group leaving an SU(N+4) ETC group and a breaking pattern

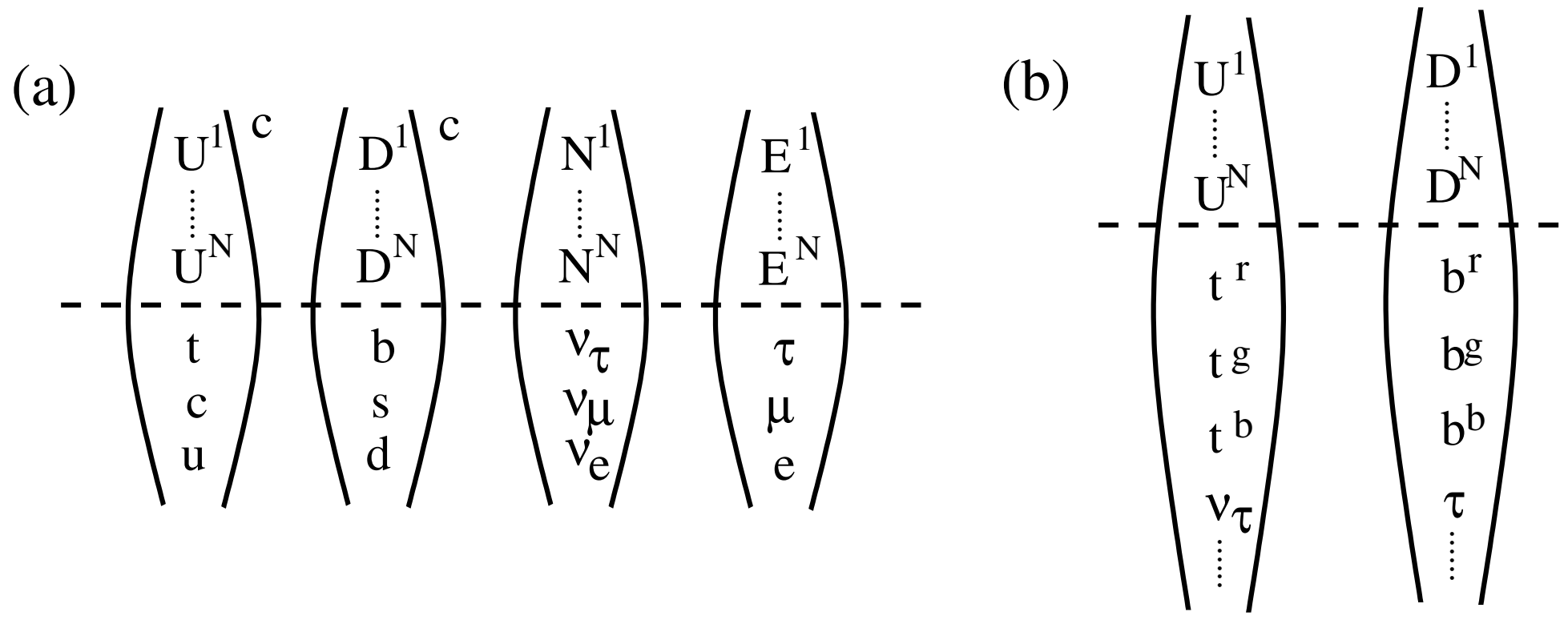

Fig. 12.8: Example ETC fermion multiplets - SU(N) technicolor acts on the techni-quarks with colors $1..N$ above the dashed line. The ETC group acts on the techni-quarks and standard model fermions so there are gauge bosons that convert one to the other. In (a) the four fermion multiplets of a one family model are shown and the ETC group is a family symmetry group. The c index indicates which mulitplets have QCD color. In (b) a one doublet model's multiplets are shown. Here the ETC group contains the QCD color symmetry acting on the $r, g, b$ colors of quarks, and leptons are the "fourth" color.

$$SU(N+4) \to SU(N+3) \to SU(N) \otimes SU(3) \tag{12.4}$$

Expanding this scenario to incorporate the full set of standard model fermions involves promoting the ETC group to SU(N+12) and a very complicated symmetry breaking pattern.

Many examples of the ETC symmetry breaking mechanisms exist in the literature including entirely dynamical sectors or less satisfactorily higgs fields. Another idea that has been proposed is that the ETC group might sequentially break itself via a "tumbling" mechanism [93–96] in which non-color singlet condensates are formed by the ETC dynamics. All these models are typically complicated since they seek to explain many different symmetry breaking sectors. Given that we still struggle to understand electroweak symmetry breaking these mechanisms are perhaps best left for the future if technicolor is discovered.

The ETC schemes sketched so far contain no dynamics capable of explaining the splitting in masses between different weak isospin partner fermions. The most marked splitting is in the top bottom quark doublet which must have the lowest ETC scale - such isospin breaking will be apparent in any initial discovery of such a sector.

One solution is to make the ETC gauge symmetries chiral (eg [87–89]) so that the right handed top and bottom quarks transform under different groups with potentially different breaking scales and couplings. The ETC symmetry breaking sector must then be further complicated as these different ETC groups are broken together to leave a single technicolor group. Extra fermions must be included in such a model to maintain the anomaly freedom of the ETC gauge groups and a mechanism must be found to give these fermions masses above current limits. Such models predict a rich structure of new physics beyond the standard model.

Alternatively the right handed top or bottom could be placed in a different representation of a single ETC group (eg [42]). This approach is a model building challenge since the top and bottom must emerge without extra fields present at low energy from the higher dimensional representation.

The generation of the fermion CKM mixing angles and CP violation are also a challenge to ETC models [97–99]. Models have been made in which these aspects of the mass spectrum are inherited from physics at higher energy scales without explanation [88, 89]. There are models that more directly attack their generation which typically try to arrange the mixings' emergence from a dynamical vacuum alignment problem (eg [99]). The small size of the neutrino masses need explanation too. Recently dynamical generation of a Majorana mass for the right handed neutrinos has been investigated leading to a see-saw mechanism to suppress their masses [19].

Finally we note that in the schemes discussed above the SU(2) gauge symmetry of the weak interactions was assumed to commute with the ETC group. In principle this need not be the case. One could imagine a chiral ETC symmetry as large as

$$SU(N+24)_R \otimes SU(N+24)_L \tag{12.5}$$

with the weak SU(2) embedded in the ETC dynamics. Such models are called non-commuting ETC [25, 100].

### *12.2.2 Indirect constraints*

A number of indirect constraints exist on ETC theories.

#### 12.2.2.1 Flavour changing neutral currents

The gauging of flavour symmetries is well known to induce flavour changing neutral currents (FCNC) since the flavour group's gauge eigenstates cannot be expected to match the standard model fermion mass eigenstates. The absence of FCNC in the standard model makes these constraints (from kaon and D-meson physics), on the first two families, very restrictive [18, 101]. In fact such a flavour gauge boson must have a mass in excess of 500 TeV. This is hard to reconcile in an ETC model with the measured values of the second family quark masses.

An early model building suggestion for overcoming the difficulty was walking technicolor [20,22]. If the technicolor gauge coupling runs close to an infra-red fixed point, so it is strong over a large energy range, then the techni-fermion condensate is enhanced pushing up the ETC scale needed to generate a particular fermion mass. Pushing the ETC scale up by an order of magnitude does though require walking dynamics over a long energy regime.

An alternative solution has been to build models that have a GIM mechanism [88, 89, 102] in the spirit of the standard model. These models have separate ETC groups acting on the left handed doublets, and on both the $+1/2$ and $-1/2$ isospin right handed doublets. The constraints on such models from FCNC are very much weaker than on normal ETC models.

#### 12.2.2.2 $\rho/T$ parameter

To generate the large top mass by a standard ETC mechanism requires the ETC scale to be as low as 1 TeV. Since this dynamics breaks isospin in order to explain the top bottom mass splitting, the light ETC gauge bosons must also violate isospin. These gauge bosons will enter into virtual contributions to the W and Z boson masses through diagrams like those in Fig 12.9(a). There will thus be contributions to the $\delta\rho$ or $T$ parameter of rough order of magnitude

$$\alpha T \sim \frac{v^2}{M_{ETC}^2} \tag{12.6}$$

where $\alpha$ is the electromagnetic coupling.

For a 1TeV ETC scale such contributions are huge ($T \sim 10!$) [103, 104]. One concludes the ETC scale must be larger than about 4TeV to stand a chance of being compatible with precision data. Some method for enhancing the top mass over the predictions from naive ETC are then needed.

The ETC isospin breaking will also generate techni-fermion masses which break isospin. Typically one might expect this mass splitting, $\Delta M$, to be of order the top bottom mass splitting. The contribution to T is given by roughly [5]

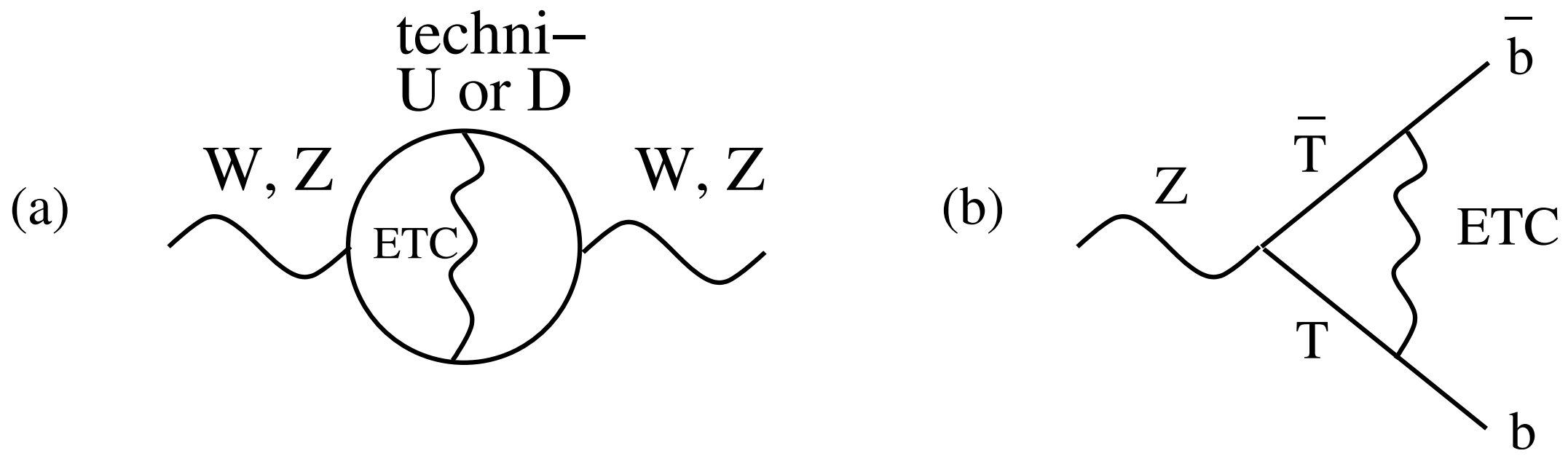

Fig. 12.9: (a) Isospin violating ETC gauge bosons contribute to the mass of the W and Z bosons through their couplings within loops of techni-fermions. The contributions to the T parameter can be very large. (b) ETC gauge bosons that couple the left handed bottom quark to the techni-quark sector generate substantial corrections to the $Zb\bar{b}$ vertex.

$$T \sim \frac{N_{TC}}{12\pi^2\alpha}\frac{\Delta M^2}{v^2} \sim 0.6N_{TC}\frac{\Delta M^2}{m_t^2} \tag{12.7}$$

with $N_{TC}$ the number of techni-colors, which lies close to the upper experimental bound.

12.2.2.3 $Zb\bar{b}$ vertex

The most strongly coupled ETC gauge bosons responsible for the top mass must also couple to the left handed bottom quark. The exchange of such a gauge boson across the $Zb\bar{b}$ vertex, as shown in Fig 12.9(b), leads to a correction to the $Zb\bar{b}$ width that has been estimated to be of the order [105]

$$\frac{\delta\Gamma}{\Gamma} \sim -6.5\% \left(\frac{m_t}{175GeV}\right) \tag{12.8}$$

Such a large contribution is not compatible with data. If a mechanism for enhancing the top mass over naive ETC estimates can be found then this contribution will fall off quadratically as the ETC scale is raised. Finally we note that positive corrections to the width are possible in non-commuting technicolor theories [25].

### *12.2.3 Strong ETC and top condensation*

It is clear from the estimates of the T parameter and the corrections to the $Zb\bar{b}$ vertex that a naive ETC model can not generate the observed top mass. The lowest ETC scale must be pushed up to of order 4 TeV or above. One mechanism for enhancing the top mass is walking technicolor which enhances the techni-fermion condensate. In such a scenario the ETC interactions will themselves be strong [106–108] and there will be further non-perturbative enhancement of the top mass. The extreme version of such a model has the ETC interactions on the top quark sufficiently large on their own that they generate a top quark condensate and the top mass independently of the rest of the ETC sector [27, 45, 46, 49]. When the ETC couplings lie close to the critical values for triggering chiral symmetry breaking on their own, small isospin violating effects, such as from an extra U(1) gauge group, which tip the combined coupling super critical may generate large top bottom splitting. This would reduce tension with the T parameter measurements. If strong ETC is the route nature has chosen it is most likely that a combination of all these ideas are responsible for the large top mass.

#### 12.2.4 Direct searches for ETC

ETC gauge bosons are flavour gauge bosons and their presence can be seen as an enhancement of a number of standard model processes [109]. They most strongly couple to the third family where they are most likely to be observed first. Models with additional gauged flavour symmetries on the quarks such as top color models [27] will generate enhanced dijet production in hadronic machines. ETC gauge bosons coupling to quarks may also mediate large enhancement of single top production processes over the standard model rate. A wider set of models was considered in [110,111] which also include couplings to leptons - the ETC gauge bosons then give new contributions in Drell Yan production.

### 12.3 Composite Higgs from higher representations

*Dennis D. Dietrich*

#### 12.3.1 The minimal walking technicolour theory

Technicolour theories [1–4] can be constructed with techniquarks in higher dimensional representations [18, 36, 38] of the technicolour gauge group. One of these theories, denoted by S($N, N_{\rm f}$)=S(2,2), with two techniflavours in the two-index symmetric representation of SU(2) [1] [10, 40, 112, 113] is found to be in agreement with electroweak precision data [39]. This variant of the model is closest to the currently available data and is the main subject of this contribution. There exists another version, S(3,2), which is consistent at the two sigma (standard deviation) level and which is discussed briefly toward the end of this contribution.

The important feature of the S(2,2) theory is that, with the smallest possible number of techni-flavours and -colours, it is quasi conformal [8, 9], that is it possesses a walking coupling constant (see Introduction, Fig. 12.1). The small number of additional degrees of freedom leads only to small corrections to the standard model at energies below $\Lambda_{\rm TC}$. The quasi-conformality allows to generate the required masses for the ordinary fermions by means of extended technicolour interactions (ETC) [42, 43, 91, 114] while avoiding flavour-changing neutral currents and lepton number violation, which would be at odds with data [35] (see below).

Furthermore, the walking of a theory has the capability to reduce the mass of the composite scalar (the Higgs) below the value of the typical scale of the underlying theory. Due to its very near conformality, the theory S(2,2) is likely to feature a remarkably light Higgs (see below). Whether a technicolour theory is nearly conformal depends on the number of technicolours and techniflavours as well as of the representation of the techniquarks. If a theory is not conformal for a given number of techniflavours it will enter the conformal phase when their number is increased. At leading order, the point where this happens can be characterised by the anomalous dimension of the quark mass operator becoming unity [114, 115] [2]. Hence, for a given number of technicolours and a given representation of the technicolour gauge group, the aforementioned criterion defines the critical number of techniflavours. Based on the two-loop beta-function, which in the t'Hooft scheme is exact, the critical number of flavours for a theory with adjoint techniquarks is found to be $N_{\rm f,crit.} = 2.075$, independent of the number of technicolours. The squared mass of the Higgs scalar is suppressed with respect to the scale of the theory by a factor of the small difference of the critical number of flavours and the actual number of flavours, $(N_{\rm f,crit.} - N_{\rm f})$ [10, 40], [3] the latter of which is necessarily an even integer. This leads to an estimated mass for the composite Higgs of about 150 GeV [10, 40].

[1] For two (techni-)colours the two-index symmetric representation coincides with the adjoint representation.

[2] At any finite higher order the criterion will usually receive corrections.

[3] Note, however, that this result might acquire corrections [29, 116]. Near the conformal phase transition other states could become light, which, in turn, could affect the argument supporting universal behavior near the phase transition [117]. That, however, need not change the result. Especially, a vanishing chiral symmetry scale does not imply that the chiral partner of the pions does not become parametrically lighter and narrow near the phase transition, as the self coupling of the scalar is also expected to vanish near a continuous phase transition [35].

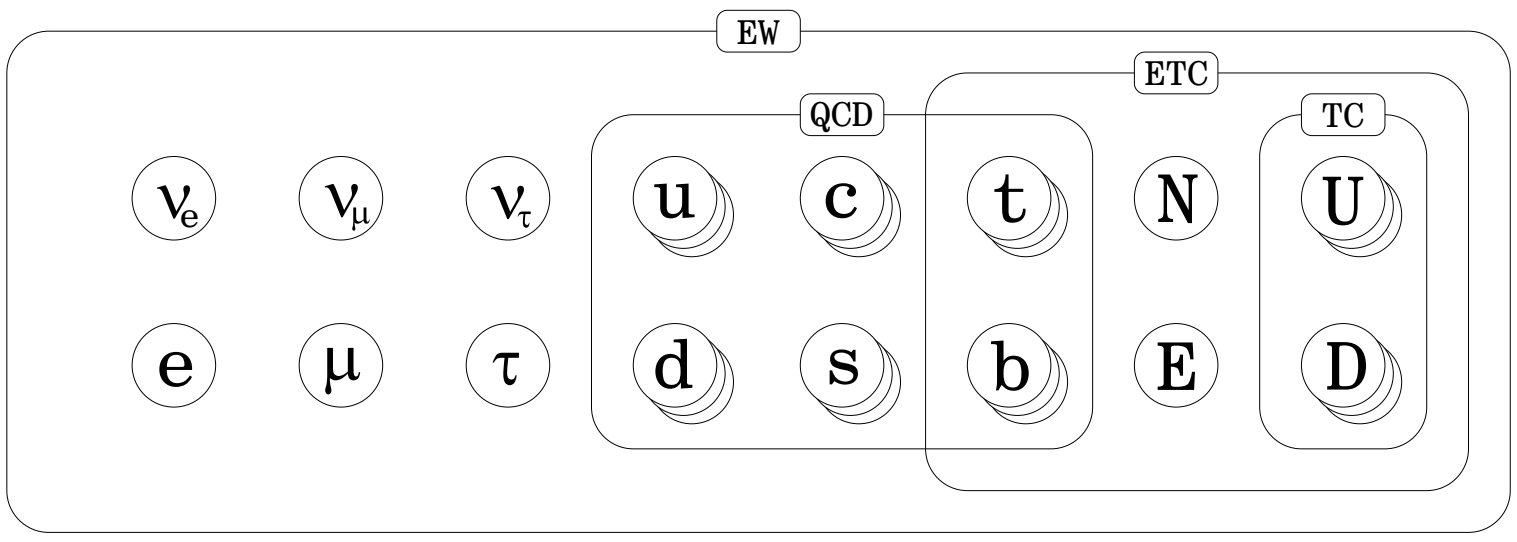

| $q$ | (I) | (II) | (III) |
|---|---|---|---|
| U | $+2/3$ | $+1/2$ | $+1$ |
| D | $-1/3$ | $-1/2$ | $0$ |
| N | $0$ | $+1/2$ | $-1$ |
| E | $-1$ | $-1/2$ | $-2$ |

Fig. 12.10: Matter content of the S(2,2) model: *Left panel:* Staggered circles represent the number of colour degrees of freedom of a particle. Boxes denote under which subset of the gauge group the particles transform. Usually ETC models comprise all massive particles. Here, however, the matter content of the residual ETC model of Ref. [35] is indicated. *Right panel:* Electric charges, $q$, of the new particles—U (techniup), D (technidown), as well as N and E (fourth family leptons)—in the three studied cases.

In the S(2,2) model the topological Witten anomaly [118] is evaded by including a fourth lepton family [10, 40]. The matter content of the model is summarised in the left panel of Fig. 12.10. At this point, only the hypercharge assignment of the new particles remains to be specified. It is constrained but not uniquely determined by requiring the absence of gauge anomalies [10, 40]. Here three possibilities for the electric charge, $q$, of the techniup (U), technidown (D), the fourth neutrino (N), and the fourth electron (E) are considered (see Fig. 12.10, right panel).

In case (I) the neutrino, N, can constitute a non-hadronic component of dark matter. The technicolour of any number of techniquarks in the adjoint representation can be neutralised by adding technigluons. With respect to the standard model sector, such bound states interact only weakly and behave like leptons. In case (III) a D techniquark whose colour has been neutralised by additional gluons can contribute to dark matter. In the same case, (III), bound states of two D techniquarks and in case (II) UD bound states represent potential technihadronic contributions to dark matter. For more details on these technihadrons and dark matter candidates see Section 12.4.

In [10, 40, 112, 113] predictions from these models have been compared to electroweak precision data. To this end the oblique parameters $S$ and $T$ [119] have been calculated and compared to data. They quantify the contribution to the vacuum polarisation of the gauge bosons from the particles which are not contained in the standard model. Therefore, per definition, the standard model with a reference Higgs mass has $S = 0 = T$. $S$ is a measure for the mixing between the photon and the $Z_0$, while $T$ measures the additional breaking of the custodial symmetry. In Fig. 12.2 of the Introduction, the areas filled in black depict the perturbatively calculated values taken by the oblique parameters for degenerate techniquarks and when the masses of the two additional leptons are varied independently between one and ten $m_Z$. The ellipse represents the 68% confidence level contour from the global fit to data in [39]. There it is assumed that the third oblique parameter, $U$, is zero, which is consistent with the values predicted in the present framework. In the fractionally charged case there is no variation in the direction of $S$ and varying the masses of the leptons gives a vertical line exactly in the opening of the area shaded in black.

Already this perturbative analysis of the oblique parameters, which can be seen as conservative, displays a sizeable overlap with the range favoured by the data. In quasi-conformal theories like the present one, the techniquarks' contribution to $S$ is lowered by non-perturbative effects [32–34, 120, 121], corresponding to a reduction of about 20% [34, 121]. The aforementioned reduction of the value of $S$ caused by the walking of the coupling has also been confirmed by a holographic analysis of technicolour theories [122]. If the reduction of 20% is taken into account the overlap between the 68% level of confidence ellipse and the values predicted for the S(2,2) technicolour model becomes even larger (see Fig. 12.11). Especially, an overlap with the right branch of the area filled in black can be achieved.

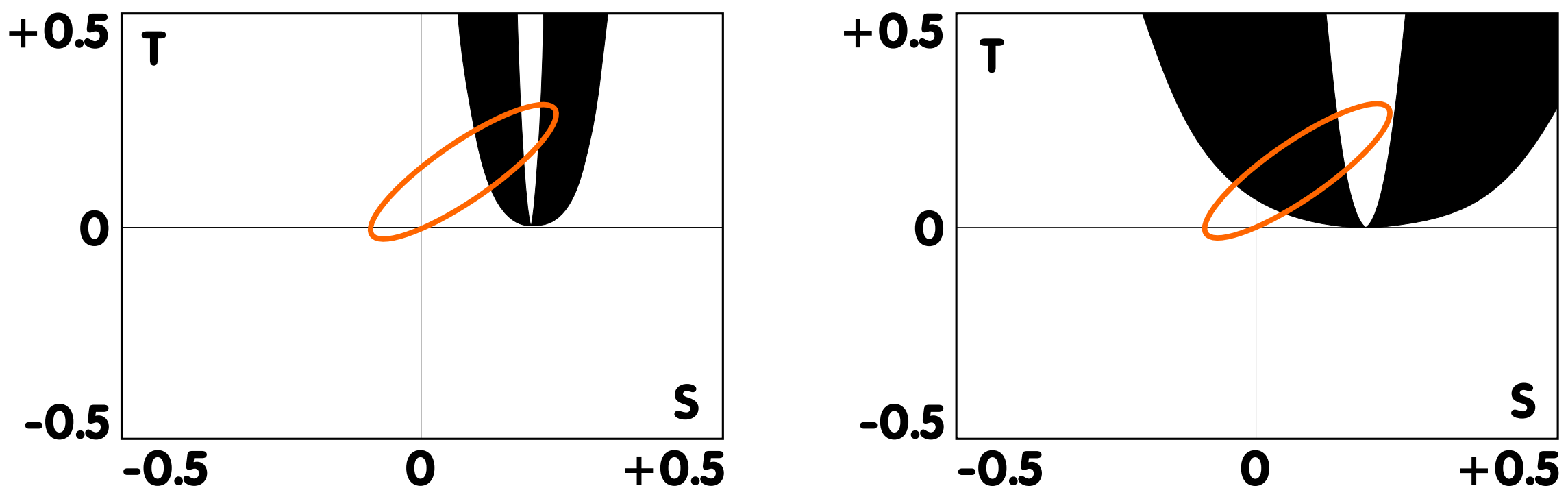

(a) Standard-model-like charge assignment (I)

(b) Integerly charged leptons (III)

Fig. 12.11: Including nonperturbative corrections. The black area represents the accessible range for the oblique parameters $S$ and $T$ for the masses of the fourth family leptons $m_E, m_N \in [m_Z; 10m_Z]$ and with degenerate techniquarks, which corresponds to a contribution of $0.8/2\pi$ to $S$ from the latter. The ellipse is the 68% confidence level contour for a global fit to electroweak precision data [39] with the third oblique parameter $U$ put to zero and for a Higgs mass of $m_H = 150$ GeV, as expected for the S(2,2) model. Putting $U$ to zero is also consistent with the S(2,2) model, where it lies typically between 0 and 0.05.

In the purely perturbative computation of the oblique parameters, the overlap between the ellipses and the black shaded areas in Fig. 12.2 of the Introduction corresponds to the masses depicted in Fig. 12.12 of the present contribution. In the cases (I) and (III) the main feature is a mass gap of the order of $m_Z$ between the new leptons by which N is lighter ($m_2$) than E ($m_1$). The mass gap is mostly determined by the limits in ($S - T$)-direction. In the case (II) an additional branch with the opposite sign for the mass gap is present. Including nonperturbative corrections to the technicolour sector, the additional overlap of the 68% level of confidence contour with the right-hand side of the black areas in Fig. 12.11 translates to a second branch also for cases (I) and (III), which is otherwise suppressed by the limit on $S$. In the present model the expected mass for the composite Higgs is 150 GeV, but even if it was as heavy as 1 TeV there would still be an overlap to the data at the 68% level of confidence.

Recently, there have been indications that the experimental constraints on the oblique parameters are much weaker than obtained in previous analyses [123,124]. In [123] it has been demonstrated that the bounds become much less stringent when the uncertainty in the triple and quartic gauge boson couplings is taken into account. Hence, the range of favoured masses tends to be even larger.

### *12.3.2 Extended technicolour*

As mentioned above, the masses of the standard model fermions are to be generated by ETC interactions, that is the direct exchange of ETC gauge bosons between the fermions of the standard model and the techniquarks. For this purpose the TC model is embedded in a larger gauge group, whose additional symmetries are broken successively. In the broken phase of the ETC group typically three types of effective four-fermion interactions occur, (a) $\bar{Q}Q\bar{Q}Q\Lambda^{-2}_{\mathrm{ETC}}$, (b) $\bar{\psi}\psi\bar{Q}Q\Lambda^{-2}_{\mathrm{ETC}}$, and (c) $\bar{\psi}\psi\bar{\psi}\psi\Lambda^{-2}_{\mathrm{ETC}}$, where $Q$ stands for any techniquark and $\psi$ represents any standard model fermion. (Different fermion types may be coupled together in this way.) In the broken phase of the TC group a chiral condensate of techniquarks develops, $\bar{Q}Q \to \langle\bar{Q}Q\rangle$, which leads to the following contributions: (a) $\bar{Q}Q\langle\bar{Q}Q\rangle\Lambda^{-2}_{\mathrm{ETC}}$, (b) $\bar{\psi}\psi\langle\bar{Q}Q\rangle\Lambda^{-2}_{\mathrm{ETC}}$, and (c) $\bar{\psi}\psi\bar{\psi}\psi\Lambda^{-2}_{\mathrm{ETC}}$. Terms of the type (c) are a byproduct of the ETC mechanism and lead to flavour changing neutral currents and lepton number violation. If no suitable GIM mechanism [88, 89, 125, 126] can be devised, the breaking scale $\Lambda_{\mathrm{ETC}}$ must be sufficiently large to suppress these terms. Terms of the type (b) provide the masses for the standard model fermions. With a running

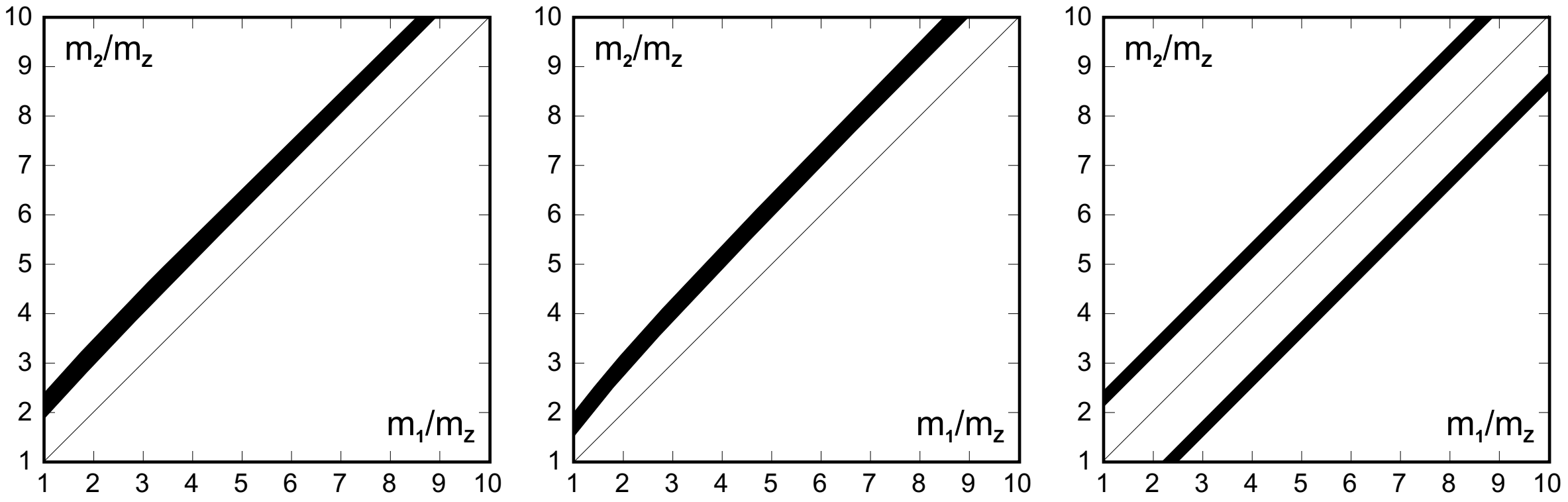

(a) Standard-model-like (I) (b) Integer charges (III) (c) Fractional charges (II)

Fig. 12.12: The black slabs mark the combination of masses favoured by electroweak precision data [39, 40] at the 68% level of confidence based on the conservative perturbative estimate. $m_1$ ($m_2$) denotes the mass of the fourth family lepton with the higher (lower) charge. The black stripes do not correspond exactly to the overlap of the parabolic area with the 68% ellipse in Fig. 12.2 of the Introduction but with a polygonal area defined by $-0.1 < S + T < +0.5$, $-0.15 < S - T < +0.025$, and $S < 0.22$.

TC sector (see Introduction, Fig. 12.1, left panel) the one-loop estimate for the mass, $m_{\rm t}$, is given by [35]

$$m_{\rm t} = \frac{N}{4\pi^2}\frac{\Sigma(0)^2}{\Lambda^2_{ETC}}\Sigma(0), \tag{12.9}$$

where $\Sigma(0)$ stands for the self energy of the techniquark at zero momentum and is of the order of 1 TeV. In a walking TC theory, where the scale $\Lambda_{\rm ETC}$ can be larger than $\Sigma(0)$ (see Introduction, Fig. 12.1, right panel), the attainable mass, $m_{\rm t}$, is enhanced by a factor of up to $\Lambda_{\rm ETC}/\Sigma(0)$ [19, 35],

$$m_{\rm t} = \frac{N}{4\pi^2}\frac{\Sigma(0)}{\Lambda_{ETC}}\Sigma(0). \tag{12.10}$$

Therefore, the top quark mass can still be generated if $\Lambda_{\rm ETC}$ = 4-8 TeV [35], whereas, in the running case, it is already difficult to reach it with $\Lambda_{\rm ETC}$ = 1 TeV. Thus, $\Lambda_{\rm ETC}$ may be sufficiently large to suppress terms of type (c). Terms of type (a) are even more enhanced than terms of type (b) and provide very large masses for additional (pseudo) Goldstone bosons, which do not become the longitudinal degrees of freedom of the gauge bosons and which might be present depending on the symmetry breaking pattern. Thereby their direct detection is impeded.

After these general considerations we consider below a concrete ETC model [35], which generates the mass of the top quark. It contains the type (I) S(2,2) model. The matter fields are arranged in a SU(7) multiplet and a SU(4) right-handed part [35],

$$\mathrm{SU}(7): \left\{\left[\begin{pmatrix} U^a \\ D^a \end{pmatrix}_L, \begin{pmatrix} N \\ E \end{pmatrix}_L, \begin{pmatrix} t^c \\ b^c \end{pmatrix}_L\right], [U^a_R\,,\ N_R\,,\ t^c_R]\right\} \quad \times \quad \mathrm{SU}(4): \{[D^a_R\,, E_R]\}. \tag{12.11}$$

$a$ and $c$ are the proto TC and QCD indices, respectively. The symmetry breaking pattern is

$$\mathrm{SU}(7)\times\mathrm{SU}(4)\times\mathrm{SU}(3)\to\mathrm{SO}(3)_{\rm TC}\times\mathrm{SU}(3)_{\rm QCD}, \tag{12.12}$$

which exploits the fact that the fundamental representation of SO(3) coincides with the adjoint of SU(2). Here the bottom quark is excluded from the second multiplet because its mass is negligible compared to the top mass. It will be generated by the breaking of a larger ETC group at higher energies, which will certainly lie outside the reach of experiments of the near future. The specific arrangement of the particles implies that, if the top mass is generated, the mass gap between U and D as well as between N and E must be of the same order. This is consistent with the size of the mass gap favored by data presented in reference [39]. Especially, in view of the findings of [123] the above scheme represents a viable candidate for a residual ETC model. The same conclusion is obtained from direct calculation of the oblique parameters in [35].

#### *12.3.3 A different variant, S(3,2)*

The S(2,2) model with two technicolours and two techniflavours in the two-index symmetric representation represents the scenario favoured by data. The runner-up is the S(3,2) model (three technicolours, two techniflavours). Its perturbative contribution to the $S$ parameter equals $1/\pi$, which is exactly twice the value of the S(2,2) model. Thereby, according to [39], S(3,2) lies still within two standard deviations from the mean value even before taking into account non-perturbative reductions. The set-up S(3,2) is not as close to the conformal window as S(2,2), whence a heavier Higgs with $m_H = 170 - 300$ GeV [10] is expected. The symmetric representation of SU(3) is of dimension six. Hence, from the point of view of the electroweak sector, the family of techniquarks comes in an even number of (colour-)copies, which does not trigger a topological Witten anomaly even without a fourth family of leptons. It does not feature the enhanced flavour symmetry $\mathrm{SU}(2\mathrm{N}_\mathrm{f})$ of the adjoint S(2,2), whence the breaking of S(3,2) only leads to the three Goldstone bosons, which represent the longitudinal degrees of freedom of the electroweak gauge bosons. In the case of S(2,2) one has a total of nine, that is six unabsorbed Goldstone bosons [127], because of the enhanced symmetry [SU(2$N_f$=4)] for the adjoint matter (for more details see again Section 12.4). Hence, for energies below the technicolour scale $\Lambda_{\mathrm{TC}}$, which might be larger than 1 TeV, S(3,2) looks identical to the standard model; apart from a higher but not inconsistent contribution to the $S$ parameter.

### 12.4 Minimal walking technicolor: effective theories and dark matter

*Sven Bjarke Gudnason and Chris Kouvaris*

In this contribution we examine the phenomenological implications of the technicolor theory with two techniquarks transforming according to the adjoint representation of $SU(2)$ [127] (see Section 12.3). The theory predicts Goldstone bosons that carry nonzero technibaryon number. These technibaryons must acquire a mass from some, yet unspecified, theory at a higher scale. Since we assume a bottom up approach we postpone the problem of producing the underlying theory providing these masses, but we expect it to be similar to the ETC type theory proposed in [35] (see Sections 12.2 and 12.3). If the technibaryon number is left intact by the ETC interactions, the lightest technibaryon (LTB) is stable and the hypercharge assignment can be chosen in a way that the LTB is also electrically neutral. The mass of the LTB is expected to be of the order of the electroweak scale. Therefore it has many features required for a dark matter component.

In the first part of the contribution, we provide the associated linear and non-linear effective theories. In the second part of this contribution we consider the scenario of one of the Goldstone bosons to be a dark matter component. We assume one of the Goldstone bosons to be neutral and we calculate its contribution to the dark matter density.

#### *12.4.1 The model*

The new dynamical sector underlying the Higgs mechanism we consider is an $SU(2)$ technicolor gauge group with two adjoint technifermions. The theory is asymptotically free if the number of flavors $N_f <$ 2.75.

As it is shown in Ref. [8], the number of flavors $N_f = 2$ lies sufficiently close to the critical value for which an infrared stable fixed point emerges so the theory is a perfect candidate for a walking technicolor theory. Although the critical number of flavors is independent of the number of colors when keeping the underlying fermions in the adjoint representation of the gauge group, the electroweak precision measurements do depend on it. Since the lowest number of colors is privileged by data [10, 40] we choose the two-technicolor theory.

Then the two adjoint fermions may be written as

$$T_L^a = \begin{pmatrix} U^a \\ D^a \end{pmatrix}_L , \qquad U_R^a \, , \quad D_R^a \, , \qquad a = 1, 2, 3 \, , \tag{12.13}$$

with $a$ the adjoint technicolor index of $SU(2)$. The left fields are arranged in three doublets of the $SU(2)_L$ weak interactions in the standard fashion. The condensate is $\langle \bar{U}U + \bar{D}D \rangle$ which breaks spontaneously the electroweak symmetry.

Our additional matter content is essentially a copy of a standard model fermion family with quarks (here transforming in the adjoint of $SU(2)$) and the following lepton doublet in order to cancel Witten's global anomaly

$$\mathcal{L}_L = \begin{pmatrix} N \\ E \end{pmatrix}_L , \qquad N_R \, , \, E_R \, . \tag{12.14}$$

Since we do not wish to disturb the walking nature of the technicolor dynamics, the doublet (12.14) must be a technicolor singlet [10]. In general, the gauge anomalies cancel using the following generic hypercharge assignment

$$Y(T_L) = \frac{y}{2} \, , \qquad Y(U_R, D_R) = \left( \frac{y+1}{2}, \frac{y-1}{2} \right) \, , \tag{12.15}$$

$$Y(\mathcal{L}_L) = -\, 3\frac{y}{2} \, , \qquad Y(N_R, E_R) = \left( \frac{-3y+1}{2}, \frac{-3y-1}{2} \right) \, , \tag{12.16}$$

where the parameter $y$ can take any real value. In our notation the electric charge is $Q = T_3 + Y$, where $T_3$ is the weak isospin generator. One recovers the SM hypercharge assignment for $y = 1/3$. In [35], the SM hypercharge has been investigated in the context of an extended technicolor theory. Another interesting choice of the hypercharge is $y = 1$, which has been investigated, from the point of view of the electroweak precision measurements, in [10, 40]. In this case

$$Q(U) = 1 \, , \quad Q(D) = 0 \, , \quad Q(N) = -1 \, , \quad \text{and} \quad Q(E) = -2 \, , \quad \text{with} \quad y = 1 \, . \tag{12.17}$$

Notice that in this particular hypercharge assignment, the $D$ technidown is electrically neutral.

Since we have two Dirac fermions in the adjoint representation of the gauge group, the global symmetry is $SU(4)$. To discuss the symmetry properties of the theory it is convenient to use the Weyl base for the fermions and arrange them in the following vector transforming according to the fundamental representation of $SU(4)$

$$Q = \begin{pmatrix} U_L \\ D_L \\ -i\sigma^2 U_R^* \\ -i\sigma^2 D_R^* \end{pmatrix} , \tag{12.18}$$

where $U_L$ and $D_L$ are the left handed techniup and technidown respectively and $U_R$ and $D_R$ are the corresponding right handed particles. Assuming the standard breaking to the maximal diagonal subgroup, the $SU(4)$ symmetry breaks spontaneously down to $SO(4)$. Such a breaking is driven by the following condensate

$$\langle Q_i^\alpha Q_j^\beta \epsilon_{\alpha\beta} E^{ij} \rangle = -2 \langle \overline{U}_R U_L + \overline{D}_R D_L \rangle \, , \tag{12.19}$$

where the indices $i, j = 1, \ldots, 4$ denote the components of the tetraplet of $Q$, and the Greek indices indicate the ordinary spin. The matrix $E$ is a $4 \times 4$ matrix defined as $E = \sigma^1 \otimes \mathbb{1}$ , where $\mathbb{1}$ is the 2-dimensional unit matrix. We have used $\epsilon_{\alpha\beta} = -i\sigma^2_{\alpha\beta}$ and $\langle U_L^{\alpha} {U_R}^{*\beta} \epsilon_{\alpha\beta}\rangle = -\langle \overline{U}_R U_L\rangle$. A similar expression holds for the $D$ techniquark. The above condensate is invariant under an $SO(4)$ symmetry.

In terms of the underlying degrees of freedom, and focusing only on the techniflavor symmetries, there are nine Goldstone bosons, three of which, transforming like

$$\overline{D}_R U_L \; , \qquad \overline{U}_R D_L \; , \qquad \frac{1}{\sqrt{2}}(\overline{U}_R U_L - \overline{D}_R D_L) \; , \tag{12.20}$$

will be eaten by the longitudinal components of the massive electroweak gauge bosons. The electric charge is respectively one, minus one and zero. For the other six Goldstone bosons we have

$$U_L U_L, \quad D_L D_L, \quad U_L D_L \; , \qquad \text{with electric charges} \quad y+1, \; y-1, \; y \; , \tag{12.21}$$

together with the associated anti-particles. These Goldstone bosons (Eq. (12.21)) are di-technibaryons carrying technibaryon number. The technibaryon generator can be identified with one of the generators of $SU(4)$.

### *12.4.2 Effective theories*

While the leptonic sector can be described within perturbation theory since it interacts only via electroweak interactions, the situation for the techniquarks is more involved since they combine into composite objects interacting strongly among themselves. It is therefore useful to construct low energy effective theories encoding the basic symmetry features of the underlying theory. We construct the linearly and nonlinearly realized low energy effective theories for our underlying theory. The theories we will present can be used to investigate relevant processes of interest at LHC and LC.

#### 12.4.2.1 The linear realization

The relevant effective theory for the Higgs sector at the electroweak scale consists, in our model, of a light composite Higgs and nine Goldstone bosons. These can be assembled in the matrix

$$M = \left(\frac{\sigma}{2} + i\sqrt{2}\Pi^a\, X^a\right) E \; , \tag{12.22}$$

which transforms under the full $SU(4)$ group according to $M \rightarrow uMu^T$, with $u \in SU(4)$ , and $X^a$ are the generators of the $SU(4)$ group which do not leave invariant the vacuum expectation value $\langle M\rangle = vE/2$ .

It is convenient to separate the fifteen generators of $SU(4)$ into the six that leave the vacuum invariant ($S^a$) and the other nine that do not ($X^a$). One can show that the $S^a$ generators of the $SO(4)$ subgroup satisfy the following relation

$$S^a\, E + E\, S^{aT} = 0 \; , \qquad \text{with} \qquad a = 1, \ldots, 6 \; . \tag{12.23}$$

The electroweak subgroup can be embedded in $SU(4)$, as explained in detail in [33]. The main difference here is that we have a more general definition of the hypercharge. The electroweak covariant derivative is

$$D_\mu M = \partial_\mu M - i\, g \left[G_\mu M + M G_\mu^T\right] \; , \tag{12.24}$$

with

$$G_\mu = \left( \begin{array}{cc} W_\mu & 0 \\ 0 & -\frac{g'}{g} B_\mu^T \end{array} \right) + \frac{y}{2}\frac{g'}{g} B_\mu \left( \begin{array}{cc} 1 & 0 \\ 0 & -1 \end{array} \right) \; , \quad W_\mu = W_\mu^a \frac{\tau^a}{2} \; , \quad B_\mu^T = B_\mu \frac{\tau^{3^T}}{2} = B_\mu \frac{\tau^3}{2} \; , \tag{12.25}$$

where $\tau^a$ are the Pauli matrices. The generators satisfy the normalization conditions $\text{Tr}[X^a\, X^b] = \delta^{ab}/2$, $\text{Tr}[S^a\, S^b] = \delta^{ab}/2$ and $\text{Tr}[SX] = 0$. Three of the Goldstone bosons in the unitary gauge, are absorbed in the longitudinal degrees of freedom of the massive weak gauge bosons while the extra six Goldstone bosons will acquire a mass due to extended technicolor interactions as well as the electroweak interactions per se. Assuming a bottom up approach we will introduce by hand a mass term for the Goldstone bosons. The new Higgs Lagrangian is then

$$\begin{aligned} L &= \frac{1}{2}\text{Tr}\left[D_\mu M D^\mu M^\dagger\right] + \frac{m^2}{2}\text{Tr}[MM^\dagger] \\ &- \frac{\lambda}{4}\text{Tr}\left[MM^\dagger\right]^2 - \widetilde{\lambda}\text{Tr}\left[MM^\dagger MM^\dagger\right] - \frac{1}{2}\,\Pi_a(M^2_{\text{ETC}})^{ab}\Pi_b\,, \end{aligned} \tag{12.26}$$

with $m^2 > 0$ and $a$ and $b$ running over the six uneaten Goldstone bosons. The matrix $M^2_{ETC}$ is dynamically generated and parametrizes our ignorance about the underlying extended technicolor model, yielding the specific mass texture. The pseudo Goldstone bosons are expected to acquire a mass of the order of a TeV. Direct and computable contributions from the electroweak corrections break $SU(4)$ explicitly down to $SU(2)_L \times SU(2)_R$ yielding an extra contribution to the uneaten Goldstone bosons. However the main contribution comes from the ETC interactions.

We stress that the expectation of a light composite Higgs relies on the assumption that the quantum chiral phase transition as function of number of flavors near the nontrivial infrared fixed point is smooth and possibly of second order [4]. The composite Higgs Lagrangian is a low energy effective theory and higher dimensional operators will also be phenomenologically relevant.

#### 12.4.2.2 The non-linearly realized effective theory

One can always organize the low energy effective theory in a derivative expansion. The best way is to make use of the exponential map

$$U = \exp\left(i\frac{\Pi^a X^a}{F}\right)E\,, \tag{12.27}$$

where $\Pi^a$ represent the 9 Goldstone bosons and $X^a$ are the 9 generators of $SU(4)$ that do not leave the vacuum invariant. To introduce the electroweak interactions one simply adopts the same covariant derivative used for the linearly realized effective theory, see Eqs. (12.24)–(12.25).

The associated non-linear effective Lagrangian reads

$$L = \frac{F^2}{2}\text{Tr}\left[D_\mu U D^\mu U^\dagger\right] - \frac{1}{2}\Pi_a(M^2_{\text{ETC}})^{ab}\Pi_b\,. \tag{12.28}$$

Still the mass squared matrix parametrizes our ignorance about the underlying ETC dynamics.

A common ETC mass for all the pseudo Goldstone bosons carrying baryon number can be provided by adding the following term to the previous Lagrangian

$$2C\text{Tr}\left[UBU^\dagger B\right] + C = \frac{C}{4F^2}\sum_{i=1}^{6}\Pi^i_B\Pi^i_B\,, \quad \text{with} \quad B = \frac{1}{2\sqrt{2}}\begin{pmatrix} \mathbb{1} & 0 \\ 0 & -\mathbb{1}\end{pmatrix}. \tag{12.29}$$

Dimensional analysis requires $C \propto \Lambda^6_{TC}/\Lambda^2_{ETC}$. A similar term can be added to the linearly realized version of our theory.

[4] We have provided supporting arguments for this picture in [10] where the reader will find also a more general discussion of this issue and possible pitfalls.

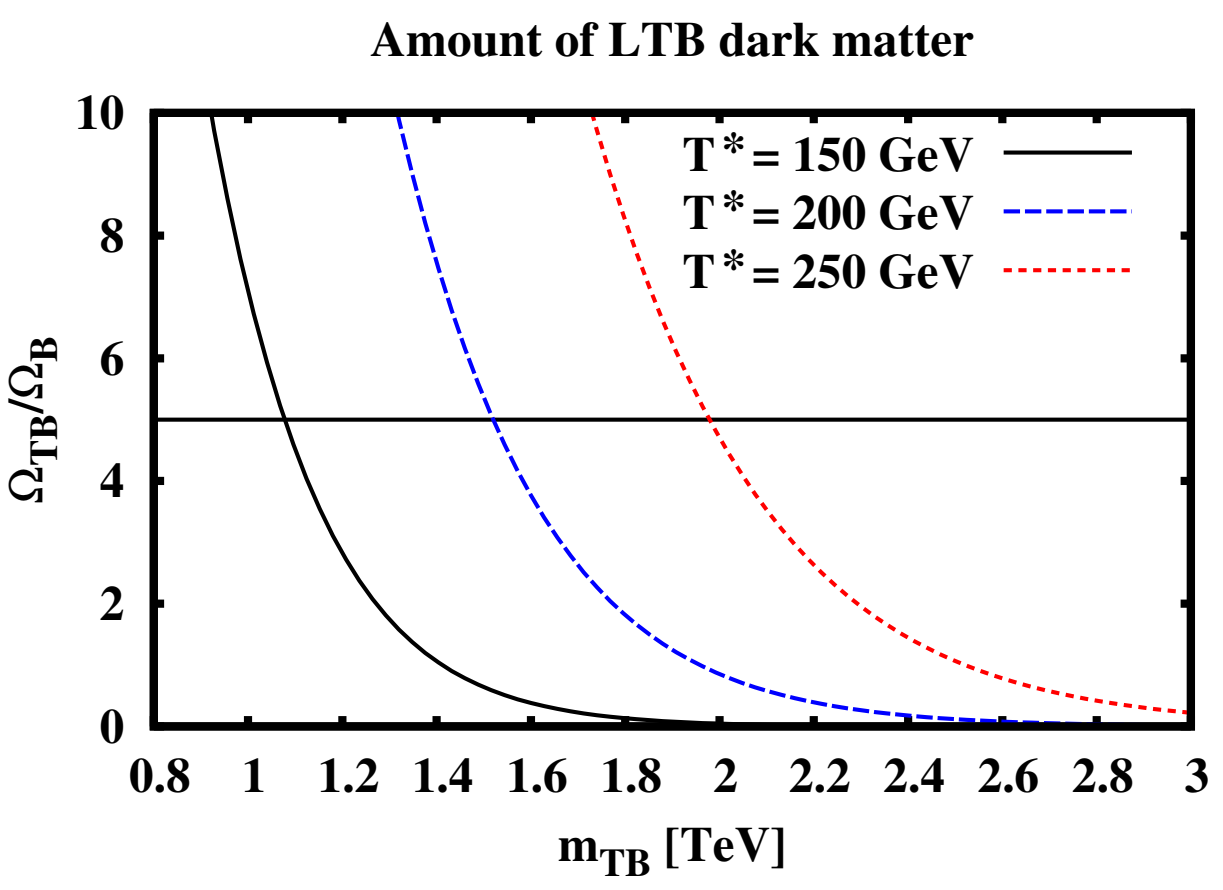

Fig. 12.13: The fraction of technibaryon matter density over the baryonic one as function of the technibaryon mass. The desired value of $\Omega_{TB}/\Omega_B \sim 5$ depends on the lightest technibaryon mass and the value of $T^*$.

### *12.4.3 The dark-side of the $5^{\rm th}$ force*

According to the choice of the hypercharge there are two distinct possibilities for a dark matter candidate. If we assume the SM-like hypercharge assignment for the techniquarks and the new lepton family, the new heavy neutrino can be an interesting dark matter candidate. For that, it must be made sufficiently stable by requiring no flavor mixing with the lightest generations and be lighter than the unstable charged lepton [10]. This possibility is currently under investigation [128]. However, we can also consider another possibility. We can choose the hypercharge assignment in such a way that one of the pseudo Goldstone bosons does not carry electric charge. The dynamics providing masses for the pseudo Goldstone bosons may be arranged in a way that the neutral pseudo Goldstone boson is the LTB. If conserved by ETC interactions the technibaryon number protects the lightest baryon from decaying. Since the masses of the technibaryons are of the order of the electroweak scale, they may constitute interesting sources of dark matter. Some time ago in a pioneering work Nussinov [129] suggested that, in analogy with the ordinary baryon asymmetry in the Universe, a technibaryon asymmetry is a natural possibility. A new contribution to the mass of the Universe then emerges due to the presence of the LTB. It is useful to compare the fraction of technibaryon mass $\Omega_{TB}$ to baryon mass $\Omega_B$ in the Universe

$$\frac{\Omega_{TB}}{\Omega_B} = \frac{TB}{B}\frac{m_{TB}}{m_p} , \tag{12.30}$$

where $m_p$ is the proton mass, $m_{TB}$ is the mass of the LTB. $TB$ and $B$ are the technibaryon and baryon number densities, respectively.

Knowing the distribution of dark matter in the galaxy, earth based experiments can set stringent limits on the physical features of the dominant component of dark matter [130]. Such a distribution, however, is not known exactly [131] and it depends on the number of components and type of dark matter. In order to determine few features of our LTB particle we make the oversimplified approximation in which our LTB constitutes the whole dark matter contribution to the mass of the Universe. In this limit the previous ratio should be around 5 [132]. By choosing in our model the hypercharge assignment $y = 1$ the lightest neutral Goldstone boson is the state consisting of the $DD$ techniquarks. The fact that it is charged under $SU(2)_L$ makes it probably detectable in Ge detectors [133].

It is well known that weak anomalies violate the baryon and the lepton number. More precisely, weak processes violate $B + L$, while they preserve $B - L$. Similarly, the weak anomalies violate also the technibaryon number, since technibaryons couple weakly. The weak technibaryon-, lepton- and baryon- number violating effects are highly suppressed at low temperatures while they are enhanced at temperatures comparable to the critical temperature of the electroweak phase transition [134]. With $T^*$ we define the temperature below which the sphaleron processes cease to be important. This temperature is not exactly known but it is expected to be in the range between $150 - 250$ GeV [134].

Following early analysis [135, 136] we have performed a careful computation of $\Omega_{TB}/\Omega_B$ within our model. Imposing thermal equilibrium, electric neutrality condition and the presence of a continuous electroweak phase transition ($T^*$ now is below the critical temperature) we find:

$$\frac{TB}{B} = \frac{11}{36}\sigma_{TB}\left(\frac{m_{TB}}{T^*}\right) , \tag{12.31}$$

with $\sigma_{TB}$ the statistical weight function

$$\sigma_{TB}\left(\frac{m_{TB}}{T^*}\right) = \frac{3}{2\pi^2}\int_0^\infty dx\, x^2 \sinh^{-2}\left(\frac{1}{2}\sqrt{x^2 + \left(\frac{m_{TB}}{T^*}\right)^2}\right) . \tag{12.32}$$

In the previous estimate (12.31) the LTB is taken to be lighter then the other technibaryons and the new lepton number is violated. We have, however, considered different scenarios and various limits which will be reported in [137]. Our results are shown in Fig. 12.13. The desired value of the dark matter fraction in the Universe can be obtained for a LTB mass of the order of a TeV for quite a wide range of values of $T^*$. The only free parameter in our analysis is essentially the mass of the LTB which is ultimately provided by ETC interactions.

## 12.5 Associate production of a light composite Higgs at the LHC

*Alfonso R. Zerwekh*

Very recently a new kind of technicolor models has been proposed [10] whose main characteristic is that technifermions are not in the fundamental representation of the technicolor group. In these models the walking behavior of the coupling constant appears naturally and they are not in conflict with the current limits on the oblique parameters. But the most remarkable feature of these models is that they predict the existence of a light composite Higgs with a mass around 150 GeV.

Inspired by such models, we write down an effective Lagrangian which describes the Standard Model with a light Higgs and vector resonances [78] which are a general prediction of dynamical symmetry breaking models [138]. The model is minimal in the sense that we assume that any other composite state would be heavier than the vector resonances, and so they are not taken into account, and there are no physical technipions in the spectrum.

We start by noticing that, in general, dynamical electroweak symmetry breaking models predict the existence of composite vector particles (the so called technirho and techniomega) that mix with the gauge bosons of the Standard Model. In order to describe this mixing, we use a generalization of Vector Meson Dominance [139] introduced in [63] and developed in [140]. In this approach we choose a representation where all vector fields transform as gauge fields and mix through a mass matrix. On the other hand, gauge invariance imposes that the mass matrix has a null determinant. In our case, the Lagrangian for the gauge sector can be written as:

$$\begin{aligned}\mathcal{L} \;=\; & -\frac{1}{4}W^a_{\mu\nu}W^{a\mu\nu} - \frac{1}{4}\tilde{\rho}^a_{\mu\nu}\tilde{\rho}^{a\mu\nu} + \frac{M^2}{2}\left(\frac{g^2}{g_2^2}W^a_\mu W^{a\mu} + \tilde{\rho}^a_\mu\tilde{\rho}^{a\mu} - \frac{2g}{g_2}W^a_\mu\tilde{\rho}^{a\mu}\right) \\ & -\frac{1}{4}B_{\mu\nu}B^{\mu\nu} - \frac{1}{4}\tilde{\omega}_{\mu\nu}\tilde{\omega}^{\mu\nu} + \frac{M'^2}{2}\left(\frac{g'^2}{g_2'^2}B_\mu B^\mu + \tilde{\omega}_\mu\tilde{\omega}^\mu - \frac{2g'}{g_2'}B_\mu\tilde{\omega}^\mu\right)\end{aligned} \tag{12.33}$$

Notice that our Lagrangian is written in terms of non-physical fields. The physical ones will be obtained by diagonalizing the mass matrix.

By construction, Lagrangian (12.33) is invariant under $SU(2)_{\rm L} \times U(1)_{\rm Y}$. The symmetry breaking to $U(1)_{\rm em}$ will be described by mean of the vacuum expectation value of a scalar field, as in the Standard Model. In other words, we will use an effective gauged linear sigma model as a phenomenological description of the electroweak symmetry breaking.

As usual, fermions are minimally coupled to gauge bosons through a covariant derivative. Because in our scheme all the vector bosons transform as gauge fields, it is possible to include the proto-technirho and the proto-techniomega in an "extended" covariant derivative [140], resulting in the following Lagrangian for the fermion sector:

$$\mathcal{L} = \bar{\psi_L} i\gamma^\mu D_\mu \psi_L + \bar{\psi_R} i\gamma^\mu \tilde{D}_\mu \psi_R \tag{12.34}$$

with

$$D_\mu = \partial_\mu + i\tau^a g(1-x_1)W^a_\mu + i\tau^a g_2 x_1 \tilde{\rho}^a_\mu + i\frac{Y}{2}g'(1-x_2)B_\mu + i\frac{Y}{2}g'_2 x_2 \tilde{\omega}_\mu$$

and

$$\tilde{D}_\mu = \partial_\mu + i\frac{Y}{2}g'(1-x_3)B_\mu + i\frac{Y}{2}g'_2 x_3 \tilde{\omega}_\mu \tag{12.35}$$

Although a direct coupling between fermions and the vector resonances can appear naturally in technicolor due to extended technicolor interactions we will set $x_i = 0$ ($i = 1, 2, 3$), for simplicity.

In our effective model the Higgs sector is assumed to be the same as in the Standard Model. We avoid the possibility of a direct coupling between the vector resonances and the Higgs because, in principle, it can introduce dangerous tree level corrections to the $\rho$ parameter.

Once the electroweak symmetry has been broken, the mass matrix of the vector bosons takes contributions from (12.33) and from the Higgs mechanism. For the neutral vector bosons, the resulting mass matrix is (written in the basis $(W^3, \tilde{\rho}^3, B, \tilde{\omega})$ and assuming $M' = M$ and $g'_2 = g_2$):

$$\mathcal{M}_{\rm neutral} = \frac{v^2}{4}\begin{bmatrix} (1+\alpha)g^2 & -\alpha g g_2 & -gg' & 0 \\ -\alpha g g_2 & \alpha g_2^2 & 0 & 0 \\ -gg' & 0 & (1+\alpha)g'^2 & -\alpha g' g_2 \\ 0 & 0 & -\alpha g' g_2 & \alpha g'^2 \end{bmatrix} \tag{12.36}$$

where $\alpha = \frac{4M^2}{v^2 g_2^2}$.

On the other hand, the mass matrix for the charged vector bosons can be written as(written in the basis $(\tilde{W}^+, \tilde{\rho}^+)$ where $\tilde{W}^+ = (W^1 - iW^2)/\sqrt{2}$ and $\tilde{\rho}^+ = (\tilde{\rho}^1 - i\tilde{\rho}^2)/\sqrt{2}$):

$$\mathcal{M}_{\rm charged} = \frac{v^2}{4}\begin{bmatrix} (1+\alpha)g^2 & -\alpha g g_2 \\ -\alpha g g_2 & \alpha g_2^2 \end{bmatrix} \tag{12.37}$$

After diagonalizing the mass matrix we can write the interactions in terms of the physical fields. The relevant Feynman rules for the associate production of a Higgs and a gauge boson, in the limit $g/g_2 \ll 1$, can be found in Table 12.1.

We compute the cross section of the associate production of a Higgs and a gauge boson at the LHC. We choose to work with $\alpha = 0.1$ because for values of $\alpha$ of this order, the vector resonances can be light (i.e. $M_\rho \approx 250$ GeV) while $g_2/g$ is still much bigger than one. As $\alpha$ approaches unity, the vector resonances became too heavy and their observation increasingly difficult. On the other hand, if $\alpha$ is too small the coupling of the vector resonances to the SM fields are suppressed.

In Fig. 12.14 we show the value of $(\sigma - \sigma_{\rm SM})/\sigma_{\rm SM}$ as a function of the mass of the technirho ($M_\rho$) for three values of the Higgs mass ($M_H = 115$ GeV (solid line),150 GeV (dashed line) and 200 GeV (dotted line)) for the process $pp \to HW^+$ at the LHC. Observe that in this case, the cross section is significatively enhanced with respect to the Standard Model when the technirho has a mass between

Table 12.1: Feynman Rules for the relevant couplings of the vector resonances for the associated production of a Higgs and a gauge bosons. The couplings of the $W^{\pm}$ and $Z$ to the quarks are identical, in our limit, to the SM.

| Fields in the vertex | Variational derivative of Lagrangian by fields |
|---|---|
| $H\ \omega^0_\mu\ Z_\nu$ | $\frac{1}{2}\frac{e^2 M_W\sqrt{\alpha}v}{{c_w}^3 M_\rho}g^{\mu\nu}$ |
| $H\ \rho^0_\mu\ Z_\nu$ | $-\frac{1}{2}\frac{e^2 M_W\sqrt{\alpha}v}{{c_w}^2 M_\rho s_w}g^{\mu\nu}$ |
| $H\ \rho^+_\mu\ W^-{}_\nu$ | $-\frac{1}{2}\frac{e^2 M_W\sqrt{\alpha}v}{M_\rho {s_w}^2}g^{\mu\nu}$ |
| $\bar{u}\ d\ \rho^+_\mu$ | $\frac{1}{8}\frac{e^2\sqrt{2}\sqrt{\alpha}V_{ud}v}{M_\rho {s_w}^2}(1-\gamma^5)\gamma^\mu$ |
| $\bar{u}\ u\ \omega^0_\mu$ | $\frac{1}{24}\frac{e^2\sqrt{\alpha}v}{{c_w}^2 M_\rho}\gamma^\mu\big((1-\gamma^5)+4(1+\gamma^5)\big)$ |
| $\bar{u}\ u\ \rho^0_\mu$ | $\frac{1}{8}\frac{e^2\sqrt{\alpha}v}{c_w M_\rho s_w}(1-\gamma^5)\gamma^\mu$ |

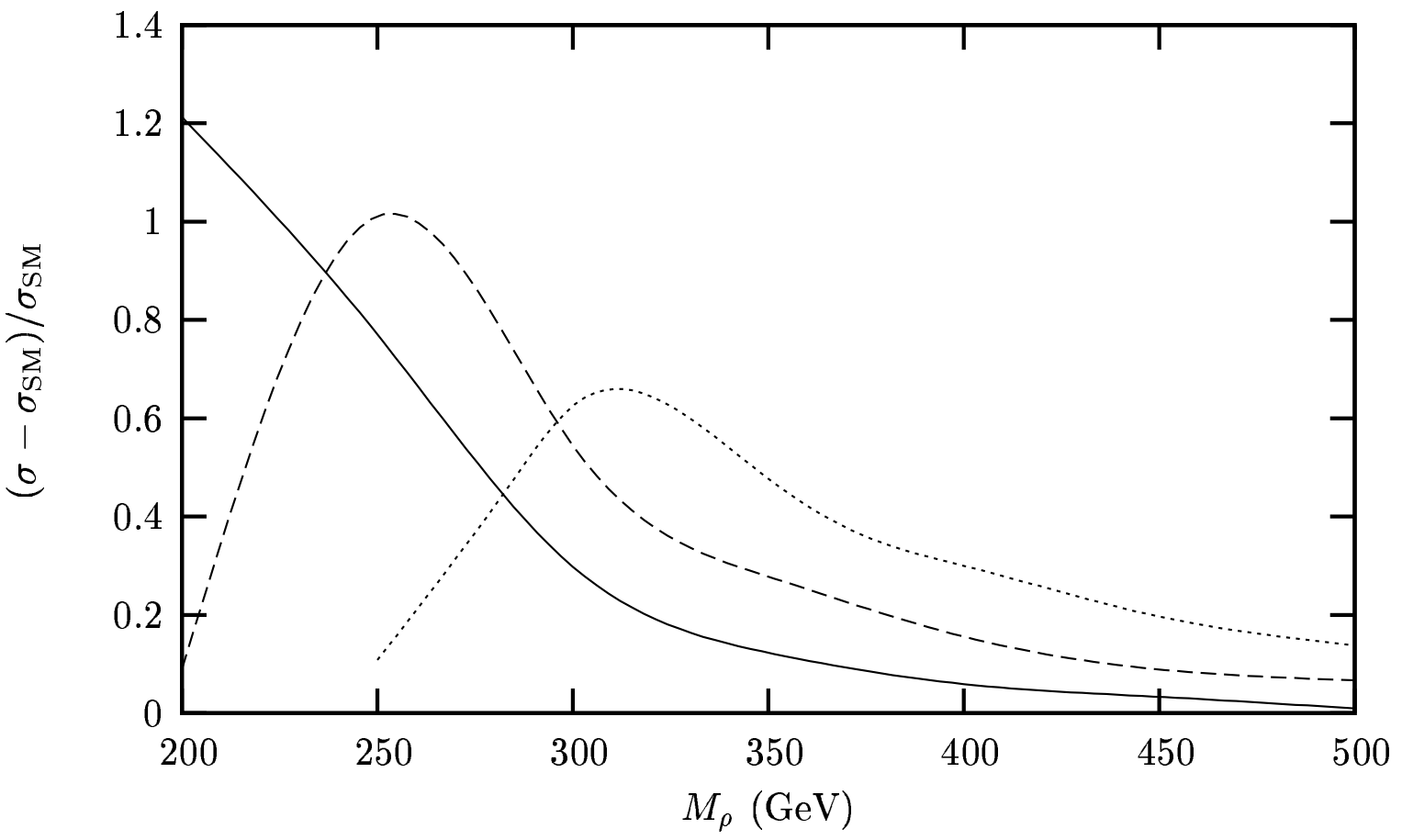

Fig. 12.14: Enhancement of the cross section, $(\sigma-\sigma_{\rm SM})/\sigma_{\rm SM}$, in the process $pp \to W^+H$ at the LHC for three values of the Higgs mass: $M_H = 115$ GeV (solid line), 150 GeV (dashed line) and 200 GeV (dotted line). In all cases we took $\alpha = 0.1$.

200 GeV and 350 GeV. The variation of this enhancement as a function of $\alpha$ is shown in Fig. 12.15 for $M_H = 150$ GeV and $g_2/g = 10$. On the other hand, when a Higgs and a $Z$ are produced (Fig. 12.16), the cross section is less enhanced and we expect that this channel will not be sensible to the presence of the vector resonances.

The point in the parameter space we use for studying our model was chosen in order to maximize the deviation from the Standard Model for the selected channel. This procedure allows us to evaluate the possibility of testing the model. Unfortunately, this point is disfavored by precision measurements. Nevertheless, we can be consistent with the constrains imposed by precision data by choosing $x_1 = 2(g/g_2)^2$ in (12.34). In this case, our results on $\sigma - \sigma_{\rm SM}$ are modified by a factor 0.60 and an important enhancement remains in the channel $pp \to HW^+$ for $M_\rho$ around 250 GeV.

In conclusion, we have constructed an effective Lagrangian which represents the Standard Model with a light (composite) Higgs boson and vector resonances that mix with the gauge bosons. We fixed the parameter of the model that connects the mass of the new vector bosons with their coupling constant, in such a way that the model is compatible with light resonances. The most obvious process for searching differences between our model and the predictions of the Standard Model is the associate production of a Higgs and a gauge boson. We found that the most sensitive channel is the production of the Higgs and $W$.

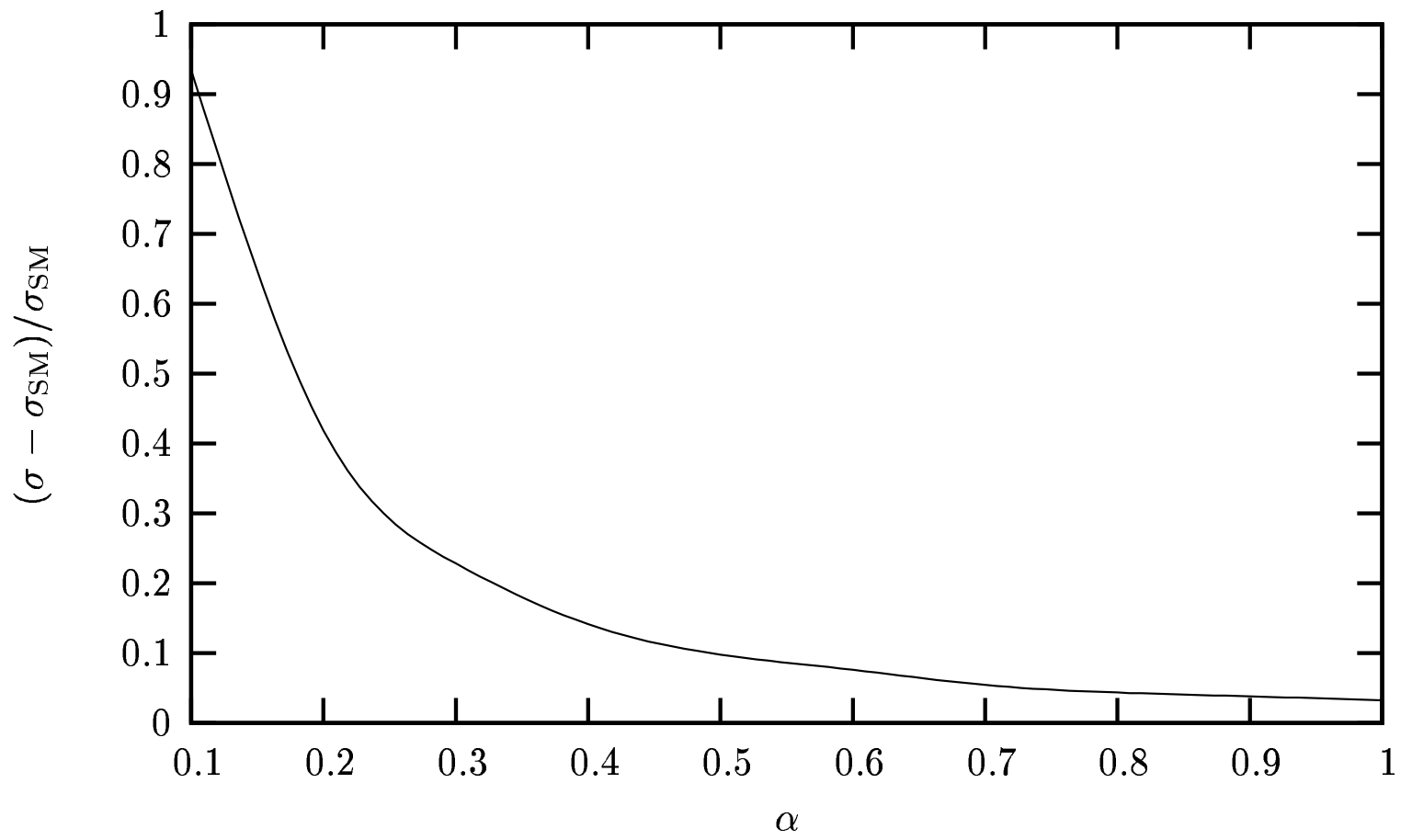

Fig. 12.15: Enhancement of the cross section in the process $p\bar{p} \to W^+H$ at the LHC as a function of $\alpha$ for $M_H = 150$ GeV and $g/g_2 = 0.1$.

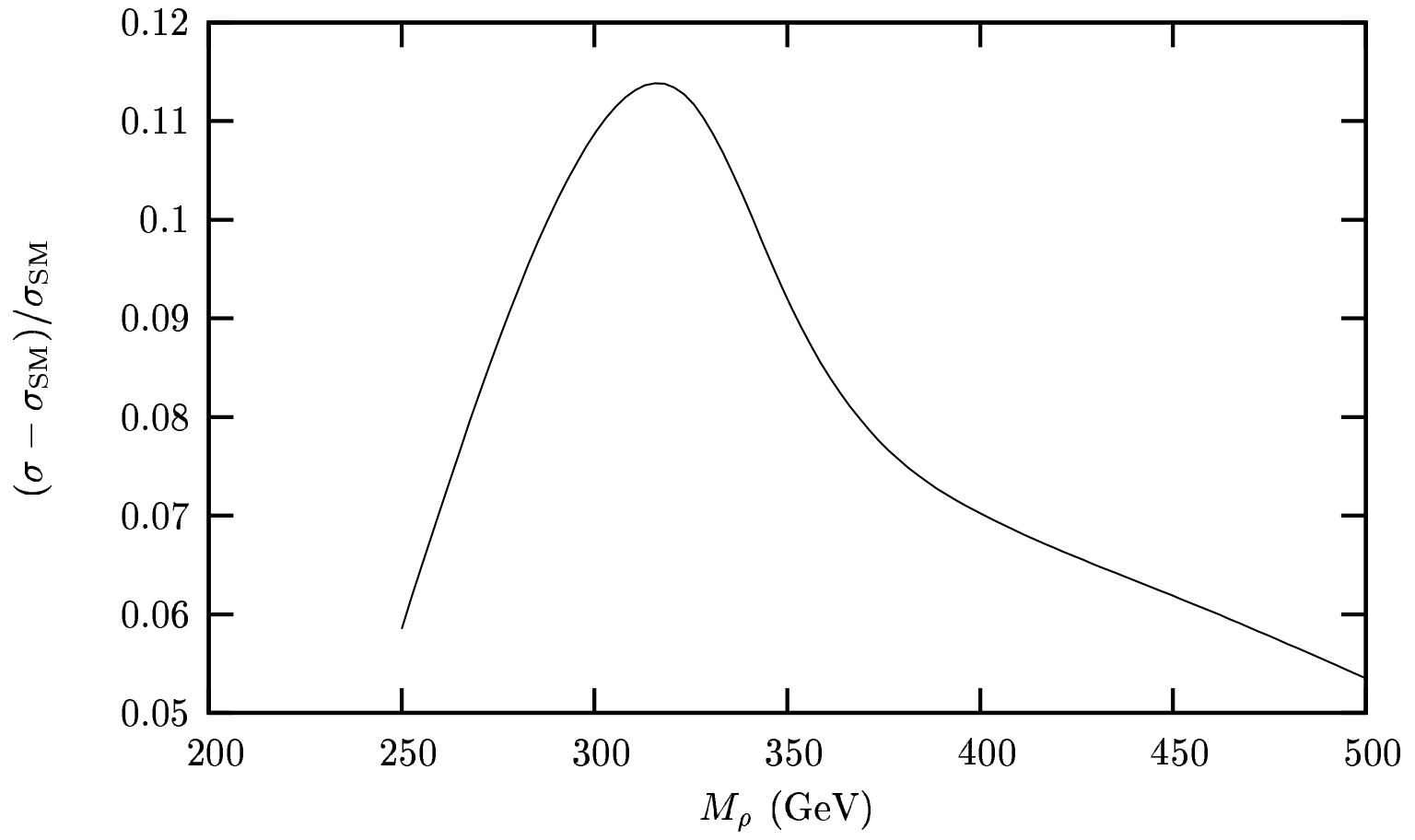

Fig. 12.16: Enhancement of the cross section in the process $pp \to ZH$ at the LHC for $M_H = 200$ GeV.

For a range of resonance's mass between 200 GeV and 350 GeV the enhancement of the cross section is significant at the LHC, depending on the Higgs mass. A relative light Higgs ($M_H = 115$ GeV) would be sensible to a technirho with mass between 200 GeV and 270 GeV while the production of a heavier Higgs ($M_H = 200$ GeV) would be enhanced by the presence of a technirho with a mass around 320 GeV.

### 12.6 Towards understanding the nature of electroweak symmetry breaking at hadron colliders: distinguishing technicolor and supersymmetry.

*Alexander Belyaev*

Many alternative models of electroweak symmetry breaking have spectra that include new scalar or pseudoscalar states whose masses could easily lie in the range to which Run II is sensitive. The new scalars tend to have cross-sections and branching fractions that differ from those of the SM Higgs. Here we discuss how to extract information about non-Standard theories of electroweak symmetry breaking from searches for a light SM Higgs at Tevatron Run II and CERN LHC. Ref. [141] studied the potential of Tevatron Run II to augment its search for the SM Higgs boson by considering the process $gg \to h_{SM} \to \tau^+\tau^-$. Authors determined what additional enhancement of scalar production and branching rate would enable a scalar to become visible in the $\tau^+\tau^-$ channel alone at Tevatron Run II. Similar work has been done for $gg \to h_{MSSM} \to \tau^+\tau^-$ at the LHC [142] and for $gg \to h_{SM} \to \gamma\gamma$ at the Tevatron [143] and LHC [144].

This contribution builds on these results and studies enhanced signals from (pseudo)scalar production in dynamical electroweak symmetry and supersymmetry considering an additional production mechanism (b-quark annihilation), more decay channels ($b\bar{b}$, $W^+W^-$, $ZZ$, and $\gamma\gamma$). We suggest the mass reach of the standard Higgs searches for each kind of non-standard scalar state. We also compare the key signals for the non-standard scalars across models and also with expectations in the SM, to show how one could identify which state has actually been found.

### *12.6.1 Models of electroweak symmetry breaking*

#### 12.6.1.1 Supersymmetry

One interesting possibility for addressing the hierarchy and triviality problems of the Standard Model is to introduce supersymmetry. In order to provide masses to both up-type and down-type quarks, and to ensure anomaly cancellation, the MSSM contains two Higgs complex-doublet superfields: $\Phi_d = (\Phi_d^0, \Phi_d^-)$ and $\Phi_u = (\Phi_u^+, \Phi_u^0)$ which aquire two vacuum expectation values $v_1$ and $v_2$ respectively. Out of the original 8 degrees of freedom, 3 serve as Goldstone bosons, absorbed into longitudinal components of the $W^\pm$ and $Z$, making them massive. The other 5 degrees of freedom remain in the spectrum as distinct scalar states, namely two neutral CP-even states($h$, $H$), one neutral, CP-odd state ($A$) and a charged pair ($H^\pm$). It is conventional to choose $\tan\beta = v_1/v_2$ and $M_A = \sqrt{M_{H^\pm}^2 - M_W^2}$ to define the SUSY Higgs sector. There are following tree-level relations between Higgs masses which will be useful for understanding enhanced Higgs boson interactions with fermions:

$$M_{h,H}^2 = \frac{1}{2}\left[(M_A^2 + M_Z^2) \mp \sqrt{(M_A^2 + M_Z^2)^2 - 4M_A^2 M_Z^2 \cos^2 2\beta}\right],$$
$$\cos^2(\beta - \alpha) = \frac{M_h^2(M_Z^2 - M_h^2)}{M_A^2(M_H^2 - M_h^2)}, \tag{12.38}$$

where $\alpha$ is the mixing angle of CP-even Higgs bosons. The Yukawa interactions of the Higgs fields with the quarks and leptons can be written as:

$$Y_{ht\bar{t}}/Y_{ht\bar{t}}^{SM} = \cos\alpha/\sin\beta, \quad Y_{Ht\bar{t}}/Y_{ht\bar{t}}^{SM} = \sin\alpha/\sin\beta, \quad Y_{At\bar{t}}/Y_{ht\bar{t}}^{SM} = \cot\beta,$$
$$Y_{hb\bar{b}}/Y_{hb\bar{b}}^{SM} = -\sin\alpha/\cos\beta, \quad Y_{Hb\bar{b}}/Y_{hb\bar{b}}^{SM} = \cos\alpha/\cos\beta, \quad Y_{Ab\bar{b}}/Y_{hb\bar{b}}^{SM} = \tan\beta, \tag{12.39}$$

relative to the Yukawa couplings of the Standard Model ($Y_{hf\bar{f}}^{SM} = m_f/v$). Once again, the same pattern holds for the tau lepton's Yukawa couplings as for those of the $b$ quark. There are several circumstances under which various Yukawa couplings are enhanced relative to Standard Model values. For high $\tan\beta$ (small $\cos\beta$), Eqs. (12.39) show that the interactions of all neutral Higgs bosons with the down-type fermions are enhanced by a factor of $1/\cos\beta$. In the decoupling limit, where $M_A \to \infty$, applying Eq. (12.38) to Eqs. (12.39) shows that the $H$ and $A$ Yukawa couplings to down-type fermions are enhanced by a factor of $\simeq \tan\beta$. Conversely, for low $M_A \simeq M_h$, one can check that $Y_{hb\bar{b}}/Y_{hb\bar{b}}^{SM} = Y_{h\tau\bar{\tau}}/Y_{h\tau\bar{\tau}}^{SM} \simeq \tan\beta$ and that $h$ and $A$ Yukawas are enhanced instead.

#### 12.6.1.2 Technicolor

Another intriguing class of theories, dynamical electroweak symmetry breaking (DEWSB), supposes that the scalar states involved in electroweak symmetry breaking could be manifestly composite at scales not much above the electroweak scale $v \sim 250$ GeV. In these theories, a new asymptotically free strong gauge interaction (technicolor [1, 2, 145]) breaks the chiral symmetries of massless fermions $f$ at a scale $\Lambda \sim 1$ TeV. If the fermions carry appropriate electroweak quantum numbers (e.g. left-hand (LH) weak doublets and right-hand (RH) weak singlets), the resulting condensate $\langle \bar{f}_L f_R \rangle \neq 0$ breaks the electroweak symmetry as desired. Three of the Nambu-Goldstone Bosons (technipions) of the chiral symmetry breaking become the longitudinal modes of the $W$ and $Z$. The logarithmic running of the strong gauge coupling renders the low value of the electroweak scale natural. The absence of fundamental scalars obviates concerns about triviality. For details, we refer the reader to section 12.1.

Many models of DEWSB have additional light neutral pseudo Nambu-Goldstone bosons which could potentially be accessible to a standard Higgs search; these are called "technipions" in technicolor models. Our analysis will assume, for simplicity, that the lightest PNGB state is significantly lighter than other neutral (pseudo) scalar technipions, so as to heighten the comparison to the SM Higgs boson. The specific models we examine are: 1) the traditional one-family model [146] with a full family of techniquarks and technileptons, 2) a variant on the one-family model [147] in which the lightest technipion contains only down-type technifermions and is significantly lighter than the other pseudo Nambu-Goldstone bosons, 3) a multiscale walking technicolor model [37] designed to reduce flavor-changing neutral currents, and 4) a low-scale technicolor model (the Technicolor Straw Man model) [53] with many weak doublets of technifermions, in which the second-lightest technipion $P'$ is the state relevant for our study (the lightest, being composed of technileptons, lacks the anomalous coupling to gluons required for $gg \to P$ production). For simplicity the lightest relevant neutral technipion of each model will be generically denoted $P$; where a specific model is meant, a superscript will be used. One of the key differences among these models is the value of the technipion decay constant $F_P$, which is related to the number $N_D$ of weak doublets of technifermions that contribute to electroweak symmetry breaking. We refer the reader to [148] for details.

### *12.6.2 Results for each model*

#### 12.6.2.1 Supersymmetry

Let us consider how the signal of a light Higgs boson could be changed in the MSSM, compared to expectations in the SM. There are several important sources of alterations in the predicted signal, some of which are interconnected. First, the MSSM includes three neutral Higgs bosons $\mathcal{H} = (h, H, A)$ states. The apparent signal of a single light Higgs could be enhanced if two or three neutral Higgs species are nearly degenerate. Second, the alterations of the couplings between Higgs bosons and ordinary fermions in the MSSM can change the Higgs decay widths and branching ratios relative to those in the SM. Radiative effects on the masses and couplings can substantially alter decay branching fractions in a non-universal way. For instance, $B(h \to \tau^+\tau^-)$ could be enhanced by up to an order of magnitude due to the suppression of $B(h \to b\bar{b})$ in certain regions of parameter space [149, 150]. However, this gain in branching fraction would be offset to some degree by a reduction in Higgs production through channels involving $Y_{\mathcal{H}b\bar{b}}$ [141]. Third, a large value of $\tan\beta$ enhances the bottom-Higgs coupling (Eqs. (12.39) ), making gluon fusion through a $b$-quark loop significant, and possibly even dominant over the top-quark loop contribution. Fourth, the presence of superpartners in the MSSM gives rise to new squark-loop contributions to Higgs boson production through gluon fusion. Light squarks with masses of order 100 GeV have been argued to lead to a considerable universal enhancement (as much as a factor of five) [151–154] for MSSM Higgs production compared to the SM. Finally, enhancement of the $Y_{\mathcal{H}b\bar{b}}$ coupling at moderate to large $\tan\beta$ makes $b\bar{b} \to \mathcal{H}$ a significant means of Higgs production in the MSSM – in contrast to the SM where it is negligible. To include both production channels when looking for a Higgs decaying as $\mathcal{H} \to xx$, we define a combined enhancement factor

$$\kappa^{\mathcal{H}}_{total/xx} = \frac{\sigma(gg \to \mathcal{H} \to xx) + \sigma(bb \to \mathcal{H} \to xx)}{\sigma(gg \to h_{SM} \to xx) + \sigma(bb \to h_{SM} \to xx)} \equiv [\kappa^{\mathcal{H}}_{gg/xx} + \kappa^{\mathcal{H}}_{bb/xx} R_{bb:gg}]/[1 + R_{bb:gg}]. \tag{12.40}$$

Here $R_{bb:gg}$ is the ratio of $b\bar{b}$ and $gg$ initiated Higgs boson production in the Standard Model, which can be calculated using HDECAY.

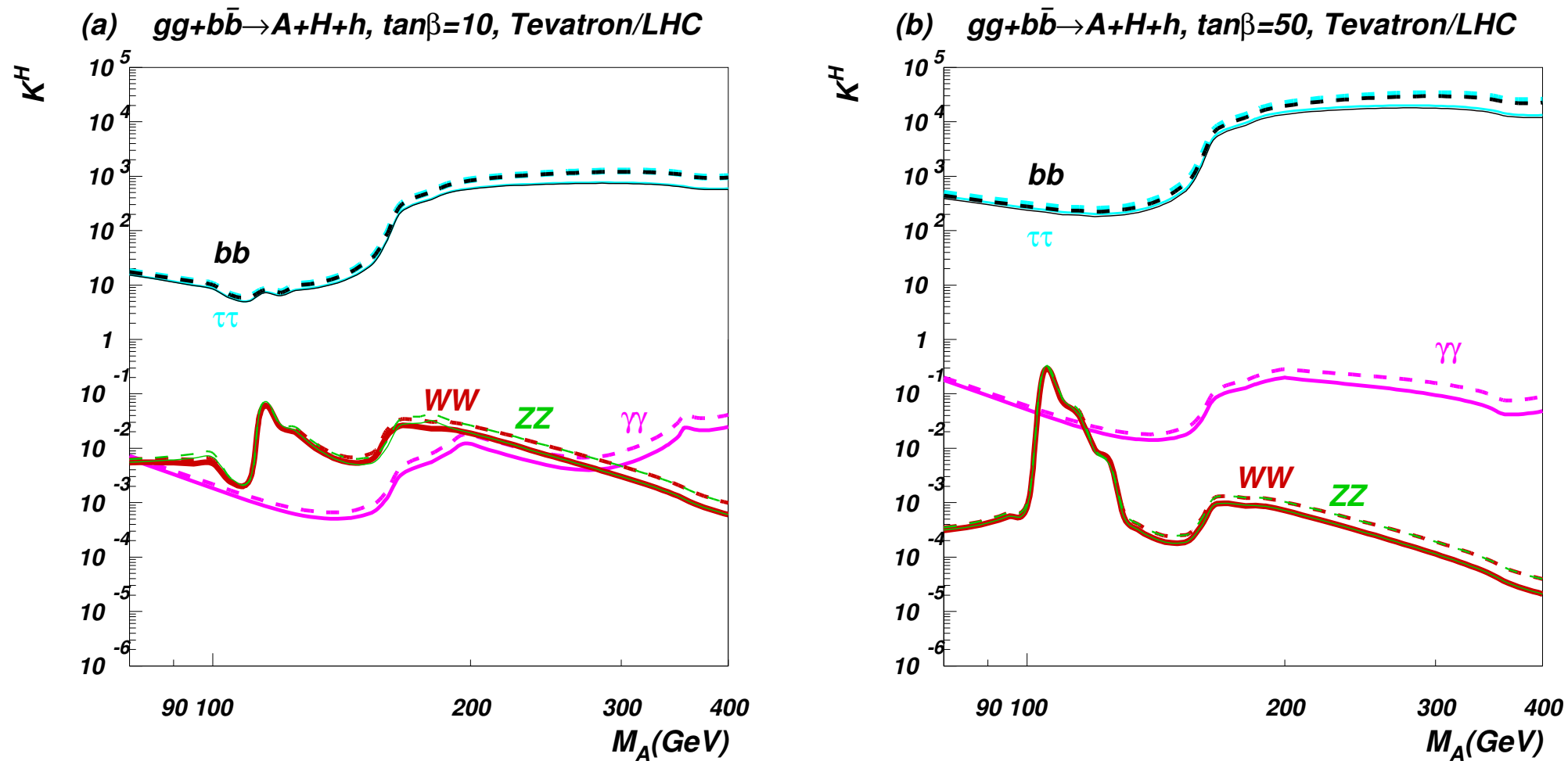

Fig. 12.17: Enhancement factor $\kappa^{\mathcal{H}}_{tot/xx}$ for final states $xx = b\bar{b},\ \tau^+\tau^-,\ WW,\ ZZ,\ \gamma\gamma$ when both $gg \to \mathcal{H}$ and $b\bar{b} \to \mathcal{H}$ are included and the signals of all three MSSM Higgs states are combined. Frames (a) and (b) correspond to $\tan\beta = 10$ and 50, respectively, at the Tevatron (solid lines) and at the LHC (dashed lines).

In the MSSM the contribution from $b\bar{b} \to \mathcal{H}$ becomes important even for moderate values of $\tan\beta \sim 10$ and both, $b\bar{b} \to \mathcal{H}$ and $gg \to \mathcal{H}$ productions are significantly enchanced with $\tan\beta$ compared to the SM rates [148]. For $M_{\mathcal{H}} < 110 - 115$ GeV the contribution from $gg \to \mathcal{H}$ process is a bit bigger than that from $b\bar{b} \to \mathcal{H}$, while for $M_{\mathcal{H}} > 115$ GeV $b$-quark-initiated production begins to outweigh gluon-initiated production. Results for LHC are qualitatively similar, except the rate, which is about two orders of magnitude higher compared to that at the Tevatron. Using the Higgs branching fractions with these NLO cross sections for $gg \to H$ and $b\bar{b} \to H$ allows us to derive $\kappa^{\mathcal{H}}_{total/xx}$, as presented in Fig. 12.17 for the Tevatron and LHC. There are several "physical" kinks and peaks in the enhancement factor for various Higgs boson final states related to $WW$, $ZZ$ and top-quark thresholds which can be seen for the respective values of $M_A$. At very large values of $\tan\beta$ the top-quark threshold effect for the $\gamma\gamma$ enhancement factor is almost gone because the b-quark contribution dominates in the loop. One can see from Fig. 12.17 that the enhancement factors at the Tevatron and LHC are very similar. In contrast to strongly enhanced $b\bar{b}$ and $\tau\bar{\tau}$ signatures, the $\gamma\gamma$ signature is always strongly suppressed! This particular feature of SUSY models, as we will see below, may be important for distinguishing supersymmetric models from models with dynamical symmetry breaking. It is important to note that combining the signals from $A, h, H$ has the virtue of making the enhancement factor independent of the degree of top squark mixing (for fixed $M_A$, $\mu$ and $M_S$ and medium to high values of $\tan\beta$), which greatly reduces the parameter-dependence of our results.

#### 12.6.2.2 Technicolor

Single production of a technipion can occur through the axial-vector anomaly which couples the technipion to pairs of gauge bosons. For an $SU(N_{TC})$ technicolor group with technipion decay constant $F_P$, the anomalous coupling between the technipion and a pair of gauge bosons is given, in direct analogy with the coupling of a QCD pion to photons, by [155–157]. Comparing a PNGB to a SM Higgs boson of the same mass, we find the enhancement in the gluon fusion production is

$$\kappa_{gg\ prod} = \frac{\Gamma(P \to gg)}{\Gamma(h \to gg)} = \frac{9}{4} N_{TC}^2 \mathcal{A}_{gg}^2 \frac{v^2}{F_P^2} \tag{12.41}$$

The main factors influencing $\kappa_{gg\ prod}$ for a fixed value of $N_{TC}$ are the anomalous coupling to gluons and the technipion decay constant. The value of $\kappa_{prod}$ for each model (taking $N_{TC} = 4$) is given in

Table 12.2: Enhancement Factors for 130 GeV technipions produced at the Tevatron and LHC, compared to production and decay of a SM Higgs Boson of the same mass. The slight suppression of $\kappa^P_{prod}$ due to the b-quark annihilation channel has been included. The rightmost column shows the cross-section (pb) for $p\bar{p}/pp \to P \to xx$ at Tevatron Run II/LHC.

| Model | Decay mode | $\kappa^P_{prod}$ | $\kappa^P_{dec}$ | $\kappa^P_{tot/xx}$ | $\sigma$(pb) Tevatron/LHC |
|---|---|---|---|---|---|
| 1) one family | $b\bar{b}$ | 47 | 1.1 | 52 | 14 / 890 |
| | $\tau^+\tau^-$ | 47 | 0.6 | 28 | 0.77 / 48 |
| | $\gamma\gamma$ | 47 | 0.12 | 5.6 | $6.4 \times 10^{-3}$ / 0.4 |
| 2) variant one family | $b\bar{b}$ | 5.9 | 1 | 5.9 | 1.8 / 100 |
| | $\tau^+\tau^-$ | 5.9 | 5 | 30 | 0.84 / 52 |
| | $\gamma\gamma$ | 5.9 | 1.3 | 7.7 | $8.7 \times 10^{-3}$ / 0.55 |
| 3) multiscale | $b\bar{b}$ | 1100 | 0.43 | 470 | 130 / 8000 |
| | $\tau^+\tau^-$ | 1100 | 0.2 | 220 | 6.1 / 380 |
| | $\gamma\gamma$ | 1100 | 0.27 | 300 | 0.34 /22 |
| 4) low scale | $\tau^+\tau^-$ | 120 | 0.6 | 72 | 2/120 |
| | $\gamma\gamma$ | 120 | 2.9 | 350 | 0.4/25 |

Table 12.2. One should note, that the value of $\kappa_{bb\ prod}$ is at least one order of magnitude smaller than $\kappa_{gg\ prod}$ in each model. From the $\kappa_{gg\ prod}/\kappa_{bb\ prod}$ ratio which reads as

$$\frac{\kappa_{gg\ prod}}{\kappa_{bb\ prod}} = \frac{9}{4} N_{TC}^2 \mathcal{A}_{gg}^2 \lambda_b^{-2} \left(1 - \frac{4m_b^2}{m_h^2}\right)^{\frac{3-s}{2}}, \tag{12.42}$$

we see that the larger size of $\kappa_{gg\ prod}$ is due to the factor of $N_{TC}^2$ coming from the fact that gluons couple to a technipion via a techniquark loop. Technicolor models with a large number of techniquarks are in quite a tension with the precision data. However the recent Technicolor models with two technicolors and only two Dirac technifermions in the adjoint representation of the Technicolor gauge group [35] are in better agreement with the precision electroweak measurements. The extended technicolor (ETC) interactions coupling $b$-quarks to a technipion have no such enhancement. With a smaller SM cross-section and a smaller enhancement factor, it is clear that technipion production via $b\bar{b}$ annihilation is essentially negligible at these hadron colliders. Comparing the technicolor and SM branching ratios in, we see immediately that all decay enhancements are of the order of one. Model 2 is an exception; its unusual Yukawa couplings yield a decay enhancement in the $\tau^+\tau^-$ channel of order the technipion's (low) production enhancement. In the $\gamma\gamma$ channel, the decay enhancement strongly depends on the group-theoretical structure of the model, through the anomaly factor. Our results for the Tevatron Run II and LHC production enhancements (including both $gg$ fusion and $b\bar{b}$ annihilation), decay enhancements, and overall enhancements of each technicolor model relative to the SM are shown in Table 12.2 for a technipion or Higgs mass of 130 GeV. Multiplying $\kappa^P_{tot/xx}$ by the cross-section for SM Higgs production via gluon fusion [158] yields an approximate technipion production cross-section, as shown in the right-most column of Table 12.2.

In each technicolor model, the main enhancement of the possible technipion signal relative to that of an SM Higgs arises at production, making the size of the technipion decay constant the most critical factor in determining the degree of enhancement for fixed $N_{TC}$.

### *12.6.3 Interpretation*

We are ready to put our results in context. The large QCD background for $q\bar{q}$ states of any flavor makes the tau-lepton-pair and di-photon final states the most promising for exclusion or discovery of the Higgs-like states of the MSSM or technicolor.

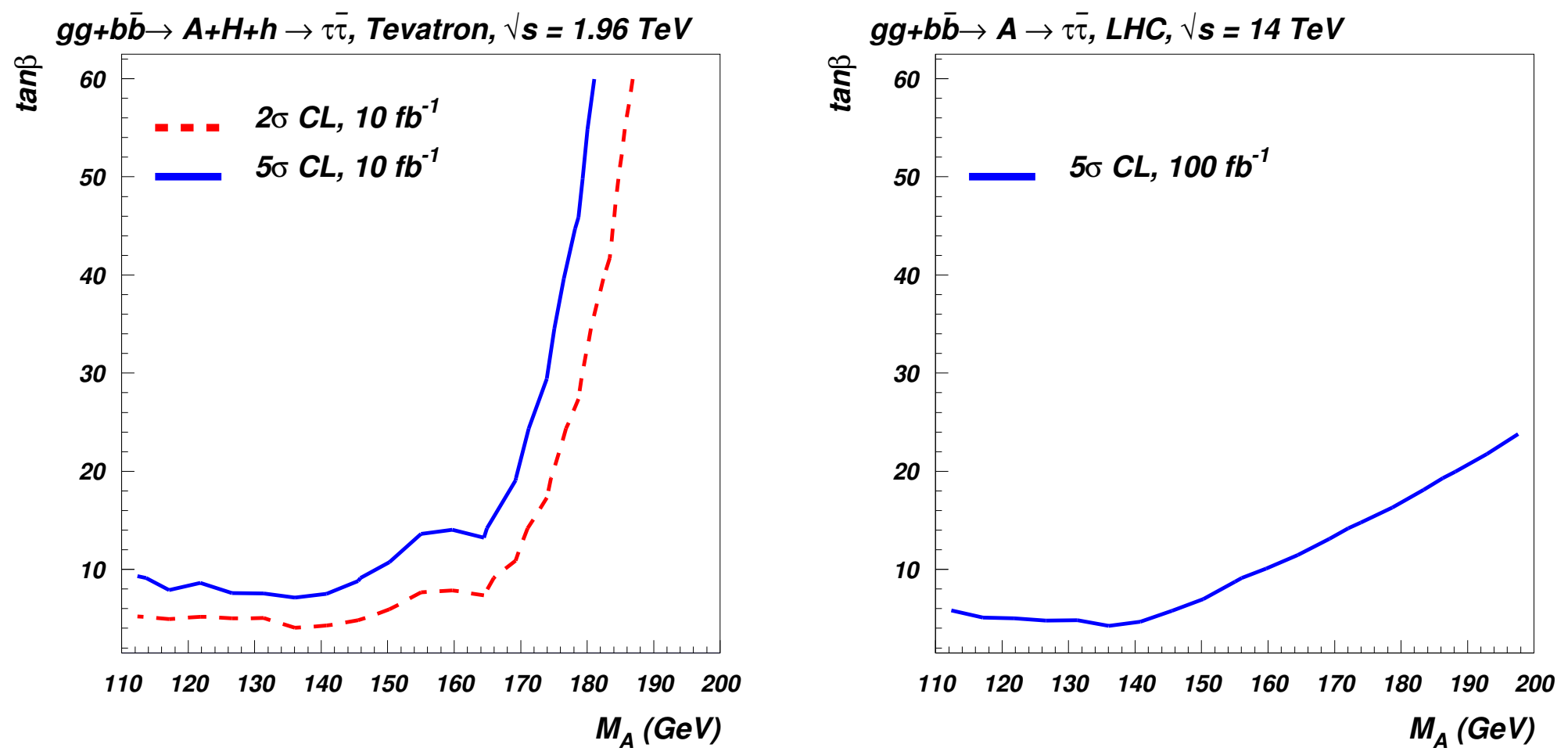

Fig. 12.18: Observability of Supersymmetric Higgs boson production $gg+b\bar{b} \to h+H+A \to \tau^+\tau^-$ for Tevatron (left) and LHC (right). For nearly degenerate Higgs bosons signal is combined in the mass window around $M_A$ (see [148] for details). The parameter space above curves is covered with a given confidence level (CL). The collider reach is based on the $h_{SM} \to \tau^+\tau^-$ studies of [142], in the MSSM parameter space.

In Figure 12.18 we summarize the ability of Tevatron (left) and LHC (right) to explore the MSSM parameter space (in terms of both a $2\sigma$ exclusion curve and a $5\sigma$ discovery curve) using the process $gg + b\bar{b} \to h + A + H \to \tau^+\tau^-$. Translating the enhancement factors into this reach plot draws on the results of [141]. As the $M_A$ mass increases up to about 140 GeV, the opening of the $W^+W^-$ decay channel drives the $\tau^+\tau^-$ branching fraction down, and increases the $\tan\beta$ value required to make Higgses visible in the $\tau^+\tau^-$ channel. At still larger $M_A$, a very steep drop in the gluon luminosity (and the related $b$-quark luminosity) at large $x$ reduces the phase space for $\mathcal{H}$ production. Therefore for $M_A$ >170 GeV, Higgs bosons would only be visible at very high values of $\tan\beta$. The pictures for Tevatron and LHC are qualitatively similar, the main differences compared to the Tevatron are that the required value of $\tan\beta$ at the LHC is lower for a given $M_A$ and it does not climb steeply for $M_A$ >170 GeV because there is much less phase space suppression.

It is important to notice that both, Tevatron and LHC, could observe MSSM Higgs bosons in the $\tau^+\tau^-$ channel even for moderate values of $\tan\beta$ for $M_A \lesssim$ 200 GeV, because of significant enhancement of this channel. However the $\gamma\gamma$ channel is so suppressed that even the LHC will not be able to observe it in any point of the $M_A <$ 200 GeV parameter space studied in this paper! [5]

The Figure 12.19 presents the Tevatron and LHC potentials to observe technipions. For the Tevatron, the observability is presented in terms of enhancement factor, while for the LHC we present signal rate in term of $\sigma \times Br(P \to \tau\tau/\gamma\gamma)$. At the Tevatron, the available enhancement is well above what is required to render the $P$ of any of these models visible in the $\tau^+\tau^-$ channel. Likewise, the right frame of that figure shows that in the $\gamma\gamma$ channel at the Tevatron the technipions of models 3 and 4 will be observable at the $5\sigma$ level while model 2 is subject to exclusion at the $2\sigma$ level. The situation at the LHC is even more promising: all four models could be observable at the $5\sigma$ level in both the $\tau^+\tau^-$ (left frame) and $\gamma\gamma$ (right frame) channels.

Once a supposed light "Higgs boson" is observed in a collider experiment, an immediate important task will be to identify the new state more precisely, i.e. to discern "the meaning of Higgs" in this context. Comparison of the enhancement factors for different channels will aid in this task. Our study has shown that comparison of the $\tau^+\tau^-$ and $\gamma\gamma$ channels can be particularly informative in distinguishing supersymmetric from dynamical models. In the case of supersymmetry, when the $\tau^+\tau^-$ channel is enhanced, the $\gamma\gamma$ channel is suppressed, and this suppression is strong enough that even the LHC would not observe the $\gamma\gamma$ signature. In contrast, for the dynamical symmetry breaking models studied we expect *simultaneous* enhancement of both the $\tau^+\tau^-$ and $\gamma\gamma$ channels. The enhancement of the $\gamma\gamma$ channel is so significant, that even at the Tevatron we may observe technipions via this signature at the $5\sigma$ level for Models 3 and 4, while Model 2 could be excluded at 95% CL at the Tevatron. The LHC collider, which will have better sensitivity to the signatures under study, will be able to observe all four models of dynamical symmetry breaking studied here in the $\gamma\gamma$ channel, and can therefore distinguish more conclusively between the supersymmetric and dynamical models.

[5]In the decoupling limit with large values of $M_A$ and low values of $\tan\beta$, the lightest MSSM Higgs could be dicovered in the $\gamma\gamma$ mode just like the SM model Higgs boson

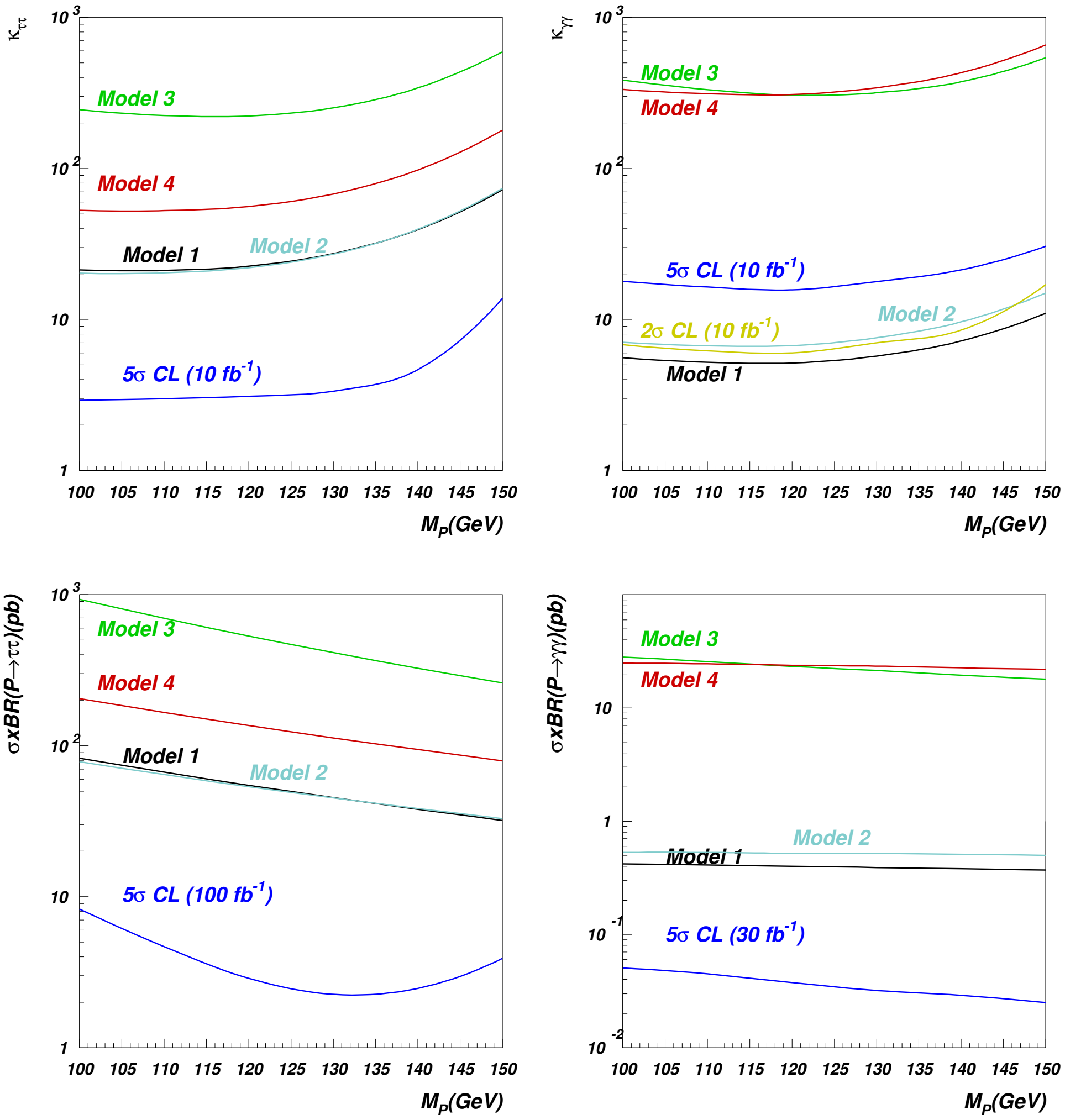

Fig. 12.19: Observability of technipions as a function of technipion mass for tau pair (left frame) or photon pair (right frame) at the Tevatron and LHC in the final state. Top: enhancement factors required for a $5\sigma$ discovery and $2\sigma$ exclusion for a Higgs-like particle at Tevatron Run II. Bottom: the lowest curve is the $\sigma \times Br$ required to make a Higgs-like particle visible in $\tau^+\tau^-$ or in $\gamma\gamma$ at LHC.

*12.6.4 Conclusions*

We have shown that searches for a light Standard Model Higgs boson at Tevatron Run II and CERN LHC have the power to provide significant information about important classes of physics beyond the Standard Model. We demonstrated that the new scalar and pseudo-scalar states predicted in both supersymmetric and dynamical models can have enhanced visibility in standard $\tau^+\tau^-$ and $\gamma\gamma$ search channels, making them potentially discoverable at both the Tevatron Run II and the CERN LHC. In comparing the key signals for the non-standard scalars across models we investigated the likely mass reach of the Higgs search in $pp/p\bar{p} \to \mathcal{H} \to \tau^+\tau^-$ for each kind of non-standard scalar state, and we demonstrated that $pp\ \ p\bar{p} \to \mathcal{H} \to \gamma\gamma$ may cleanly distinguish the scalars of supersymmetric models from those of dynamical models.

## 12.7 Dynamical breakdown of an Abelian gauge chiral symmetry by strong Yukawa couplings [6]

*Petr Beneš, Tomáš Brauner and Jiří Hošek*

The standard technicolour scenarios [1, 2] dispense with the elementary Higgs and, instead of its vacuum expectation value, generate the order parameter for electroweak symmetry breaking by a bilinear condensate of new fermions. This is bound together by a new strong gauge interaction.

In this contribution we suggest a different mechanism for dynamical electroweak symmetry breaking. The central idea may be summarised as follows. We retain the elementary scalar, but with a positive mass squared so that the usual particle interpretation is preserved even in the absence of interactions. Our basic assumption is the existence of a strong Yukawa interaction between the scalar and the massless fermions. We show that, provided the Yukawa coupling is large enough, the fermion masses may be generated spontaneously as a self-consistent solution of the Schwinger–Dyson equations. In order to make the proposed mechanism more transparent, we demonstrate it on the dynamical breaking of an Abelian chiral symmetry. The extension to the full electroweak $SU(2)_L \times U(1)_Y$ gauge invariance is currently being worked on.

Our plan is the following. First, we recall our previous results and show how a global chiral symmetry may be dynamically broken by a strong Yukawa interaction, thus generating the fermion masses [159]. Second, the axial part of the symmetry is gauged and a sum rule for the gauge boson mass is derived. Last, we explicitly work out the one-loop triple gauge boson vertex as a genuine prediction of the broken symmetry. The extension to the electroweak theory is discussed in the conclusions.

*12.7.1 Global chiral symmetry*

Following closely our recent paper [159], we consider a model of two Dirac fermions $\psi_{1,2}$ and a complex scalar $\phi$, defined by the Lagrangian,

$$\mathcal{L} = \sum_{j=1,2} \left(\bar{\psi}_{jL} i\partial\!\!\!/ \psi_{jL} + \bar{\psi}_{jR} i\partial\!\!\!/ \psi_{jR}\right) + \partial_\mu \phi^\dagger \partial^\mu \phi - M^2 \phi^\dagger \phi - \frac{1}{2}\lambda(\phi^\dagger\phi)^2 + \\ + y_1(\bar{\psi}_{1L}\psi_{1R}\phi + \bar{\psi}_{1R}\psi_{1L}\phi^\dagger) + y_2(\bar{\psi}_{2R}\psi_{2L}\phi + \bar{\psi}_{2L}\psi_{2R}\phi^\dagger). \quad (12.43)$$

The Yukawa couplings $y_1, y_2$ are assumed real. Note that the Lagrangian (12.43) is invariant under the global Abelian group $U(1)_{V_1} \times U(1)_{V_2} \times U(1)_A$. The two vector $U(1)$'s correspond to independent phase transformations of $\psi_1$ and $\psi_2$ i.e., are generated by the operators of the number of fermions of the respective type. The axial $U(1)_A$, on the other hand, relates all the fields included. It consists of transformations of the type

$$\psi_1 \to e^{+i\theta\gamma_5}\psi_1, \quad \psi_2 \to e^{-i\theta\gamma_5}\psi_2, \quad \phi \to e^{-2i\theta}\phi.$$

[6] Although this contribution is not related to technicolor, it is included in this section. It constitutes an exotic example of dynamical breaking of an abelian gauge theory which may one day be used to break the electroweak theory.

The fact that the scalar carries nonzero axial charge will be crucial in the following. Note also that the fermions have opposite charges in order to remove the axial anomaly. While this is a convenience at this stage, where the considered symmetry is global, it will become a theoretical necessity later when it is gauged.

Our goal is to show that at sufficiently strong Yukawa interaction, the axial $U(1)_A$ is spontaneously broken. In the fermion sector this means, of course, that nonzero Dirac masses are generated. Analogously in the scalar sector we find the 'anomalous' axial-charge-violating two-point function $\langle\phi\phi\rangle$. Both these effects are related to each other through one-loop Feynman graphs and both thus have to be analysed simultaneously.

To account for the symmetry breaking in the scalar sector, we introduce the formal doublet

$$\Phi = \begin{pmatrix} \phi \\ \phi^\dagger \end{pmatrix},$$

and work with the matrix propagator $iD(x-y) = \langle 0|T\{\Phi(x)\Phi^\dagger(y)\}|0\rangle$. We use the method of the Schwinger–Dyson equations. For the sake of simplicity we neglect all symmetry-preserving radiative corrections to the fermion and scalar propagators, making the following Ansatz,

$$S^{-1}_{1,2}(p) = \not{p} - \Sigma_{1,2}(p^2), \quad D^{-1}(p) = \begin{pmatrix} p^2 - M^2 & -\Pi(p^2) \\ -\Pi^*(p^2) & p^2 - M^2 \end{pmatrix}.$$

The functions $\Sigma_{1,2}(p^2)$ are the Lorentz-scalar fermion proper self-energies that are responsible for the nonzero masses. Likewise, $\Pi(p^2)$ is the anomalous scalar proper self-energy. The scalar spectrum then consists of two real particles with masses given self-consistently by $M^2_{1,2} = M^2 \pm |\Pi(M^2_{1,2})|^2$.

For the purpose of the numerical computation of the self-energies we perform two additional simplifications. First, we abandon the $\lambda(\phi^\dagger\phi)^2$ interaction. This is because the dynamical symmetry breaking is assumed to happen due to the Yukawa interaction, while the $\lambda$ term in the Lagrangian serves merely as a perturbative counterterm. Second, we neglect all vertex corrections, thus closing the Schwinger–Dyson hierarchy self-consistently at the propagator level. The fermion and scalar propagators are then determined by the solution of the one-loop equations,

$$\begin{aligned}
\Sigma_{1,p} &= iy_1^2 \int \frac{d^4k}{(2\pi)^4} \frac{\Sigma_{1,k}}{k^2 - \Sigma^2_{1,k}} \frac{\Pi_{k-p}}{[(k-p)^2 - M^2]^2 - |\Pi_{k-p}|^2}, \\
\Sigma_{2,p} &= iy_2^2 \int \frac{d^4k}{(2\pi)^4} \frac{\Sigma_{2,k}}{k^2 - \Sigma^2_{2,k}} \frac{\Pi^*_{k-p}}{[(k-p)^2 - M^2]^2 - |\Pi_{k-p}|^2}, \\
\Pi_p &= -\sum_{j=1,2} 2iy_j^2 \int \frac{d^4k}{(2\pi)^4} \frac{\Sigma_{j,k}}{k^2 - \Sigma^2_{j,k}} \frac{\Sigma_{j,k-p}}{(k-p)^2 - \Sigma^2_{j,k-p}}.
\end{aligned}$$

This set of equations have been solved iteratively in the Euclidean space. We found that a nontrivial solution exists only if the Yukawa couplings are large enough. For $y_1 = y_2$ the critical value is about 80. The typical form of the solution is shown in Fig. 12.20. Our model has the remarkable property that a moderate change of the ratio $y_1/y_2$ transforms into a tremendous change of the mass ratio. For instance, for $y_1 = 77.4$ and $y_2 = 88$ we find $m_1^2/m_2^2 \approx 10^{-3}$ and the mass ratio seems to fall down to zero as $y_1$ approaches a critical value about 77. This is alluring and gives us a hope that, when applied to the $SU(2)_L \times U(1)_Y$ gauge theory of electroweak interactions, the present mechanism of dynamical symmetry breaking could provide a natural explanation of the hierarchy of the fermion masses.

### *12.7.2 Gauge axial symmetry*

The next step in our analysis is the gauging of the axial part of the global symmetry. Formally, this is done by replacing the ordinary derivatives in the Lagrangian (12.43) with the covariant ones, $D_\mu\psi_1 =$

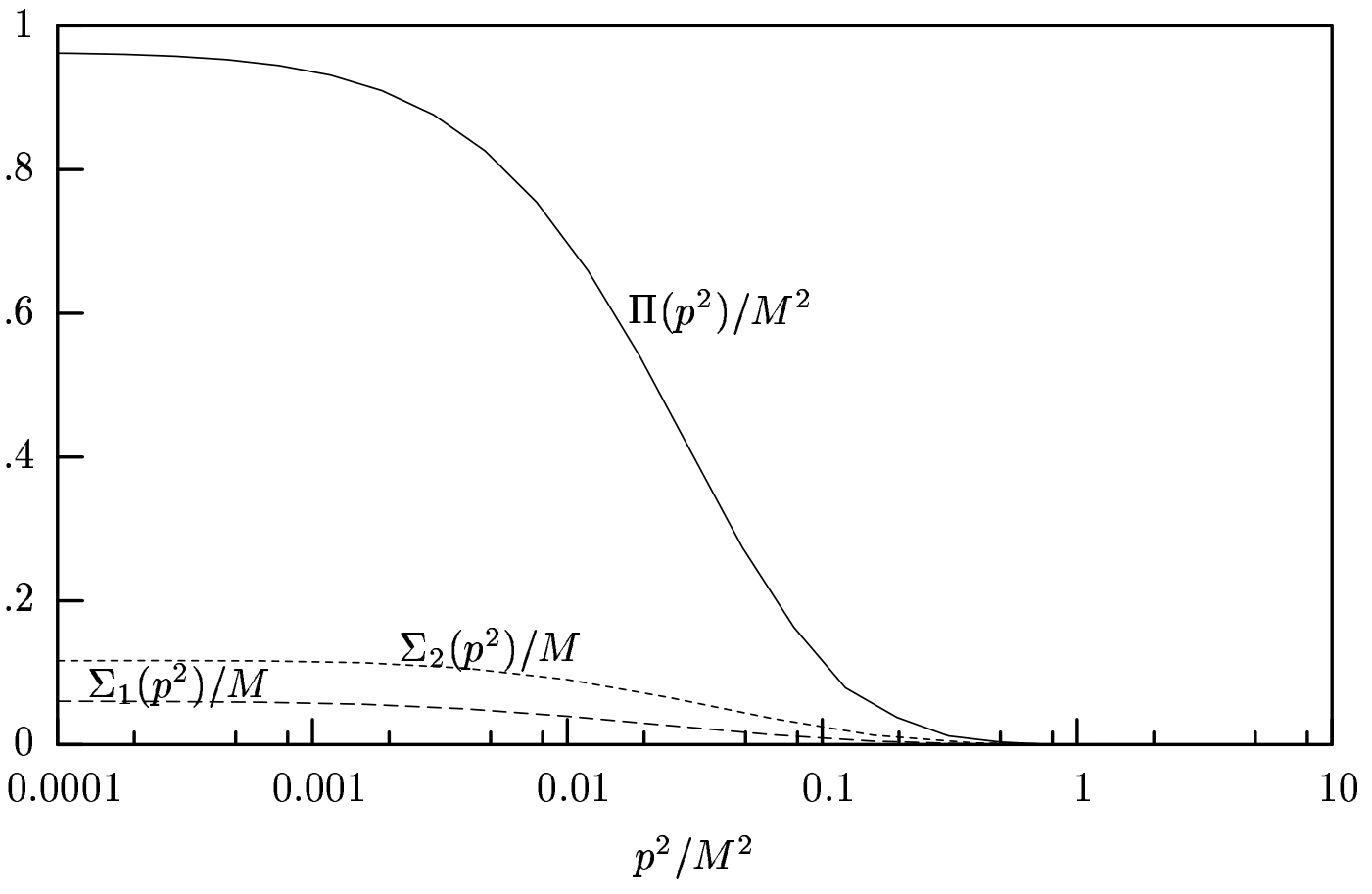

Fig. 12.20: The fermion and scalar proper self-energies for $y_1 = 79$ and $y_2 = 88$.

$(\partial_\mu - igA_\mu\gamma_5)\psi_1$, $D_\mu\psi_2 = (\partial_\mu + igA_\mu\gamma_5)\psi_2$ and $D_\mu\phi = (\partial_\mu + 2igA_\mu)\phi$, and adding the kinetic term for the Abelian gauge field $A_\mu$. We emphasise that the gauge interaction may be switched on perturbatively since it does not play any role in the proposed dynamical mechanism for chiral symmetry breaking.

Once the axial symmetry is spontaneously broken, the gauge boson acquires a nonzero mass through the Schwinger mechanism [160]. Technically, the mass is given by the residue of the massless pole in the gauge boson polarisation tensor.[7] This pole in turn arises from the propagation of the (composite) Goldstone boson of the broken symmetry.

To determine the pole part of the polarisation tensor, one has to know the effective coupling of the Goldstone boson to the axial current or to the gauge boson. First, the axial Ward identity is used to calculate the pole part of the axial current vertex functions, and thus the effective couplings of the Goldstone to the fermions and the scalar [159]. The Goldstone–gauge boson coupling then arises through the fermion and scalar loops. (Recall that the interaction of the fermions and the scalar with the gauge boson is perturbative.) The resulting formula for the dynamically generated gauge boson mass reads $m_A^2 = g^2[I_{\psi_1}(0) + I_{\psi_2}(0) + I_\phi(0)]$, where the one-loop integrals are given by

$$iq^\nu I_{\psi_j}(q^2) = 4\int \frac{d^4k}{(2\pi)^4} \frac{[(k+q)^\nu \Sigma_{j,k} - k^\nu \Sigma_{j,k+q}]\,[\Sigma_{j,k+q} + \Sigma_{ji,k}]}{\left[(k+q)^2 - \Sigma^2_{j,k+q}\right]\left[k^2 - \Sigma^2_{j,k}\right]},$$

$$iq^\nu I_\phi(q^2) = -2\int \frac{d^4k}{(2\pi)^4} \frac{(2k+q)^\nu \left\{(\Pi_{k+q} + \Pi_k)\left[[(k+q)^2 - M^2]\Pi_k^* - (k^2 - M^2)\Pi^*_{k+q}\right] + \text{c.c.}\right\}}{\left\{[(k+q)^2 - M^2]^2 - |\Pi_{k+q}|^2\right\}[(k^2 - M^2)^2 - |\Pi_k|^2]}.$$

The sample of numerical results is presented in Fig. 12.21.

Note that the gauge boson mass is expressed in terms of the fermion and scalar self-energies by the above one-loop integrals [162]. This clearly distinguishes our model of dynamical symmetry breaking from the standard Higgs mechanism where the fermion and gauge boson masses are generated independently.

### *12.7.3 Triple gauge boson vertex*

The sheer existence of the Goldstone boson and the generation of the gauge boson mass are robust, model-independent predictions of the broken symmetry. In order to achieve deeper insight into the dynamical origin of the symmetry breaking, it is necessary to work out in detail the consequences of the particular model. In the electroweak interactions this will, of course, be of crucial importance in the search for the signatures of new physics.

[7] In fact, this procedure gives only an approximate value of the gauge boson mass, because of the neglected finite part of the polarisation tensor [161]. Such an approximation is justified provided the generated mass is small enough.

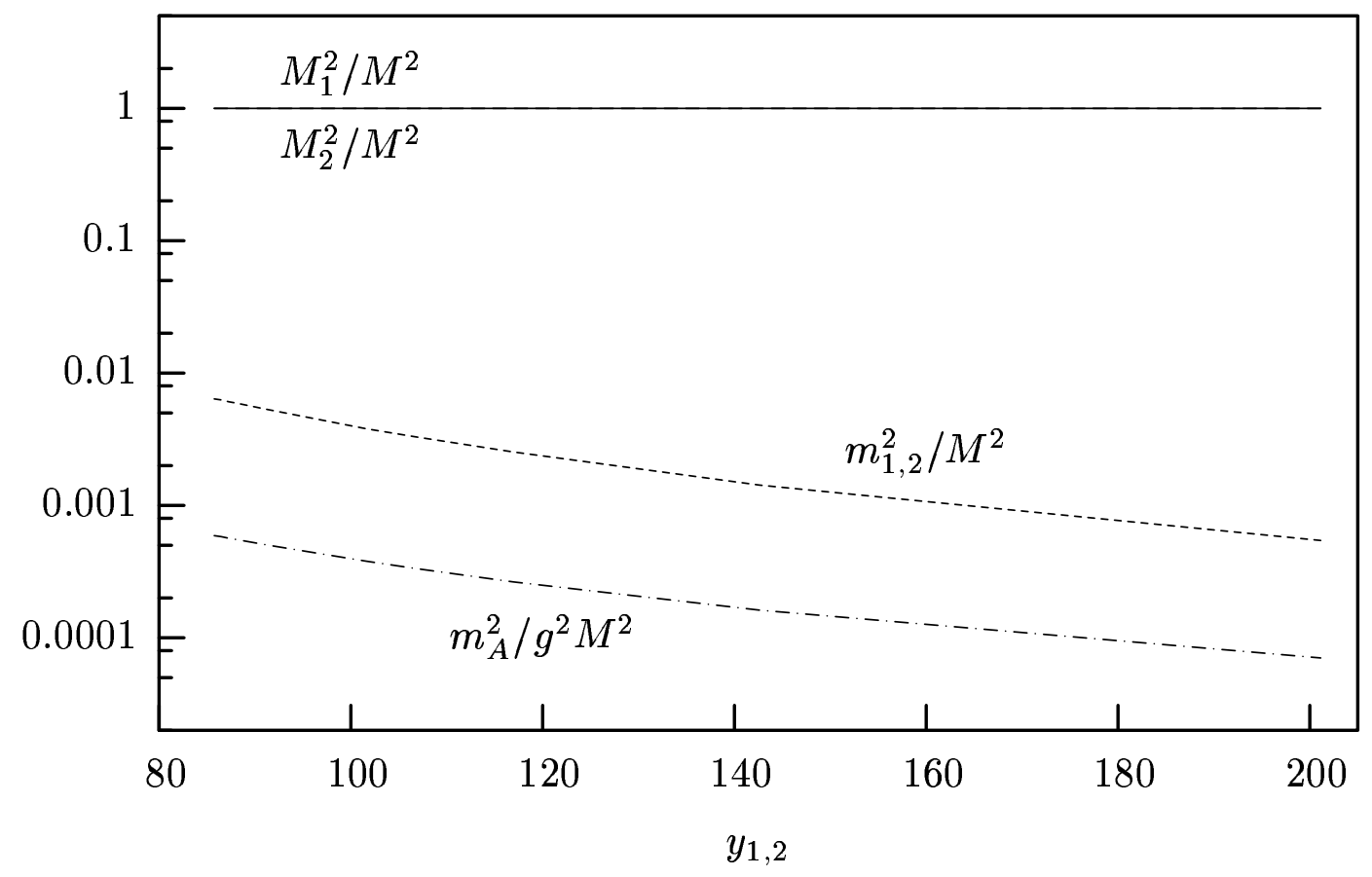

Fig. 12.21: The mass spectrum as a function of $y_1 = y_2$. Here $M_{1,2}, m_{1,2}, m_A$ are the masses of the heavy scalars, the fermions, and the gauge boson, respectively.

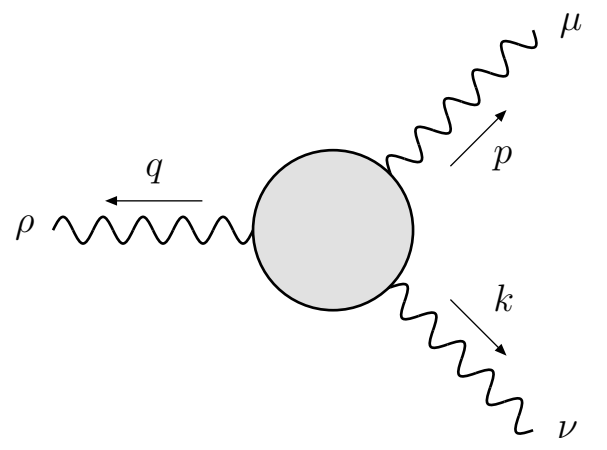

Fig. 12.22: The $A^3$ vertex, denoted as $iT^{\mu\nu\rho}(p,k)$. There are only two independent momentum variables due to the conservation $p + k + q = 0$.

Here we calculate the axial-vector $A^3$ vertex, see Fig. 12.22 for notation. To order $g^3$, it is given by the sum of one-loop diagrams with either the fermions or the scalar circulating in the loop. Explicitly we find

$$T^{\mu\nu\rho}(p,k) = \sum_j \big[T^{\mu\nu\rho}_{\psi_j}(p,k) + T^{\nu\mu\rho}_{\psi_j}(k,p)\big] + T^{\mu\nu\rho}_{\Phi}(p,k) + T^{\nu\mu\rho}_{\Phi}(k,p) + \\ + 2\big[T_4^{\mu\nu\rho}(p,k) + T_4^{\mu\rho\nu}(p,-p-k) + T_4^{\nu\rho\mu}(k,-p-k)\big],$$

where $T^{\mu\nu\rho}_{\psi_j}$ denote the fermion triangle loops, $T^{\mu\nu\rho}_{\Phi}$ the analogous scalar loops, and $T_4^{\mu\nu\rho}$ the scalar loops with the insertion of the vertex $\phi^\dagger\phi A^\mu A_\mu$.

We do not write down the lengthy explicit expression for $T^{\mu\nu\rho}(p,k)$. It cannot be calculated analytically anyway since the self-energies $\Sigma_j$ and $\Pi$ have been computed only numerically. (In fact, in our approximation the scalar loops do not contribute.)

The $A^3$ vertex can, however, be determined analytically in the special case when the external gauge bosons are on-shell i.e., $p^2 = k^2 = q^2 = m_A^2$ and the fermion self-energies are set constant, $\Sigma_j(p^2) = m_j$. It is then found that

$$iT^{\mu\nu\rho}(p,k) = G\left[(q^\mu k^\alpha - k^\mu q^\alpha)p^\beta \epsilon^{\nu\rho}{}_{\alpha\beta} + (p^\nu q^\alpha - q^\nu p^\alpha)k^\beta \epsilon^{\rho\mu}{}_{\alpha\beta} + (k^\rho p^\alpha - p^\rho k^\alpha)q^\beta \epsilon^{\mu\nu}{}_{\alpha\beta}\right]. \tag{12.44}$$

The effective coupling constant $G$ is expressed as $G = -g^3 \sum_{j=1,2}(-1)^j f(m_j^2, m_A^2)$, where

$$f(x,y) = \frac{2}{\pi^2 y}\int_0^1 dz \frac{z(1-z)}{\sqrt{z(3z-4)+\frac{4x}{y}-i\epsilon}} \arctan\frac{z}{\sqrt{z(3z-4)+\frac{4x}{y}-i\epsilon}}.$$

For very small coupling $g$ the gauge boson mass is small as well and the value $f(m^2, 0) = 1/24\pi^2 m^2$ is then of particular interest. Note also that the $A^3$ vertex (12.44) may be obtained from the effective Lagrangian

$$\mathcal{L}_{\text{eff}} = G\epsilon_{\alpha\beta\gamma\delta}(\partial_\sigma A^\alpha)(\partial^\beta A^\sigma)(\partial^\gamma A^\delta).$$

### *12.7.4 Conclusions*

In the framework of a simple Abelian model we have shown that a sufficiently strong Yukawa interaction may induce the spontaneous breakdown of the chiral symmetry. Our ultimate goal is, however, to apply this idea to the gauge theory of electroweak interactions. The upcoming LHC machine at CERN is going to explore the new physics underlying the electroweak symmetry breaking and we hope that our proposal might provide a reasonable alternative to the existing models.

Our hope is based on the following observations. First, with the elementary scalars kept in the Lagrangian one explicitly breaks the interfamily symmetries. Models without scalars struggle with guaranteeing unobservability of physical consequences of these unwanted symmetries. Further, our numerical results suggest that, unlike in the Higgs mechanism, we are able to generate the huge hierarchy of fermion masses without introducing vastly different Yukawa couplings. Another notable fact is that the gauge boson masses are tied to the fermion spectrum in terms of sum rules.

To date, the most serious obstacle to a direct application of the present mechanism to the electroweak interactions is its nonperturbative nature. We have already shown that the scalar spectrum may be adapted to fit the symmetries of the standard model [163]. On the other hand, much work is still in order to make phenomenological predictions that could be compared with experiment. Not only the flavour structure of the standard model makes the set of coupled equations to be solved much larger and thus more complicated, but also the approximations used here to generate the Schwinger–Dyson equations have to be revised. Nevertheless, we are convinced that our goal is worthy of the effort required by this task.

## 13 HIGGS TRIPLETS

### 13.1 Introduction

*John F. Gunion and Chris Hays*

Even though the Standard Model (SM) of the strong and electroweak interactions has proven enormously successful, it need not be the case that the Higgs sector consists solely of a single $SU(2)_L$ Higgs doublet field, that is a field with total weak isospin $T = \frac{1}{2}$ having two members with $T_3 = +\frac{1}{2}$ and $T_3 = -\frac{1}{2}$ and $U(1)$ hypercharge $Y = 1$. The inclusion of additional doublets as well as singlets (i.e., fields with $T = Y = 0$) is a frequently considered possibility (and at least two doublets are required in the supersymmetric context). The next logical step is to consider the inclusion of one or more triplet $SU(2)_L$ representations (i.e. a $T = 1$, three-component field with $T_3 = +1, 0, -1$ members). The purpose of this section is to review the phenomenology of a Higgs sector which contains both doublet and triplet fields. Surprisingly attractive models, fully consistent with all existing experimental constraints, can be constructed. These yield many exotic features and unusual experimental signatures. In exploring the physics of electroweak symmetry breaking at future colliders, it will be important to consider the alternative possibilities characteristic of this and other non-minimal Higgs sectors. Our discussion here will focus primarily on models in which only the Higgs sector of the SM is extended via the addition of triplet representation(s) with hypercharge $Y = 0$ and/or $Y = \pm 2$ (which are real and complex, respectively).

There are many models in which both the Higgs sector and the gauge sector are expanded that provide a natural setting for Higgs triplet fields, as for example the left-right (LR) symmetric models with extended gauge group $SU(2)_L \times SU(2)_R \times U(1)_{B-L}$ [1–6] (see also Section 6.1). Supersymmetric left-right (SUSYLR) symmetric models with triplets can also be constructed and have many attractive features [7–14]. One of the primary motivations for left-right symmetric models with triplets in the Higgs sector is that they provide a natural setting for the see-saw mechanism of neutrino mass generation. The minimal Higgs sector contains a bi-doublet Higgs field, an $SU(2)_R$ triplet (employed for the see-saw) and its LR partner $SU(2)_L$ triplet. For the phenomenology of such models in the absence of supersymmetry, useful starting references are [15–17]. A brief outline of the SUSYLR Higgs scenarios is given at the end of this review.

We will focus on the $Y = 0$ and $Y = \pm 2$ triplet models. Only the $Y = 0$ and $Y = \pm 2$ triplet representations have a neutral member which, if it acquires a non-zero vacuum expectation value (vev), can influence electroweak symmetry breaking and give rise to non-zero $VV'$-Higgs vertices (where $V, V' = W, Z, \gamma$). Triplet models with still larger even-integer values of $Y$ do not have a neutral member. We also do not consider triplet models with an odd-integer value of $Y$. The Higgs fields of such a model would have fractional charge. Table 13.1 lists the triplet models we consider and establishes some notation for the Higgs bosons appearing in the various models, including those with custodial or left-right symmetry.

Before zeroing in on triplet models, we make a few more general remarks regarding Higgs representations with weak isospin $T \geq 1$, generically denoting the Higgs fields and bosons by $\delta$. Let us focus on two particular hallmark signatures, both of which require the presence of a Higgs representation with $T \geq 1$ that contains a neutral field member *with non-zero vacuum expectation value (vev)*.

(1) Verification of the presence of a non-zero $\delta^{++}W^-W^-$ tree-level strength interaction. In the $T = 1$ context, this would require the $Y = \pm 2$ representation.
This vertex would be detected by observation of the single-$\delta^{++}$ fusion production process $W^+W^+$ $\delta^{++}$ and/or via the presence of $\delta^{++} \to W^+W^+$ decays.

(2) Verification of the existence of a tree-level $\delta^-W^+Z$ vertex. For $T = 1$, this can occur for either a $Y = 0$ or $Y = \pm 2$ representation (or if both are present).
Detection of this vertex would be via single $\delta^+$ production, e.g. $Z^* \to \delta^+W^-$, or decays such as

Table 13.1: Notation for different Higgs representations and models. Without superscripts, $T$ and $T_3$ refer to the SM $SU(2)_L$ group. $SU(2)_C$ refers, when relevant, to the custodial symmetry group of the model. For left-right symmetric models, we distinguish $T^L$ and $T_3^L$ of $SU(2)_L$ from $T^R$ and $T_3^R$ of $SU(2)_R$. In the absence of the $SU(2)_R$ group the charge formula is $Q = T_3 + \frac{1}{2}$. In left-right symmetric models $Q = T_3^L + T_3^R + \frac{1}{2}(B - L)$. The listed couplings are those present at tree-level. Without subscripts $V$ and $V'$ refer to the usual $W^\pm, Z$. For left-right symmetric models, the $V_L = W_L, Z_L$ of $SU(2)_L$ and the $V_R = W_R, Z_R$ of $SU(2)_R$ are distinguished. In the limit where no triplet neutral field has a vev, the doublet and triplet fields separate and the couplings of the triplets to $VV'$ are absent.

Generic Higgs field

| General $(T, Y)$ | $\phi_{T,Y}$ | Couplings |
|---|---|---|

Complex doublet Higgs field with neutral member

| | | |
|---|---|---|
| $(T, Y) = (1/2, \pm 1)$ | $\phi^\pm, \phi^0$ | $VV'$ and $f\overline{f}$ |

Generic Triplet Higgs fields

| | | |
|---|---|---|
| $T = 1$, $Y$ arbitrary, $\delta$ fields | $\delta^{\pm\pm}, \delta^\pm, \delta^0, \ldots$ | |

Complex triplet Higgs representation with neutral member

| | | |
|---|---|---|
| $(T, Y) = (1, \pm 2)$, $\chi$ fields | $\chi^{\pm\pm}, \chi^\pm, \chi^0$ | $VV'$ and $\ell\ell$ |

Real triplet Higgs representation with neutral member

| | | |
|---|---|---|
| $(T, Y) = (1, 0)$, $\xi$ fields | $\xi^\pm, \xi^0$ | $VV'$ only |

Model with one doublet and one real triplet

| | | |
|---|---|---|
| mix of $\phi = (1/2, \pm 1)$ and $\xi = (1, 0)$ | $\begin{pmatrix} \phi^+ \\ \phi^0 \end{pmatrix}, \begin{pmatrix} \xi^+ \\ \xi^0 \\ \xi^- \end{pmatrix}$ | |
| Mass eigenstates for $\langle \xi^0 \rangle \neq 0$ | $h^\pm, h^0, k^0$ | $VV'$ and $f\overline{f}$ |

Triplet model with tree-level custodial $SU(2)_C$ symmetry

| | | |
|---|---|---|
| mix of $\phi = (1/2, \pm 1)$, $\xi = (1, 0)$ and $\chi = (1, \pm 2)$ | $\begin{pmatrix} \phi^+ \\ \phi^0 \end{pmatrix}, \begin{pmatrix} \chi^{0*} & \xi^+ & \chi^{++} \\ \chi^- & \xi^0 & \chi^+ \\ \chi^{--} & \xi^- & \chi^0 \end{pmatrix}$ | |
| $SU(2)_C$ decomposition for $\langle \xi^0 \rangle = \langle \chi^0 \rangle \neq 0$ | Mass eigenstates | Couplings |
| $SU(2)_C$ 5-plet: $(1, \pm 2)$ and $(1, 0)$ mix | $H_5^{\pm\pm}, H_5^\pm, H_5^0$ | $VV'$ and $\ell\ell$ |
| $SU(2)_C$ 3-plet: $(1/2, \pm 1)$, $(1, \pm 2)$ and $(1, 0)$ mix | $H_3^\pm, H_3^0$ | $f\overline{f}$ only |
| $SU(2)_C$ singlet #1: pure $(1/2, 0)$ | $H_1^0$ | $VV'$ and $f\overline{f}$ |
| $SU(2)_C$ singlet #2: $(1, \pm 2), (1, 0)$ mixture | $H_1^{0\prime}$ | $VV'$ and $\ell\ell$ |

Left-right symmetric models

| | | |
|---|---|---|
| Bi-doublet, $\phi$, $(T^L, T^R, B - L) = (\frac{1}{2}, \frac{1}{2}, 0)$ | $\begin{pmatrix} \phi_1^0 & \phi_1^+ \\ \phi_2^- & \phi_2^0 \end{pmatrix}$ | $V_{L,R} V'_{R,L}$ and $f\overline{f}$ |
| Left triplet, $\Delta_L$, $(T^L, T^R, B - L) = (1, 0, 2)$ | $\begin{pmatrix} \frac{\delta_L^+}{\sqrt{2}} & \delta_L^{++} \\ \delta_L^0 & -\frac{\delta_L^+}{\sqrt{2}} \end{pmatrix}$ | $V_L V'_L$ and $\ell\ell$ |
| Right triplet, $\Delta_R$, $(T^L, T^R, B - L) = (0, 1, 2)$ | $\begin{pmatrix} \frac{\delta_R^+}{\sqrt{2}} & \delta_R^{++} \\ \delta_R^0 & -\frac{\delta_R^+}{\sqrt{2}} \end{pmatrix}$ | $V_R V'_R$ and $\ell\ell$ |

$\delta^+ \to W^+ Z$.

Neither signature would be observed for any doublet model. For example, although a doublet representation with $Y = 3$ has a doubly-charged Higgs boson, it does not contain a neutral member and the doubly-charged Higgs could only be pair produced at a hadron or $e^+e^-$ collider. (Note that to preserve $U(1)_{EM}$, any Higgs potential involving such exotic representations must be constructed so that vevs do not develop for the charged fields.) Nonetheless, it should also be kept in mind that such exotic doublet representations *can* influence gauge coupling running and, therefore, coupling unification.

An example of a Higgs sector with $T > 1$ that can give rise to the two hallmark signatures is the $T = 2$, $Y = 0$ Higgs representation, which contains $\delta^{\pm\pm}$, $\delta^{\pm}$ and $\delta^0$ fields. If the neutral member has non-zero vev then non-zero $\delta^{\pm\pm}W^{\mp}W^{\mp}$ and $\delta^{\pm}W^{\mp}Z$ vertices will both be generated at tree-level. Both signatures also arise for non-zero neutral field vev in the case of the $T = 3, Y = 4$ representation that is the next simplest beyond the doublet representation to have a built-in custodial symmetry that guarantees a tree-level value of $\rho = \frac{m_W{}^2}{m_Z{}^2 \cos^2\theta_W} = 1$ with *finite* radiative corrections.

For any $T \geq 1$ model, it is entirely possible and, in the absence of a built-in custodial symmetry, probably most natural for the vev of the neutral field to be zero. In particular, only in this way can we guarantee $\rho = 1$ at tree-level and that radiative corrections to $\rho$ will be finite. Thus, we will devote considerable discussion in our triplet review to the dramatic alterations in phenomenology that arise in such a case as compared to the case when the neutral-field vev is non-zero. The most obvious phenomenological consequence is that for zero neutral-field vev all $VV'\delta$ vertices are zero at tree-level. The single $\delta$ production processes and $\delta \to VV'$ decays that rely on such vertices are then highly suppressed or absent altogether.

Of course, couplings of Higgs bosons to the SM fermions are also a crucial ingredient for phenomenology. In fact, subject to the exception discussed below, only doublet Higgs bosons can have couplings to SM fermions. If a $T = 1$ representation has a non-zero neutral field vev, then the physical neutral and singly-charged Higgs eigenstates will typically be mixtures of doublet and triplet members and will have fermionic couplings proportional to their doublet components. If the neutral field vev is zero, then the triplet fields will not mix with the doublet fields and they will form their own separate set of physical mass eigenstates and these will not have couplings to SM fermions. The only exception to this statement is the following. In the case of the $Y = \pm 2$ triplet representation there is the possibility of $\delta^{++}l^-l^-$, $\delta^+l^-\nu_l$ and $\delta^0\nu_l\nu_l$ (Majorana-like) couplings, where the $l^-$'s and $\nu_l$'s are the left-handed objects with $Y = -1$. Analogous couplings are not possible in the case of the $Y = 0$ triplet representation since the right-handed leptons and neutrinos have $T = 0$. If Higgs-lepton-lepton couplings are present for the $T = 1, Y = \pm 2$ case, they play a particularly prominent phenomenological role when the vev for the neutral field is zero (or very small).

We wish to note that the discussions presented here for purely Higgs sector additions to the SM require some modification in the context of the Little Higgs models which also contain Higgs triplets. In particular, custodial symmetry issues become much more complicated, and their implications are closely tied to the ultraviolet cutoff of the effective theory. For example, in a model with an ultraviolet cutoff it is natural to allow for the presence of "non-renormalizable" effective operators suppressed by some inverse power of the cutoff that could affect such observables as $\rho$ at tree-level.

### *13.1.1 Model Considerations*

Higgs triplets can, in principle, carry any hypercharge $Y$. The real triplet with $Y = 0$ and the complex triplet with $Y = \pm 2$ both contain a neutral Higgs field, namely that component with $T_3 + \frac{1}{2}Y = 0$. For these cases, a non-zero vev for the neutral component (generically denoted by $\phi^0_{T=1}$) leads to a deviation in the tree-level prediction of $\rho = \frac{m_W{}^2}{m_Z{}^2 \cos^2\theta_W} \sim 1$, whereas $\rho \sim 1$ is automatic for doublets (plus possible singlets). For triplet models, the simplest possibility will therefore be that $\langle \phi^0_{T=1} \rangle = 0$. If $\langle \phi^0_{T=1} \rangle \neq 0$, one can avoid large corrections to the electroweak $\rho$ parameter at tree level by either (i) choosing $\langle \phi^0_{T=1} \rangle$ very small compared to the vevs of neutral doublet fields or (ii) arranging the triplet fields and the vevs of their neutral members so that a custodial $SU(2)$ symmetry, $SU(2)_C$, is maintained. A number of models of type (ii), with a custodial $SU(2)_C$ symmetry, have been proposed in the literature. The most popular is that containing a real $Y = 0$ triplet and a $Y = \pm 2$ complex triplet as proposed in [18] and later considered in greater depth in [19, 20], with further follow-up in [21]. The Higgs potential for the model can be constructed in such a way that it preserves the tree-level $SU(2)_C$ symmetry. For such a model, the $SU(2)_C$ is maintained after higher-order loop corrections from Higgs self-interactions.

However, in all triplet models with $\langle \phi^0_{T=1} \rangle \neq 0$, the presence of interactions of the $U(1)$ $B$ field with the Higgs sector necessarily violates $SU(2)_C$. This is because the $U(1)$ hypercharge operator corresponds to the $T_3^C$ of the would-be custodial $SU(2)_C$ and the non-zero vev then explicitly breaks the $SU(2)_C$ $T_3^C$ symmetry. As a result, loop corrections to the $W$ and $Z$ masses are infinite [22]. In fact, corrections to $\rho$ are quadratically divergent and achieving $\rho = 1$ is at least as unnatural in the presence of such quadratic divergences as is achieving a low mass for the Higgs boson in the presence of quadratic divergences due to SM particle loops. Just as in the SM, these quadratic divergences can be ignored and computations can be carried out using standard renormalization procedures, where a set of experimental observables (measured in some appropriate way) are input and other observables are computed in terms of them. In the case of triplet models with $\langle \phi^0_{T=1} \rangle \neq 0$, $\rho$ must be renormalized, implying that $\rho = 1$ is no longer a prediction of the theory but rather the observed value of $\rho$ (or some other related electroweak parameter, often chosen to be $\sin\theta_W$ as defined via the $Zee$ coupling, $-i\bar{e}(v_e + \gamma_5 a_e)\gamma_\mu e Z^\mu$, where $1 - 4\sin^2\theta_W = \mathrm{Re}v_e/\mathrm{Re}a_e$, $\theta_W$ being the Weinberg angle) must be considered as an additional experimental input. Other observables in the electroweak sector can then be computed in terms of the observed value of $\rho$ [23–25]. To obtain $\rho = 1$ up to only *finite* radiative corrections, implying that $\rho \sim 1$ is a natural prediction of the model, $\langle \phi^0_{T=1} \rangle = 0$ is required. This implies an extra custodial symmetry of the theory such that the triplet fields generate only finite loop corrections to $\rho$. The phenomenology associated with this class of triplet models is very different from that which arises in models with $\langle \phi^0_{T=1} \rangle \neq 0$. We will consider the two possibilities in turn.

### *13.1.2 Phenomenology when a neutral triplet field has non-zero vev ( $\langle \phi^0_{T=1} \rangle \neq 0$)*

The general tree-level expression for $\rho$ is [26]

$$\rho = \frac{\sum_{T,Y}[4T(T+1) - Y^2]|V_{T,Y}|^2 c_{T,Y}}{\sum_{T,Y} 2Y^2 |V_{T,Y}|^2}\,, \tag{13.1}$$

where $\langle \phi^0_{T,Y} \rangle = V_{T,Y}$ ($\phi^0_{T,Y}$ being the neutral field in a given $T, Y$ representation) and $c_{T,Y} = 1$ $(1/2)$ for a complex (real) representation. If we consider a Higgs sector with one $Y = 1$ doublet and one $Y = 0$ or $Y = \pm 2$ triplet, and define $r_{1,0} = V_{1,0}/V_{1/2,1}$ and $r_{1,2} = V_{1,\pm 2}/V_{1/2,1}$, we obtain

$$\rho = \begin{cases} 1 + 2r_{1,0}^2\,, & Y = 0 \\ (1 + 2r_{1,2}^2)(1 + 4r_{1,2}^2)^{-1}\,, & Y = \pm 2 \end{cases} \tag{13.2}$$

so that $\rho - 1 > 0$ $(< 0)$ for the $T, Y = 1, 0$ $(1, \pm 2)$ case. If there is more than one $Y = 1$ doublet field, the above results can be generalized by replacing $V_{1/2,1}^2 \to \sum_k V_{k\,1/2,1}^2$. The notation we will employ for the $Y = 1$ doublet and the $Y = 0$ and $Y = \pm 2$ triplets is:

$$\phi_{T=1/2,Y=1} = \begin{pmatrix} \phi^+ \\ \phi^0 \end{pmatrix}\,, \quad \phi_{T=1,Y=0} = \begin{pmatrix} \xi^+ \\ \xi^0 \\ \xi^- \end{pmatrix}\,, \quad \phi_{T=1,Y=2} = \begin{pmatrix} \chi^{++} \\ \chi^+ \\ \chi^0 \end{pmatrix}\,. \tag{13.3}$$

Of course, if the neutral fields have non-zero vevs, the above refers to the quantum fluctuations of the fields relative to these vevs. In the conventions we employ, $\phi^{+\,*} = -\phi^-$, $(\xi^0)^* = \xi^0$, $(\xi^+)^* = -\xi^-$, $(\chi^{++})^* = \chi^{--}$ and $(\chi^+)^* = -\chi^-$.

We have already noted the two experimental signatures that would immediately signal a Higgs sector with representations beyond the usual $T = 1/2, Y = 1$ doublets (at least one of which will always be assumed to be present) and $T = 0, Y = 0$ singlets. First, as exemplified by the $T = 1, Y = \pm 2$ representation, there can be doubly-charged Higgs bosons. More exotic choices of $T$ and $Y$ can yield Higgs bosons with still larger integer charge or even fractional charge. Second, triplet models that have a non-zero vev for a neutral field member typically predict a non-zero $ZW^\pm H^\mp$ vertex, where $H^\pm$ is some charged Higgs (or linear combination of charged Higgs) of the model. The general result (allowing for any $T, Y$ and assuming only neutral fields have vevs) is[1]

$$\mathcal{L}_{H^\pm W^\mp Z} = -\frac{g^2}{2c_W}\kappa\left[W^+{}_\mu Z^\mu H^- + \text{h.c.}\right] , \tag{13.4}$$

where

$$\kappa^2 = \sum_{T,Y} Y^2\left[4T(T+1) - Y^2\right]|V_{T,Y}|^2 - \frac{\left\{\sum_{T,Y} 2Y^2|V_{T,Y}|^2\right\}^2}{\sum_{T,Y}\left[4T(T+1) - Y^2\right]|V_{T,Y}|^2 c_{T,Y}} \tag{13.5}$$

This formula has many implications. For instance, if $\rho = 1$ we can use Eq. (13.1) to simplify Eq. (13.5) to obtain:

$$\kappa^2 \overset{\rho=1}{=} \frac{\sum_{T,Y}\left[4T(T+1) - Y^2 - 2\right]|V_{T,Y}|^2}{\sum_{T,Y} 2Y^2|V_{T,Y}|^2} \geq \frac{\sum_{T,Y}\left[4T - 2\right]|V_{T,Y}|^2}{\sum_{T,Y} 2Y^2|V_{T,Y}|^2} , \tag{13.6}$$

where to obtain the 2nd equality one must note that $|Y/2| = |T_3|$ is required for the neutral field with non-zero $V_{T,Y}$ and that $|T_3| \leq T$. Eq. (13.6) shows that if $\rho = 1$ then $\kappa^2 = 0$ is only possible if all representations with non-zero $V_{T,Y}$ have $T = 1/2$ (i.e., if they are doublets). In other words, any model containing triplet or higher Higgs representations with a neutral field member that has a non-zero vacuum expectation value, and that simultaneously yields $\rho = 1$ at tree-level, must have at least one charged Higgs with non-zero coupling to the $WZ$ channel [26].

It will be useful to discuss several specific triplet models in order to illustrate some of the many subtleties. We will consider two models: a) the model with one $T = 1/2, Y = 1$ doublet plus one $T = 1, Y = 0$ triplet; and b) the model with one $T = 1/2, Y = 1$ doublet, one $T = 1, Y = 0$ triplet and one $T = 1, Y = \pm 2$ triplet with Higgs potential such that $\langle\xi^0\rangle = \langle\chi^0\rangle$ so that there is a custodial symmetry at tree-level implying $\rho(\text{tree}) = 1$. We will end with a discussion of the implications of Higgs-lepton-lepton couplings for a $Y = 2$ triplet when $\langle\chi^0\rangle \neq 0$.

#### 13.1.2.1 The model with one $T = 1/2, Y = 1$ doublet plus one $T = 1, Y = 0$ triplet

The model with one $T = 1/2, Y = 1$ doublet plus one $T = 1, Y = 0$ triplet model illustrates many important aspects of triplet models. For this model, we will write, using the notation of Eq. (13.3), $\langle\phi^0\rangle = v/\sqrt{2}$ and $\langle\xi^0\rangle = v' = \frac{1}{2}v\tan\beta$, where $\beta$ is the mixing angle that isolates the charged Goldstone boson absorbed in giving mass to the $W$ (see below). Then at tree level, $\rho = 1/c_\beta^2$. (We will write $t_\beta \equiv \tan\beta$, $c_\beta = \cos\beta$, and so forth). The physical Higgs bosons comprise the $h^0$, the $k^0$ and the $h^\pm$. In general, the physical eigenstate $h^0$ is a mixture of $\text{Re}\phi^0/\sqrt{2}$ and $\xi^0$ (since $\xi^0$ is a real field, the Goldstone boson eaten by the $Z$ is $\text{Im}\phi^0/\sqrt{2}$) and $h^+$ is a mixture of $\phi^+$ and $\xi^+$ (the orthogonal $g^+$ being eaten by the $W^+$). The mixings are specified by two angles, $\beta$ and $\gamma$, according to:

$$\begin{pmatrix} g^+ \\ h^+ \end{pmatrix} = \begin{pmatrix} c_\beta & s_\beta \\ -s_\beta & c_\beta \end{pmatrix}\begin{pmatrix} \phi^+ \\ \xi^+ \end{pmatrix} , \quad \begin{pmatrix} h^0 \\ k^0 \end{pmatrix} = \begin{pmatrix} c_\gamma & s_\gamma \\ -s_\gamma & c_\gamma \end{pmatrix}\begin{pmatrix} \text{Re}\phi^0/\sqrt{2} \\ \xi^0 \end{pmatrix} . \tag{13.7}$$

In general, $m_{h^0}$, $m_{k^0}$ and $m_{h^\pm}$ can be adjusted independently of one another. However, if the Higgs potential trilinear and quartic couplings are kept finite (they should not be too big in order to avoid a non-perturbative regime) then in the limit of $t_\beta \to 0$ the $h^0$ approaches the SM Higgs boson while the $k^0$ and $h^\pm$ have $m_{k^0} \sim m_{h^\pm}$ and decouple. The limit $t_\beta \to 0$ requires $\lambda_4 \to 0$ in the $\lambda_4 \phi^\dagger \sigma^a \phi \xi^a$ [$a = 1, 2, 3$, where, e.g., $\xi^+ = \frac{1}{\sqrt{2}}(\xi^1 + i\xi^2)$] Higgs potential term that explicitly mixes the doublet and triplet fields. It is also possible to choose the Higgs potential so that it has an extra custodial $SU(2)_C$ symmetry from the beginning by setting $\lambda_4 = 0$. In this case, $v' = 0$ and $m_{h^\pm} = m_{k^0}$ and the triplet Higgs sector gives no correction to $\rho$ at any order. As a final theoretical point, we note that the unitarity constraints for $W^+W^- \to W^+W^-$ scattering and/or perturbativity for the Higgs potential parameters imply that if $\tan\beta$ is not small then all the physical Higgs states should have mass below $\sim 1$ TeV.

[1] We follow [26], correcting a small error.

We now turn to the prediction for $\rho$ in the $T = 1, Y = 0$ model just described. The prediction must be compared to the standard precision electroweak constraints as encapsulated, for instance, in the $S, T, U$ parameters of [27]. In particular, it is useful to note that $\alpha T = \rho - 1$. The current $S, T$ plot, based purely on $Z$-pole data, from the LEP Electroweak Working Group (LEWWG) [28] is shown as the left, top plot in Fig. 13.1. Note that the data have a somewhat positive $S$ and positive $T$ relative to the $m_{h_{SM}} = 114$ GeV prediction. If one includes NuTeV, atomic parity violation and SLAC results on Moller scattering as well, one finds the $S, T$ ellipse [29] (with updated $S, T$ plot provided by P. Langacker) appearing at the bottom of Fig. 13.1. The center of the ellipse shifts to slightly negative $S$ and $T$ values. Assuming $U = 0$ the center is at (assuming $m_{h_{SM}} = 117$ GeV and $m_t = 172.6 \pm 2.9$ GeV)

$$S = -0.07 \pm 0.09\,, \quad T = -0.03 \pm 0.09\,. \tag{13.8}$$

These latter values are completely consistent with $S = T = 0$ for new physics contributions. The value of $\rho$ corresponding to the above $T$ is $\rho = 0.9990 \pm 0.0009$, leaving very little room for new physics effects. At tree-level, the fit to the pure $Z$-pole LEWWG ellipse can be improved by using a fairly heavy SM Higgs with $m_{h_{SM}} \sim 500$ GeV along with a $T = 1, Y = 0$ triplet with $r_{1,0} \sim 0.03$, corresponding to $\beta \sim 0.045$ radians. The large $m_{h_{SM}}$ value moves the prediction towards positive $S$ and negative $T$. This is compensated by the correction from the above $r_{1,0}$ value which gives a positive $T$ contribution that places the net prediction more or less at the center of the ellipse. If one employs all available data as represented by the bottom-center $S, T$ ellipse, $\beta \sim 0$ is preferred, but $\beta \sim 0.045$ and a heavy Higgs is still a possibility, giving a prediction in the upper right-hand corner of the PDG ellipse.

Of course, loop corrections from the triplet sector should also be included [23, 30]. For small $\beta$ (as above), and assuming for simplicity that parameters are chosen so that there is no mixing between the doublet $\mathrm{Re}\phi^0/\sqrt{2}$ and the triplet $\xi^0$ ($\gamma = 0$), the one-loop triplet contributions give [30]

$$S_{1,0} = 0, \quad T_{1,0} \sim \frac{1}{6\pi}\frac{1}{s_W^2 c_W^2}\frac{\Delta m^2}{m_Z{}^2}, \quad U_{1,0} = \frac{\Delta m}{3\pi m_{h^\pm}}, \tag{13.9}$$

where $\Delta m = m_{k^0} - m_{h^\pm}$, and $s_W$ and $c_W$ are the sine and cosine of the standard electroweak angle, respectively. If the $\lambda_i$ parameters of the most general Higgs potential are kept fixed (in particular, with $\lambda_4 \neq 0$) and $r_{1,0} \to 0$, then $m_{k^0}, m_{h^\pm} \to \infty$ while $\Delta m \to 0$. In this limit, the triplet decouples from the precision electroweak parameters. A small value for $\Delta m$ is thus a natural possibility.

In a fully general treatment of $\rho$ at one loop, we have already noted that $\rho$ (or some related parameter) must be input to the renormalization scheme as an additional observable. In [25, 31], a study of the $m_t$ dependence of the precision electroweak constraints deriving from $G_\mu = \frac{\pi\alpha}{\sqrt{2} m_W{}^2 \sin^2\theta_W}(1 + \Delta R)$ and $\Gamma_Z$ is performed in the $\tan\beta \neq 0$ context. In their approach, the value of $\sin\theta_W$ from the $Zee$ vertex is input as the additional observable. An interesting phenomenon emerges: the sensitivity of $\Delta R$ (through fixing $\sin\theta_W$) to $m_t$ is greatly reduced. In the SM, $\Delta R$ depends on $m_t$ quadratically, whereas in the triplet model with $\sin\theta_W$ as the additional experimental input parameter, renormalization proceeds differently, and $\Delta R$ is only logarithmically sensitive to $m_t$. This weak dependence is shown in the left plot of Fig. 13.2. Additional flexibility in the predictions arises if one allows for a range of $m_{k^0}$ and $m_{h^\pm}$ values — a large selection of models are consistent with current data.

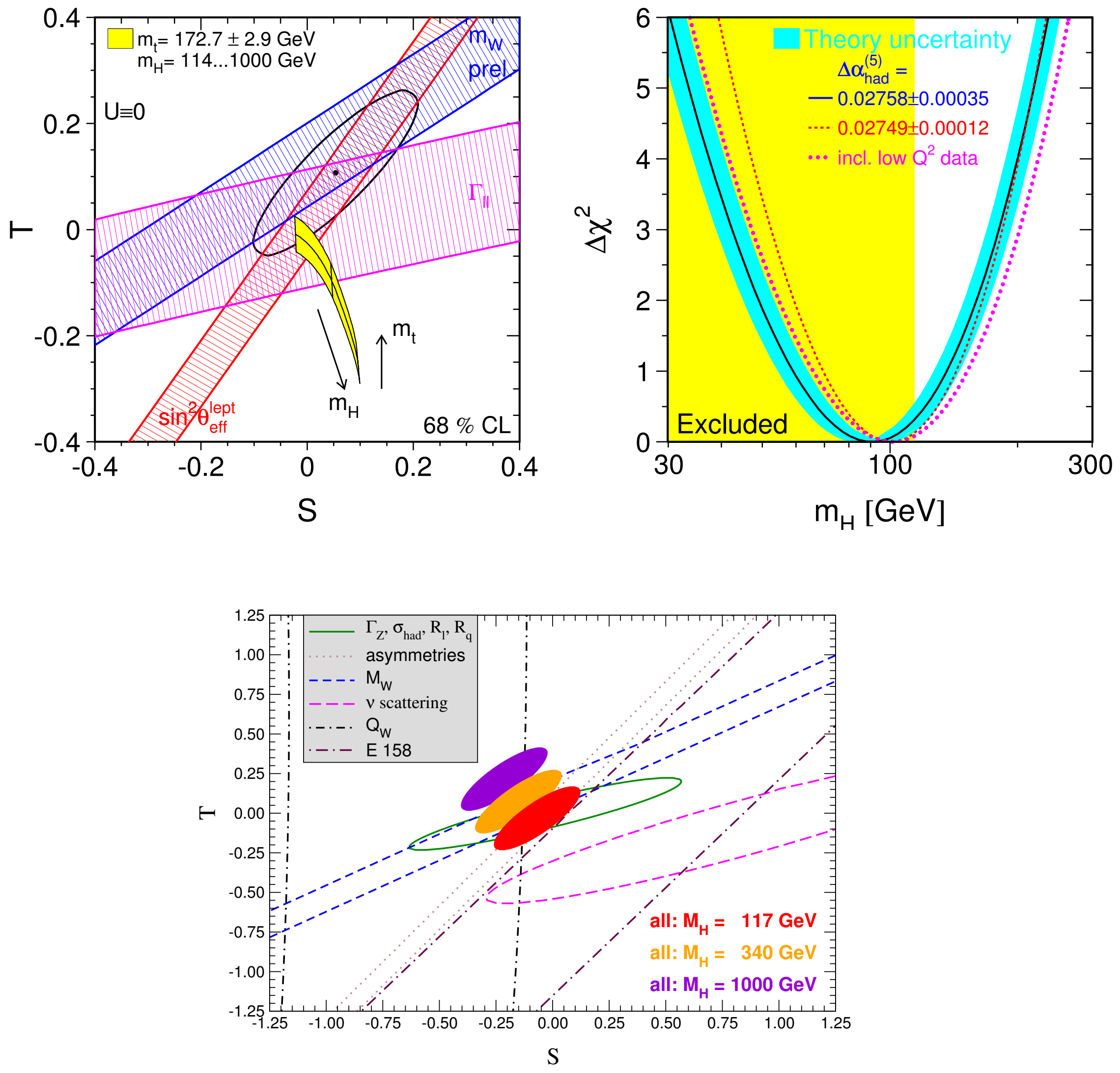

Fig. 13.1: Top left: The latest $S,T$ ellipse for $Z$-pole data only from the LEWWG is shown. The reference SM Higgs and top masses employed are $m_{h_{SM}} = 150$ GeV and $m_t = 175$ GeV and $U = 0$ is assumed. The SM prediction as a function of $m_{h_{SM}}$ from $m_{h_{SM}} = 114$ GeV to 1 TeV with $m_t = 172.7 \pm 2.9$ GeV is shown. Top right: The blueband plot showing that $m_{h_{SM}} \sim 100$ GeV provides the best SM fit. Bottom center: The latest $S,T$ ellipse to appear in the PDG (thanks to P. Langacker) which includes $Z$-pole data as well as parity violation and NuTeV data.

Other schemes are also possible. Instead of inputting $\sin\theta_W$ as the extra observable, one could directly input $T$ as the extra observable and simply fix it to agree with the value at the center of the ellipse. Then, $\sin\theta_W$ would acquire logarithmic sensitivity to $m_t$.

Of course, other observables also depend upon $m_t$, most notably the $Z$ boson width $\Gamma_Z$ which has a strong $m_t$ dependence from vertex corrections to the $Z \to b\bar{b}$ decay width — $\Gamma_Z$ decreases rapidly with increasing $m_t$ as shown in the right-hand plot of Fig. 13.2. Combining the $m_W$ and $\Gamma_Z$ sensitivity to $m_t$ gives a prediction for $m_t$ within about 30 GeV. Whatever scheme is employed, the important consequence is that the SM *prediction* of the top mass from precision electroweak data is considerably weakened in triplet models with non-zero vev for the neutral field.

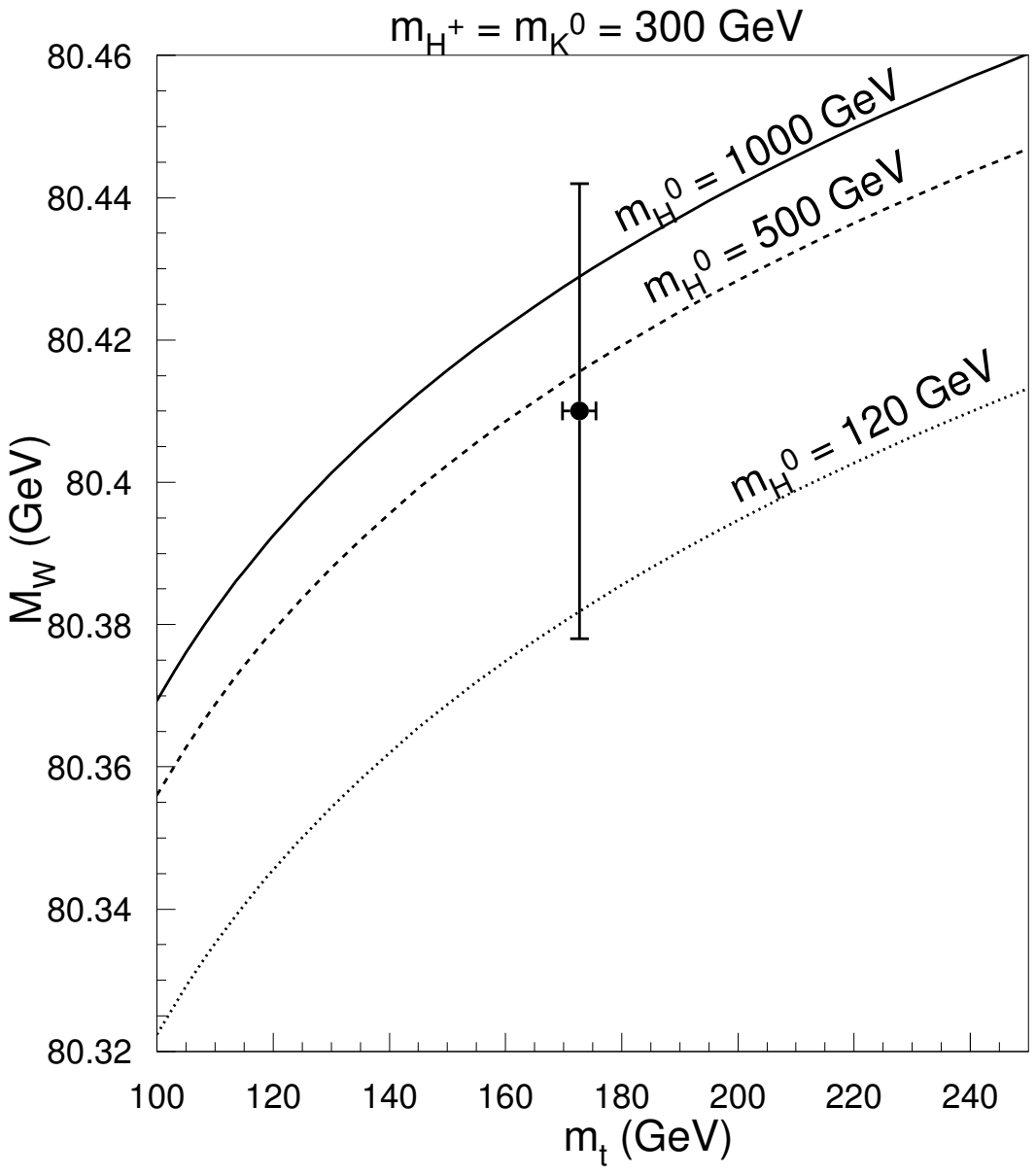

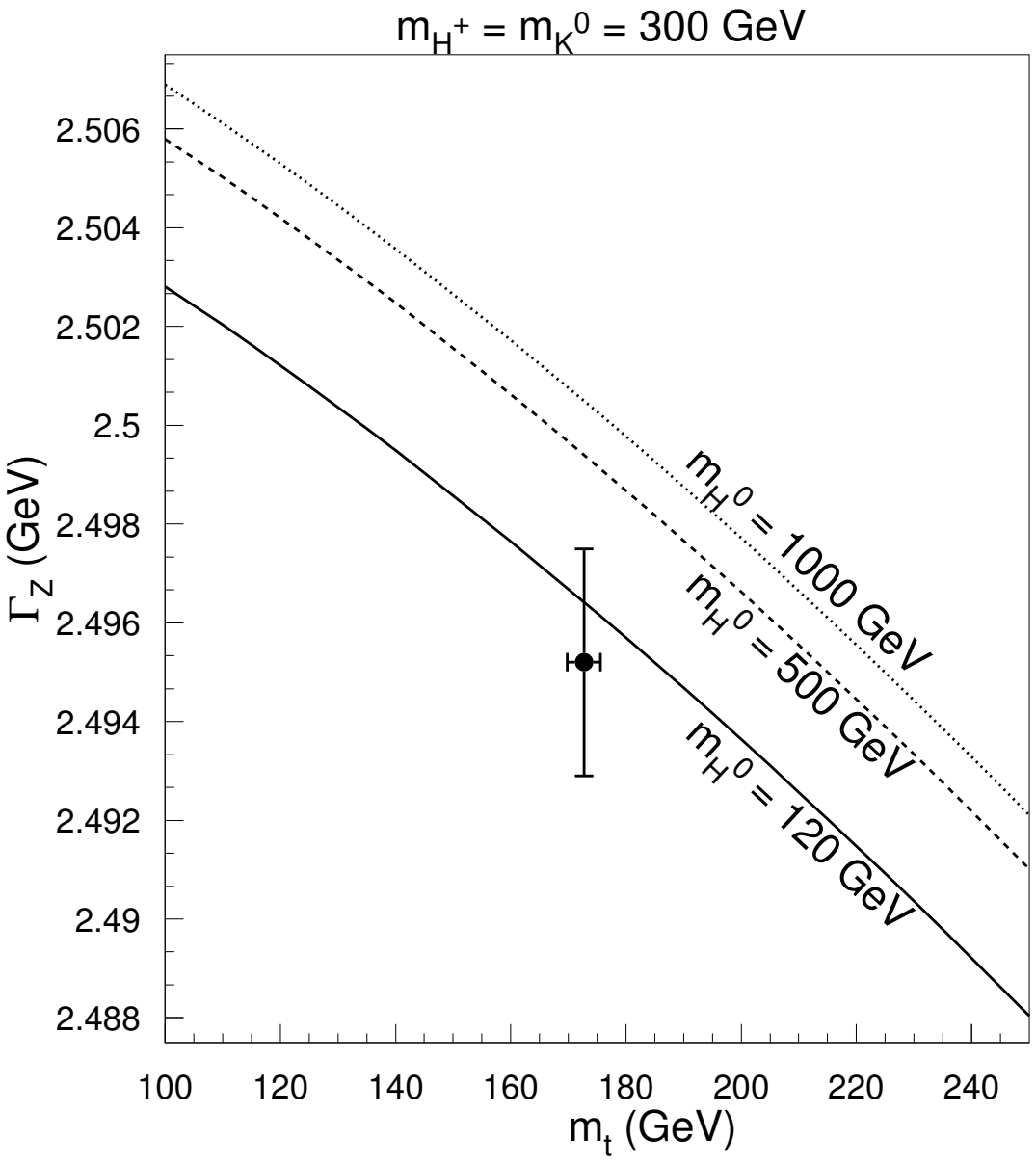

Fig. 13.2: Left: Prediction for $m_W$ as a function of $m_t$ in the $Y = 0$ triplet model (TM). Right: The prediction of the TM for $\Gamma_Z$ as a function of $m_t$. The $1\sigma$ error bars are shown in both plots. Results shown are for $m_{h^+} = m_{k^0} = 300$ GeV and for $m_{h^0}$ values as indicated by the different lines.

Of course, we now have a very accurate measurement of $m_t$. We have already noted that this measurement plus the various LEP and other precision measurements very strongly constrain $\rho$, allowing only very small deviations from $\rho = 1$ corresponding to a small non-zero vev for $\xi^0$ and requiring $m_{h^0}$ to be near 100 GeV.

Some final notes are the following. Although there is no doubly-charged Higgs boson in this model, for $\tan\beta \neq 0$ there is a non-zero $H^+W^-Z$ vertex specified by setting $\kappa = v\sin\beta$ in Eq. (13.4). Experimental probes of this vertex will be discussed shortly. It is also quite amusing to note [32] that a model containing two $Y = 1$ doublets and one $Y = 0$ triplet leads to gauge coupling unification at a scale of $M_U \sim 1.6 \times 10^{14}$ GeV, for $\alpha_s(m_Z) \sim 0.115$. While full gauge group unification cannot occur at this $M_U$ without encountering difficulties with proton decay, coupling unification without gauge group unification is a feature of certain string models.

#### 13.1.2.2 The model with one $T = 1/2, Y = 1$ doublet, one $T = 1, Y = 0$ triplet and one $T = 1, Y = \pm 2$ triplet with $\rho(\text{tree}) = 1$

If one wishes to construct a triplet model with non-zero neutral triplet vevs and $\rho = 1$ at tree-level, despite the fact that at the one-loop level $\rho$ is infinitely renormalized and simply must be treated as an input parameter, then one can consider the model of [18] containing one doublet, one $Y = 0$ triplet and one $Y = 2$ triplet. The following outlines the full analysis of this model presented in [20]. A convenient representation of the Higgs boson sector is

$$\phi = \begin{pmatrix} \phi^+ \\ \phi^0 \end{pmatrix}, \quad \chi = \begin{pmatrix} \chi^{0*} & \xi^+ & \chi^{++} \\ \chi^- & \xi^0 & \chi^+ \\ \chi^{--} & \xi^- & \chi^0 \end{pmatrix}. \tag{13.10}$$

By taking $\langle\chi^0\rangle = \langle\xi^0\rangle = b$ and $\langle\phi^0\rangle = a/\sqrt{2}$, at tree-level one finds $\rho = 1$ with ${m_W}^2 = {m_Z}^2\cos^2\theta_W = \frac{1}{4}g^2v^2$ where $v^2 \equiv a^2 + 8b^2$. The amount of vev carried by the doublet sector is then characterized by $c_H \equiv \cos\theta_H = a/v$ and that in the triplet sector is then given by $s_H \equiv \sin\theta_H = 2\sqrt{2}b/v$. Thus, $t_H \equiv \tan\theta_H$ characterizes the amount of the $W$ mass coming from the doublet vs. the triplet fields. To fit a deviation from $\rho = 1$ at tree-level, the $\langle\chi^0\rangle = \langle\xi^0\rangle$ equality could be slightly broken. The one-loop corrections to $S, T, U$ have not been computed in this model, but the Higgs sector parameters could probably be adjusted to allow for $\rho = 1$ at tree level along with a fairly heavy SM-like Higgs and an appropriate positive addition to $T$ so as to remain within the current $S, T$ ellipse of the last plot in Fig. 13.1.

If the Higgs potential preserves the custodial $SU(2)_C$, as desirable to minimize deviations from $\rho \sim 1$ arising from Higgs loops, then the physical states of this model can be classified according to their transformation properties under the tree-level custodial $SU(2)_C$. One finds a five-plet $H_5^{\pm\pm}, H_5^{\pm}, H_5^0$, a three-plet $H_3^{\pm}, H_3^0$, and two singlets, $H_1^0$ and $H_1^{0\prime}$. All members of a given multiplet are degenerate in mass at tree-level and only the $H_1^0$ and $H_1^{0\prime}$ can mix. For simplicity, we will present a discussion in which this mixing is absent and the $H_1^0$ and $H_1^{0\prime}$ are mass eigenstates. The phenomenology of the model reveals many new features and corresponding phenomenological possibilities.

Several features of the $VV$ and $f\overline{f}$ couplings should be noted. First, ignoring the $HV$ and $HH$ type channels, at tree level the $H_5$'s couple and decay only to vector boson pairs (we return later to the possibility of $U(1)$-conserving Higgs-lepton-lepton couplings), while the $H_3$'s couple and decay only to fermion-antifermion pairs. Second, we observe that the SM is regained in the limit where $s_H \to 0$, in which case the $H_1^0$ plays the role of the SM Higgs and has SM couplings and the $H_3$'s decouple from fermions. However, in this model with custodial $SU(2)_C$ symmetry, there is no intrinsic need for $s_H$ to be small in order to keep $\rho$ near 1 at tree-level. When $s_H \neq 0$, many new phenomenological features emerge. First, using the setup giving $\rho = 1$ at tree-level, there is a non-zero $H_5^+ W^- Z$ coupling specified by $\kappa = s_H v$, where $\frac{1}{2} g v = m_W$. The second highly distinctive feature is the presence of a doubly-charged $H_5^{++}$ with coupling to $W^+W^+$ given by $\sqrt{2} g m_W s_H$. The effects of both these new couplings are maximized in the $c_H \to 0$ limit where all the electroweak symmetry breaking resides in the triplets. We note that if $c_H$ is small then the couplings of the doublet to the fermions must be much larger than in the SM in order to obtain the experimentally determined quark masses. Then, the Higgs bosons that do couple to fermions have much larger fermion-antifermion pair couplings and decay widths than in the SM.

Constraints on the model are significant. First, there is unitarity for vector-boson scattering. For $t_H \neq 0, \infty$, many of the Higgs bosons contribute to preserving unitarity for the various $VV$ scattering processes. For example, unitarity for $ZW^- \to ZW^-$ requires the presence of $H_1^0$, $H_1^{0\prime}$, and $H_5^0$ in $t$-channel graphs and $H_5^-$ in $s$- and $u$-channel graphs. The masses of all four must lie below $\sim 1$ TeV in order to avoid unitarity violation. In $W^+W^+ \to W^+W^+$, the $H_1^0$, $H_1^{0\prime}$ and $H_5^0$ appear in $t$-channel and $u$-channel graphs while the $H_5^{++}$ appears in the $s$-channel. Again, all masses must lie below $\sim 1$ TeV to avoid unitarity violation. Note that for $s_H \to 1$ the $H_1^{0\prime}$ can have $W^+W^-, ZZ$ couplings that are larger than in the SM so long as canceling contributions from exchanges of one of the $H_5$ states are present. Of course, if $s_H \sim 0$, then the masses of the triplet Higgs states (i.e., the $H_1^{0\prime}$ and the $H_5$ states) are unconstrained by unitarity, while if $c_H \sim 0$ then $m_{H_1^0}$ can be arbitrarily large.

We now discuss the general phenomenology when $s_H$ is not near zero. In this case, the Higgs-lepton-lepton couplings discussed later do not play a role. Due to space limitations, we mainly restrict the discussion to a very brief outline of the phenomenology of the $H_5^{++}$, the hallmark state of this model. Quantum numbers imply that the only tree-level decays of the $H_5^{++}$ are to virtual or real $H_3^+W^+$, $W^+W^+$ or $H_3^+H_3^+$ channels. When the $H_5$'s are sufficiently light both of the bosons must be virtual. This results in four-body states with two non-resonant fermion-antifermion pairs. In practice, in the four-body region only $W^{+*}W^{+*}$ diagrams are important due to the weak coupling of the $H_3^+$ to $f\bar{f}$ states. As a result, the $H_5^{++}$ has a very long lifetime at low mass. For larger $m_{H_5}$, mixed real-virtual channels take over. The final state then consists of a real $V$ or $H$ plus a $f\bar{f}$ pair. The possibilities include $H_3^+W^{+*}$ ($W^+H_3^{+*}$ and $H_3^+H_3^{+*}$ contributions are much smaller) and $W^+W^{+*}$. At still higher $m_{H_5}$, we enter the region where at least one of the two-body modes — $W^+W^+$, $H_3^+W^+$ and $H_3^+H_3^+$ — is allowed. In all cases, secondary $H_3^+$ and $W^+$ bosons typically decay to an $f\bar{f}$ final state. In Fig. 13.3 we present the lifetime $c\tau$ for the $H_5^{++}$ as a function of $m_{H_5}$. We have chosen $m_{H_3} = m_W$ for this plot. For $\tanh = 0.1$, the $W^+W^+ \to H_5^{++}$ coupling, given by $\sqrt{2}gm_W s_H$, is suppressed and $c\tau$ can be larger than a $\mu m$ for $m_{H_5} \lesssim 45$ GeV.

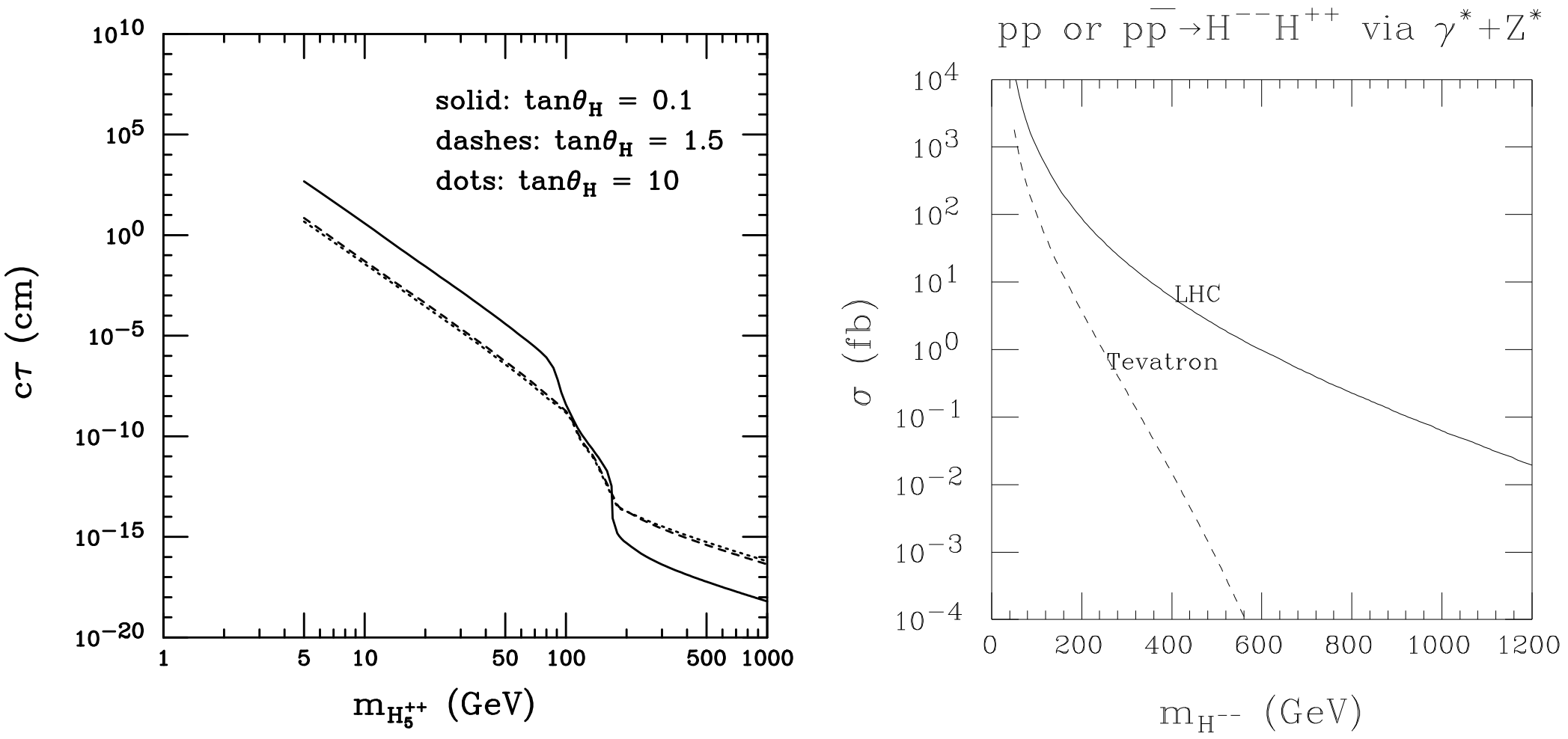

Fig. 13.3: Left: The lifetime $c\tau$ (in $cm$) of the $H_5^{++}$ as a function of $m_{H_5}$, assuming $m_{H_3} = m_W$. Right: The Drell-Yan cross section for $H_5^{++}H_5^{--}$ pair production as a function of $m_{H_5}$ (labeled as $m_{H^{--}}$) at the LHC and Tevatron, taken from [33].

Regarding the experimental implications, we note that LEP $Z$-pole data would have revealed an extra width contribution coming from $Z \to H_5^{++}H_5^{--}$ for $m_{H_5} \lesssim 40$ GeV. For higher $m_{H_5}$, continuum pair production is the most relevant process, yielding (independent of $t_H$)

$$\sigma(e^+e^- \to H_5^{++}H_5^{--}) \overset{\sqrt{s}\gg m_{H_5}}{\to} \frac{1 + 4\sin^4\theta_W}{2\sin^4 2\theta_W}\,\sigma(e^+e^- \to \mu^+\mu^-) \qquad (13.11)$$

which corresponds to slightly more than 1 unit of $R$. For $m_{H_5} \gtrsim 45$ GeV, the $H_5^{++}$ decay becomes prompt and only the two fermion-antifermion pairs emerging from the decay are visible. Although the cross section above is substantial, we are not aware of LEP $\sqrt{s} \sim 200 - 210$ GeV analyses that exclude $H_5^{++}H_5^{--}$ production with cascade decays to such complicated final states.

Turning to $H_5^{++}$ production at hadron colliders, there are two basic processes. First, there is $H_5^{++}H_5^{--}$ Drell-Yan pair production via $\gamma^*, Z^*$. The cross section for this process is independent of $t_H$ and is presented for the LHC and Tevatron in Fig. 13.3. These tree-level cross sections are increased significantly (20% to 30%) by QCD radiative corrections [34]. Second, there is $W^+W^+ \to H_5^{++}$ fusion production, which is large for large $t_H$, but is quickly suppressed as $t_H \to 0$. An estimate for the cross section is easily obtained starting with the fact that

$$\gamma(H_5^{++} \to W^+W^+) = 2s_H^2\gamma(h_{SM} \to W^+W^-)\,. \qquad (13.12)$$

For $pp$ collisions, the $W^+W^+$ luminosity is slightly larger than the $W^+W^-$ luminosity, but after adding in the $ZZ \to h_{SM}$ fusion processes one obtains at LHC energies and moderate $m_{H_5}$

$$\sigma(W^+W^+ \to H_5^{++}) \sim s_H^2\,\sigma(W^+W^-, ZZ \to h_{SM}) \qquad (13.13)$$

for $m_{H_5} = m_{h_{SM}}$. As regards the $H_5^{++}$ decays, the main decay other than $H_5^{++} \to W^+W^+$ is likely to be $H_5^{++} \to H_3^+W^+$. In the limit where $m_{H_5} \gg m_{H_3}, m_W$, the widths of the two modes are in the ratio $\frac{\gamma(H_5^{++}\to W^+W^+)}{\gamma(H_5^{++}\to H_3^+W^+)} \to 2t_H^2$ . Thus, for $t_H \gtrsim 1$ a highly distinctive signature for the $H_5^{++}$ would arise via the process $W^+W^+ \to W^+W^+$ with $W^+ \to l^+\nu_l$ decays [35]. To cleanly observe the $s$-channel $H_5^{++}$ exchange as a peak in $M_{W^+W^+}$, given the presence [20] of $t$- and $u$-channel graphs with exchanges of $H_5^0$, $H_1^0$ and $H_1^{0\prime}$, would require using the mode where one $W^+ \to l^+\nu_l$ while the second decays via $W^+ \to q'\overline{q}$. The charge conjugate process, $W^-W^- \to W^-W^-$, will also be present at a somewhat lower rate. At the Tevatron, the rate for $W^+W^+ \to H_5^{++}$ will be rather small. Only searches based on Drell-Yan production are likely to be fruitful. There are currently no Tevatron searches for $H_5^{++}H_5^{--}$ pair production with $m_{H_5} > 2m_W$ based on $H_5^{++} \to W^+W^+$ and $H_5^{--} \to W^-W^-$.

At a linear collider it is possible to operate in the $e^-e^-$ mode, in which case $W^-W^- \to W^-W^-$ scattering will take place [20, 36]. Using the $W^- \to q\overline{q}'$ decay modes, the $W^-W^-$ mass can be reconstructed. If there is an $H_5^{--}$ present in the $s$-channel, sizable bumps in the $M_{W^-W^-}$ distribution will emerge for $t_H = 1$ if $m_{H_5} \sim 200 - 300$ GeV, assuming $\sqrt{s} = 500$ GeV. Another interesting possibility is $W^-W^- \to H_3^-H_3^-$, with $H_5^{--}$ exchange in the $s$-channel [37]. The reaction $W^-W^- \to H_5^-H_5^-$ occurs via $t$- and $u$-channel Higgs exchanges. Although there is no $s$-channel resonance, the size of the cross section depends strongly on $t_H$ and the masses of the exchanged $H_5^0$ and $H_3^0$.

It is also interesting to note that the $H_1^{0\prime}$ can be quite light and at tree-level would only decay via the $s_H$ suppressed $H_1^{0\prime} \to W^{-*}W^{+*} \to fermions$. As pointed out in [38], see also [21], the $\gamma\gamma$ loop-induced decay can be quite competitive in such an instance and some experimental limits may be applicable [39], depending on the $t_H$ value.

As already noted, triplet models with $\rho = 1$ at tree-level and non-zero neutral field vevs will yield a non-zero charged-Higgs-$ZW$ vertex. In general, observation of such an interaction would be an immediate signal for a Higgs sector with $SU(2)_L$ representations beyond the doublet. In the present model, for $s_H \neq 0$ there is a non-zero $H_5^+ZW^-$ vertex given by $\kappa = s_H v$. In the $T = 1/2, Y = 1$ plus $T = 1, Y = 0$ model there was a non-zero $h^+ZW^-$ vertex with $\kappa = s_\beta v$. This kind of coupling, especially if suppressed by small $s_H$, $s_\beta$ or their equivalents, is not easy to probe experimentally. Possibilities include $e^+e^- \to Z^* \to \chi^\mp W^\pm$ [40, 41], $pp \to Z^* \to \chi^\pm W^\mp$, and $pp \to W^{\pm *} \to Z\chi^\pm$ [42]. Constraints on a charged-Higgs-$ZW$ vertex from the static electromagnetic properties of the $W$ boson are discussed in [43].

#### 13.1.2.3 $Y = 2$ triplets with non-zero Higgs-lepton-lepton coupling

Let us finally return to the Higgs-lepton-lepton couplings of a $Y = 2$ triplet. These can be written in the form

$$\mathcal{L} = ih_{ij}\left(\psi_{i\,L}^T C\tau_2 \Delta \psi_{j\,L}\right) + h.c. \;, \tag{13.14}$$

where $\psi_{i\,L}$ is the usual two-component leptonic doublet field, $\psi_{i\,L} = \begin{pmatrix} \nu_{l_i\,L} \\ l_{i\,L} \end{pmatrix}$, $\Delta$ is a $2 \times 2$ representation of the $Y = 2$ complex triplet field,

$$\Delta \equiv \begin{pmatrix} \frac{\chi^+}{\sqrt{2}} & \chi^{++} \\ \chi^{0*} & -\frac{\chi^+}{\sqrt{2}} \end{pmatrix} \;, \tag{13.15}$$

and $i, j$ are family indices. Expanding out this Yukawa interaction, we find Majorana mass terms for the neutrinos of the form

$$m_{ij} = 2h_{ij}\langle\chi^0\rangle = \frac{h_{ij}s_H v}{\sqrt{2}}. \tag{13.16}$$

If we assume that this matrix is diagonal, then the strongest limit on the Majorana mass is that for $\nu_e$ deriving from neutrinoless double-beta decay ($\beta\beta_{0\nu}$). From this we obtain $h_{ee} \lesssim 5.75 \times 10^{-12}/s_H$. For the muon and tau neutrinos, there are the usual limits from $\mu$ and $\tau$ decays. But, WMAP data, especially in combination with results from SDSS and/or 2dFGRS, imply [44] a much stronger upper bound of roughly 1 eV on the largest of the neutrino masses, corresponding to $h \lesssim 1 \times 10^{-11}/s_H$. Neutrino oscillation data provide further constraints on the $h$'s. Indeed, we know that there is mixing among the neutrinos and that at least two of the neutrinos must have some Majorana mass. This could arise entirely from a see-saw mechanism and all the $h_{ij}$ could be zero. Lower bounds on the $h_{ij}$ arise if the entire Majorana neutrino mass is assumed to come from the triplet vev. The combination of $\Delta m^2_{\rm atm} \sim 2.4 \times 10^{-3}$ eV$^2$ and $\Delta m^2_{\rm solar} \sim 8 \times 10^{-5}$ eV$^2$ imply that at least one of the neutrinos (which depends upon whether we have a normal or inverted hierarchy) should have a Majorana mass of order $0.05$ eV and a second should have a mass of at least $0.009$ eV for normal hierarchy or again of order $0.05$ eV for the inverted hierarchy case. A Majorana mass of $\sim 0.05$ eV corresponds to $h \sim 3 \times 10^{-13}/s_H$. Whether or not couplings that saturate these limits can be phenomenologically relevant is determined by the extent to which lepton-lepton channels can be of significance in the decays of the Higgs bosons. (The limits above clearly imply that the couplings are not useful for Higgs boson production.) For any $Y = \pm 2$ triplet Higgs boson with decay mediated by an $h_{ij}$, such as the decay $H_5^{++} \to l^+ l^+$, the relevant Feynman rule coupling for the decay is easily obtained from Eq. (13.14) and takes the form $-2h_{ll}v^T(k)CP_L v(l)$, where $P_L \equiv (1-\gamma_5)/2$, $C$ is the usual charge conjugation matrix, and $k$ and $l$ are the momenta of the two final state leptons. The resulting decay width for a generic $\chi$ is

$$\Gamma(\chi \to l l') = \frac{|h_{ll'}|^2}{8\pi} m_\chi, \tag{13.17}$$

where $l, l'$ might be either charged leptons or neutrinos. In the present $Y = 0$ plus $Y = \pm 2$ triplet model, the small size of the $h_{ll'}$ imply that these decays are rather unlikely to be phenomenologically important unless $s_H$ is very small, a limit to which we will now turn.

This completes our summary of results applicable when the neutral member of a triplet has a substantial vev and thus makes a substantial contribution to electroweak symmetry breaking. We next turn to triplet models in which the triplet(s) play little or no role in electroweak symmetry breaking.

### *13.1.3 Triplet models with no or forbidden triplet vev ($\langle \phi^0_{T=1} \rangle = 0$)*

From the perspective of the preceding section, this would seem a very special case. However, the $\langle \phi^0_{T=1} \rangle = 0$ limit of a triplet model is the point at which custodial $SU(2)_C$ is an unbroken symmetry to all orders. One obtains $\rho = 1$ at tree-level with *finite* radiative corrections. It is no longer necessary to input $\rho$ as an additional observable as part of the renormalization procedure. However, at least in the $Y = 0$ TM, the SM Higgs must then be fairly light. The $\langle \phi^0_{T=1} \rangle = 0$ choice also has the advantage of restoring the prediction that $m_t \sim 174$ GeV in order to agree with precision electroweak data.

If $\langle \phi^0_{T=1} \rangle = 0$, then the triplet Higgs boson(s) will not have any couplings to purely SM particle final states (leaving aside the lepton-lepton coupling possibility for the moment). In addition, all couplings of triplets to the SM-like Higgs will be ones in which two triplet Higgs of the same type appear — these also do not allow for decay to the SM Higgs which would in turn decay as usual. To explore the non-Higgs-diagonal Higgs-Higgs-$V$ couplings, we first turn to the model containing one $Y = 0$ and one $Y = \pm 2$ triplet. An important issue is whether the Higgs-Higgs-$V$ couplings could allow a cascade decay of the $H_5^{++}$. For the model in question, for $s_H = 0$ there are non-zero $H_5^{++} H_3^- W^-$ and $H_3^+ H_1^{0\prime} W^-$ couplings. Further, for $s_H = 0$ we have ${m_{H_5}}^2 = 3{m_{H_3}}^2$ and $m_{H_1^0{}'} = 0$ (at tree-level). As a result, there will be a rapid cascade of $H_5^{++} \to H_3^+ W^+ \to H_1^{0\prime} W^+ W^+$. The $H_1^{0\prime}$, being stable and having no interactions with SM particles, would lead to missing energy. (Of course, one or more of the above particles could be virtual.) Thus, we would have a very distinctive $H_5^{++}$ decay chain. A final state of four $W$'s plus missing energy coming from the production of an $H_5^{++} H_5^{--}$ pair would be hard to miss if the rate is adequate.

In the case of the simpler single $Y = 0$ triplet, $\langle \xi^0 \rangle = 0$ implies that the $k^0$ and $h^\pm$ are degenerate. Presumably this degeneracy would be slightly broken by electromagnetic interactions, resulting in a larger mass for the $h^\pm$. Generically speaking, these corrections would be expected to yield $m_{h^\pm} - m_{k^0}$ of order few$\times m_\pi$, in which case the $h^\pm$ decay would eventually take place, but perhaps not in a typical detector (see below). The $k^0$ would be stable.

If all triplet neutral Higgs fields have zero vev, then the lightest of the associated Higgs bosons could well be absolutely stable and would then provide an excellent dark matter candidate. For example, in the case of the $T = 1, Y = 0$ representation if $\langle \xi^0 \rangle = 0$ then the $k^0$ is expected to be lighter than the $h^\pm$ and would be absolutely stable against decay to a purely SM particle state by virtue of the custodial $SU(2)_C$ (a direct $\nu_L \nu_L$ coupling being forbidden by $Y$ conservation). Annihilation would proceed via $k^0 k^0 \to h^0$. In the model with one $Y = 0$ and one $Y = \pm 2$ triplet, the very light $H_1^{0\prime}$ (which is massless at tree-level) would be stable. Annihilation in the early universe would proceed via $H_1^{0\prime} H_1^{0\prime} \to H_1^0$, where $H_1^0$ is the SM Higgs boson when $s_H = 0$. A consistent description of the observed dark matter density would require an appropriate choice of $m_{H_1^0}$ relative to $2m_{H_1^{0\prime}}$. In the case where only a $T = 1, Y = 2$ Higgs representation is added to the SM, the $\chi^0$ would similarly provide a good dark matter candidate if the (allowed by $Y$) coupling to $\nu_L \nu_L$ is absent, i.e., $h_{\nu\nu} = 0$.

### *13.1.4 Triplet Higgs bosons with large $c\tau$*

Let us first discuss the situation when there are no Majorana couplings leading to decays of $Y = 2$ triplet Higgs bosons to lepton-lepton channels. In general, see the examples above, any triplet model with $\langle \phi^0_{T=1} \rangle = 0$ will have a custodial $SU(2)_C$ which guarantees the absence of all other decays for the lightest neutral triplet Higgs boson. Custodial $SU(2)_C$ is easily imposed in the models considered above by requiring that the Higgs potential be invariant under an appropriate discrete symmetry. Avoiding decays of a charged Higgs boson, at least within a detector size, is a far trickier business, For example, we have already noted that in the $T = 1, Y = 0$ plus $T = 1, Y = \pm 2$ Higgs model described above, ${m_{H_5}}^2 = 3{m_{H_3}}^2$ and the $H_5^+$ and $H_5^{++}$ would quickly chain decay down to the $H_1^{0\prime}$. Single triplet representation theories are safer against such chain decays. For example, the $h^+$ of the single $Y = 0$ representation triplet model would be split from the $k^0$ by electromagnetic radiative corrections by an amount of order a few times $m_\pi$. Thus, it would decay via the far off-shell doubly virtual $h^+ \to k^{0*} W^{+*}$ process which would yield a long path length (given the small mass splitting) resulting in stability of the $h^+$ within the detector. Similarly, if one employed a single $Y = 2$ triplet representation, the $\chi^{++}$ and $\chi^+$ would be split from one another and from the $\chi^0$ by electromagnetic amounts only (for the custodial $SU(2)_C$ symmetry limit of $\langle \chi^0 \rangle = 0$) and would be stable within the detector.

For $Y = \pm 2$ triplet Higgs bosons, the $h_{ll'}$ couplings can dramatically alter the above conclusions. (In the following, we do not assume, except where noted, that the $h_{ll}$, $h_{l\nu_l}$ and $h_{\nu_l \nu_l}$ couplings are related by the Clebsch-Gordon factors predicted by Eq. (13.14). This allows for model independent statements. However, the reader should keep in mind that they are most probably fixed relative to one another.) For non-zero $h_{ll'}$, we would have neutral triplet Higgs decaying to $\nu\nu'$, singly $+$-charged triplet Higgs decaying to $\nu' l^+$ and doubly $++$-charged triplet Higgs decaying to $l^+ l'^+$. Rewriting Eq. (13.17) in terms of the corresponding $c\tau$ yields (here $l$ and $l'$ refer to either charged leptons or neutrinos that could be in the same family or different families)

$$c\tau(\chi \to ll') = 0.5\,\mu\text{m} \left( \frac{100 \text{ GeV}}{m_\chi} \right) \left( \frac{10^{-5}}{h_{ll'}} \right)^2 = 1\text{ m} \left( \frac{100 \text{ GeV}}{m_\chi} \right) \left( \frac{0.7 \times 10^{-8}}{h_{ll'}} \right)^2 . \quad (13.18)$$

For $m_\chi = 100$ GeV, the decay is very prompt unless all $h_{ll'}$'s are considerably smaller than $10^{-5}$, detector sized decay lengths being reached for $h < 0.7 \times 10^{-8}$. What are the constraints? First, *for zero triplet vev, there are no constraints on the $h_{ll'}$ couplings arising from neutrino mass limits.* Limits on diagonal $ee$ and $\mu\mu$ couplings come from $e^+e^- \to e^+e^-$, $\frac{1}{2}(g-2)_\mu$, and muonium to anti-muonium conversion; limits on lepton flavor-violating couplings derive from $\mu \to e\gamma$, $\mu \to ee\bar{e}$ and $\tau \to l_i l_j \bar{l}_j (i, j = e, \mu)$. Theoretical formalism for these decays focused on triplet Higgs models appears in [16,45]. It is particularly interesting to note that the contributions to $\Delta a_\mu$ from a $\chi^-$ and a $\chi^{--}$ are

$$\Delta a_\mu(\chi^-) = -\frac{m_\mu^2}{48\pi m_{\chi^-}^2} \sum_{j=e,\mu,\tau} h_{\mu j}^2 , \quad \Delta a_\mu(\chi^{--}) = -\frac{m_\mu^2}{6\pi m_{\chi^{--}}^2} \sum_{j=e,\mu,\tau} h_{\mu j}^2 , \quad (13.19)$$

i.e., both are opposite in sign to the observed experimental deviation. Current experimental limits on the $h$'s are reviewed in the separate experimental section. One finds that all of the diagonal limits are well above the $h = 10^{-5}$ prompt decay range. Thus, in direct collider searches all possibilities ranging from prompt to long-lived decays must be considered.

As reviewed in the experimental section, there are substantial direct limits from LEP and Tevatron experiments on doubly-charged Higgs bosons that either have a very long path length or decay to like-sign dileptons. A long-lived heavily ionizing $\chi^{++}$ track is easily seen, while in the prompt decay limit the $l^+l^+ + l^-l^-$ events have very small background. LEP limits include those from [46–48]. Tevatron limits have been obtained in [49–51]. Very roughly, current limits are of order 120 GeV, and will be extended to $\sim 250$ GeV by the end of the Tevatron running. At the LHC, the heavily ionizing track or $l^+l^+ + l^-l^-$ events would again stand out and limits on $m_{\chi^{++}}$ of order 1 TeV will be achieved [33,52]. Backgrounds for singly-charged Higgs bosons that decay to $l\nu$ are much larger and Tevatron results for this case have not been presented. The neutral triplet Higgs bosons are produced entirely by $Z^* \to \chi^0\chi^0$ and decay only to $\nu\nu$. At hadron colliders, the events would only be observable through the initial state radiation of photons or gluons. Such Tevatron analyses have been performed, but have not been interpreted in this context.

At a linear collider, one can probe the triplet Higgs with only $ll'$ couplings via $e^+e^- \to \gamma^*, Z^* \to \chi\overline{\chi}$ pair production. (Recall that for $\langle\phi^0_{T=1}\rangle = 0$ there are no $ZZ\chi^0$ or $ZW^+\chi^-$ coupling.) Searches for long-lived track pairs or $l^+l^+ + l^-l^-$ events will be sensitive to $m_\chi$ values up to nearly $\frac{1}{2}\sqrt{s}$. Should a doubly-charged Higgs boson decaying to like-sign leptons be seen either at the LHC or in $e^+e^-$ collisions, operation of the linear collider in the $e^-e^-$ collision mode will be very highly motivated. Very small values of $h_{ee}$ can be probed using $s$-channel resonance production $e^-e^- \to \chi^{--} \to l^-l^-$. This would provide the best means for actually determining $h_{ee}$. This is reviewed in [32,53]. The alternative processes of $e^-e^- \to \chi^{--}Z^0$ and $e^-e^- \to \chi^-W^-$ are much less sensitive [54], as are $\gamma e^- \to l^+\chi^{--}$ and $e^+e^- \to e^+l^+\chi^{--}$ [55–57].

### *13.1.5 Left-Right Symmetric (LR) and Supersymmetric Left-Right (SUSYLR) Models*

In this section, we provide a very brief overview of models based on left-right symmetry with an extended gauge group $SU(2)_L \times SU(2)_R \times U(1)_{B-L}$ [1–6] and the role therein of triplet Higgs bosons. Supersymmetric left-right symmetric models have some especially attractive features [7–14, 58]. One of the primary motivations for left-right symmetric models is that they provide a natural setting for the see-saw mechanism of neutrino mass generation. For many years, the preferred means of implementing the see-saw has been to employ a $SU(2)_R$ Higgs triplet representation (which requires the presence also of an $SU(2)_L$ triplet in order to implement the LR symmetry). We shall denote our triplet members in the LR case by $\delta^{++}_{L,R}$, $\delta^+_{L,R}$ and $\delta^0{}_{L,R}$. In addition to the triplet Higgs fields, the LR models typically contain a bi-doublet Higgs representation for generating the usual Dirac quark and lepton masses. The minimal Higgs field content of the LR model is then

$$\phi = \begin{pmatrix} \phi_1^0 & \phi_1^+ \\ \phi_2^- & \phi_2^0 \end{pmatrix}, \quad \Delta_L = \begin{pmatrix} \delta_L^+/\sqrt{2} & \delta_L^{++} \\ \delta_L^0 & -\delta_L^+/\sqrt{2} \end{pmatrix}, \quad \Delta_R = \begin{pmatrix} \delta_R^+/\sqrt{2} & \delta_R^{++} \\ \delta_R^0 & -\delta_R^+/\sqrt{2} \end{pmatrix}. \tag{13.20}$$

A non-zero neutral field triplet vev, $\langle\delta^0_R\rangle = v_R/\sqrt{2}$, breaks $SU(2)_R \times U(1)_{B-L}$ down to $U(1)_Y$. The non-zero vevs in the bi-doublet, $\langle\phi^0_{1,2}\rangle = \kappa_{1,2}/\sqrt{2}$, break the $SU(2)_L \times U(1)_Y$ to $U(1)_{EM}$ as usual. The neutrino see-saw operates as follows. First, LR symmetry requires the presence of a $\nu_R$ as well as the usual $\nu_L$. Quantum numbers allow a Majorana style $\delta^0_R\nu_R\nu_R$ coupling, see Eq. (13.14). For $\langle\delta^0_R\rangle = v_R/\sqrt{2}$ a Majorana mass of order $v_R h_{\nu\nu}$ (LR symmetry requires $h_{\nu_R\nu_R} = h_{\nu_L\nu_L}$ and so we drop the $R, L$ subscript). For large $v_R$ and small Dirac neutrino masses, if $h_{\nu\nu}$ is not extremely small then the see-saw mechanism takes place. The LR models can be constructed either with the requirement that $\langle\delta^0_L\rangle = v_L/\sqrt{2}$ be zero or non-zero. In general, at least some of the additional Higgs bosons of the LR models can be light; their phenomenology was first studied in [15–17] and basic results are summarized in [26]. Further results appear in [21, 55, 57, 59–73]. Here, we confine ourselves to a very few remarks regarding the status of Higgs triplets in LR and SUSYLR models.

Precision electroweak constraints are most easily satisfied if $v_L = 0$, but can also be satisfied with $v_L \neq 0$ [25, 74, 75]. Renormalization proceeds much as in the $T = 1/2, Y = 1$ plus $T = 1, Y = 0$ triplet model; for $v_L \neq 0$ the experimental value of $\sin\theta_W$ (or $\rho$ or, equivalently, $T$) becomes an input rather than a prediction. Using the input-$\sin\theta_W$ scheme, one again finds a greatly reduced sensitivity of $\Delta R$ to $m_t$, in this case because the $\delta \sin^2\theta_W/\sin^2\theta_W$ contribution to $\Delta R$, though quadratic in $m_t$, is proportional to $m_{W_1}^2/m_{W_2}^2$, where $m_{W_1}$ must be very close to the observed $W$ mass. Thus, this contribution to $\Delta R$ vanishes as $m_{W_2} \to \infty$.

In the simplest LR model, with Higgs content sketched above, minimization of the Higgs potential results in surprisingly strong constraints. In [5, 16, 17] it was shown that there is a "vev see-saw" relation that reads (following [17])

$$(2\rho_1 - \rho_3)v_L v_R = \beta_2\kappa_1^2 + \beta_1\kappa_1\kappa_2 + \beta_3\kappa_2^2 \tag{13.21}$$

where the $\beta_i$ and $\rho_i$ are certain Higgs potential parameters. Thus, for generic Higgs potential parameter choices, if $v_L \ll \kappa_{1,2}$ (where $\kappa_1^2 + \kappa_2^2$ is of order the usual 246 GeV), then $v_R \gg \kappa_{1,2}$ is required by the minimization. In [17], it was shown that the only phenomenologically acceptable solutions are $\beta_2 = 0$ (as required if we demand that the Higgs potential be invariant under $\phi \to i\phi$, which also cures certain FCNC problems of the model) with $v_L = \kappa_2 = 0$, $\kappa_1, v_R \neq 0$ and $\rho_{diff} \equiv 2\rho_1 - \rho_3 \neq 0$. If $\rho_{diff}$ is of order 1, then all Higgs bosons other than a single SM-like Higgs boson will have masses of order $v_R$. Interesting new Higgs phenomenology at the TeV-scale would require a very small value for $\rho_{diff}$. In fact, an additional symmetry can be imposed on the Higgs potential that guarantees $\rho_{diff} = 0$. However, $\rho_{diff} = 0$ implies that the Higgs bosons residing in the real and imaginary parts of $\delta_L^0$ are massless at tree-level; this is inconsistent with constraints from the $Z$. Thus, this symmetry must be slightly broken at the $v_R$ scale by effective operators. Assuming very small $\rho_{diff}$, we would have the following. After removing the usual Goldstone bosons, the $\delta_L^{++}$ and $\delta_L^+$ states have masses of order $v_L$, the $\mathrm{Im}\delta_L^0/\sqrt{2}$ and $\mathrm{Re}\delta_L^0/\sqrt{2}$ states have masses of order $\sqrt{\rho_{diff}}v_R$, while the $\delta_R^{++}$, $\delta_R^+$ and $\mathrm{Im}\phi_2^0/\sqrt{2}$ states have masses of order $v_R$. The residual $h^+$ state that is a combination of $\phi_1^+$ and $\delta_R^+$ is heavy, as is one combination, called $H^0$, of $\mathrm{Re}\phi_1^0/\sqrt{2}$ and $\delta_R^0$, while the orthogonal combination ($h^0$) plays the role of a light SM-like Higgs boson. In the end, the TeV-scale phenomenology has many similarities to that of a one-doublet + one $Y = 2$ triplet model, including the presence of $ll$ couplings for the $\delta_L$ states (a remnant of the LR symmetry and the see-saw neutrino mass generation mechanism). The detailed phenomenology of this model can be found in [15–17].

The SUSYLR models have some important attractive features. In particular, it is possible to construct them so that both the strong CP problem and the SUSY CP problem (i.e., the generic problem of SUSY phases giving large EDM's unless cancellations are carefully arranged) are automatically solved [7–9, 11]. If LR symmetry and SUSY are implemented in the triplet-Higgs context, then one needs additional triplet fields; in the SUSY extension of the triplet model discussed above, these would be the conjugates of the $\Delta_R$ and $\Delta_L$. In addition, as we sketch below, one also needs to include singlet superfields. Before symmetry breaking, both the $\delta_R^{++}$ Higgs bosons (there are now two) and their higgsino partners are massless due to the existence of a flat direction associated with rotations in $\langle\delta_R^0\rangle$–$\langle\delta_R^{++}\rangle$ space. If the scale of supersymmetry breaking, $m_{\mathrm{SUSY}}$ is above $v_R$, then after SU$\not{S}$Y the theory lives in a charge-violating vacuum unless non-renormalizable operators suppressed by $1/M_{\mathrm{P}}$ involving the singlet field(s) are included. After including these operators, the $\delta_R^{++}$ Higgs bosons typically acquire only a small mass. If $m_{\mathrm{SUSY}} < v_R$, then the renormalizable theory may live in a charge-conserving vacuum and the $\delta_R^{++}$ Higgs bosons pick up a mass of order $m_{\mathrm{SUSY}}$. Now, however, the corresponding higgsinos are very light since the breaking of supersymmetry is assumed to be soft. Non-renormalizable operators are now needed to give the higgsinos sufficient mass to avoid current experimental constraints. Minimization of the Higgs potential after including the $\Delta_L$ and $\overline{\Delta}_L$ fields, and the associated Higgs bosons, have not be carefully studied in this context.

More recently, an alternative SUSYLR model has emerged in which Higgs triplet fields do not play a role [58]. The $SU(2)_R \times U(1)_{B-L}$ symmetry is broken down to $U(1)_Y$ in the supersymmetric limit by $B-L = \pm 1$ doublet scalar fields, namely the right-handed doublet denoted by $\chi^c(1,2,-1)$ accompanied by its left-handed partner $\chi(2,1,1)$, where the items in parenthesis indicate the $SU(2)_L$, $SU(2)_R$ and $B-L$ representations, respectively. Anomaly cancellation requires the presence of the charge conjugate fields, $\overline{\chi}^c(1,2,1)$ and $\overline{\chi}(2,1,-1)$, as well. The vevs $\langle \chi^c \rangle = \langle \overline{\chi}^c \rangle = v_R$ break the left-right symmetry group down to the MSSM gauge symmetry. The only Higgs bosons of the resulting model with masses at the TeV scale (rather than at scale $v_R$) correspond to the $H_u$ and $H_d$ doublet fields of the MSSM. There are some additional singlet Higgs fields with masses of order $v_R$, but no triplet Higgs fields are employed. The main advantage of this model over SUSYLR models with triplets is that the SUSY phase problem is solved based on requiring LR parity symmetry alone as opposed to requiring charge conjugation symmetry as well. In addition, introduction of non-renormalizable effective interactions is not required in order to guarantee a charge-conserving vacuum. However, non-renormalizable operators suppressed by $1/M_{\rm P}$, as well as both a visible sector singlet field and a hidden sector singlet field, *are* required in order to generate an effective soft-supersymmetry breaking $B\mu$ term. A non-renormalizable operator form $(fLL\chi\chi + f^* L^c L^c \chi^c \chi^c)/M_{\rm P}$ is also employed to produce Majorana masses for the $\nu_R$'s of size $v_R^2/M_{\rm P}$. For $v_R \sim 10^{14} - 10^{16}$ GeV, the predicted Majorana masses are in the right ball park to explain solar and atmospheric oscillation data. Overall, the model is not very simple and the canonical see-saw with Majorana masses of order $(246~{\rm GeV})^2/v_R$ is totally abandoned. Thus, the SUSYLR models with triplet Higgs should certainly not be ignored.

### *13.1.6 Experiment*

A wide variety of experimental searches and standard model tests probe the existence of Higgs triplets. The possibilities depend strongly on the type of triplet model. In this section, we will focus on a single $T = 1, Y = \pm 2$ triplet addition to the SM doublet with Higgs bosons denoted as previously by $\chi^0$, $\chi^\pm$, and $\chi^{\pm\pm}$. For this case, if we ignore loop corrections then $V_{1,\pm 2} \equiv \langle \chi^0 \rangle$ is constrained to be small ($\langle \chi^0 \rangle < 8$ GeV) from the measurements of the $W$ mass and other electroweak parameters — see, for example, Eq. (13.2) and note that $\rho < 1$ is predicted at tree-level whereas data suggest a small positive value for $\rho - 1$. Radiative corrections for this case have not been worked out, but it seems safe to say that even after their inclusion $\langle \chi^0 \rangle$ would have to be quite small. Our discussion will be based on this approximation. Among other things, it implies that there is rather small mixing between the doublet Higgs bosons and the triplet Higgs bosons. Thus, we will speak of the $\chi^0$ and $\chi^\pm$ as though they are unmixed states with phenomenology determined by perturbative corrections associated with the small non-zero value of $\langle \chi^0 \rangle$. Of course, the $\chi^{++}$ and $\chi^{--}$ are pure states.

The parameters determining the sensitivity of a given experiment are the Higgs mass and couplings. The expected phenomena depend on whether the lightest triplet member is $\chi^0$, $\chi^\pm$ or $\chi^{\pm\pm}$. If there were no mixing between the $\chi^0$ and the $\phi^0$ nor between the $\chi^+$ and the $\phi^+$, then it would be most probable that the $\chi^0$ would be the lightest state. The effects of introducing mixing terms into the Higgs Lagrangian have not been worked out for this case, but it seems possible for the mass ordering of the states to be altered. We will discuss various signatures for each of the states in turn assuming that the state in question is the lightest of the triplet states.

We will begin with the $\chi^{\pm\pm}$. We focus on various extreme possibilities.

- The $\chi^{\pm\pm}$ has significant couplings to leptons.
- The $\chi^{\pm\pm}$ has negligible couplings to leptons and $W$ bosons.
- The $\chi^{\pm\pm}$ has negligible couplings to leptons and small but significant couplings to $W$ bosons.
- The $\chi^{\pm\pm}$ is a member of a supersymmetric triplet with R-parity-conserving interactions.

#### 13.1.6.1 $\chi^{\pm\pm}$ with leptonic couplings

The richest triplet phenomenology occurs for light doubly charged Higgs bosons with significant leptonic couplings. This is possible without conflicting with neutrino masses if either $\langle\chi^0\rangle = 0$ (as would be more or less required if $h_{ll}$, $l$ being the charged lepton, is substantial and $h_{\nu\nu} \sim h_{ll}$, as predicted by Eq. (13.14)) or if only $h_{ll}$ is substantial. In any case, it is important to obtain limits on $h_{ll}$ without introducing any model-dependent inputs. The effects of the $\chi^{++}$ through $h_{ll}$ couplings can be observed indirectly through rare leptonic decays or conversion processes, or directly through production at lepton and hadron colliders.

Including flavor changing possibilities, there are six $\chi^{\pm\pm}$ leptonic couplings, which we denote by $h_{ij}$. These are undetermined parameters, so there is no theoretical guidance to whether a particular leptonic decay is preferred, and if so, which one. Off-diagonal couplings lead to lepton-flavor-violating processes such as $\mu \to 3e$ [16], $\tau \to 3l$ [16], and $\mu \to e\gamma$ [76], while diagonal couplings contribute to the Bhabha scattering cross section [16, 77, 78] the muon anomalous magnetic moment [16, 79], and muonium to antimuonium conversion (Figs. 13.4 and 13.5) [77]. Table 13.2 shows the coupling limits for $m_{\chi^{\pm\pm}} = 100$ GeV from searches for, or measurements of, these processes. Future data to be taken by the BELLE and MEG collaborations will improve coupling sensitivity by about an order of magnitude from $\tau \to 3l$ and $\mu \to e\gamma$ searches, respectively.

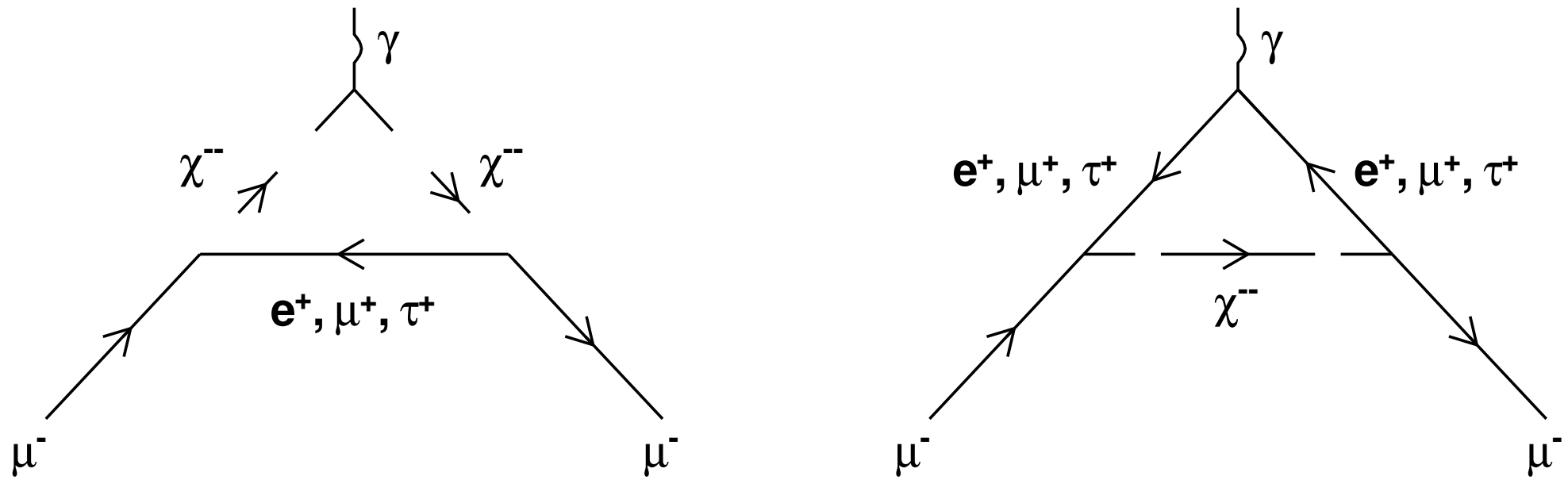

Fig. 13.4: The $\chi^{\pm\pm}$ contributions to the $\mu^-$ anomalous magnetic moment. Equivalent charge-conjugate diagrams exist for the $\mu^+$. The same diagrams with the outgoing muons replaced by electrons result in the decay $\mu \to e\gamma$.

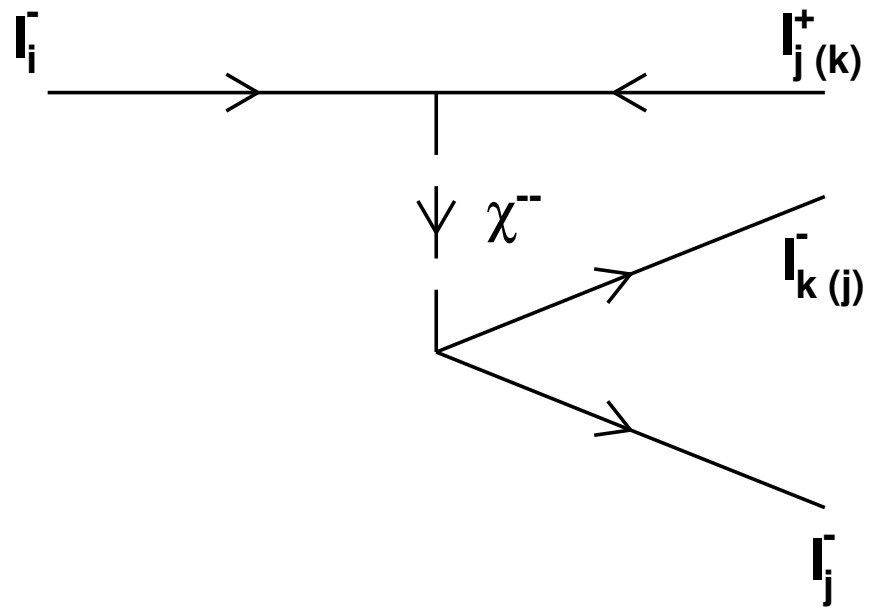

Fig. 13.5: $\chi^{--}$-mediated $l_i^- \to l_j^- l_{k(j)}^- l_{j(k)}^+$ decay. A corresponding charge-conjugate diagram mediates $l_i^+ \to l_j^+ l_{k(j)}^+ l_{j(k)}^-$ decay.

While indirect studies probe the lepton-coupling-to-$\chi^{\pm\pm}$-mass ratio in the form $c_{ij} \equiv h_{ij}^2/m_{\chi^{\pm\pm}}^2$, direct searches are sensitive to a given mass for couplings spanning several orders of magnitude. Pairs of $\chi^{\pm\pm}$ bosons are produced in $e^+e^-$ and hadron collisions through $Z/\gamma^*$ exchange and for $m_{\chi^{++}} \gtrsim 100$ GeV decay promptly ($c\tau < 10$ microns) if $\sum h_{ij} > 10^{-5}$ — see Eq. (13.18) — even if $\langle\chi^0\rangle = 0$ (so that $\chi^{\pm\pm} \to W^\pm W^\pm$ decays are absent). Searches for $\chi^{++}\chi^{--}$ pair production have excluded the $\chi^{\pm\pm}$ if its mass is below 100-135 GeV (Fig. 13.6), provided decay channels other than the dilepton channels have small branching ratio (as would be true if $\langle\chi^0\rangle$ is tiny or zero or if one or more of the $h_{ij}$ are large). The limits depend on the dominant $h_{ij}$ coupling and on whether the $\chi^{\pm\pm}$ has left-handed or right-handed[2] couplings. The limits also assume $m_{\chi^\pm} >> m_{\chi^{\pm\pm}}$, and become stronger if $m_{\chi^\pm} \approx m_{\chi^{\pm\pm}}$ (see Section 13.2). Ongoing $p\bar{p}$ data collected at the Tevatron will increase the mass sensitivity to $\sim 250$ GeV, and future $pp$ data from the LHC will further increase the sensitivity to $\sim 1$ TeV. For an analysis of doubly charged Higgs bosons in the left-right symmetric model at the LHC, see Section 6.4.

| Process | Limit |
|---|---|
| $e^+e^- \to e^+e^-$ | $h_{ee} < 0.15$ [80] |
| $\frac{1}{2}(g-2)_\mu$ | $h_{\mu\mu} < 0.22$ [81] |
| $M \to \bar{M}$ | $h_{ee}h_{\mu\mu} < 2.0 \times 10^{-3}$ [82] |
| $\mu \to e\gamma$ | $h_{e\mu}, h_{e\tau}, h_{\tau\mu} < 4.5 \times 10^{-3}$ [83] |
| $\mu \to eee$ | $h_{ee}h_{\mu e} < 2 \times 10^{-7}$ [84] |
| $\tau \to l_i l_j l_j,\ i,j = e,\mu$ | $h_{ij}h_{j\tau} < 6 \times 10^{-2}$ [85, 86] |

Table 13.2: The Yukawa coupling limits on $\chi^{\pm\pm}$ for $m_{\chi^{\pm\pm}} = 100$ GeV/$c^2$. The $h_{ij}$ limits increase linearly with increasing $\chi^{\pm\pm}$ mass. Any assumptions made on the relative couplings have been chosen to produce conservative (i.e., higher) limits.

#### 13.1.6.2 Long-Lived $\chi^{\pm\pm}$

If the $\chi^{\pm\pm}$ leptonic couplings are significantly suppressed, and it has no other significant decay channels (requiring very small or zero $\langle\chi^0\rangle$ and very small or negative $m_{\chi^{++}} - m_{\chi^+}$), then the $\chi^{\pm\pm}$ is likely to be long-lived ($c\tau > 10$ m). In this case, $\chi^{\pm\pm}$ phenomena will be limited to direct production at lepton and hadron colliders. Current mass limits range from 110-135 GeV (Fig. 13.6), depending on whether the doubly charged Higgs has Majorana couplings to the left- or right-handed leptons. The full Tevatron data set will extend the sensitivity to $\sim$ 250 GeV, and the LHC $pp$ collisions will make observation of $\sim 1$ TeV $\chi^{\pm\pm}$ bosons possible.

#### 13.1.6.3 $\chi^{\pm\pm}$ with $W$ couplings

Doubly charged Higgs couplings to $W$ bosons are determined by the vacuum expectation value of the neutral Higgs field in the triplet. If $\langle\chi^0\rangle$ takes its maximum allowed value, then $pp$ collisions at the LHC will produce an observable rate of single $\chi^{++}$ bosons produced both in association with a $W$ boson and also via $WW$ fusion processes such as $W^+W^+ \to \chi^{++}$. The sensitivity of the Tevatron data to these topologies has not yet been determined.

The $W^-\chi^{++} + W^+\chi^{--}$ final state would result in low-background signatures of like-sign leptons $+\not{E}_T + X$ in hadron and $e^+e^-$ colliders assuming that the $h_{ll}$ couplings are large enough that the $\chi^{++} \to l^+l^+$ decay is dominant over decays such as $\chi^{++} \to W^+\phi^+$ which would typically be present for $\langle\chi^0\rangle \neq 0$. Note, however, that this scenario with both large $h_{ll}$ and large $\langle\chi^0\rangle$ is inconsistent with neutrino masses unless $h_{\nu\nu} \ll h_{ll}$, in contradiction to the $SU(2)_L$ invariant interaction form of Eq. (13.14).

#### 13.1.6.4 $\chi^{\pm\pm}$ with SUSY couplings

In the supersymmetric extension of the $T = 1, Y = \pm 2$ model, many additional possibilities emerge. First, there would be SUSY-determined analogues of the $h_{ij}$ specifying the coupling of the $\chi^{++}$ to slepton pairs $\widetilde{l}^+\widetilde{l}^+$. These would give rise to a like-sign slepton signal for the $\chi^{++}$. In addition, there would also be doubly-charged Higgsinos $\tilde{\chi}^{\pm\pm}$, and SUSY-determined analogues of the $h_{ij}$ couplings that would lead to like-sign slepton-lepton decays, e.g. $\tilde{\chi}^{++} \to l^+\tilde{l}^+$. Assuming dominance of these decays (as most easily arranged if $\langle\chi^0\rangle = 0$), these topologies are probably observable at the Tevatron and the LHC, although the associated discovery reaches will certainly depend on the slepton and LSP masses and have not been explored in detail. For either $\chi^{++}\chi^{--}$ production followed by $\chi^{\pm\pm} \to \tilde{l}^\pm\tilde{l}^\pm$ or $\tilde{\chi}^{++}\tilde{\chi}^{--}$ production with $\tilde{\chi}^{\pm\pm} \to \tilde{l}^\pm l^\pm$, the final state signature at both hadron and $e^+e^-$ colliders will be that of like-sign leptons+$\not{E}_T + X$.

[2]In considering the case of couplings to right-handed charged leptons, we are implicitly extending our considerations to a left-right symmetric model with a $\chi^{\pm\pm}$ $SU(2)_R$ triplet member.

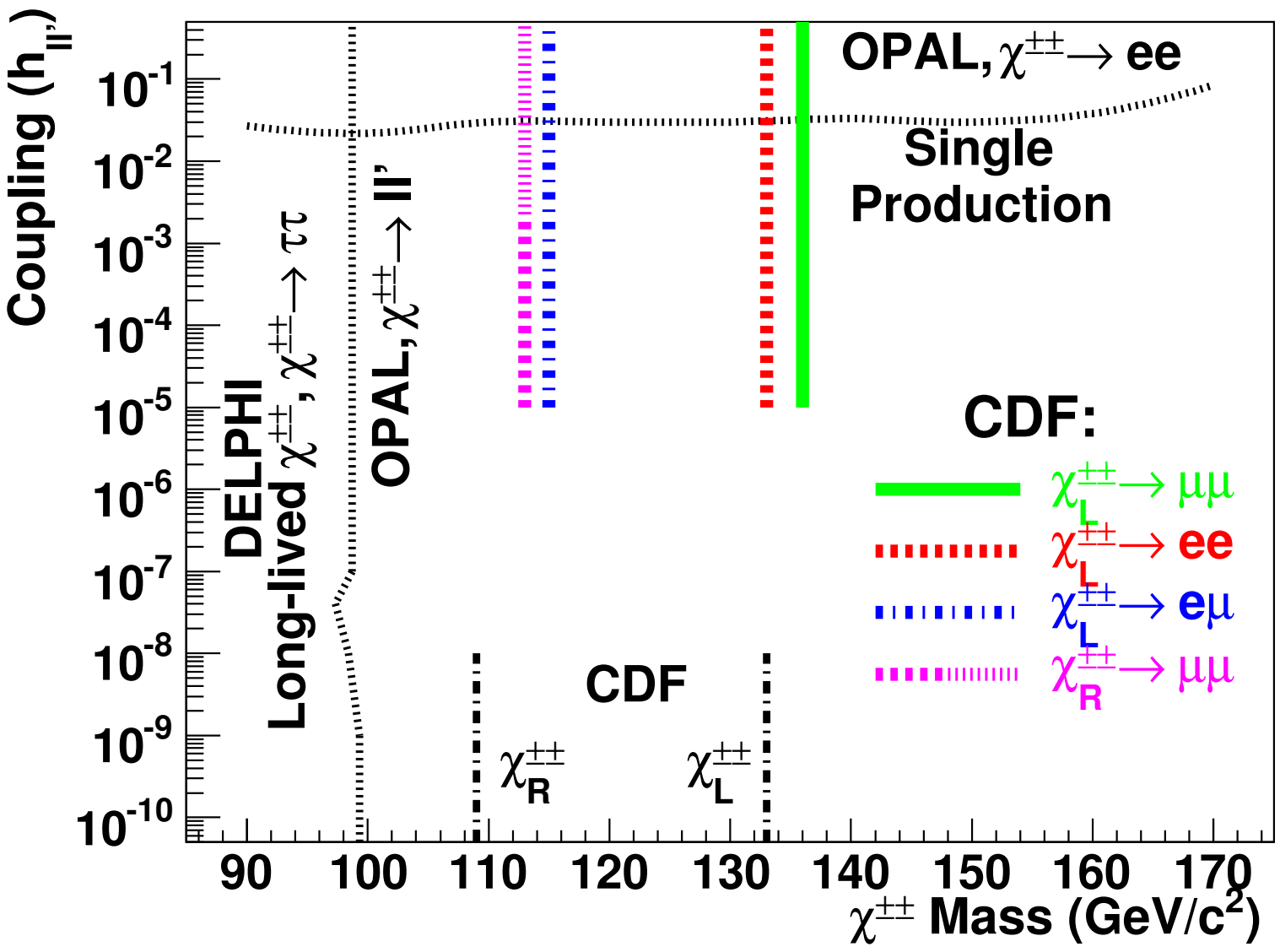

Fig. 13.6: Exclusions in the $\chi^{\pm\pm}$ mass-Yukawa coupling plane from direct searches for $\chi^{++}\chi^{--}$ pair production, assuming exclusive decays to a given dilepton pair [87]. The CDF searches have undetermined sensitivity beyond the published coupling ranges [49, 50]. The DØ Collaboration has also excluded $\chi^{\pm\pm} \to \mu\mu$ assuming left-handed couplings below a mass of 118 GeV/$c^2$ for $h_{\mu\mu} > 10^{-7}$ [51]. Also shown is the limit on $h_{ee}$ coming from searches for single $\chi^{\pm\pm}$ production through the $h_{ee}$ coupling, $e^+e^- \to e^+e^+\chi^{--} + e^-e^-\chi^{++}$, followed by $\chi^{\pm\pm} \to l^\pm l^\pm$ [80].

#### 13.1.6.5 $\chi^\pm$

Singly charged members of Higgs triplets would contribute indirectly to the same leptonic processes mediated by $W$ bosons, for example muon decay. No limits on the coupling-to-$\chi^\pm$-mass ratio have been derived for these processes, but the consistency of the electroweak theory at NLO suggests that the contributions must be small.

For small or zero $\langle\chi^0\rangle$, the $\chi^\pm$ bosons will be mainly pair-produced in leptonic and hadronic collisions. Assuming the presence of the relevant $h_{l\nu}$ coupling, the signature would consist of two leptons with missing energy from the neutrinos, which is similar to that of $W$ pair production. The consistency of the measured $WW$ cross sections with SM predictions suggests that the $\chi^\pm$ mass is larger than that of the $W$ boson, though no explicit limits have been derived. Ongoing $p\bar{p}$ collisions at the Tevatron and future $pp$ collisions at the LHC offer opportunities to observe $\chi^\pm$ in the ~TeV mass range. Of course, if the $h_{l\nu}$ couplings are very tiny and $\langle\chi^0\rangle = 0$, then there is a chance that the $\chi^\pm$ would be stable within the detector.

13.1.6.6 $\chi^0$

We first note that if the $h_{\nu_i\nu_j}$ couplings are off-diagonal, then the $\chi^0$ could cause neutrino decay, contributing to neutrino disappearance experiments. Neutrino data have not yet been interpreted in this context.

For $\langle\chi^0\rangle = 0$, it might well be that the $\chi^0$ state is the lightest of the triplet states. The only possibility for it to decay would be if one or more of the $h_{\nu_i\nu_j}$'s are non-zero. If $\chi^0$-$W$-$W$ couplings are sufficiently large, the $\chi^0$ can decay to two photons via $W$ loop diagrams (see Section 13.3). If $\chi^0 \to \nu\nu$ is its only decay, there are significant experimental challenges as sketched below.

In $e^+e^-$ collisions, $e^+e^- \to Z^* \to \chi^0\chi^0$ would not be directly observable, but the $e^+e^- \to Z^* \to \gamma\chi^0\chi^0$ photon-tag leads to a striking signature of an isolated photon plus missing energy in the detector. In hadron collisions, $\chi^0\chi^0$ pair production could be tagged by either a radiated photon or radiated gluon jet. Either would be much harder to separate from backgrounds than in the case of $e^+e^- \to \gamma\chi^0\chi^0$.

Of course, if $\langle\chi^0\rangle \neq 0$, then associated $W\chi^0$ and $Z\chi^0$ production becomes possible. However, neutrino mass limits directly forbid a large enough $h_{\nu\nu}$ coupling in this case for $\chi^0 \to \nu\nu$ to dominate such decays as $\chi^0 \to \phi^0\gamma, Z$ or $\chi^0 \to W^+W^-$ (with the final state particles being either real or virtual). In the above, the $\phi^0$ would decay more or less like a SM Higgs boson of equivalent mass. While collaborations at LEP and the Tevatron have not performed explicit searches for $\chi^0$ based on the $W\chi^0$ and $Z\chi^0$ production modes followed by such decays, they observe no excesses in these channels.

## 13.2 Single and Pair production of $\chi^{\pm\pm}$ at Hadron Colliders

*A.G. Akeroyd and Mayumi Aoki*

In 2003 the Fermilab Tevatron performed the first search for $\chi^{\pm\pm}$ (a doubly charged scalar from a $T = 1, Y = 2$ Higgs triplet) at a hadron collider. D0 [51] searched for $\chi^{\pm\pm} \to \mu^\pm\mu^\pm$ while CDF [50] searched for 3 final states: $\chi^{\pm\pm} \to e^\pm e^\pm, e^\pm\mu^\pm, \mu^\pm\mu^\pm$. The assumed production mechanism for $\chi^{\pm\pm}$ is $q\overline{q} \to \gamma^*, Z^* \to \chi^{++}\chi^{--}$. This cross-section depends on only one unknown parameter, $m_{\chi^{\pm\pm}}$, and importantly is not suppressed by any potentially small factor such as the leptonic Yukawa coupling $h_{ij}$ (Eq. (13.14)) or a triplet vev. Assuming that $\chi^{\pm\pm}$ production proceeds via this pair production process, the absence of signal enables a limit to be set on the product:

$$\sigma(p\overline{p} \to \chi^{++}\chi^{--}) \times BR(\chi^{\pm\pm} \to l_i^\pm l_j^\pm) \tag{13.22}$$

Clearly the strongest constraints on $m_{\chi^{\pm\pm}}$ are obtained assuming BR$(\chi^{\pm\pm} \to l_i^\pm l_j^\pm) = 100\%$. Currently these mass limits stand at: 133,115,136 GeV for the $e^\pm e^\pm, e^\pm\mu^\pm, \mu^\pm\mu^\pm$ channels respectively [50]. Due to the low backgrounds for the same sign leptonic decay mode the search potential of the Tevatron merely depends on the signal efficiencies for the signal (currently $\approx 34\%, 34\%, 18\%$ for $\mu\mu, ee, e\mu$) and the integrated luminosity. These relatively high efficiencies together with an expected $\mathcal{L} = 4 - 8fb^{-1}$ by the year 2009 would enable discovery with $> 5$ events for $\sigma_{\chi^{++}\chi^{--}}$ of a few fb, which corresponds to a mass reach $m_{\chi^{\pm\pm}} < 200$ GeV.

The current search strategy is in fact sensitive to any *singly produced* $\chi^{\pm\pm}$, i.e. signal candidates are events with *one pair* of same sign leptons reconstructing to $m_{\chi^{\pm\pm}}$. Although single $\chi^{\pm\pm}$ production processes such as $p\overline{p} \to W^\pm \to W^\mp\chi^{\pm\pm}$ can be neglected due to the strong triplet vev suppression (if $< \chi^0 >= v_\Delta << 8$ GeV), the mechanism $p\overline{p} \to W^* \to \chi^{\pm\pm}\chi^\mp$ is potentially sizeable [88]. This latter process proceeds via a gauge coupling constant and is not suppressed by any small factor. The LO partonic cross-section is as follows:

$$\sigma_{LO}(q'\overline{q} \to \chi^{++}\chi^-) = \frac{\pi\alpha^2}{144\sin^4\theta_W Q^2} C_T^2 p_W^2 \beta_2^3 \tag{13.23}$$

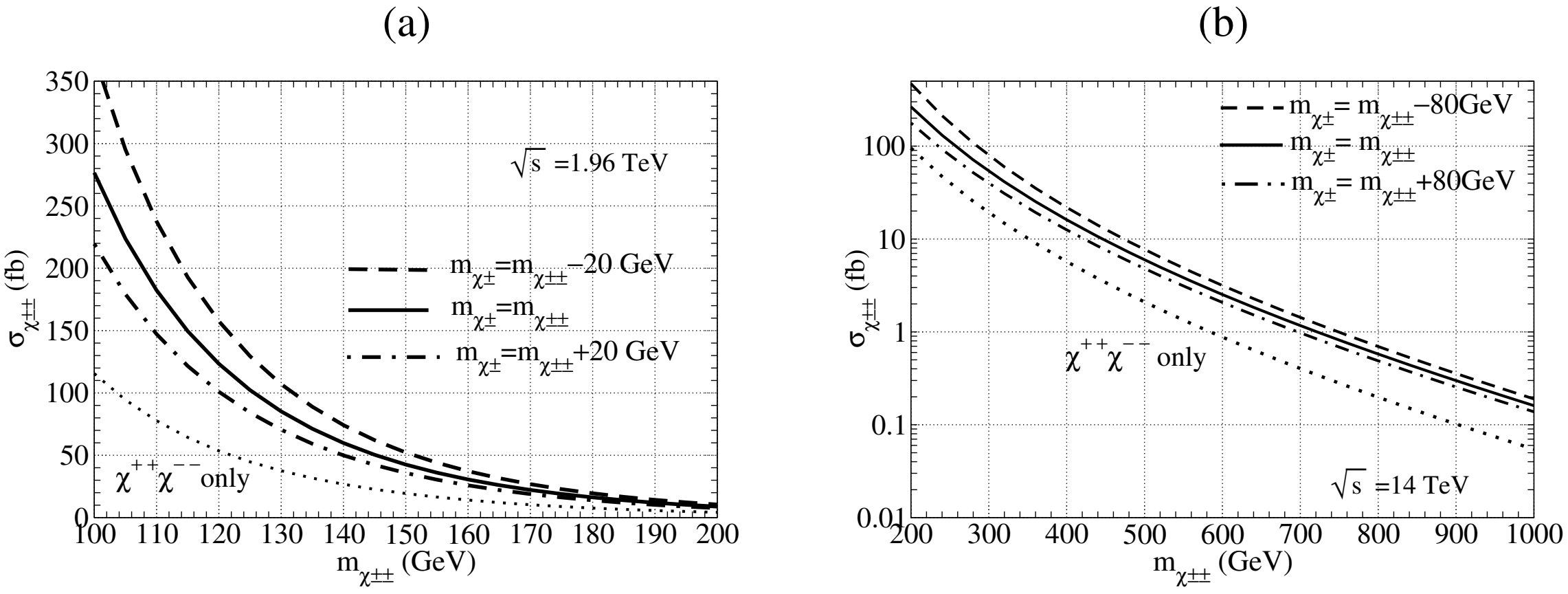

Fig. 13.7: (a) Single production cross-section of $\chi^{\pm\pm}$ ($\sigma_{\chi^{\pm\pm}}$) at the Tevatron as a function of $m_{\chi^{\pm\pm}}$ for different values of $m_{\chi^\pm}$. (b) as above but for LHC. We use CTEQ6L1 parton distribution functions.

Here $C_T$ arises from the $\chi^{\pm\pm}\chi^{\mp}W^{\mp}$ vertex and $C_T = 2$ for $T = 1$, $Y = 2$ triplet fields $\beta_2 = \sqrt{(1-(m_{\chi^\pm}+m_{\chi^{\pm\pm}})^2/Q^2)(1-(m_{\chi^\pm}-m_{\chi^{\pm\pm}})^2/Q^2)}$ and $p_W = Q^2/(Q^2 - M_W^2)$. In this contribution we will study the impact of the production channel $\sigma(q'\overline{q} \to \chi^{\pm\pm}\chi^{\mp})$ on the direct search for $\chi^{\pm\pm}$ at hadron colliders. An expanded treatment can be found in [89]. We perform our analysis for the one doublet ($\Phi$), one $T = 1$, $Y = 2$ triplet model hereafter referred to as the "Higgs Triplet Model" (HTM). The scalar potential for HTM is given in [90] and the mass splitting between $m_{\chi^{\pm\pm}}$ and $m_{\chi^\pm}$ is determined by the term $\lambda_5\Phi^\dagger\Phi\Delta^\dagger\Delta$. However, our numerical analysis is also relevant for other models which possess a left-handed $T = 1, Y = 2$ doubly charged scalar e.g. $\delta_L^{\pm\pm}$ in LR symmetric models and $H_5^{\pm\pm}$ in the triplet model with tree-level custodial $SU(2)_C$ symmetry, where both $q'\overline{q} \to H_5^{\pm\pm}H_5^{\mp}$ and $H_5^{\pm\pm}H_3^{\mp}$ are possible.

Motivated by the fact that the currently employed Tevatron search strategy is sensitive to *single production* of $\chi^{\pm\pm}$, we advocate the use of the inclusive single production cross-section ($\sigma_{\chi^{\pm\pm}}$) when comparing the experimentally excluded region with the theoretical cross-section. This leads to a strengthening of the mass bound for $m_{\chi^{\pm\pm}}$ which now carries a dependence on $m_{\chi^\pm}$. We introduce the single production cross-section as follows:

$$\sigma_{\chi^{\pm\pm}} = \sigma(p\overline{p}, pp \to \chi^{++}\chi^{--}) + \sigma(p\overline{p}, pp \to \chi^{++}\chi^{-}) + \sigma(p\overline{p}, pp \to \chi^{--}\chi^{+}) \tag{13.24}$$

At the Tevatron $\sigma(p\overline{p} \to \chi^{++}\chi^{-}) = \sigma(p\overline{p} \to \chi^{--}\chi^{+})$ while at the LHC $\sigma(pp \to \chi^{++}\chi^{-}) > \sigma(pp \to \chi^{--}\chi^{+})$. If a signal for $\chi^{\pm\pm}$ were found in the 2 lepton channel, subsequent searches could select signal events with 3 or 4 leptons, in order to disentangle $q\overline{q} \to \chi^{++}\chi^{--}$ and $q'\overline{q} \to \chi^{\pm\pm}\chi^{\mp}$.

In Fig. 13.7 (a) we plot $\sigma_{\chi^{\pm\pm}}$ as a function of $m_{\chi^{\pm\pm}}$ at the Tevatron for three different values of $m_{\chi^\pm}$. The mass splitting between $m_{\chi^{\pm\pm}}$ and $m_{\chi^\pm}$ is determined by the $\lambda_5$ term in the Higgs potential. The current excluded regions from the $e^\pm e^\pm, e^\pm\mu^\pm, \mu^\pm\mu^\pm$ searches correspond to the area above horizontal lines at roughly 40, 70, 35 fb respectively. The present mass limits for $m_{\chi^{\pm\pm}}$ are where the curve for $\chi^{++}\chi^{--}$ intersects with the above horizontal lines, and read as $133, 115, 136$ GeV respectively for BR$(\chi^{\pm\pm} \to l_i^\pm l_j^\pm) = 100\%$ . With the inclusion of the $\chi^{\pm\pm}\chi^{\mp}$ channel, these mass limits increase to 150, 130, 150 for $m_{\chi^\pm} = m_{\chi^{\pm\pm}} + 20$ GeV, strengthening to 160,140,160 for $m_{\chi^\pm} = m_{\chi^{\pm\pm}} - 20$ GeV. Clearly the search potential of the Tevatron (i.e. the mass limit on $m_{\chi^{\pm\pm}}$) increases significantly when one includes the contribution to $\sigma_{\chi^{\pm\pm}}$ from $p\overline{p} \to \chi^{\pm\pm}\chi^{\mp}$. Note that the above mass limits strictly apply to the case when $\chi^{\pm\pm}$ decays leptonically, and with BR=$100\%$ in a given channel.

In Fig. 13.7 (b) we plot the analogy of Fig. 13.7 (a) for the LHC, allowing larger mass splittings ($|m_{\chi^{\pm\pm}} - m_{\chi^{\pm}}| \leq 80$ GeV). As before, the inclusion of $q'\overline{q} \to \chi^{\pm\pm}\chi^{\mp}$ significantly increases the search potential e.g. if sensitivity to $\sigma_{\chi^{\pm\pm}} = 1$ fb is attained, the mass reach extends from $m_{\chi^{\pm\pm}} < 600$ GeV ($\chi^{++}\chi^{--}$ only) to 750 GeV for ($m_{\chi^{\pm}} = m_{\chi^{\pm\pm}} - 80$ GeV). Recently [52] performed a simulation of the detection prospects at the LHC for $q\overline{q} \to \chi^{++}\chi^{--}$ for the cases where 3 and 4 leptons are detected. With 100 fb$^{-1}$, sensitivity to $m_{\chi^{\pm\pm}} < 800$ GeV (3 leptons) and $m_{H^{\pm\pm}} < 700$ GeV (4 leptons) is expected. We are not aware of a simulation for the case where only 2 leptons are detected. Presumably even larger values of $m_{\chi^{\pm\pm}}$ ($<$ 800 GeV) could be probed.

An attractive feature of the $T = 1, Y = 2$ triplet representation is the leptonic Yukawa coupling $h_{ij}$ which leads to neutrino mass and mixing if $v_\Delta (=< \chi^0 >) \neq 0$ (see Eq. (13.16)). In the HTM $h_{ij}$ and the neutrino mass matrix are related by the following equation (where $V_{\text{MNS}}$ is the Maki-Nakagawa-Sakata matrix and $|m_i|$ are the neutrino mass eigenvalues):

$$h_{ij} = \frac{1}{\sqrt{2}v_\Delta} V_{\text{MNS}} diag(m_1, m_2, m_3) V_{\text{MNS}}^T \tag{13.25}$$

Hence BR($\chi^{\pm\pm} \to l^{\pm}l^{\pm}$) are predicted and different [90] in each of the neutrino mass scenarios: *Normal hierarchy* (=NH), *Inverted hierarchy* (=IH) and *Quasi-degenerate* (=DG). We take $\varphi_1, \varphi_2 = 0$ or $\pi$ for the Majorana phases which leads to seven distinct cases:

| | | | |
|---|---|---|---|
| NH: | $m_1 < m_2 \ll m_3$, | | |
| IH1: | $m_2 > m_1 \gg m_3$, | IH2: | $-m_2 > m_1 \gg m_3$, |
| DG1: | $m_1 \simeq m_2 \simeq m_3$, | DG2: | $m_1 \simeq m_2 \simeq -m_3$, |
| DG3: | $m_1 \simeq -m_2 \simeq m_3$, | DG4: | $m_1 \simeq -m_2 \simeq -m_3$. |

Note that such predictions of BR($\chi^{\pm\pm} \to l^{\pm}l^{\pm}$) are a feature of the HTM in which the couplings $h_{ij}$ are the sole origin of neutrino mass. This direct correlation between BR($\chi^{\pm\pm} \to l^{\pm}l^{\pm}$) and the neutrino mass matrix may not extend to $\chi^{\pm\pm}$ of other models in which neutrinos can acquire mass by other means or by a combination of mechanisms which may or may not include the $h_{ij}$ couplings.

For $m_{\chi^{\pm}} < m_{\chi^{\pm\pm}}$ the alternative decay mode $\chi^{\pm\pm} \to \chi^{\pm}W^*$ competes with $\chi^{\pm\pm} \to l^{\pm}l^{\pm}$. In Fig. 13.8(a) we show contours of BR($\chi^{\pm\pm} \to \chi^{\pm}W^*$) in the plane ($m_{\chi^{\pm\pm}}, v_\Delta$) for the NH case. Clearly BR($\chi^{\pm\pm} \to \chi^{\pm}W^*$) can be sizeable and approaches 100% for larger $v_\Delta$. In Figs. 13.8(b)–13.10 we plot $\sigma_{ll}$ as a function of $m_{\chi^{\pm\pm}}$, where $\sigma_{ll}$ is the total leptonic ($l = e, \mu, \tau$) cross-section defined by:

$$\sigma_{ll} = \sigma(p\overline{p} \to \chi^{++}\chi^{--}) \times B_{ll}(2 - B_{ll}) + 2\sigma(p\overline{p} \to \chi^{++}\chi^{-}) \times B_{ll} \tag{13.26}$$

The contribution to $\sigma_{ll}$ from $\sigma(p\overline{p} \to \chi^{++}\chi^{--})$ falls more slowly with decreasing $B_{ll}$ since signal candidates are events with at least 2 leptons. Eq. (13.26) simplifies to Eq. (13.22) in the limit where $\sigma(p\overline{p} \to \chi^{\pm\pm}\chi^{\mp}) = 0$ and $B_{ll} = 1$. In the figures we take $m_{\chi^{\pm}} = m_{\chi^{\pm\pm}} - 20$ GeV, which induces a sizeable (but not dominant) BR($\chi^{\pm\pm} \to \chi^{\pm}W^*$), and hence $\sum \sigma_{ll} < \sigma_{\chi^{\pm\pm}}$. We set $v_\Delta = 10$ eV in Figs. 13.8(b) and 13.9 and $v_\Delta = 100$ eV in Fig. 13.10. We only plot $\sigma_{ll}$ for $ee, e\mu, \mu\mu$ since the Tevatron has already performed searches in these channels. Sensitivity to $\sigma_{ll}$ of a few fb will be possible with the anticipated integrated luminosities of $4 - 8$ fb$^{-1}$. There are plans to search for the 3 leptonic decays involving $\tau$ ($e\tau, \mu\tau, \tau\tau$) although the discovery reach in $m_{\chi^{\pm\pm}}$ is expected to be inferior to that for the $ee, e\mu, \mu\mu$ channels. In all figures we take $\theta_{13} = 0°$, the smallest angle in $V_{\text{MNS}}$. From the figures it is clear that $\sigma_{ee,e\mu,\mu\mu}$ differ considerably in each of the 7 scenarios. Optimal coverage is for cases DG1 and DG4, which have $\sigma_{ee,\mu\mu} \geq 5$ fb and $\sigma_{e\mu,\mu\mu} \geq 5$ fb respectively for $m_{\chi^{\pm\pm}} < 180$ GeV. For NH, $\sigma_{\mu\mu} \geq 5$ fb for $m_{\chi^{\pm\pm}} < 190$ GeV but $\sigma_{ee}$ and $\sigma_{e\mu}$ are both unobservable. Taking $\theta_{13}$ at its largest experimentally allowed value results in minor changes to all figures, with the most noticeable effect being a significant reduction of $\sigma_{\mu\mu}$ in DG4. Clearly the Tevatron Run II not only has strong search potential for $\chi^{\pm\pm}$, but is also capable of distinguishing between the various allowed scenarios for the neutrino mass matrix.

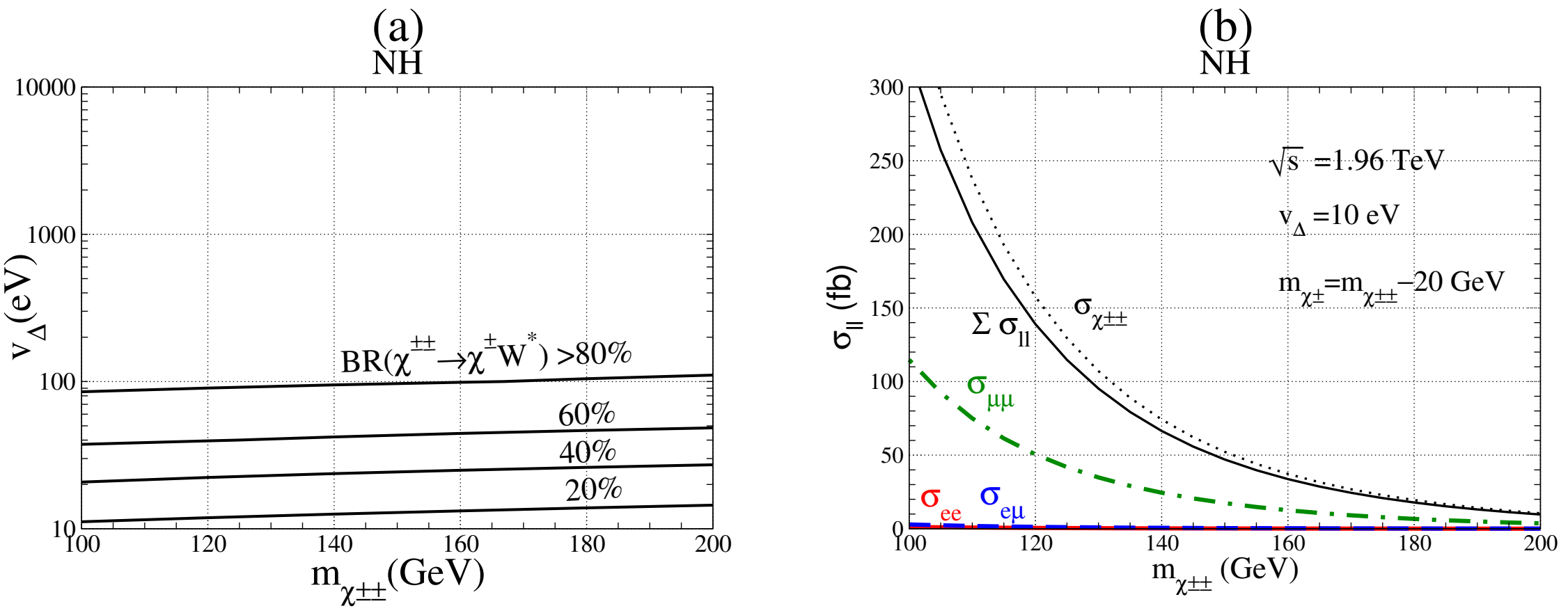

Fig. 13.8: (a) Contours of BR($\chi^{\pm\pm} \to \chi^{\pm} W^*$) for NH in the plane ($m_{\chi^{\pm\pm}}, v_\Delta$); (b) $\sigma_{ll}$ as a function of $m_{\chi^{\pm\pm}}$ for NH with $m_{\chi^\pm} = m_{\chi^{\pm\pm}} - 20$ GeV .

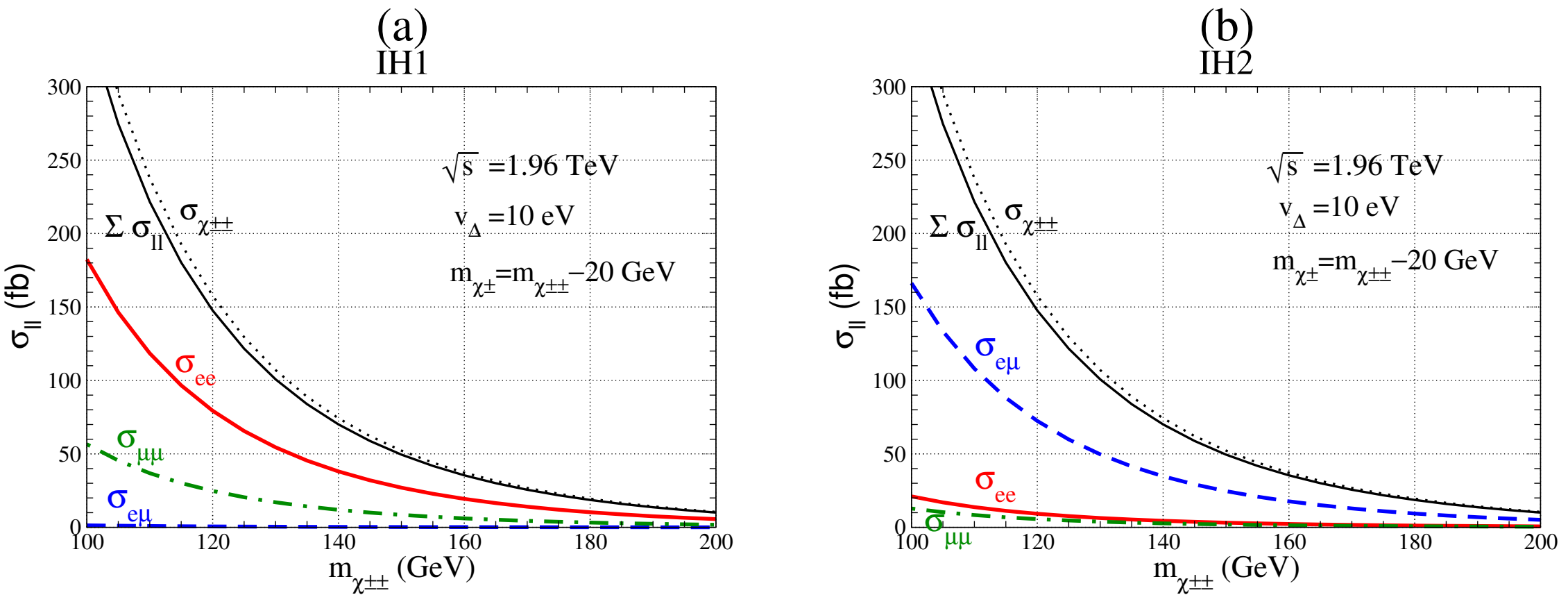

Fig. 13.9: $\sigma_{ll}$ as a function of $m_{\chi^{\pm\pm}}$ for (a) IH1 and (b) IH2.

## 13.3 Photonic decays of neutral triplet scalars

*A.G. Akeroyd, A. Alves, M.A. Diaz and O. Eboli*

Neutral Higgs bosons with very suppressed couplings to fermions, generically called "fermiophobic Higgs bosons" ($h_f$) [91], are possible in specific extensions of the SM Higgs sector. Such a $h_f$ is expected to decay dominantly to two photons $h_f \to \gamma\gamma$ for $m_{h_f} < 95$ GeV or to two massive gauge bosons, $h_f \to VV^{(*)}$, ($V = W^\pm, Z$) for $m_{h_f} > 95$ GeV [92, 93]. The large branching ratio (BR) for $h_f \to \gamma\gamma$ (a "photonic Higgs") would provide a very clear experimental signature, and observation of such a particle would strongly constrain the possible choices of the underlying Higgs sector.

In the triplet model with tree-level custodial $SU(2)_C$ symmetry the eigenstate $H_1^{0\prime}$ is entirely composed of triplet fields and is given by [20]

$$H_1^{0\prime} = \frac{1}{\sqrt{3}}(\sqrt{2}\chi^{0r} + \xi^0) \, , \tag{13.27}$$

where $\chi^{0r}$ is the real part of $\chi^0$ and $< \chi^0 >= b$. In Eq. (13.27) the $\chi^{0r}$ component in $H_1^{0\prime}$ couples to $\nu\overline{\nu}$ via the $h_{ij}$ coupling, Eq. (13.14). Since the product $h_{ij}b$ is fixed by the neutrino mass matrix $m_{ij}$ to be of order eV (Eq. (13.16)), it is apparent that the decay $H_1^{0\prime} \to \gamma\gamma$, mediated by $W$ loops proportional to $b$, will dominate over $H_1^{0\prime} \to \nu\overline{\nu}$ if $b$ is of the order of a few GeV. Thus $H_1^{0\prime}$ is a candidate for a $h_f$ since it has very suppressed fermionic (leptonic only) couplings and a non-negligible coupling to the massive vector bosons if $b$ is GeV scale. However, in general $H_1^0$ and $H_1^{0\prime}$ mix through the following mass matrix written in the ($H_1^0$,$H_1^{0\prime}$) basis [20]

$$\mathcal{M} = \begin{pmatrix} 8c_H^2(\lambda_1+\lambda_3) & 2\sqrt{6}s_H c_H \lambda_3 \\ 2\sqrt{6}s_H c_H \lambda_3 & 3s_H^2(\lambda_2+\lambda_3) \end{pmatrix} v^2 \, . \tag{13.28}$$

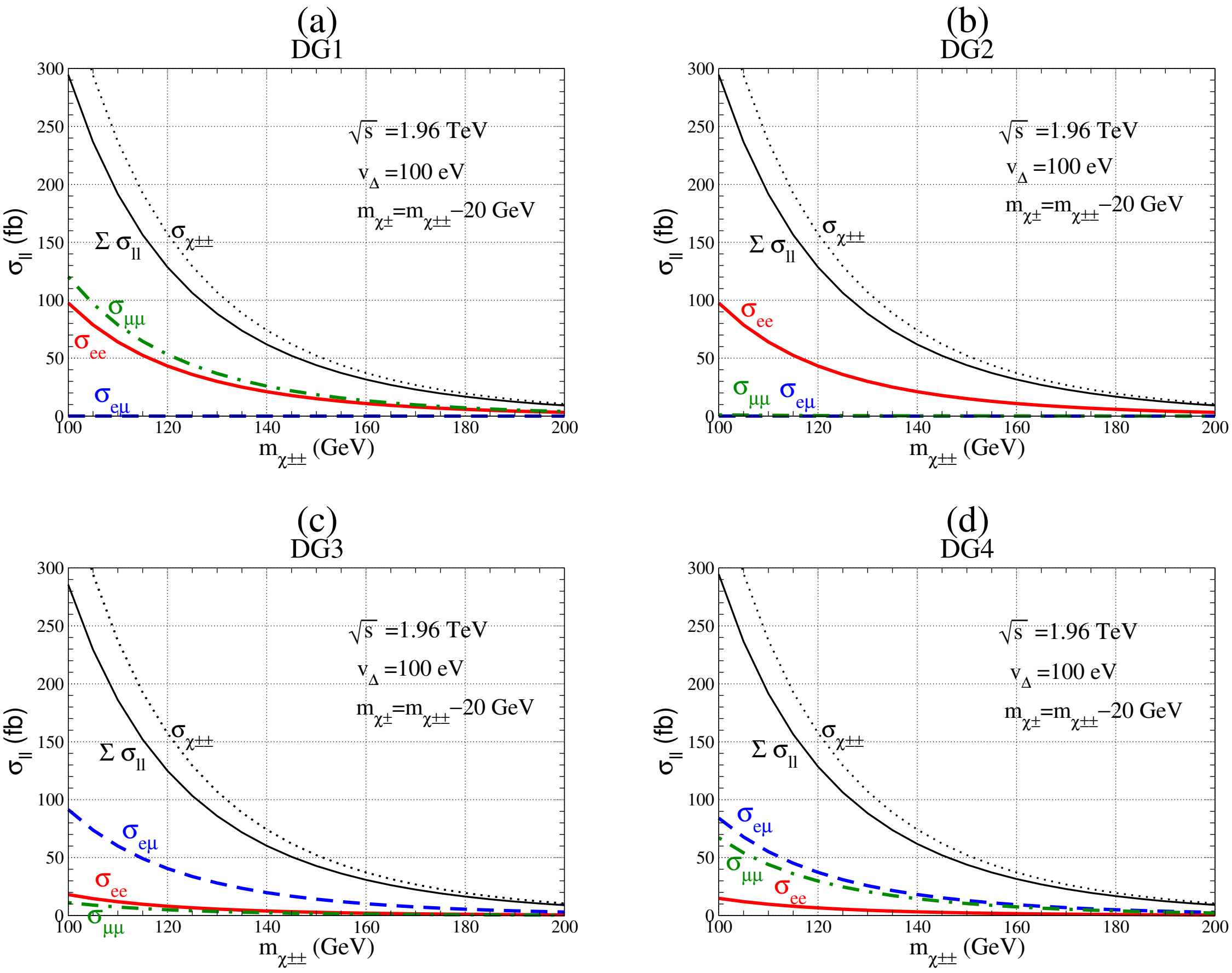

Fig. 13.10: $\sigma_{ll}$ as a function of $m_{\chi^{\pm\pm}}$ for (a) DG1, (b) DG2, (c) DG3 and (d) DG4.

Here $\lambda_i$ are dimensionless quartic couplings in the Higgs potential, $s_H = 2\sqrt{2}b/v$, $c_H = a/v$ (where $a/\sqrt{2} =< \phi^0 >$) and $v^2 = a^2 + 8b^2$. The assumption that the $\lambda_i$ couplings are roughly the same order of magnitude together with the imposition of the bound $s_H < 0.4$ (obtained from the effect of $H_3^\pm$ on $Z \to b\overline{b}$) results in very small mixing [94]. Hence $H_1^{0\prime}$ may be taken to be a $h_f$ in a sizeable region of the triplet model parameter space. Moreover, $H_1^{0\prime}$ would be the lightest Higgs boson in this model in the limit of small $s_H$, as stressed in [21].

We depict in Fig. 13.11 the branching ratios for a generic fermiophobic Higgs boson $h_f$ into $VV$ where $V$ can be either a $W$, $Z$ or $\gamma$. In this figure we assumed that the $h_f$ couplings to fermions are absent and that $h_f \to \gamma\gamma$ is mediated solely by a $W$ boson loop, giving rise to the following $h_f$ partial width into two photons:

$$\Gamma(h_f \to \gamma\gamma) = \frac{\alpha^2 g^2}{1024\pi^3}\frac{m_{h_f}^3}{m_W^2}|F_1 C_{WW}|^2 \tag{13.29}$$

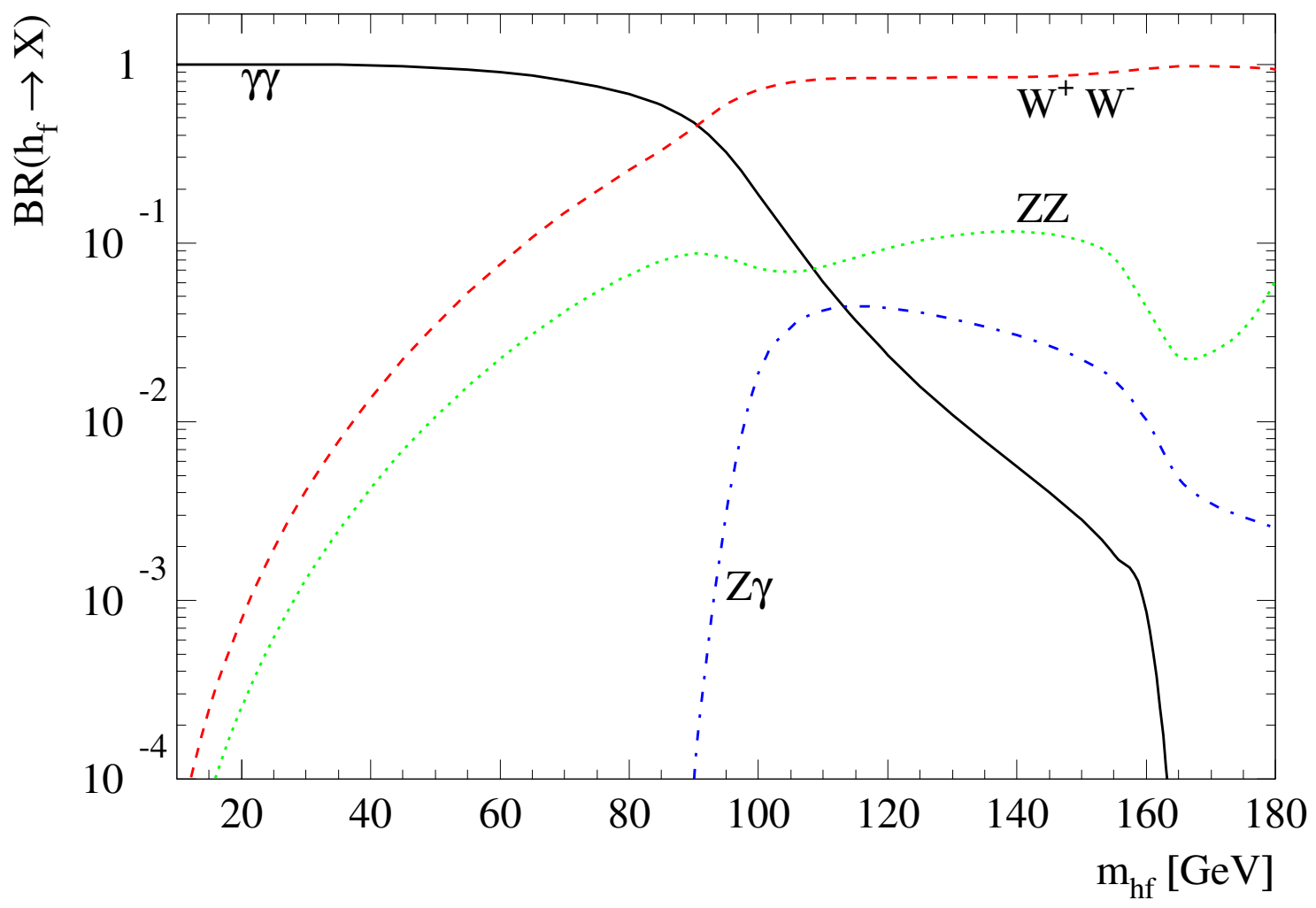

Fig. 13.11: Branching ratios of the largest decay modes of a fermiophobic Higgs boson assuming exact fermiophobia at tree-level. The branching ratio into $\gamma\gamma$ equals the $W^*W^*$ mode for $m_{h_f} \approx 90$ GeV and drops to 20% for $m_{h_f} = 100$ GeV.

where $F_1 = F_1(\tau)$, $\tau = 4m_W^2/m_{h_f}^2$, is a function given in [95]. The $h_f WW$ coupling $C_{WW}$ is normalized to the SM $\phi_0 WW$ coupling. In the triplet model one has:

$$C_{WW} = WWH_1^{0\prime} = \frac{2\sqrt{2}}{\sqrt{3}}\, s_H \; . \tag{13.30}$$

This gives rise to benchmark BRs which are used in the ongoing searches to derive mass limits on $m_{h_f}$. In practice, $h_f \to \gamma\gamma$ can also be mediated by charged scalar loops: $H_3^\pm, H_5^\pm, H_5^{\pm\pm}$ [22]. Although such contributions are suppressed relative to the $W$ loops by a phase space factor, they can be important if $C_{WW}$ is suppressed. In our numerical analysis we will assume the benchmark BRs given in Fig. 13.11. One can see from the figure that the loop induced decay mode $h_f \to \gamma\gamma$ is dominant for $m_{h_f} < 95$ GeV and drops below $0.1\%$ for $h_f$ masses above 150 GeV. On the other hand, the decay channel $h_f \to W^*W^*$ dominates for $m_{h_f} > 95$ GeV, being close to $100\%$ until the threshold for $h_f$ decay into two real $Z$'s is reached.

$H_1^{0\prime}$ can be produced at both $e^+e^-$ colliders and hadron colliders through the couplings $C_{WW}, C_{ZZ}$ Since the maximum value of $s_H$ is around 0.4, $H_1^{0\prime}$ can be produced with SM strength at best. Mass limits for $h_f$ have been obtained at both LEP and the Fermilab Tevatron assuming that the coupling $h_f VV$ has the same strength as the Standard Model (SM) Higgs coupling $VV\phi^0$, and that all fermion BRs are exactly zero. Lower bounds of the order $m_{h_f} > 100$ GeV have been obtained by the LEP collaborations OPAL [96], DELPHI [97], ALEPH [98], and L3 [99], utilizing the channel $e^+e^- \to h_f Z$, $h_f \to \gamma\gamma$. At the Tevatron Run I, the limits on $m_{h_f}$ from the DØ and CDF collaborations are respectively 78.5 GeV [100] and 82 GeV [101] at $95\%$ C.L., using the mechanism $qq' \to V^* \to h_f V$, $h_f \to \gamma\gamma$, with the dominant contribution coming from $V = W^\pm$. For an integrated luminosity of 2 fb$^{-1}$, Run II will extend the coverage of $m_{h_f}$ in the benchmark model slightly beyond that of LEP [102]. In addition, Run II will be sensitive to the region $110\,\text{GeV} < m_{h_f} < 160\,\text{GeV}$ and BR$(h_f \to \gamma\gamma) > 4\%$ which could not be probed at LEP. A preliminary search in the inclusive $2\gamma + X$ channel has been performed with $190\,pb^{-1}$ of Run II data [103]. However, a small $s_H$ would suppress the coupling $VVH_1^{0\prime}$ and consequently deplete the production rate. Hence it is of concern to consider other production mechanisms which are unsuppressed in the above scenario.

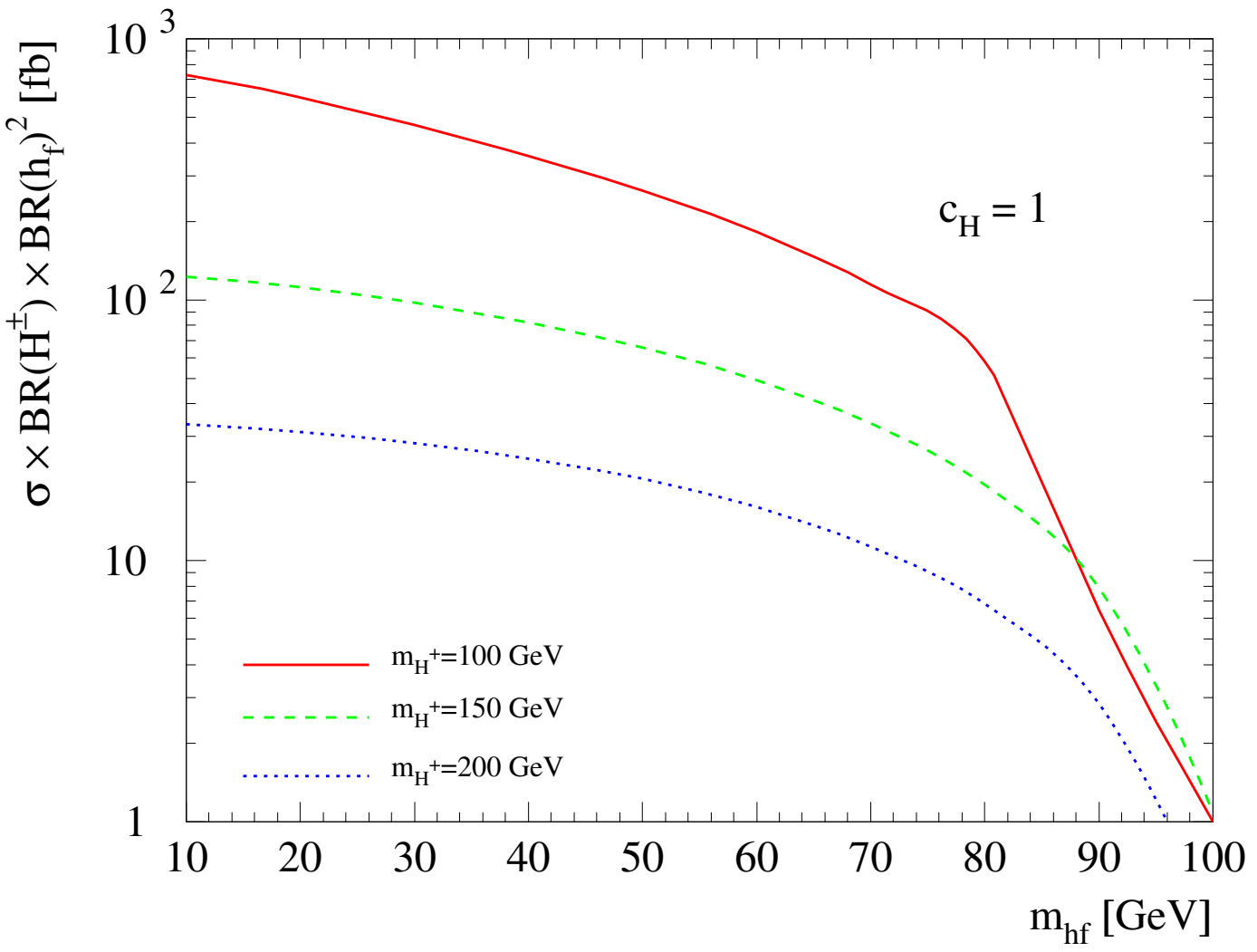

Fig. 13.12: Total production cross sections times branching ratios of $H_1^0{}' \to \gamma\gamma$ and $H_3^\pm \to W^\pm H_1^0{}'$ for $p\bar{p} \to H_1^0{}' H_3^\pm \to \gamma\gamma\gamma\gamma + W^\pm$ before cuts at the Tevatron Run II in femtobarns. The values of $c_H$ and $m_{H_3^\pm}$ are as indicated in the figure.

The production mechanism $qq' \to H_3^\pm H_1^0{}'$ is complementary to that of $qq' \to V H_1^0{}'$. This can be seen immediately from the explicit expression for the coupling:

$$W H_3^\pm H_1^0{}' \propto \frac{2\sqrt{2}}{\sqrt{3}} c_H \ . \tag{13.31}$$

Hence the mechanism $qq' \to H_3^\pm H_1^0{}'$ is unsuppressed in the region of the parameter space where the standard production mechanism $qq' \to V H_1^0{}'$ becomes ineffective. In our numerical analysis we take $c_H = 1$ as a benchmark value. From the bound $s_H < 0.4$ one obtains $c_H > 0.9$. In the exact $c_H = 1$ limit (i.e. triplet vev $b = 0$) the neutrinos would not receive a mass at tree-level. Extremely small $s_H < 10^{-9}$ would require non-perturbative values of $h_{ij}$ to generate realistic neutrino masses. We are interested in the interval $0.9 < c_H < 0.99$ (corresponding to GeV scale triplet vev) in which $H_1^0{}'$ decays primarily to photons in the detector, and neutrino mass is generated with a very small $h_{ij} \sim 10^{-10}$.

To date complementary mechanisms have not been considered in the direct fermiophobic Higgs searches at the Tevatron. As emphasized in [104], [105] a more complete search strategy for $h_f$ at hadron colliders must include such production processes in order to probe the scenario of fermiophobic Higgs bosons with a suppressed coupling $h_f VV$. Moreover, one expects $H_1^0{}'$ to be the lightest Higgs boson in the one doublet, two triplet model for small $s_H$, which further motivates a search in the complementary channel $qq' \to H_3^\pm H_1^0{}'$.

The experimental signature of the process $qq' \to H_3^\pm H_1^0{}'$ depends on the decay modes of $H_3^\pm$. If $H_3^\pm$ decays to two fermions then the signal would be of the type $\gamma\gamma + X$, which is essentially the same as that assumed in the searches which rely on the coupling $VVH_1^0{}'$. However, the decay $H_3^\pm \to H_1^0{}' W^*$ may have a very large BR [106]. This is because the decay width to the fermions ($H_3^\pm \to f'\overline{f}$) is proportional to $t_H = (s_H/c_H)$. Thus in the region of small $s_H$ the fermionic decays of $H_3^\pm$ are depleted. This enables the decay $H_3^\pm \to H_1^0{}' W^*$ to become the dominant channel *even if* the mass difference $m_{H_3^\pm} - m_{H_1^0{}'}$ is much less than $m_W$. Consequently, this scenario would give rise to double $H_1^0{}'$ production, with subsequent decay of $H_1^0{}' H_1^0{}' \to \gamma\gamma\gamma\gamma$ for light $H_1^0{}' < 90$ GeV. In Fig. 13.12 we show the production cross section times branching ratios of the complementary process under study. The rates are considerably larger than for the analogous case in the 2HDM (Model I) because of the enhancement factor $2\sqrt{2}/\sqrt{3}$ in the triplet model [107].

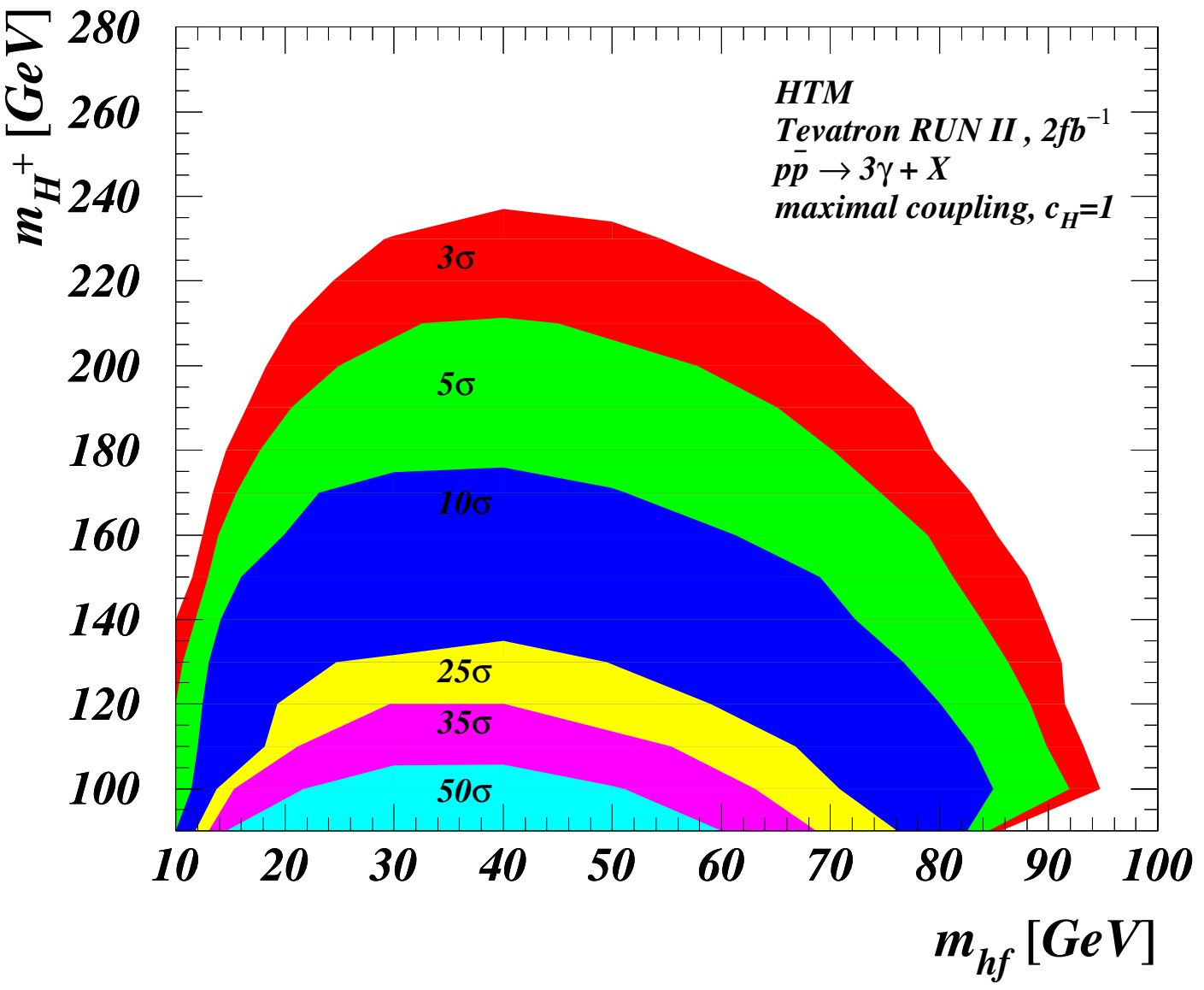

Fig. 13.13: The expected signal statistical significance (from top to bottom: $3\sigma$, $5\sigma$, $10\sigma$, $25\sigma$, $35\sigma$, $50\sigma$) of multi-photon final states in the $m_{H^\pm} \otimes m_{h_f}$ plane in the HTM frame work, assuming an integrated luminosity of 2 fb$^{-1}$ at the Tevatron RUN II.

In Ref. [107] we showed that Tevatron RUN II will be able to detect a fermiophobic Higgs boson, in the framework of the triplet model or 2HDM, in a sizable portion of the parameter space in the inclusive multi-photon channel, namely, at least three photons ($> 3$) or four photons which may or may not be accompanied by extra leptons and/or jets. A multi-photon final state has the great advantage of being extremely clean (concerning SM backgrounds) after applying a mild set of cuts:

$$E_T^\gamma > 15 \text{ GeV} \quad , \quad |\eta^\gamma| < 1.0 \quad , \quad \Delta R_{\gamma\gamma} > 0.4 \quad , \quad \Delta R_{\gamma X} > 0.4 \quad , \tag{13.32}$$

which are sufficient to guarantee the trigger of these events and proper isolation. On the other hand they are very effective against SM backgrounds from photon and/or gluon bremmstrahlung processes and parton collinear splittings.

We show in Fig. 13.13 the search potential in the $m_{H_3^\pm} \otimes m_{H_1^0{}'}$ plane accessible at the Tevatron RUN II with 2 fb$^{-1}$ of integrated luminosity. Fermiophobic Higgs bosons with masses up to 90 GeV can be detected at the $3\sigma$ statistical level (at least) for $m_{H_3^\pm} < 140$ GeV, while a 40–50 GeV $H_1^0{}'$ can be observed at $3\sigma$ with $m_{H_3^\pm}$ up to 240 GeV. More than 100 signal events are expected for 20 GeV $< m_{H_1^0{}'} < 50$ GeV and $m_{H_3^\pm} < 100$ GeV. Notably, at least 40 signal events are expected against around 4 background events inside the whole of the $10\sigma$ region in the $m_{H_3^\pm} \otimes m_{H_1^0{}'}$ plane, which should suffice to reconstruct the $\gamma\gamma$ invariant mass peak structure with a good accuracy [107]. Note that the signal gets depleted for very light fermiophobic Higgs masses ($m_{H_1^0{}'} < 20$ GeV). This is because a light neutral Higgs decays to a pair of soft photons which do not pass the trigger requirements. Heavier Higgs bosons have more severe phase space suppression as well as a decreasing branching ratio into photons.

In summary, fermiophobic Higgs bosons in the framework of the triplet model with tree-level custodial $SU(2)_C$ symmetry produced in association with charged Higgs bosons through the complementary process $p\bar{p} \to H_1^0{}' H_3^\pm$ give rise to multi-photon signatures with very low SM background. The Tevatron RUN II will be able to cover a sizable region of the triplet model parameter space and detect or exclude Higgs masses up to $m_{H_3^\pm} < 240$ GeV or $m_{H_1^0{}'} < 100$ GeV.

## Acknowledgements

The list below quotes all the funding agencies, ministries and international instances whose funding was indispensable in the work described in this report.

AUSTRIA

- Austrian Academy of Sciences, APART Grant No. 10983
- Fonds zur Förderung der wissenschaftlichen Forschung Project No. P16592-N02

BELGIUM

- IAP program from the Politique scientifique fédérale belge

BRAZIL

- Fundação de Amparo à Pesquisa do Estado de São Paulo (FAPESP)
- Conselho Nacional de Desenvolvimento Científico e Tecnológico (CNPq)
- Conicyt grant No. 1040384

CANADA

- Natural Science and Engineering Research Council

CHINA

- National Natural Science Foundation of China

CZECH REPUBLIC

- GACR grants Nos. 202/06/0734 and 202/05/H003

FRANCE

- ACI Jeunes Chercheurs contract 2068
- CNRS GDRI-ACPP

GERMANY

- Deutsche Forschungsgemeinschaft: Contract No. FR 1064/5-2;
  Forschergruppe 'Quantenfeldtheorie, Computeralgebra und Monte-Carlo-Simulation' Grant SFB/T
  Graduiertenkolleg 'High Energy Physics and Particle Astrophysics Sonderforschungsbereich/Transı
  SFB/TR-9 Computational Particle Physics'.
- Federal Ministry of Education, Science, Research and Technology (BMBF) contract No. 05HA4PD
- Studienstiftung des deutschen Volkes
- Helmholtz-Gemeinschaft Grant No. VH–NG–005

INTERNATIONAL INSTITUTIONS

- Cambridge Commonwealth Trust
- CNRS/USA grant 3503
- EU Human Potential Programme HPRN-CT-2000-00149
- EU Research Training Networks Programme MRTN contracts 2004-503369 and CT-2004-503369
- European Science Foundation Network grant N.86

– German–Israeli Project Cooperation in Future-Oriented Topics (DIP)
– INTAS grant 03-51-4007
– Marie Curie Excellence Grant MEXT-CT-2004-013510
– NATO grant PST.CLG.980066
– Royal Society Joint Project Grant with the former Soviet Union

ITALY

– INFN, Sezione di Trieste

JAPAN

– Grant-in-Aid of the Ministry of Education, Culture, Sports, Science, and Technology, Government of Japan, Nos. 17540286, 13135225, 16081211, 17043008

KOREA

– Korea Research Foundation grant KRF–2006–013–C00097
– The Korean Federation of Science and Technology Societies Grant funded by Korea Government (MOEHRD, Basic Research Promotion Fund)

NORWAY

– Research Council of Norway

POLAND

– Polish Committee for Scientific Research Grants No. 1 P03B 040 26, 115/E-343/SPB/DESY/P-03/DWM517/2003-2005, 115/E-343/SPB/5 and PR-UE 'Sieci'/DIE 367/2004–2006
– Polish Ministry of Science and Higher Education Grant No. 1 P03B 108 30 (2006–2008)

RUSSIA

– The 'Dynasty' Foundation
– Federal Agency for Science Grants NS-1685.2003.2, RFBR-04-02-17448, RFBR-04-02-16073
– Federal Program of the Russian Ministry of Industry, Science and Technology grant SS-1124.2003.2
– ICPPM

SPAIN

– CICYT grant FPA2004-02948
– DGIID-DGA grant 2005-E24/2
– Generalitat Valenciana
– MCyT Ramon y Cajal contracts
– Spanish grant FPA2005-01269

SWEDEN

– Göran Gustafsson Foundation

UNITED KINGDOM

– British Council under the ‘Alliance: Franco British Partnership Programme 2004 (Project No. PN 04.051)’
– PPARC grants PPA/G/S/2003/00096
– University of Cambridge Board of Graduate Studies, Cavendish Laboratory, Cambridge

UNITED STATES OF AMERICA

– Davis Institute for High Energy Physics
– Department of Energy grants Nos. DE-FG01-04ER04-02, DE-FG02-95ER40896, DOE-EY-76-02-3071, DE–FG02–95ER–40896, DE–FG03–91ER–40674, DE-FG03-92ER40689, DE-AC02-76SF00515 and DoE Outstanding Junior Investigator Award (grant DE-FG02-97ER41209)
– U.C. Davis Dean’s Office
– Wisconsin Alumni Research Foundation